\documentclass[12pt,letterpaper]{report}
\usepackage{epsfig}
\usepackage{setspace}
\usepackage{cite}
\usepackage{helvet}%sets the sffamily font to helvetica
\usepackage{fancyhdr}
\usepackage{amsmath} % See http://en.wikibooks.org/wiki/LaTeX/Print_version, for /tfrac and /dfrac use
\usepackage{graphicx}
\usepackage{appendix}
\usepackage{apptools}
\usepackage{titlesec}
\usepackage[titles]{tocloft}
\usepackage{calc}
\usepackage{setspace}
\usepackage{mathtools}
\usepackage[pdftex,colorlinks,linkcolor=black,citecolor=blue]{hyperref} % creates links for references (and other things)
\PassOptionsToPackage{hyphens}{url}
\titleformat
{\chapter} % command
[hang] % shape
{\normalfont\normalsize\bfseries\center} % format
{\chaptertitlename\ \thechapter} % label
{1em} % sep
{\MakeUppercase} % before code

\titleformat
{\section} % command
[hang] % shape
{\normalfont\normalsize\bfseries} % format
{\thesection} % label
{1em} % sep
{} % before code

\titleformat
{\subsection} % command
[hang] % shape
{\normalfont\normalsize\bfseries} % format
{\thesubsection} % label
{1em} % sep
{} % before code

\titlespacing{\chapter}{0in}{-0.5in}{0in}
\titlespacing{\section}{0in}{0in}{0in}
\titlespacing{\subsection}{0in}{0in}{0in}
\makeatletter
\g@addto@macro\appendix{%
  \addtocontents{toc}{%
		\protect%
		\protect%
  }%
}
\makeatother

\let\oldbibliography\thebibliography
\renewcommand{\thebibliography}[1]{%
  \oldbibliography{#1}%
  \setlength{\itemsep}{0pt}%
}
\large
\catcode`\@=11 % @ signs are allowed in definitions
\def\cs2{CS$_2$}
\sffamily
\let\@afterindentfalse\@afterindenttrue
\@afterindenttrue
\newcommand{\cmmnt}[1]{\ignorespaces}

\begin{document}

	% sets numbering upto 2nd level in chapter,section, subsection, subsubsection
	% numbering starts from 0 - so chapter = 0, section = 1, etc
	\setcounter{secnumdepth}{2} 

	% makes levels up to the 2nd (subsection) appear in the table of contents
	\setcounter{tocdepth}{2}
	
	% set page numbering for the preliminary pages to roman numerals
	\pagenumbering{roman}
	\setcounter{page}{1} 
\thispagestyle{empty}
\begin{titlepage}

	\begin{center}

	\singlespacing
	\textbf{Balance functions of charged hadron pairs $(\pi,K,p)\otimes (\pi,K,p)$ in Pb--Pb collisions at $\sqrt{s_{_{\rm NN}}} =$2.76 TeV}\\
	\doublespacing
	
	by\\
	
	\textbf{Jinjin Pan}\\
	\textbf{DISSERTATION}\\

	Submitted to the Graduate School\\
	
        of Wayne State University,\\
        Detroit, Michigan\\
	in partial fulfillment of the requirements\\
	for the degree of\\
%	\vspace{0.8cm}
	\textbf{DOCTOR OF PHILOSOPHY}\\

	2019\\
	\end{center}
	\begin{flushright}
	\makebox[8.7cm][l]{MAJOR: Physics}
	\makebox[8.7cm][l]{Approved By:}\\
   \vspace{1.0cm}
	\makebox[8.7cm][l]{$\overline { \mathrm{Advisor^{\vphantom{\frac{I}{I}}}} 
									\qquad\qquad\qquad\qquad\qquad
									\mathrm{Date}\hspace{1.375cm}}$} \\
   \vspace{0.7cm}
	\makebox[8.7cm][l]{$\overline {\hspace{7.8cm}}$} \\
   \vspace{0.9cm}
	\makebox[8.7cm][l]{$\overline {\hspace{7.8cm}}$} \\
   \vspace{0.9cm}
	\makebox[8.7cm][l]{$\overline {\hspace{7.8cm}}$} \\
% Add this if you need an extra line
%   \vspace{0.9cm}
%	\makebox[8.7cm][l]{$\overline {\hspace{7.8cm}}$} \\
	
	\end{flushright}

	\end{titlepage}

	%Include this if you intend to copyright your document
%	\include{copyright}

	% add dedication and acknowledgements
	\setcounter{page}{2} 
	\addcontentsline{toc}{chapter}{Dedication}
	\newpage
\begin{center}
{\bf  DEDICATION}
\end{center}

%\phantom{a} 
%
%\phantom{a}
%
%\phantom{a}
%
%\phantom{a}
%
%\phantom{a}
%
%\phantom{a}
%
%\phantom{a}
%
%\phantom{a}
%
%\phantom{a}
%
%\phantom{a}

{\it 
\centering 

I dedicate this dissertation to my parents, my grandparents and my wife, for giving me courage to conquer all the difficulties, for giving me strength to embrace changes in my life, and for giving me hope to chase a brighter future.

}

	\addcontentsline{toc}{chapter}{Acknowledgments}
	\newpage
{\centering \bf ACKNOWLEDGMENTS \par}

I owe a great deal of gratitude to my PhD advisor Dr. Claude Pruneau for his guidance, mentorship and patience throughout the whole process of my PhD program.

I would also like to thank the members of my PhD Committee, Dr. Sergei Voloshin, Dr. Chun Shen, Dr. Scott Pratt for their support and valuable suggestions.

And I am extremely grateful for the invaluable help from Dr. Takafumi Niida, Dr. Ron Belmont, Dr. Sumit Basu, and Victor Gonzalez, who often sit with me for hours digging the code or solving my analysis problems.

I wish to express my warm and sincere thanks to Dr. Kolja Kauder, Dr. Joern Putschke, Dr. Rosi Reed, Dr. W.J. Llope, Dr. Raghav Elayavalli, Dr. 
Maxim Konyushikhin, Dr. Shanshan Cao for all their help and suggestions.

I thank my fellow grad students Dr. Mohammad Saleh, Dr. Chris Zin, Nick Elsey, Isaac Mooney, Amit Kumar for fruitful discussions and fun time together.

Many thanks to the ALICE Collaboration for all the support and help.

This work has been supported in part by the Office of Nuclear Physics at the United States Department of Energy (DOE NP) under Grant No. DE-FOA-0001664.

\vspace*{\fill}

\begin{flushright}
{\it Jinjin Pan $\qquad \qquad$}
\end{flushright}

	% build table of contents
	\renewcommand{\contentsname}{Table of Contents}
  \tableofcontents

	% add list of figures - it requires some formatting changes
%	{
%		\newpage
%		\addcontentsline{toc}{chapter}{List of Figures}
%
%		% change the left margin to match the rest of the document
%		\setlength\cftfigindent{0cm}
%		\singlespacing
%
%		% double space between each entry - this will have to be changed 
%		% if document font size changes
%		\setlength{\parskip}{12pt}
%		\listoffigures 
%	}
	
	% start the body of the work on a new page
	\newpage

	\pagenumbering{arabic}
  \pagestyle{fancy}
  \fancyhead{} 
	\fancyfoot{} % clear all header and footer fields
  \chead[]{\thepage}
  \renewcommand{\headrulewidth}{0pt}
  \renewcommand{\footrulewidth}{0pt}
	\fancypagestyle{plain}{%
		% clear all header and footer fields
		\fancyhf{} 
		% except the center
		\fancyhead[C]{\thepage} 
		\renewcommand{\headrulewidth}{0pt}
		\renewcommand{\footrulewidth}{0pt}
	}

	% insert chapters
	% Add more chapters by copying chapters/chapter_1.tex and adding a line for it here	
	\chapter{Introduction}\label{chap:RHIP}

\section{Understanding Hadron Productions in Relativistic Heavy Ion Collisions}

The two-wave quark production scenario~\cite{PhysRevLett.108.212301} proposes that there are two waves of light quark production in relativistic heavy ion collisions. 
The first wave takes place within $\approx$1 fm/$c$ immediately after the beginning of a collision when gluons of the system thermalize and form a Quark Gluon Plasma (QGP). 
This first wave is then followed by a  period of (nearly) isentropic expansion that lasts for 5-10 fm/$c$. During this period, the newly formed QGP expands longitudinally,  to a lesser extent transversely, and cools down. Little charge production occurs during this stage but, eventually, as the QGP temperature drops to approximately 160 MeV, gluon collisions quickly yield a large burst of quark production. 
These quarks then rapidly combine to yield hadrons as the QGP transitions into a short lived hadron phase which expands further but quickly ends up streaming into free hadrons.
Pratt and collaborators argued that the vast majority of charge production, based on light $u$ and $d$ quarks and yielding mostly pions, takes place during  the second wave.
They additionally argued that the production of strange and heavier quarks should predominantly occur during the first stage.  
Pratt et al. further articulated that the two waves of charge creation can be studied with general charge balance functions of identified particle pairs such as pion pairs, kaon pairs, proton antiproton pairs, proton/$K^-$ pairs, etc. 
As such, they argued, measurements of general charge balance functions (BF) shall provide quantitative insight into the time of formation of quarks, the longitudinal and transverse expansion dynamics of the QGP, as well as its quark/hadron chemistry.

In this work, we present extensive measurements of BFs of charged hadron pairs $(\pi,K,p)\otimes (\pi,K,p)$ in Pb--Pb collisions at $\sqrt{s_{_{\rm NN}}} =$2.76 TeV. These BF results provide new and challenging constraints for theoretical models of the production and evolution of the quark gluon plasma, as well as hadron production and transport in relativistic heavy-ion collisions.

\section{Summary of the Objectives Of This Work}

The scientific objectives of this work are established based on the state of knowledge of collision dynamics and, in particular, general balance functions described in Sec.~\ref{sec:theory},~\ref{sec:experiment},~\ref{sec:THvsEX}.
They can be summarized as follows: Measure and use general balance functions to ...
\begin{itemize}
\item test the two-wave quark production scenario in relativistic heavy ion collisions,
\item study the collision dynamics of large relativistic collisional systems and better constrain models of these systems,
\item study the evolution of the hadron pairing probability vs. collision centrality in relativistic heavy ion collisions.
\end{itemize}

These generic goals are pursued with the following specific objectives and measurements
\begin{itemize}
\item establish sound procedures to measure BFs as functions of $\Delta y$  and $\Delta\varphi$, 
\item measure balance functions $B(\Delta y,\Delta\varphi)$ of charged hadron pairs $(\pi,K,p)\otimes (\pi,K,p)$ in Pb--Pb collisions for specific transverse momentum ranges of these hadrons,
\item measure the collision centrality evolution of these BFs in Pb--Pb collisions,
\item study the evolution of the longitudinal and azimuth widths of these BFs vs. collision centrality in Pb--Pb collisions,
\item study the evolution of the integral of these BFs vs. collision centrality in Pb--Pb collisions.
\end{itemize}

\section{Organization and Structure of This Dissertation}

Chapter~\ref{chap:BFct} introduces the notion of correlation functions and balance functions and presents a brief literature survey of recent theoretical and experimental works on balance functions.
The details of the analysis method, the datasets, the optimization techniques used, and the various checks performed are described in Chapter~\ref{chap:AnalysisMethods}, while Chapter~\ref{chap:SystematicUncertainties} presents a detailed discussion of systematic uncertainties. The results are presented and discussed in Chapter~\ref{chap:Results}.
The conclusions of this work are summarized in Chapter~\ref{chap:Summary}.

% Sample figure, table, and equation:

%\begin{figure}
%\begin{center}
%\includegraphics*[width=16.5cm]{plots/somefile.eps}
%\end{center}
%\caption{A caption}
%\label{fig:figurereference1}
%\end{figure}

%\begin{table}[htb]
%\caption{Table caption.
%\label{tab:tablereference1}}
%\begin{center}
%\begin{tabular}{cc}
%\hline
%Symbol & Name \\
%\hline
%{\it u} & Up \\
%{\it d} & Down \\
%\hline
%{\it c} & Charm \\
%{\it s} & Strange \\
%\hline
%{\it t} & Top \\
%{\it b} & Bottom \\
%\hline
%\end{tabular}
%\end{center}
%\end{table}

%\begin{equation}
%n \to p + e^- + \bar{\nu}_e
%\label{eq:equationreference1}
%\end{equation}

	\chapter{Balance Functions}\label{chap:BFct}

Section~\ref{sec:ObservablesDefinition} introduces the notations and observable definitions used throughout this work. 
In particular, it introduces the definition of the $R_2^{\alpha\beta}$ correlator, linear combinations of unlike- and like-sign charge combinations of this correlator, as well as the charge dependent correlator, $R_2^{\alpha\beta,CD}$, used in this work to determine general balance functions $B^{\alpha\beta}$.

Sections~\ref{sec:theory} -- \ref{sec:THvsEX} present a brief literature survey of theoretical and prior experimental works on BFs. They define the historical and practical context of this study and establish the formal scientific goals of the work.

\section{Definitions and Notations}
\label{sec:ObservablesDefinition}

%This section presents the definitions of correlation function(CF) and balance function(BF).

\subsection{Particle Number Densities}
\label{subsec:ParticleNumberDensities}

%Let $\{\vec p_{i}\}$, $i=1, \ldots$ represent the momentum vectors of particles. 
The momentum $\vec{p}$ of particles is decomposed into transverse momentum $p_{\rm T}$, rapidity $y$, and azimuthal angle $\varphi$ components:
\begin{equation}
p_{\rm T} = \sqrt{p_x^2+p_y^2},
\end{equation}
\begin{equation}
y = \frac{1}{2} \ln\left( \frac{E+p_zc}{E-p_zc}\right),
\end{equation}
\begin{equation}
\varphi = \tan^{-1}\left( \frac{p_y}{p_x} \right).
\end{equation}

Whenever the identity of particle species is unknown, the pseudorapidity, $\eta$, is used as proxy to their rapidity $y$.
\begin{equation}
\eta = - \ln\left[ \tan\left( \frac{\theta}{2}\right)\right], 
\end{equation}
where $\theta$ is the polar angle of emission of the particles
\begin{equation}
\theta =  \tan^{-1}\left(   \frac{p_{\rm T}}{p_z}  \right).  
\end{equation}

Single-particle number density $\rho_{1}^{\alpha}$ and two-particle number density $\rho_{2}^{\alpha\beta}$ are defined according to:

\begin{equation}
\rho_{1}^{\alpha}(\vec{p}^{\alpha}) = \rho_{1}^{\alpha}(y^{\alpha},\varphi^{\alpha},p_{\rm T}^{\alpha}) = \frac{1}{\sigma} \frac{d\sigma}{p_{\rm T}^{\alpha}dp_{\rm T}^{\alpha}dy^{\alpha}d\varphi^{\alpha}},
\label{eq:rho1}
\end{equation}

\begin{equation}
\rho_{2}^{\alpha\beta}(\vec{p}^{\alpha},\vec{p}^{\beta}) = \rho_{2}^{\alpha\beta}(y^{\alpha},\varphi^{\alpha},p_{\rm T}^{\alpha},y^{\beta},\varphi^{\beta},p_{\rm T}^{\beta}) = \frac{1}{\sigma} \frac{d\sigma}{p_{\rm T}^{\alpha}p_{\rm T}^{\beta}dp_{\rm T}^{\alpha}dp_{\rm T}^{\beta}dy^{\alpha}dy^{\beta}d\varphi^{\alpha}d\varphi^{\beta}},
\label{eq:rho2}
\end{equation}

where $\sigma$ is the particle production cross section, while $\alpha$ and $\beta$ represent the particle species. The densities are then integrated over a fixed $p_{\rm T}$ range according to

\begin{equation}
\rho_{1}^{\alpha}(y^{\alpha},\varphi^{\alpha}) =  \int_{p_{{\rm T},\min}^{\alpha}}^{p_{{\rm T},\max}^{\alpha}} \rho_{1}^{\alpha}(y^{\alpha}, \varphi^{\alpha}, p_{\rm T}^{\alpha}) p_{\rm T}^{\alpha} dp_{\rm T}^{\alpha},
\end{equation}

\begin{equation}
\begin{aligned}
\rho_{2}^{\alpha\beta}(y^{\alpha},\varphi^{\alpha},y^{\beta},\varphi^{\beta}) =
\int_{p_{{\rm T},\min}^{\alpha}}^{p_{{\rm T},\max}^{\alpha}}
\int_{p_{{\rm T},\min}^{\beta}}^{p_{{\rm T},\max}^{\beta}}
\rho_{2}^{\alpha\beta}(y^{\alpha},\varphi^{\alpha},p_{\rm T}^{\alpha},y^{\beta},\varphi^{\beta},p_{\rm T}^{\beta}) p_{\rm T}^{\alpha} p_{\rm T}^{\beta} dp_{\rm T}^{\alpha} dp_{\rm T}^{\beta}.
\end{aligned}
\end{equation}

The ranges $[p_{{\rm T},\min}^{\alpha},p_{{\rm T},\max}^{\alpha}]$ used in this work are selected independently for each of the hadron species in order to focus on low--$p_{\rm T}$ ``bulk" physics, maximize the yield of usable hadrons, and minimize problematic $p_{\rm T}$ ranges where significant secondary track contamination is known to arise in  Pb--Pb collisions at $\sqrt{s_{\rm NN}}=$ 2.76 TeV measured with the ALICE detector. In this work, as described in more details in sec.~\ref{subsec:EventTrackSelection}, identified $\pi^{\pm}$ and $K^{\pm}$ are selected in the low transverse momentum regime $0.2 \le p_{\rm T} \le 2.0$ GeV/$c$, while  identified $p/\bar{p}$ are selected in the range $0.5 \le p_{\rm T} \le 2.5$ GeV/$c$.

\subsection{Normalized Two-particle Differential Cumulants}
\label{subsec:Cumulant}
{\bf Normalized two-particle differential cumulants}, hereafter noted $R_2^{\alpha\beta}$, are defined (and measured) as functions of particle rapidity $y$ and azimuthal angle $\varphi$ according to:
\begin{equation}
%\begin{aligned}
%\begin{align*}
R_2^{\alpha\beta}(y^{\alpha},\varphi^{\alpha},y^{\beta},\varphi^{\beta}) = \frac{\rho_{2}^{\alpha \beta }(y^{\alpha},\varphi^{\alpha},y^{\beta},\varphi^{\beta})}{\rho_{1}^{\alpha}(y^{\alpha},\varphi^{\alpha})\rho_{1}^{\beta}(y^{\beta},\varphi^{\beta})}- 1,
%\end{align*}
%\end{aligned}
\label{eq:R2}
\end{equation}

The correlators $R_2^{\alpha\beta}(y^{\alpha},\varphi^{\alpha},y^{\beta},\varphi^{\beta})$ are  averaged over the fiducial acceptance $\bar y = (y^{\alpha}+y^{\beta})/2$ and $\bar \varphi = (\varphi^{\alpha}+\varphi^{\beta})/2$ to yield  functions expressed in terms of the relative (difference) rapidity $\Delta y = y^{\alpha}-y^{\beta}$ and azimuthal angle $\Delta \varphi = \varphi^{\alpha}-\varphi^{\beta}$ according to
\begin{equation}
\begin{aligned}
R_2^{\alpha\beta}(\Delta y,\Delta \varphi) = \frac{1}{\Omega(\Delta y)} \int_{\Omega(\Delta y)} dy^{\alpha} dy^{\beta} \int d\varphi^{\alpha} d\varphi^{\beta} R_2(y^{\alpha},\varphi^{\alpha},y^{\beta},\varphi^{\beta}) \\
 \times \delta(\Delta y -y^{\alpha}+y^{\beta}) \delta(\Delta \varphi -\varphi^{\alpha}+\varphi^{\beta}),
 \end{aligned}
\end{equation}PhysRevLett.118.162302

where the function $\Omega(\Delta y)$ represents the fiducial acceptance in $\bar y$ for a given value of $\Delta y$.

Experimentally, the determination of the densities proceeds with histograms involving finitely many bins as described in \cite{PhysRevLett.118.162302,Acharya:2018ddg}. The above integrals are then replaced by sums. The algorithms and computing codes used, in this work, to measure the densities and carry out these operations are similar to those developed in ~\cite{PhysRevLett.118.162302,Acharya:2018ddg}. By definition, $R_{2}^{\alpha\beta}$ is a robust observable 
because it involves a ratio of  two-particle densities and a product of single particle densities within which detection efficiencies cancel out.
Properties of the  $R_{2}^{\alpha\beta}$ correlator are discussed in detail elsewhere~\cite{PhysRevC.89.024906}.

\subsection{US, LS, CI and CD Combinations of $R_2^{\alpha\beta}$}

The analysis of the $R_2^{\alpha\beta}$ correlation functions is carried out in four different charge combinations, namely $\alpha^+\beta^+$, $\alpha^+\beta^-$, $\alpha^-\beta^+$, and $\alpha^-\beta^-$, where $\alpha$ and $\beta$ represent the reference and associate particle species, respectively. 

Opposite-sign particle pairs are combined to yield {\bf unlike-sign (US)} combinations of $R_{2}$ correlators, hereafter noted $R_{2}^{\alpha\beta,{\rm US}}$, and defined according to
\begin{equation}
R_{2}^{\alpha\beta,{\rm US}} = \frac{1}{2} \left( R_{2}^{\alpha^+\beta^-}  + R_{2}^{\alpha^-\beta^+} \right).
\end{equation}
Similarly, same-sign particle pairs are combined to yield {\bf like-sign (LS)} combinations of $R_{2}$ correlators, hereafter noted  $R_{2}^{\alpha\beta,{\rm LS}}$, and defined according to 
\begin{equation}
R_{2}^{\alpha\beta,{\rm LS}} = \frac{1}{2} \left( R_{2}^{\alpha^+\beta^+} + R_{2}^{\alpha^-\beta^-} \right).
\end{equation}
The US and LS correlators are  combined to form {\bf charge independent (CI)} and {\bf charge dependent (CD)} correlators according to 
\begin{equation}
R_{2}^{\alpha\beta,{\rm CI}} = R_{2}^{\alpha\beta,{\rm US}} + R_{2}^{\alpha\beta,{\rm LS}},
\end{equation}
\begin{equation}
R_{2}^{\alpha\beta,{\rm CD}} = R_{2}^{\alpha\beta,{\rm US}} - R_{2}^{\alpha\beta,{\rm LS}}.
\end{equation}

By construction, the $R_{2}^{\alpha\beta,{\rm CI}}$ correlator measures the charge independent correlation between particle species $\alpha$ and $\beta$, while $R_{2}^{\alpha\beta,{\rm CD}}$ quantifies the charge dependent correlation between particle species $\alpha$ and $\beta$.

Within this document, the notations $R_{2}^{\alpha\beta,{\rm CD}}$ and $R_{2}^{{\rm CD}}$ are used interchangeably depending on the context.

\subsection{General Balance Functions}
\label{sec:GBF}

The {\bf balance function}  ({\bf BF}) was originally defined as~\cite{PhysRevLett.85.2689}
\begin{equation}
\begin{aligned}
B^{\alpha\beta}(\vec{p}^{\alpha},\vec{p}^{\beta}) \equiv \frac{1}{2} \left[ \frac{\rho_2^{\alpha^+\beta^-}(\vec{p}^{\alpha},\vec{p}^{\beta})}{\rho_1^{\alpha^+}(\vec{p}^{\alpha})} - \frac{\rho_2^{\alpha^+\beta^+}(\vec{p}^{\alpha},\vec{p}^{\beta})}{\rho_1^{\alpha^+}(\vec{p}^{\alpha})} \right. 
 + \left. \frac{\rho_2^{\alpha^-\beta^+}(\vec{p}^{\alpha},\vec{p}^{\beta})}{\rho_1^{\alpha^-}(\vec{p}^{\alpha})} - \frac{\rho_2^{\alpha^-\beta^-}(\vec{p}^{\alpha},\vec{p}^{\beta})}{\rho_1^{\alpha^-}(\vec{p}^{\alpha})} \right],
\label{eq:BF_Bass}
\end{aligned}
\end{equation}
where $\alpha$ stands for the reference species  and $\beta$ represents the associate species. Ratios of the form  
$\frac{\rho_2^{\alpha^+\beta^-}}{\rho_1^{\alpha^+}}$ are known as conditional densities expressing the number (i.e., the density) of (negative) particles of species $\beta$ observed at $\vec{p}^{\beta}$ given a (positive)  particle of species $\alpha$ is observed $\vec{p}^{\alpha}$.

The above  definition of the balance function misses some important details. In addition, by construction, it is not a robust quantity given it depends on conditional densities. It must then be explicitly corrected for detection efficiencies. By contrast, the correlation functions $R_{2}^{\alpha\beta}$ defined in the previous section are robust and thus independent of efficiencies (to first order). A more detailed BF definition derived from $R_{2}^{\alpha\beta}$ correlators  is thus used in this work.

We first introduce the notion of {\bf two-particle conditional cumulant}, hereafter noted  $A_2^{\alpha\beta}$, and  defined according to:
\begin{equation}
A_2^{\alpha\beta}(\vec{p}^{\alpha},\vec{p}^{\beta}) \equiv \rho_{1}^{\beta}(\vec{p}^{\beta}) \cdot R_2^{\alpha\beta}(\vec{p}^{\alpha},\vec{p}^{\beta})
 = \frac{\rho_{2}^{\alpha\beta}(\vec{p}^{\alpha},\vec{p}^{\beta})}{\rho_{1}^{\alpha}(\vec{p}^{\alpha})}- \rho_{1}^{\beta}(\vec{p}^{\beta}),
\label{eq:A2}
\end{equation}
The functions $A_2^{\alpha\beta}(\vec{p}^{\alpha},\vec{p}^{\beta})$ defined in Eq.~(\ref{eq:A2}) have  also been referred to as balance function in ~\cite{VOLOSHIN2006490}. They should be interpreted  as distributions of the ``associate" particles $\beta$ to be found at momentum  $\vec{p}^{\beta}$ given (under the condition)  a ``reference" particle $\alpha$ is observed at momentum $\vec{p}^{\alpha}$.

Opposite-sign particle pairs are combined to yield unlike-sign combinations of $A_{2}$ correlators, hereafter noted  $A_{2}^{\alpha\beta,{\rm US}}$, and defined according to
\begin{equation}
\begin{aligned}
\label{eq:A_US}
A_2^{\alpha\beta,US}(\vec{p}^{\alpha},\vec{p}^{\beta}) =  \frac{1}{2} \left[ A_2^{\alpha^+\beta^-}(\vec{p}^{\alpha},\vec{p}^{\beta}) + A_2^{\alpha^-\beta^+}(\vec{p}^{\alpha},\vec{p}^{\beta}) \right] \\
= \frac{1}{2} \left[ \frac{\rho_{2}^{\alpha^+\beta^-}(\vec{p}^{\alpha},\vec{p}^{\beta})}{\rho_{1}^{\alpha^+}(\vec{p}^{\alpha})} - \rho_{1}^{\beta^-}(\vec{p}^{\beta}) + \frac{\rho_{2}^{\alpha^-\beta^+}(\vec{p}^{\alpha},\vec{p}^{\beta})}{\rho_{1}^{\alpha^-}(\vec{p}^{\alpha})} - \rho_{1}^{\beta^+}(\vec{p}^{\beta}) \right].
\end{aligned}
\end{equation}
Similarly, same-sign particle pairs are combined to yield like-sign combinations of $A_{2}$ correlators, hereafter noted $A_{2}^{\alpha\beta,{\rm LS}}$, and defined according to 
\begin{equation}
\begin{aligned}
\label{eq:A_LS}
A_2^{\alpha\beta,LS}(\vec{p}^{\alpha},\vec{p}^{\beta}) =  \frac{1}{2} \left[ A_2^{\alpha^+\beta^+}(\vec{p}^{\alpha},\vec{p}^{\beta}) + A_2^{\alpha^-\beta^-}(\vec{p}^{\alpha},\vec{p}^{\beta}) \right] \\
= \frac{1}{2} \left[ \frac{\rho_{2}^{\alpha^+\beta^+}(\vec{p}^{\alpha},\vec{p}^{\beta})}{\rho_{1}^{\alpha^+}(\vec{p}^{\alpha})} - \rho_{1}^{\beta^+}(\vec{p}^{\beta}) + \frac{\rho_{2}^{\alpha^-\beta^-}(\vec{p}^{\alpha},\vec{p}^{\beta})}{\rho_{1}^{\alpha^-}(\vec{p}^{\alpha})} - \rho_{1}^{\beta^-}(\vec{p}^{\beta}) \right].
\end{aligned}
\end{equation}

{\bf Balance functions of positive charged reference particle $\alpha^+$} are defined according to
\begin{equation}
\begin{aligned}
\label{eq:BF_experiment_plus}
B^{\alpha^+\beta}(\vec{p}^{\alpha},\vec{p}^{\beta}) = A_2^{\alpha^+\beta^-}(\vec{p}^{\alpha},\vec{p}^{\beta}) - A_2^{\alpha^+\beta^+}(\vec{p}^{\alpha},\vec{p}^{\beta}) \\
= \frac{\rho_{2}^{\alpha^+\beta^-}(\vec{p}^{\alpha},\vec{p}^{\beta})}{\rho_{1}^{\alpha^+}(\vec{p}^{\alpha})} - \rho_{1}^{\beta^-}(\vec{p}^{\beta}) - \frac{\rho_{2}^{\alpha^+\beta^+}(\vec{p}^{\alpha},\vec{p}^{\beta})}{\rho_{1}^{\alpha^+}(\vec{p}^{\alpha})} + \rho_{1}^{\beta^+}(\vec{p}^{\beta}).
%= \rho_1^{\beta^-} R_2^{\alpha^+\beta^-}(\Delta y,\Delta \varphi) - \rho_1^{\beta^+} {R_2^{\alpha^+\beta^+}(\Delta y,\Delta \varphi)},
\end{aligned}
\end{equation}
The $B^{\alpha^+\beta}$ describes the conditional probability that a particle of reference species $\alpha$ with positive charge, in the bin $\vec{p}^{\alpha}$, is accompanied by a particle of associate species $\beta$ of negative charge in the bin $\vec{p}^{\beta}$.

The integral of $B^{\alpha^+\beta}(\vec{p}^{\alpha},\vec{p}^{\beta})$ over the full momentum space and all possible associate species $\beta$ is unity.
This normalization comes from the fact that for every produced positive charge, there must be a negative charge produced at approximately the same space-time, due to charge conservation.
Experimentally, detectors cover a finite acceptance. Only a fraction of the particles produced is observed. It is thus impossible to measure BFs for all possible associate species $\beta$. Consequently, for any given reference species $\alpha$ with positive charge, the BF integral measured experimentally is smaller than unity.
%if only charged kaons are measured and the hyperons with strangeness are excluded, the balance function would sum to the fraction of strangeness held in kaons.

The {\bf balance function of negative charged reference particle $\alpha^-$} is defined similarly and features similar properties
\begin{equation}
\begin{aligned}
\label{eq:BF_experiment_minus}
B^{\alpha^-\beta}(\vec{p}^{\alpha},\vec{p}^{\beta}) = A_2^{\alpha^-\beta^+}(\vec{p}^{\alpha},\vec{p}^{\beta}) - A_2^{\alpha^-\beta^-}(\vec{p}^{\alpha},\vec{p}^{\beta}) \\
= \frac{\rho_{2}^{\alpha^-\beta^+}(\vec{p}^{\alpha},\vec{p}^{\beta})}{\rho_{1}^{\alpha^-}(\vec{p}^{\alpha})} - \rho_{1}^{\beta^+}(\vec{p}^{\beta}) - \frac{\rho_{2}^{\alpha^-\beta^-}(\vec{p}^{\alpha},\vec{p}^{\beta})}{\rho_{1}^{\alpha^-}(\vec{p}^{\alpha})} + \rho_{1}^{\beta^-}(\vec{p}^{\beta})
\end{aligned}
\end{equation}
The functions $B^{\alpha^+\beta}$ and $B^{\alpha^-\beta}$ are also referred as {\bf signed balance functions} in ~\cite{PhysRevC.99.044901}, which are key observables to probe the initial strong magnetic field produced in high-energy nuclear collisions.

Computing the average of $B^{\alpha^+\beta}$ and $B^{\alpha^-\beta}$, we arrive at the same definition of BF as that given in Eq.~(\ref{eq:BF_Bass}), according to
\begin{equation}
\begin{aligned}
B^{\alpha\beta}(\vec{p}^{\alpha},\vec{p}^{\beta}) =  \frac{1}{2} \left[ B^{\alpha^+\beta}(\vec{p}^{\alpha},\vec{p}^{\beta}) + B^{\alpha^-\beta}(\vec{p}^{\alpha},\vec{p}^{\beta}) \right] \\
=  \frac{1}{2} \left[ \frac{\rho_{2}^{\alpha^+\beta^-}(\vec{p}^{\alpha},\vec{p}^{\beta})}{\rho_{1}^{\alpha^+}(\vec{p}^{\alpha})} - \frac{\rho_{2}^{\alpha^+\beta^+}(\vec{p}^{\alpha},\vec{p}^{\beta})}{\rho_{1}^{\alpha^+}(\vec{p}^{\alpha})} \right.
 + \left. \frac{\rho_{2}^{\alpha^-\beta^+}(\vec{p}^{\alpha},\vec{p}^{\beta})}{\rho_{1}^{\alpha^-}(\vec{p}^{\alpha})} - \frac{\rho_{2}^{\alpha^-\beta^-}(\vec{p}^{\alpha},\vec{p}^{\beta})}{\rho_{1}^{\alpha^-}(\vec{p}^{\alpha})} \right],
\end{aligned}
\end{equation}
which may also be written in terms of correlators $A_2^{\alpha\beta}$ and $R_2^{\alpha\beta}$
\begin{equation}
\begin{aligned}
\label{eq:BF_experiment}
B^{\alpha\beta}(\vec{p}^{\alpha},\vec{p}^{\beta}) = A_2^{\alpha\beta,US}(\vec{p}^{\alpha},\vec{p}^{\beta}) - A_2^{\alpha\beta,LS}(\vec{p}^{\alpha},\vec{p}^{\beta}) \\
= \frac{1}{2} \left[ A_2^{\alpha^+\beta^-}(\vec{p}^{\alpha},\vec{p}^{\beta}) - A_2^{\alpha^+\beta^+}(\vec{p}^{\alpha},\vec{p}^{\beta}) \right.
 + \left. A_2^{\alpha^-\beta^+}(\vec{p}^{\alpha},\vec{p}^{\beta}) - A_2^{\alpha^-\beta^-}(\vec{p}^{\alpha},\vec{p}^{\beta})  \right] \\
= \frac{1}{2} \left[ \rho_1^{\beta^-}(\vec{p}^{\beta}) \cdot R_2^{\alpha^+\beta^-}(\vec{p}^{\alpha},\vec{p}^{\beta}) - \rho_1^{\beta^+}(\vec{p}^{\beta}) \cdot R_2^{\alpha^+\beta^+}(\vec{p}^{\alpha},\vec{p}^{\beta}) \right. \\
 + \left. \rho_1^{\beta^+}(\vec{p}^{\beta}) \cdot R_2^{\alpha^-\beta^+}(\vec{p}^{\alpha},\vec{p}^{\beta}) - \rho_1^{\beta^-}(\vec{p}^{\beta}) \cdot R_2^{\alpha^-\beta^-}(\vec{p}^{\alpha},\vec{p}^{\beta})  \right]
\end{aligned}
\end{equation}
Experimentally, in this work, we measure BFs in terms of correlators $R_2^{\alpha\beta}$, defined as in Eq.~(\ref{eq:BF_experiment}). Given the correlators $R_2^{\alpha\beta}$ are robust against efficiency losses, 
 detection efficiency corrections are primarily needed for single particle densities $\rho_1^{\beta^{\pm}}(\vec{p}^{\beta})$.

%where the ratios  $\frac{\rho_2^{\alpha\beta}(\vec{p}^{\alpha},\vec{p}^{\beta})}{\rho_1^{\alpha}(\vec{p}^{\alpha})}$ correspond to conditional number densities, also known as per trigger yield correlator.
%Here, $\alpha$ is the species of the particle of interest, and $\beta$ is the species of the associated particle.
Measurements of balance functions can exploit any conserved charge including electric charge, strangeness, baryon number, charm, or other quantum numbers. In this context, the balance function is then termed {\bf General Balance Function (GBF)}. For instance, one can treat the ``$+$" and ``$-$" signs in the BF definition as strangeness and anti-strangeness, respectively, thereby yielding  strangeness BFs. Similarly, one can treat the  ``$+$" and ``$-$" signs  as baryon  and anti-baryon numbers, respectively, to obtain   baryon number BFs. In this work, the BFs of kaon--kaon ($KK$) account for  both electric charge and strangeness, while the BFs of proton-proton ($pp$) account for both electric charge and baryon number.

BFs are first computed as functions of $y_{\alpha}$, $\varphi_{\alpha}$, $y_{\beta}$, and $\varphi_{\beta}$ according to Eq.~(\ref{eq:BF_experiment}). 
The BFs are next averaged across $\bar y=(y_{\alpha}+y_{\beta})/2$ and  $\bar \varphi=(\varphi_{\alpha}+\varphi_{\beta})/2$ to yield
functions of  $\Delta y$ and $\Delta\varphi$ as follows
\begin{equation}
\begin{aligned}
\label{eq:BF_experiment_dydphi}
B^{\alpha\beta}(\Delta y,\Delta \varphi) = \frac{1}{2} \left[ \rho_1^{\beta^-} \cdot R_2^{\alpha^+\beta^-}(\Delta y,\Delta \varphi) - \rho_1^{\beta^+} \cdot R_2^{\alpha^+\beta^+}(\Delta y,\Delta \varphi) \right. \\
+ \left. \rho_1^{\beta^+} \cdot R_2^{\alpha^-\beta^+}(\Delta y,\Delta \varphi) - \rho_1^{\beta^-} \cdot R_2^{\alpha^-\beta^-}(\Delta y,\Delta \varphi)  \right]
\end{aligned}
\end{equation}
where the densities $\rho_1^{\beta^{\pm}}$ are calculated based on the published $p_{\rm T}$ spectra of $\pi^{\pm}$, $K^{\pm}$ and $p/\bar p$ in Pb--Pb collisions at $\sqrt{s_{_{\rm NN}}} =$2.76 TeV~\cite{PhysRevC.88.044910}, since $\rho_1^{\beta^{\pm}}$ are independent of rapidity $y^{\beta}$ and azimuth $\varphi^{\beta}$ in the fiducial volume of interest. 

\clearpage

\section{A Brief Survey of Balance Function Literature }
\label{sec:Survey}

This section presents a survey of the recent literature on (general) balance functions. It should be clear at the outset that the literature on balance functions is quite abundant. This  review is thus  focusing on what we believe are essential landmark works.

\subsection{Theoretical Works on Balance Functions}
\label{sec:theory}

\subsubsection{Balance function -- Definition and Motivations}

Bass, Danielewicz, and Pratt~\cite{PhysRevLett.85.2689} introduced the notion of balance function  to test the hypothesis that there exists a novel state of matter produced in relativistic heavy ion collisions, with normal hadrons not appearing until several fm/$c$ after the start of the collision. They proposed that charge dependent correlations be evaluated with the use of BFs. They argued that late-stage hadronization could be identified as tightly correlated charge vs. anti-charge pairs  as a function of relative rapidity.

Pratt~\cite{PRATT2002531} later introduced and discussed measurements of balance functions as a means to study the dynamics of the separation of balancing conserved charges. He argued that charges produced later in the collisions are more tightly correlated in (relative) rapidity. He also argued that late-stage production of charges signals the existence of a long-lived novel state of matter, the Quark Gluon Plasma.

\subsubsection{Sensitivity to radial flow}
Voloshin~\cite{VOLOSHIN2006490} broke down the balance function definition by~\cite{PhysRevLett.85.2689}, and introduced a more basic definition of charge balance function. He argued that the width of the BF is roughly inversely proportional to the transverse mass, which is consistent with the experimentally observed  narrowing of the charge balance function by STAR~\cite{PhysRevLett.90.172301}.
He also argued that because the charge BF is normalized to unity, the narrowing of the BF means an increase in the magnitude of the BF. 
In turn, this means an enhancement of the net charge multiplicity fluctuations if measured in a rapidity region comparable or smaller than the correlation length (1--2 units of rapidity). 
This observation might be an explanation for the centrality dependence of the net charge fluctuations measured at RHIC~\cite{PhysRevC.68.044905}. 
He also showed that the azimuthal correlations generated by transverse expansion could be a major contributor to the non-flow azimuthal correlations. 
And more importantly, this contribution would depend on centrality (following the development of radial flow), unlike many other non-flow effects. Note, however, that the typical elliptic flow centrality dependence (rise and fall) is different from that of the correlations due to transverse radial expansion.

\subsubsection{Sensitivity to transverse flow} 

Building on ideas by Pratt et al., Bozek~\cite{BOZEK2005247} presented a theoretical study of charge and  baryon number balance function vs. the relative azimuthal angle of pairs of particles emitted in ultrarelativistic heavy ion collisions. The $\pi\pi$ and $pp$ balance functions are computed using thermal models with two different set of parameters, corresponding to a large freeze-out temperature with a moderate transverse flow or a low temperature with a large transverse flow. The single-particle spectra including pions from resonance decays are similar for the two scenarios, but the azimuthal BFs are very different and could serve as an independent measure of transverse flow at freeze-out.

Bozek and Broniowski~\cite{BOZEK2013479c} presented a calculation of two-dimensional correlation functions in $\Delta\eta$--$\Delta\varphi$ for charged hadrons emitted in heavy ion collisions using an event-by-event hydrodynamics model. With the Glauber model for the initial density distributions in the transverse plane and elongated density profiles in the longitudinal direction, they reproduced the measured flow patterns in azimuthal angle of the two-dimensional correlation function. They showed that the additional fall-off of the same-side ridge in the longitudinal direction, an effect first seen in two-particle correlation measurements in Au--Au collisions at 200 GeV, can be explained as an effect of local charge conservation at a late stage of the evolution. They then argued that this additional non-flow effect increases the harmonic flow coefficients for unlike-sign particle pairs.

\subsubsection{General balance function}

Based on the canonical picture of the evolution of the QGP during a high-energy heavy ion collision in which quarks are produced in two waves, Pratt~\cite{PhysRevC.85.014904} introduced the notion of GBF. In his model, the first wave takes place during the first fm/$c$ of the collision, when gluons thermalize into the QGP. After a period of isentropic expansion, during which the number of quarks is approximately conserved, a second wave ensues at hadronization, about 5-10 fm/$c$ into the collision. Since entropy conservation requires the number of quasi-particles to stay roughly constant, and since each hadron contains at least two quarks, the majority of quark production occurs at this later time. For each quark produced in a heavy ion collision, an anti-quark of the same flavor must be created at the same point in space-time. Given the picture above one expects the distribution in relative rapidity of balancing charges to be characterized by two scales. To test this idea, STAR~\cite{PhysRevC.82.024905} presented BF measurements of identified charged-pion pairs and charged-kaon pairs in Au-Au collisions at $\sqrt{s_{\rm NN}}$ = 200 GeV.

Pratt~\cite{PhysRevC.85.014904} also showed  how BFs can be defined using any pair of hadronic states, and how one can identify and study both processes of quark production and transport. By considering BFs of several hadronic species, and by performing illustrative calculations, he showed that BF measurements hold the prospect of providing the field's most quantitative insight into the chemical evolution of the QGP. This dissertation is the first application of this idea by measuring the full charged hadron pair matrix $(\pi,K,p)\otimes (\pi,K,p)$ in relativistic heavy ion collisions.

Pratt~\cite{Pratt:2013xca} additionally discussed that correlations from charge conservation are affected by the time at which  charge/anticharge pairs are created during the course of a relativistic heavy ion collision. For charges created early, balancing charges are typically separated by about one unit of spatial rapidity by the end of the collision, whereas charges produced later in the collision are found to be more closely correlated in rapidity. By analyzing correlations from STAR for different species, he showed that one can distinguish the two separate waves of charge creation expected in a high-energy collision, one at early times when the QGP is formed and a second at hadronization. Furthermore, he extracted the density of up, down, and strange quarks in the QGP and found agreement at the 20\% level with expectations for a chemically thermalized plasma.

\subsubsection{Connection to other observables}

Jeon and Pratt~\cite{PhysRevC.65.044902} showed there is a simple connection between charge BFs, charge fluctuations, and correlations. In particular, they showed that charge fluctuations can be directly expressed in terms of BFs under certain assumptions. They discussed, more specifically, the distortions of charge BFs due to experimental acceptance and the effects of identical boson interference.

\subsubsection{Quark coalescence predictions}

Bialas~\cite{BIALAS200431} presented a quark antiquark coalescence mechanism for pion production which he uses to explain the small pseudorapidity width of BF observed for central heavy ion collisions. This model includes effects of the finite acceptance region and of the transverse flow. In contrast, the standard hadronic cluster model is not compatible with data.

Bialas and Rafelski~\cite{BIALAS2006488} presented a study of charge and baryon BFs based on coalescence hadronization mechanism of the QGP. Assuming that in the plasma phase, the $q\bar q$ pairs form uncorrelated clusters whose decay is also uncorrelated, one can understand the observed small width of the charge balance function in the Gaussian approximation. The coalescence model predicts even smaller width of the baryon-antibaryon BF relative to charge BF: 
$\sigma_{B\bar B} /\sigma_{+-} = \sqrt{2}/3$.

\subsubsection{Thermal model predictions}

Florkowski et al. ~\cite{Florkowski:2004em} presented a calculation of the $\pi^+\pi^-$ invariant-mass correlations and the pion BFs in the single-freeze-out model. A satisfactory agreement with the data measured in Au-Au collisions by the STAR Collaboration was found.

\subsubsection{Distortions from other physical processes}

Pratt and Cheng~\cite{PhysRevC.68.014907} showed that distortions from residual interactions and unbalanced charges may impair measurements of BFs. They estimated, within the context of simple models, the significance of these effects by constructing BFs in both relative rapidity and invariant relative momentum. While these considerations are not strictly relevant at LHC energies, because of the transparency of the colliding nuclei.

\subsubsection{Hydrodynamic models}

Ling, Springer, and Stephanov~\cite{PhysRevC.89.064901} applied a stochastic hydrodynamics model to study charge-density fluctuations in QCD matter undergoing Bjorken expansion. They found that the charge-density correlations are given by a time integral over the history of the system, with   dominant contribution coming from the QCD crossover region where the change of susceptibility per entropy, $\chi_T/s$, is most significant. They studied the rapidity and azimuthal angle dependence of the resulting charge balance function using a simple analytic model of heavy ion collision evolution. Their results are in agreement with experimental measurements, indicating that hydrodynamic fluctuations contribute significantly to the measured charge correlations in high-energy heavy ion collisions. The sensitivity of the balance function to the value of the charge diffusion coefficient, $D$, allowed them to estimate that the typical value of this coefficient in the crossover region is rather small, of the order of $(2\pi T )^{-1}$, characteristic of a strongly coupled plasma.

In general, hydrodynamic models do not handle conserved quantum numbers correctly within the Cooper-Frye prescription used to convert the energy into particles at freeze-out.

\subsubsection{Statistical models}

Cheng et al. ~\cite{PhysRevC.69.054906} investigated charge BFs with microscopic hadronic models and thermal models. The microscopic models give results which are contrary to STAR BF of pions whereas the thermal model roughly reproduces the experimental results. This suggests that charge conservation is local at breakup, which is in line with expectations for a delayed hadronization.

\subsubsection{Other models}

Song et al. ~\cite{PhysRevC.86.064903} calculated the charge BF of the bulk quark system before hadronization and those for the directly produced and the final hadron system in high-energy heavy ion collisions. They used the variance coefficient to describe the strength of the correlation between the momentum of the quark and that of the antiquark if they are produced in a pair and fix the parameter by comparing the results for hadrons with the available data. They studied the hadronization effects and decay contributions by comparing the results for hadrons with those for the bulk quark system. Their results indicate that while hadronization via a quark combination mechanism slightly increases the width of the charge BFs, it preserves the main features of these functions such as the longitudinal boost invariance and scaling properties in rapidity space. The influence from resonance decays on the width of the BF is more significant but it does not destroy its boost invariance and scaling properties in rapidity space either. Based on these considerations, it makes sense to consider BF measurements averaged over $\bar y = (y^{\alpha}+y^{\beta})/2$ across the acceptance of the ALICE detector as shown in Sec.~\ref{subsec:Cumulant}.

Li, Li, and Wu~\cite{PhysRevC.80.064910} used the PYTHIA  and AMPT models to study the longitudinal boost invariance of charge BF and its transverse momentum dependence. They found that within the context of these models, the charge BF is boost invariant in both pp and Au-Au collisions, in agreement with experimental data. The BF properly scaled by the width of the pseudorapidity window is independent of the position or the size of the window and is corresponding to the BF of the whole pseudorapidity range. 
They found that widths of BF also holds for particles in small transverse momentum ranges in the PYTHIA and the AMPT default models, but is violated in the AMPT with string melting.

\subsection{Experimental Works on Balance Functions}
\label{sec:experiment}

Several experiments have undertaken or completed measurements of BF. We here present a summary of the most important results.

%\subsubsection{EHS/NA22 Results}
%
%The EHS/NA22 collaboration~\cite{EHS/NA22:2006BF} reported a study of the boost invariance and multiplicity dependence of the charge BF in $\pi^+$p and $K^+$p collisions at 250 GeV/{\it c} incident beam momentum with full acceptance coverage. Charge balance, as well as charge fluctuations, are found to be boost invariant over the whole rapidity region, but both depend on the size of the rapidity window. It is also found that the BF becomes narrower with increasing multiplicity, which is consistent with the narrowing of the BF with increasing centrality and/or system size, as observed in current relativistic heavy ion experiments.

\subsubsection{STAR Results}

The STAR collaboration~\cite{PhysRevLett.90.172301} presented first BF measurements at RHIC for unidentified charged particle pairs and identified charged pion pairs in Au-Au collisions at $\sqrt{s_{\rm NN}}$ = 130 GeV. BFs for peripheral collisions have widths consistent with model predictions based on a superposition of nucleon-nucleon scattering. The measured BF widths in central collisions are smaller, which the STAR collaboration concluded is consistent with trends predicted by models incorporating late hadronization.

The STAR collaboration~\cite{Abelev2010239} presented BF measurements of unidentified charged particles, for diverse pseudorapidity and transverse momentum ranges in Au + Au collisions at $\sqrt{s_{\rm NN}}$ = 200 GeV. They observed that the BF is boost-invariant within the pseudorapidity coverage [-1.3,1.3]. The BF properly scaled by the width of the observed pseudorapidity window does not depend on the position or size of the pseudorapidity window. This scaling property also holds for particles in different transverse momentum ranges. In addition, they found that the BF width decreases monotonically with increasing transverse momentum for all centrality classes.

The STAR collaboration~\cite{PhysRevC.82.024905} presented BF measurements for charged-particle pairs, identified charged-pion pairs, and identified charged-kaon pairs in Au-Au, d-Au, and pp collisions at $\sqrt{s_{\rm NN}}$ = 200 GeV at RHIC. STAR observed that for charged-particle pairs, the BF width in $\Delta\eta$ scales smoothly with the number of participating nucleons, while HIJING and UrQMD model calculations show no dependence on centrality or system size. For charged-particle and charged-pion pairs, the BF widths in $\Delta\eta$ and $\Delta y$, are narrower in central Au-Au collisions than in peripheral collisions. The width for central collisions is consistent with thermal blast-wave models where the balancing charges are highly correlated in coordinate space at breakup. This strong correlation might be explained by either delayed hadronization or limited diffusion during the reaction. Furthermore, the narrowing trend is consistent with the lower kinetic temperatures inherent to more central collisions. In contrast, the BF width in $\Delta y$ of charged-kaon pairs shows little centrality dependence, which may signal a different production mechanism for kaons. The BF widths in $q_{\rm inv}$ for charged pions and kaons narrow in central collisions compared to peripheral collisions, which may be driven by the change in the kinetic temperature.

The STAR collaboration~\cite{PhysRevC.94.024909} reported BF measurements in terms of $\Delta\eta$ for unidentified charged particle pairs in Au-Au collisions with energies ranging from $\sqrt{s_{\rm NN}}$ = 7.7 GeV to 200 GeV. These results are compared with BFs measured in Pb-Pb collisions at $\sqrt{s_{\rm NN}}$ =  2.76 TeV by the ALICE Collaboration. The BF width decreases as the collisions become more central and as the beam energy is increased. In contrast, the BF widths calculated using shuffled events show little dependence on centrality or beam energy and are larger than the observed widths. The BF widths calculated using events generated by UrQMD are wider than the measured widths in central collisions and show little centrality dependence. STAR concluded that the measured BF widths in central collisions are consistent with the delayed hadronization of a deconfined QGP. They also stated that the narrowing of the BF in central collisions at $\sqrt{s_{\rm NN}}$ = 7.7 GeV implies that a QGP is still being created at this relatively low energy.

\subsubsection{ALICE Results}

The ALICE collaboration~\cite{2013267} reported the first LHC measurements of electric charge BF as functions of $\Delta\eta$ and $\Delta\varphi$ in Pb-Pb collisions at $\sqrt{s_{\rm NN}}$ = 2.76 TeV. The BF widths decrease for more central collisions in both projections. This centrality dependence is not reproduced by HIJING, while AMPT, a model which incorporates strings and parton rescattering, exhibits qualitative agreement with the measured correlations in $\Delta\varphi$ but fails to describe the correlations in $\Delta\eta$. A thermal blast-wave model incorporating local charge conservation and tuned to describe the $p_{\rm T}$ spectra and $v_2$ measurements reported by ALICE, is used to fit the centrality dependence of the BF width and to extract the average separation of balancing charges at freeze-out. The comparison of their results with measurements at lower energies reveals an ordering with $\sqrt{s_{\rm NN}}$: the BFs become narrower with increasing energy for all centralities. This is consistent with the effect of larger radial flow at the LHC energies but also with the late stage creation scenario of balancing charges. However, the relative decrease of the BF widths in $\Delta\eta$ and $\Delta\varphi$ with centrality from the highest SPS to the LHC energy exhibits only small differences. This observation cannot be interpreted solely within the framework where the majority of the charge is produced at a later stage in the evolution of the heavy ion collision.

The ALICE collaboration~\cite{Adam:2015gda, Pan:2017vrs} reported BF measurements in pp, p-Pb, and Pb-Pb collisions as a function of $\Delta\eta$ and $\Delta\varphi$. They presented the dependence of the BF on the event multiplicity as well as on the reference and associated particle $p_{\rm T}$ in pp, p-Pb, and Pb-Pb collisions at $\sqrt{s_{\rm NN}}$ = 7, 5.02, and 2.76 TeV, respectively. In the low transverse momentum region, for 0.2 $< p_{\rm T} <$ 2.0 GeV/c, the BF becomes narrower in both $\Delta\eta$ and $\Delta\varphi$ directions in all three systems for events with higher multiplicity. The experimental findings favor models that either incorporate some collective behavior (e.g., AMPT) or different mechanisms that lead to effects that resemble collective behavior (e.g. PYTHIA8 with color reconnection). For higher $p_{\rm T}$ values the BF becomes even narrower but exhibits no multiplicity dependence, indicating that the observed narrowing with increasing multiplicity at low $p_{\rm T}$ is a feature of bulk particle production.

\subsection{Theory vs. Experimental results}
\label{sec:THvsEX}

Pratt, McCormack, and Ratti~\cite{PhysRevC.92.064905} analyzed preliminary experimental measurements of charge BFs from the STAR Collaboration. They found that scenarios in which balancing charges are produced in a single surge, and therefore separated by a single length scale, are inconsistent with data. In contrast, a model that assumes two surges, one associated with the formation of a thermalized QGP and a second associated with hadronization, provides a far superior reproduction of the data. A statistical analysis of the model comparison finds that the two-surge model best reproduces the data if the charge production from the first surge is similar to expectations for equilibrated matter taken from lattice gauge theory. The charges created in the first surge appear to separate by approximately one unit of spatial rapidity before emission, while charges from the second wave appear to have separated by approximately a half unit or less.

%Pratt et al. presented an analysis of preliminary charge balance functions from the STAR Collaboration at the Relativistic Heavy Ion Collider (RHIC) with a model where quarks are produced in two waves. If a chemically equilibrated quark-gluon plasma (QGP) is created the strength and diffusive spread of the first wave should be governed by the chemical composition of the QGP, while the second wave should be determined by the increased number of quarks required to make the observed final-state hadrons. A simple model parameterizes the chemistry of the super-hadronic matter and the two correlation lengths for the two waves. Calculations are compared to preliminary data from the STAR Collaboration. The chemistry of the super-hadronic matter appears to be within 20\% of expectations from lattice gauge theory.

\subsection{The state of balance functions}

%Here, it would be nice to have a very short statement of the status of the field and what this thesis can contribute to.

As shown by the above literature review, the balance function is a key observable to learn about general charge creation mechanisms, the time scales of quark production, and the collective motion of the QGP, which are still open questions in relativistic heavy ion physics.
It is thus critical for experiments to measure BFs of various identified particle pairs with great precision. 
This dissertation presents the first BF measurement of the full charged hadron pair matrix $(\pi,K,p)\otimes (\pi,K,p)$ in relativistic heavy ion collisions, and provides new and challenging constraints for theoretical models of hadron production and transport.
To further deepen the understanding of the properties of the QGP, future BF measurement results in various collision systems with different energies are very much needed by the field, including BF of other particle pairs (e.g. lambda baryon), BF of heavy flavor particle pairs, BF as a function of $p_{\rm T}$, BF in terms of collision event plane, and signed BFs.

\clearpage

	\chapter{Experimental Setup}\label{chap:Setup}

In this chapter, we briefly describe the characteristics of the LHC accelerator and the ALICE experiment that are relevant for the balance function measurements presented in this work. 
Section~\ref{sec:LHC} presents a very brief description of the large hadron collider (LHC), while Sec.~\ref{ALICE} describes the features and components of the ALICE experiment relevant for the measurements presented.

\section{The Large Hadron Collider}
\label{sec:LHC}

\begin{figure}[h!]
\centering
\includegraphics[width=0.75\linewidth]{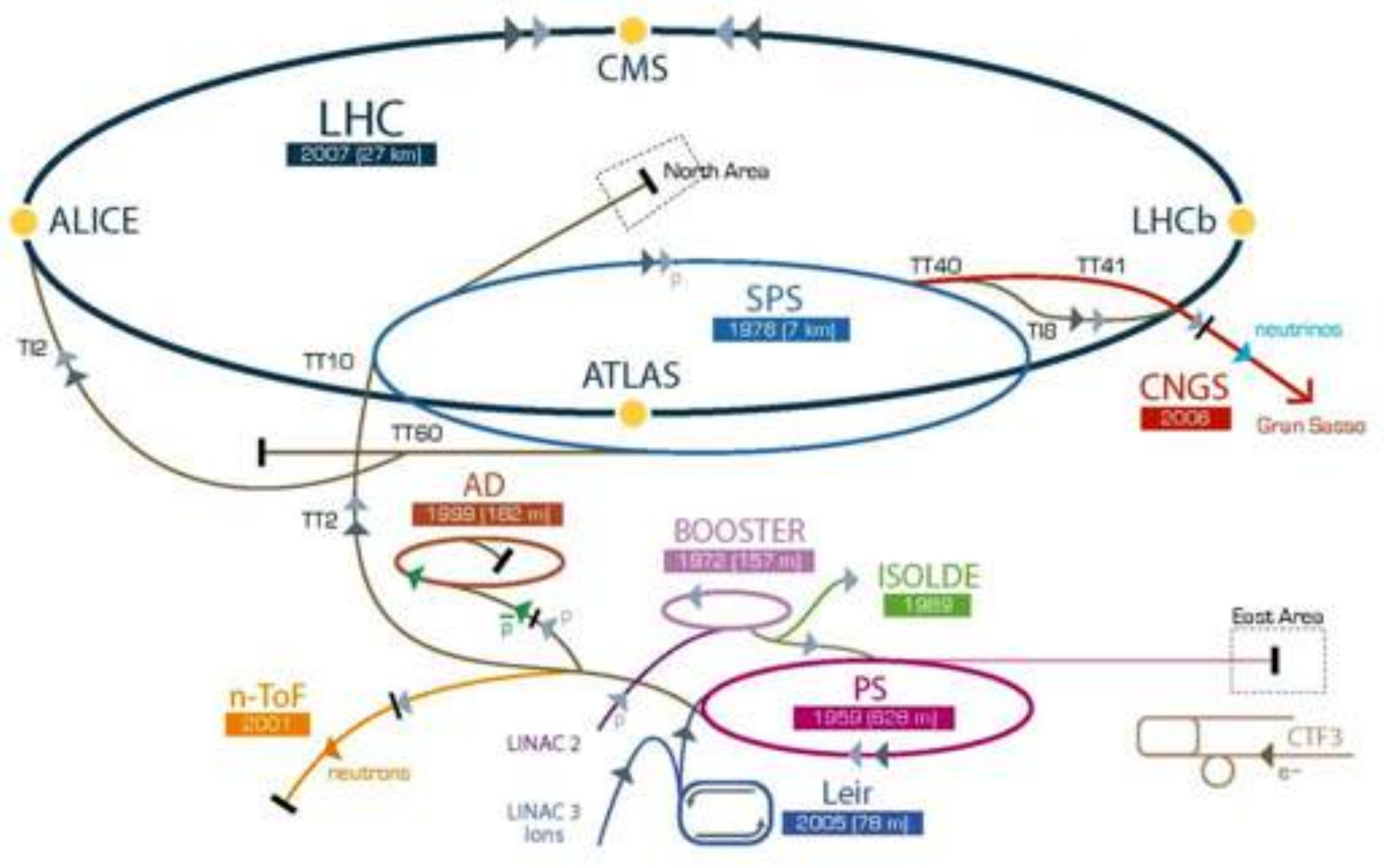}
\caption{The schematic illustration of the Large Hadron Collider and its major components.}
\label{fig:LHCparts}
\end{figure}

The Large Hadron Collider (LHC), as schematically illustrated in Figure~\ref{fig:LHCparts}, is the largest and most powerful particle accelerator ever built. It is located in a 27 km long circular underground tunnel across the border between France and Switzerland. It has the ability to accelerate charged particles to relativistic energies in excess of 1 TeV per nucleon. 

This PhD work is based on measurements carried out with the ALICE detector at the LHC.

\section{A Large Ion Collider Experiment}
\label{ALICE}

\begin{figure}[h!]
\centering
\includegraphics[width=0.75\linewidth]{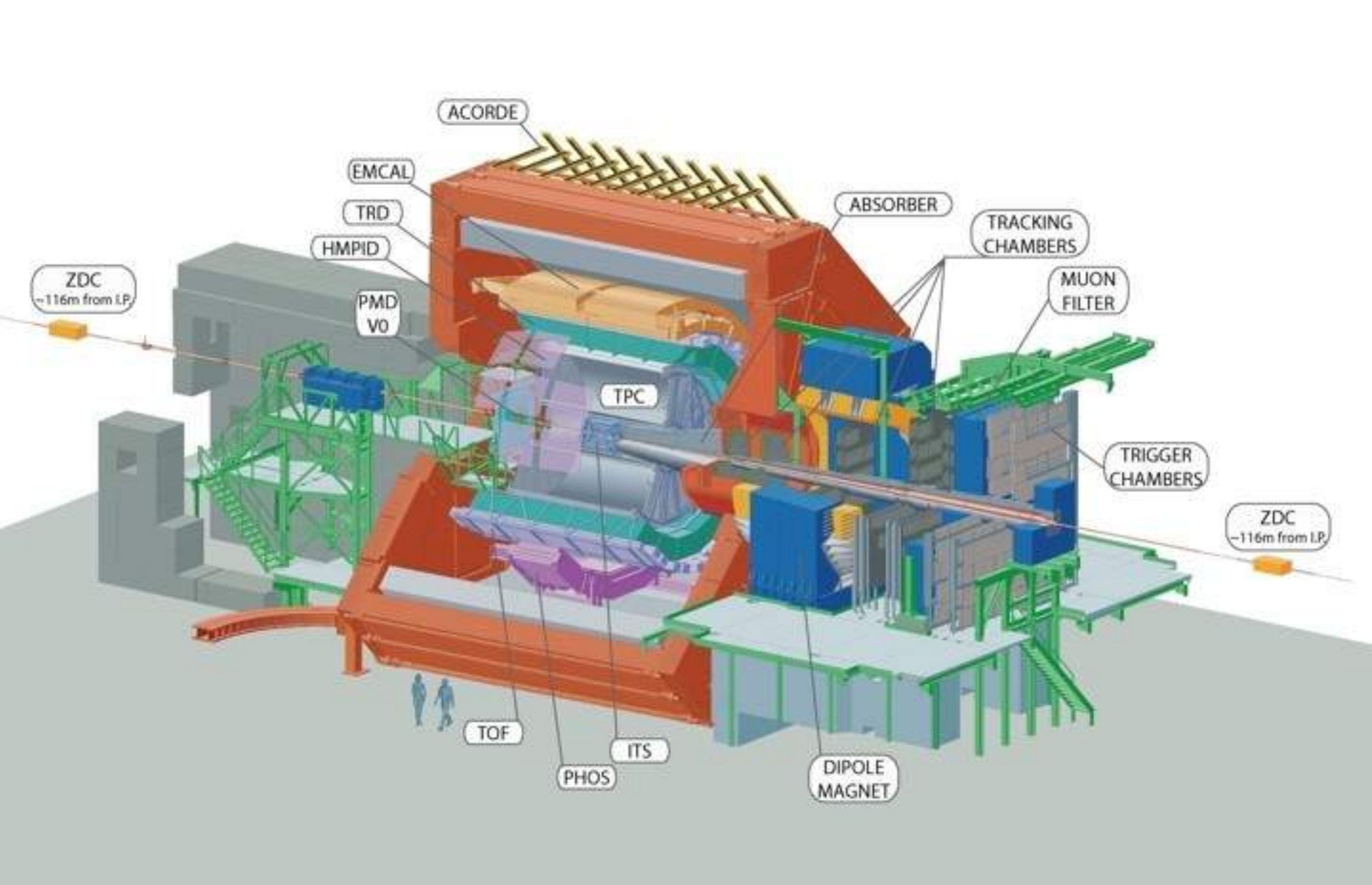}
\caption{The ALICE detectors.}
\label{fig:ALICEdetectors}
\end{figure}

A Large Ion Collider Experiment (ALICE)~\cite{Aamodt:2008zz} is a large multi-purpose experiment located at one of the collision points (P2) of the LHC. The ALICE collaboration involves about a thousand scientists and engineers from more than 100 institutes around the world. It has built and operates a dedicated heavy-ion detector to exploit the unique physics potential of nucleus-nucleus interactions at LHC energies. The aim is to study the physics of strongly interacting matter at extreme energy densities, where the formation of a new phase of matter, the quark-gluon plasma, is expected. The existence of such a phase and its properties are key issues in QCD for the understanding of confinement and of chiral-symmetry restoration. 

ALICE has an efficient and robust tracking system that enables the detection of particles over a large momentum range, from tens of MeV/c (soft physics) to over 100 GeV/c (jet physics). A specificity of the ALICE detector relative to other LHC experiments is its large focus on particle identification (PID). PID is achieved over a large momentum range based on a variety of techniques, including specific ionization energy loss dE/dx, time-of-flight, transition and Cherenkov radiation, electromagnetic calorimetry, muon filters, and topological decay reconstruction.

This work utilizes the TPC and ITS for charged hadron reconstruction. PID is achieved with signals from the TPC and TOF detectors, while triggering and event classification is largely based on signals from the V0 detectors. A brief overview of these detectors is provided in the following paragraphs, but a more complete and detailed description of these devises and their performances are provided in~\cite{Alessandro:2006yt}. The geometry and position of the detectors are illustrated in Figure~\ref{fig:ALICEdetectors}.

\subsection{ITS}
\label{ITS}

The Inner Tracking System (ITS) consists of six cylindrical layers of silicon detectors for precision tracking in the pseudorapidity range $|\eta|< 0.9$. The layers surround the collision point and measure the properties of particles emerging from the collision, pin-pointing their positions to a fraction of a millimeter. 
More details about the design, construction, calibration, and performance of the ITS can be found in~\cite{Dellacasa:1999kf,Alessandro:2006yt}.

\subsection{TPC}
\label{TPC}

The ALICE Time Projection Chamber (TPC) is the main detector used for the detection man momentum determination of charged particles. 
The ALICE TPC is a conventional but large TPC optimized for extreme track densities. The design, operation, and performance of the TPC are reported in~\cite{ALME2010316}.

\subsection{TOF}
\label{TOF}

The ALICE Time-Of-Flight (TOF) detector consists of is a large area array of Multigap Resistive Plate Chambers (MRPC), positioned at 370-399 cm from the beam axis. It covers the full azimuth and the pseudorapidity range $|\eta| < 0.9$. The TOF can separate $\pi/K$ and K/p up to approximately 2.5 GeV/c. The performance of the ALICE TOF is described in~\cite{Akindinov:2013tea}.

%\subsection{V0}
%
%Consider saying something about the Z-Cal and V0 detectors which are used collision centrality determination.

\clearpage

	\chapter{Analysis Methods}\label{chap:AnalysisMethods}

This chapter describes the analysis methods and techniques used in the determination of the balance functions presented in this work. The analyzed  datasets   are described  in Sec.~\ref{sec:DataSamples} while the specific  techniques used in this analysis  are detailed in Secs.~\ref{subsec:EventTrackSelection} -- \ref{sec:OtherFilterBits}. Section~\ref{subsec:EventTrackSelection} discusses event and track selection criteria used for inclusion of events and tracks in the determination of the correlation functions and balance functions. Section~\ref{sec:Particle Identification} presents a summary of the particle identification techniques used towards the identification of charged pions and kaons as well as protons and anti-protons.
Section~\ref{sec:Efficiency_Correction} discusses techniques and issues associated with track reconstruction and efficiency losses, whereas Sec.~\ref{subsec:TrackSplitting} elaborates on potential issues associated with splitting and merging in the reconstruction of long or close-by tracks. The roles and impacts of $\phi$-meson and $\Lambda$-baryon decays are discussed in Secs.~\ref{subsec:PhiDecay} and \ref{subsec:LambdaDecay}, respectively. Section~\ref{sec:OtherFilterBits} presents a comparative analysis of balance functions obtained with alternative track selection criteria.

\clearpage

\section{Data Samples}
\label{sec:DataSamples}

\subsection{Real Data Samples}
\label{subsec:RealDataSamples}

The results reported in this dissertation are based on Pb--Pb collisions at $\sqrt{s_{\mathrm{NN}}}=$ 2.76 TeV acquired during the 2010 LHC run. The analysis is specifically based on the data production {\it LHC10h-pass2-AOD160}, and involves runs taken with the ALICE magnet operated with positive and negative polarity. The data sample for this analysis is composed of the following list of runs.

Selected positive polarity runs are:\\
139510, 139507, 139505, 139503, 139465, 139438, 139437, 139360, 139329,
139328, 139314, 139310, 139309, 139173, 139107, 139105, 139038, 139037,
139036, 139029, 139028, 138872, 138871, 138870, 138837, 138732, 138730,
138666, 138662, 138653, 138652, 138638, 138624, 138621, 138583, 138582,
138579, 138578, 138534, 138469, 138442, 138439, 138438, 138396, 138364,

while negative polarity runs are:\\
138275, 138225, 138201, 138197, 138192, 138190, 137848, 137844, 137752,
137751, 137724, 137722, 137718, 137704, 137693, 137692, 137691, 137686,
137685, 137639, 137638, 137608, 137595, 137549, 137546, 137544, 137541,
137539, 137531, 137530, 137443, 137441, 137440, 137439, 137434, 137432,
137431, 137430, 137366, 137243, 137236, 137235, 137232, 137231, 137230,
137162, 137161, 137135.

These data are acquired with a minimum bias trigger based on the V0 detector and the SPD detector. A description of the triggering system and its performance are recorded in~\cite{Abelev:2014ffa}.

\clearpage

\subsection{Monte Carlo Data Samples}
\label{subsec:MCDataSamples}

In this work, we also utilize Monte Carlo (MC) simulation data of the Alice detector performance, specifically the production {\it LHC11a10a-bis-AOD162} with option ``PIDResponseTuneOnData", which anchors the real data production {\it LHC10h-pass2-AOD160} in Pb--Pb collisions at $\sqrt{s_{\mathrm{NN}}}=$ 2.76 TeV.
The MC production {\it LHC11a10a-bis-AOD162} is produced with the HIJING model~\cite{Gyulassy:1994ew}, with the same run numbers as in the real data production {\it LHC10h-pass2-AOD160}.

The MC data is analyzed to determine particle identification purities reported in Sec.~\ref{sec:Particle Identification}, and the MC closure tests reported in Sec.~\ref{subsubsec:MCHIJINGClosureTest}.

\clearpage

\section{Event and Track Selection}
\label{subsec:EventTrackSelection}

This section presents a detailed description of the analysis techniques used towards event selection, collision centrality definition and selection~\cite{Abelev:2013qoq}, as well as track selection.

Measurements of the correlation functions are restricted to primary charged hadrons, which are reconstructed with the ALICE ITS and TPC.
%The track reconstruction and momentum determination are based on a Kalman filter.
% described in Ref.~\cite{xxxxx}.
Events are in general acquired with minimum bias triggers primarily based on the V0 counters and the Silicon Pixel Detector (SPD). 
Tracks are selected based on their length determined based on the number of reconstructed space points ($N_{\rm TPC clusters}$), as well as their longitudinal ($DCA_z$) and radial ($DCA_{xy}$) distance of closest approach (DCA) to the collision main vertex.
In ALICE, AOD tracks have a filter-bit mask, which stores the information about whether the track satisfies standard sets of quality criteria.
Each filter-bit corresponds to a given set of cuts.
In this work, the nominal analysis is performed on TPC only tracks corresponding to filter-bit 1.

Kinematic ranges and selection criteria were tuned to optimize the statistics (number of tracks), quality, and more specifically the purity of particle identification. Studies performed towards such optimizations are presented in Sec.~\ref{subsubsec:RangeOptimization}.
The data taking conditions can be summarized as follows:

\begin{itemize}
\renewcommand{\labelitemi}{$\bullet$}
%%%%%%%%%%%%%%%%%%%%%%%%%%%%%%%
\item Pb--Pb collisions at $\sqrt{s_{\mathrm{NN}}}=$ 2.76 TeV
\begin{itemize}
\item Accepted events: $1 \times 10^7$ collisions,
\item Trigger:  {\it AliVEvent::kMB},
\item Centrality selection:  V0-M detector,
\item Longitudinal event vertex position: $|V_z| \le 6$ cm,
\item Track selection criteria (for inclusion in this analysis),
\begin{itemize}
  \item Charged Pion $\pi^{\pm}$
  \begin{itemize}
  \item Transverse momentum: $0.2 \le p_{\rm T} \le 2.0$ GeV/$c$,
  \item Rapidity: $|y| \le 0.8$ for $\pi\pi$ pair; $|y| \le 0.7$ for cross-species pairs,
  \item $N_{\rm TPC clusters} \ge 70$ out of a maximum of 159,
  \item DCA: $DCA_z \le 2.0$ cm, $DCA_{xy} \le 0.04$ cm,
%  \item $\chi^2 \le XXXXX$
%  \item PID:
  \end{itemize}

  \item Charged Kaon $K^{\pm}$
  \begin{itemize}
  \item Transverse Momentum: $0.2 \le p_{\rm T} \le 2.0$ GeV/$c$,
  \item Rapidity: $|y| \le 0.7$,
  \item $N_{\rm TPC clusters} \ge 70$ out of a maximum of 159,
  \item DCA: $DCA_z \le 2.0$ cm, $DCA_{xy} \le 2.0$ cm,
%  \item $\chi^2 \le XXXXX$
%  \item PID:
  \end{itemize}

  \item (Anti-)Proton $p/\bar{p}$
  \begin{itemize}
  \item Transverse Momentum: $0.5 \le p_{\rm T} \le 2.5$ GeV/$c$,
  \item Rapidity: $|y| \le 0.6$ for $pp$ pair; $|y| \le 0.7$ for cross-species pairs,
  \item $N_{\rm TPC clusters} \ge 70$ out of a maximum of 159,
  \item DCA: $DCA_z \le 2.0$ cm, $DCA_{xy} \le 0.04$ cm,
%  \item $\chi^2 \le XXXXX$
%  \item PID:
  \end{itemize}
\end{itemize}
\end{itemize}
\end{itemize}

\clearpage

\section{Analysis Software}

This analysis is conducted within the context of the Physics Working Group of Correlation Function (PWG--CF) in the ALICE Collaboration, with the computing software initially developed by C. Pruneau and P. Pujahari for the study of unidentified hadron number and $p_{\rm T}$ correlations. I have further developed and generalized the code to extend the analysis to identified particle pairs with the freedom of choosing reference and associate particle species separately.

The open source analysis codes are available on Github:\\
\sloppy\url{https://github.com/alisw/AliPhysics/blob/master/PWGCF/Correlations/DPhi/AliAnalysisTaskGeneralBF.cxx} \\
\url{https://github.com/alisw/AliPhysics/blob/master/PWGCF/Correlations/DPhi/AliAnalysisTaskGeneralBF.h} \\
\url{https://github.com/alisw/AliPhysics/blob/master/PWGCF/Correlations/macros/PIDBFDptDpt/AddTaskGeneralBF.C}

\clearpage

\section{Primary Charged Hadron Identification}
\label{sec:Particle Identification}

%\subsubsection{Charged Hadron Identification}
%\label{ChargedHadronIdentification}

\begin{figure}
\centering
  \includegraphics[width=1.0\linewidth]{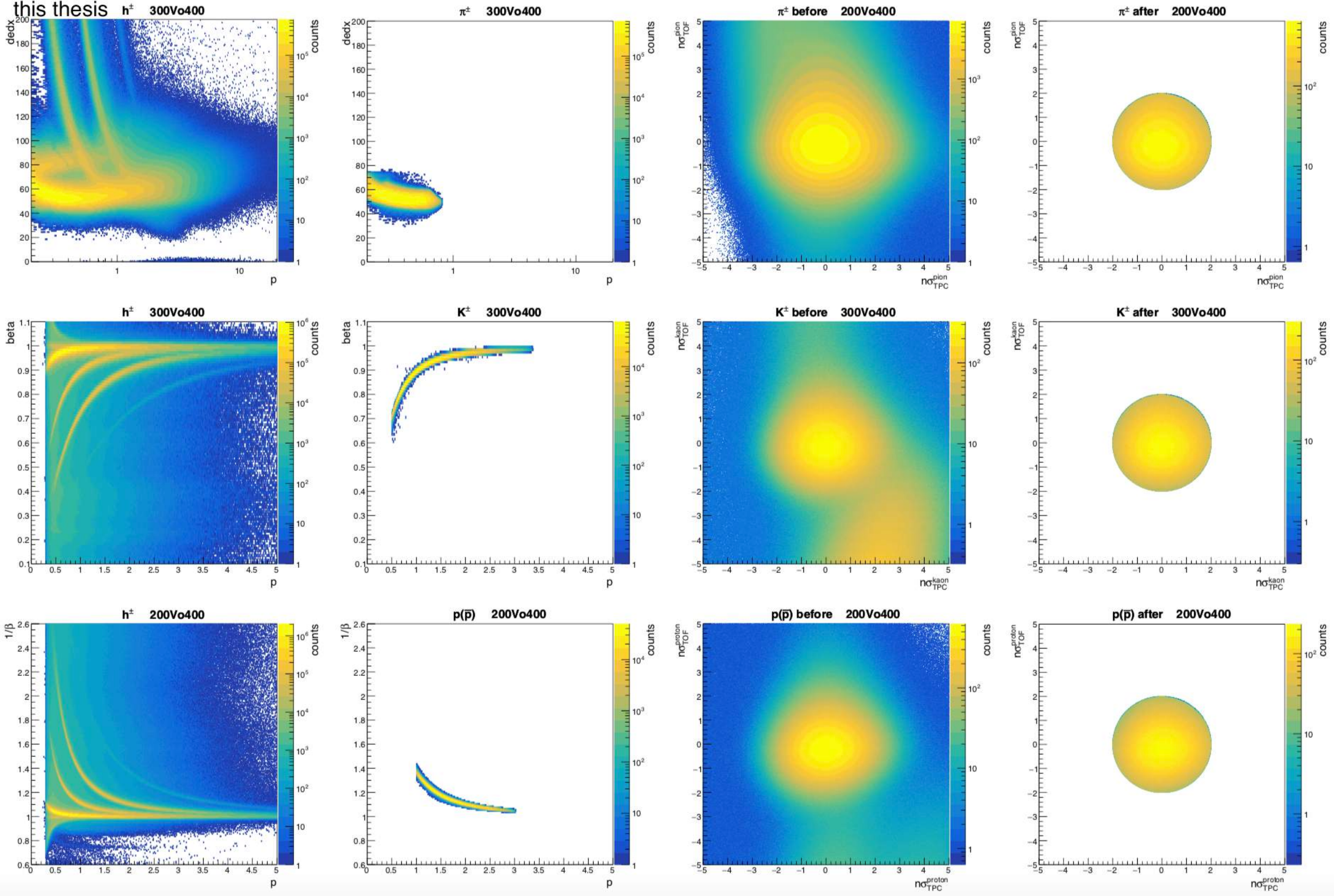}
  \caption{Examples of PID quality assurance plots used in this work: TPC dE/dx distribution of all charged hadrons (upper row $1^{st}$ column) and after $\pi^{\pm}$ PID cuts (upper row $2^{nd}$ column); TPC and TOF $n\sigma$ of $\pi^{\pm}$ before (upper row $3^{rd}$ column) and after PID cuts (upper row $4^{th}$ column); TOF $\beta$ distribution of all charged hadrons (middle row $1^{st}$ column) and after $K^{\pm}$ PID cuts (middle row $2^{nd}$ column); TPC and TOF $n\sigma$ of $K^{\pm}$ before (middle row $3^{rd}$ column) and after PID cuts (middle row $4^{th}$ column); TOF $1/\beta$ distribution of all charged hadrons (bottom row $1^{st}$ column) and after $p/\bar{p}$ PID cuts (bottom row $2^{nd}$ column); TPC and TOF $n\sigma$ of $p/\bar{p}$ before (bottom row $3^{rd}$ column) and after PID cuts (bottom row $4^{th}$ column).}
  \label{fig:PIDQA}
\end{figure}

\begin{figure}
\centering
  \includegraphics[width=0.99\linewidth]{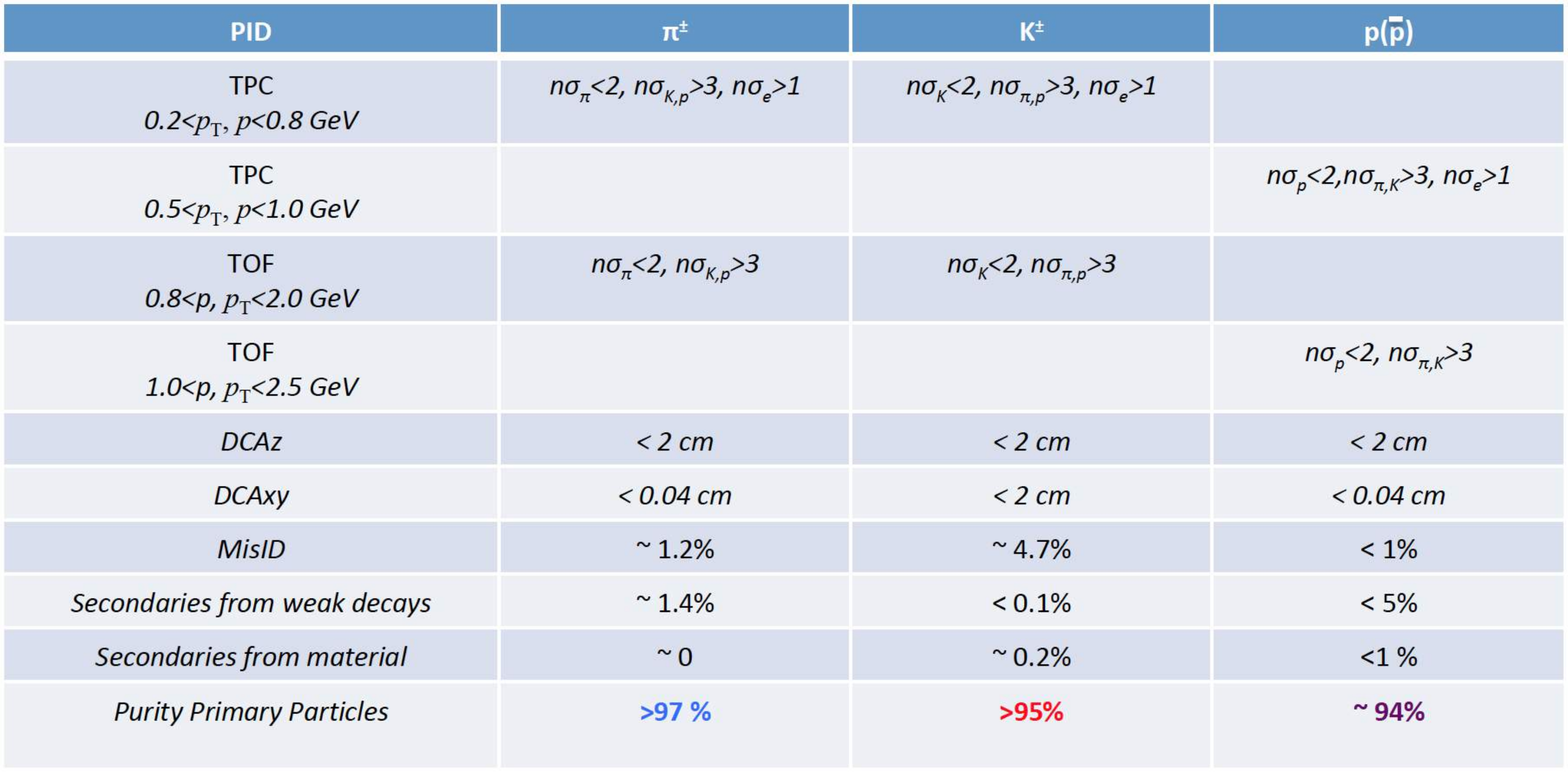}
  \caption{Detailed PID $n\sigma$ selection and veto cuts in TPC and TOF regions used towards the identification of $\pi^{\pm}$ ($2^{nd}$ column), $K^{\pm}$ ($3^{rd}$ column) and $p/\bar{p}$ ($4^{th}$ column), along with the DCA cuts used to reduce secondary particles from weak decays and interaction of primary particles with detector materials. The purities of identified primary hadrons along with detailed contamination percentages from different sources are also listed.}
  \label{fig:PID_Cuts}
\end{figure}

The identification of charged hadrons is performed using the $n\sigma$ method based on TPC dE/dx and TOF signals.
Fig.~\ref{fig:PIDQA} presents illustrative examples of quality assurance (QA) plots used in this work, towards  the identification and separation of $\pi^{\pm}$, $K^{\pm}$, and $p/\bar{p}$.
The detailed PID cuts used for the final results are listed in Fig.~\ref{fig:PID_Cuts}.

\begin{figure}
\centering
  \includegraphics[width=0.32\linewidth]{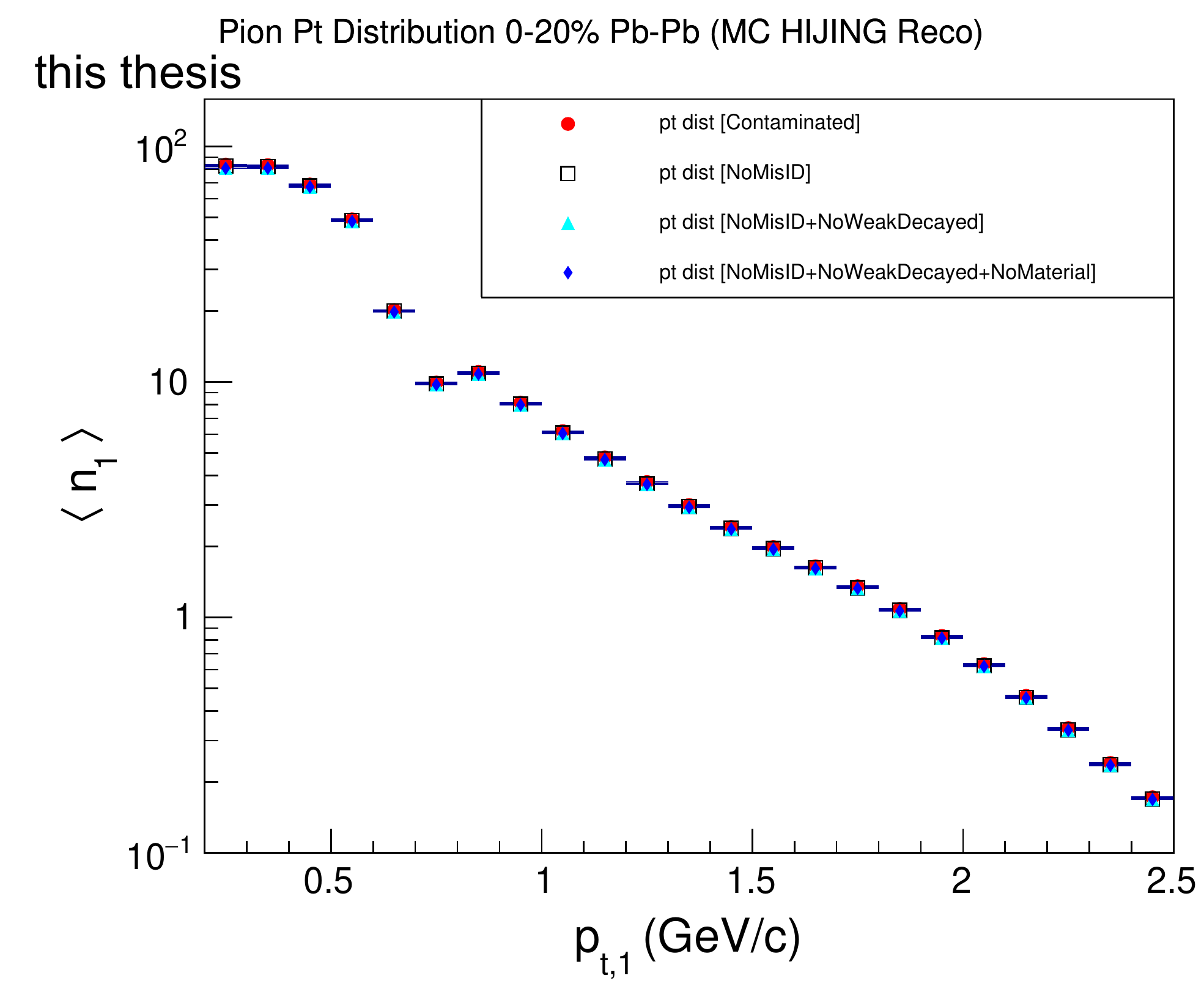}
  \includegraphics[width=0.32\linewidth]{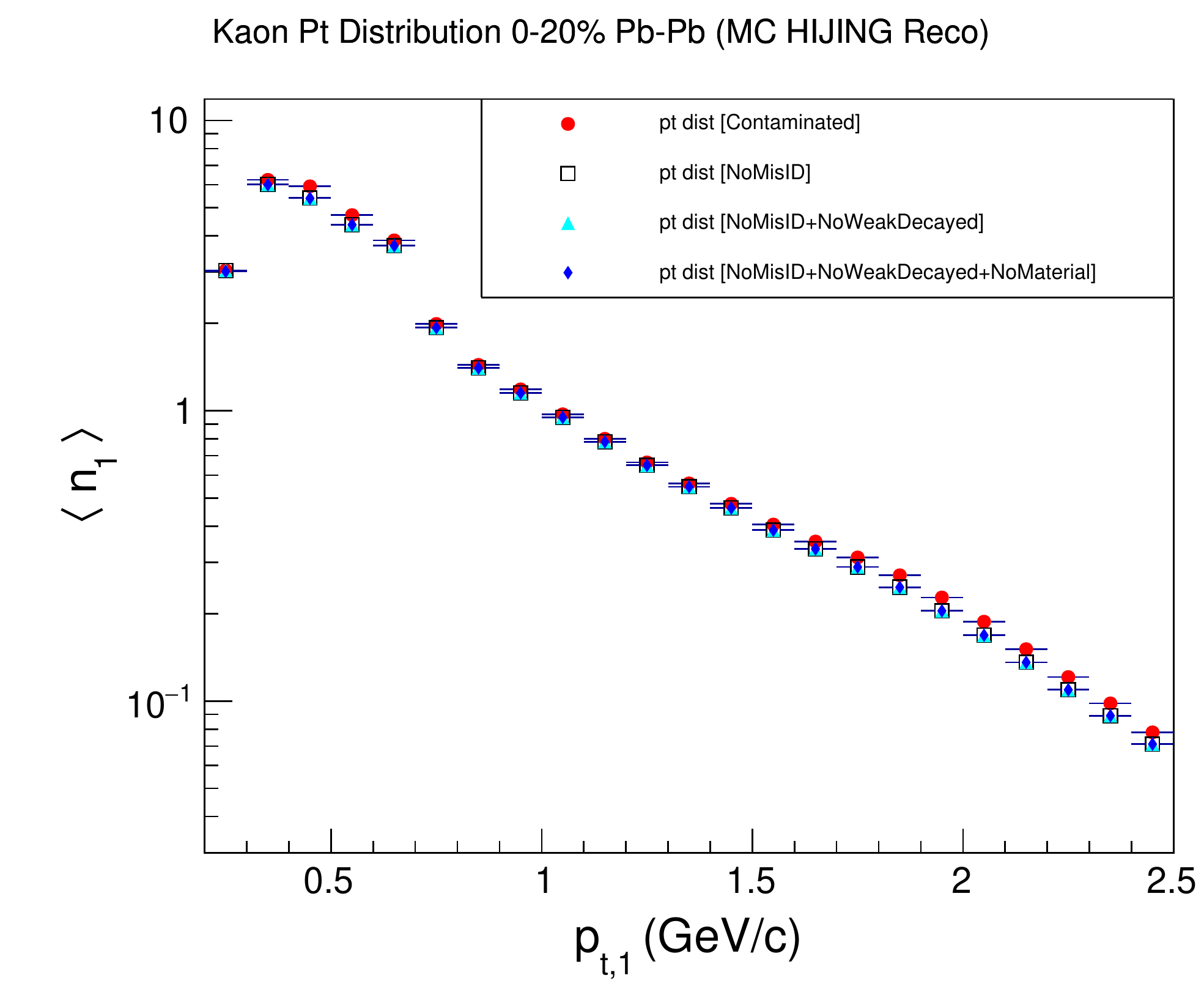}
  \includegraphics[width=0.32\linewidth]{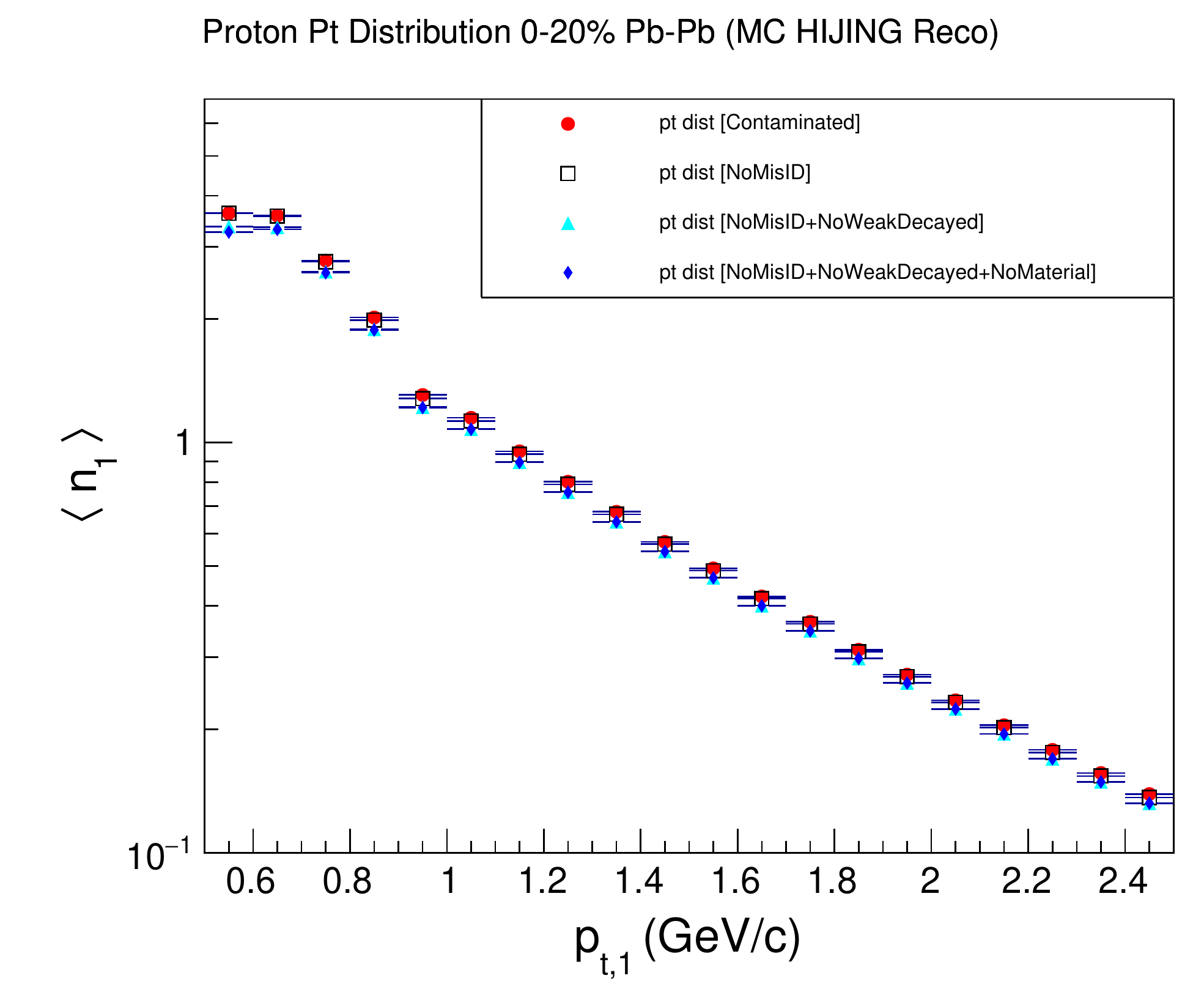}
  \includegraphics[width=0.32\linewidth]{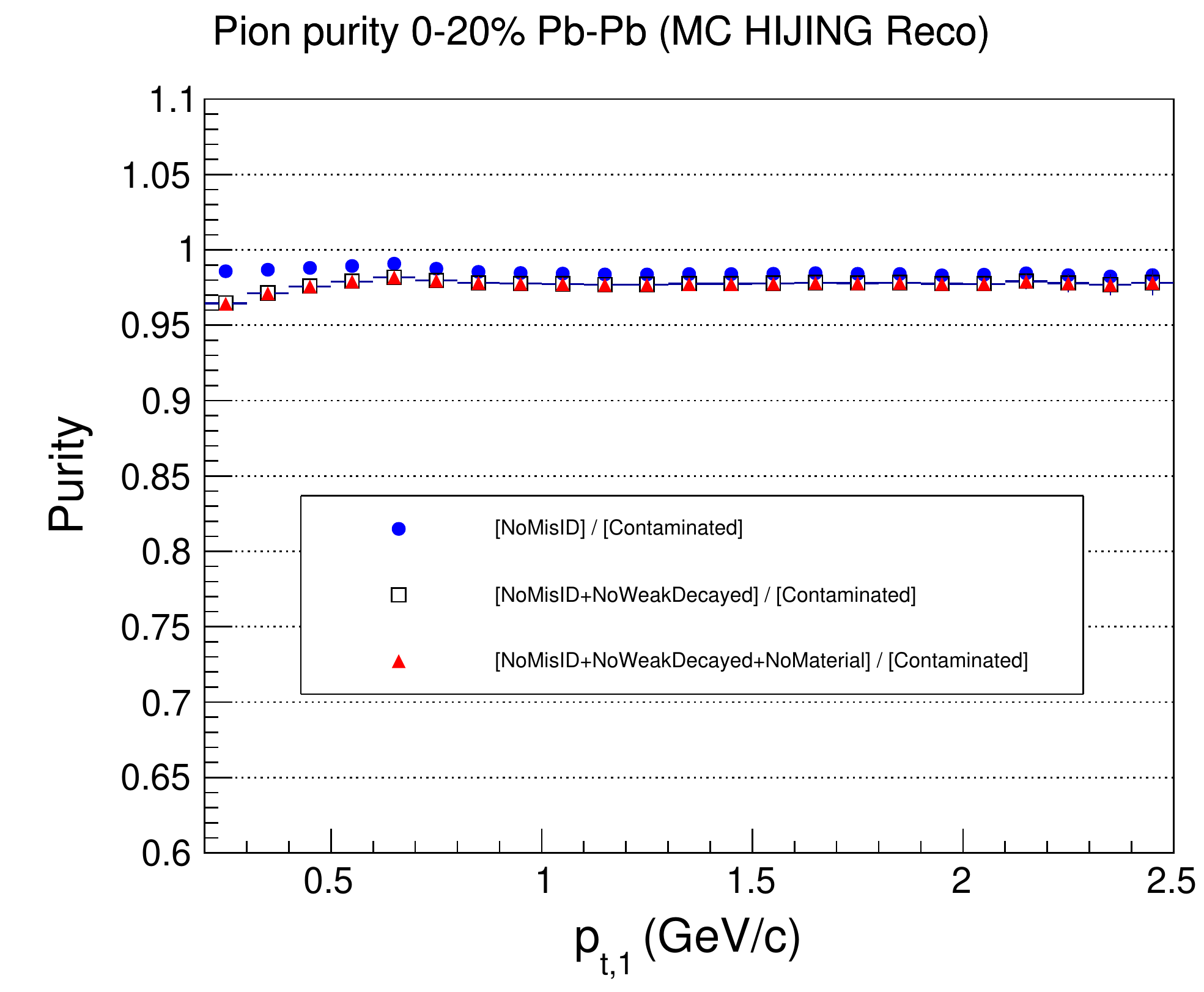}
  \includegraphics[width=0.32\linewidth]{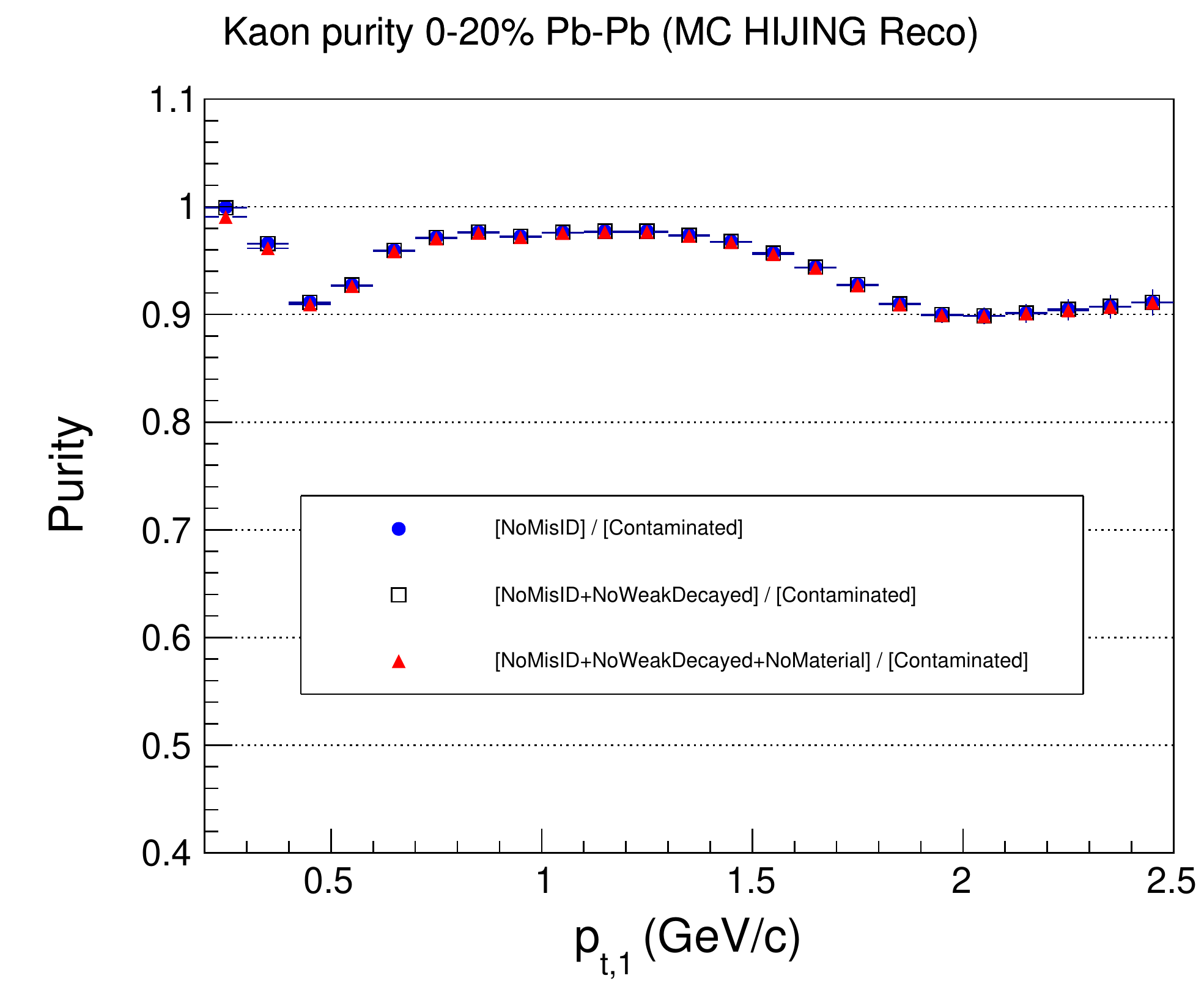}
  \includegraphics[width=0.32\linewidth]{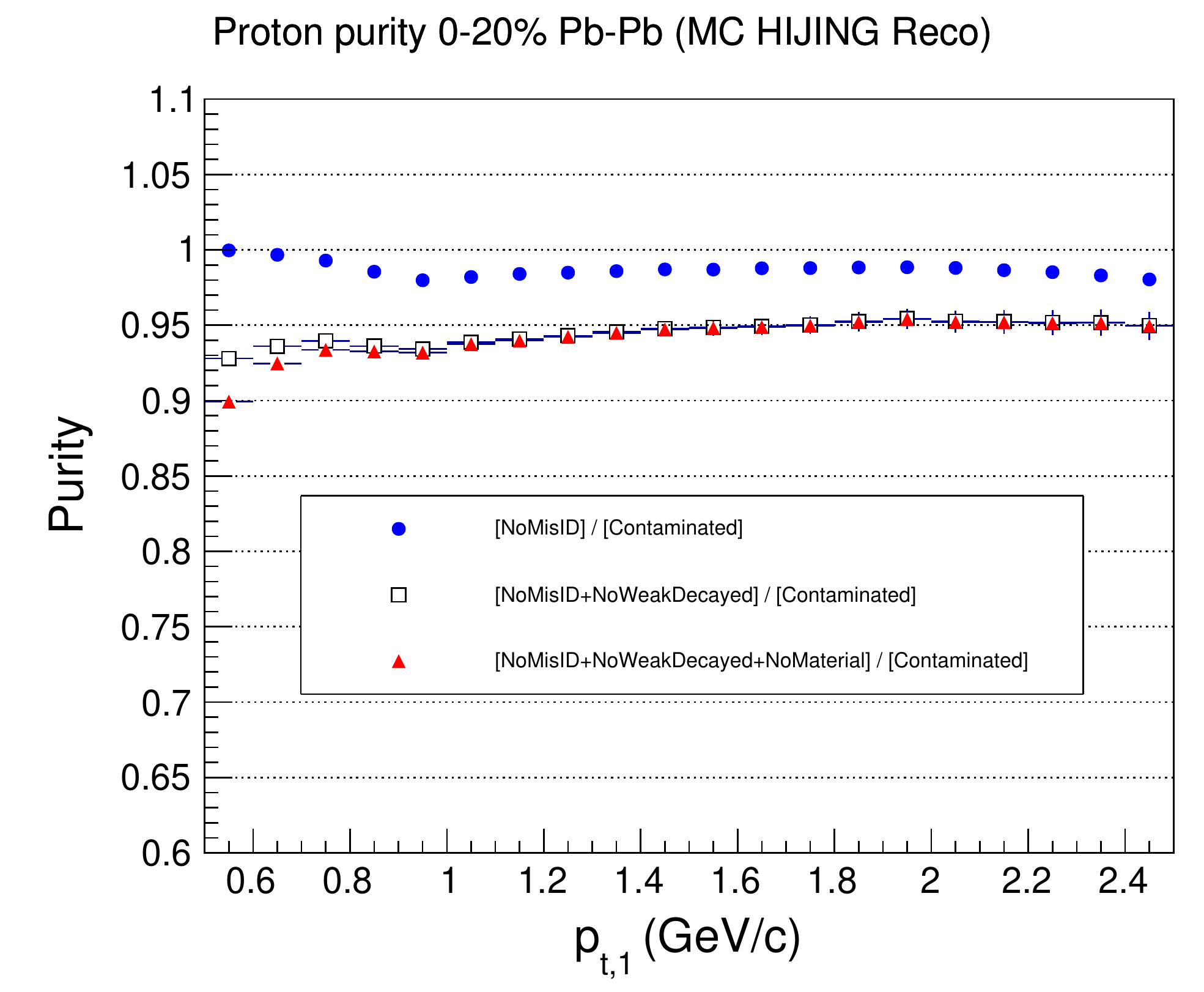}
  \caption{Purities (bottom row) of primary $\pi^{\pm}$ ($1^{st}$ column), $K^{\pm}$ ($2^{nd}$ column) and $p/\bar{p}$ ($3^{rd}$ column), obtained   using  MC reconstructed data simulations of  0-20\% Pb--Pb collisions. The $p_{\rm T}$ distributions (upper row) quantify the contributions from primary particles, secondary particles from weak decays and interactions with detector materials, and mis-identified particles. Purities have similar values in other centrality ranges.}
   \label{fig:Purity_pion_kaon_proton}
\end{figure}

Secondary particles are produced from weak decays and interactions of primary particles with the detector materials.
Their contributions to the measured BFs reported in this work must be eliminated or at the very least suppressed. This is achieved with the use of a tight $DCA_{xy}$ cut ( $DCA_{xy} \le 0.04$ cm ) for $\pi^{\pm}$ and $p/\bar{p}$ to remove secondary charged pions and protons. A wider $DCA_{xy}$ cut ( $DCA_{xy} \le 2$ cm ) is applied for $K^{\pm}$, since there are fewer secondary tracks for $K^{\pm}$.
Figures~\ref{fig:PID_Cuts} and \ref{fig:Purity_pion_kaon_proton} show that after the DCA cuts are applied, the fraction  of secondary particles remaining in the $\pi^{\pm}$ sample is about 1.4\%, while in the $K^{\pm}$ and $p/\bar{p}$  samples, they are  about 0.2\%, and  5\%, respectively.

The purity and contamination level of $\pi^{\pm}$, $K^{\pm}$ and $p/\bar{p}$ are plotted as a function of $p_{\rm T}$ in Fig.~\ref{fig:Purity_pion_kaon_proton}. The PIDResponseTuneOnData option in MC is used given it provides good dE/dx and TOF response simulations tuned to actual data. It thus provides reliable estimates of the purity of identified charged hadrons. The purities and contamination contributions from different sources are presented in Fig.~\ref{fig:PID_Cuts}. In summary, the purities of primary $\pi^{\pm}$, $K^{\pm}$ and $p/\bar{p}$ obtained in the determination of the balance 
functions reported in this work (i.e., final results)  are about 97\%, 95\%, and 94\%, respectively. The effects of remaining contamination are studied with various PID and DCA cuts, and used to estimate   systematic uncertainties.

%\clearpage

%%%%%%%%%%%%%%%%%%%%%%%%%%%%%%%%%%%%%%%%%%%%%%%%%%%%%%%%%%%%%%%%%
\section{Photon Conversion Study}

Photon conversions in the beam pipe and inner layers of the ALICE detector yields $e^{+} + e^{-}$ pairs ($\gamma + X \rightarrow e^{+} + e^{-}$). Electrons may be mis-identified as pions and kaons, and thus constitute a potential source of contamination in the study of the balance functions reported in this work. Additionally, given $e^{+} + e^{-}$ pairs resulting from photon conversions are emitted at small relative angles, such pairs may actually explicitly bias the $R_2^{US}$ correlation functions measured in this work. Such pairs would lead to a sharp and narrow spike at the origin of $R_2^{CD}$ correlation functions (as well as BFs). Suppressing $e^{+} + e^{-}$ pairs contamination is thus essential. 
  
\begin{figure}
\centering
    \includegraphics[width=0.49\textwidth]{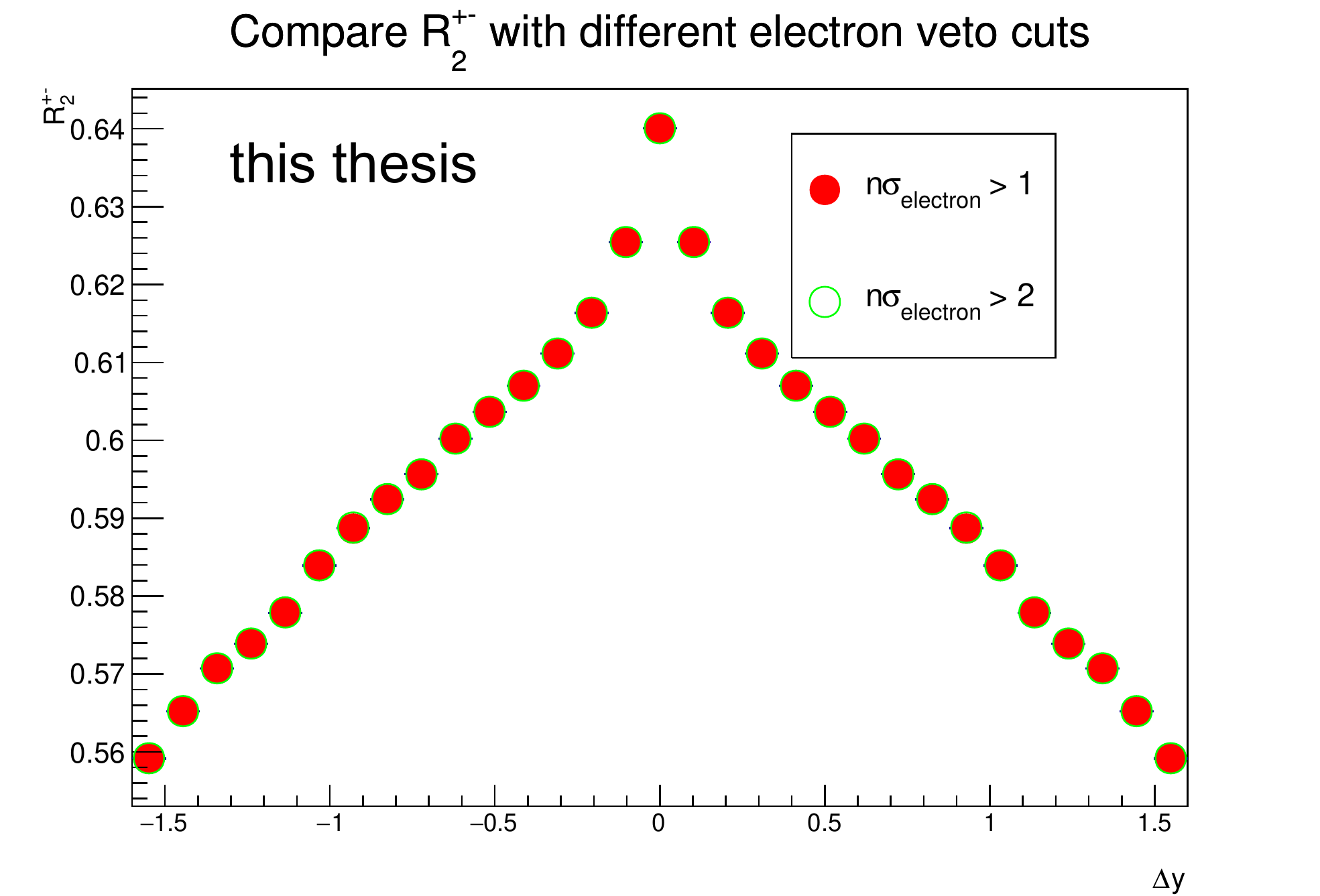}
    \includegraphics[width=0.49\textwidth]{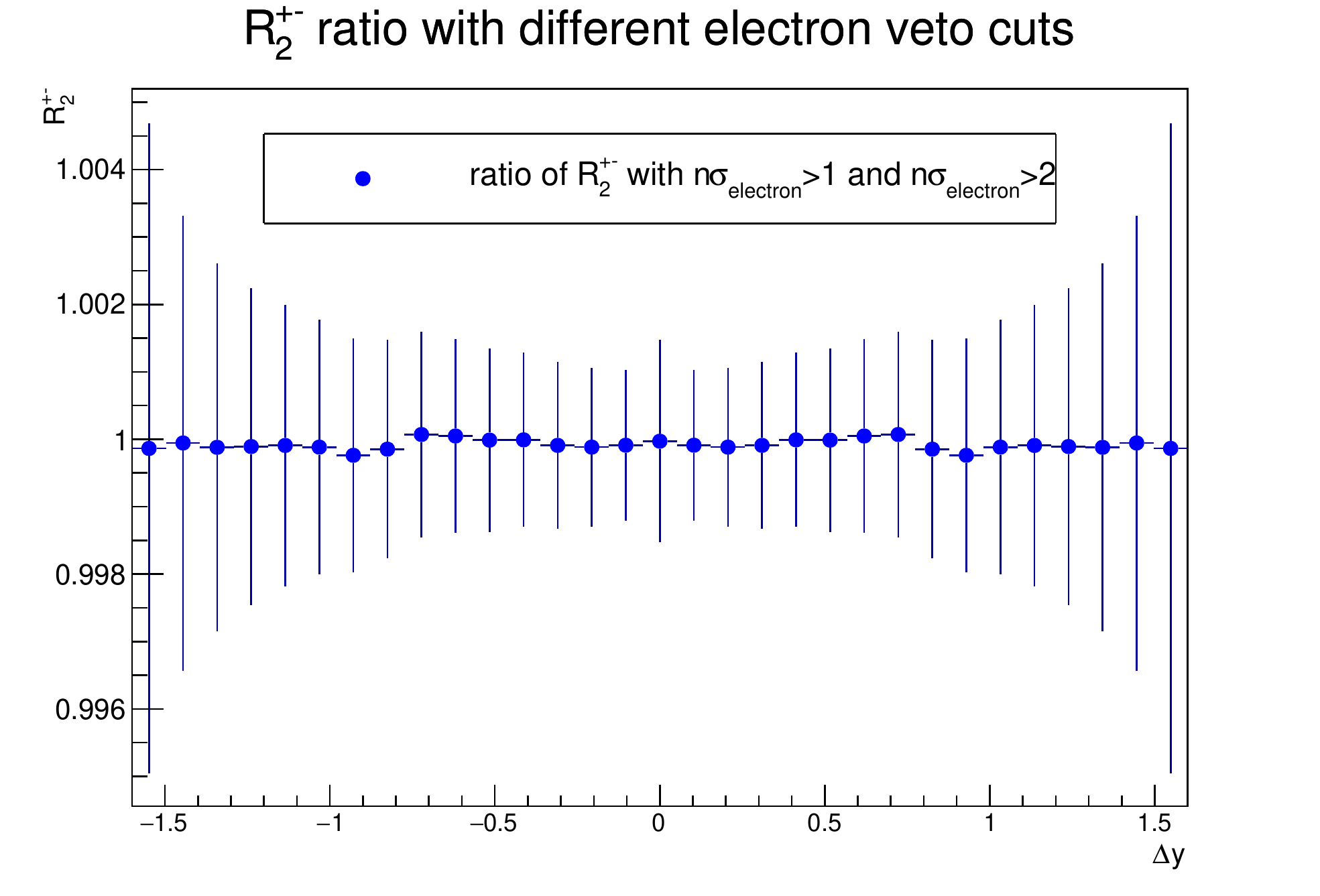}
  \caption{Study of the impact of photon conversions on correlation functions measured in Pb-Pb collisions: (left)  Comparison of  $R_{2}^{+-}$($\Delta$y) correlation functions obtained with and without an electron veto cut. (right) Ratio of the projections shown on the left. Note: Vertical bars correpond to statistical uncertainties not corrected for correlation between the $n\sigma_{electron}<1$ and $n\sigma_{electron}<2$ datasets.}
  \label{fig:Esigma1Esigma2}
\end{figure}
Particles are rejected if within 1$\sigma$ of the electron band in the TPC dE/dx region.
In the TOF region, no electron rejection cut is used for the sake of saving statistics since the electron band and the pion band overlap in the TOF $\beta$ vs momentum distribution, as shown in Fig.~\ref{fig:PIDQA}.
Figure~\ref{fig:Esigma1Esigma2} displays a comparison of $R_{\pi\pi}^{+-}$($\Delta$y) obtained with electron veto cuts $n\sigma_{electron}<1$ and $n\sigma_{electron}<2$ in 20-30\% Pb-Pb collisions. The two projections are nearly indistinguishable. 
%The right panel of the figure displays the ratio of the two projections and one finds the two are indeed essentially identical. 
Photon conversions thus appear to yield a very small contamination when a $n\sigma_{electron}<1$ cut is used. This less restrictive cut is thus used to minimize false rejections of pions and kaons.

%NOTE: the errors in the ratio are INCORRECT and must be fixed.

\clearpage

\section{Efficiency Corrections}
\label{sec:Efficiency_Correction}

\subsection{Basic Issue}

The correlation function $R_2$ is, by construction, robust against particle losses if the track detection efficiency is invariant across the experimental acceptance and independent of event parameters or detector conditions. Unfortunately, such dependences are observed in practice: The detection efficiency is a rather complicated function of the transverse momentum, the rapidity, and the azimuthal angle of the particles. It changes with detector occupancy and the position of the event primary-vertex.
%evaluated in sec.~\ref{sec:efficiency} 
This work thus uses a weight technique, described in~\cite{Ravan:2013lwa}, to correct for such dependences.
%The optimization of the acceptance and the kinematic ranges is carried out in this work, described in Sec.~\ref{subsubsec:RangeOptimization}.

\subsection{Weight Correction Method and Algorithm}

We describe the computation of the weights used in the study of correlation functions
involving  $\pi^{\pm}$ as an example. Weight calculations for other species are handled in a similar fashion.
The efficiency correction weights $w^{\pm}(y,\varphi,p_{\rm T},V_z)$ are computed using 24 bins in the primary vertex range $|V_{z}|\le6$ cm, 18 bins in the particle $p_{\rm T}$ range $0.2 \leq p_{\rm T} \leq 2.0 $ GeV, 28 bins in the azimuth range $0\le \varphi \le 2\pi$, and 16 bins in the rapidity range $|y|\le0.8$. This fine binning choice enables corrections for the efficiency dependence in 4 dimensions. The procedure used to establish the weights is as follows:

We first measure the uncorrected average transverse momentum yield, $N^{\pm}_{\rm avg}(p_{\rm T})$, of the positive and negative tracks separately.
This average, though not corrected for detection efficiency, is used as the reference yield vs $p_{\rm T}$. We next measure the positive and negative track yields, $N^{\pm}_{\rm avg}(y,\varphi,p_{\rm T},V_z)$, as a function of rapidity, y, azimuth angle, $\varphi$, and transverse momentum, $p_{\rm T}$, and the vertex position index, $V_z$, of the events.

The weights are thus calculated according to
\begin{equation}
w^{\pm}(y,\varphi,p_{\rm T},V_z) = \frac{N^{\pm}_{\rm avg}(p_{\rm T})}{N^{\pm}(y,\varphi,p_{\rm T},V_z)}
\label{eq:weights}
\end{equation}

The full dataset is then reprocessed using these weights. Single particle and two-particle yield histograms are incremented by
$w^{\pm}(y^\alpha,\varphi^\alpha,p_{\rm T}^\alpha,V_z^\alpha)$
and \\
$w^{\pm}(y^\alpha,\varphi^\alpha,p_{\rm T}^\alpha,V_z) \cdot w^{\pm}(y^\beta,\varphi^\beta,p_{\rm T}^\beta,V_z^\beta)$
respectively. It is verified that the use of the weights produces flat distributions in $y$, $\varphi$, and $V_z$.

\clearpage

\subsection{Optimization of $y$ and $V_{z}$ Ranges}
\label{subsubsec:RangeOptimization}

\begin{figure*}[h!]
\begin{center}
  \includegraphics[width=0.7\linewidth]{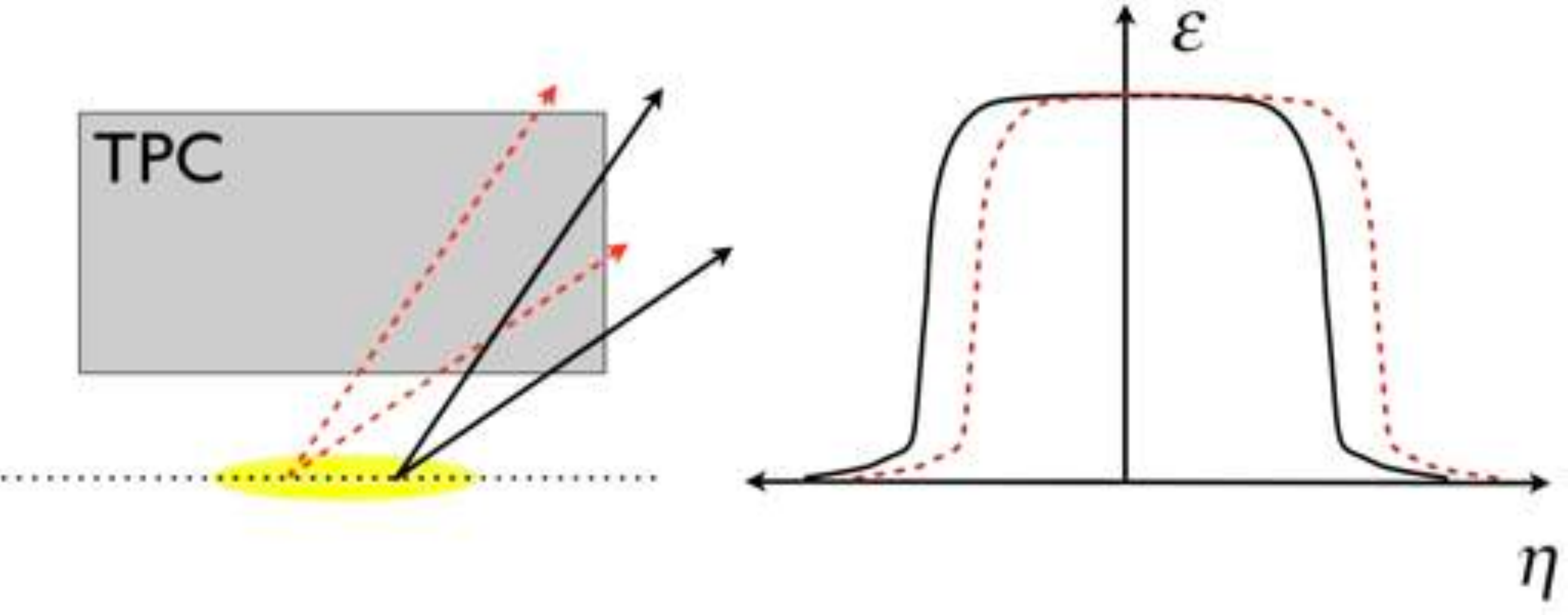}
  \caption{Schematic illustration of the TPC acceptance (left) and the longitudinal dependence (right) of the particle reconstruction efficiency on the vertex position $V_{z}$.}
  \label{fig:AcceptanceCartoon}
\end{center}
\end{figure*}

The LHC produces Pb--Pb collisions within a wide interaction diamond. In ALICE, the position  of interaction is characterized in terms of a primary vertex position $(V_{x},V_{y},V_{z})$, where $V_{z}$ represents the longitudinal position of the interaction, i.e. the position of the vertex in the beam direction.
While $V_{x}$ and $V_{y}$ represent the vertex position in the plane transverse to the beam axis.  
The distributions of $V_{x}$ and $V_{y}$ are typically relatively narrow and fully accepted in this analysis. The $V_{z}$ distribution, on the other hand, is rather wide, and characterized by  an approximately Gaussian distribution with a standard deviation of the order of 10 cm.
In the interest of maximizing the size of the statistical sample, one would wish to maximize the range of $V_{z}$-vertex position used in the analysis. 
However, the detector design provides for particle detection and identification optimization for collisions taking place near the nominal center of the detector. 
The rapidity acceptance and particle detection/identification efficiency, in particular, are found to depend on the $V_{z}$-vertex position, as schematically illustrated in Fig.~\ref{fig:AcceptanceCartoon}. 
While such dependence may be partly corrected with the weight technique used in this analysis, we found that the PID acceptance becomes severely compromised for very large $|V_{z}|$ values. 
We thus conducted an optimization study to determine what $V_{z}$ and $y$ ranges would produce the most reliable and robust correlation functions.

\begin{figure}
\centering
  \includegraphics[width=0.48\linewidth]{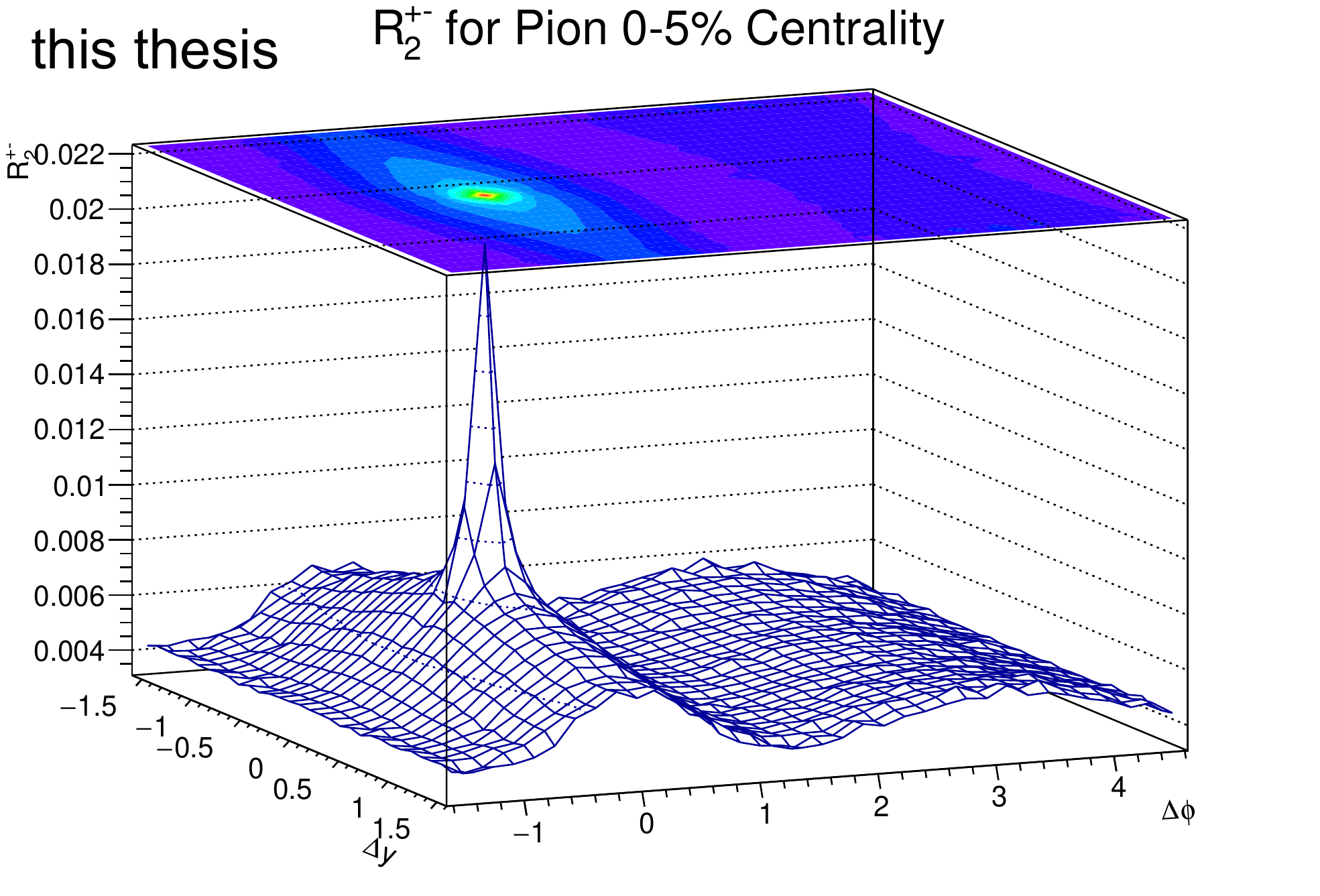}
  \includegraphics[width=0.48\linewidth]{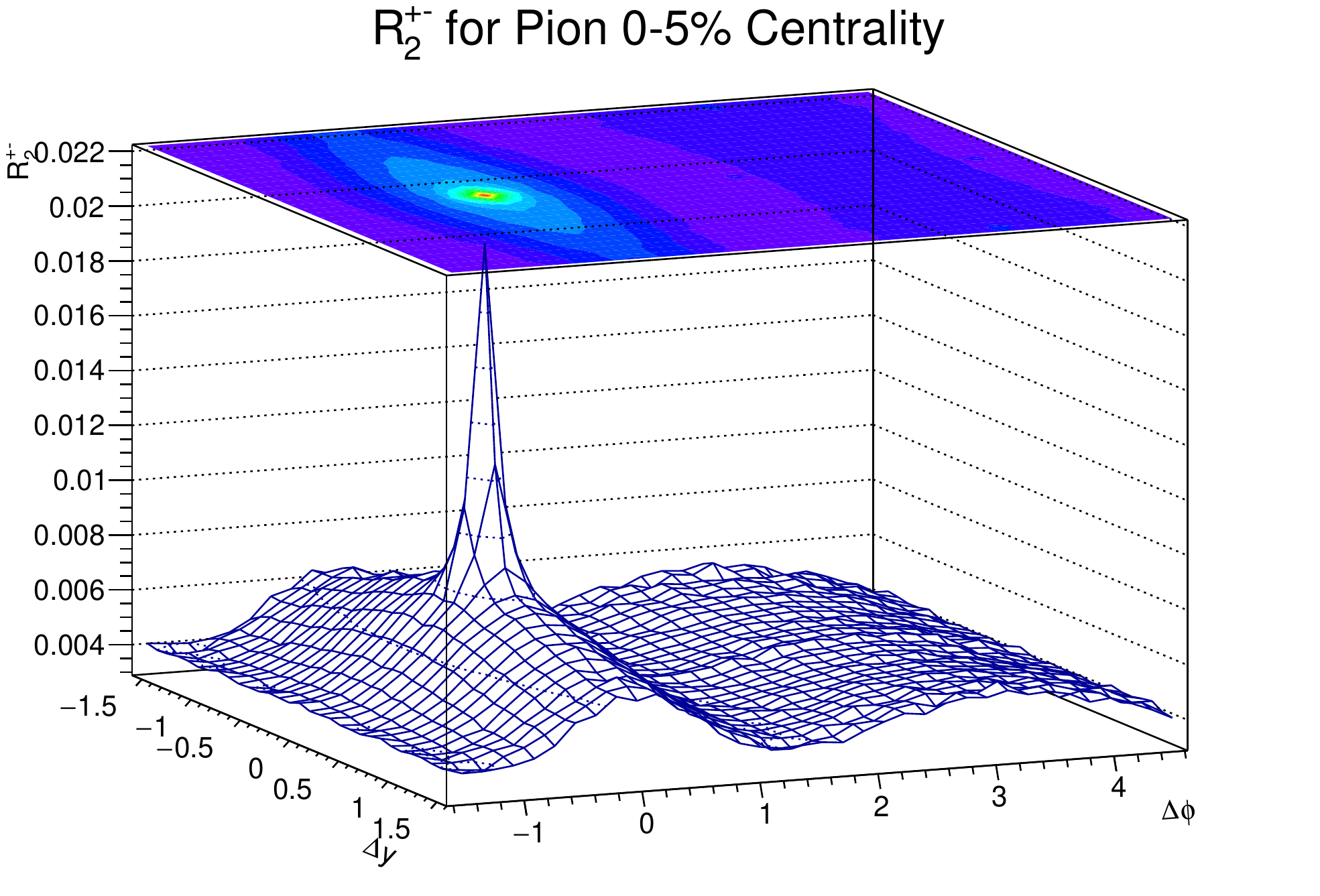}
  \caption{Comparison of correlation functions $R_{\pi\pi}^{+-}(\Delta y, \Delta\varphi)$ obtained with ranges $|V_{z}|<6$ cm (left) and $|V_{z}|<8$ cm (right) measured in 0-5\% Pb--Pb collisions with ++ B-Field.}
\label{fig:Pos_R2_PM_2d_Pion_6vZ24_0Cent5}  
\end{figure}

The previously published ALICE BF~\cite{2013267,Adam:2015gda} and $R_{2}^{CD}$~\cite{Acharya:2018ddg} papers reporting results on CFs of unidentified hadrons used the $|V_{z}|<10$ cm range.
However, Fig.~\ref{fig:Pos_R2_PM_2d_Pion_6vZ24_0Cent5} demonstrates that the use of large $V_{z}$ ranges could cause a little ridge structure along $\Delta y=0$ in 2D $R_2$ correlation functions for most central events, due to the effect illustrated in Fig.~\ref{fig:AcceptanceCartoon}. 
This little ridge is a charge independent effect and is present in both US and LS correlation functions with approximately the same strength. 
It nearly cancels out in the calculation of CD combination of correlation functions.
Thus, $|V_{z}|<6$ cm is used in the final result. 
For most central events of $\pi\pi$ and $KK$, the remaining little ridge is also corrected for US and LS correlation functions by taking the difference between $\Delta y=0$ bins and neighbor bins.
This correction is just for showing correct US and LS correlation functions, and has no impact on BFs.

\begin{figure}
\centering
  \includegraphics[width=1.0\linewidth]{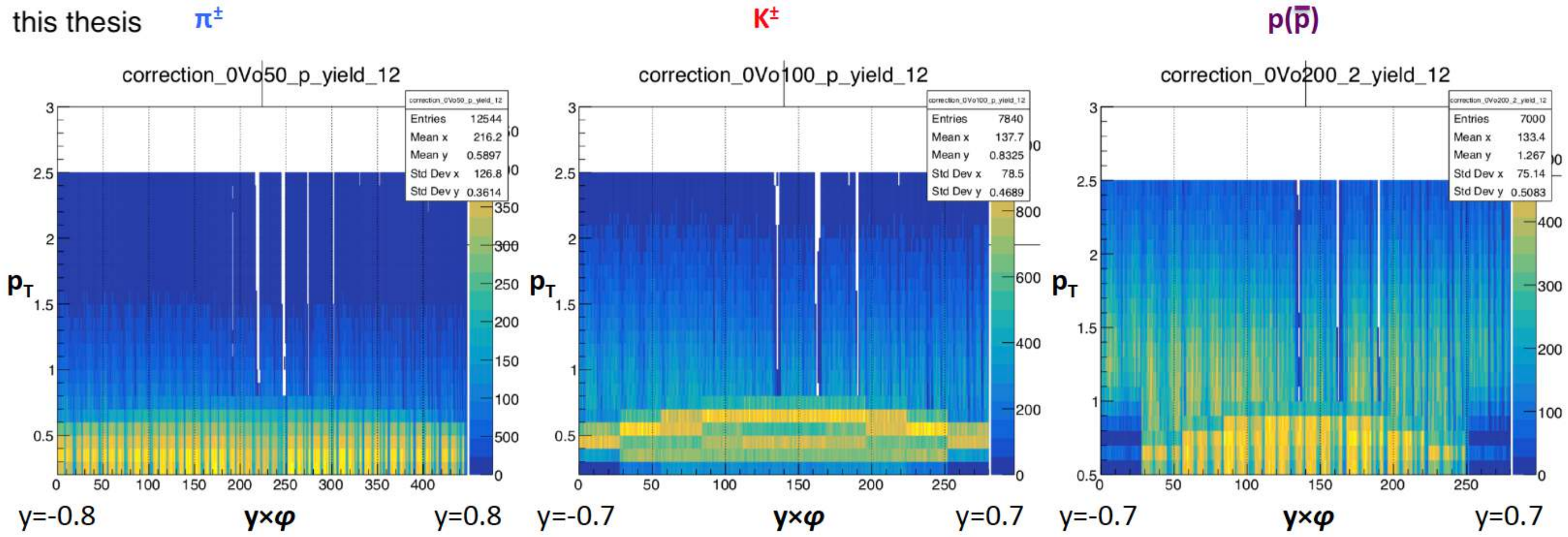}
  \caption{Uncorrected  single-particle yield distributions of (left)  $\pi^{\pm}$,  (middle) $K^{\pm}$,  and  (right)  $p/\bar{p}$, plotted as a function of rapidity $y$, azimuth $\varphi$, and $p_{\rm T}$,  obtained with $0<V_{z}<0.5$ cm in central Pb--Pb collisions, measured  with the  ++ B-field setting.}
  \label{fig:y_phi_pt_Distribution_2010PbPb_Pos_central_PiKPr}
\end{figure}

Correlation functions  of unidentified charged hadron, $R_{2}^{CD}$, previously reported by the ALICE collaboration~\cite{Acharya:2018ddg}, were obtained within the pseudorapidity range $|\eta|<1.0$.
However, for the determination of correlation functions of identified $\pi^{\pm}$, $K^{\pm}$ and $p/\bar{p}$, reported  in this work, it was necessary to narrow the the rapidity $y$ considered in order  to avoid  biases caused by kinematic bins featuring  very small statistics (i.e., very few charged particle track entries).
Figure~\ref{fig:y_phi_pt_Distribution_2010PbPb_Pos_central_PiKPr} shows that for the $p_{\rm T}$ ranges of interest in this work, rapidity ranges must be limited to $|y|\le0.8$, $|y|\le0.7$ and $|y|\le0.7$ for single $\pi^{\pm}$, $K^{\pm}$ and $p/\bar{p}$, respectively.
%Thus, these rapidity ranges are chosen to maximize the rapidity acceptance used in the determination of BFs.

\clearpage

\subsection{MC Closure Tests}
\label{subsubsec:MCHIJINGClosureTest}

MC closure tests of $\pi\pi$ and $KK$ correlation functions (CFs) were performed to examine the robustness of the analysis procedure used in this work. 
The closure tests are performed for CFs of US and LS, and BFs separately, based on the MC events described in Sec.~\ref{subsec:MCDataSamples}.
Similar to real data, the analysis of simulated data is performed separately for $++$ and $--$ magnetic polarity runs to account for known differences in the detection efficiency achieved with these two magnetic field configurations.
The final CF results are the weighted average of $++$ and $--$ magnetic polarity results.
The same event and track selection cuts as those used with real data are employed in the closure tests.
The correlators $A^{US}$ and $A^{LS}$ are first calculated at generator level separately. Primary physics tracks are used as MC Truth towards the computation of generator level correlators. 
They are next computed with simulated reconstructed data obtained with MC events processed with GEANT 3.0 and the ALICE detector simulation software.
Weights used to construct the MC correlators are calculated using the same method and software used for real data.
The closure tests are considered successful if the (simulated) reconstructed correlators match the generator level correlators with reasonable precision.

%%%%%%%%%%%%%%%%%%%%%%%%%%%%%%%%%%%%%%%%%%%%%%%%%%%%%%%%
% PionPion DCAxy004
%%%%%%%%%%%%%%%%%%%%%%%%%%%%%%%%%%%%%%%%%%%%%%%%%%%%%%%%
\subsubsection{MC Closure Test of $\pi\pi$ Correlation Functions}
\label{subsubsec:MCCTpipi}

\begin{figure}
\centering
  \includegraphics[width=0.32\linewidth]{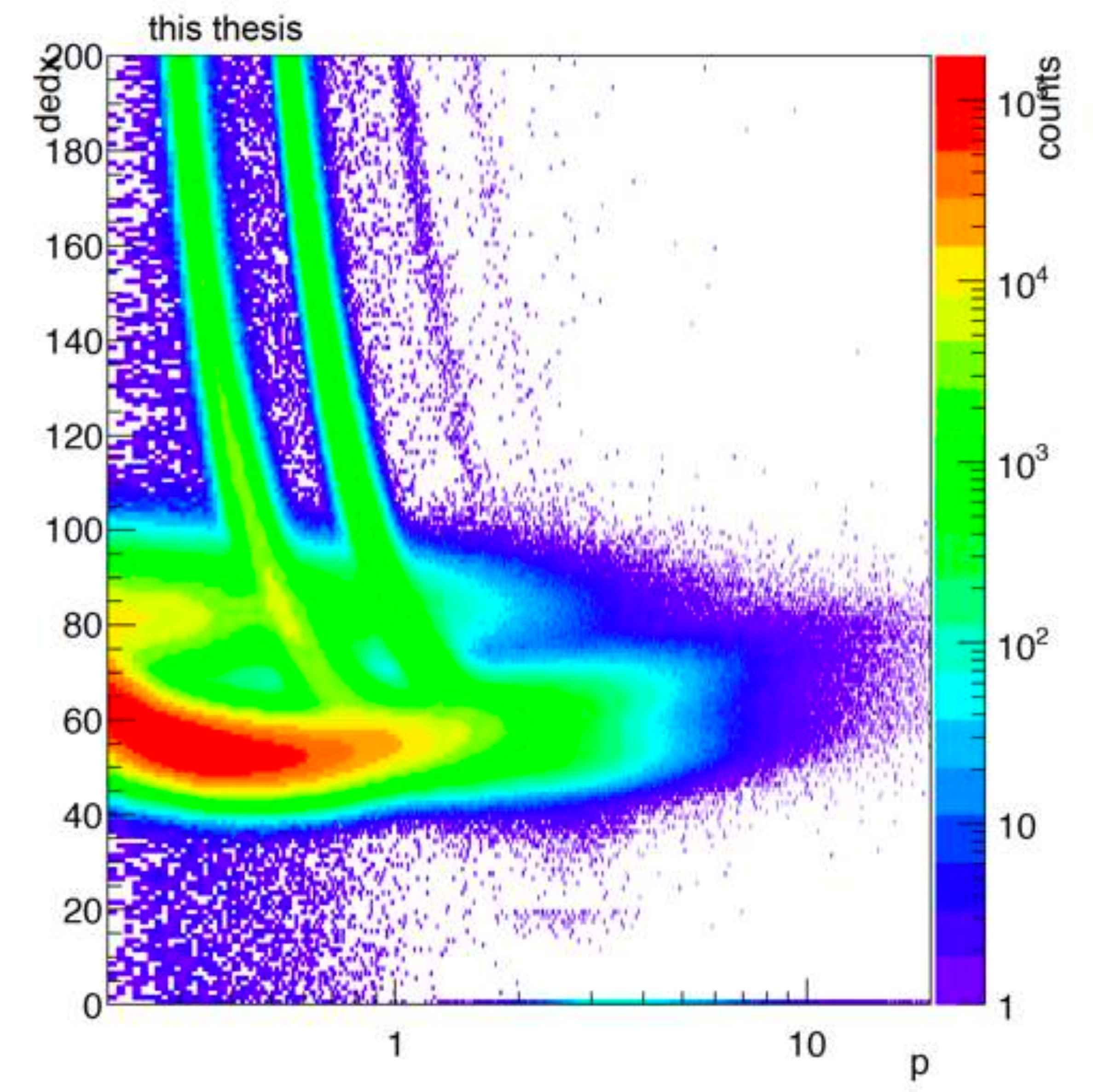}
  \includegraphics[width=0.32\linewidth]{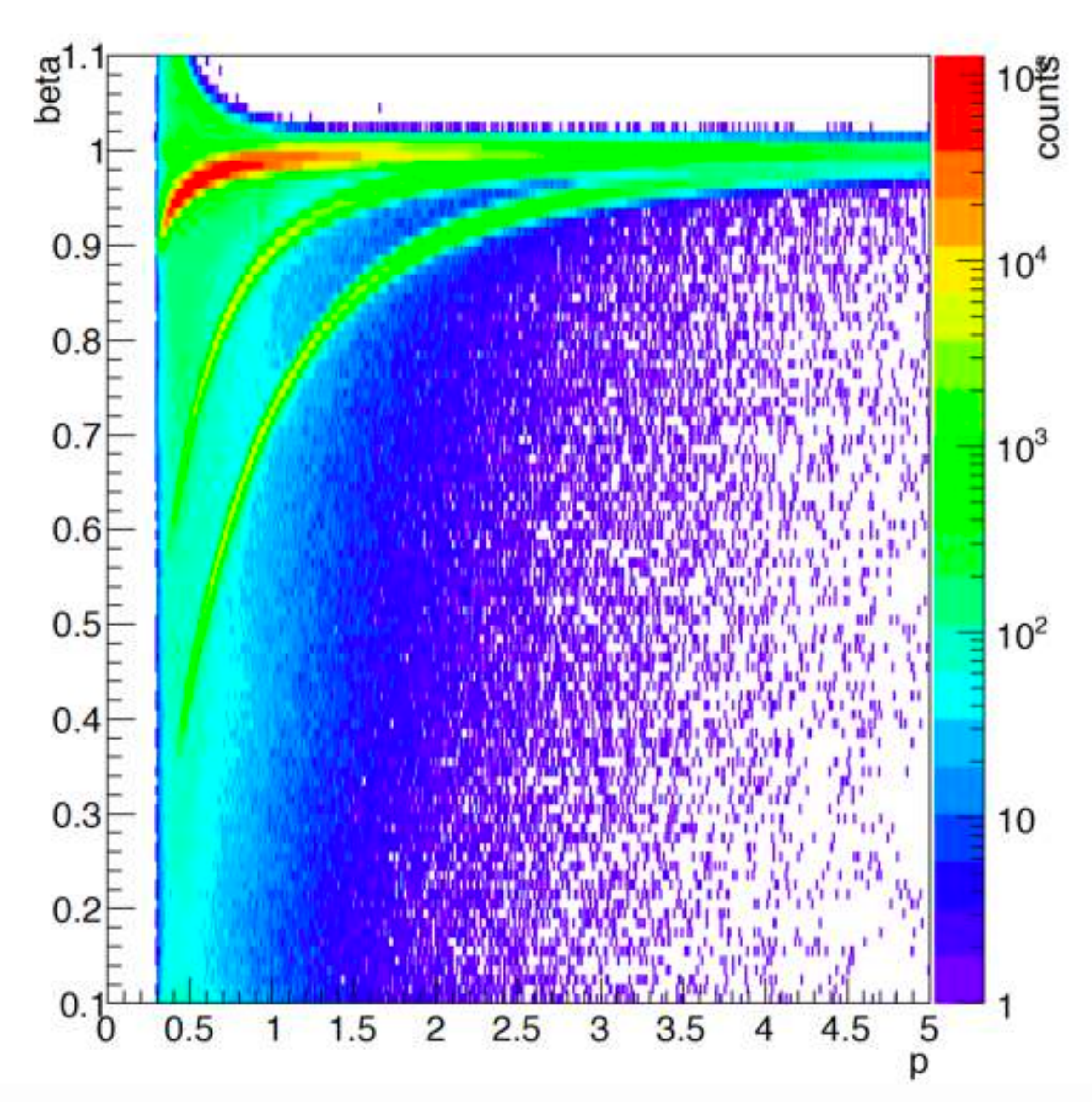}
  \includegraphics[width=0.32\linewidth]{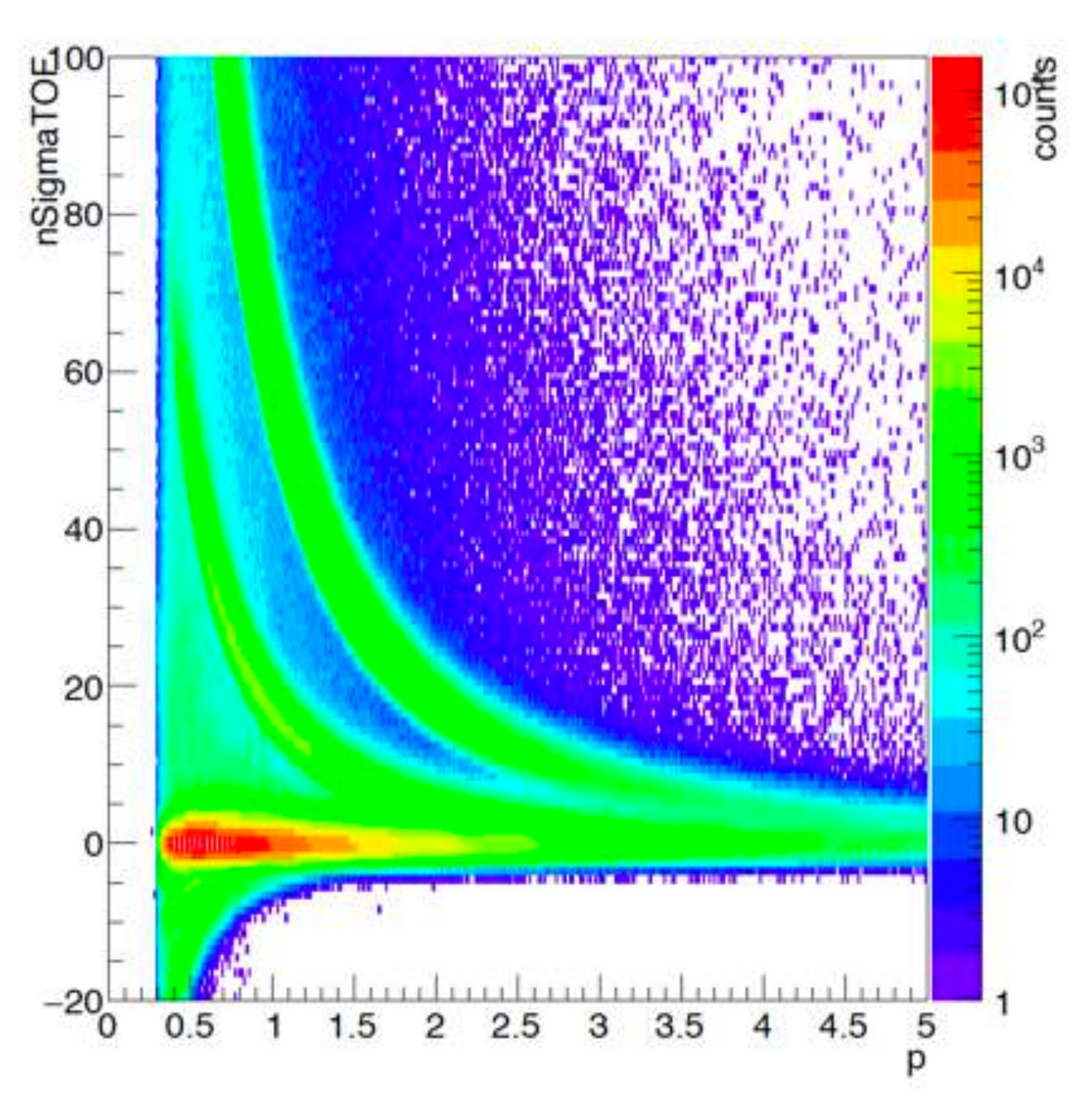}
  \includegraphics[width=0.32\linewidth]{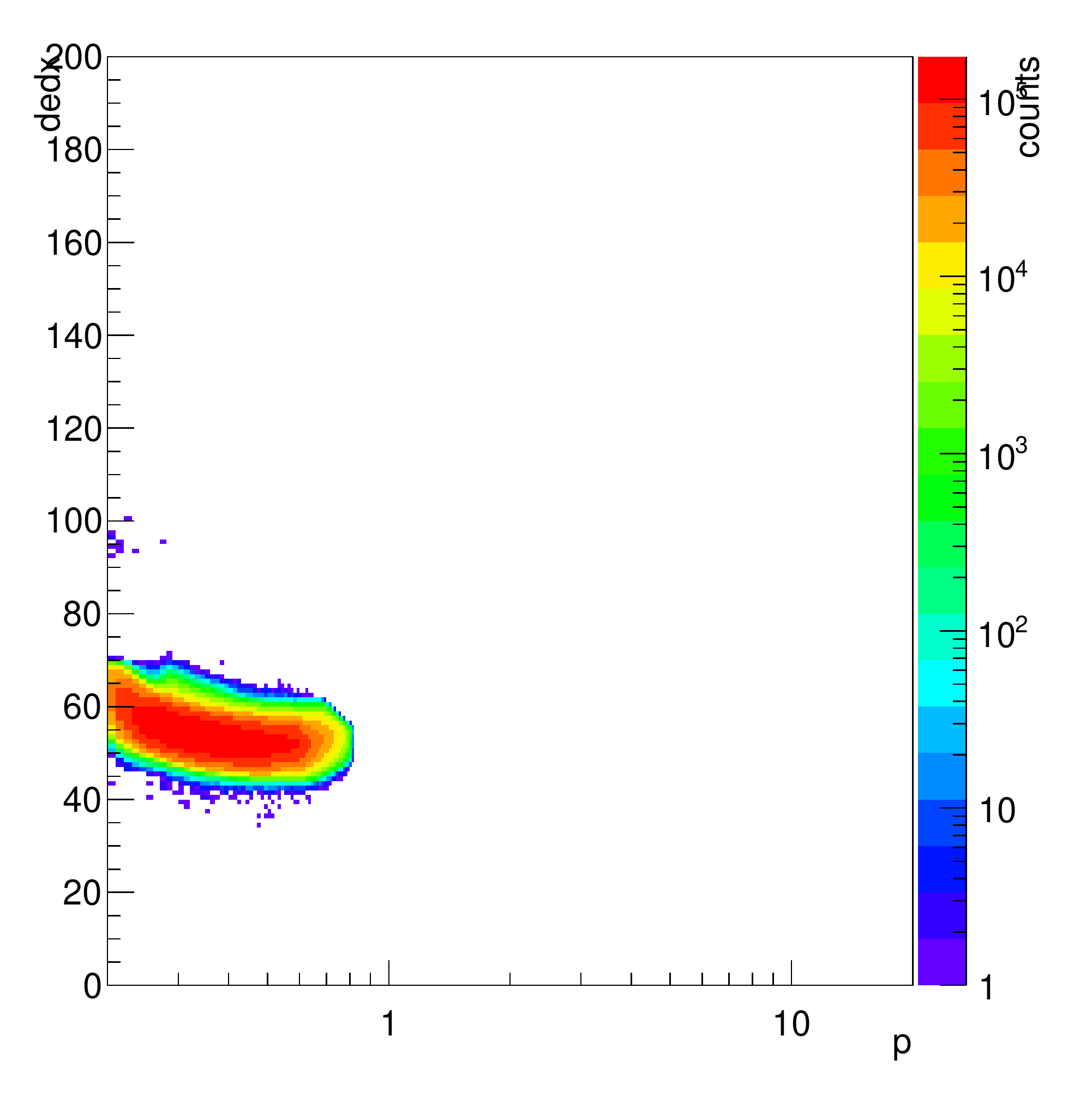}
  \includegraphics[width=0.32\linewidth]{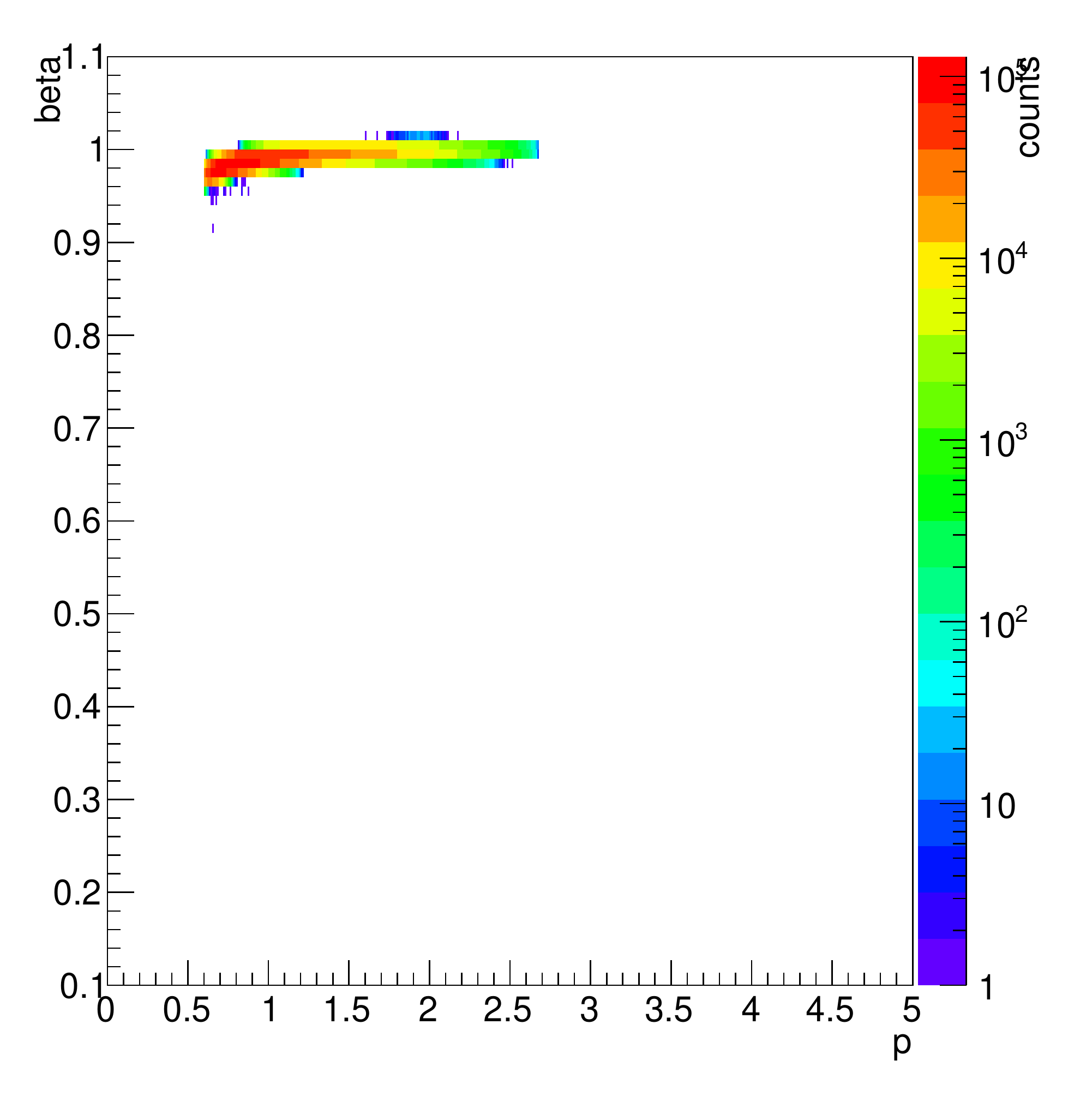}
  \includegraphics[width=0.32\linewidth]{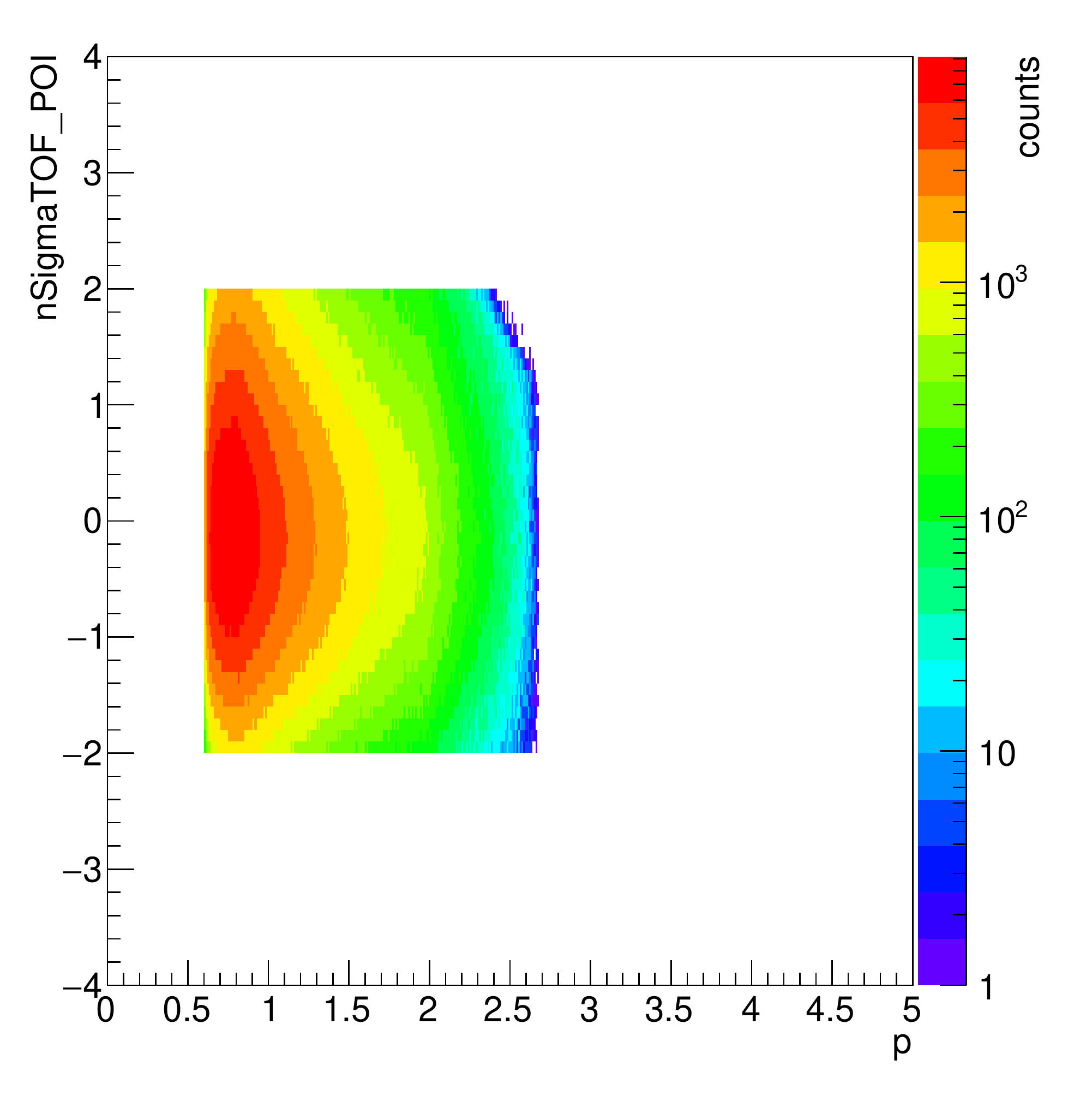}
  \caption{Illustration of hadron species identification based on the reconstructed level 20-40\% centrality Pb--Pb MC events. 
  Comparisons of TPC dE/dx (left column), TOF $\beta$ (middle column), and TOF $n\sigma$ (right column), as a function of momentum, before (upper row) and after (lower row) $\pi^{\pm}$ selection cuts.}
   \label{fig:HIJING_Reco_pion_before_after_PID_Pos_20Cent40}
\end{figure}

For $\pi\pi$ pair, given the limited size of the MC data sample, wider centrality bins corresponding to 0-20\%, 20-40\%, 40-60\%, and 60-80\% of the interaction cross section are used. Figure~\ref{fig:HIJING_Reco_pion_before_after_PID_Pos_20Cent40} shows that the reconstructed level PID plots obtained with MC simulations are similar to those obtained with real data.

\begin{figure}
\centering
  \includegraphics[width=0.3\linewidth]{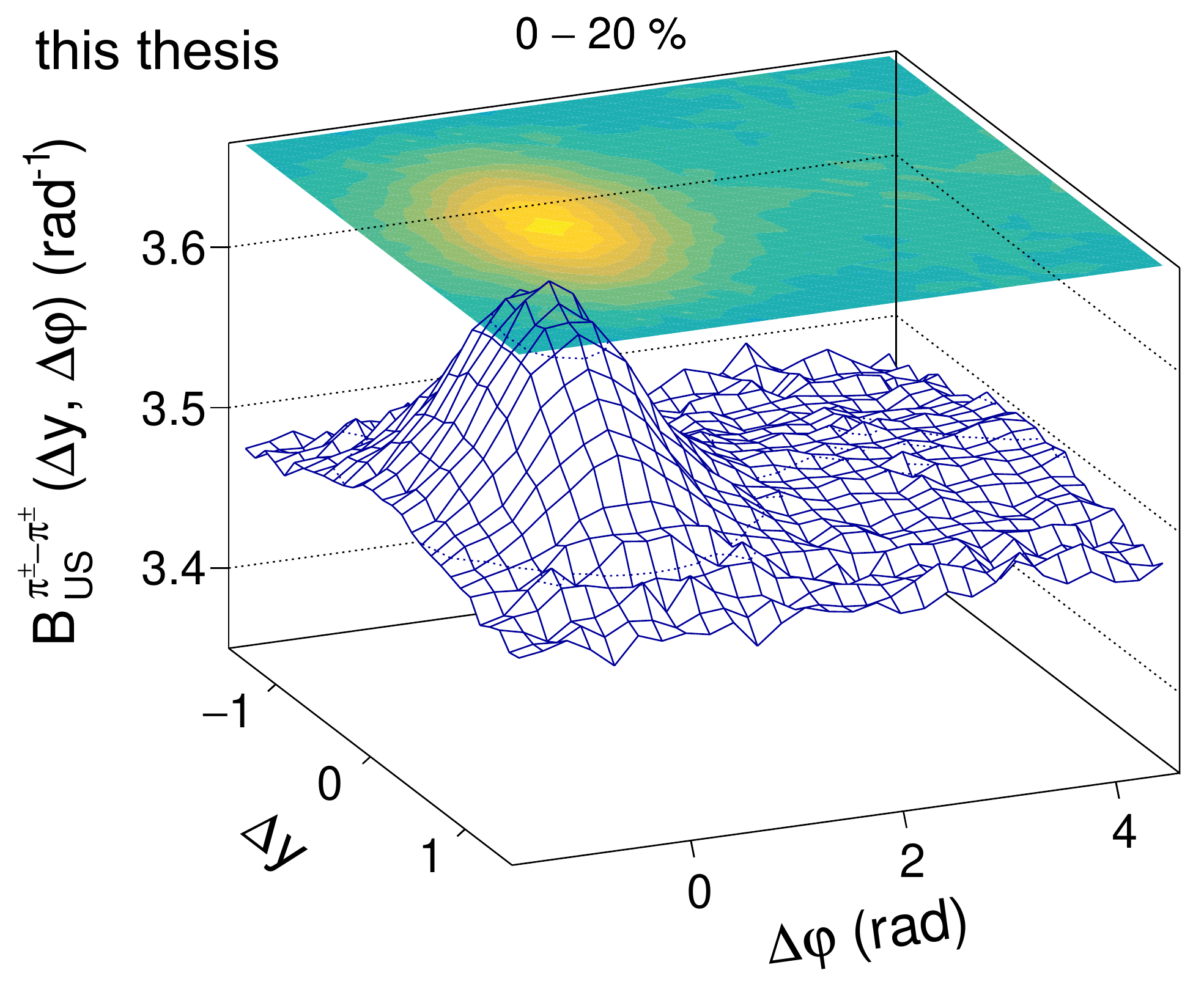}
  \includegraphics[width=0.3\linewidth]{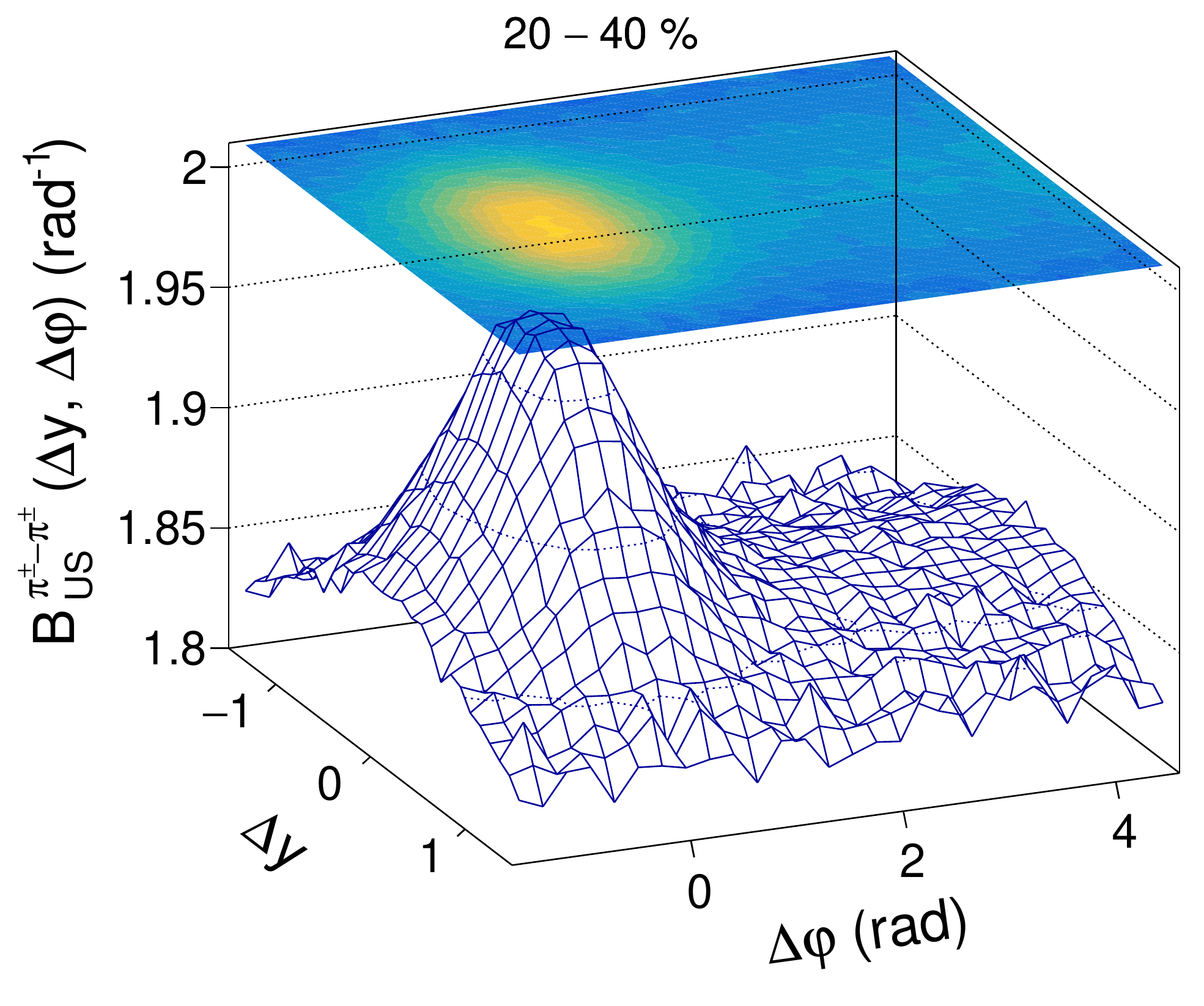}
  \includegraphics[width=0.3\linewidth]{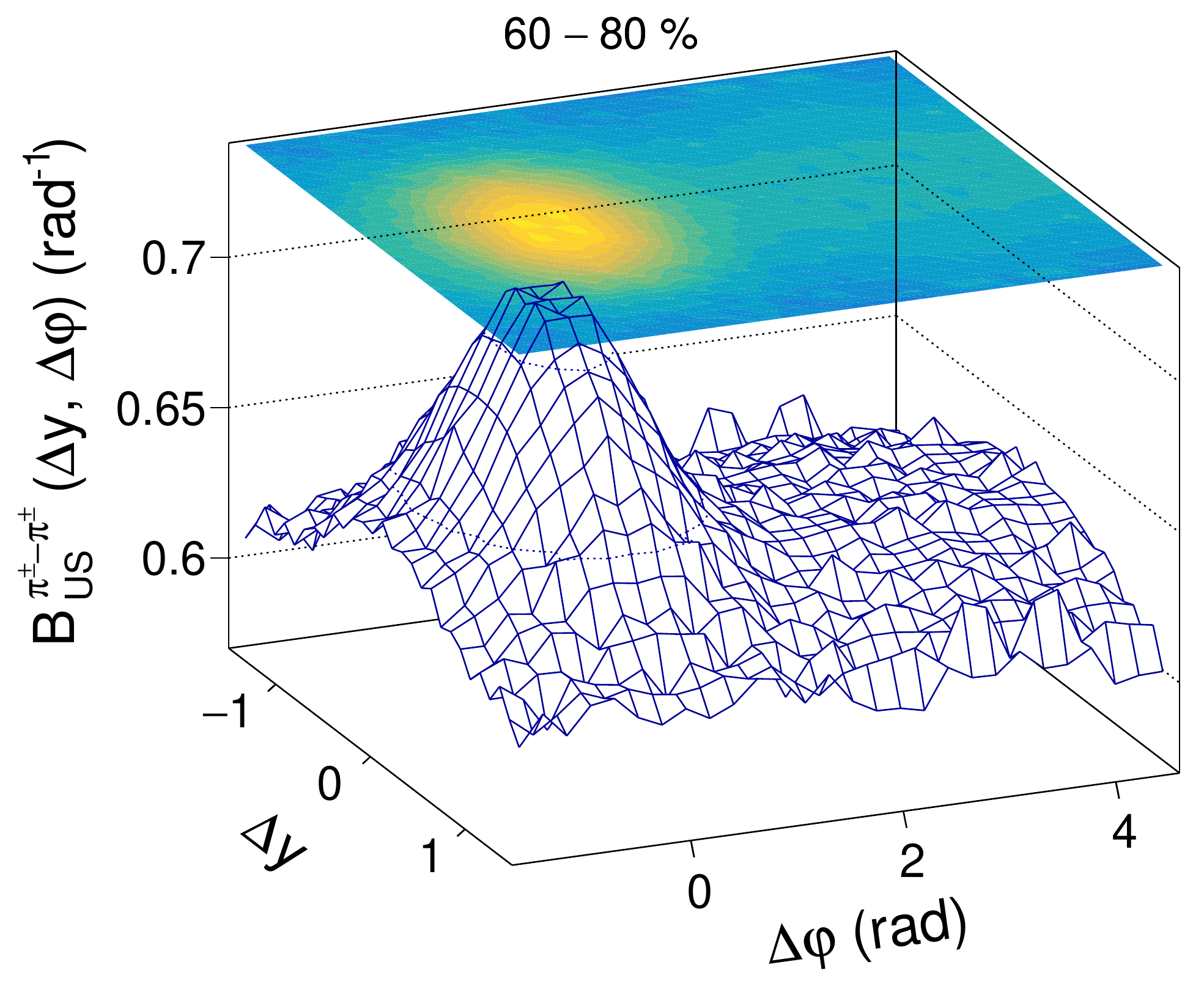}
  \includegraphics[width=0.3\linewidth]{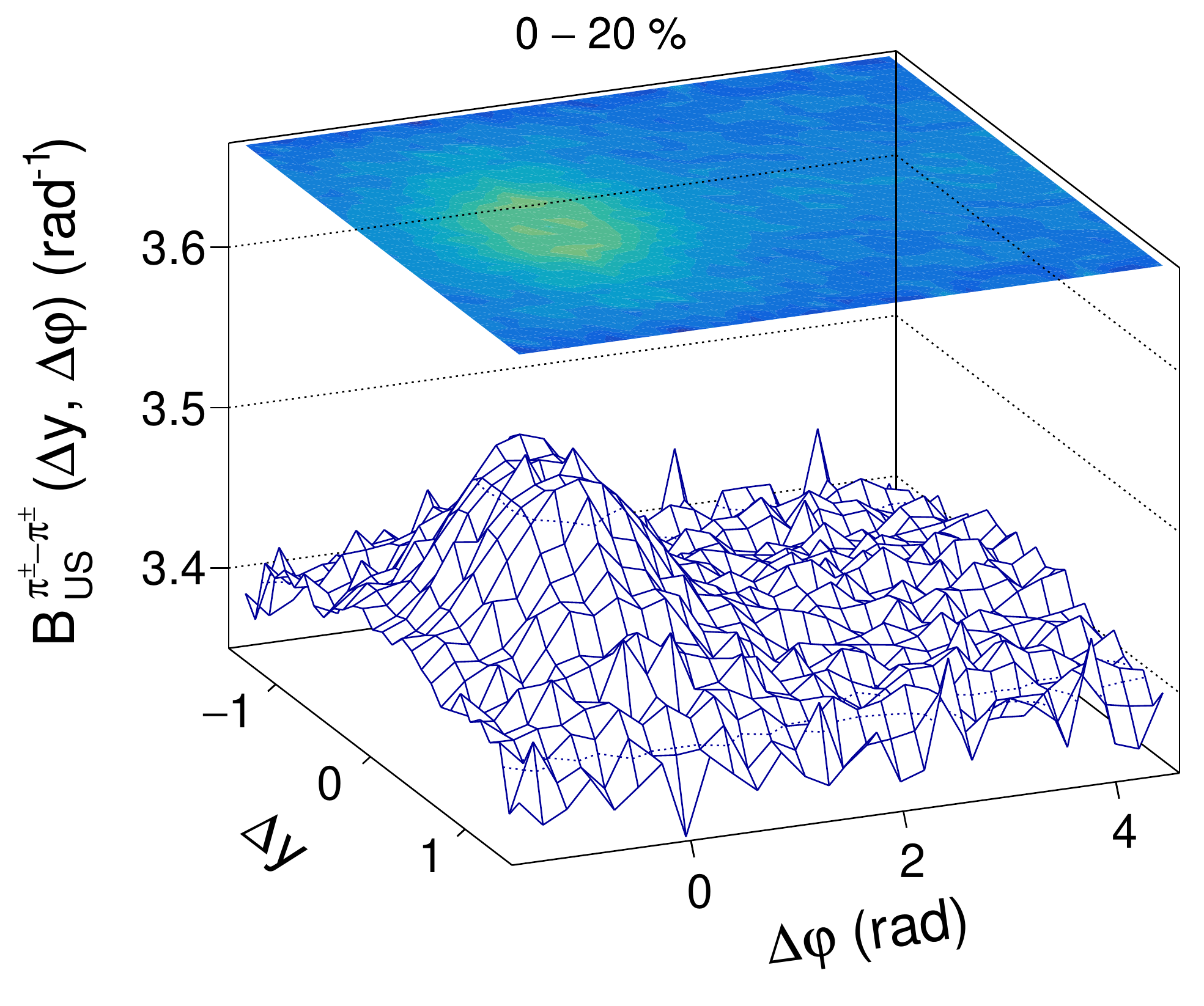}
  \includegraphics[width=0.3\linewidth]{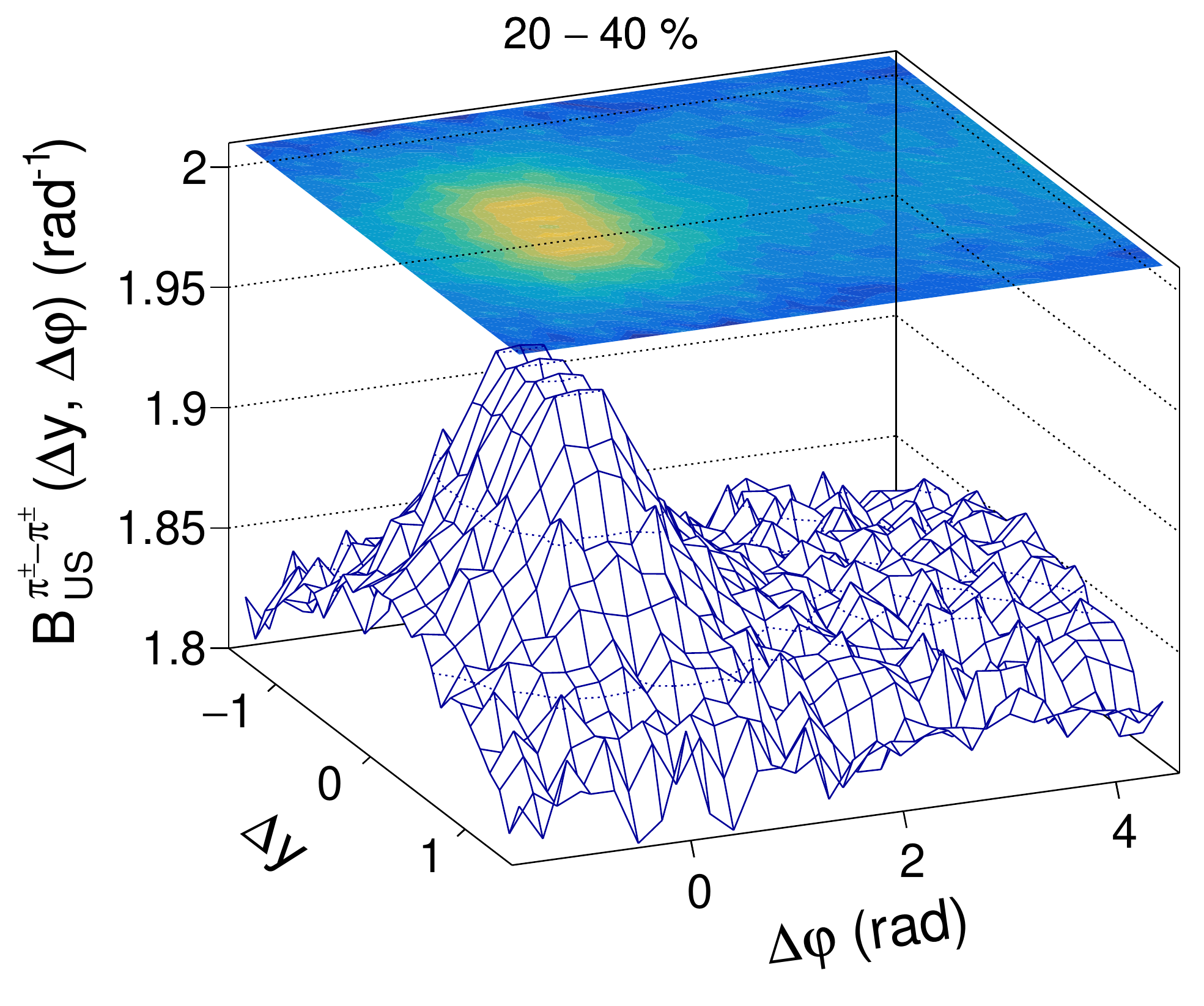}
  \includegraphics[width=0.3\linewidth]{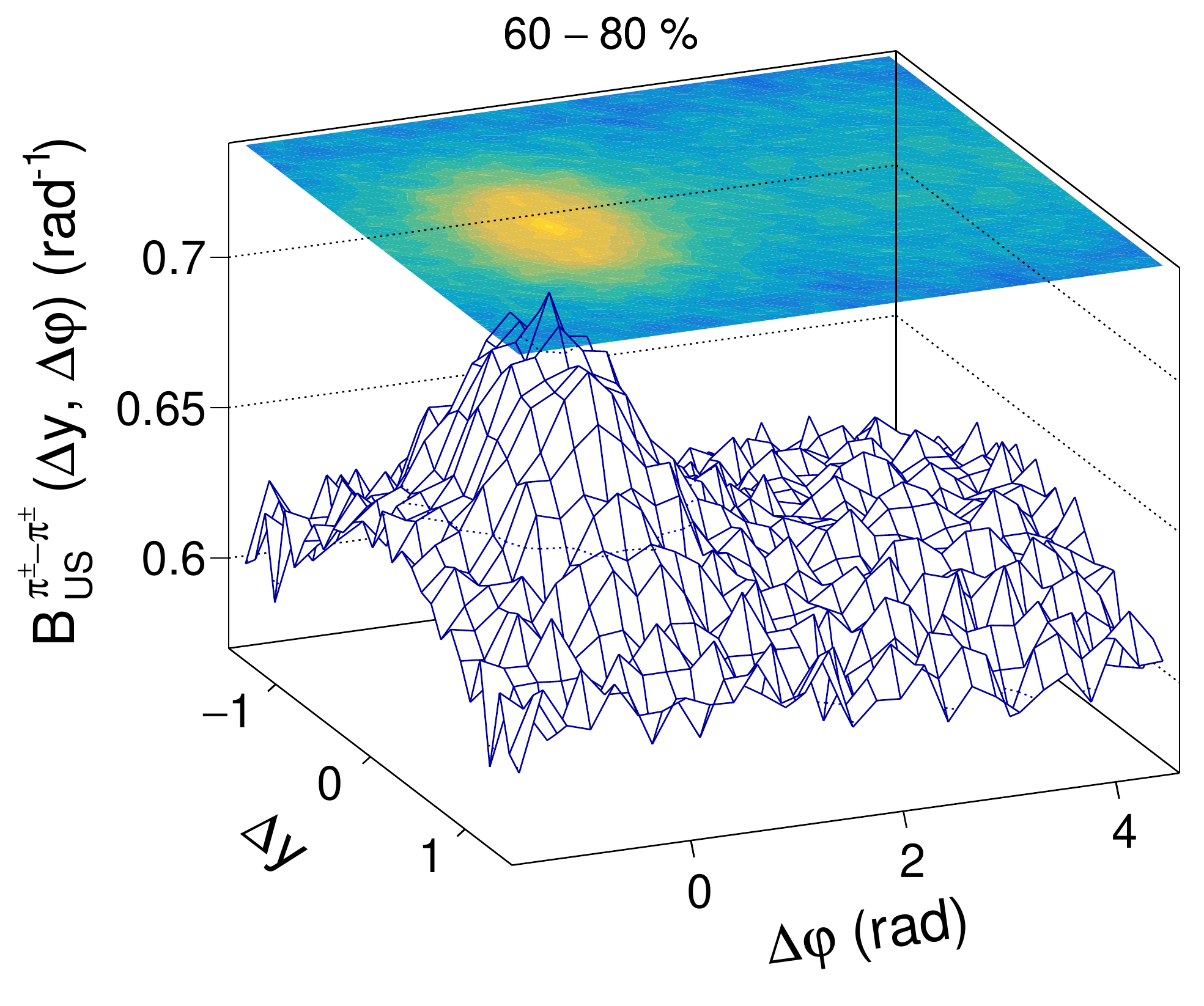}
  \includegraphics[width=0.3\linewidth]{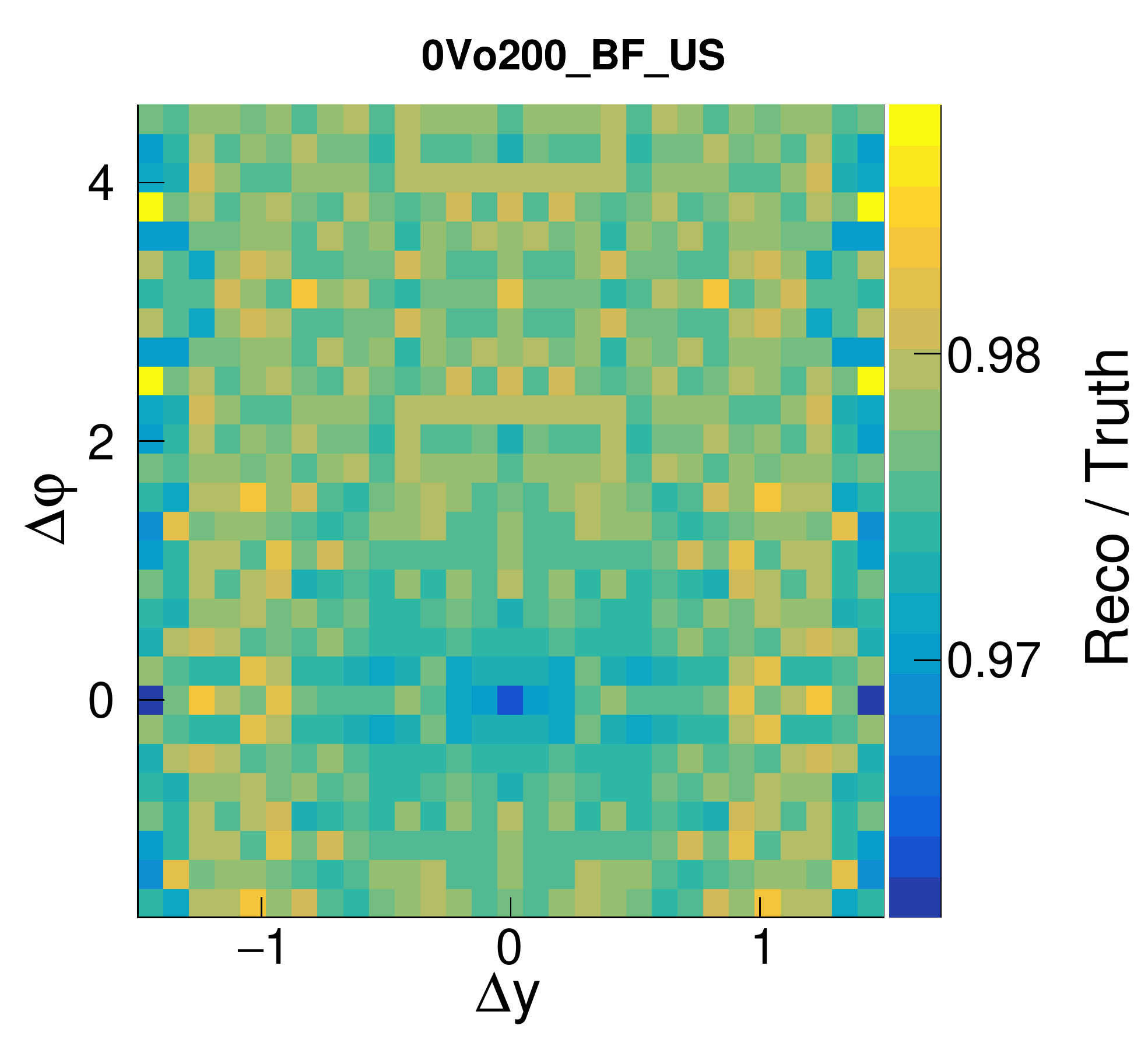}
  \includegraphics[width=0.3\linewidth]{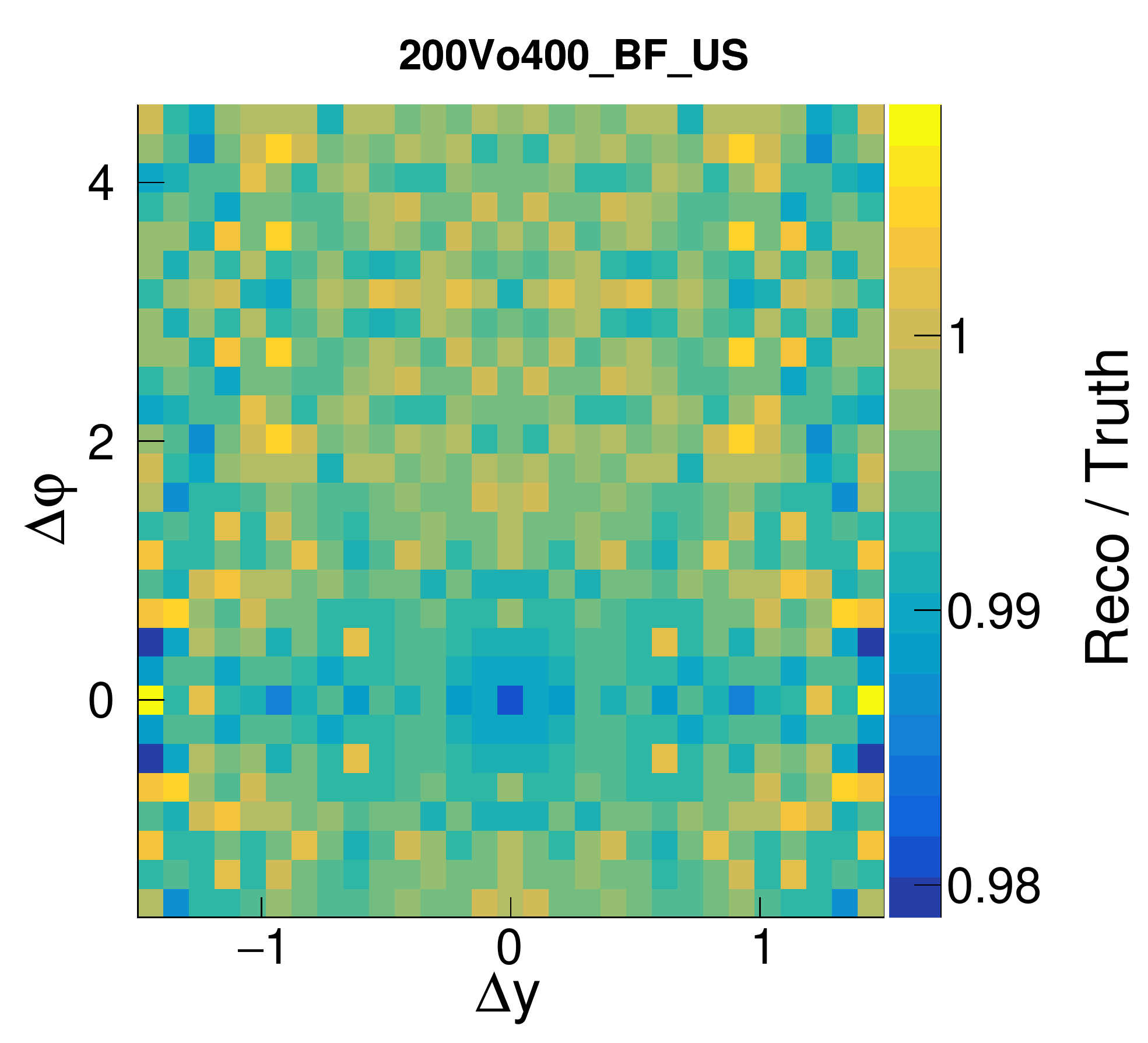}
  \includegraphics[width=0.3\linewidth]{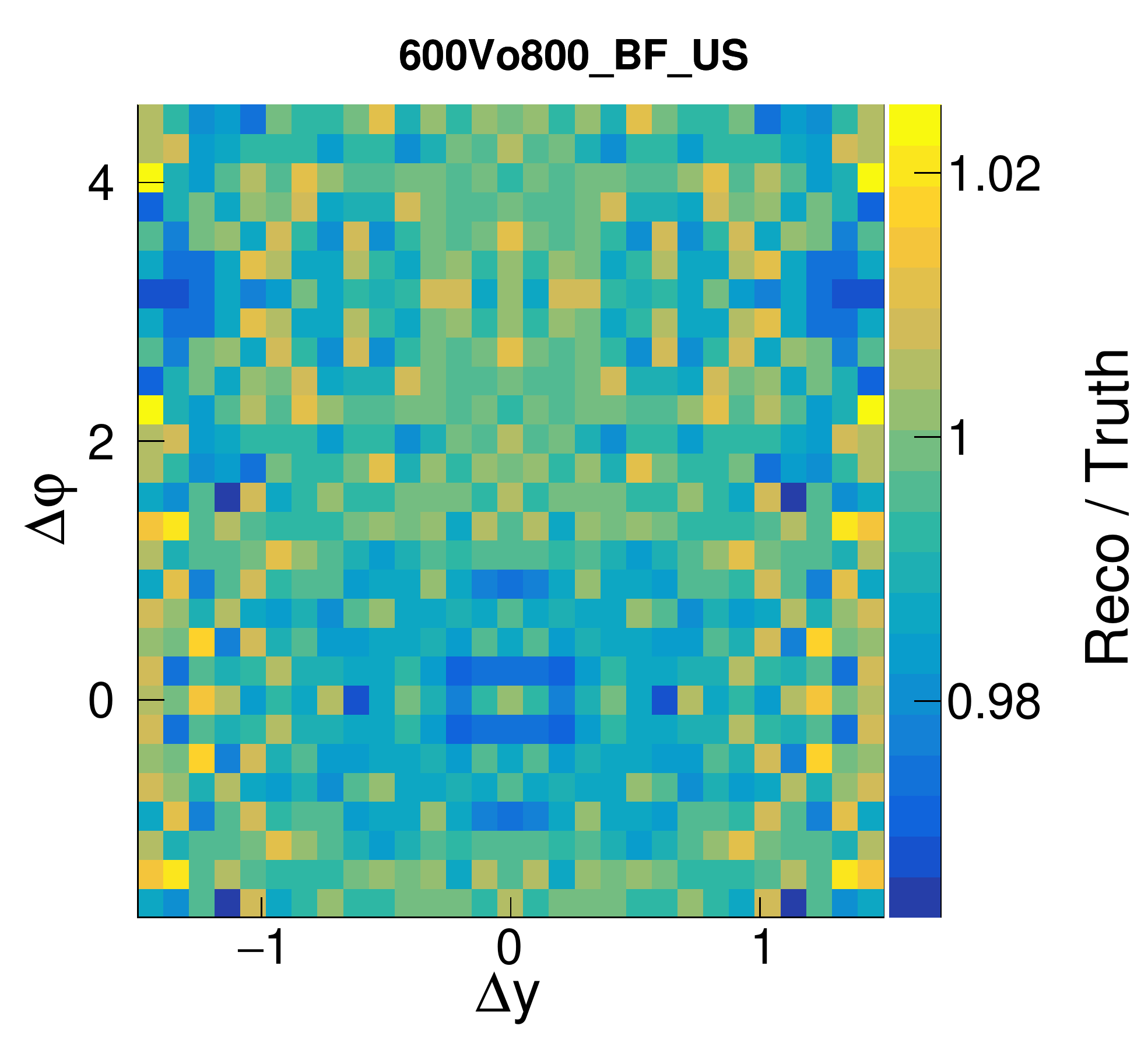}
  \includegraphics[width=0.3\linewidth]{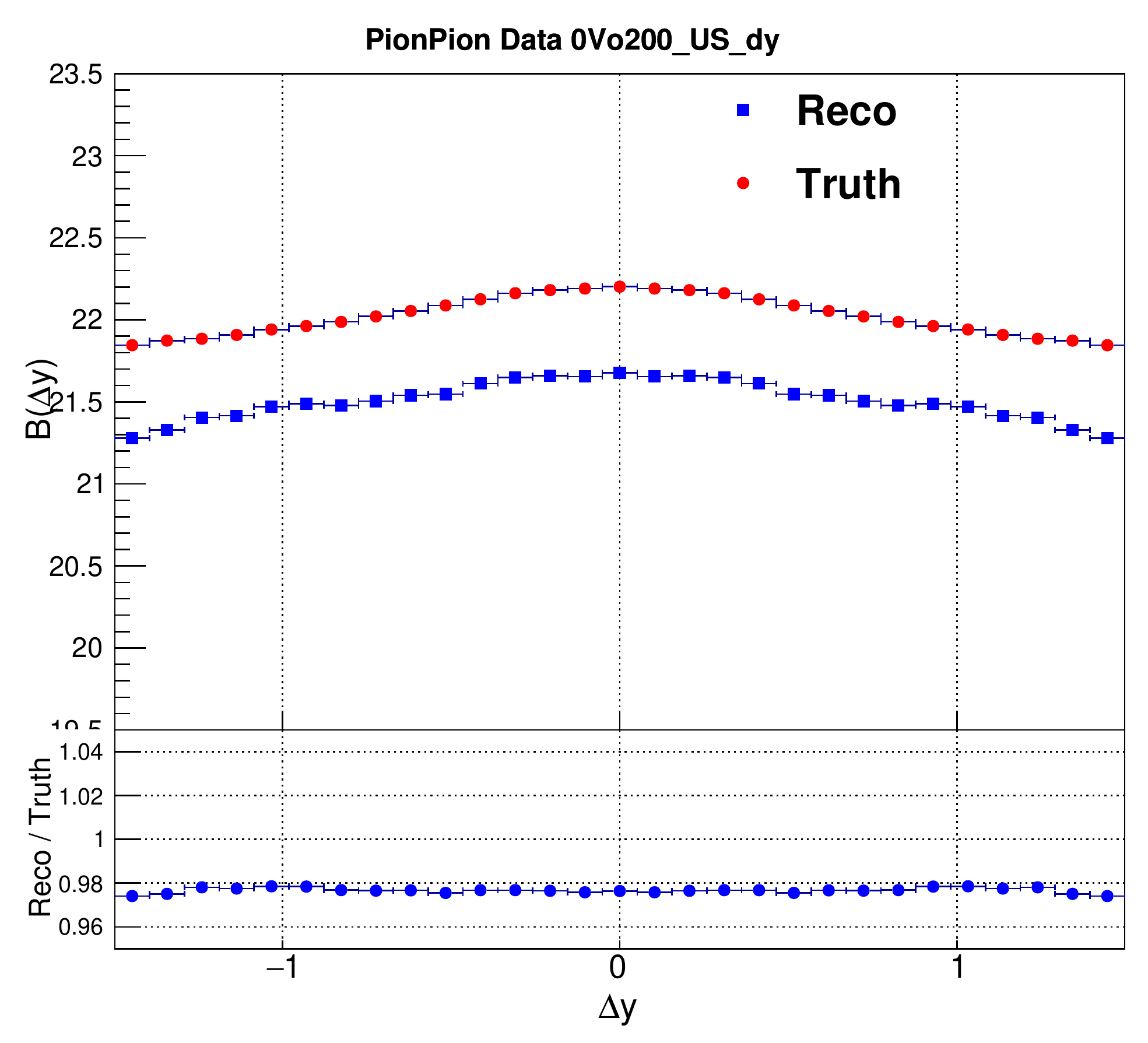}
  \includegraphics[width=0.3\linewidth]{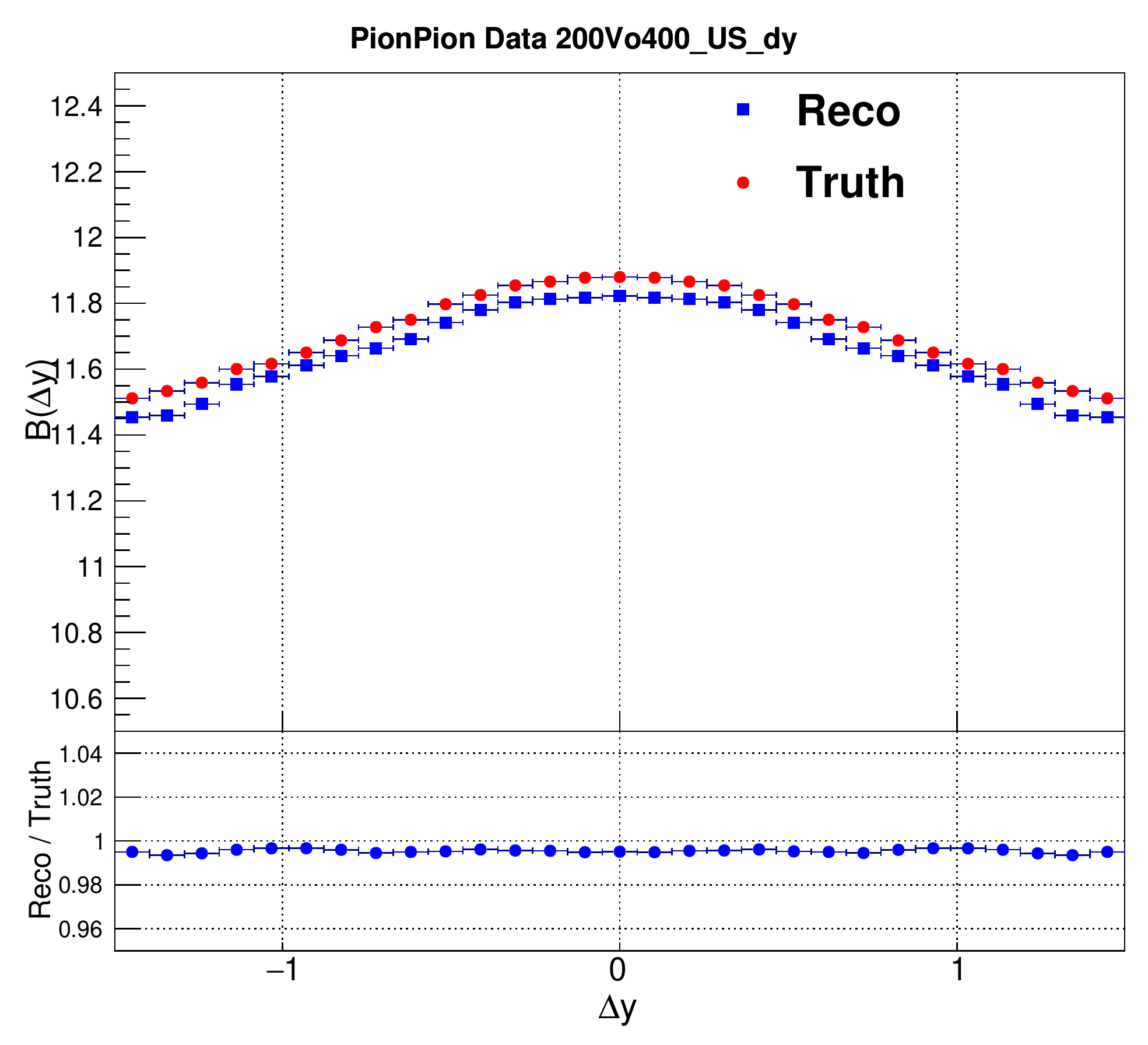}
  \includegraphics[width=0.3\linewidth]{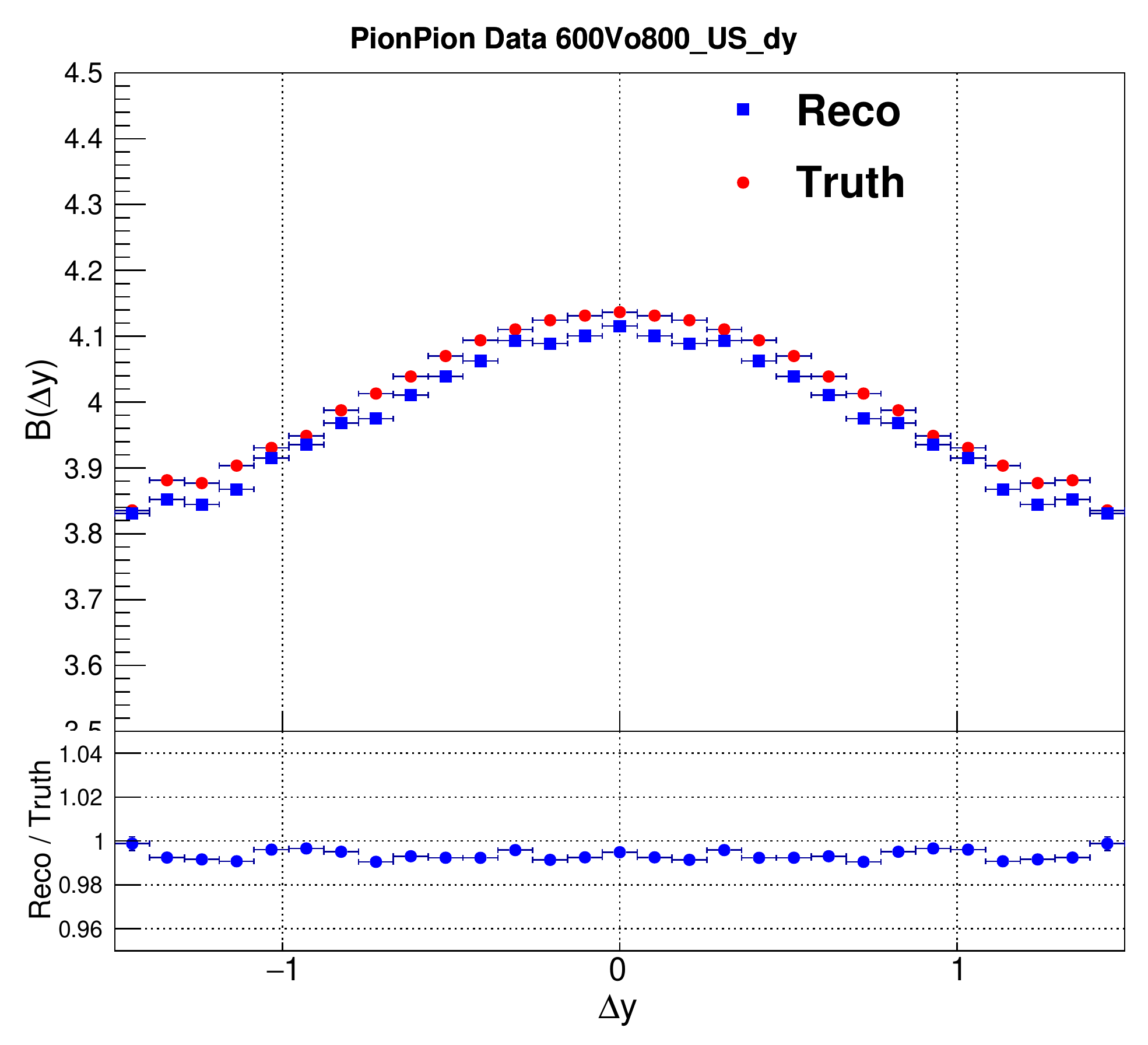}
  \includegraphics[width=0.3\linewidth]{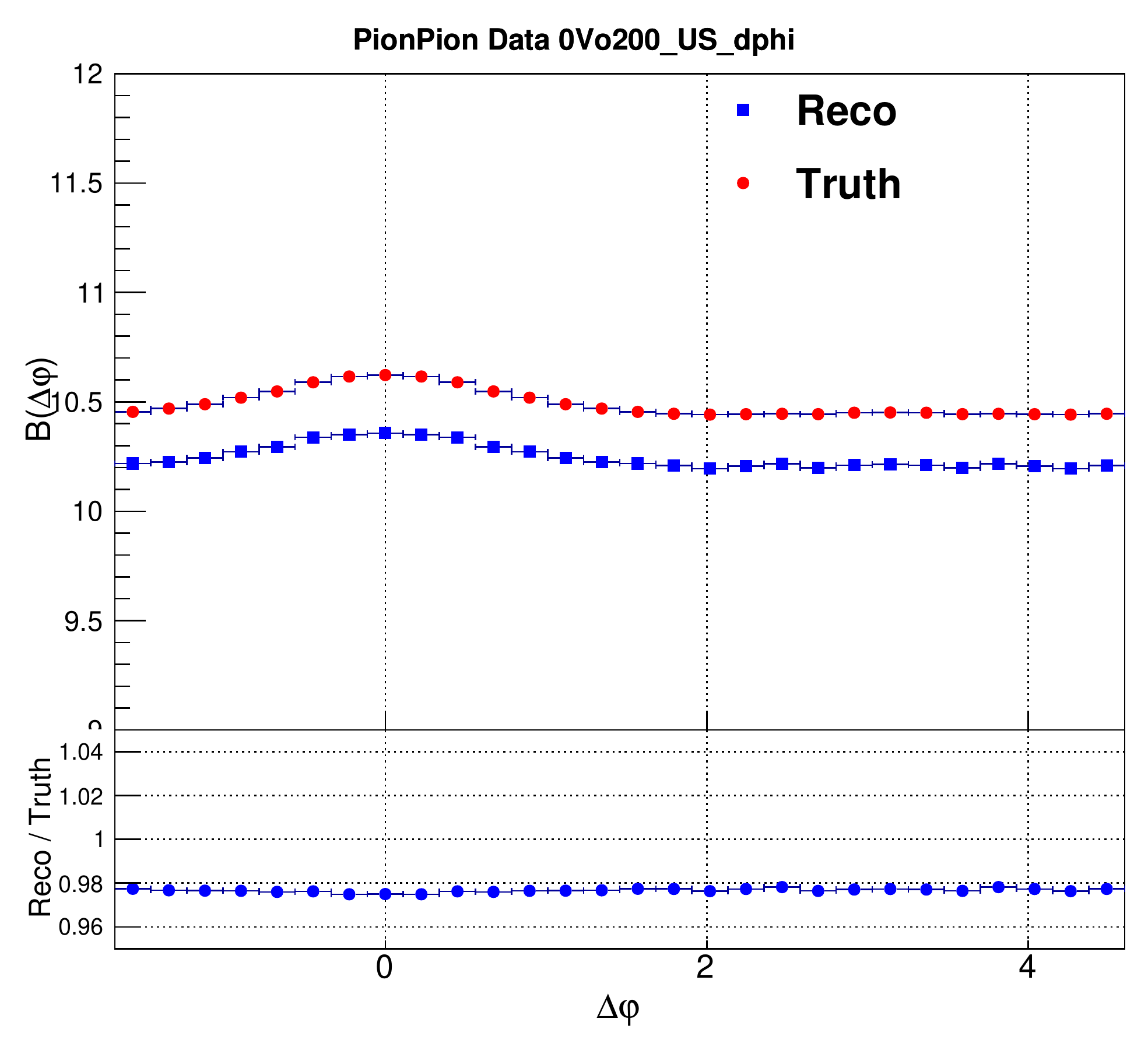}
  \includegraphics[width=0.3\linewidth]{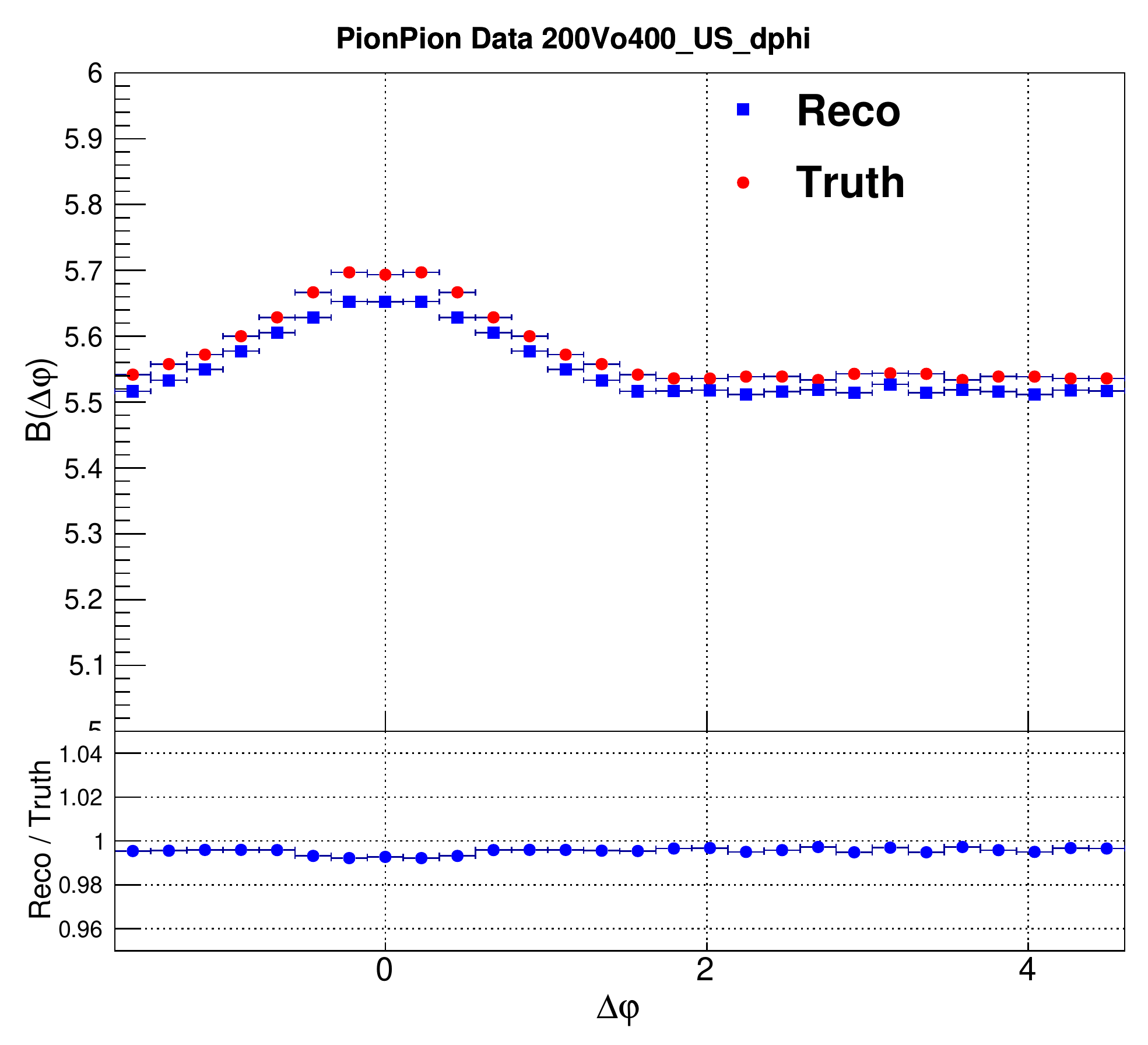}
  \includegraphics[width=0.3\linewidth]{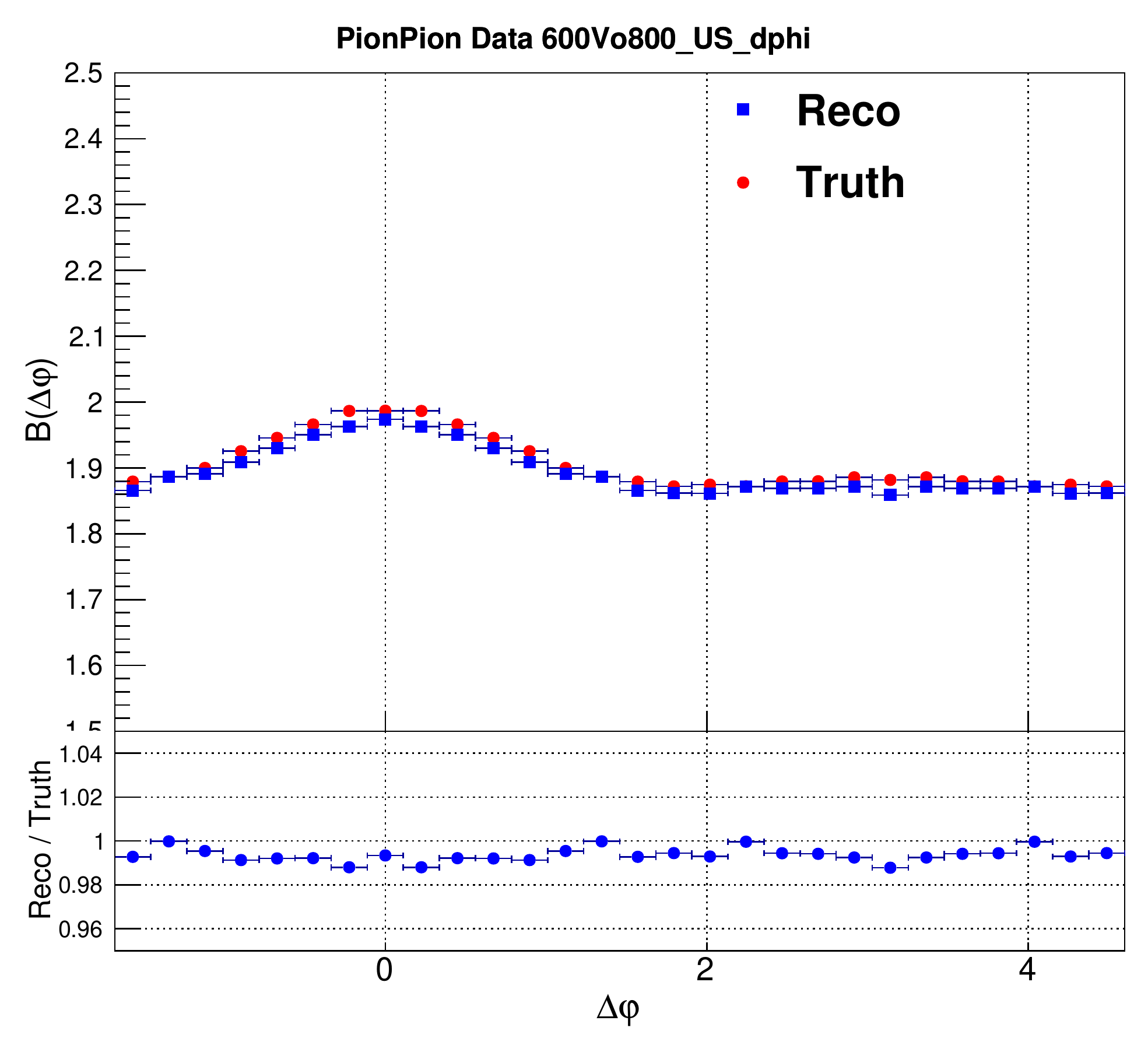}
  \caption{MC closure test of $\pi\pi$ pair. Comparisons of 2D CFs of US obtained with MC truth ($1^{st}$ row) and reconstructed ($2^{nd}$ row) events for selected collision centralities, and their ratios ($3^{rd}$ row). $4^{th}$ and $5^{th}$ rows: comparisons of the $\Delta y$ and $\Delta\varphi$ projections, respectively.}
  \label{fig:HIJING_Truth_Reco_DCAxy004_BF_US_PionPion}
\end{figure}

\begin{figure}
\centering
  \includegraphics[width=0.3\linewidth]{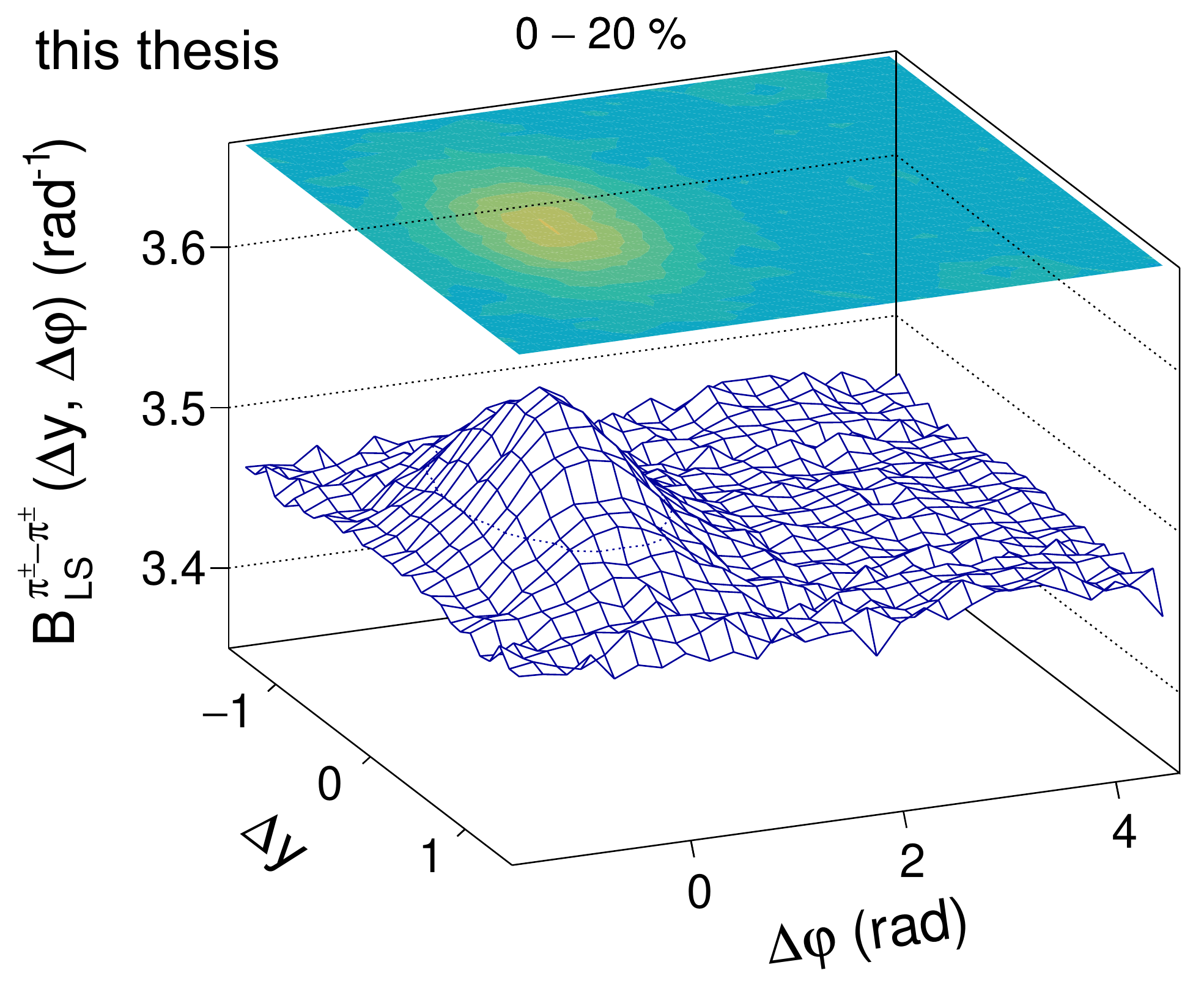}
  \includegraphics[width=0.3\linewidth]{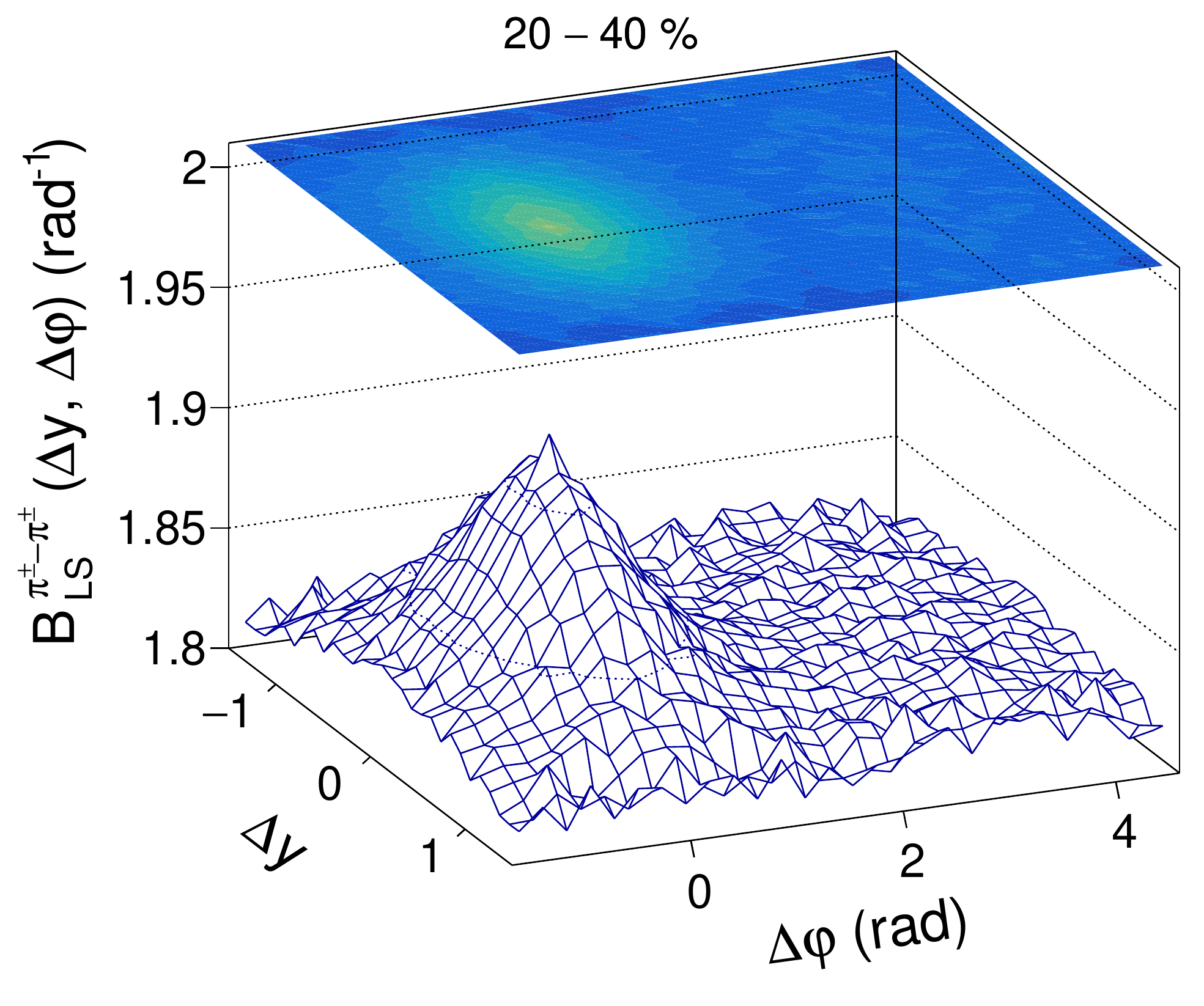}
  \includegraphics[width=0.3\linewidth]{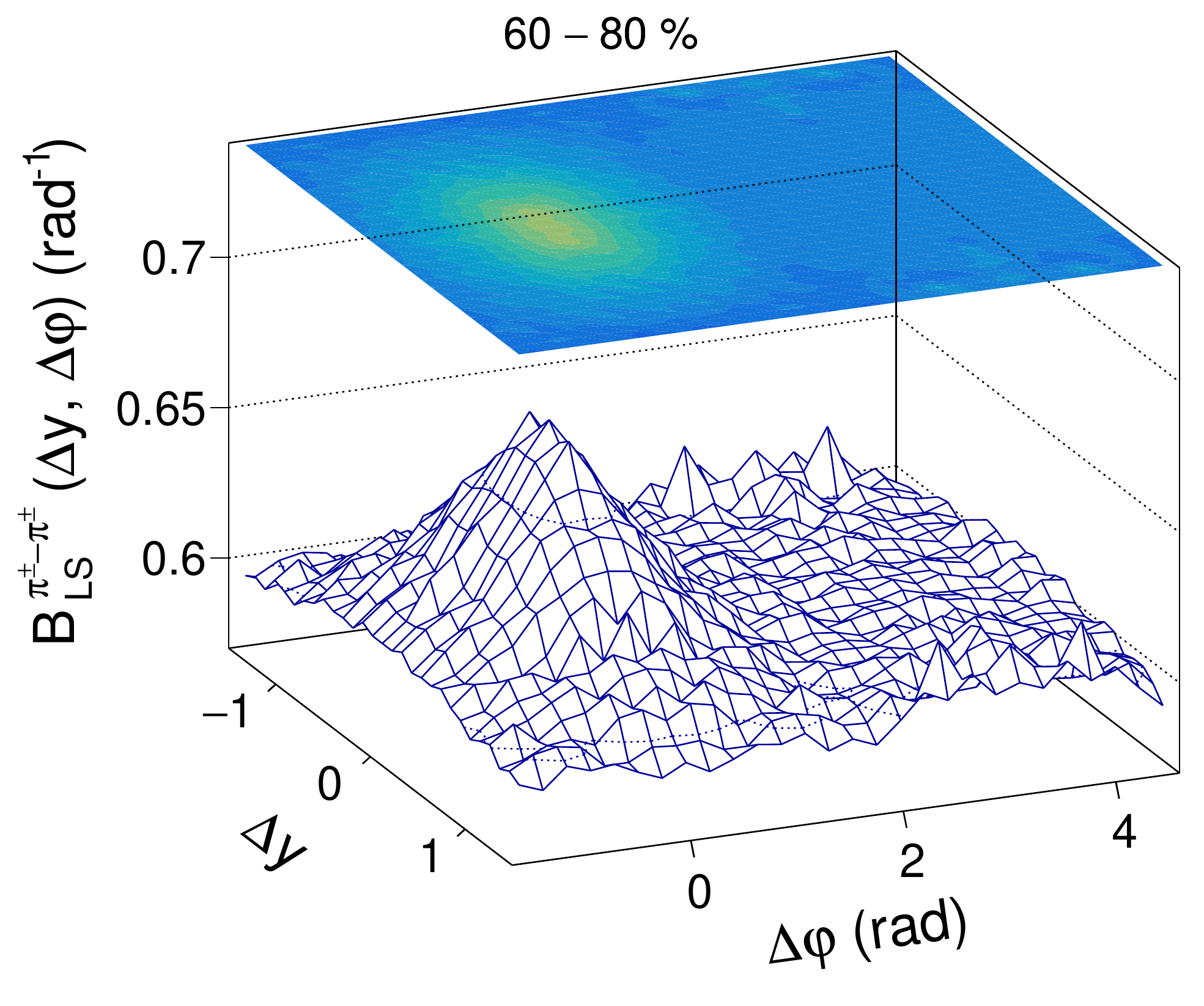}
  \includegraphics[width=0.3\linewidth]{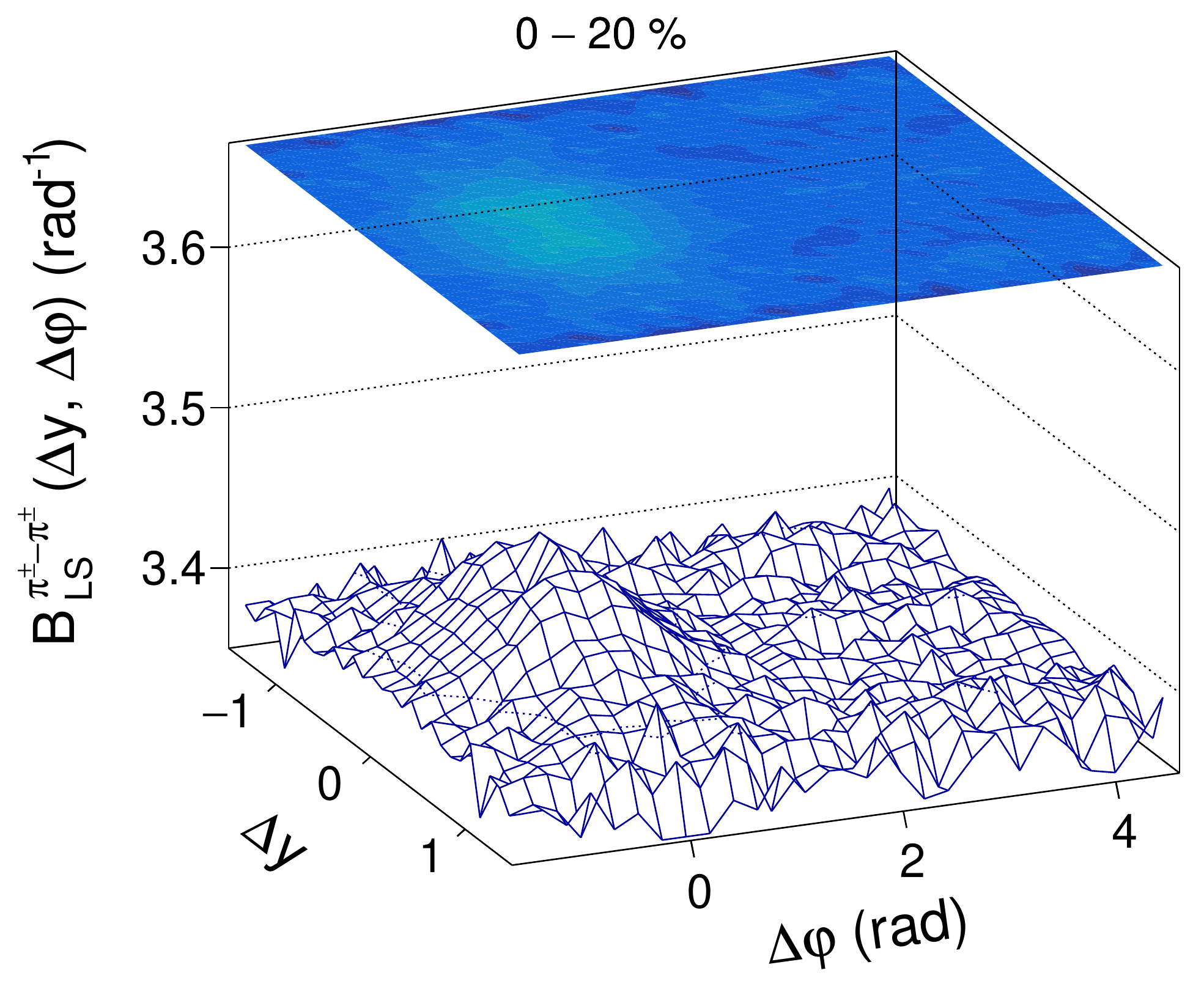}
  \includegraphics[width=0.3\linewidth]{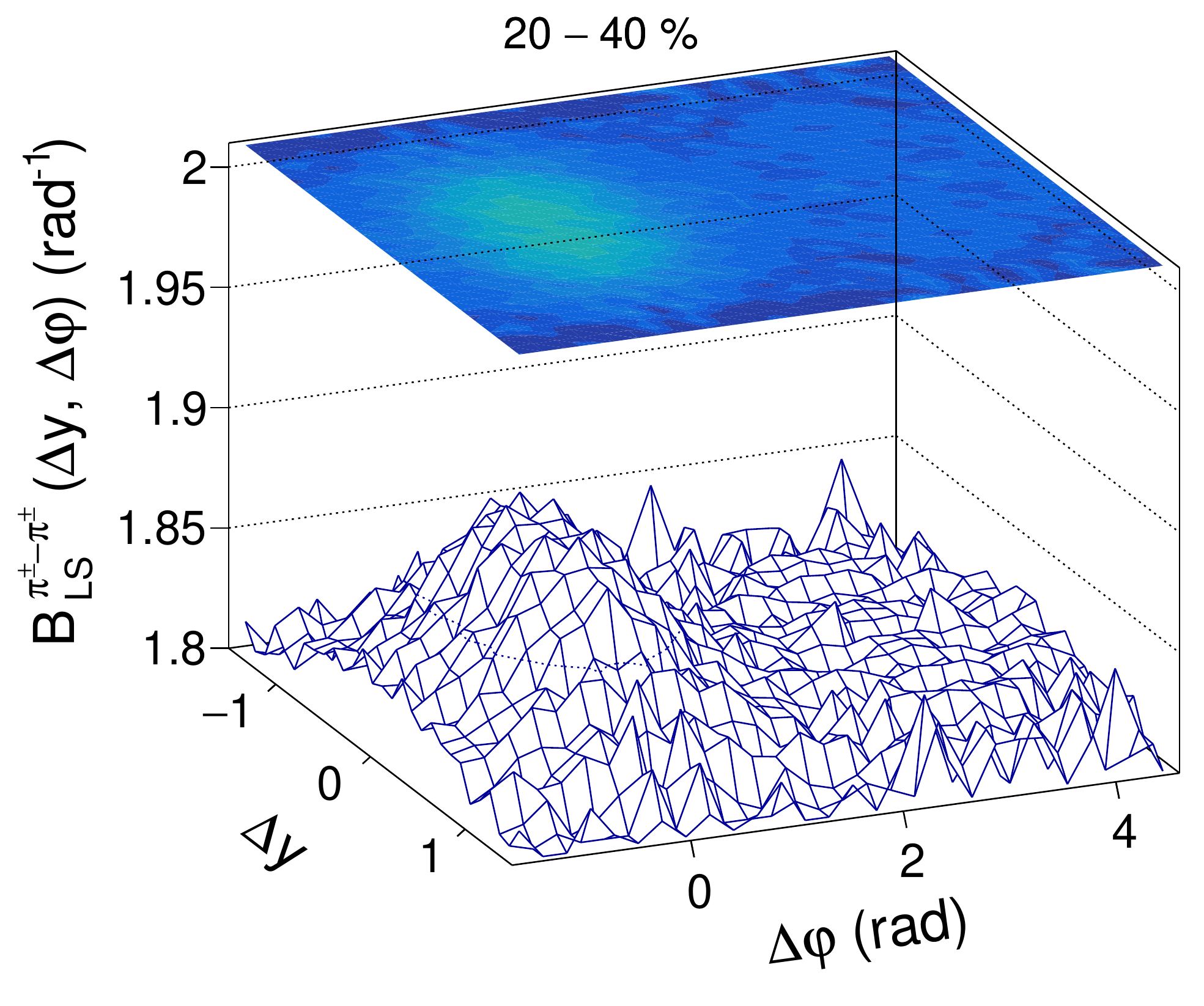}
  \includegraphics[width=0.3\linewidth]{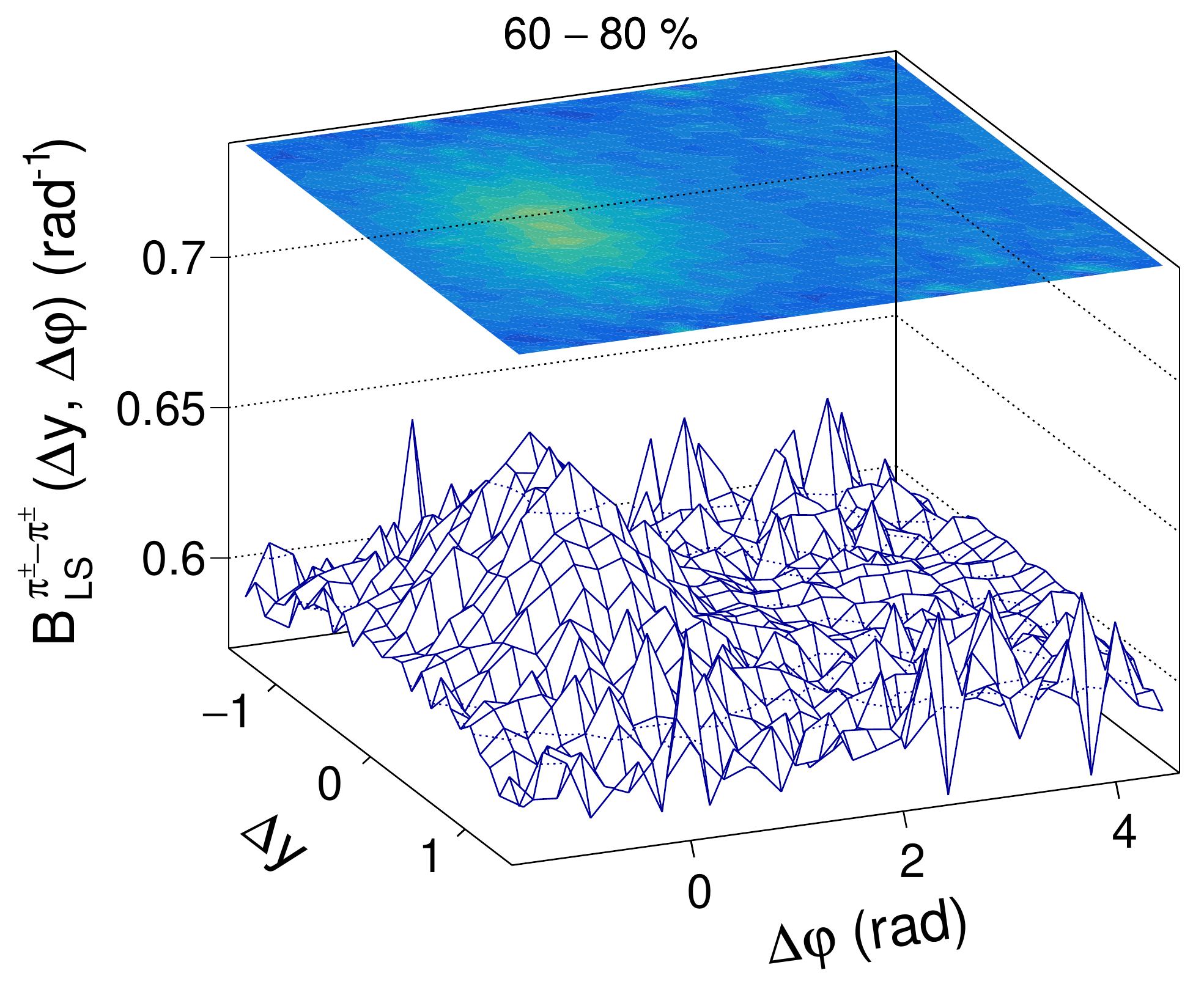}
  \includegraphics[width=0.3\linewidth]{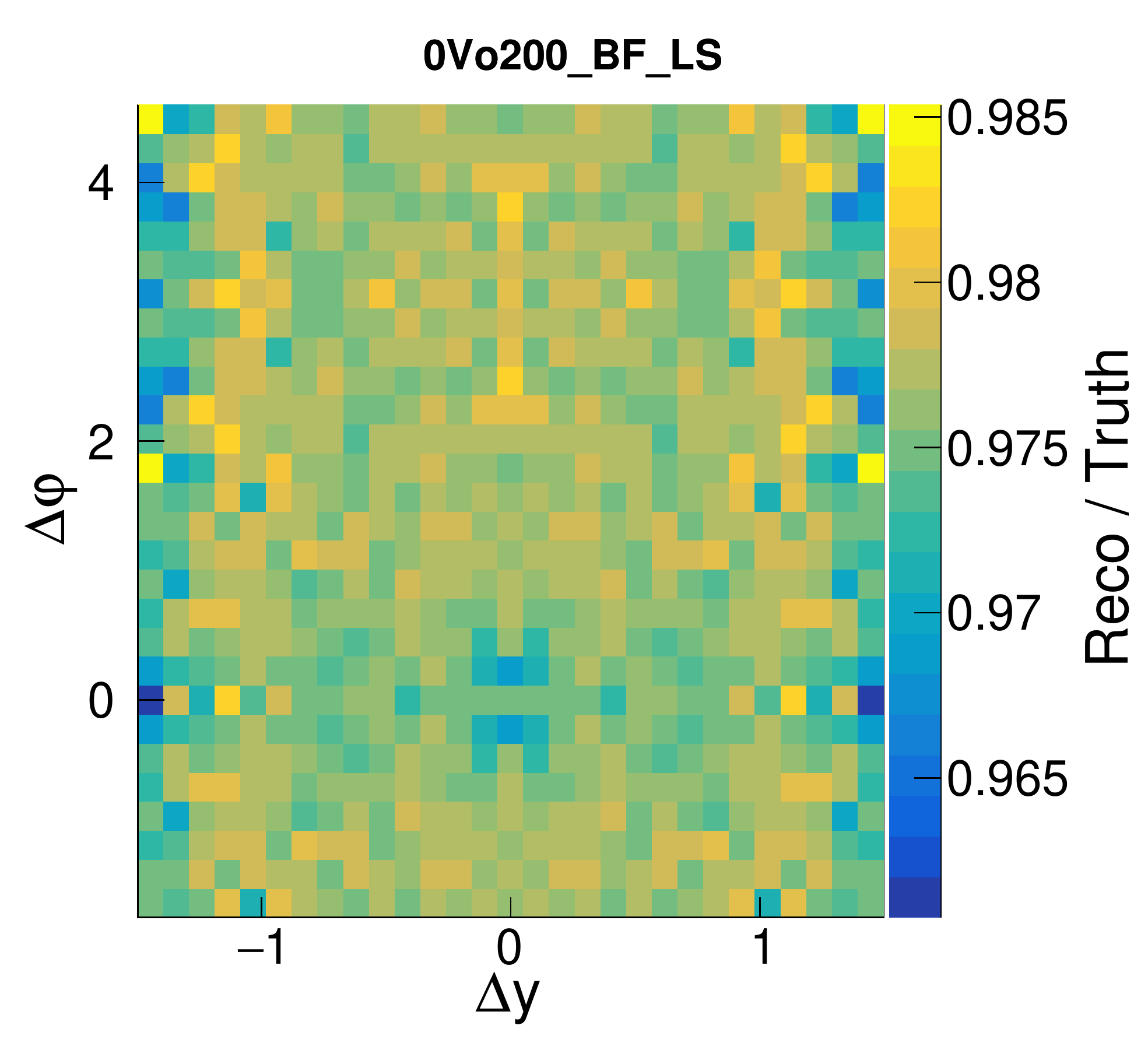}
  \includegraphics[width=0.3\linewidth]{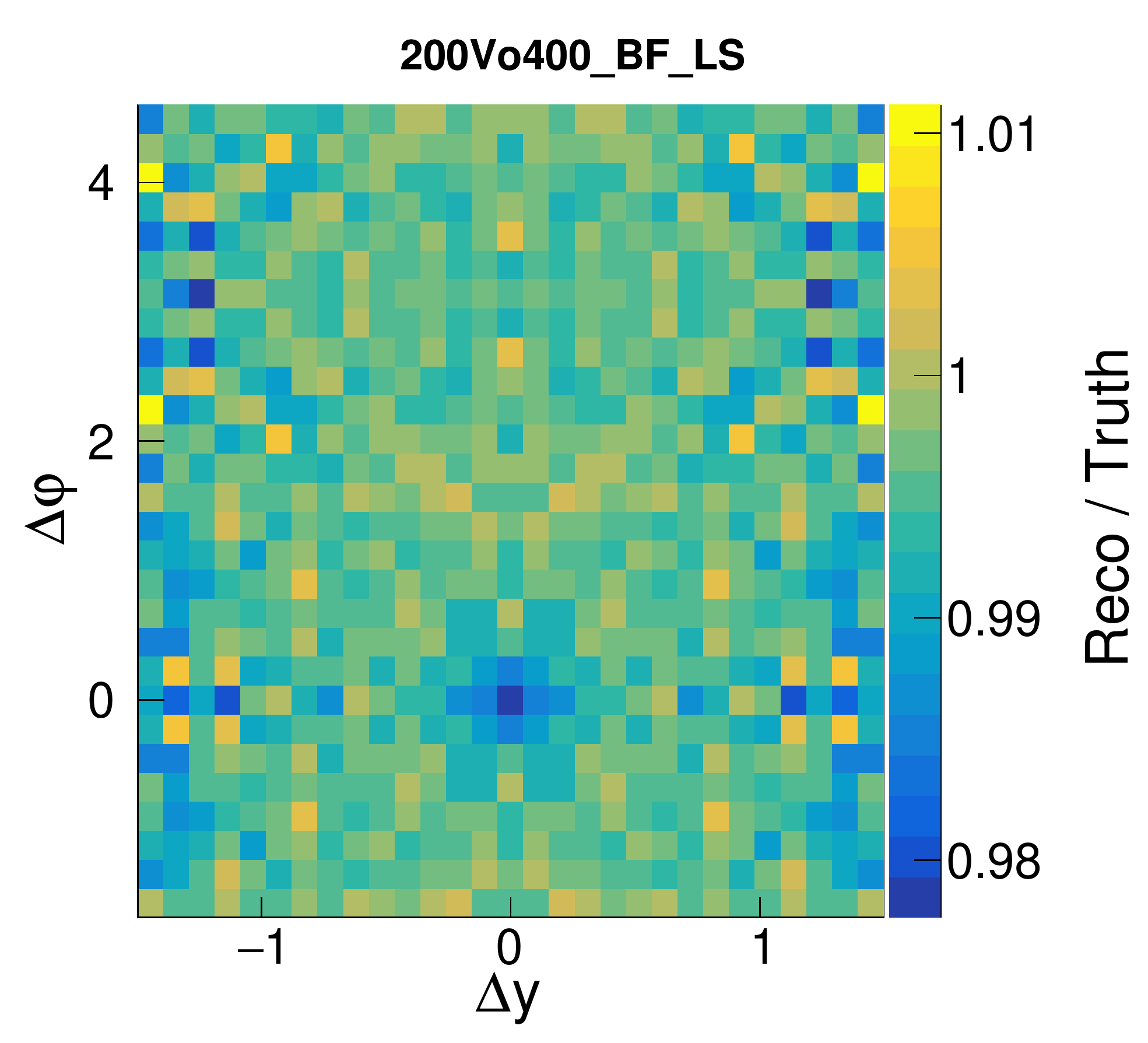}
  \includegraphics[width=0.3\linewidth]{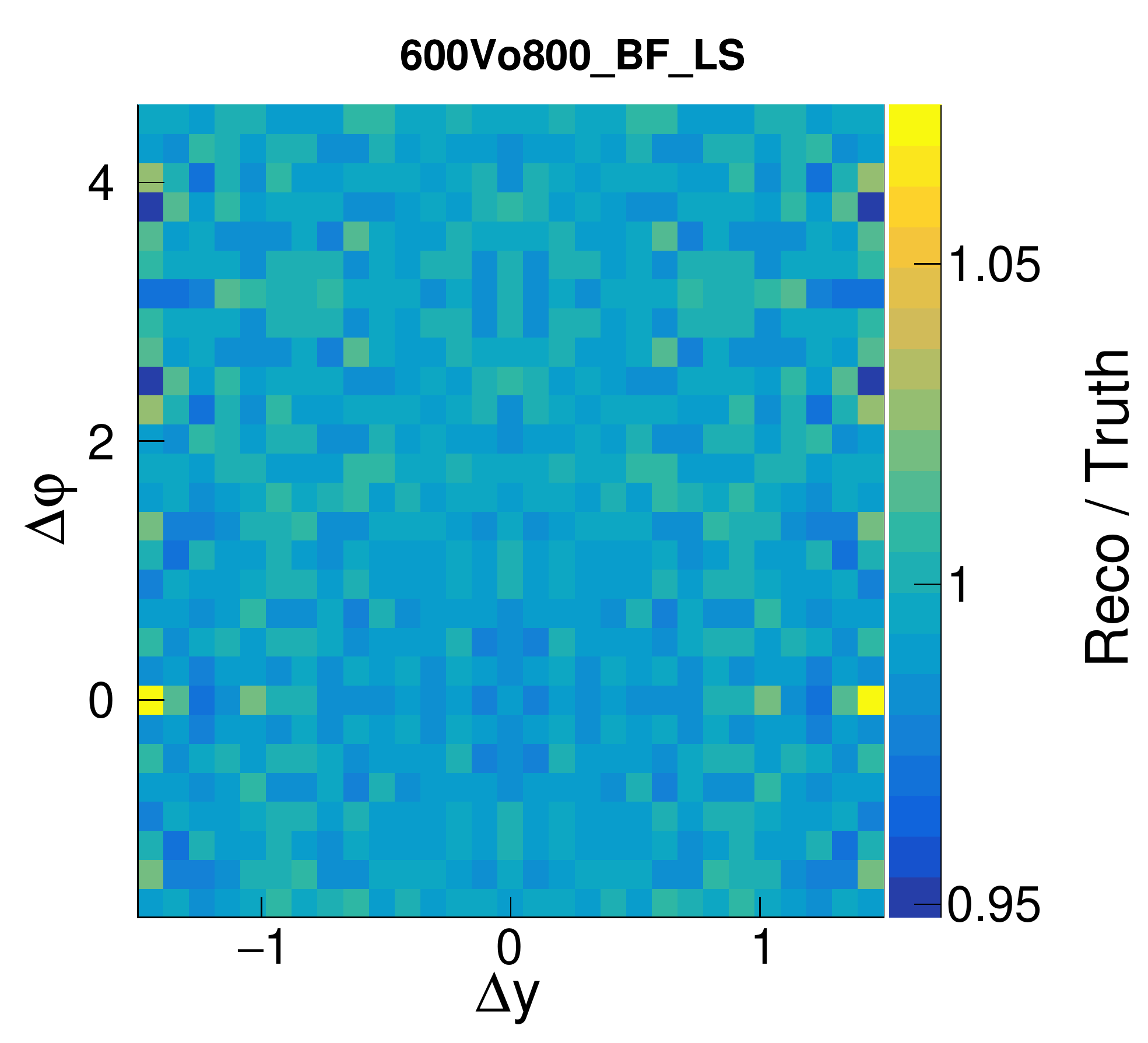}
  \includegraphics[width=0.3\linewidth]{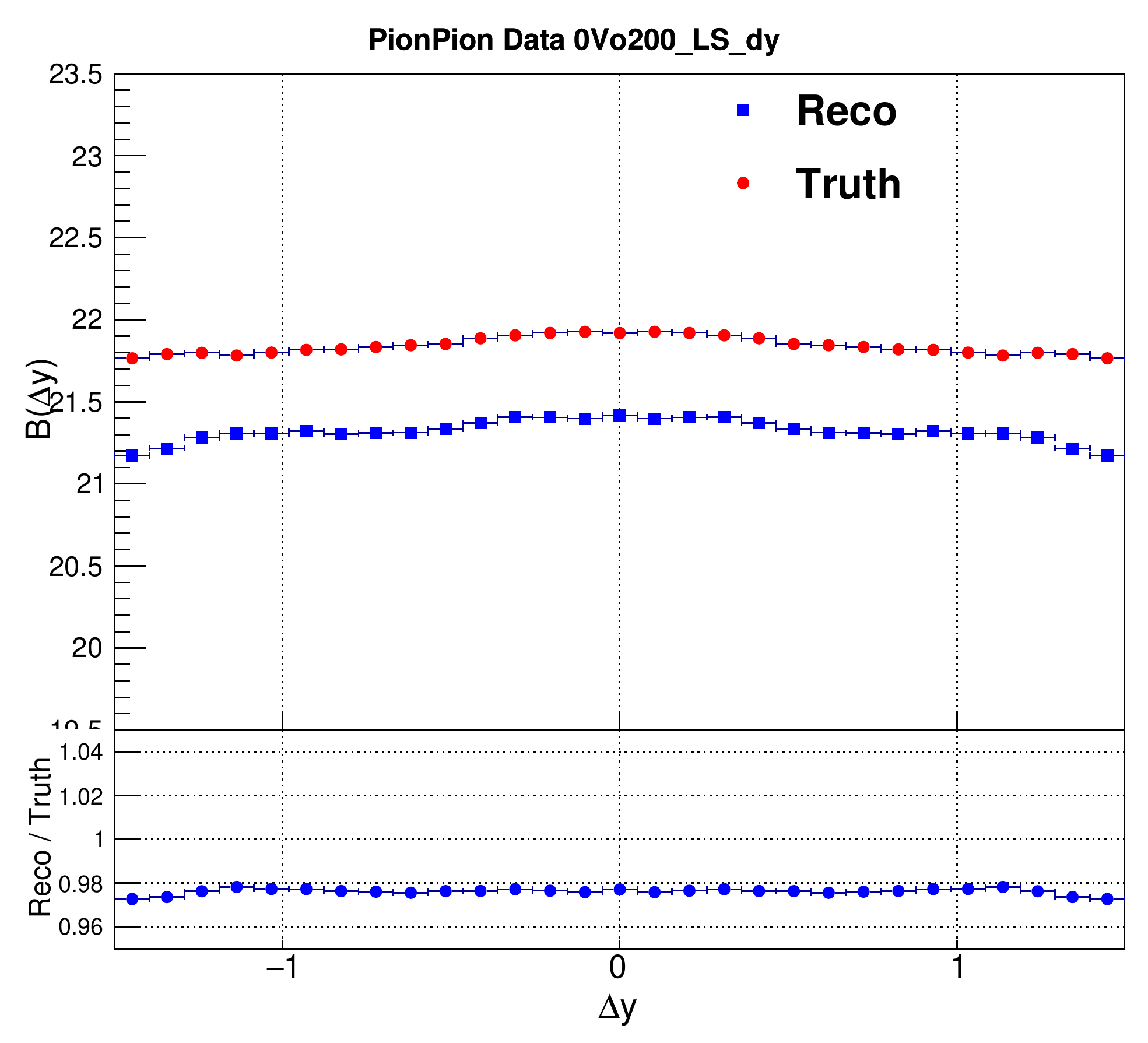}
  \includegraphics[width=0.3\linewidth]{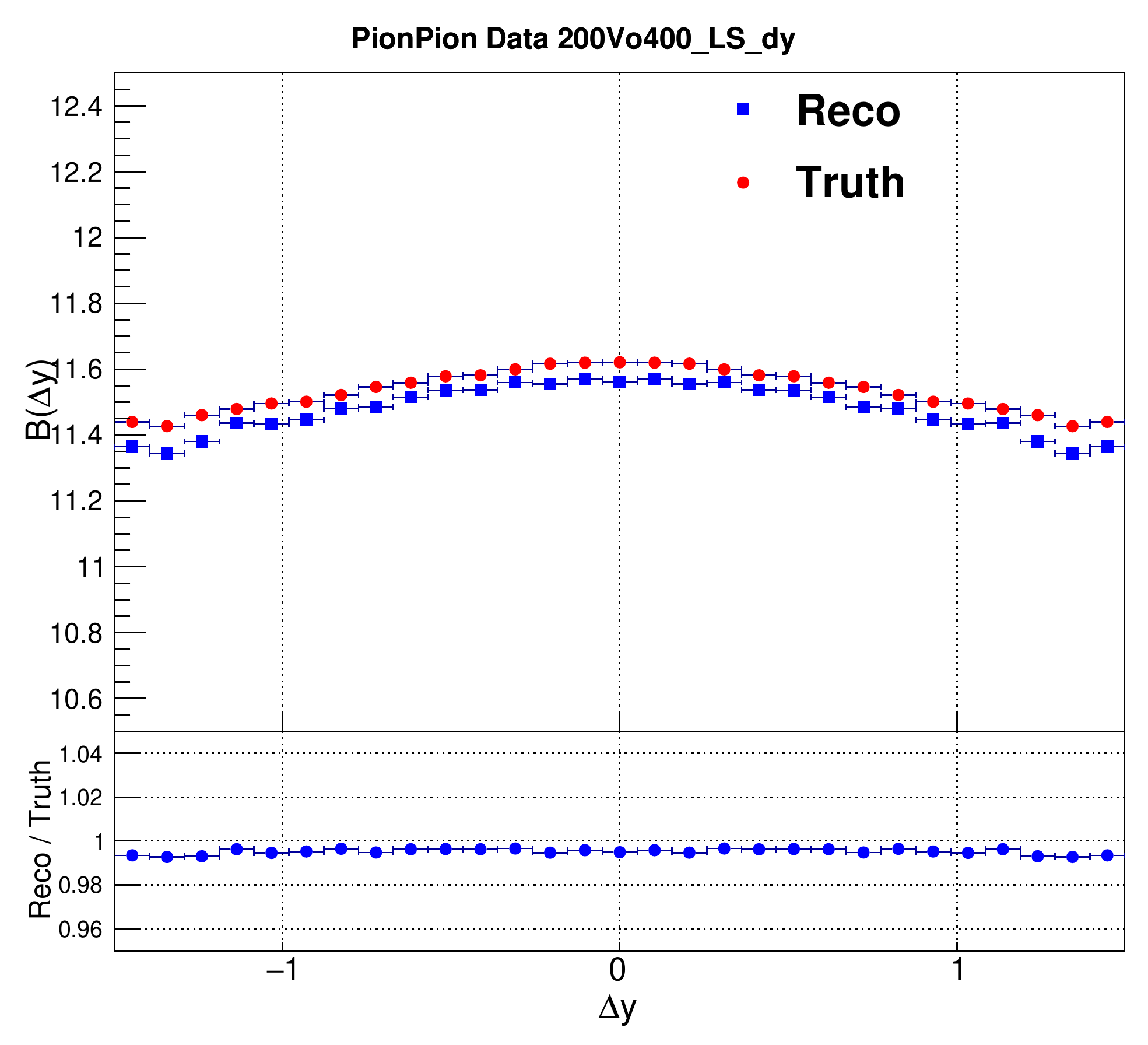}
  \includegraphics[width=0.3\linewidth]{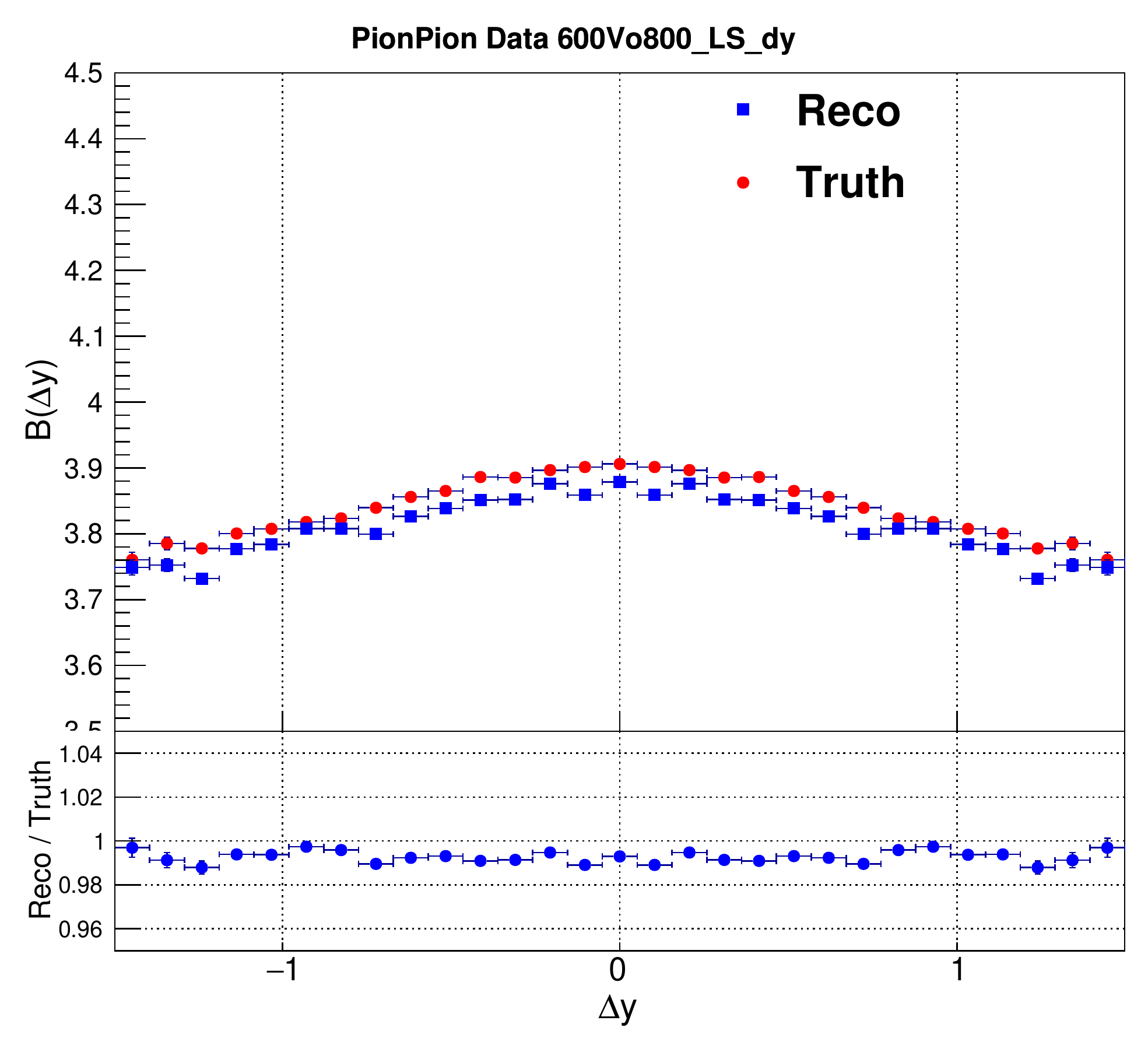}
  \includegraphics[width=0.3\linewidth]{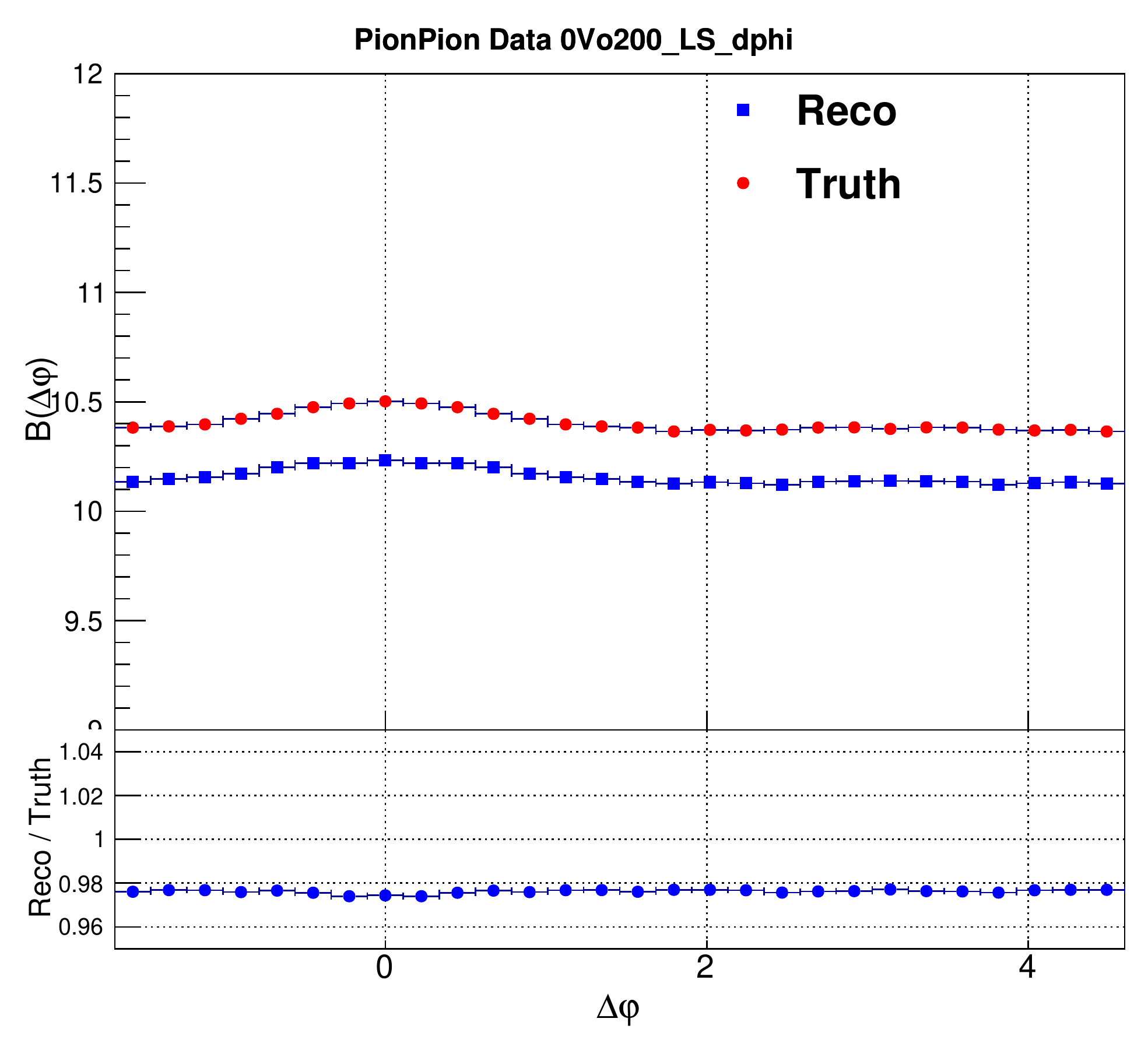}
  \includegraphics[width=0.3\linewidth]{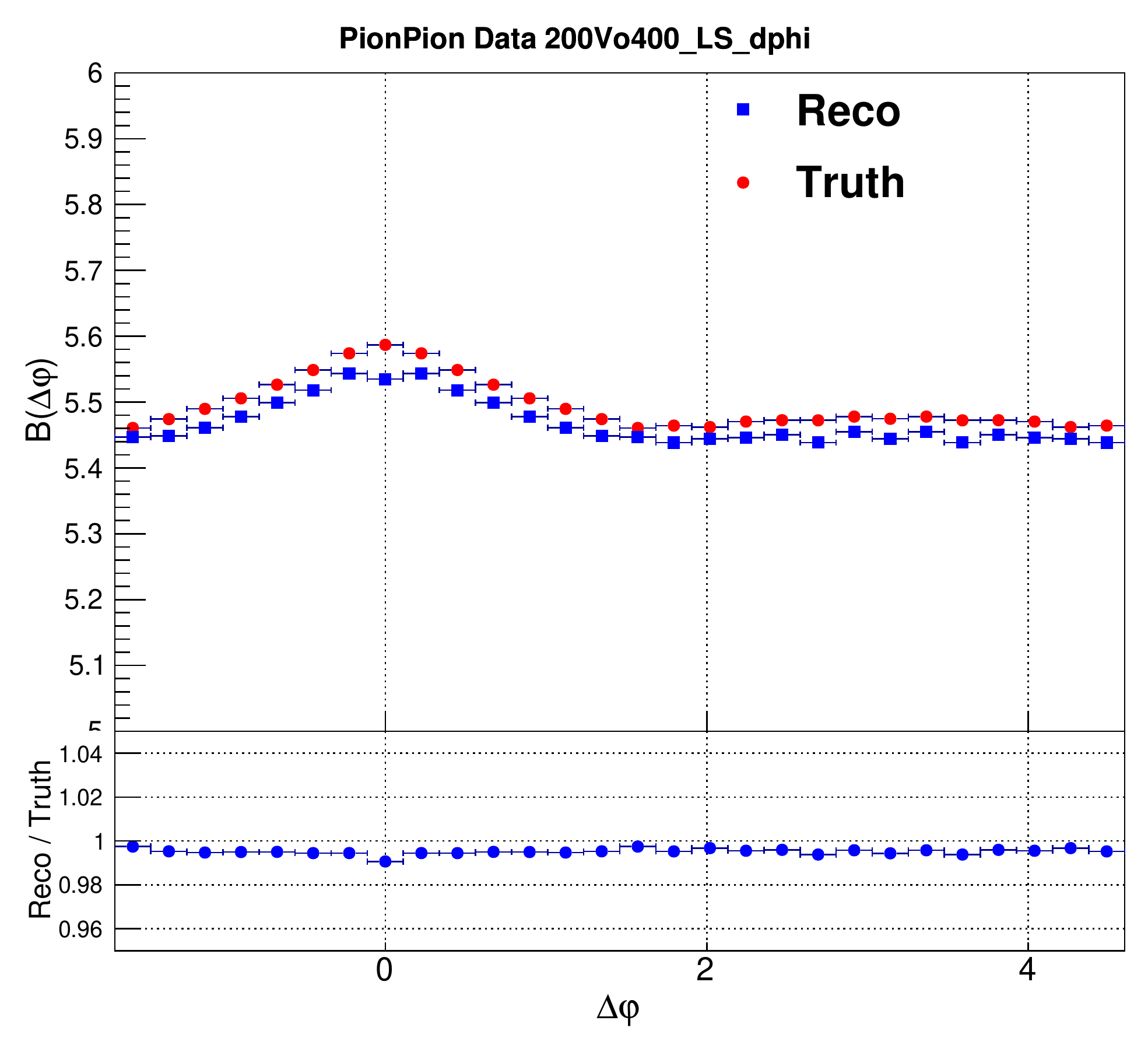}
  \includegraphics[width=0.3\linewidth]{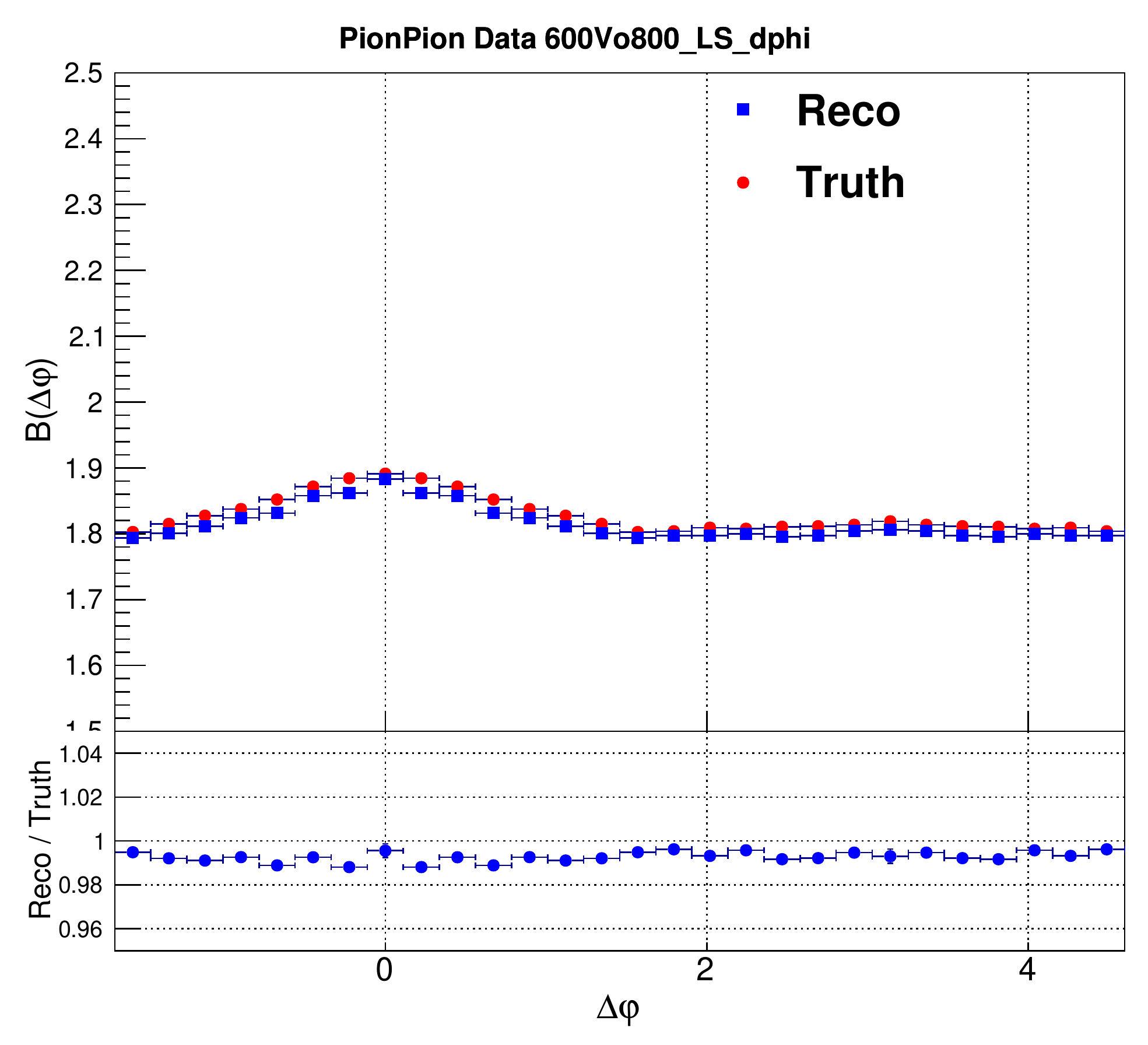}
  \caption{MC closure test of $\pi\pi$ pair. Comparisons of 2D CFs of LS obtained with MC truth ($1^{st}$ row) and reconstructed ($2^{nd}$ row) events for selected collision centralities, and their ratios ($3^{rd}$ row). $4^{th}$ and $5^{th}$ rows: comparisons of the $\Delta y$ and $\Delta\varphi$ projections, respectively.}
  \label{fig:HIJING_Truth_Reco_DCAxy004_BF_LS_PionPion}
\end{figure}

Figures~\ref{fig:HIJING_Truth_Reco_DCAxy004_BF_US_PionPion} and~\ref{fig:HIJING_Truth_Reco_DCAxy004_BF_LS_PionPion} show that the generator and reconstructed two-dimensional US and LS correlators are similar. The reconstructed $\Delta y$ and $\Delta\varphi$ projections are about 0.5-2\% (for different centralities) lower than the generator results.
On one hand, this may be explained in part by  pair loss in the reconstructed MC data, that are larger in central than peripheral events.
On the other hand, this also could be due to the existence of about 3\% mis-identified and secondary $\pi^{\pm}$ in the reconstructed $\pi^{\pm}$ sample, as shown in Fig.~\ref{fig:PID_Cuts}.
The effects are present in CFs of both US and LS with approximately the same strength, thus cancels in the BF.

%Furthermore, in order to understand the reason of bigger discrepancy for central events ($\sim2\%$), normalized single pion $p_{\rm T}$ spectra are compared for different centralities between HIJING Reconstructed and Truth. Fig.~\ref{fig:CompareMCReco_m_pt_spectra} are the single negative pion $p_{\rm T}$ spectra for HIJING Reconstructed and Truth, respectively. Fig.~\ref{fig:ratio_Reco_pt_m_Comparison} are the normalized (by dividing the $p_{\rm T}$ spectrum for peripheral $60-80\%$ centrality) single negative pion $p_{\rm T}$ spectra for HIJING Reconstructed and Truth, respectively. \\

%Fig.~\ref{fig:Compare_MC_RatioofRatio_Reco_Truth_m_pt_spectra} is the comparison of normalized single negative pion $p_{\rm T}$ spectra between HIJING Reconstructed and Truth, which shows the efficiency is lower for central events ($0-20\%$) than for mid-central events ($20-40\%$ and $40-60\%$) by about $4\%$ for $0.6\le p_{\rm T}\le2.0$ GeV/c (the TOF PID region). The efficiency correction used in this closure test only corrects efficiency in terms of $y$ and $\varphi$ but not $p_{\rm T}$, since the efficiency as a function of $p_{\rm T}$ is approximately constant for most centralities except for the central events where there is about $5\%$ difference for the TPC and TOF PID region.

\begin{figure}
\centering
  \includegraphics[width=0.3\linewidth]{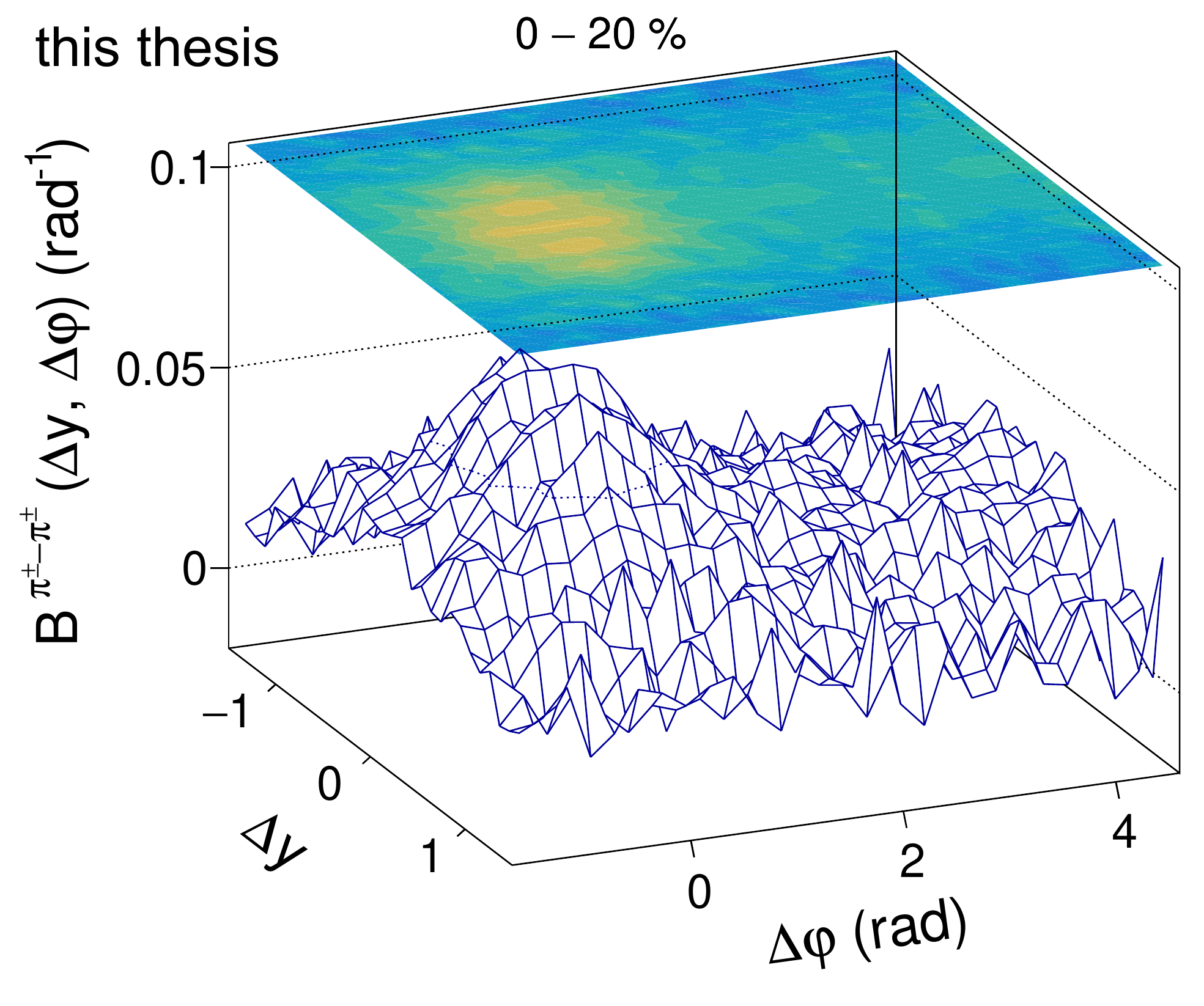}
  \includegraphics[width=0.3\linewidth]{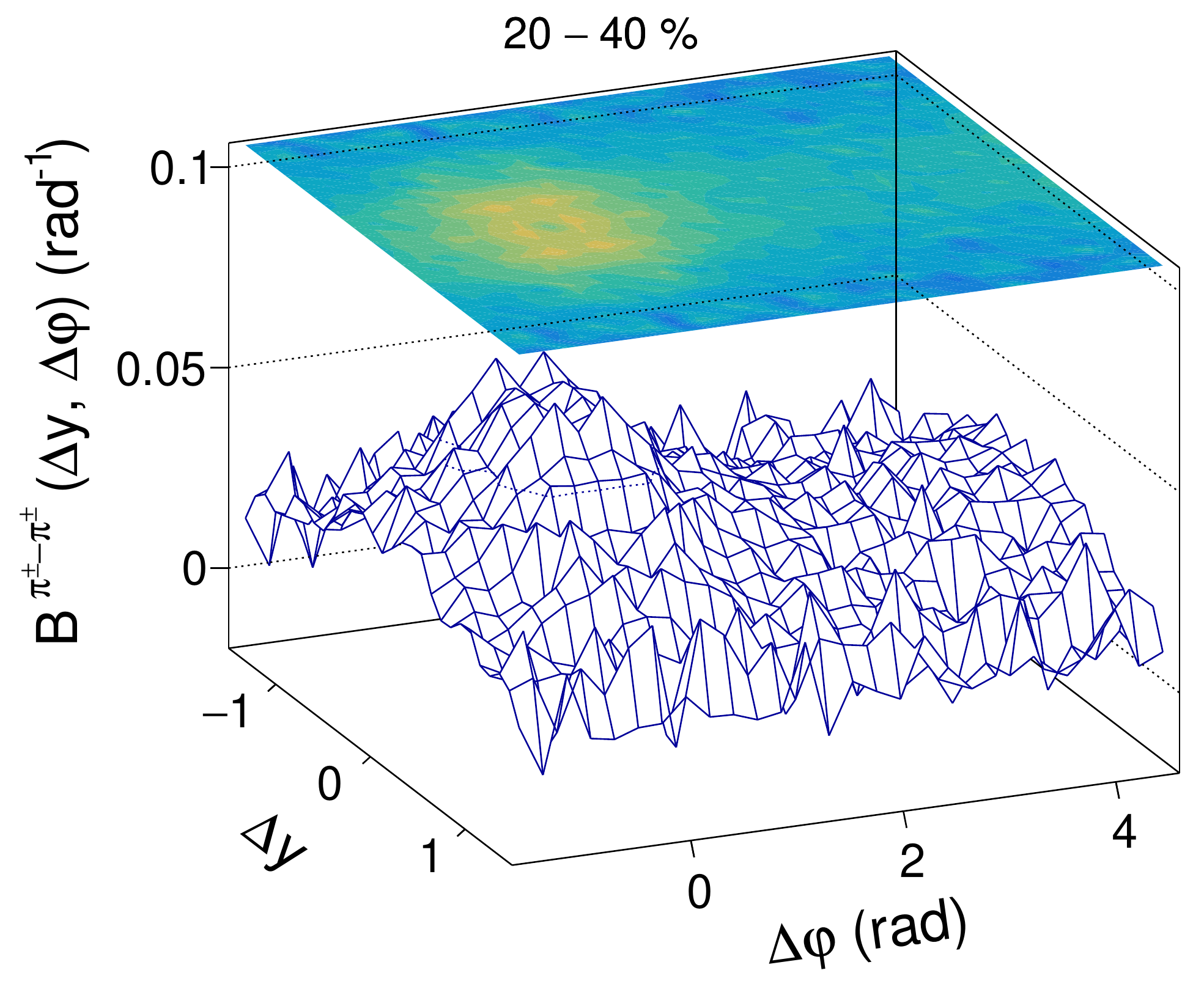}
  \includegraphics[width=0.3\linewidth]{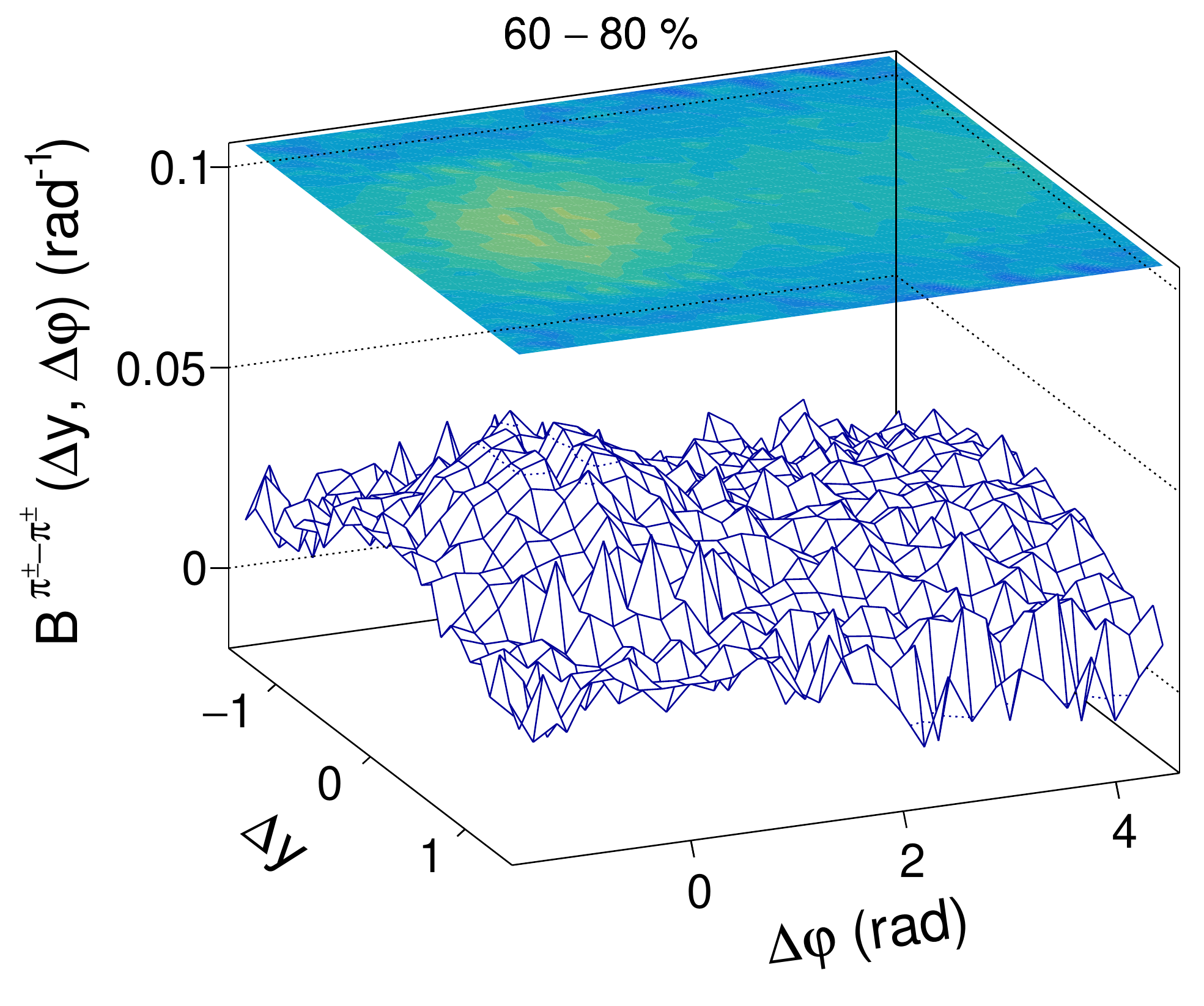}
  \includegraphics[width=0.3\linewidth]{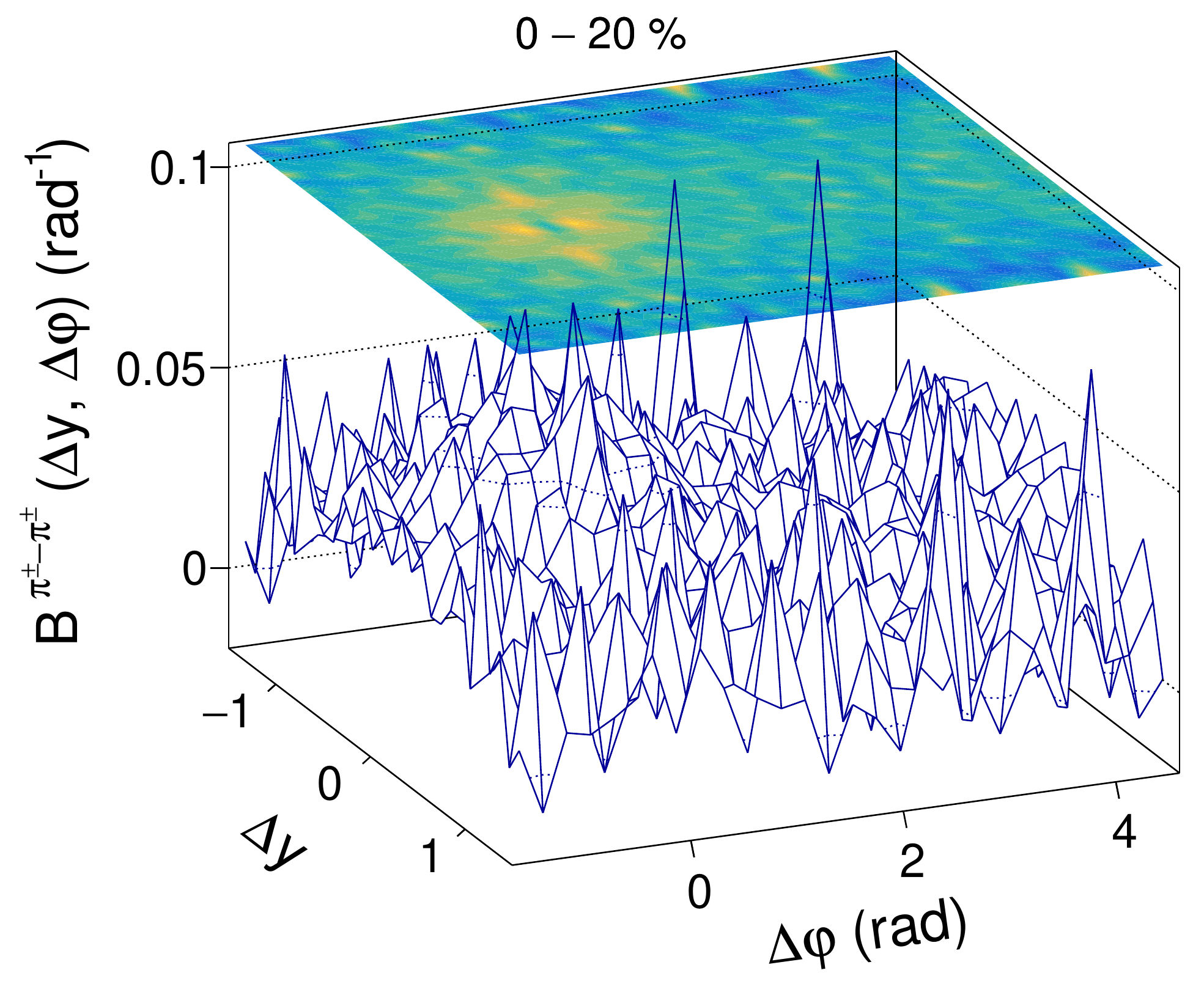}
  \includegraphics[width=0.3\linewidth]{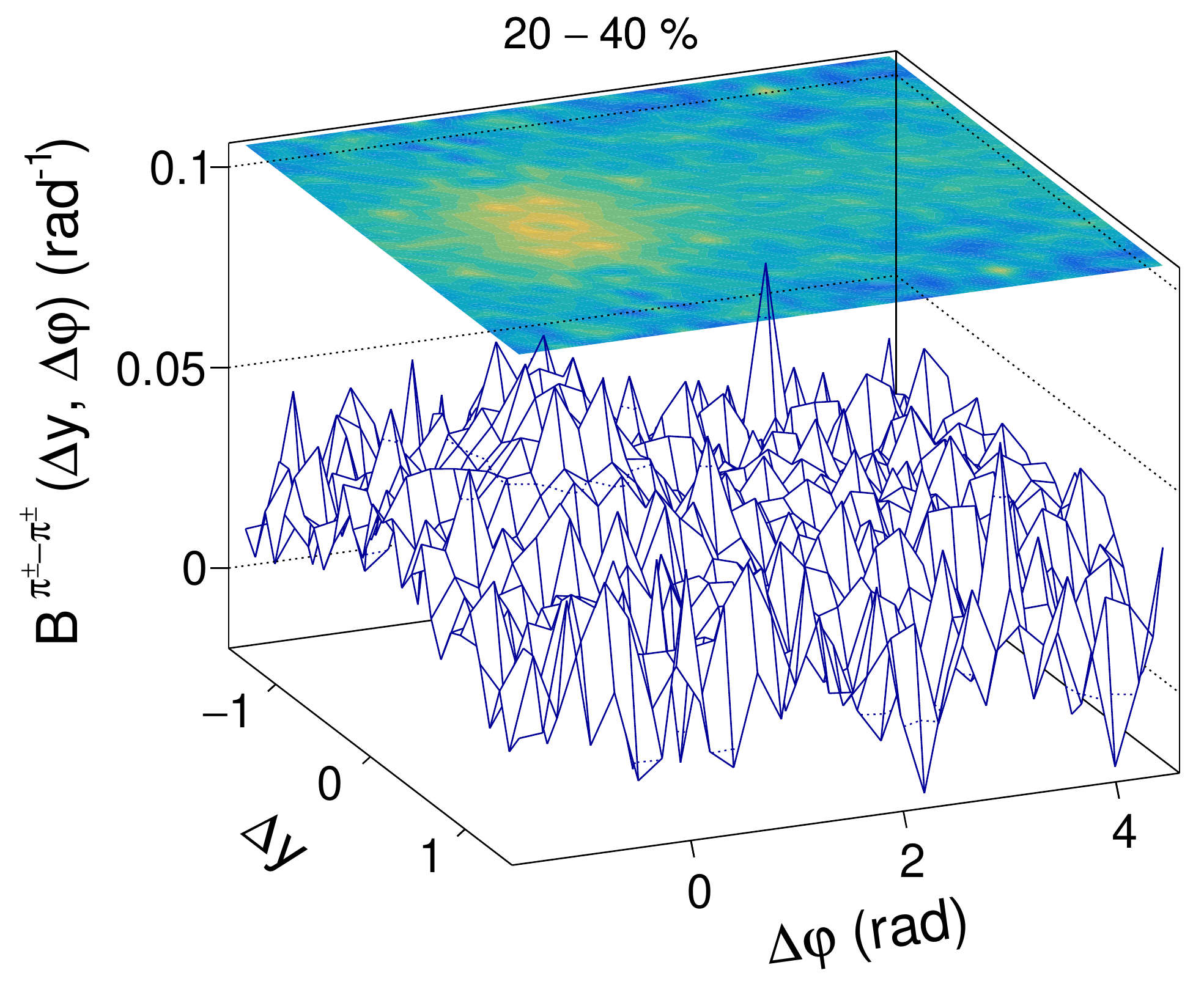}
  \includegraphics[width=0.3\linewidth]{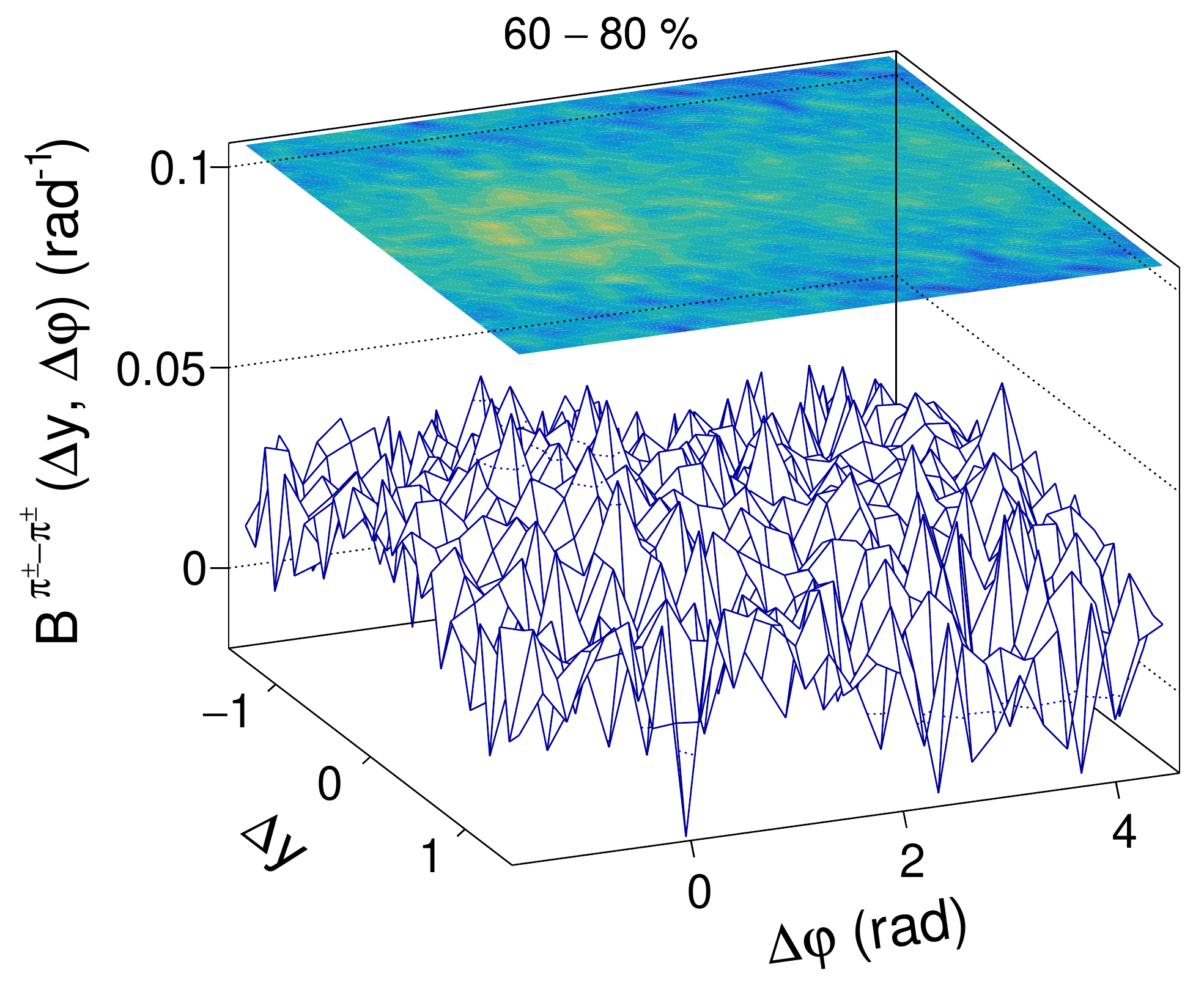}
\includegraphics[width=0.3\linewidth]{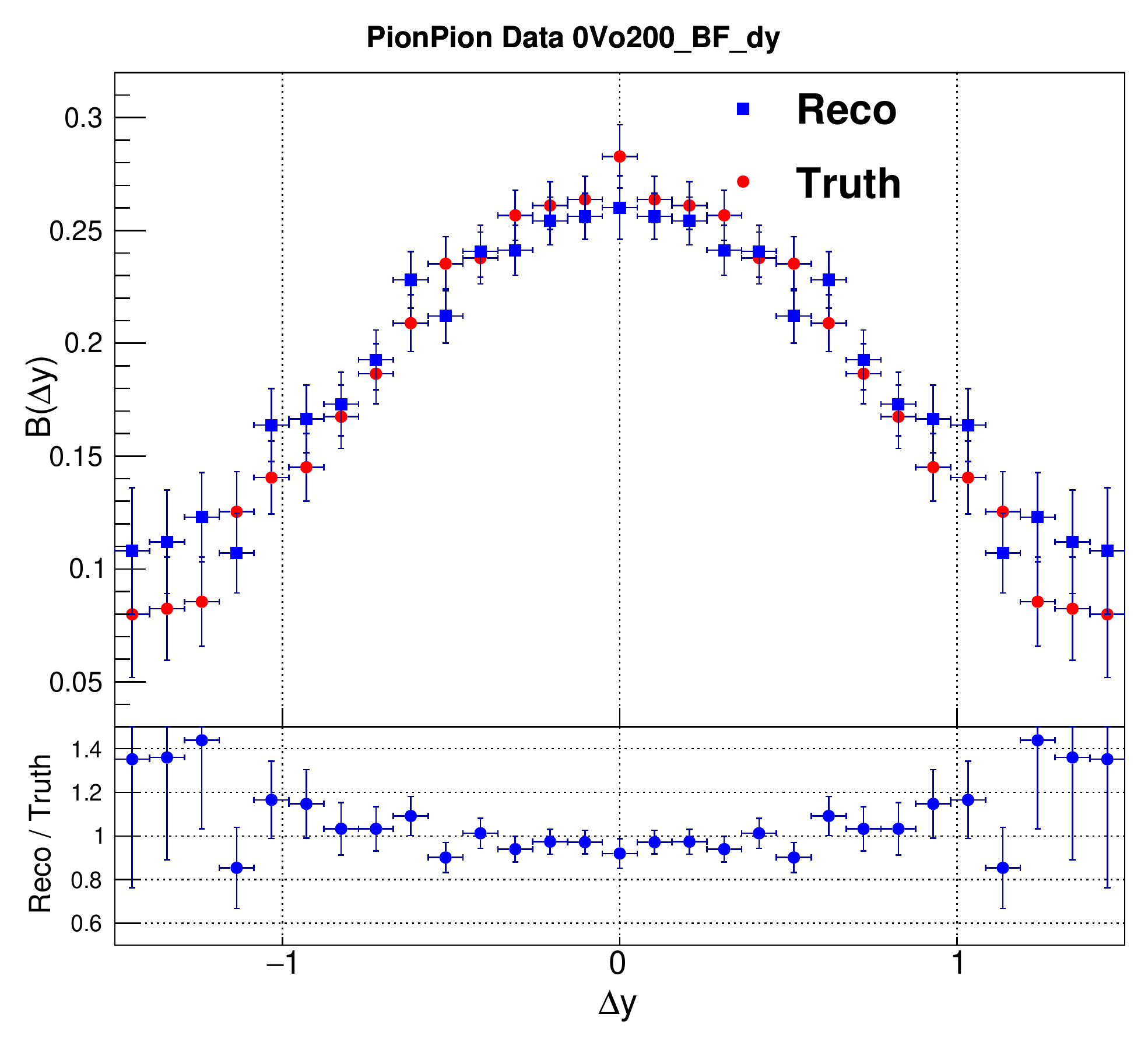}
  \includegraphics[width=0.3\linewidth]{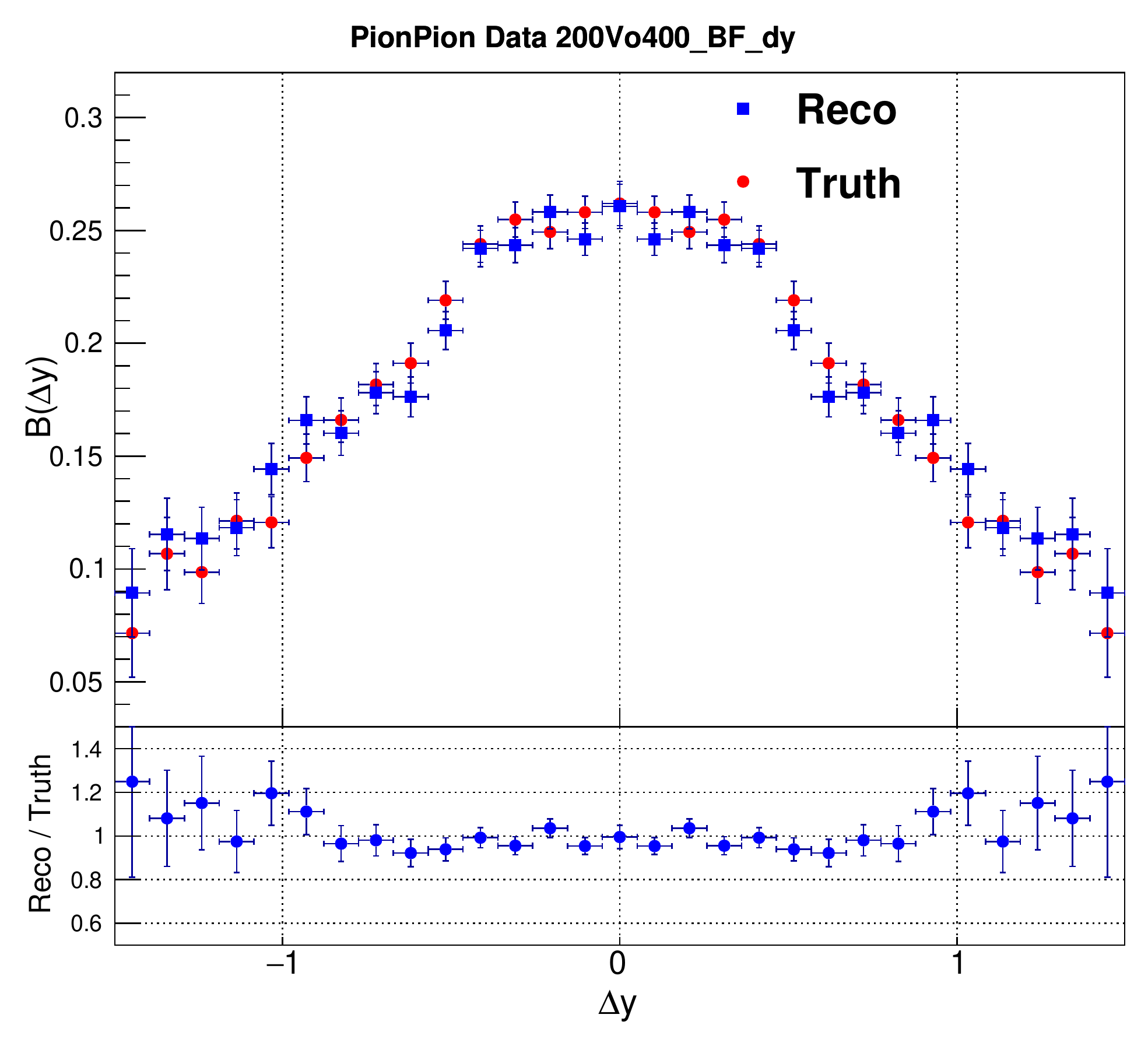}
  \includegraphics[width=0.3\linewidth]{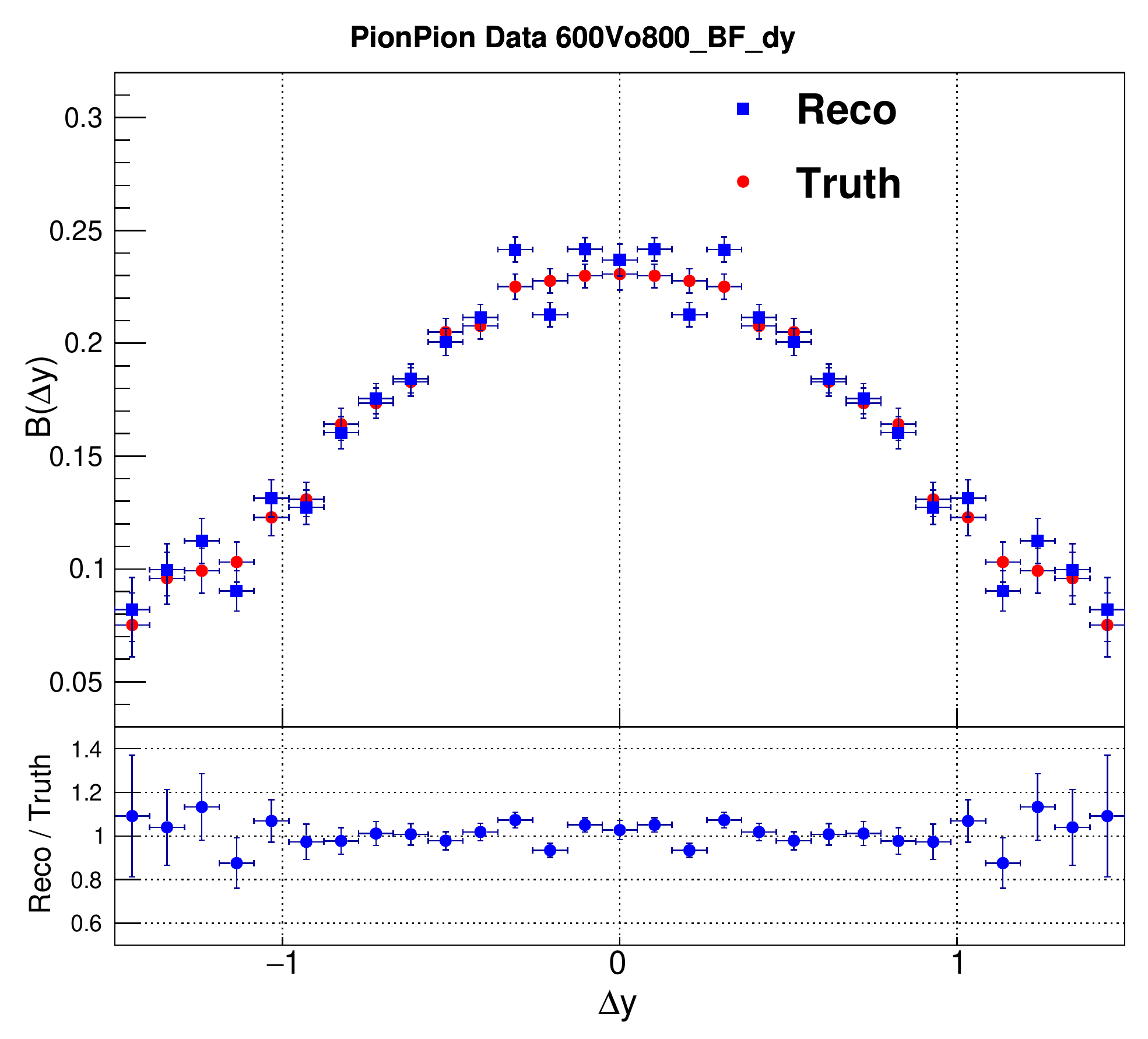}
  \includegraphics[width=0.3\linewidth]{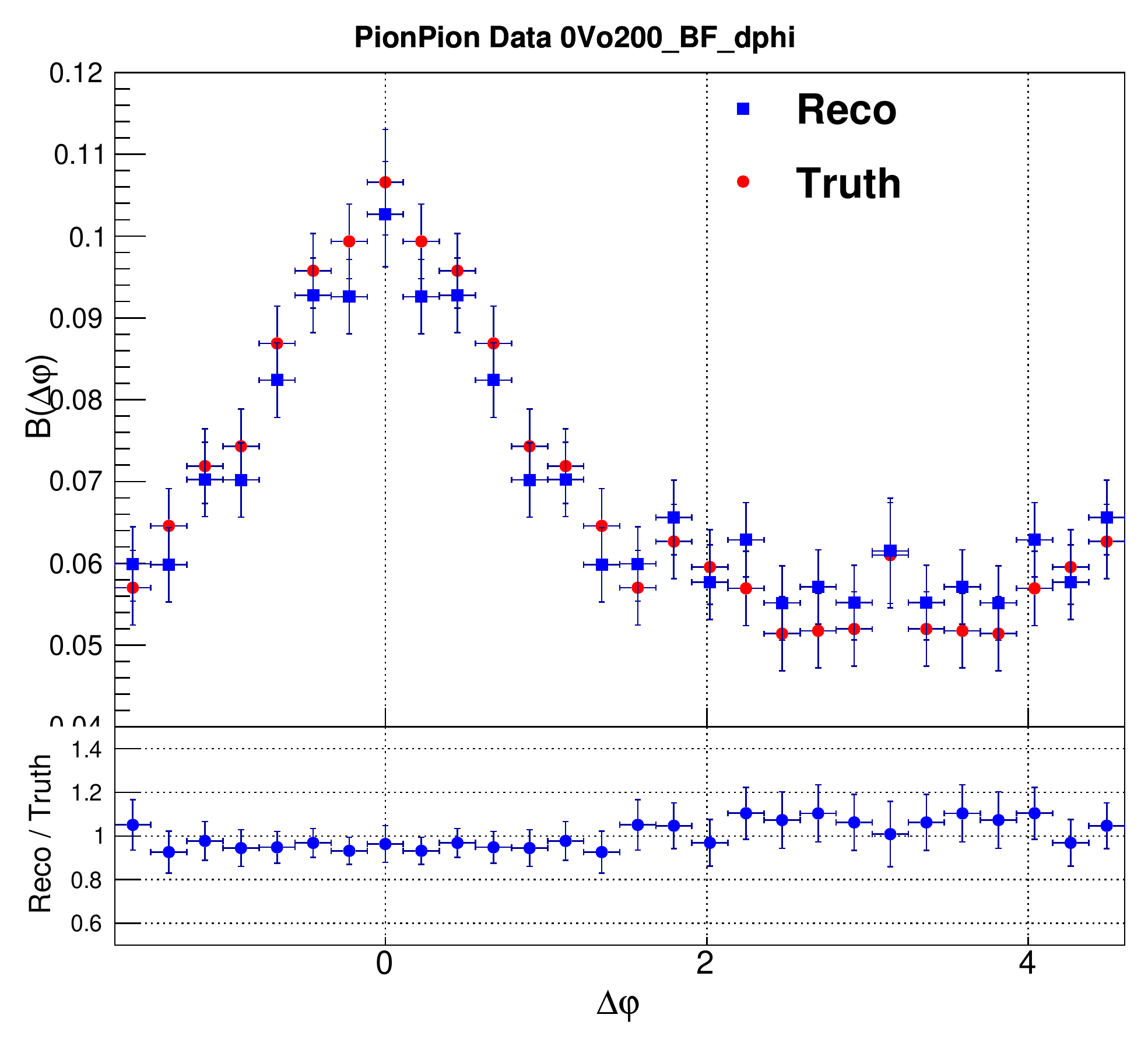}
  \includegraphics[width=0.3\linewidth]{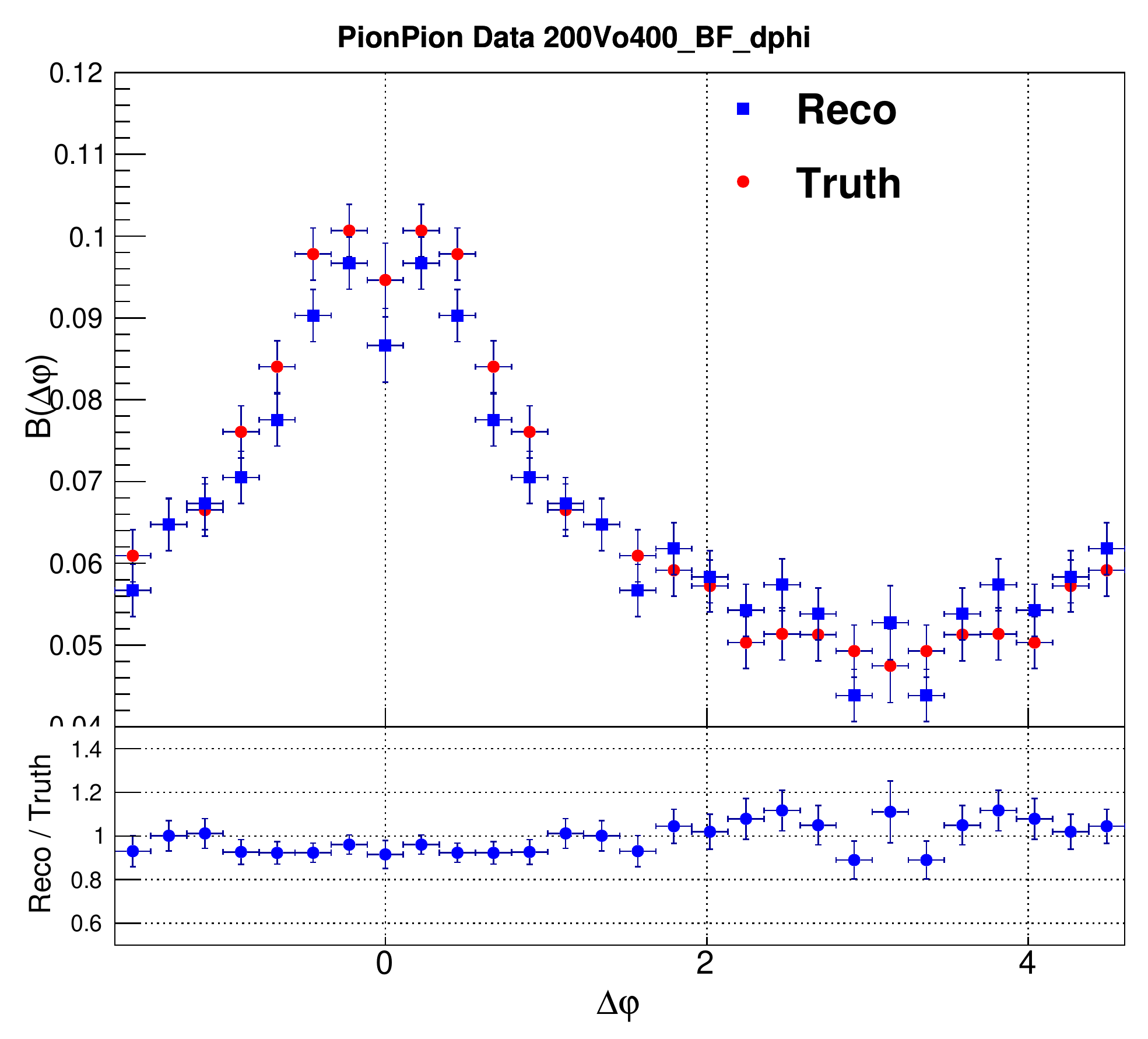}
  \includegraphics[width=0.3\linewidth]{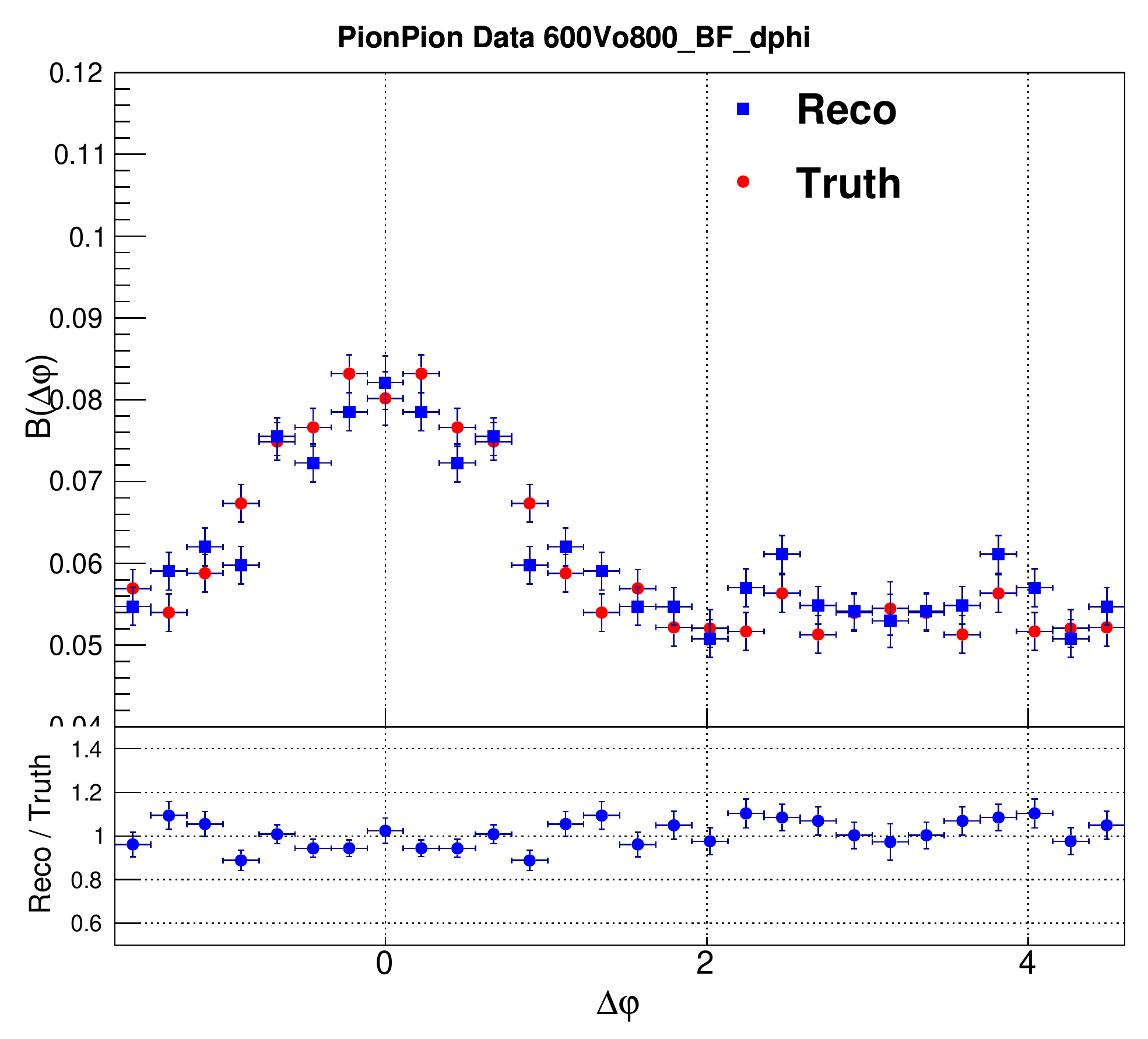}
  \caption{MC closure test of $\pi\pi$ pair. Comparisons on 2D BFs between MC truth ($1^{st}$ row) and reconstructed ($2^{nd}$ row) results for different centralities, along with their $\Delta y$ ($3^{rd}$ row) and $\Delta\varphi$ ($4^{th}$ row) projections. The $\Delta\varphi$ projections are taken for range $|\Delta y|\le0.9$ to avoid large fluctuations at large $\Delta y$.}
  \label{fig:HIJING_Truth_Reco_DCAxy004_BF_PionPion}
\end{figure}

Figure~\ref{fig:HIJING_Truth_Reco_DCAxy004_BF_PionPion} presents a comparison between the generator and reconstructed 2D BFs.
The two BFs are found to be very similar, despite the fact that there are large fluctuations in reconstructed results due to low statistics. The differences between the reconstructed and generator projections (onto both $\Delta y$ and $\Delta\varphi$) are within two standard deviation of statistical uncertainties for almost all points.

%%%%%%%%%%%%%%%%%%%%%%%%%%%%%%%%%%%%%%%%%%%%%%%%%%%%%%%%
% KaonKaon DCAxy2
%%%%%%%%%%%%%%%%%%%%%%%%%%%%%%%%%%%%%%%%%%%%%%%%%%%%%%%%
\subsubsection{MC Closure Test of $KK$ Correlation Functions}
\label{subsubsec:MCCTKK}

\begin{figure}
\centering
  \includegraphics[width=0.32\linewidth]{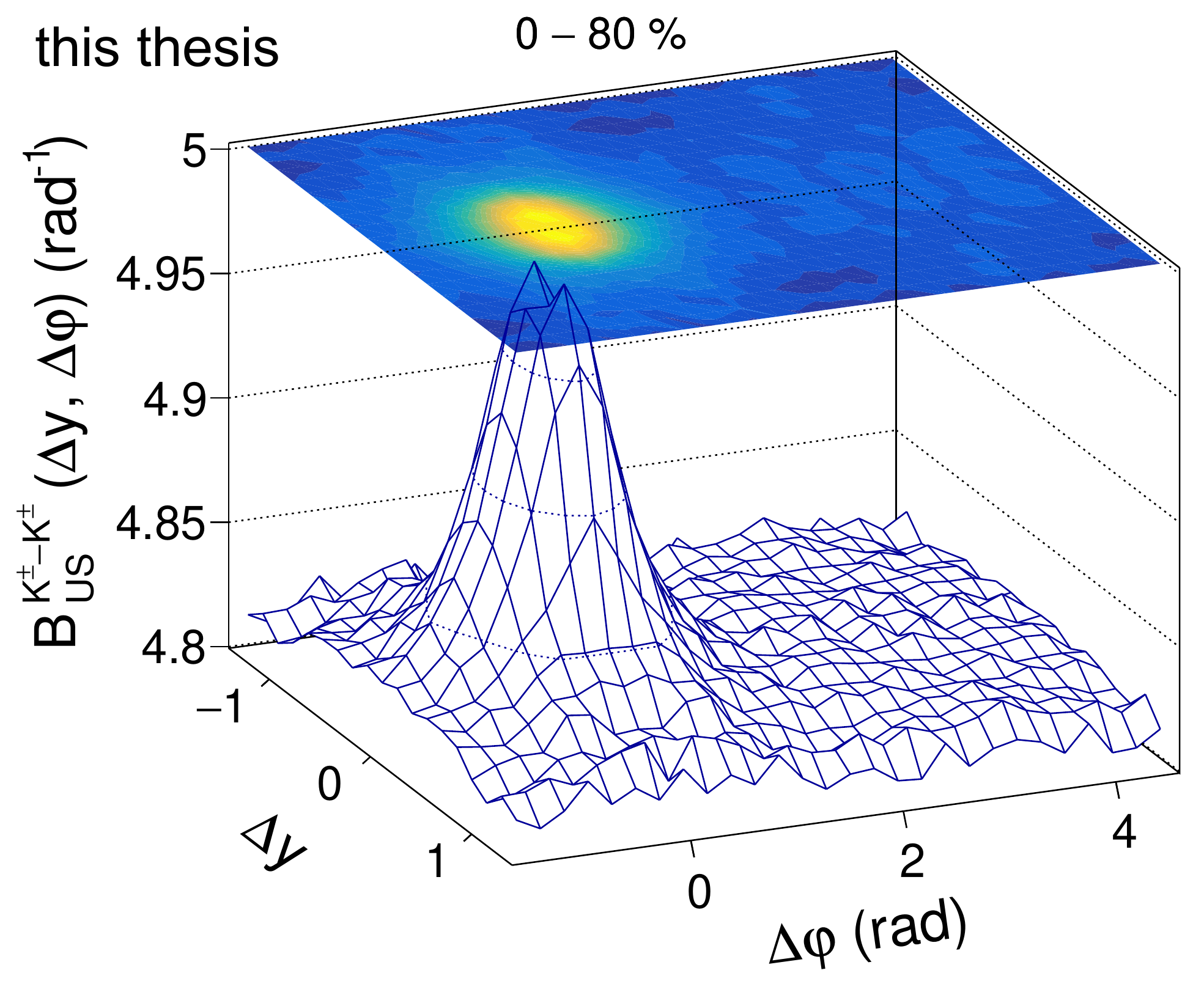}
  \includegraphics[width=0.32\linewidth]{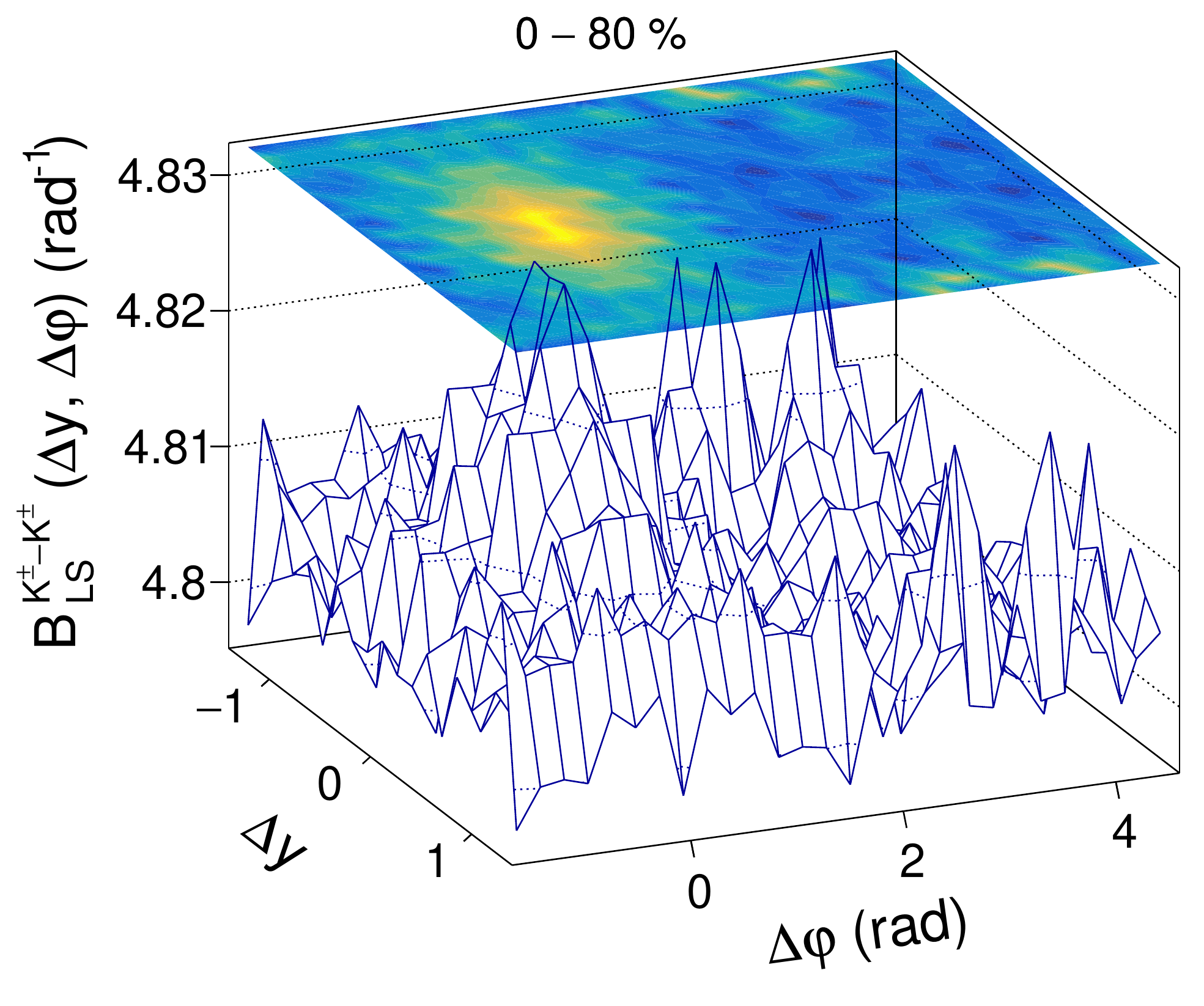}
  \includegraphics[width=0.32\linewidth]{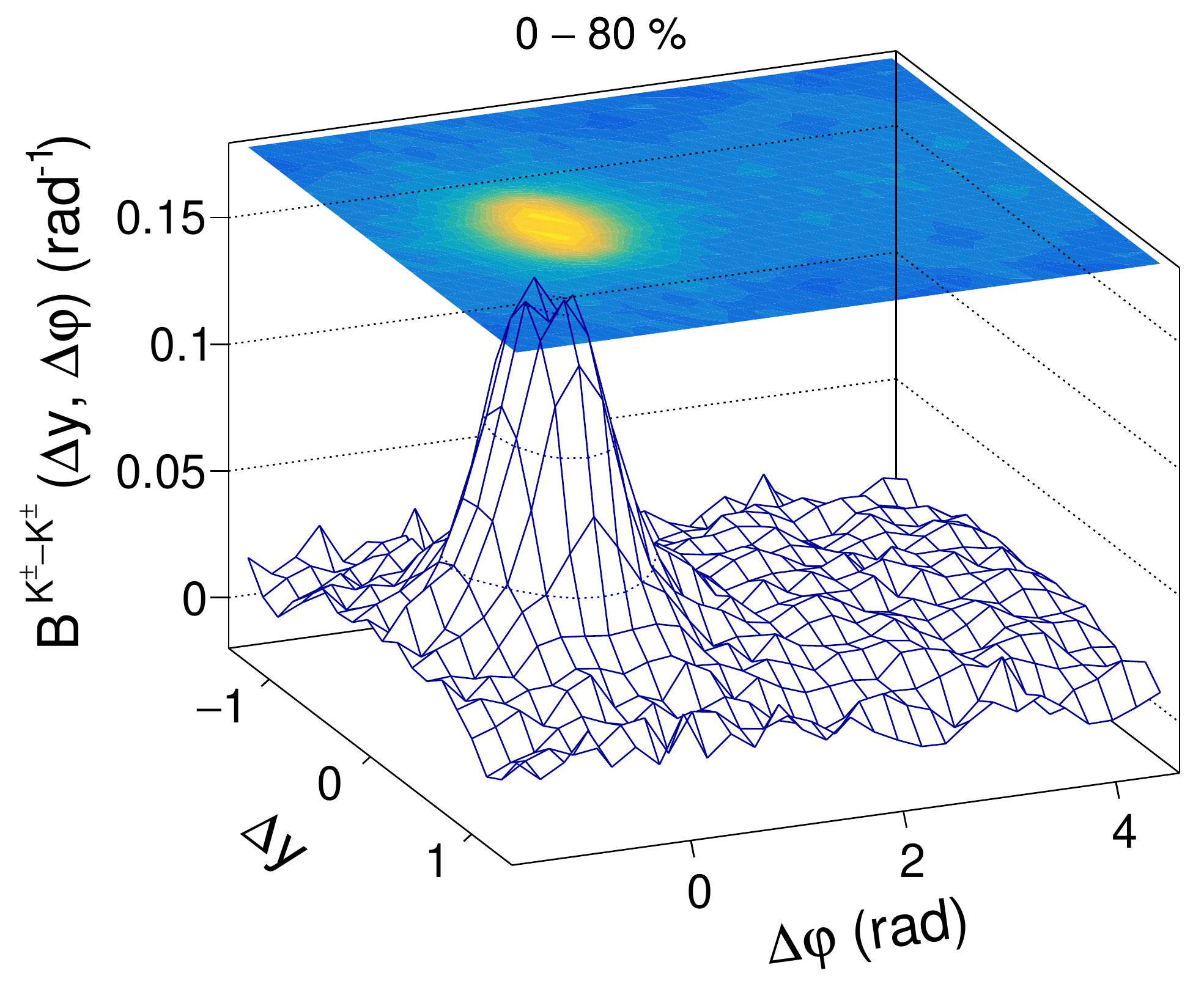}
  \includegraphics[width=0.32\linewidth]{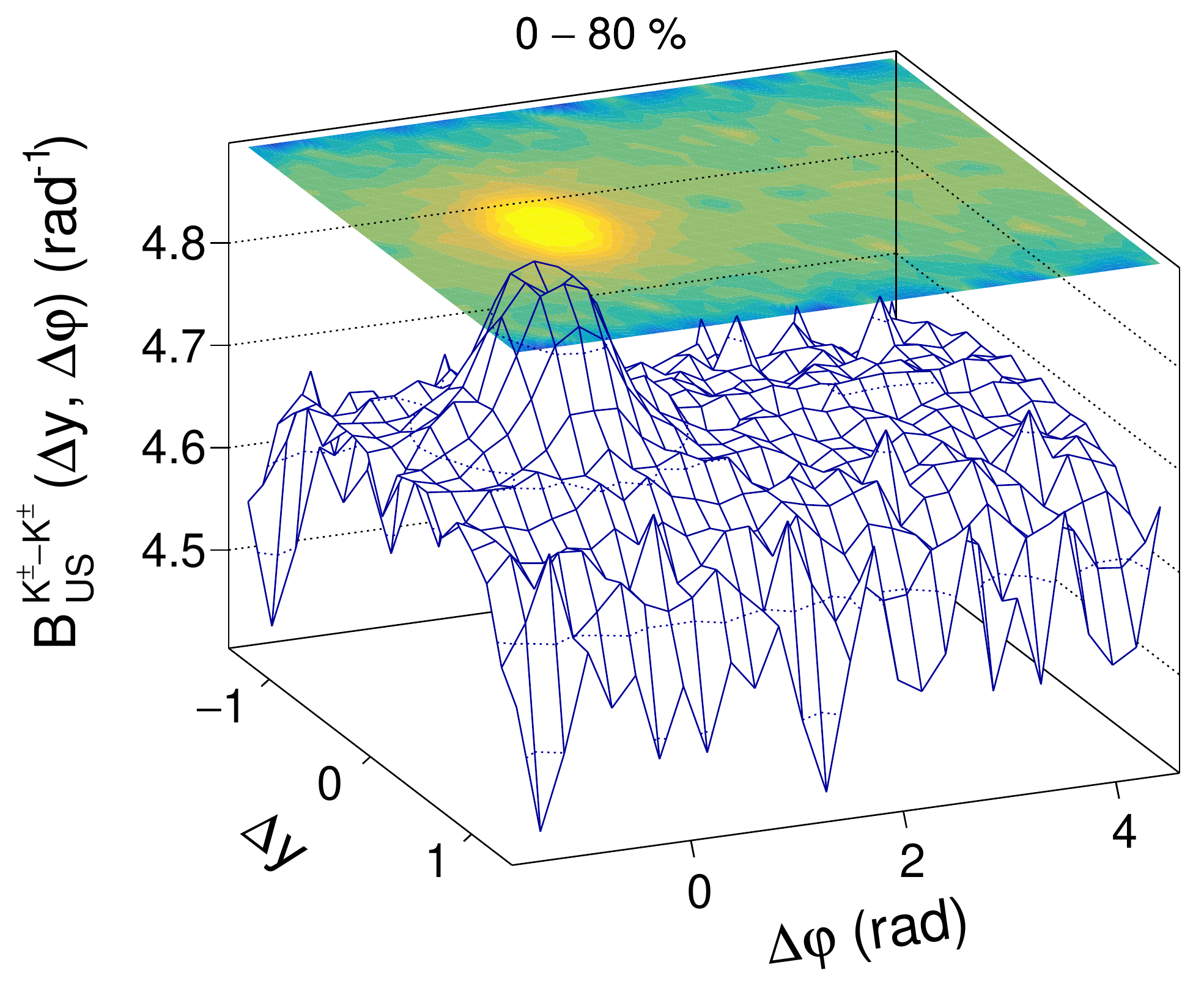}
  \includegraphics[width=0.32\linewidth]{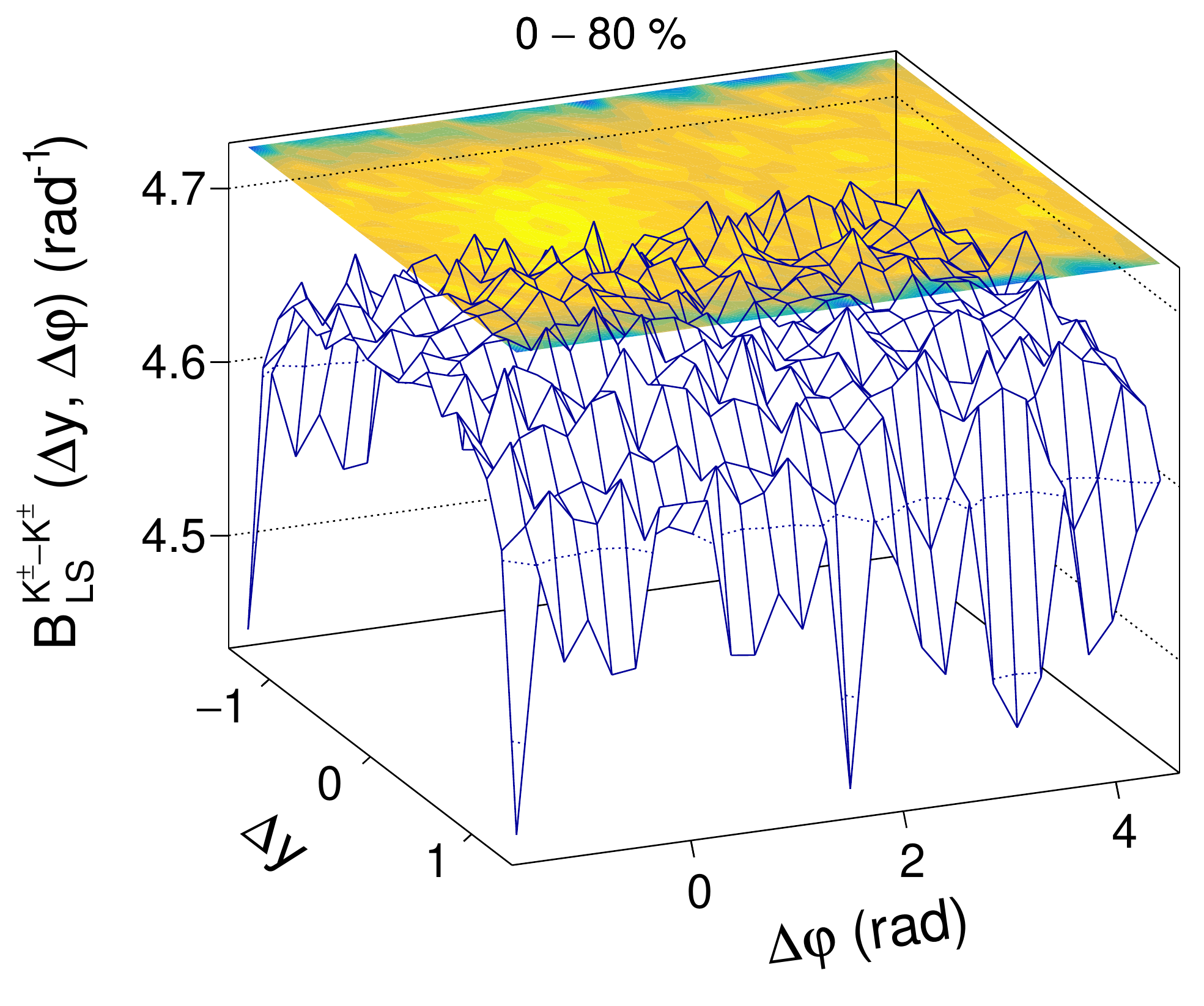}
  \includegraphics[width=0.32\linewidth]{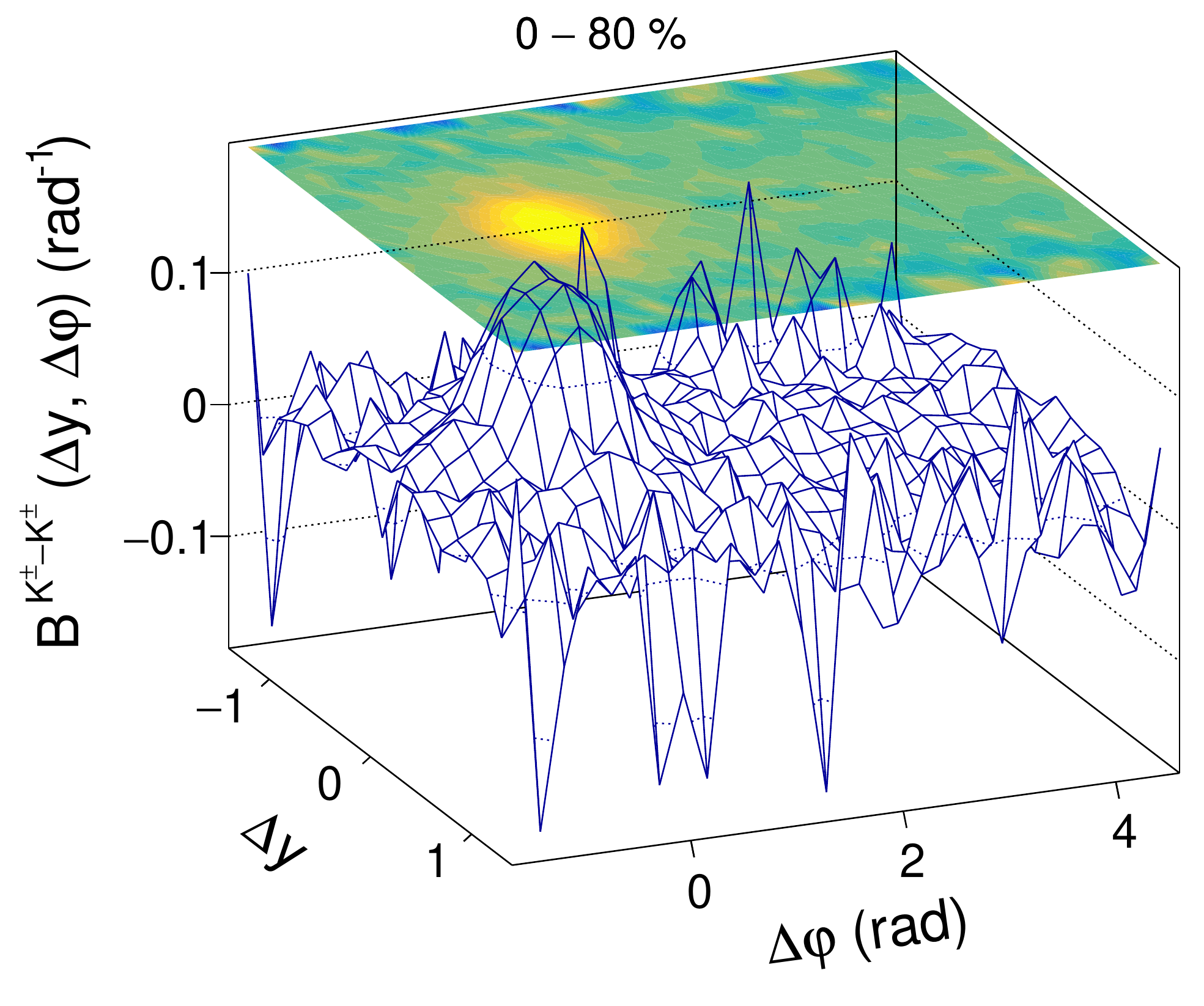}
  \includegraphics[width=0.32\linewidth]{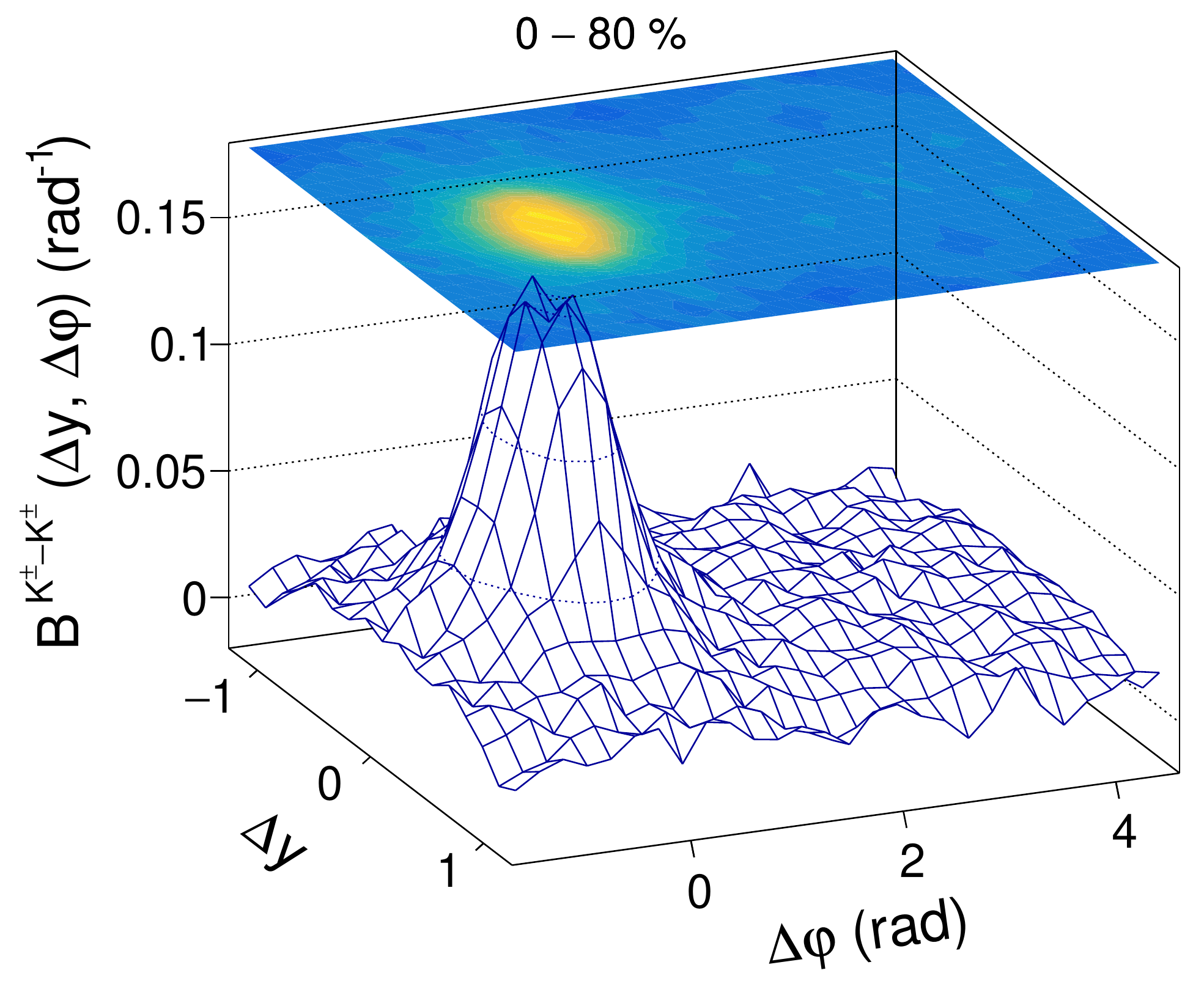}
  \includegraphics[width=0.32\linewidth]{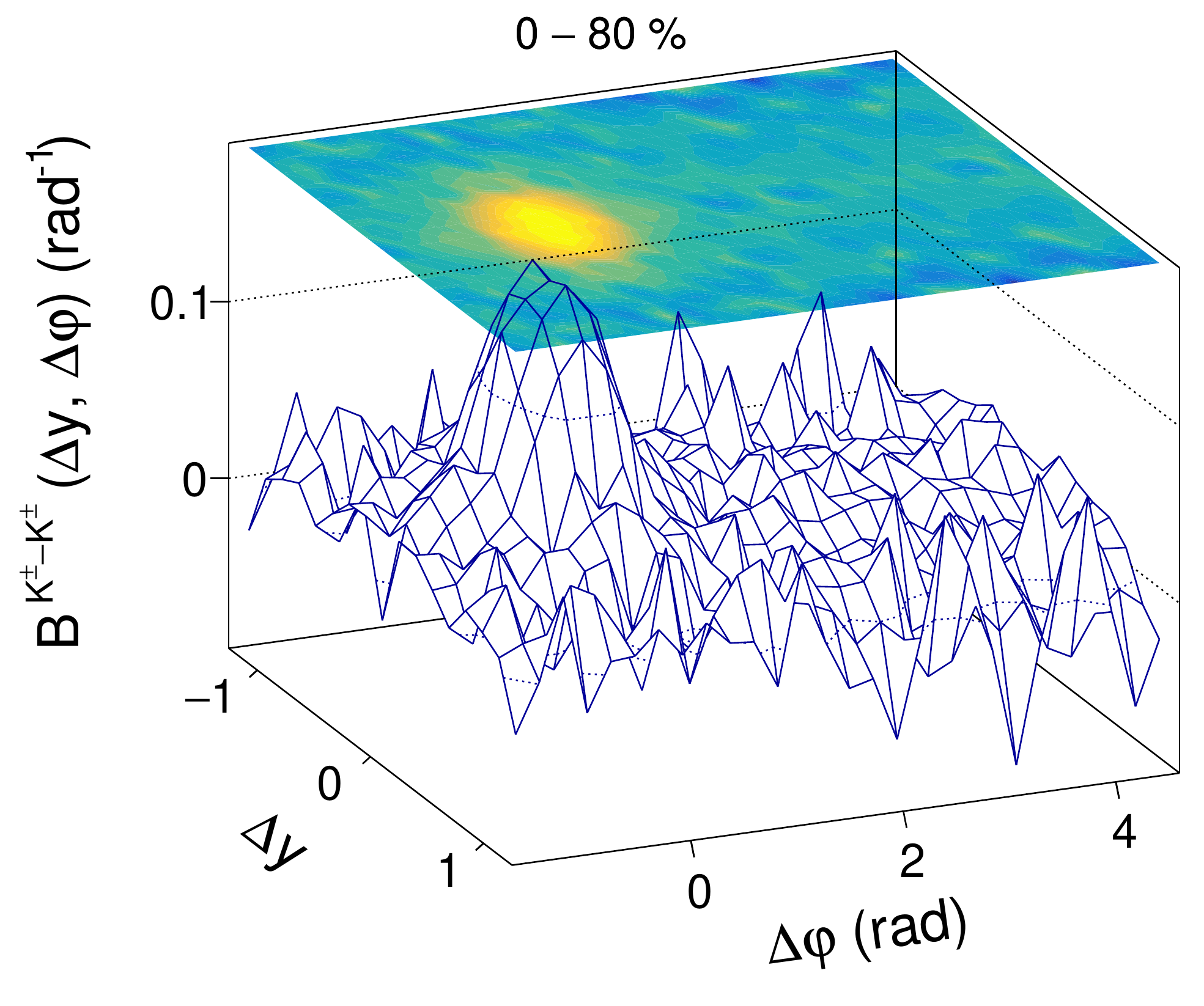}
  \includegraphics[width=0.32\linewidth]{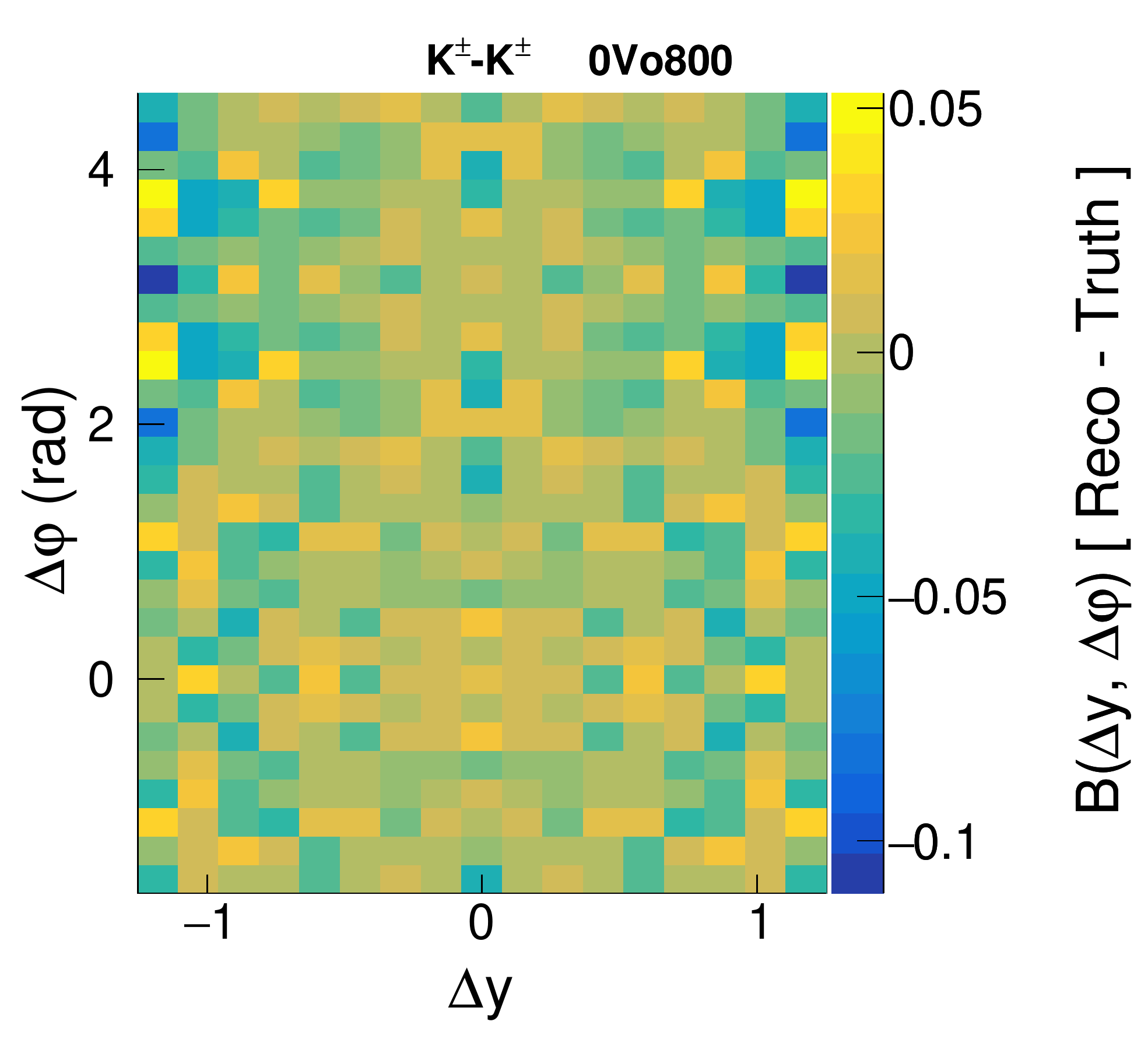}
  \includegraphics[width=0.32\linewidth]{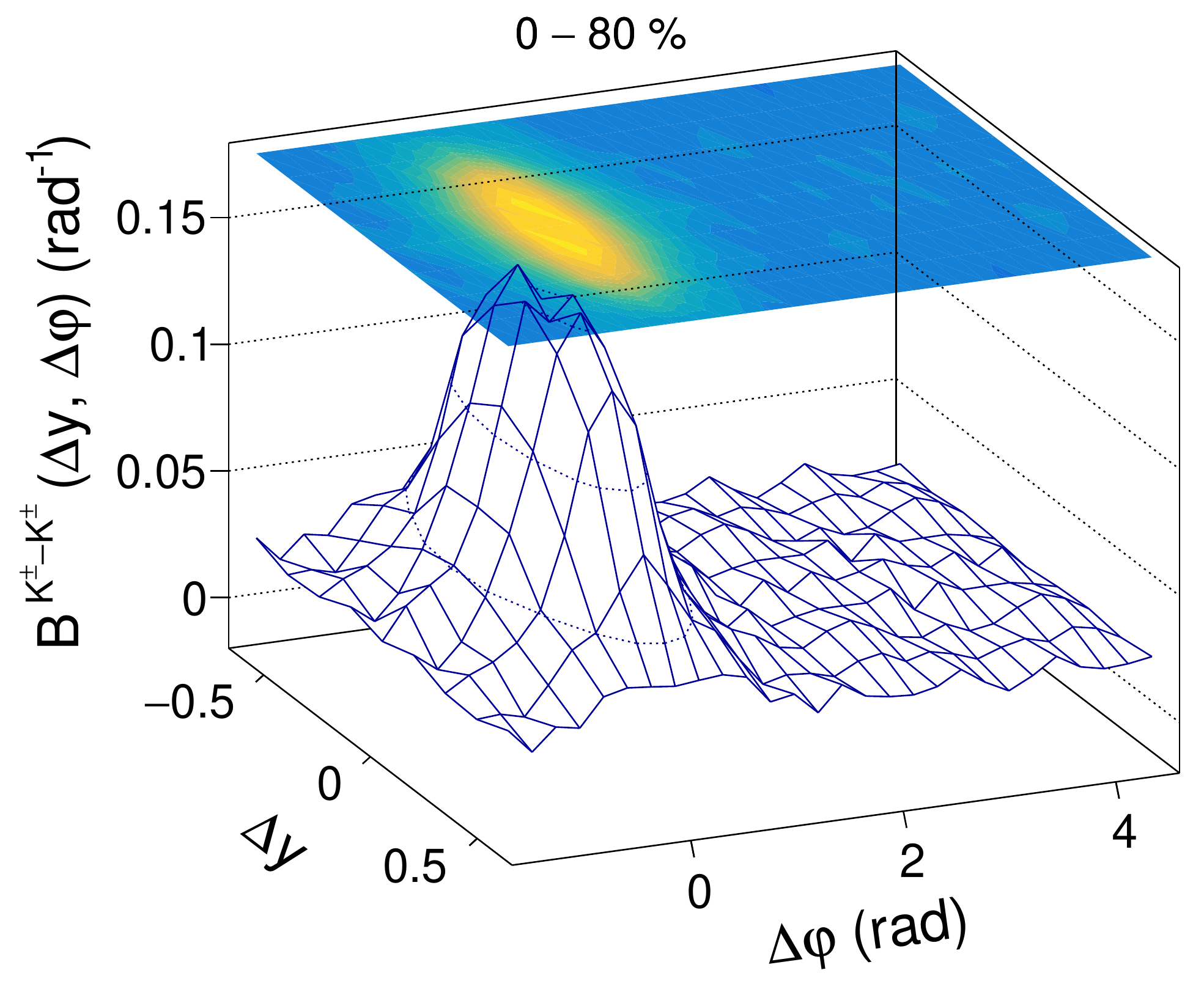}
  \includegraphics[width=0.32\linewidth]{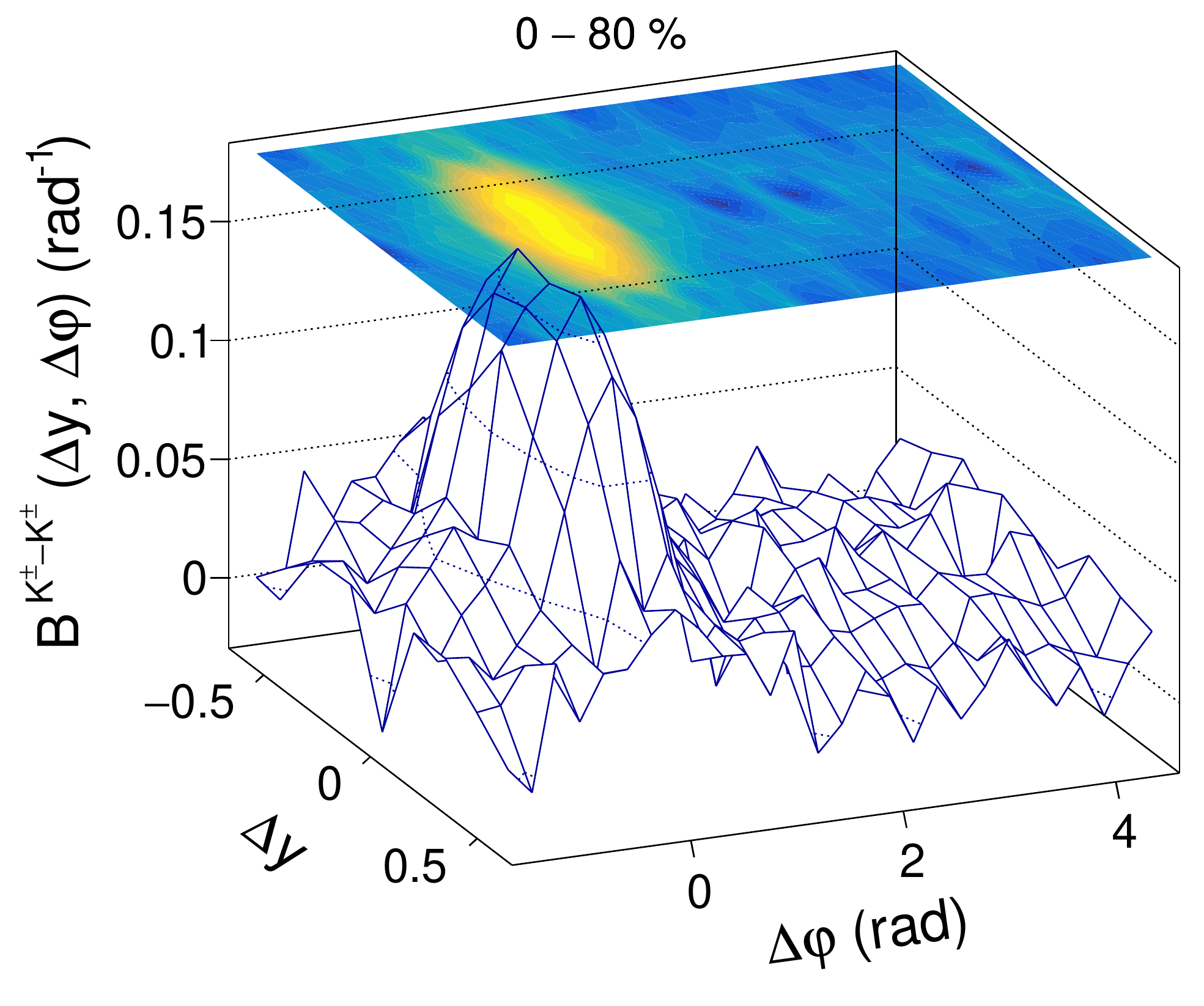}
  \includegraphics[width=0.32\linewidth]{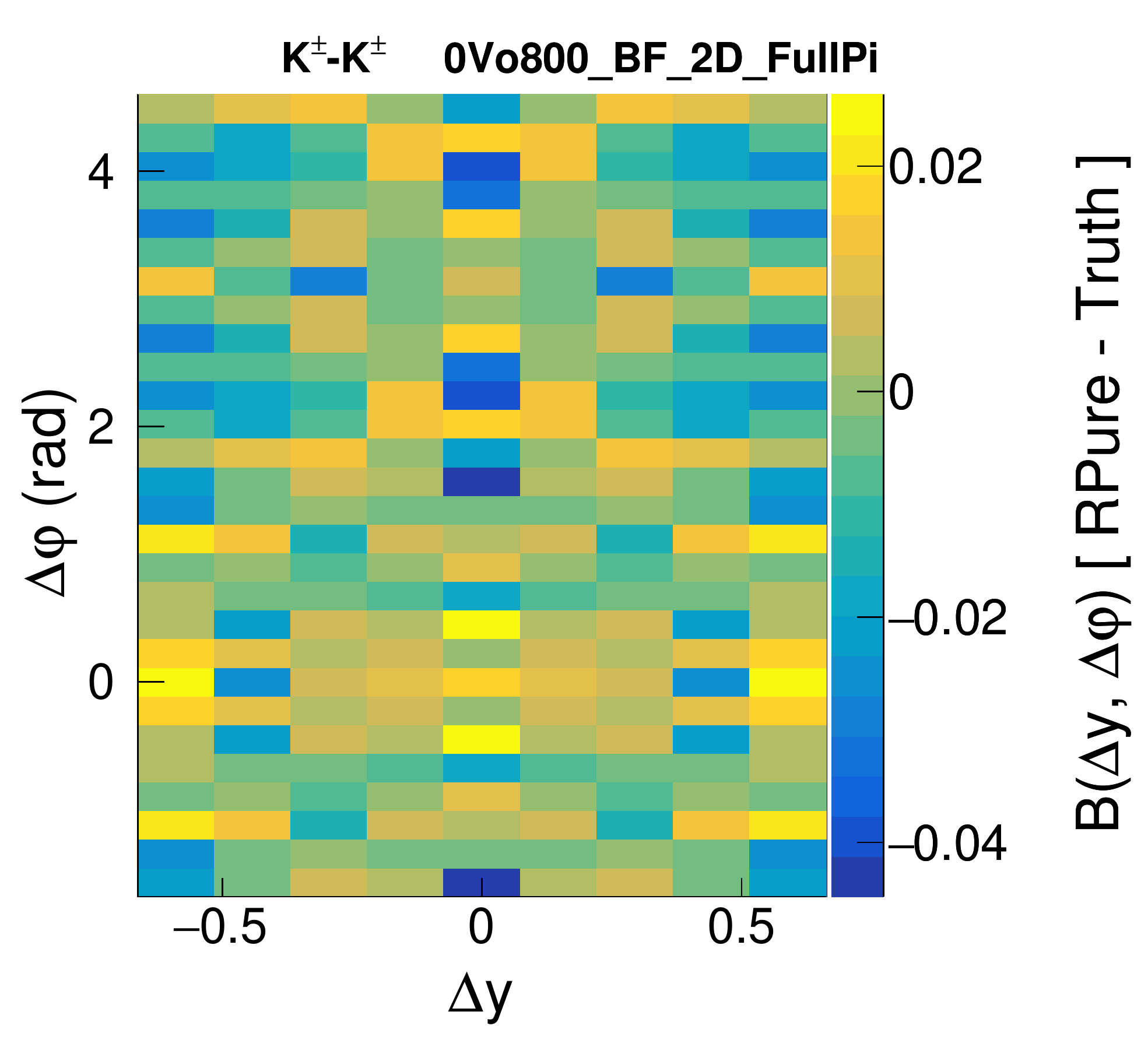}
  \caption{MC closure test of $KK$ pair for 0-80\% centrality. $1^{st}$ row for Truth (generator level) with $|\Delta y|\le1.4$: US (left) and LS (middle) CFs, and BF (right).
  $2^{nd}$ row for $R_{\rm pure}$ (reconstructed without mis-identified and secondaries from weak decays and detector material) with $|\Delta y|\le1.4$: US (left) and LS (middle) CFs, and BF (right).
  $3^{rd}$ row for BF with $|\Delta y|\le1.2$: Truth (left), $R_{\rm pure}$ (middle), and their difference (right).  
  $4^{th}$ row for BF with $|\Delta y|\le0.6$: Truth (left), $R_{\rm pure}$ (middle), and their difference (right).}
  \label{fig:HIJING_Truth_RPure_2D_KaonKaon}
\end{figure}

\begin{figure}
\centering
  \includegraphics[width=0.32\linewidth]{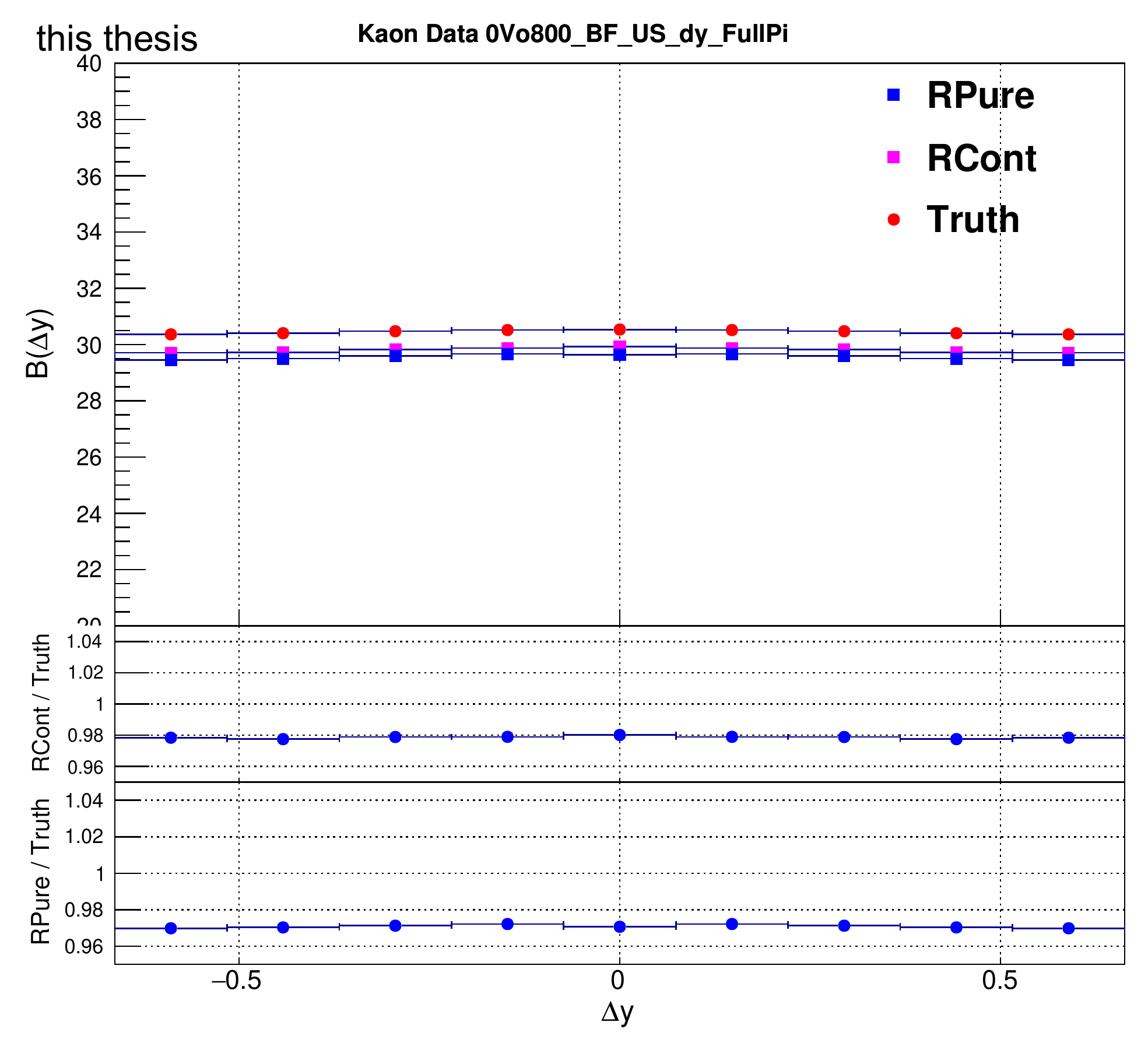}
  \includegraphics[width=0.32\linewidth]{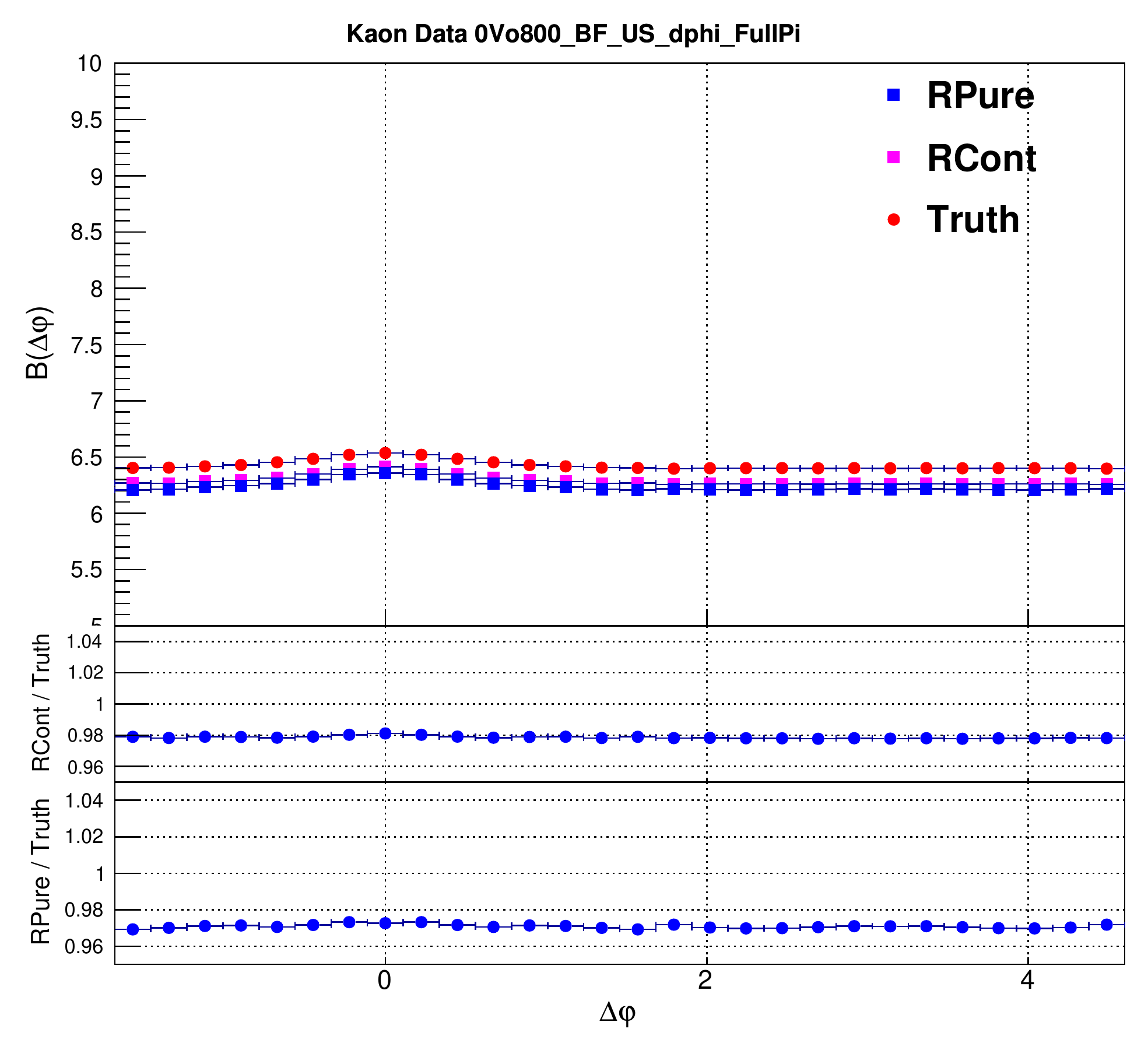}
  \includegraphics[width=0.32\linewidth]{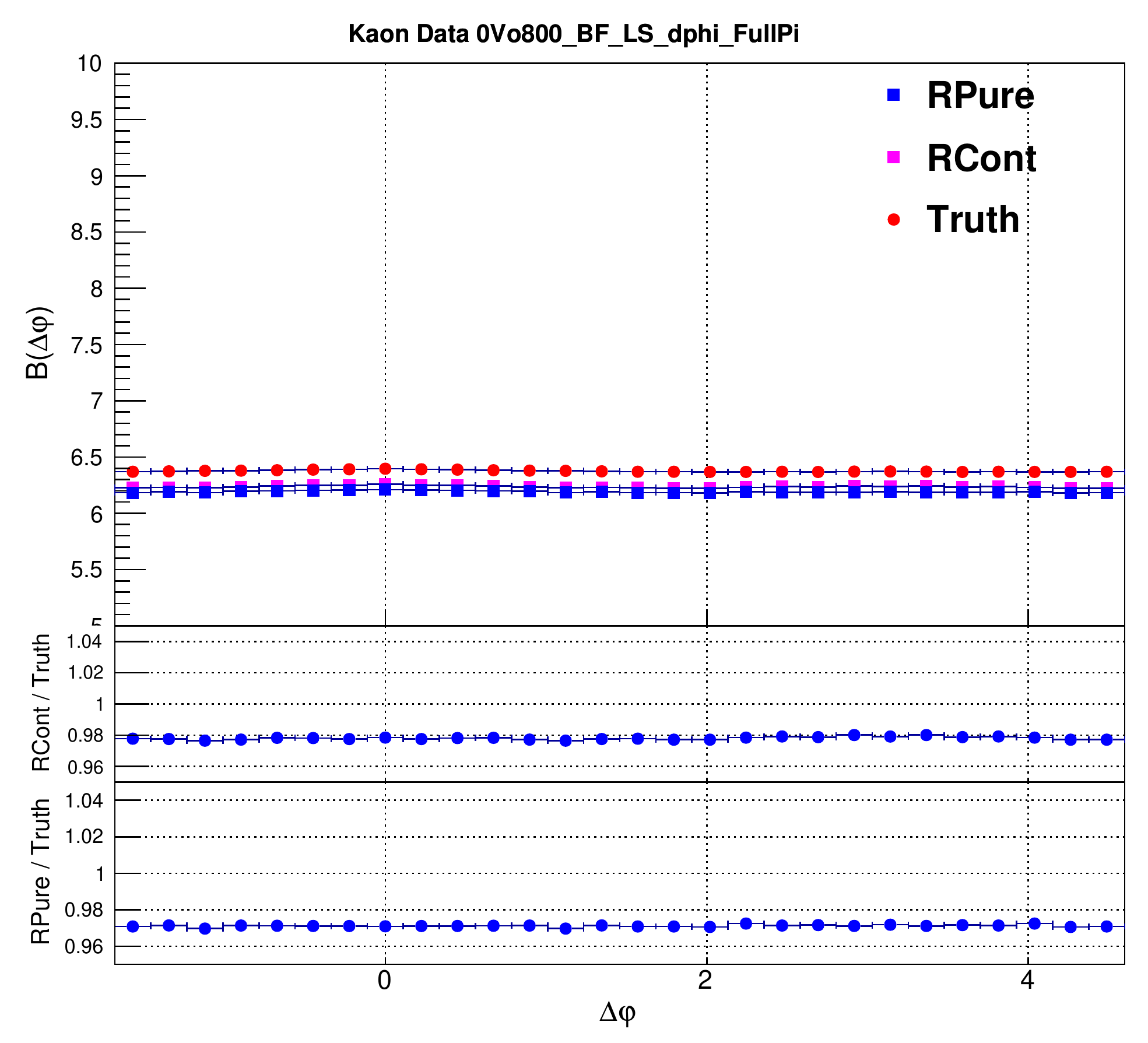}
  \includegraphics[width=0.32\linewidth]{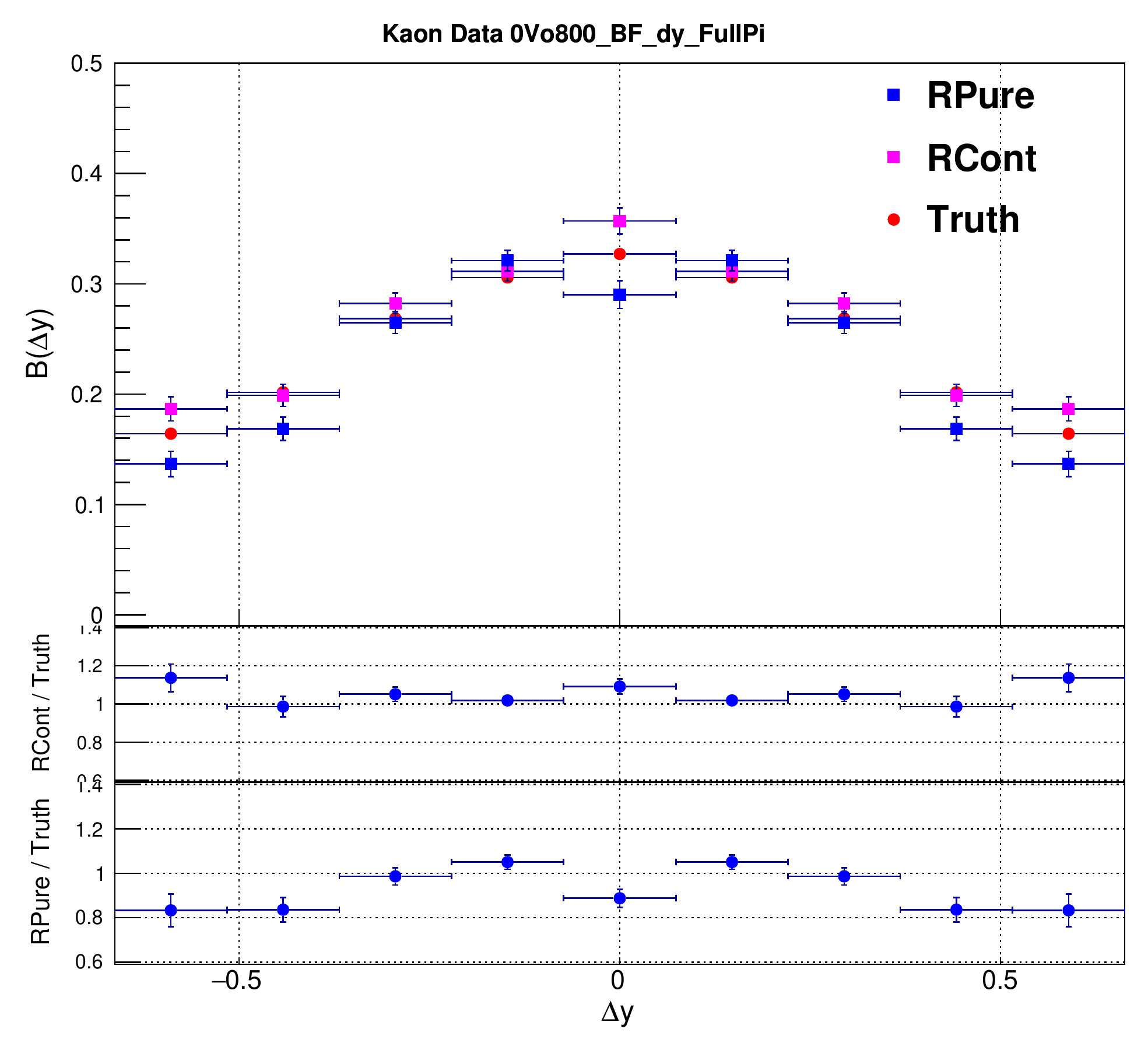}
  \includegraphics[width=0.32\linewidth]{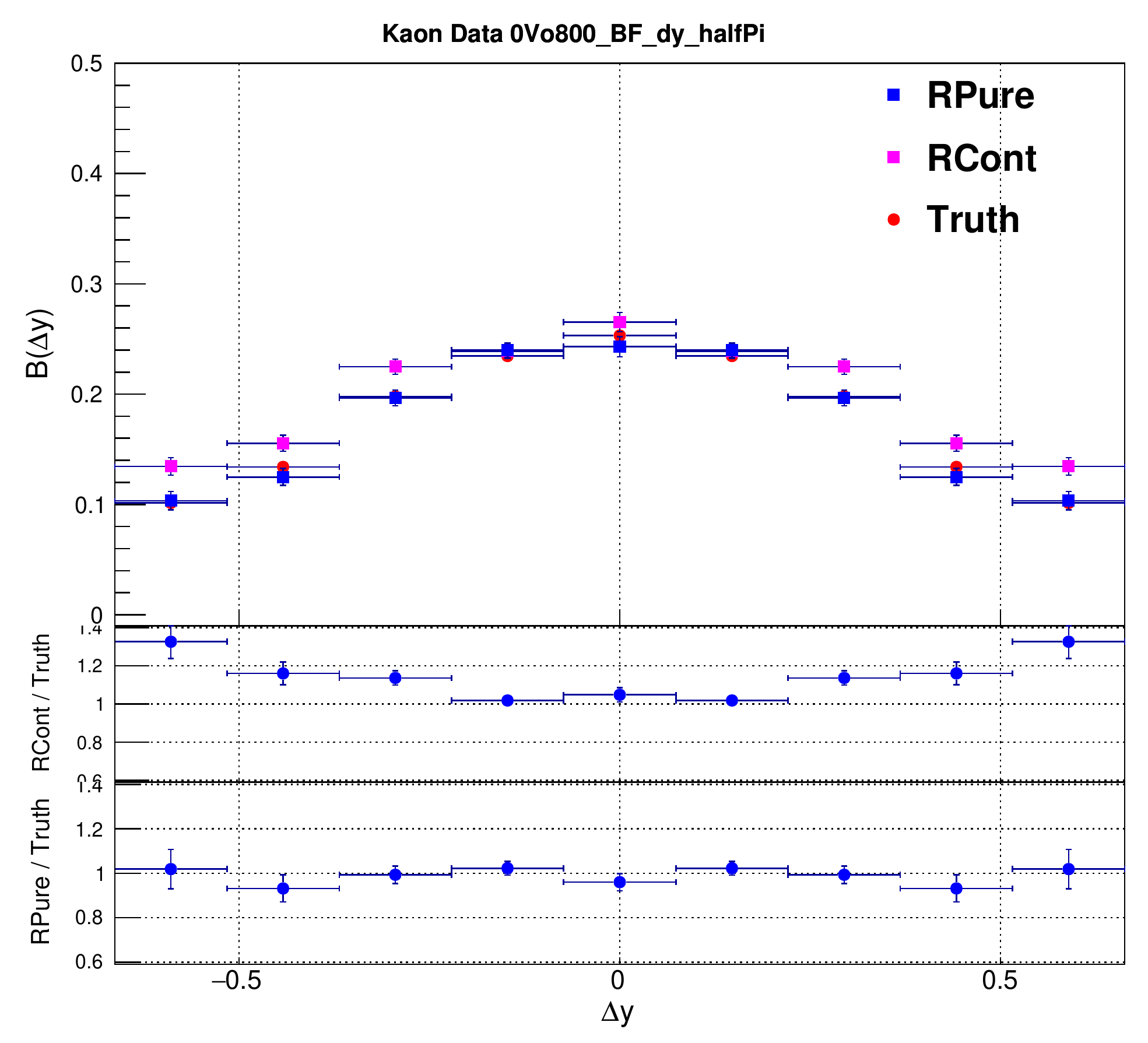}
  \includegraphics[width=0.32\linewidth]{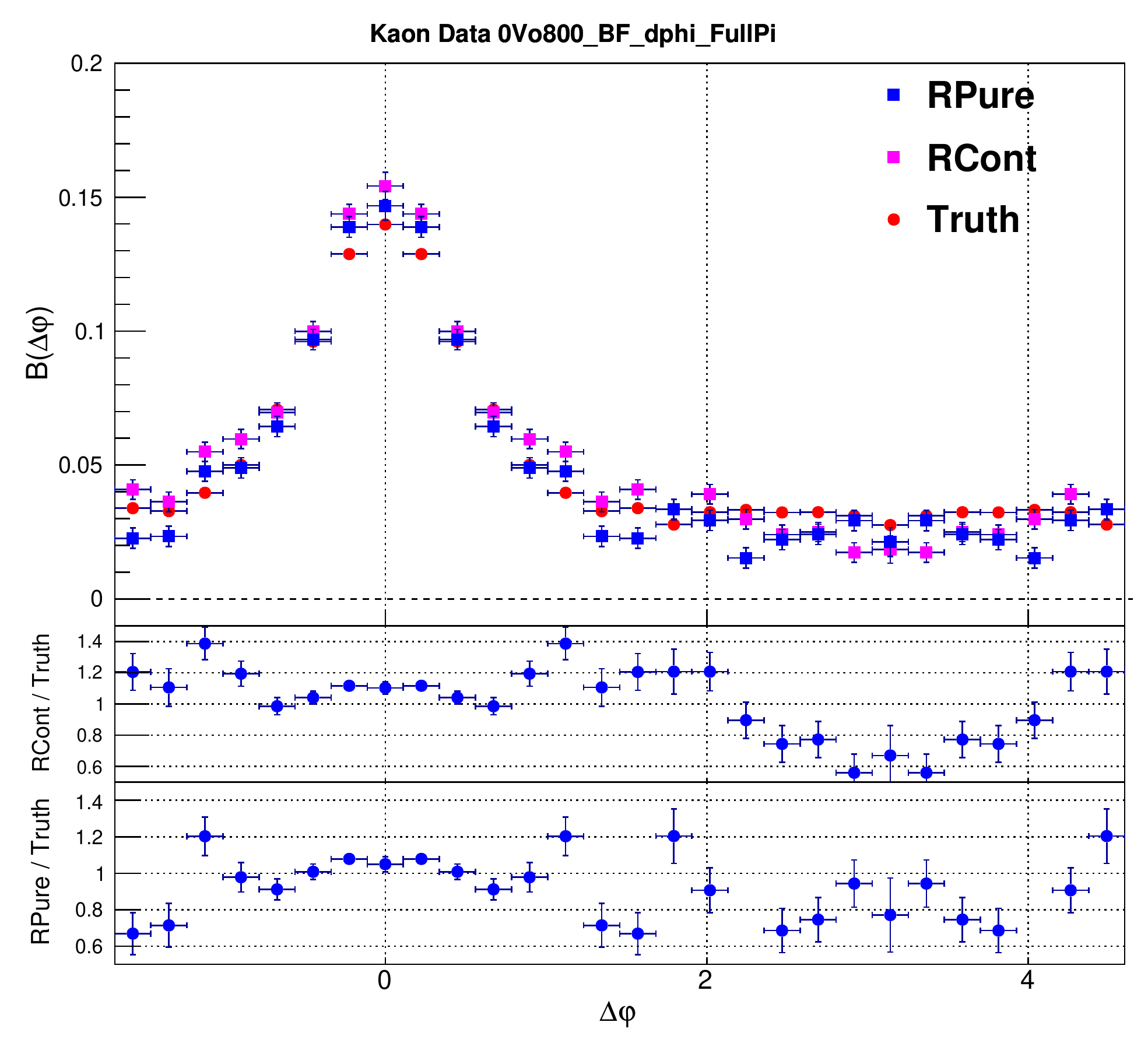}
  \includegraphics[width=0.32\linewidth]{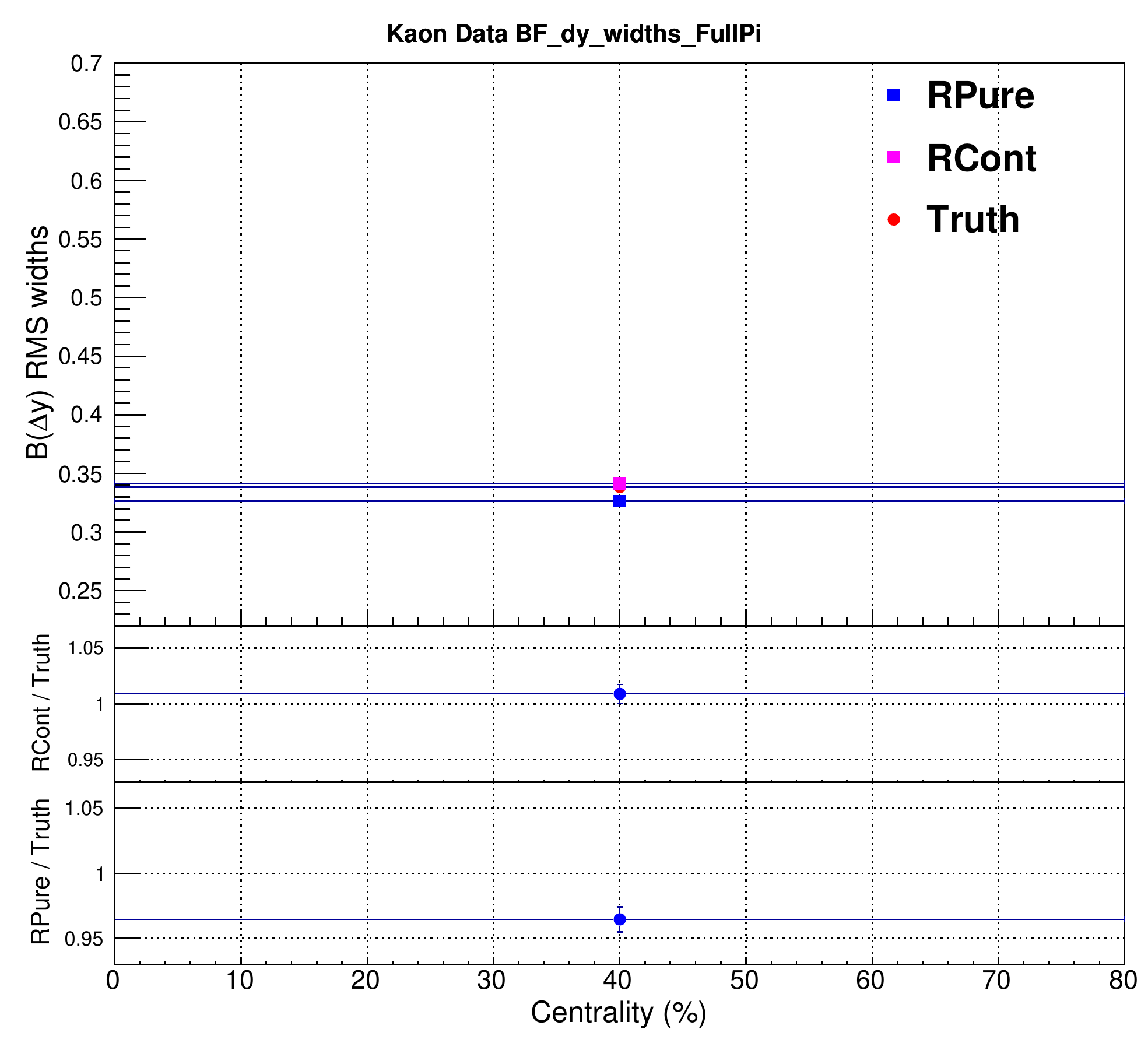}
  \includegraphics[width=0.32\linewidth]{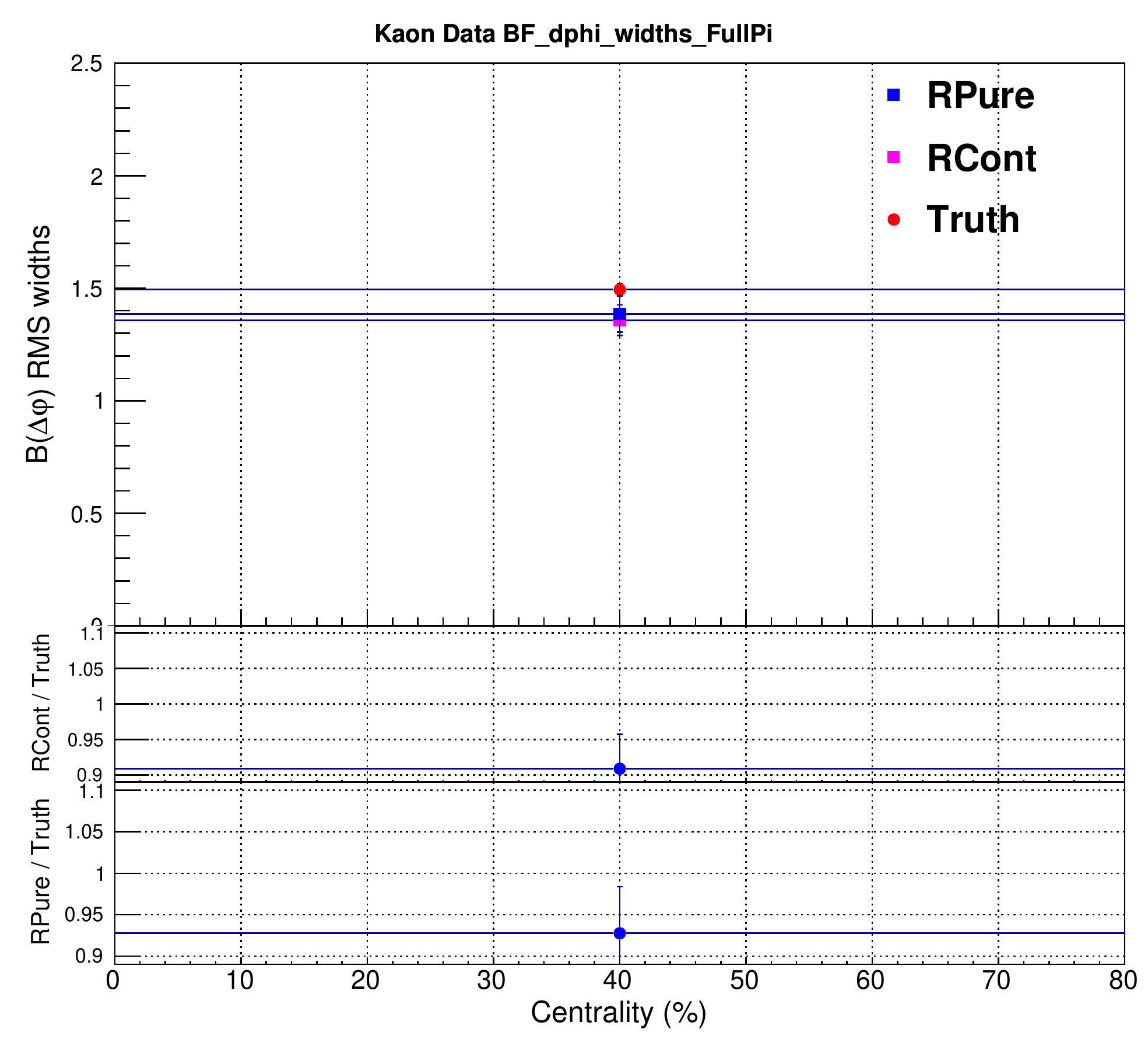}
  \includegraphics[width=0.32\linewidth]{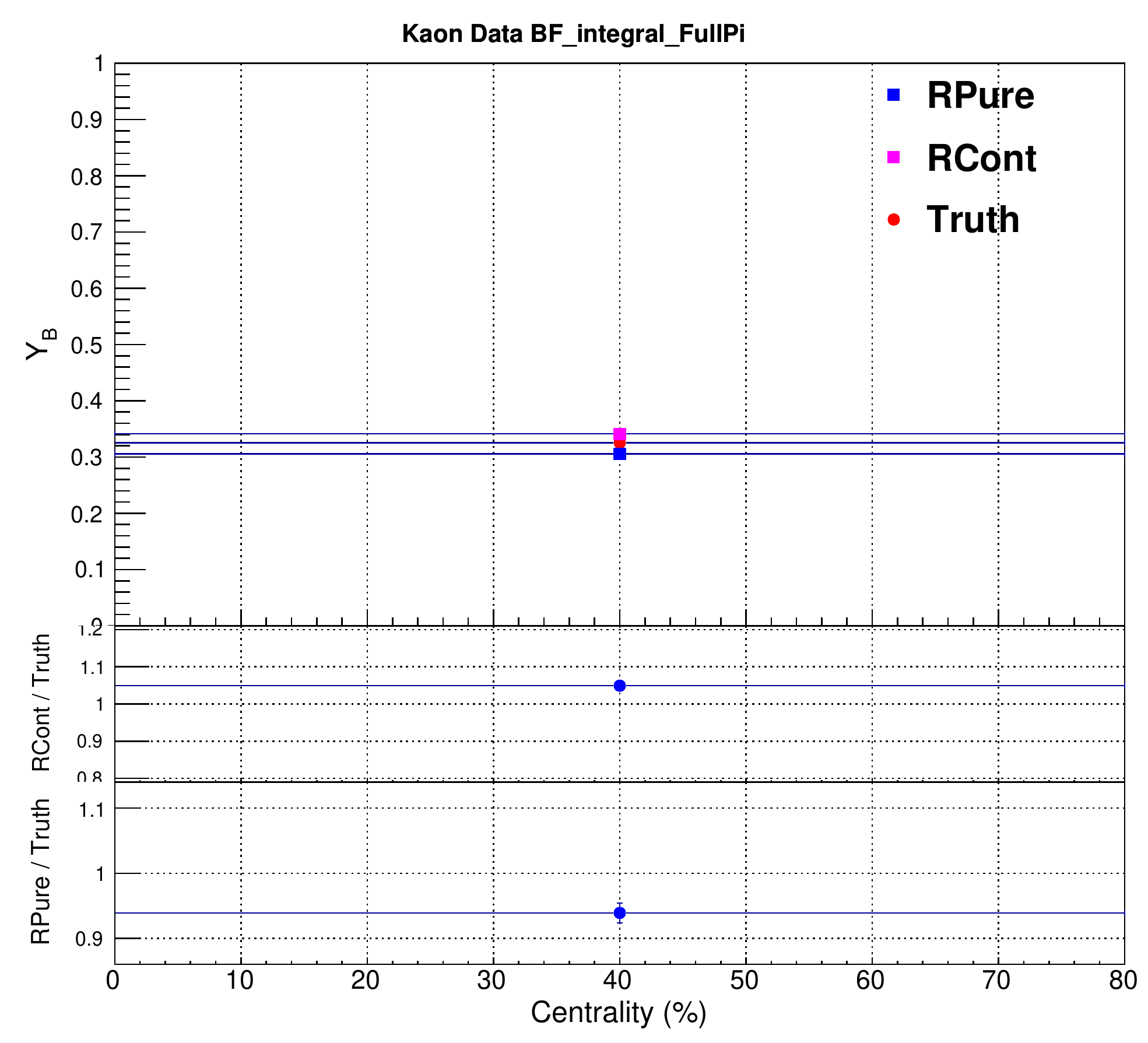}
  \includegraphics[width=0.32\linewidth]{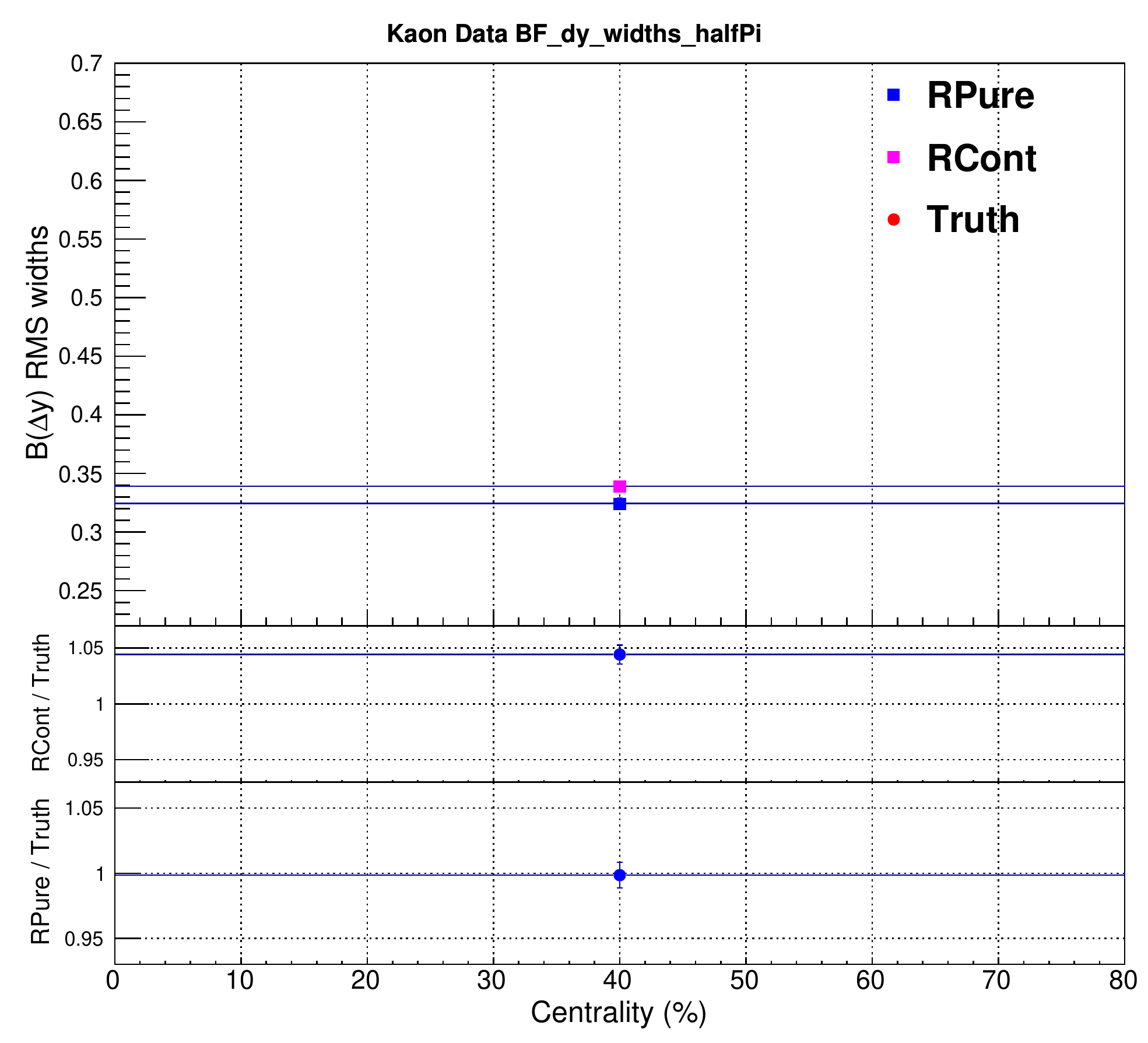}
  \includegraphics[width=0.32\linewidth]{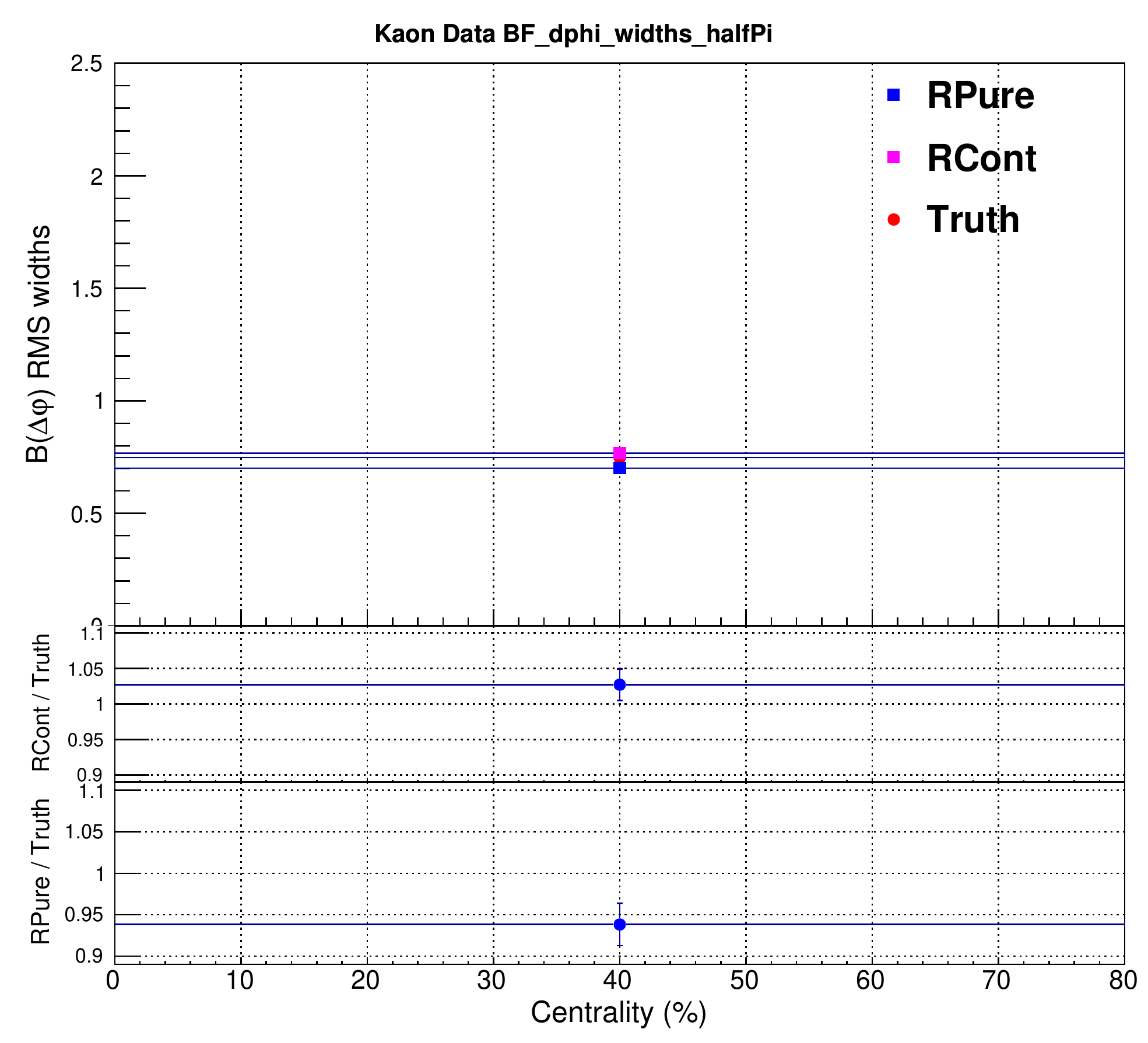}
  \includegraphics[width=0.32\linewidth]{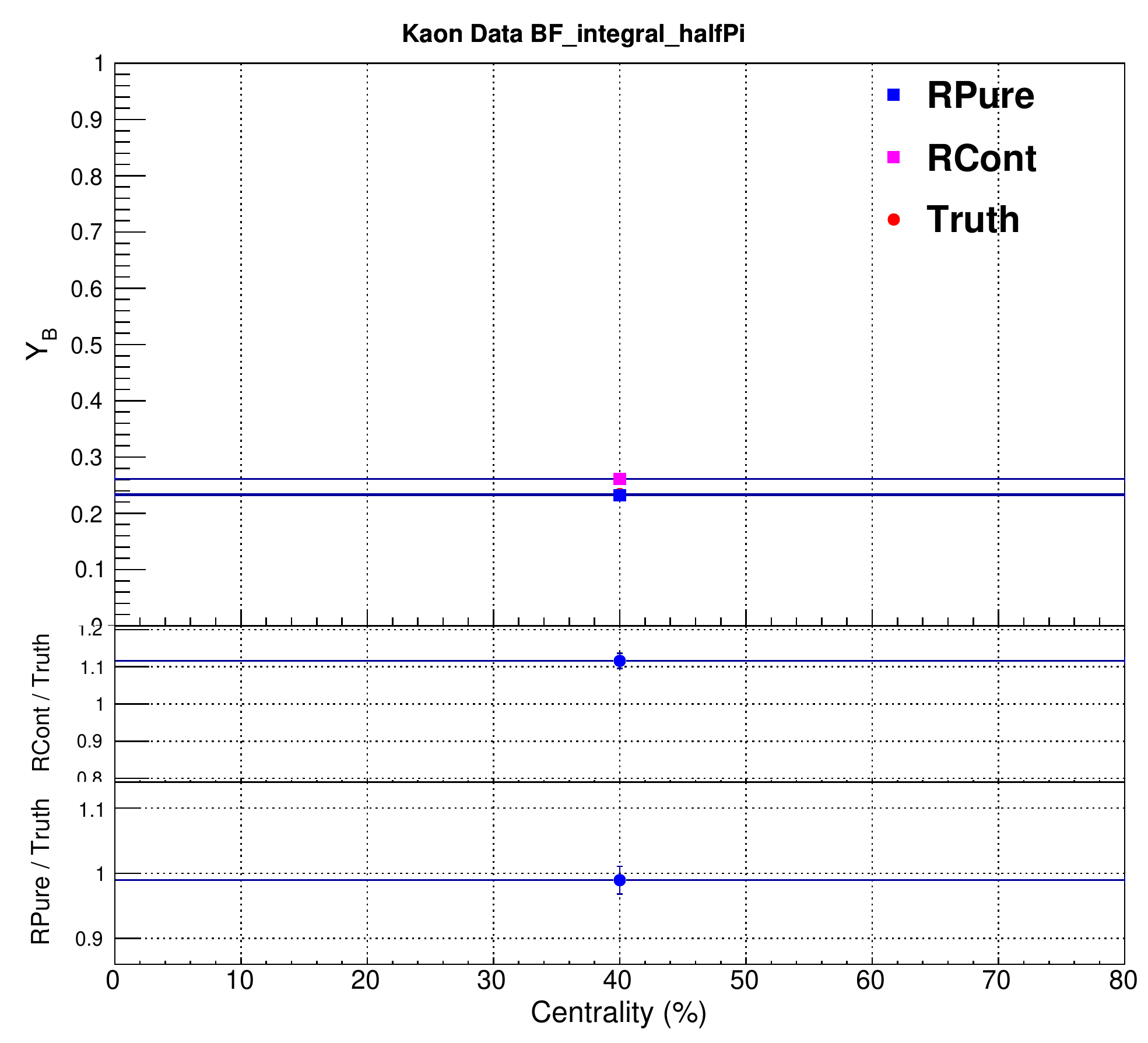}  
  \caption{MC closure test of $KK$ pair for 0-80\% centrality. Comparisons between Truth (generator level), $R_{\rm pure}$ (reconstructed without mis-identified and secondaries from weak decays and detector material), and $R_{\rm cont}$  (reconstructed with mis-identified and secondaries) with $|\Delta y|\le0.6$.
  $1^{st}$ row for CF projections: US $\Delta y$ projections for $|\Delta\varphi|\le\pi$ (left), and US (middle) and LS (right) $\Delta\varphi$ projections.
  $2^{nd}$ row for BF projections: $\Delta y$ projections for $|\Delta\varphi|\le\pi$ (left), $\Delta y$ projections for $|\Delta\varphi|\le\pi/2$ (middle), and $\Delta\varphi$ projections (right).    
  $3^{rd}$ row for BF with $|\Delta\varphi|\le\pi$: $\Delta y$ widths (left), $\Delta\varphi$ widths (middle), and integrals (right).
  $4^{th}$ row for BF with $|\Delta\varphi|\le\pi/2$: $\Delta y$ widths (left), $\Delta\varphi$ widths (middle), and integrals (right).}
  \label{fig:HIJING_Truth_RPure_RCont_1D_KaonKaon}
\end{figure}

The MC closure test of $B^{KK}$ suffers from low statistics.
Even MC Truth (generator level) CF of LS for 0-80\% centrality do not provide a clean signal, especially for large $\Delta y$, as shown in Figure~\ref{fig:HIJING_Truth_RPure_2D_KaonKaon}.
And it is worth to mention that Gonzalez et al. has shown that large centrality bin width does not bias the $R_{2}^{CD}$ correlator~\cite{PhysRevC.99.034907}.
Thus, reasonable MC closure test could only be accomplished for $B^{KK}$ within a large centrality bin (0-80\%) and narrow relative rapidity range ($|\Delta y|\le0.6$), as shown in Figures~\ref{fig:HIJING_Truth_RPure_2D_KaonKaon}, ~\ref{fig:HIJING_Truth_RPure_RCont_1D_KaonKaon}.
A few \% of differences on $B^{KK}$ projections, widths and integrals between MC Truth (generator level), $R_{\rm pure}$ (reconstructed without mis-identified and secondaries from weak decays and detector material), and $R_{\rm cont}$  (reconstructed with mis-identified and secondaries) are consistent with expected systematics and low statistics.

The $B^{\pi\pi}$ and $B^{KK}$ results show that the MC closure tests are successful.
We conclude that the data analysis techniques used in this analysis are sufficiently robust for measuring BFs of charged hadron pairs $(\pi,K,p)\otimes (\pi,K,p)$.

\clearpage

\subsection{Additional $p_{\rm T}$-dependent Efficiency Correction}
\label{subsec:PtDependentEfficiencyCorrection}

We carried out an additional   $p_{\rm T}$-dependent efficiency correction study  to insure that the efficiency correction method applied in this work, and described in Sec.~\ref{sec:Efficiency_Correction}, produces robust CF results.

\begin{figure}
\centering
  \includegraphics[width=0.32\linewidth]{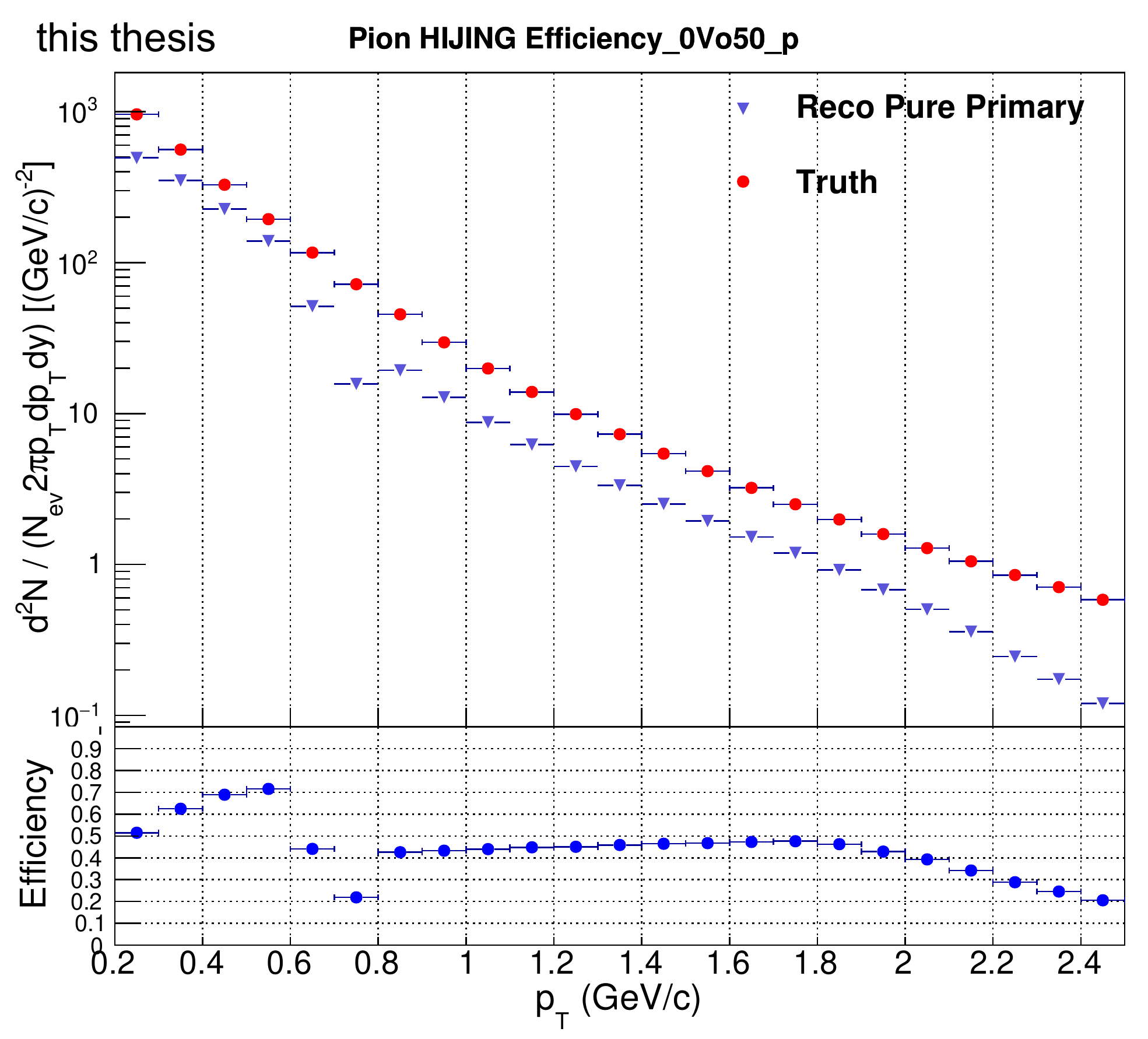}
  \includegraphics[width=0.32\linewidth]{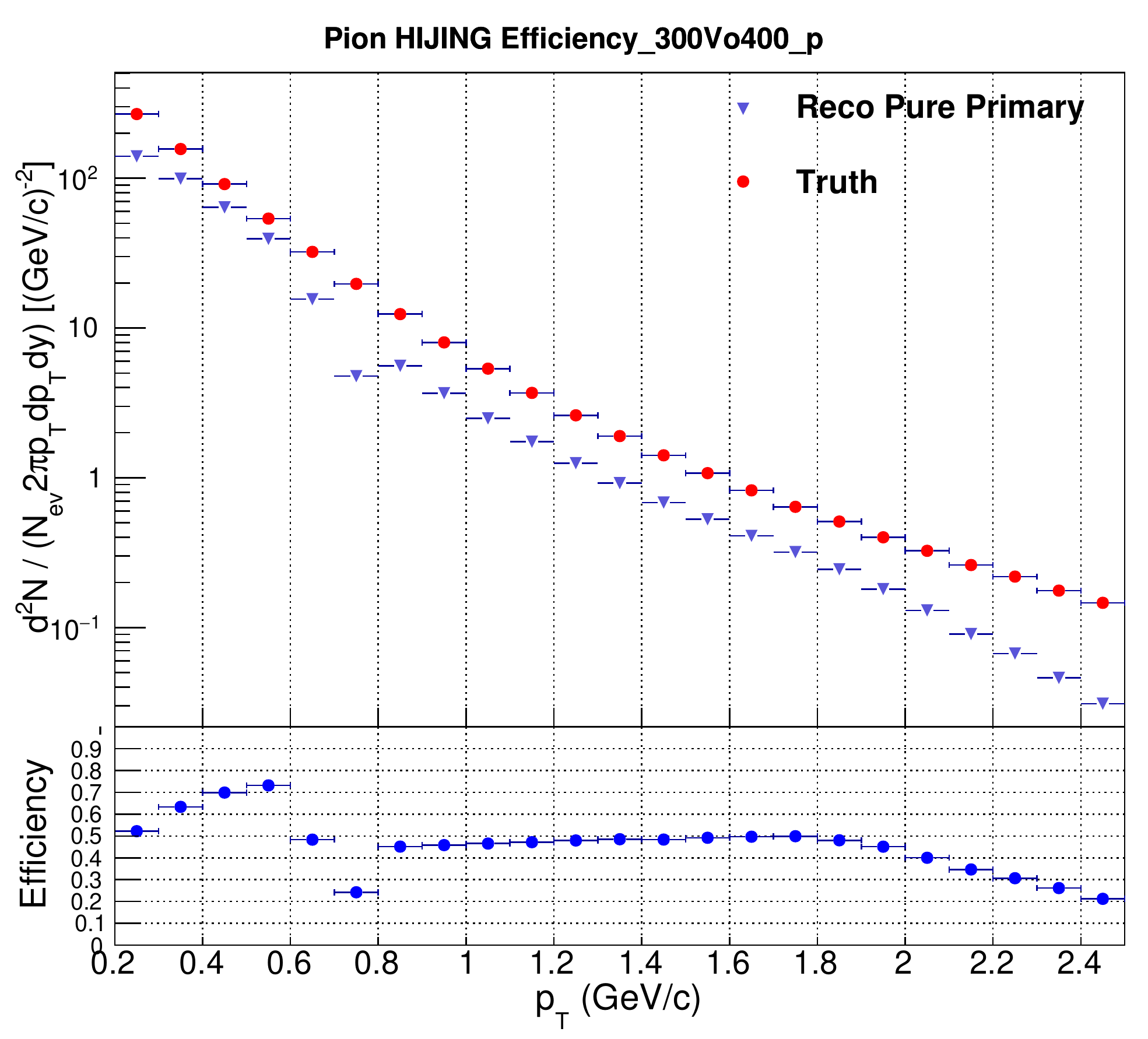}
  \includegraphics[width=0.32\linewidth]{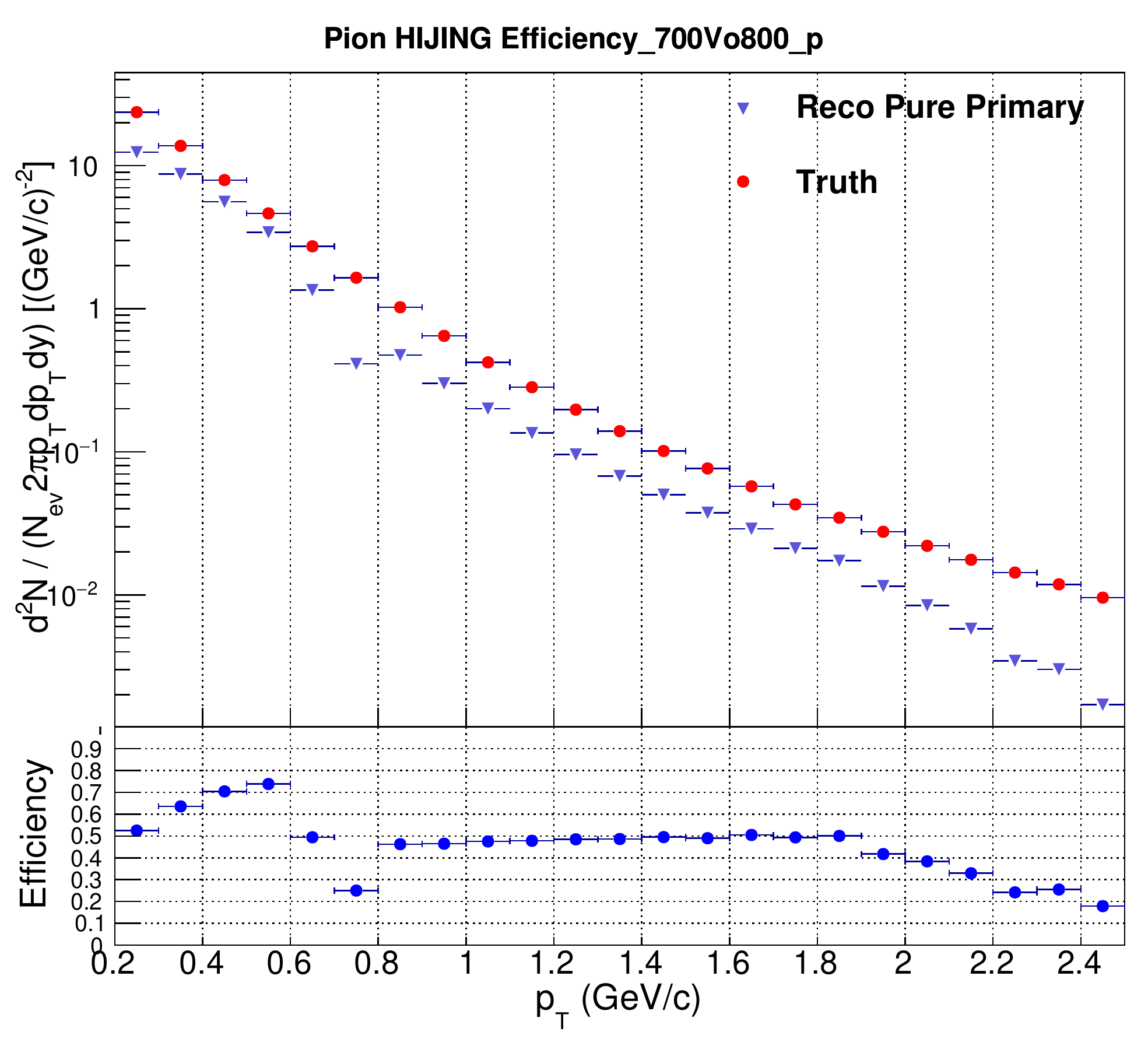}
  \includegraphics[width=0.32\linewidth]{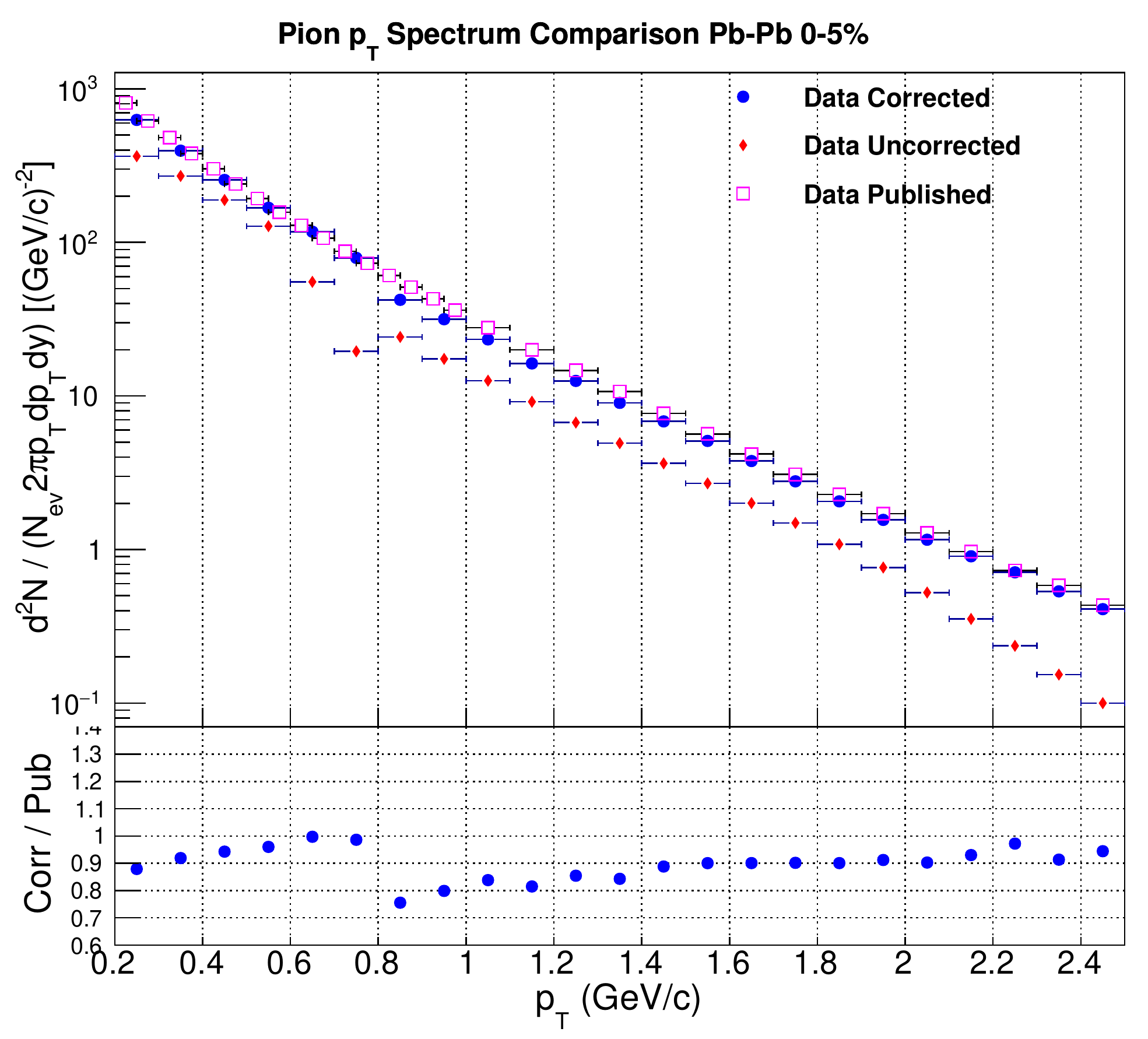}
  \includegraphics[width=0.32\linewidth]{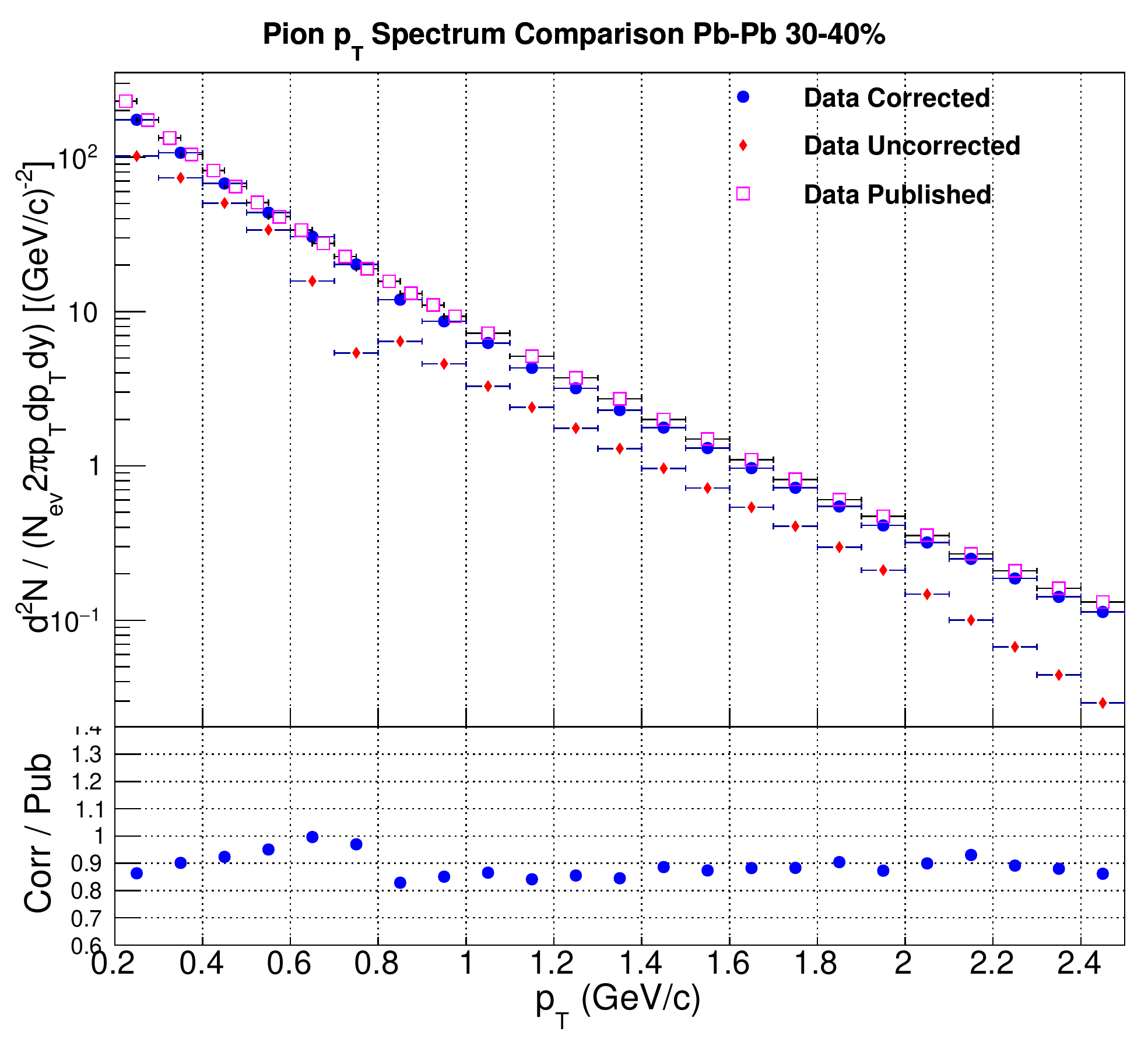}
  \includegraphics[width=0.32\linewidth]{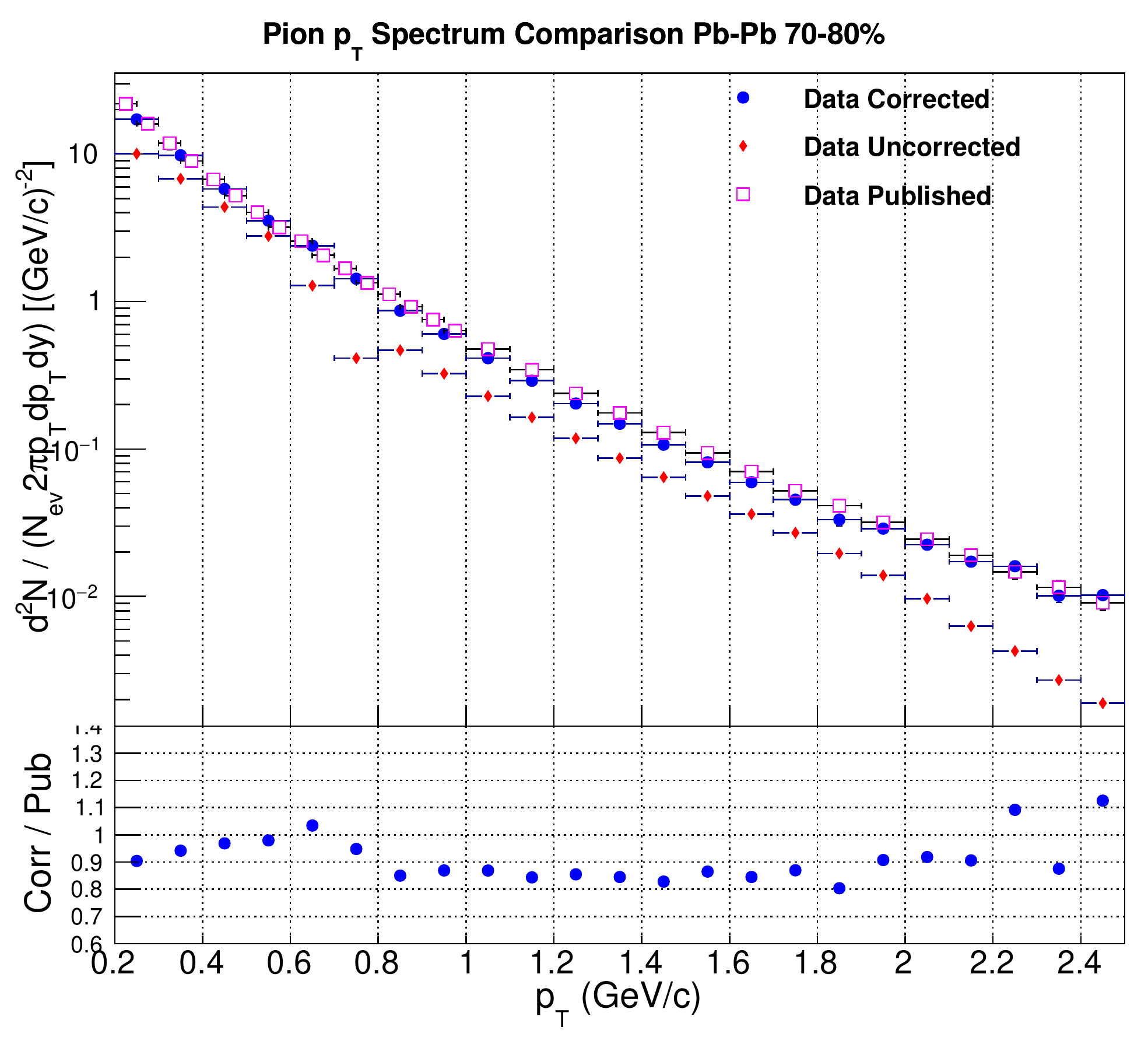}
  \caption{For $\pi^{\pm}$.
 Upper row for MC: efficiency estimation by taking the ratio of reconstructed divided by generator $p_{\rm T}$ spectra.
 Lower row for real data: comparison of $p_{\rm T}$ spectra obtained with and without MC efficiency correction, and the ALICE published, in central (left), mid-central (middle) and peripheral (right) collisions.}
   \label{fig:HIJING_Corrected_pt_pion_vs_Published}
\end{figure}

\begin{figure}
\centering
  \includegraphics[width=0.32\linewidth]{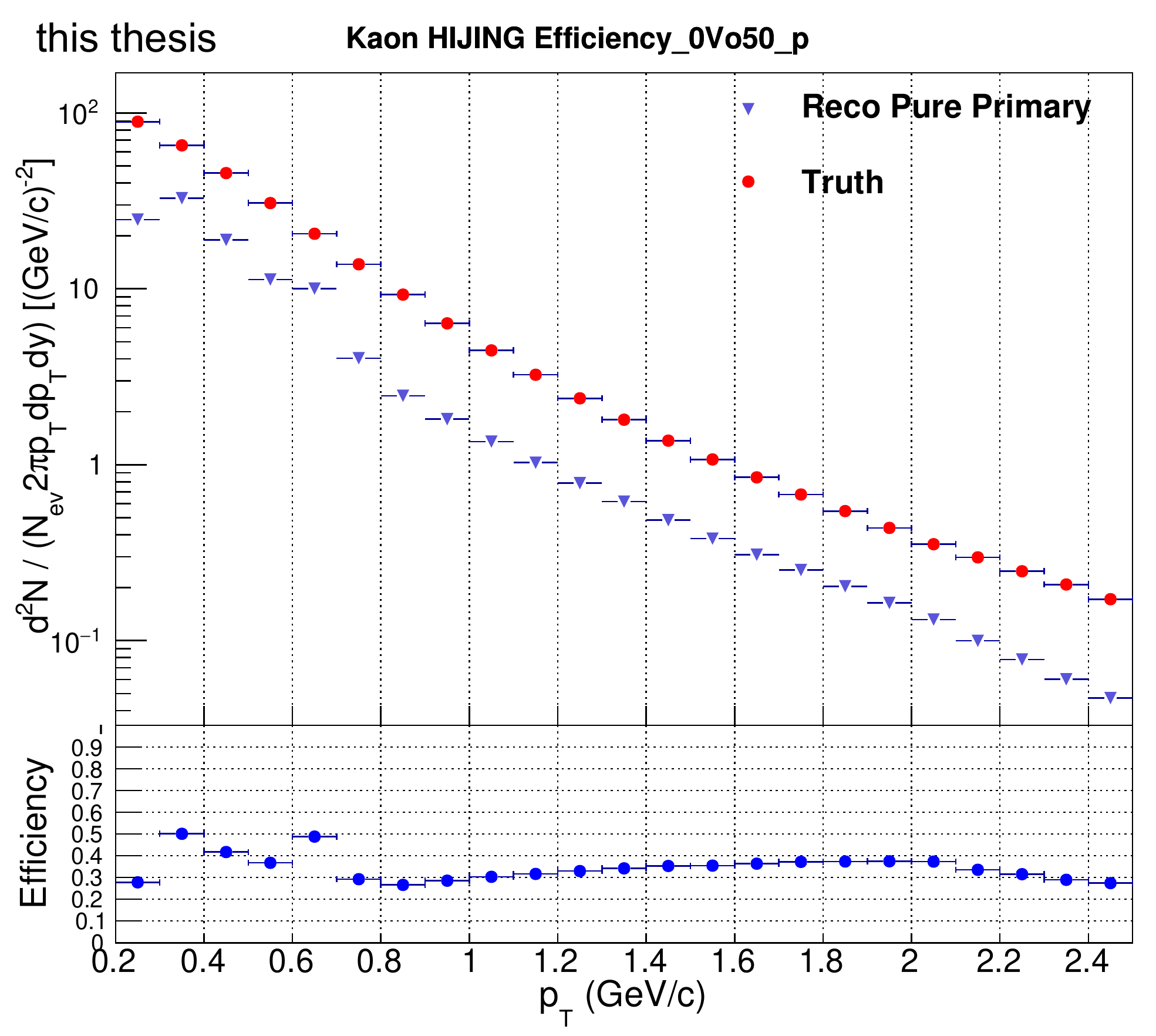}
  \includegraphics[width=0.32\linewidth]{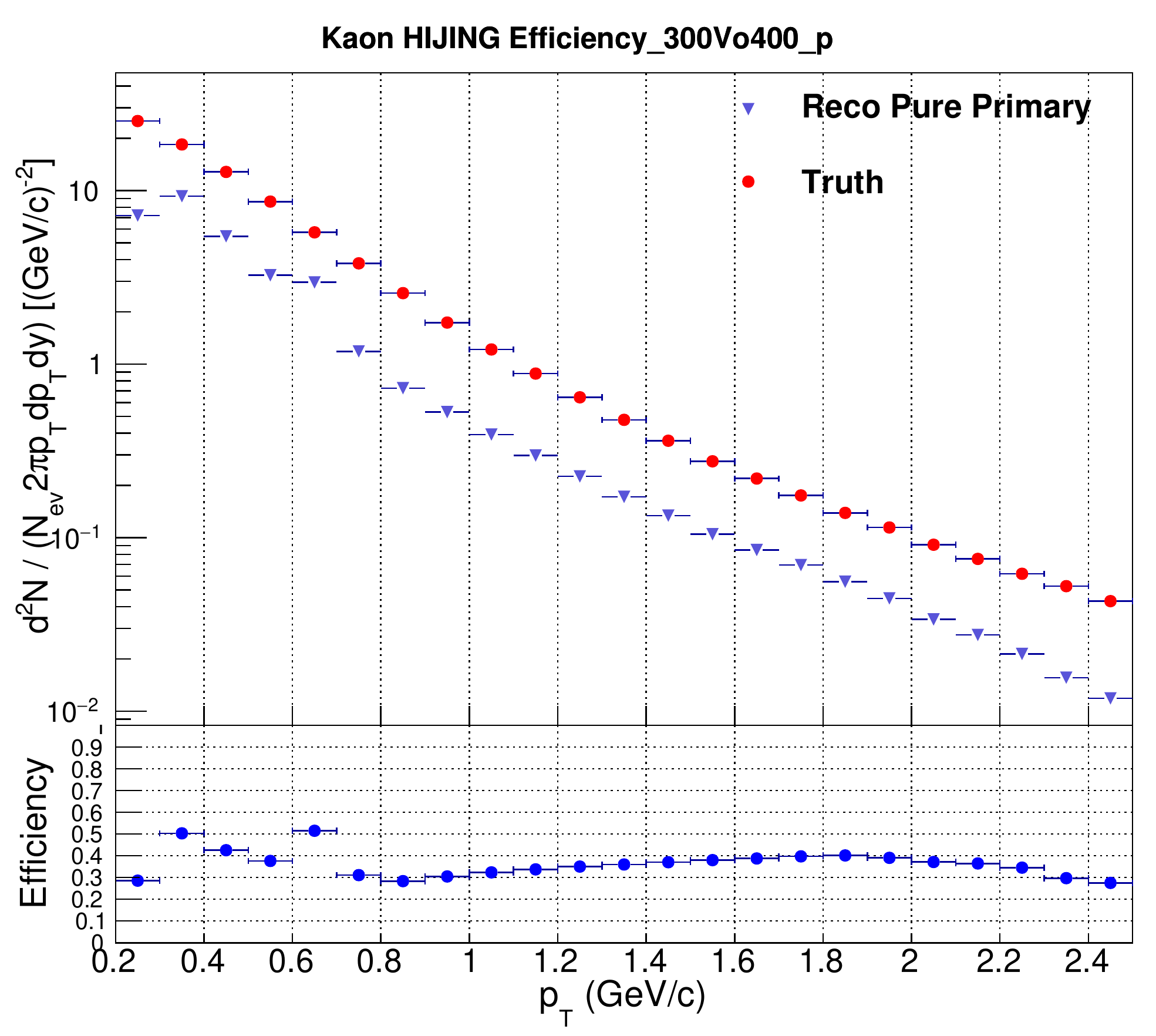}
  \includegraphics[width=0.32\linewidth]{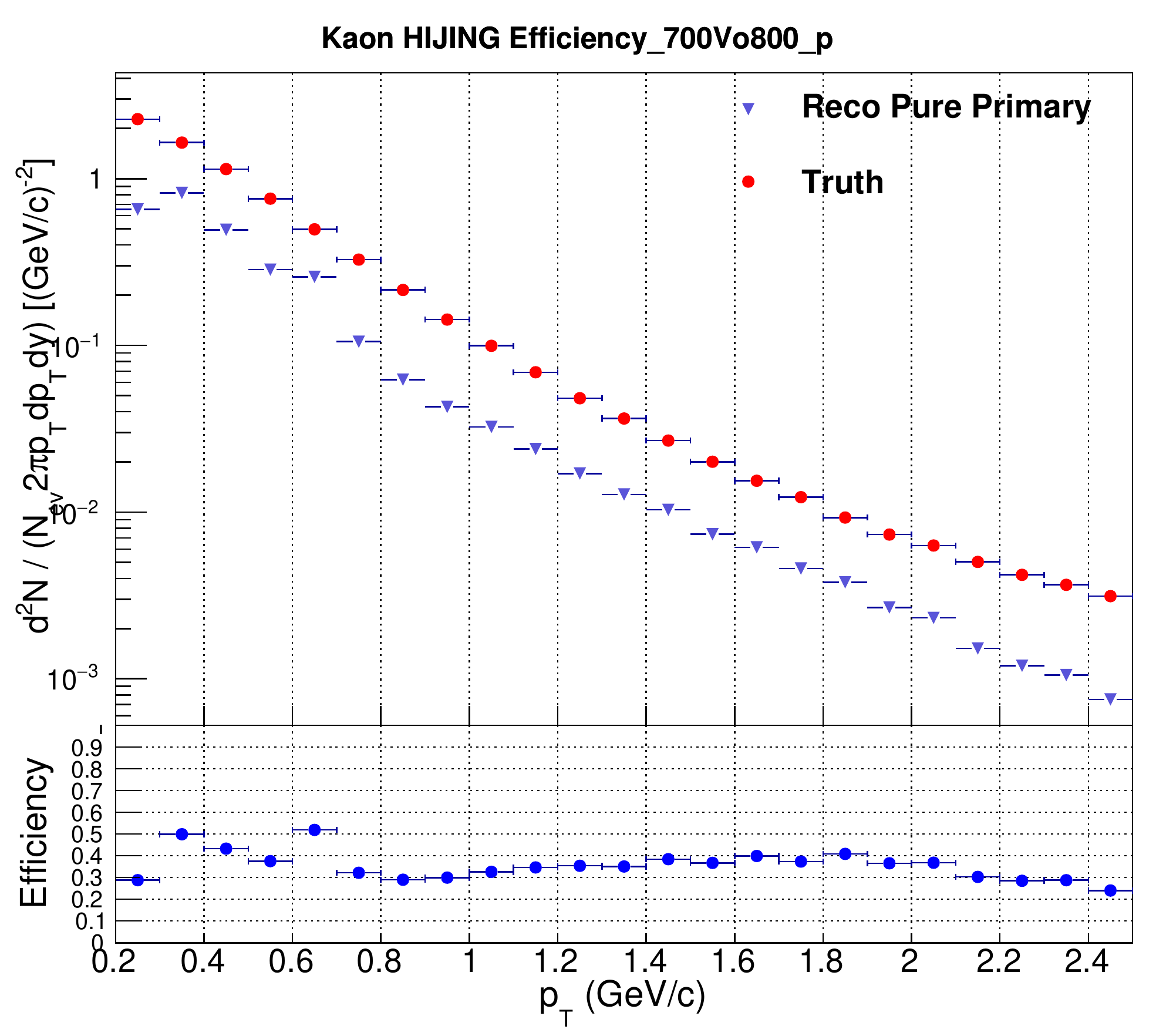}
  \includegraphics[width=0.32\linewidth]{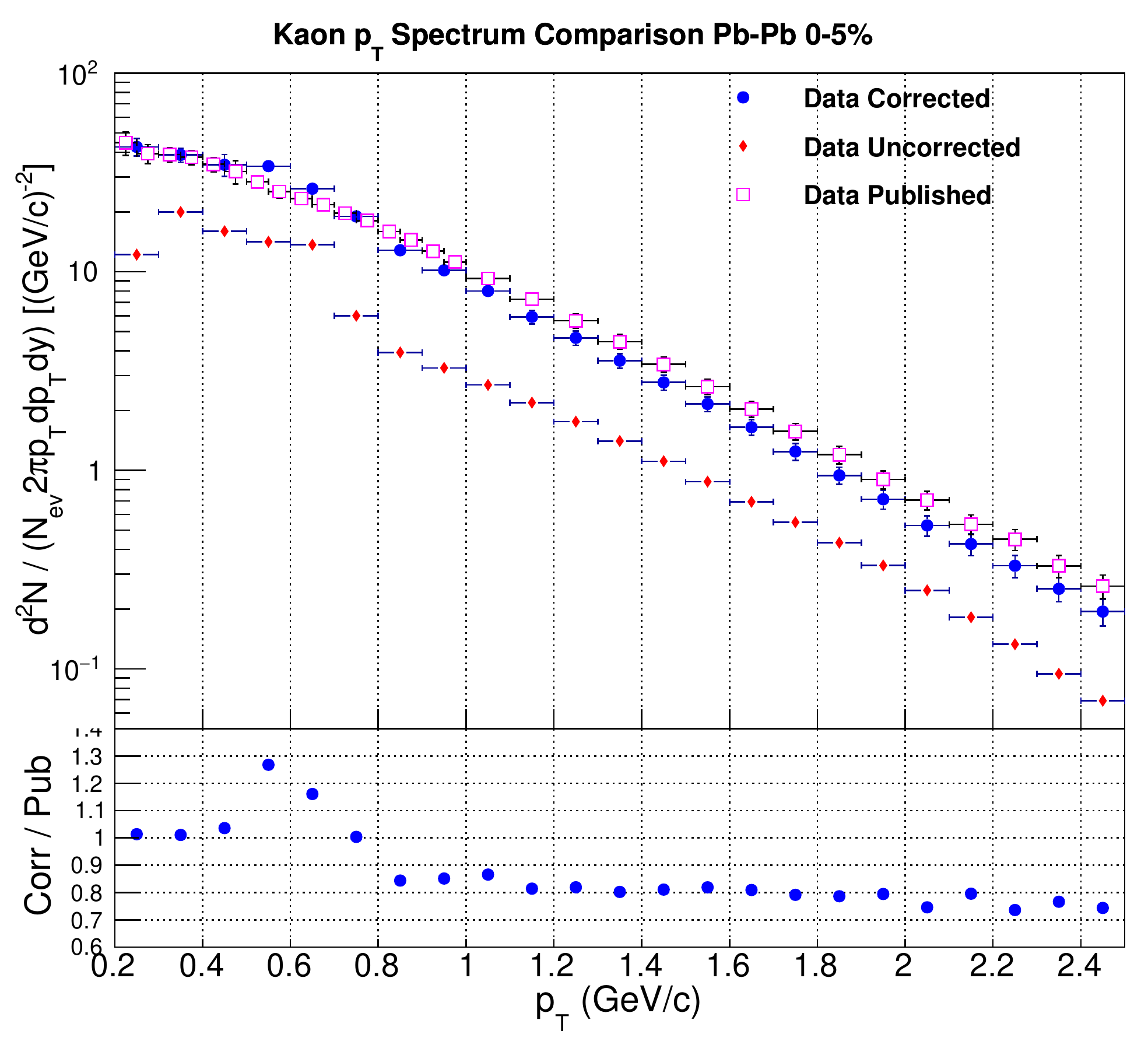}
  \includegraphics[width=0.32\linewidth]{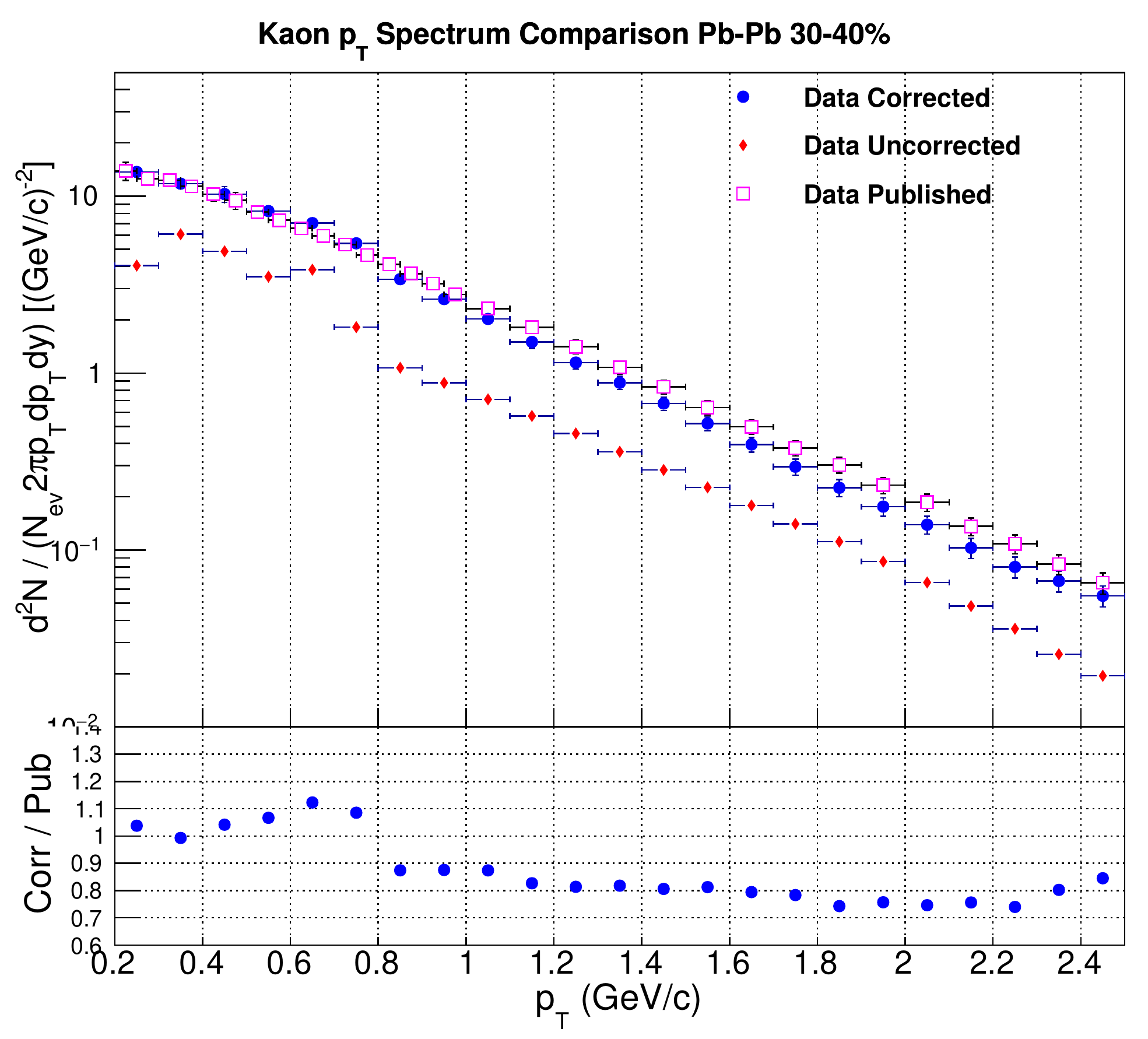}
  \includegraphics[width=0.32\linewidth]{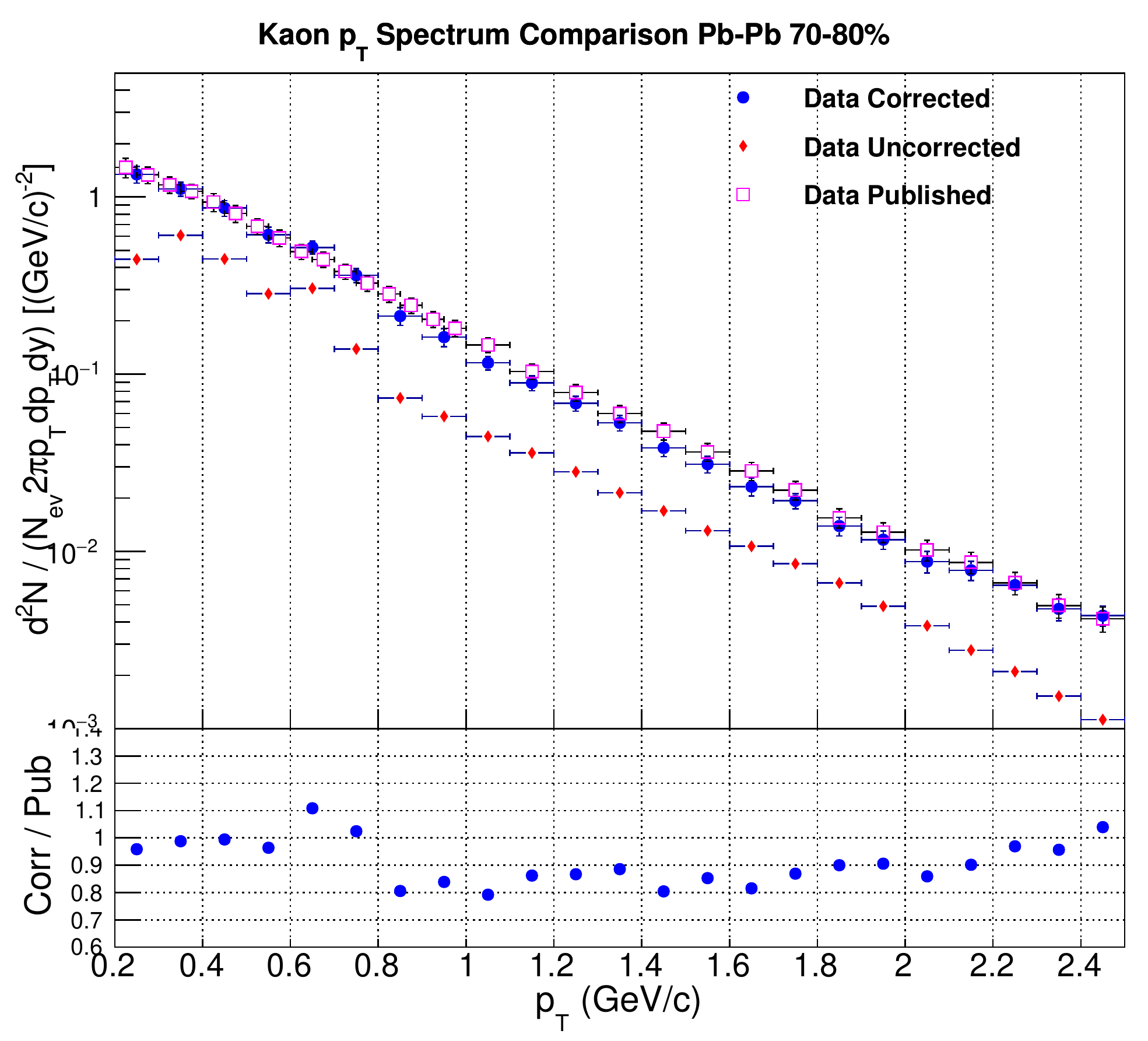}
  \caption{For $K^{\pm}$.
 Upper row for MC: efficiency estimation by taking the ratio of reconstructed divided by generator $p_{\rm T}$ spectra.
 Lower row for real data: comparison of $p_{\rm T}$ spectra obtained with and without MC efficiency correction, and the ALICE published, in central (left), mid-central (middle) and peripheral (right) collisions.}
   \label{fig:HIJING_Corrected_pt_kaon_vs_Published}
\end{figure}

\begin{figure}
\centering
  \includegraphics[width=0.32\linewidth]{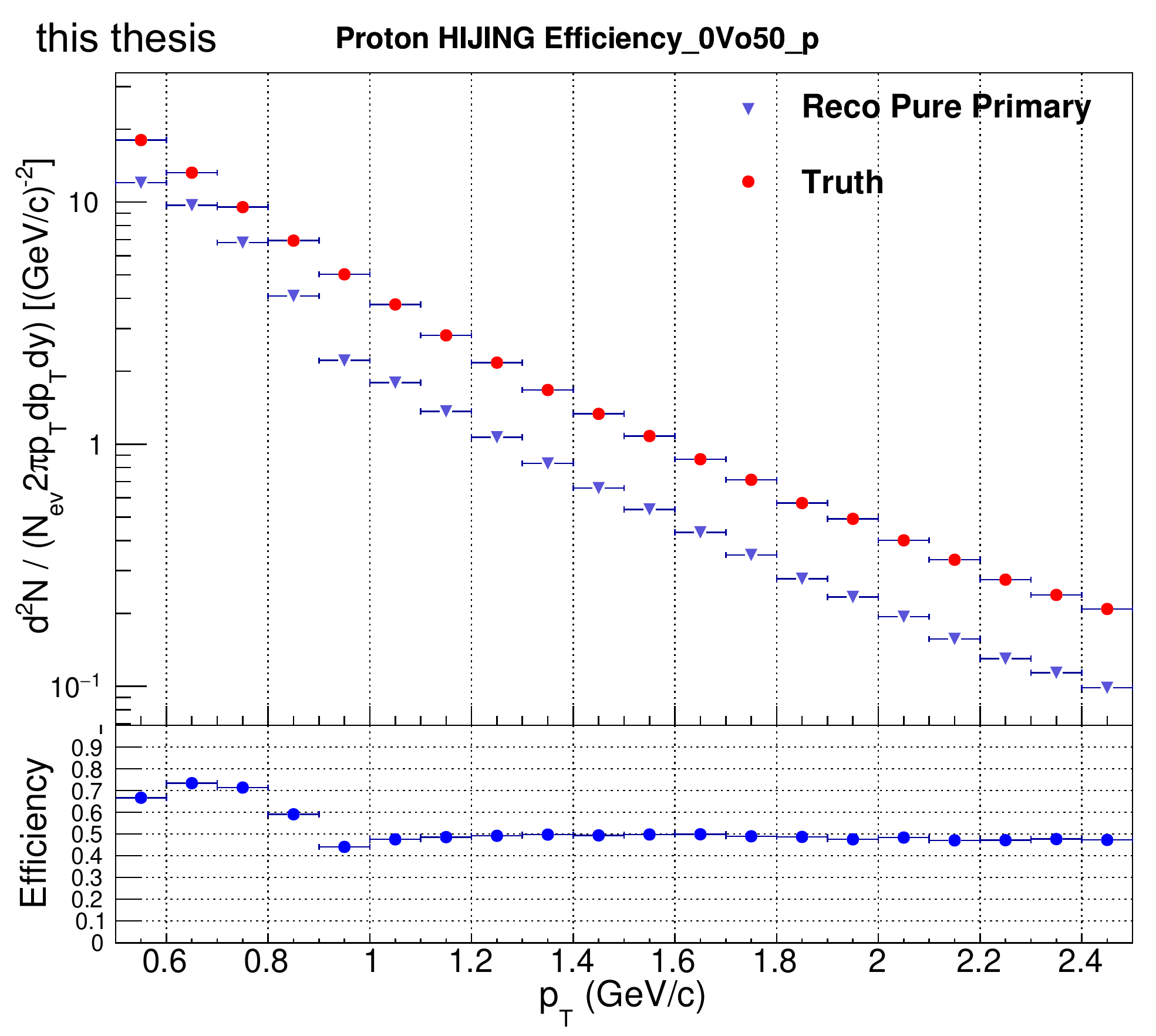}
  \includegraphics[width=0.32\linewidth]{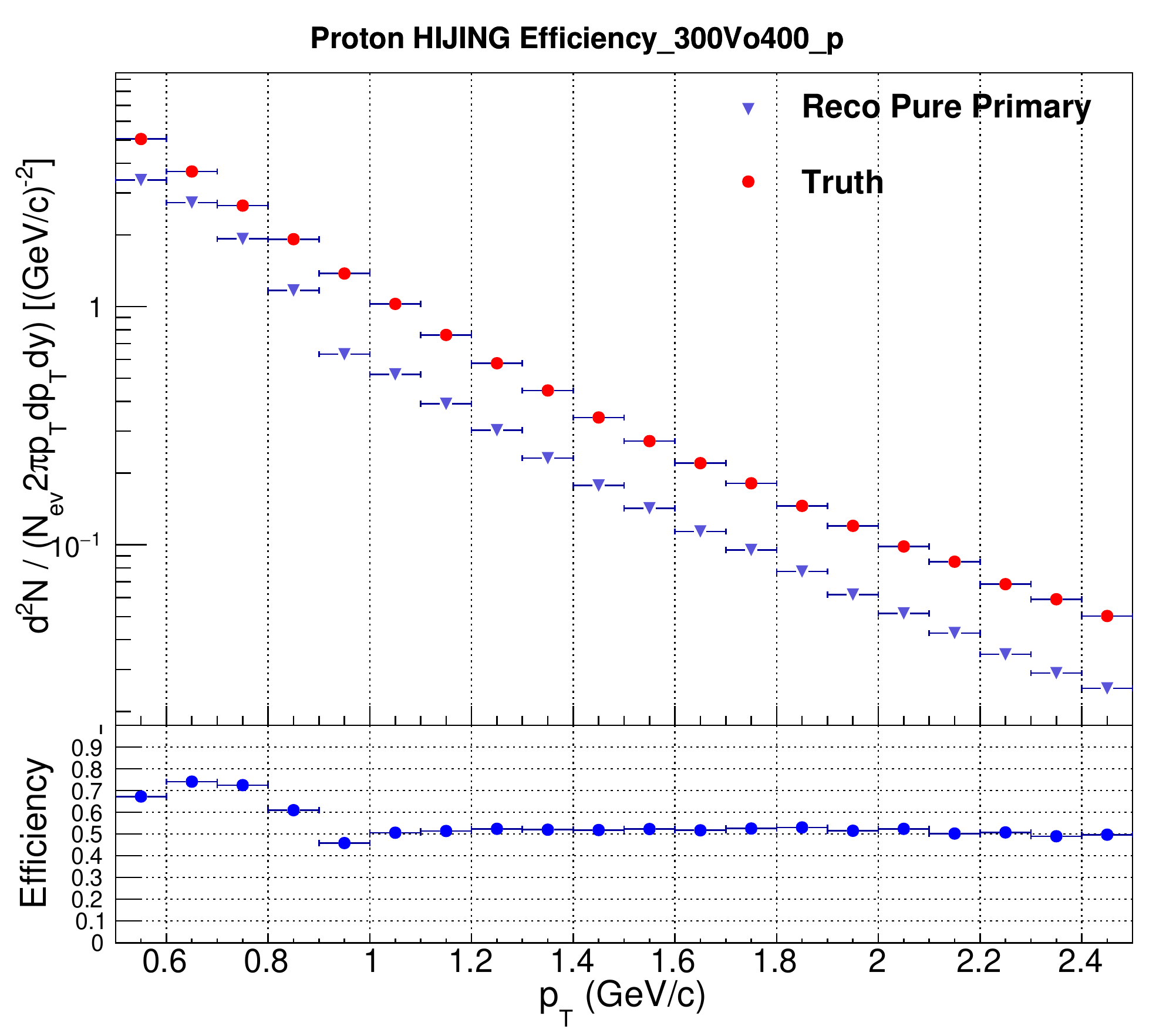}
  \includegraphics[width=0.32\linewidth]{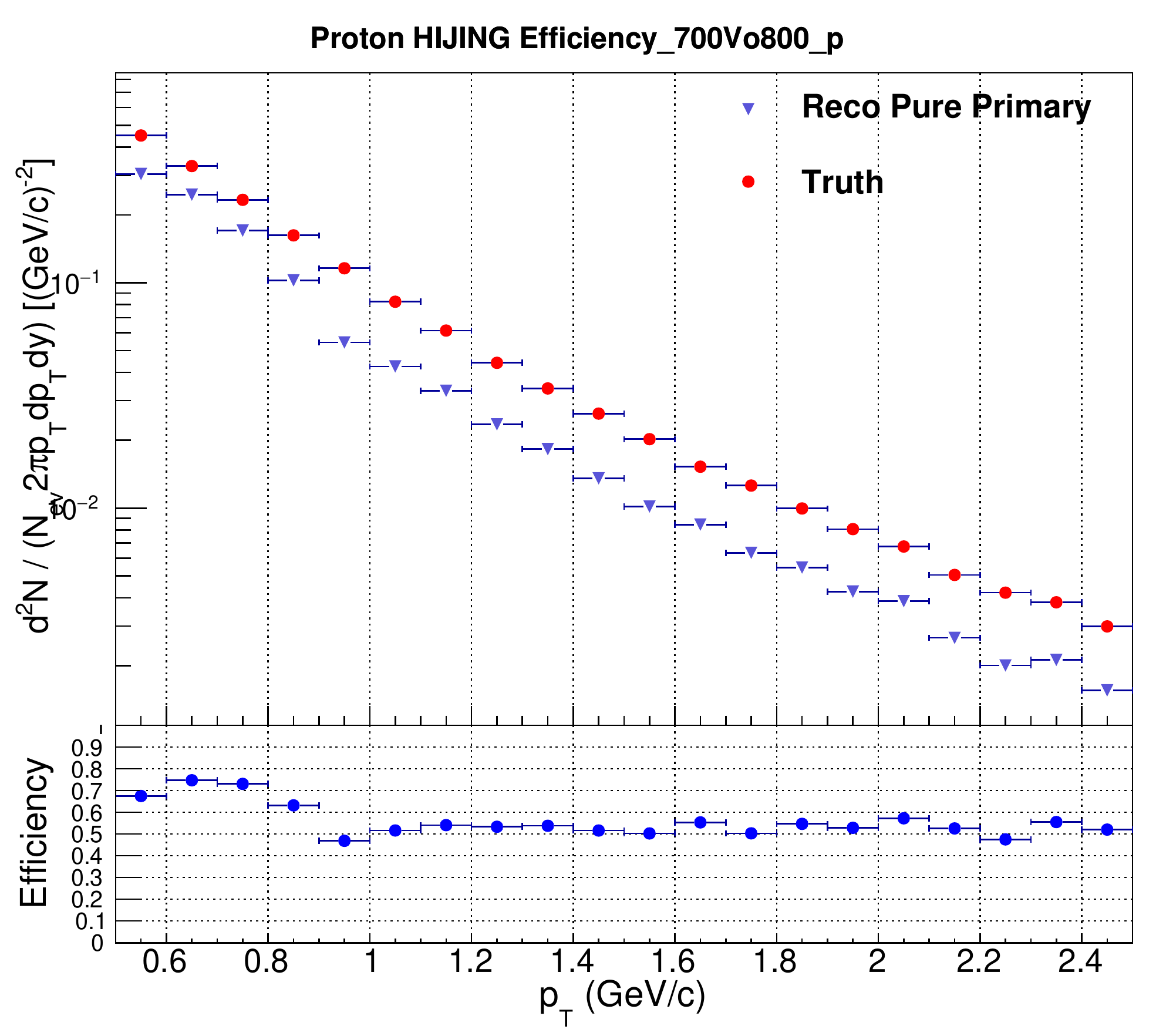} 
  \includegraphics[width=0.32\linewidth]{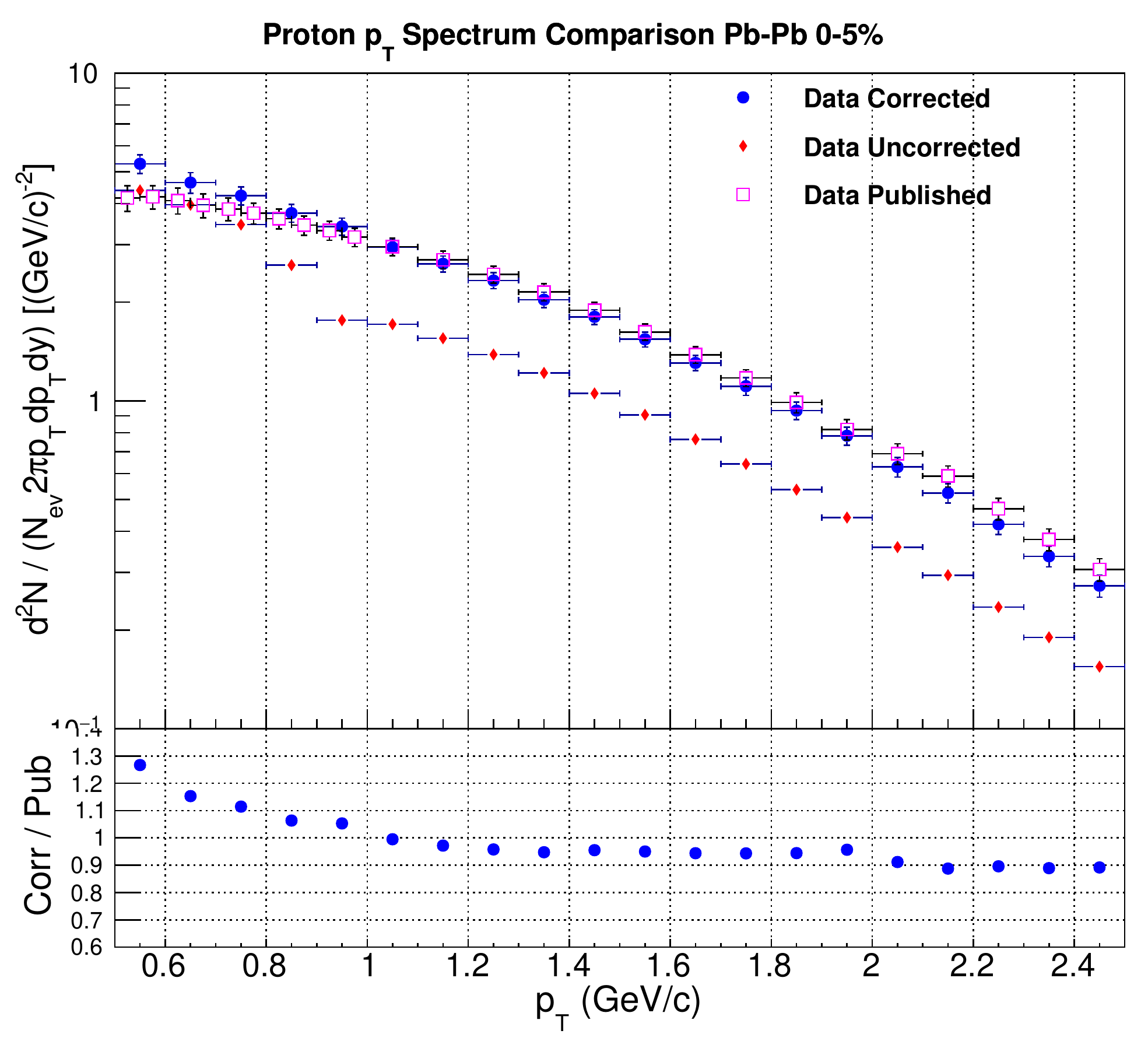}
  \includegraphics[width=0.32\linewidth]{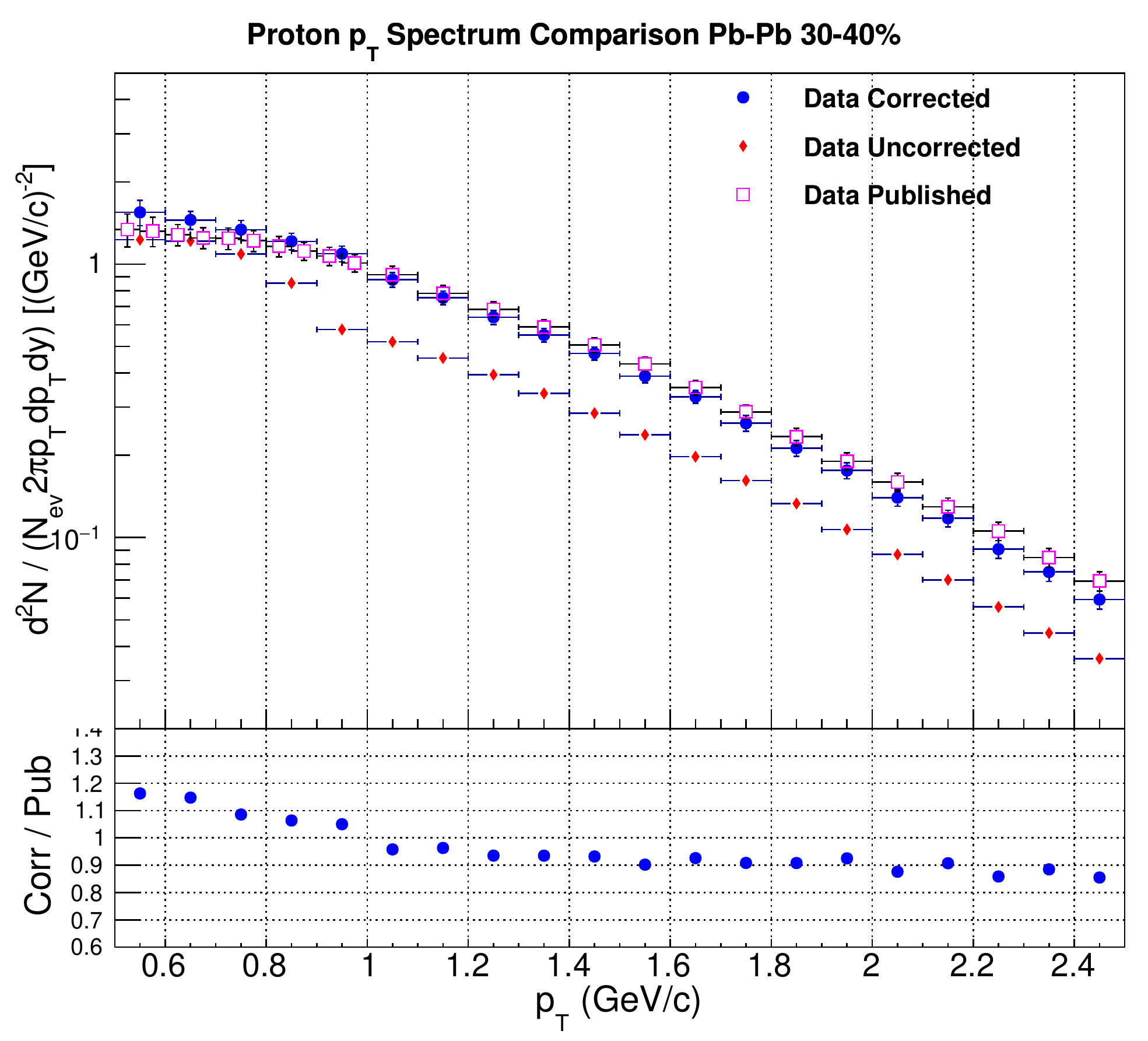}
  \includegraphics[width=0.32\linewidth]{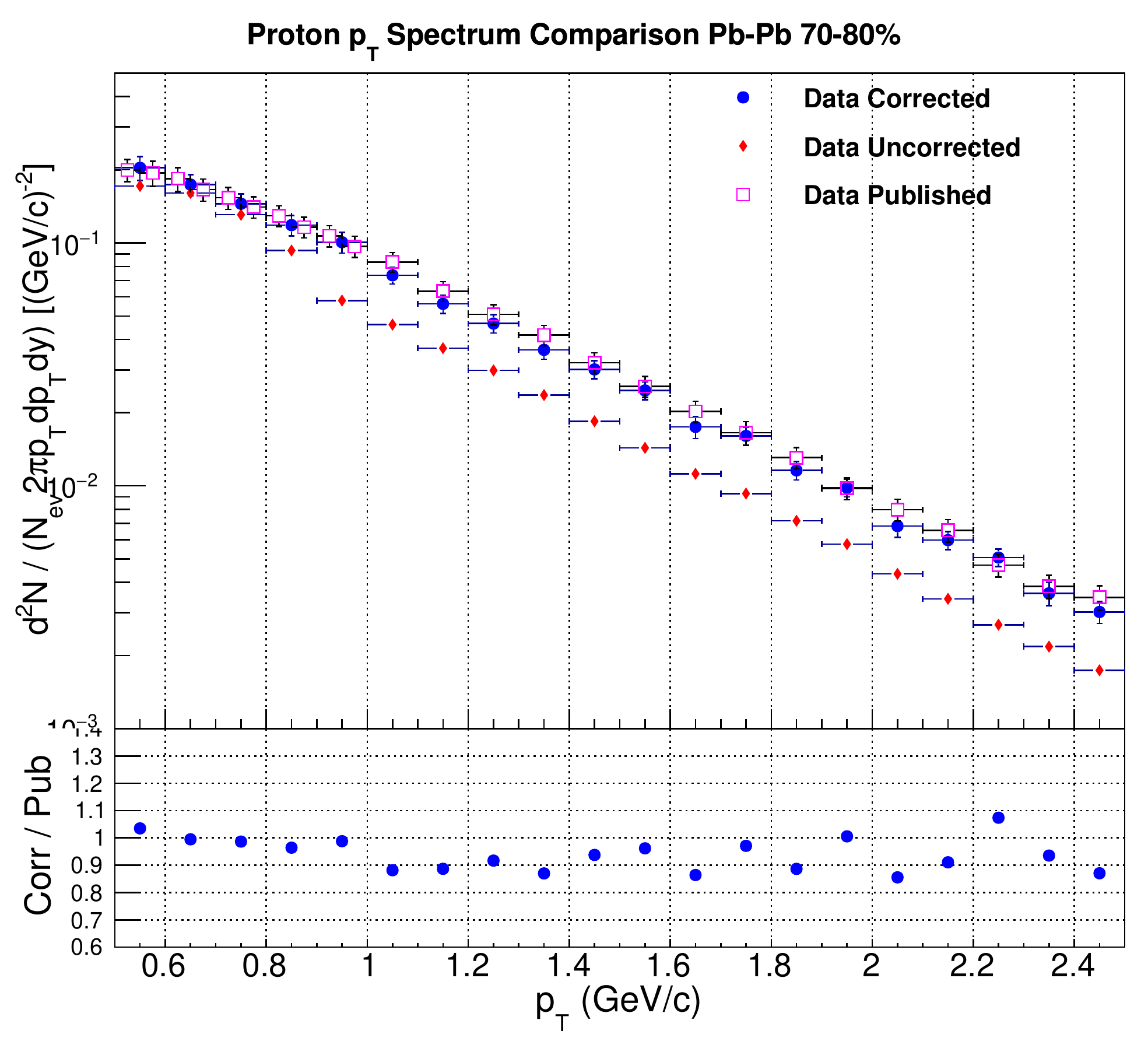}
  \caption{For $p/\bar{p}$.
 Upper row for MC: efficiency estimation by taking the ratio of reconstructed divided by generator $p_{\rm T}$ spectra.
 Lower row for real data: comparison of $p_{\rm T}$ spectra obtained with and without MC efficiency correction, and the ALICE published, in central (left), mid-central (middle) and peripheral (right) collisions.}
   \label{fig:HIJING_Corrected_pt_proton_vs_Published}
\end{figure}

\begin{figure}
\centering
  \includegraphics[width=0.50\linewidth]{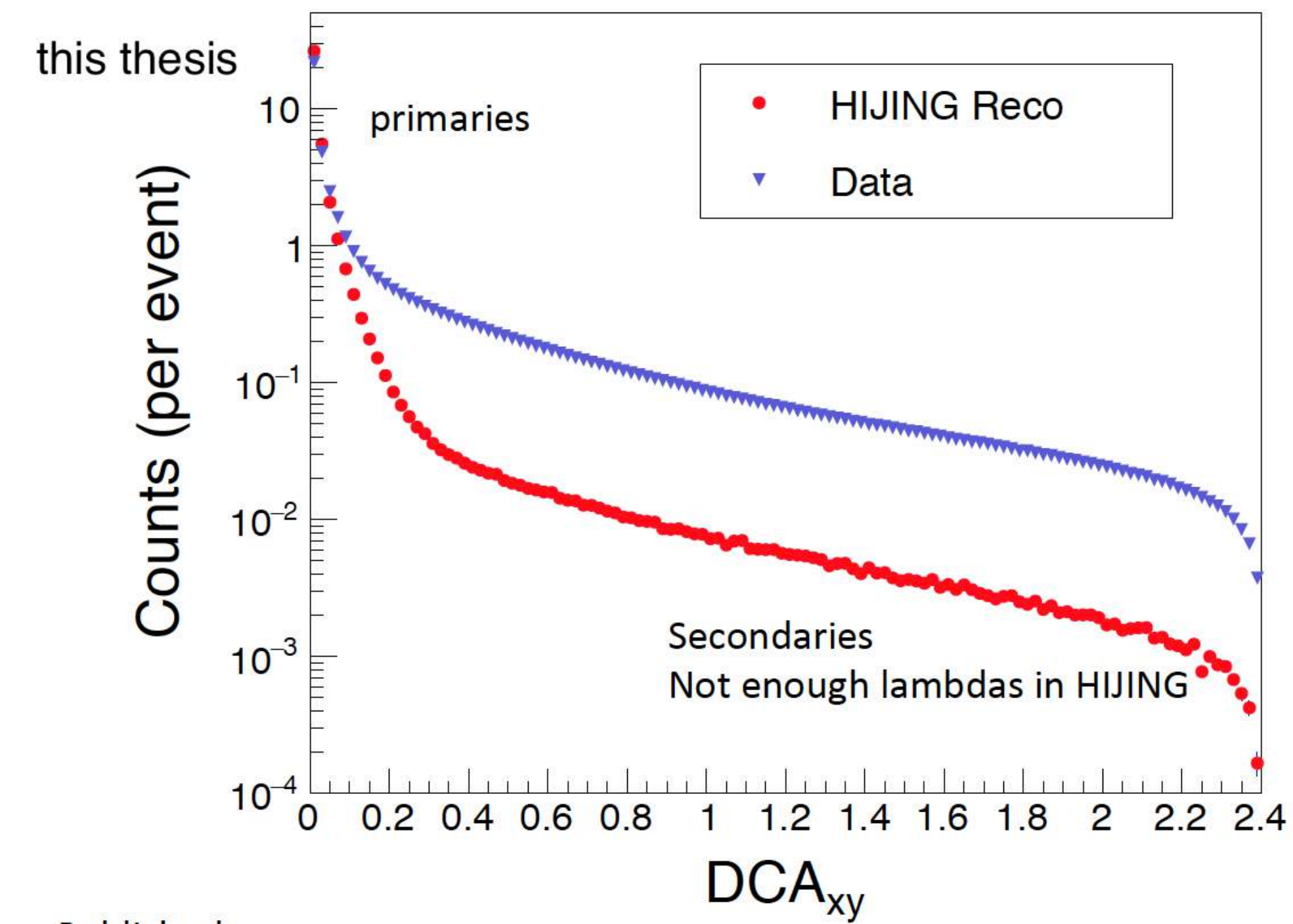}
    \caption{For $p/\bar{p}$. Comparison on $DCA_{xy}$ distributions between data and reconstructed HIJING MC in 0-20\% centrality of Pb-Pb collisions. }
   \label{fig:DCAxyDataMC}
\end{figure}

The use of both TPC and TOF for PID in this work results in $p_{\rm T}$-dependent detection efficiencies for $\pi^{\pm}$, $K^{\pm}$, and $p/\bar{p}$. 
In Figures~\ref{fig:HIJING_Corrected_pt_pion_vs_Published},~\ref{fig:HIJING_Corrected_pt_kaon_vs_Published},~\ref{fig:HIJING_Corrected_pt_proton_vs_Published}, the $p_{\rm T}$-dependent efficiencies are estimated based on the ratio of HIJING MC truth and reconstructed $p_{\rm T}$ spectra obtained with full GEANT simulated Pb--Pb collisions. 
The MC corrected $p_{\rm T}$ spectra of real data are very similar to the ALICE published $p_{\rm T}$ spectra~\cite{PhysRevC.88.044910}, with only a 10-15\% discrepancies on average between the two, which in principle are good enough for CFs and will not bias the results. 
The discrepancies are due to the fact that HIJING reconstructed events could not reproduce the real data $DCA_{xy}$ distribution, as shown in Figure~\ref{fig:DCAxyDataMC}.
Thus the MC detection efficiency estimations by HIJING reconstructed data are not robust for tight DCA cuts, which makes it difficult to perfectly reproduce the published $p_{\rm T}$ spectra.

\begin{figure}
\centering
  \includegraphics[width=0.32\linewidth]{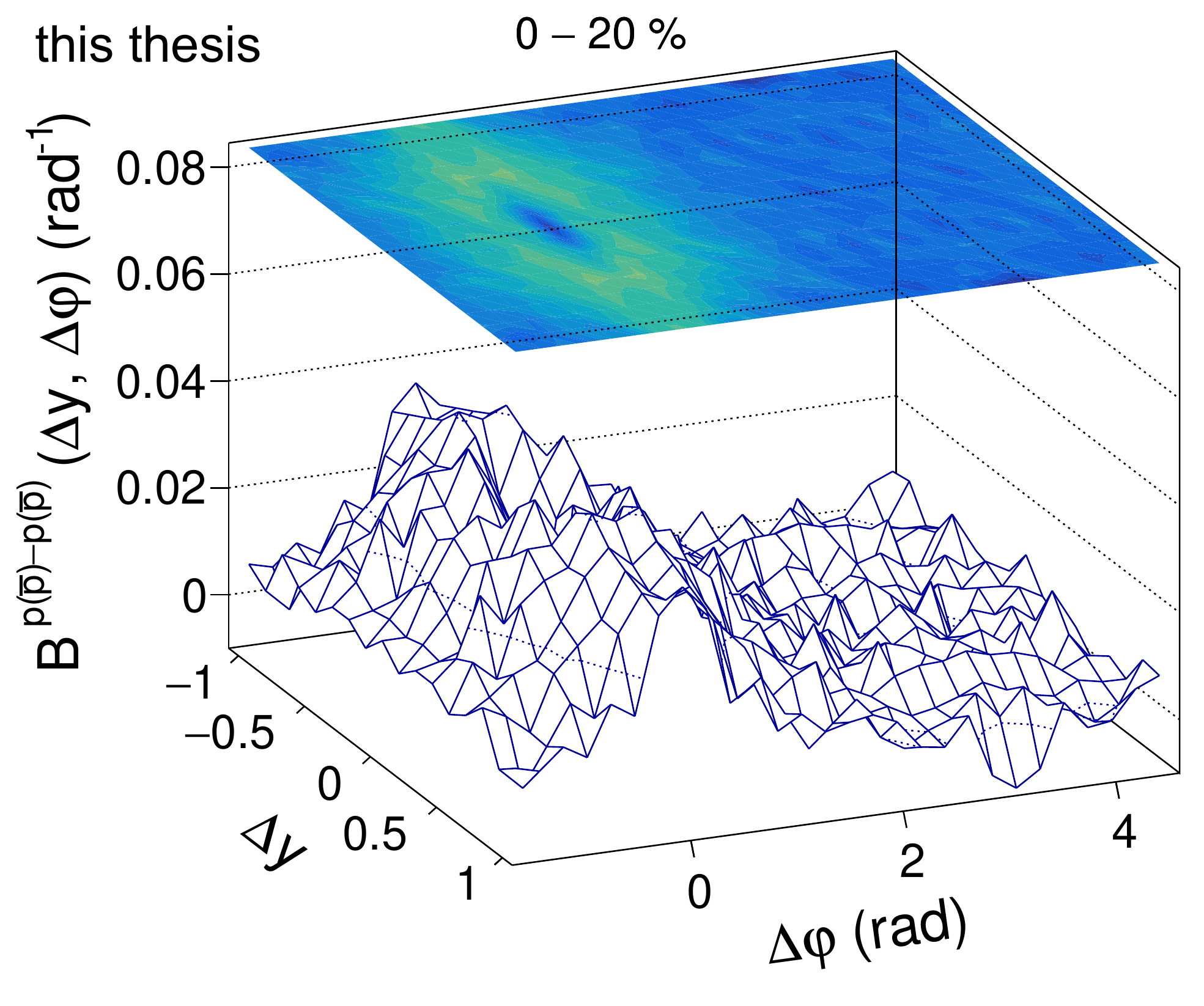}
  \includegraphics[width=0.32\linewidth]{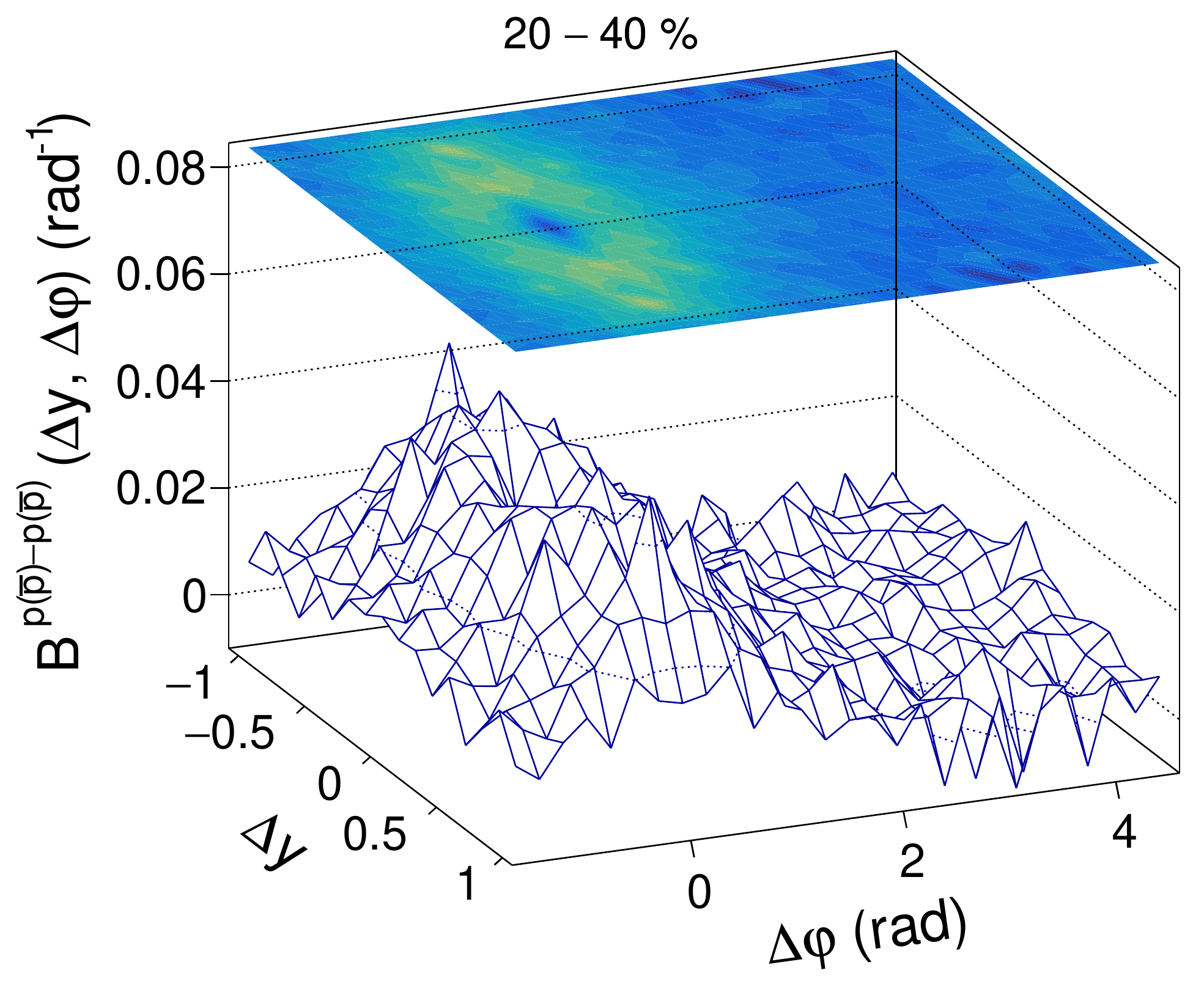}
  \includegraphics[width=0.32\linewidth]{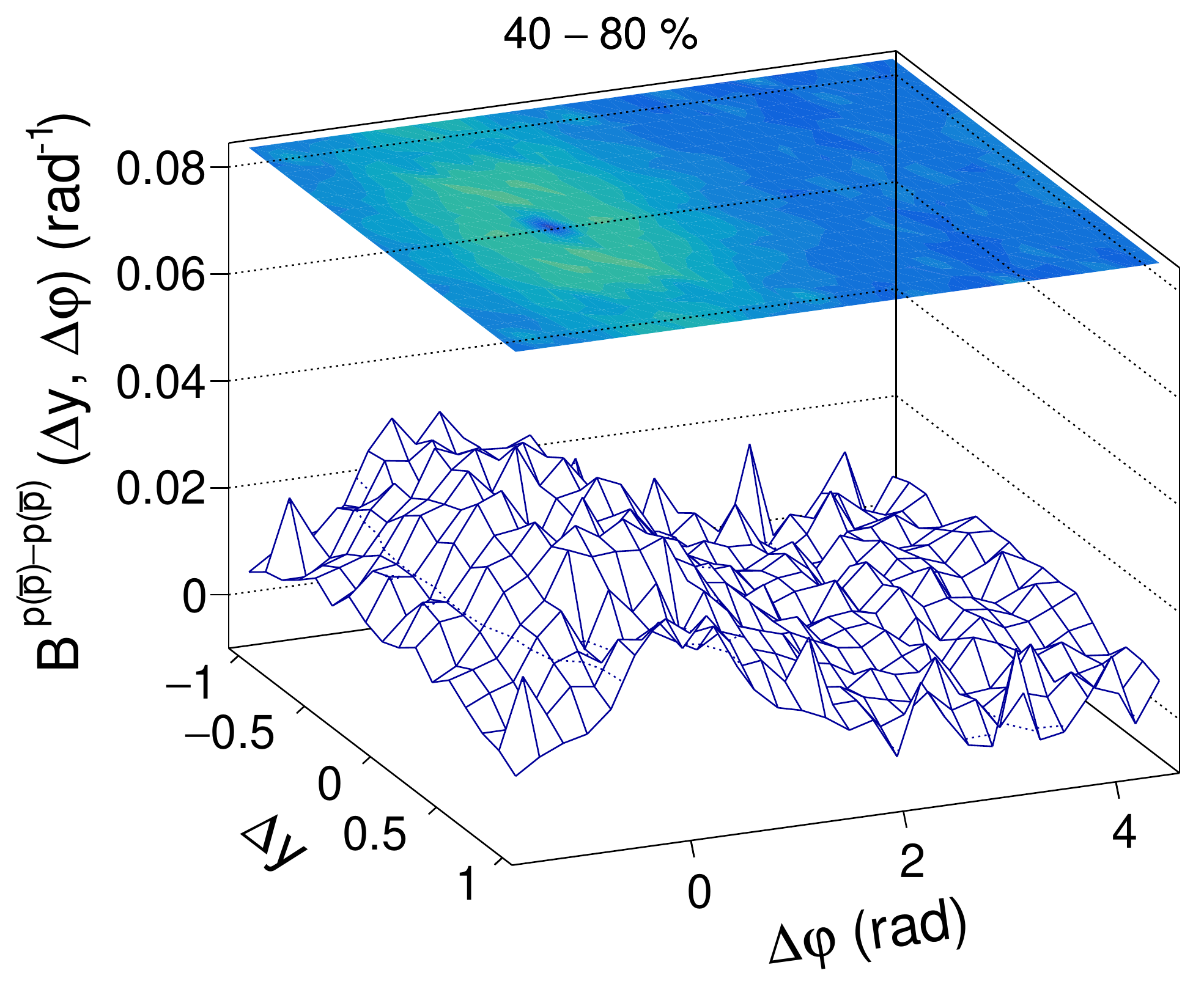} 
  \includegraphics[width=0.32\linewidth]{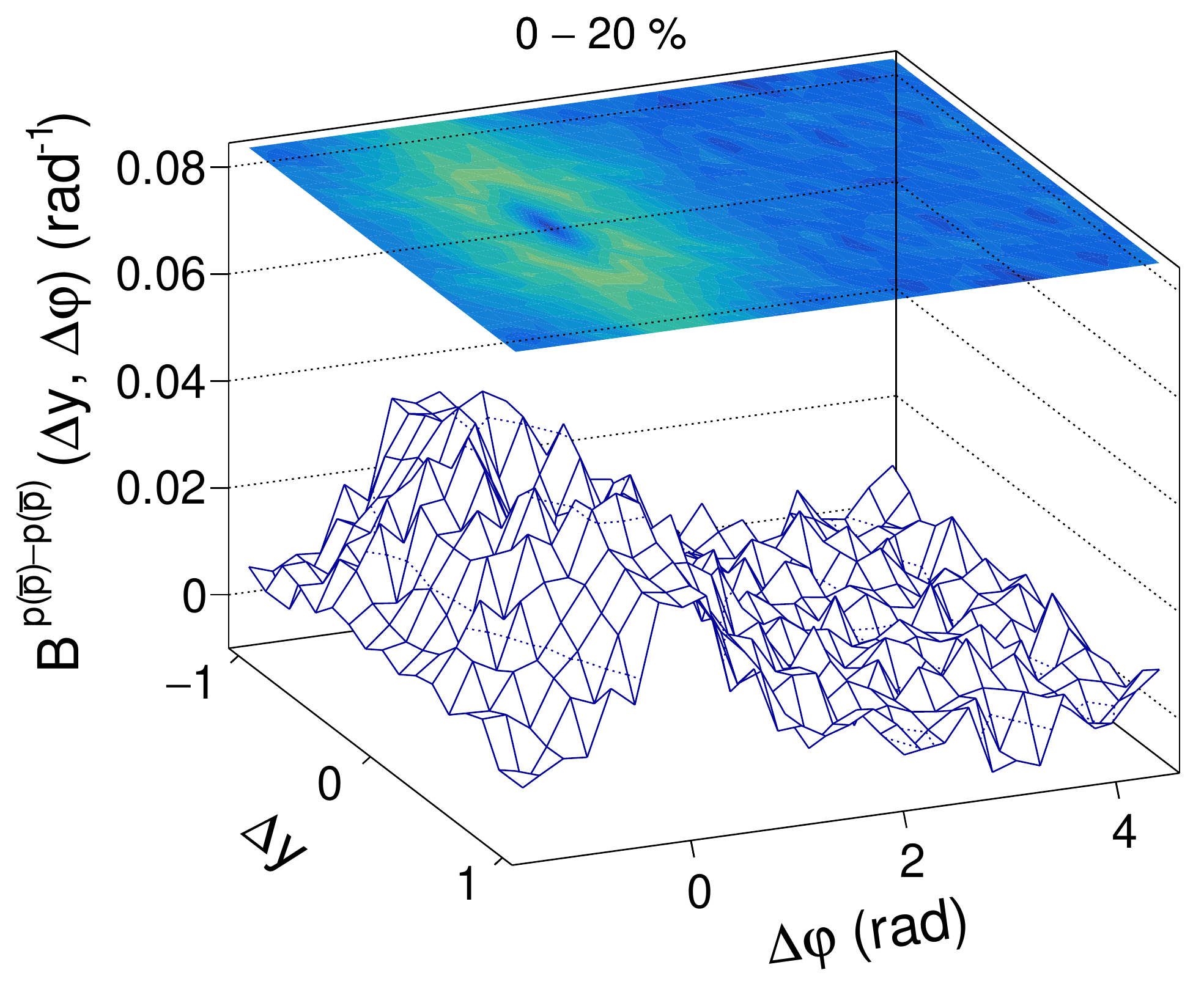}
  \includegraphics[width=0.32\linewidth]{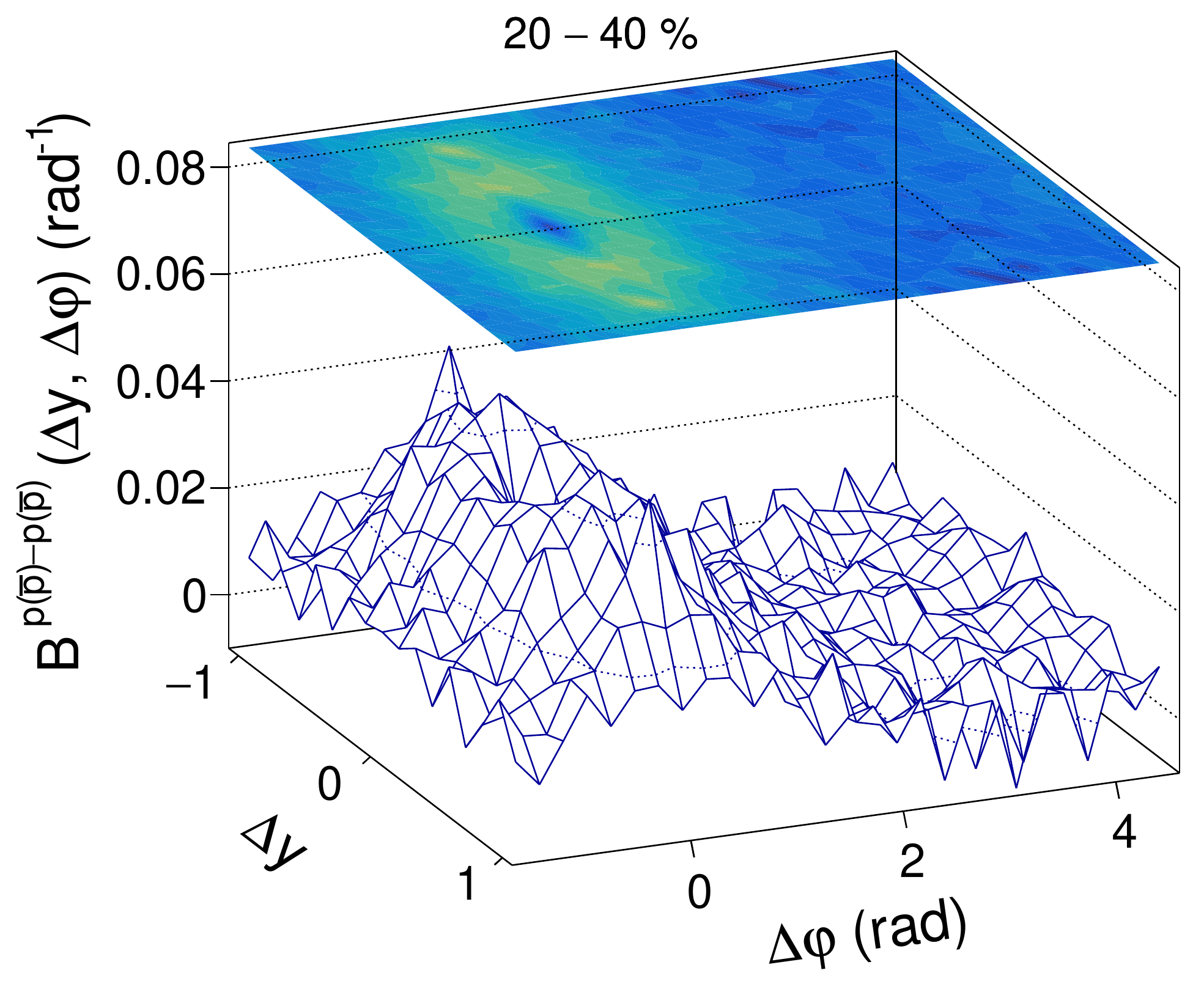}
  \includegraphics[width=0.32\linewidth]{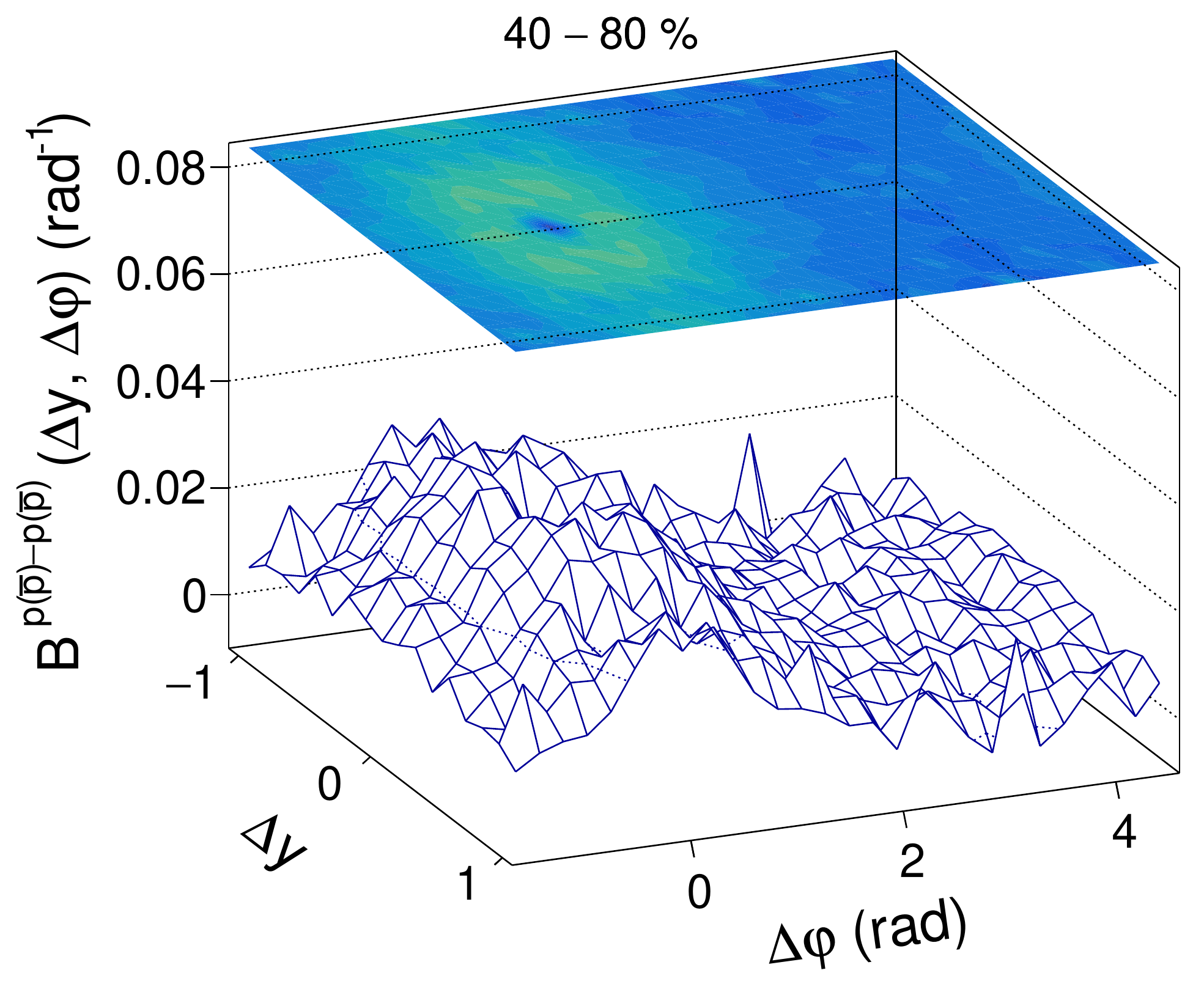} 
  \includegraphics[width=0.32\linewidth]{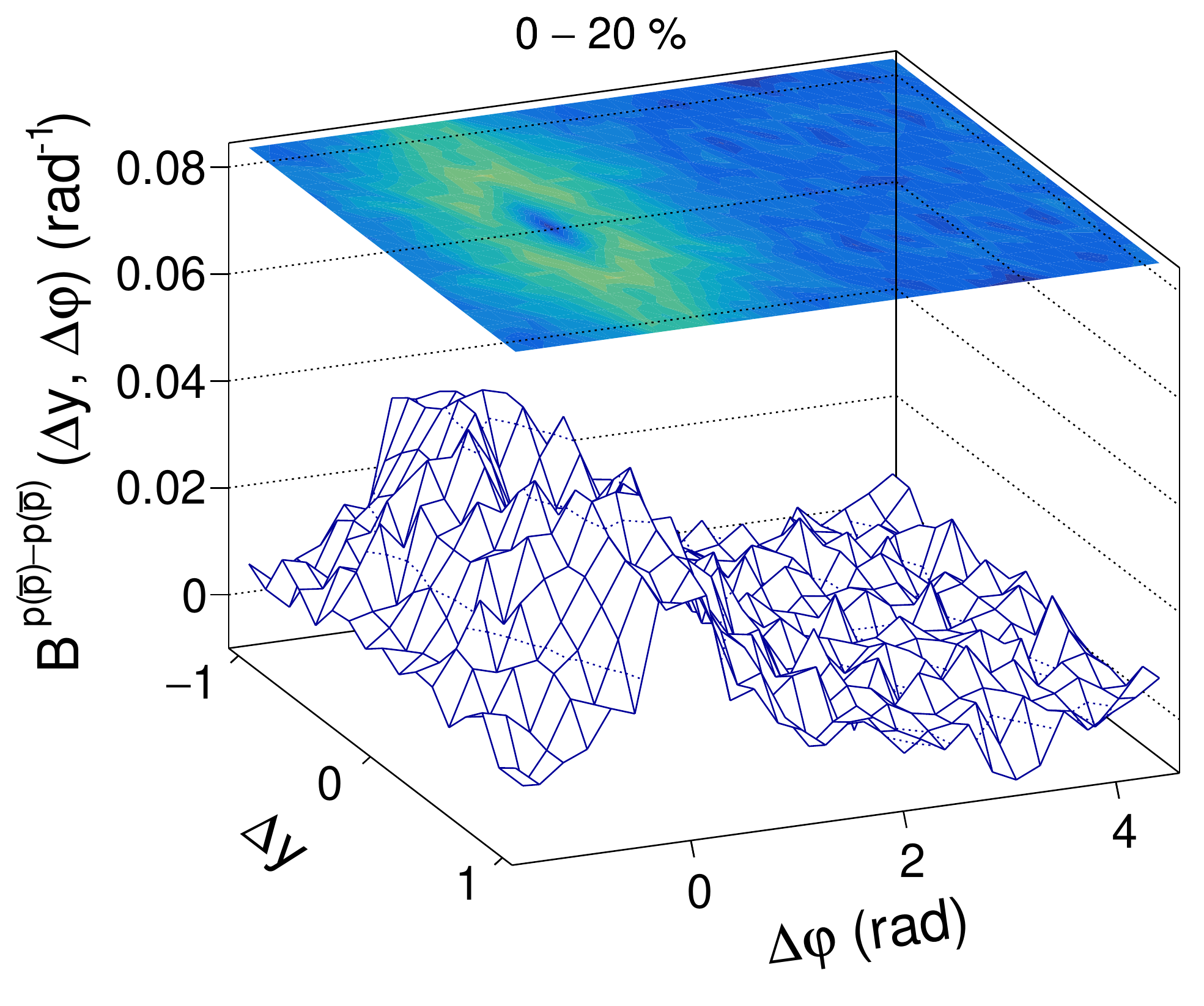}
  \includegraphics[width=0.32\linewidth]{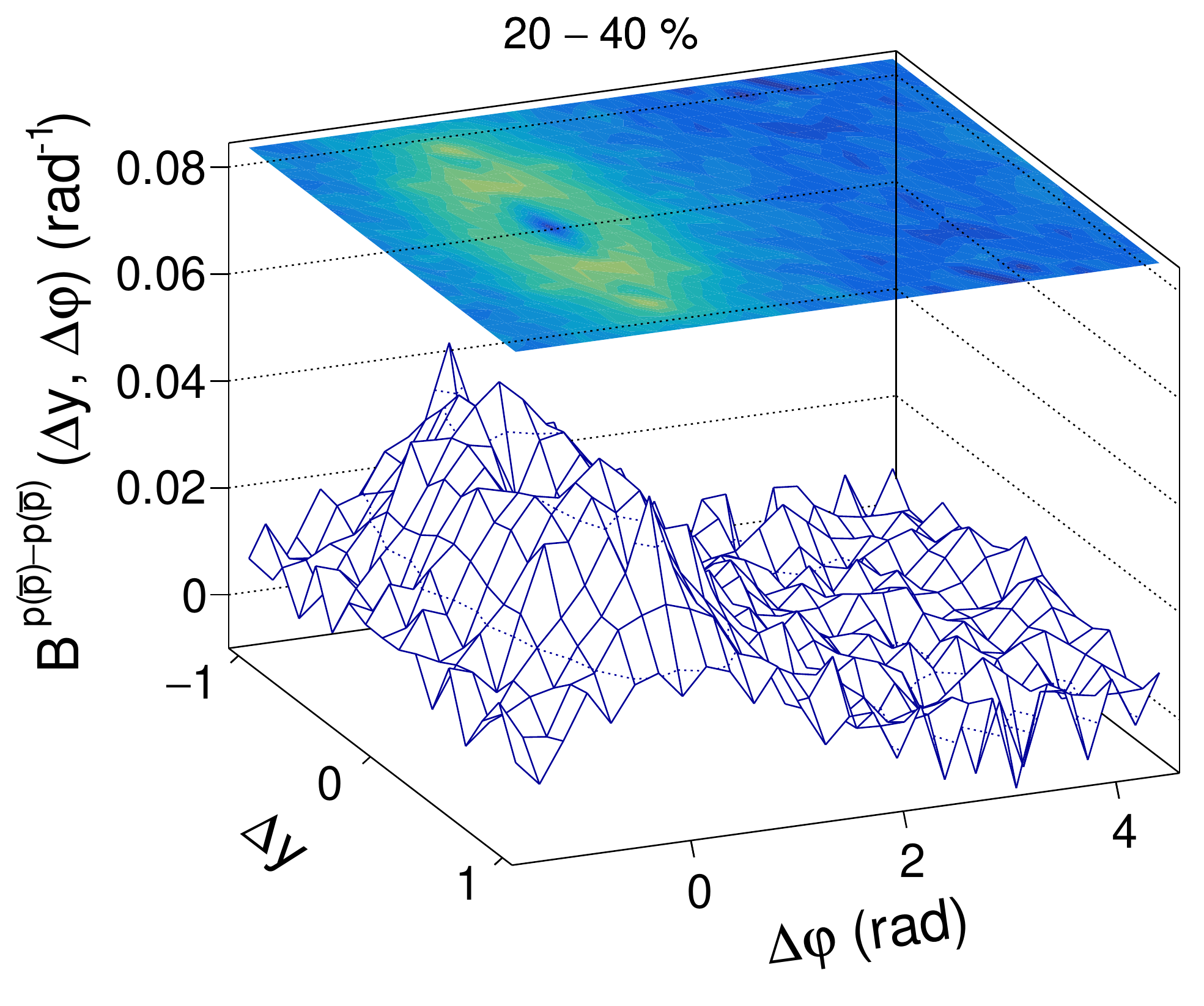}
  \includegraphics[width=0.32\linewidth]{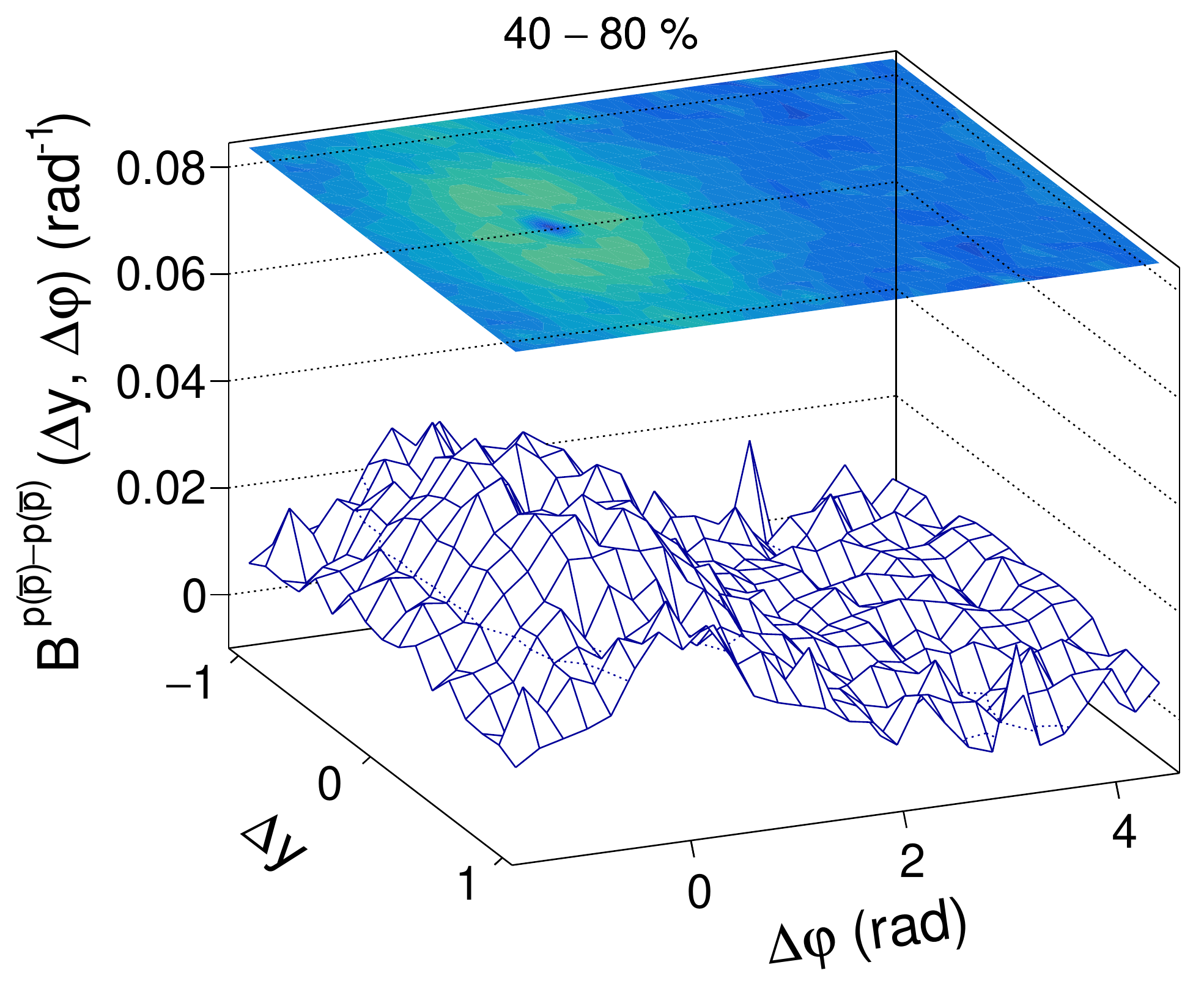}
  \includegraphics[width=0.32\linewidth]{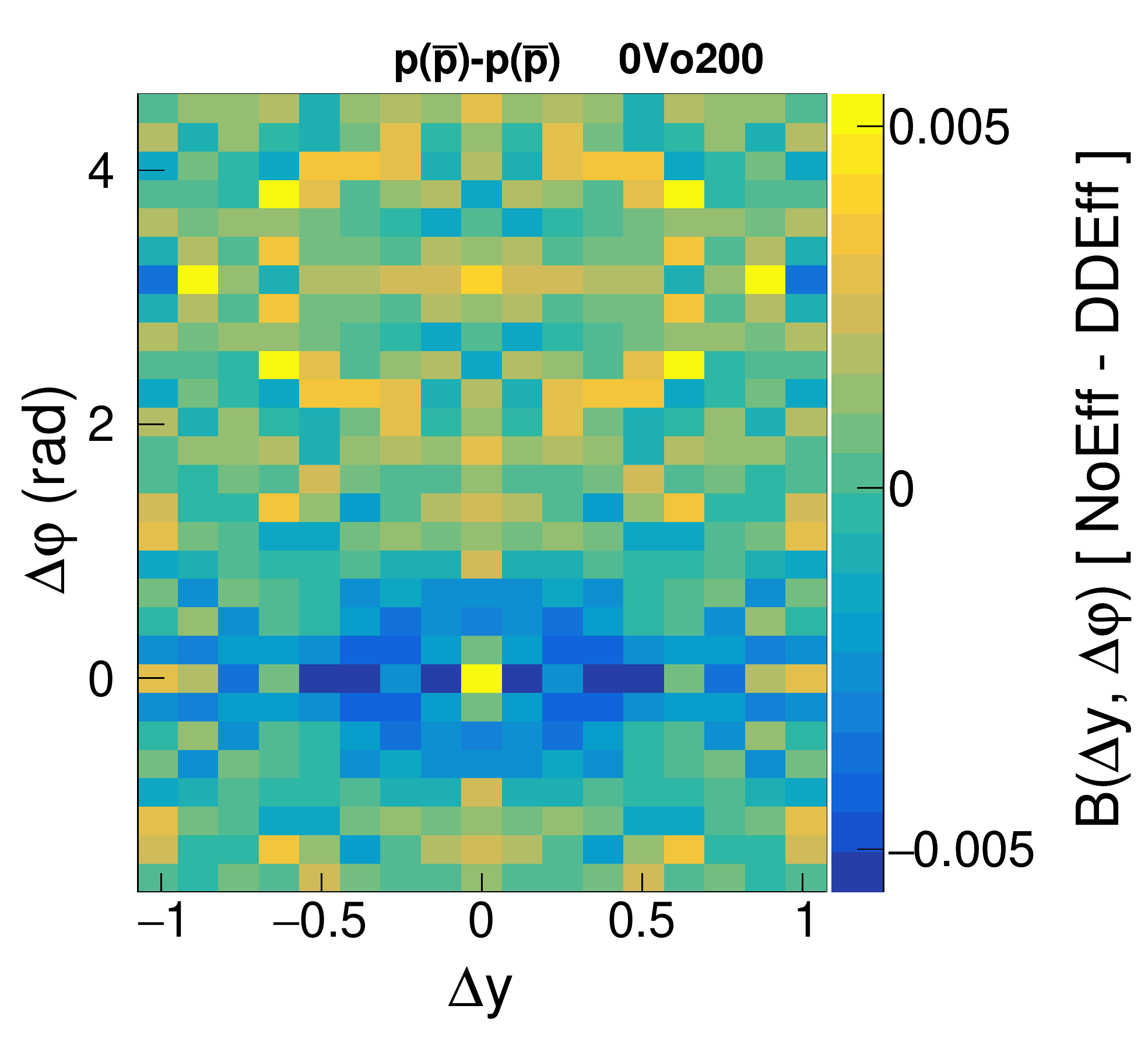}
  \includegraphics[width=0.32\linewidth]{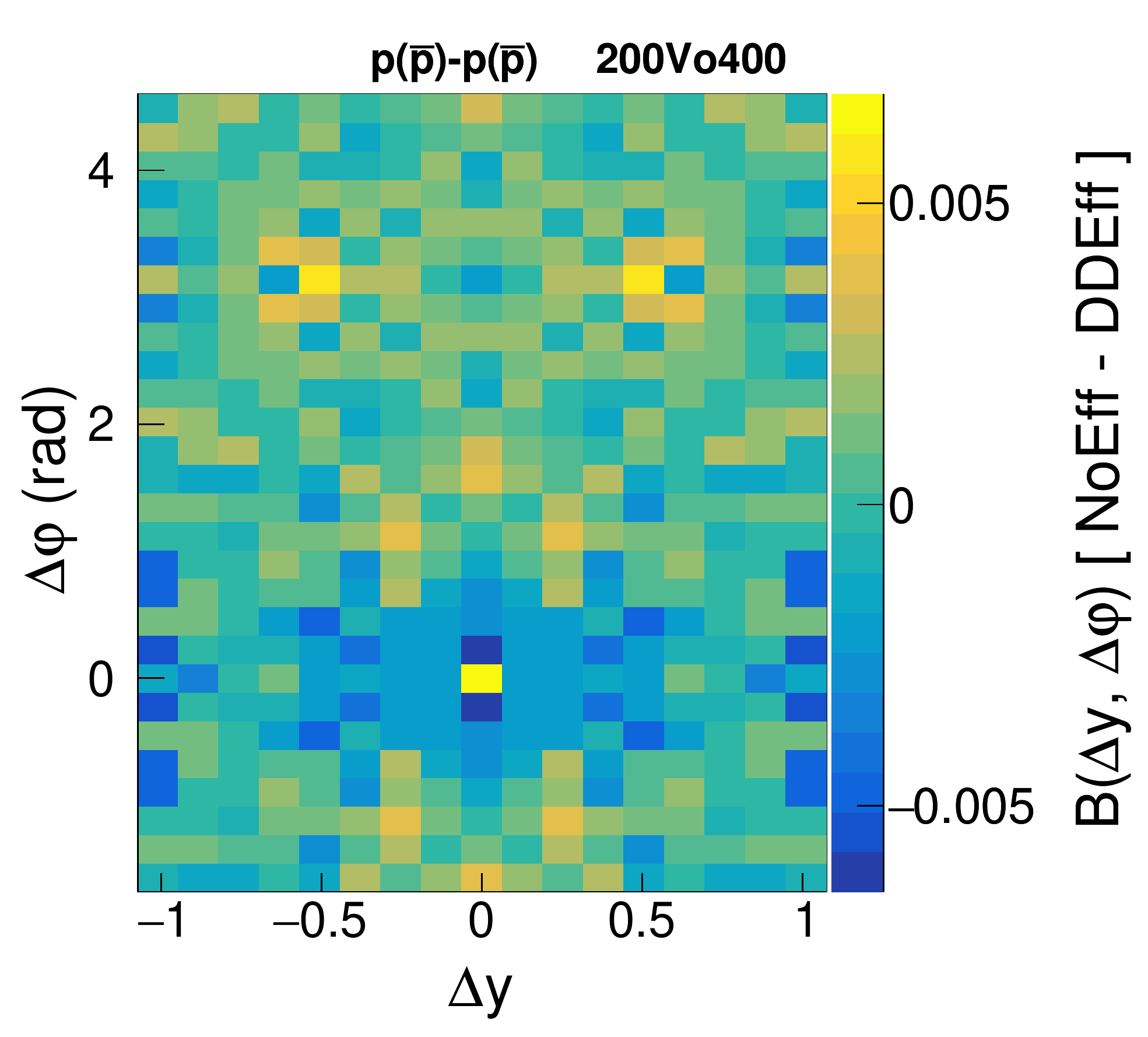}
  \includegraphics[width=0.32\linewidth]{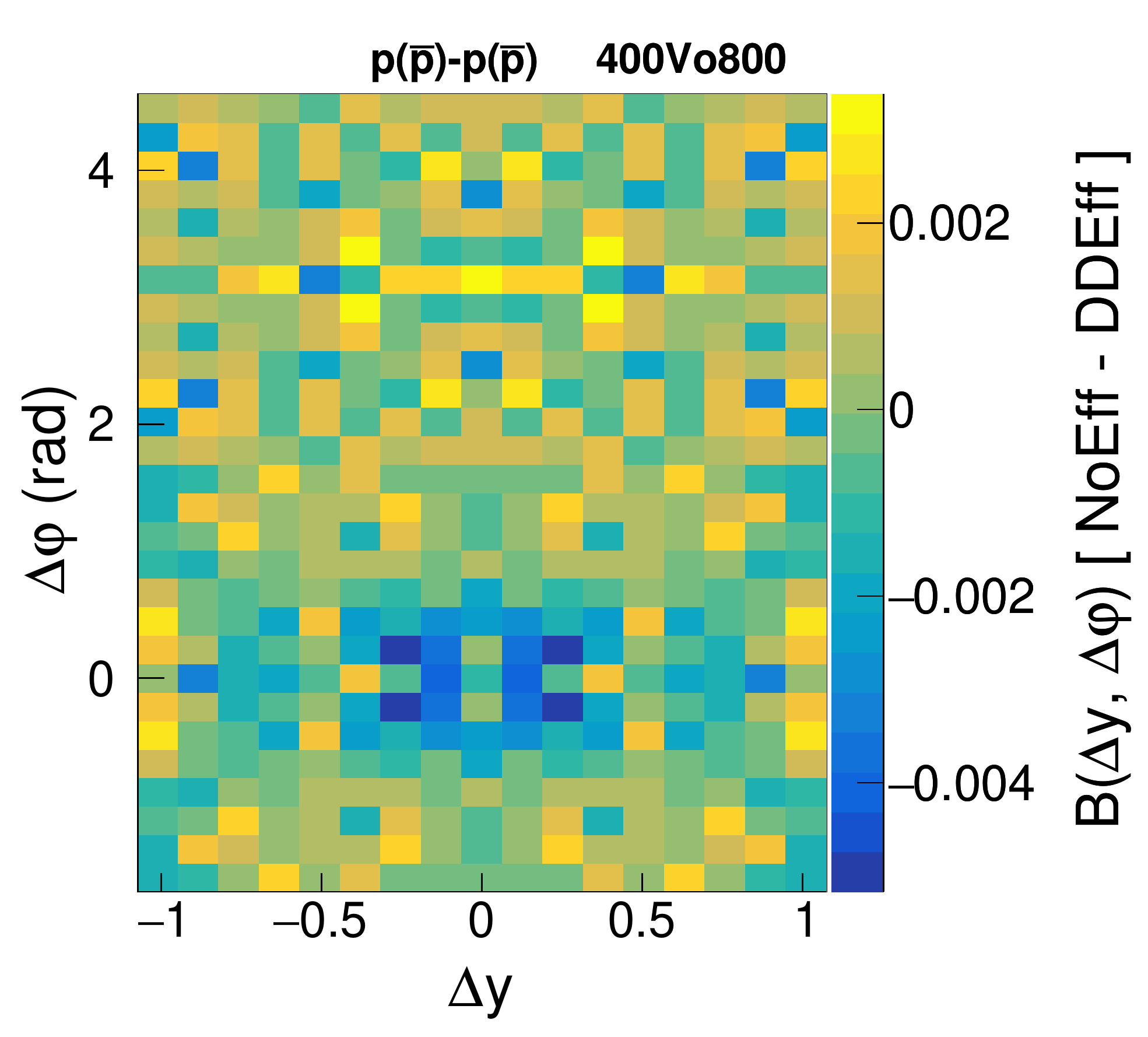} 
  \caption{Comparison of 2D $B^{pp}$ obtained without (1st row), with MC (2nd row) and Data Driven (3rd row) efficiency correction.
  4th row: differences between without and with Data Driven $p_{\rm T}$-dependent efficiency correction.}
%  5th row: differences between with MC and Data Driven $p_{\rm T}$-dependent efficiency correction.}
   \label{fig:Compare_DDEffCorr_MCEffCorr_NoEffCorr_BF_2d_ProtonProton}
\end{figure}

\begin{figure}
\centering
  \includegraphics[width=0.32\linewidth]{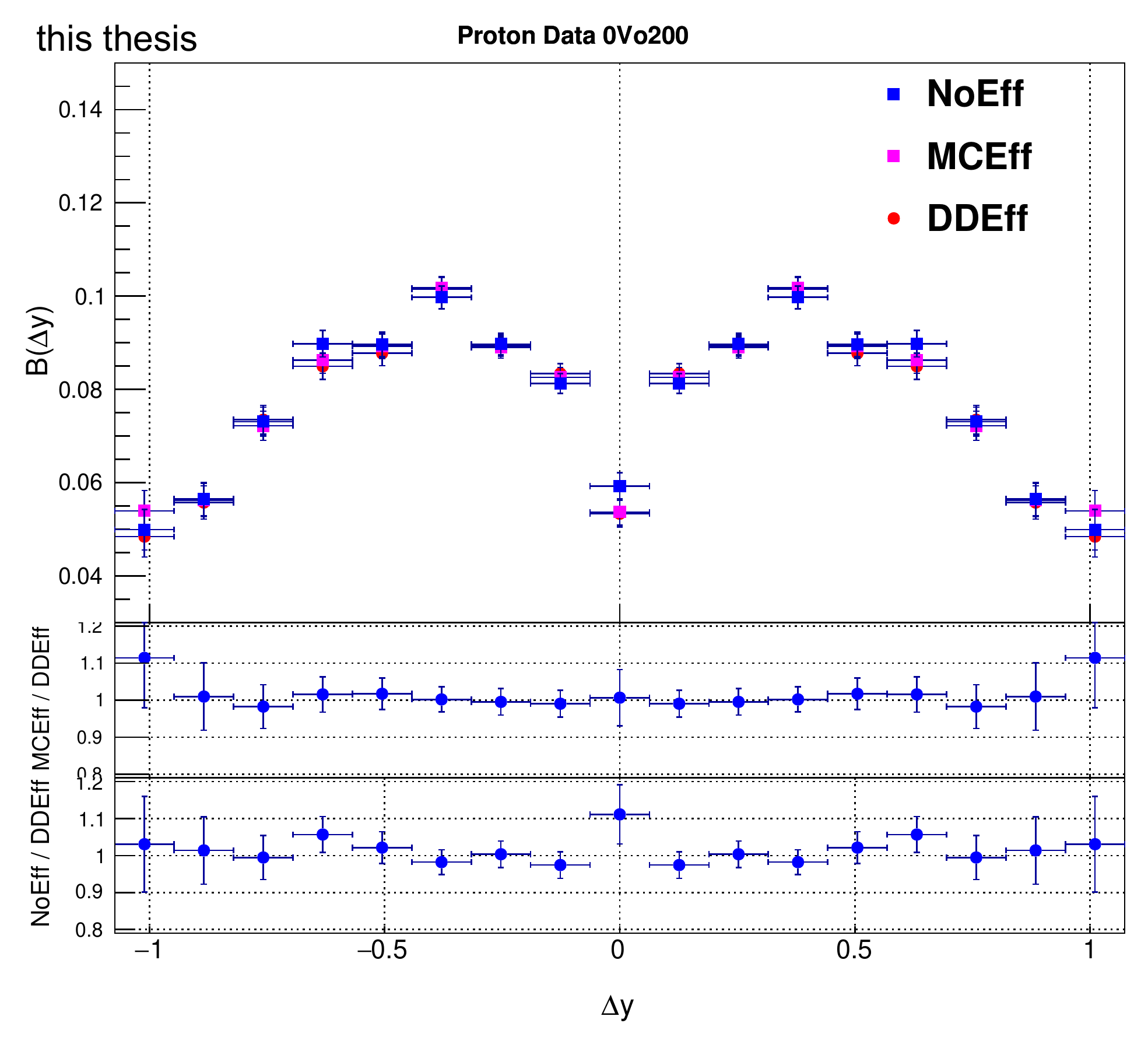}
  \includegraphics[width=0.32\linewidth]{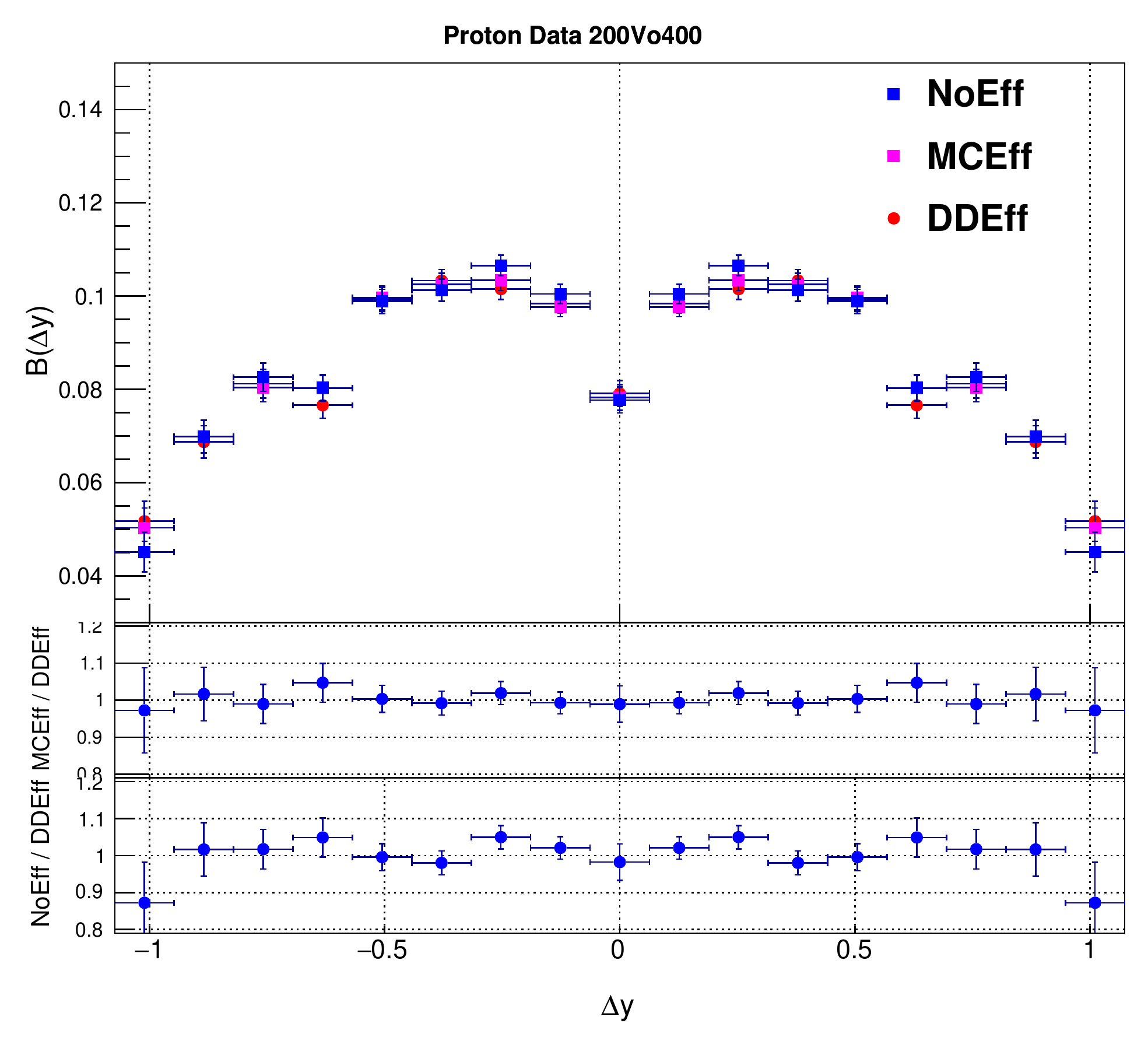}
  \includegraphics[width=0.32\linewidth]{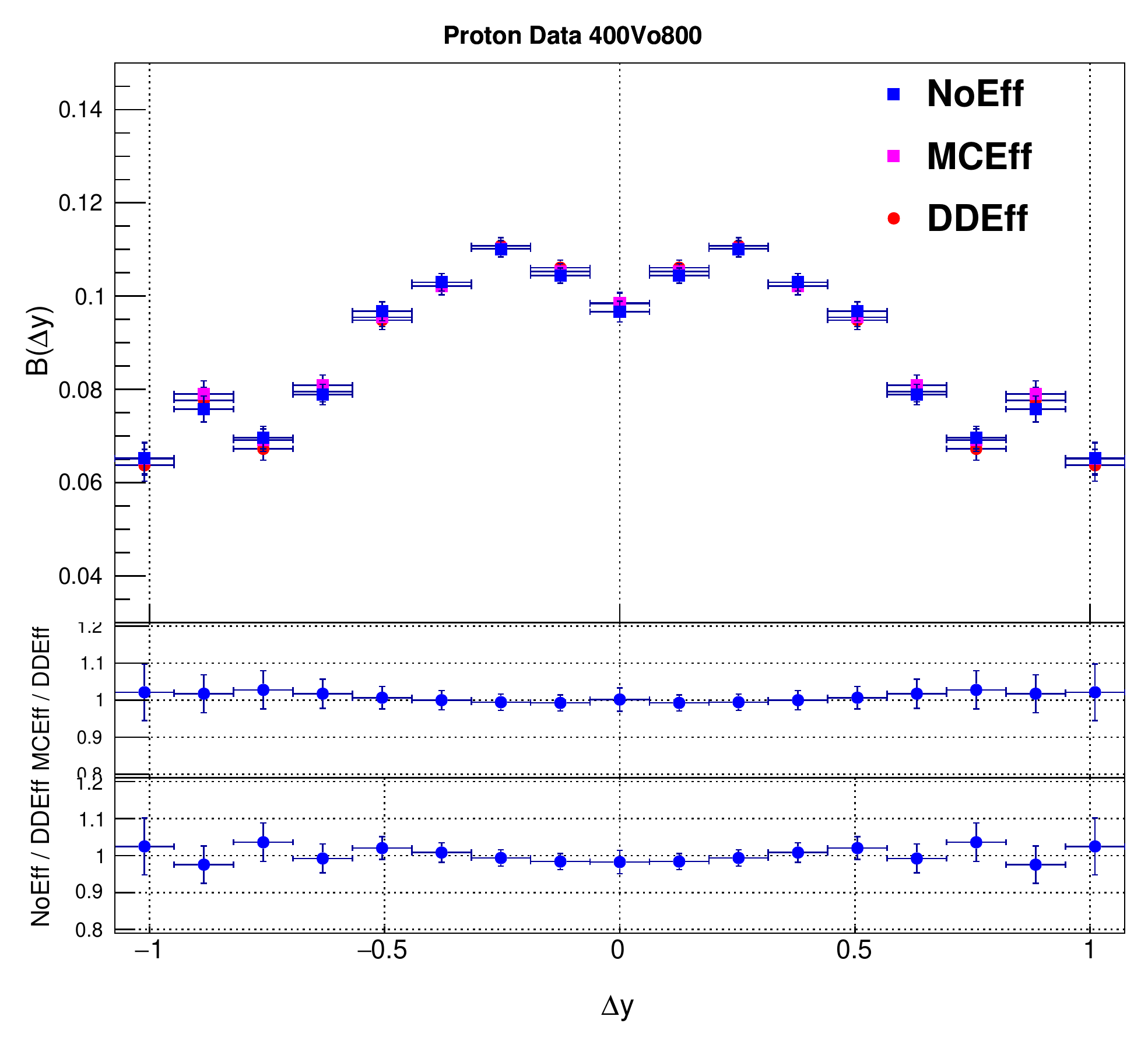} 
  \includegraphics[width=0.32\linewidth]{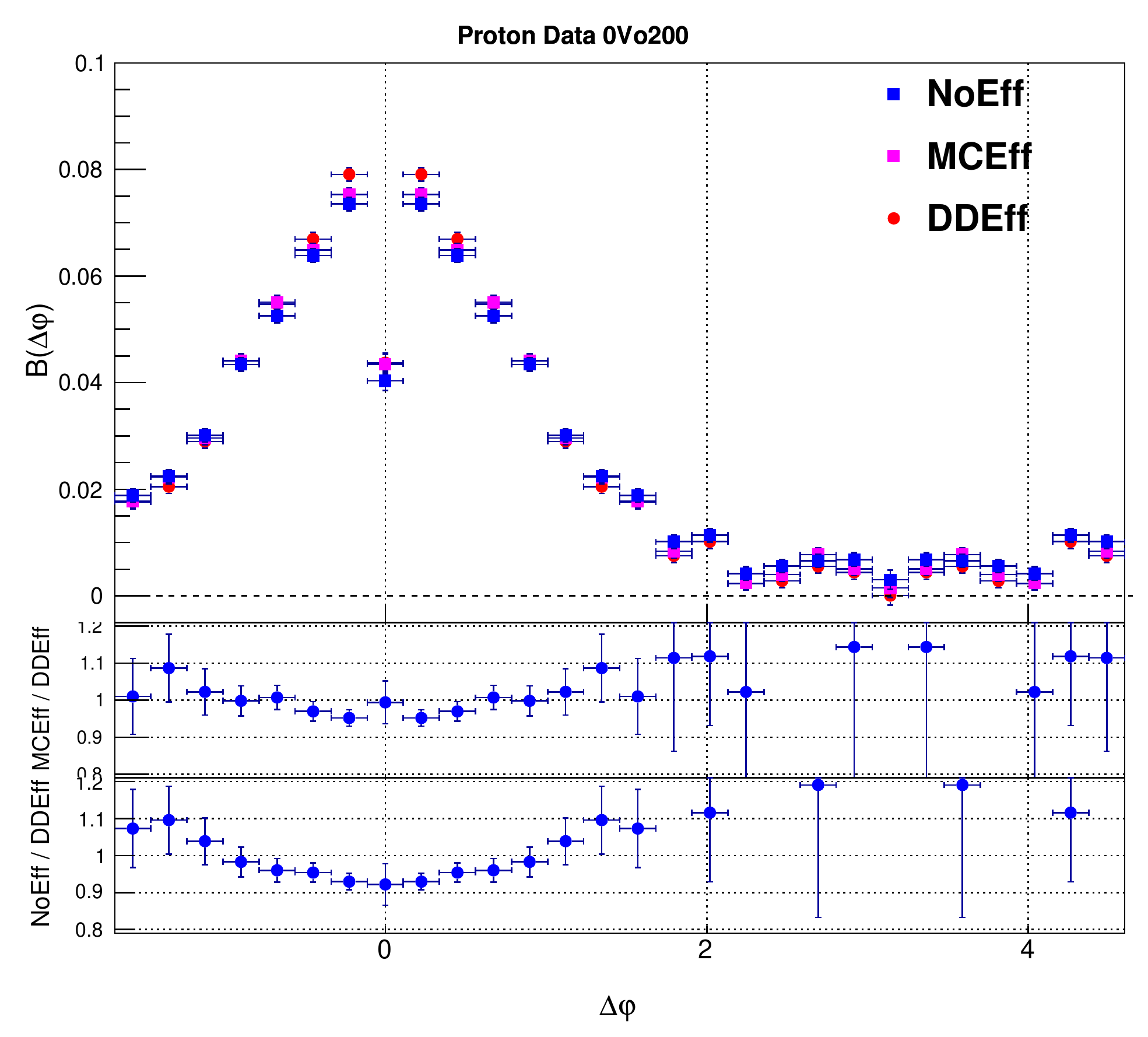}
  \includegraphics[width=0.32\linewidth]{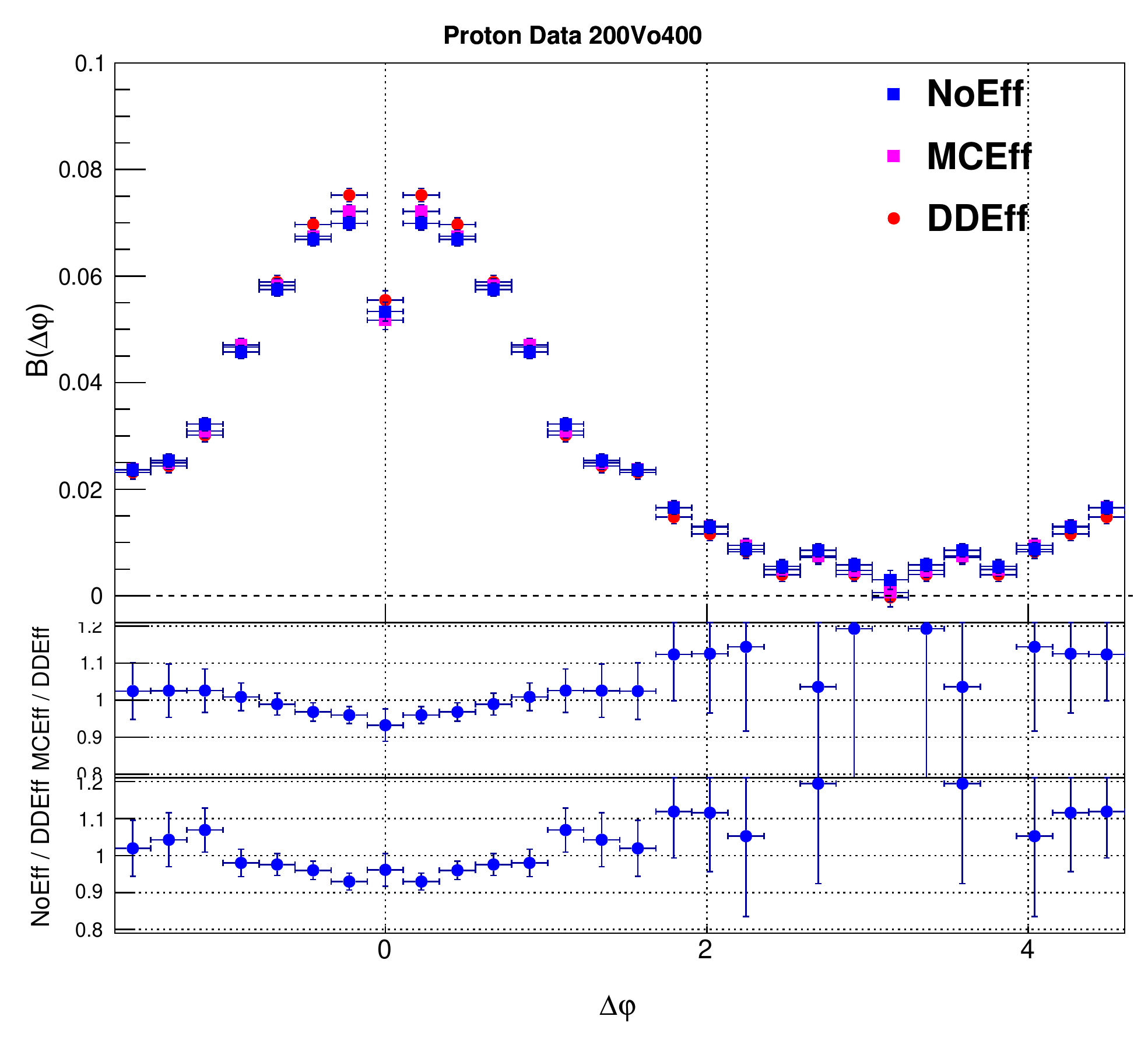}
  \includegraphics[width=0.32\linewidth]{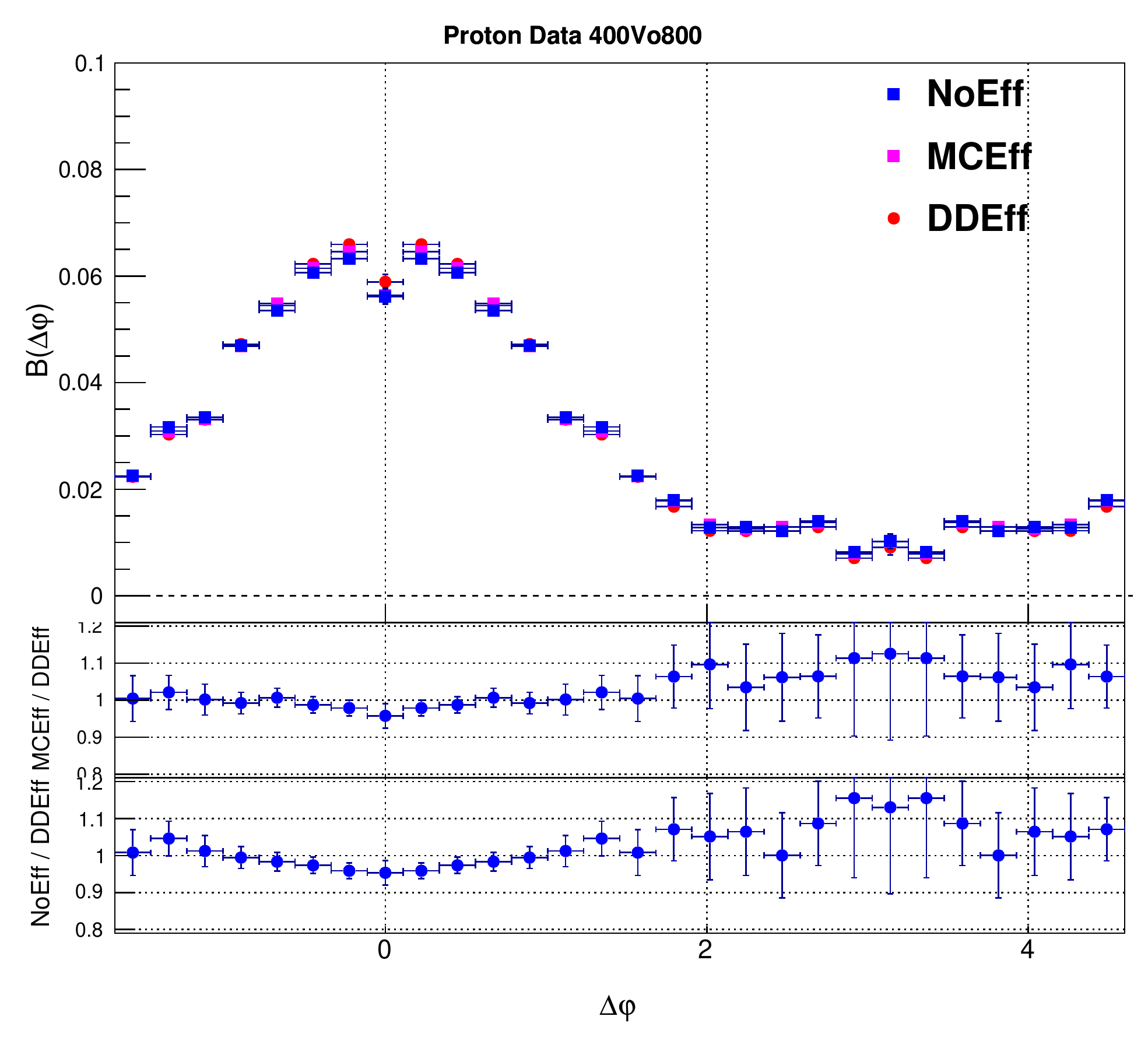} 
  \includegraphics[width=0.32\linewidth]{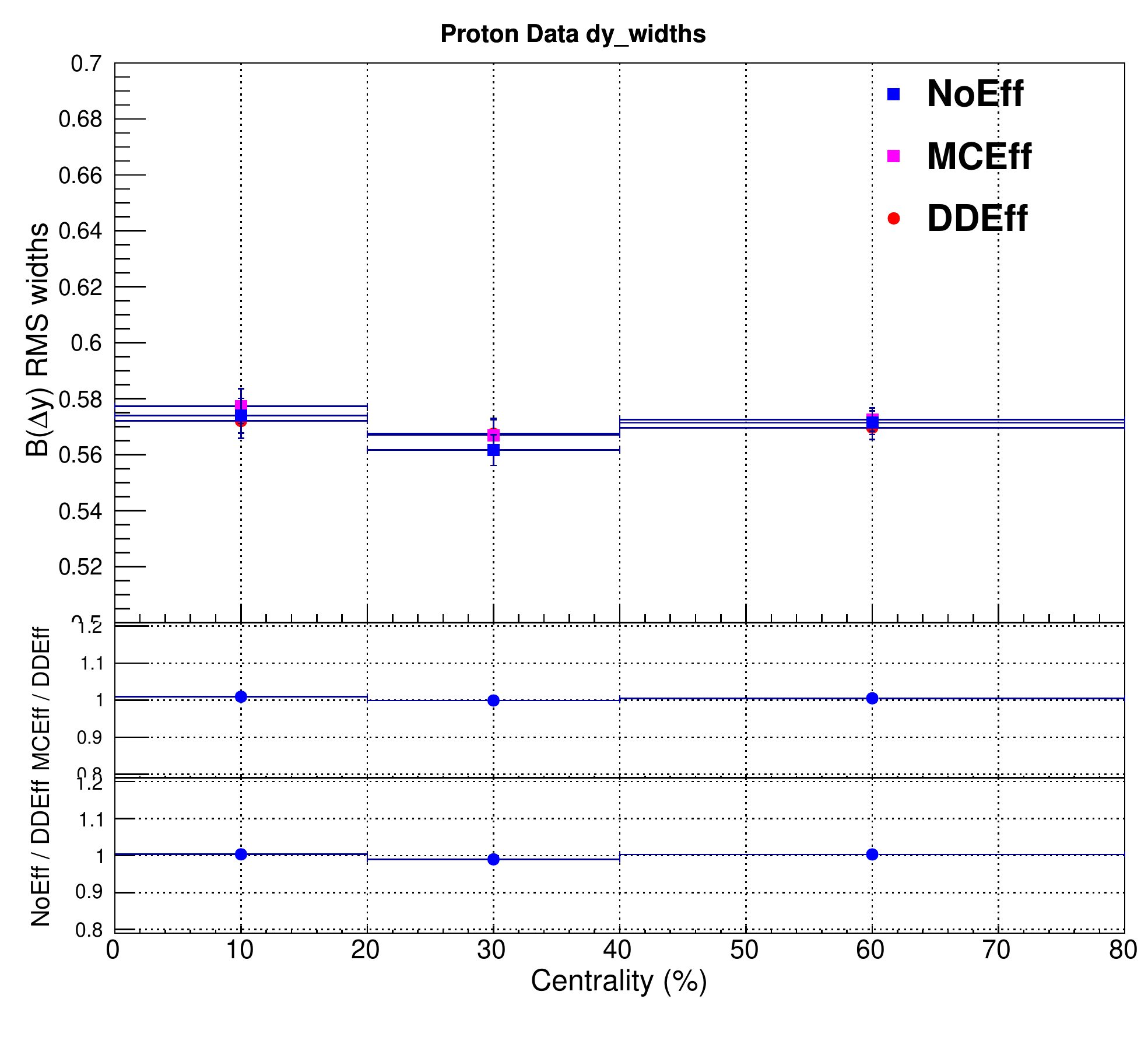}
  \includegraphics[width=0.32\linewidth]{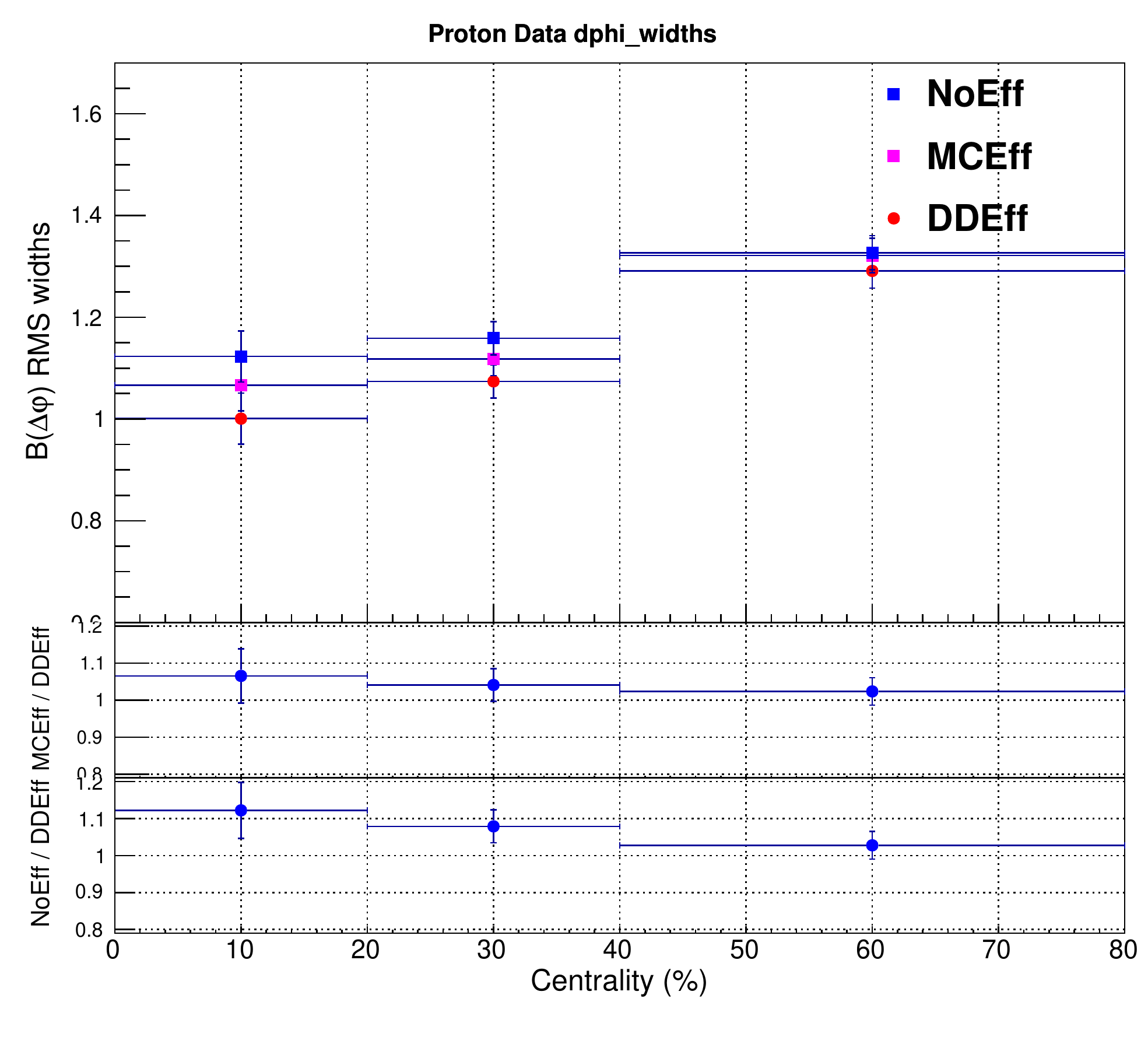}
  \includegraphics[width=0.32\linewidth]{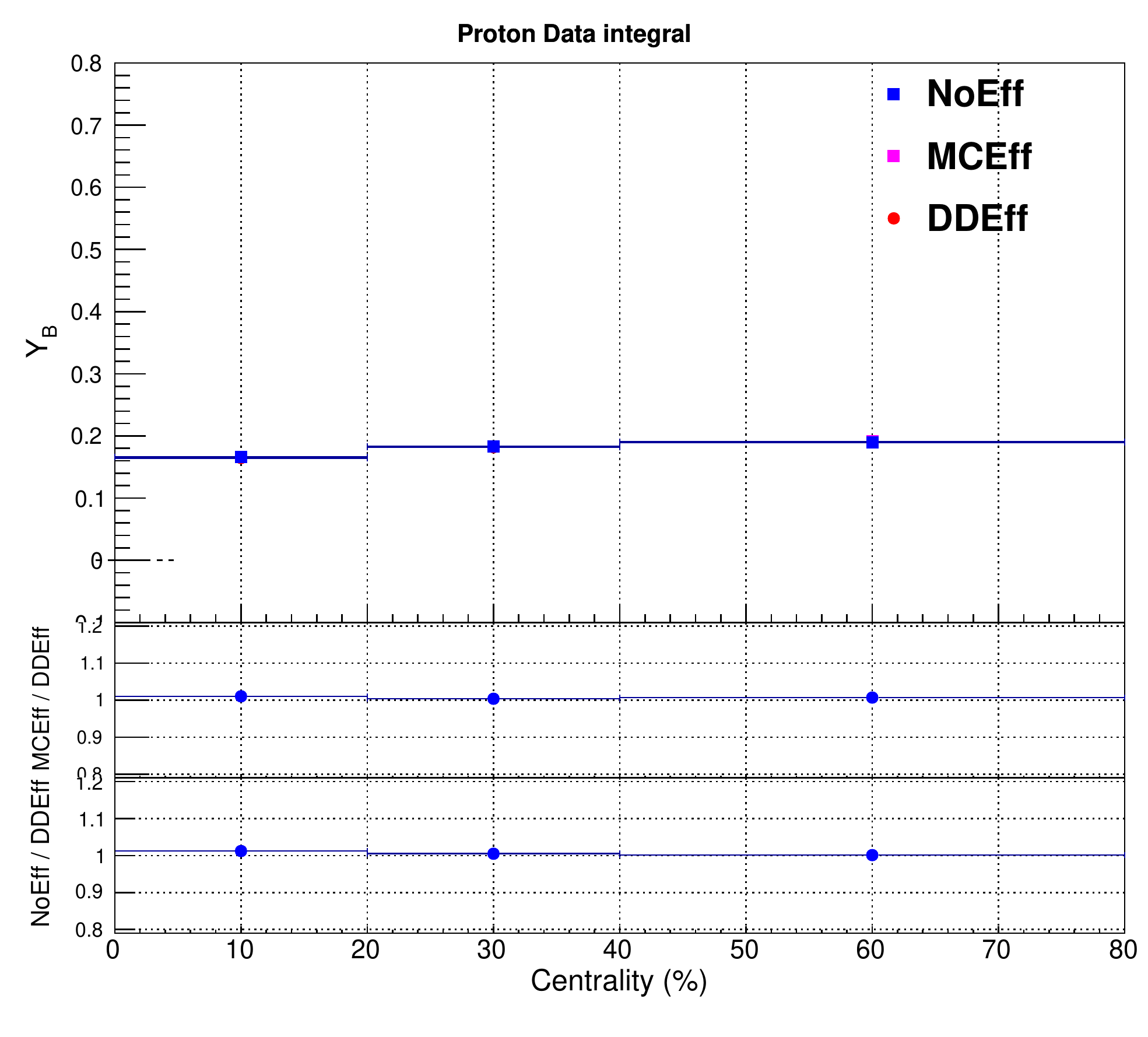} 
  \caption{Comparisons of $B^{pp}$ $\Delta y$ projections (top row), $\Delta \varphi$ projections (middle row), and $\Delta y$ and $\Delta \varphi$ widths, and integrals (bottom row), obtained without, with MC and Data Driven $p_{\rm T}$-dependent efficiency correction.
  In each plot, the second pad presents the ratio of $B^{pp}$ obtained with MC and Data Driven $p_{\rm T}$-dependent efficiency correction, while the third pad exhibits the ratio of $B^{pp}$ obtained without and with Data Driven $p_{\rm T}$-dependent efficiency correction.}
   \label{fig:Compare_DDEffCorr_MCEffCorr_NoEffCorr_BF_dy_dphi_widths_integral_ProtonProton}
\end{figure}

Figures~\ref{fig:Compare_DDEffCorr_MCEffCorr_NoEffCorr_BF_2d_ProtonProton},~\ref{fig:Compare_DDEffCorr_MCEffCorr_NoEffCorr_BF_dy_dphi_widths_integral_ProtonProton} show three sets of $B^{pp}$ results obtained with different $p_{\rm T}$-dependent efficiency corrections. The first set is obtained without a $p_{\rm T}$-dependent efficiency correction.
The second set is based on a MC $p_{\rm T}$-dependent efficiency correction, which uses the difference between the raw and the MC corrected $p_{\rm T}$ spectra as a correction factor.
The third set is obtained with a data driven $p_{\rm T}$-dependent efficiency correction, which uses the difference between the raw and the published $p_{\rm T}$ spectra as a correction factor.
The three sets of BF results are in  good agreement with each other. Differences between the three sets of results  are within 1\% for $\Delta y$ widths and BF integrals, whereas the differences between $\Delta\varphi$ widths are within a few percent, although  both the MC and data driven corrections tend to make the BF slightly narrower in $\Delta\varphi$. 
This bias may in part be explained by the fact that   the MC and data driven corrections add more weight to higher $p_{\rm T}$ (TOF region) particles in the sample. High-$p_{\rm T}$ particles, particularly those produced by fragmentation of jets, are in general more likely to be emitted at smaller relative angle  $\Delta\varphi$. Increasing the relative weight of the high-$p_{\rm T}$ pairs in  the sample should
thus produce a slight narrowing of the correlation functions and balance functions averaged over the range measured in this work. 
Since the ``MC corrected"  results  lie in-between the ``uncorrected"  and ``data driven corrected" results, we use 
the difference between the latter  two sets to estimate  systematic uncertainties associated with a $p_{\rm T}$-dependent efficiency correction.

\begin{figure}
\centering
  \includegraphics[width=0.3\linewidth]{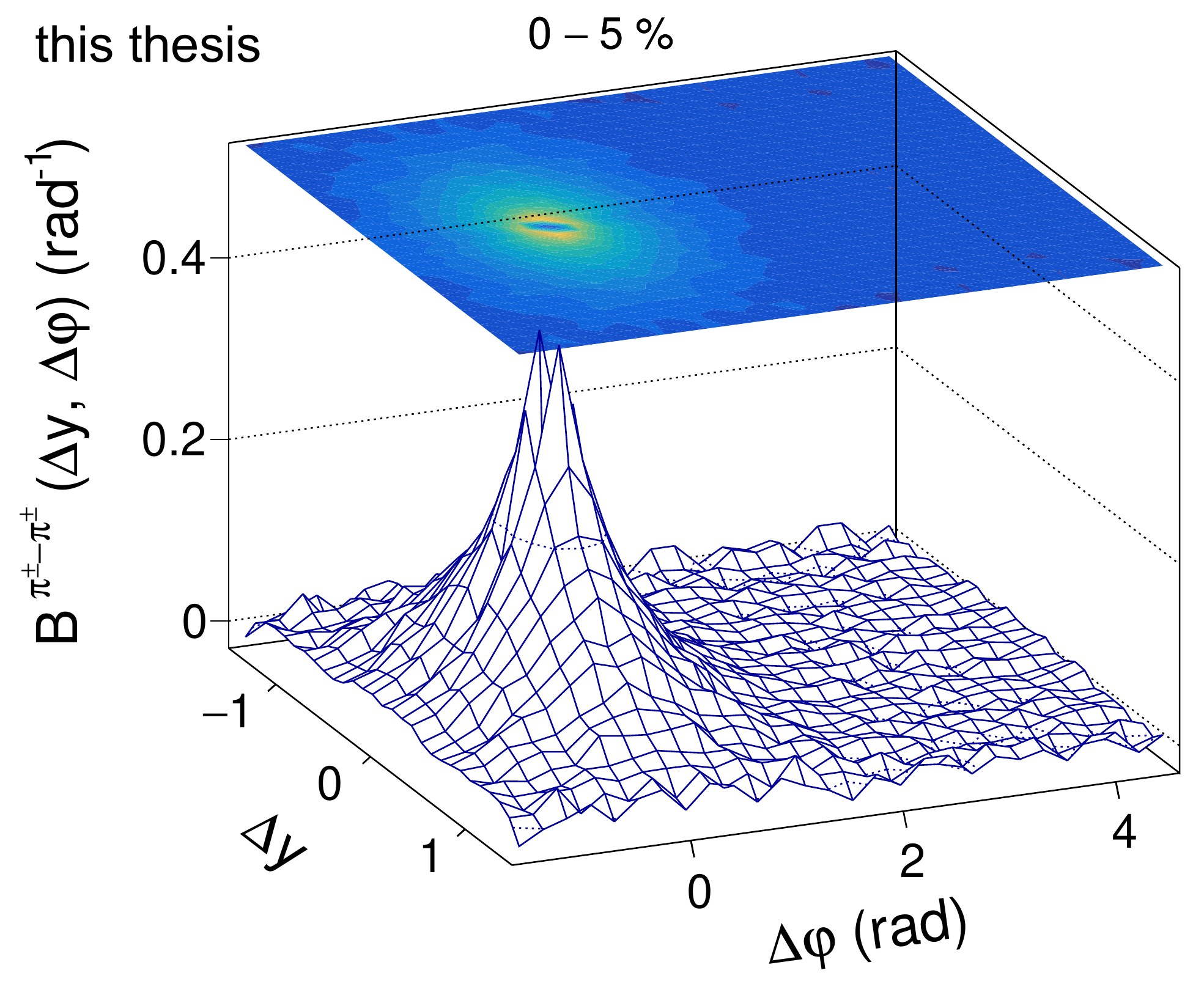}
  \includegraphics[width=0.3\linewidth]{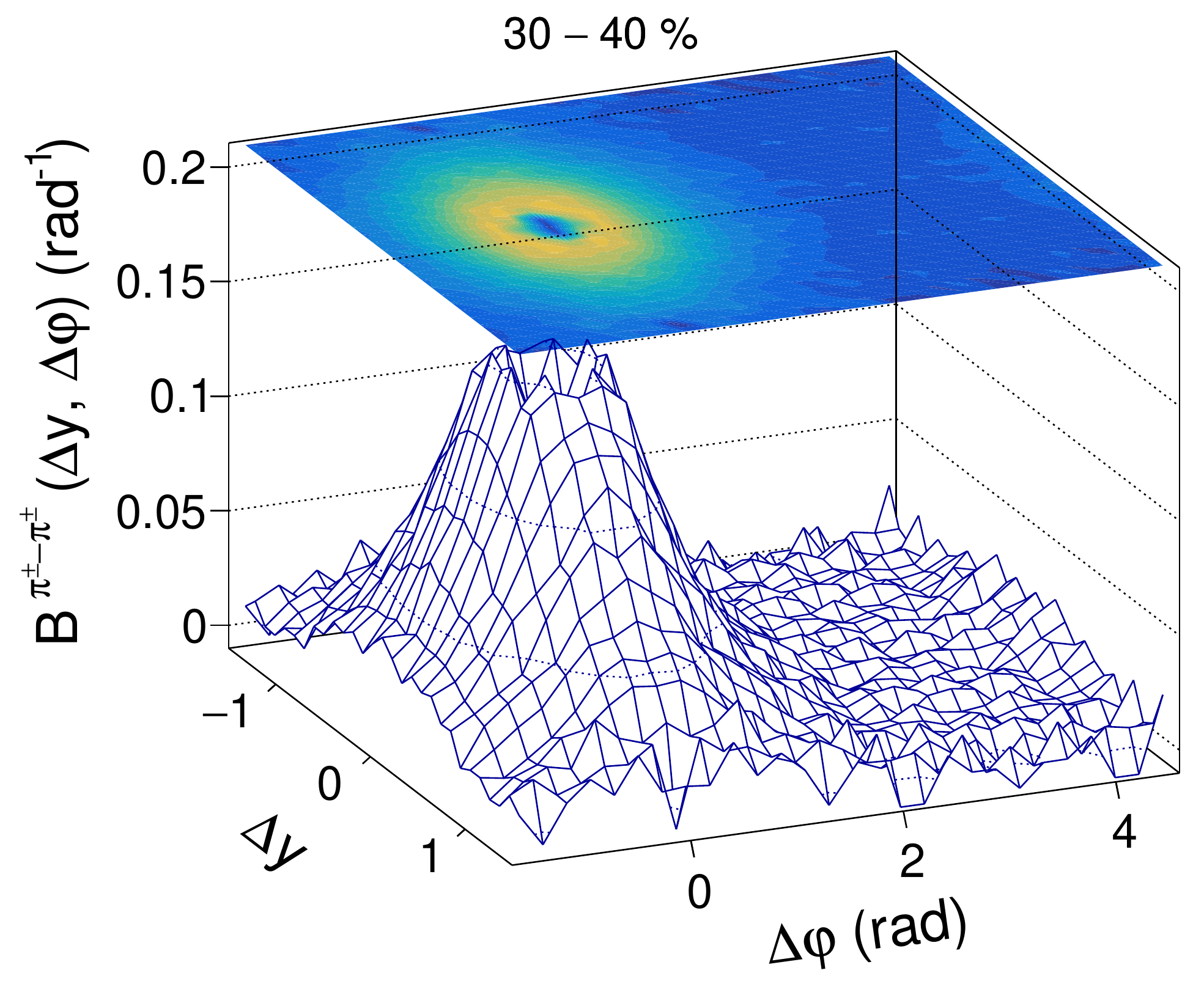}
  \includegraphics[width=0.3\linewidth]{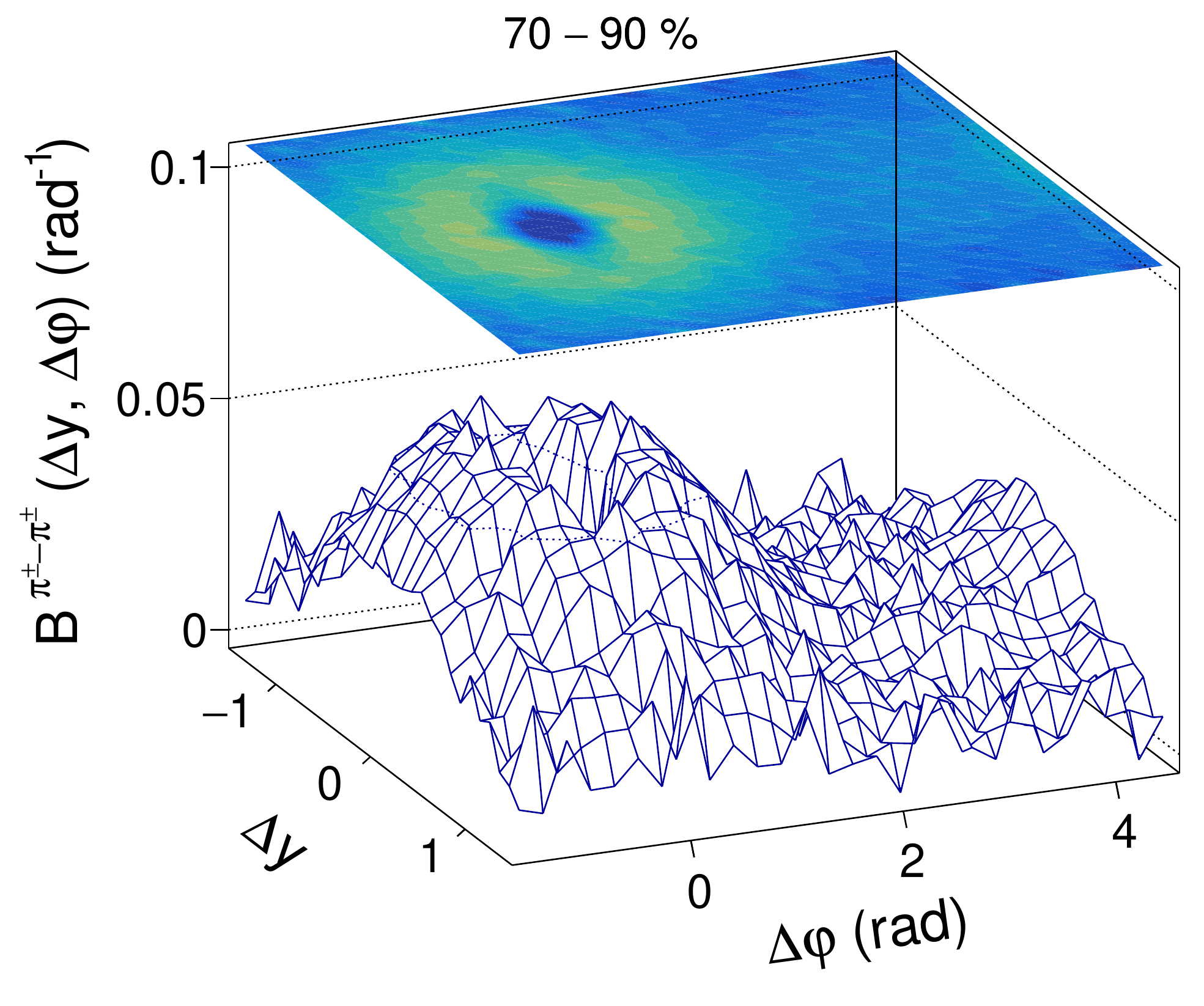}
  \includegraphics[width=0.3\linewidth]{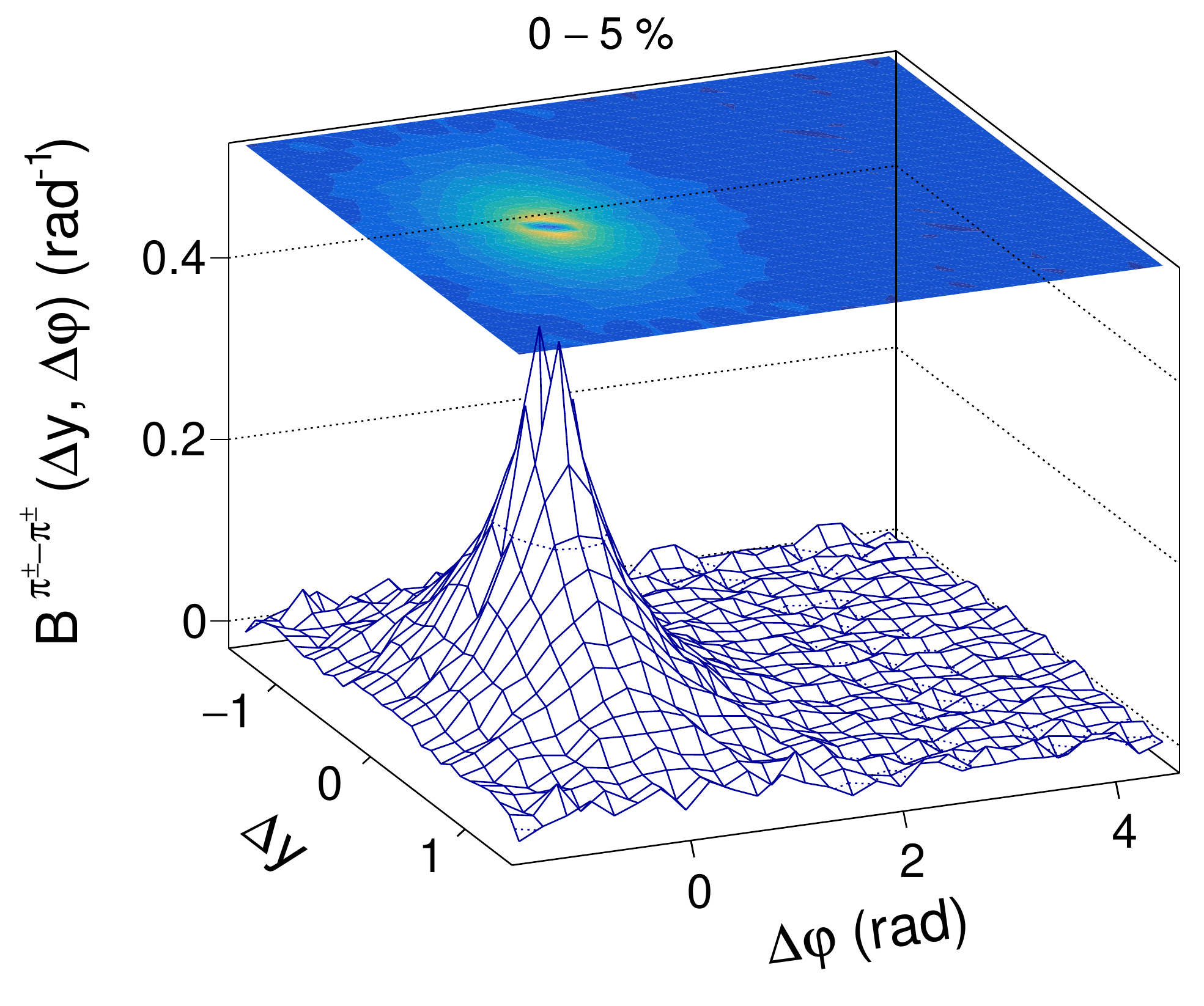}
  \includegraphics[width=0.3\linewidth]{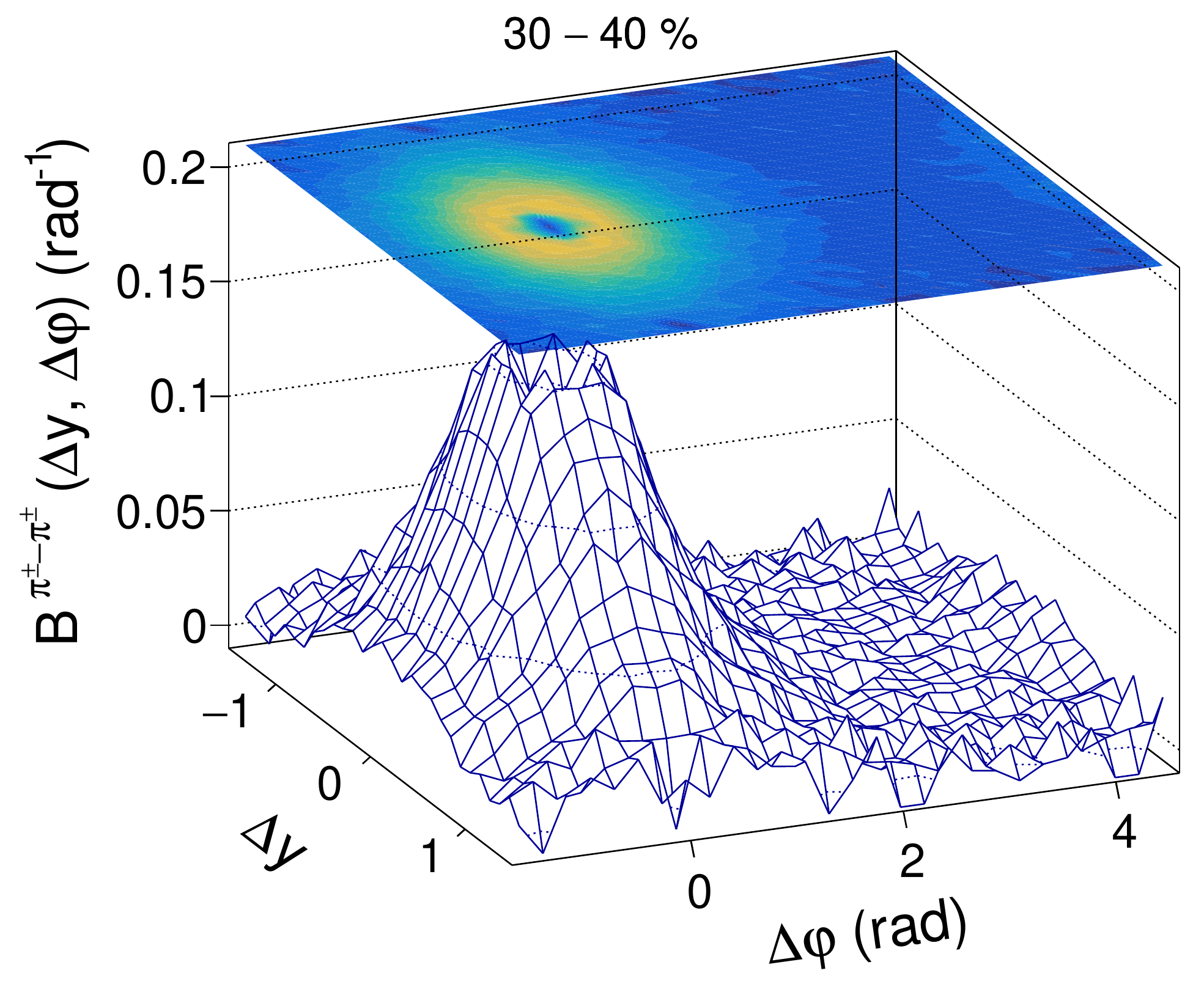}
  \includegraphics[width=0.3\linewidth]{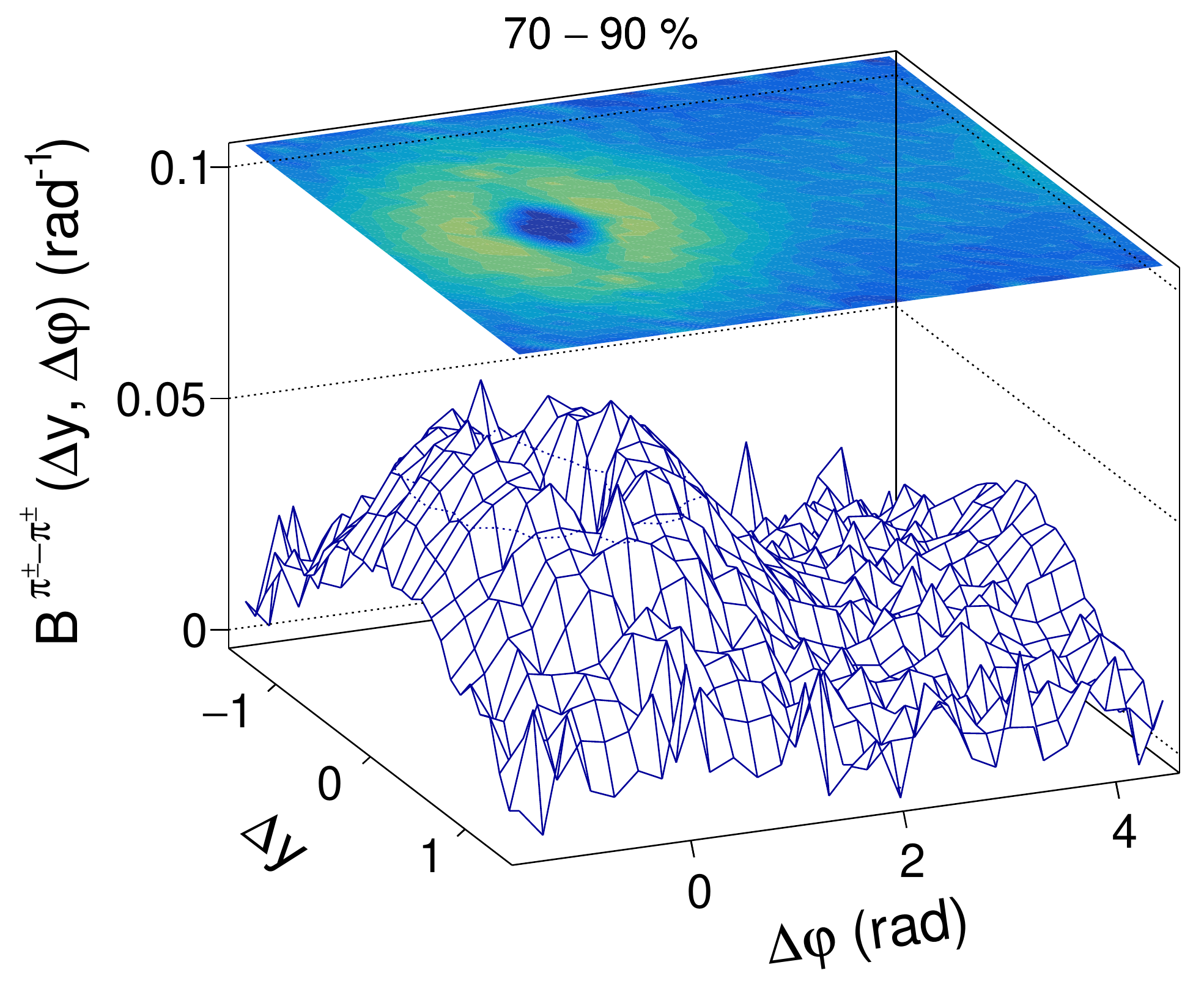}
  \includegraphics[width=0.3\linewidth]{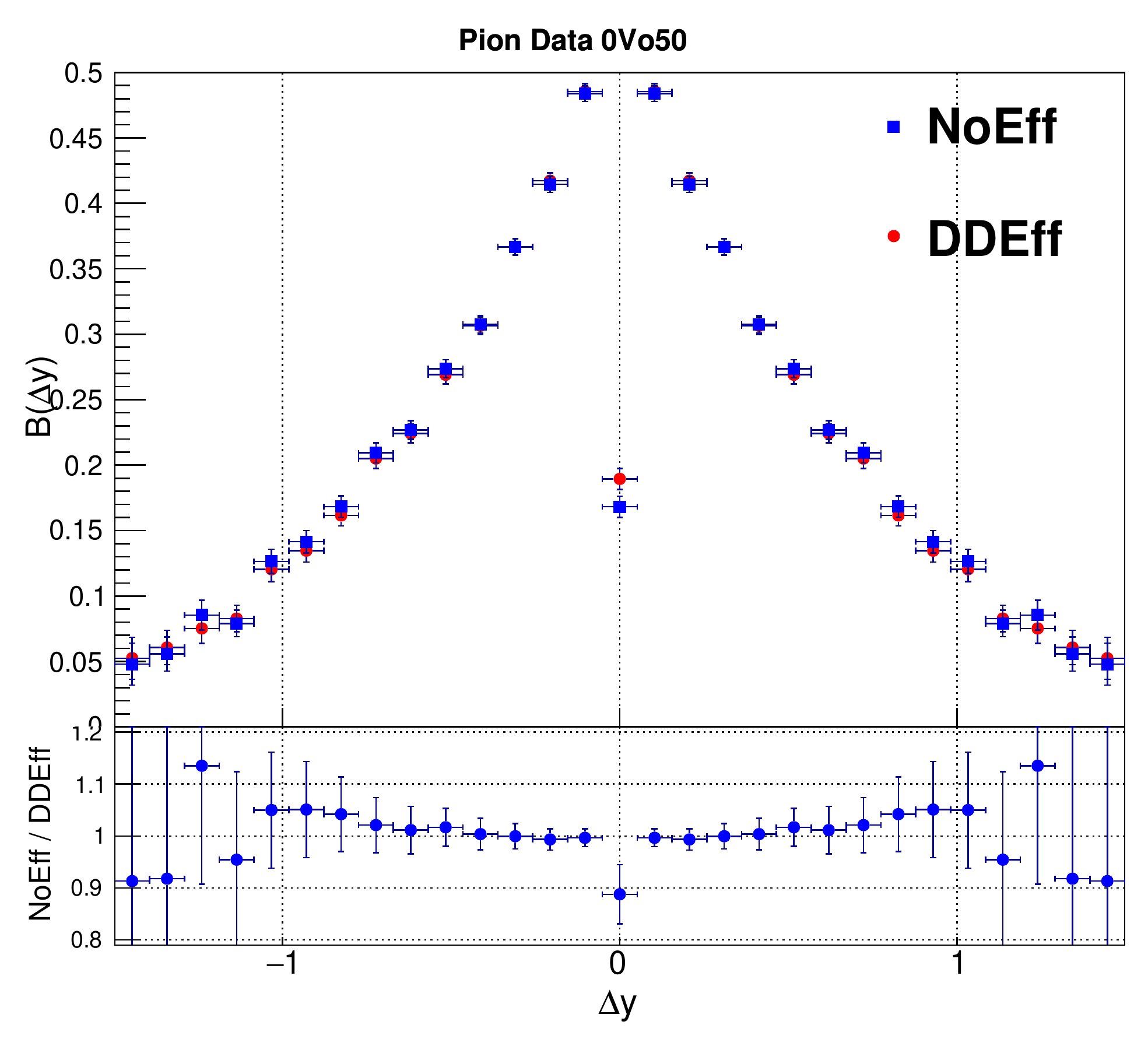}
  \includegraphics[width=0.3\linewidth]{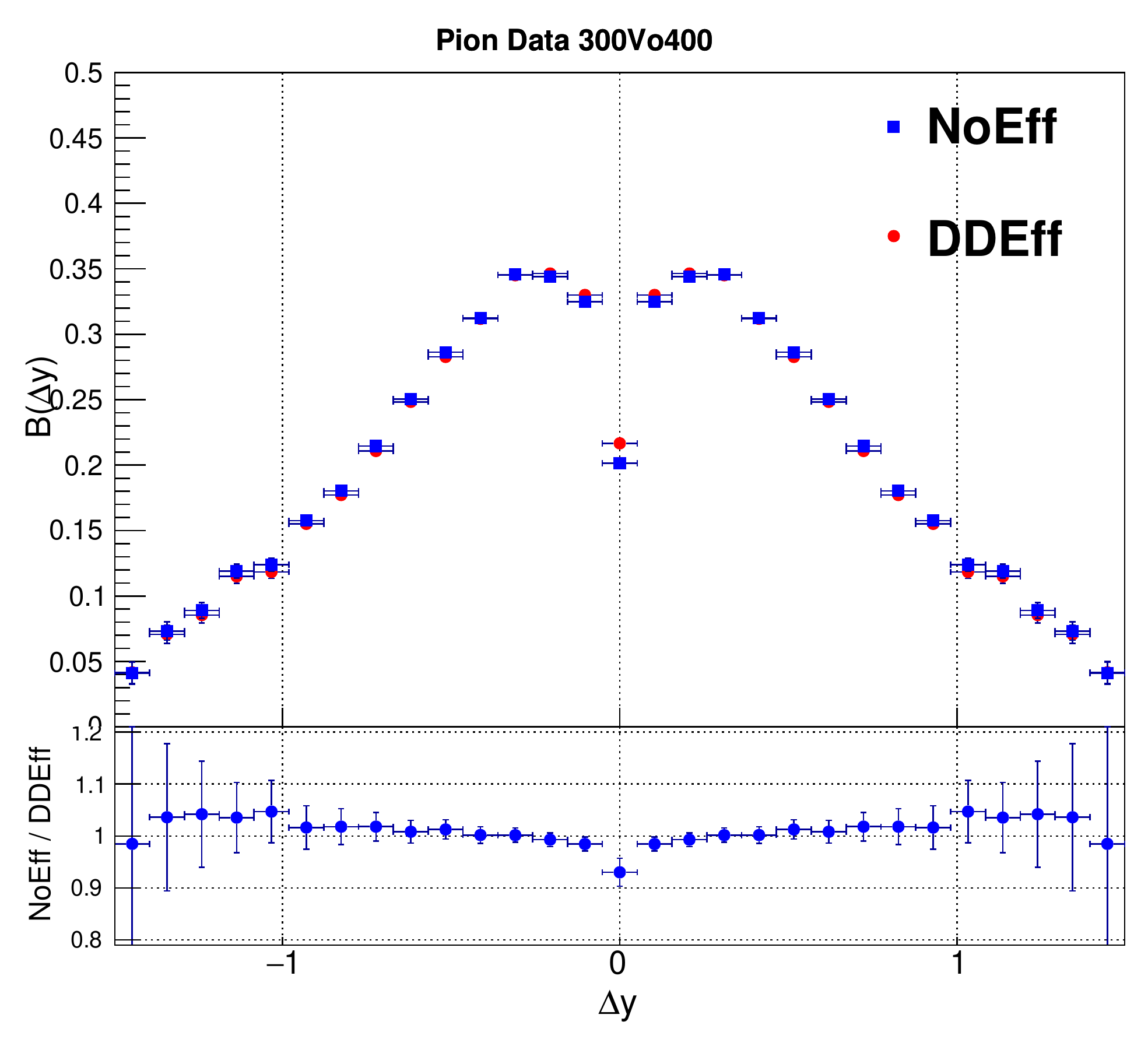}
  \includegraphics[width=0.3\linewidth]{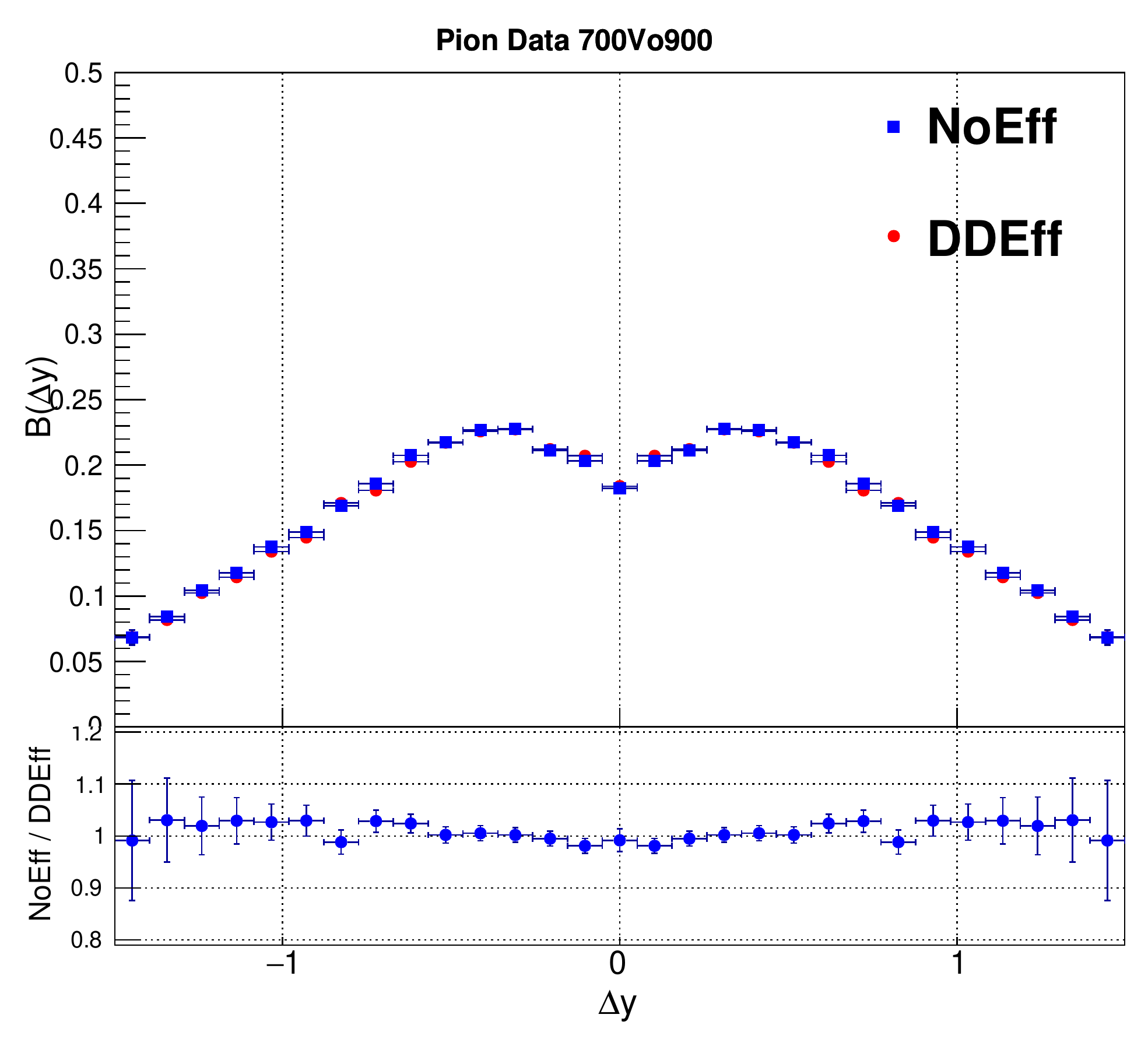} 
  \includegraphics[width=0.3\linewidth]{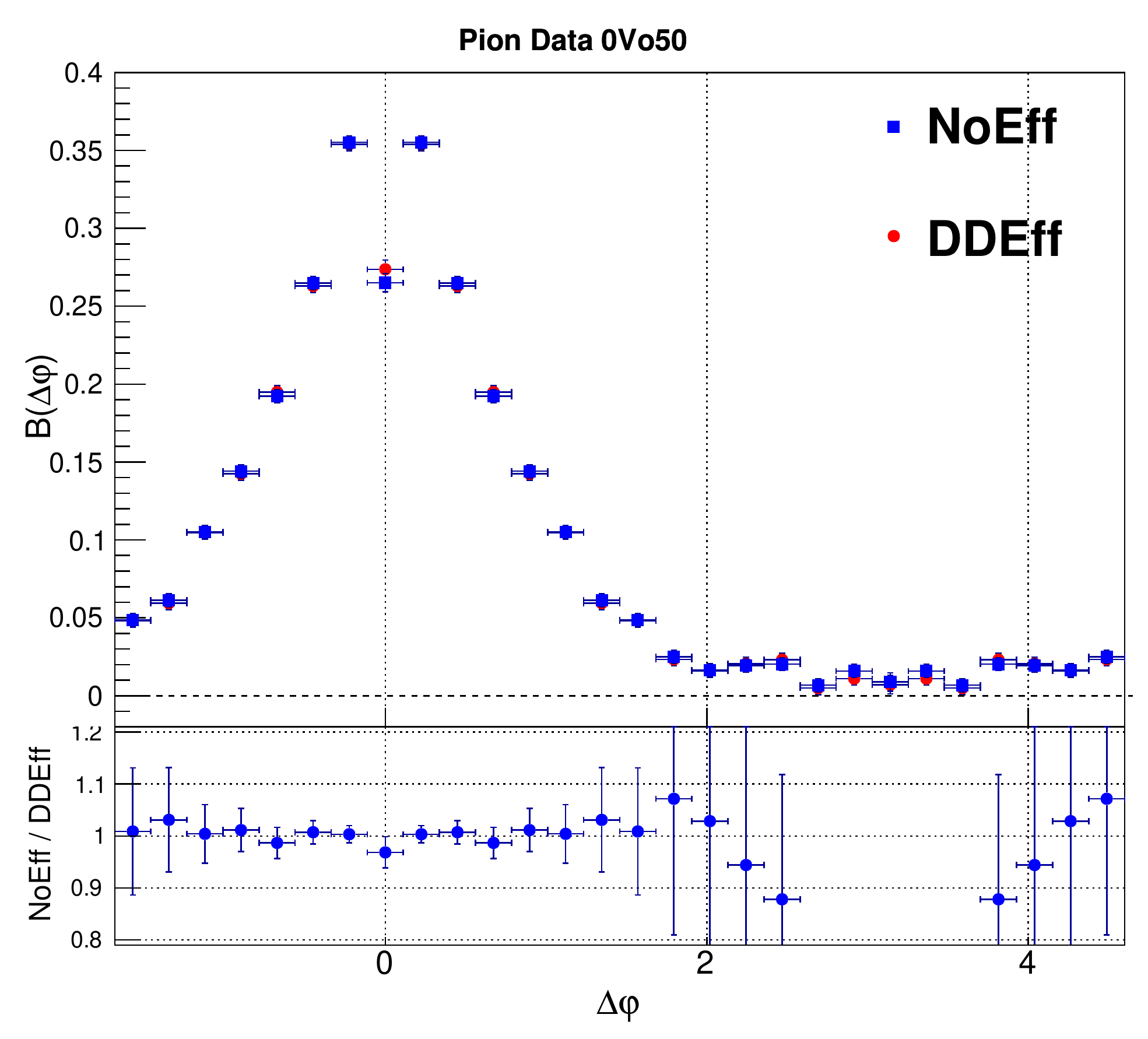}
  \includegraphics[width=0.3\linewidth]{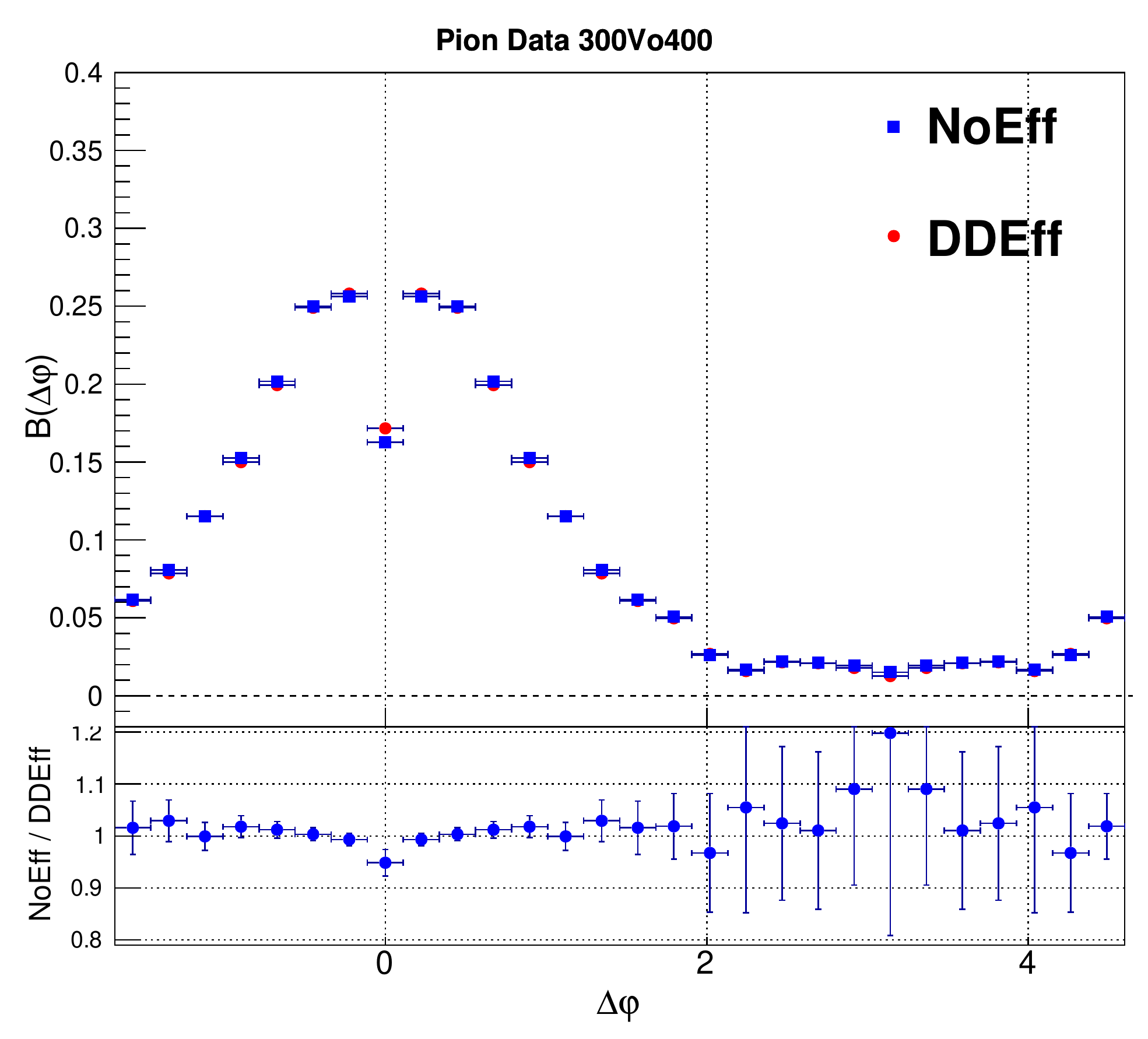}
  \includegraphics[width=0.3\linewidth]{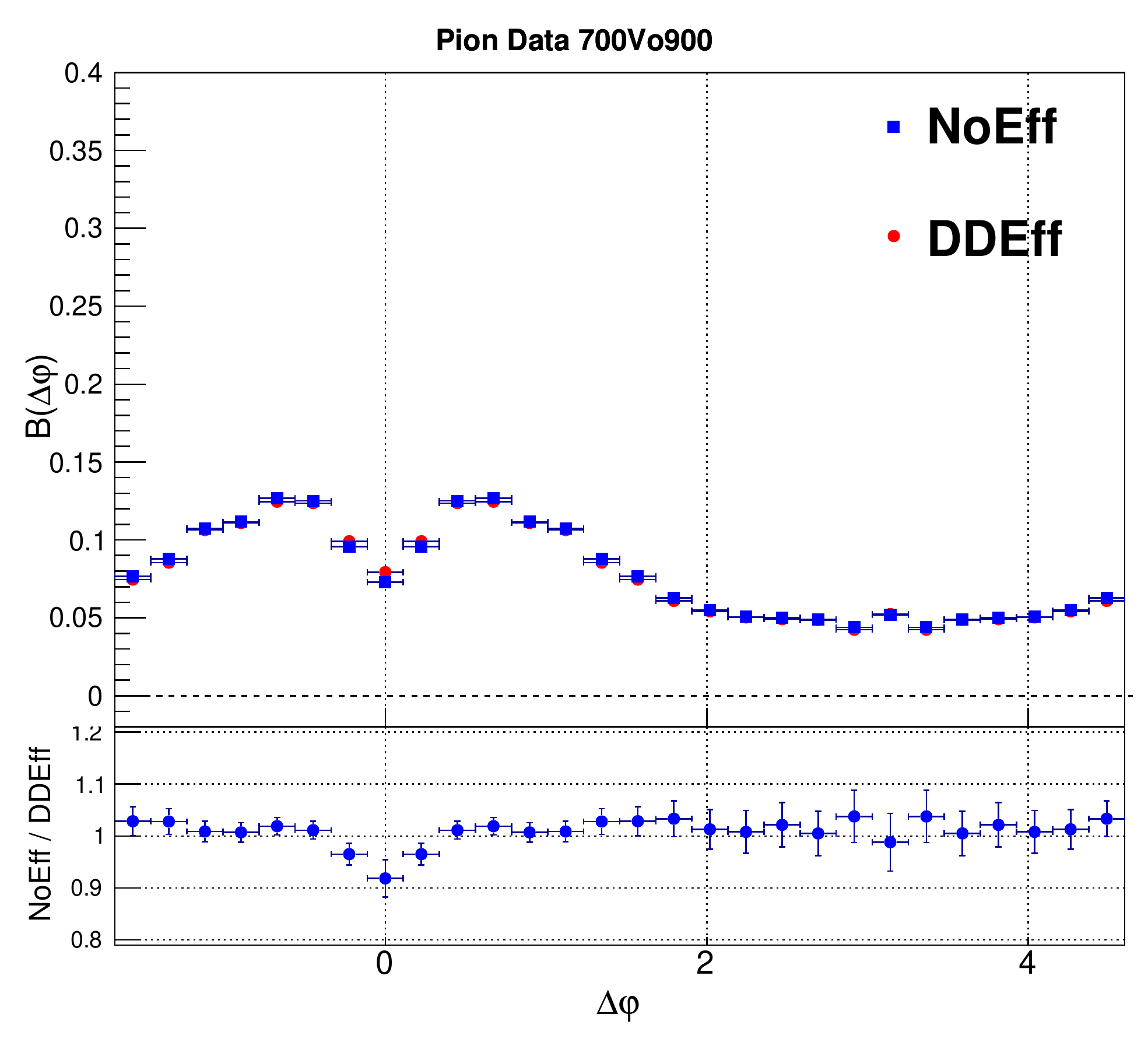} 
  \includegraphics[width=0.3\linewidth]{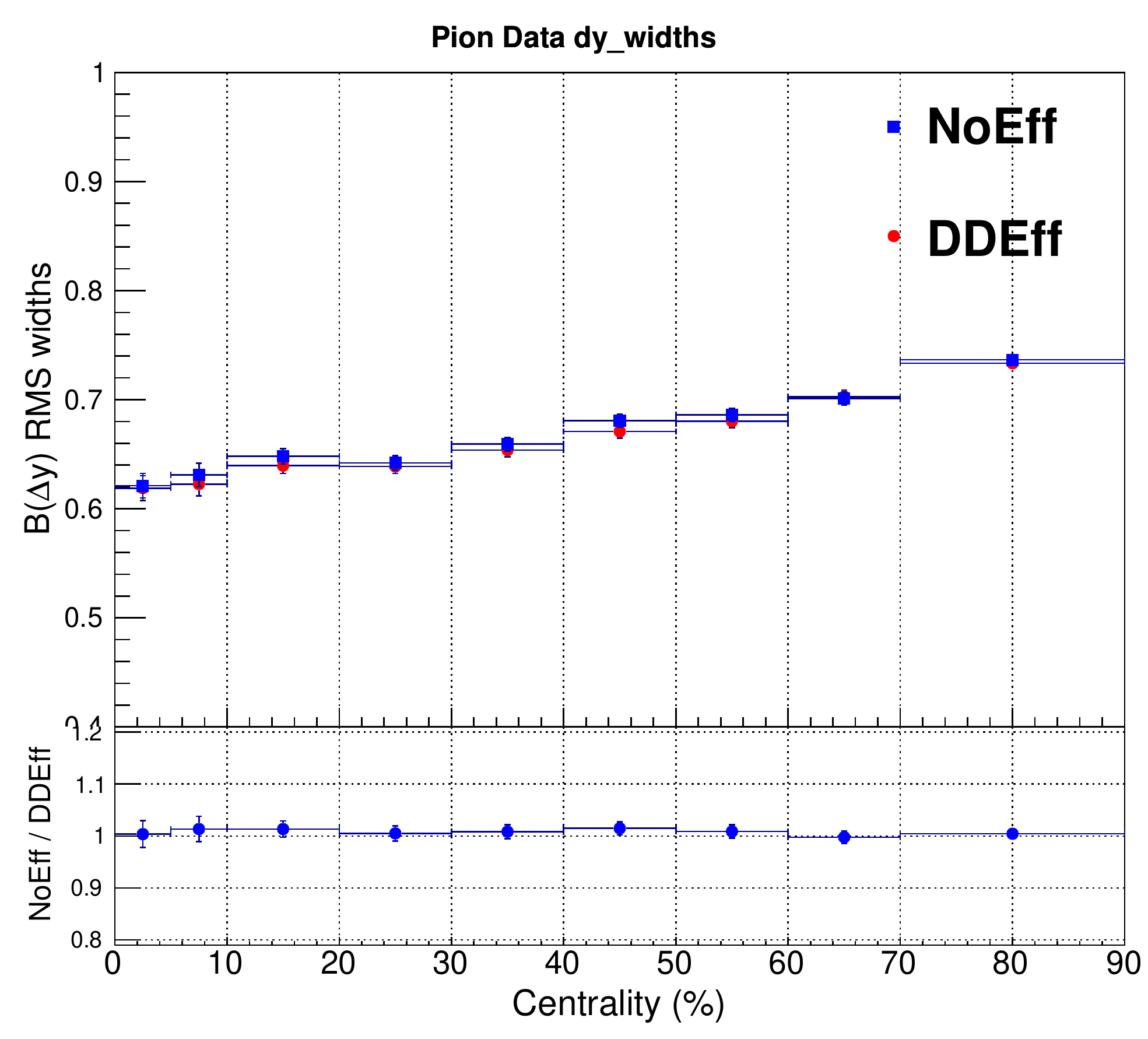}
  \includegraphics[width=0.3\linewidth]{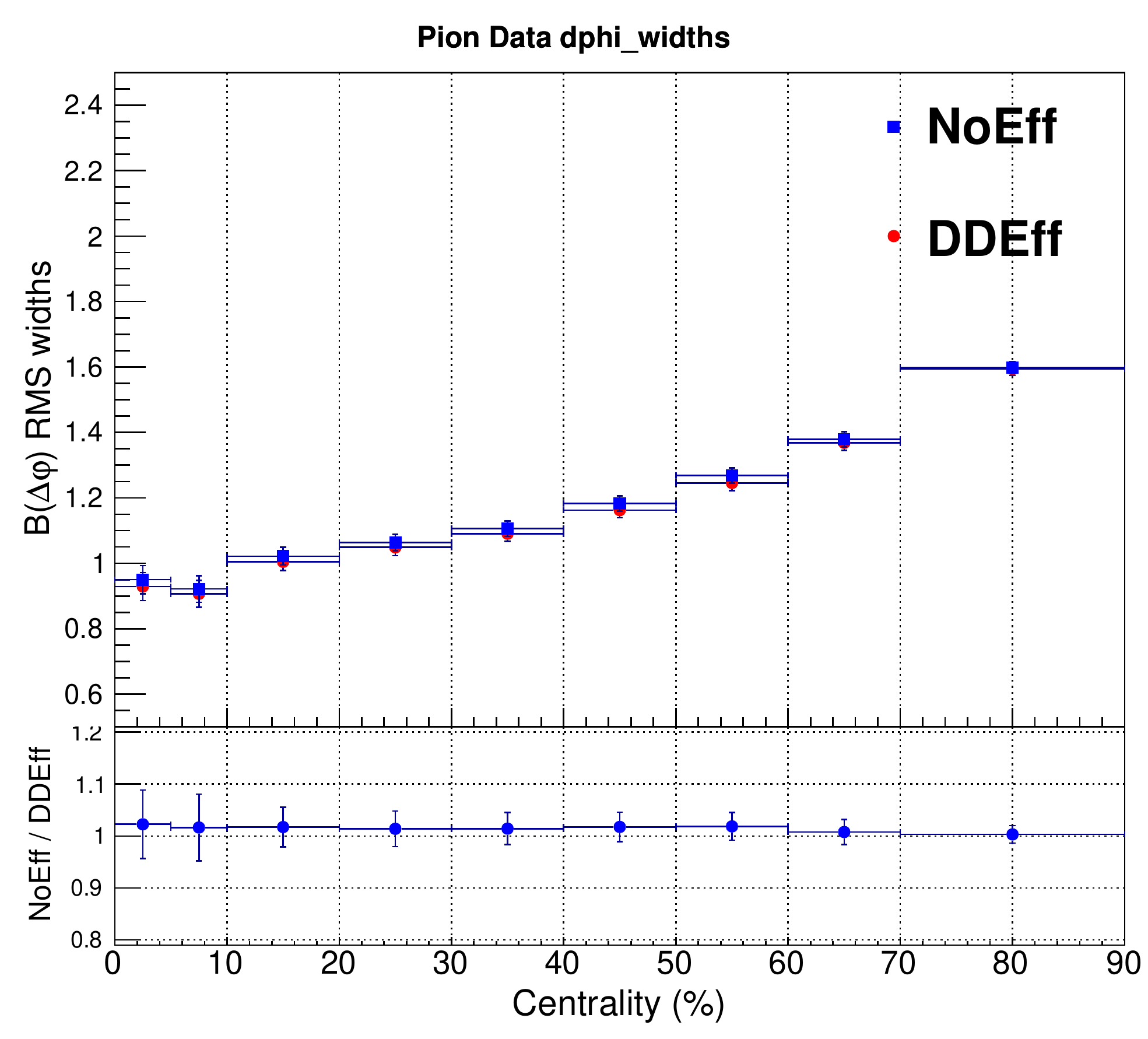}
  \includegraphics[width=0.3\linewidth]{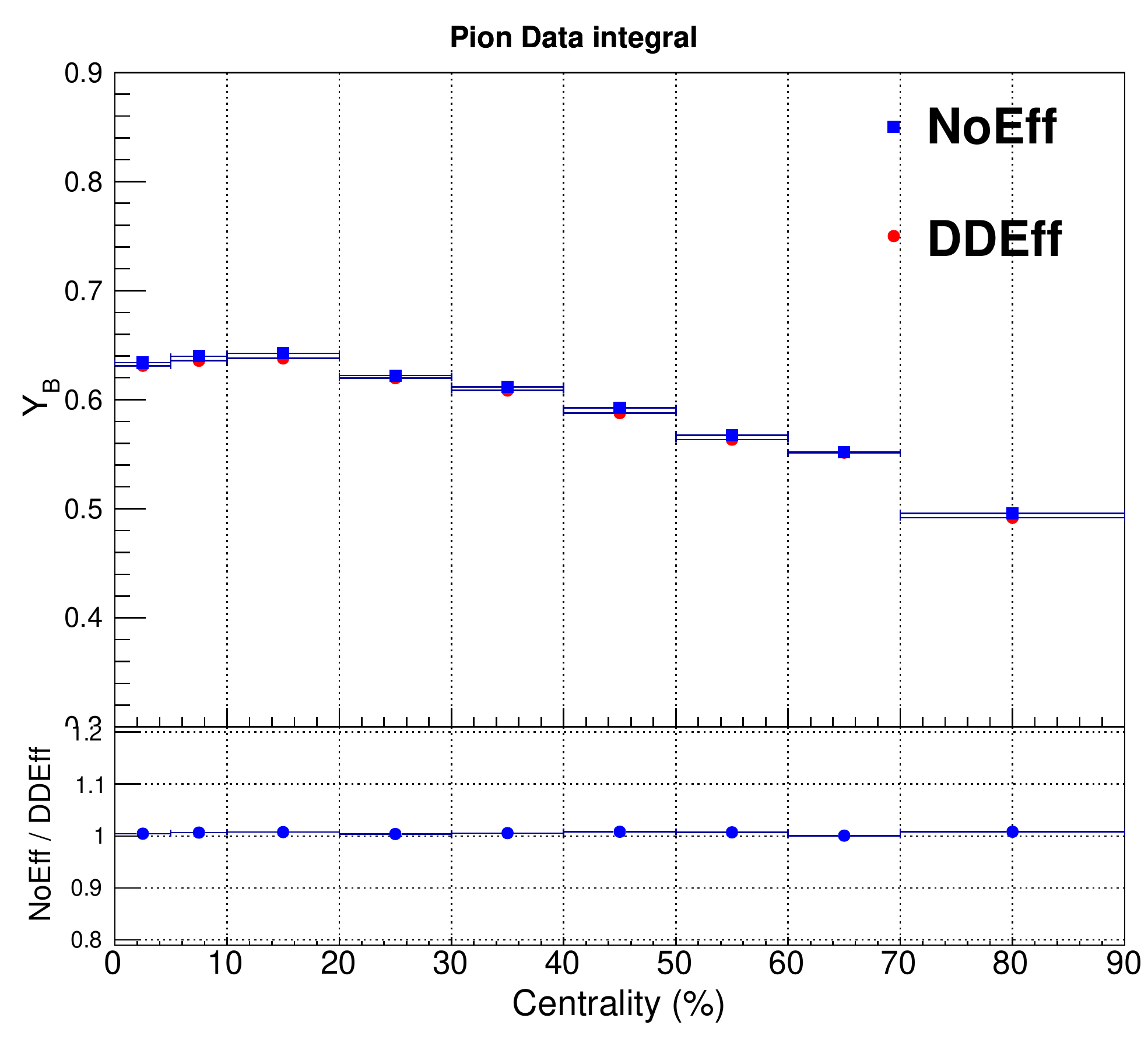}
  \caption{Comparisons of 2D $B^{\pi\pi}$ obtained without (1st row) and with Data Driven (2nd row) $p_{\rm T}$-dependent efficiency correction for selected centralities, along with their $\Delta y$ (3rd row) and $\Delta \varphi$ projections (4th row), $\Delta y$ and $\Delta \varphi$ widths, and integrals (5th row).}
   \label{fig:Compare_DDEffCorr_NoEffCorr_BF_PionPion}
\end{figure}

\begin{figure}
\centering
  \includegraphics[width=0.3\linewidth]{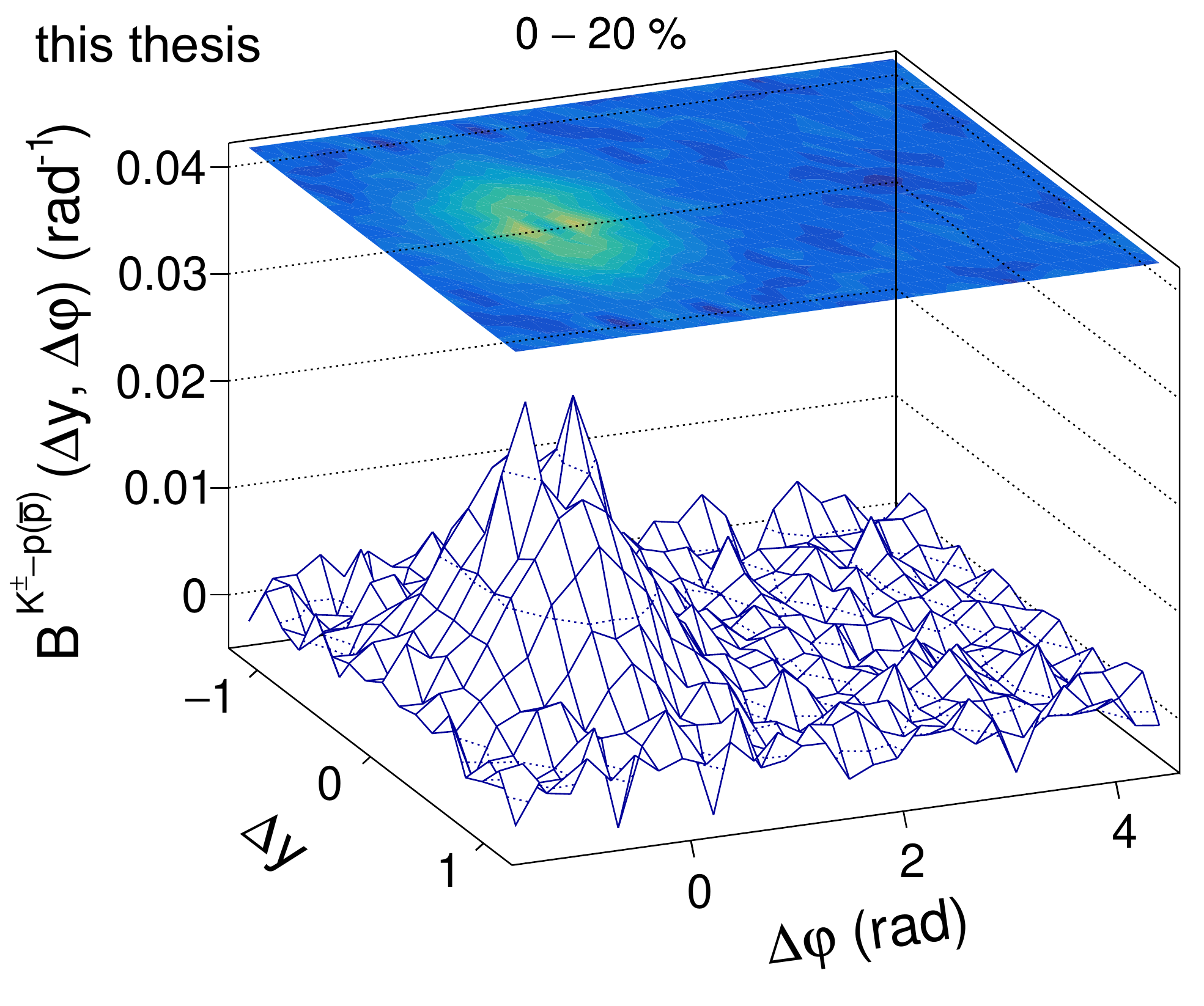}
  \includegraphics[width=0.3\linewidth]{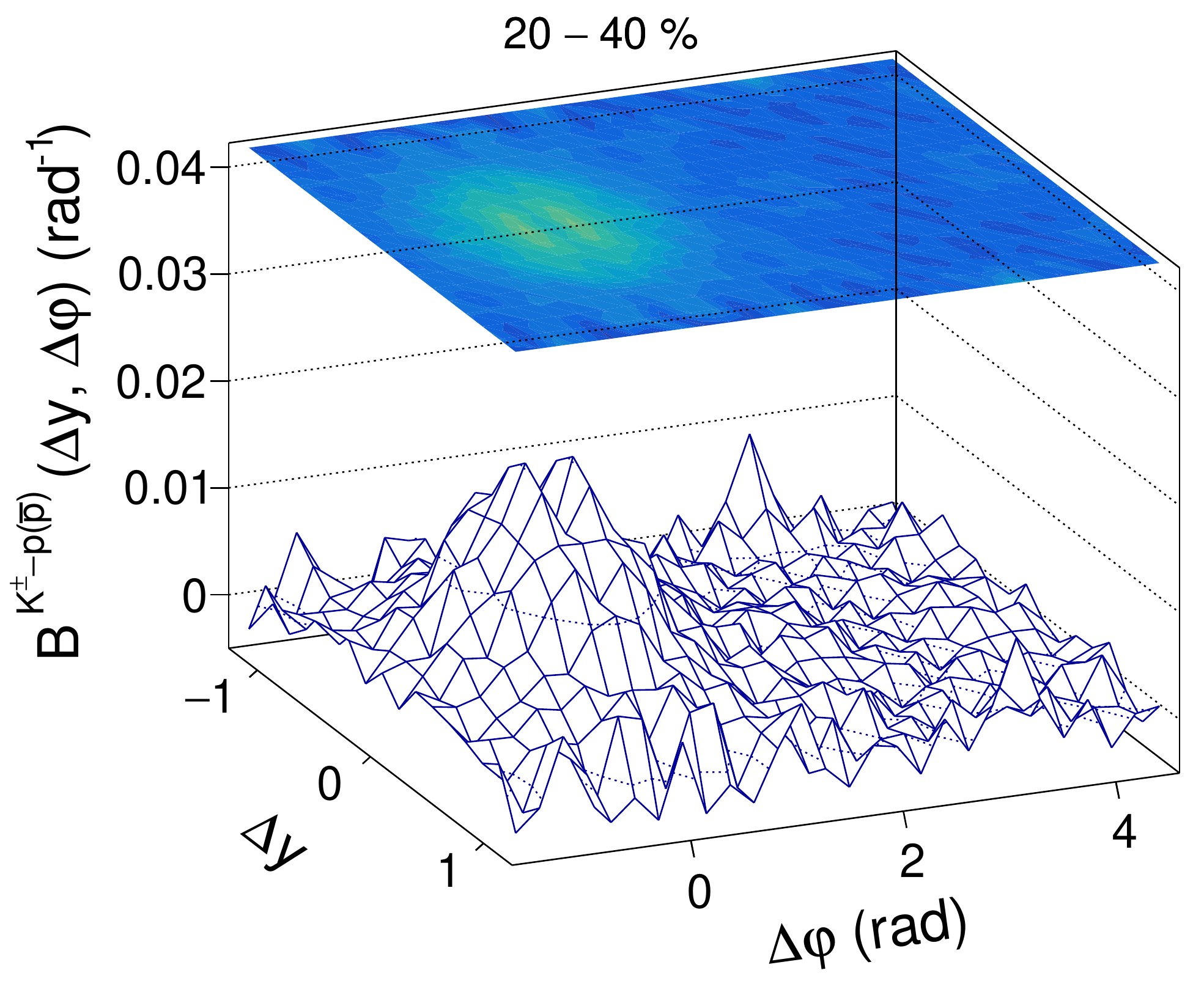}
  \includegraphics[width=0.3\linewidth]{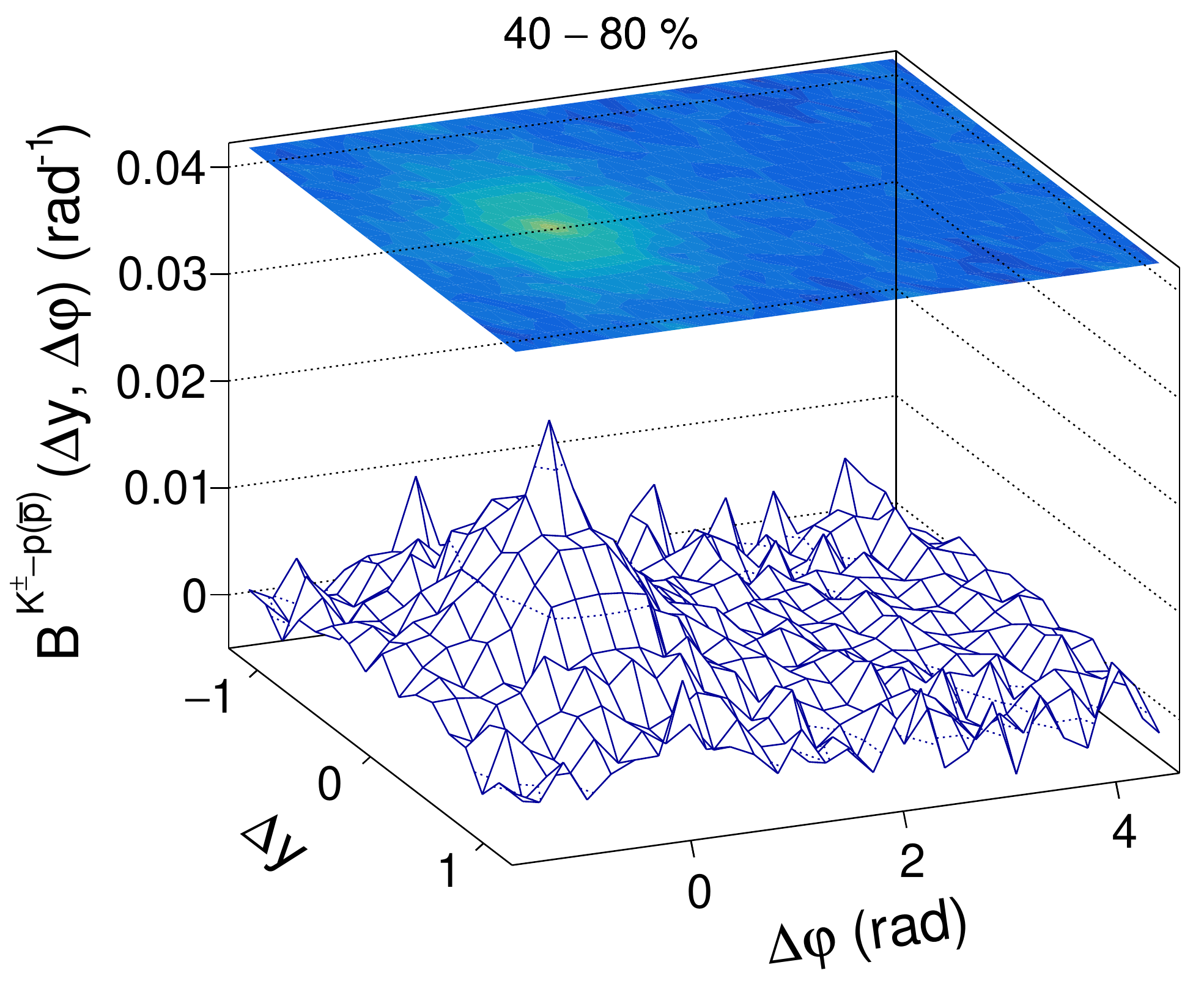}
  \includegraphics[width=0.3\linewidth]{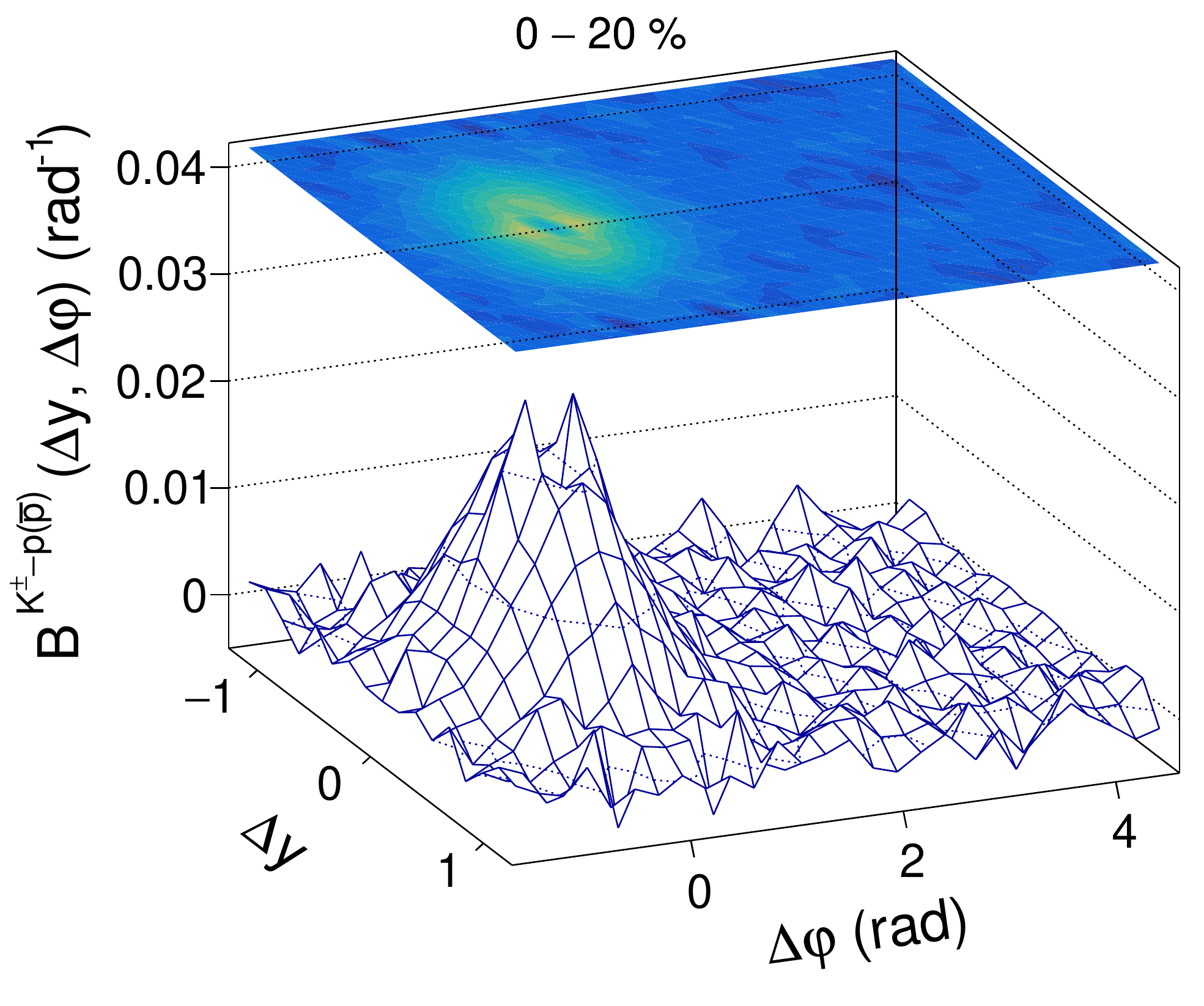}
  \includegraphics[width=0.3\linewidth]{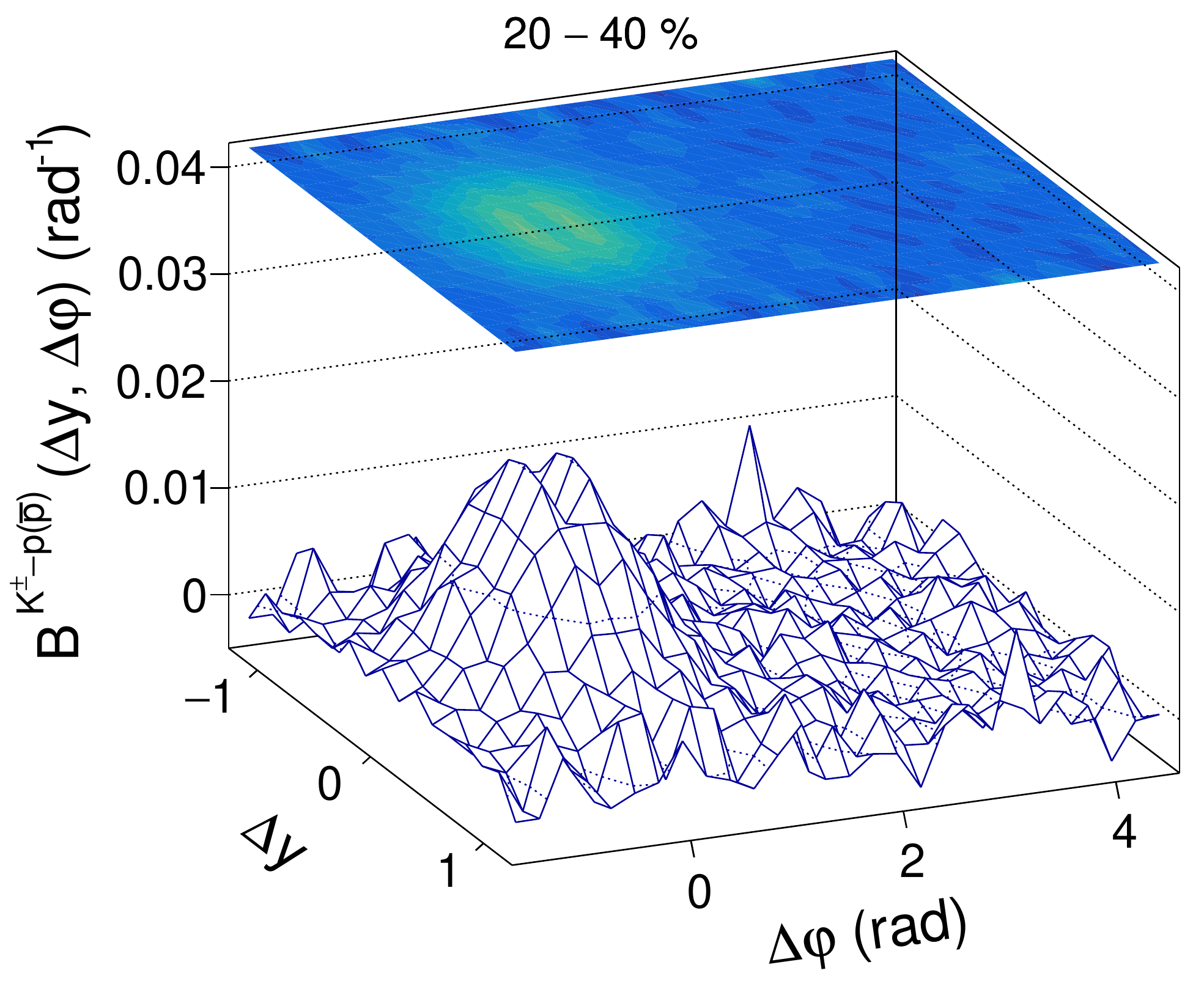}
  \includegraphics[width=0.3\linewidth]{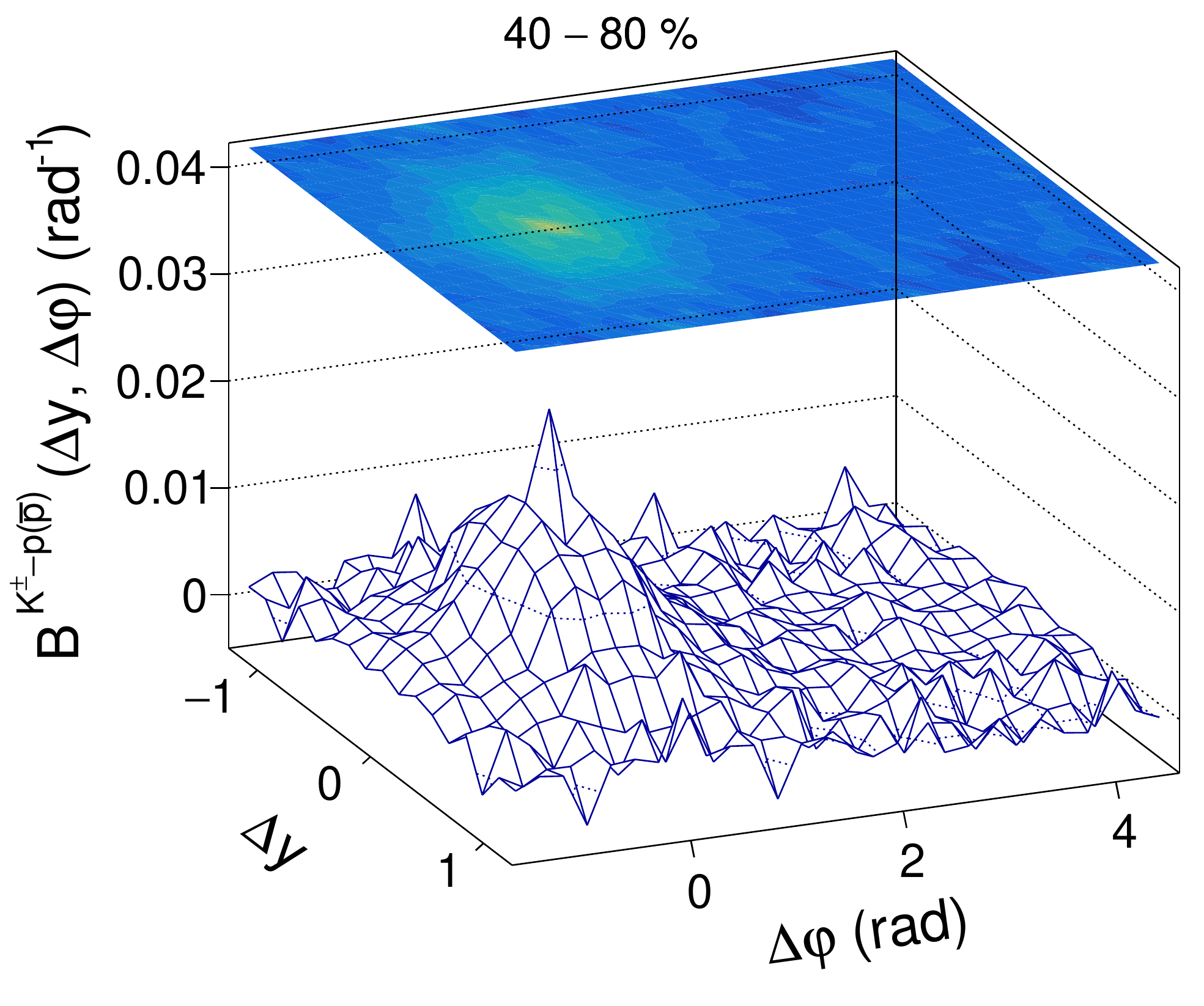}
  \includegraphics[width=0.3\linewidth]{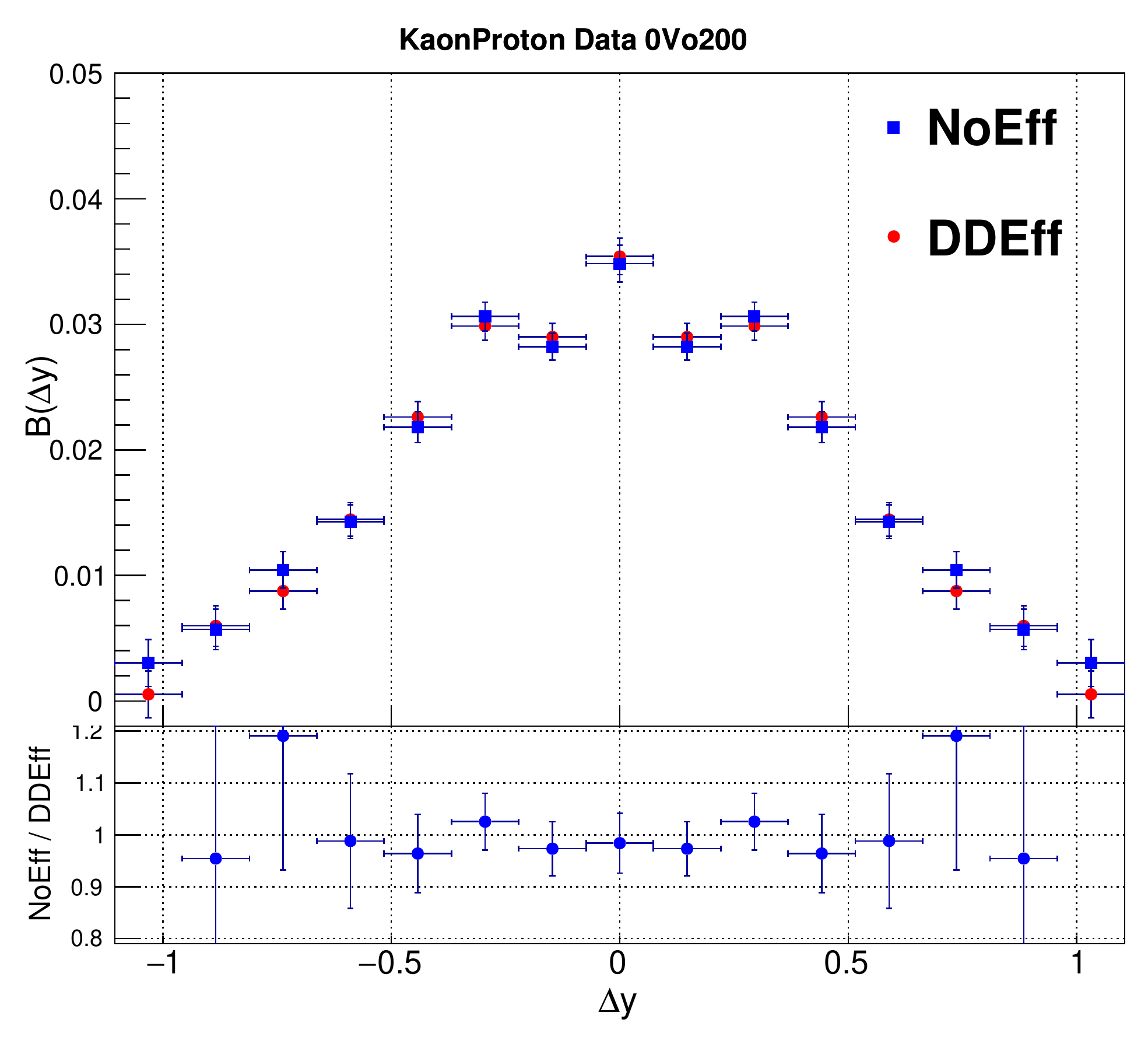}
  \includegraphics[width=0.3\linewidth]{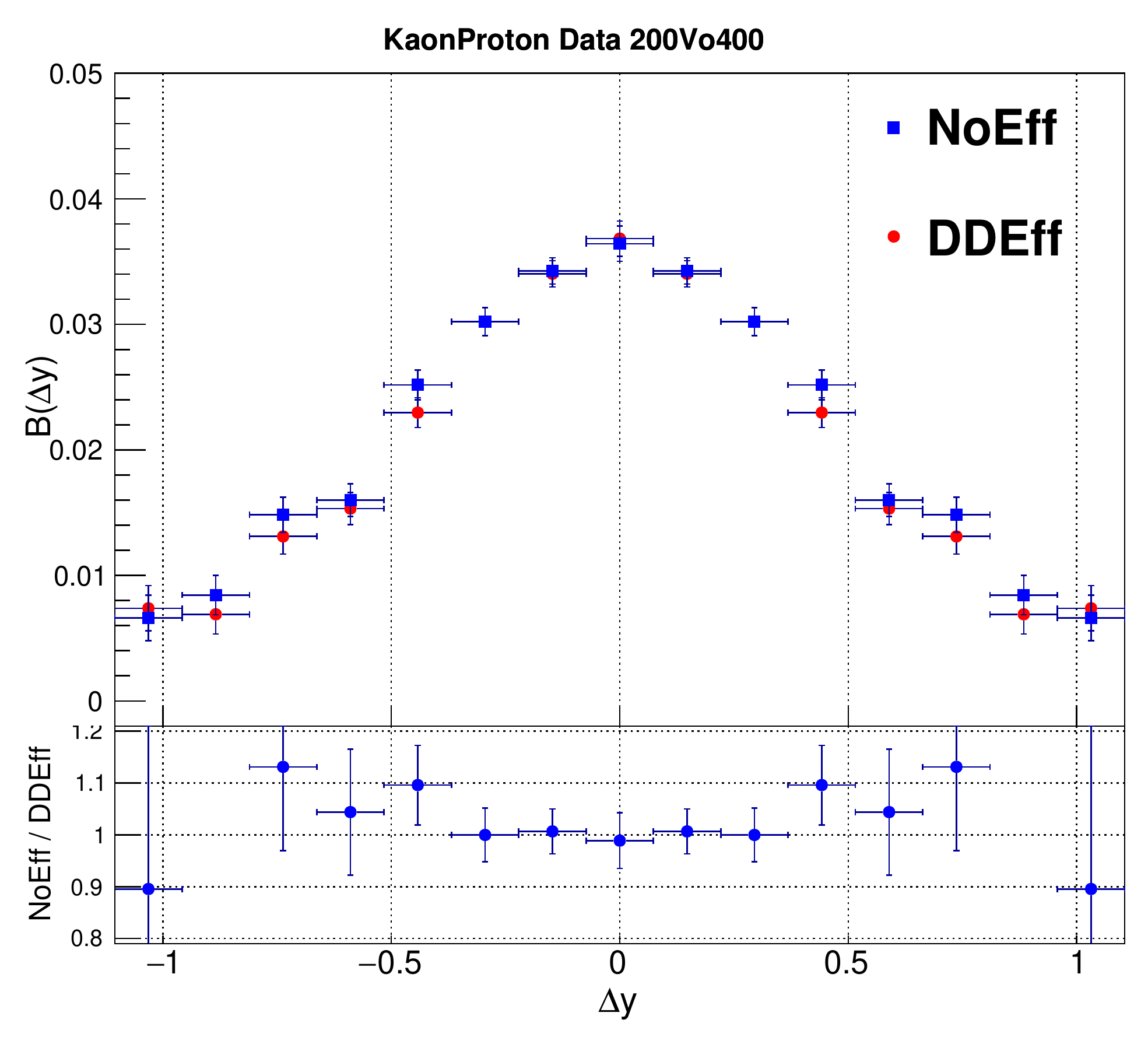}
  \includegraphics[width=0.3\linewidth]{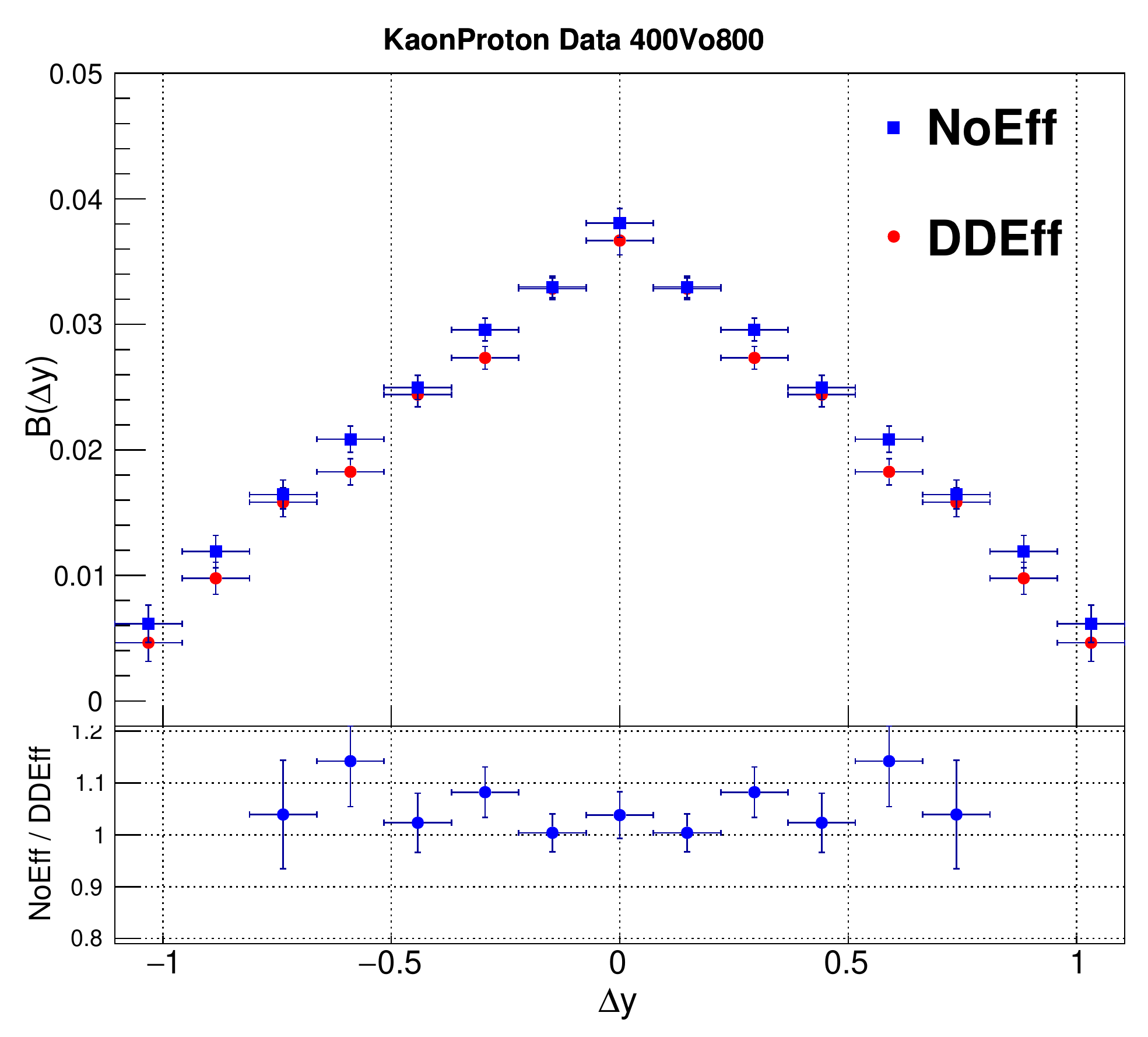} 
  \includegraphics[width=0.3\linewidth]{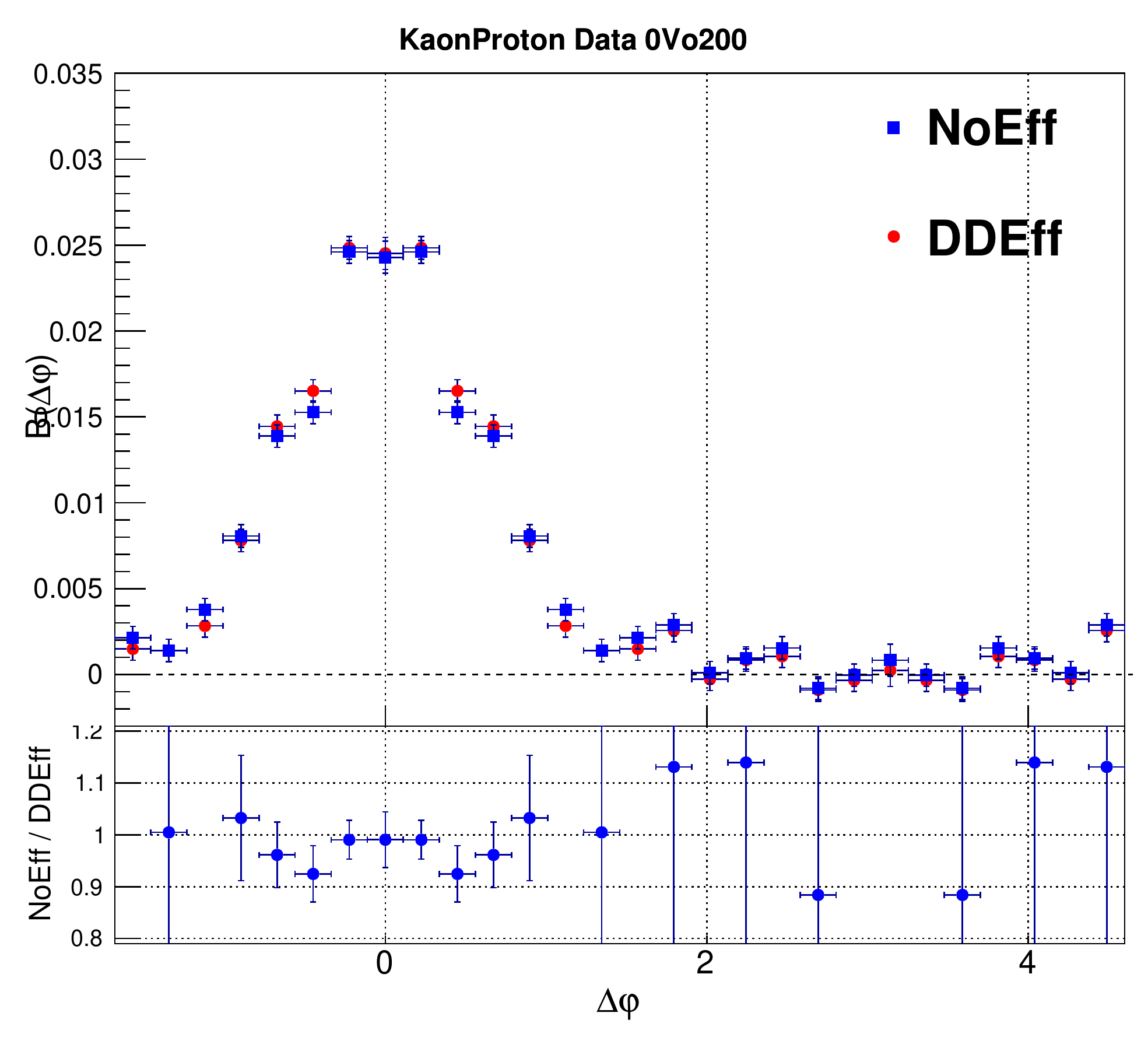}
  \includegraphics[width=0.3\linewidth]{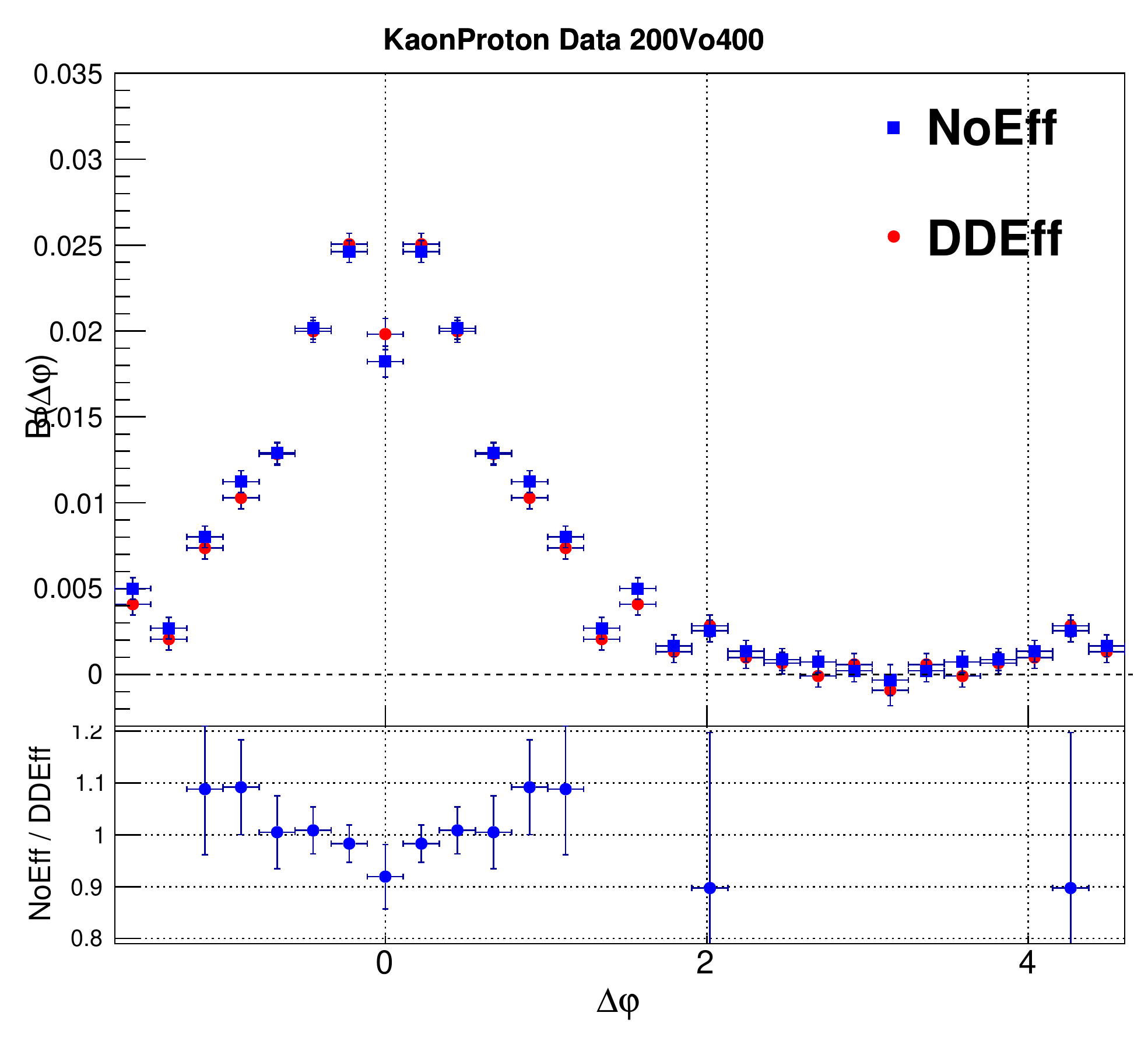}
  \includegraphics[width=0.3\linewidth]{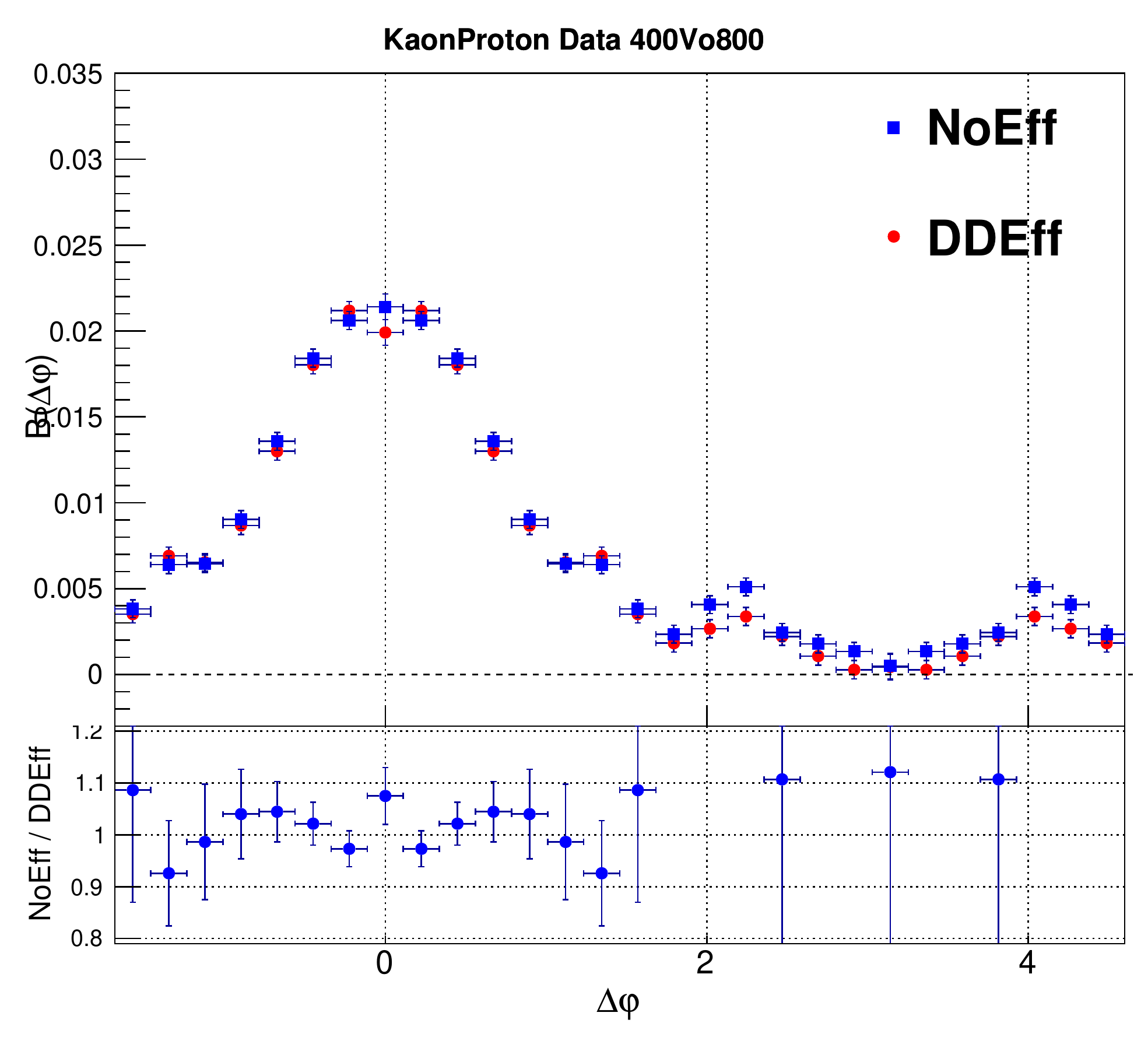} 
  \includegraphics[width=0.3\linewidth]{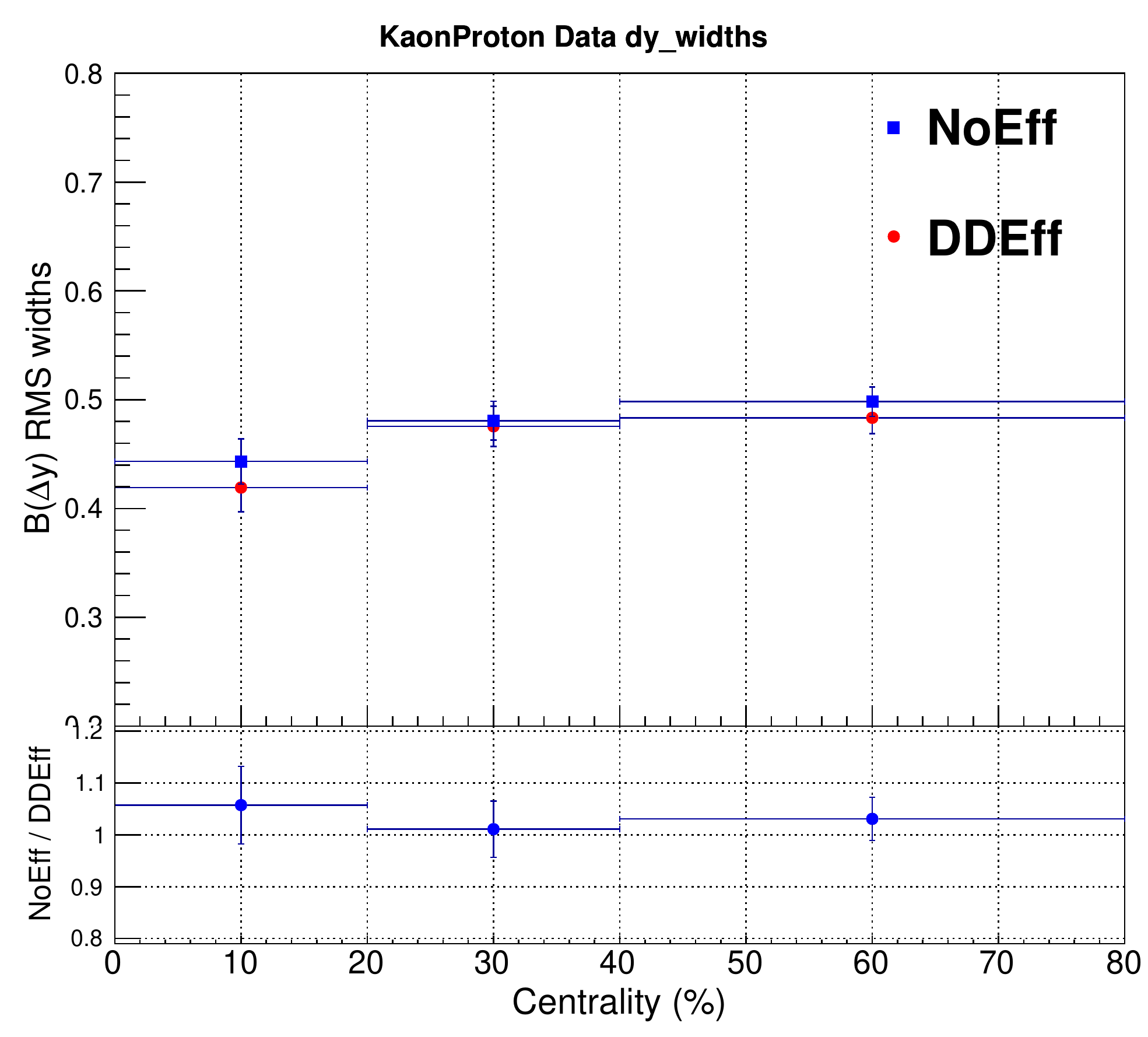}
  \includegraphics[width=0.3\linewidth]{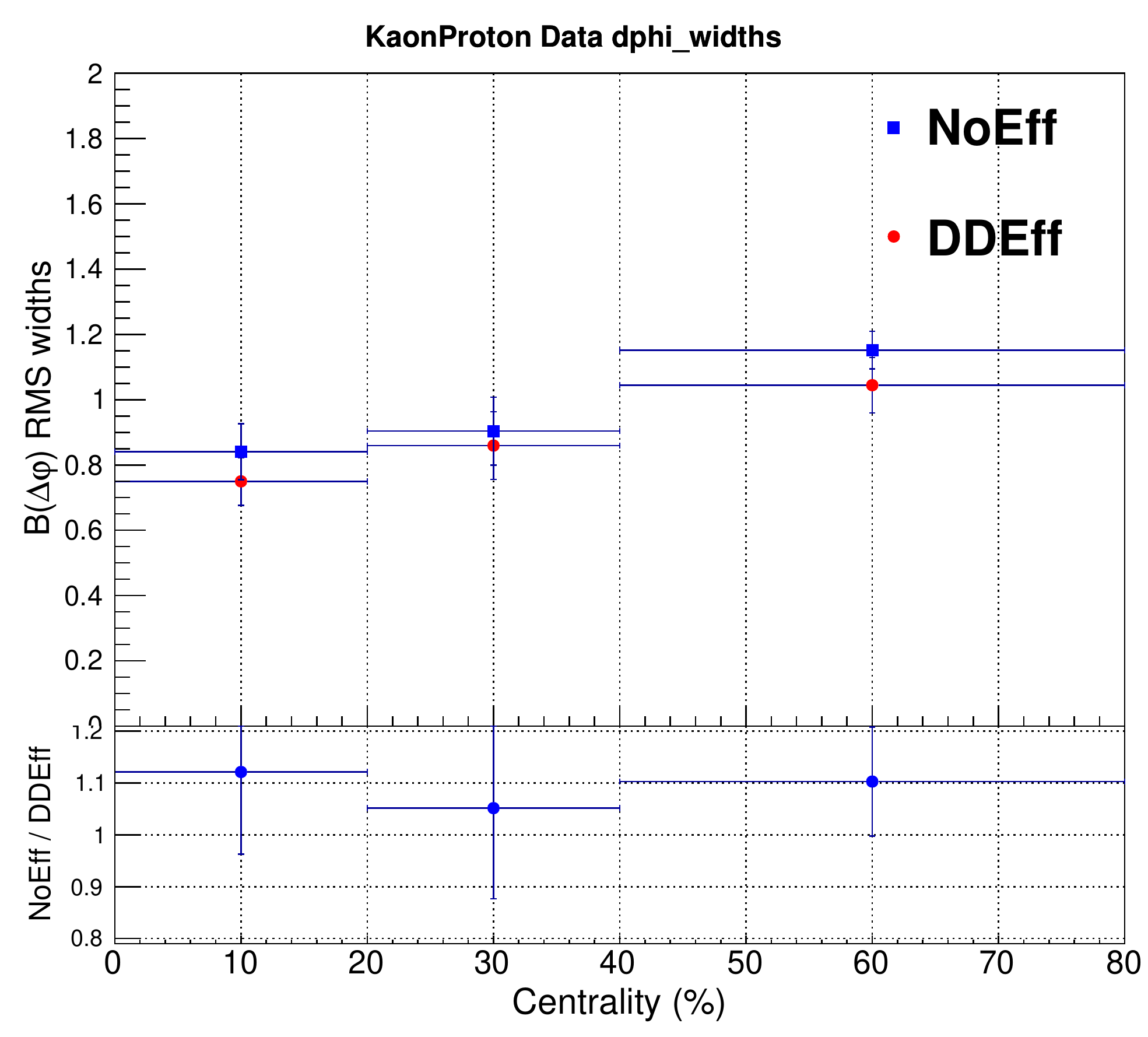}
  \includegraphics[width=0.3\linewidth]{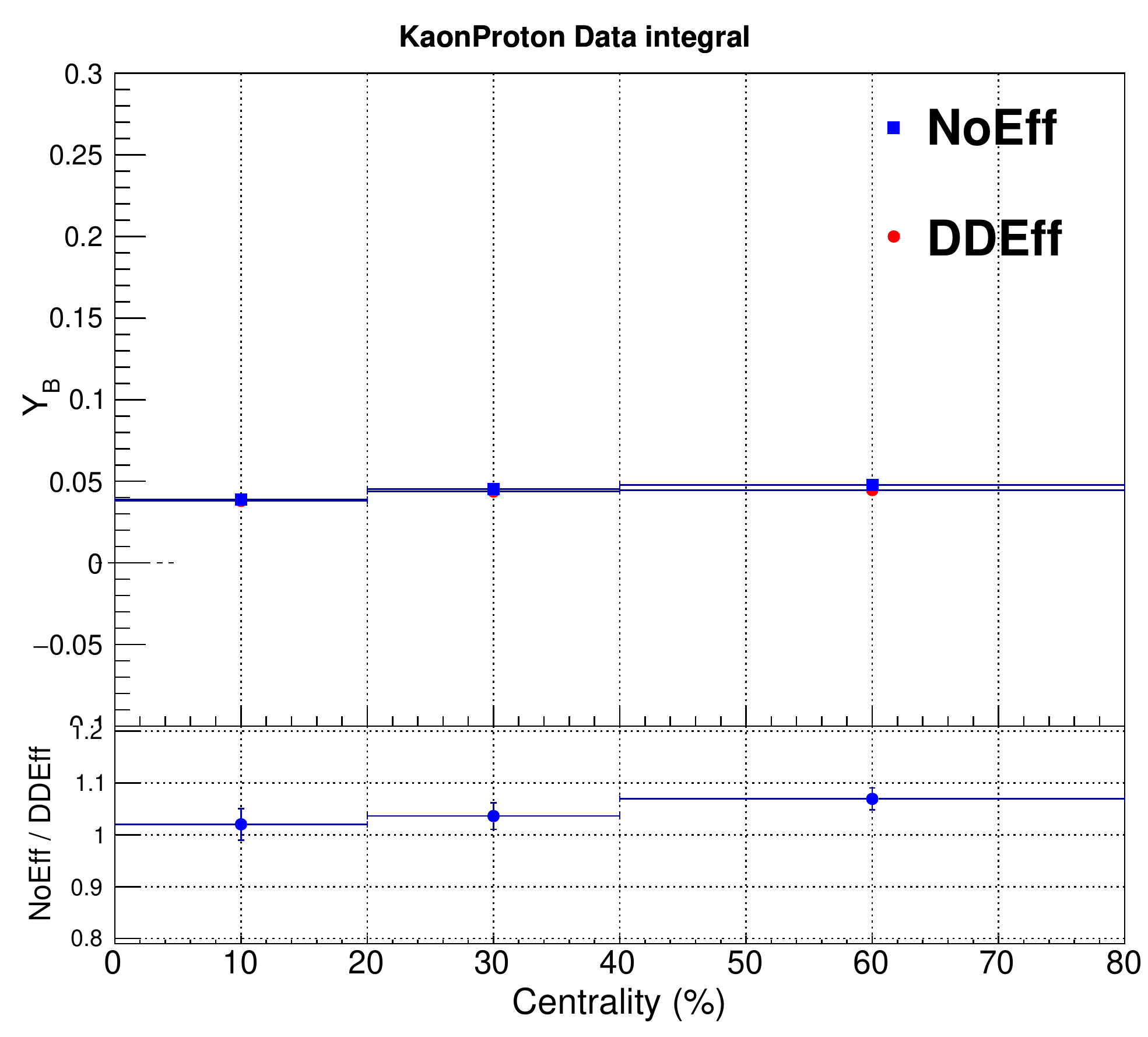}
  \caption{Comparisons of 2D $B^{Kp}$ obtained without (1st row) and with Data Driven (2nd row) $p_{\rm T}$-dependent efficiency correction for selected centralities, along with their $\Delta y$ (3rd row) and $\Delta \varphi$ projections (4th row), $\Delta y$ and $\Delta \varphi$ widths, and integrals (5th row).}
   \label{fig:Compare_DDEffCorr_NoEffCorr_BF_KaonProton}
\end{figure}

Figures~\ref{fig:Compare_DDEffCorr_NoEffCorr_BF_PionPion} --  %~\ref{fig:Compare_DDEffCorr_NoEffCorr_BF_KaonKaon},~\ref{fig:Compare_DDEffCorr_NoEffCorr_BF_PionKaon},~\ref{fig:Compare_DDEffCorr_NoEffCorr_BF_PionProton},
\ref{fig:Compare_DDEffCorr_NoEffCorr_BF_KaonProton} show that the ``uncorrected" BF results and the ``data driven"  $p_{\rm T}$-dependent efficiency corrected results are in good agreement for both  same- and cross-species pairs. 
Differences between the two sets of results are with 1$\sigma$ of statistical uncertainties at essentially all $\Delta y$ and $\Delta\varphi$ separations. In this work, we thus report ``final results" for balance functions based on  the ``data driven" $p_{\rm T}$-dependent efficiency correction method for  all  species pairs. 
The differences between the ``uncorrected" results and the ``data driven corrected" results are taken as  systematic uncertainty associated with this data driven $p_{\rm T}$-dependent efficiency correction.

%KaonKaon
\begin{figure}
\centering
  \includegraphics[width=0.3\linewidth]{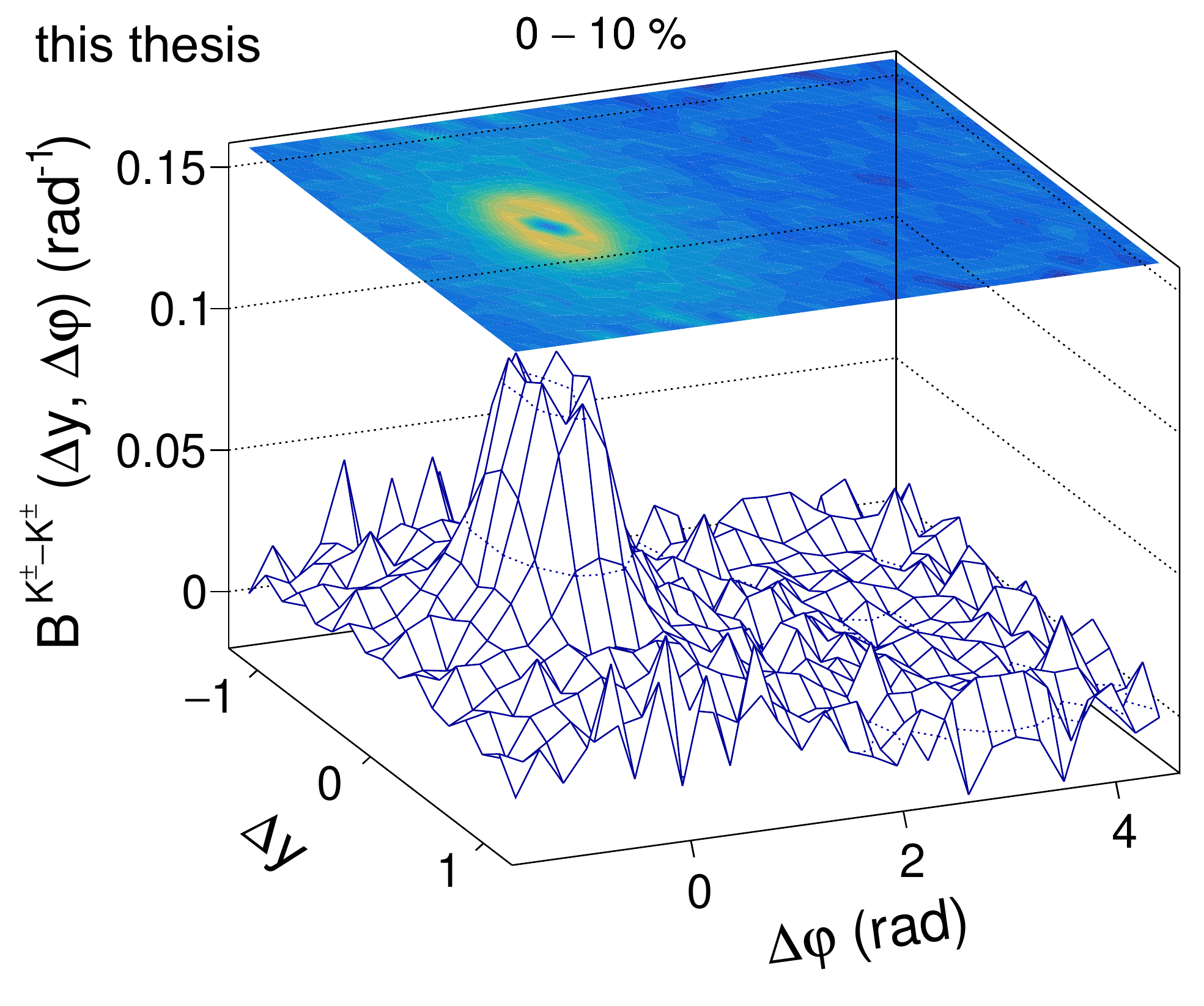}
  \includegraphics[width=0.3\linewidth]{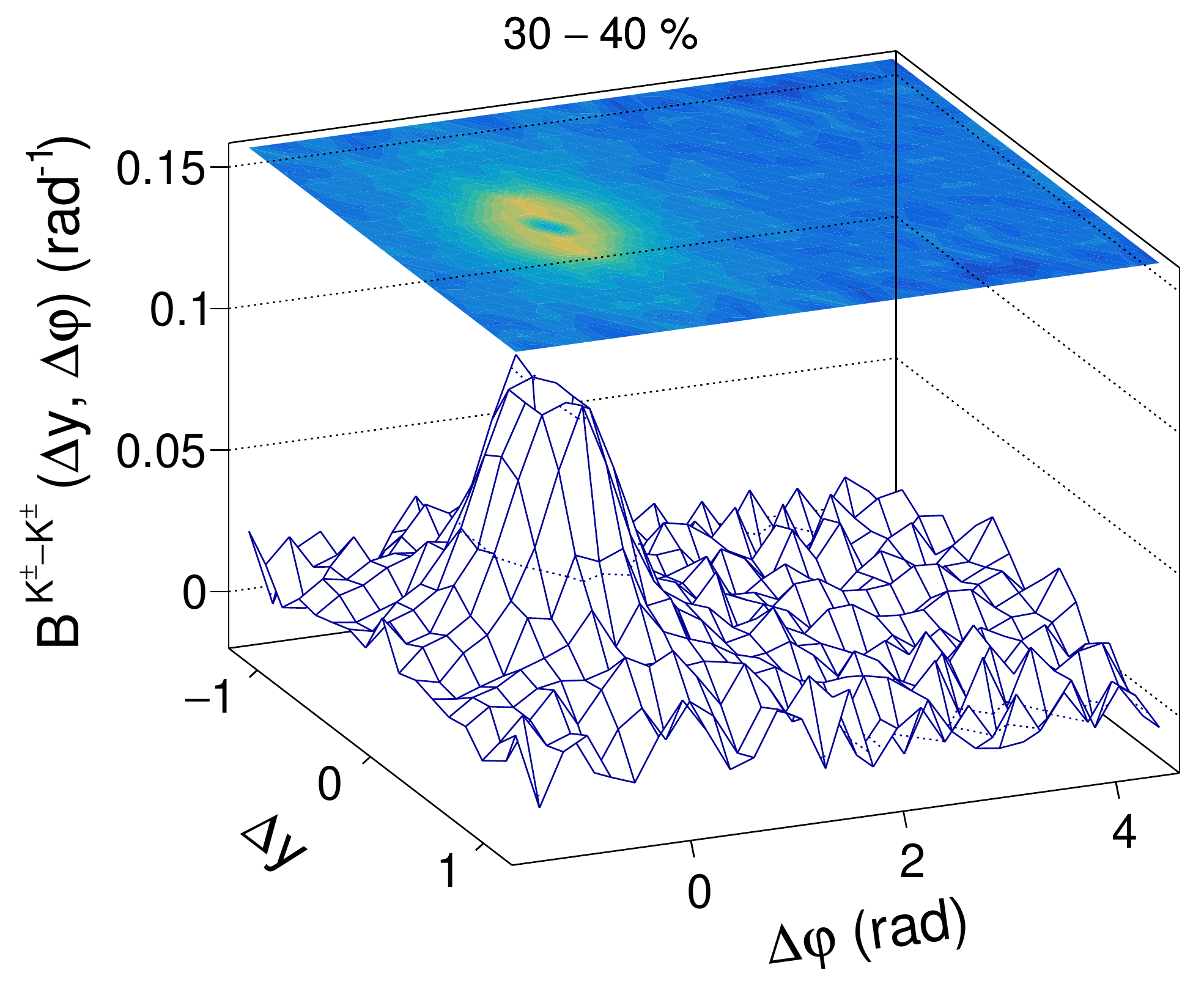}
  \includegraphics[width=0.3\linewidth]{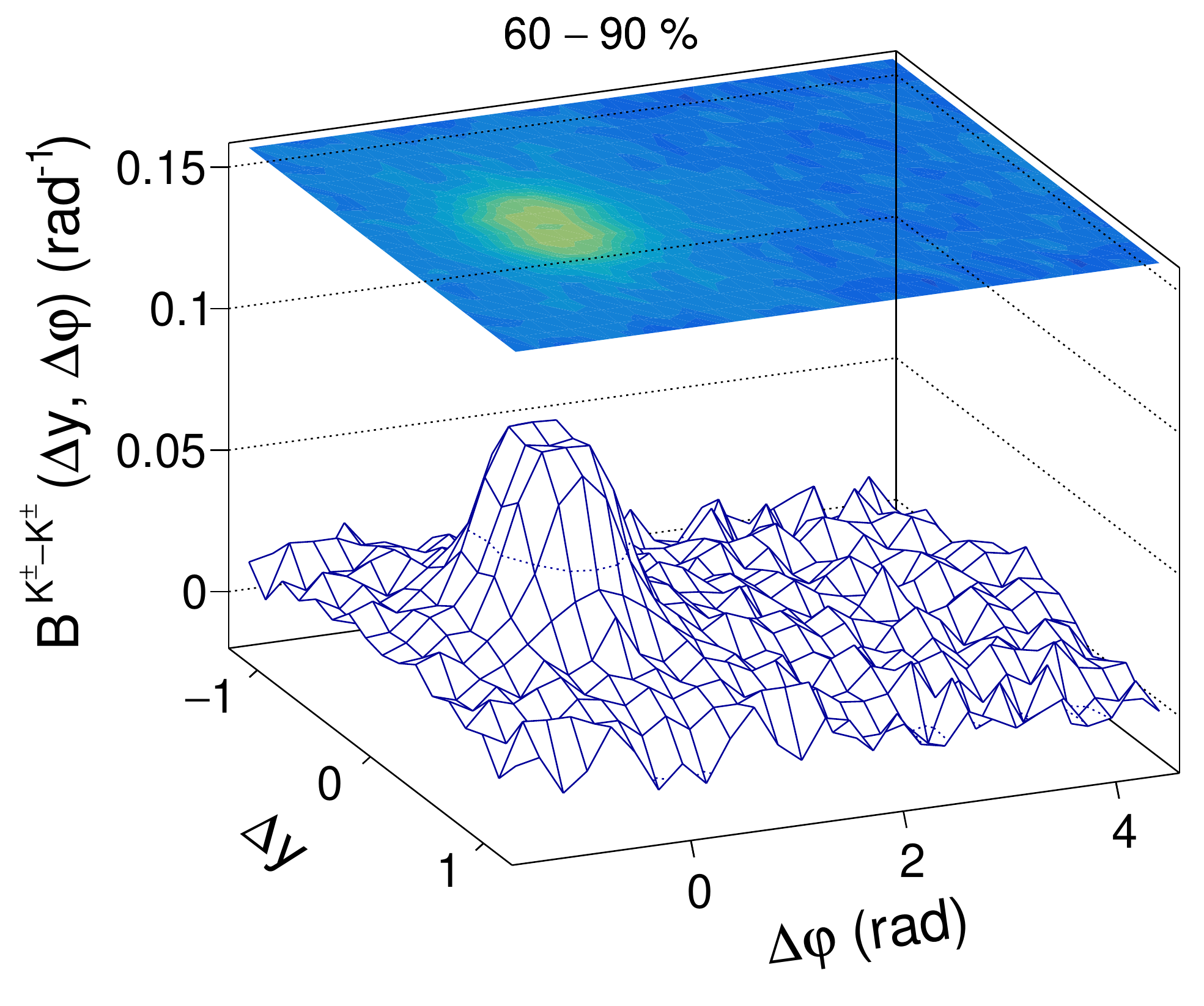}
  \includegraphics[width=0.3\linewidth]{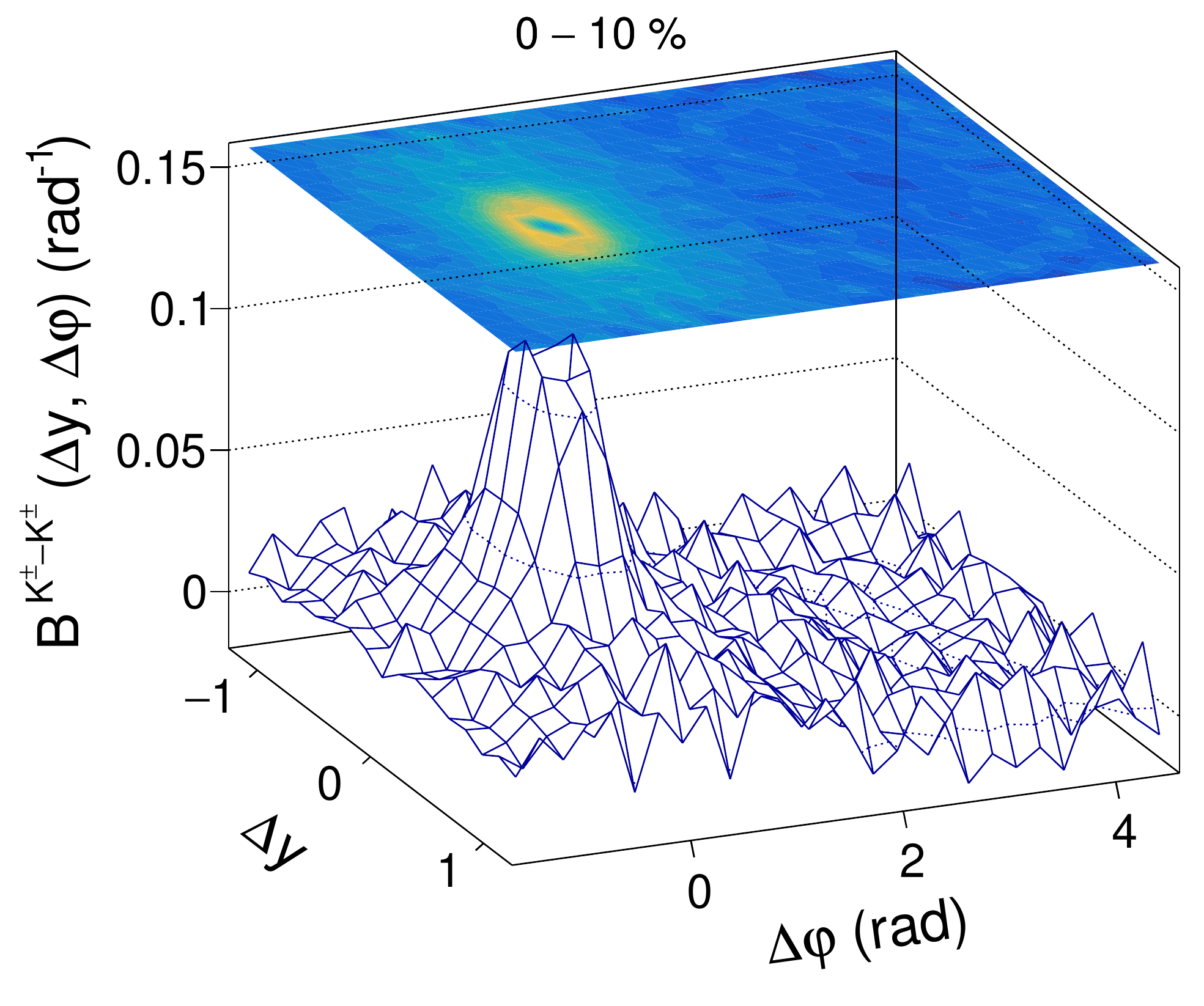}
  \includegraphics[width=0.3\linewidth]{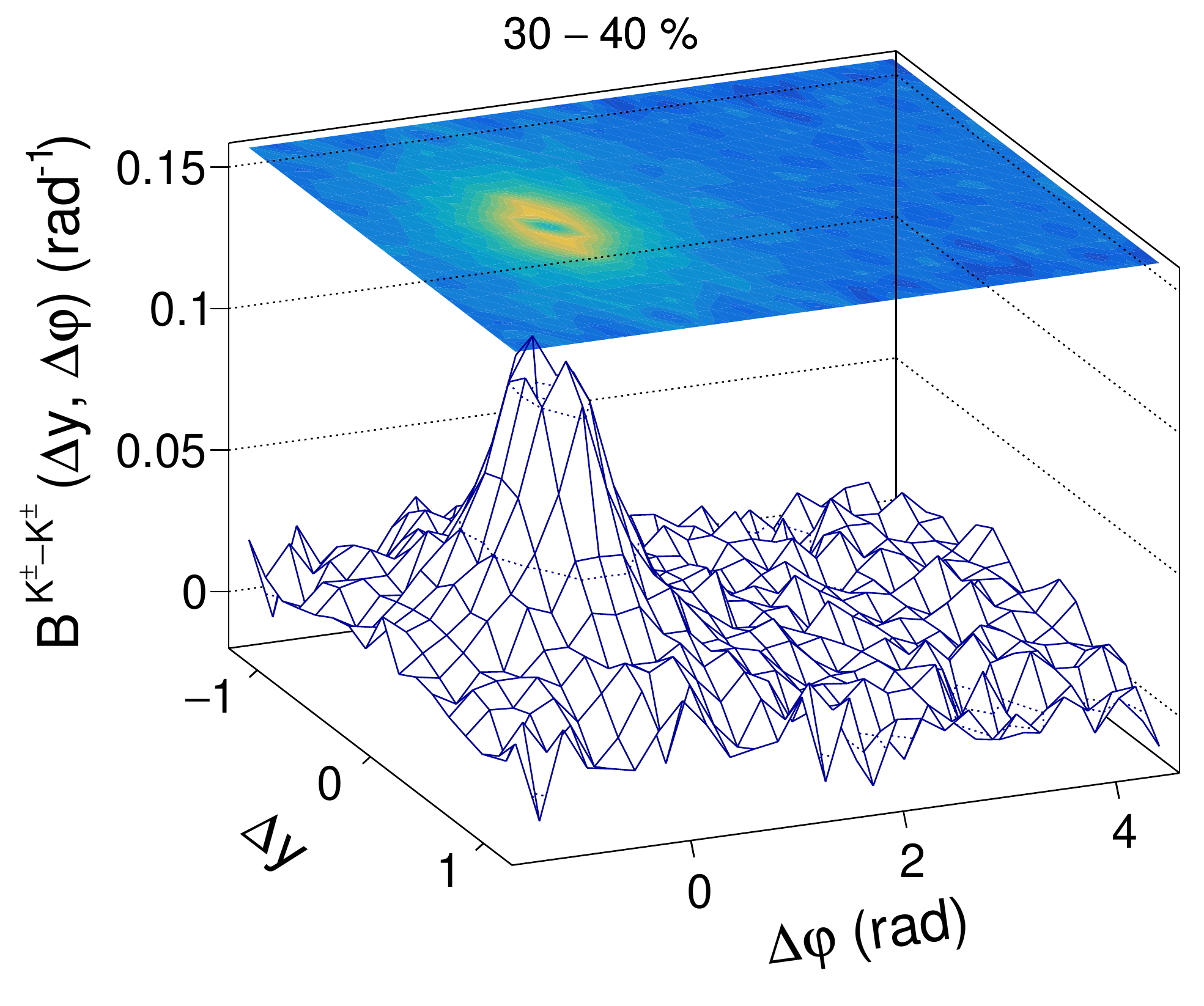}
  \includegraphics[width=0.3\linewidth]{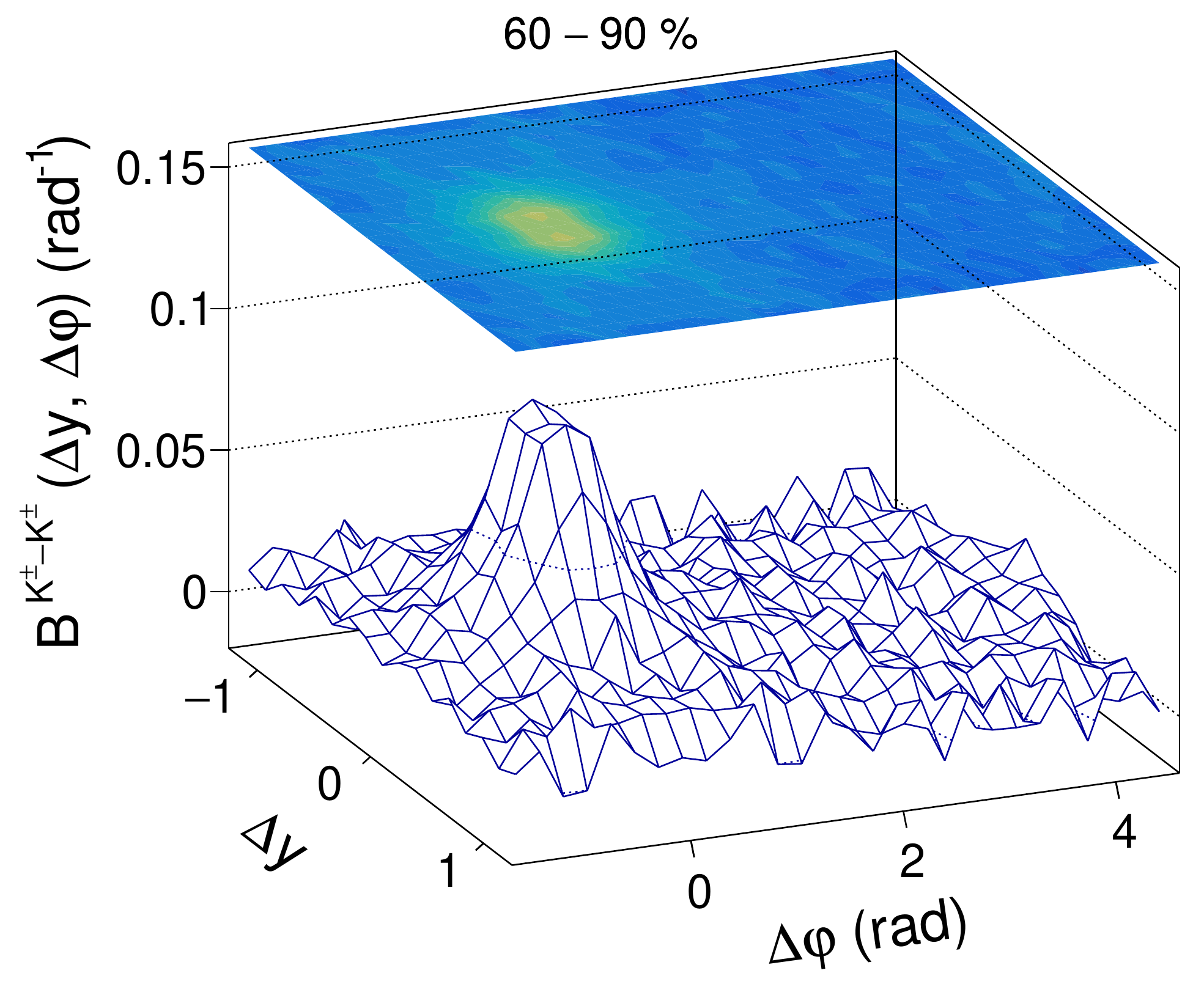}
  \includegraphics[width=0.3\linewidth]{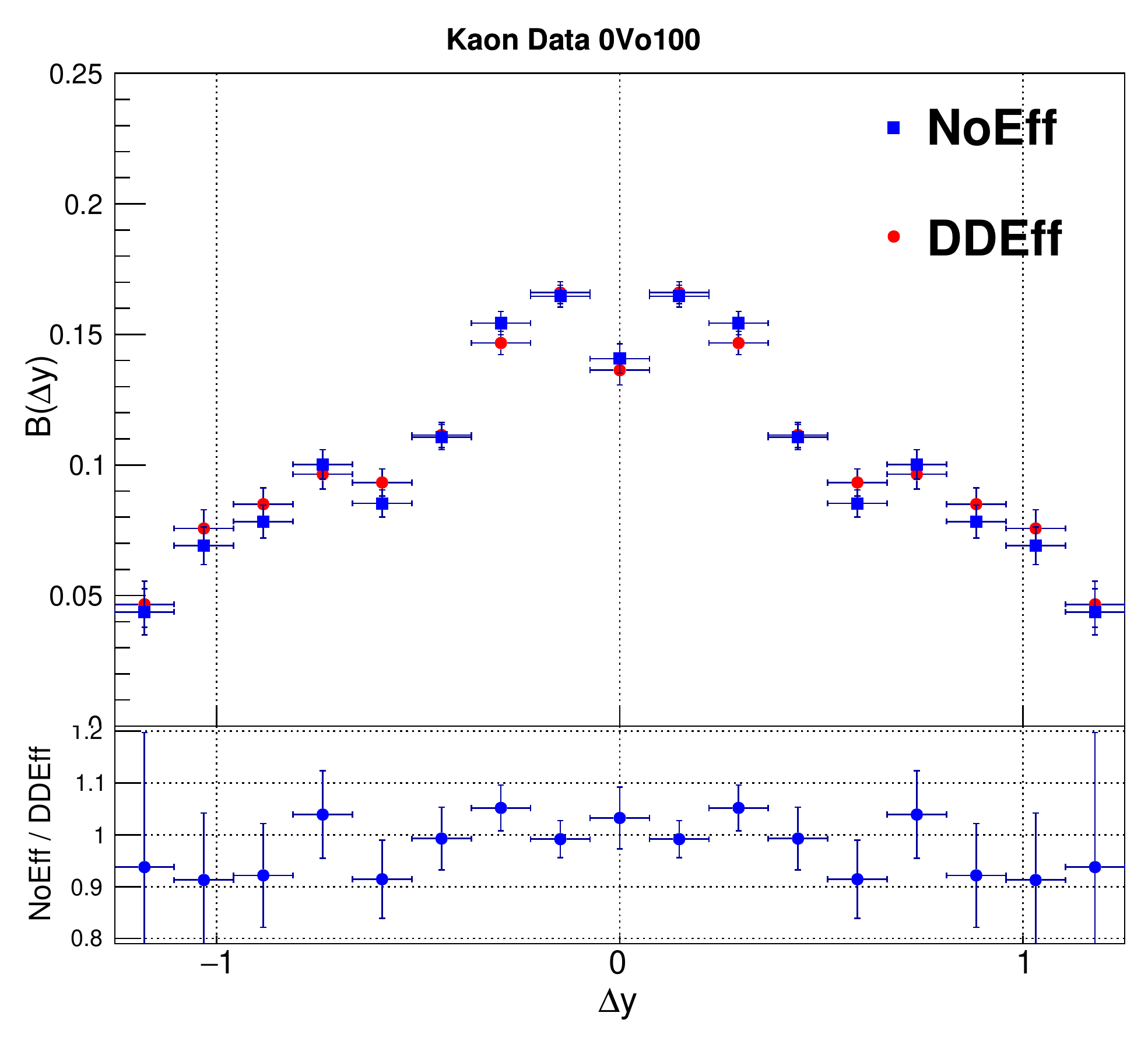}
  \includegraphics[width=0.3\linewidth]{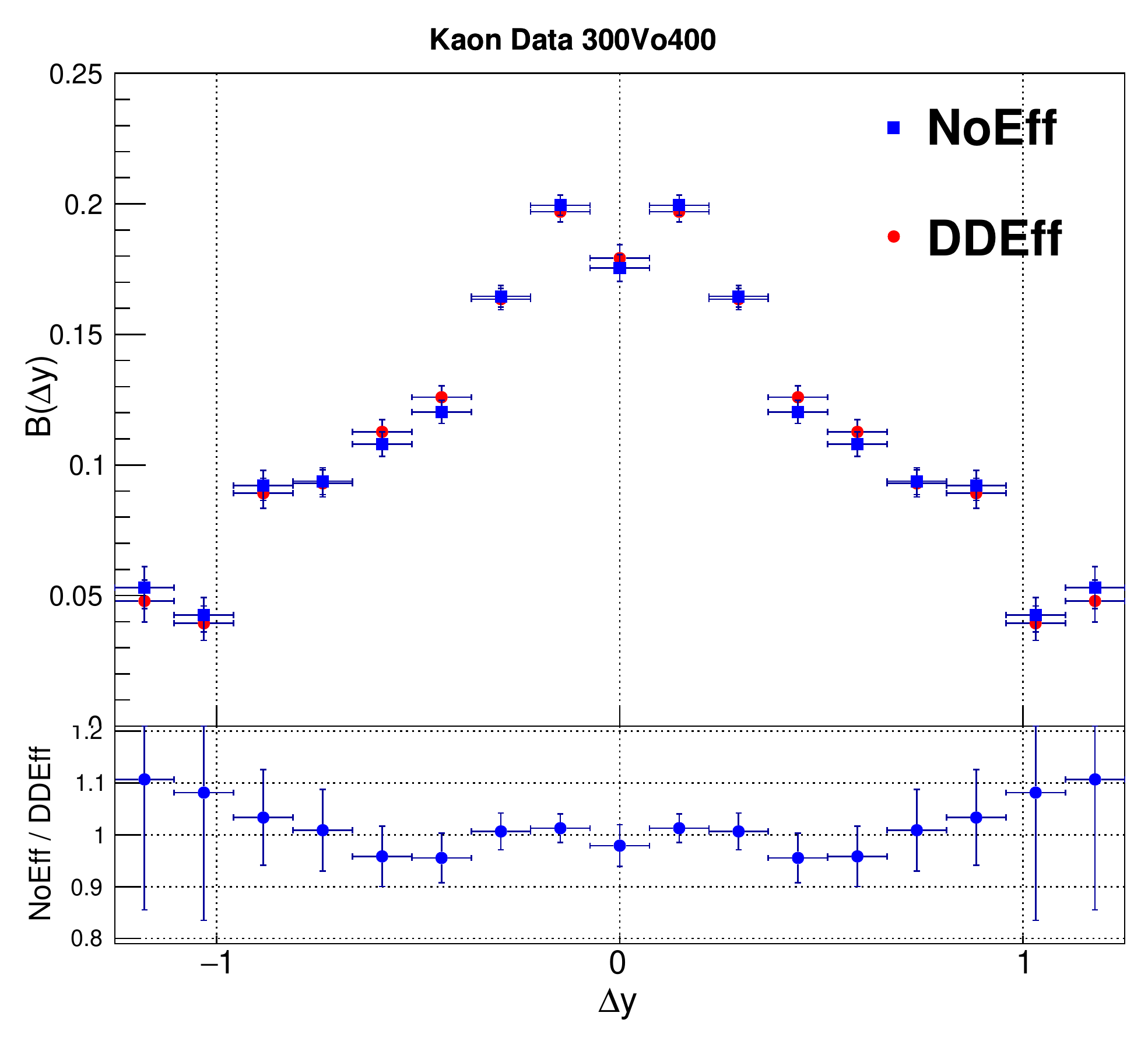}
  \includegraphics[width=0.3\linewidth]{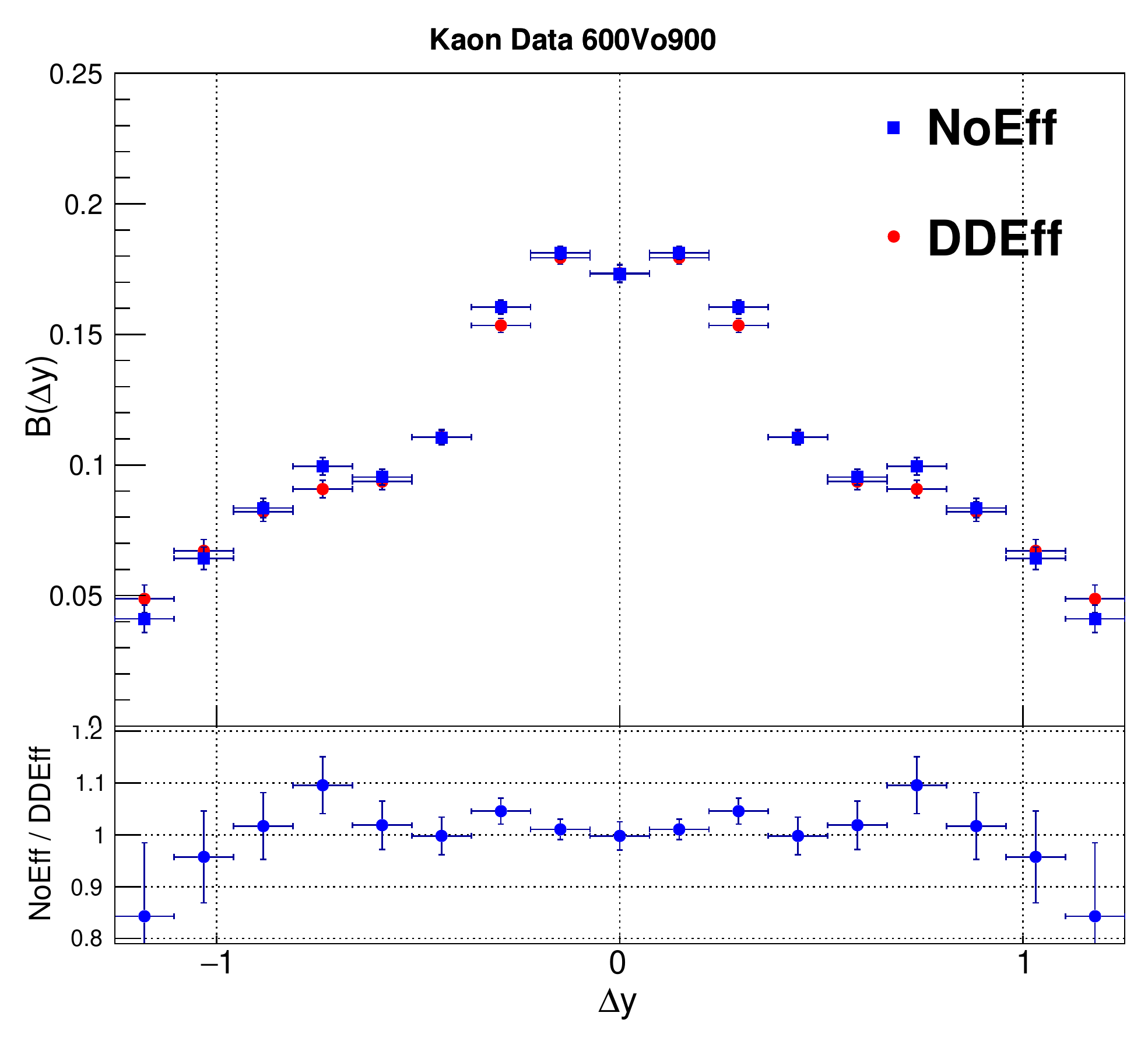} 
  \includegraphics[width=0.3\linewidth]{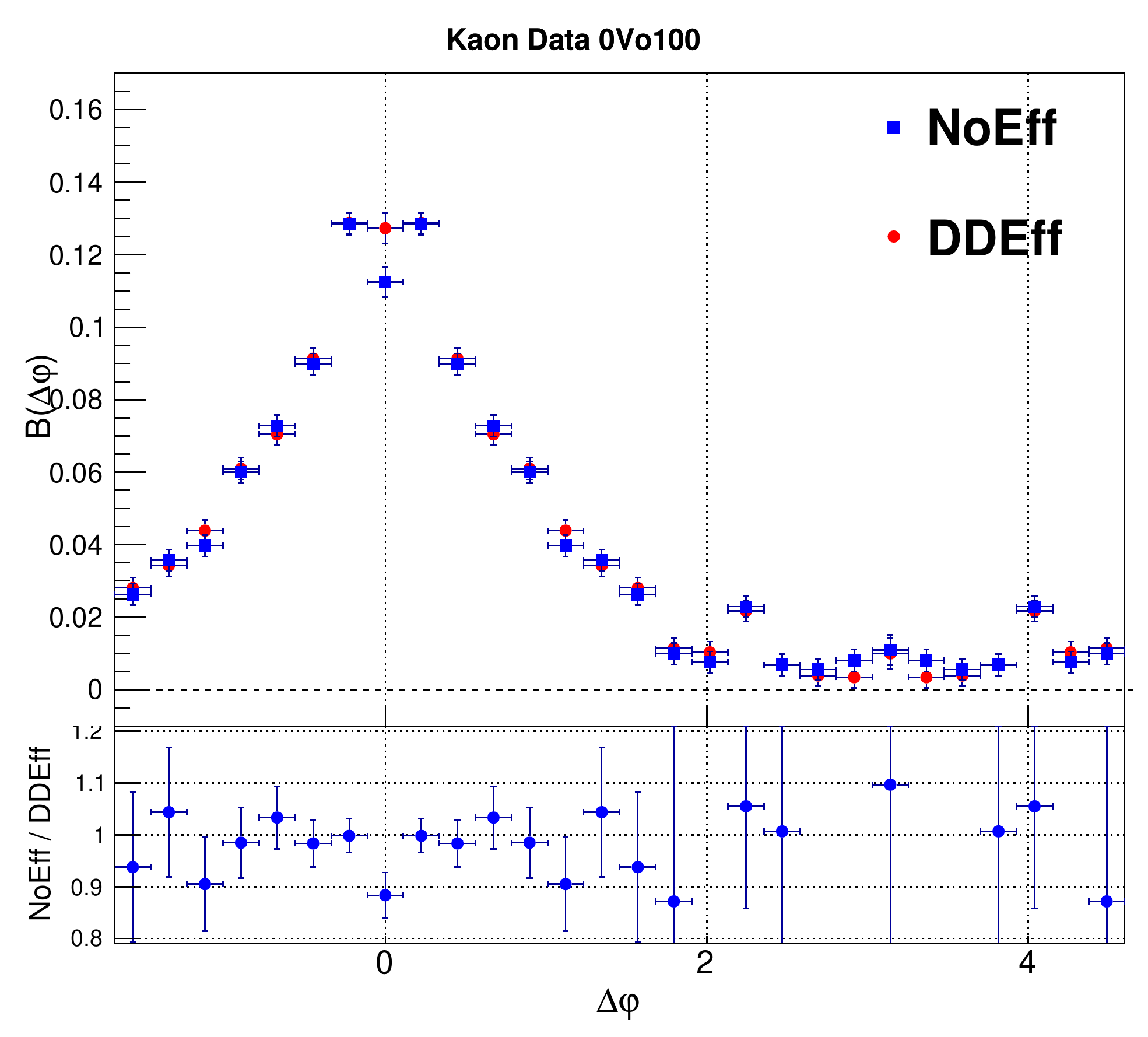}
  \includegraphics[width=0.3\linewidth]{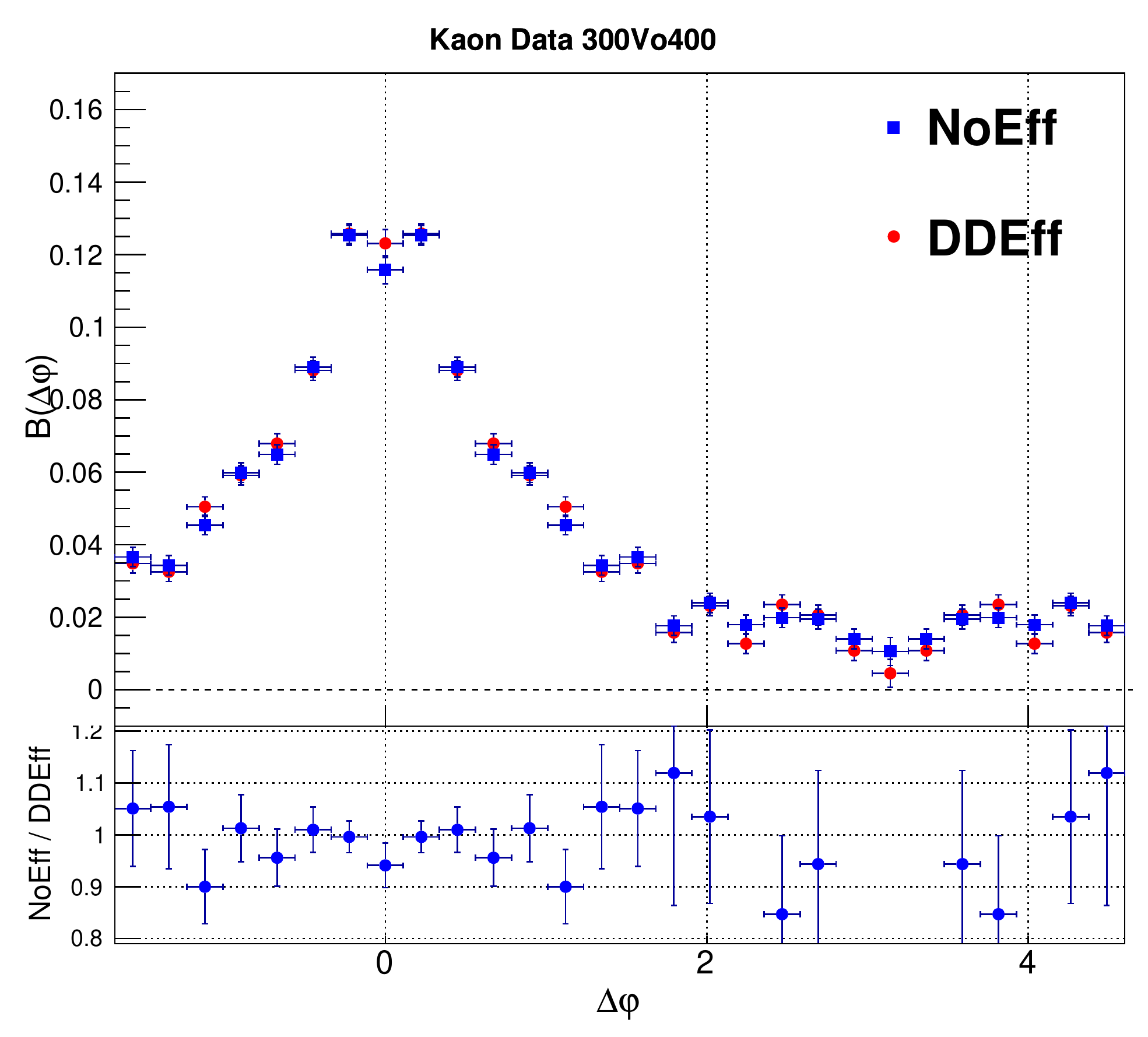}
  \includegraphics[width=0.3\linewidth]{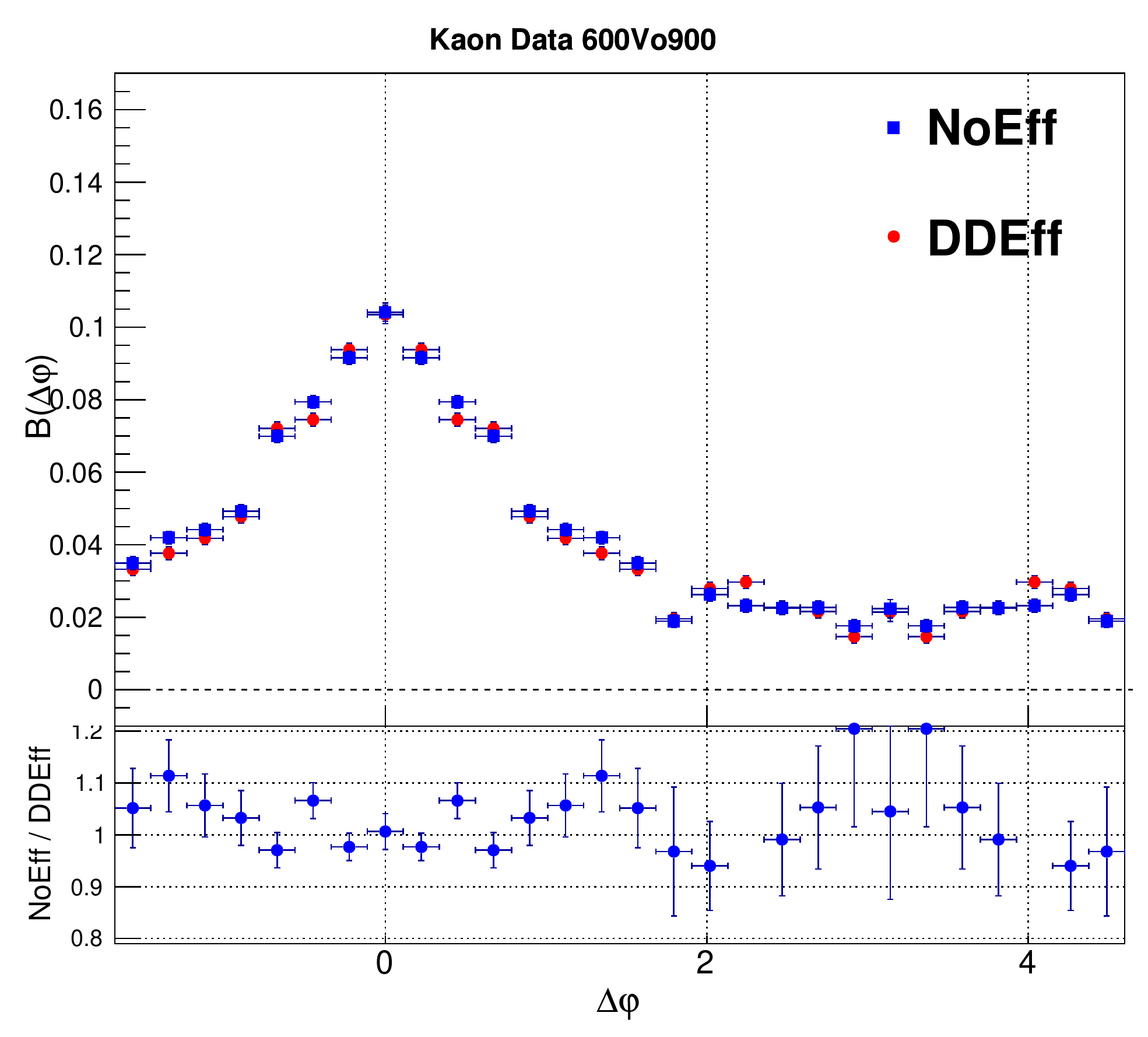} 
  \includegraphics[width=0.3\linewidth]{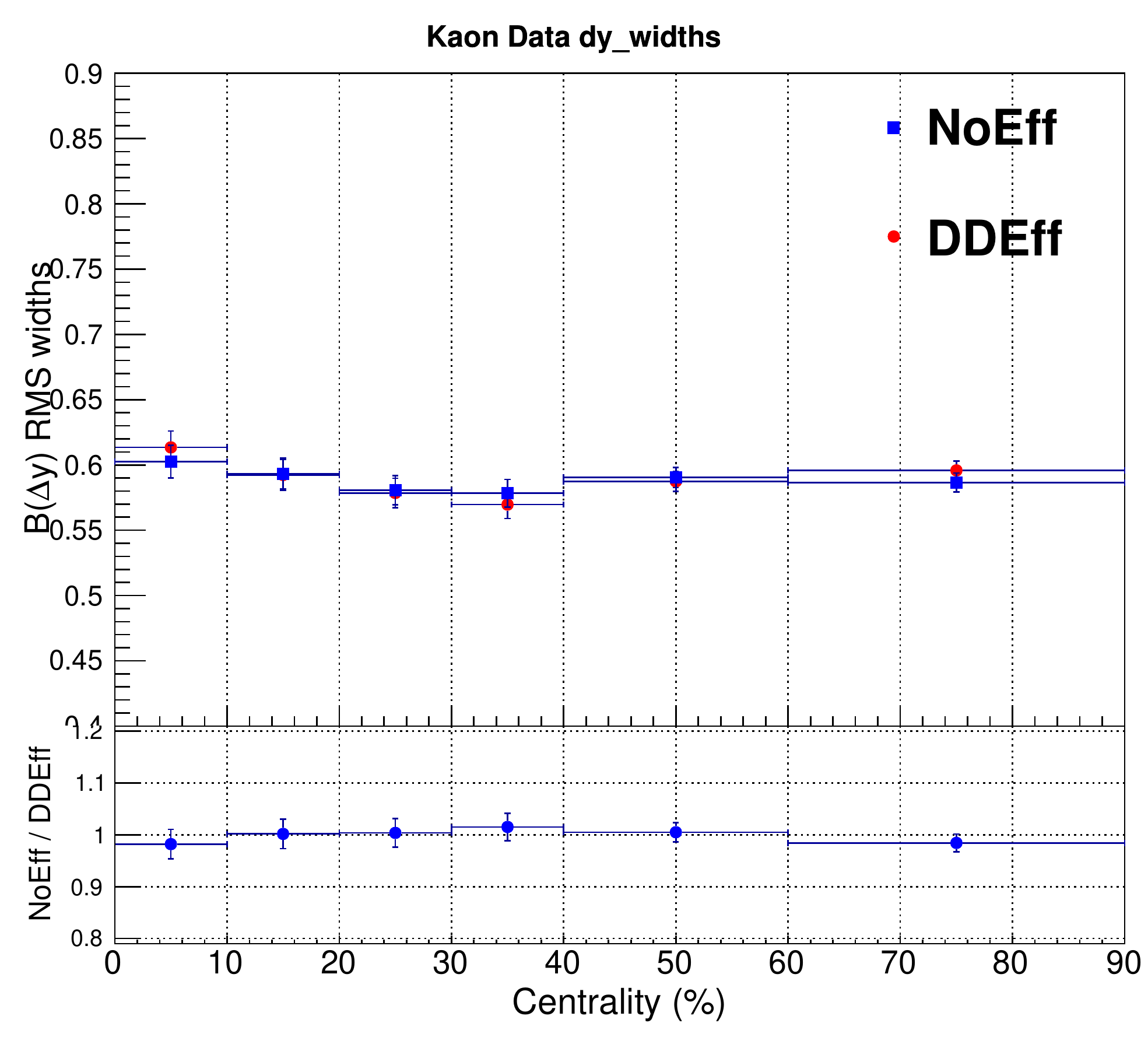}
  \includegraphics[width=0.3\linewidth]{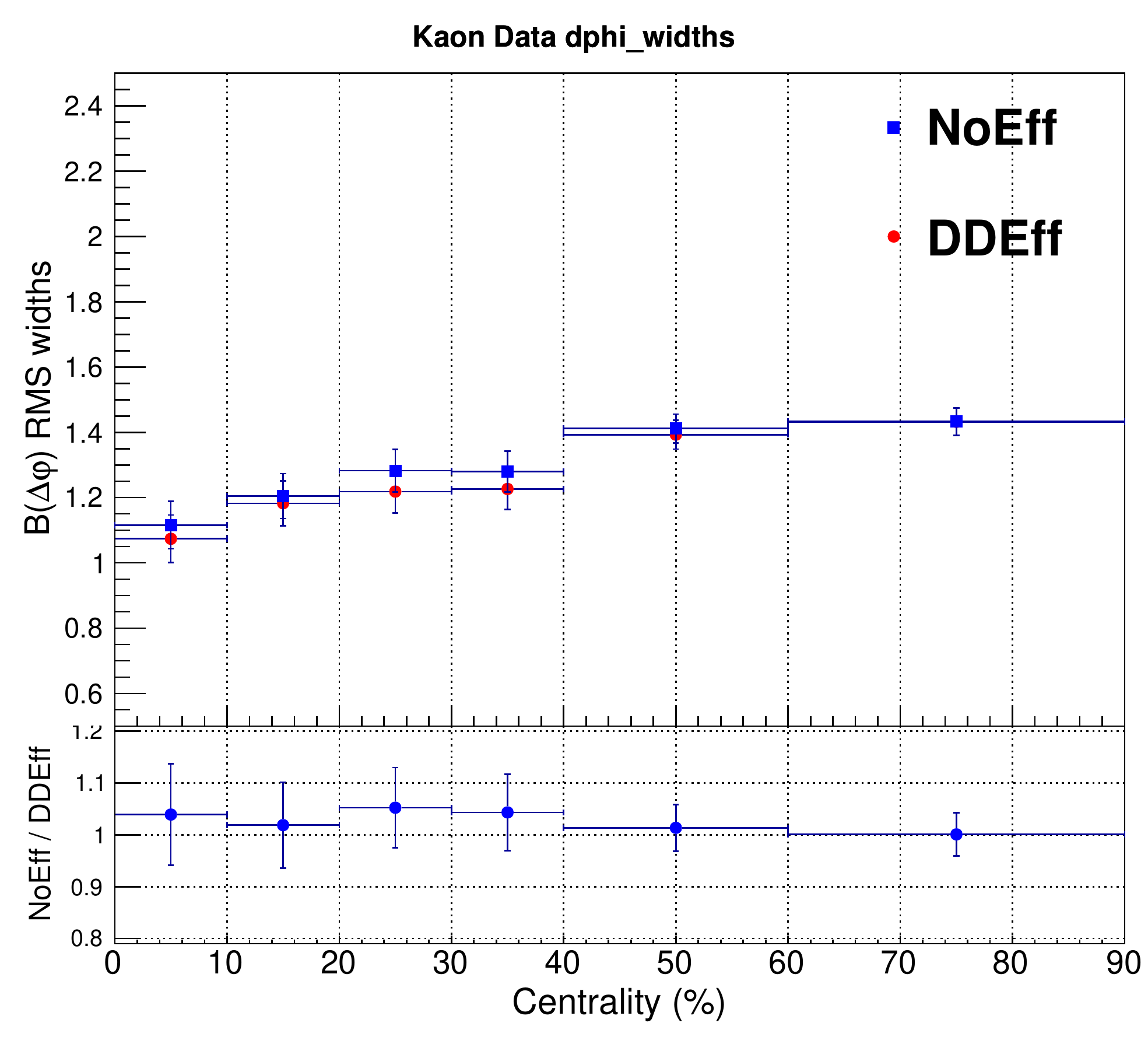}
  \includegraphics[width=0.3\linewidth]{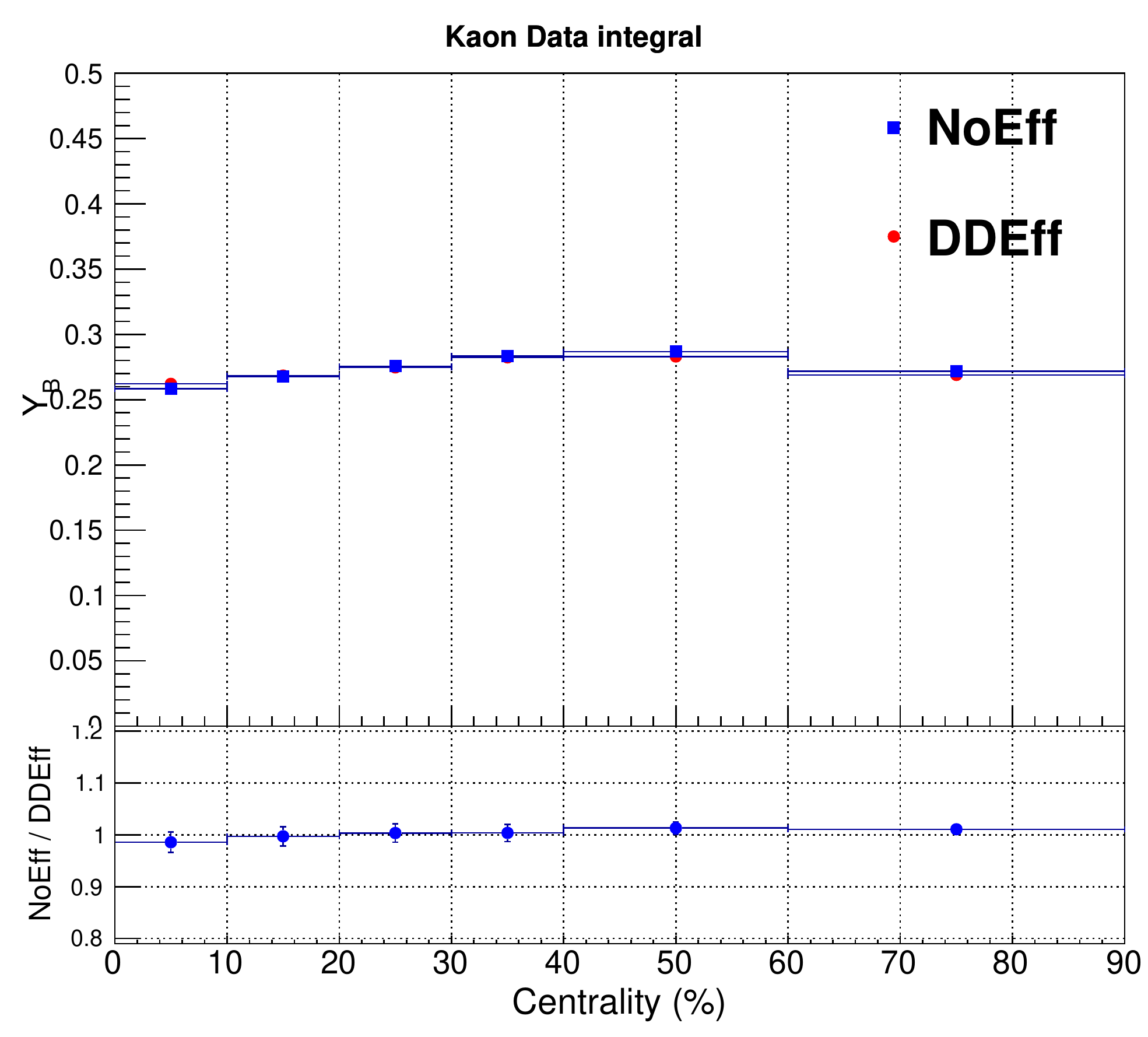} 
  \caption{Comparisons of 2D $B^{KK}$ obtained without (1st row) and with Data Driven (2nd row) $p_{\rm T}$-dependent efficiency correction for selected centralities, along with their $\Delta y$ (3rd row) and $\Delta \varphi$ projections (4th row), $\Delta y$ and $\Delta \varphi$ widths, and integrals (5th row).}
   \label{fig:Compare_DDEffCorr_NoEffCorr_BF_KaonKaon}
\end{figure}

%PionKaon
\begin{figure}
\centering
  \includegraphics[width=0.3\linewidth]{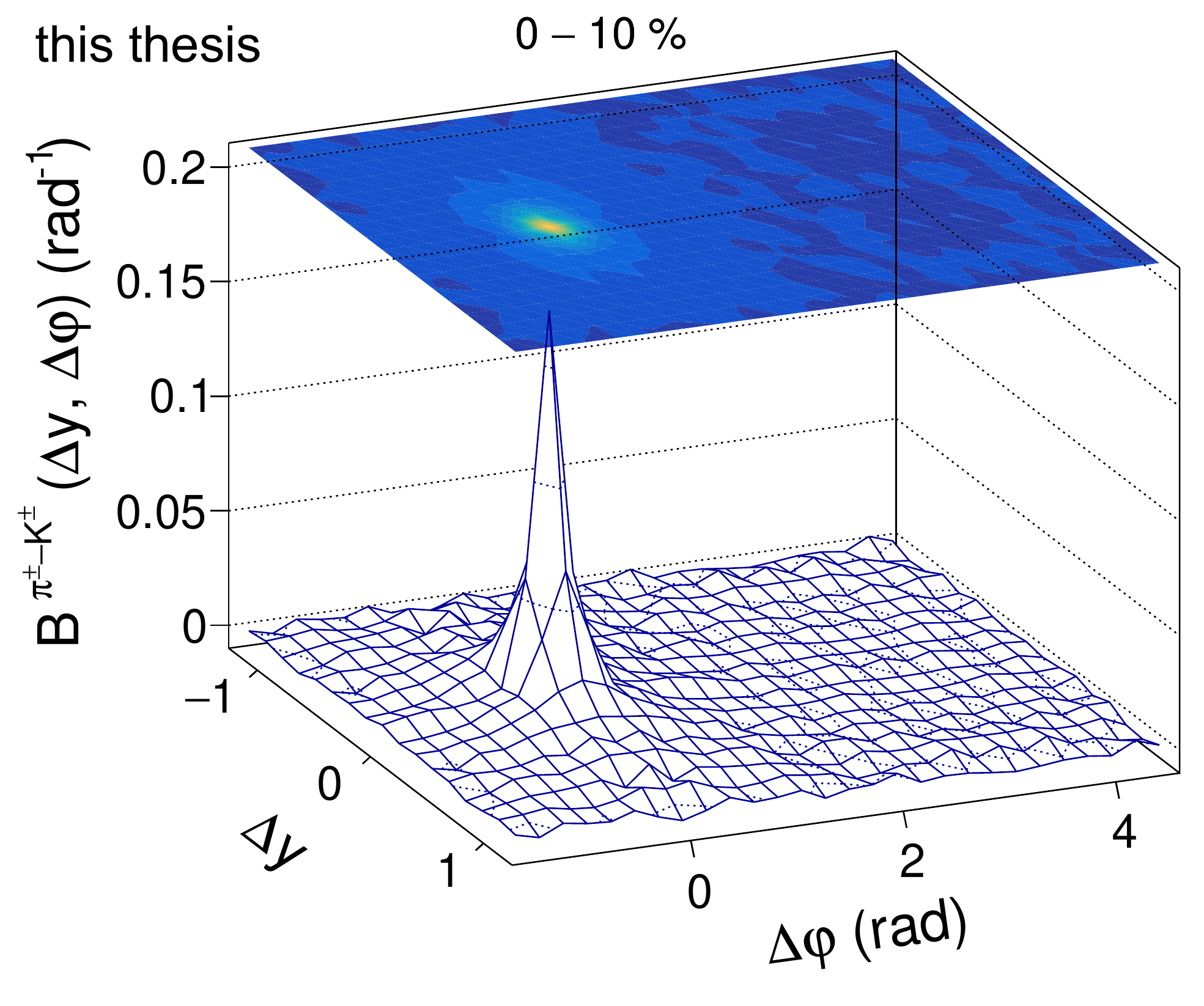}
  \includegraphics[width=0.3\linewidth]{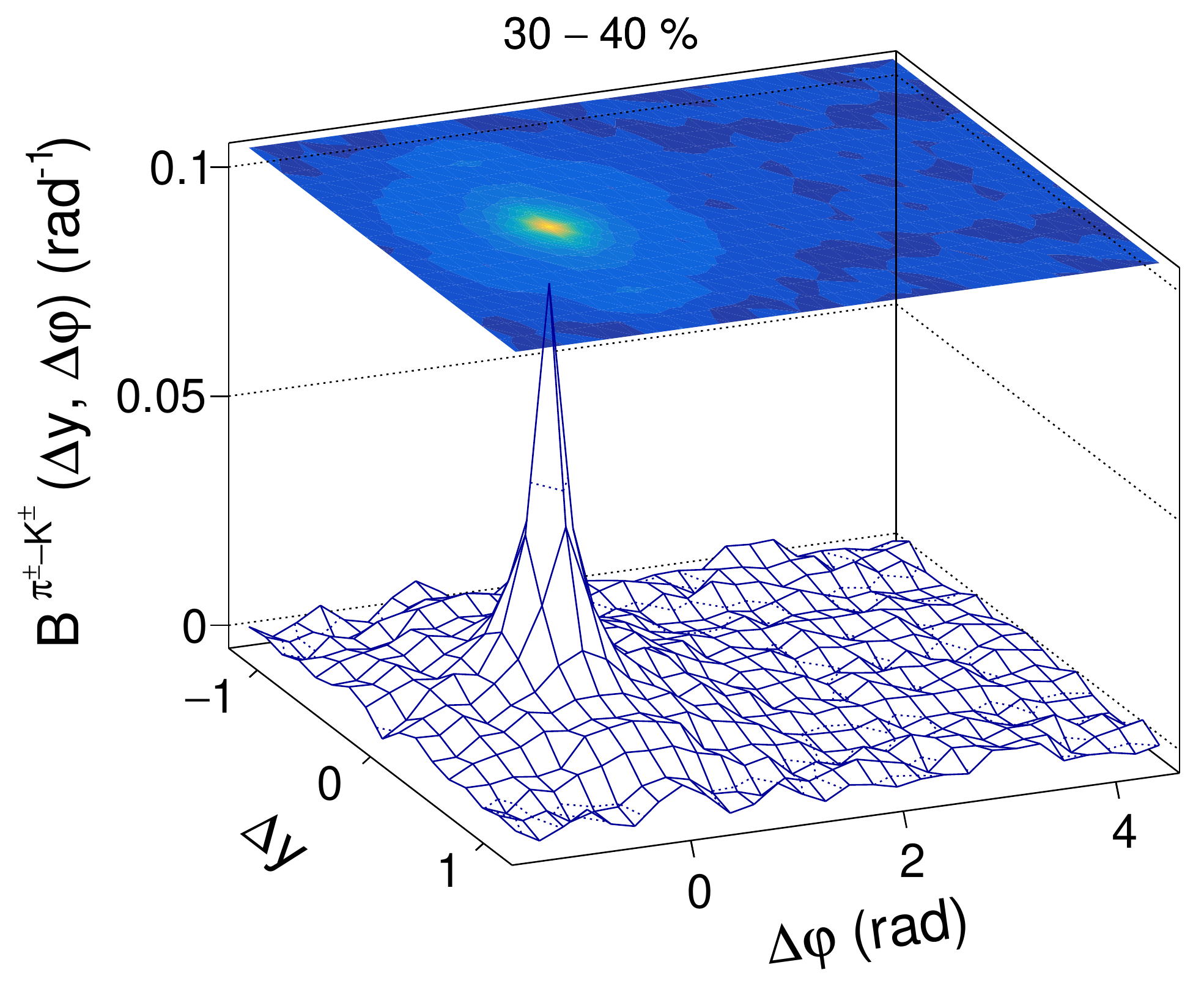}
  \includegraphics[width=0.3\linewidth]{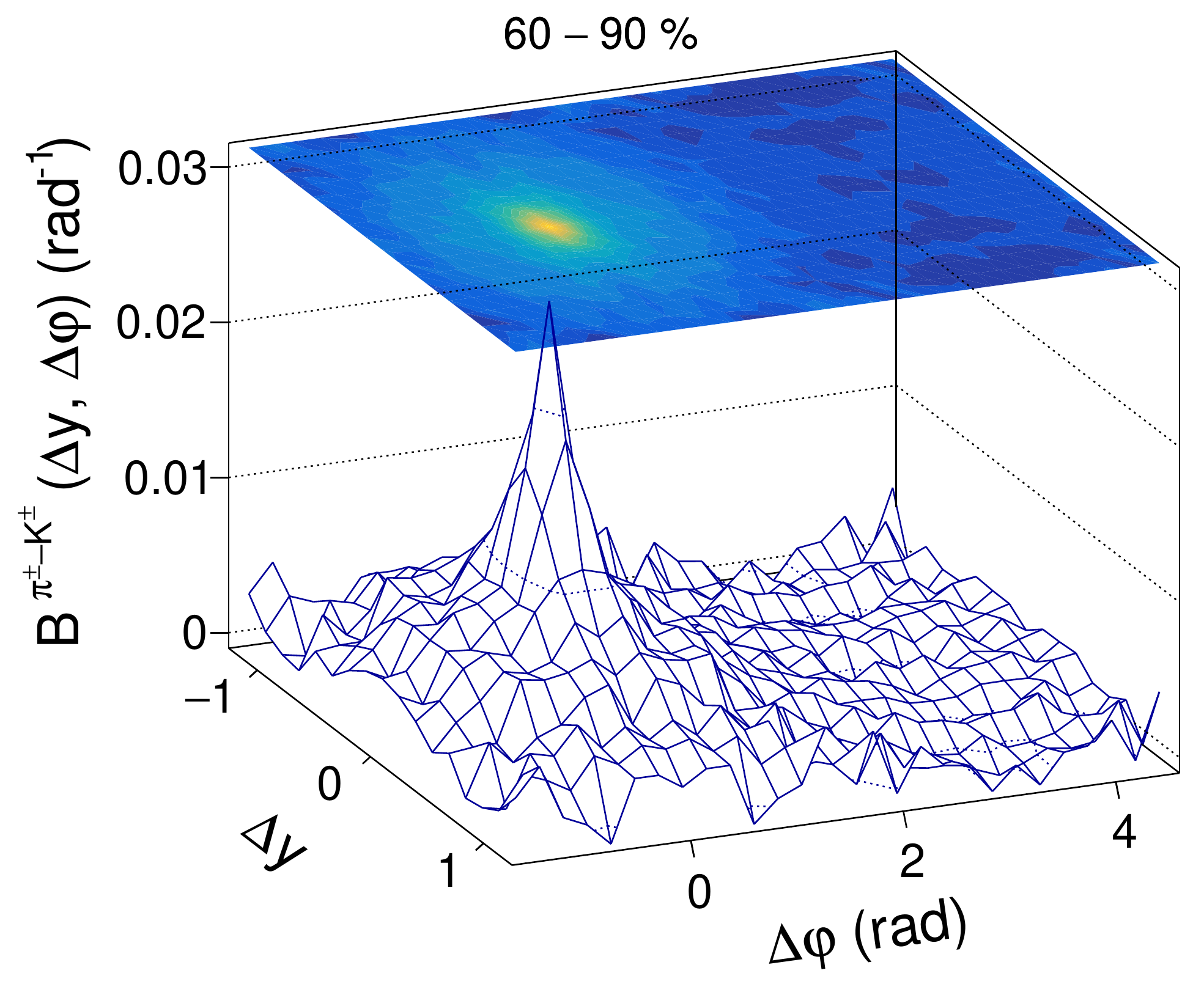}
  \includegraphics[width=0.3\linewidth]{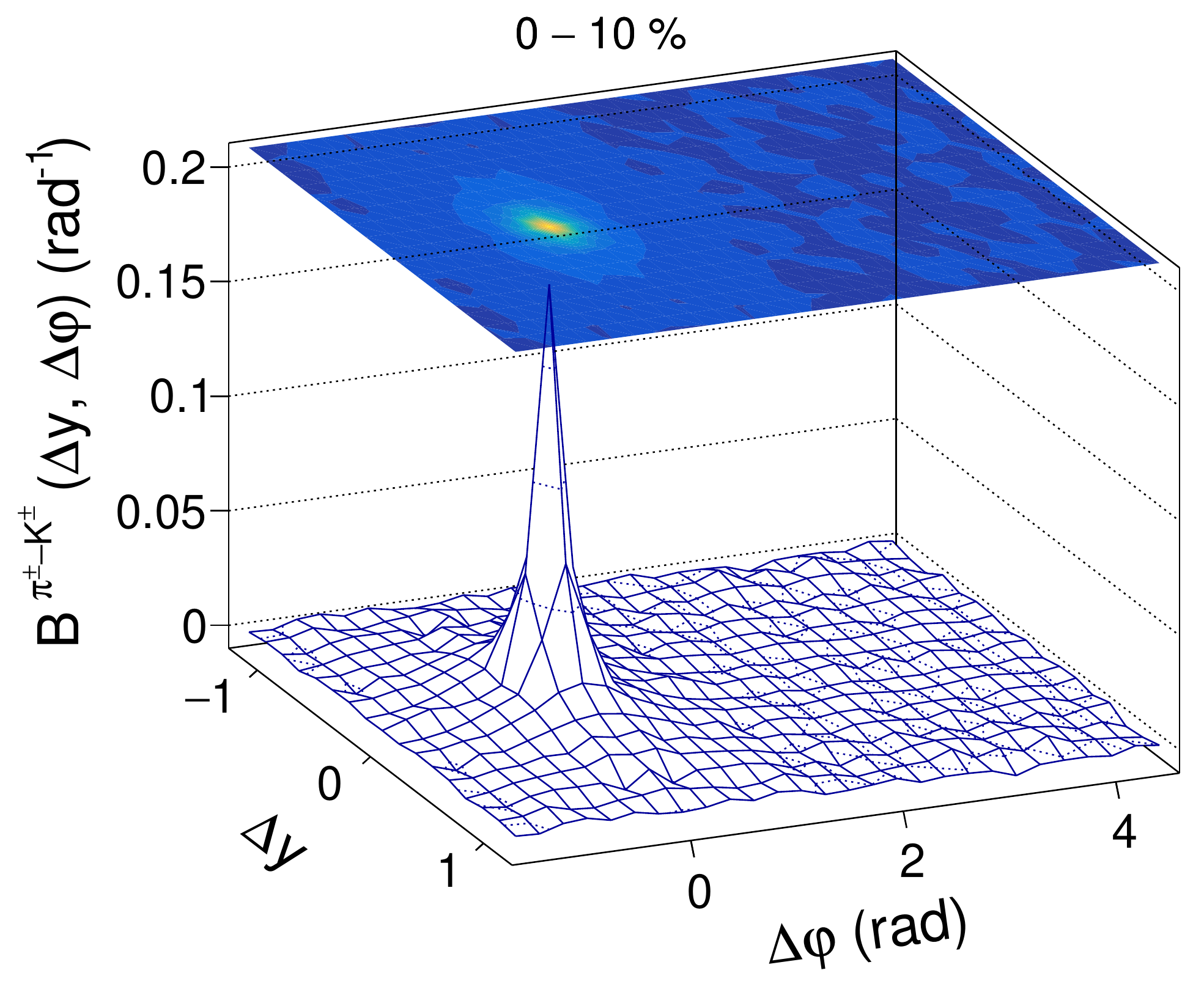}
  \includegraphics[width=0.3\linewidth]{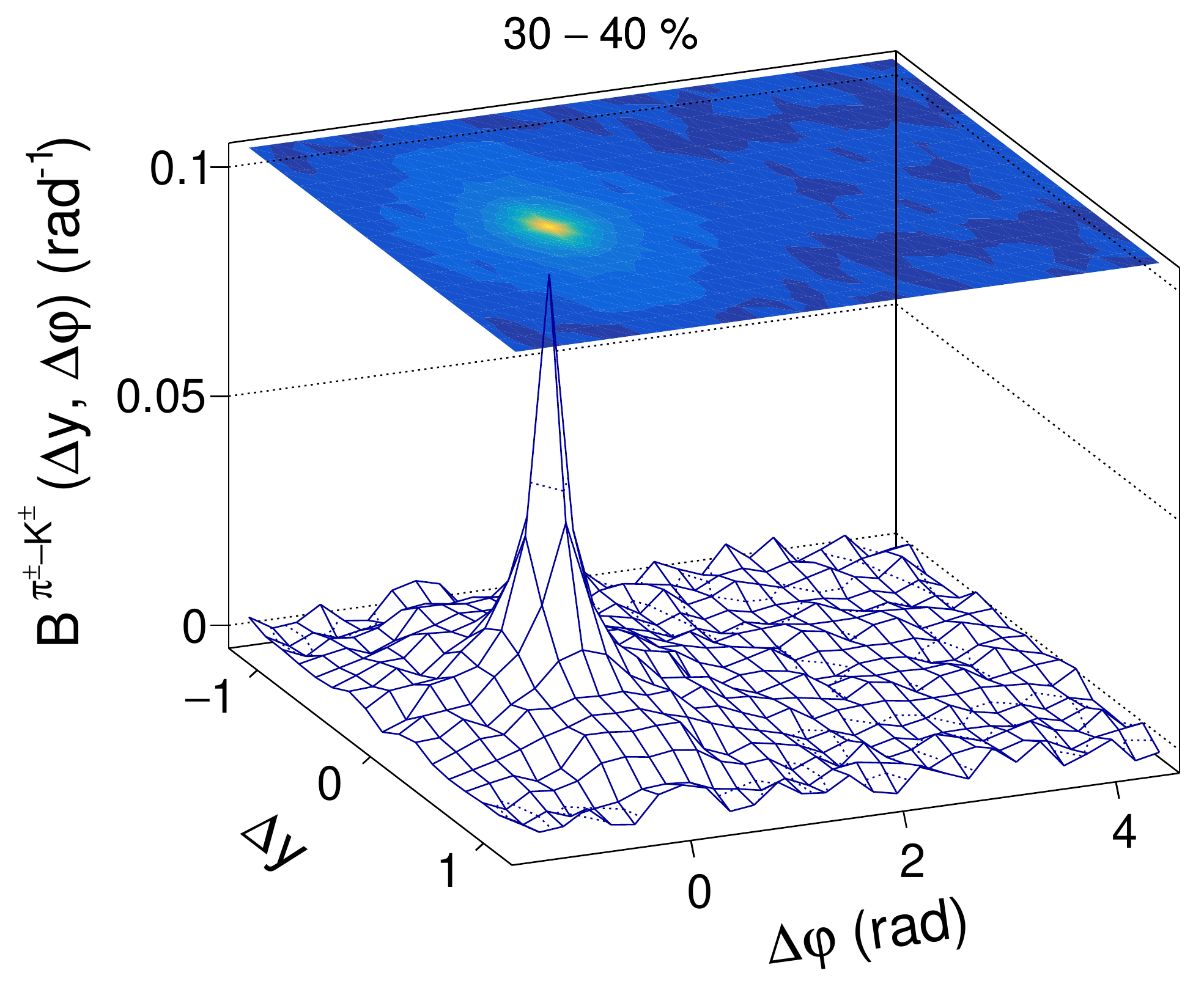}
  \includegraphics[width=0.3\linewidth]{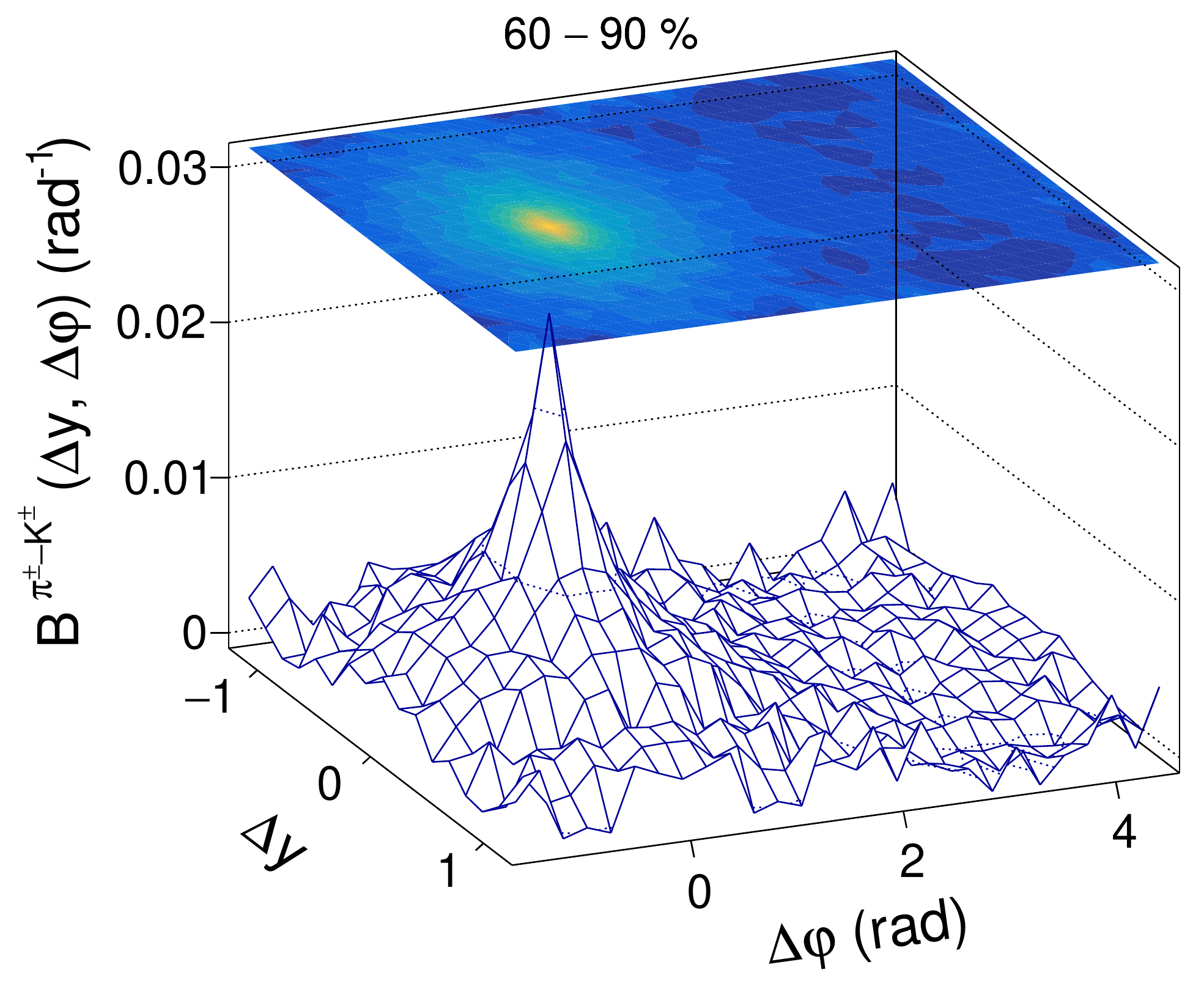}
  \includegraphics[width=0.3\linewidth]{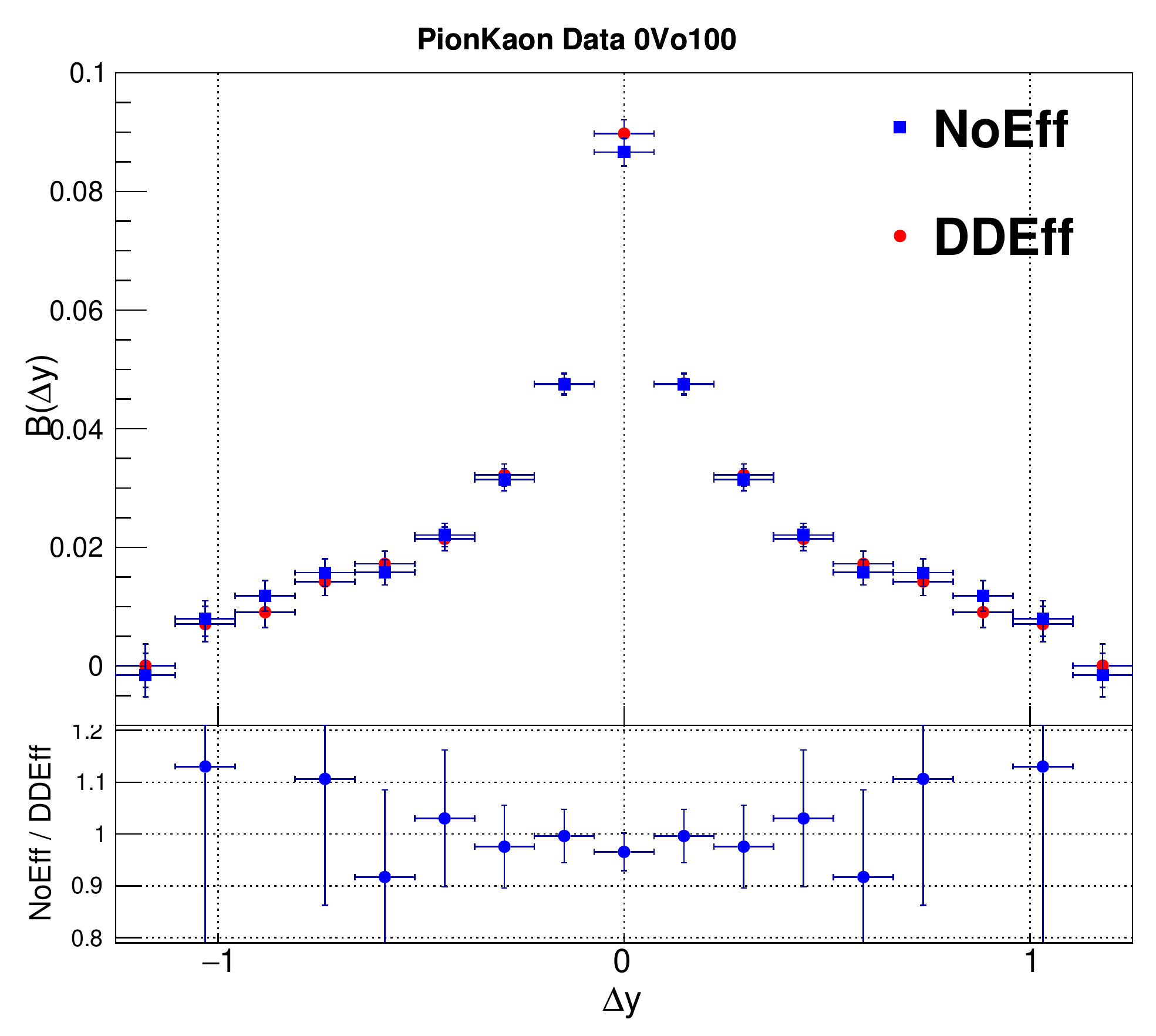}
  \includegraphics[width=0.3\linewidth]{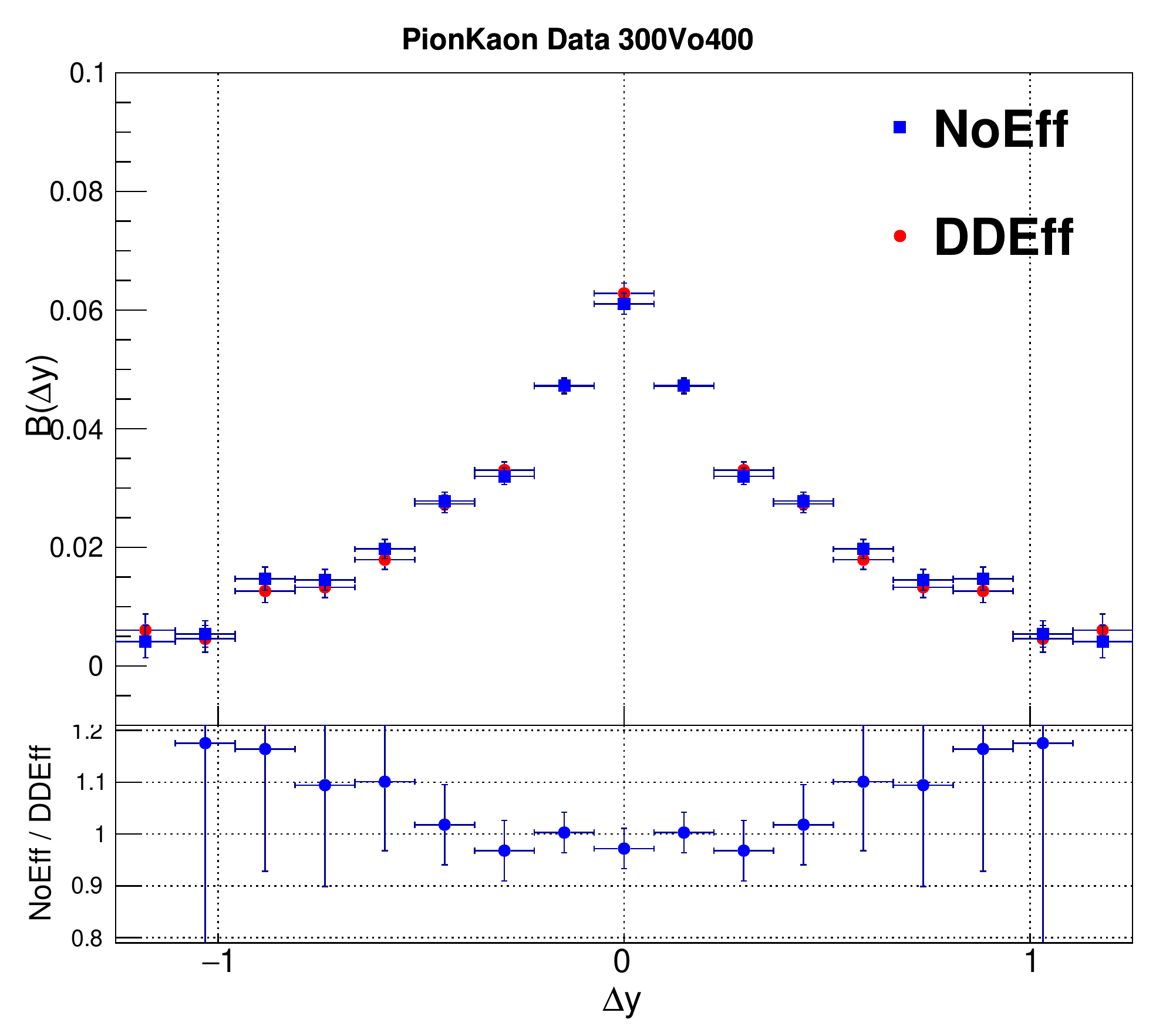}
  \includegraphics[width=0.3\linewidth]{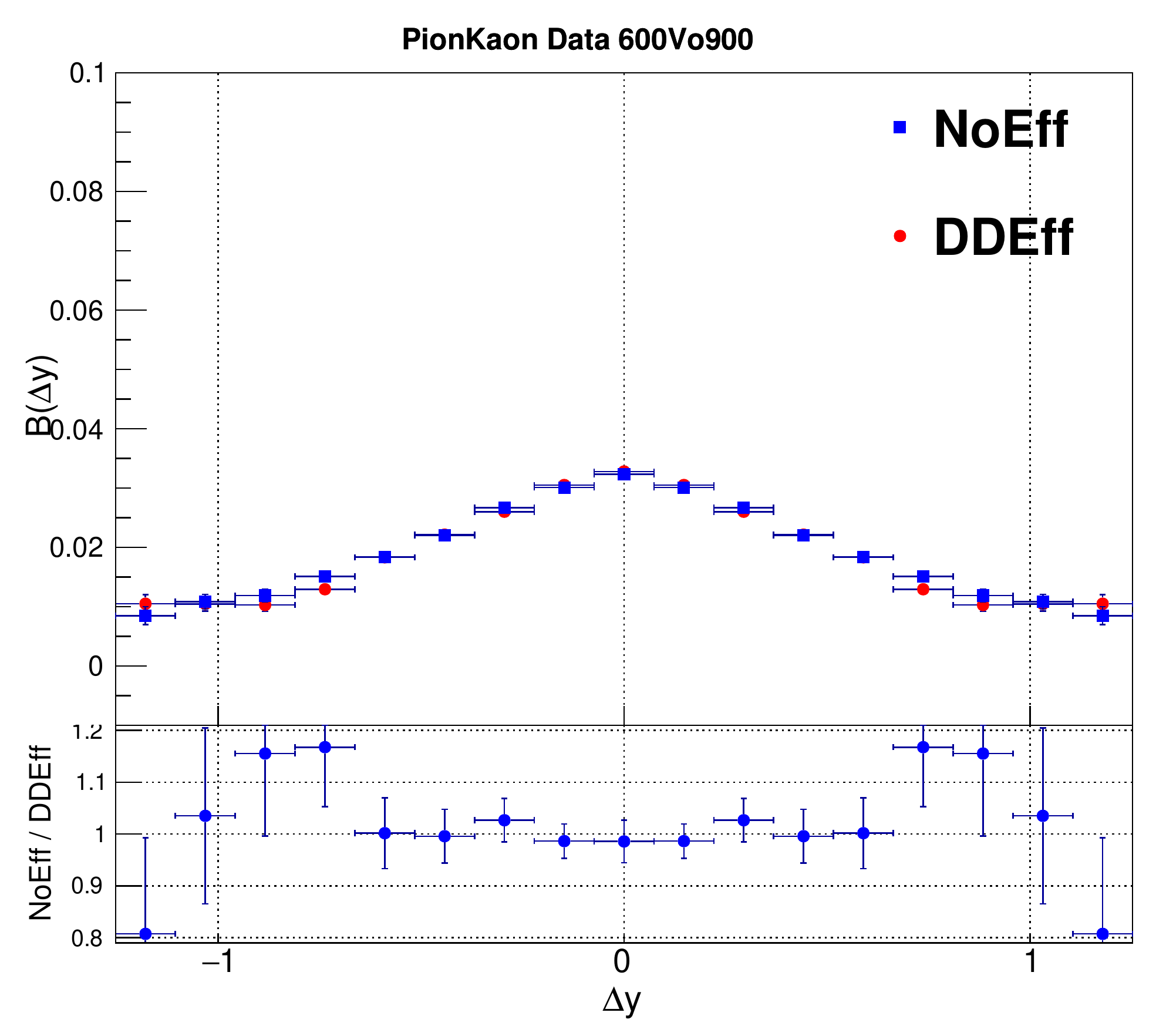} 
  \includegraphics[width=0.3\linewidth]{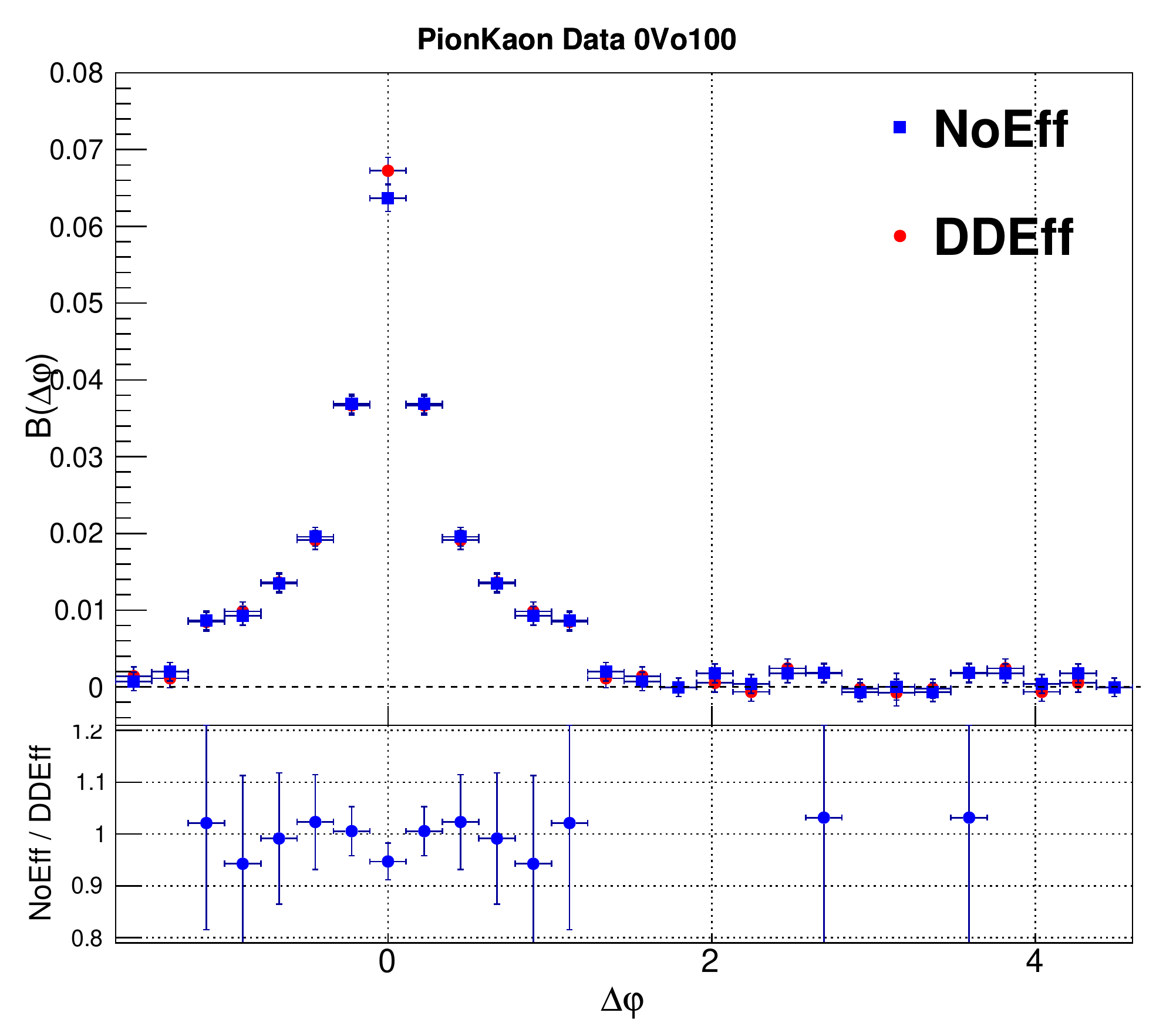}
  \includegraphics[width=0.3\linewidth]{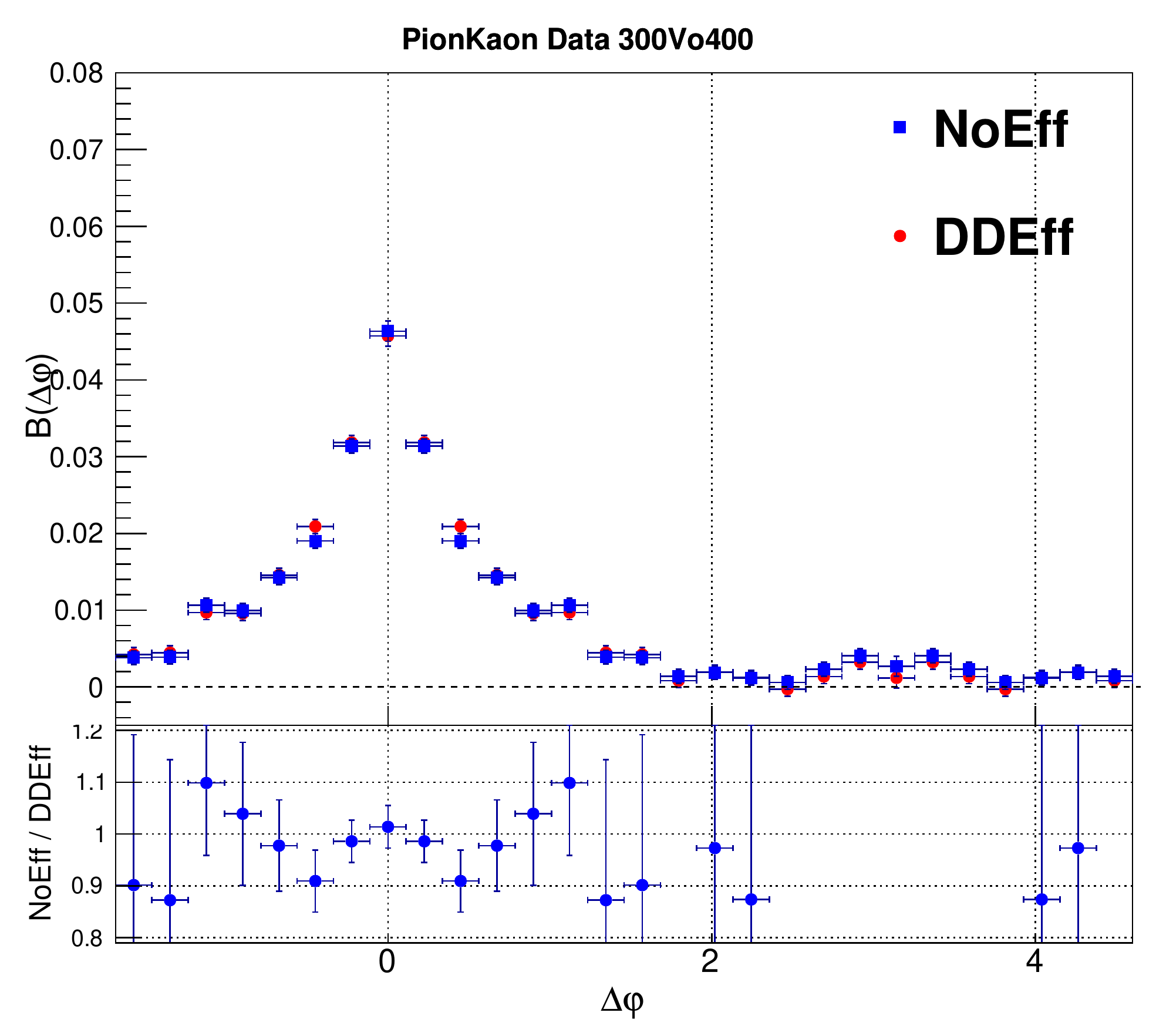}
  \includegraphics[width=0.3\linewidth]{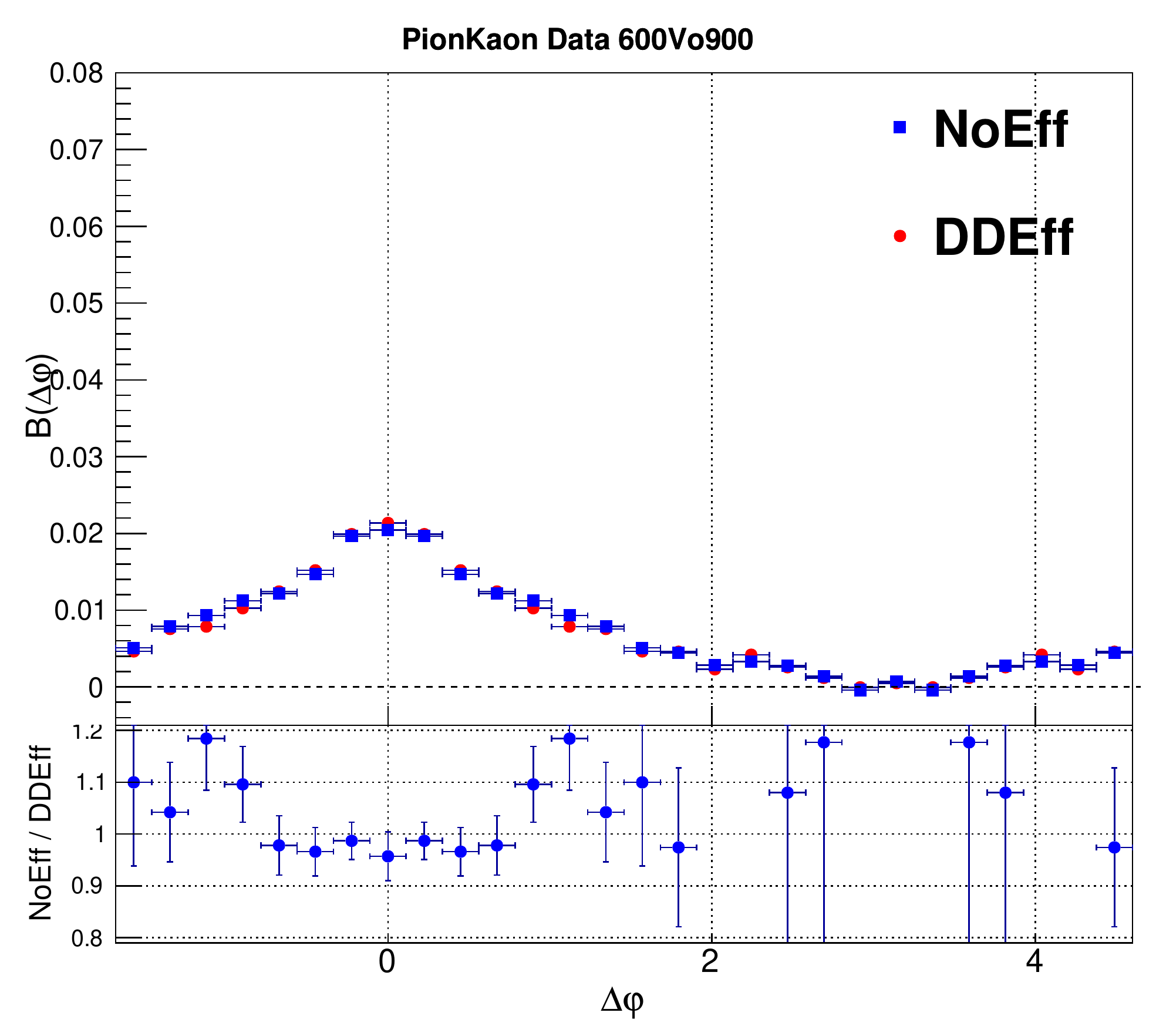} 
  \includegraphics[width=0.3\linewidth]{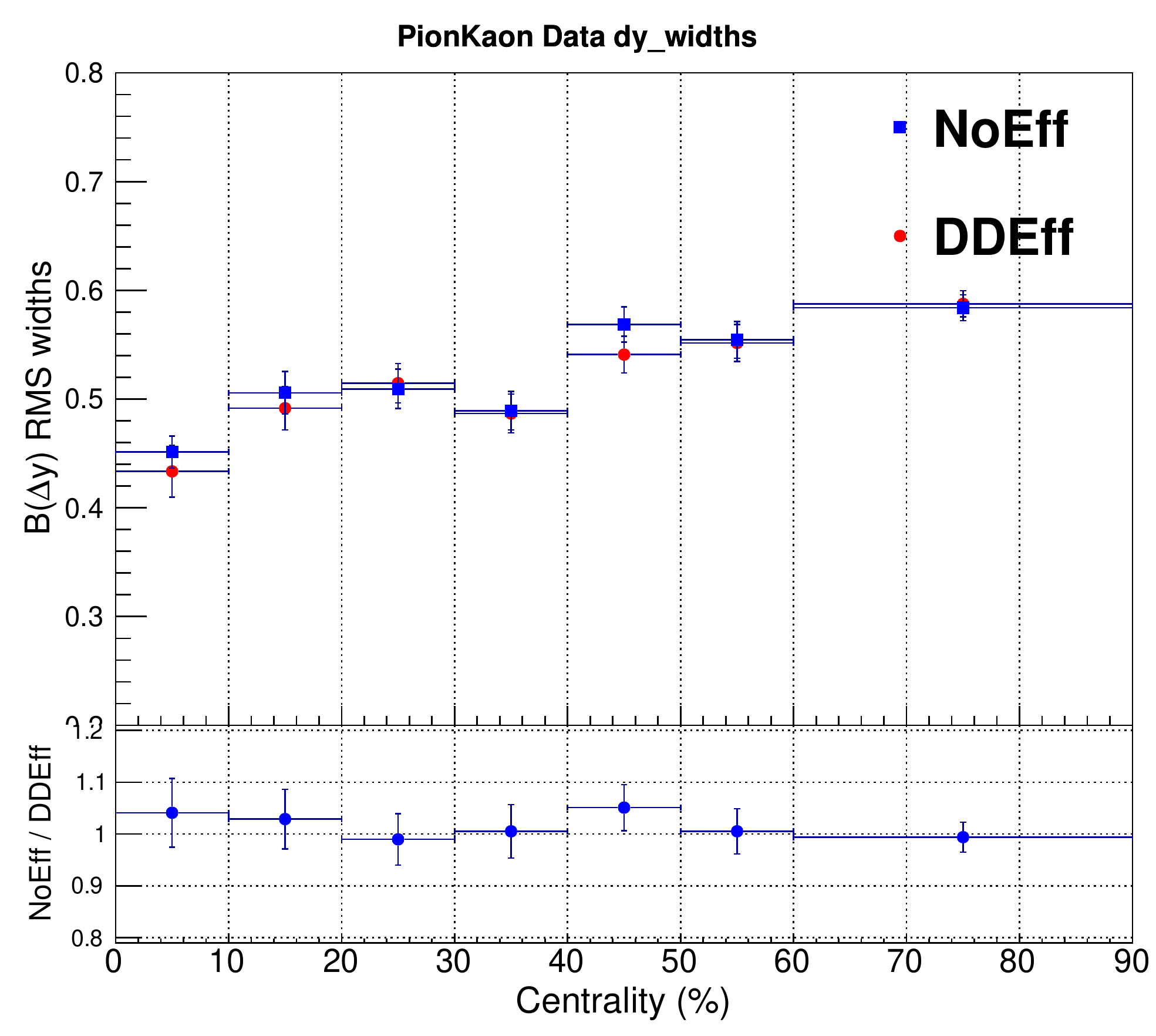}
  \includegraphics[width=0.3\linewidth]{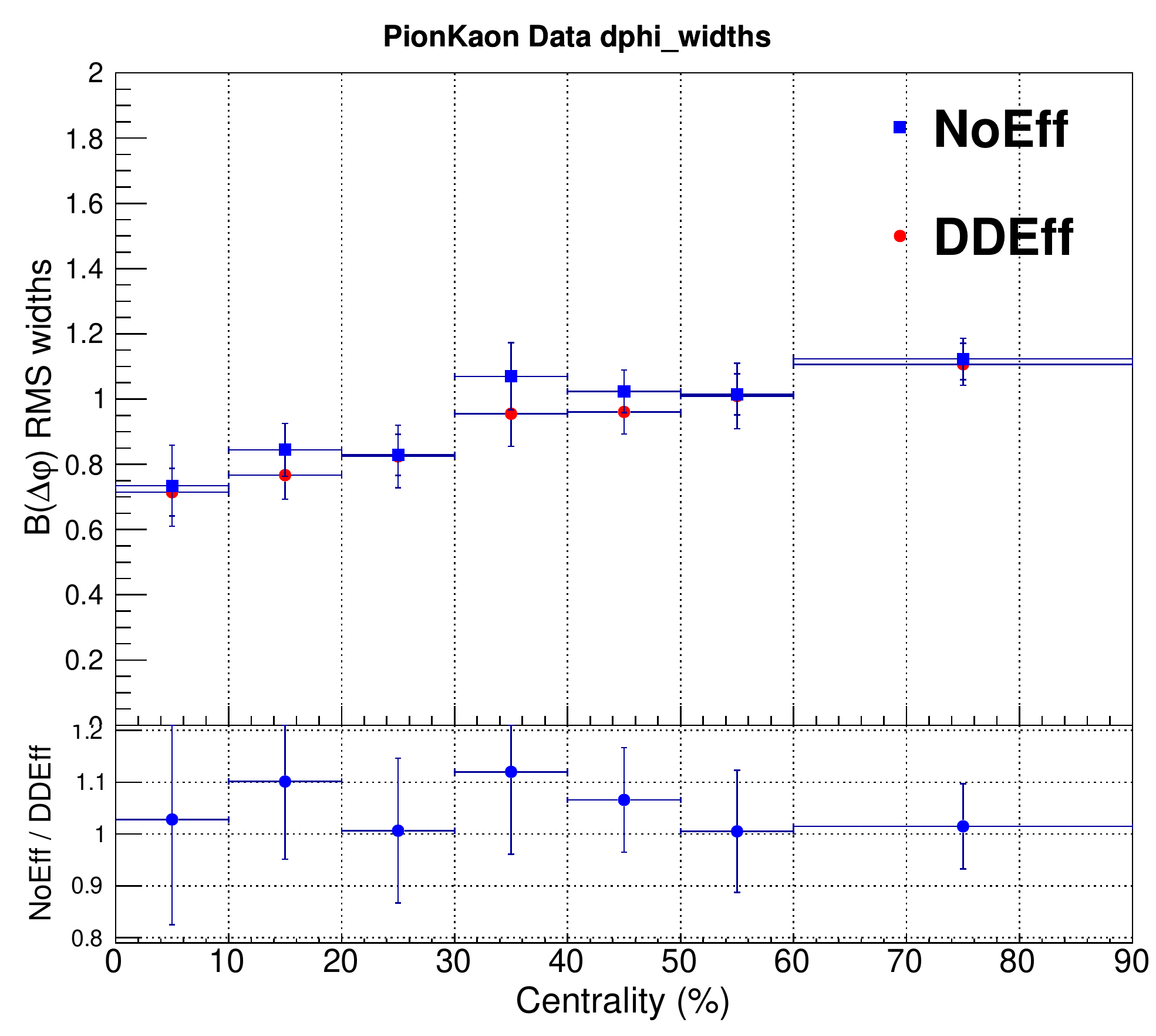}
  \includegraphics[width=0.3\linewidth]{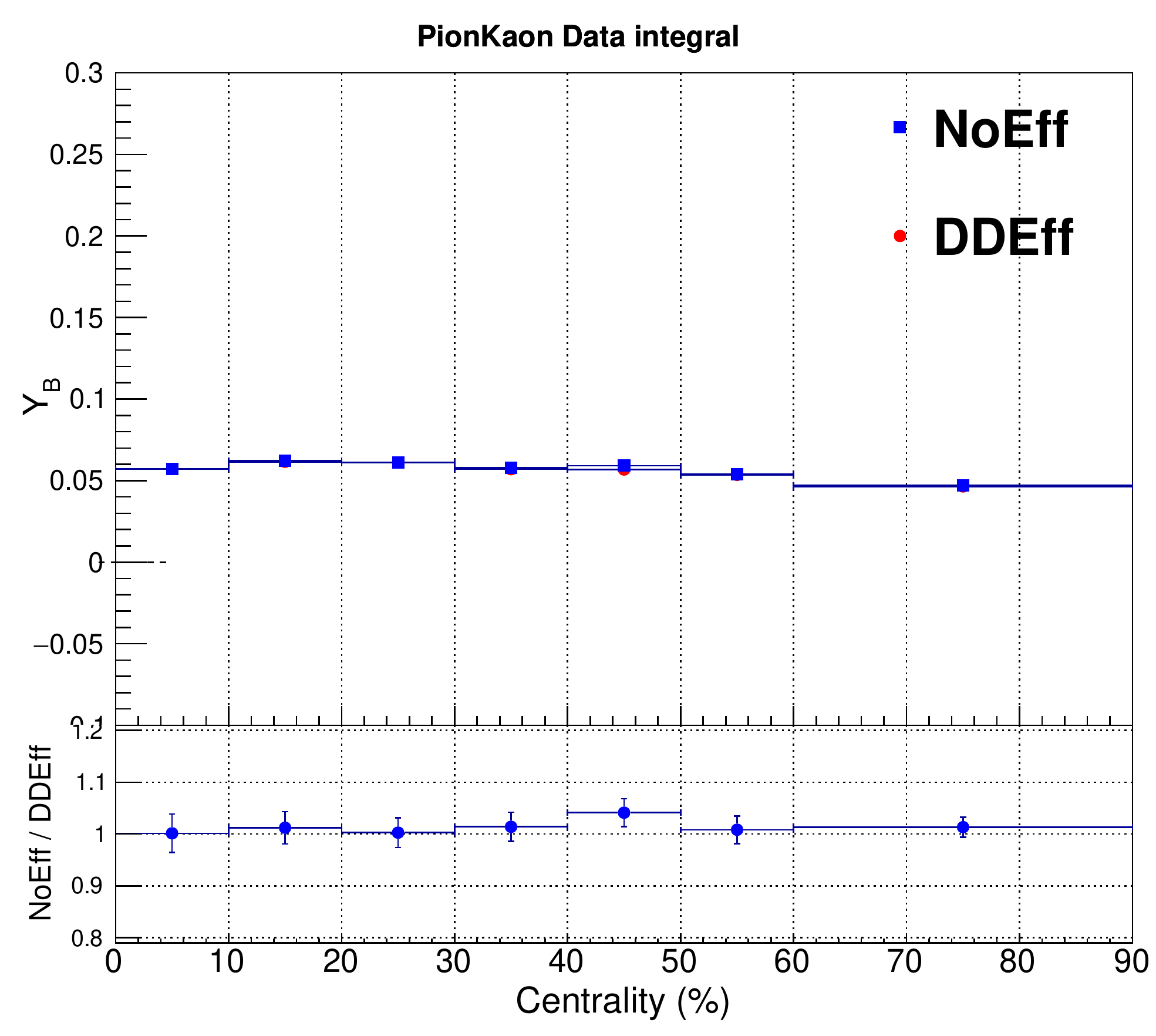} 
  \caption{Comparisons of 2D $B^{\pi K}$ obtained without (1st row) and with Data Driven (2nd row) $p_{\rm T}$-dependent efficiency correction for selected centralities, along with their $\Delta y$ (3rd row) and $\Delta \varphi$ projections (4th row), $\Delta y$ and $\Delta \varphi$ widths, and integrals (5th row).}
   \label{fig:Compare_DDEffCorr_NoEffCorr_BF_PionKaon}
\end{figure}

%PionProton
\begin{figure}
\centering
  \includegraphics[width=0.3\linewidth]{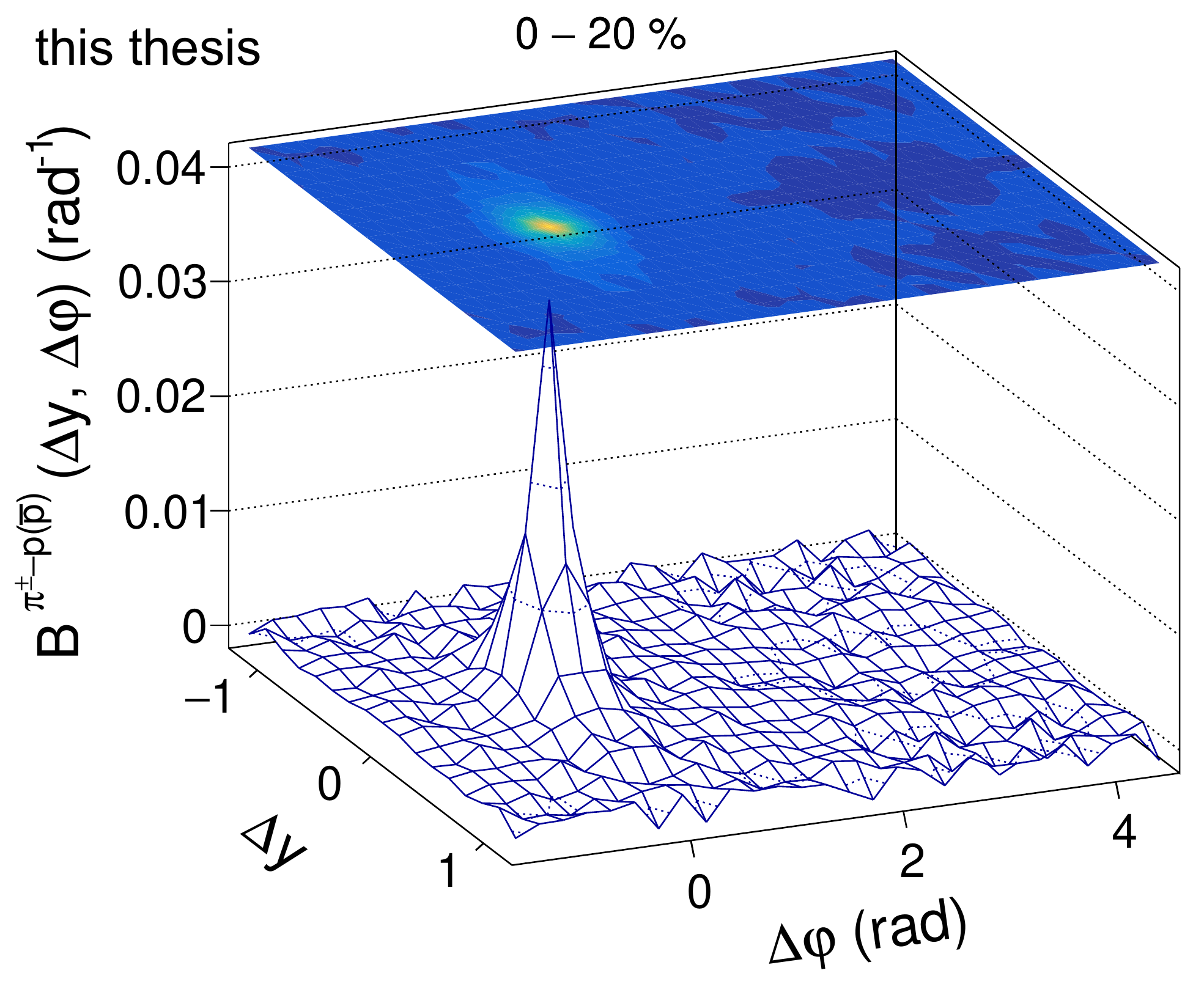}
  \includegraphics[width=0.3\linewidth]{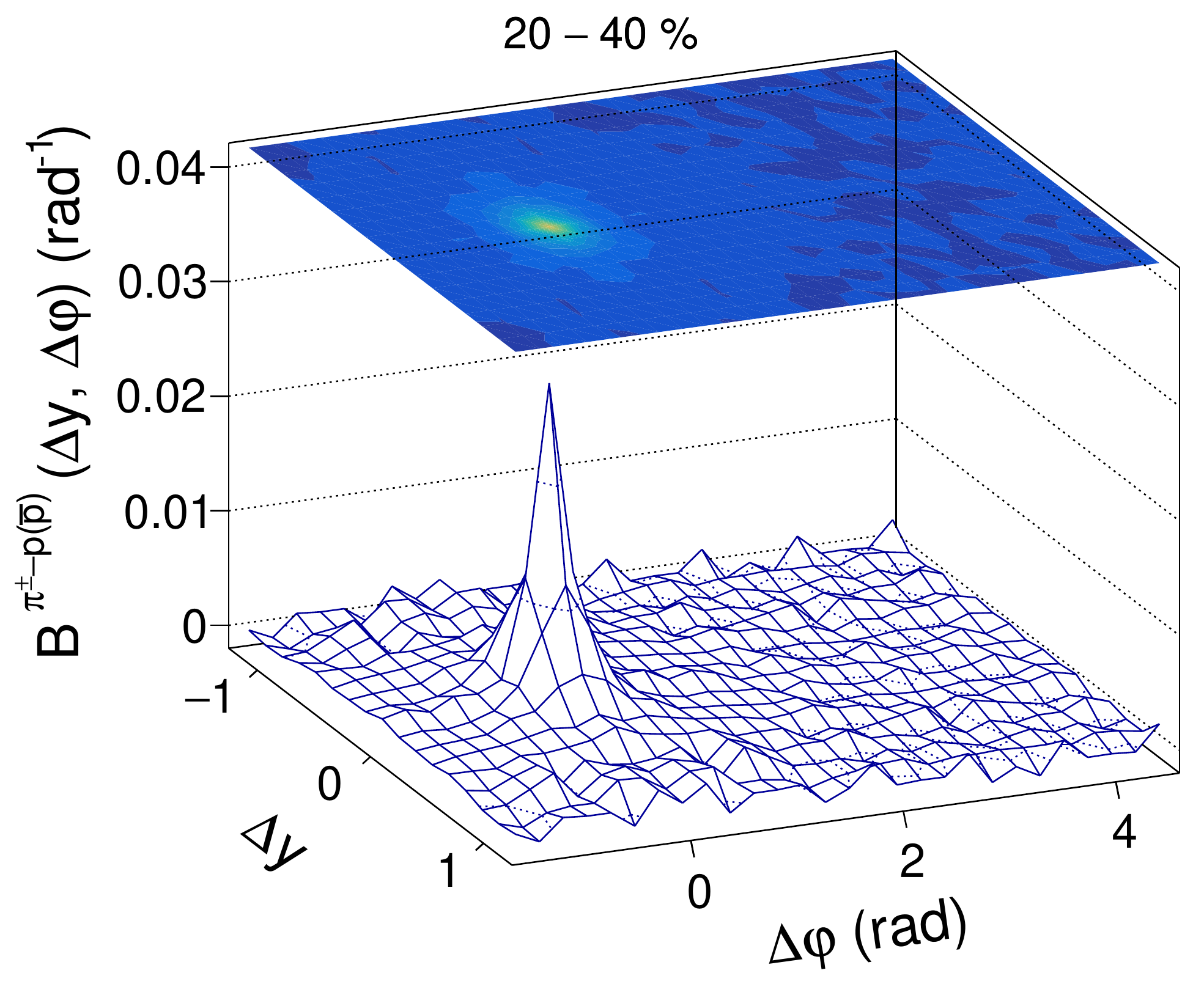}
  \includegraphics[width=0.3\linewidth]{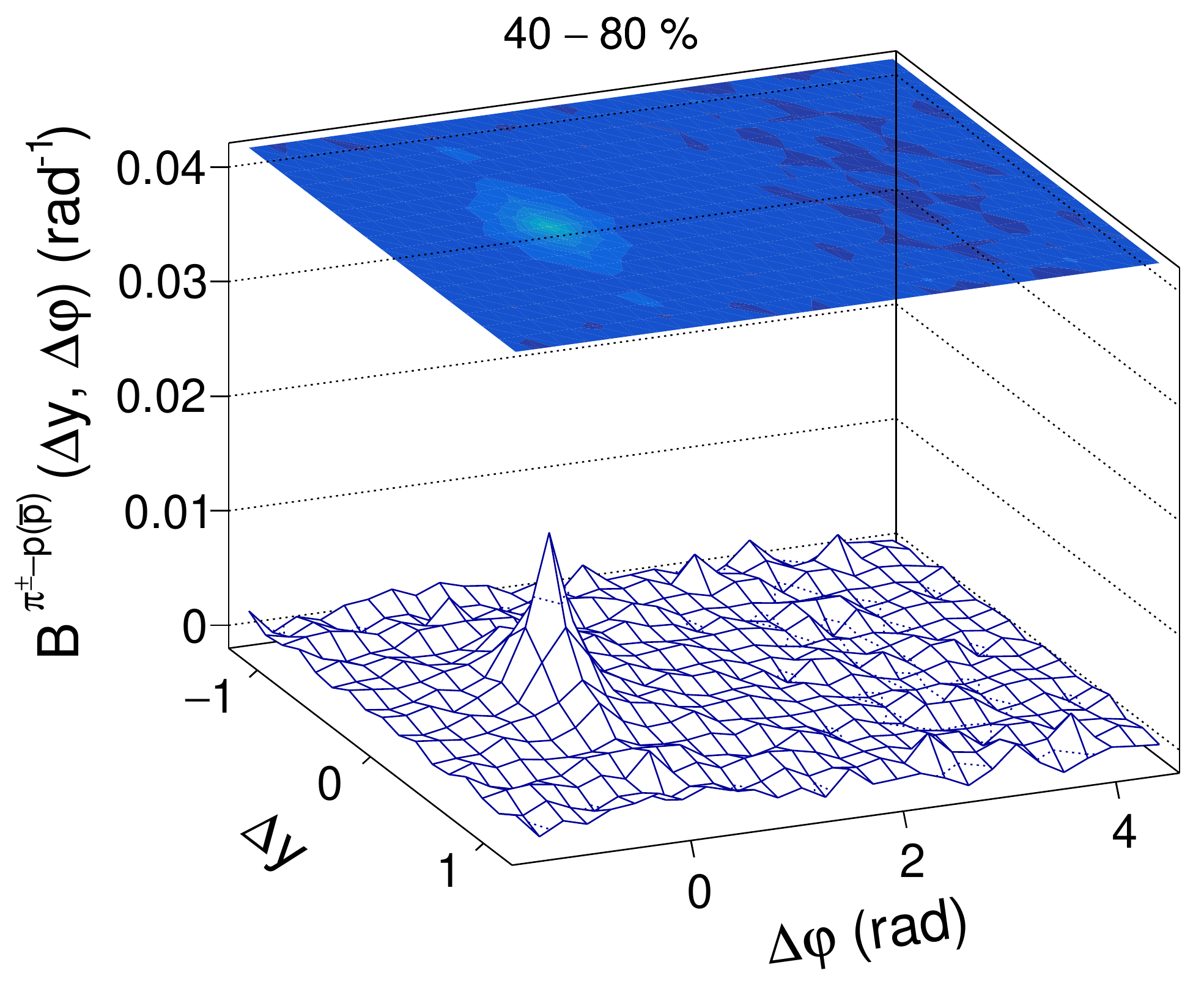}
  \includegraphics[width=0.3\linewidth]{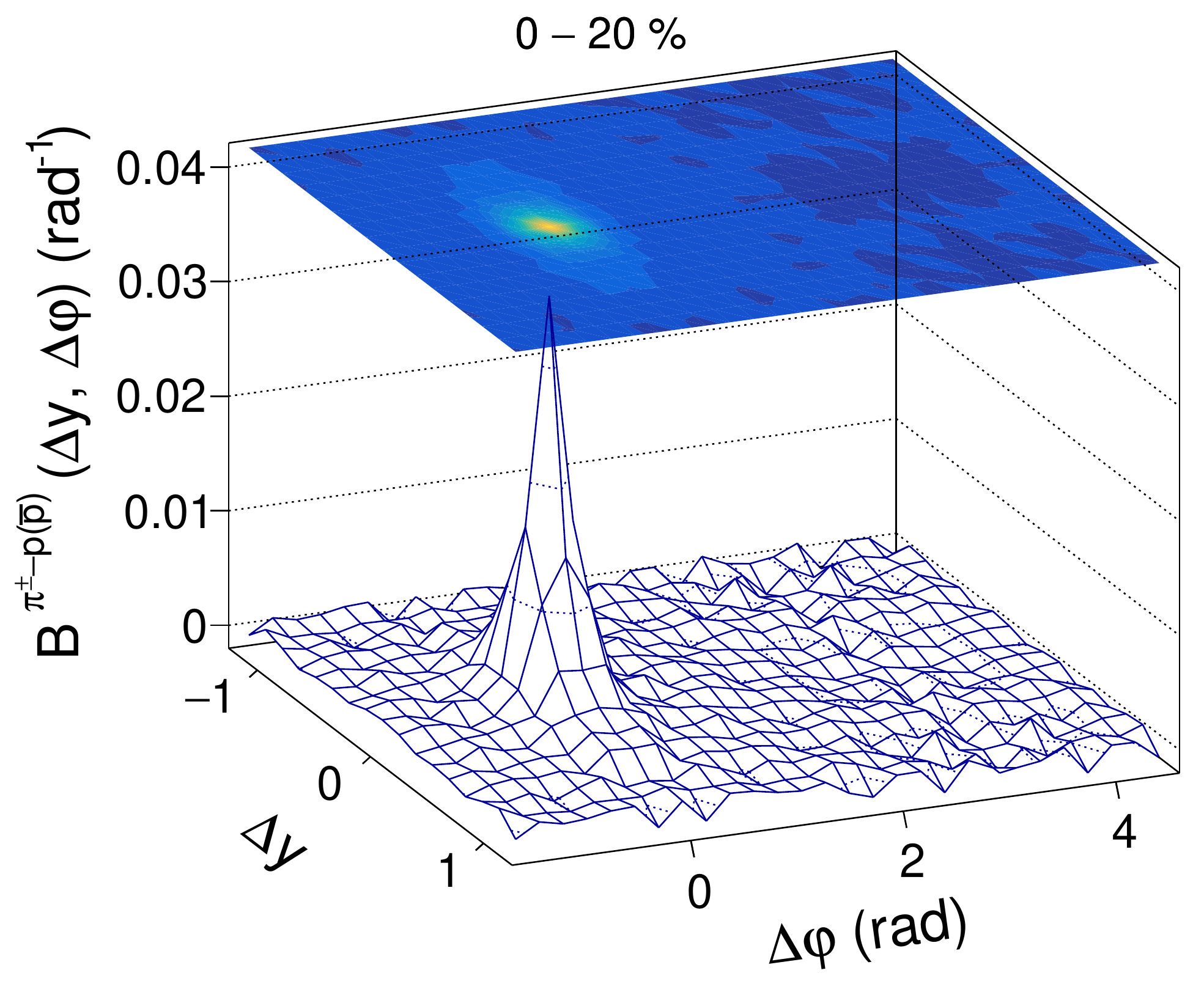}
  \includegraphics[width=0.3\linewidth]{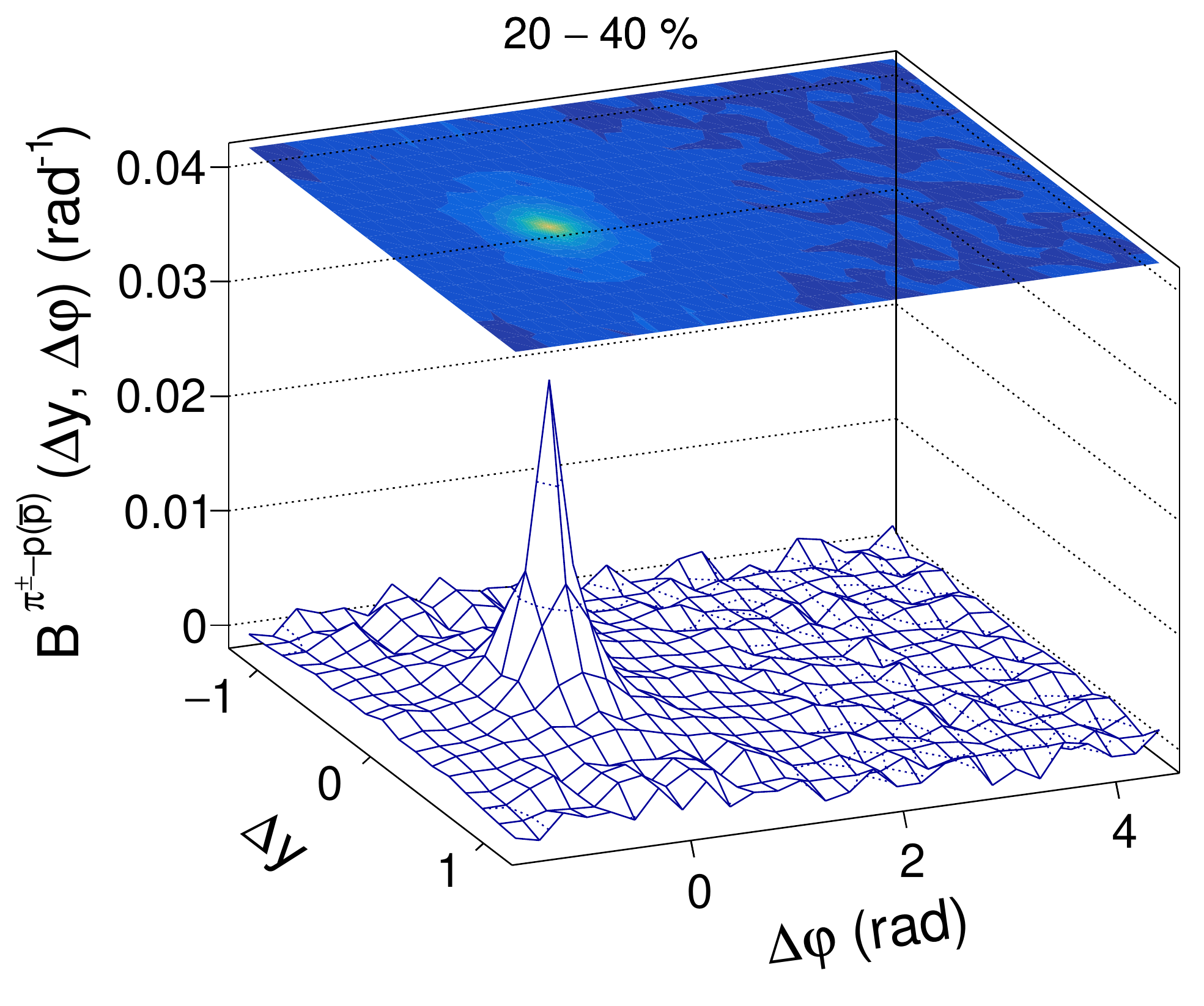}
  \includegraphics[width=0.3\linewidth]{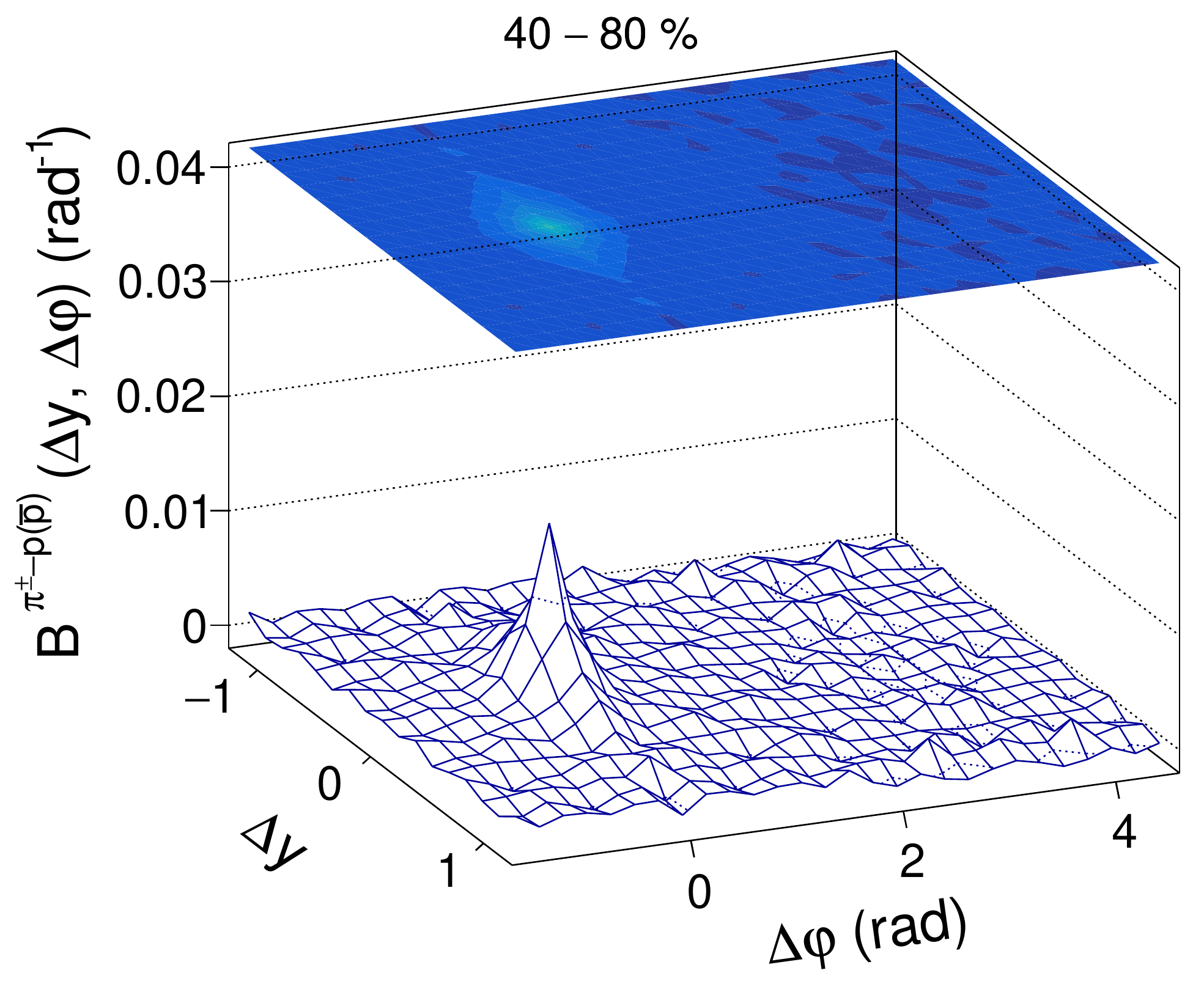}
  \includegraphics[width=0.3\linewidth]{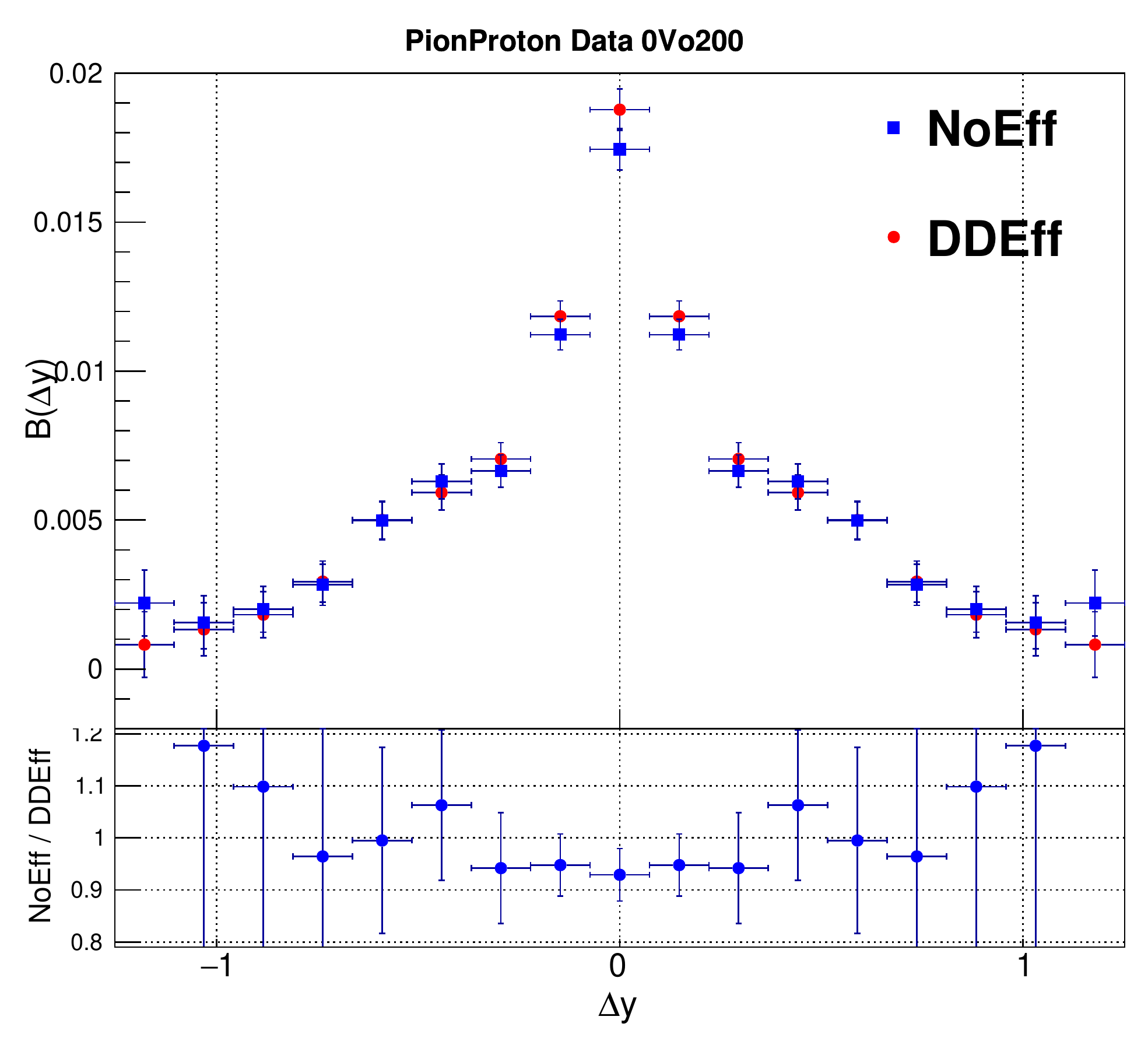}
  \includegraphics[width=0.3\linewidth]{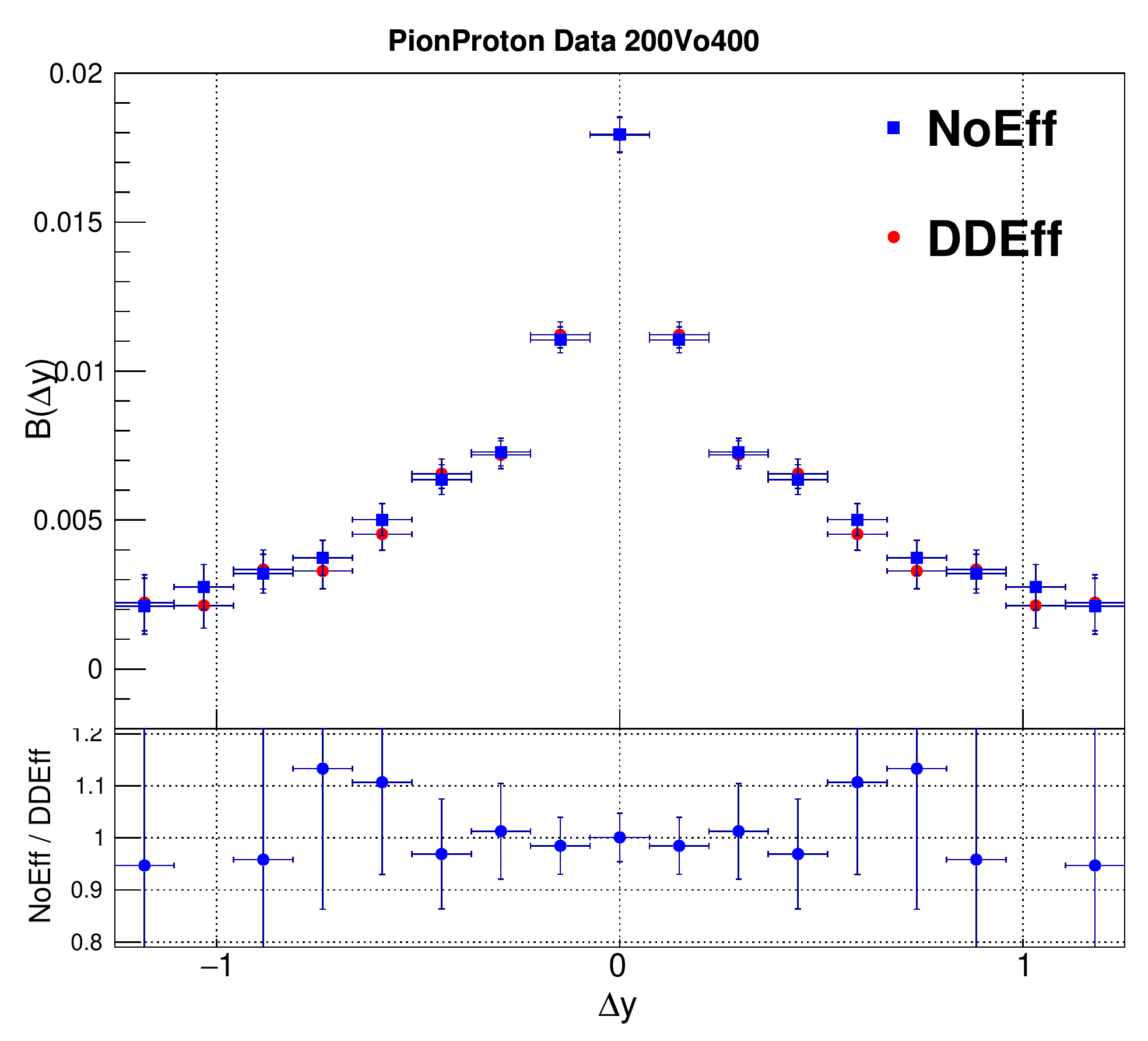}
  \includegraphics[width=0.3\linewidth]{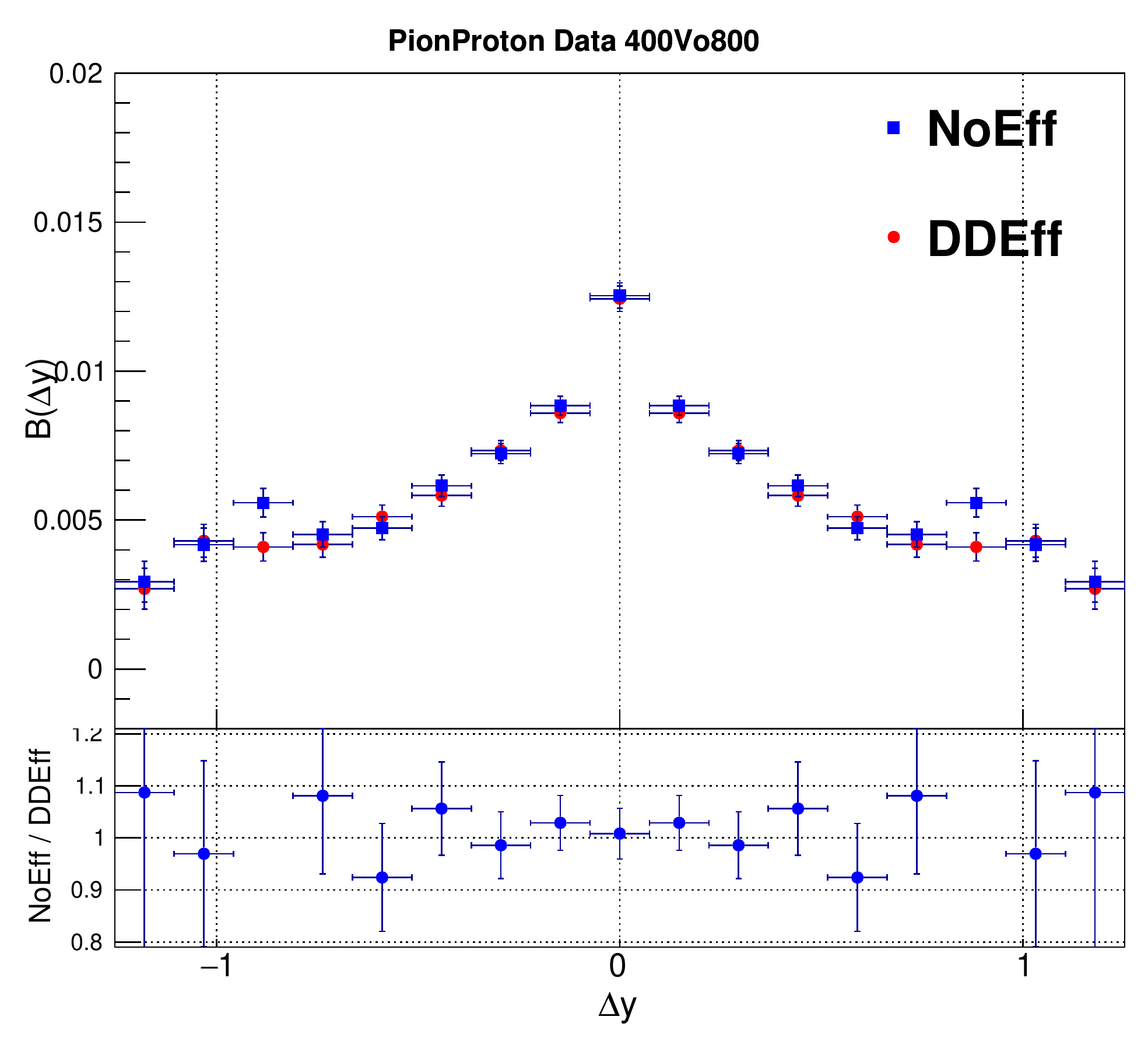} 
  \includegraphics[width=0.3\linewidth]{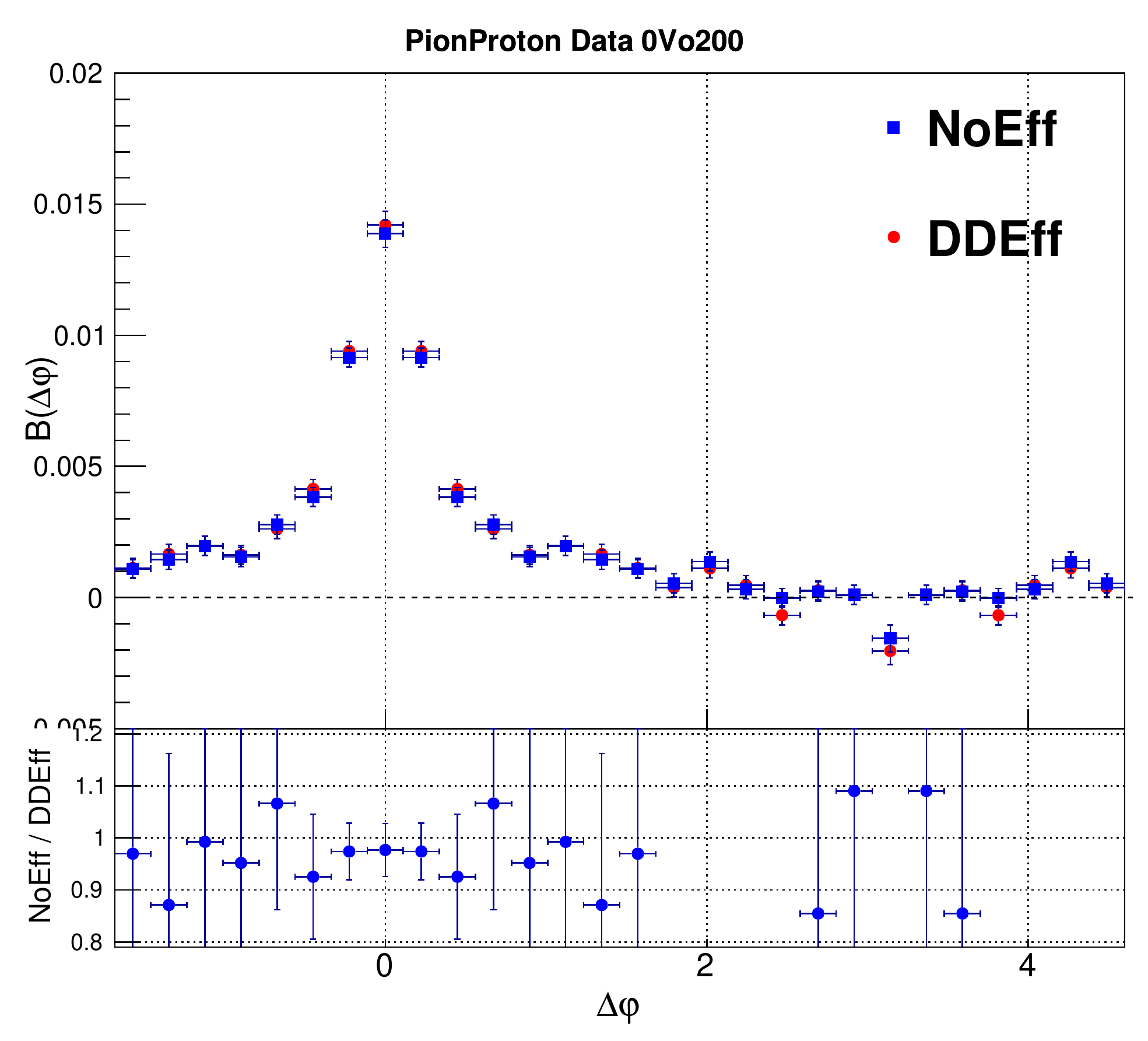}
  \includegraphics[width=0.3\linewidth]{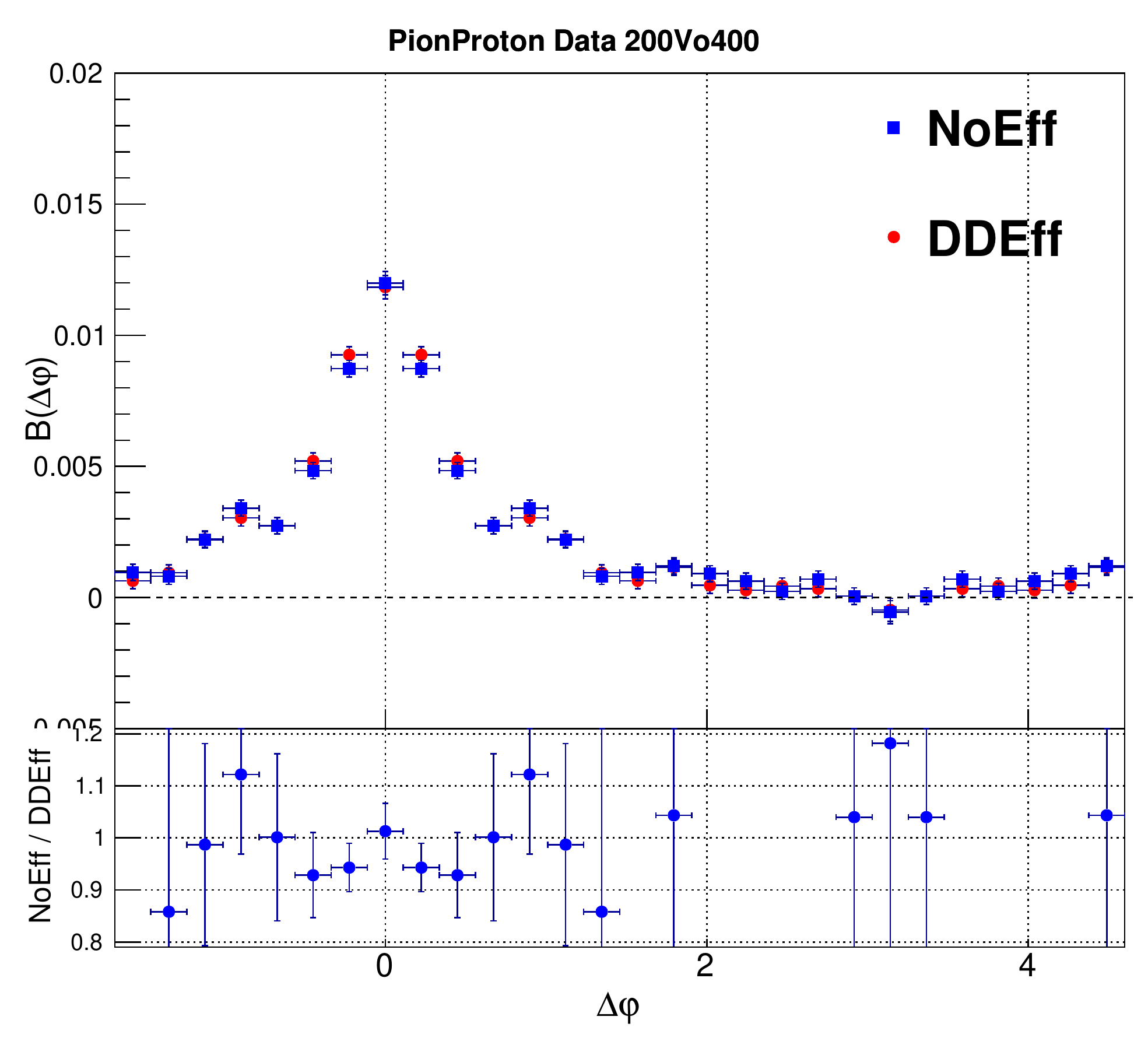}
  \includegraphics[width=0.3\linewidth]{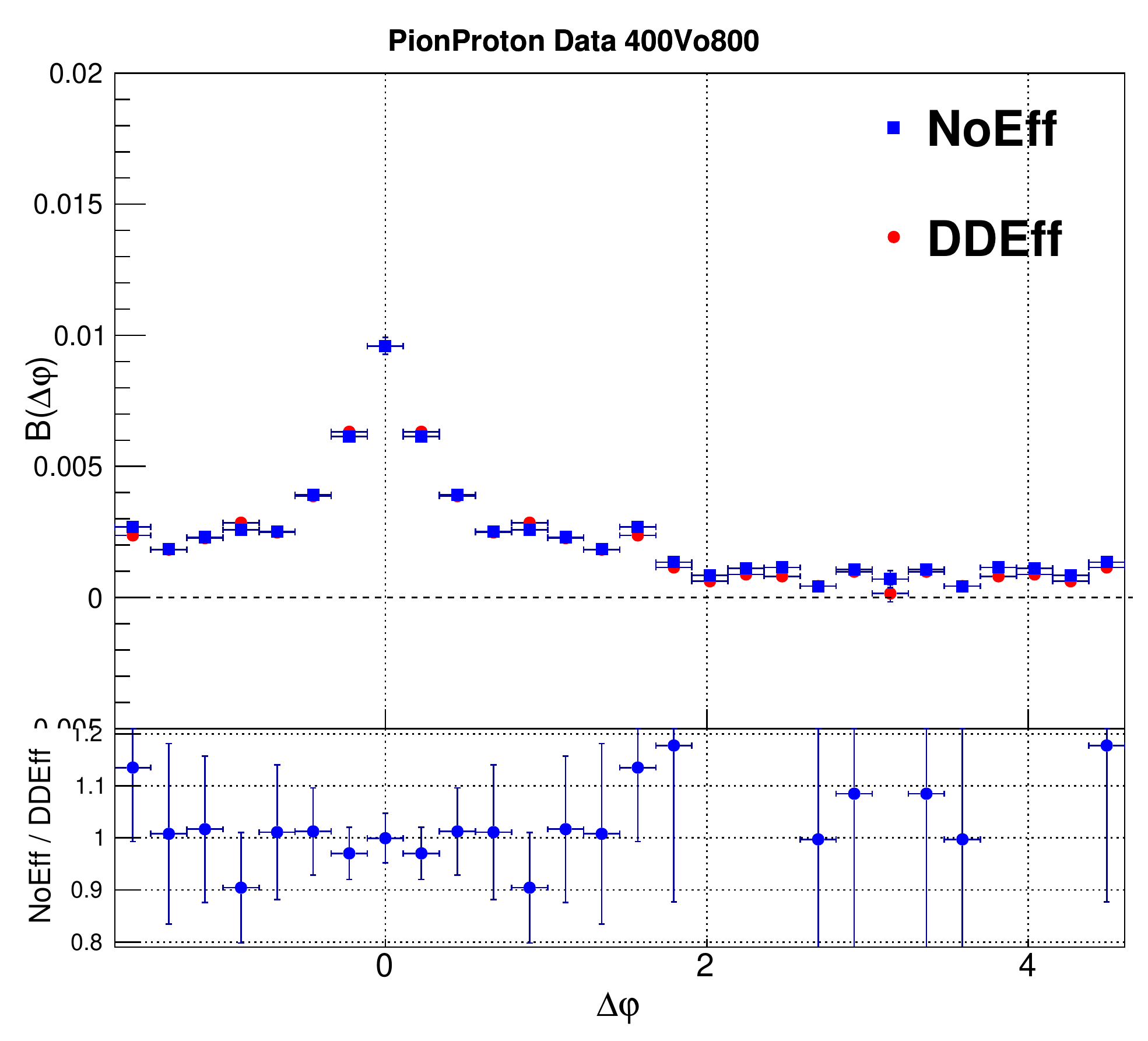} 
  \includegraphics[width=0.3\linewidth]{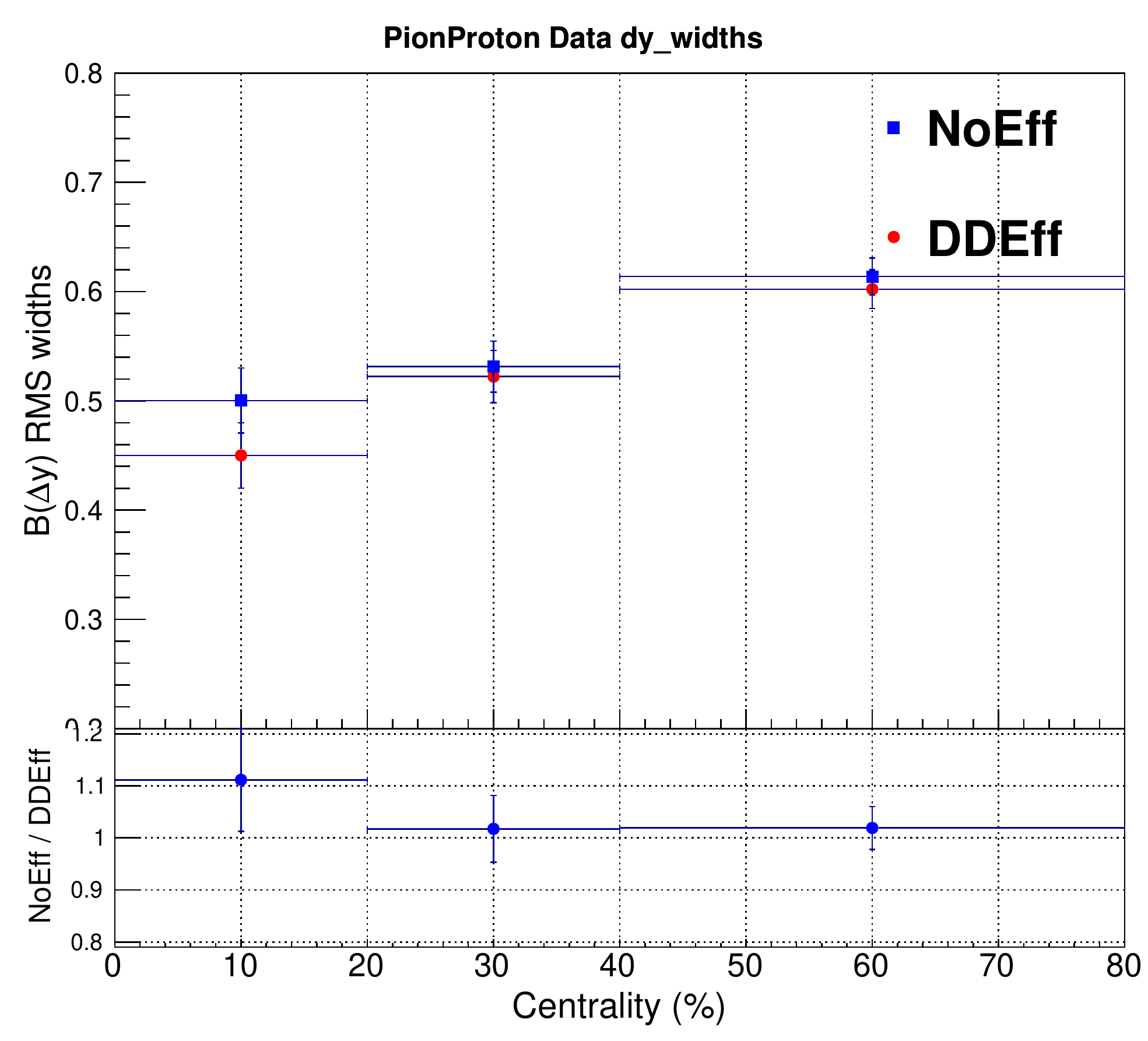}
  \includegraphics[width=0.3\linewidth]{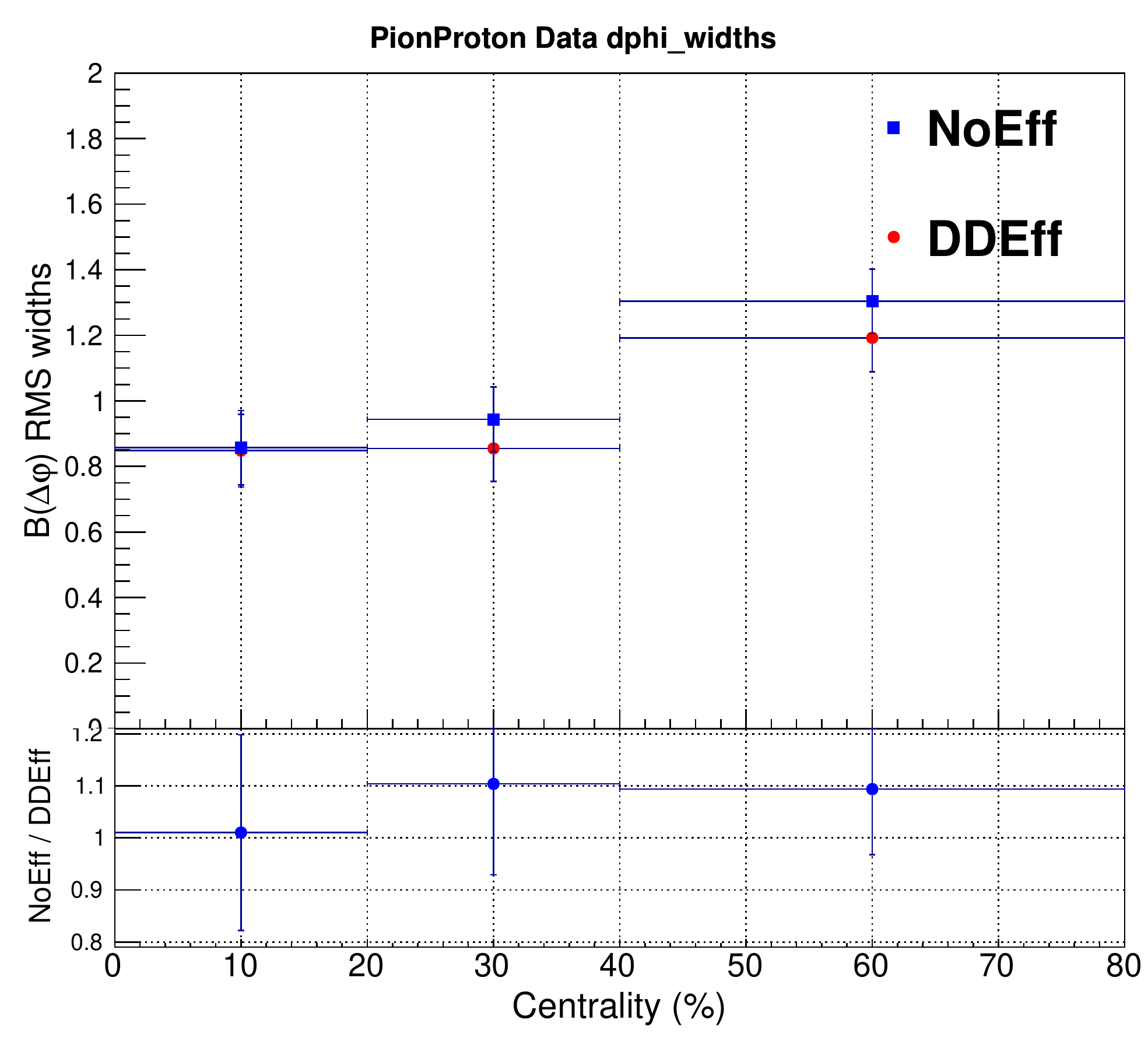}
  \includegraphics[width=0.3\linewidth]{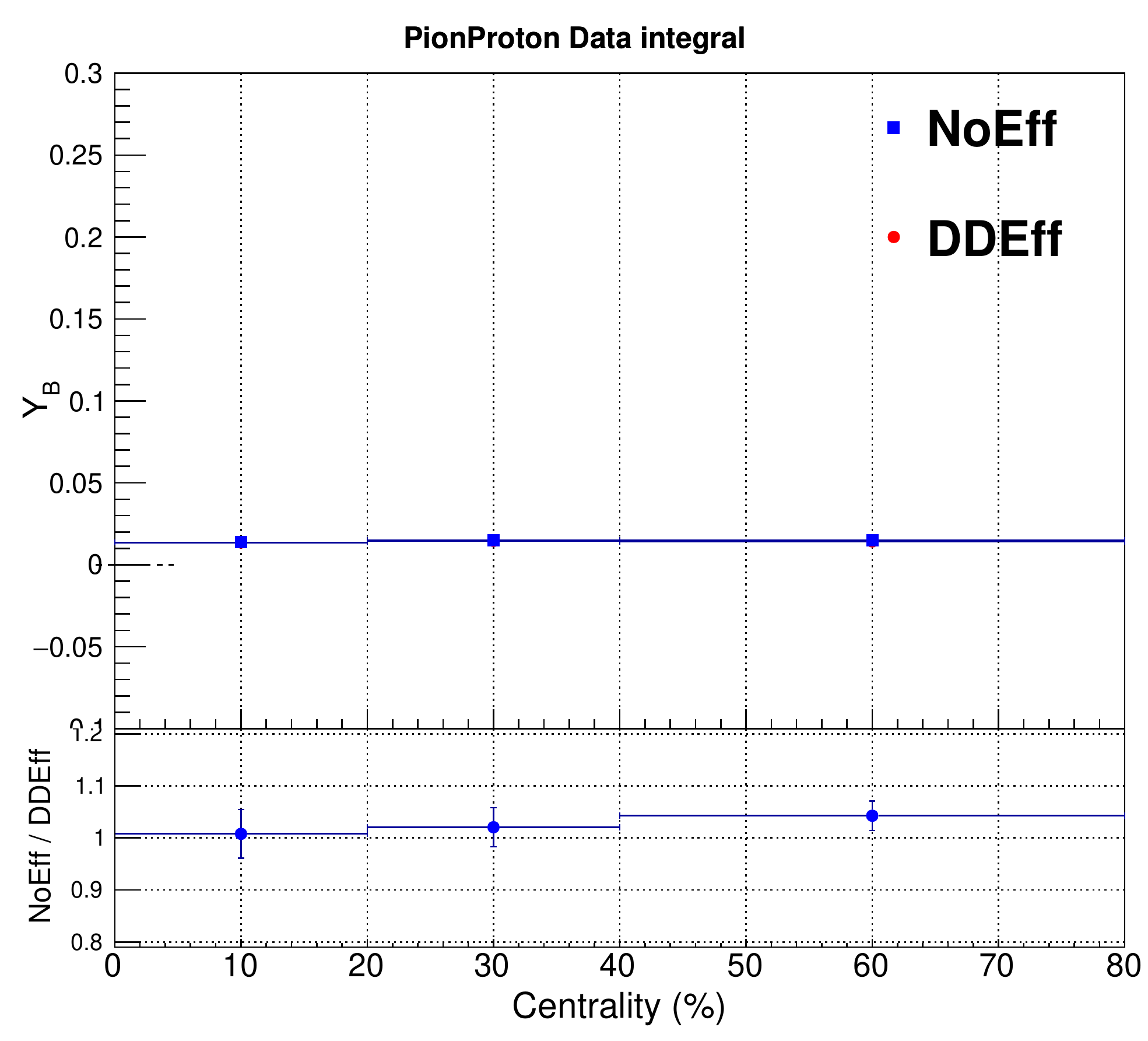} 
  \caption{Comparisons of 2D $B^{\pi p}$ obtained without (1st row) and with Data Driven (2nd row) $p_{\rm T}$-dependent efficiency correction for selected centralities, along with their $\Delta y$ (3rd row) and $\Delta \varphi$ projections (4th row), $\Delta y$ and $\Delta \varphi$ widths, and integrals (5th row).}
   \label{fig:Compare_DDEffCorr_NoEffCorr_BF_PionProton}
\end{figure}

%KaonProton

\clearpage

\section{Track Splitting Study}
\label{subsec:TrackSplitting}

Track splitting occurs when the track reconstruction software fails to properly connect track segments that belong together thereby  producing two reconstructed tracks instead of one. Such failures occur when hits belonging to  two track segments are not properly aligned. Track splitting may thus yield a narrow spike at the origin of 2D $B_{LS}(\Delta y,\Delta \varphi)$ of  same-species pairs, namely  $\pi\pi$, $KK$, and $pp$ pairs. 

Track splitting may nominally be suppressed by requiring all tracks included in the analysis are reconstructed with more than half of the possible number of hits they can have according to their geometry. However, imposing a ``long track" requirement greatly reduces the number of accepted tracks and thus the statistical quality of the measured CFs. It is thus important not to use too stringent a cut. In this work, a cut $N_{\rm TPC clusters} \ge 70$ out of a maximum of 159 is used for selecting tracks, which in principle suppresses a majority of the split tracks.  In addition, a comparison with a set of BF results with a cut $N_{\rm TPC clusters} \ge 85$ is taken as an independent systematic uncertainty, the details of which are in Chapter~\ref{chap:SystematicUncertainties}. The differences of BF results between $N_{\rm TPC clusters} \ge 70$ and $N_{\rm TPC clusters} \ge 85$ are smaller than 1\%. The impact of split tracks on $R_2^{LS}$ correlation function reported in this work is thus expected to be essentially negligible.  

\begin{figure}
\centering
    \includegraphics[width=0.32\textwidth]{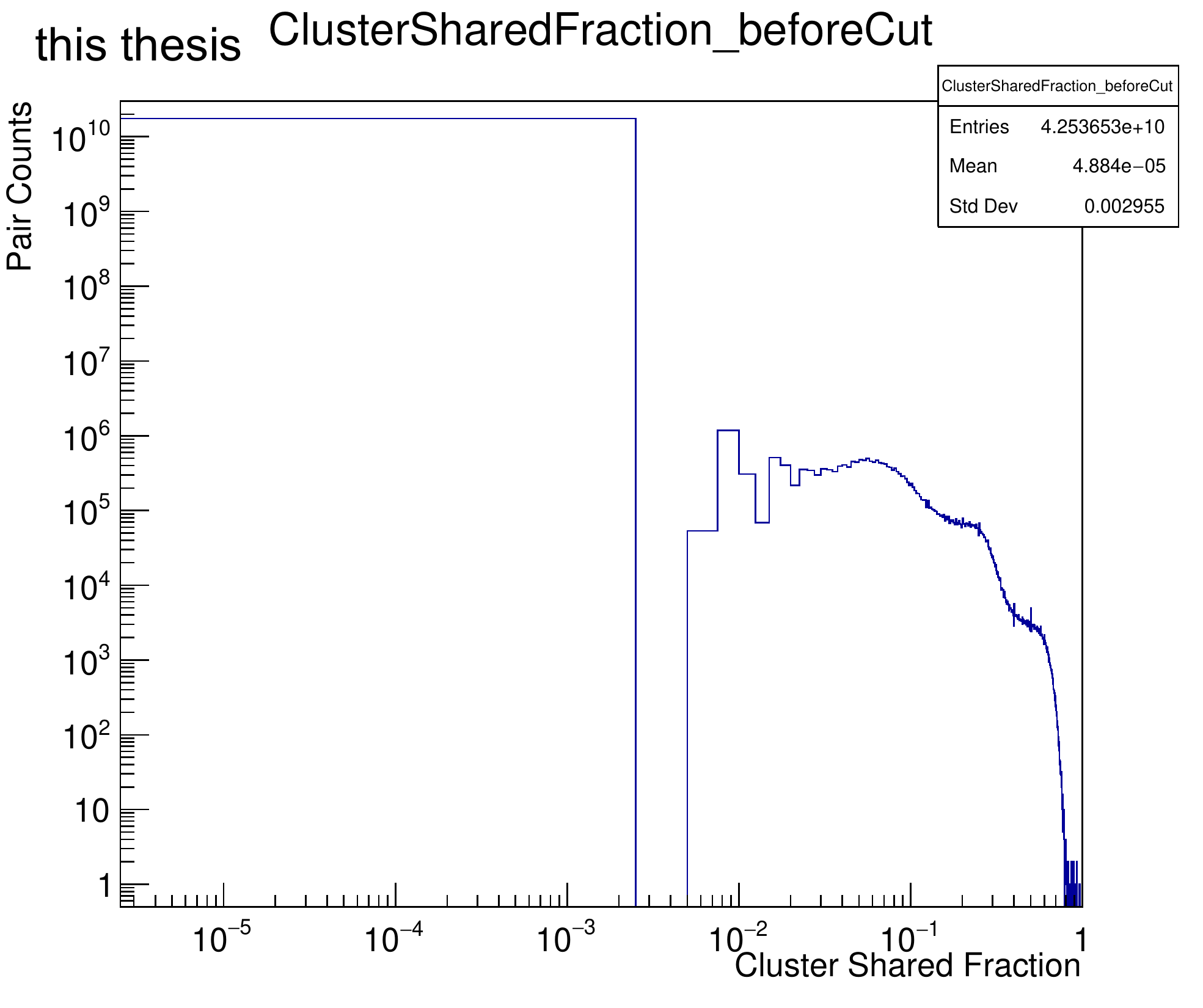}
    \includegraphics[width=0.32\textwidth]{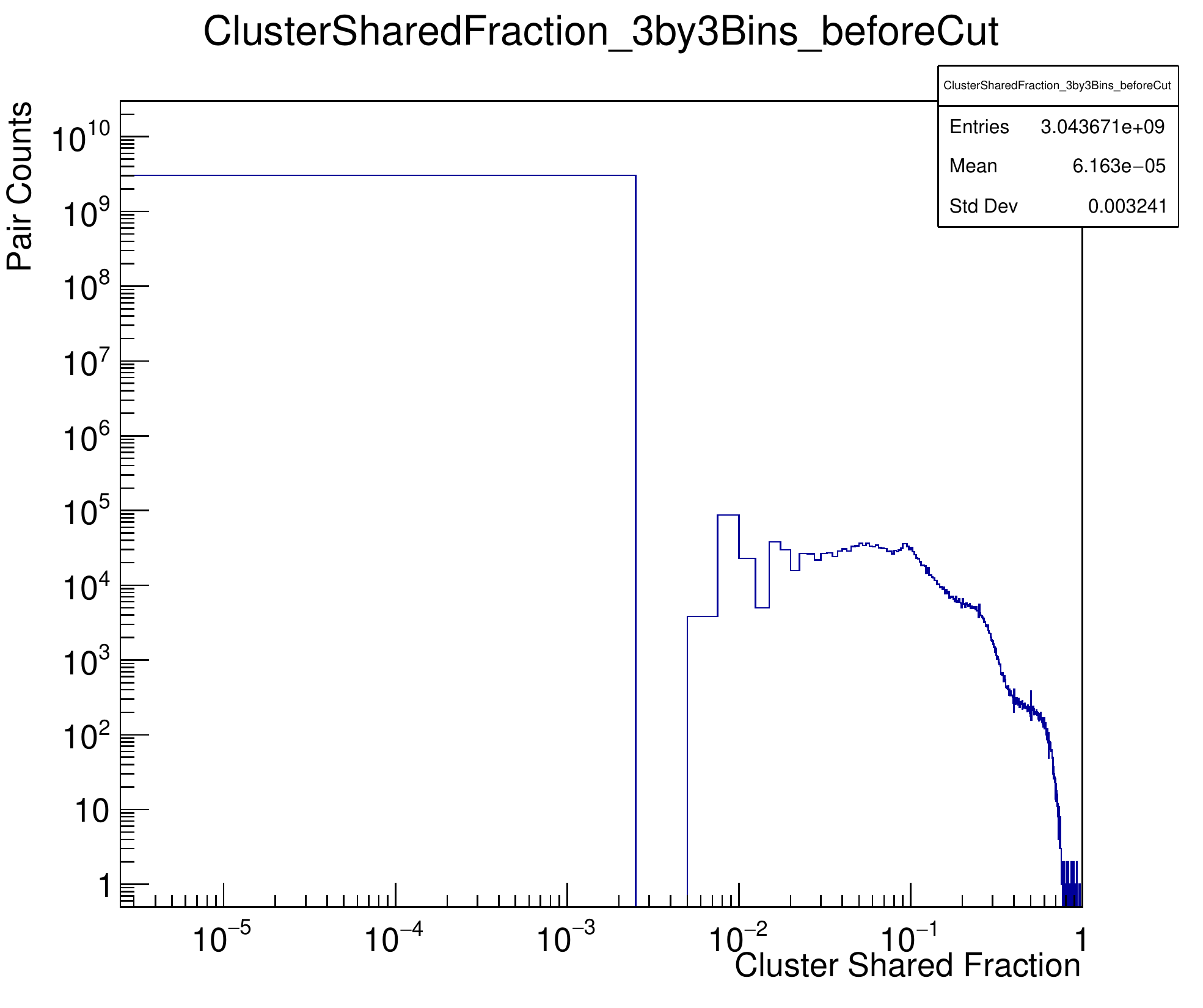}
    \includegraphics[width=0.32\textwidth]{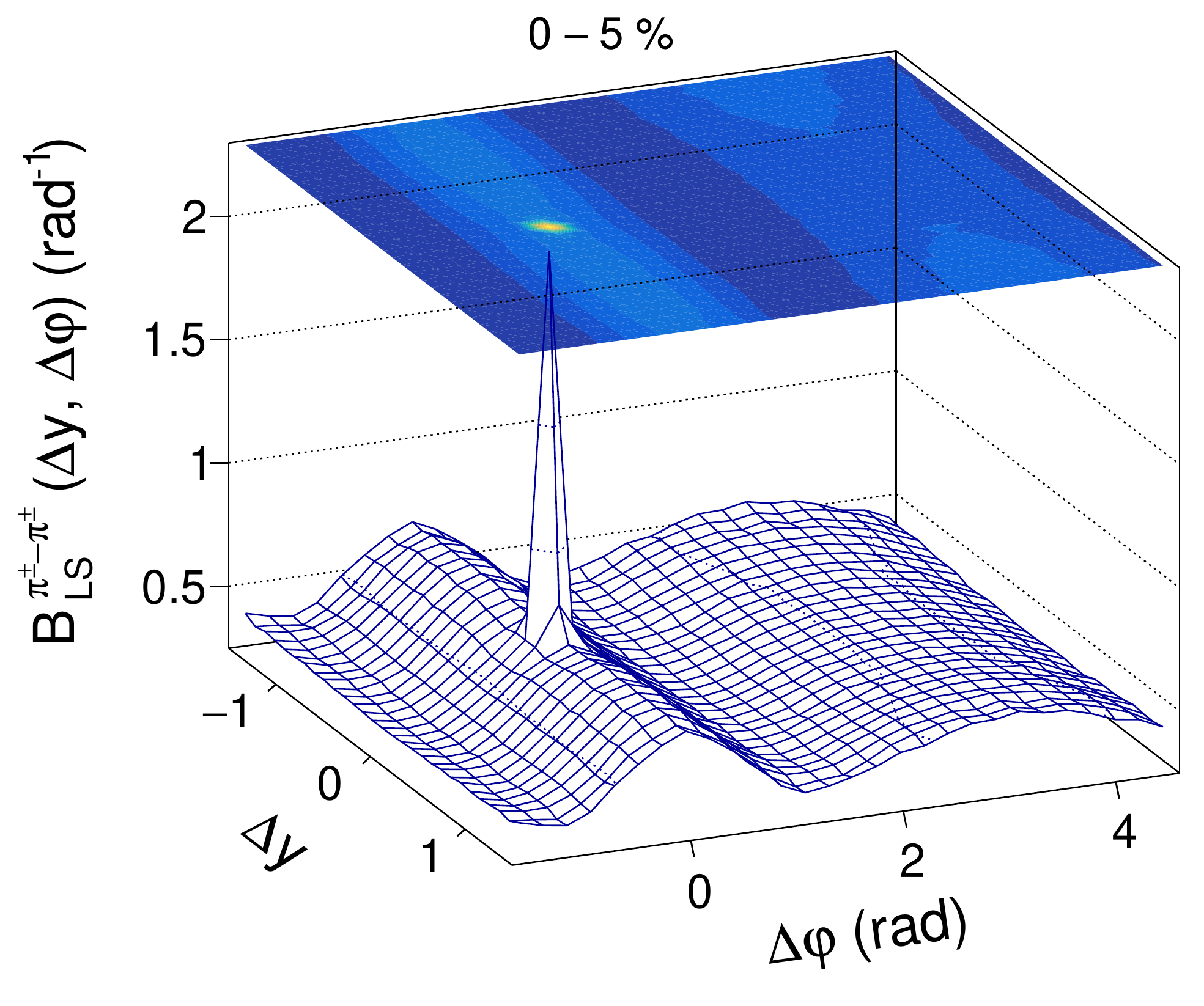}
    \includegraphics[width=0.32\textwidth]{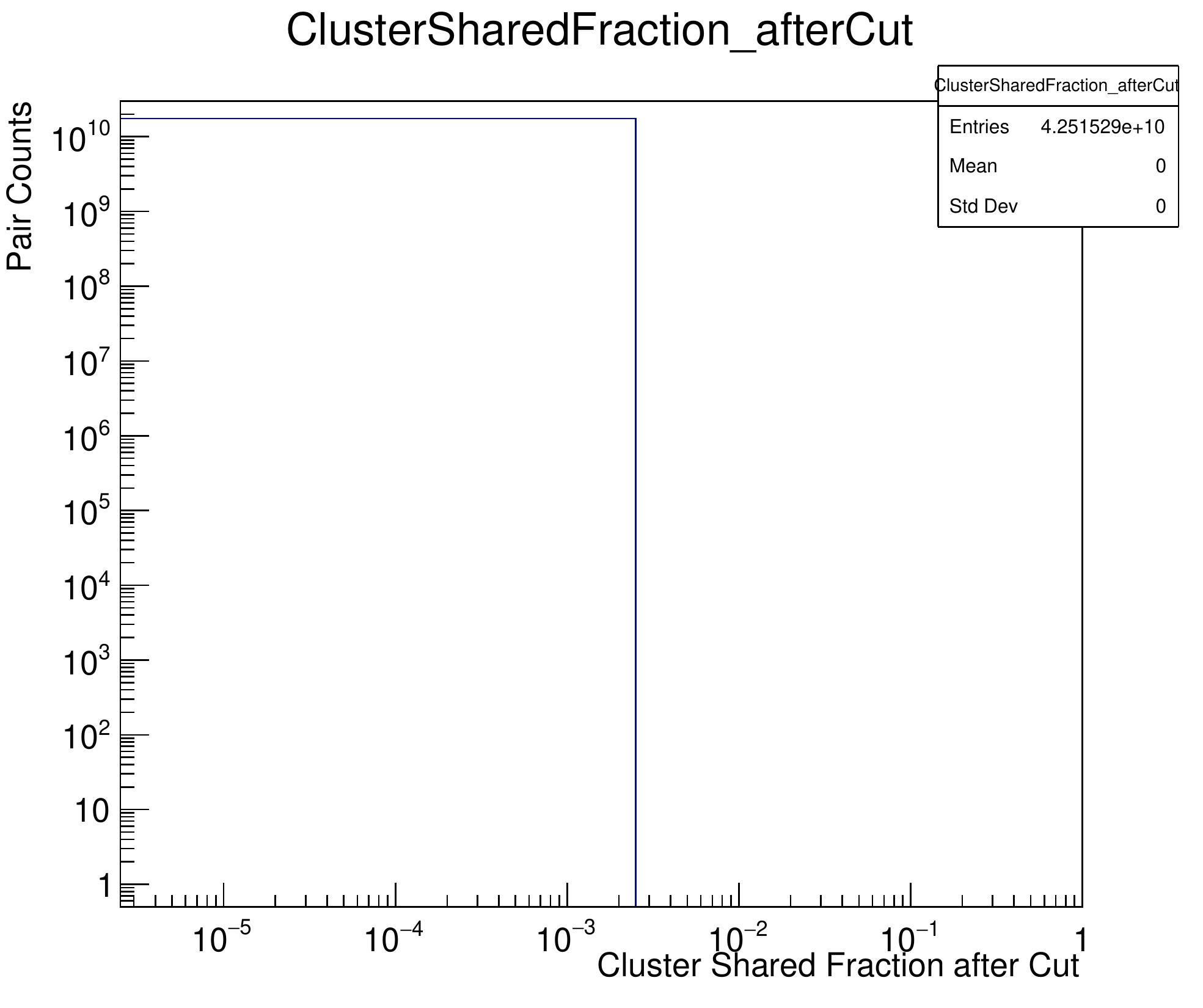}
    \includegraphics[width=0.32\textwidth]{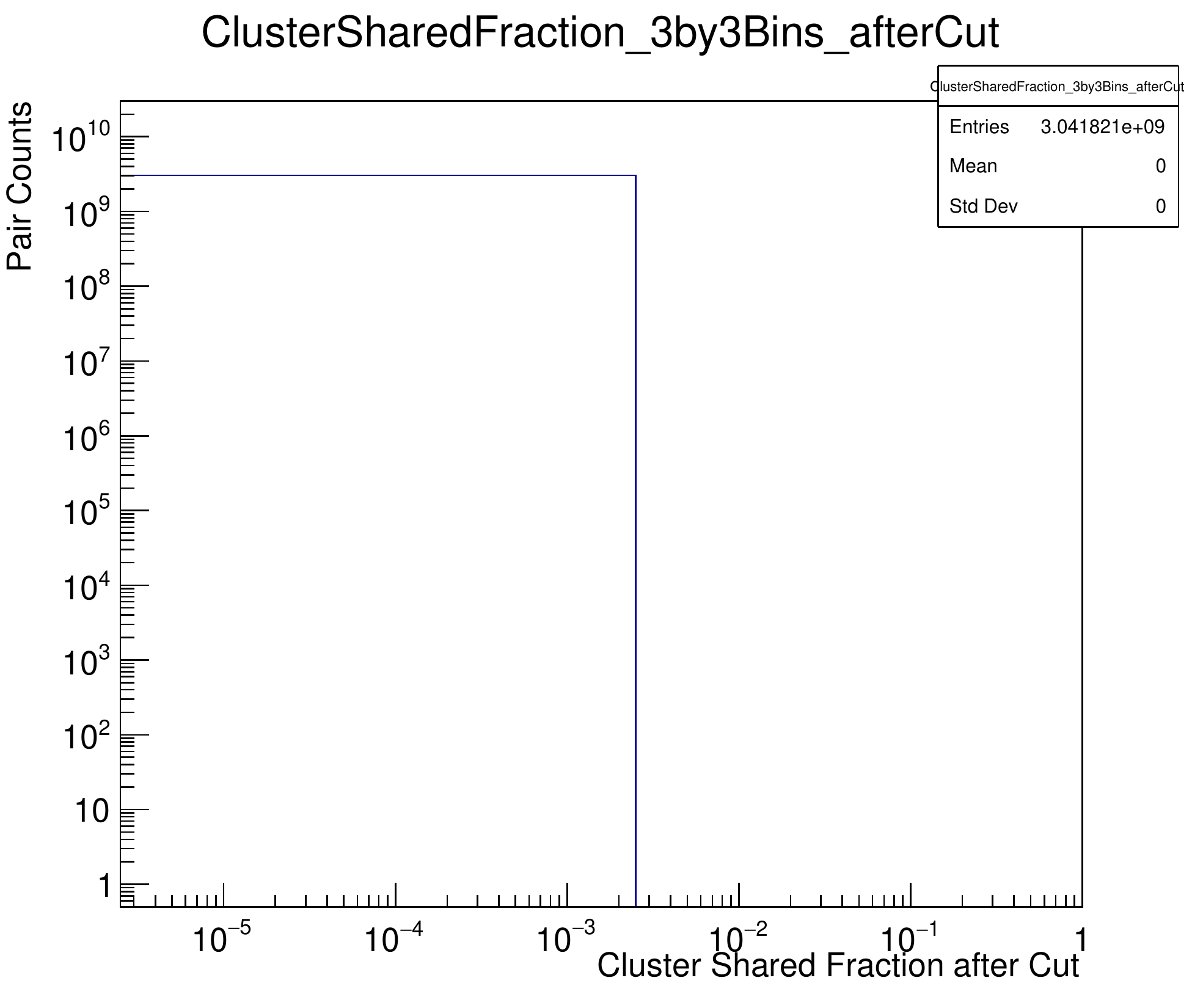}
    \includegraphics[width=0.32\textwidth]{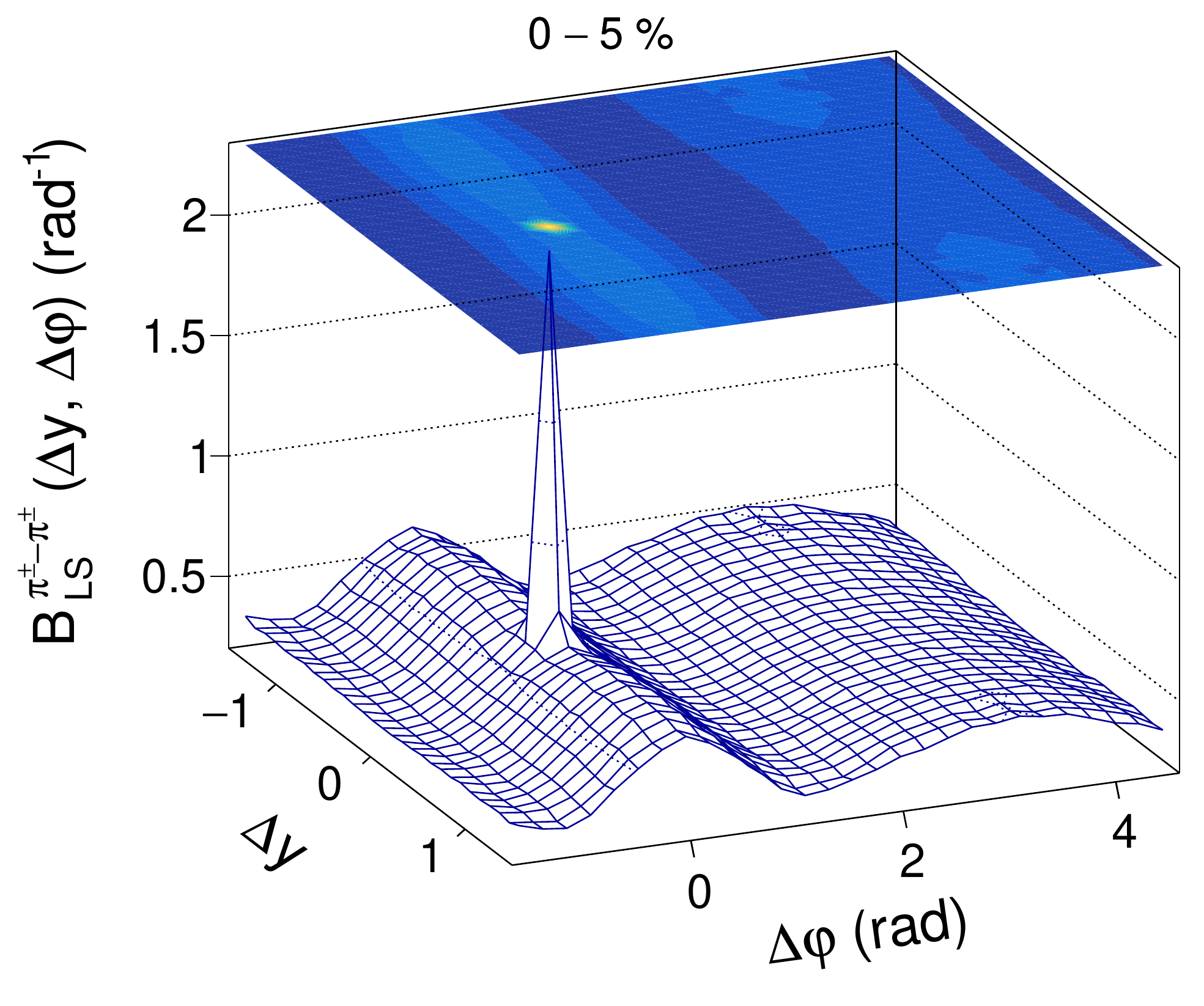}
  \caption{Comparison of $B_{LS}^{\pi\pi}$($\Delta y$,$\Delta\varphi$) of 0-5\% centrality obtained without (top row) and with (bottom row) the TPC cluster sharing cut ($SharingFraction<0.25\%$). 
  Left column plots are TPC cluster shared fraction in the full $\Delta y$ and $\Delta\varphi$ acceptance. 
  Middle column plots are TPC cluster shared fraction in the 3$\times$3 bins around $\Delta y$=0,$\Delta\varphi$=0. 
  Right column plots are the $B_{LS}^{\pi\pi}$($\Delta y$,$\Delta\varphi$) of 0-5\% results. }
\label{fig:CompareClusterSharing_PiPi}
\end{figure}

\begin{figure}
\centering
    \includegraphics[width=0.49\textwidth]{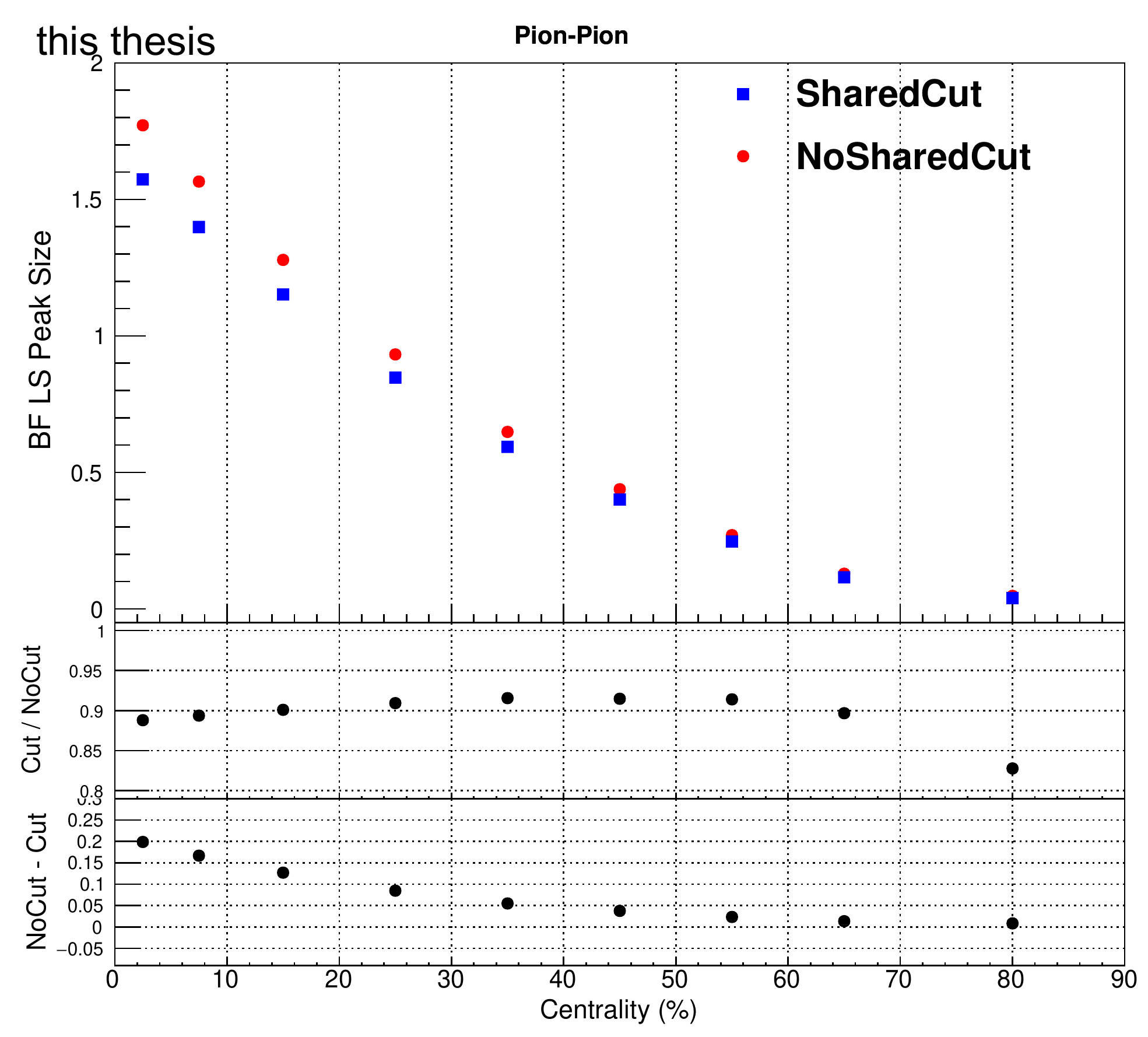}
    \includegraphics[width=0.49\textwidth]{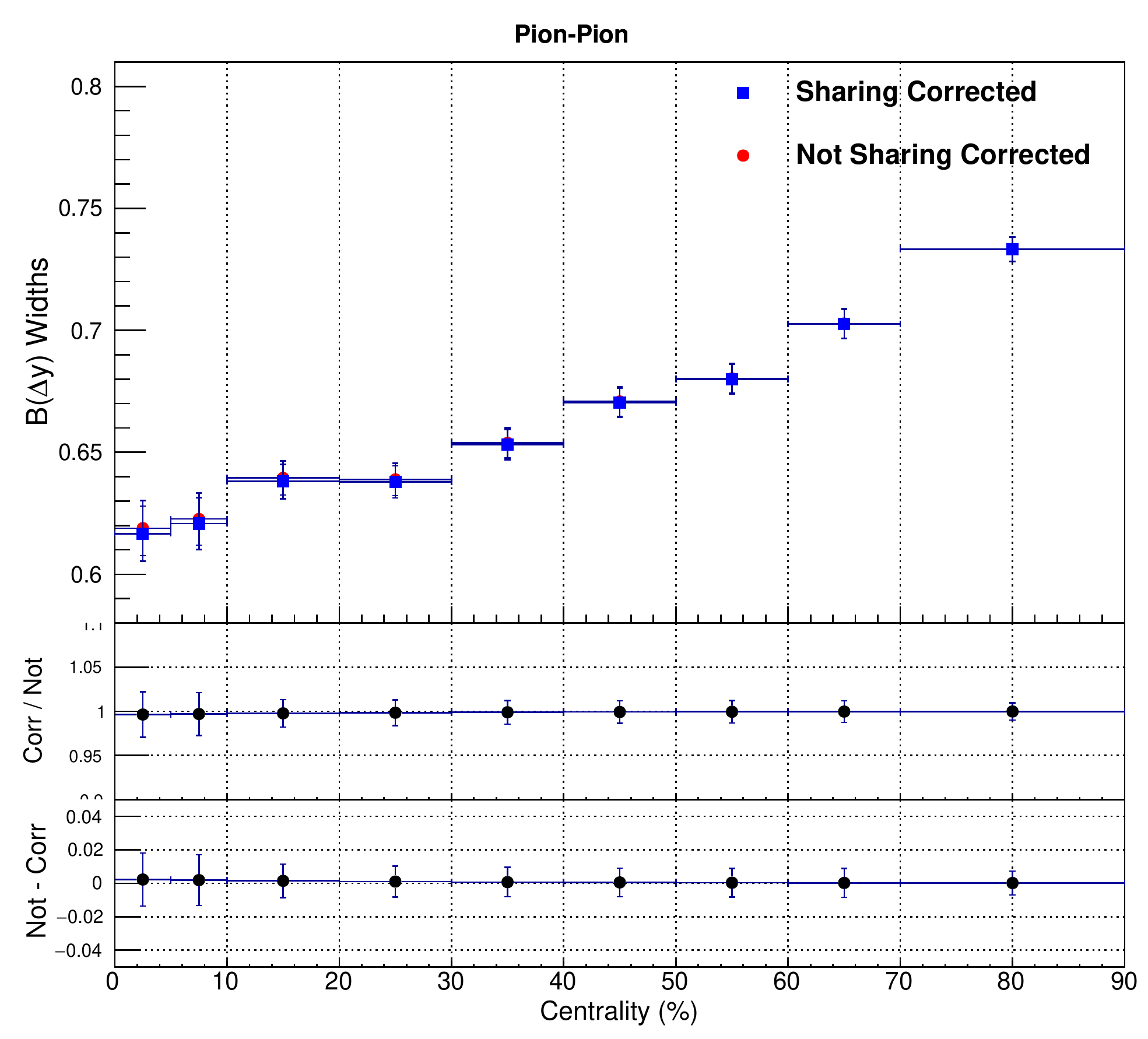}
    \includegraphics[width=0.49\textwidth]{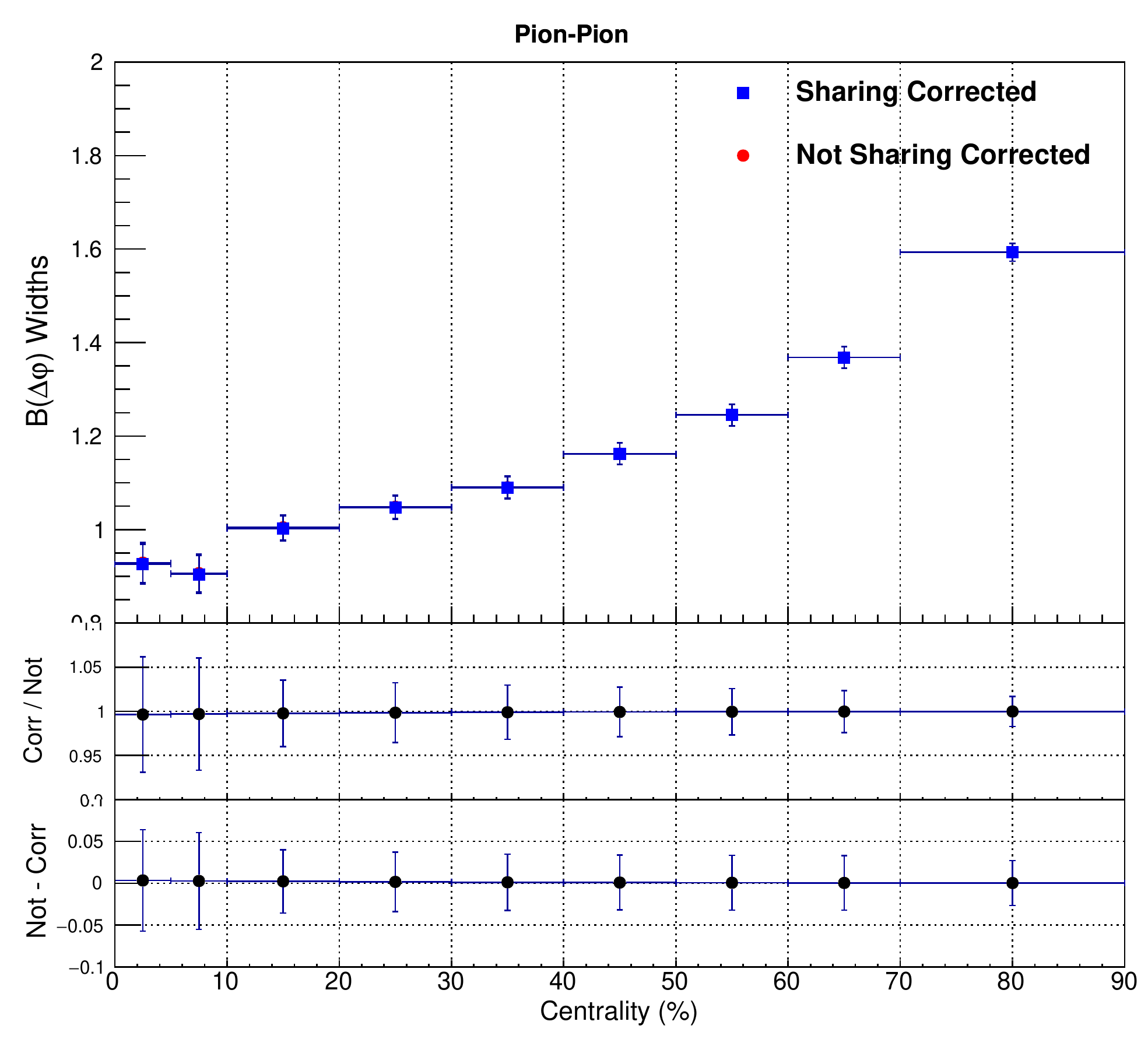}
     \includegraphics[width=0.49\textwidth]{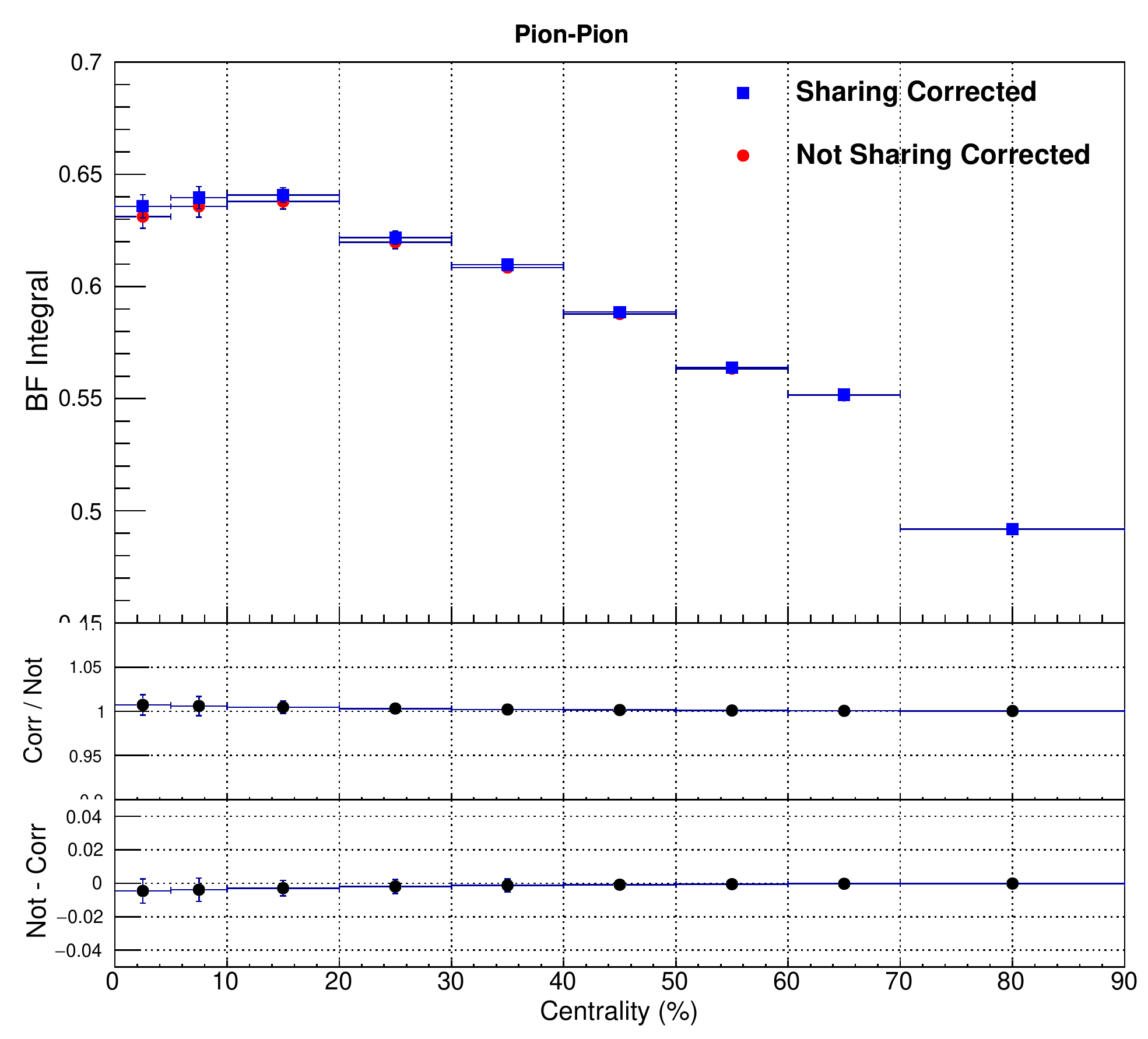}  
  \caption{Comparison of the peak size at (0,0) bin of $B_{LS}^{\pi\pi}(\Delta y,\Delta \varphi)$ obtained with and without TPC sharing cut (upper left), which serves as a correction for $B^{\pi\pi}$.
   And comparison of the $B^{\pi\pi}$ $\Delta y$ RMS widths (upper right), $\Delta\varphi$ RMS widths (lower left), and integrals (lower right) obtained with and without TPC sharing correction.}
\label{fig:CompareBF_LS_Peak_Size_PiPi}
\end{figure}

\begin{figure}
\centering
    \includegraphics[width=0.32\textwidth]{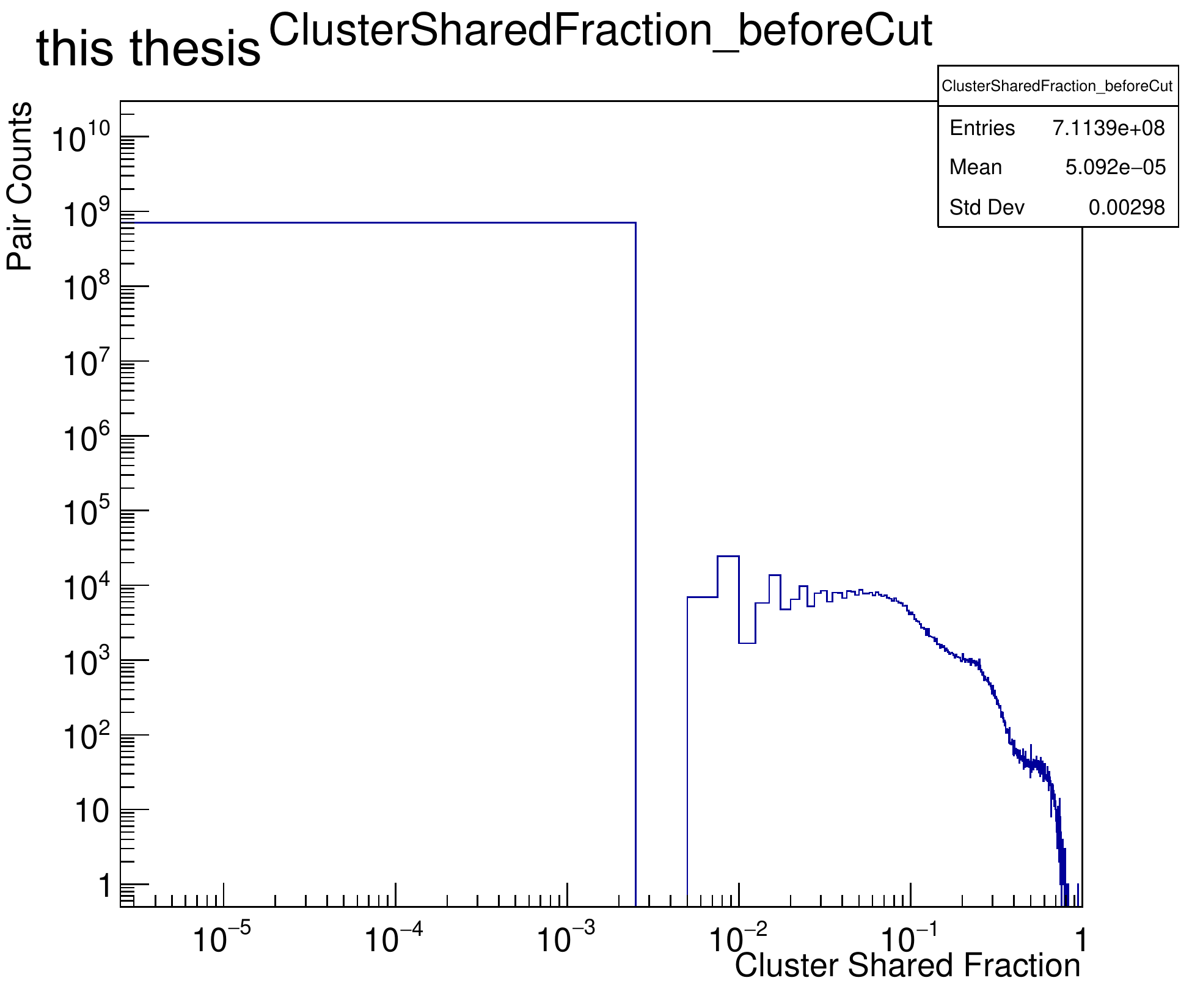}
    \includegraphics[width=0.32\textwidth]{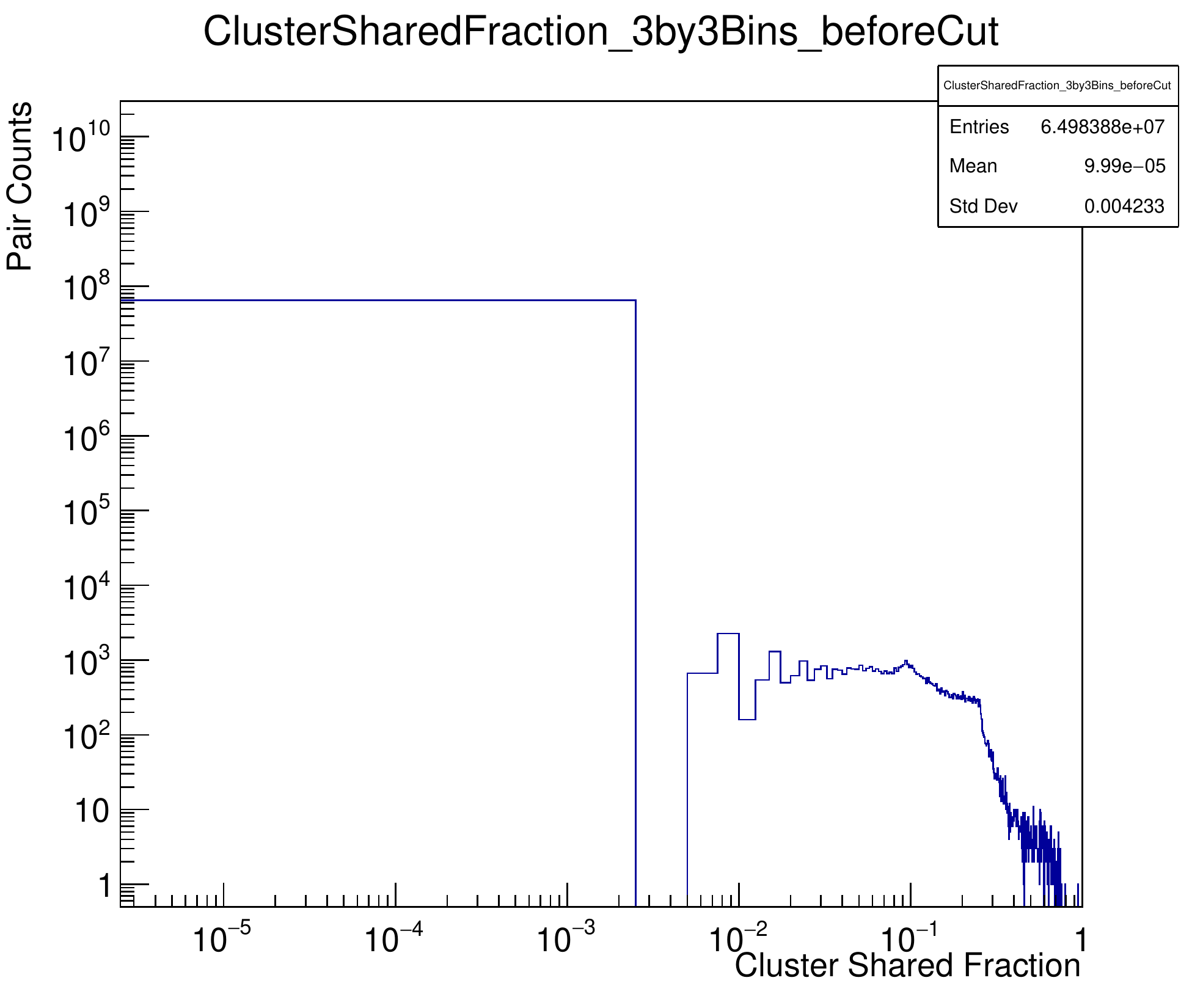}
    \includegraphics[width=0.32\textwidth]{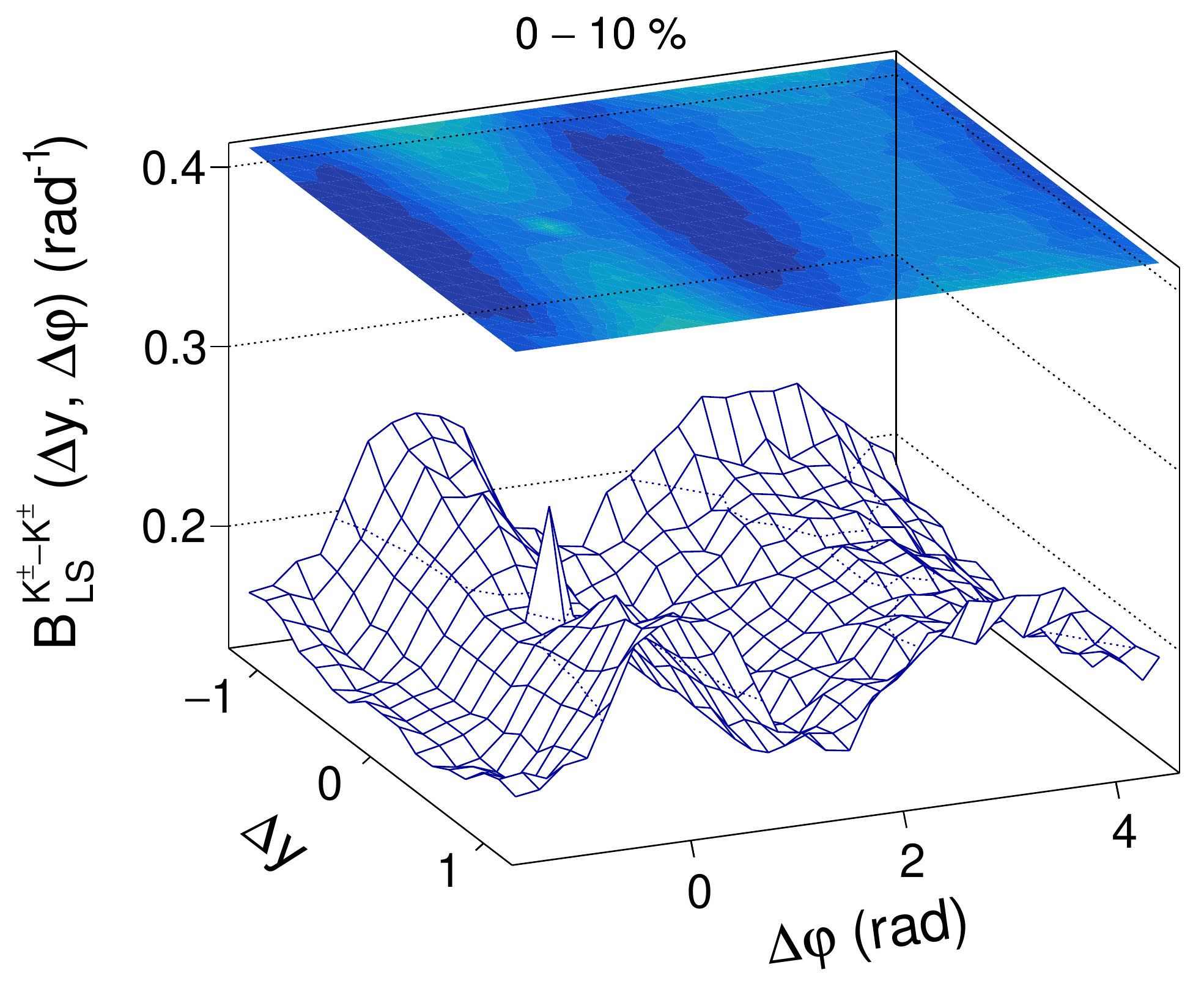}
    \includegraphics[width=0.32\textwidth]{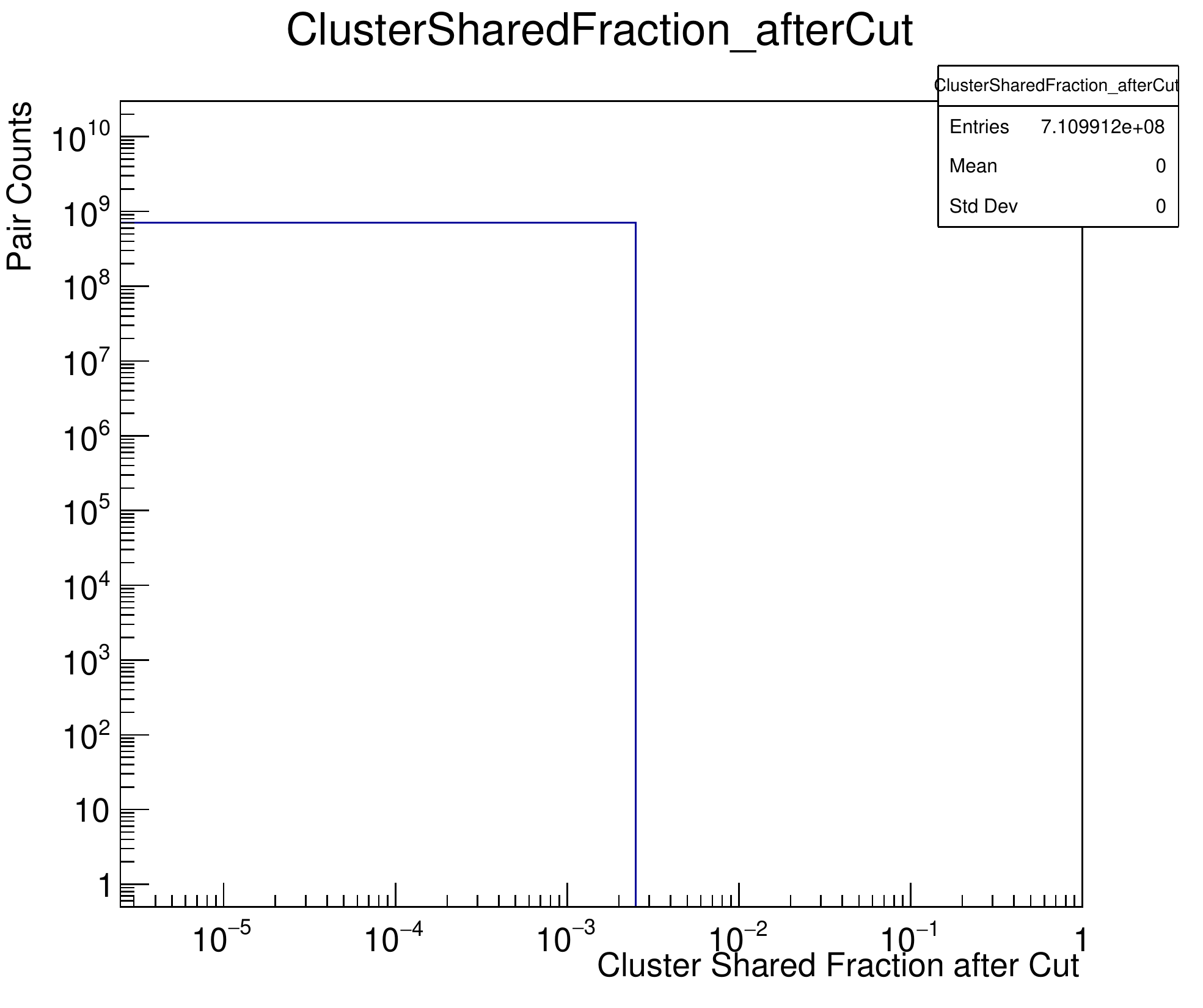}
    \includegraphics[width=0.32\textwidth]{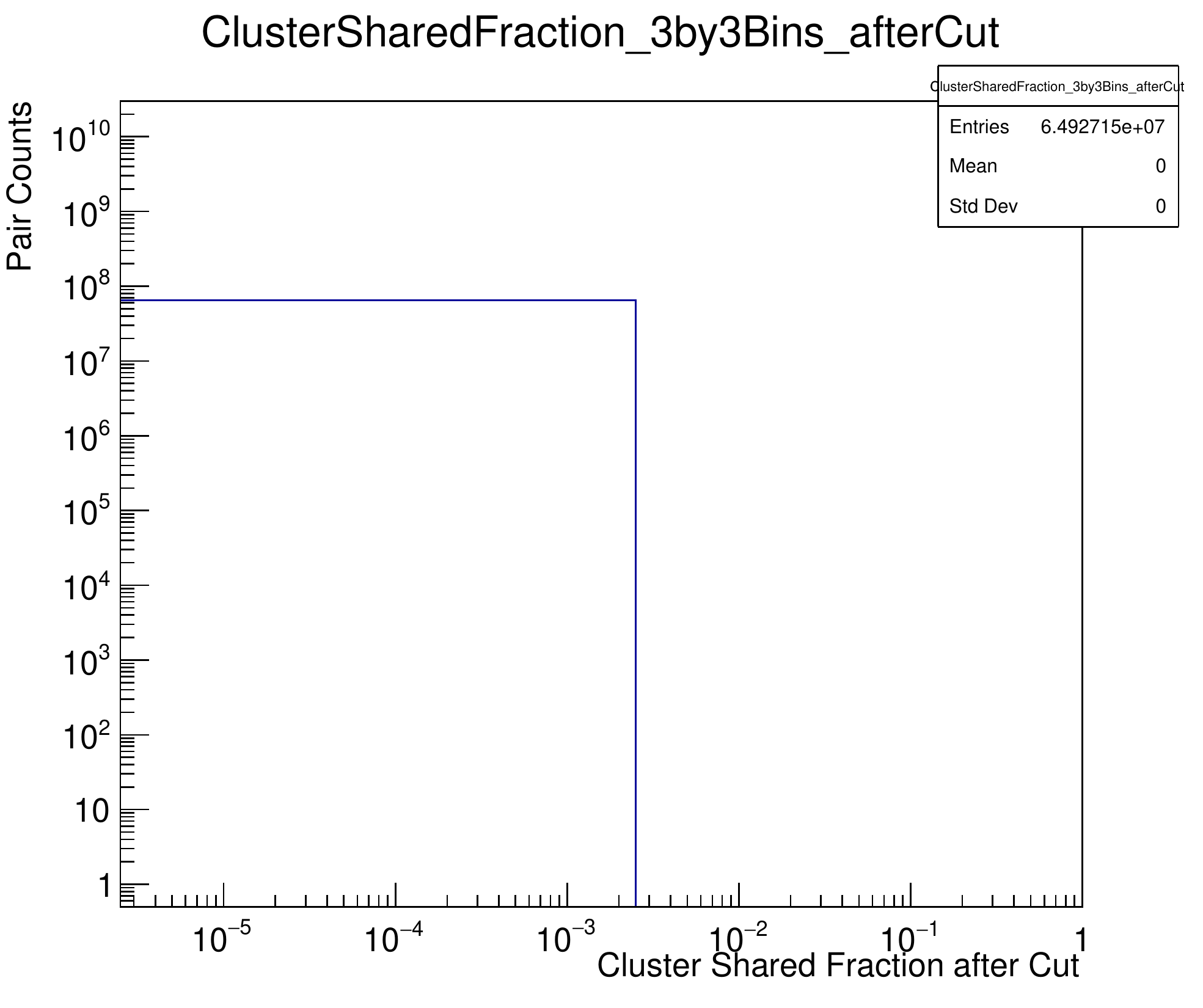}
    \includegraphics[width=0.32\textwidth]{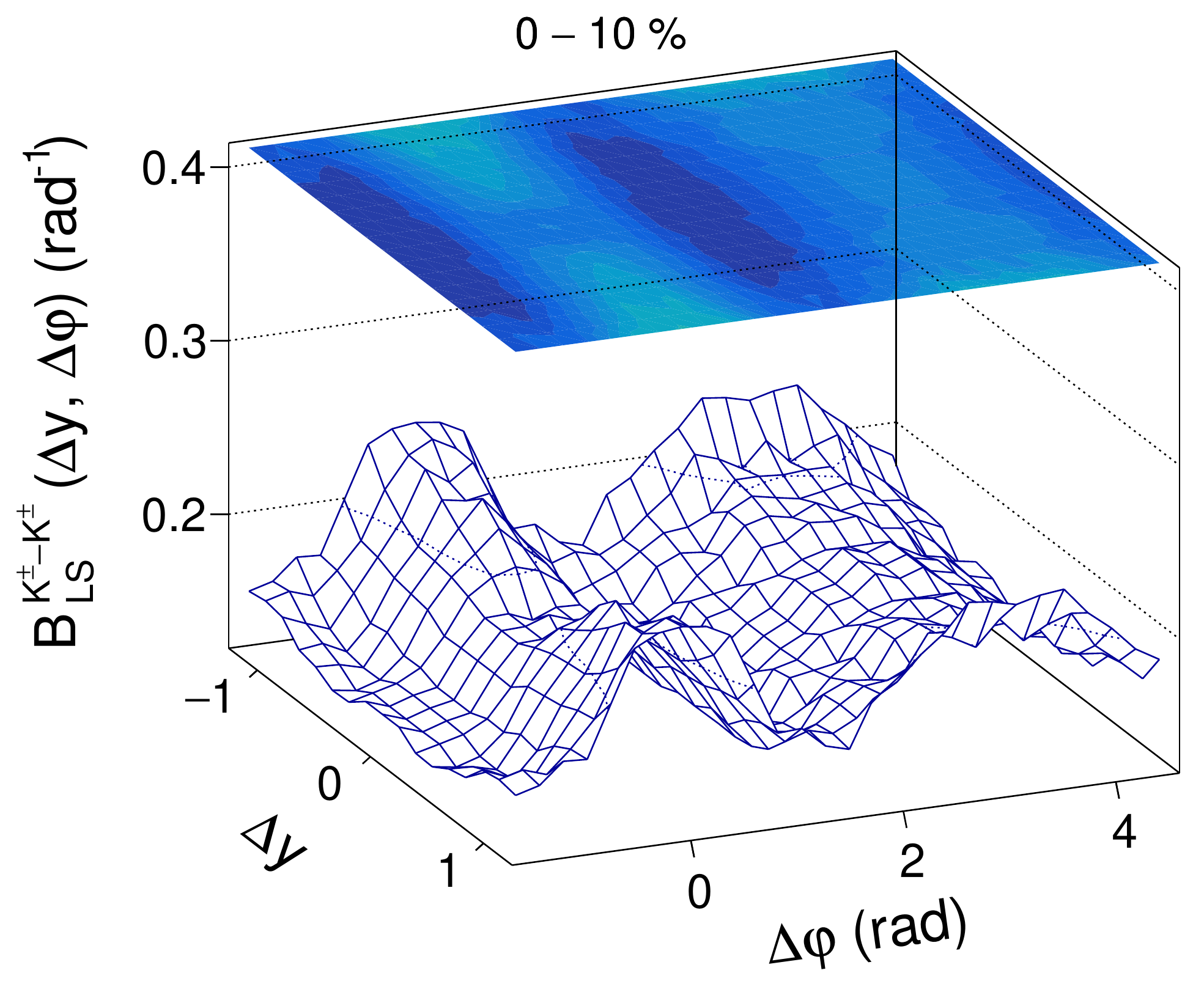}
  \caption{ Comparison of $B_{LS}^{KK}$($\Delta y$,$\Delta\varphi$) of 0-10\% centrality obtained without (top row) and with (bottom row) the TPC cluster sharing cut ($SharingFraction<0.25\%$). Left column plots are TPC cluster shared fraction in the full $\Delta y$ and $\Delta\varphi$ acceptance. Middle column plots are TPC cluster shared fraction in the 3$\times$3 bins around ($\Delta y$=0,$\Delta\varphi$=0). Right column plots are the $B_{LS}^{KK}$($\Delta y$,$\Delta\varphi$) of 0-10\% results.}
\label{fig:CompareClusterSharing_KK}
\end{figure}

\begin{figure}
\centering
    \includegraphics[width=0.49\textwidth]{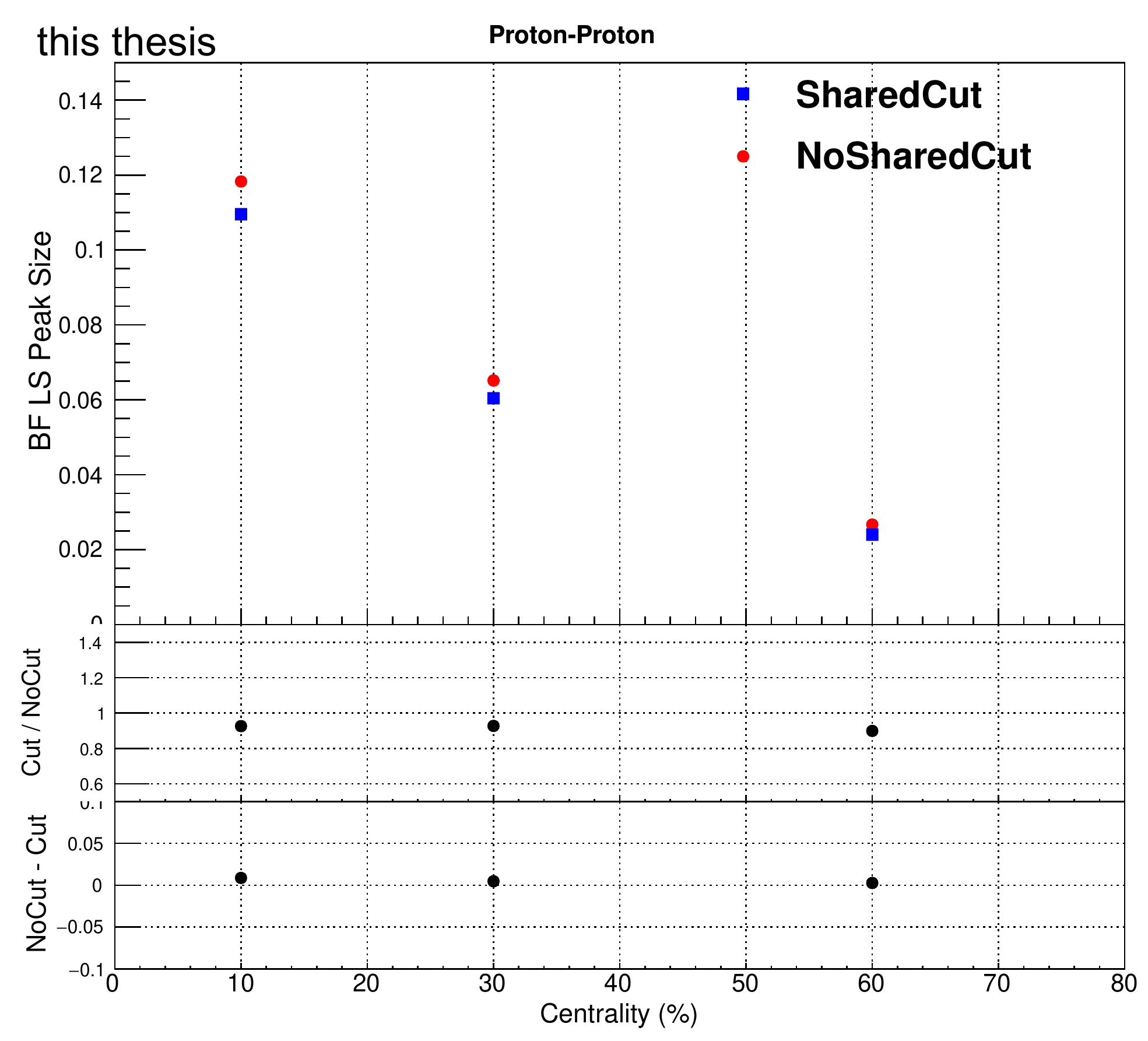}
    \includegraphics[width=0.49\textwidth]{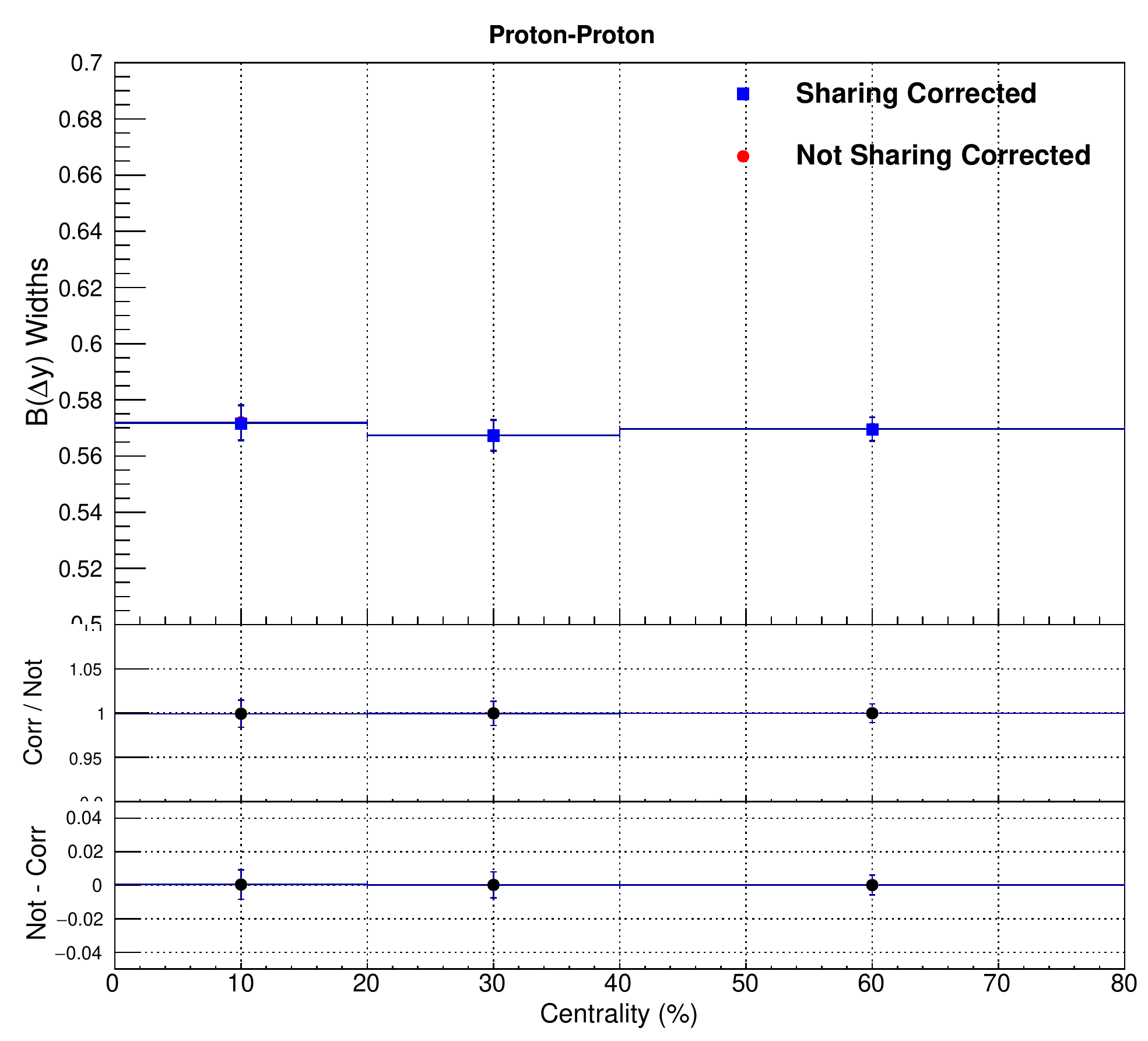}
    \includegraphics[width=0.49\textwidth]{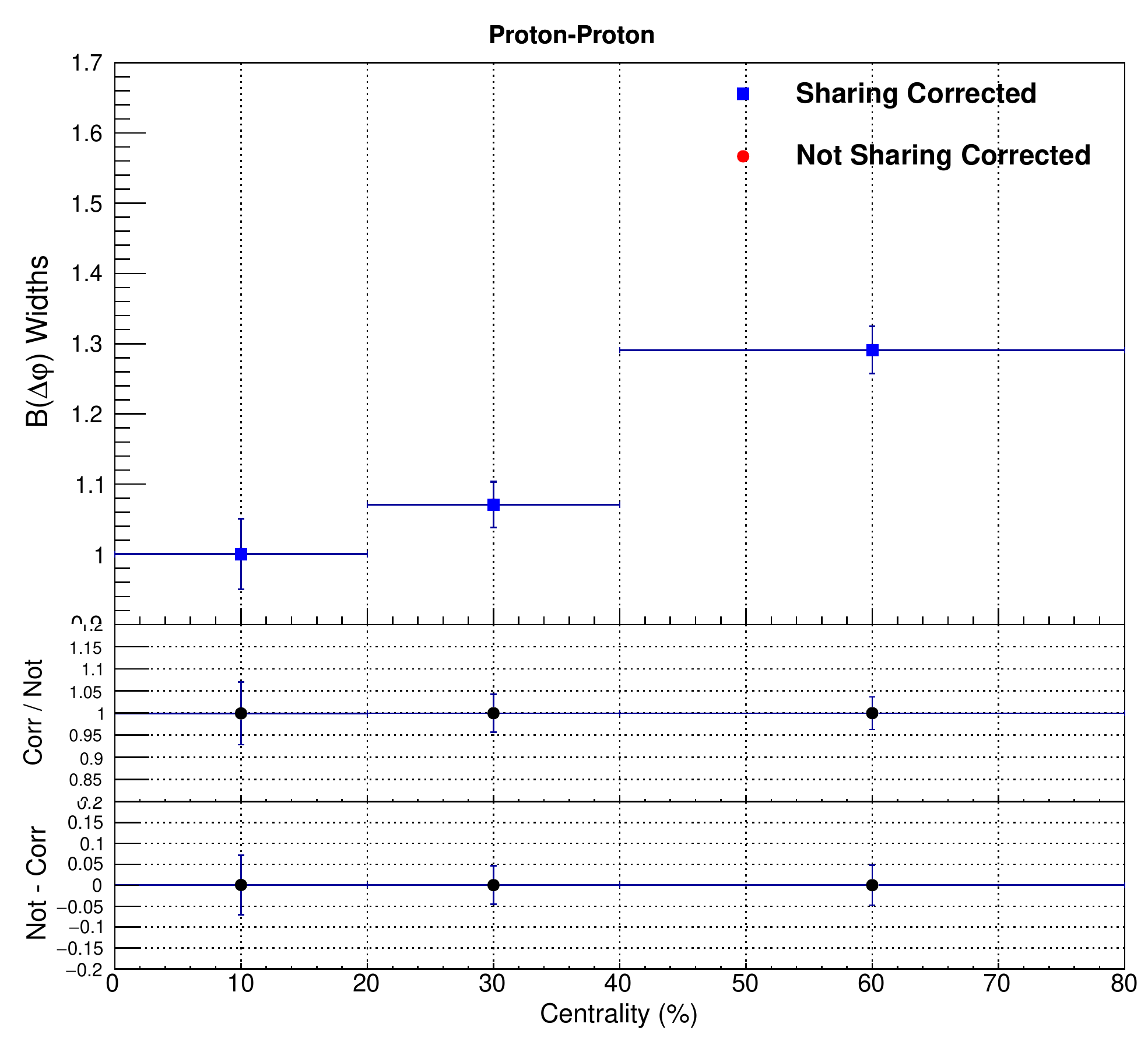}
     \includegraphics[width=0.49\textwidth]{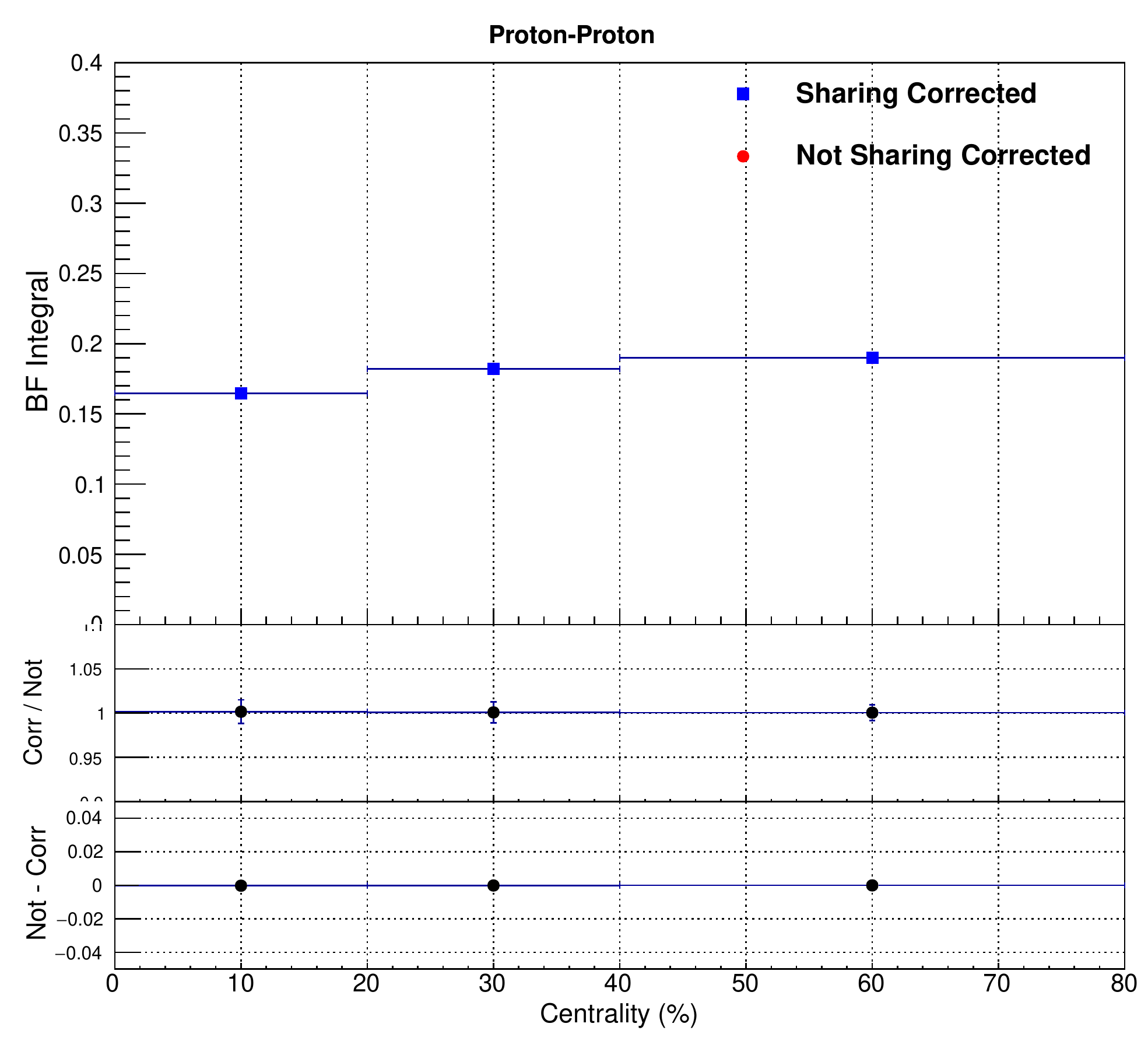}  
  \caption{Comparison of the peak size at (0,0) bin of $B_{LS}^{pp}(\Delta y,\Delta \varphi)$ obtained with and without TPC sharing cut (upper left), which serves as a correction for $B^{pp}$.
   And comparison of the $B^{pp}$ $\Delta y$ RMS widths (upper right), $\Delta\varphi$ RMS widths (lower left), and integrals (lower right) obtained with and without TPC sharing correction.}
\label{fig:CompareBF_LS_Peak_Size_PrPr}
\end{figure}

We nonetheless further explored the possibility of split tracks by means of a study of TPC shared track fraction. The study is based 
on a technique borrowed from the HBT working group. We used a function named  CalculateSharedFraction derived from the source code AliFemtoShareQualityPairCut.cxx, and which was  extensively applied for checking the presence of TPC shared hits and split tracks in many other ALICE analyses. Figure~\ref{fig:CompareClusterSharing_PiPi} shows the TPC shared track fraction and compares the $B_{LS}^{\pi\pi}$($\Delta y$,$\Delta\varphi$) correlator measured in  the 0-5\% collision centrality before and after a strict TPC shared track fraction cut.
One finds that the narrow near-side peak at ($\Delta y$=0,$\Delta\varphi$=0) is not due to the track splitting.
The size difference of the near-side peak at ($\Delta y$=0,$\Delta\varphi$=0) of $B_{LS}^{\pi\pi}$($\Delta y$,$\Delta\varphi$) is used to correct the TPC cluster sharing effect. 
Figure~\ref{fig:CompareBF_LS_Peak_Size_PiPi} shows that the differences between TPC sharing corrected and not corrected for BF $\Delta y$ and $\Delta\varphi$ RMS widths, and integral are smaller than 1\%. This TPC cluster sharing correction is included in the final BF results reported in this work.
Results presented in Figs~\ref{fig:CompareClusterSharing_KK} --  
%~\ref{fig:CompareBF_LS_Peak_Size_KK},~\ref{fig:CompareClusterSharing_PrPr},
\ref{fig:CompareBF_LS_Peak_Size_PrPr} show that TPC cluster sharing effects observed with $KK$ and $pp$ pairs are similar to those observed for $\pi\pi$ pair.

\begin{figure}
\centering
    \includegraphics[width=0.49\textwidth]{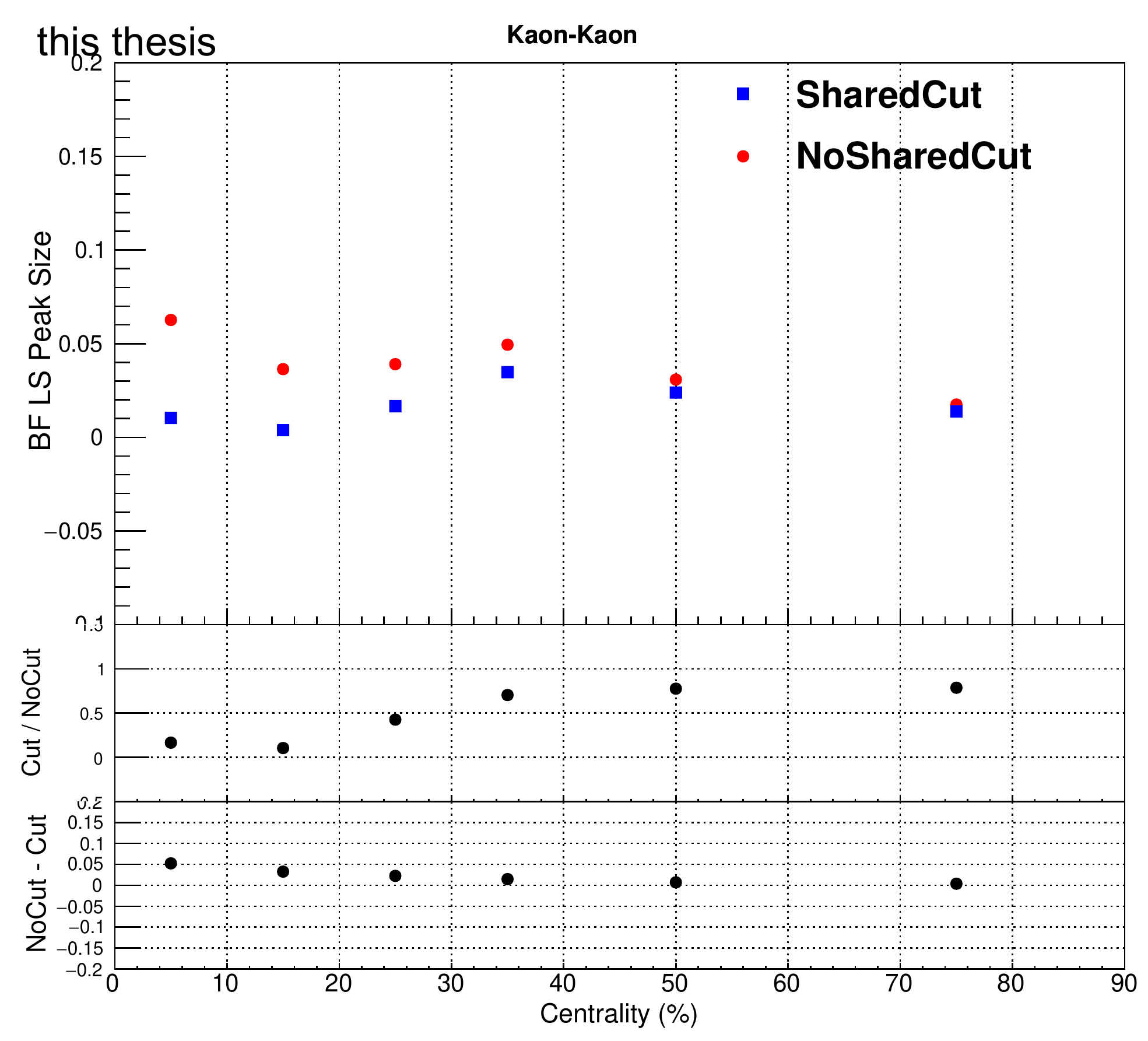}
    \includegraphics[width=0.49\textwidth]{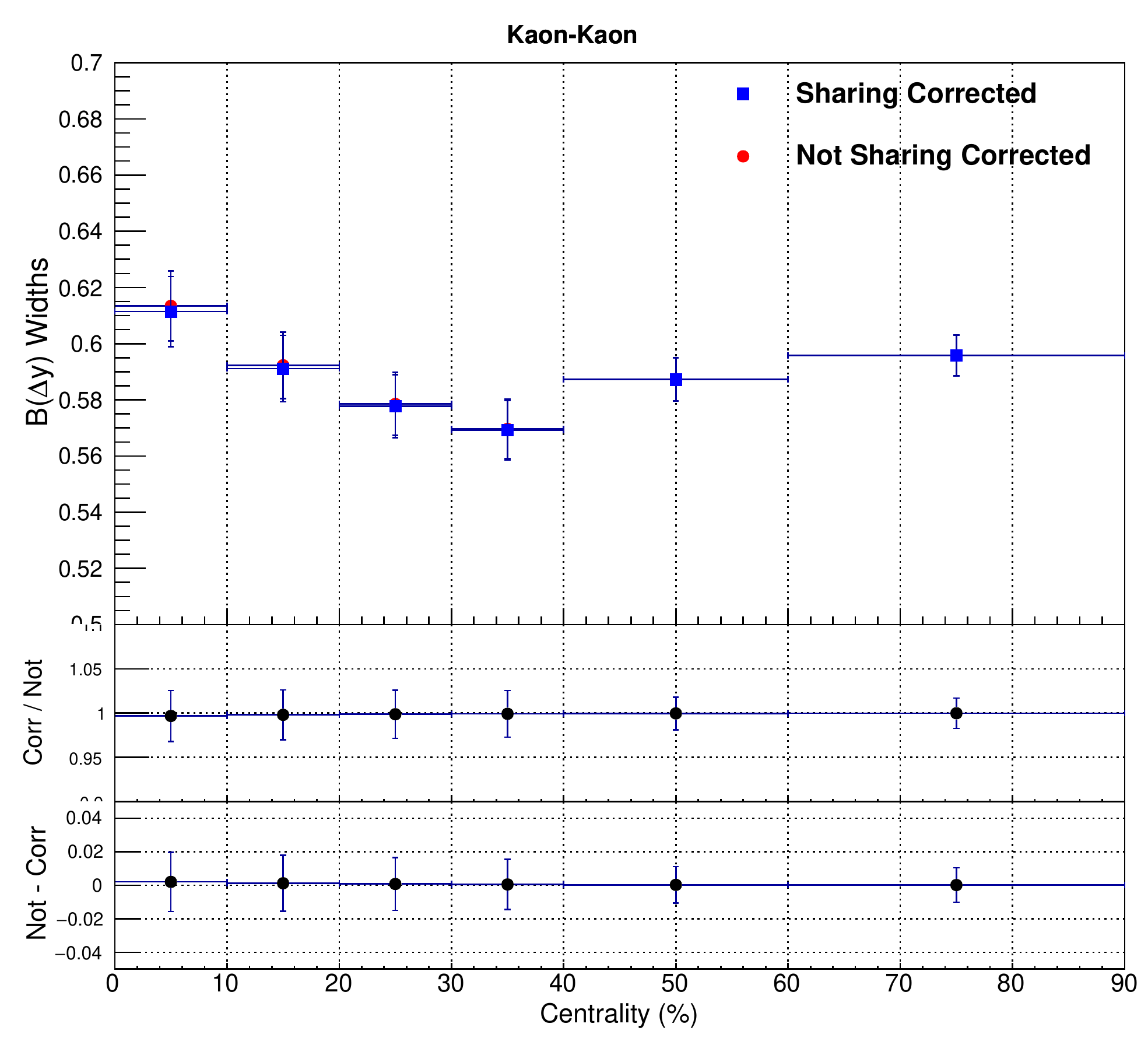}
    \includegraphics[width=0.49\textwidth]{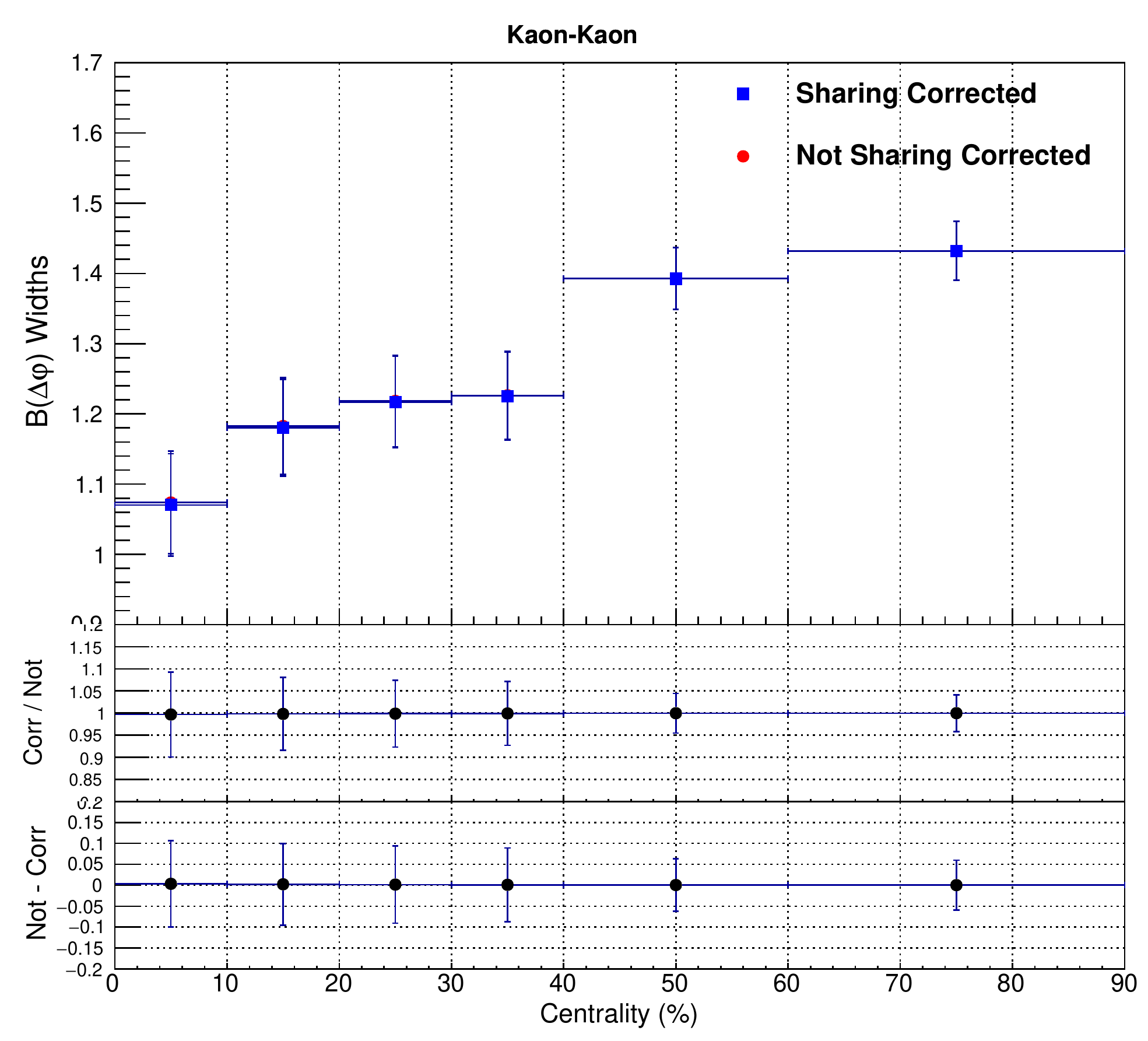}
     \includegraphics[width=0.49\textwidth]{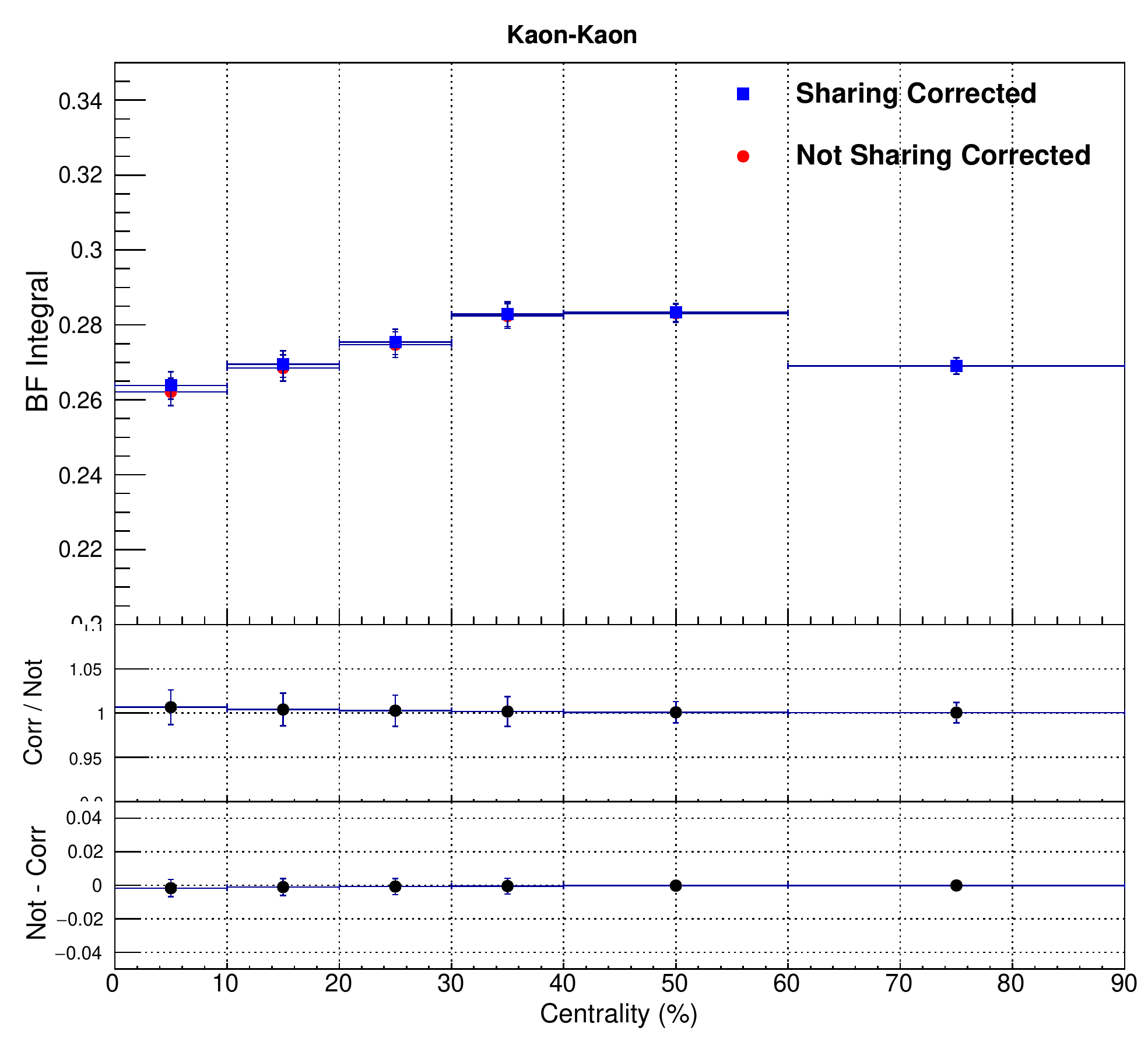}  
  \caption{Comparison of the peak size at (0,0) bin of $B_{LS}^{KK}(\Delta y,\Delta \varphi)$ obtained with and without TPC sharing cut (upper left), which serves as a correction for $B^{KK}$.
   And comparison of the $B^{KK}$ $\Delta y$ RMS widths (upper right), $\Delta\varphi$ RMS widths (lower left), and integrals (lower right) obtained with and without TPC sharing correction.}
\label{fig:CompareBF_LS_Peak_Size_KK}
\end{figure}
\begin{figure}
\centering
    \includegraphics[width=0.32\textwidth]{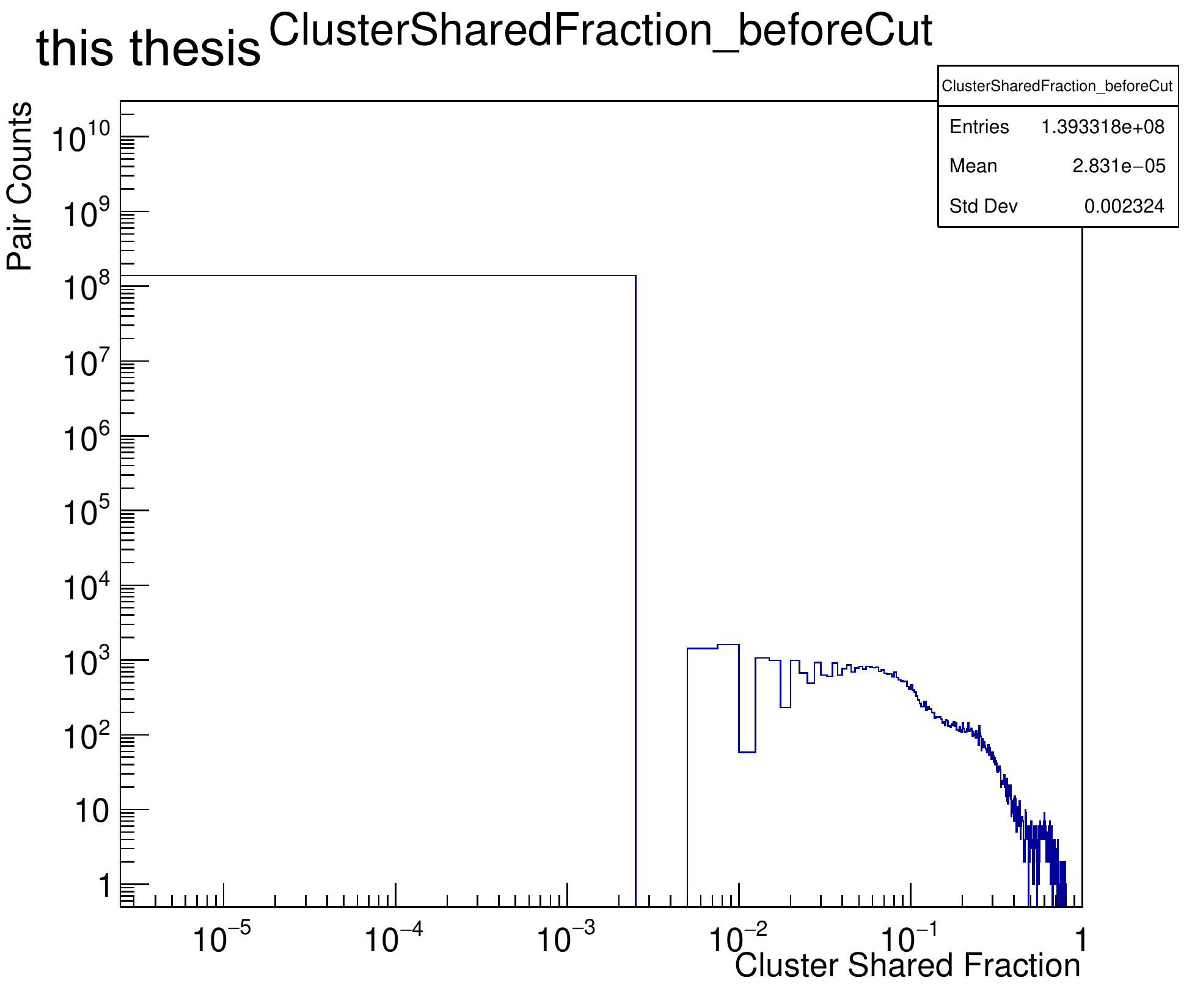}
    \includegraphics[width=0.32\textwidth]{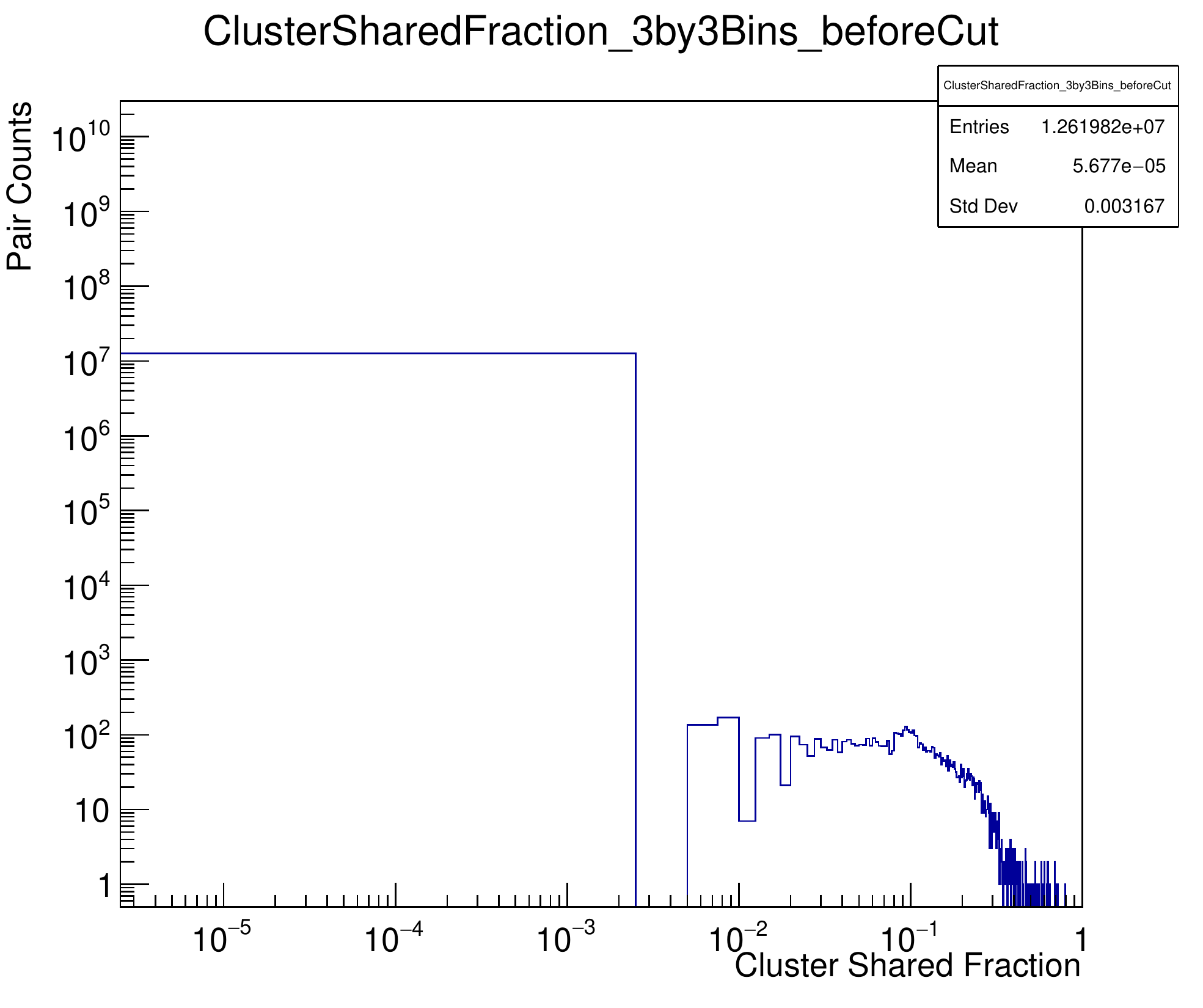}
    \includegraphics[width=0.32\textwidth]{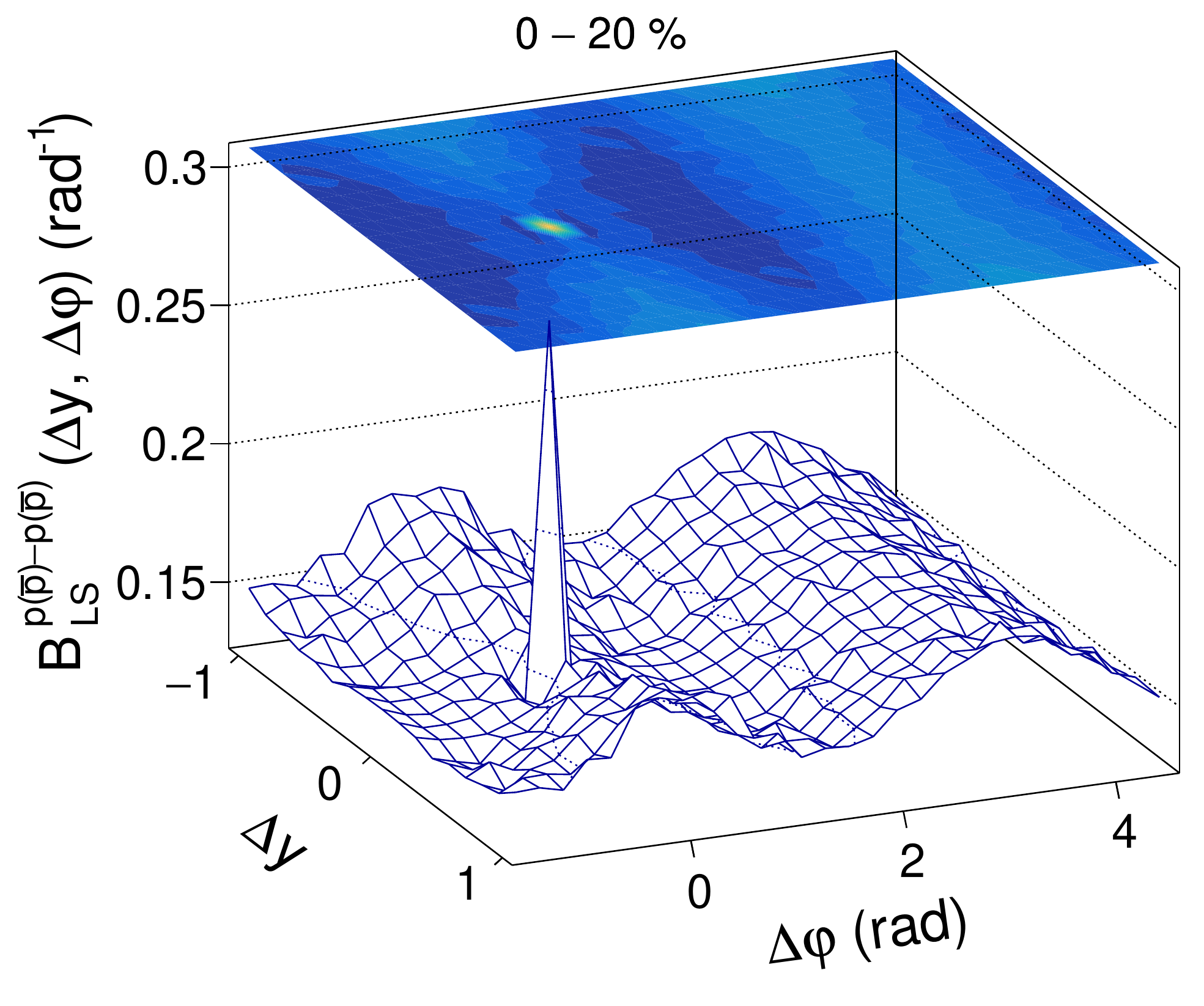}
    \includegraphics[width=0.32\textwidth]{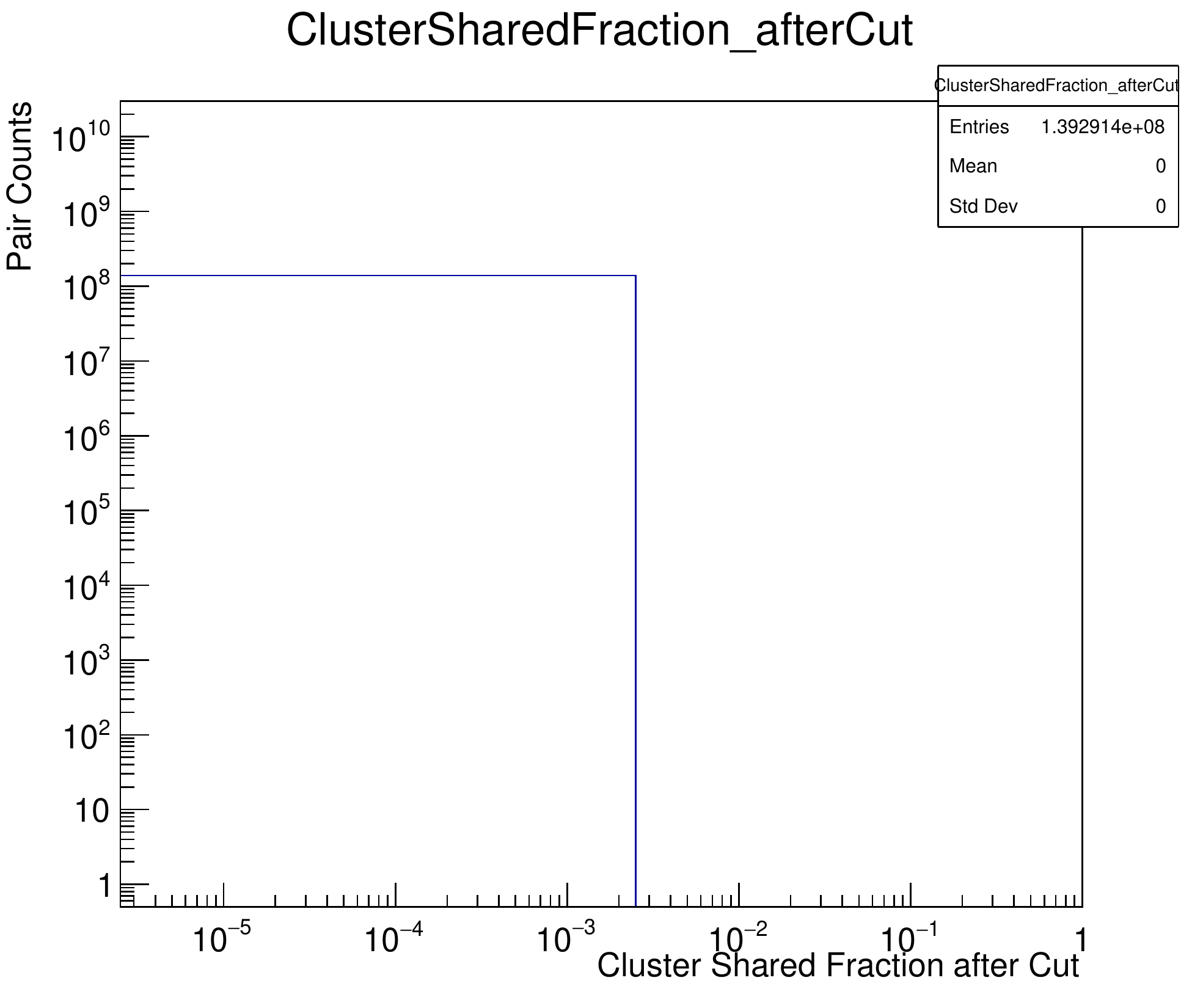}
    \includegraphics[width=0.32\textwidth]{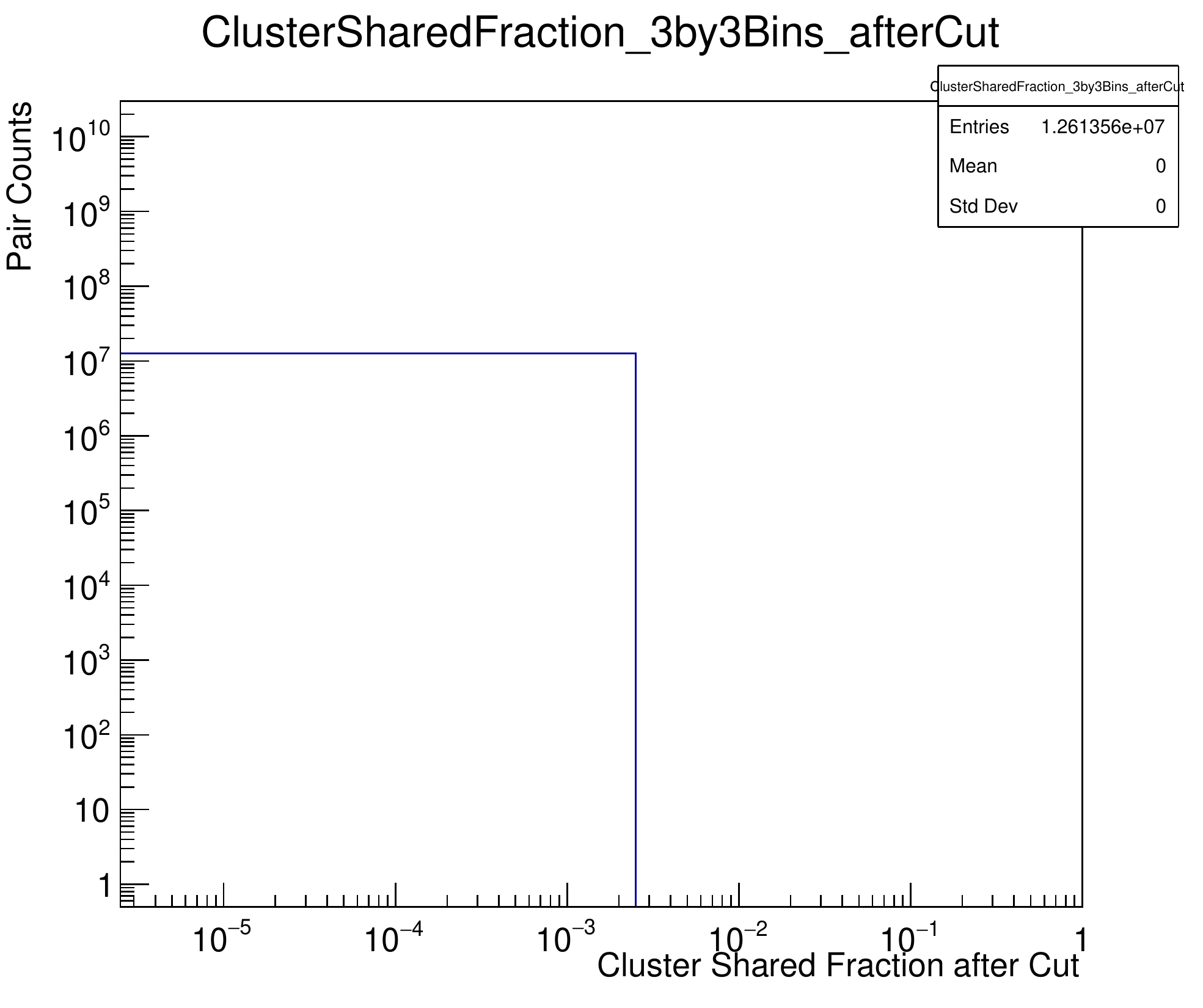}
    \includegraphics[width=0.32\textwidth]{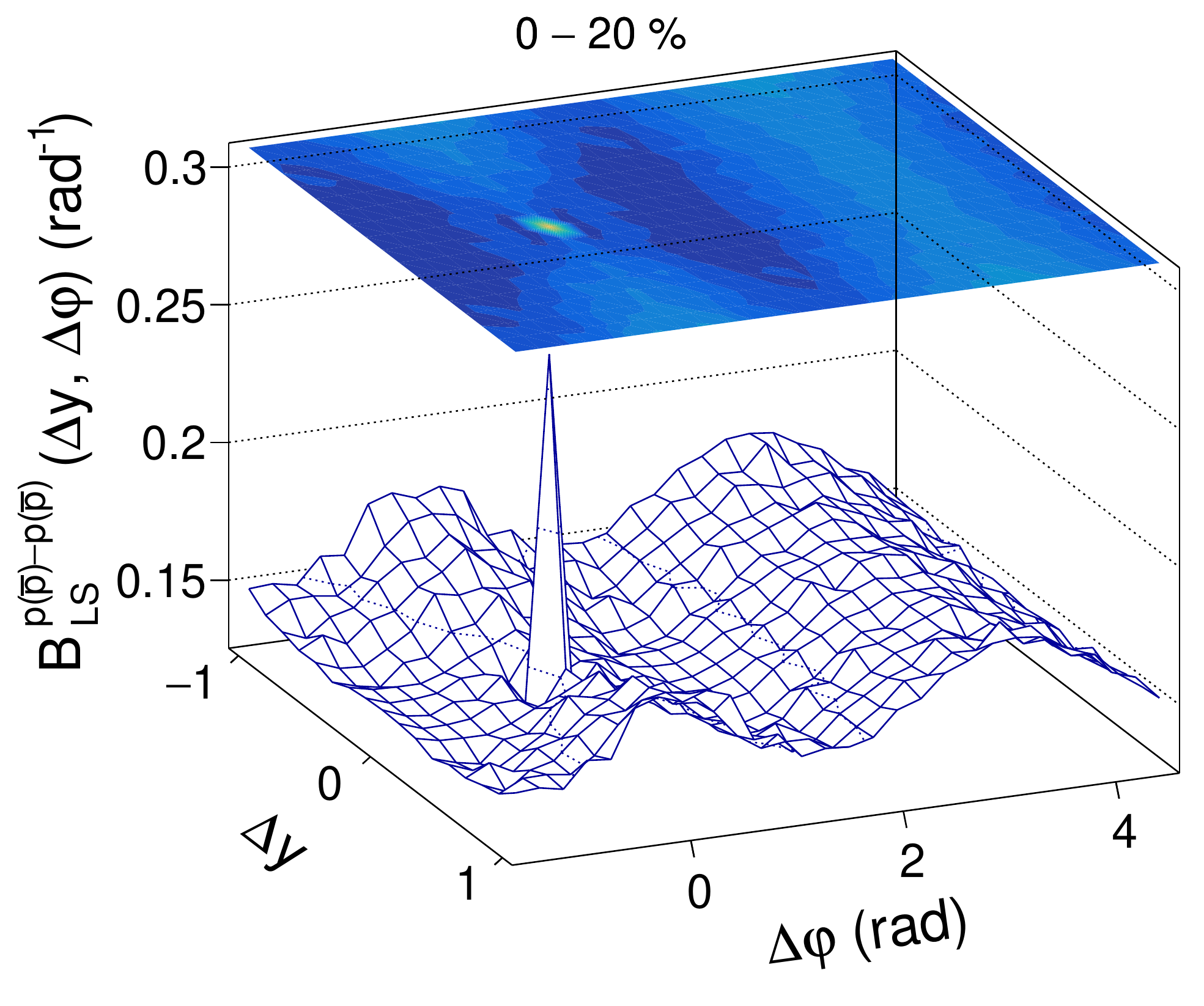}
  \caption{Comparison of $B_{LS}^{pp}$($\Delta y$,$\Delta\varphi$) of 0-20\% centrality obtained without (top row) and with (bottom row) the TPC cluster sharing cut ($SharingFraction<0.25\%$). Left column plots are TPC cluster shared fraction in the full $\Delta y$ and $\Delta\varphi$ acceptance. Middle column plots are TPC cluster shared fraction in the 3$\times$3 bins around ($\Delta y$=0,$\Delta\varphi$=0). Right column plots are the $B_{LS}^{pp}$($\Delta y$,$\Delta\varphi$) of 0-20\% results.}
 \label{fig:CompareClusterSharing_PrPr}
\end{figure}

\clearpage

\section{$\phi$-Meson Decay in $B^{KK}$}
\label{subsec:PhiDecay}

%In this section, we study the importance of $\phi$-meson decays in $B^{KK}$.
Nuclear collisions   produce  a relatively small but non-negligible amount of $\phi$-mesons. These mesons
have a lifetime of $1.55\pm 0.01\times 10^{-22}$ s, which is of  the order of the lifespan of the fireball produced in Pb--Pb collisions. The particles $\phi$-mesons decay into, particularly pairs of $K^+$ and $K^-$ strange mesons, are thus considered  primary particles in the context of this analysis. By virtue of their origin, such $K^+ + K^-$ pairs are 
intrinsically correlated and thus expected to form observable correlation structures in $KK$ balance functions. But given these kaons are considered primary particles (they do not result from weak-interaction decays), no correction for their presence in the data sample is required. However, the production and transport properties of $\phi$-mesons may differ from those of ``ordinary" kaons, i.e., those not originating from $\phi$-mesons. It is thus of interest to investigate whether the presence of $\phi$-meson decays influence the shape and evolution of $KK$ balance functions with collision centrality.

\subsection{$\phi$-Meson Yield}
\label{subsubsec:PhiMesonCounts}

\begin{figure}
\centering
  \includegraphics[width=0.32\linewidth]{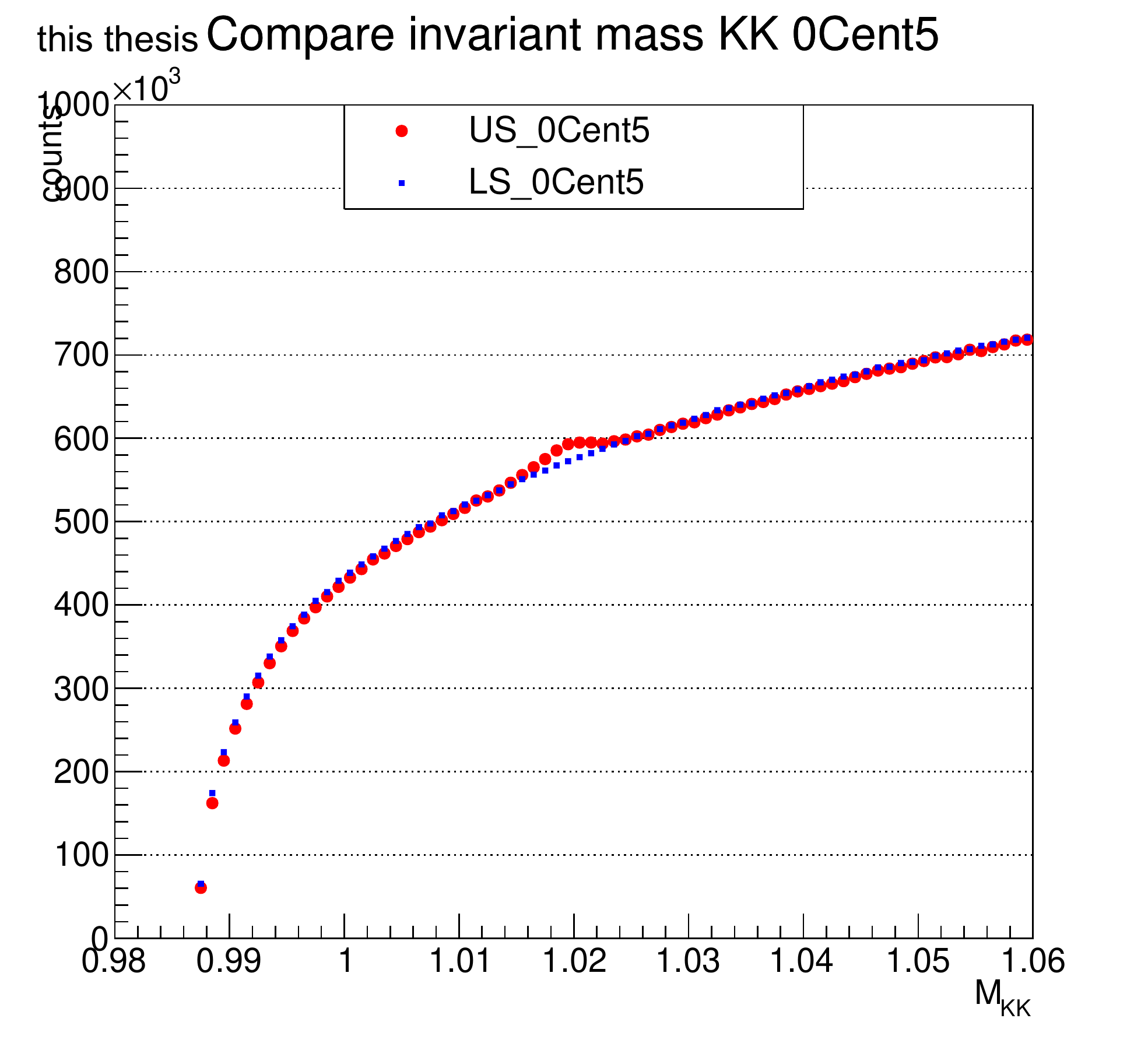}
  \includegraphics[width=0.32\linewidth]{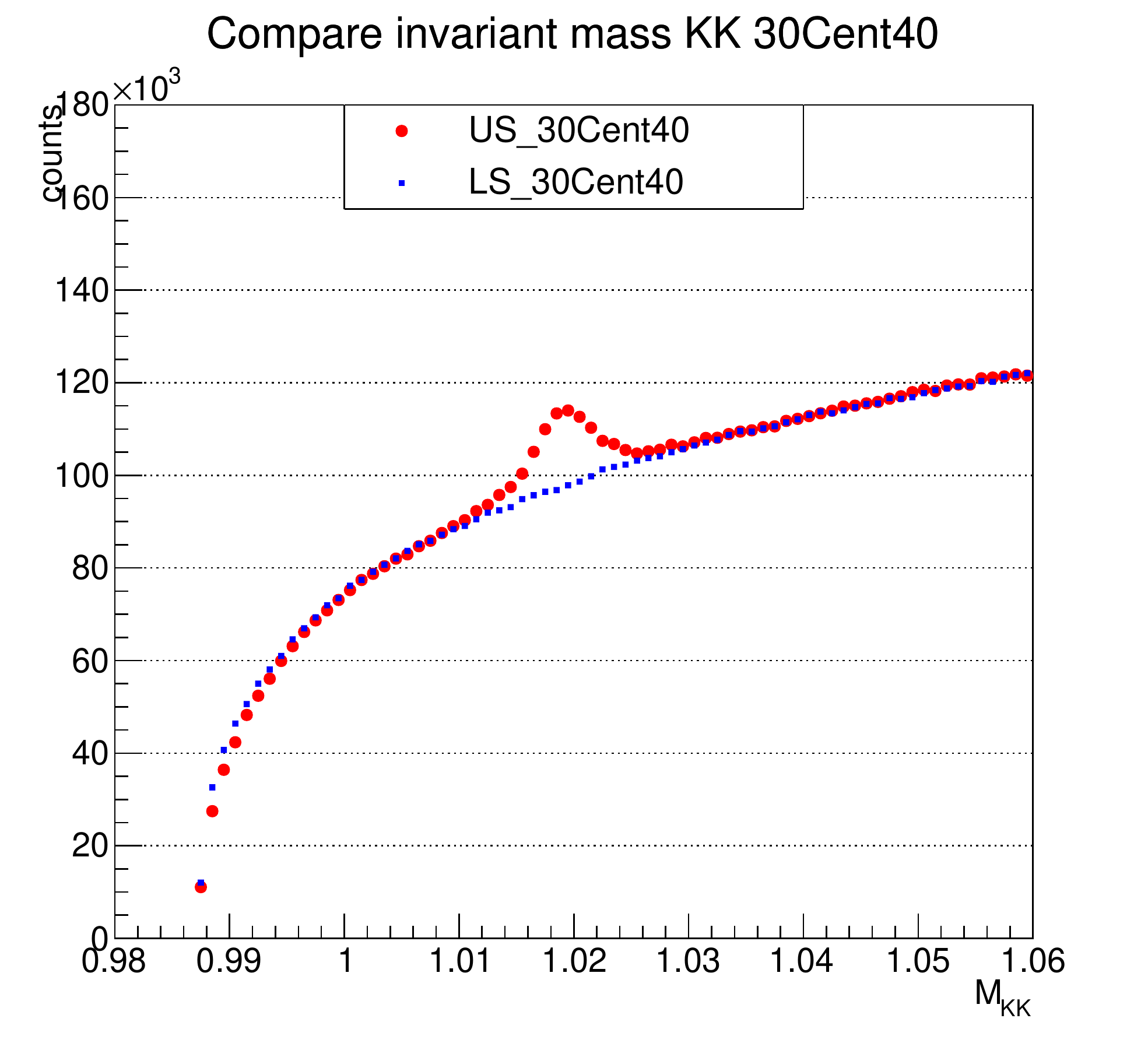}
  \includegraphics[width=0.32\linewidth]{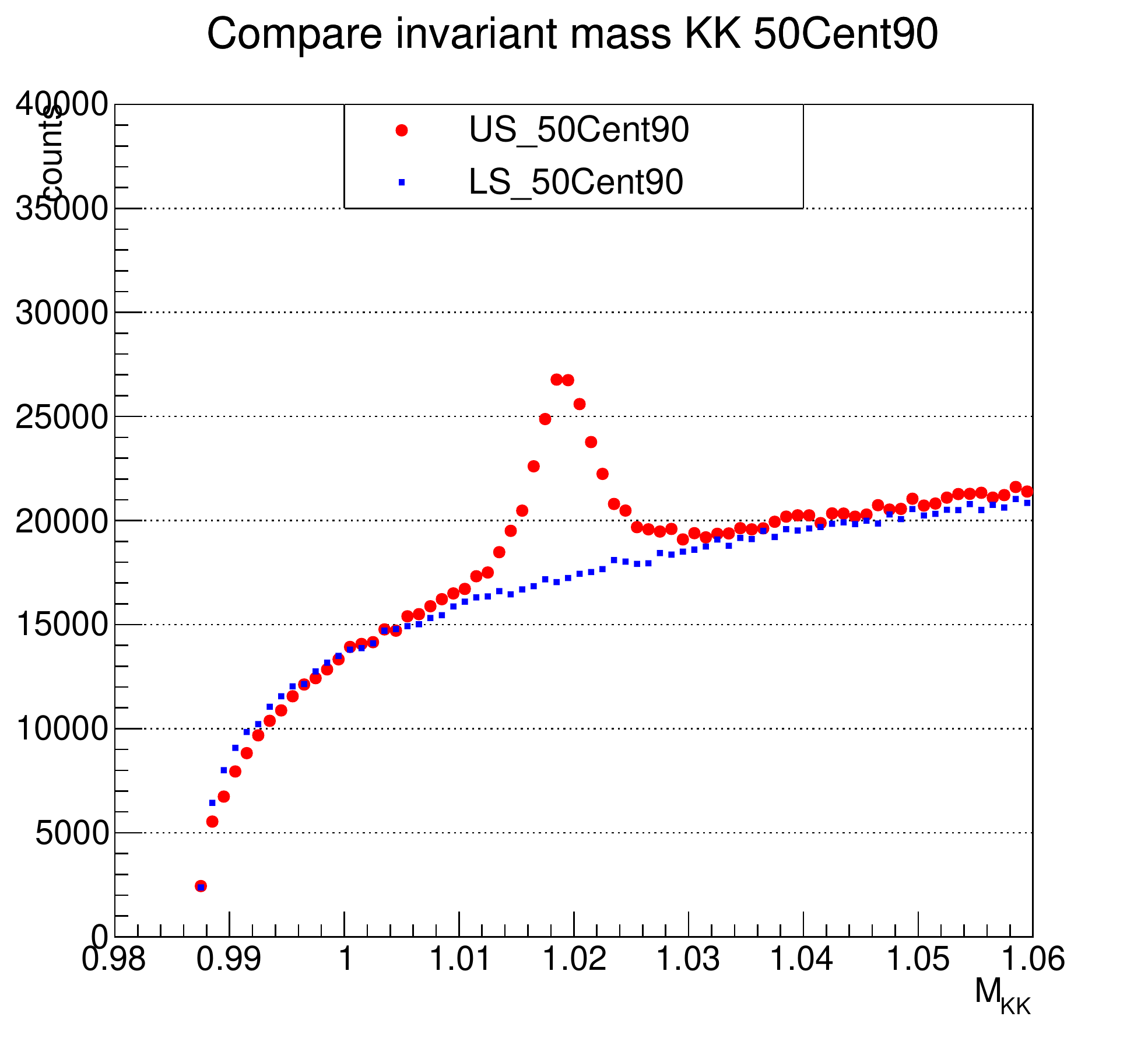}
  \caption{Comparison of US and LS kaon pairs invariant mass distributions obtained   in (left) central , (middle) mid-central, and (right) peripheral  Pb--Pb collisions. }
  \label{fig:PhiInvMass}
\end{figure}

We evaluated the yield of $\phi$-meson decays into $K^{+}$+$K^{-}$ pairs based on the $\phi$-meson (two charged kaons) invariant mass distributions of US and LS kaon pairs in Fig.~\ref{fig:PhiInvMass}. 
We found that the number of $K^{+}$+$K^{-}$ pairs from $\phi$-meson decays is smaller than 3 \% of the total number of $K^{+}$+$K^{-}$ pairs observed in the  near-side peak region  of US correlations.  However, because they tend to be emitted at small relative angle and rapidity, due in  part to  kinematical focusing associated with radial flow, $K^{+}$+$K^{-}$ pairs from $\phi$-meson may nonetheless contribute a
sizable fraction of the near-side peak of $KK$ balance functions. We thus further explored contributions to
the $KK$ BFs based on MC studies.

\subsection{MC Study of $\phi$-Meson Decay in $B^{KK}$}
\label{subsubsec:MCPhiMesonDecay}

A MC study of $\phi$ meson decayed $K^{+}$+$K^{-}$ pair contribution in $B^{KK}$ was performed using the HIJING generator level data described in Sec.~\ref{subsec:MCDataSamples}.
Unfortunately, a similar study using simulation data produced with the  AMPT model, which features  more realistic $\phi$-meson yields,  is not possible because  the  AMPT dataset produced by the ALICE collaboration  does not contain information of decaying mother particles. Producing an independent AMPT dataset suitable for this study is beyond the scope and resources of this work. 

\begin{figure}
\centering
  \includegraphics[width=0.32\linewidth]{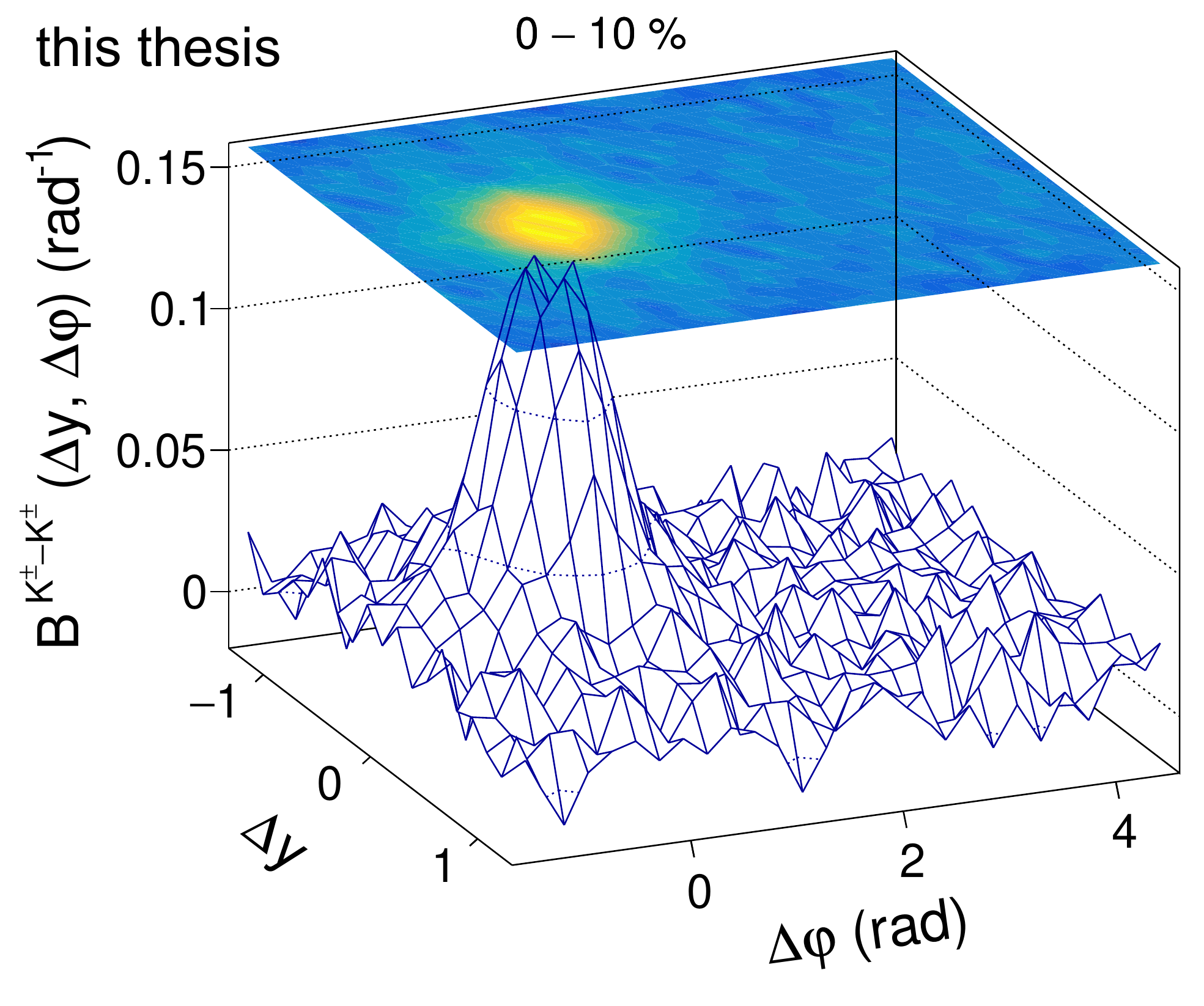}
  \includegraphics[width=0.32\linewidth]{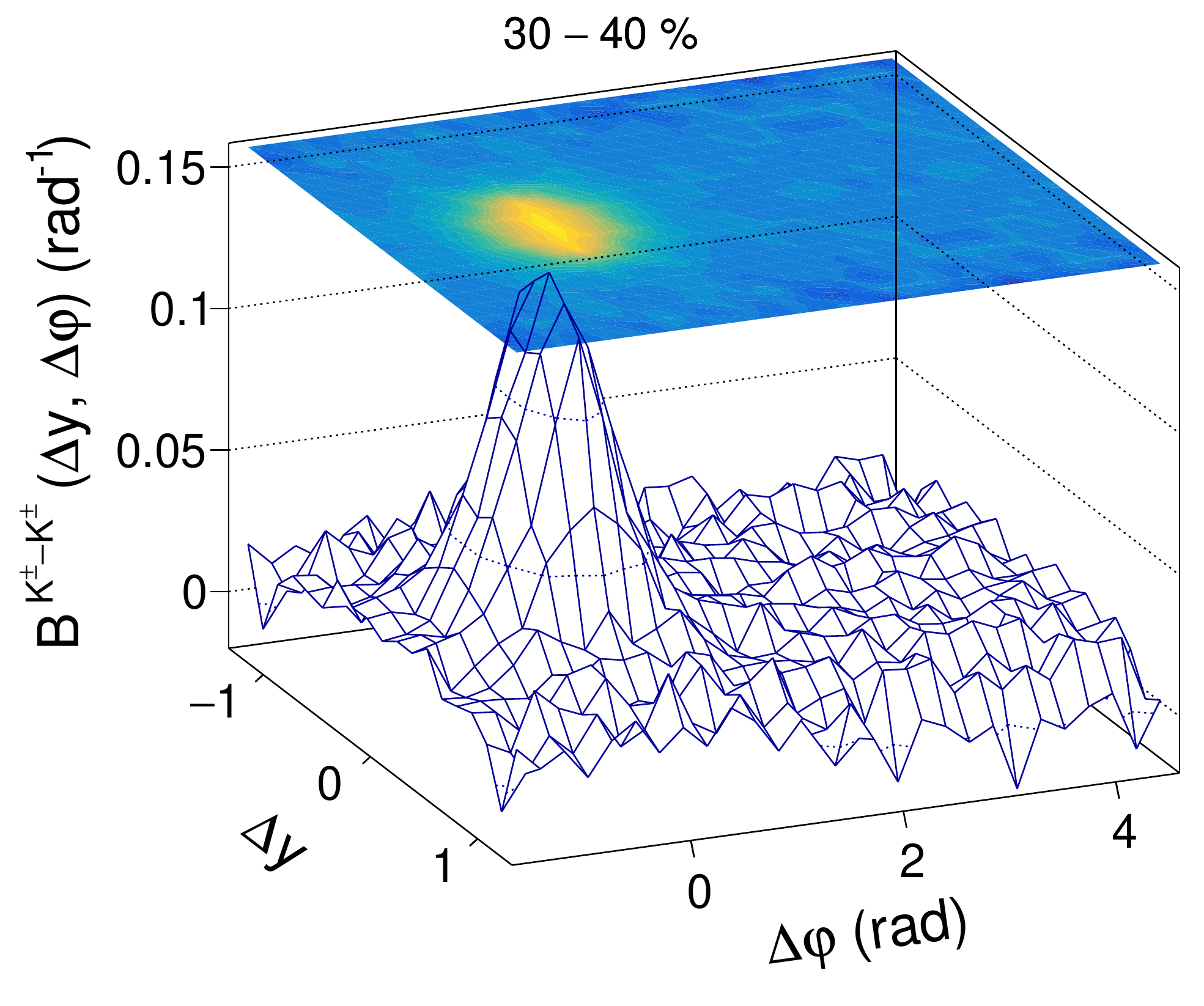}
  \includegraphics[width=0.32\linewidth]{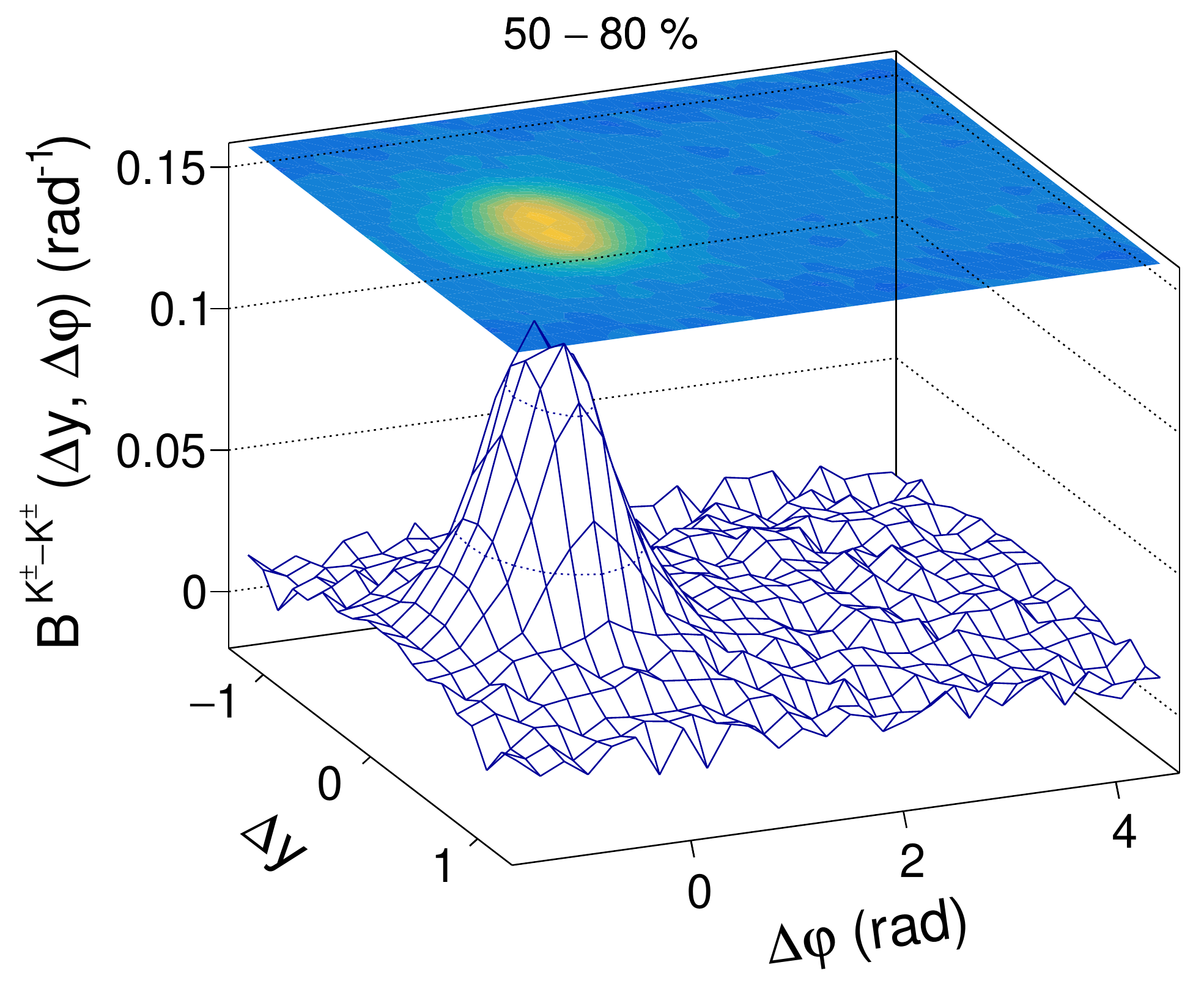}
  \includegraphics[width=0.32\linewidth]{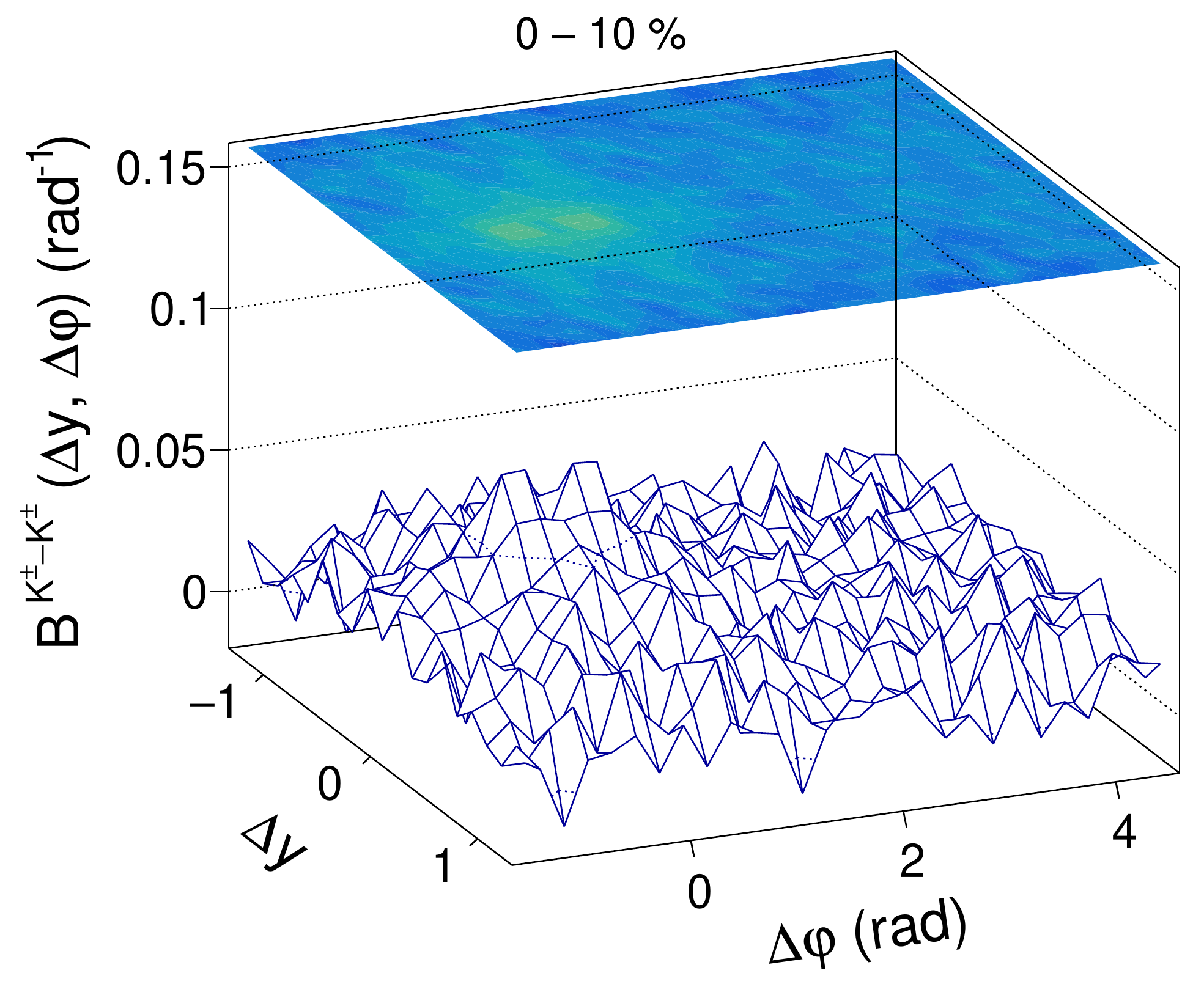}
  \includegraphics[width=0.32\linewidth]{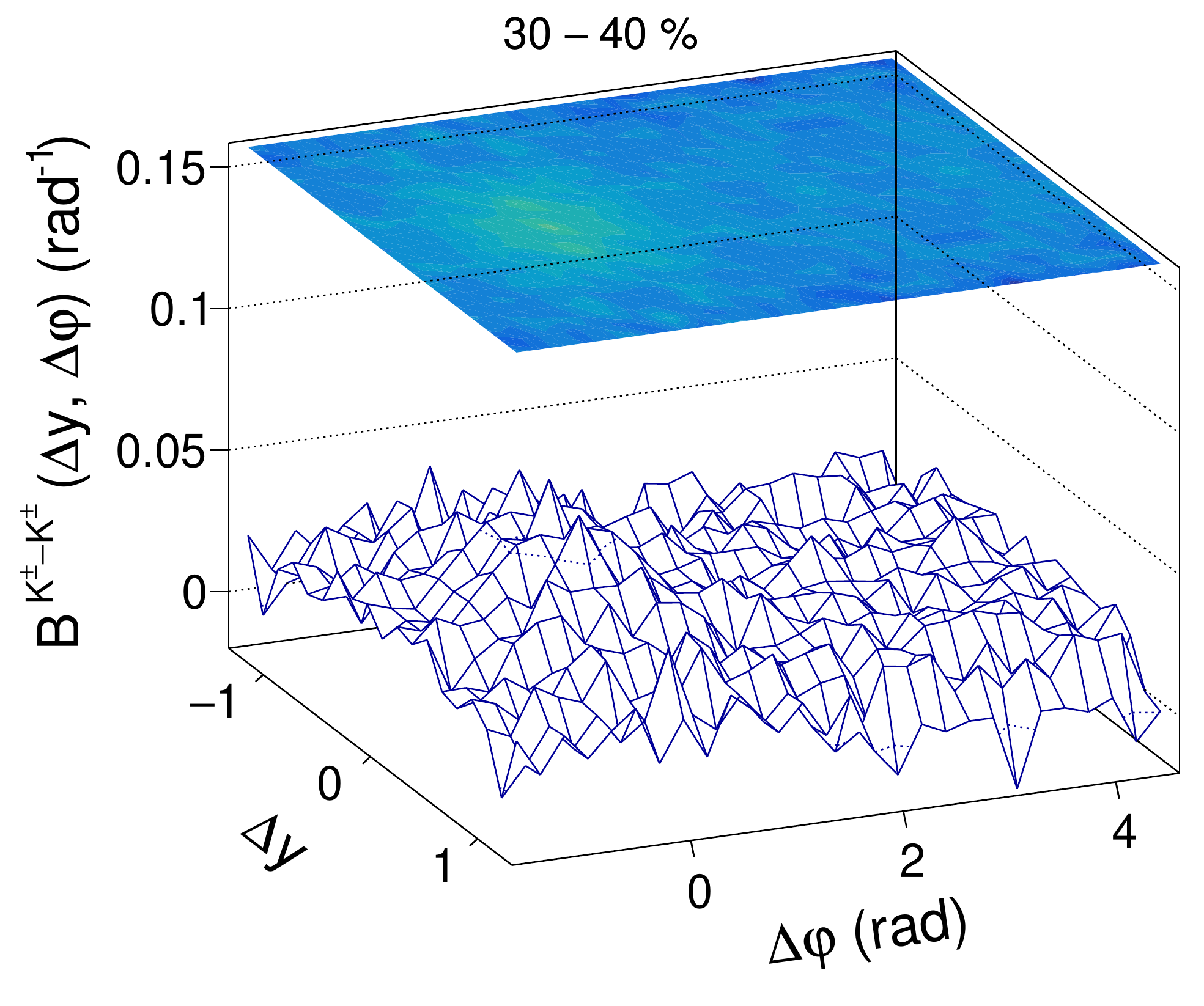}
  \includegraphics[width=0.32\linewidth]{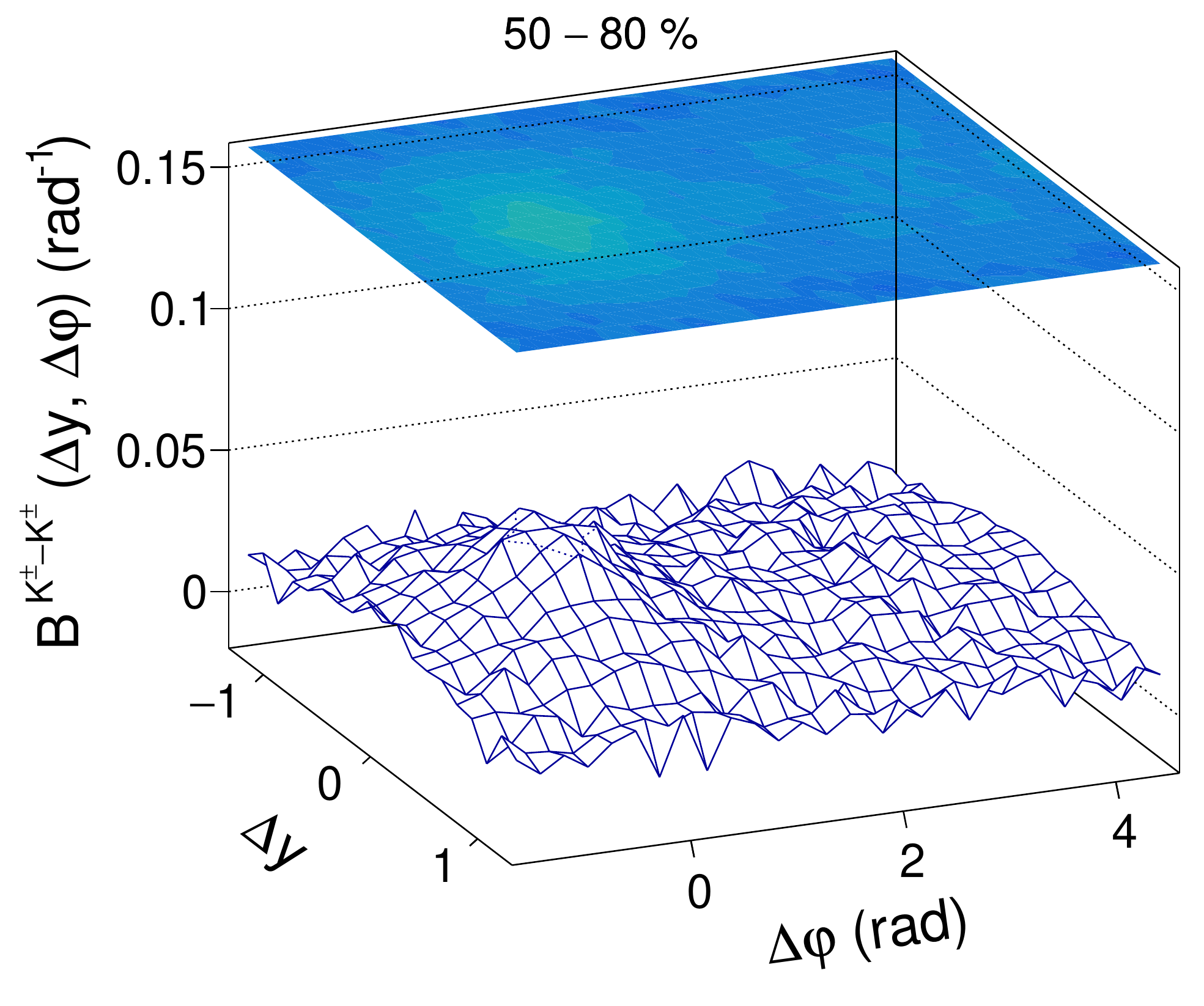} 
  \caption{Comparison of $KK$ BFs obtained when including   (top row) and excluding (bottom row) $\phi$-meson decays from  MC HIJING generator level data  in three selected ranges of Pb-Pb collision centralities.}
  \label{fig:BF_KaonKaon_HIING_Truth_Phi_Decay_2D}
\end{figure}

\begin{figure}
\centering
  \includegraphics[width=0.32\linewidth]{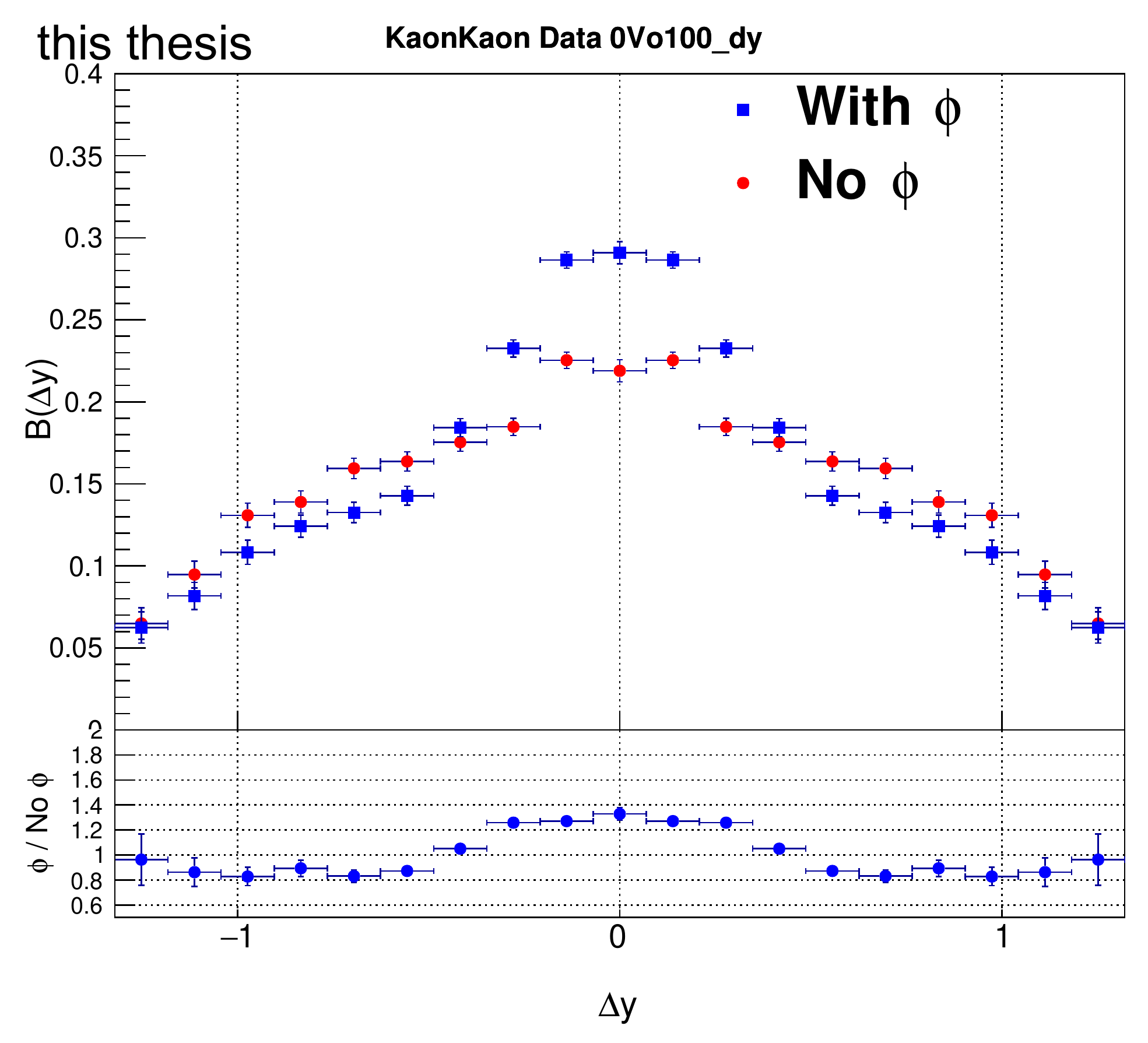}
  \includegraphics[width=0.32\linewidth]{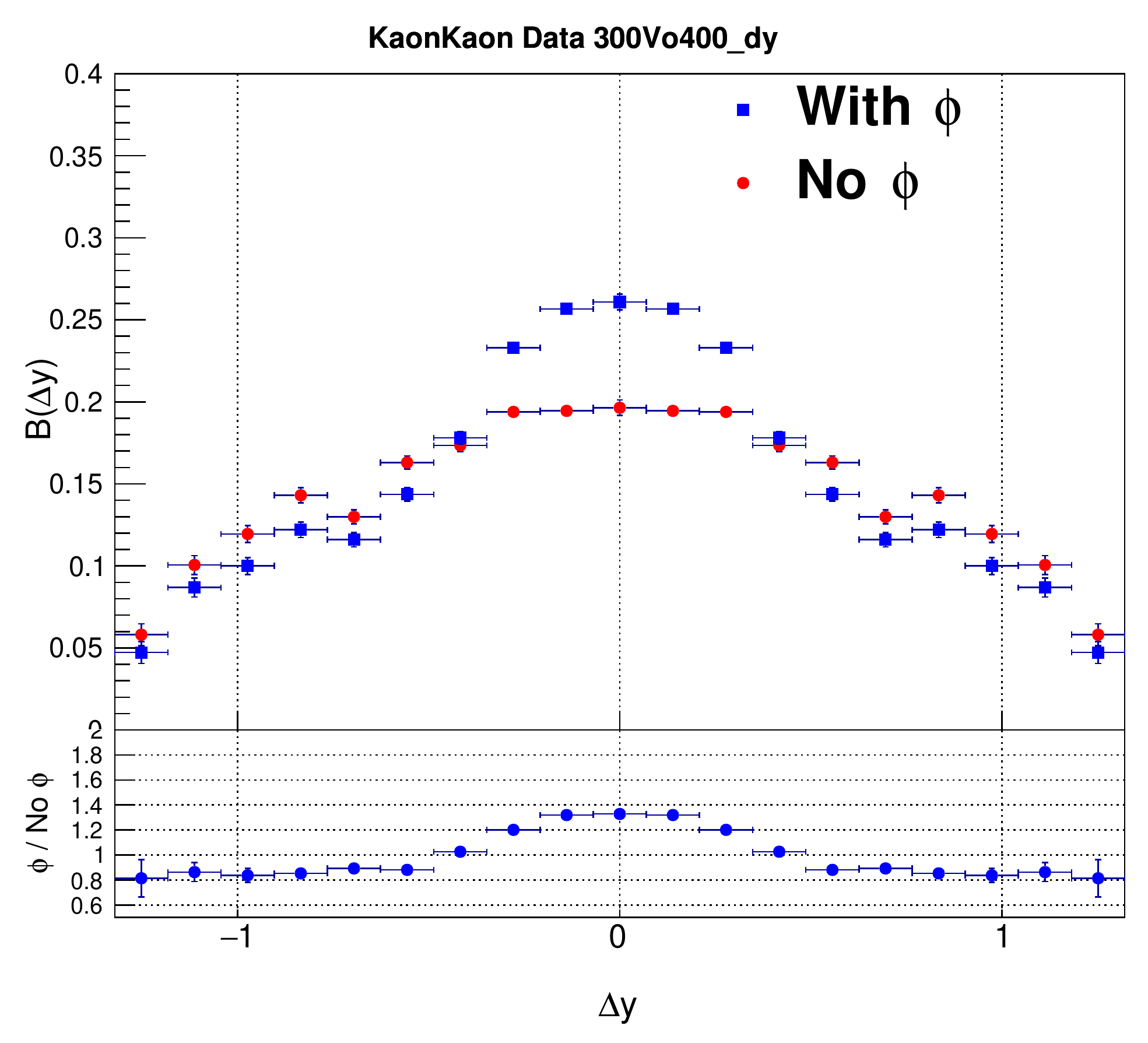}
  \includegraphics[width=0.32\linewidth]{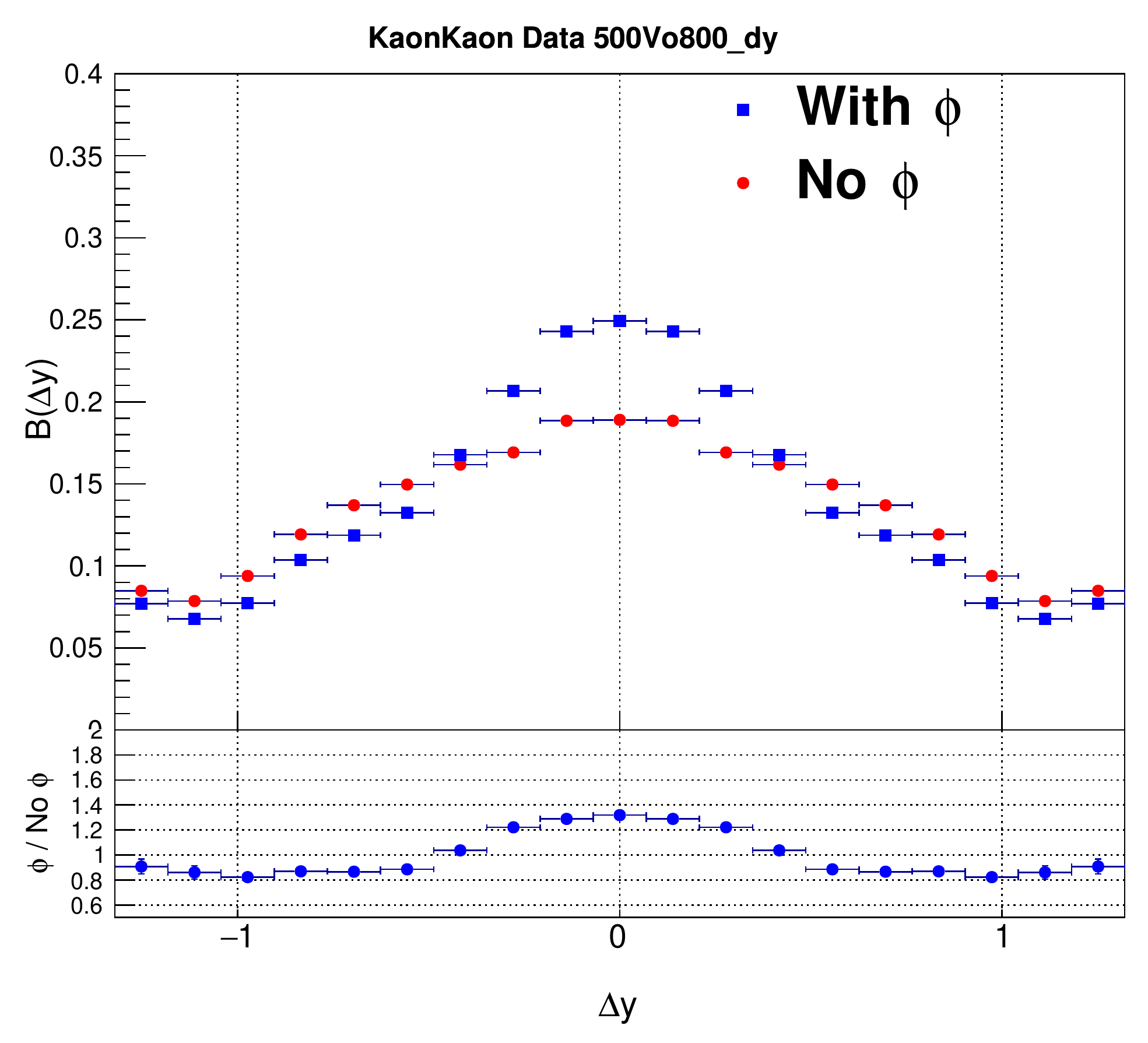}
  \includegraphics[width=0.32\linewidth]{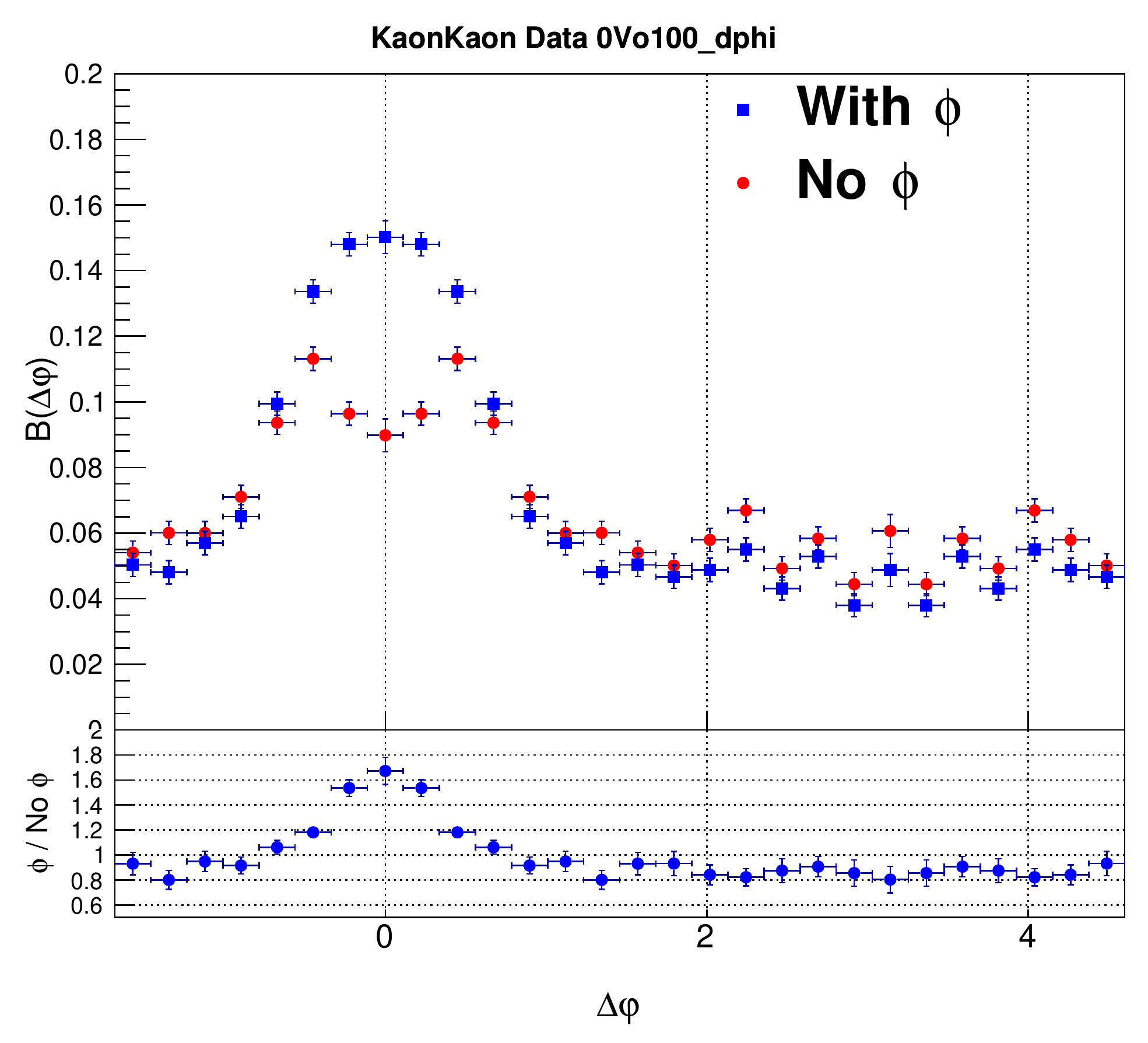}
  \includegraphics[width=0.32\linewidth]{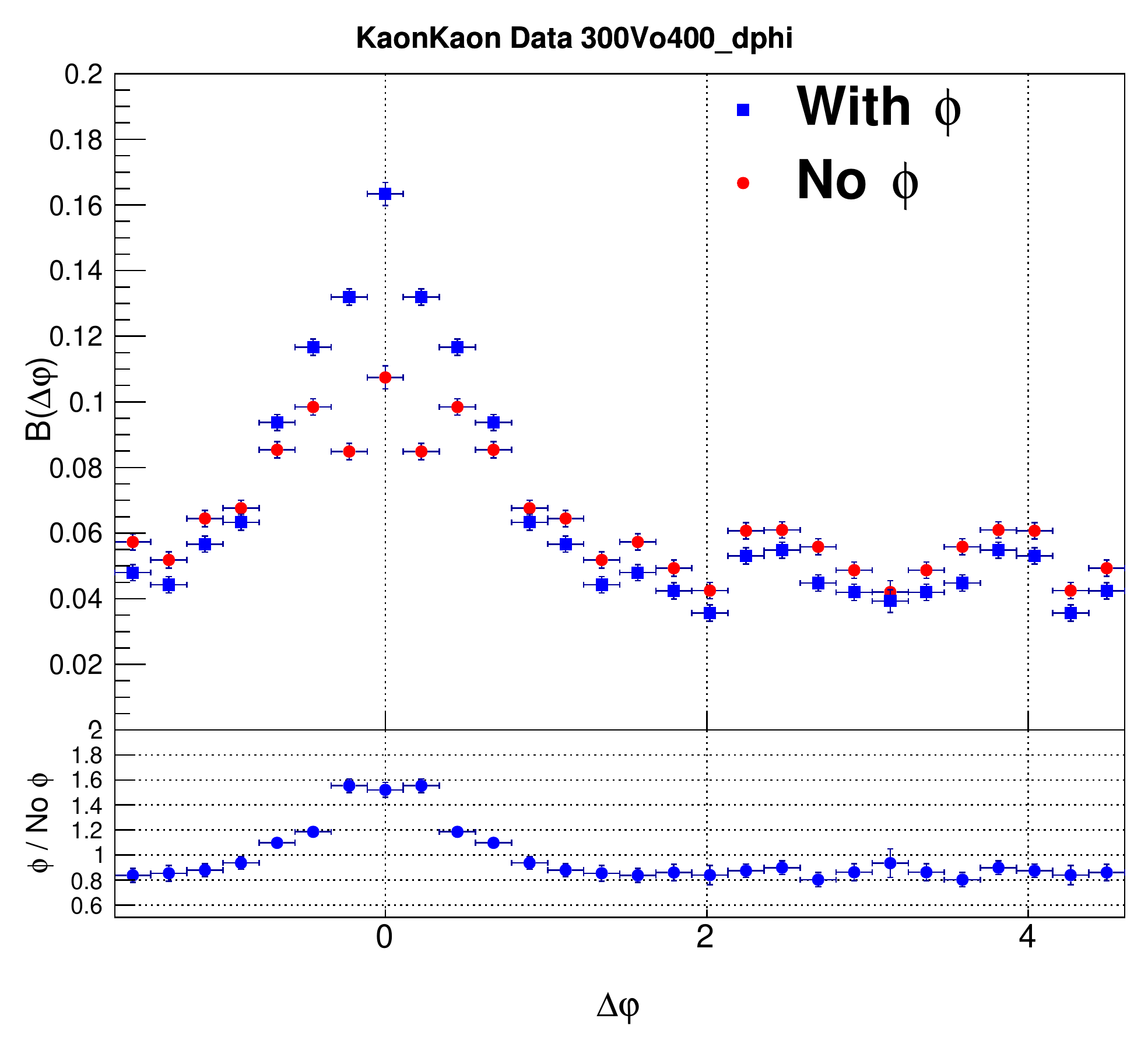}
  \includegraphics[width=0.32\linewidth]{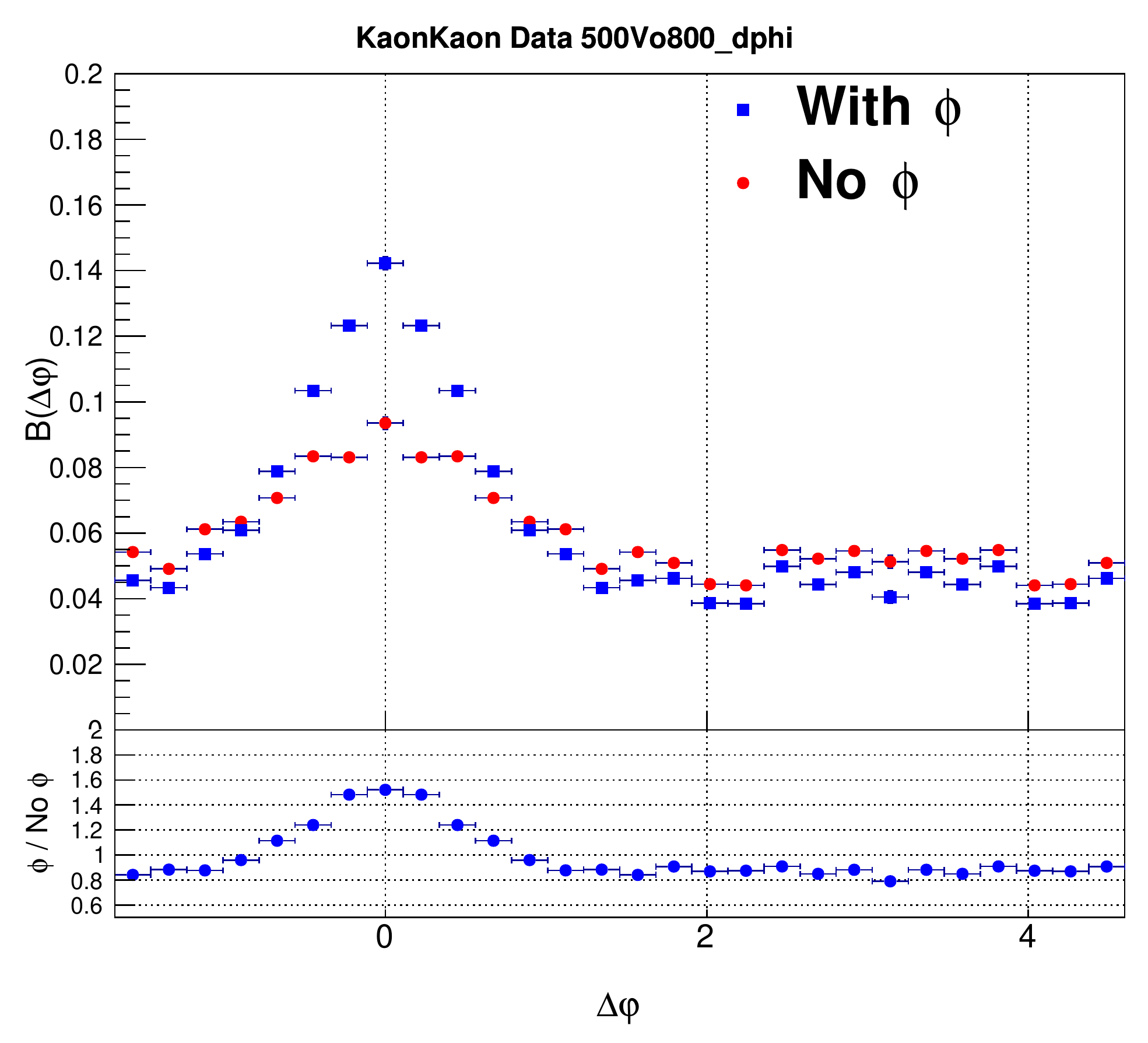}
  \includegraphics[width=0.32\linewidth]{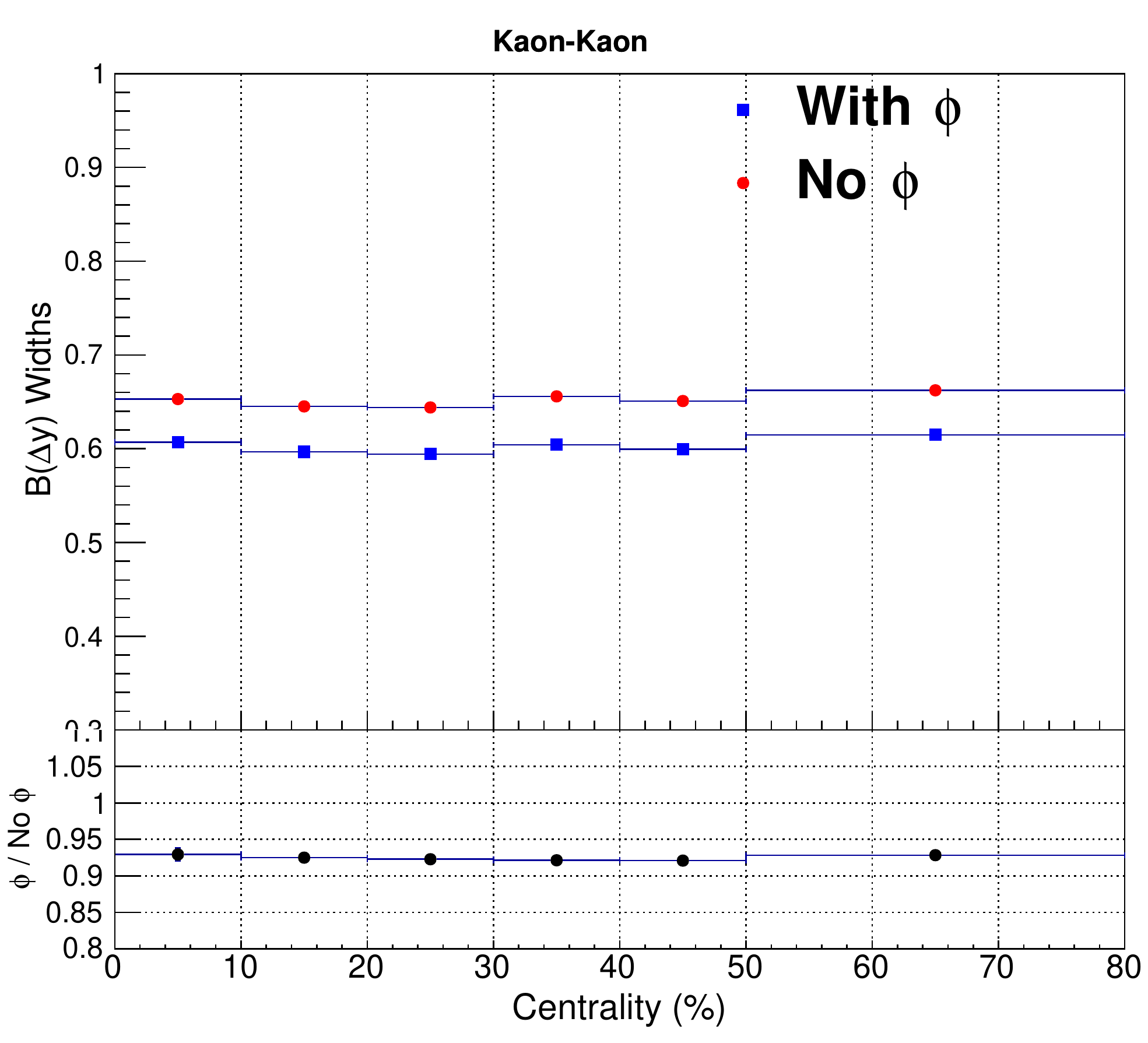}
  \includegraphics[width=0.32\linewidth]{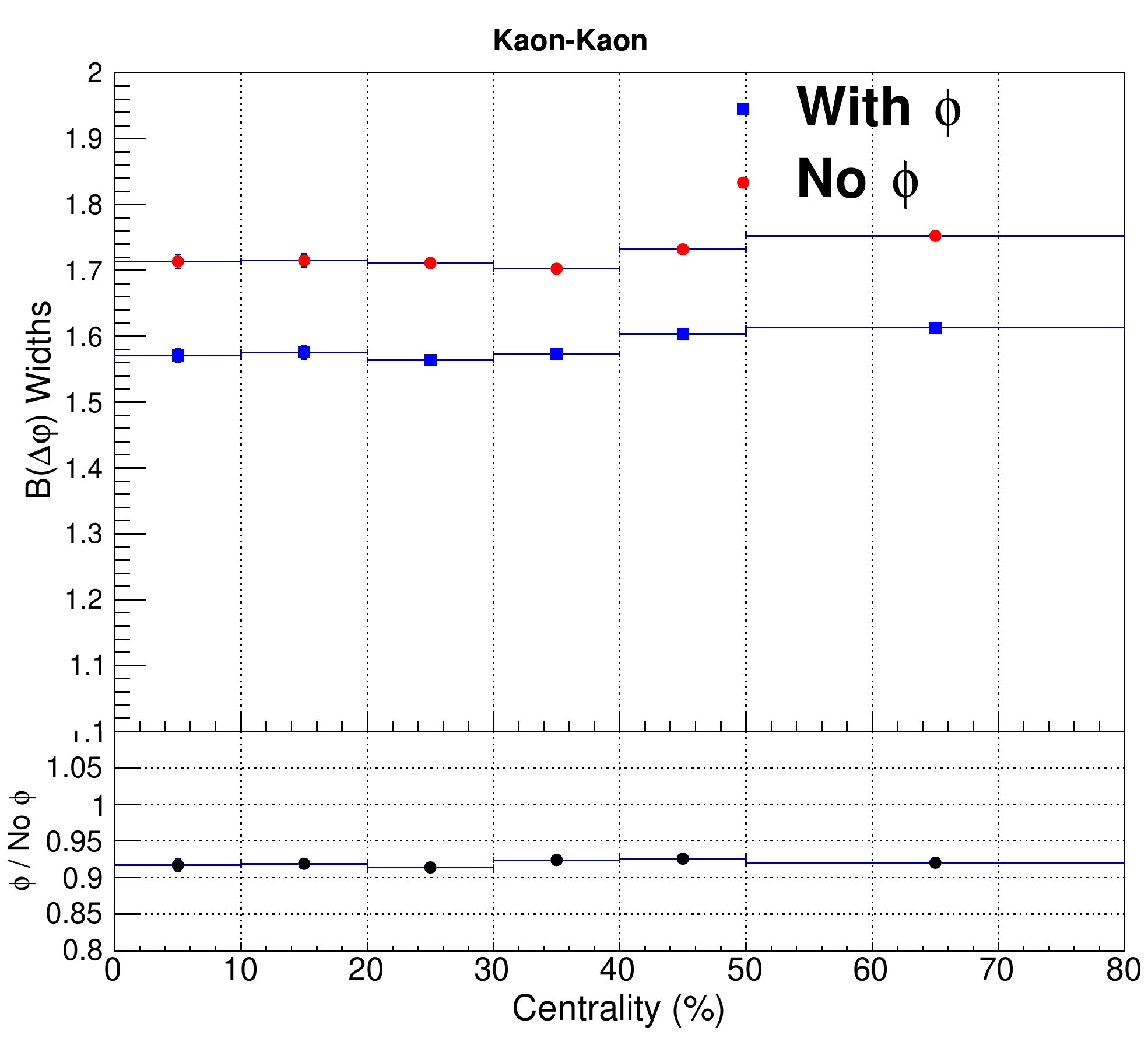}
  \includegraphics[width=0.32\linewidth]{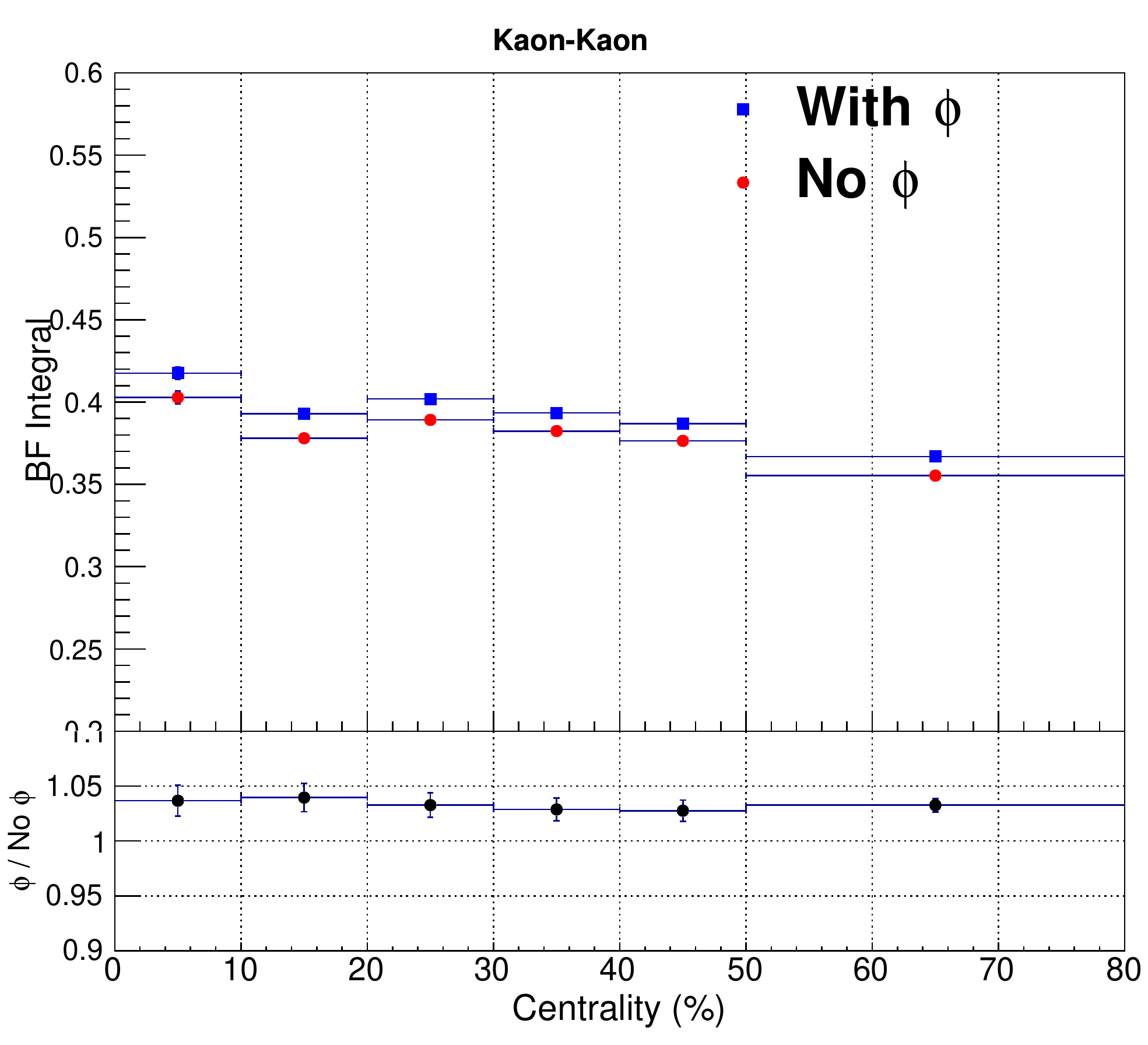} 
  \caption{ For $KK$ using MC HIJING generator level data. Comparisons on BF $\Delta y$ (top row) and $\Delta\varphi$ (middle row) projections of different centralities, and $\Delta y$ widths (lower left), $\Delta\varphi$ widths (lower middle), and BF integrals (lower right) between phi meson decay included and excluded.}
  \label{fig:BF_KaonKaon_HIING_Truth_Phi_Decay_widths_integral}
\end{figure}

%A comparative analysis of the 2D $B^{KK}$ balance functions shown in 
Figure~\ref{fig:BF_KaonKaon_HIING_Truth_Phi_Decay_2D} presents 
a comparison of two-dimensional $B^{KK}$ balance functions obtained with the inclusion (top) and the exclusion (bottom) of charged kaon pairs originating from $\phi$-mesons in three ranges of Pb--Pb collision centralities. Projections of these balance functions onto   the $\Delta y$ and $\Delta\varphi$ axes are
displayed in  Fig.~\ref{fig:BF_KaonKaon_HIING_Truth_Phi_Decay_widths_integral}. One finds that the amplitude of the near-side peak of the balance function is suppressed by about 30\% when contributions from $\phi$-meson decays are explicitly excluded.
Figure~\ref{fig:BF_KaonKaon_HIING_Truth_Phi_Decay_widths_integral} further shows that $\phi$-meson decays could possibly lower the $B^{KK}$ $\Delta y$ and $\Delta\varphi$ widths by about 7-8\%, and increase the integrals by about 3-4\%. 
%In addition, Figure~\ref{fig:BF_KaonKaon_HIING_Truth_Phi_Decay_widths_integral} shows that $\phi$-meson decays do not change the centrality dependence of $B^{KK}$ $\Delta y$ and $\Delta\varphi$ widths, and integrals.

The STAR Collaboration measured $B^{KK}$ in terms of $q_{inv}$ in Au--Au collisions at $\sqrt{s_{_{\rm NN}}} =$200 GeV in nine centrality bins~\cite{PhysRevC.82.024905}.
They concluded that the $\phi$-meson decay contribution in $B^{KK}(q_{inv})$ is approximately 50\%, independent of centrality.

\clearpage

\section{Lambda Weak Decay in $B^{\pi p}$ and $B^{p\pi}$}
\label{subsec:LambdaDecay}

We investigated the presence of contamination from Lambda baryon ($\Lambda$/$\bar{\Lambda}$) decays in $B^{\pi p}$ and $B^{p\pi}$ balance functions. Lambda baryon decays occur on a weak-interaction time scale of $(2.631\pm 0.020)\times 10^{-10}$ s (in their rest frame). Their daughter products, protons and pions (or anti-particles), are thus considered secondary particles in the context of this work. Secondary pions and protons (anti-protons) must thus be suppressed in  measurements of $\pi \pi$, $\pi p$, and  $p p$ balance functions. In this work, this was accomplished by means of tight DCA cuts. Achieved estimated purities were already discussed in Sec.~\ref{sec:Particle Identification}. The discussion presented in this section focuses on the impact of secondary pions and protons on the $B^{\pi p}$ balance function, which is most susceptible to 
contamination from $\Lambda$/$\bar{\Lambda}$ decays.

\subsection{Tight $DCA_{xy}$ Cut}
\label{subsubsec:TightDCAxyCut}

\begin{figure}
\centering
  \includegraphics[width=0.49\linewidth]{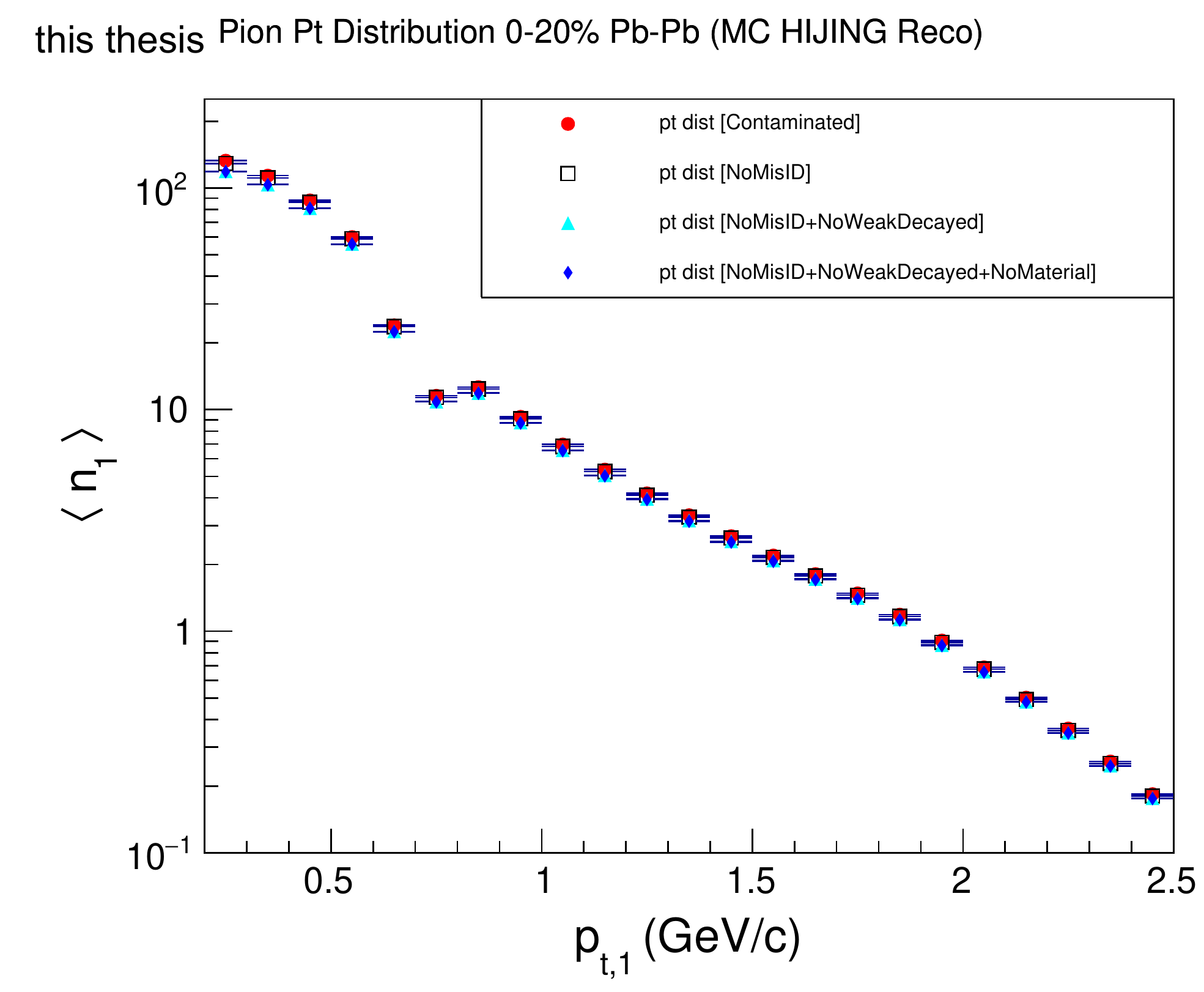}
  \includegraphics[width=0.49\linewidth]{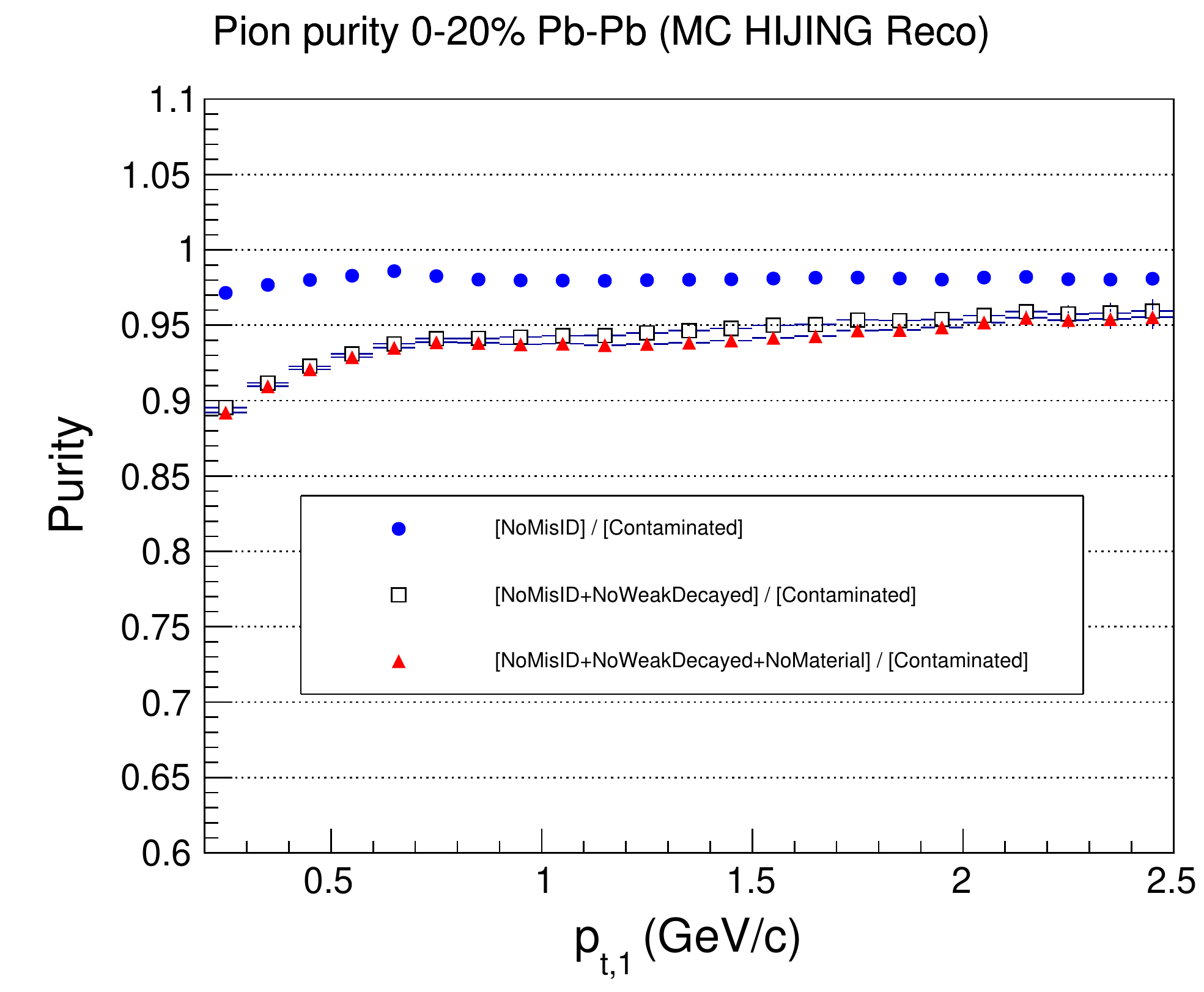}
  \includegraphics[width=0.49\linewidth]{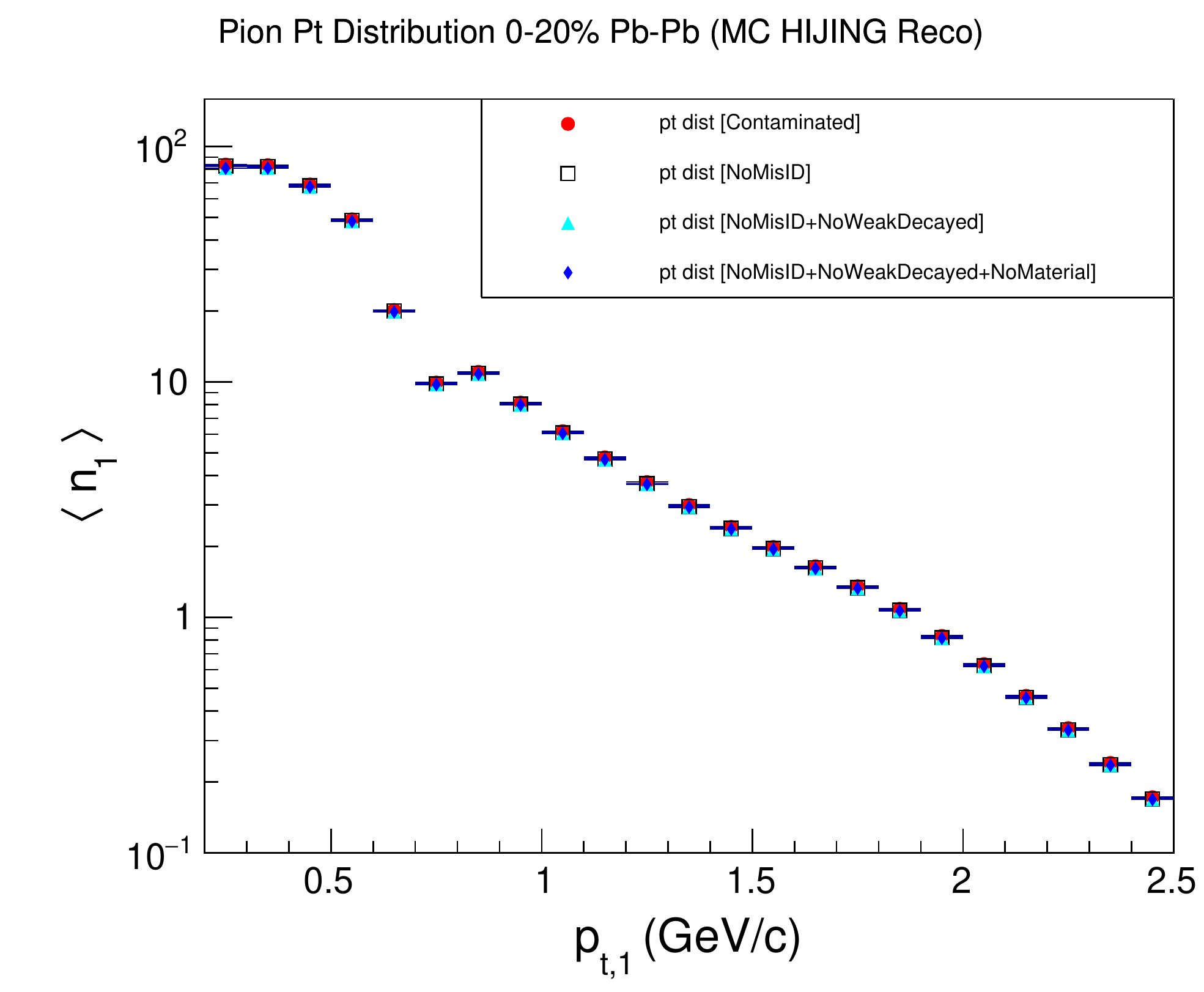}
  \includegraphics[width=0.49\linewidth]{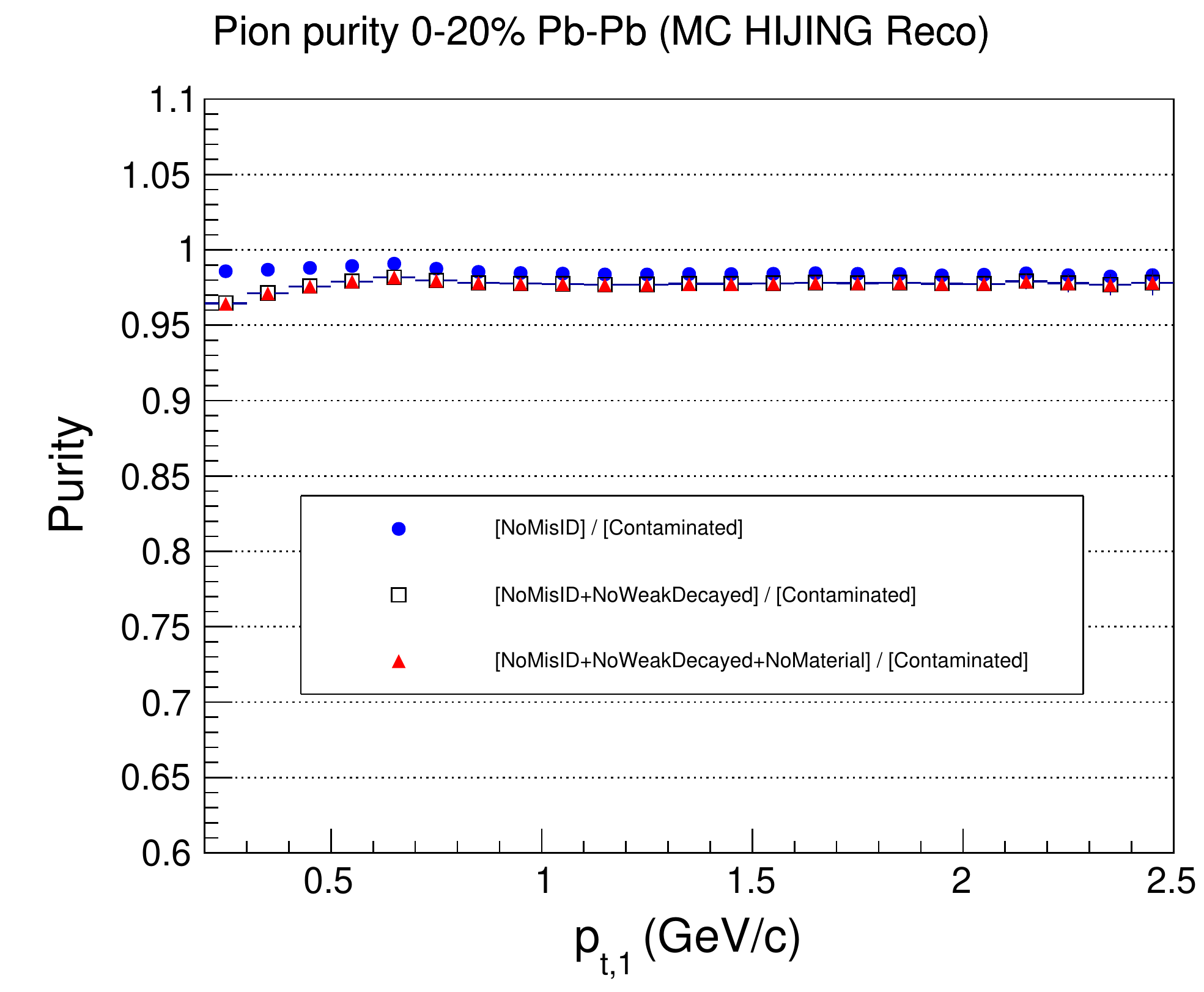}
  \caption{Comparison of primary $\pi^{\pm}$ purities (right column) obtained with cuts $DCA_{xy}<2.4$ cm (top row) and $DCA_{xy}<0.04$ cm (bottom row) in 0-20\% centrality Pb--Pb collisions. The purities are calculated using MC reconstructed data, and are based on the $p_{\rm T}$ distributions of $\pi^{\pm}$ (left column) with contributions from primary particles, secondary particles from weak decays and interaction with detector materials, and mis-identified particles, separately. Purities have similar values for other centralities.}
  \label{fig:Purity_DCAxy24_DCAxy004_Pion_0Cent20}
\end{figure}

\begin{figure}
\centering
  \includegraphics[width=0.49\linewidth]{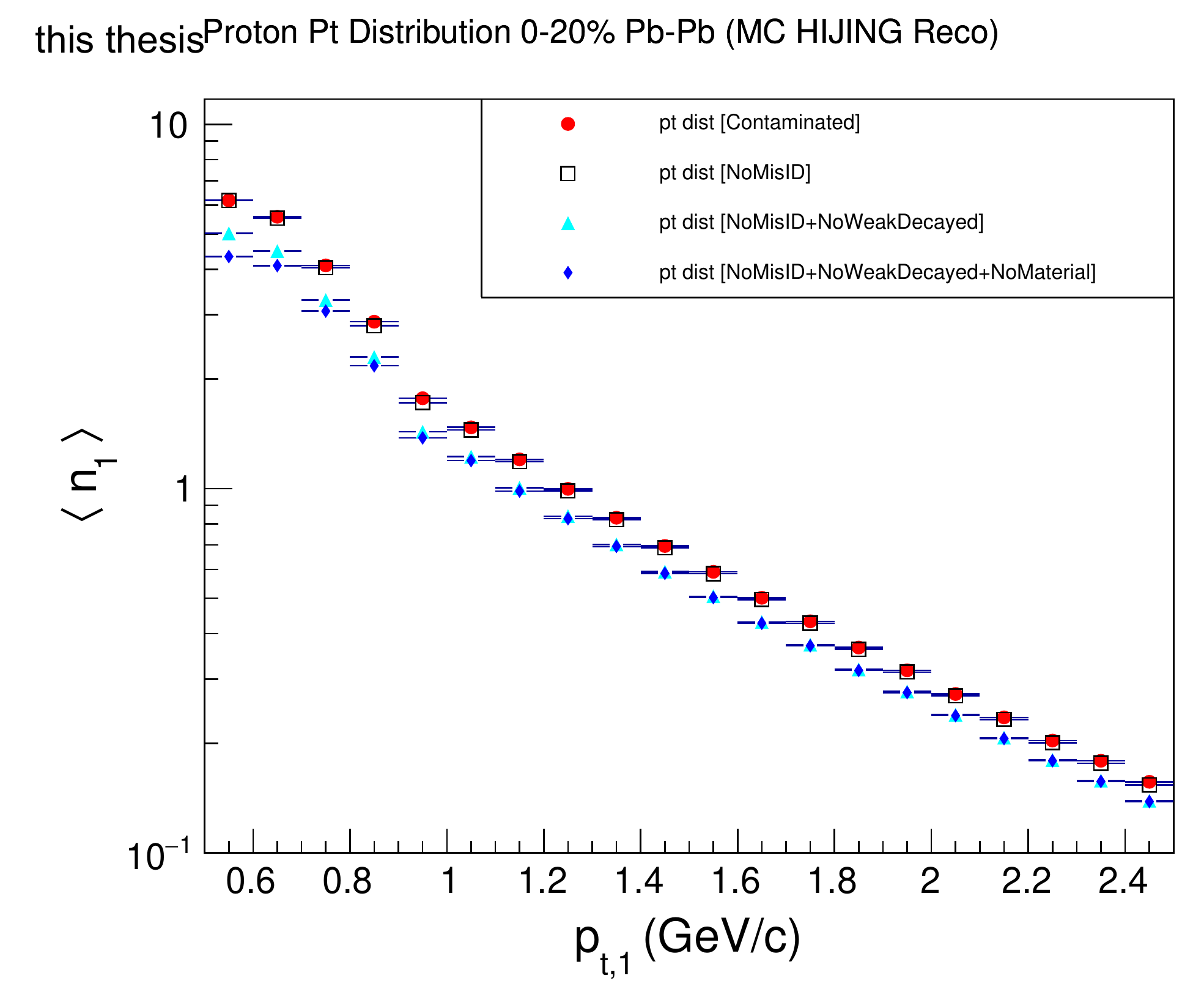}
  \includegraphics[width=0.49\linewidth]{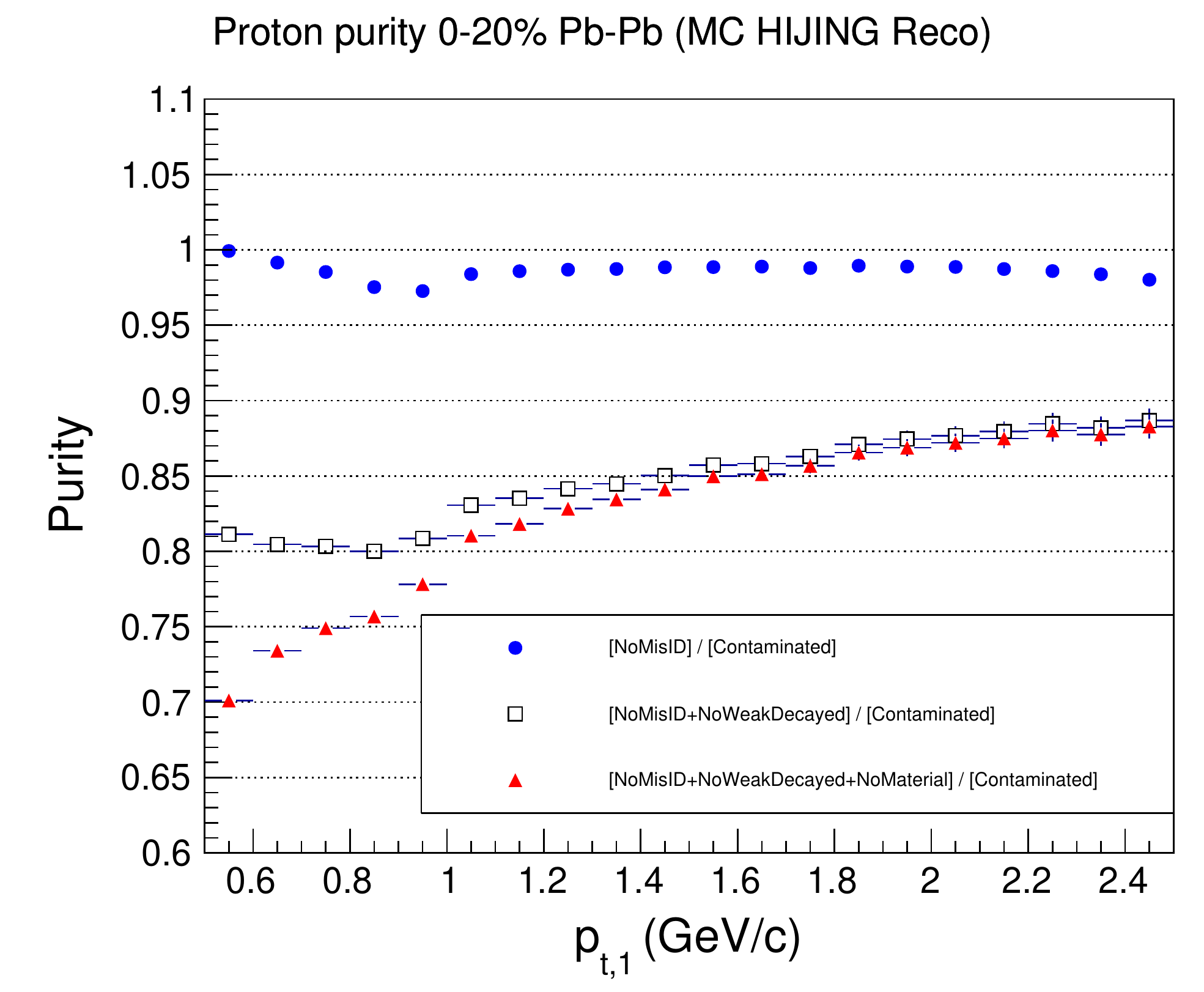}
  \includegraphics[width=0.49\linewidth]{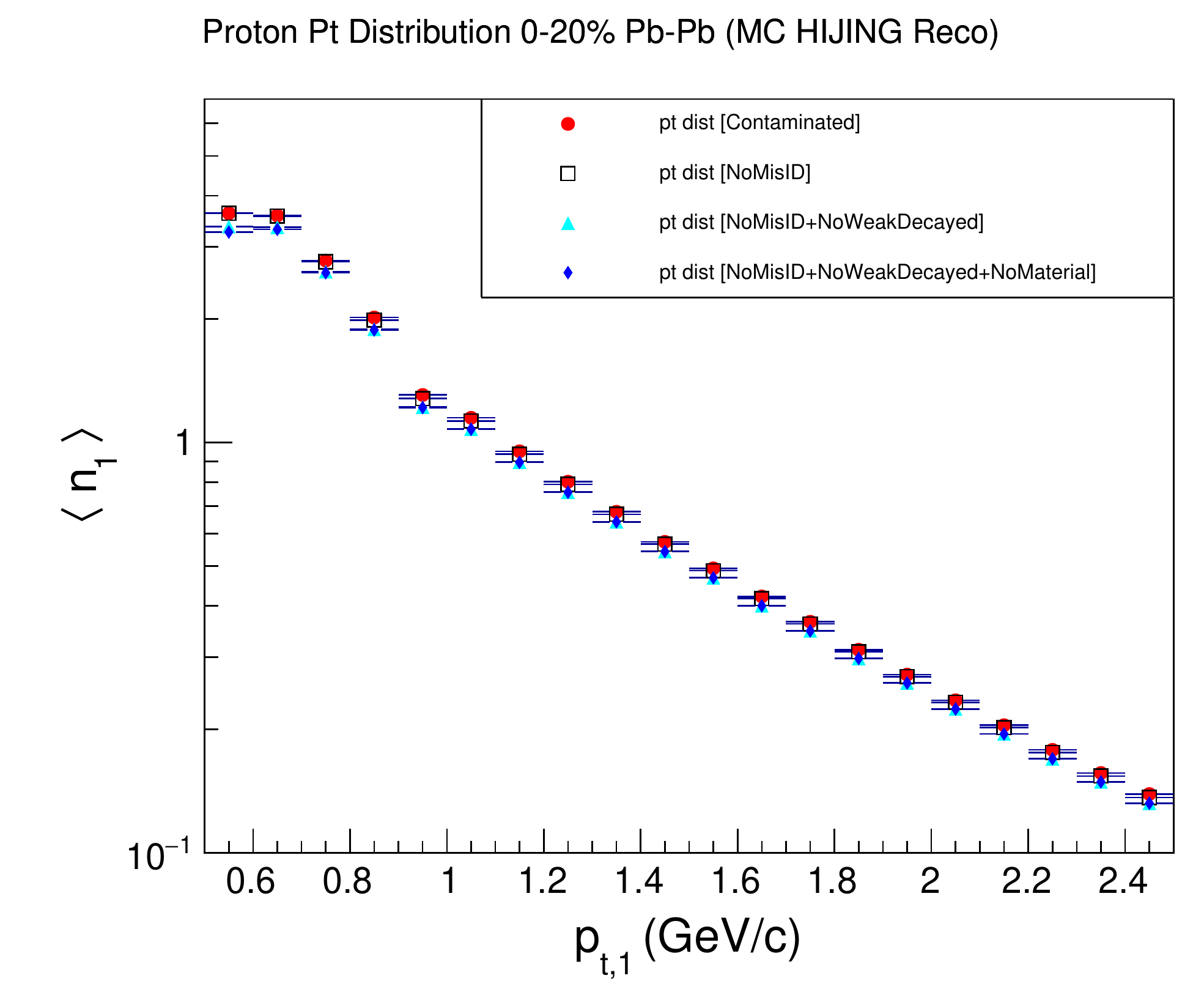}
  \includegraphics[width=0.49\linewidth]{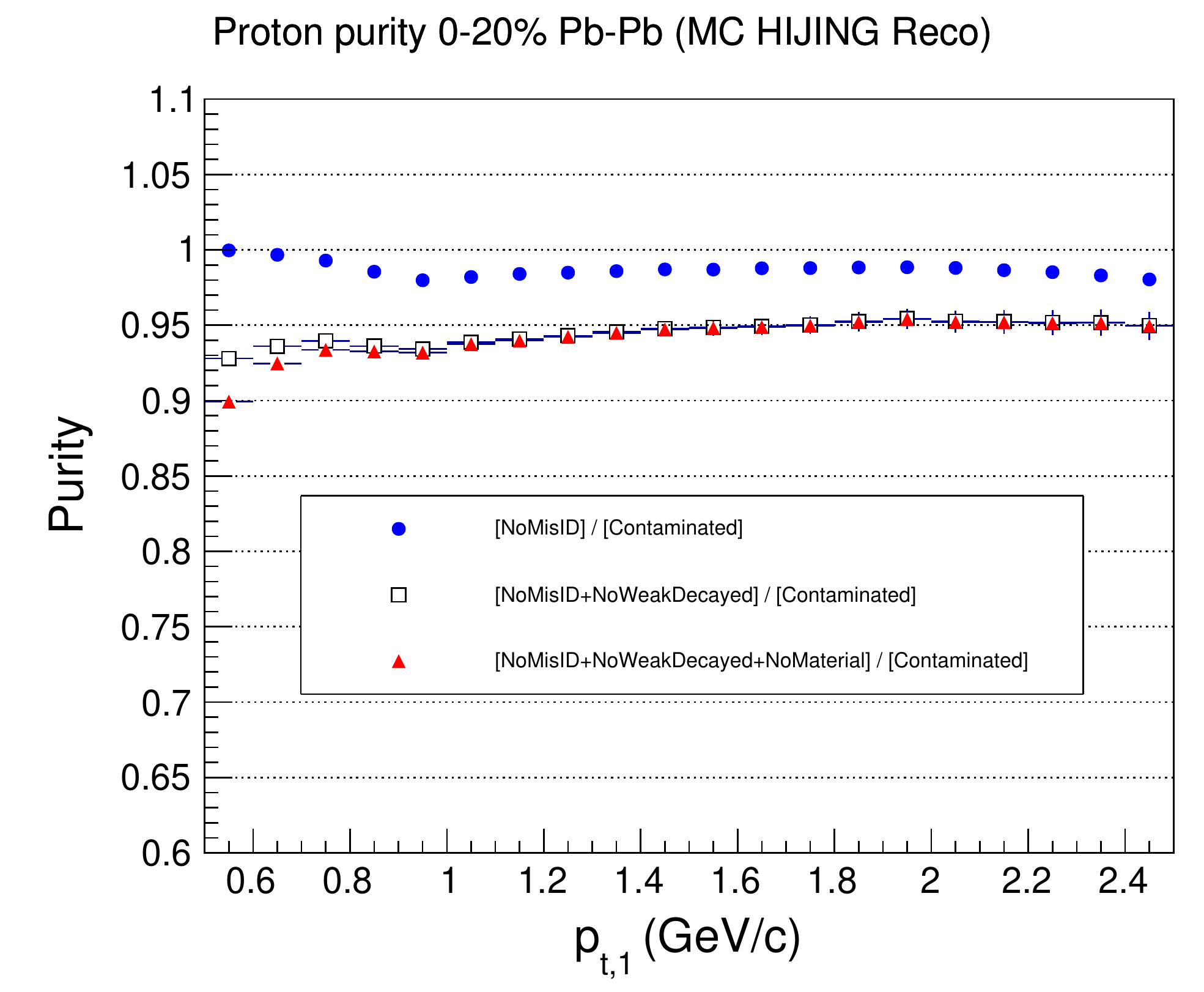}   
  \caption{Comparison of primary $p/\bar{p}$ purities (right column) obtained with cuts $DCA_{xy}<2.4$ cm (top row) and $DCA_{xy}<0.04$ cm (bottom row) in 0-20\% centrality Pb--Pb collisions. The purities are calculated using MC reconstructed data, and are based on the $p_{\rm T}$ distributions of $p/\bar{p}$ (left column) with contributions from primary particles, secondary particles from weak decays and interaction with detector materials, and mis-identified particles, separately. Purities have similar values for other centralities.}
  \label{fig:Purity_DCAxy24_DCAxy004_Proton_0Cent20}
\end{figure}

\begin{figure}
\centering
  \includegraphics[width=0.32\linewidth]{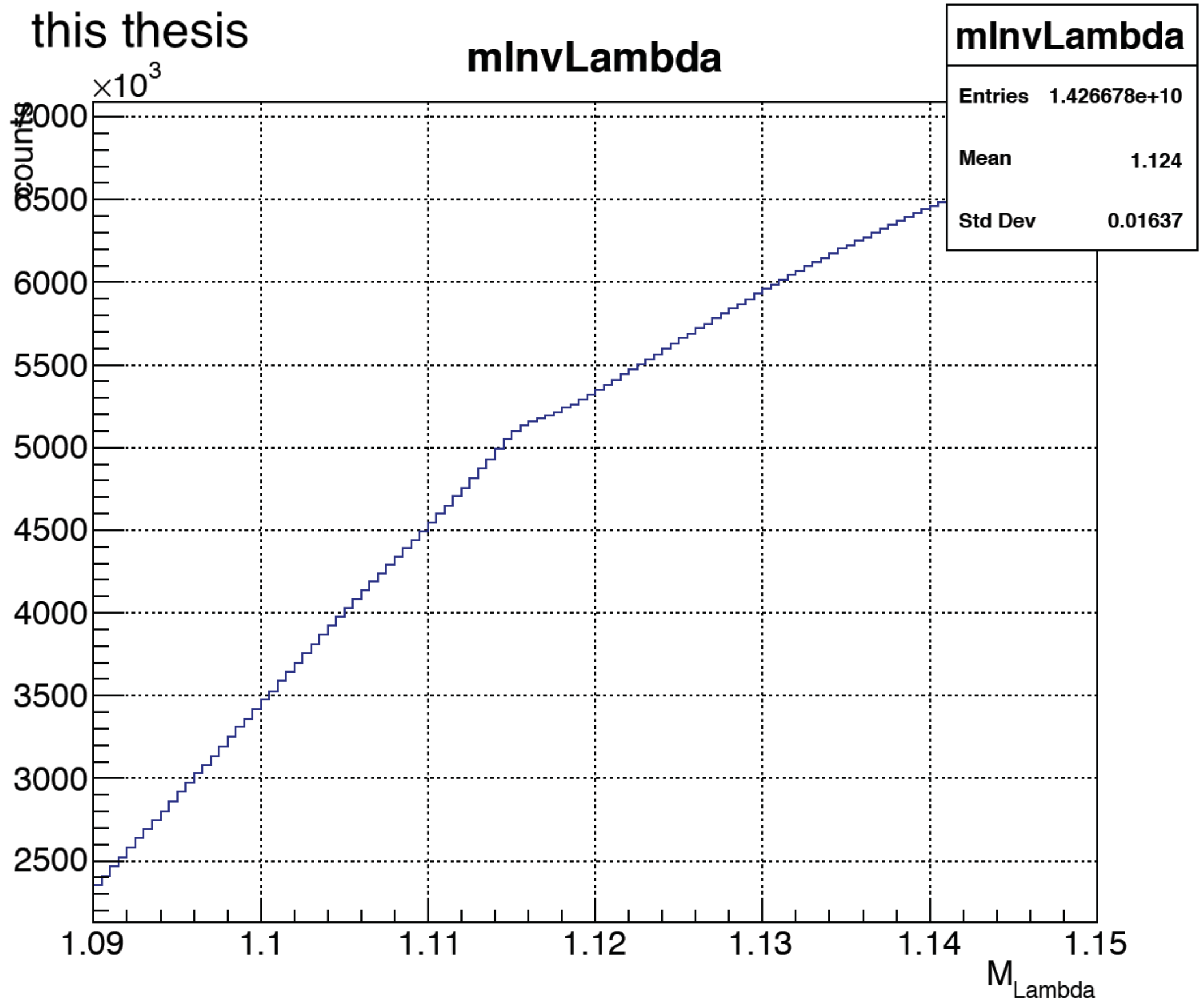}
  \includegraphics[width=0.32\linewidth]{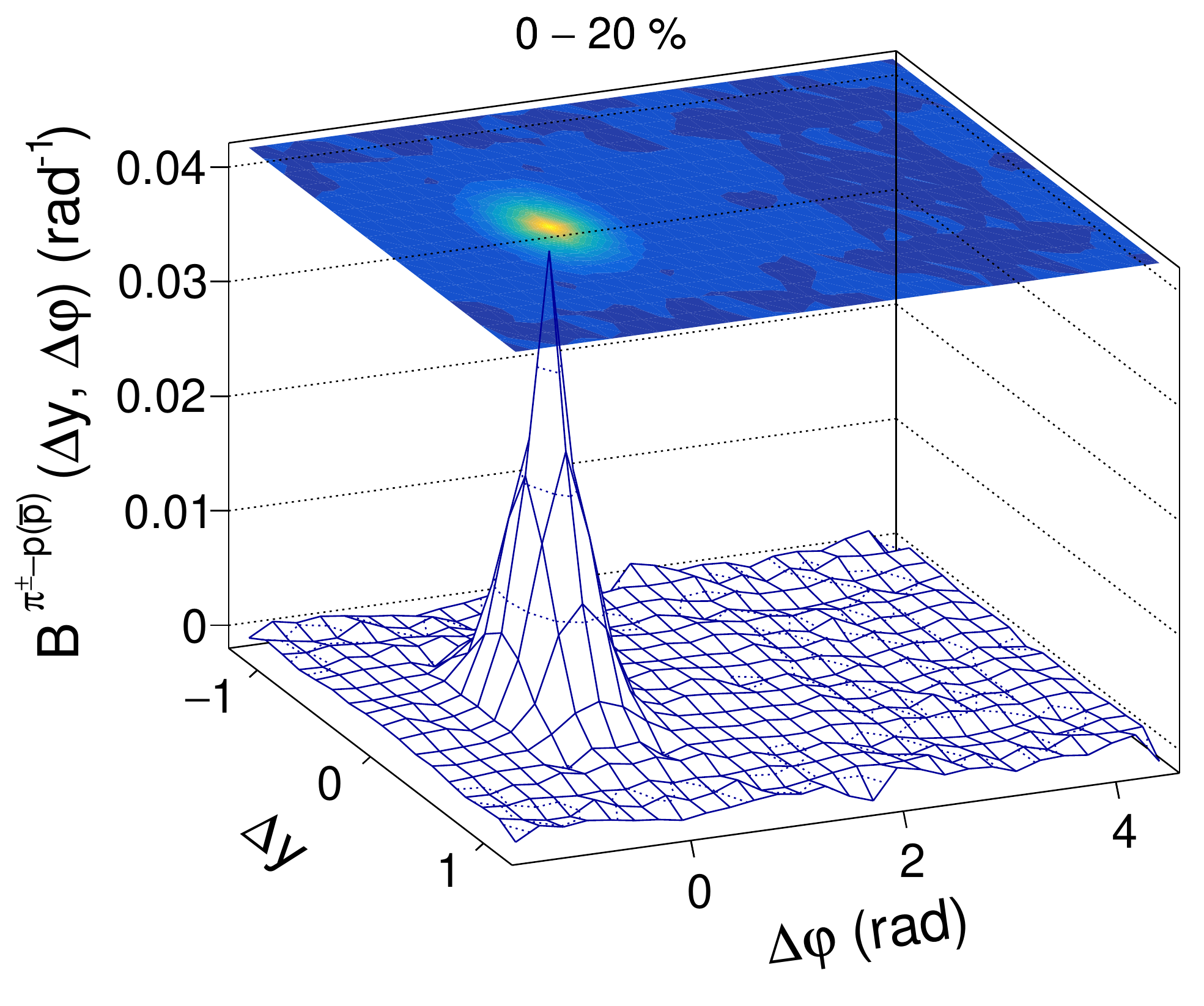}
  \includegraphics[width=0.32\linewidth]{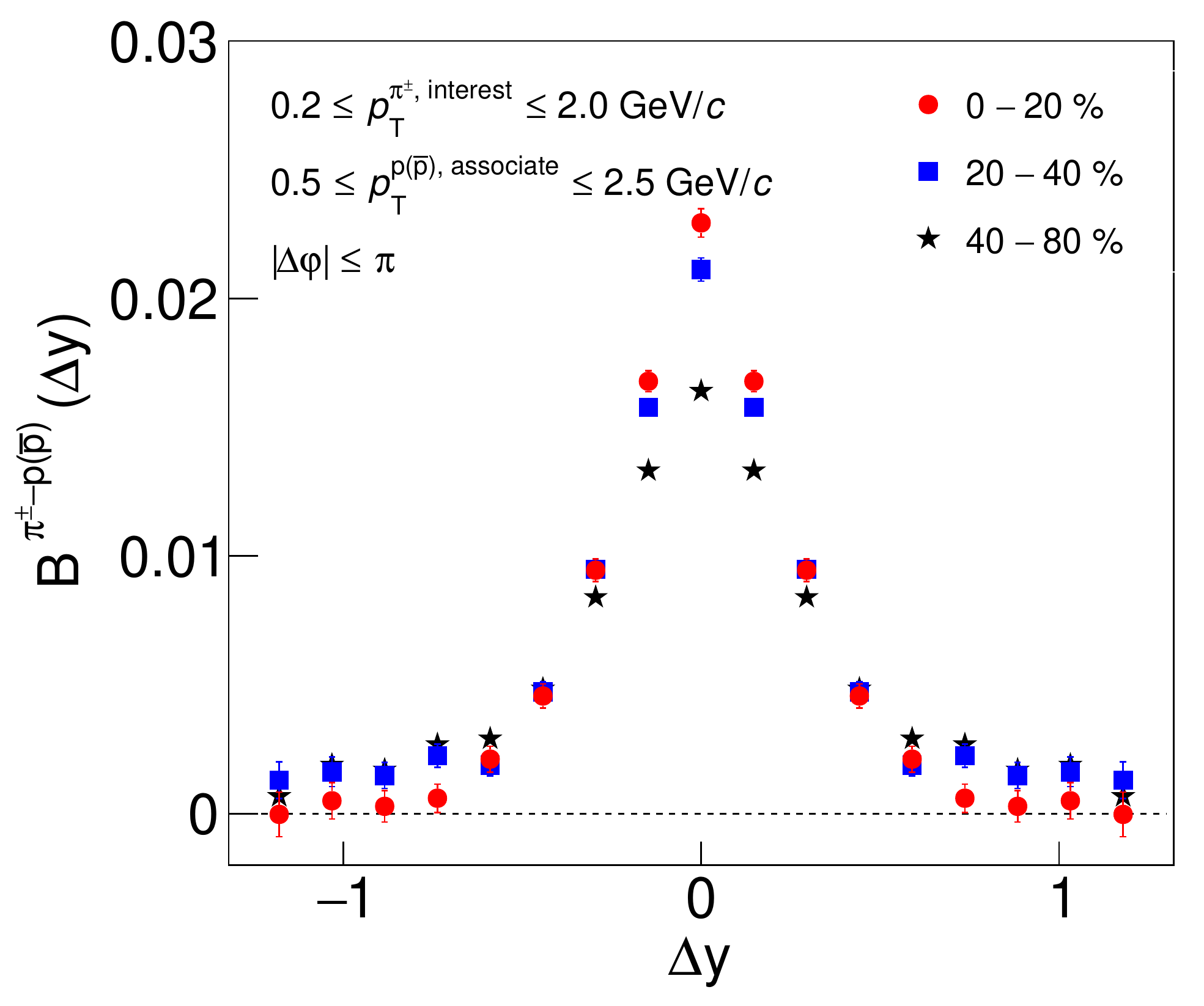}
  \includegraphics[width=0.32\linewidth]{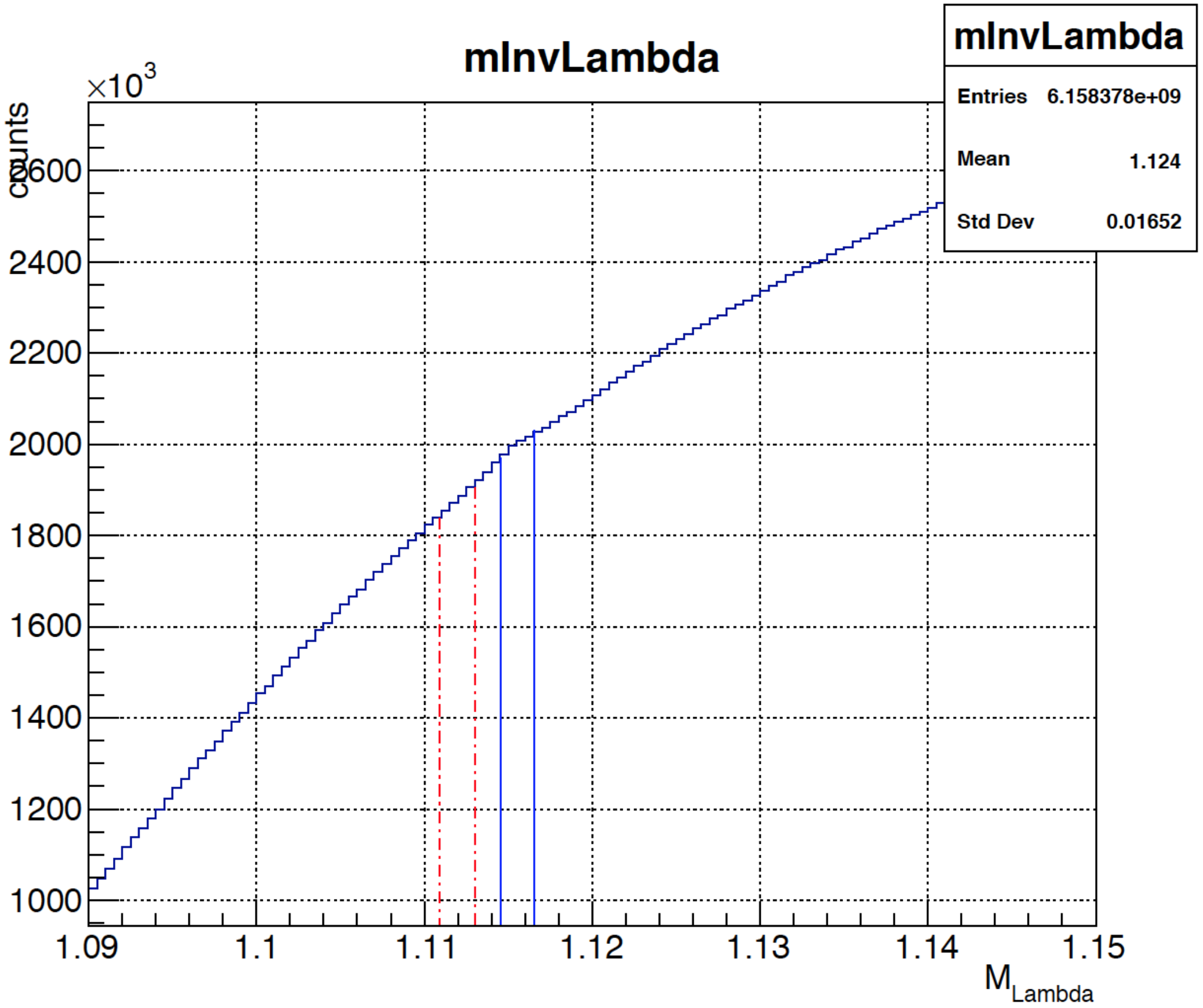}
  \includegraphics[width=0.32\linewidth]{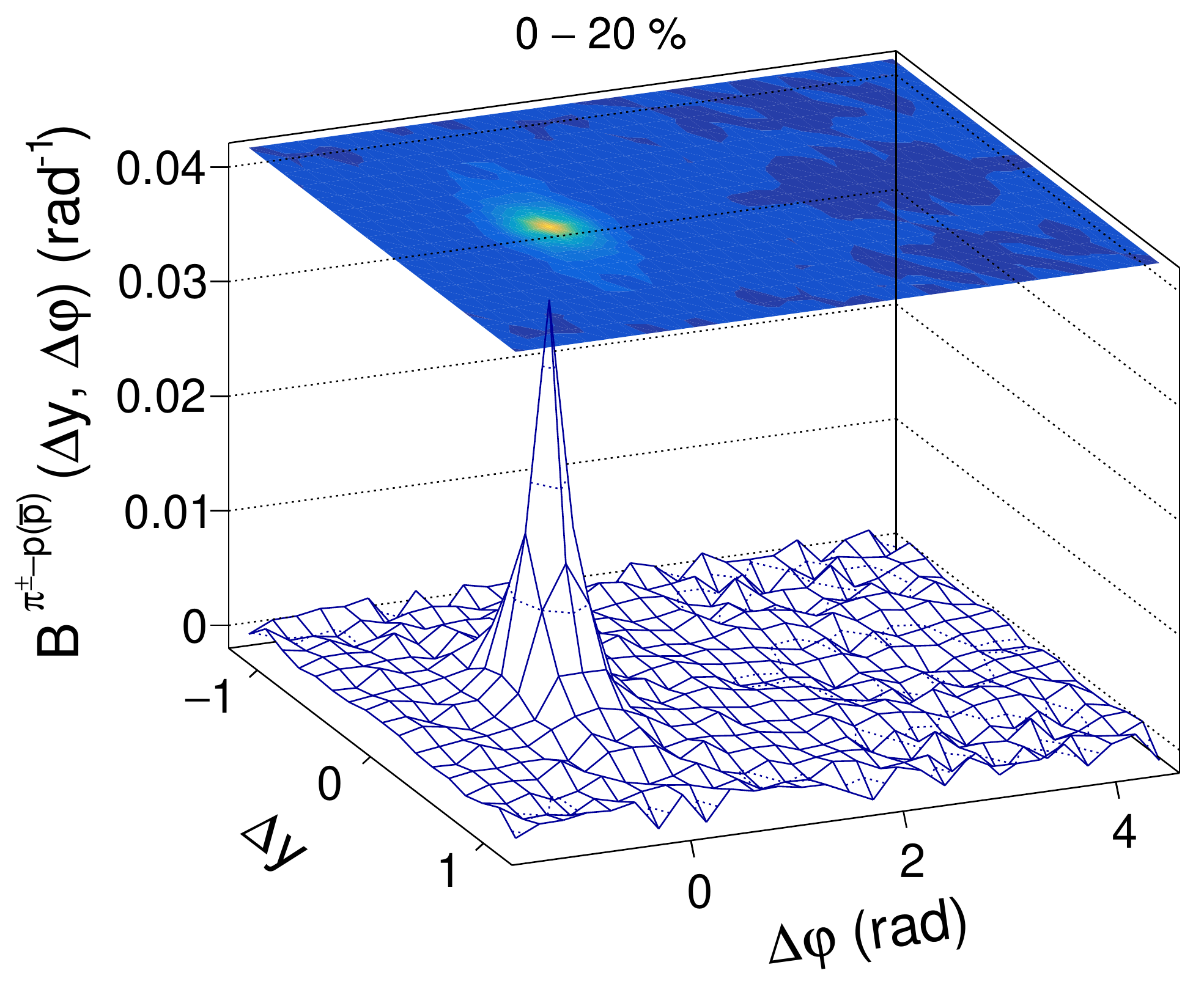}
  \includegraphics[width=0.32\linewidth]{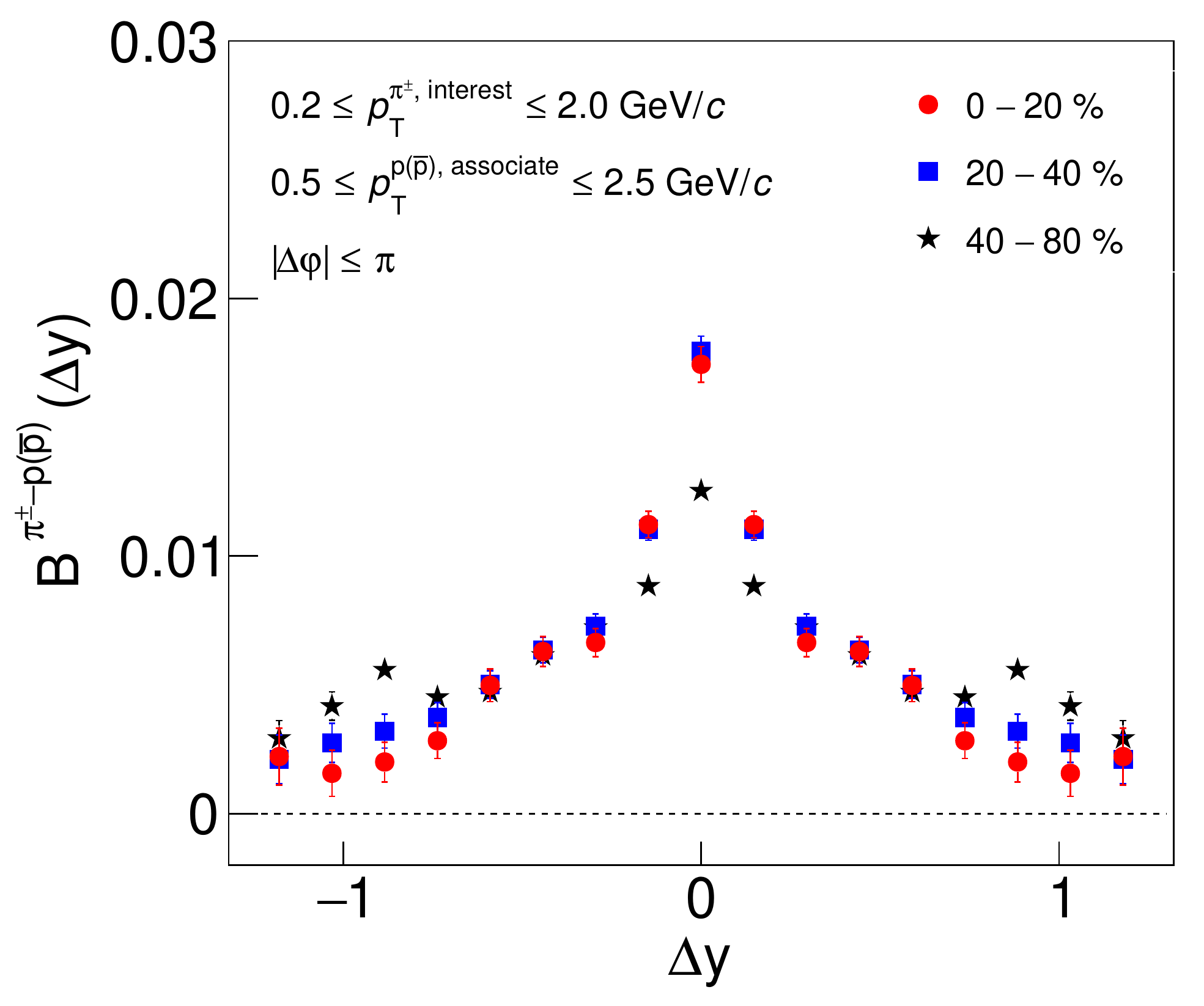} 
  \caption{Study of the impact of $\Lambda$ weak decay contamination in measurements of $B^{\pi p}$.
  The $\Lambda$ invariant mass of US pairs (left column), 2D $B^{\pi p}$ (middle column), and $\Delta y$ projections (right column) obtained with cuts $DCA_{xy}<2.4$ cm (top row) and $DCA_{xy}<0.04$ cm (bottom row) in 20-40\% centrality Pb--Pb collisions.} 
  \label{fig:BF_PionProton_DCAxy24_DCAxy004_2D_1D}
\end{figure}

The $\Lambda$/$\bar{\Lambda}$ decay daughter particles  \footnote{$\pi^{\pm}$ and $p/\bar{p}$,  ($\Lambda \rightarrow \pi^{-} + p$ and $\bar{\Lambda} \rightarrow \pi^{+} + \bar{p}$)} may cause feed down contamination in $B^{\pi p}$ and $B^{p\pi}$.
Figures~\ref{fig:Purity_DCAxy24_DCAxy004_Pion_0Cent20} and \ref{fig:Purity_DCAxy24_DCAxy004_Proton_0Cent20} show that a tight $DCA_{xy}<0.04$ cm cut does a much better job at removing the secondary particles from both weak decays and interaction of produced particles with detector materials than a wide $DCA_{xy}<2.4$ cm cut. 
Furthermore,  in Fig.~\ref{fig:BF_PionProton_DCAxy24_DCAxy004_2D_1D}, the $\Lambda$ invariant mass plots show that the $\Lambda$ signal is better suppressed by the tight $DCA_{xy}<0.04$ cm cut than the wide $DCA_{xy}<2.4$ cm cut.
%Fig.~\ref{fig:BF_PionProton_DCAxy24_DCAxy004_2D_1D} also shows the near-side peak of $B^{\pi p}$ is wider and taller for the wide $DCA_{xy}<2.4$ cm cut than the tight $DCA_{xy}<0.04$ cm cut.
Thus in this work, a tight $DCA_{xy}<0.04$ cm cut is used to reduce secondary $\pi^{\pm}$ and $p/\bar{p}$.
Note that a similar tight $DCA_{xy}$ cut was also used to remove secondary particles in other published ALICE PID CF papers~\cite{Adam:2016iwf}.
After this tight $DCA_{xy}<0.04$ cm cut, we estimate that only about 1.4\% of $\pi^{\pm}$ and 4\% of $p/\bar{p}$ are from weak decays.
This should lead to only minor contamination in $B^{\pi p}$ and $B^{p\pi}$.

%Narrow $DCA_{xy}$ cut suppresses the secondary particles from weak decays and interaction with detector materials.

\subsection{Lambda Invariant Mass Check}
\label{subsubsec:LambdaInvariantMassCheck}

\begin{figure}
\centering
  \includegraphics[width=0.32\linewidth]{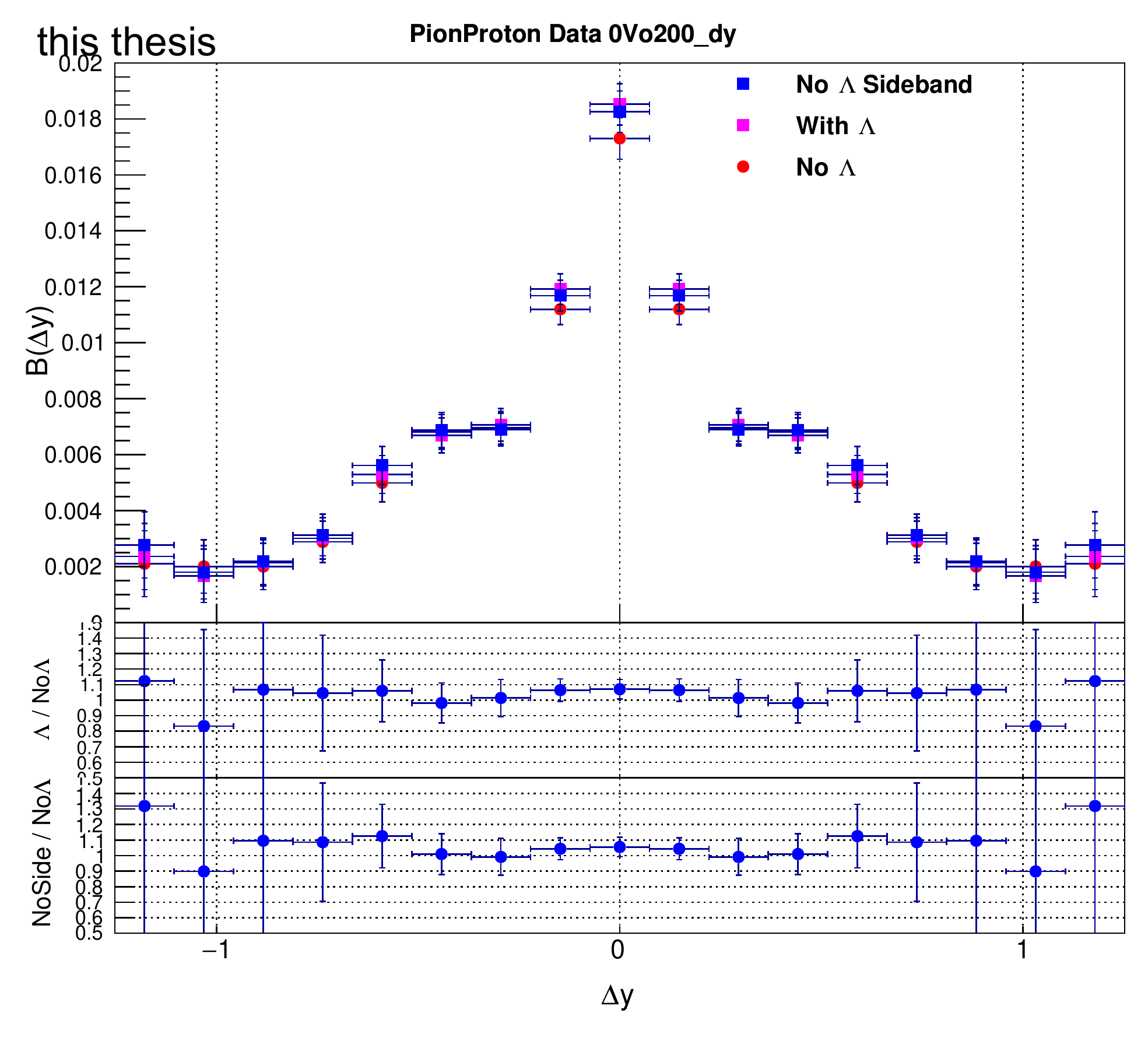}
  \includegraphics[width=0.32\linewidth]{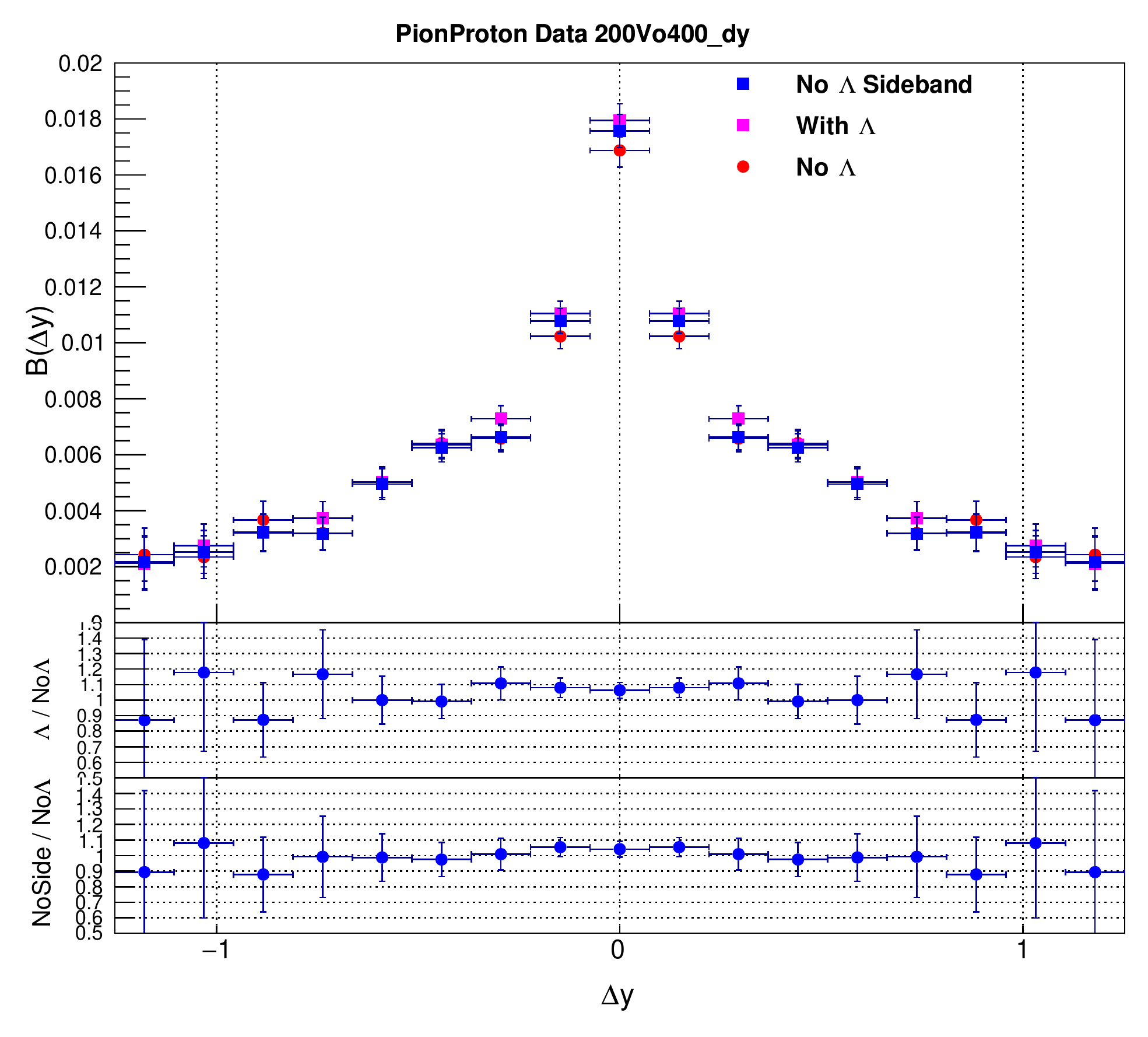}
  \includegraphics[width=0.32\linewidth]{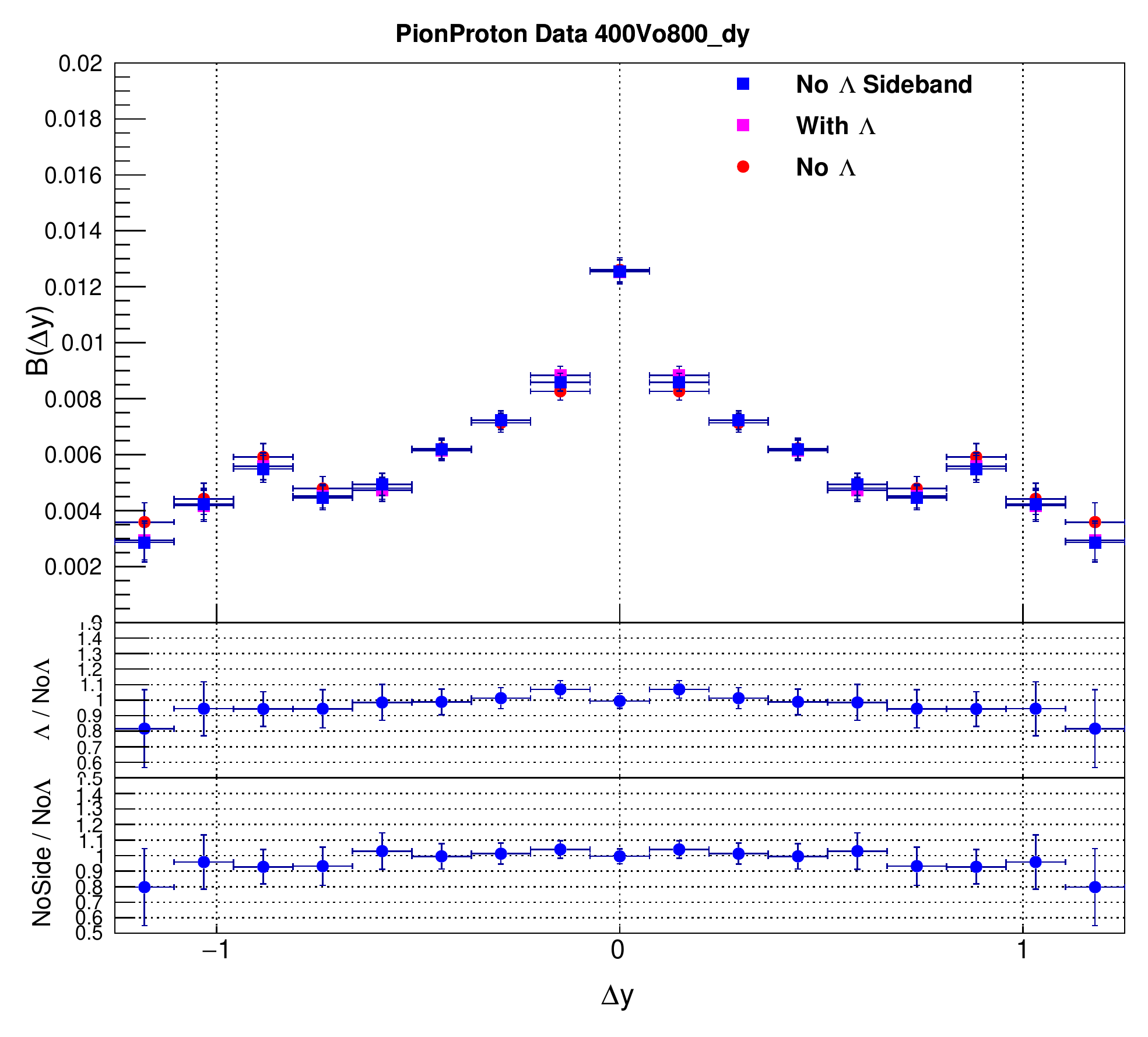}
  \includegraphics[width=0.32\linewidth]{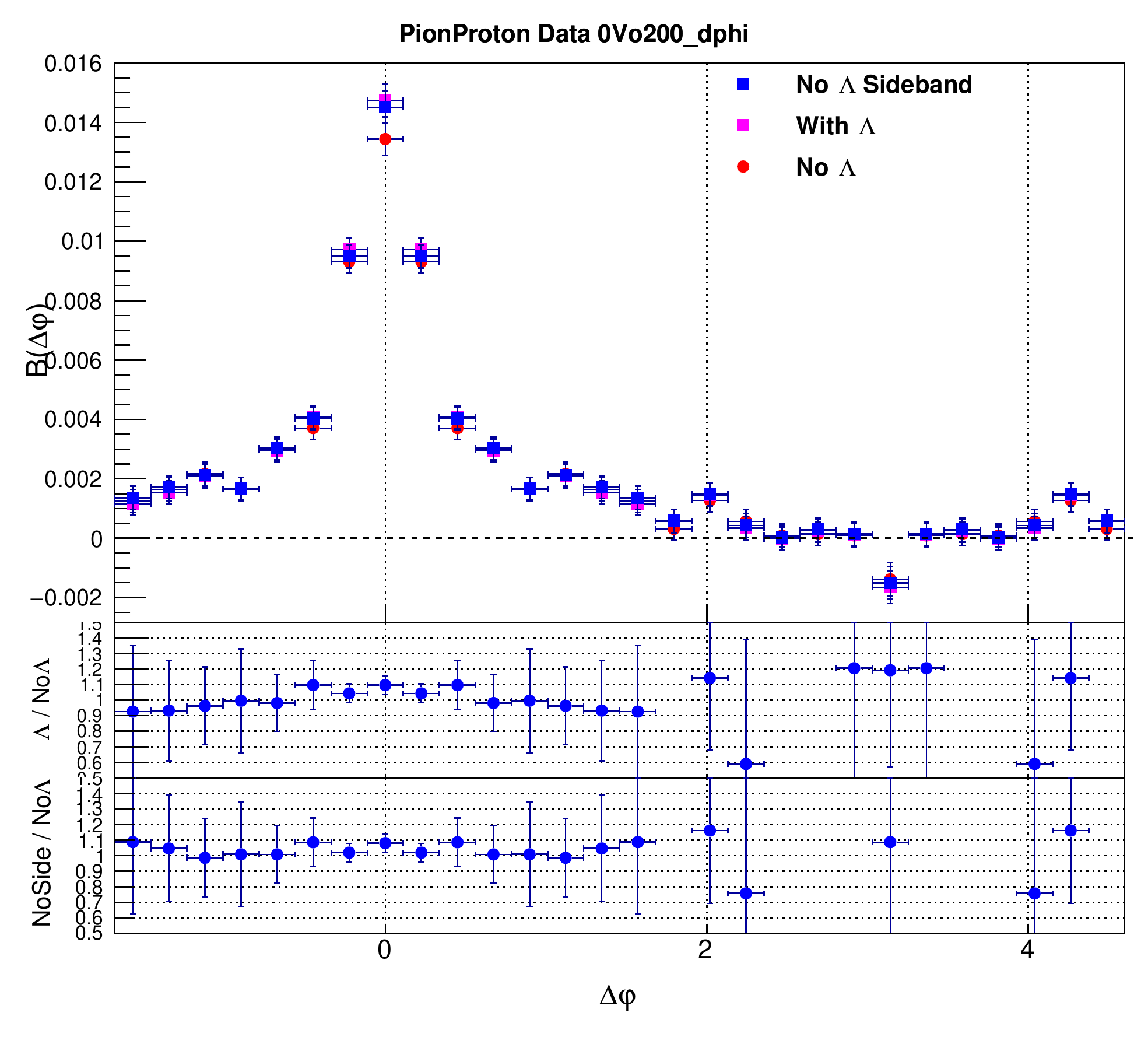}
  \includegraphics[width=0.32\linewidth]{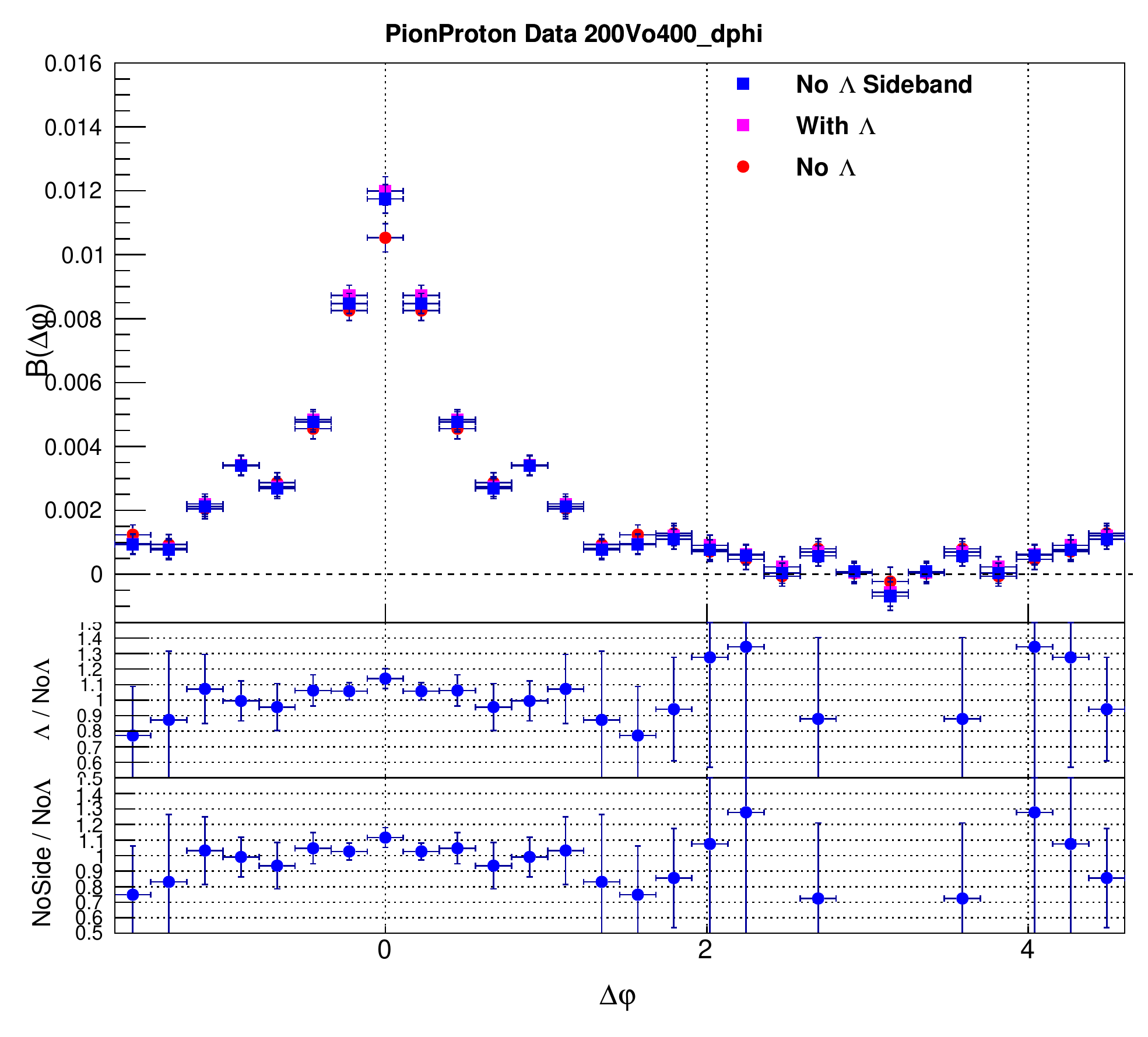}
  \includegraphics[width=0.32\linewidth]{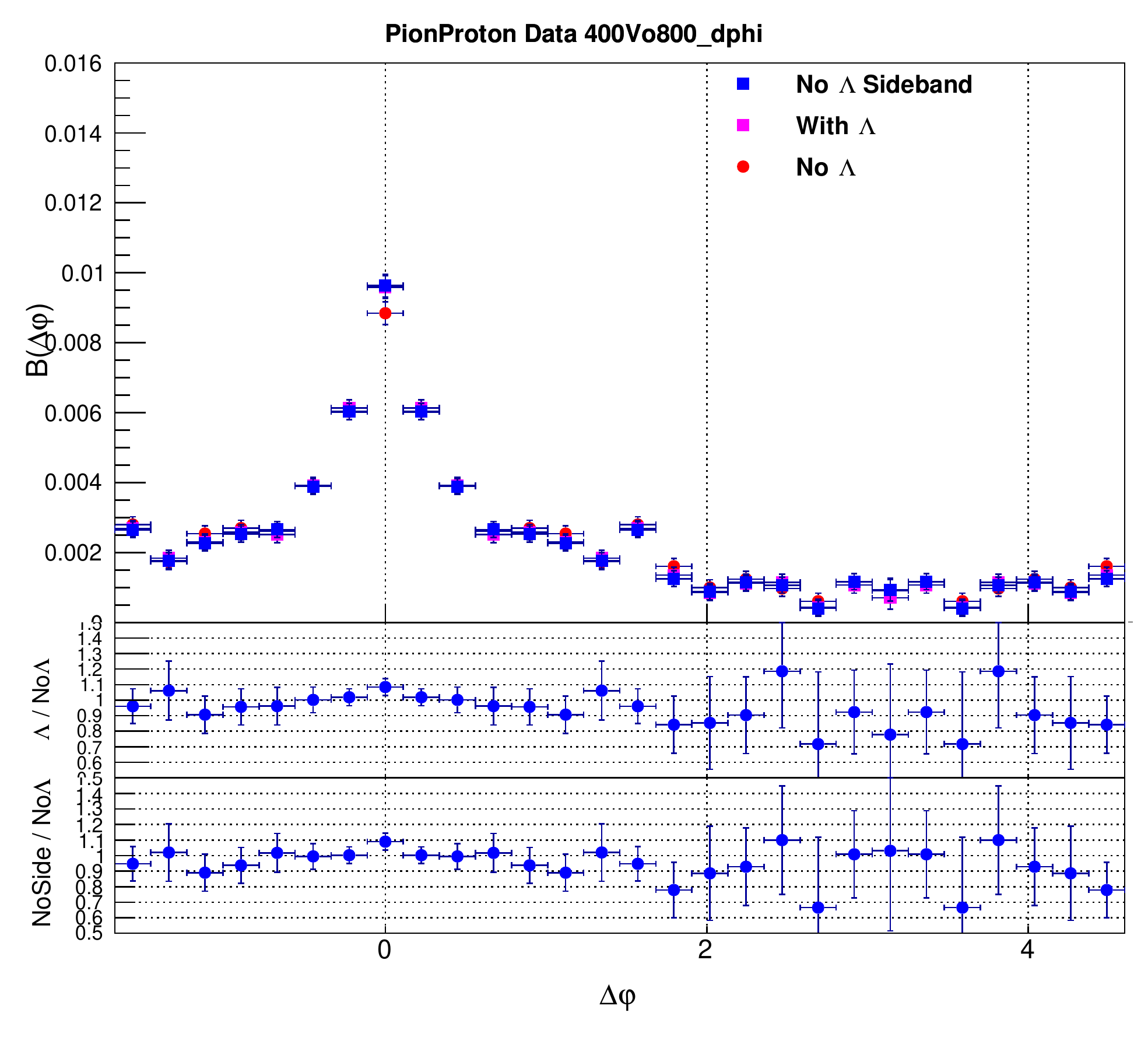}
  \includegraphics[width=0.32\linewidth]{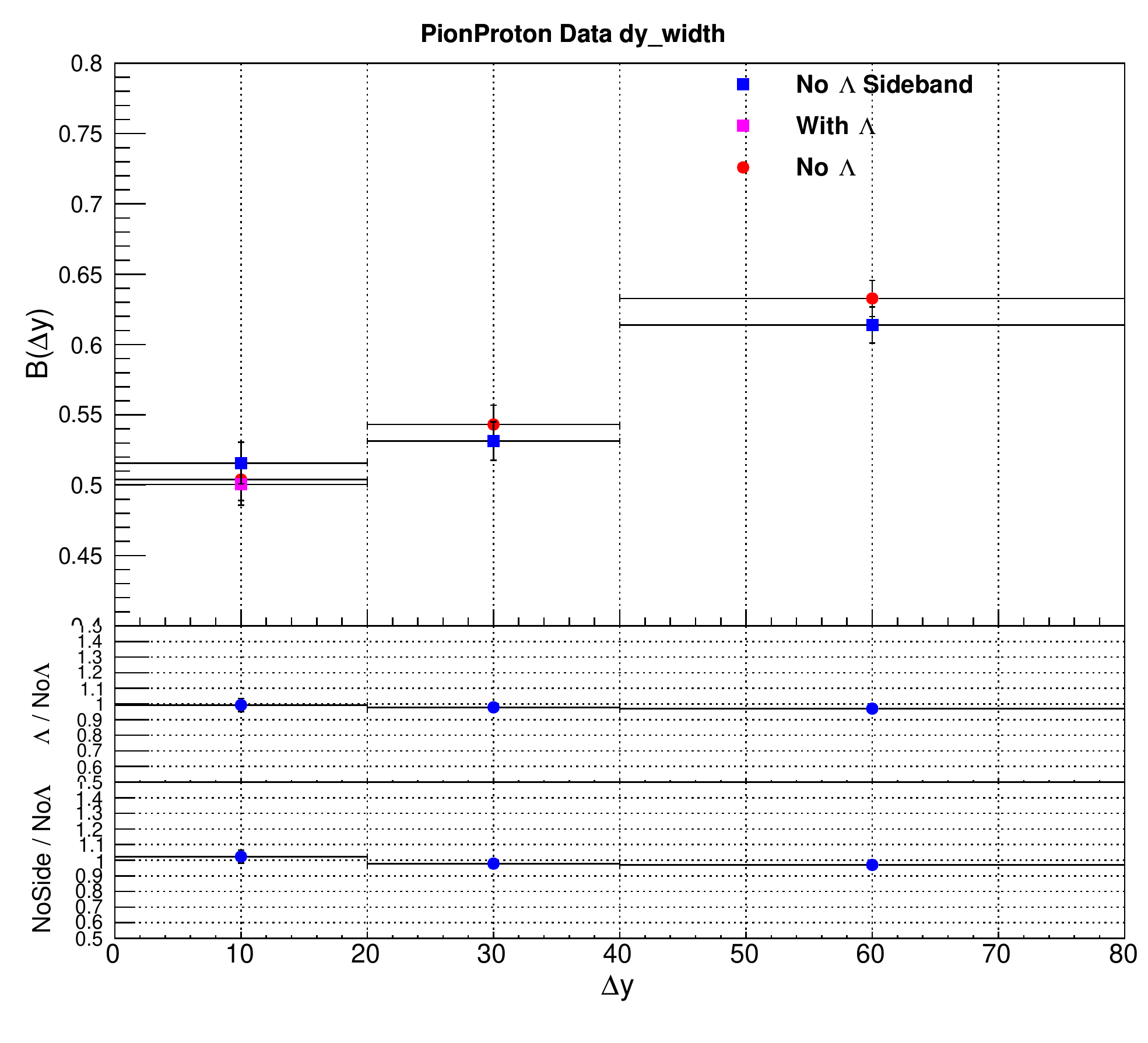}
  \includegraphics[width=0.32\linewidth]{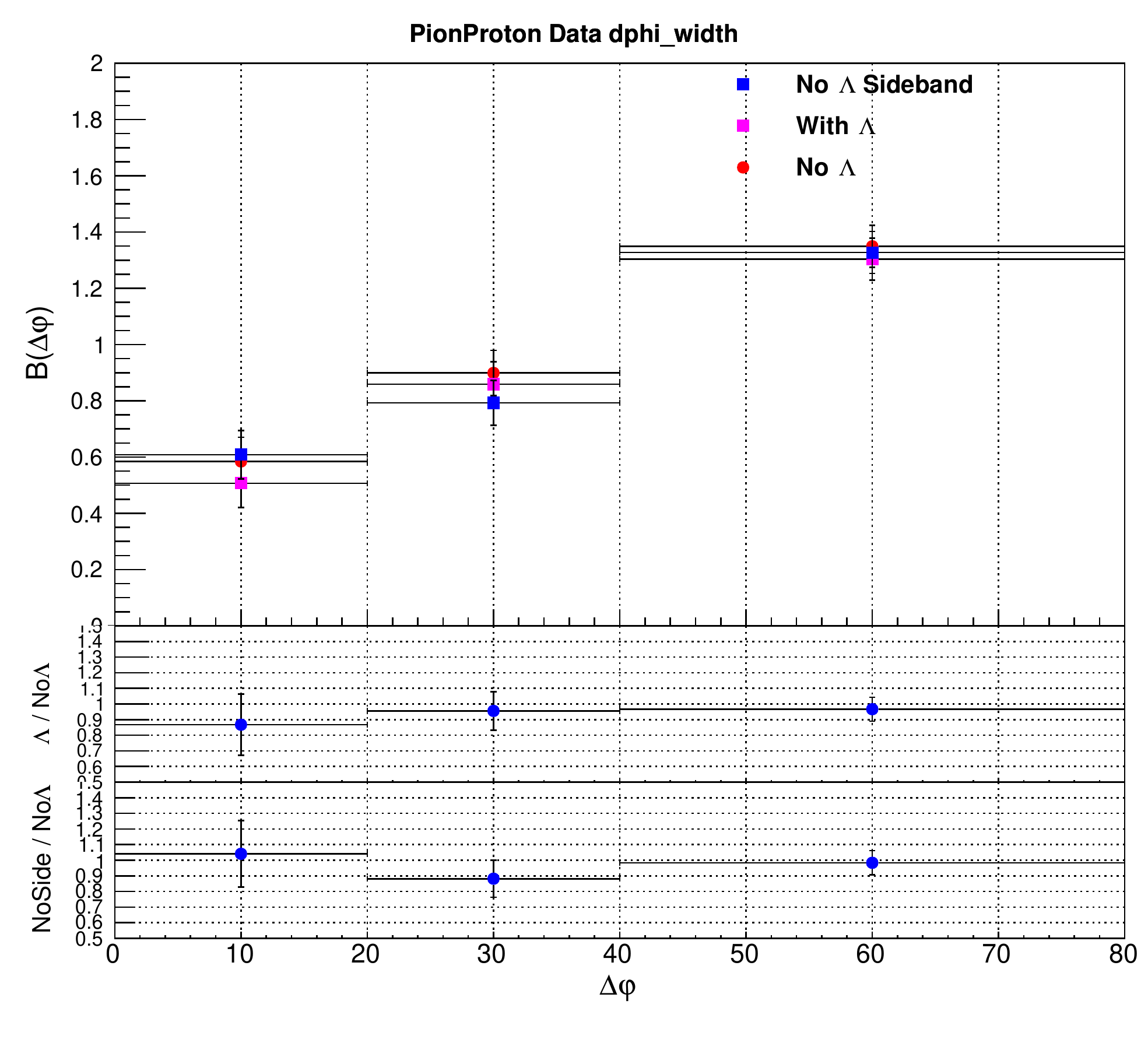}
  \includegraphics[width=0.32\linewidth]{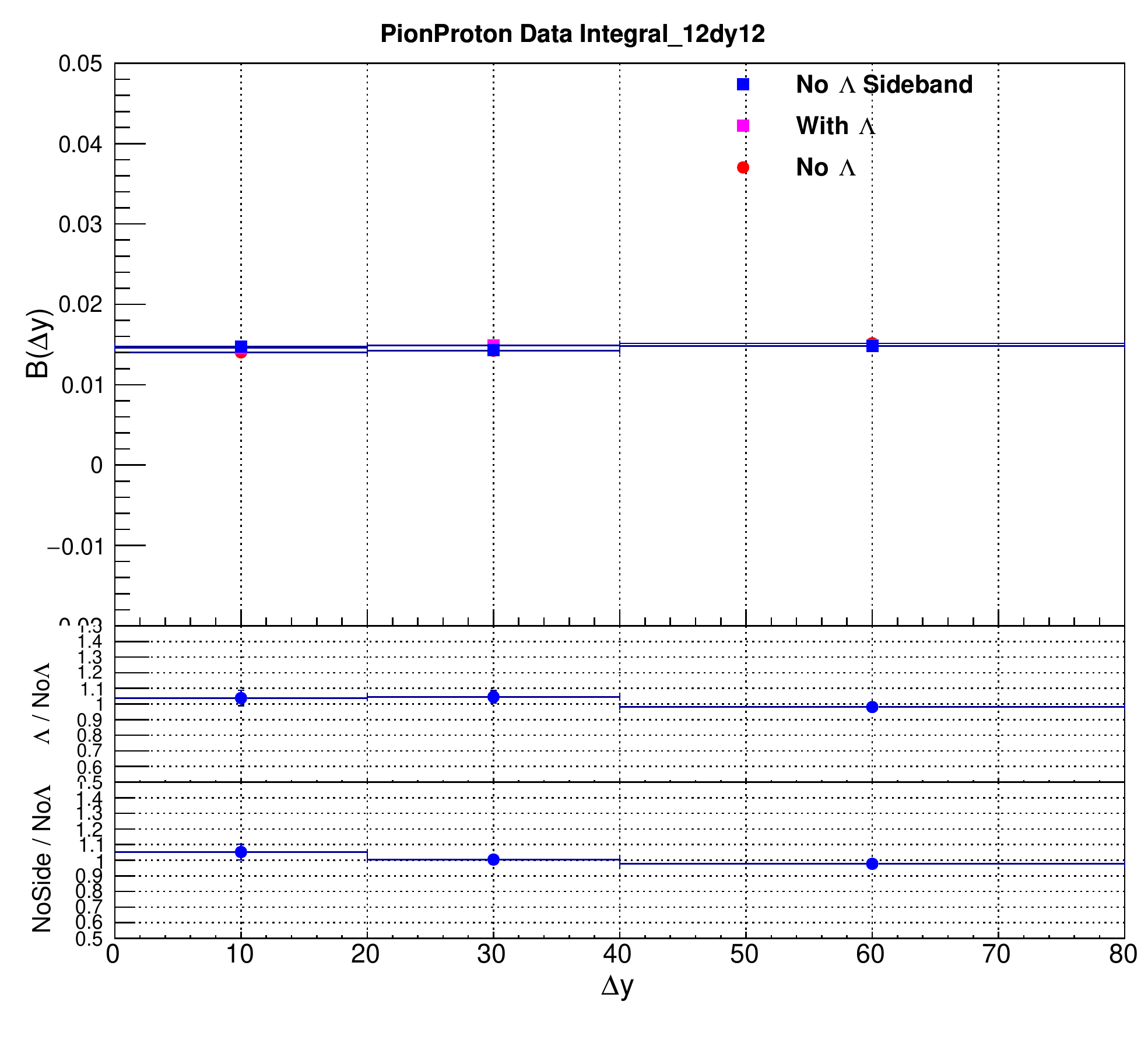} 
  \caption{Study of the impact of $\Lambda$ weak decay contamination in measurements of $B^{\pi p}$ with the tight $DCA_{xy}<0.04$ cm cut. 
  Comparisons on BF $\Delta y$ projections (top row), $\Delta\varphi$ projections of different centralities (middle row), and $\Delta y$ widths (lower left), $\Delta\varphi$ widths (lower middle), and BF integrals (lower right) between without and with a $1\sigma$ $\Lambda$ invariant mass cut ($1.114683<mInv_{\Lambda}<1.116683$), and with a $\Lambda$ invariant mass sideband cut ($1.1109<mInv_{\Lambda}<1.1131$). The $\Lambda$ invariant mass cut (corresponds to the blue lines in lower left plot in Fig.~\ref{fig:BF_PionProton_DCAxy24_DCAxy004_2D_1D}) and the sideband cut (corresponds to the red lines in lower left plot in Fig.~\ref{fig:BF_PionProton_DCAxy24_DCAxy004_2D_1D}) remove approximately same number of US pairs.}
  \label{fig:BF_PionProton_DCAxy004_dy_dphi_widths_integral}
\end{figure}

In addition, in order to make sure that contributions to $B^{\pi p}$ from $\Lambda$ decays (contamination) is negligible after 
 a tight $DCA_{xy}<0.04$ cm cut is applied, we performed a  $\Lambda$  invariant mass cut study.  
Figure~\ref{fig:BF_PionProton_DCAxy004_dy_dphi_widths_integral} presents $\Delta y$ and $\Delta \varphi$ projections of  $\pi p$ balance functions in 
three collision centrality ranges. BFs used to calculate these projections were obtained with three 
distinct sets of $\pi p$ pairs invariant mass criteria. The first set (labelled ``With $\Lambda$") is obtained with all  $\pi p$ pairs, i.e., without
selection based on the invariant mass of the pair. The second set (labelled No $\Lambda$) involves the use of a mass cut to eliminate all
particle pairs with $1\sigma$ $\Lambda$ invariant mass cut ($1.114683<mInv_{\Lambda}<1.116683$). 
The third set involves removal of pairs with a $\Lambda$ invariant mass sideband cut ($1.1109<mInv_{\Lambda}<1.1131$). This cut is selected to remove approximately 
the same number of particle pairs but is off the $\Lambda$ mass peak. It is thus possible to compare balance function obtain with all pairs, pairs that exclude the $\Lambda$ mass region, and a set of pairs from which $\Lambda$ are not removed but an equivalent number of pairs is. The top and middle rows of Fig.~\ref{fig:BF_PionProton_DCAxy004_dy_dphi_widths_integral} show projections along $\Delta y$ and $\Delta \varphi$, respectively, as well 
as ratios of these projections. One notes that the amplitude and shape of the three projections are nearly identical, thereby confirming that
the explicit removal of pairs with a  $\Lambda$ mass has a very small impact on the balance functions and their projections. Furthermore, the bottom row of the figure displays  a comparison of the $\Delta y$ and $\Delta \varphi$ rms widths and the integral of the balance functions vs. the Pb--Pb collision centrality. One finds that the widths and BF integrals obtained with and without mass cuts
are essentially identical. One then concludes from these comparisons that the residual contamination of $\Lambda$ decays into the $B^{\pi p}$ balance function reported in this work is essentially negligible.

%\subsection{Compare with Delta Baryon Decay}
%\label{subsubsec:DeltaBaryonDecay}

\clearpage

\section{Study of BF with Other Filter-bits}
\label{sec:OtherFilterBits}

%The detailed specifics of different filterbits are listed in Fig.~\ref{fig:FilterbitsTable}.

The final BF results reported in this work are obtained with TPC only tracks corresponding to filter-bit 1, as described in Sec.~\ref{subsec:EventTrackSelection}.
However, we have also studied BFs obtained with global tracks with tight DCA cut, corresponding to filter-bit 96.

\begin{figure}
\centering
  \includegraphics[width=0.32\linewidth]{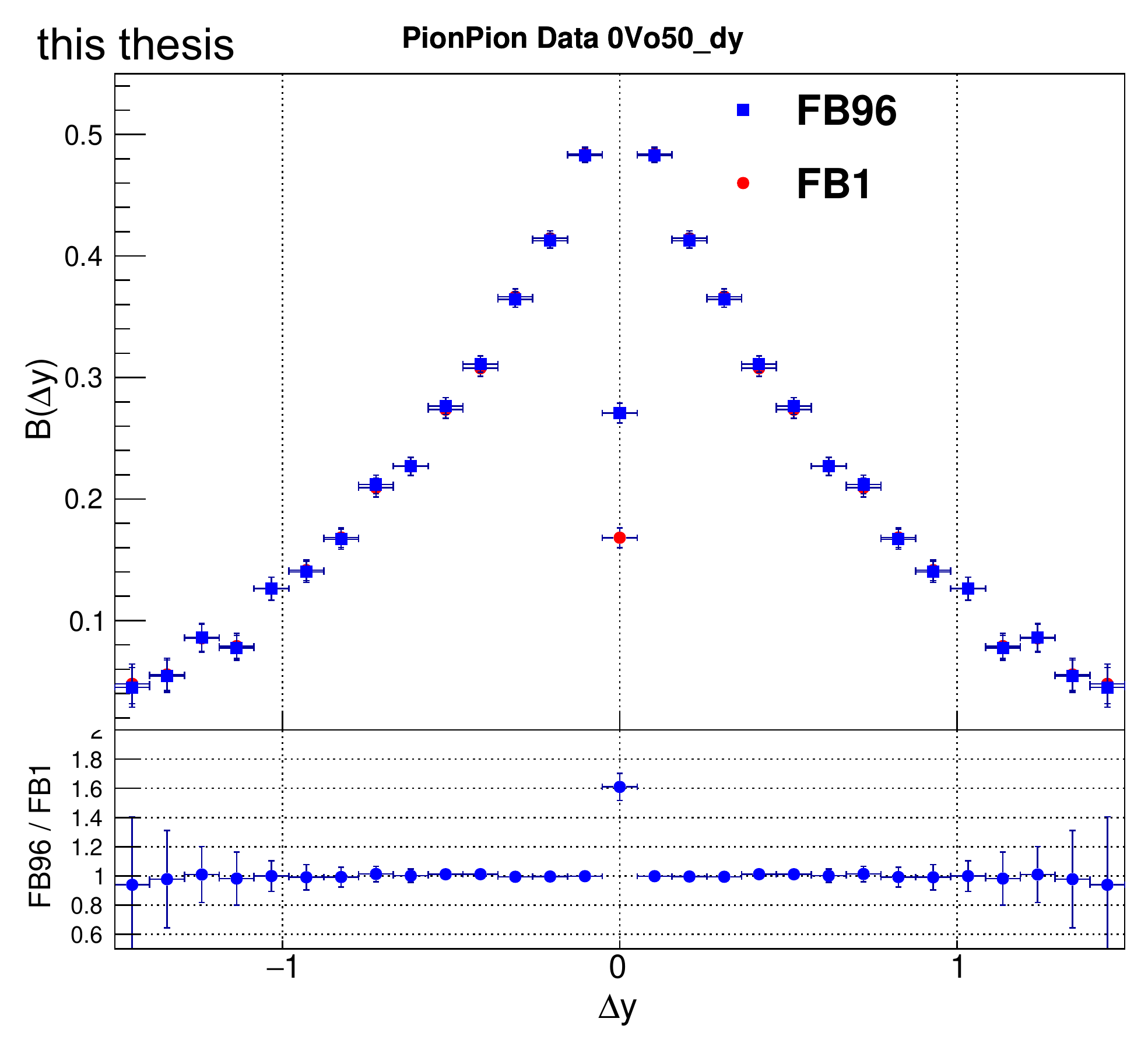}
  \includegraphics[width=0.32\linewidth]{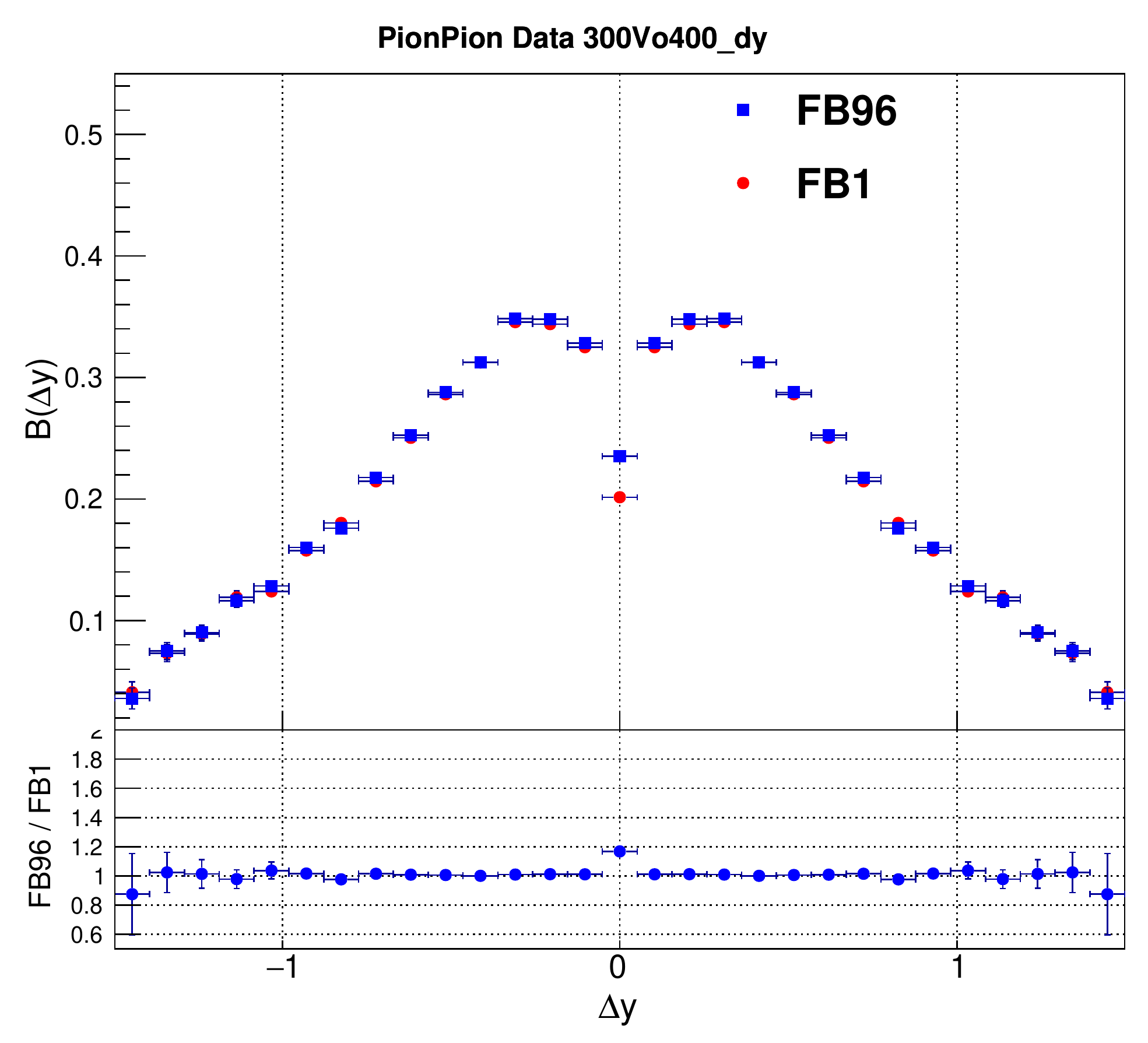}
  \includegraphics[width=0.32\linewidth]{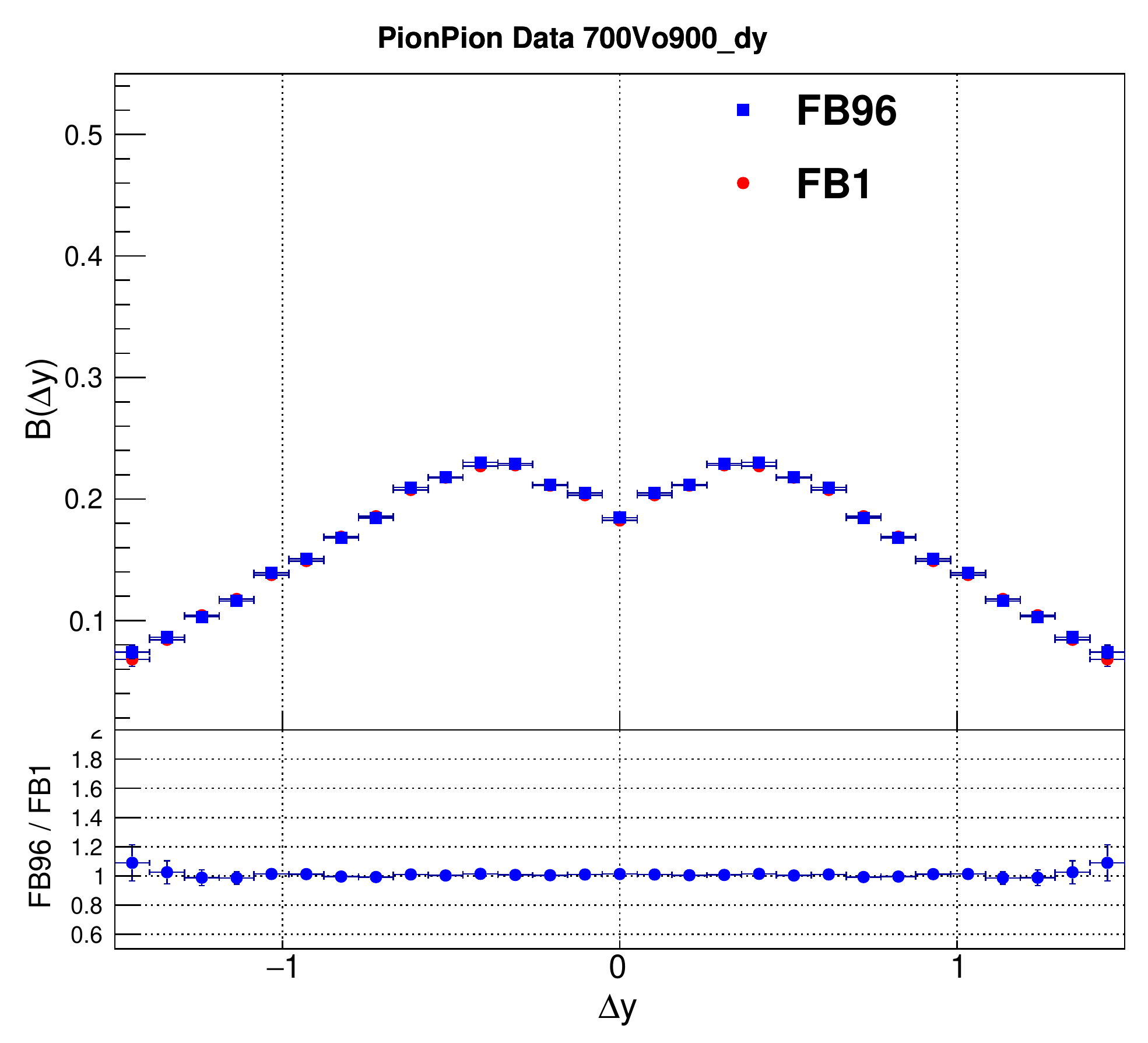}
  \includegraphics[width=0.32\linewidth]{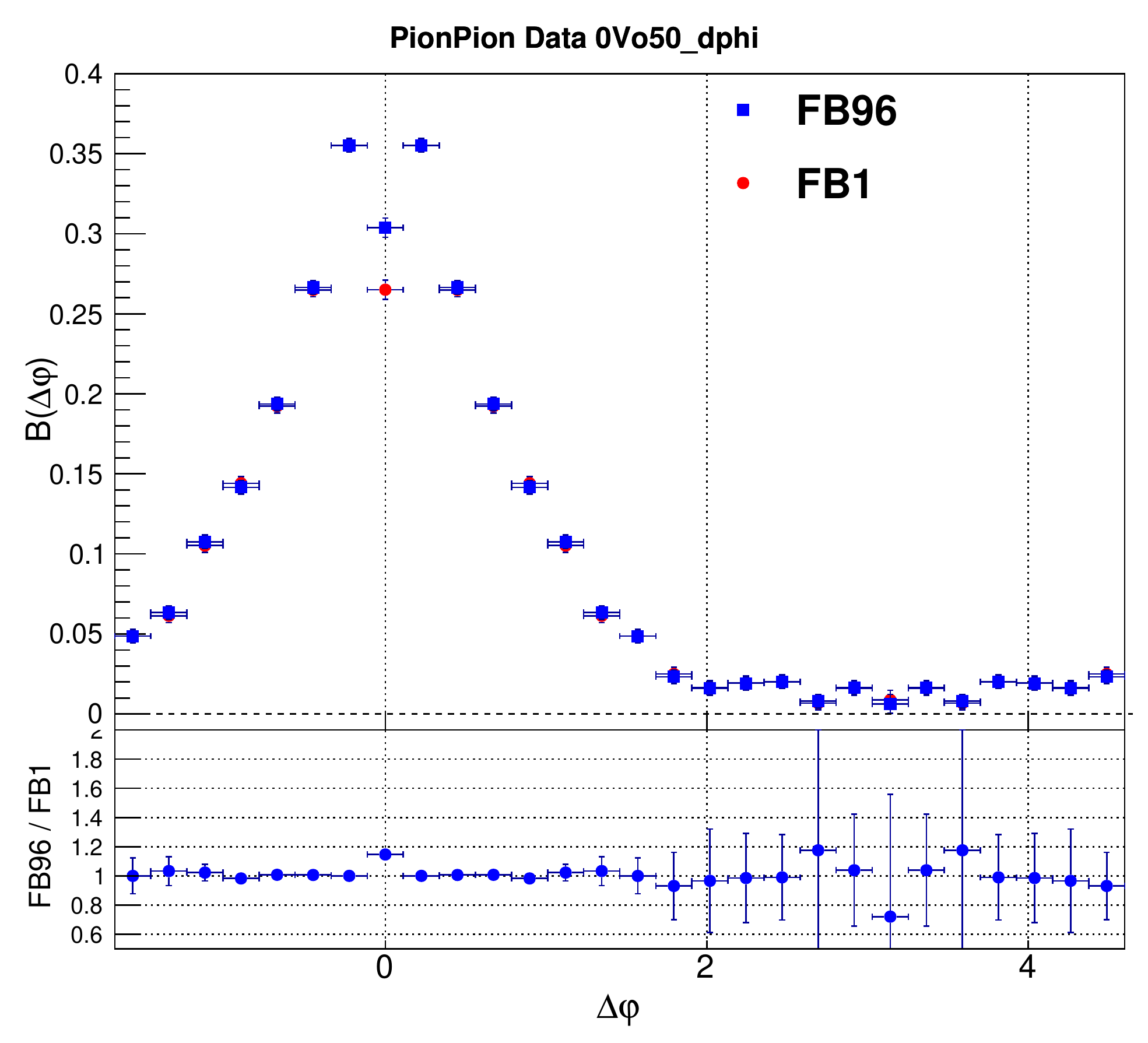}
  \includegraphics[width=0.32\linewidth]{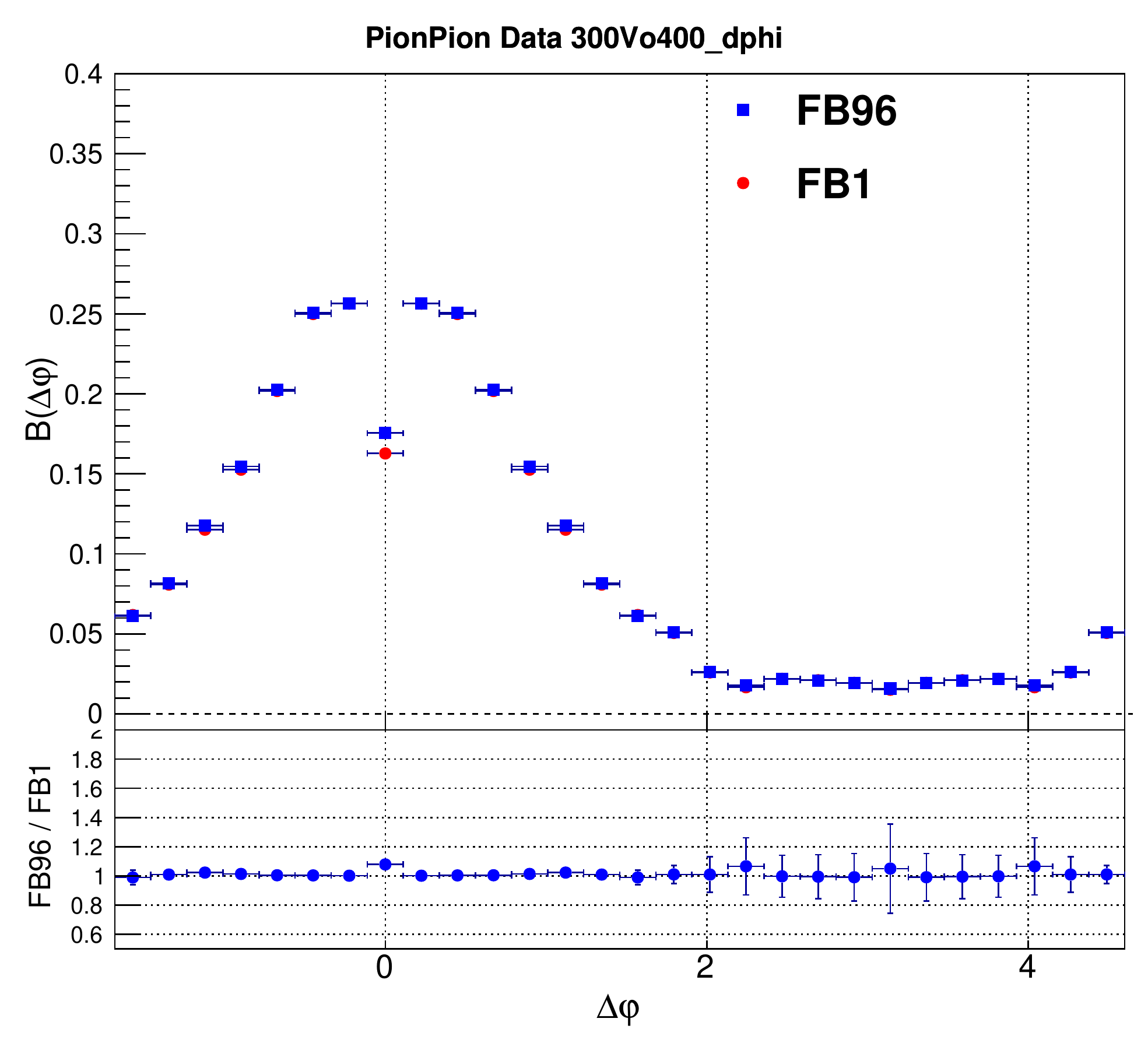}
  \includegraphics[width=0.32\linewidth]{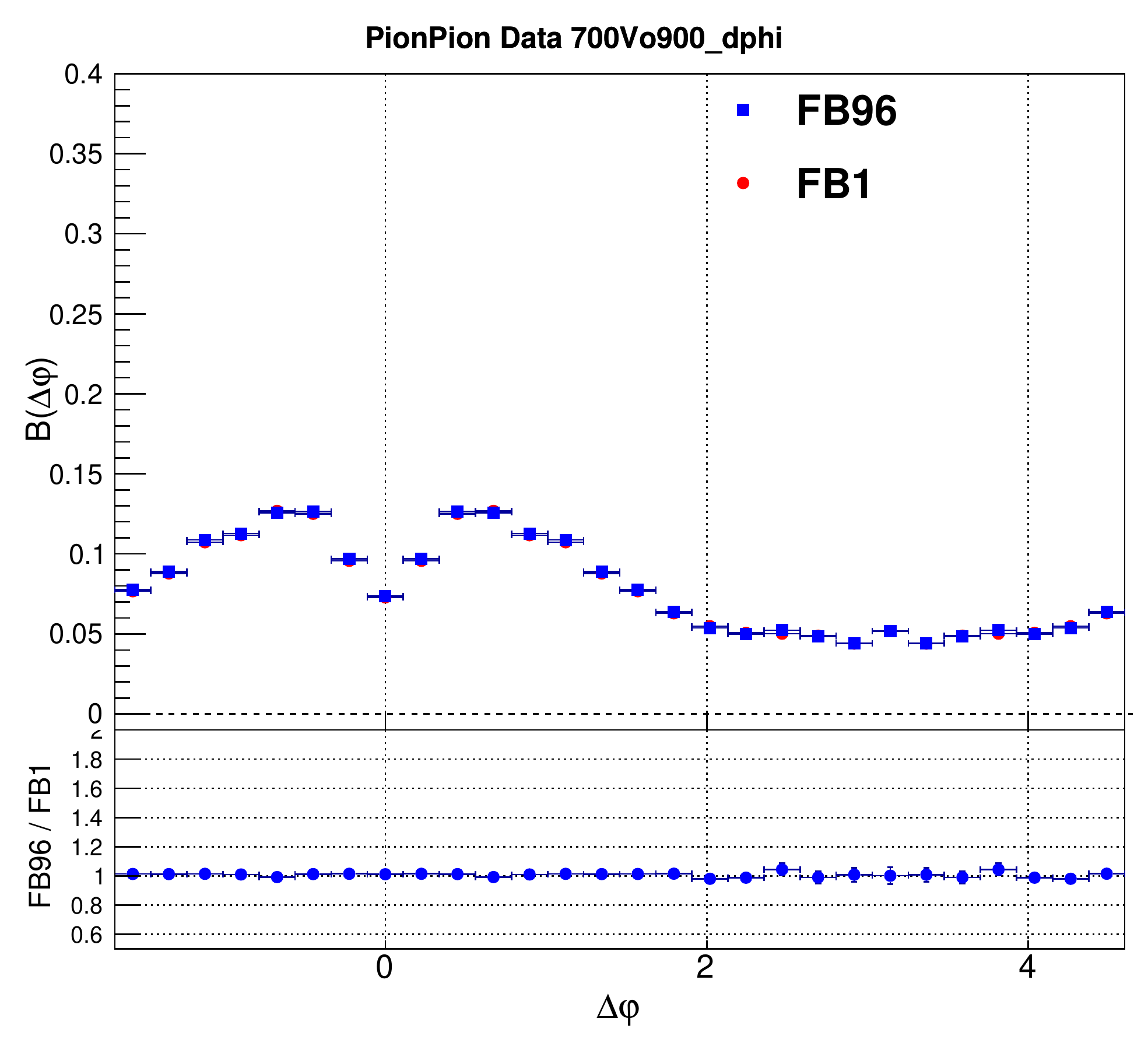}
  \includegraphics[width=0.32\linewidth]{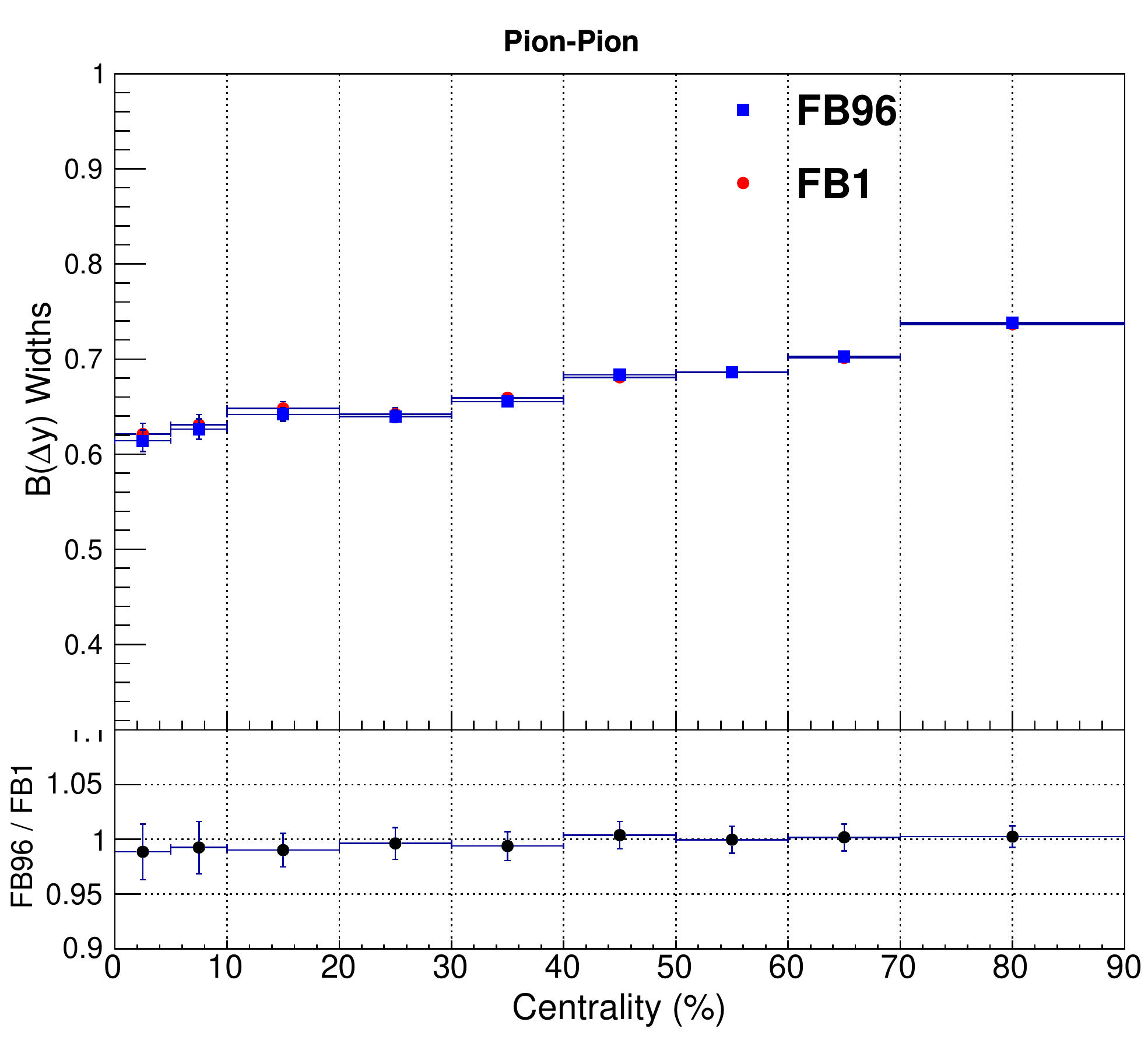}
  \includegraphics[width=0.32\linewidth]{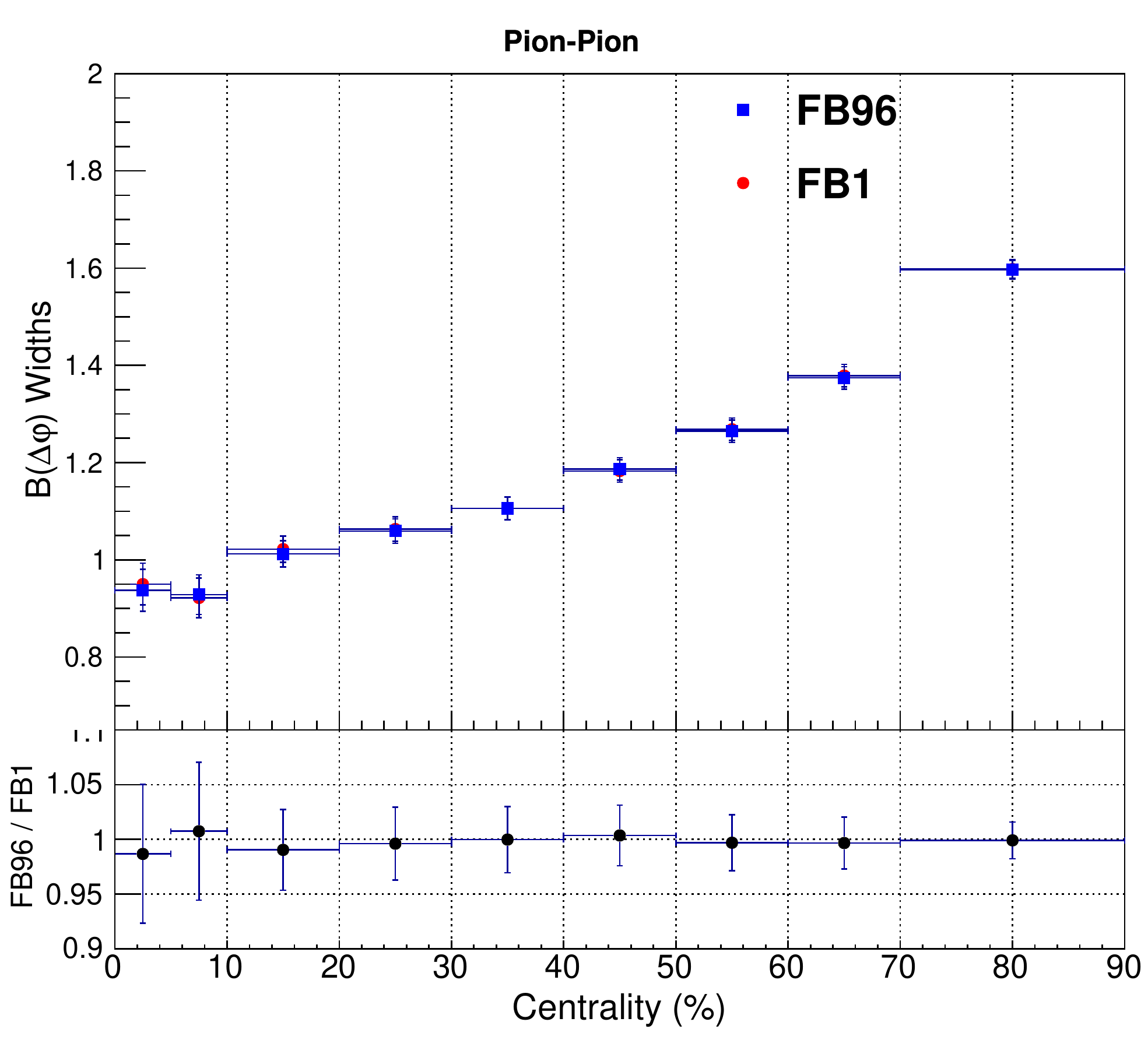}
  \includegraphics[width=0.32\linewidth]{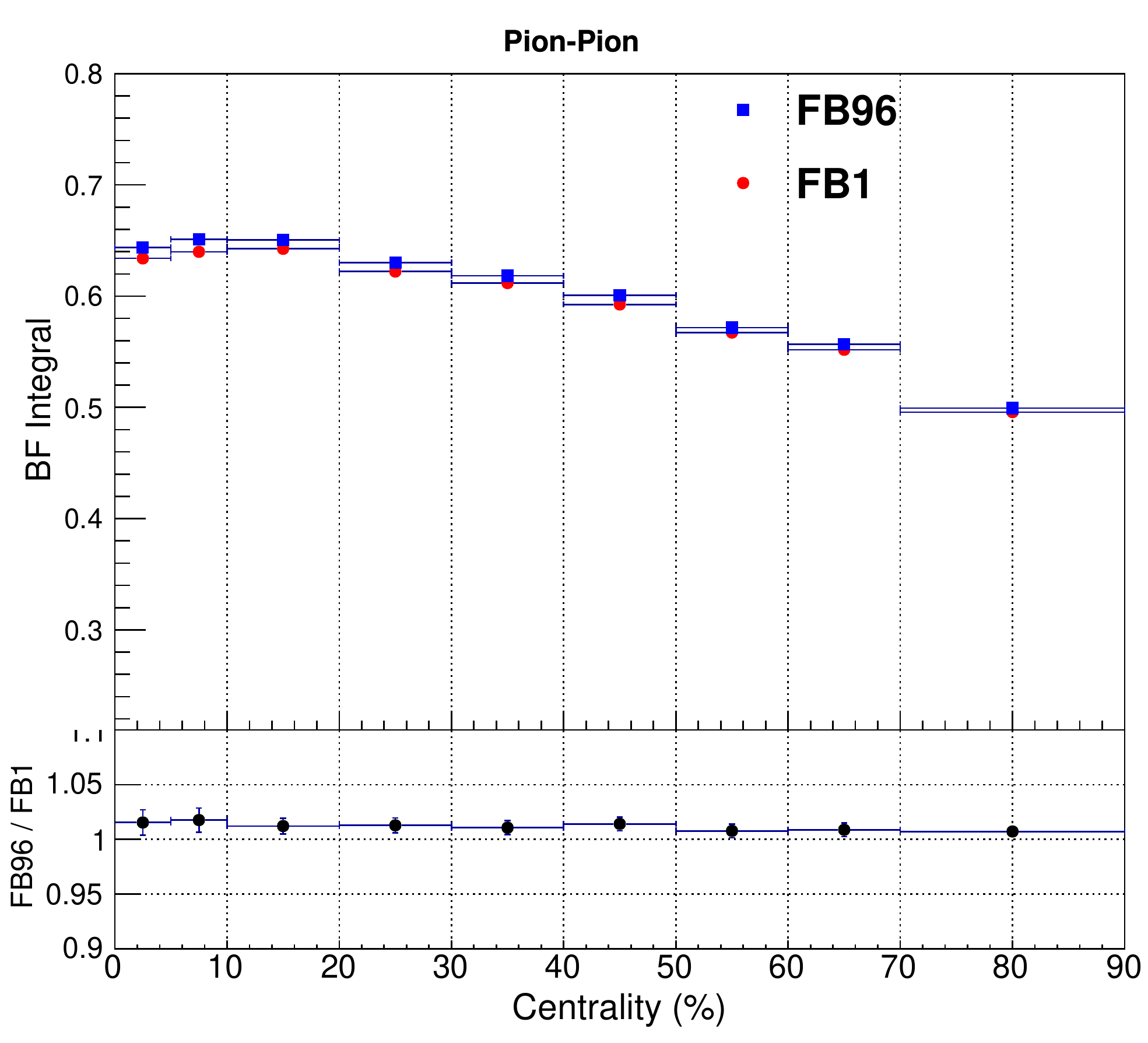}
  \caption{Comparisons of $B^{\pi\pi}$ projections onto $\Delta y$ (top row) and $\Delta\varphi$ (middle row) axis for selected collision centralities.
  Bottom row: comparisons of $B^{\pi\pi}$ $\Delta y$ widths (left), $\Delta\varphi$ widths (middle), and integrals (right) as a function of collision centrality between filter-bit 1 and 96.}
  \label{fig:BF_PionPion_FB1_FB96_Comparison}
\end{figure}

\begin{figure}
\centering
  \includegraphics[width=0.32\linewidth]{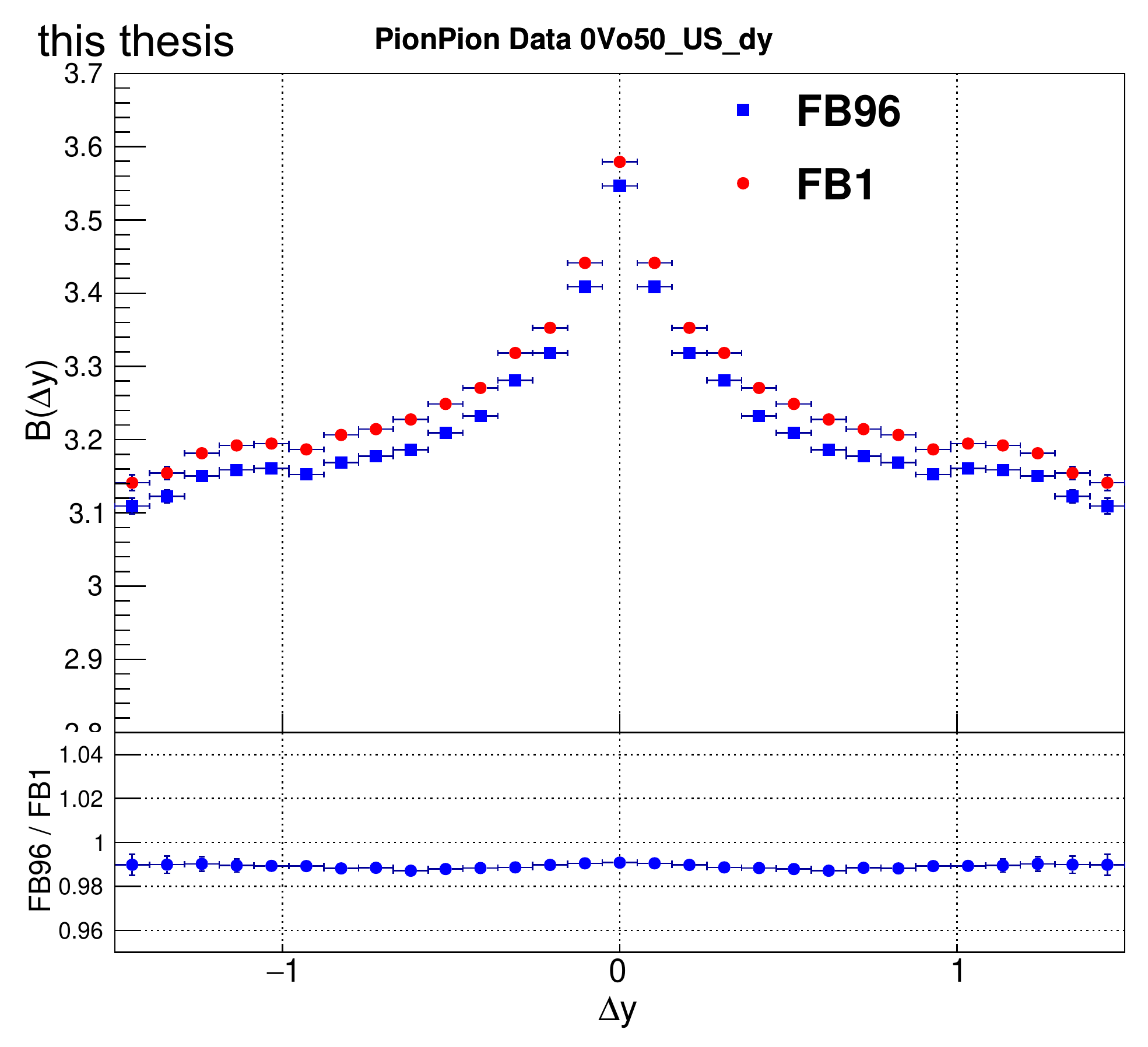}
  \includegraphics[width=0.32\linewidth]{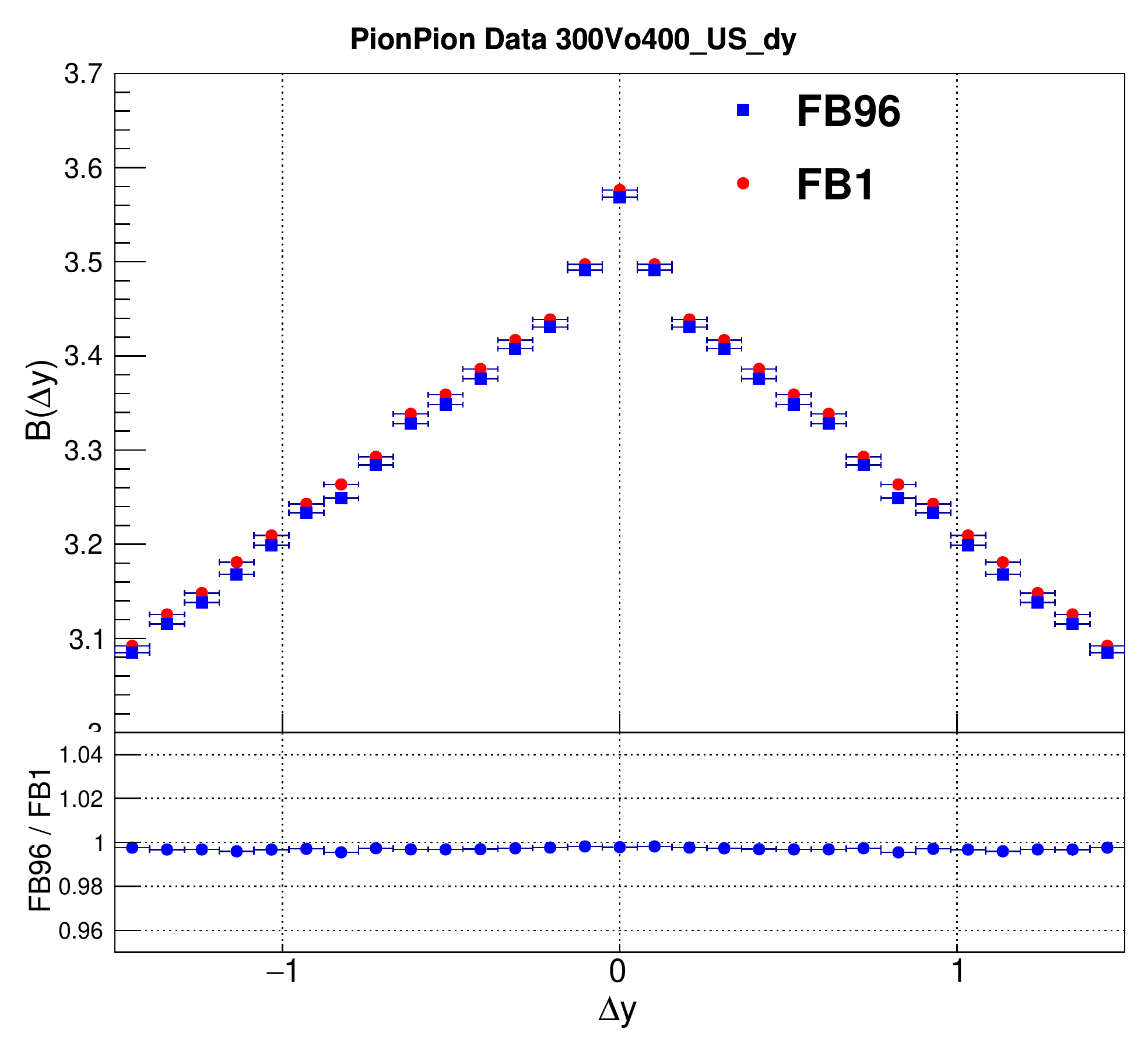}
  \includegraphics[width=0.32\linewidth]{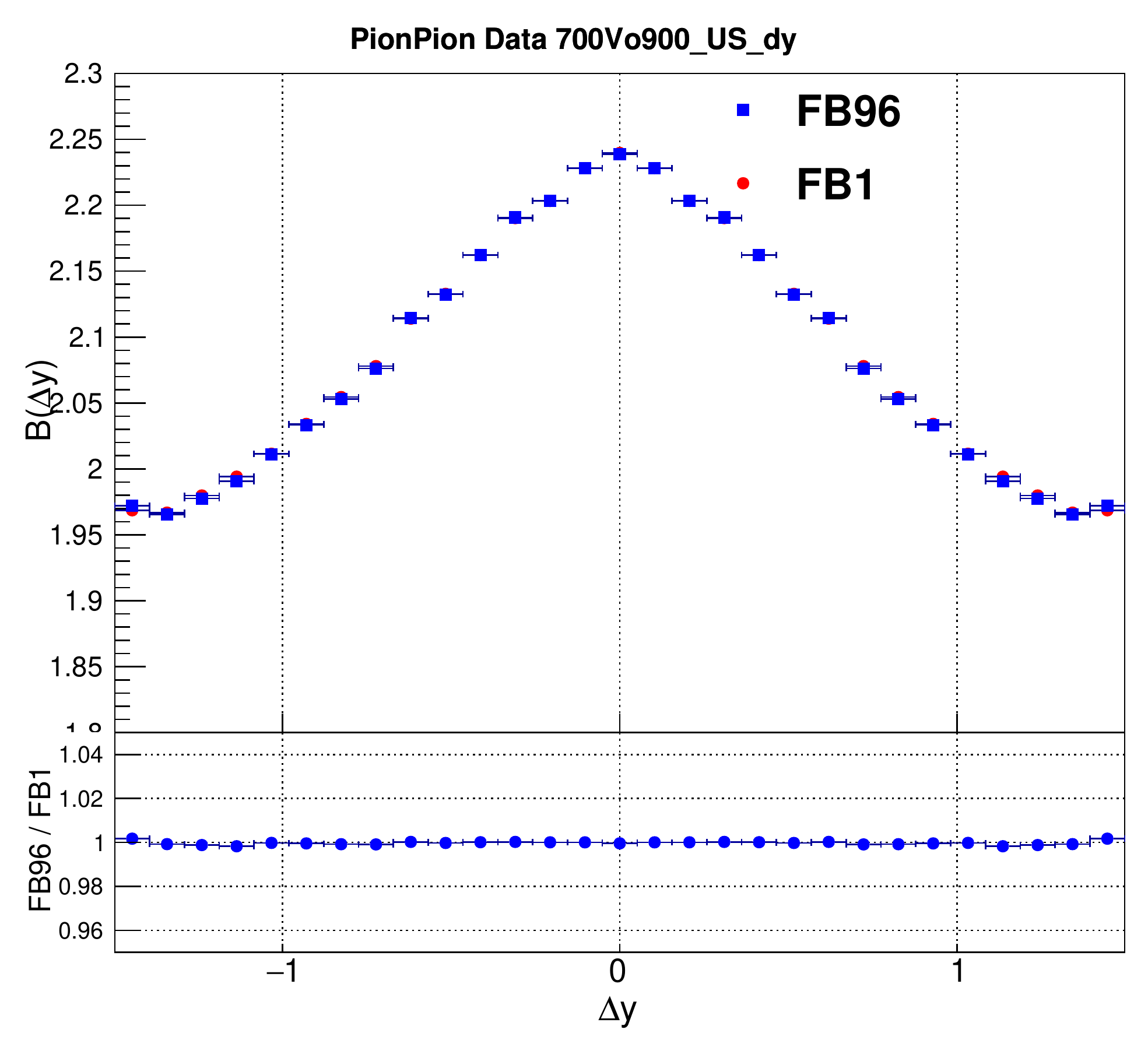}
  \includegraphics[width=0.32\linewidth]{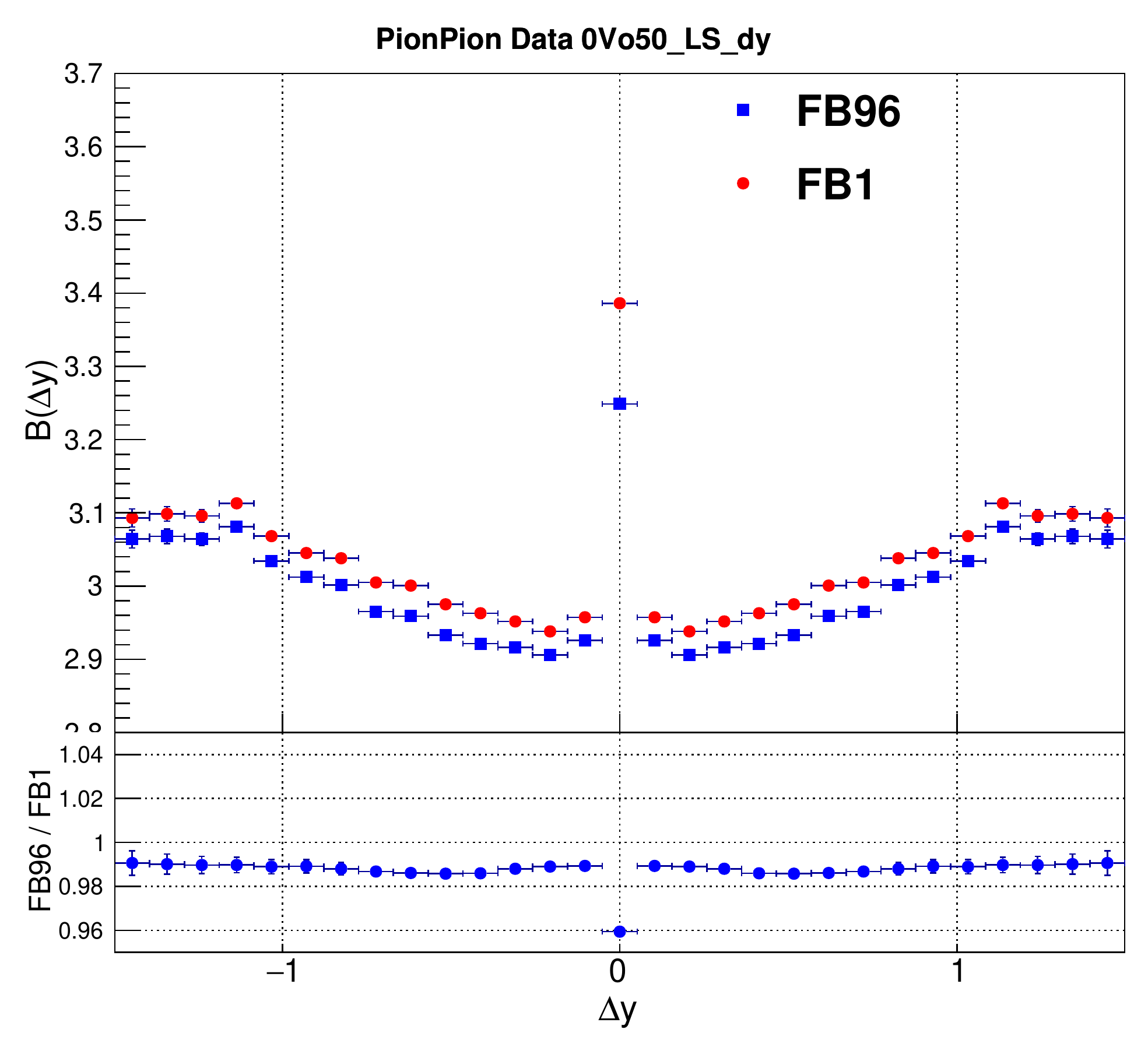}
  \includegraphics[width=0.32\linewidth]{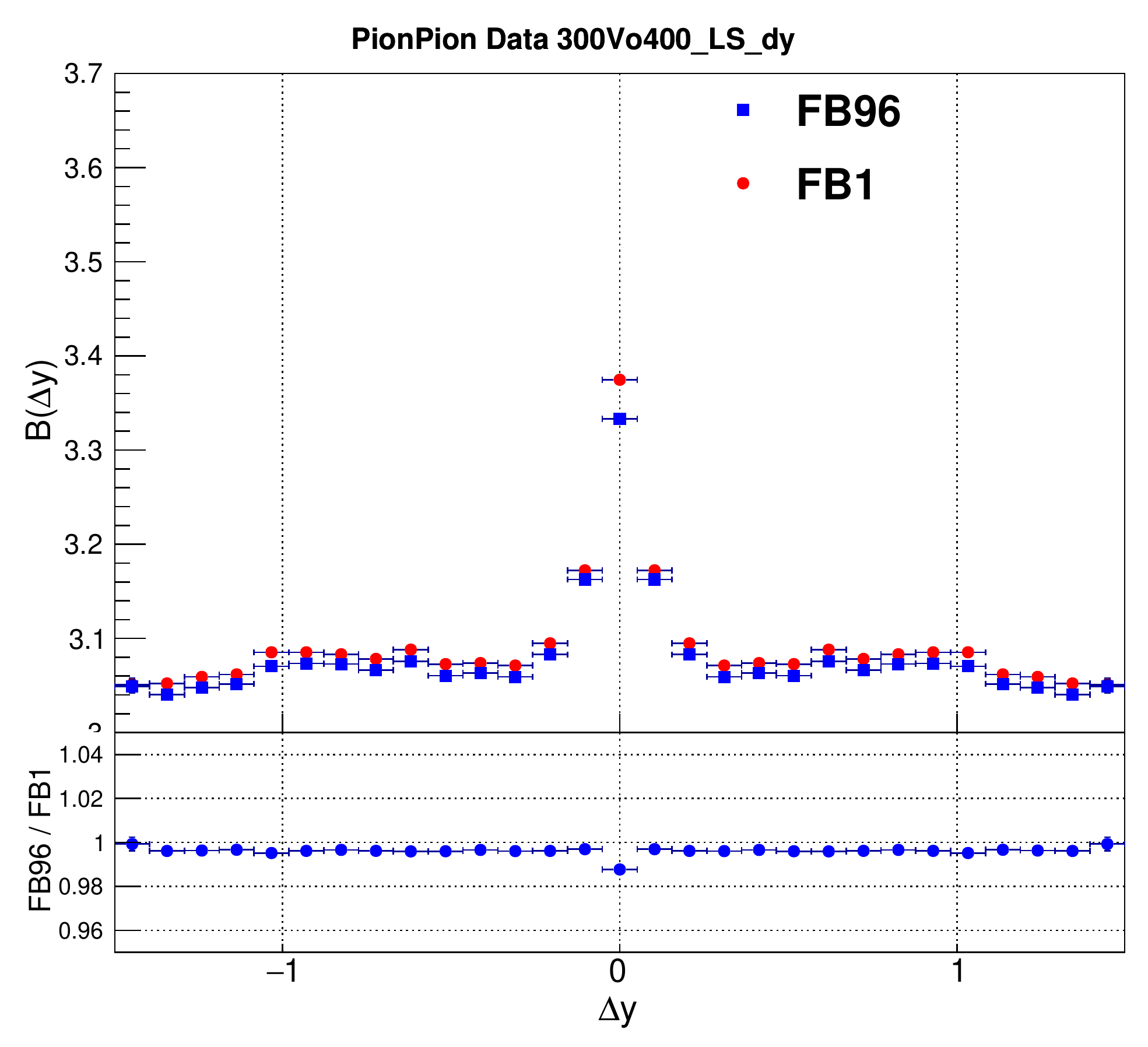}
  \includegraphics[width=0.32\linewidth]{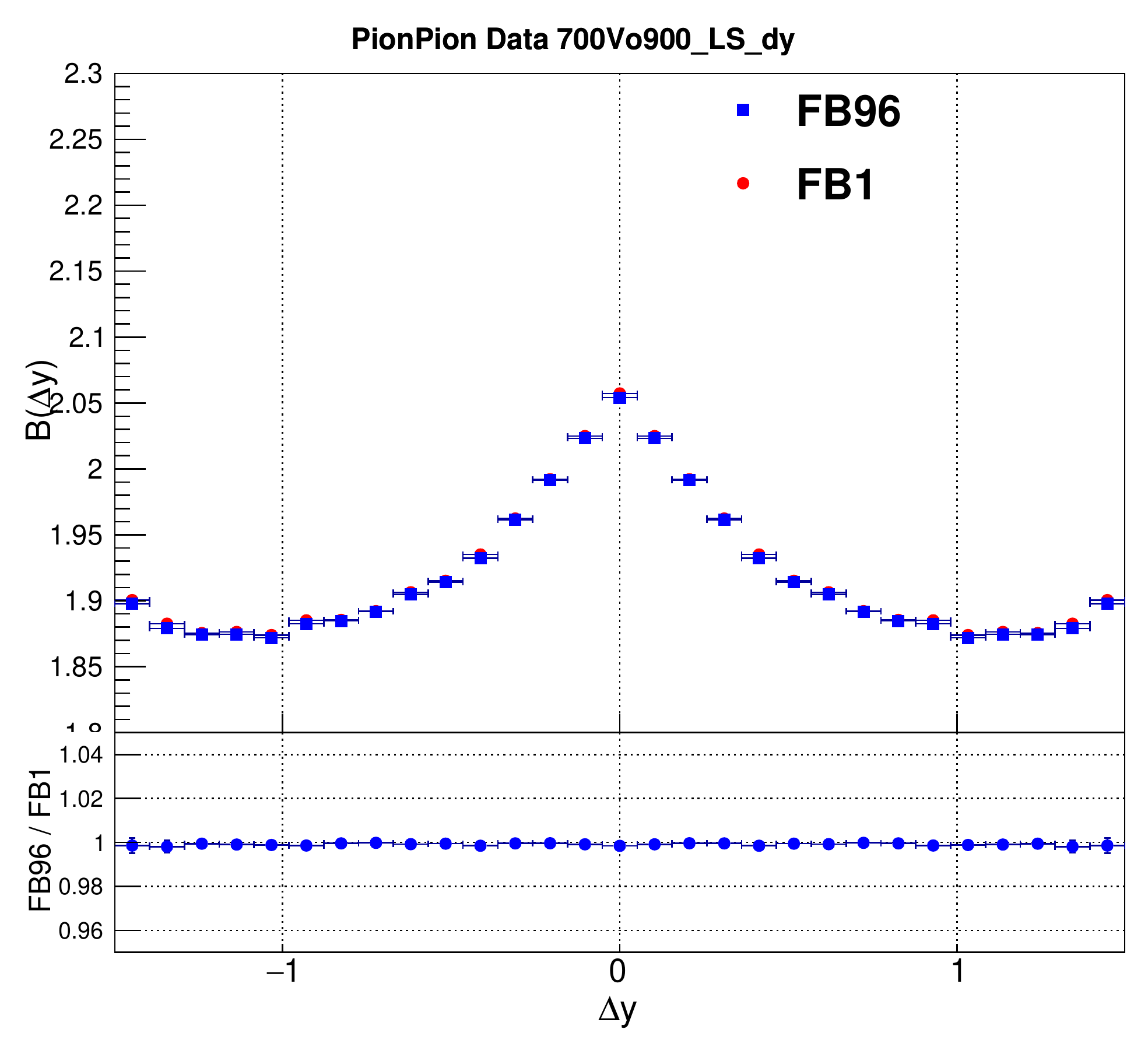}
  \includegraphics[width=0.32\linewidth]{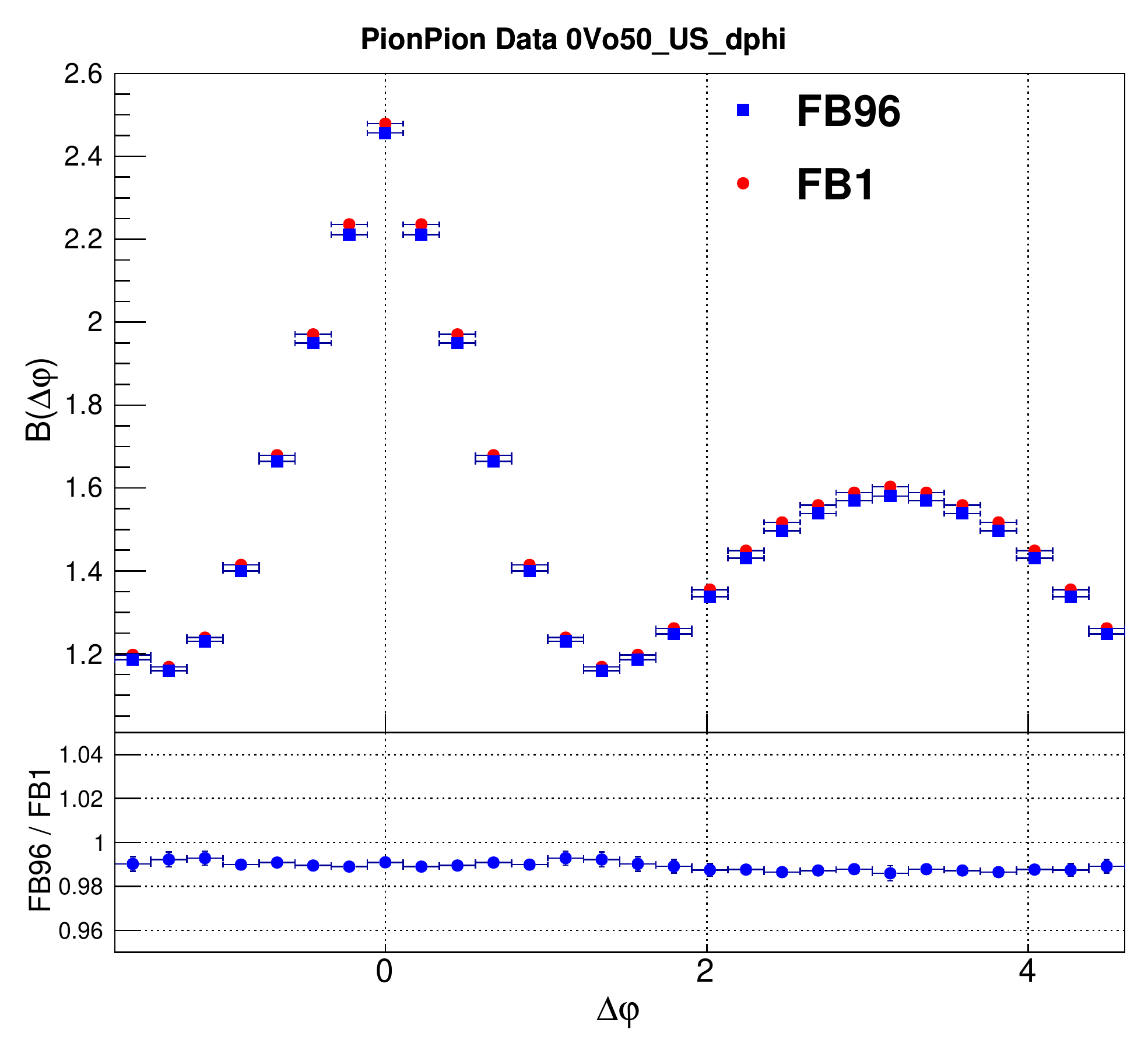}
  \includegraphics[width=0.32\linewidth]{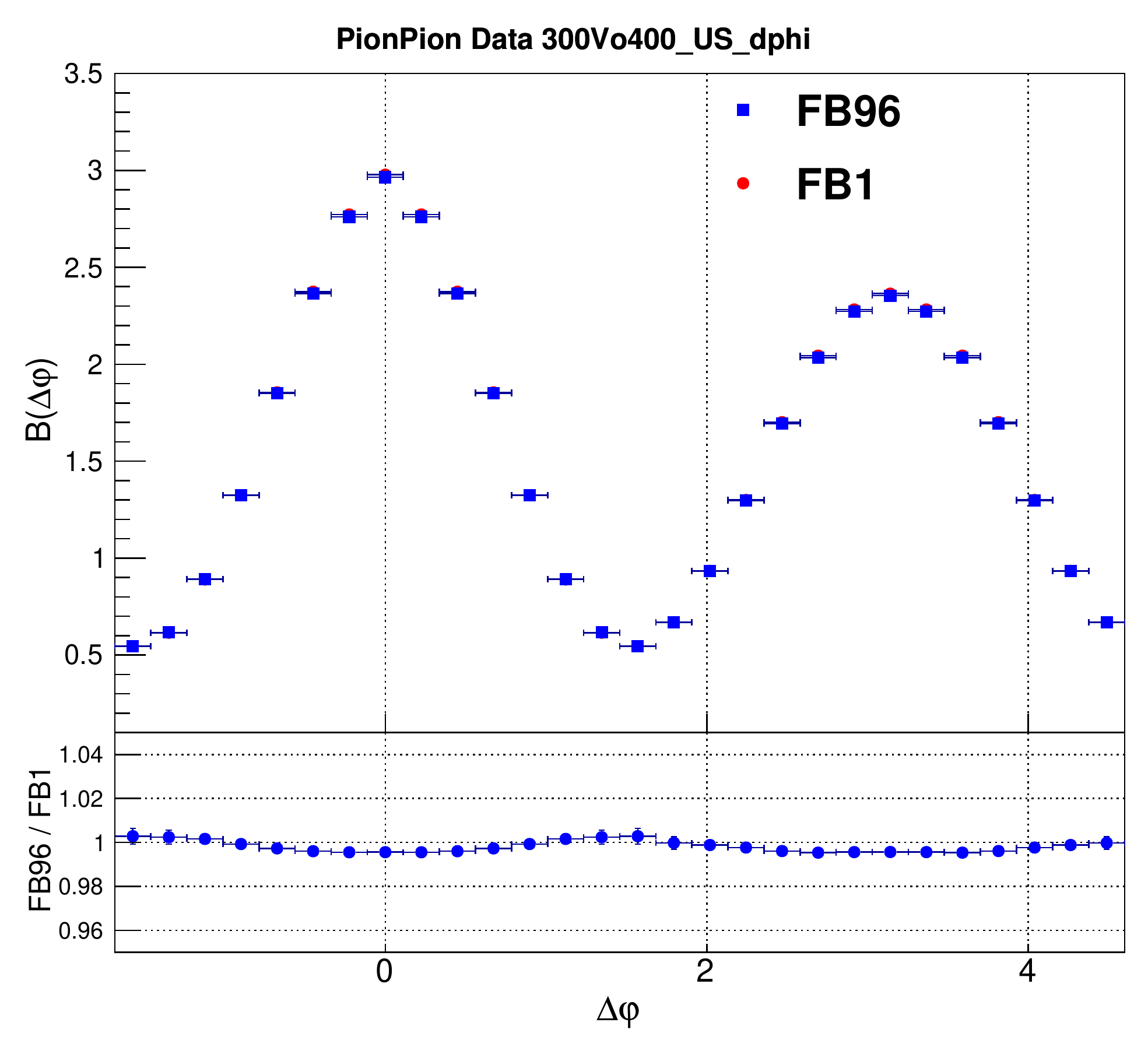}
  \includegraphics[width=0.32\linewidth]{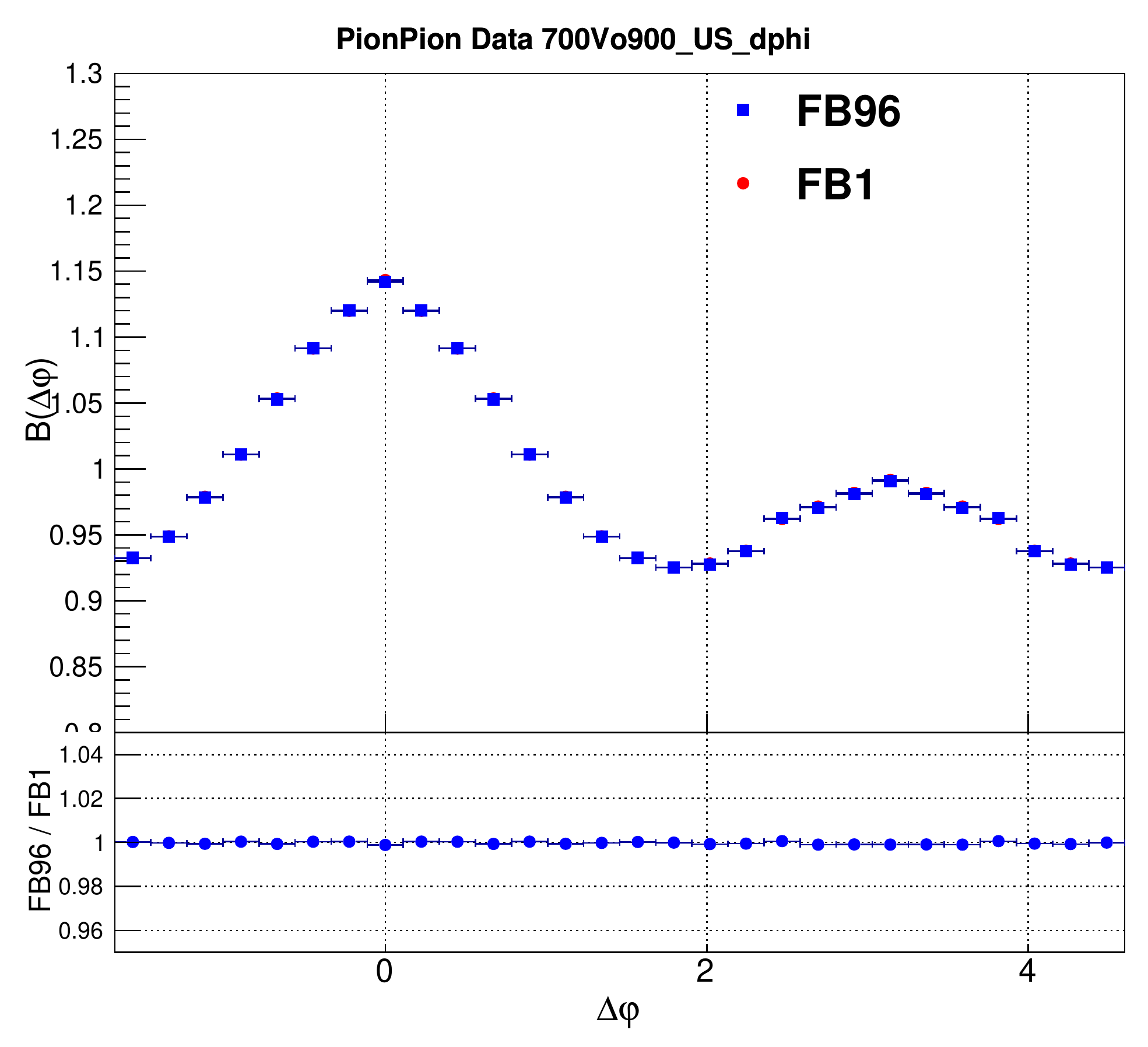}
  \includegraphics[width=0.32\linewidth]{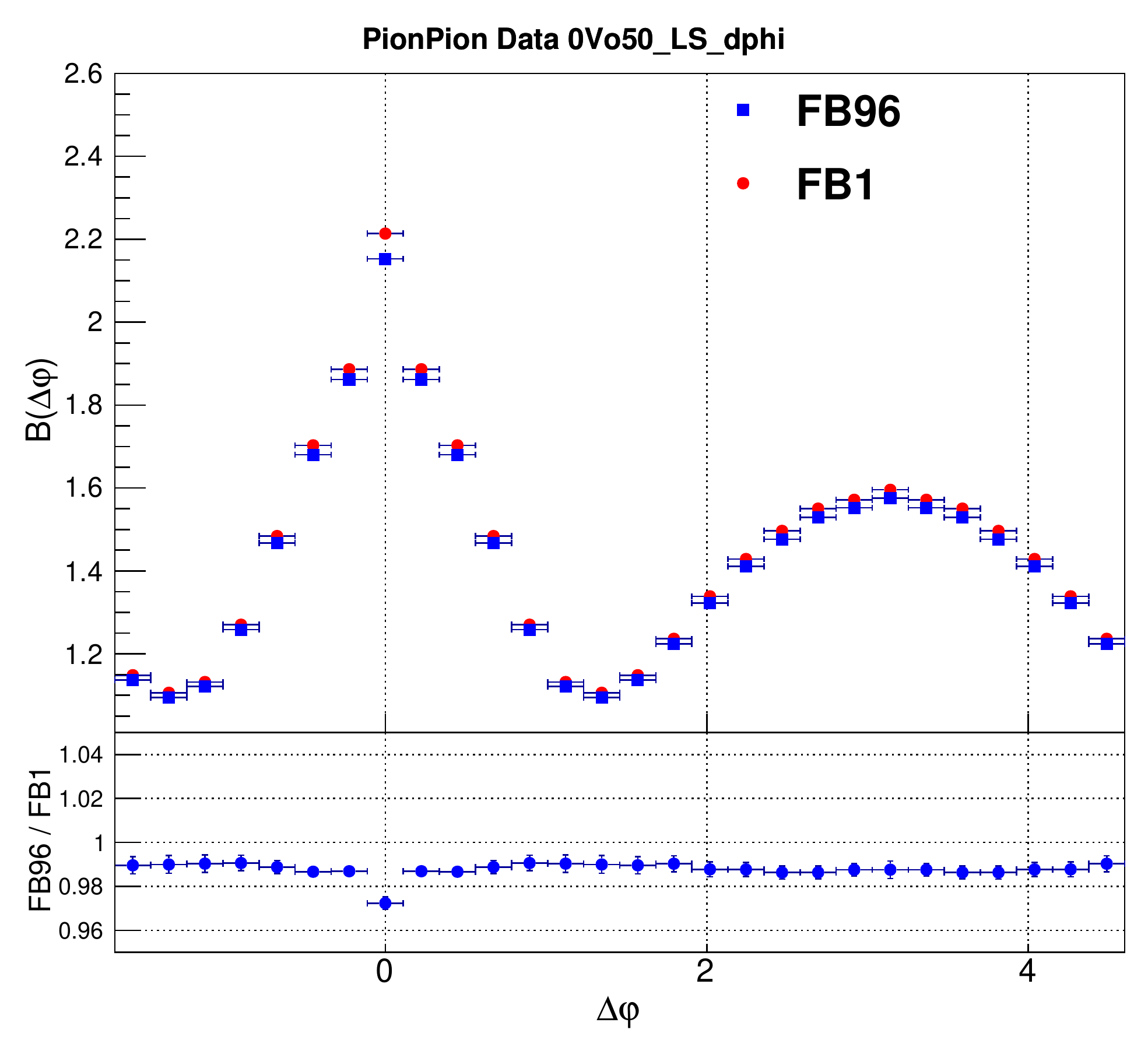}
  \includegraphics[width=0.32\linewidth]{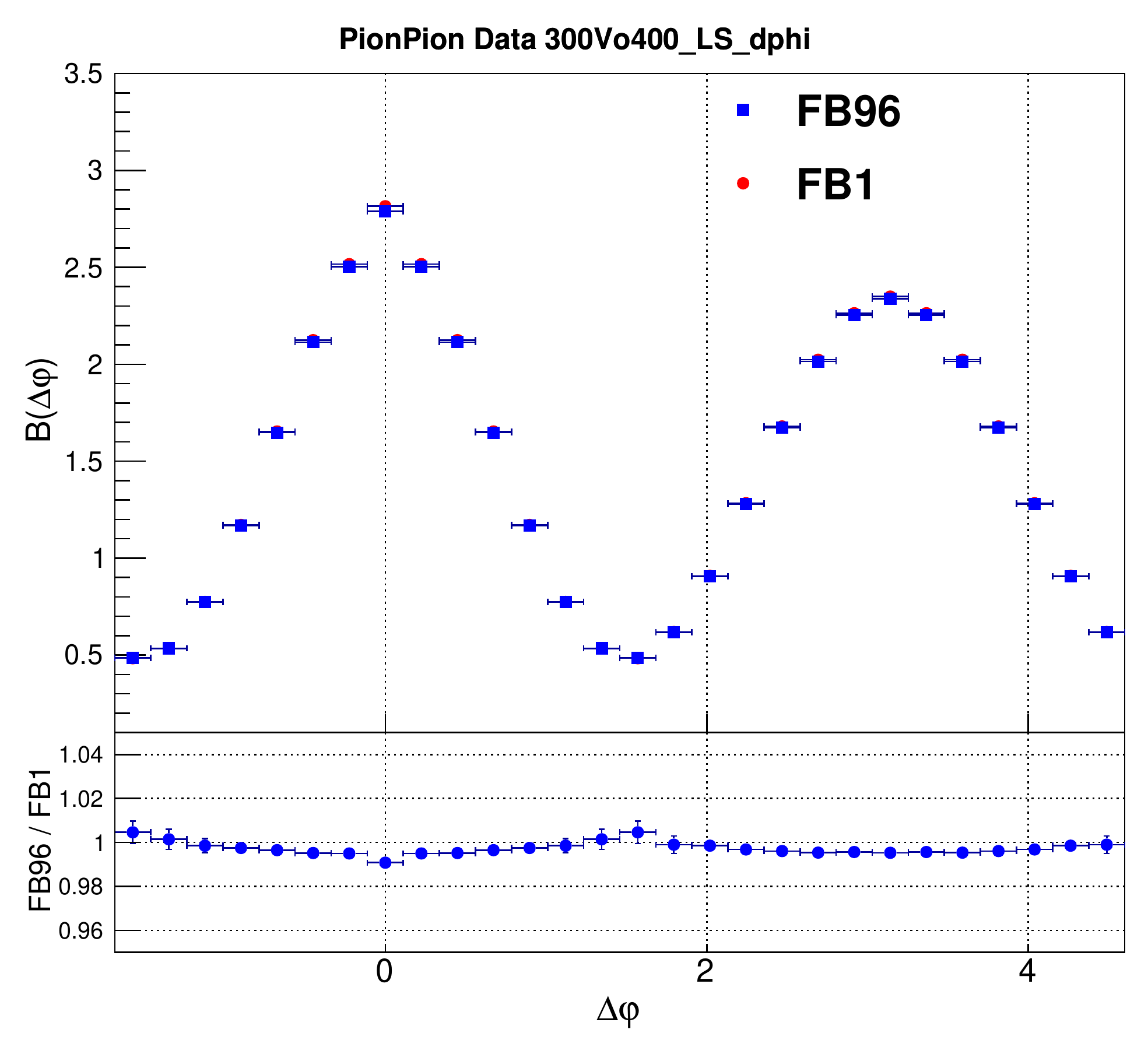}
  \includegraphics[width=0.32\linewidth]{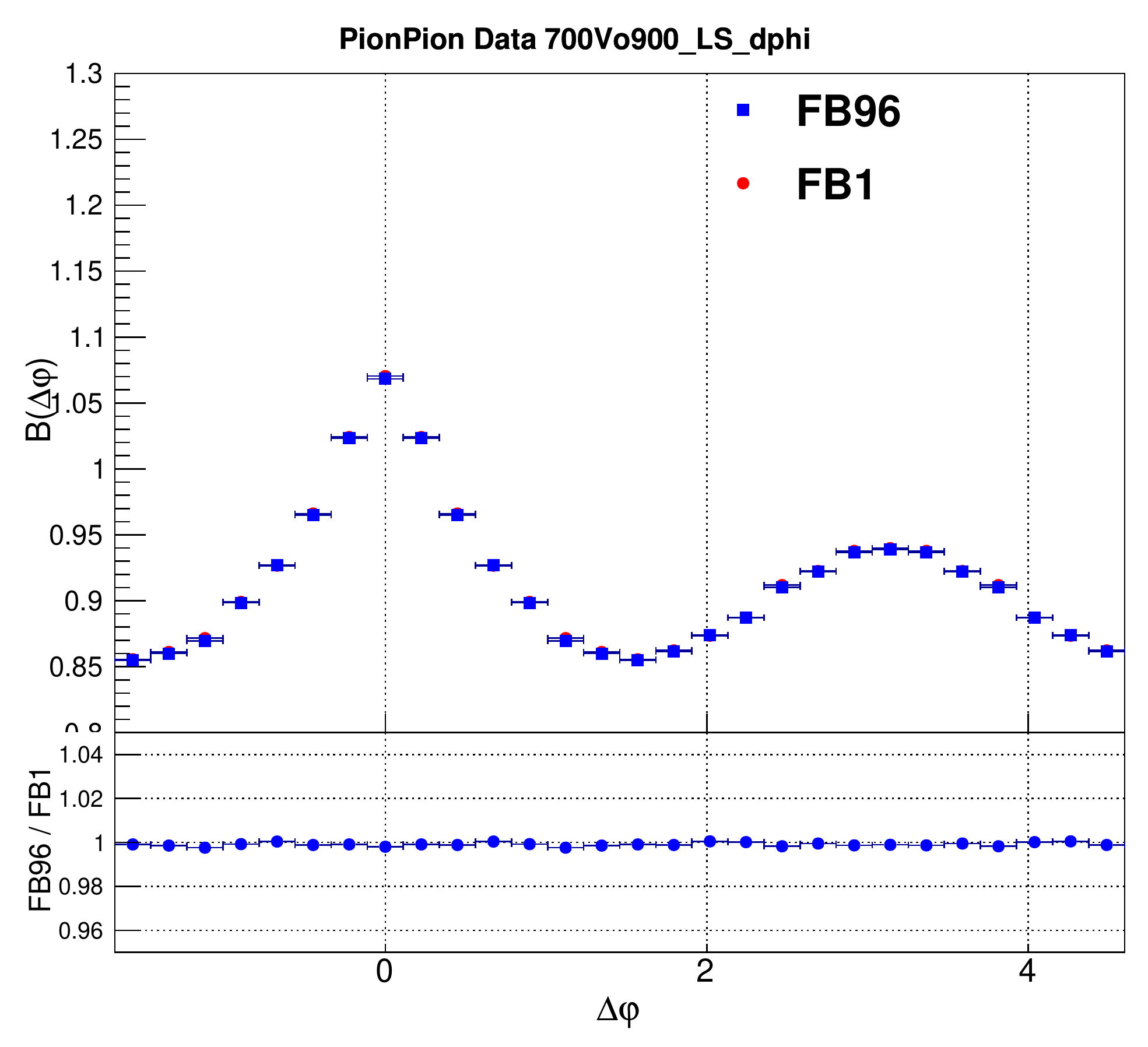}
  \caption{For $\pi\pi$ pair. Comparisons of US CF $\Delta y$ (1st row) and $\Delta\varphi$ (3rd row) projections obtained with filter-bit 1 and 96 for 0-5\% (left column), 30-40\% (middle column), and 70-90\% (right column) collision centralities. 
  Comparisons of LS CF $\Delta y$ (2nd row) and $\Delta\varphi$ (4th row) projections obtained with filter-bit 1 and 96 for the same collision centralities.}
  \label{fig:BF_PionPion_FB1_FB96_Comparison_US_LS}
\end{figure}

\begin{figure}
\centering
  \includegraphics[width=0.32\linewidth]{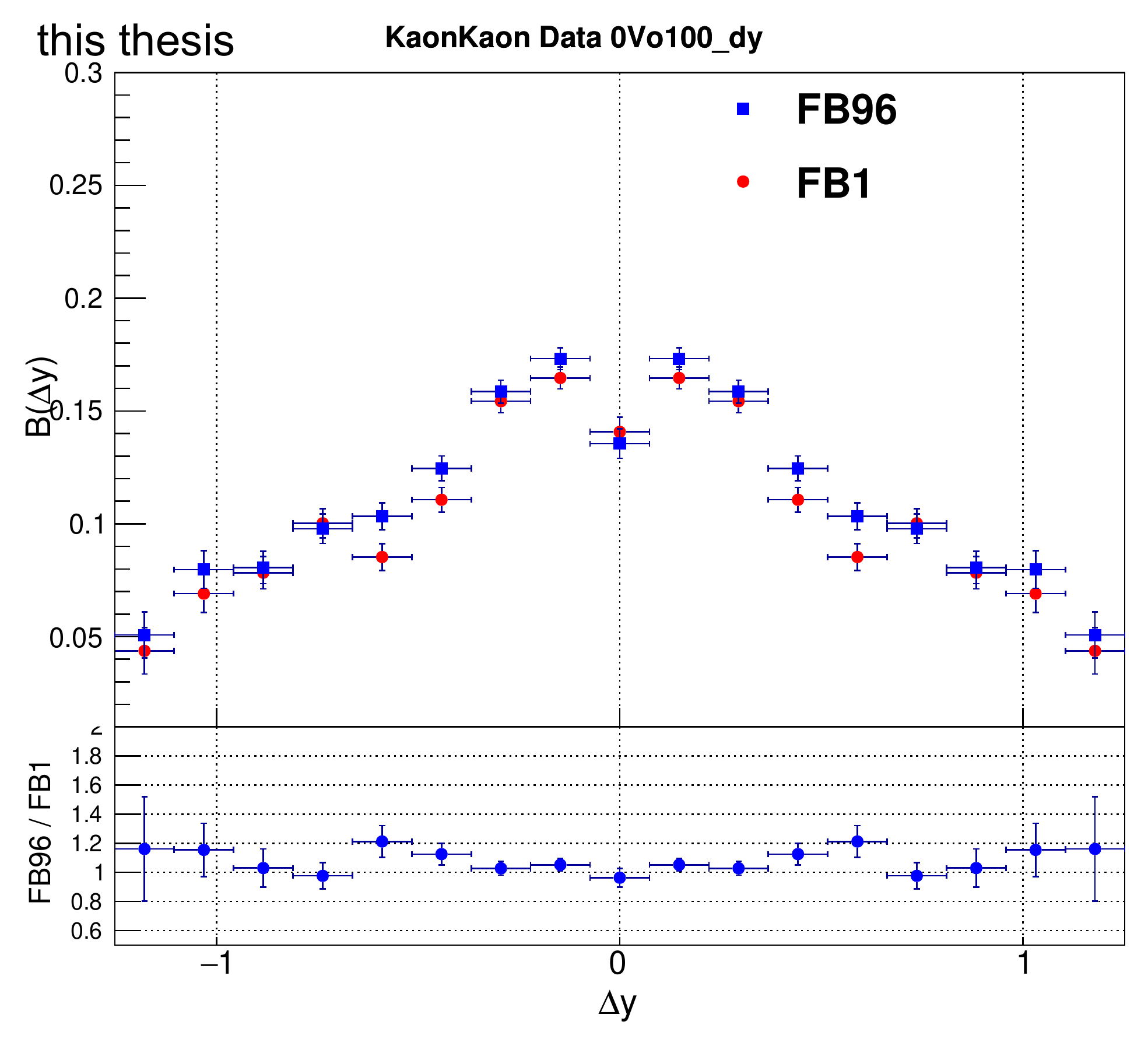}
  \includegraphics[width=0.32\linewidth]{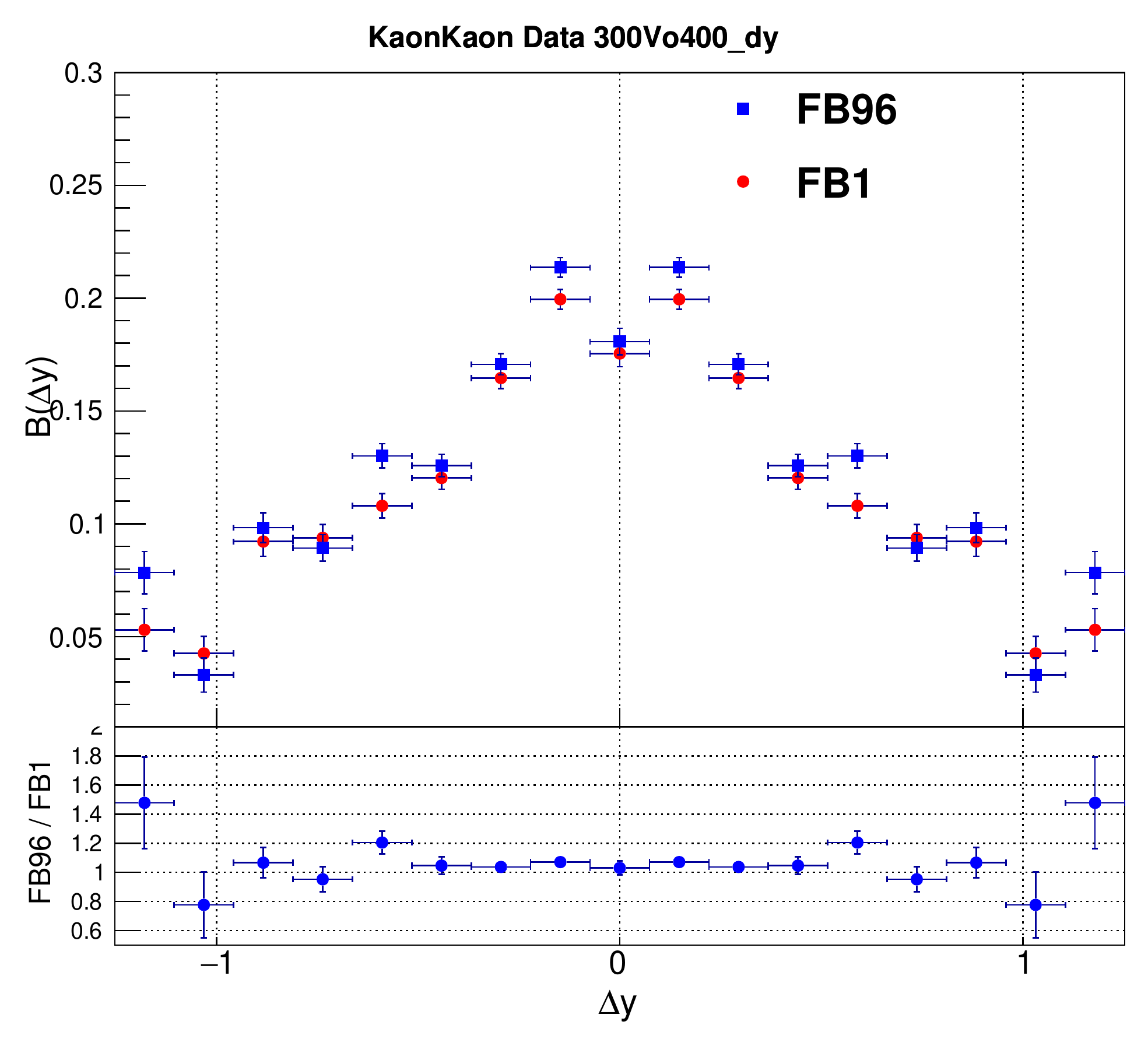}
  \includegraphics[width=0.32\linewidth]{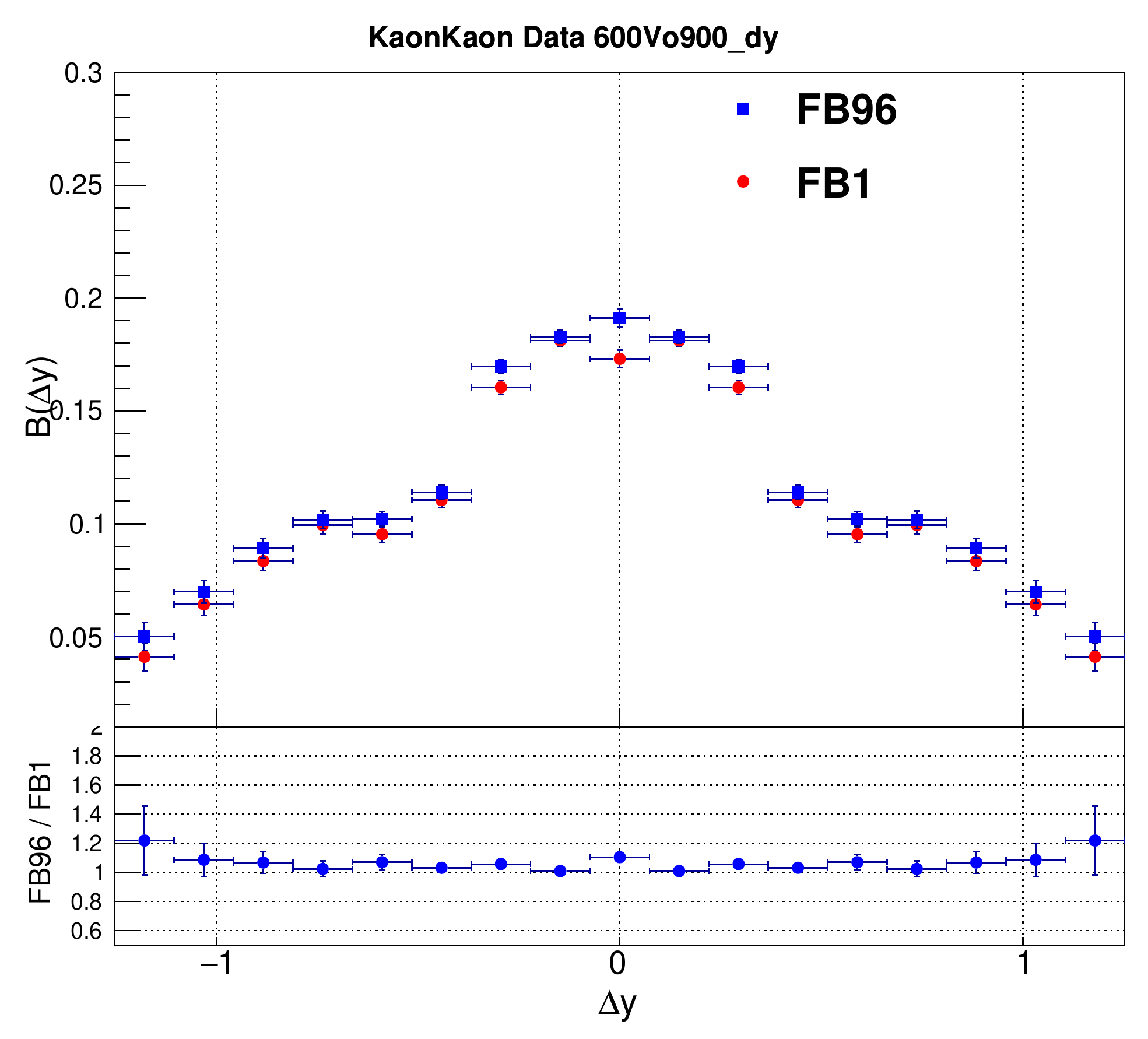}
  \includegraphics[width=0.32\linewidth]{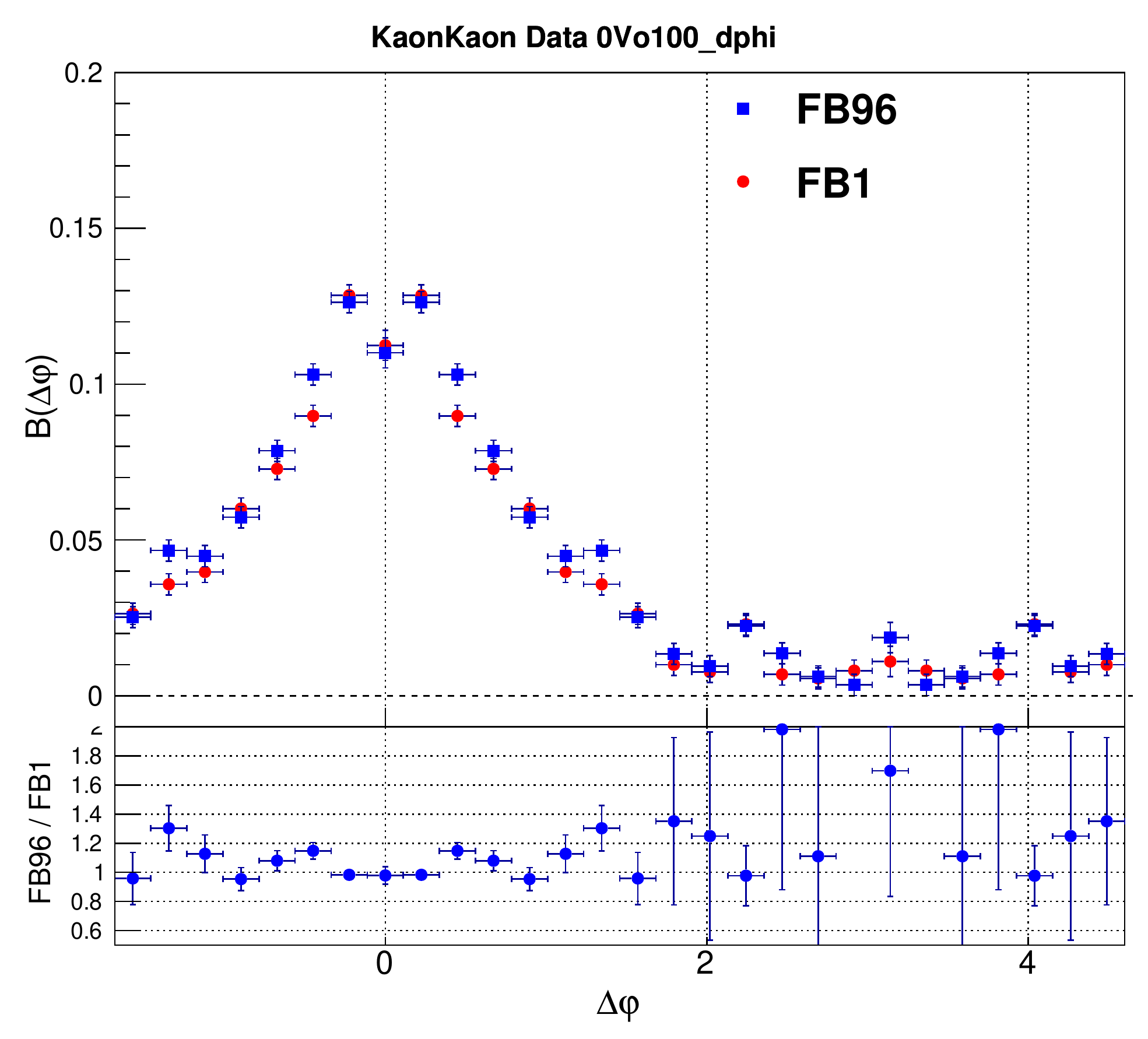}
  \includegraphics[width=0.32\linewidth]{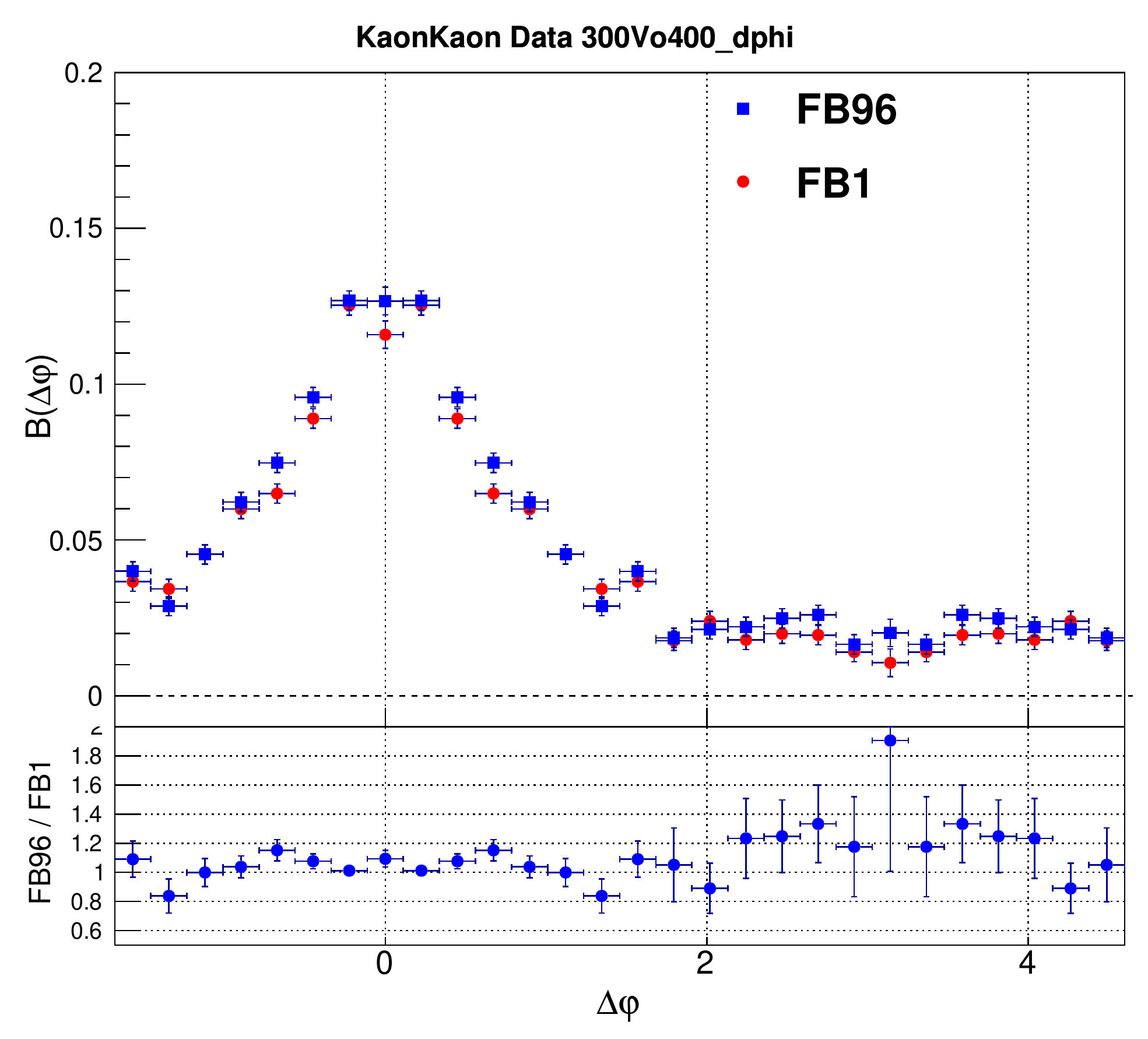}
  \includegraphics[width=0.32\linewidth]{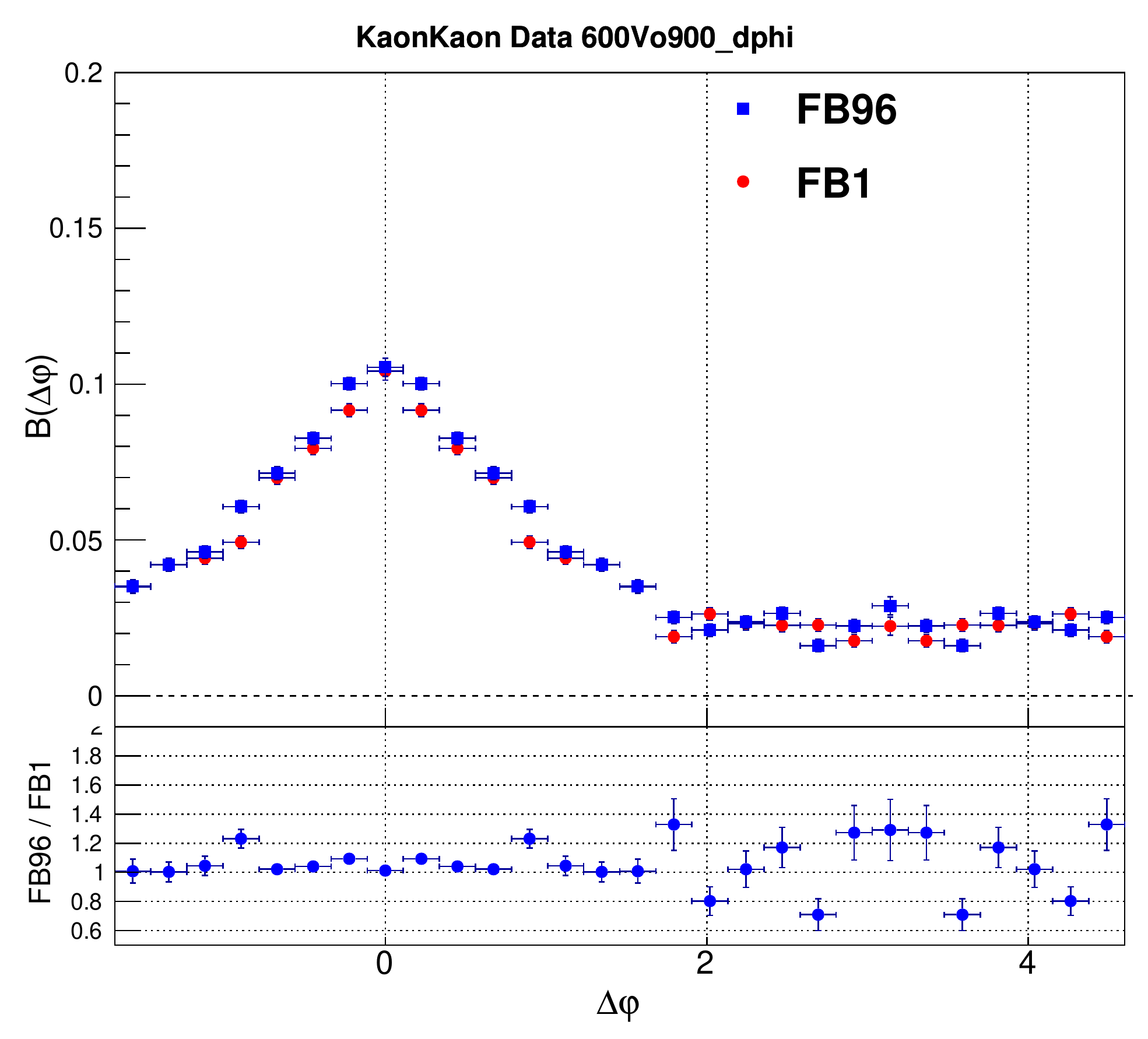}
  \includegraphics[width=0.32\linewidth]{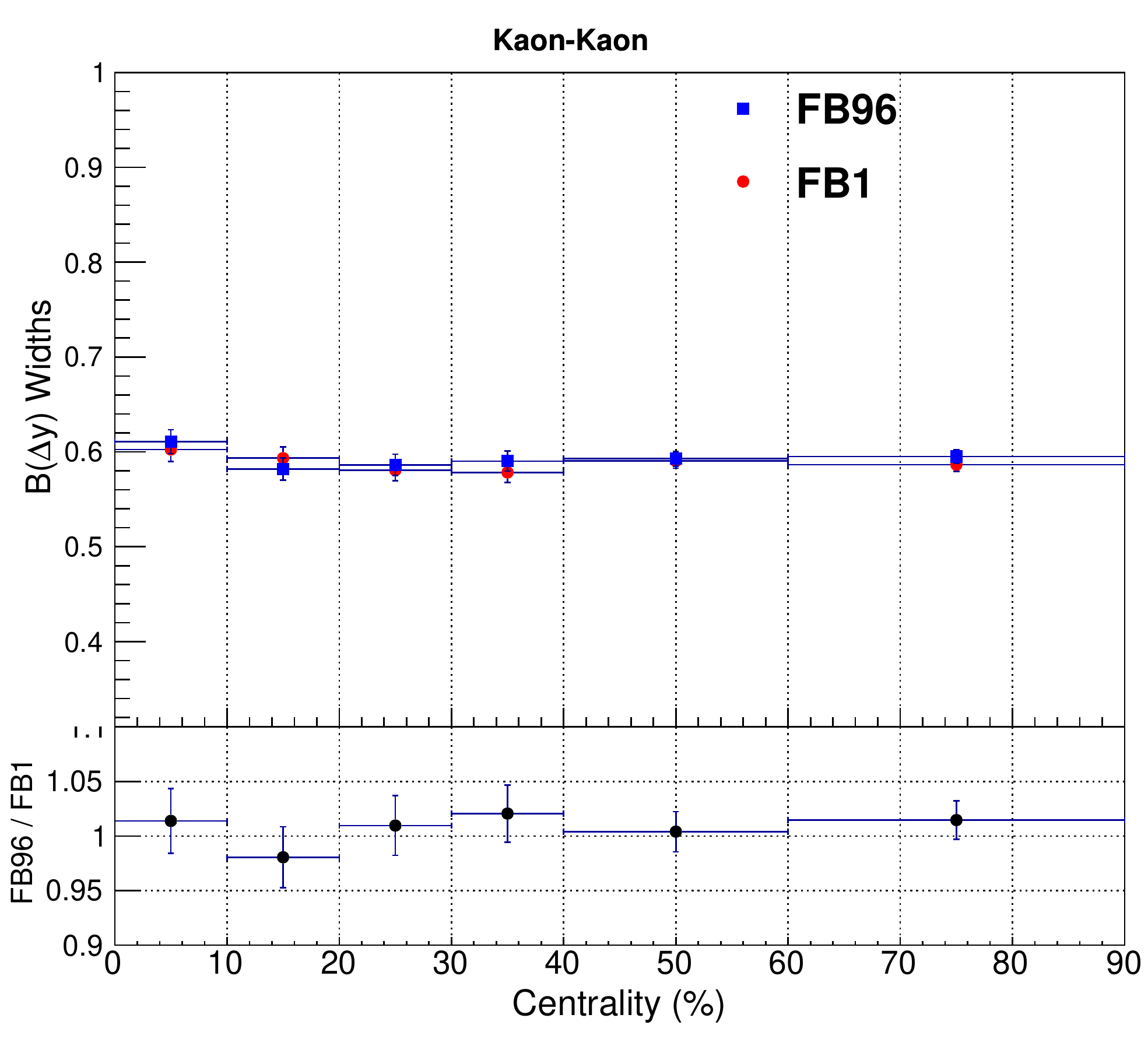}
  \includegraphics[width=0.32\linewidth]{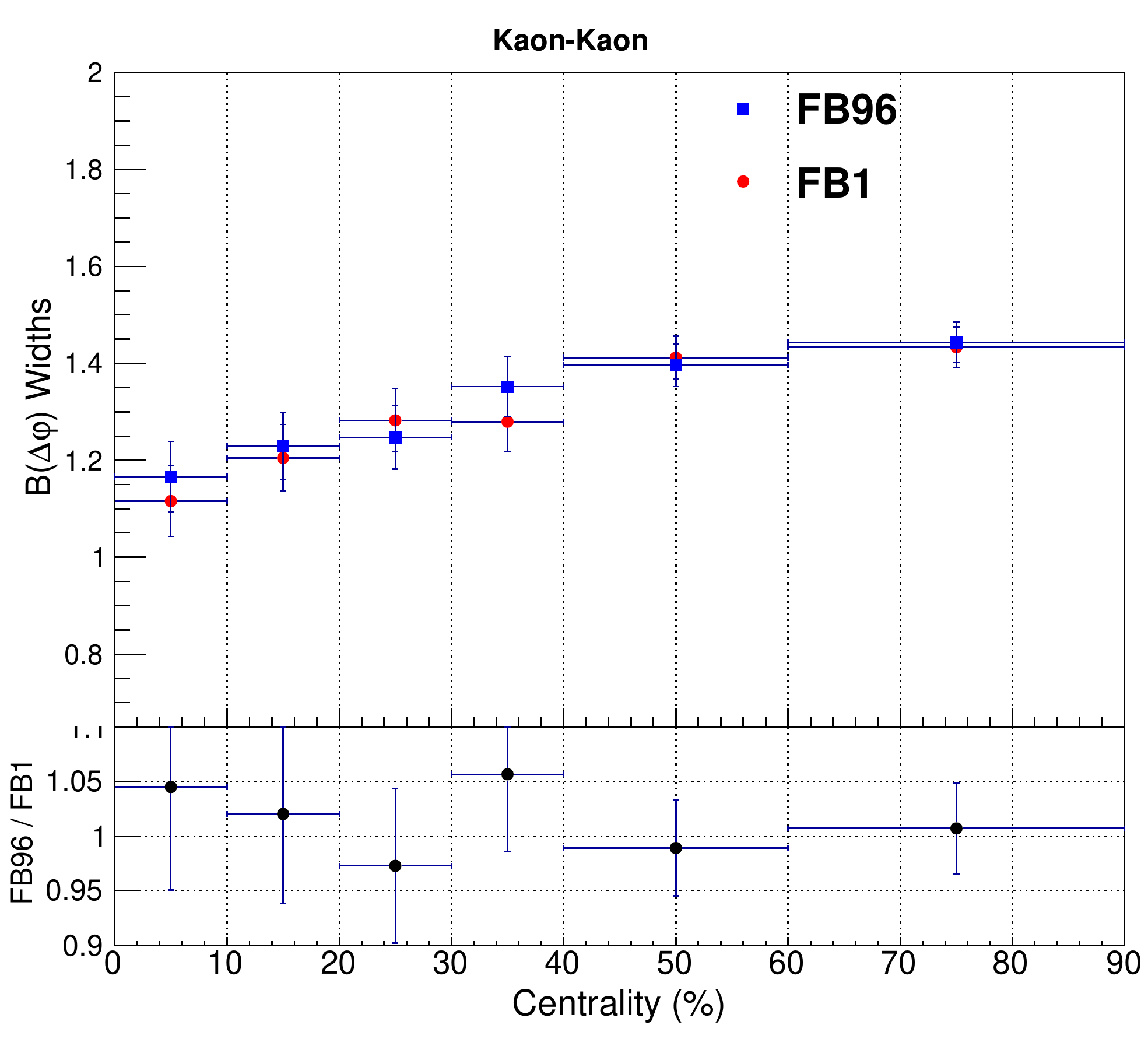}
  \includegraphics[width=0.32\linewidth]{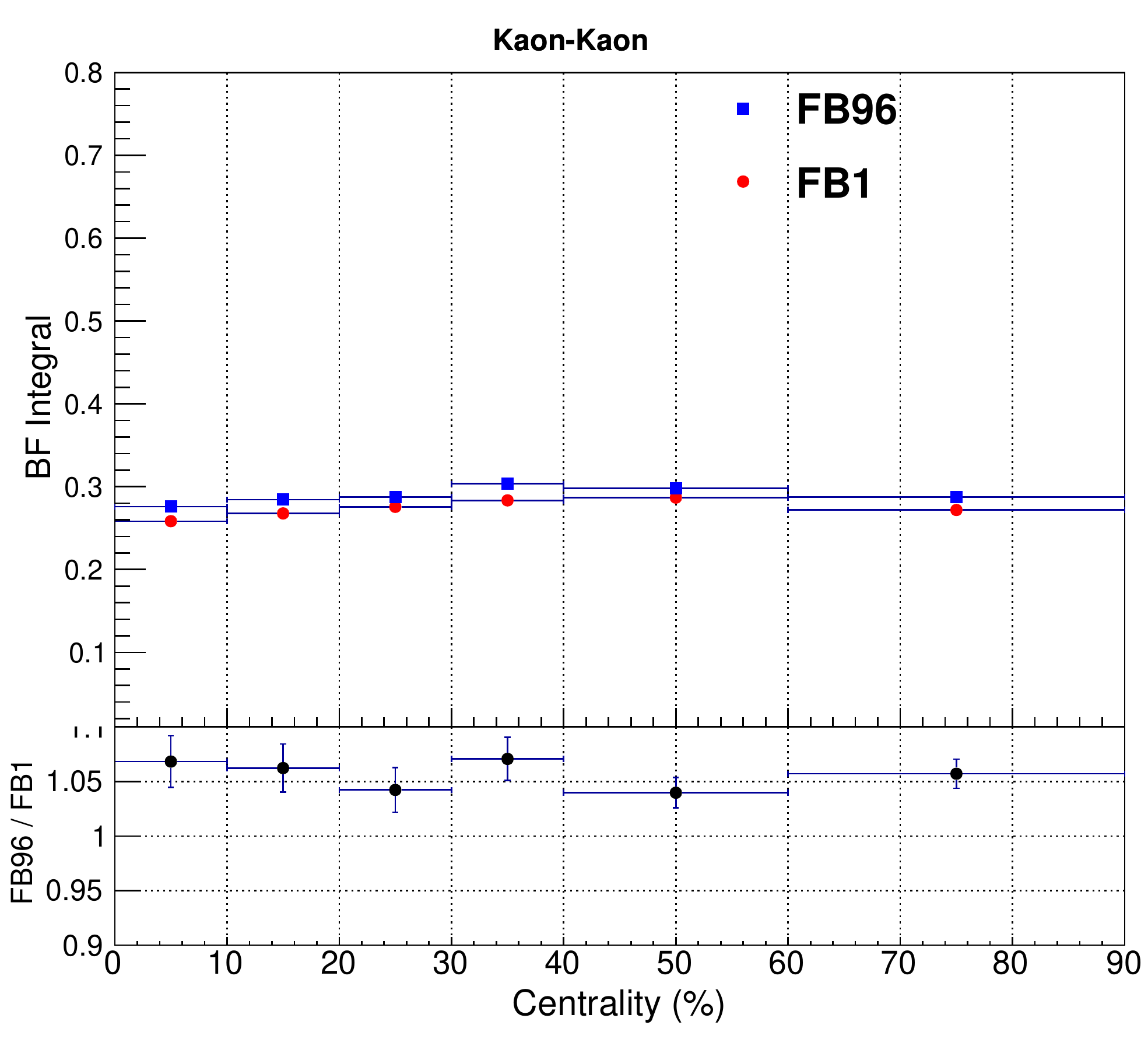}
  \caption{Comparisons of $B^{KK}$ projections onto $\Delta y$ (top row) and $\Delta\varphi$ (middle row) axis for selected collision centralities.
  Bottom row: comparisons of $B^{KK}$ $\Delta y$ widths (left), $\Delta\varphi$ widths (middle), and integrals (right) as a function of collision centrality between filter-bit 1 and 96.}
  \label{fig:BF_KaonKaon_FB1_FB96_Comparison}
\end{figure}

\begin{figure}
\centering
  \includegraphics[width=0.32\linewidth]{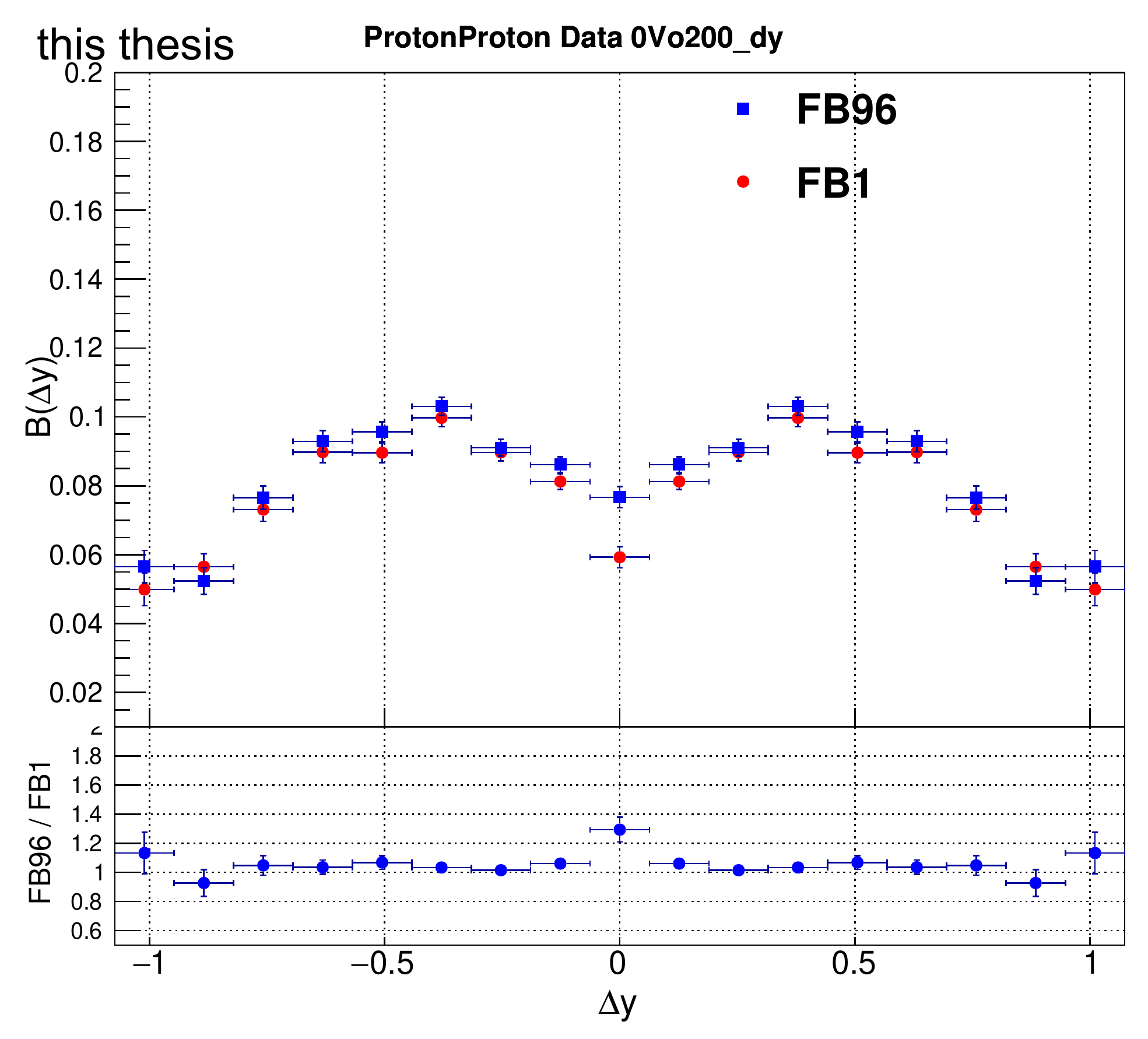}
  \includegraphics[width=0.32\linewidth]{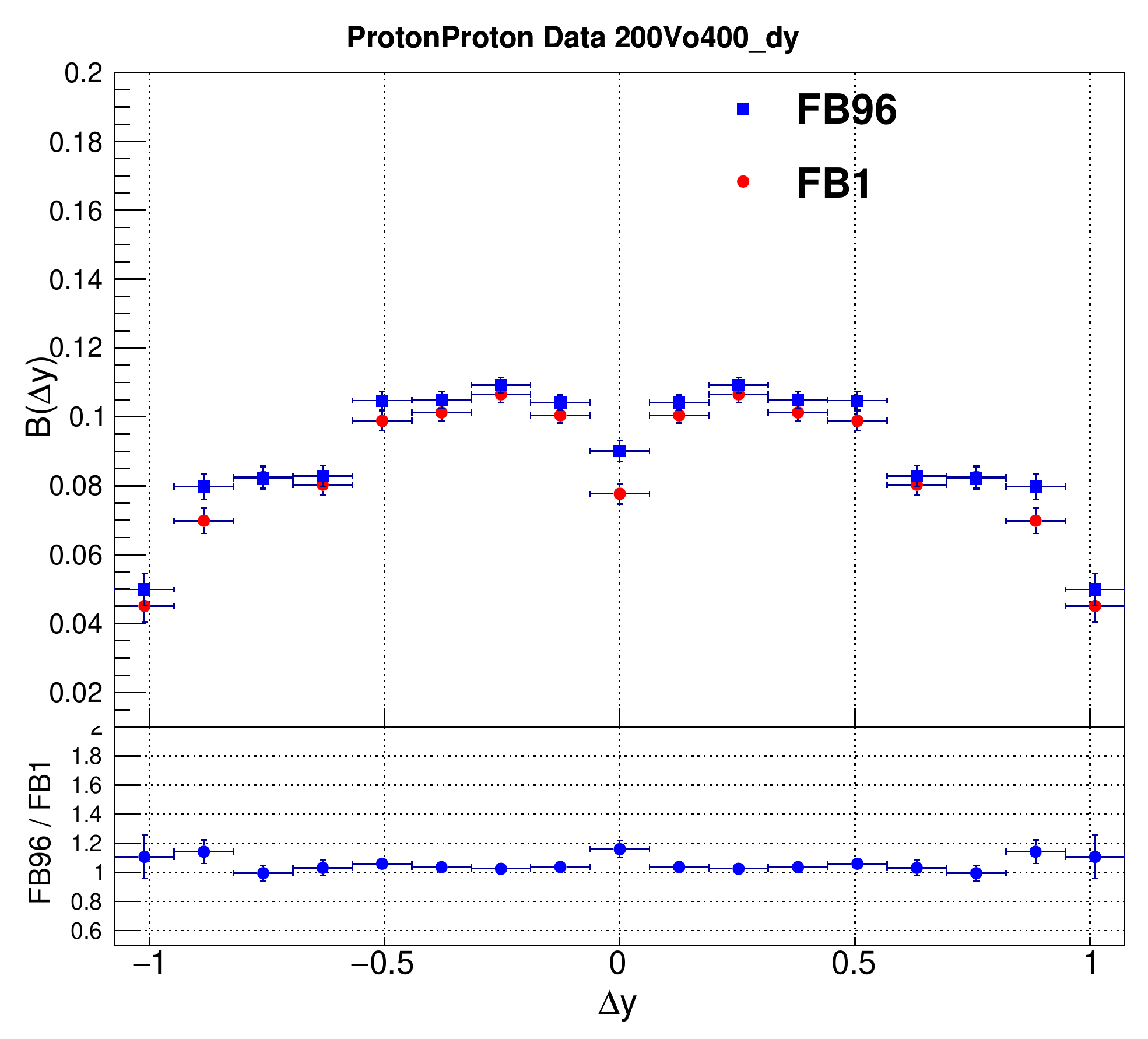}
  \includegraphics[width=0.32\linewidth]{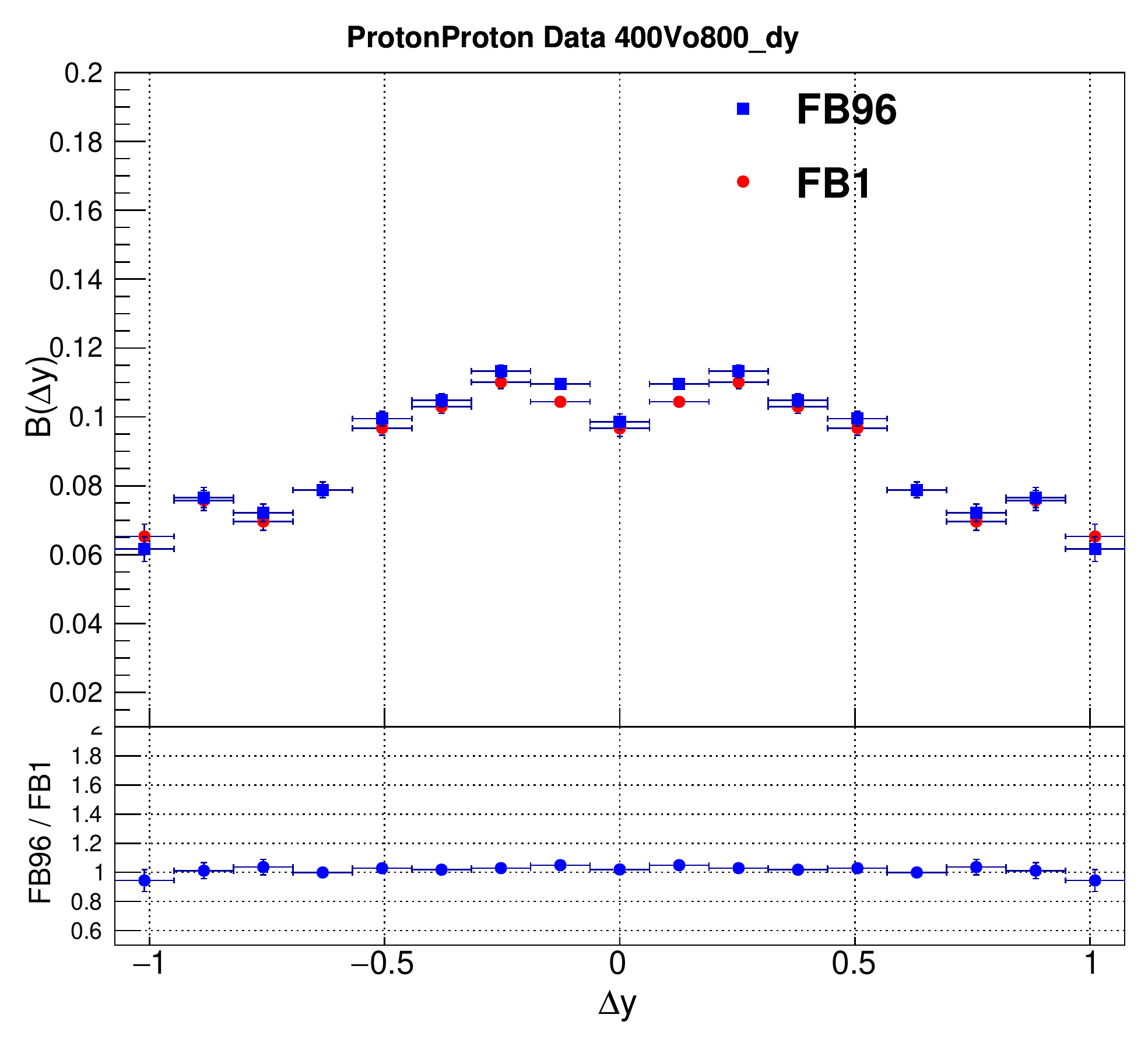}
  \includegraphics[width=0.32\linewidth]{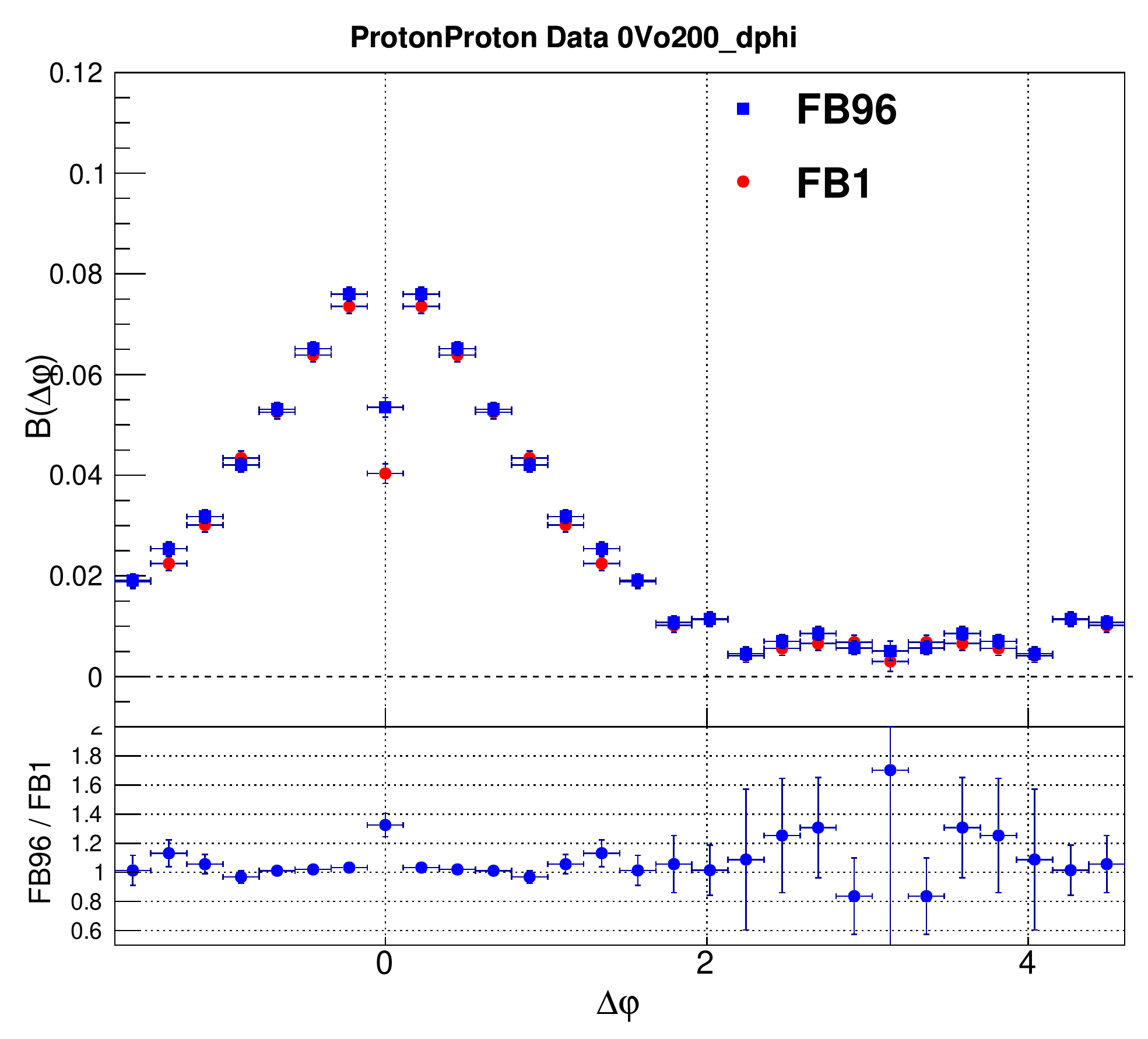}
  \includegraphics[width=0.32\linewidth]{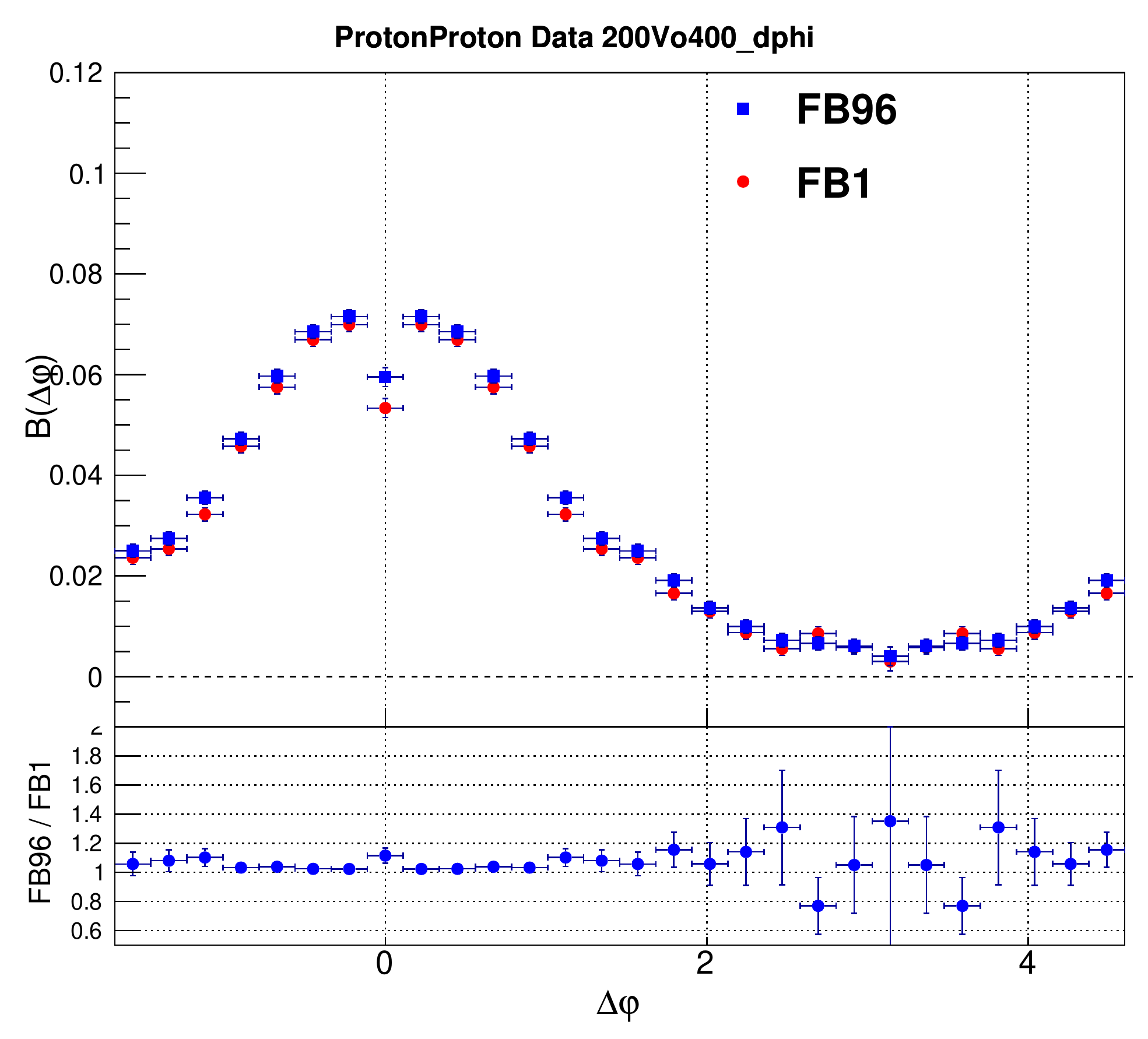}
  \includegraphics[width=0.32\linewidth]{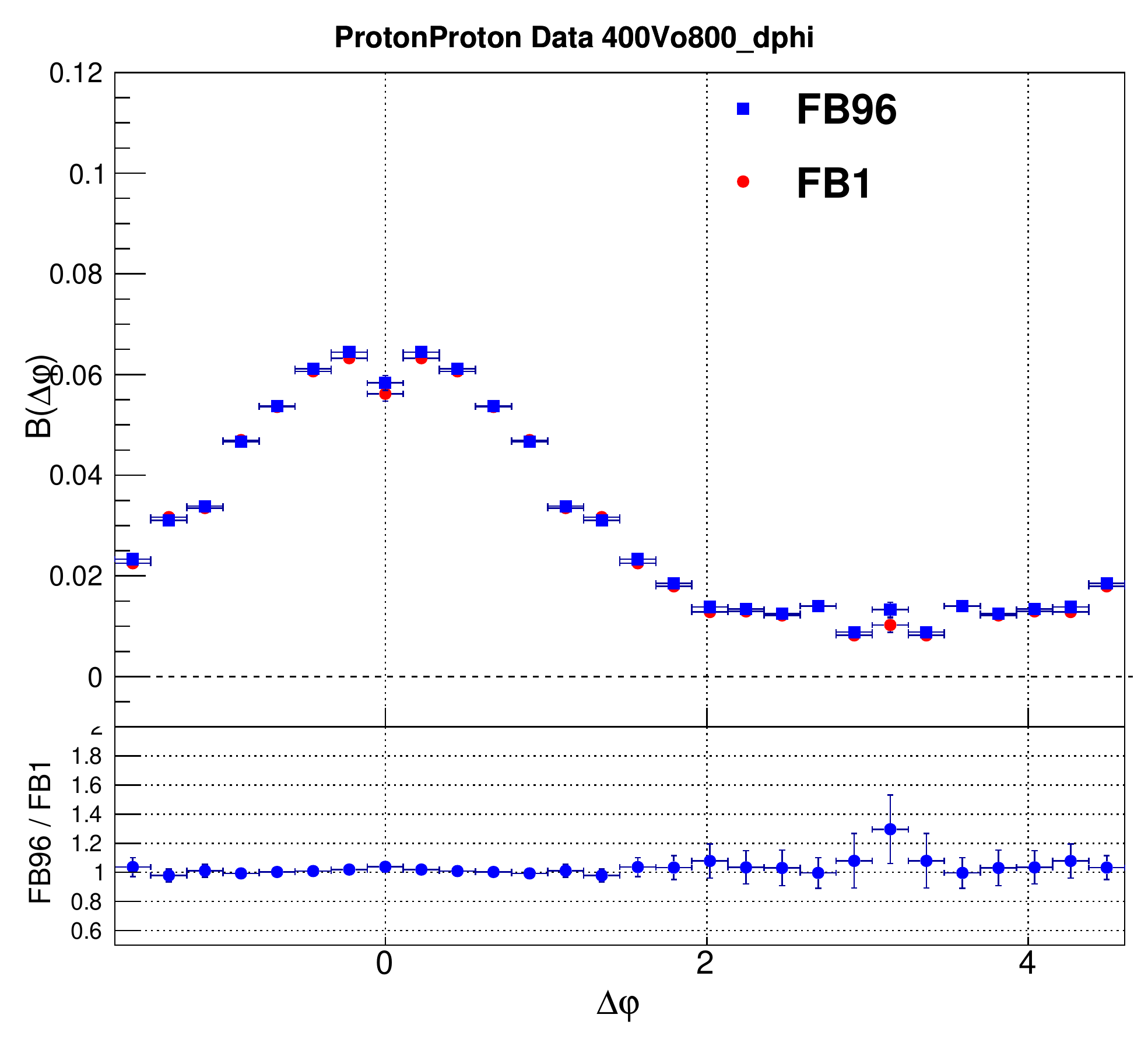}
  \includegraphics[width=0.32\linewidth]{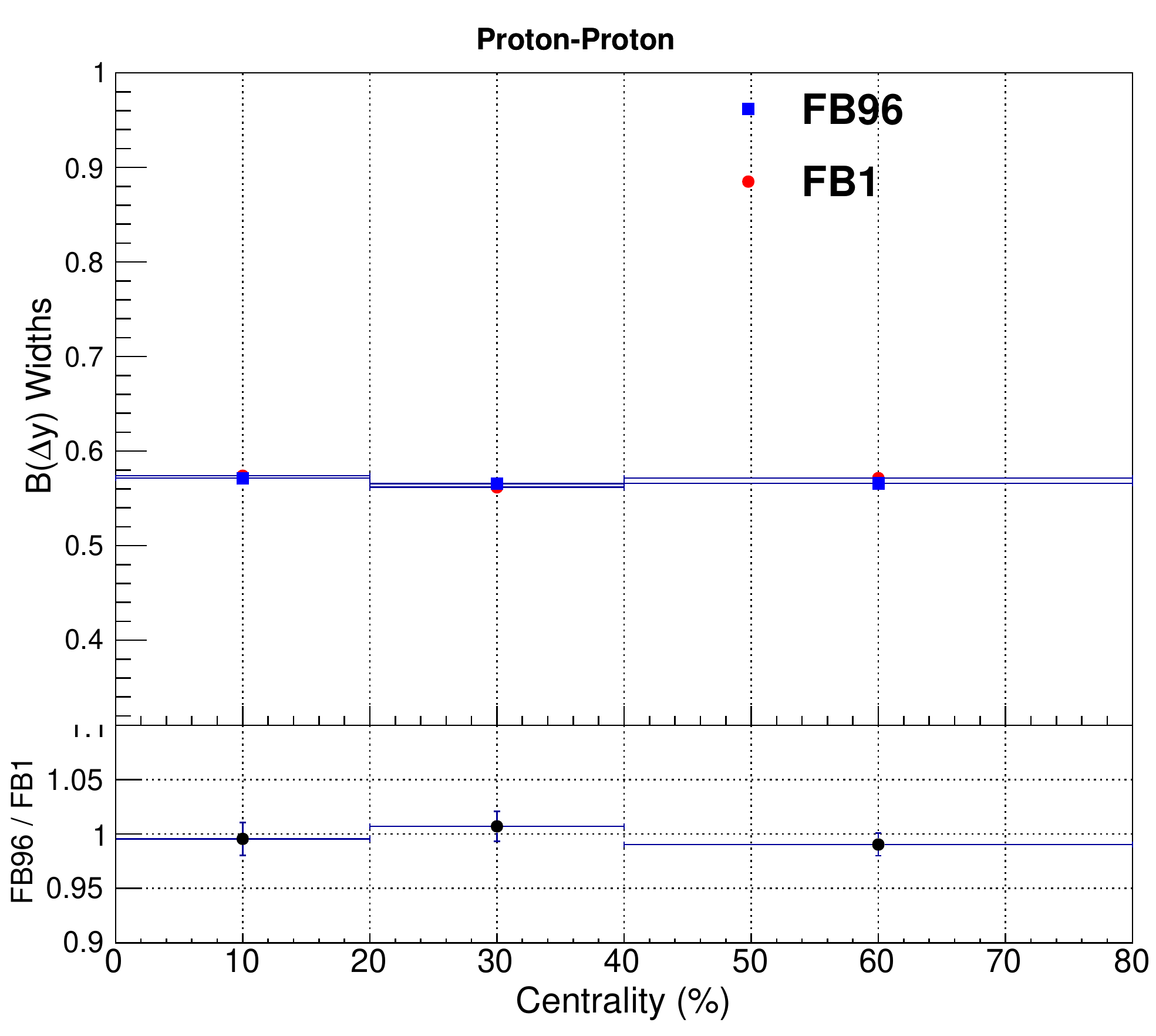}
  \includegraphics[width=0.32\linewidth]{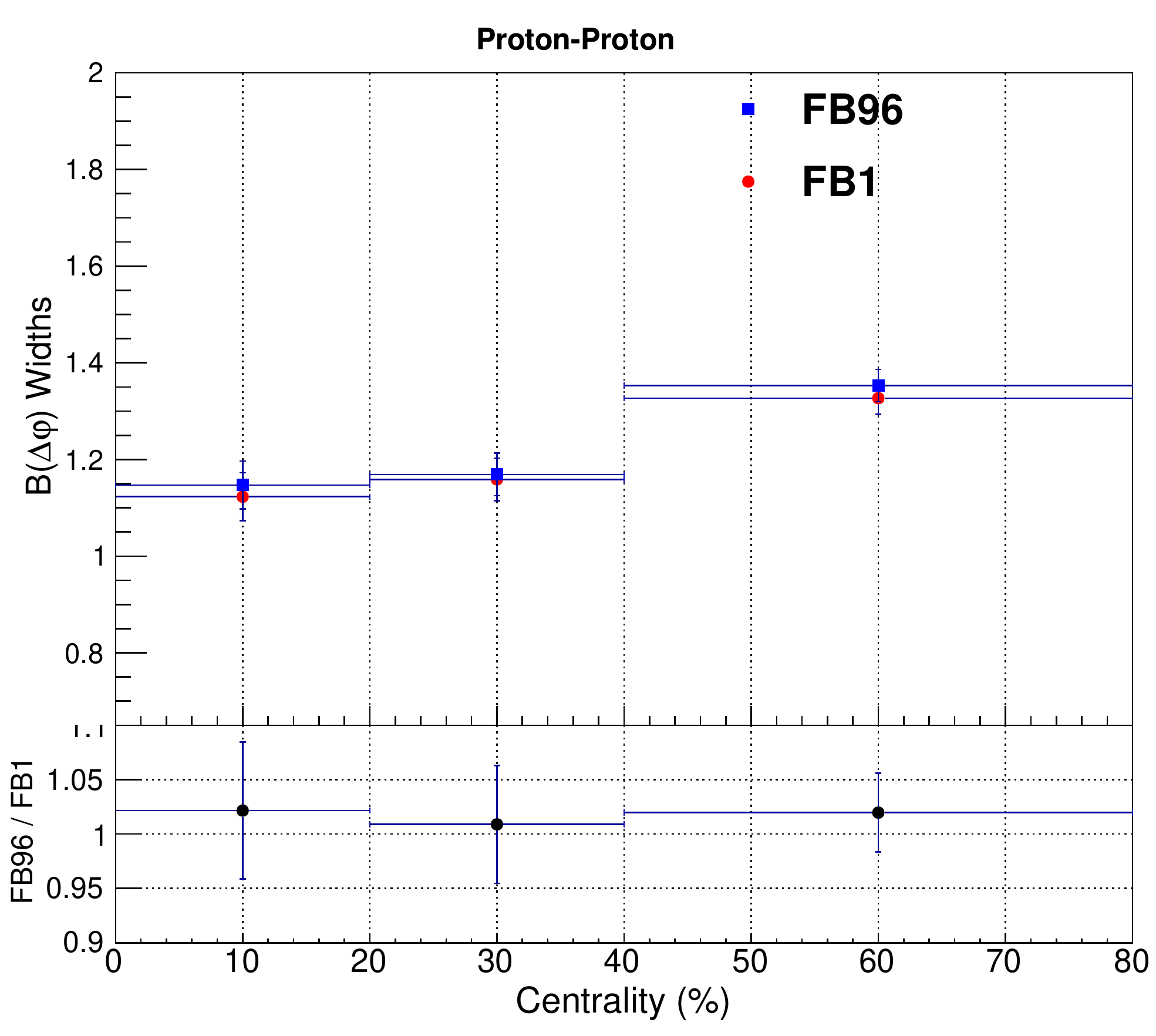}
  \includegraphics[width=0.32\linewidth]{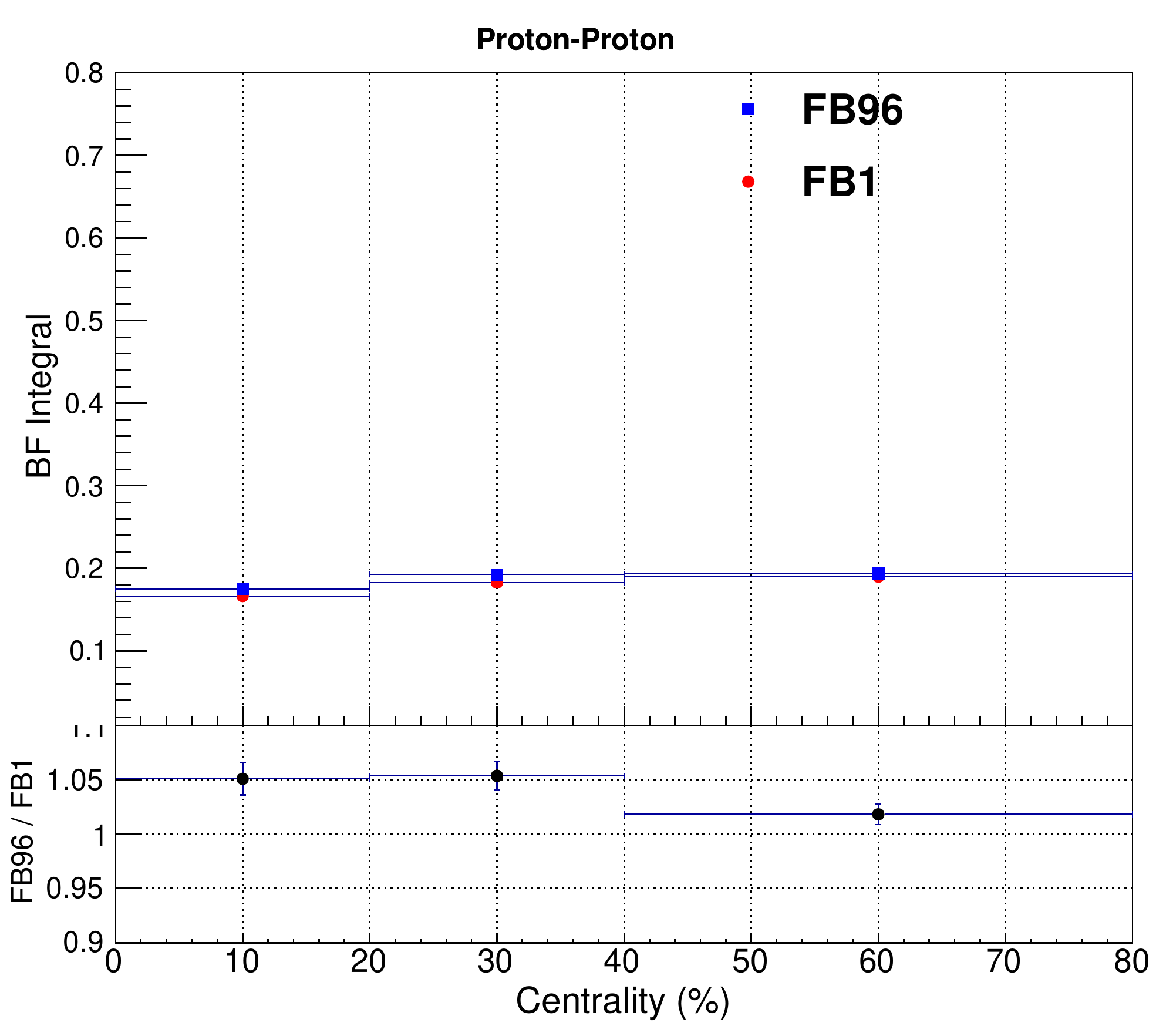}
  \caption{Comparisons of $B^{pp}$ projections onto $\Delta y$ (top row) and $\Delta\varphi$ (middle row) axis for selected collision centralities.
  Bottom row: comparisons of $B^{pp}$ $\Delta y$ widths (left), $\Delta\varphi$ widths (middle), and integrals (right) as a function of collision centrality between filter-bit 1 and 96.}
  \label{fig:BF_ProtonProton_FB1_FB96_Comparison}
\end{figure}

\begin{figure}
\centering
  \includegraphics[width=0.32\linewidth]{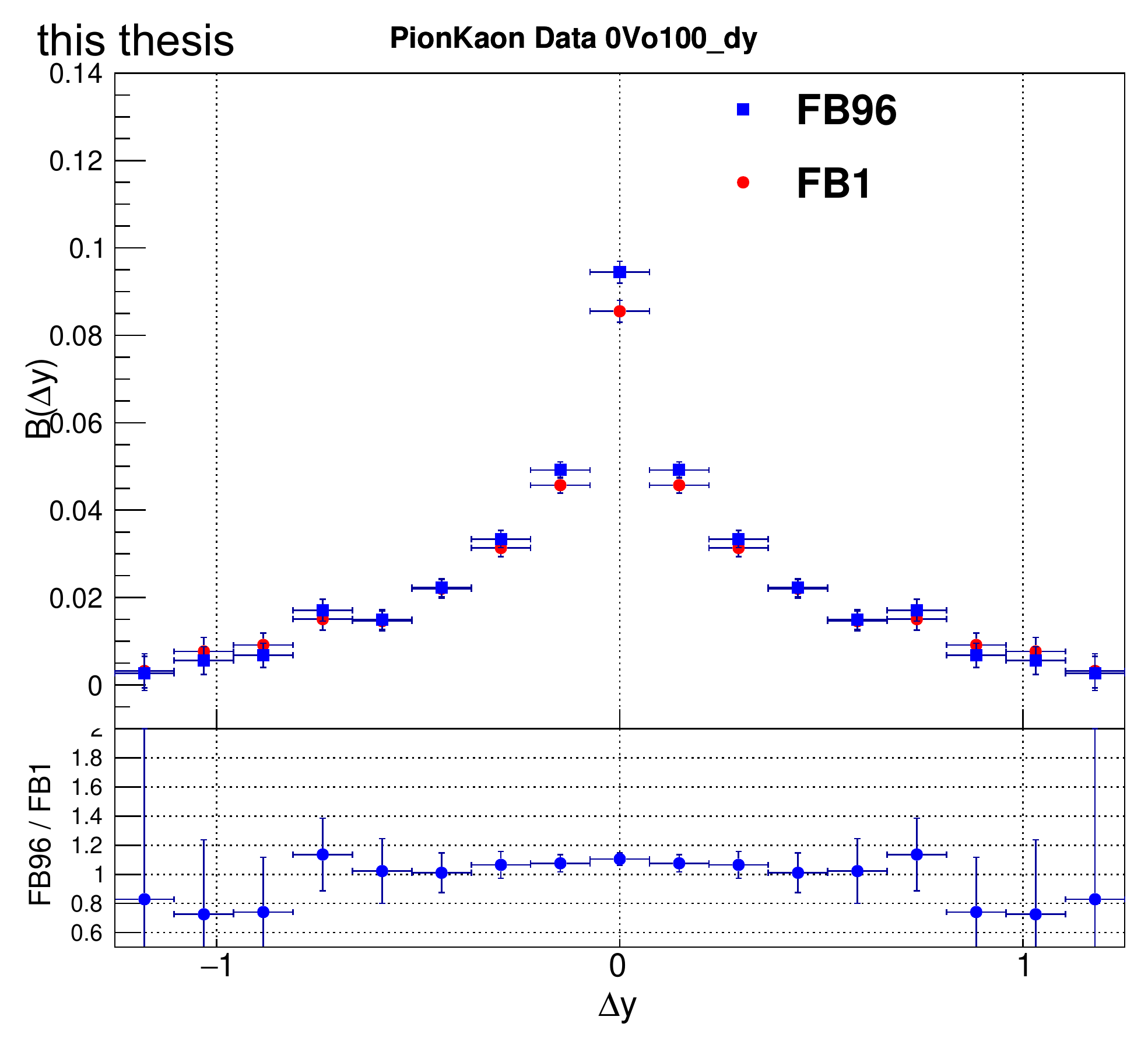}
  \includegraphics[width=0.32\linewidth]{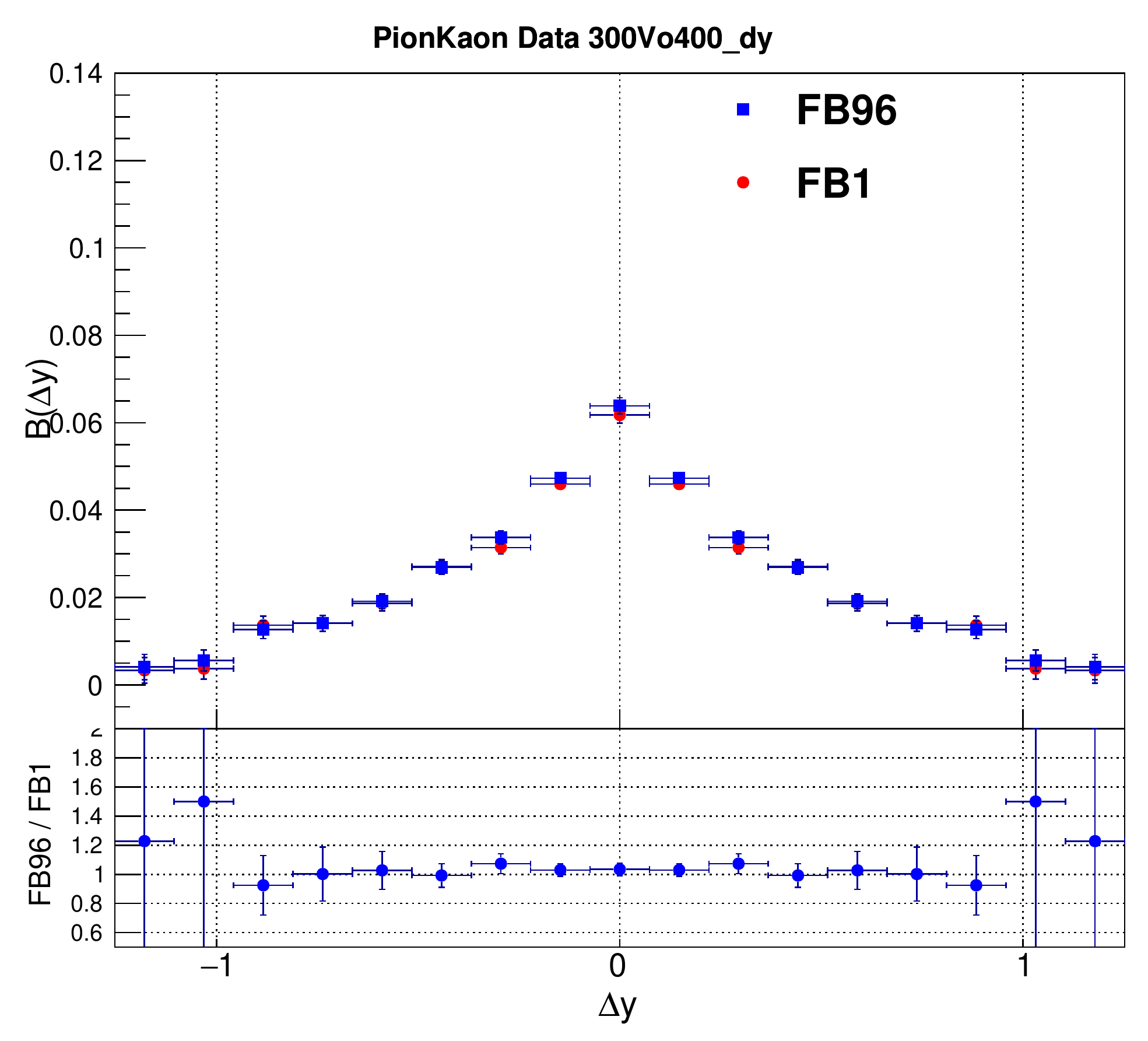}
  \includegraphics[width=0.32\linewidth]{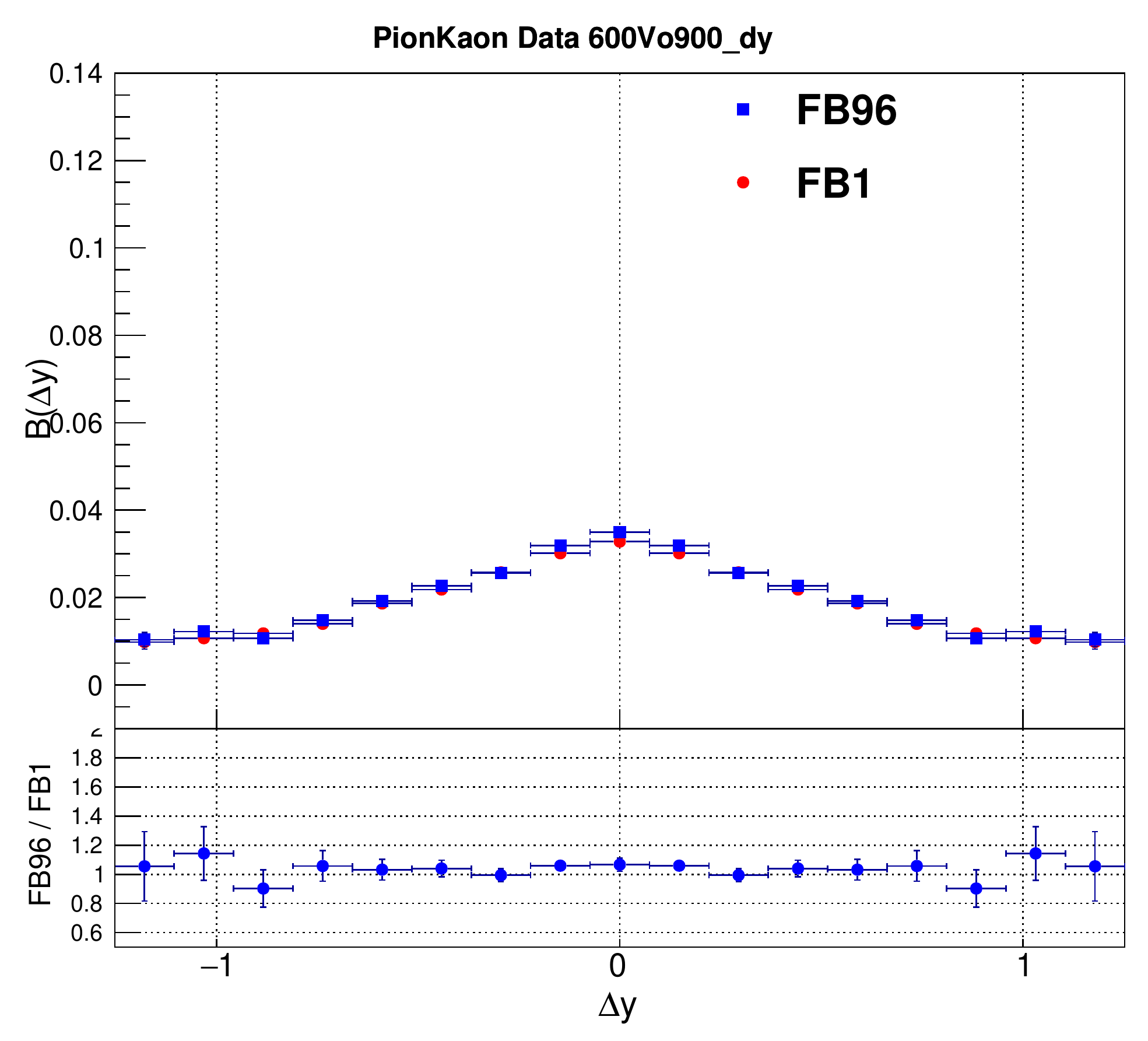}
  \includegraphics[width=0.32\linewidth]{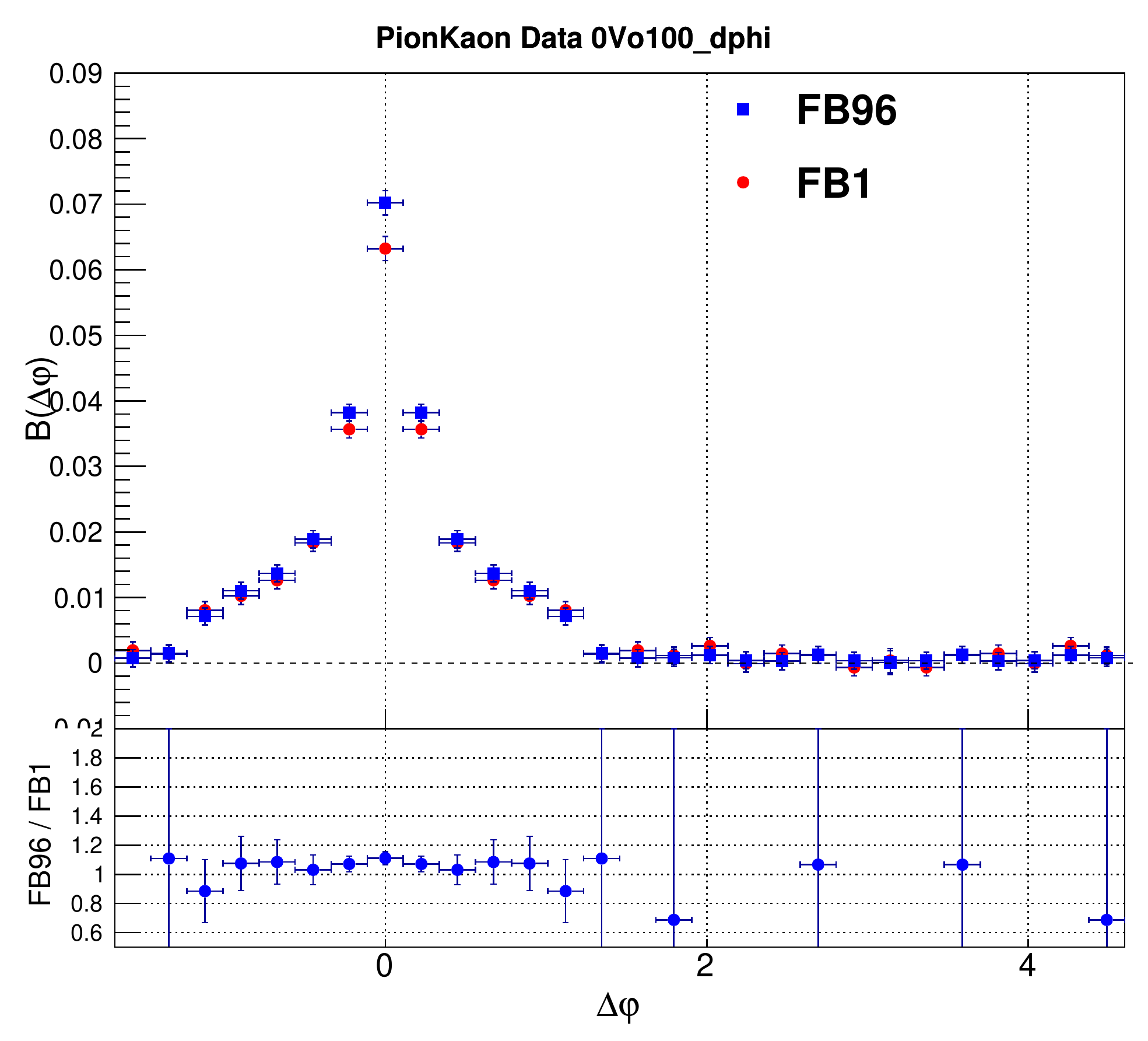}
  \includegraphics[width=0.32\linewidth]{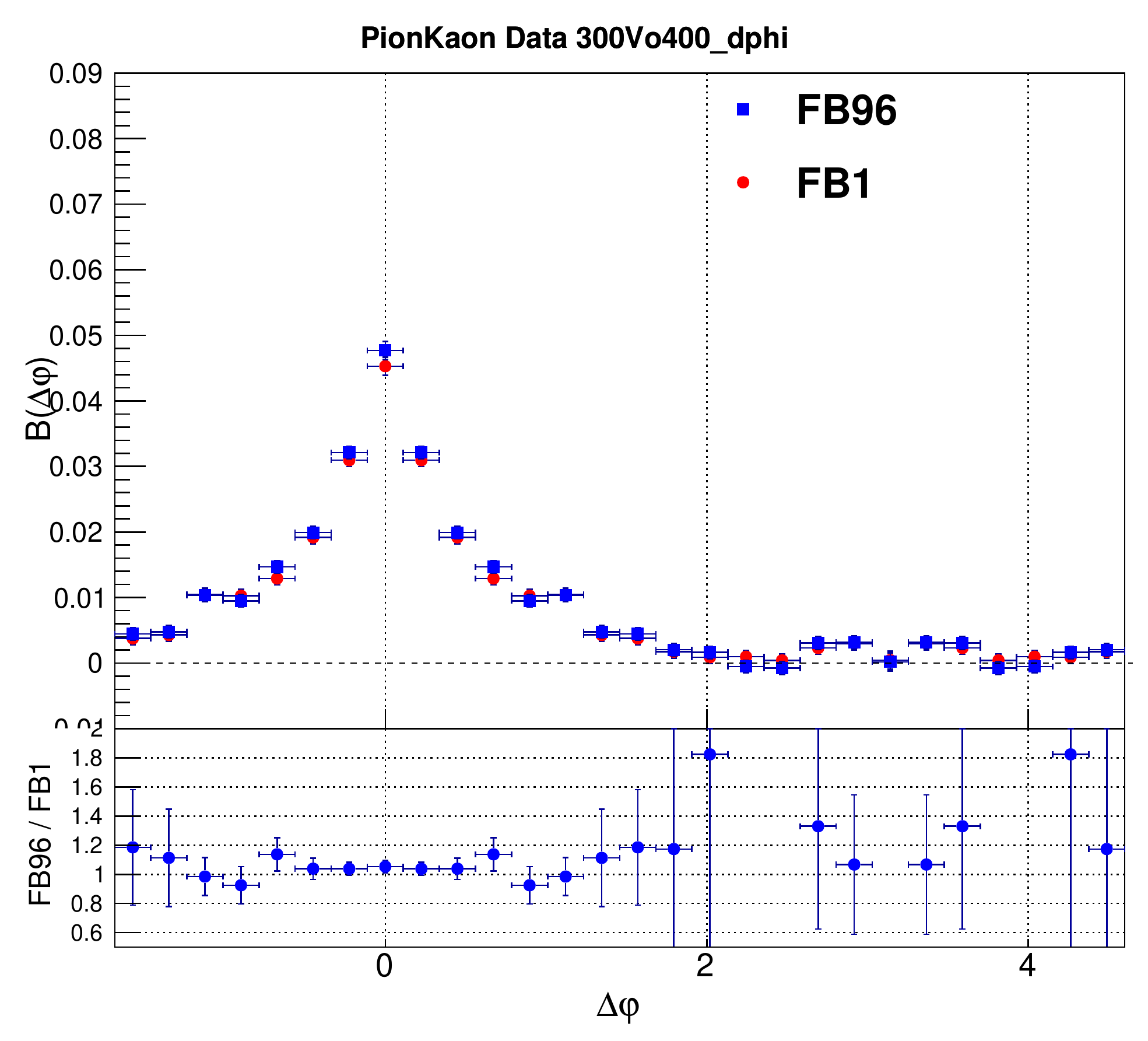}
  \includegraphics[width=0.32\linewidth]{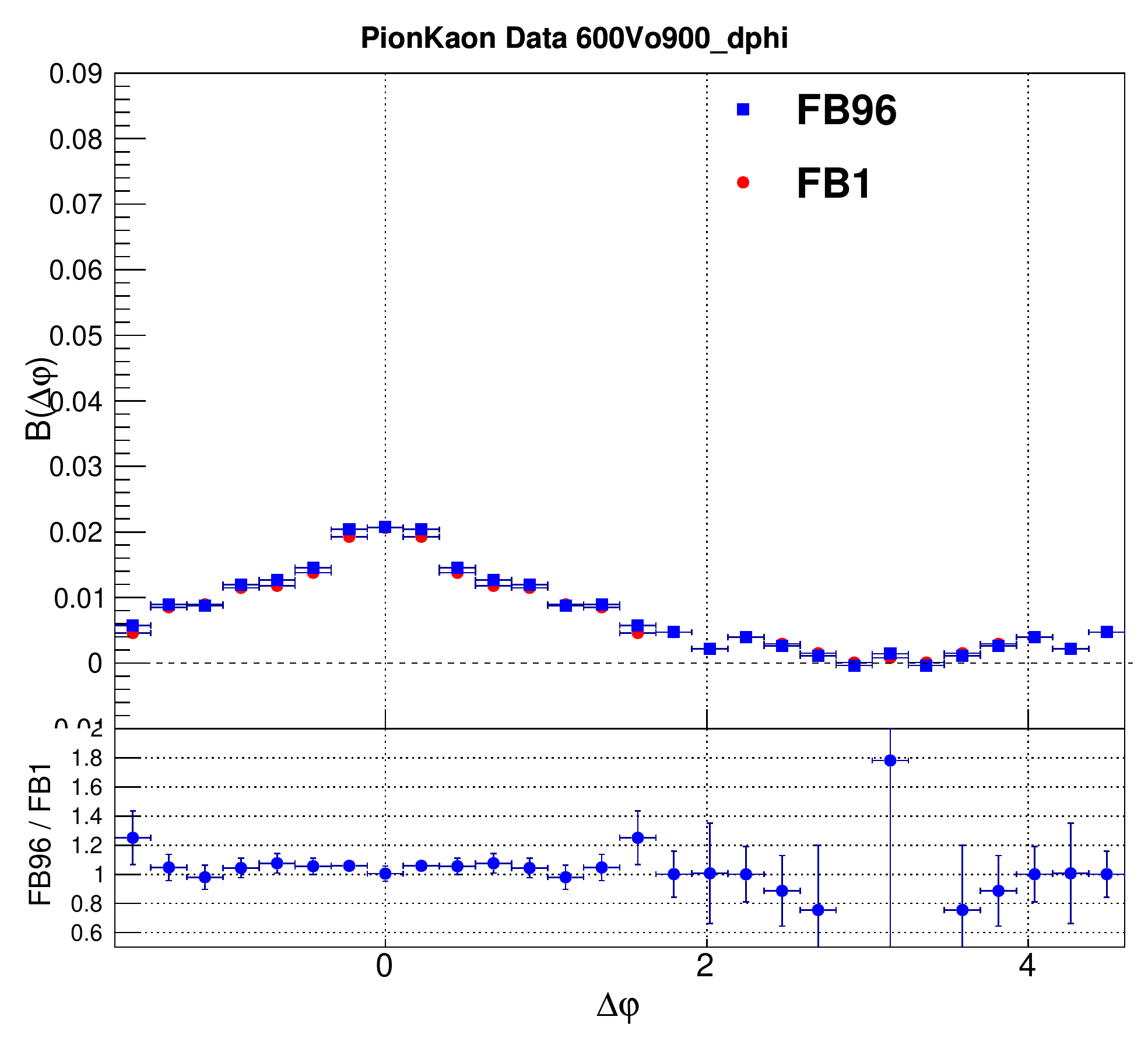}
  \includegraphics[width=0.32\linewidth]{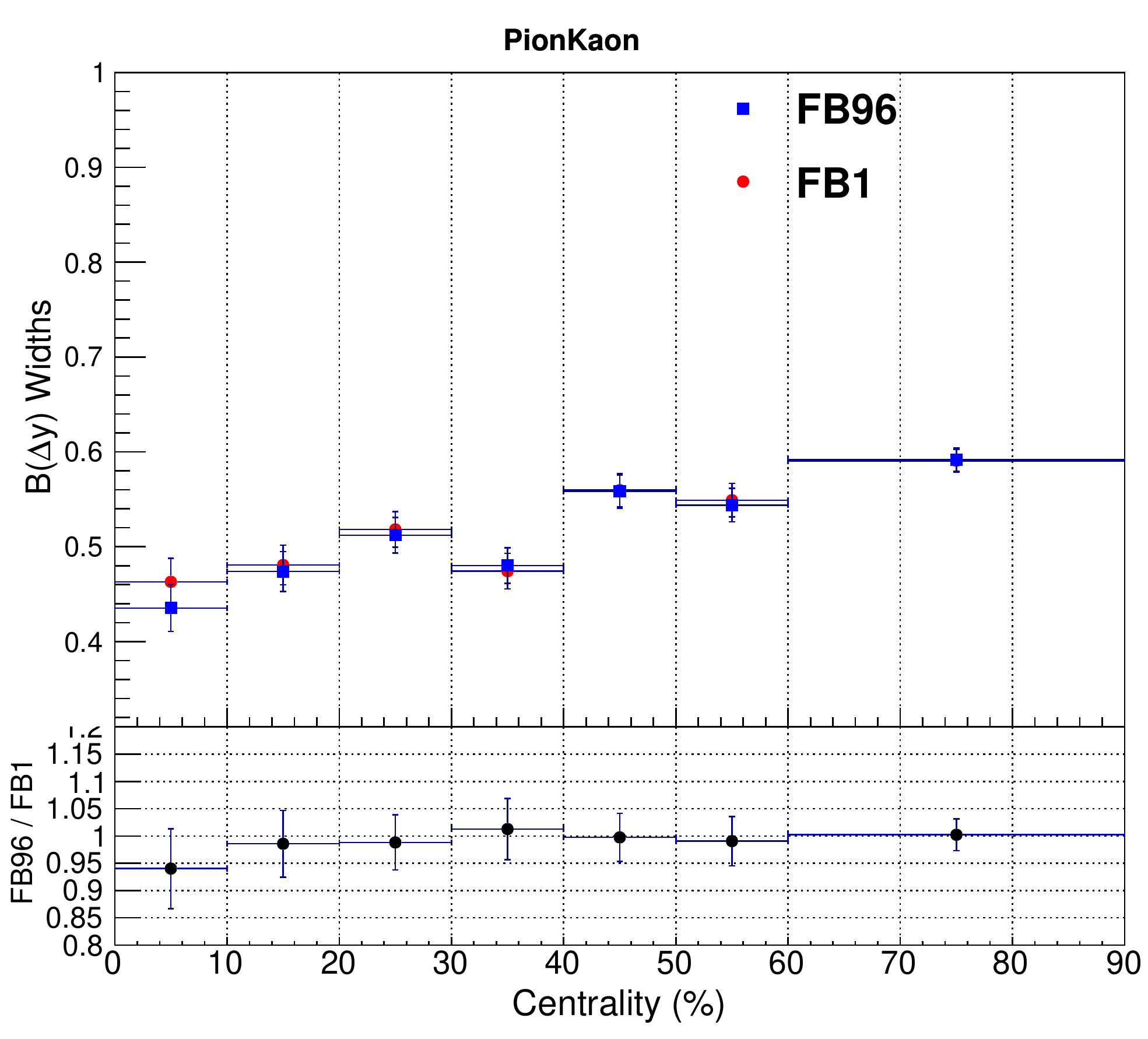}
  \includegraphics[width=0.32\linewidth]{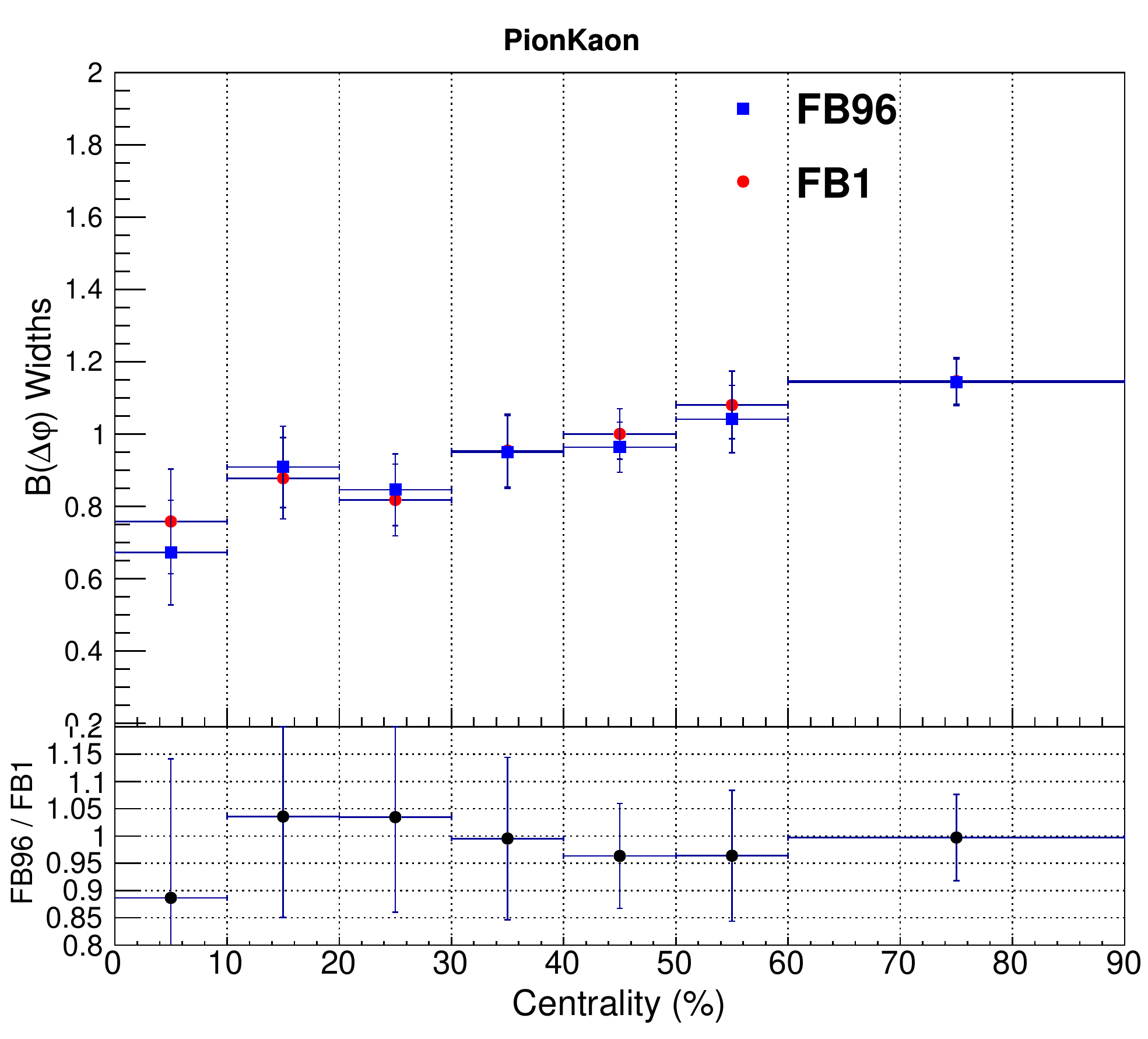}
  \includegraphics[width=0.32\linewidth]{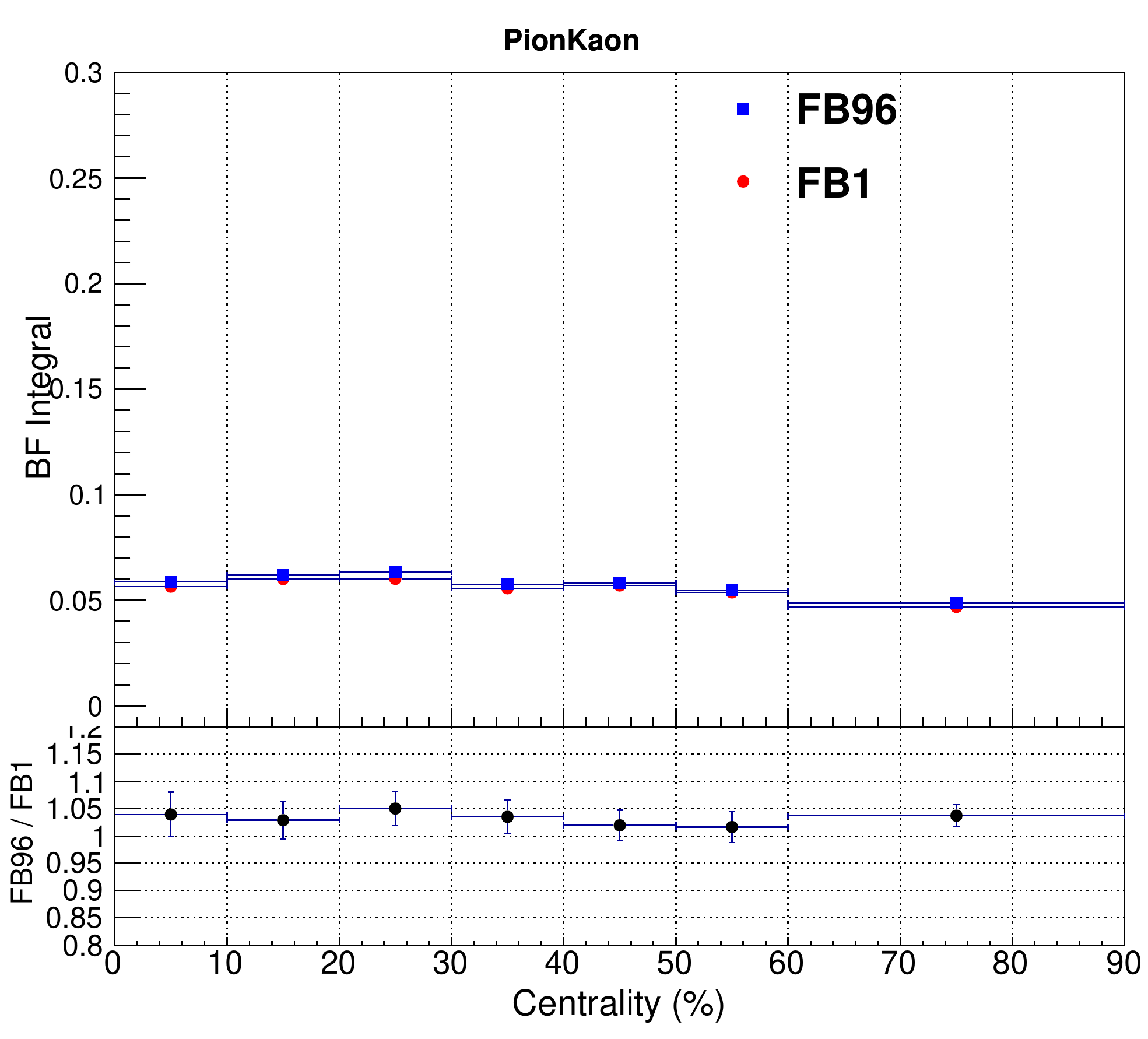}
  \caption{Comparisons of $B^{\pi K}$ projections onto $\Delta y$ (top row) and $\Delta\varphi$ (middle row) axis for selected collision centralities.
  Bottom row: comparisons of $B^{\pi K}$ $\Delta y$ widths (left), $\Delta\varphi$ widths (middle), and integrals (right) as a function of collision centrality between filter-bit 1 and 96.}
  \label{fig:BF_PionKaon_FB1_FB96_Comparison}
\end{figure}

\begin{figure}
\centering
  \includegraphics[width=0.32\linewidth]{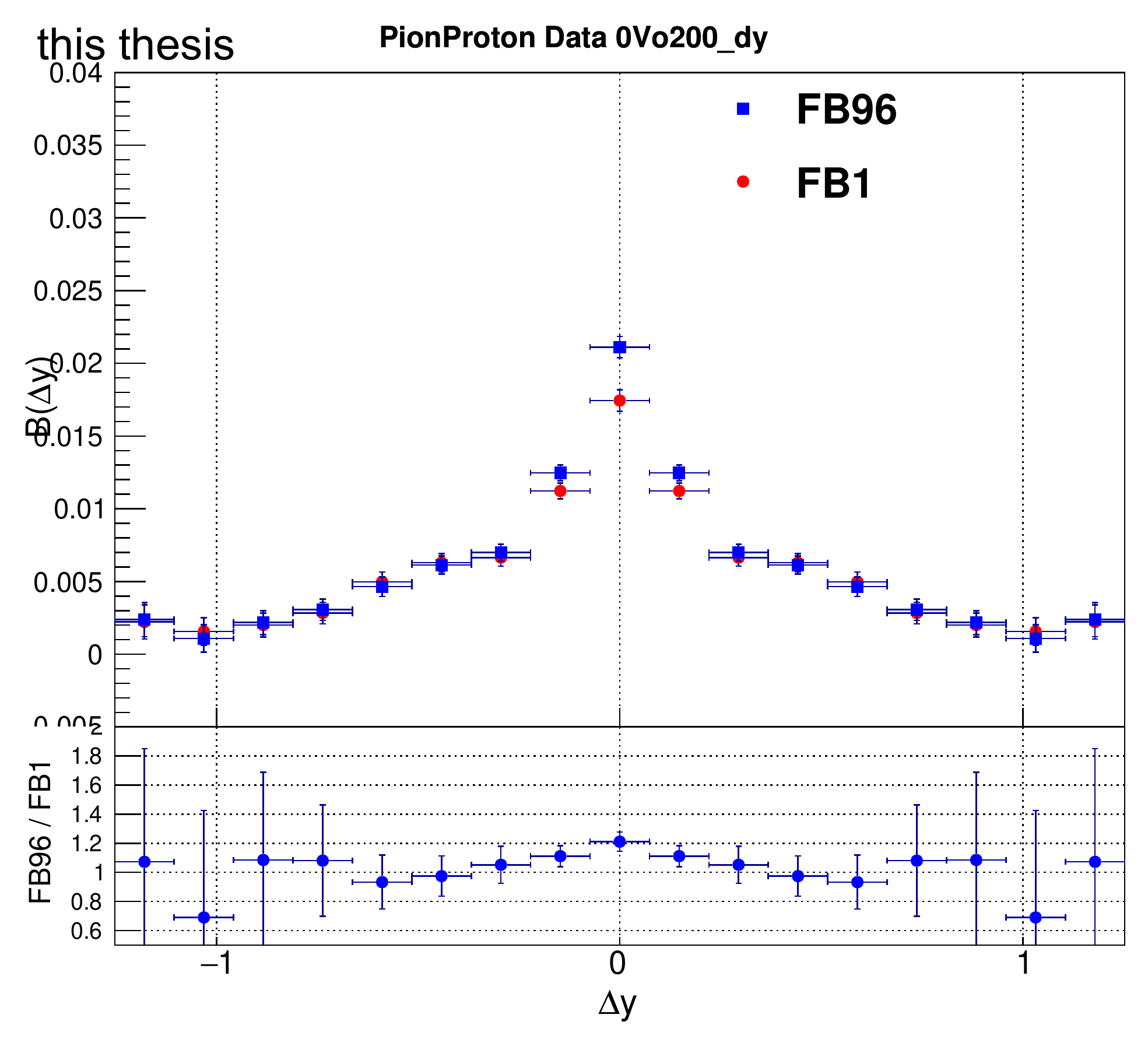}
  \includegraphics[width=0.32\linewidth]{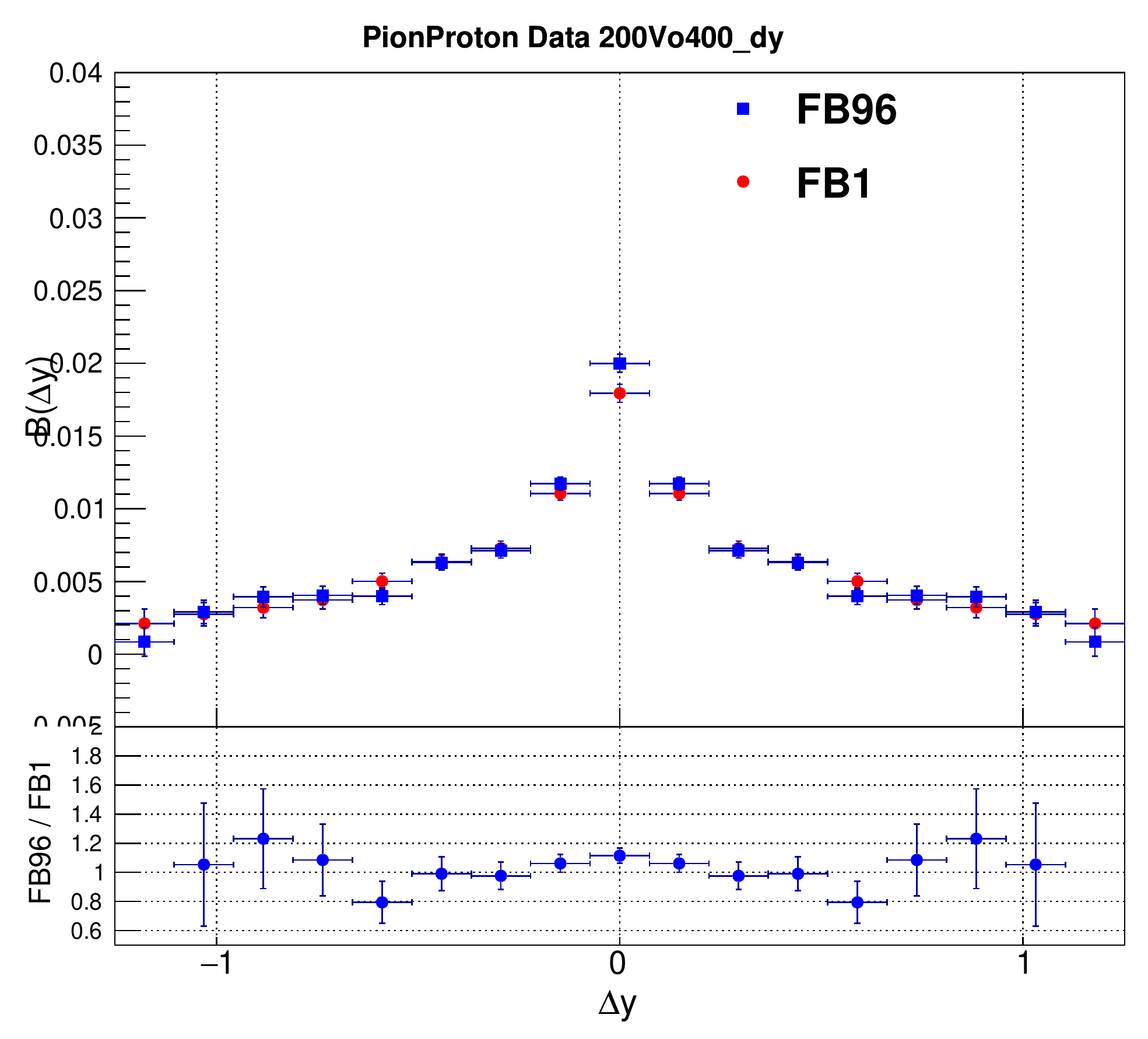}
  \includegraphics[width=0.32\linewidth]{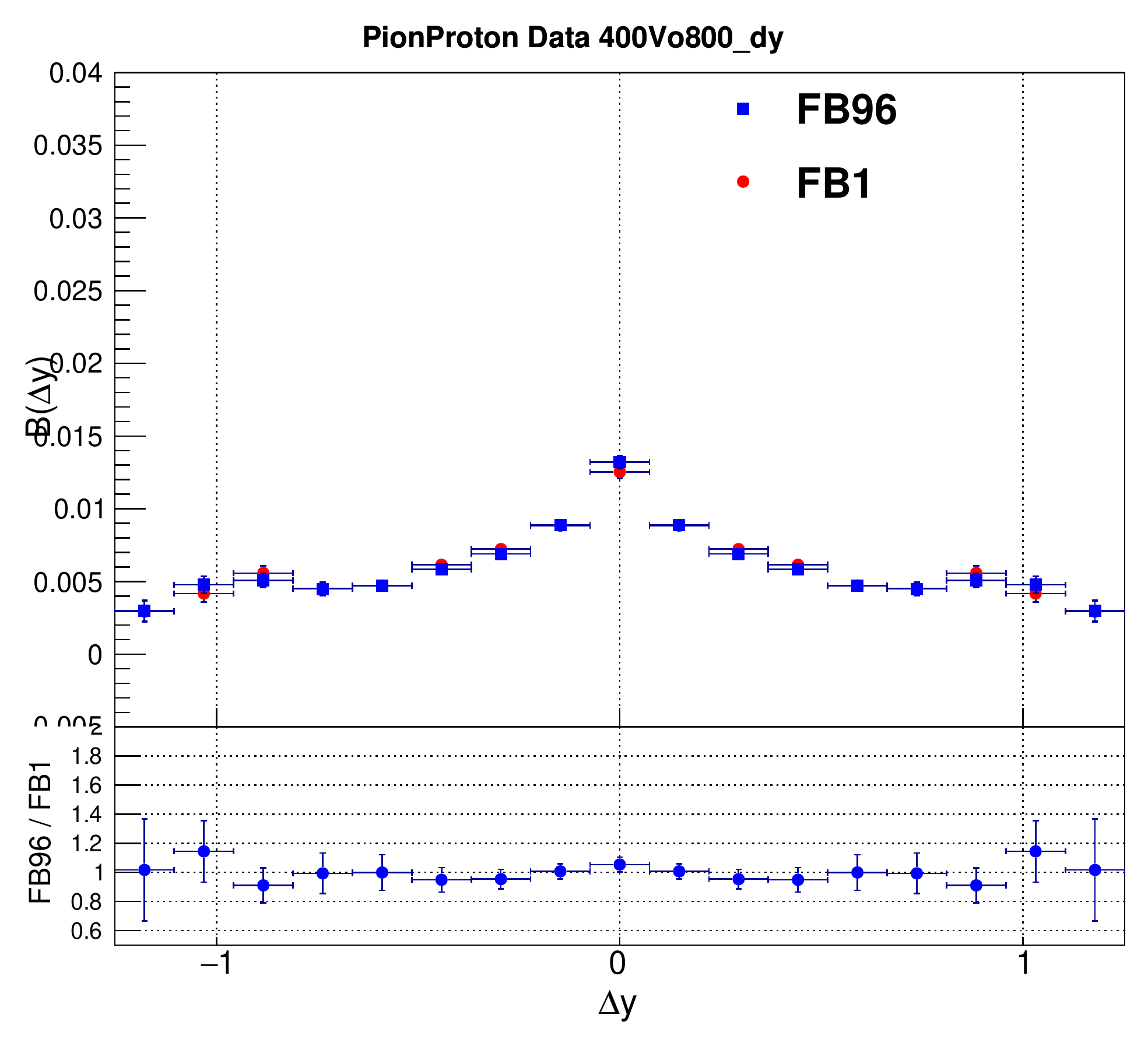}
  \includegraphics[width=0.32\linewidth]{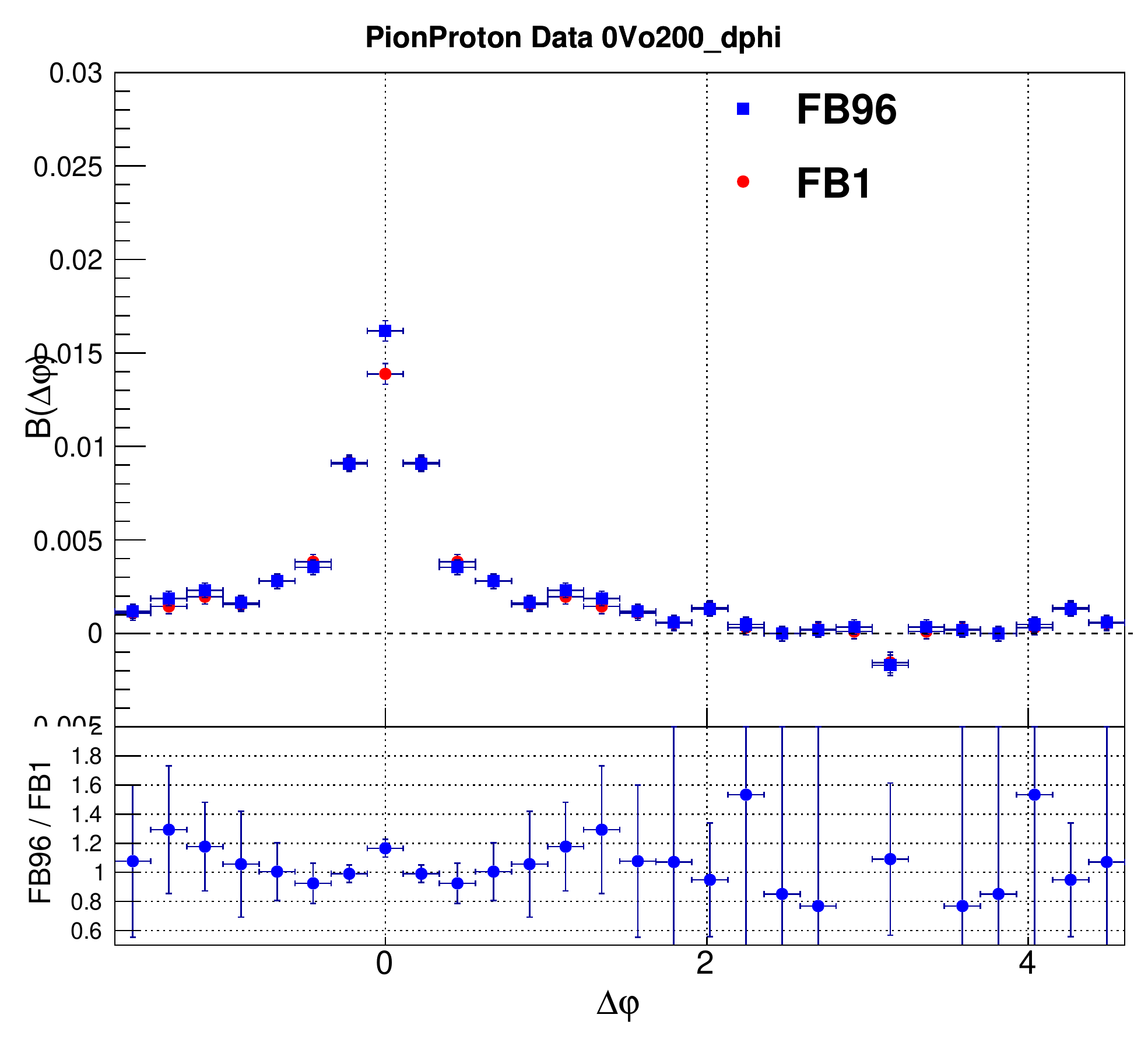}
  \includegraphics[width=0.32\linewidth]{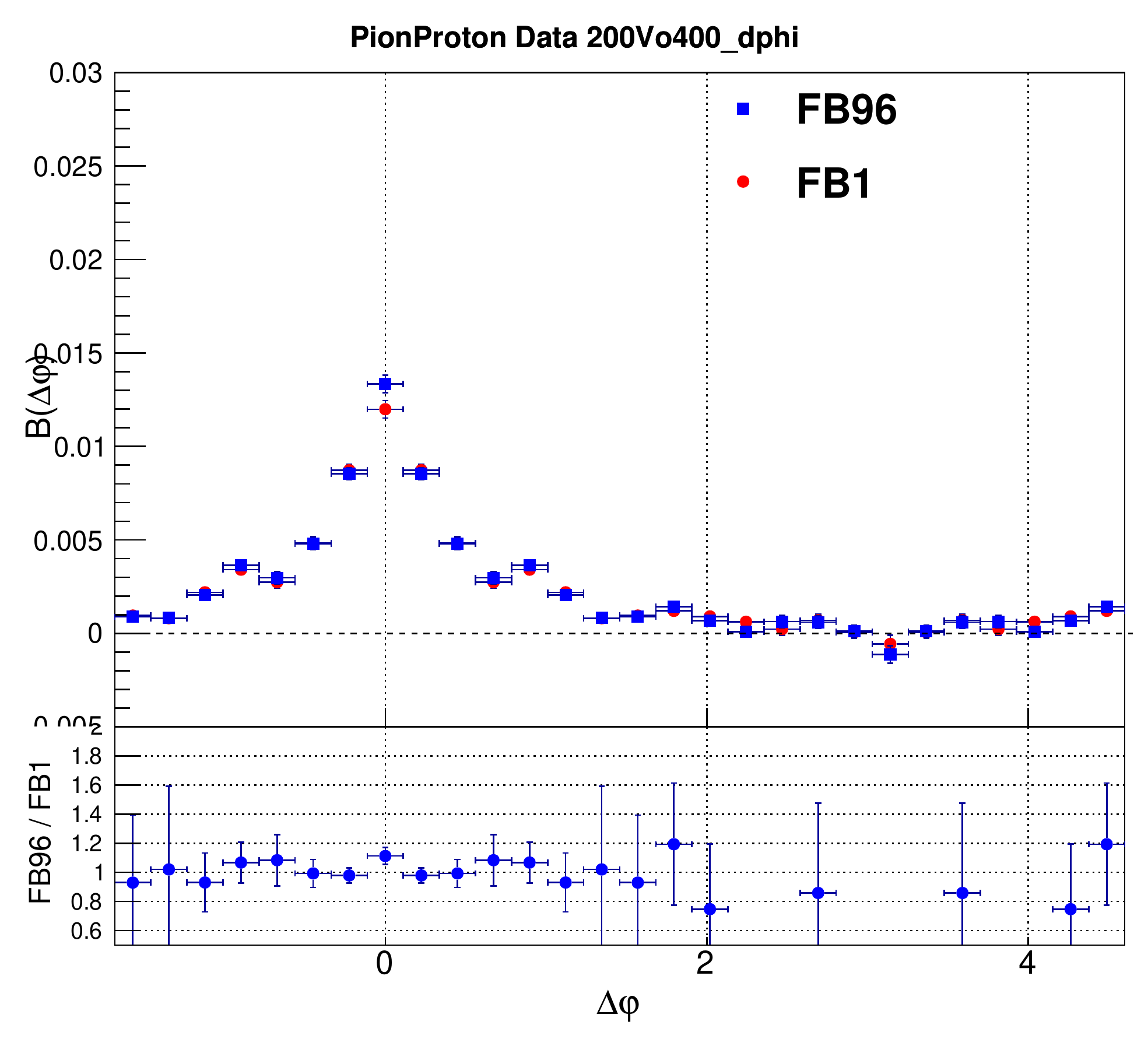}
  \includegraphics[width=0.32\linewidth]{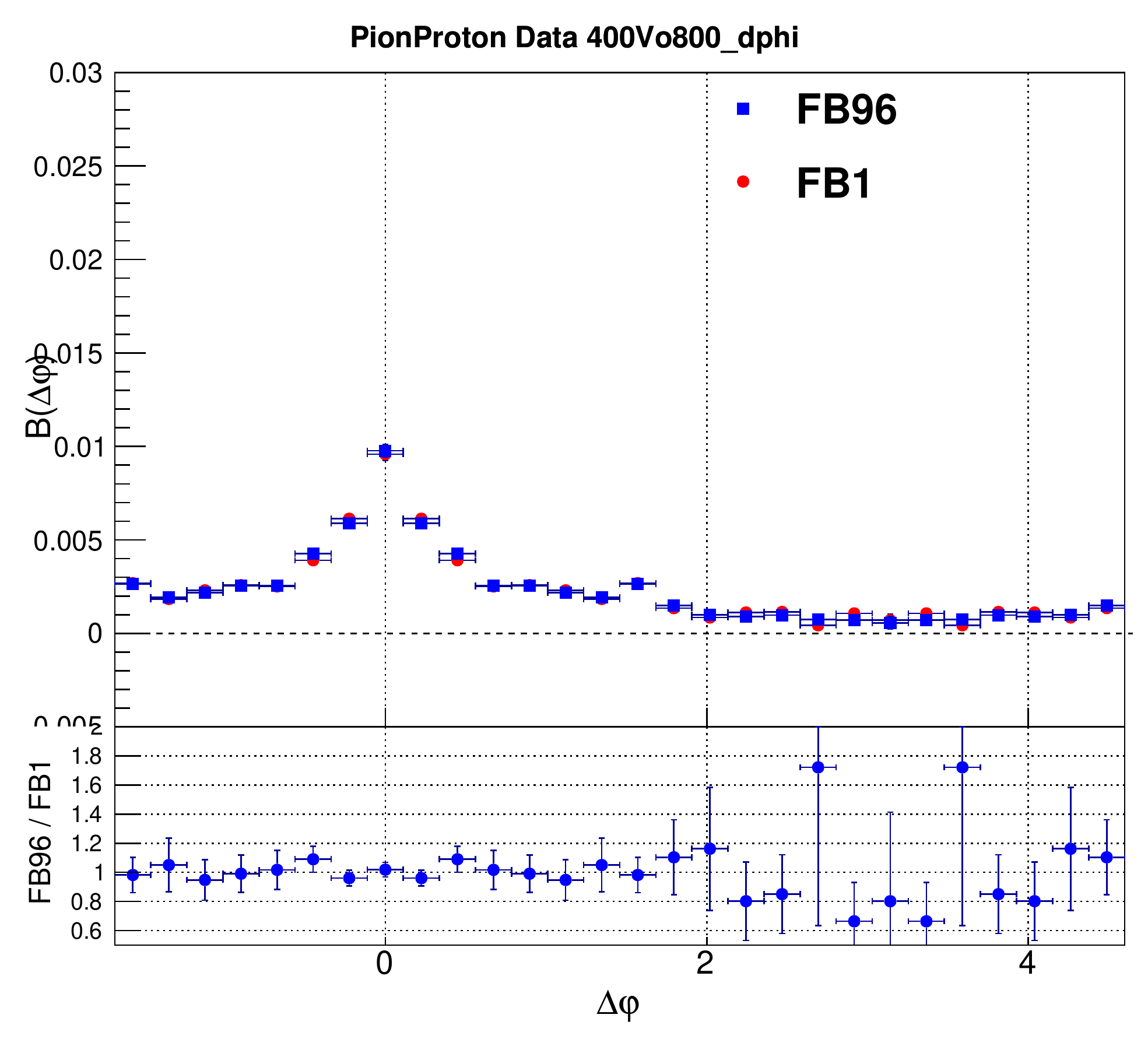}
  \includegraphics[width=0.32\linewidth]{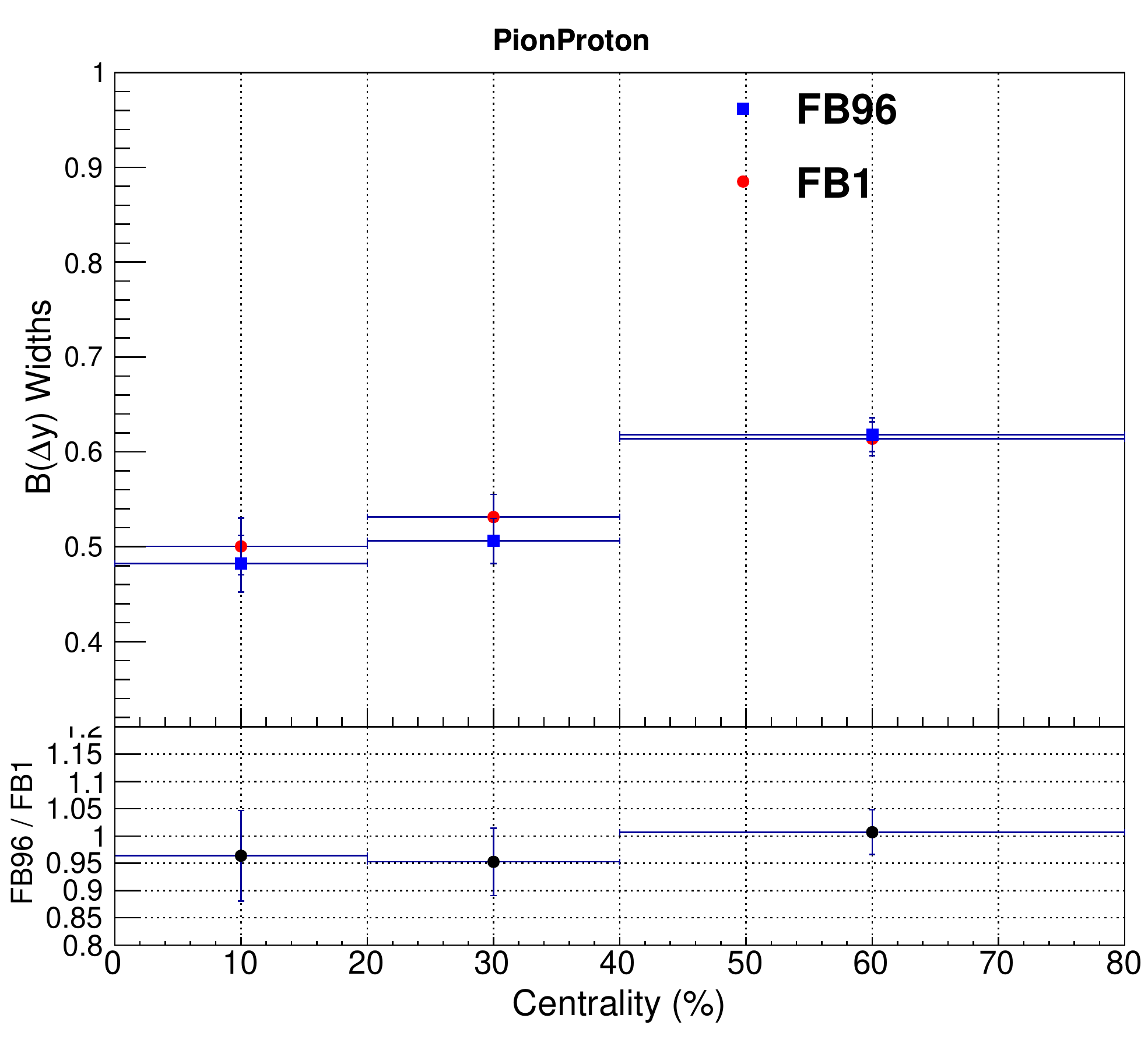}
  \includegraphics[width=0.32\linewidth]{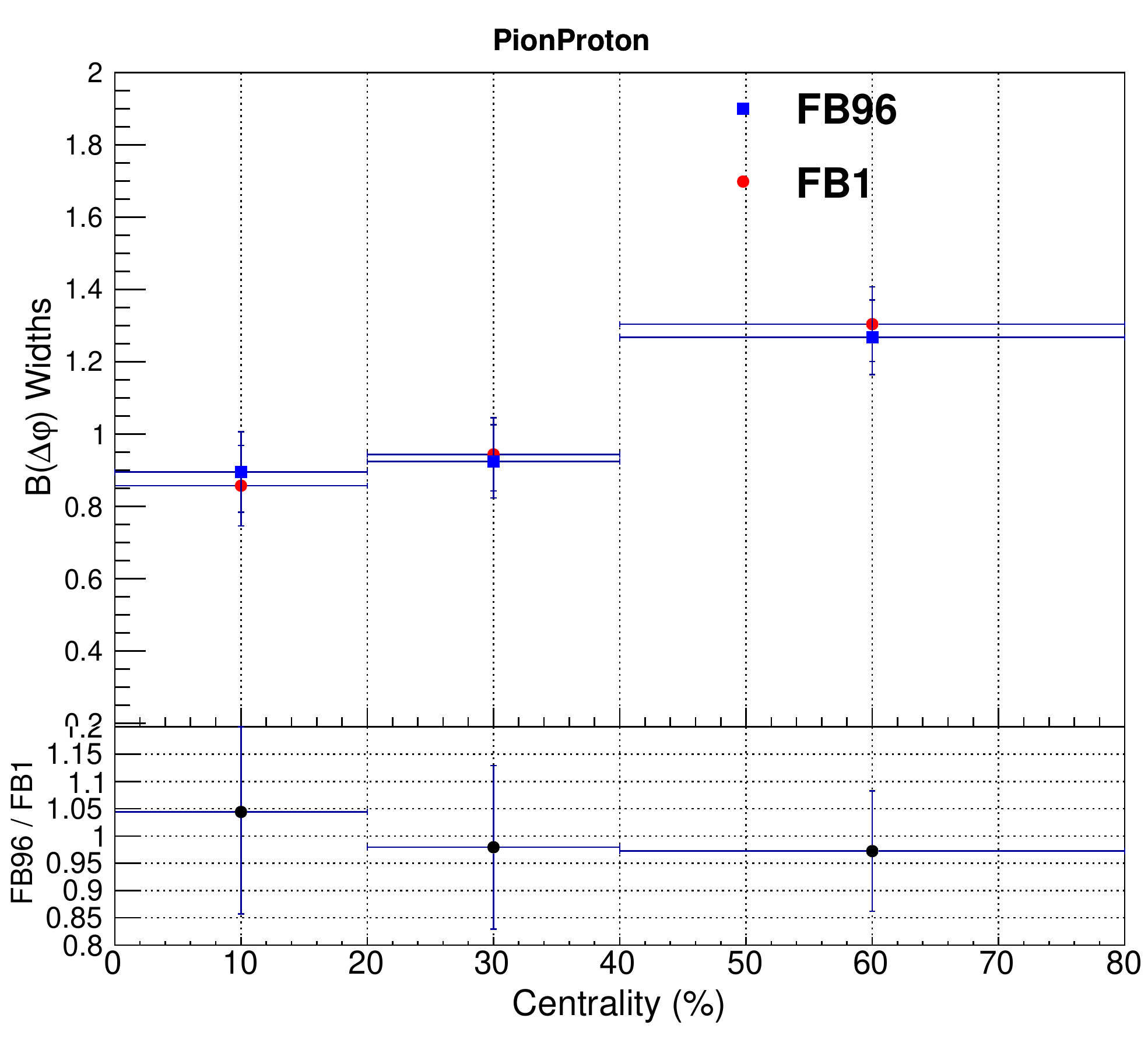}
  \includegraphics[width=0.32\linewidth]{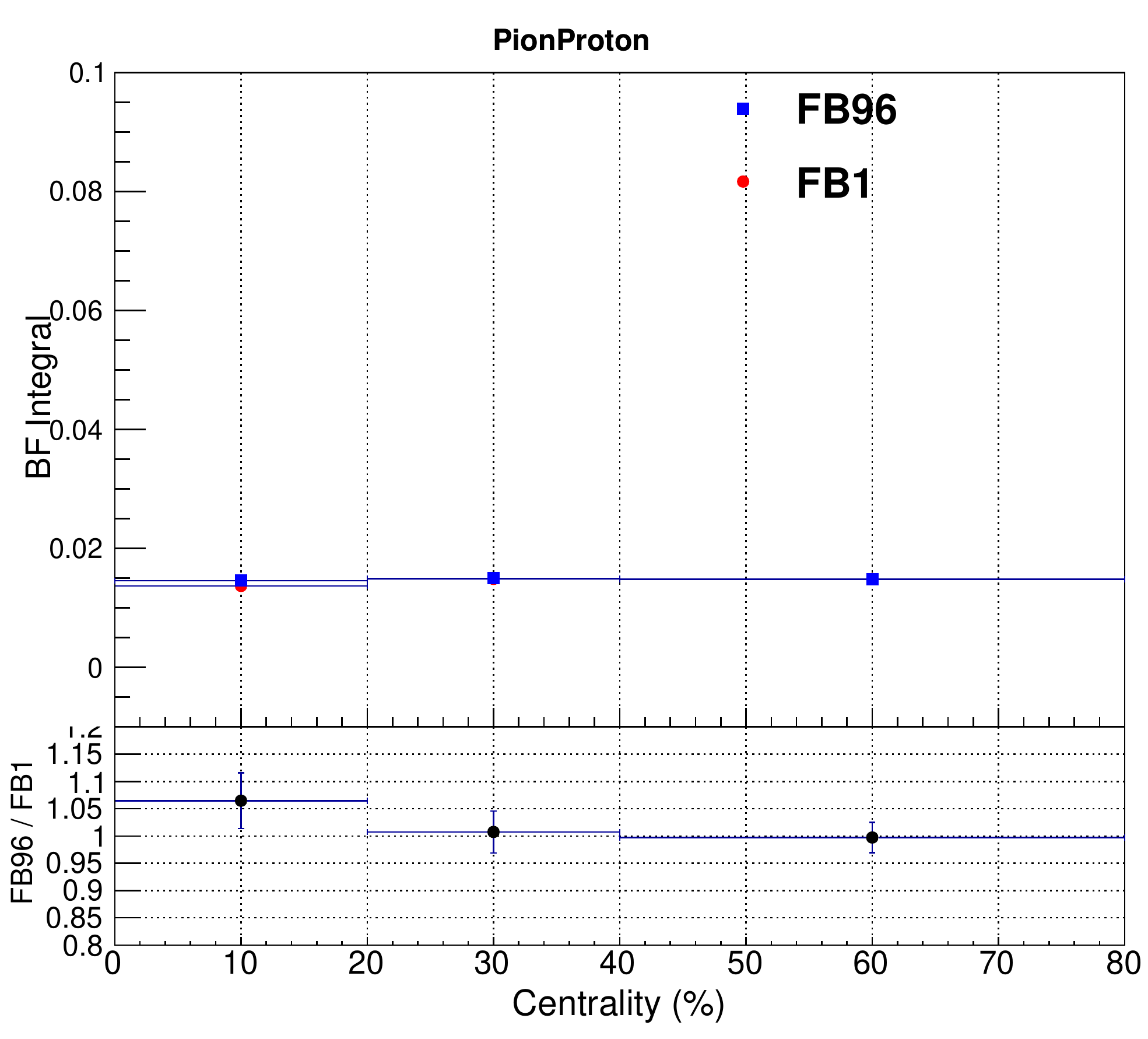}
  \caption{Comparisons of $B^{\pi p}$ projections onto $\Delta y$ (top row) and $\Delta\varphi$ (middle row) axis for selected collision centralities.
  Bottom row: comparisons of $B^{\pi p}$ $\Delta y$ widths (left), $\Delta\varphi$ widths (middle), and integrals (right) as a function of collision centrality between filter-bit 1 and 96.}
  \label{fig:BF_PionProton_FB1_FB96_Comparison}
\end{figure}

\begin{figure}
\centering
  \includegraphics[width=0.32\linewidth]{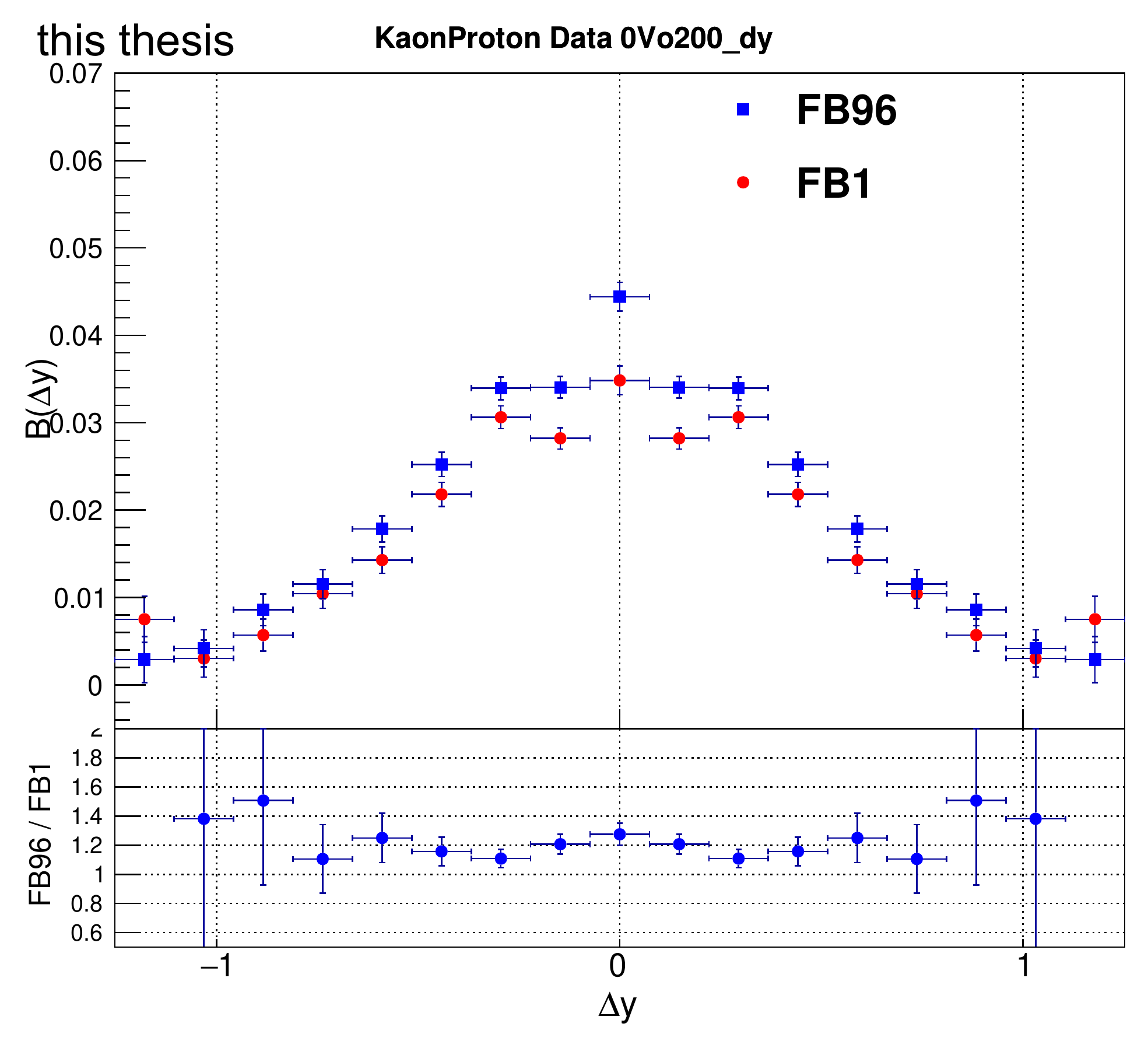}
  \includegraphics[width=0.32\linewidth]{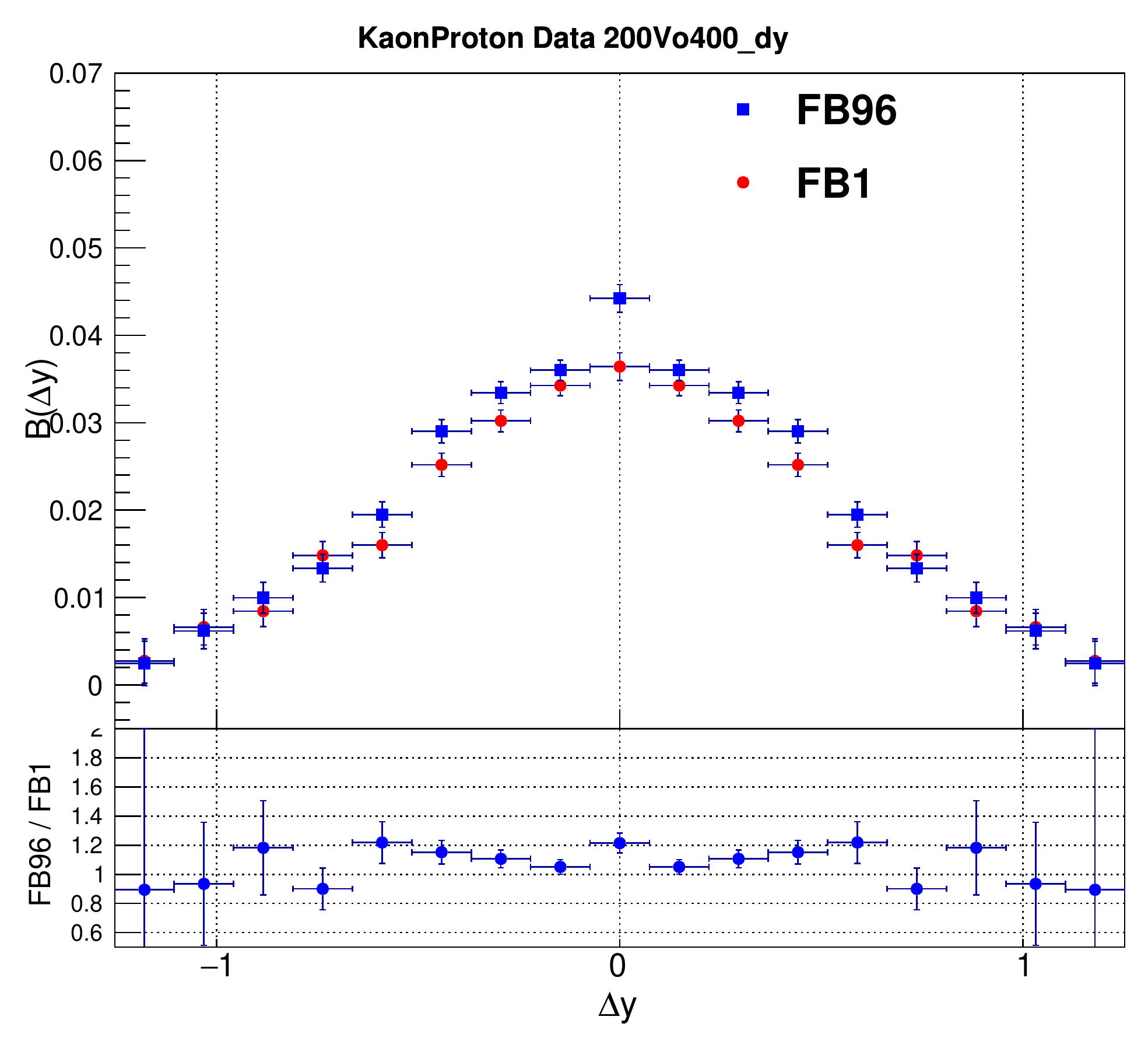}
  \includegraphics[width=0.32\linewidth]{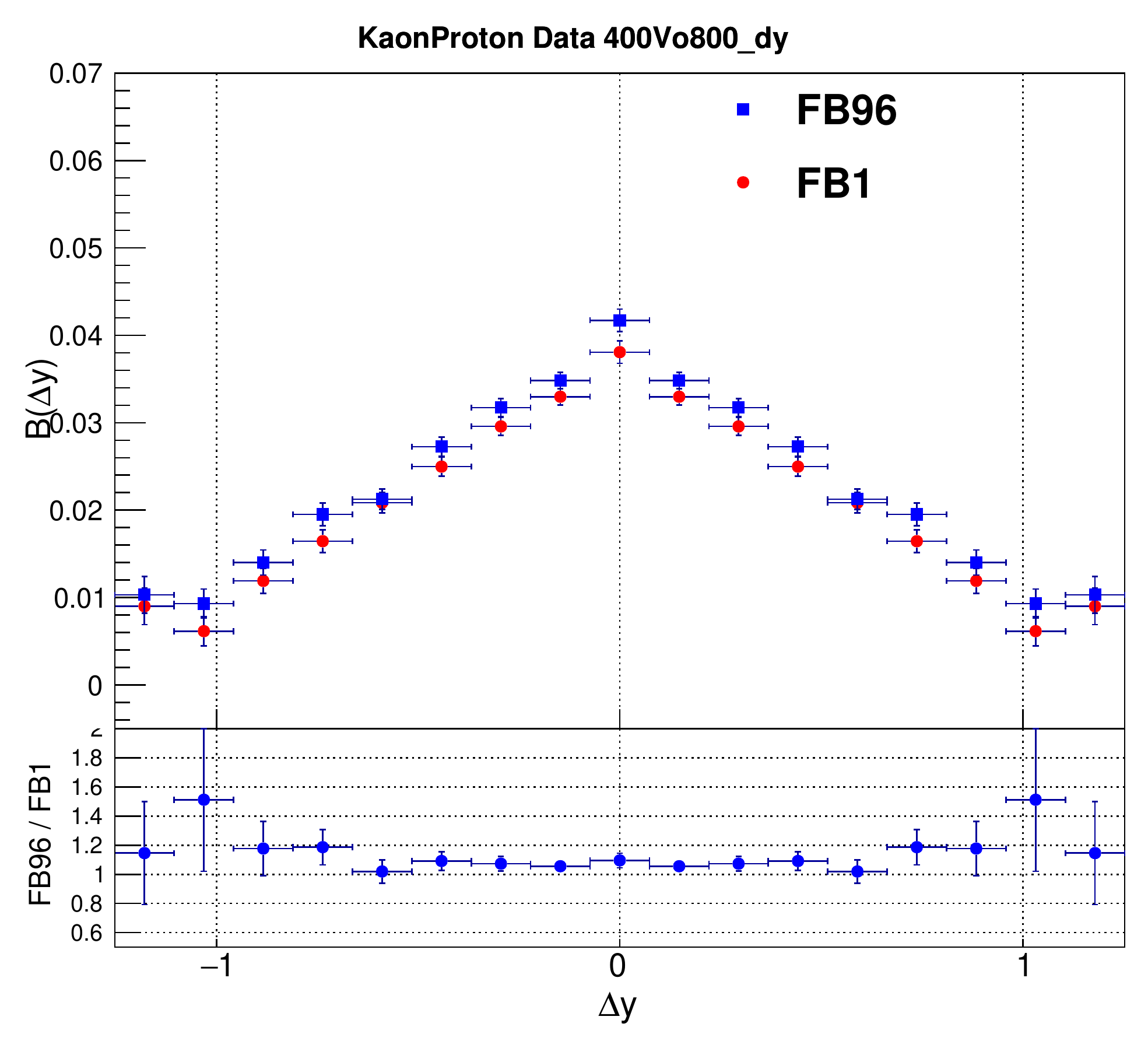}
  \includegraphics[width=0.32\linewidth]{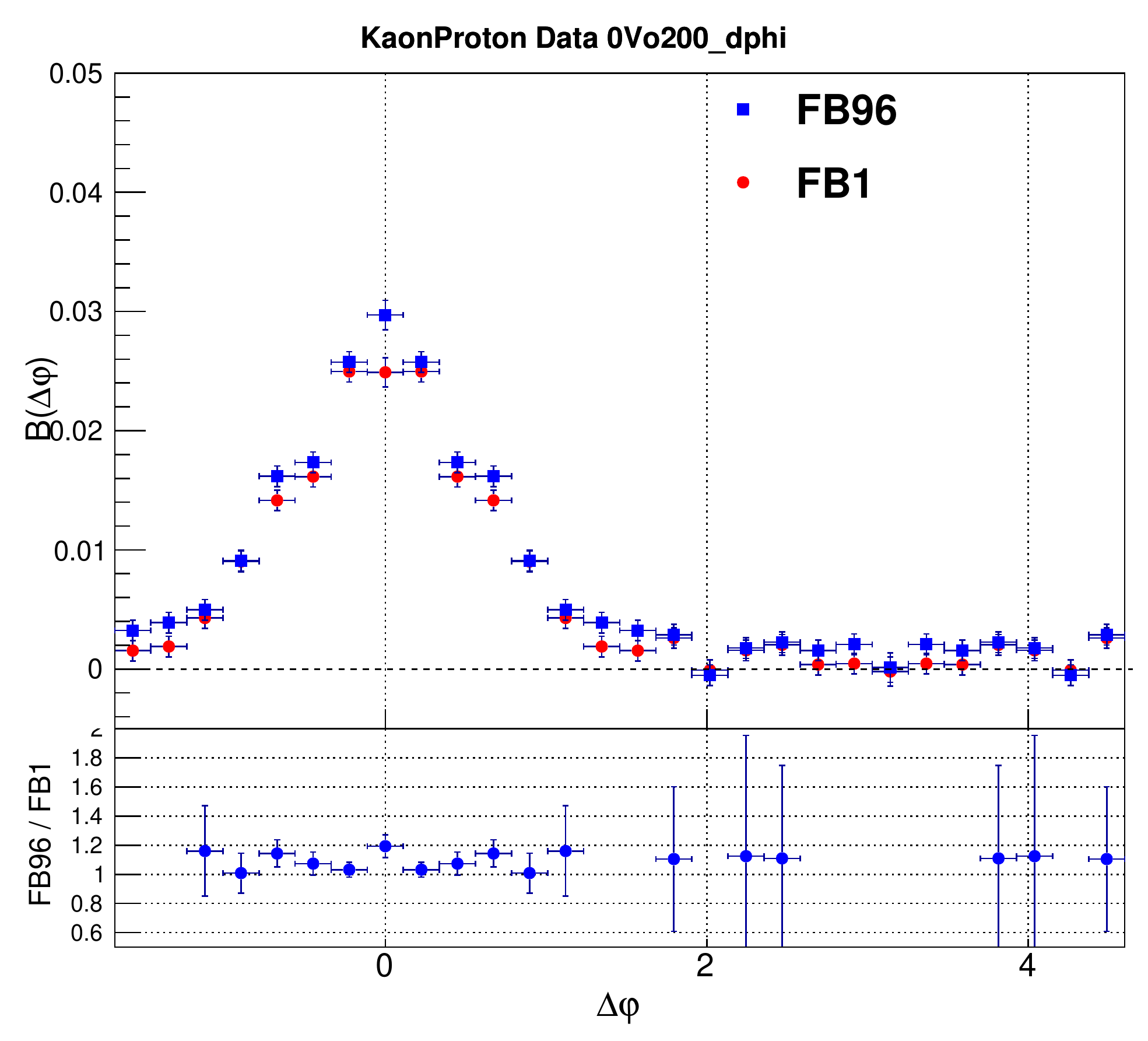}
  \includegraphics[width=0.32\linewidth]{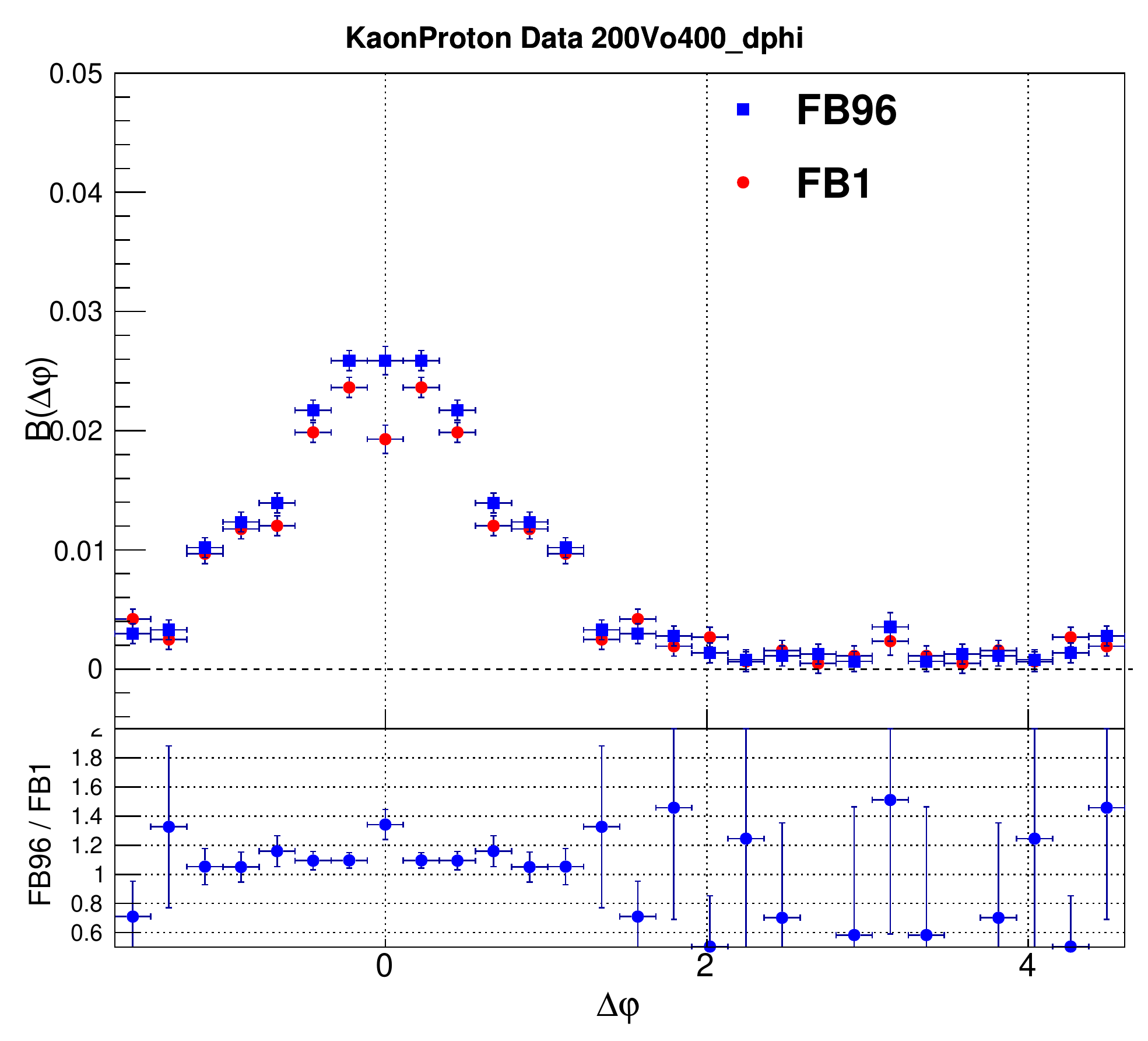}
  \includegraphics[width=0.32\linewidth]{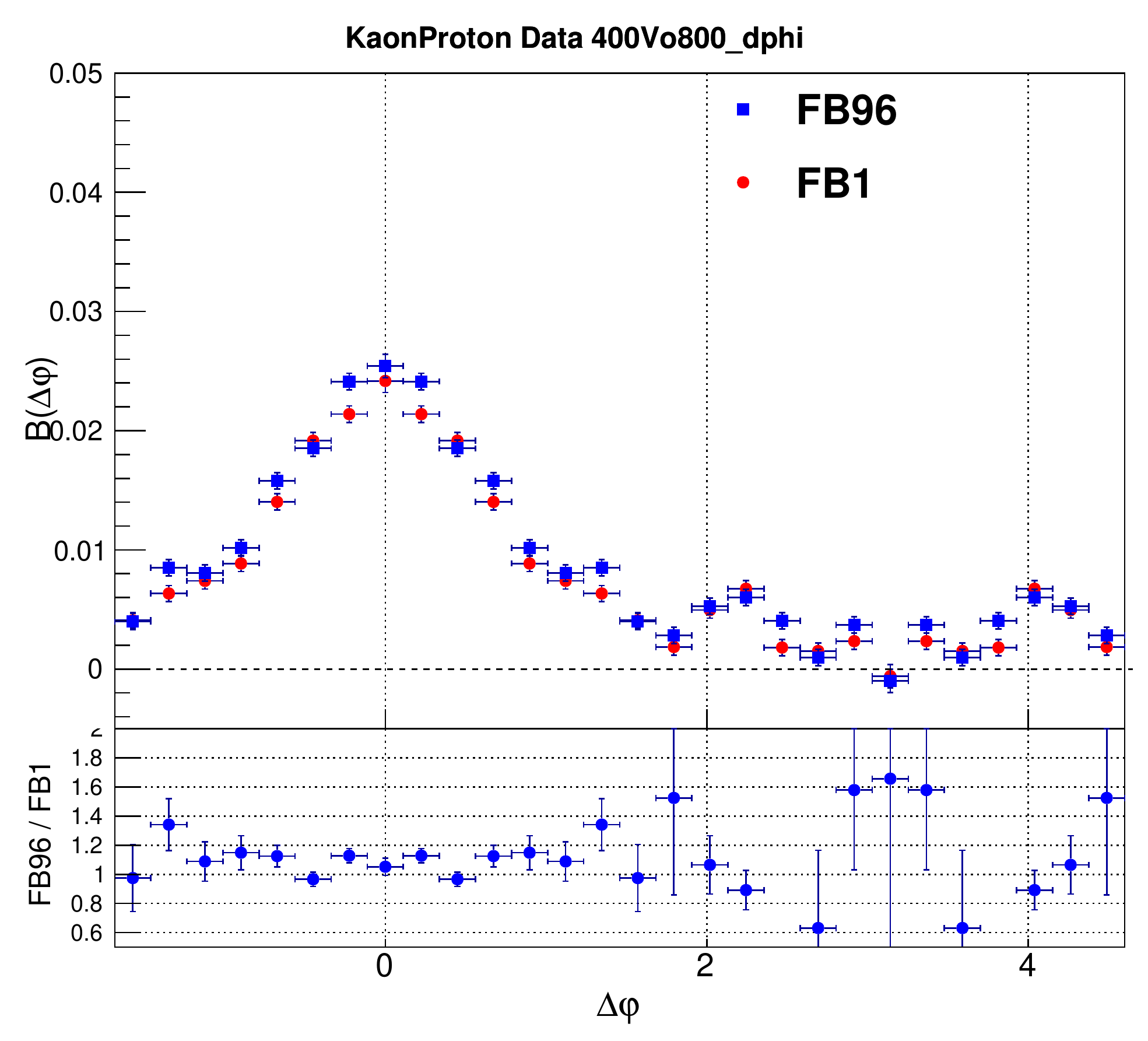}
  \includegraphics[width=0.32\linewidth]{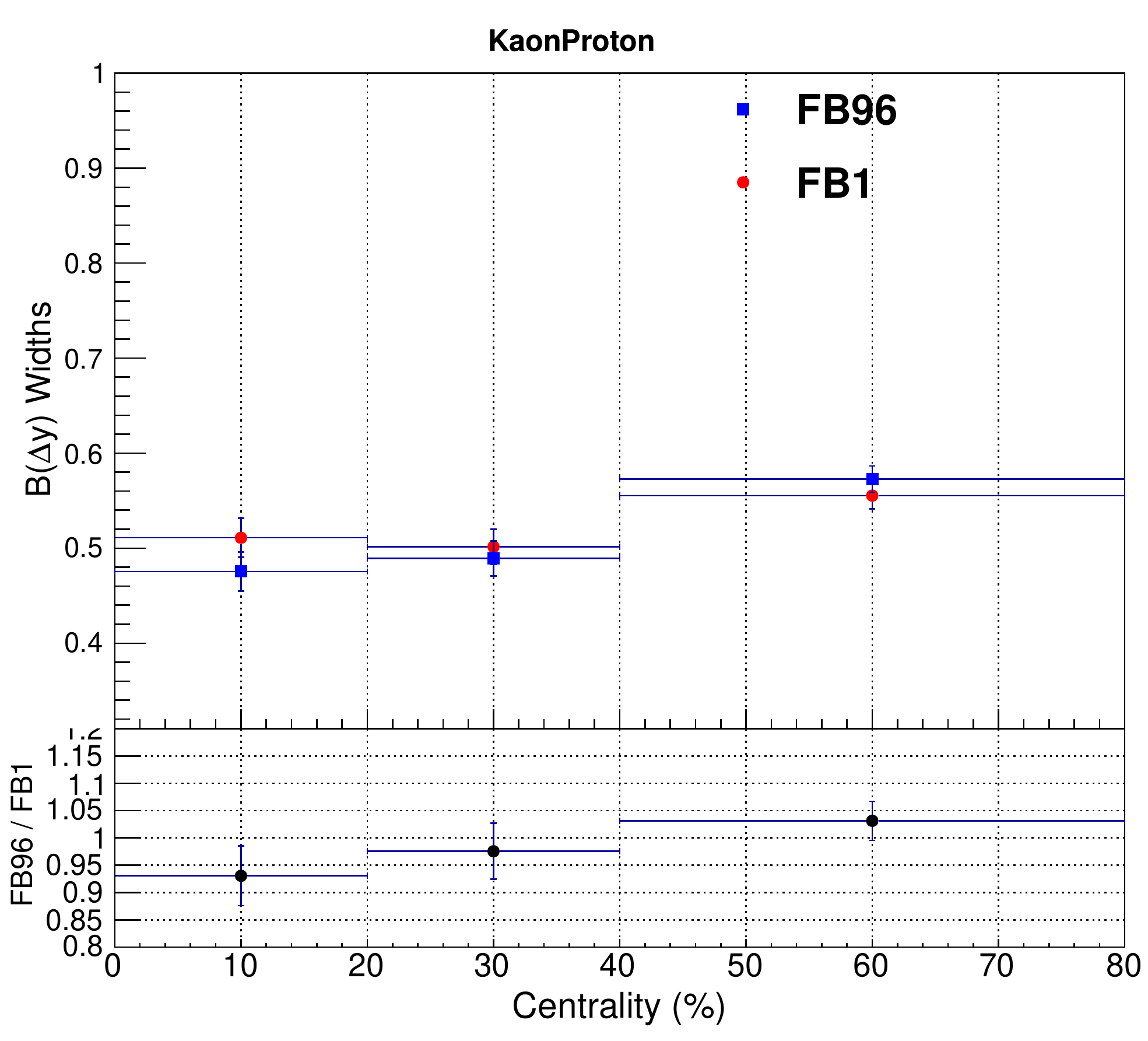}
  \includegraphics[width=0.32\linewidth]{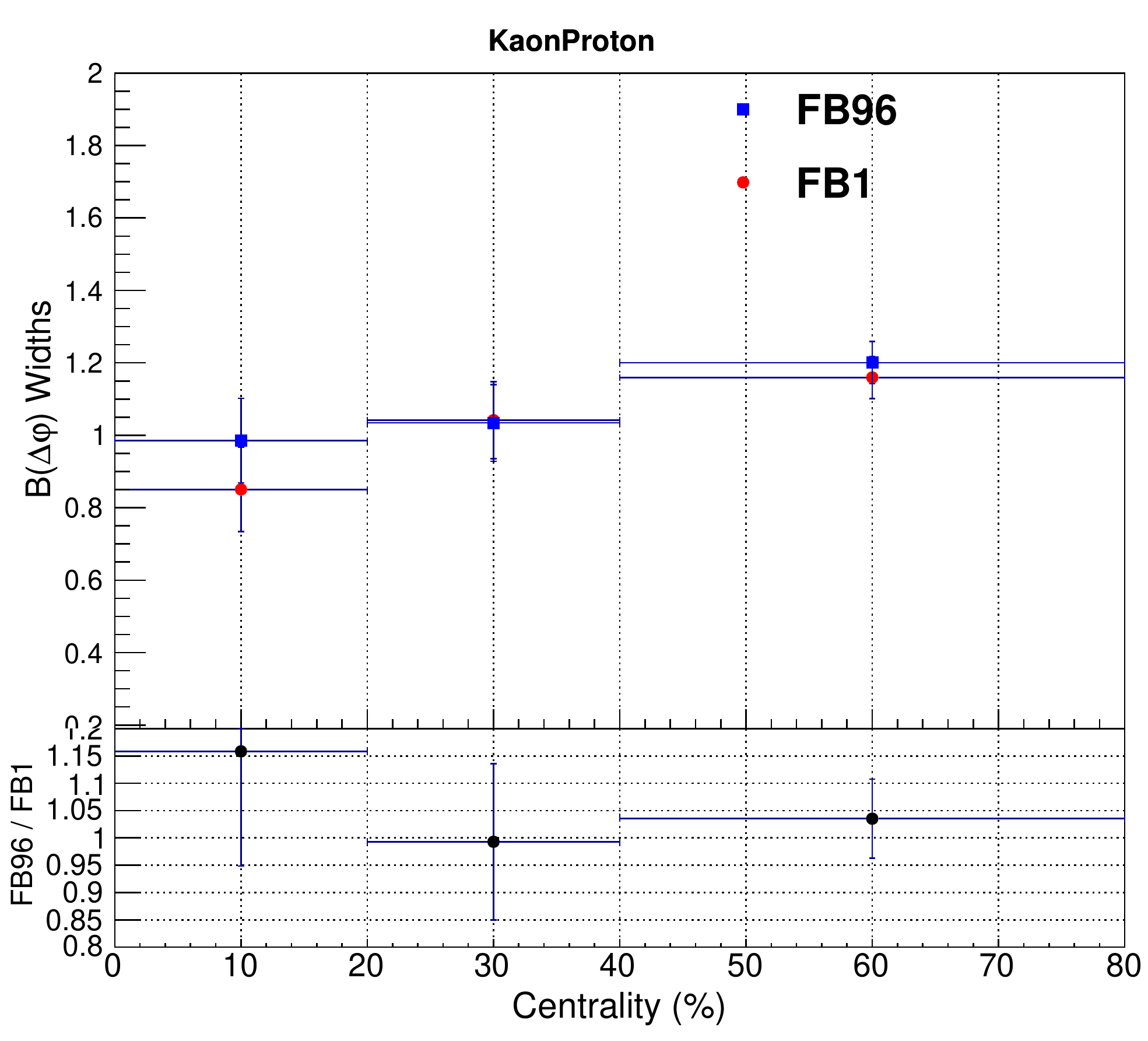}
  \includegraphics[width=0.32\linewidth]{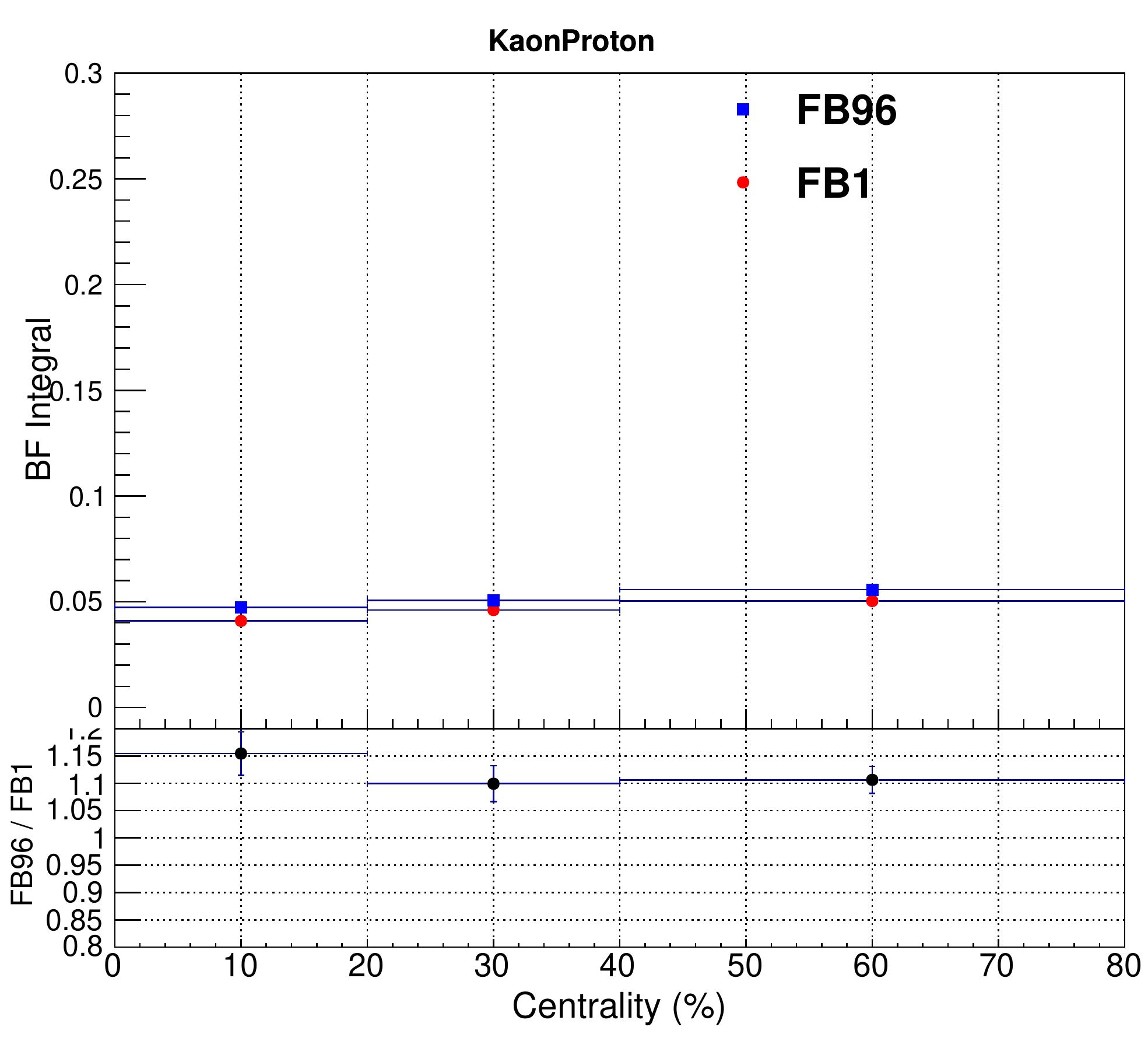}
  \caption{Comparisons of $B^{Kp}$ projections onto $\Delta y$ (top row) and $\Delta\varphi$ (middle row) axis for selected collision centralities.
  Bottom row: comparisons of $B^{Kp}$ $\Delta y$ widths (left), $\Delta\varphi$ widths (middle), and integrals (right) as a function of collision centrality between filter-bit 1 and 96.}
  \label{fig:BF_KaonProton_FB1_FB96_Comparison}
\end{figure}

The differences between BFs obtained with filter-bit 1 and 96, as shown in \\
Figures~\ref{fig:BF_PionPion_FB1_FB96_Comparison},~\ref{fig:BF_KaonKaon_FB1_FB96_Comparison},~\ref{fig:BF_ProtonProton_FB1_FB96_Comparison},~\ref{fig:BF_PionKaon_FB1_FB96_Comparison},~\ref{fig:BF_PionProton_FB1_FB96_Comparison},~\ref{fig:BF_KaonProton_FB1_FB96_Comparison}, where most points in the projection, width and integral plots comparing BFs of filter-bit 1 and 96 are within two standard deviation of statistical uncertainties.
Larger differences between BFs obtained with filter-bits 1 and 96 are observed at $\Delta y=0$ and $\Delta\varphi=0$ in the projections of $B^{\pi\pi}$, especially for most central events, as shown in Fig.~\ref{fig:BF_PionPion_FB1_FB96_Comparison}. These are due to differences on CFs of LS at $\Delta y=0$ and $\Delta\varphi=0$ between filter-bit 1 and 96, as shown in Figure~\ref{fig:BF_PionPion_FB1_FB96_Comparison_US_LS}, which are probably because filter-bit 96 incorporates the refit towards the ITS. Thus, in the final $B^{\pi\pi}$ and $B^{pp}$ results, the differences on 
$(\Delta y, \Delta\varphi)=(0,0)$ bin between filter-bit 1 and 96 are taken as an additional systematic uncertainty for $(\Delta y, \Delta\varphi)=(0,0)$ bin.

\clearpage
	\chapter{Systematic Uncertainties}\label{chap:SystematicUncertainties}

\begin{figure}
\centering
  \includegraphics[width=0.32\linewidth]{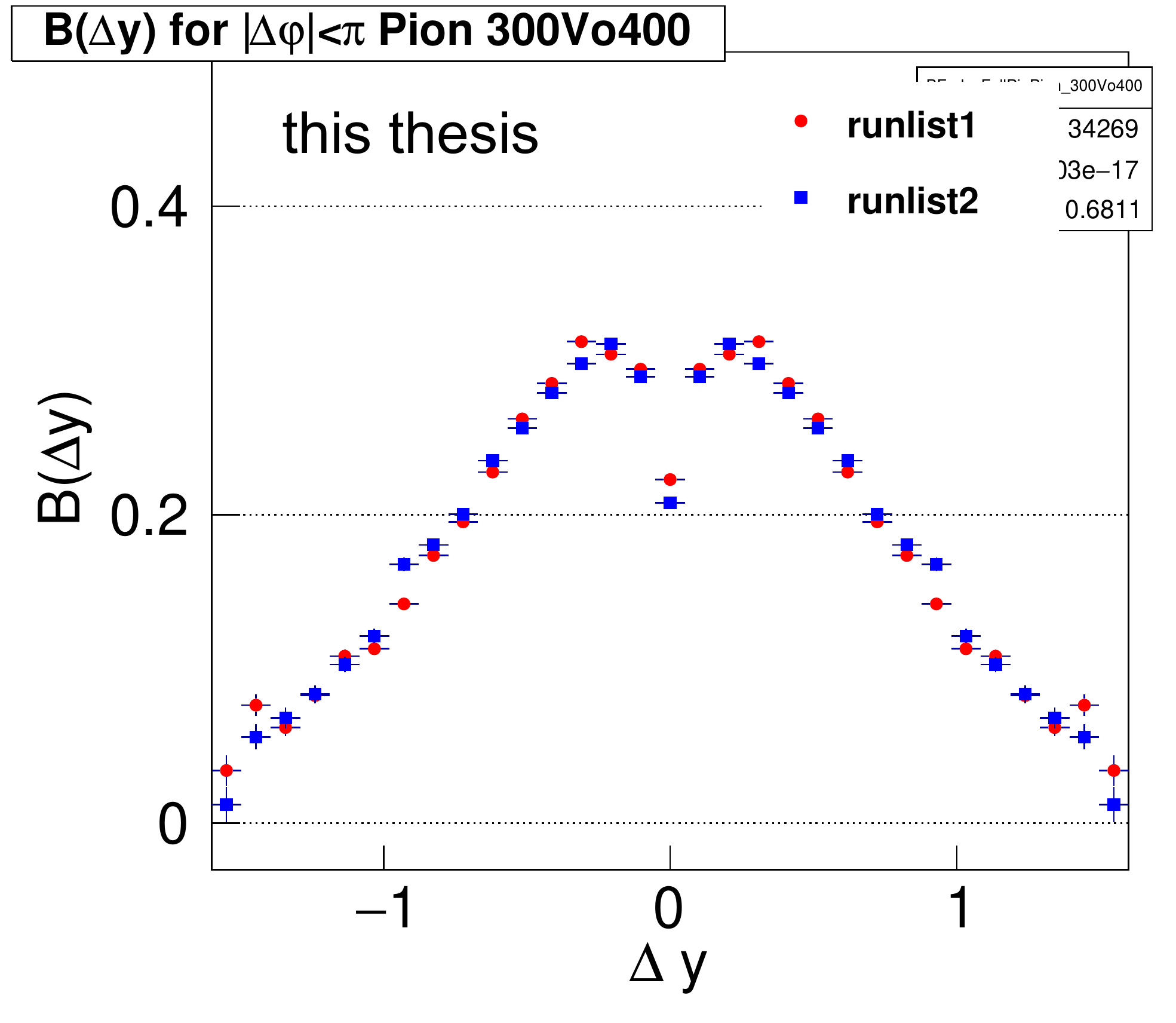}
  \includegraphics[width=0.32\linewidth]{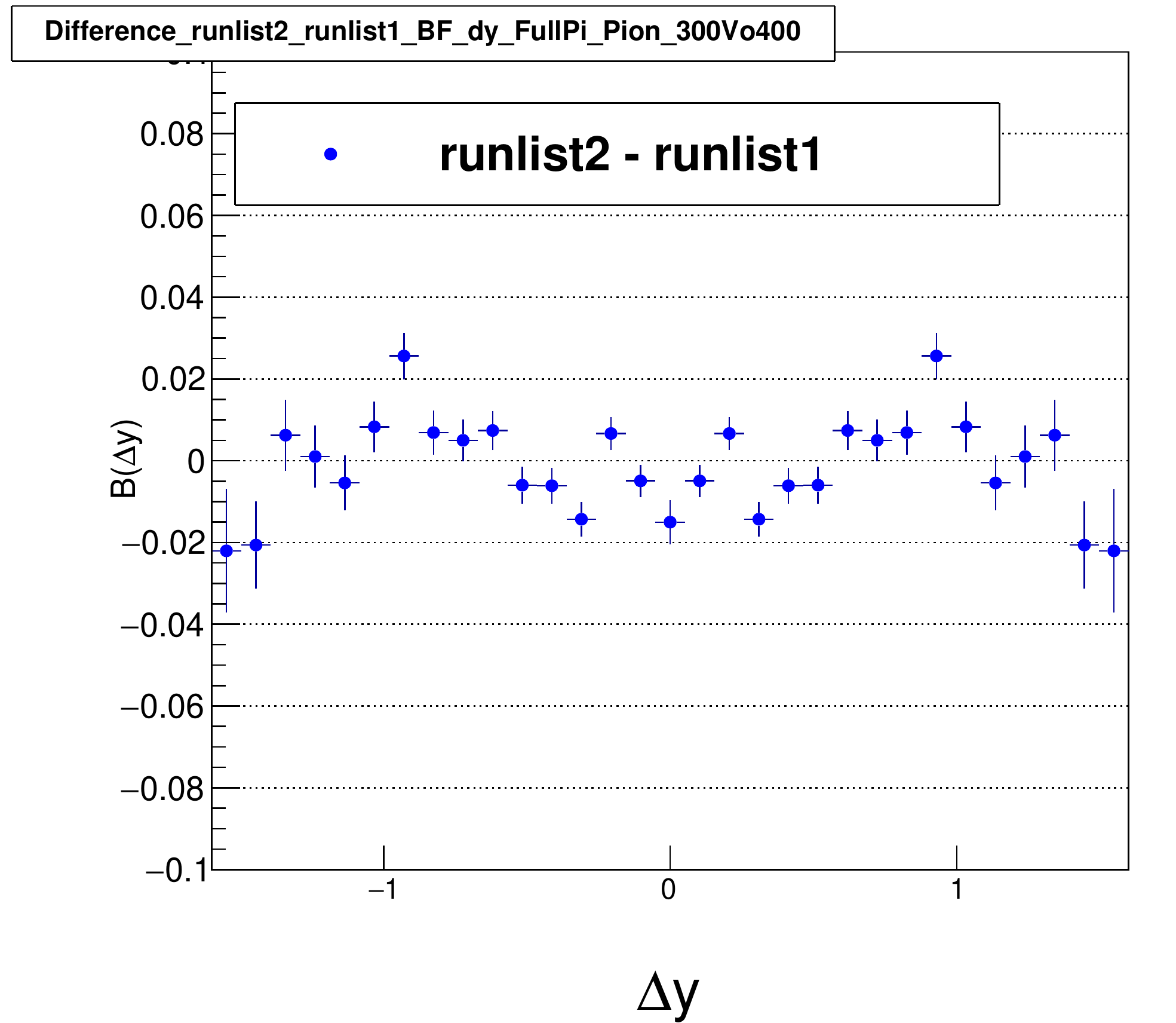}
  \includegraphics[width=0.32\linewidth]{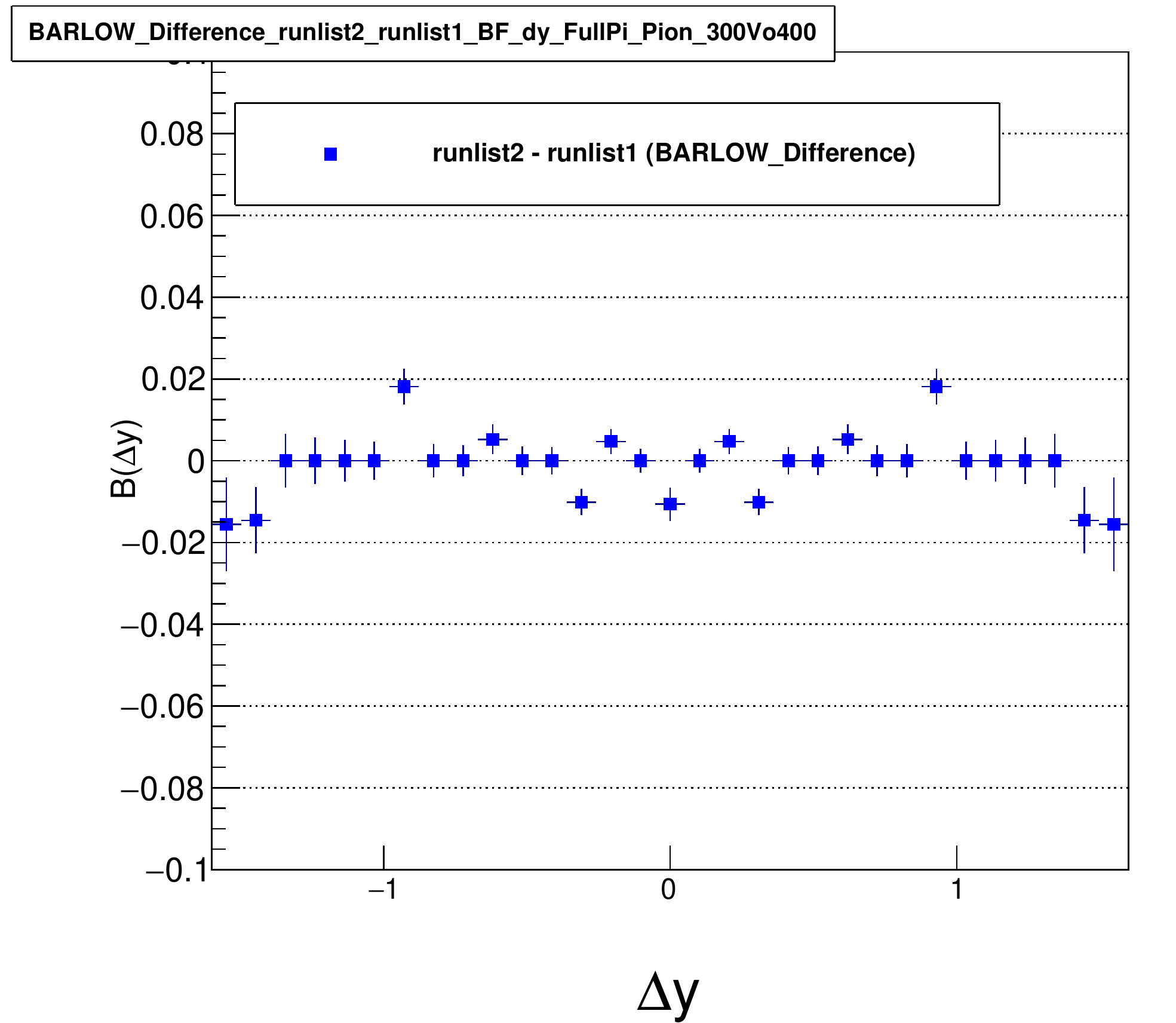}
  
  \includegraphics[width=0.32\linewidth]{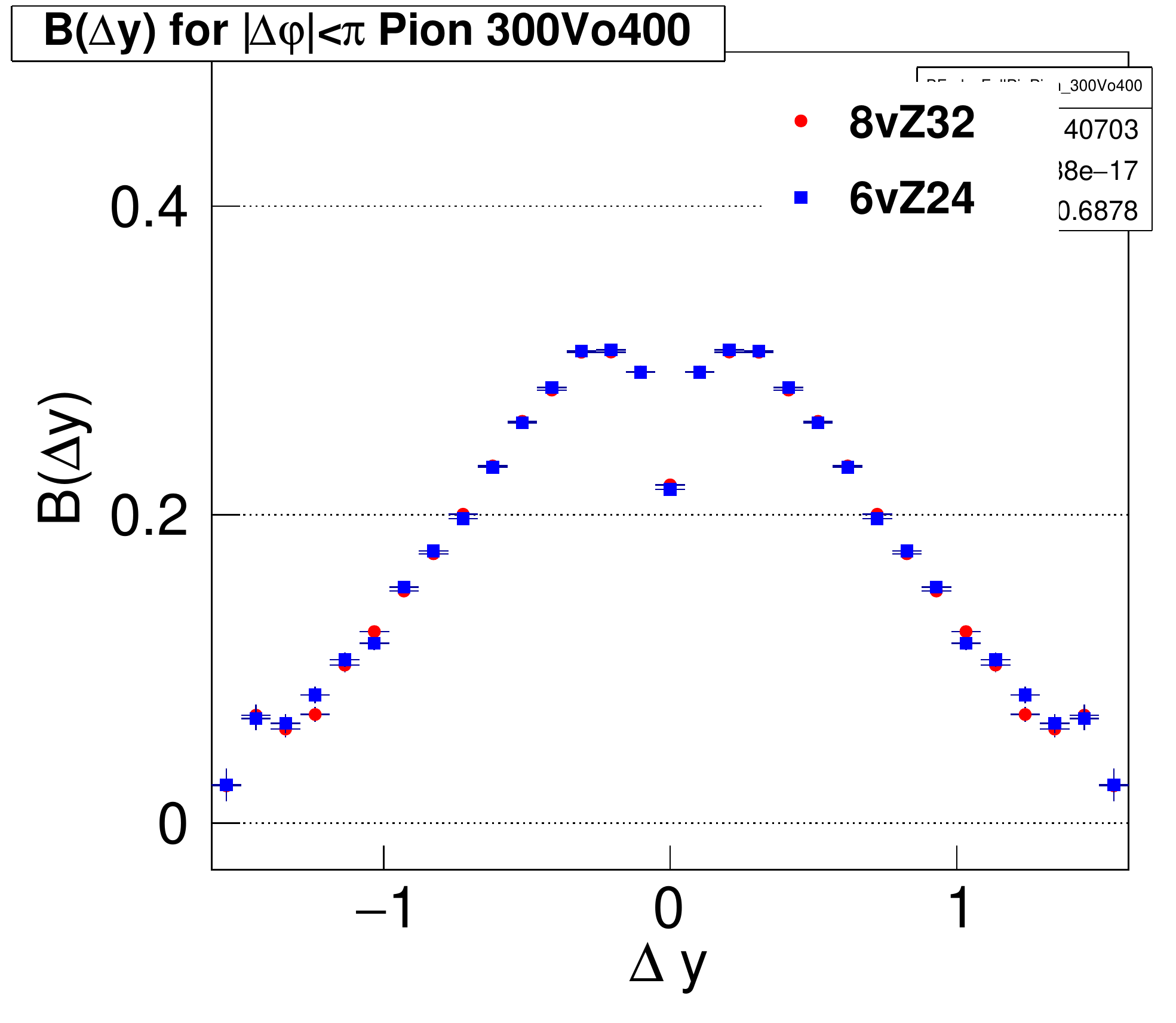}
  \includegraphics[width=0.32\linewidth]{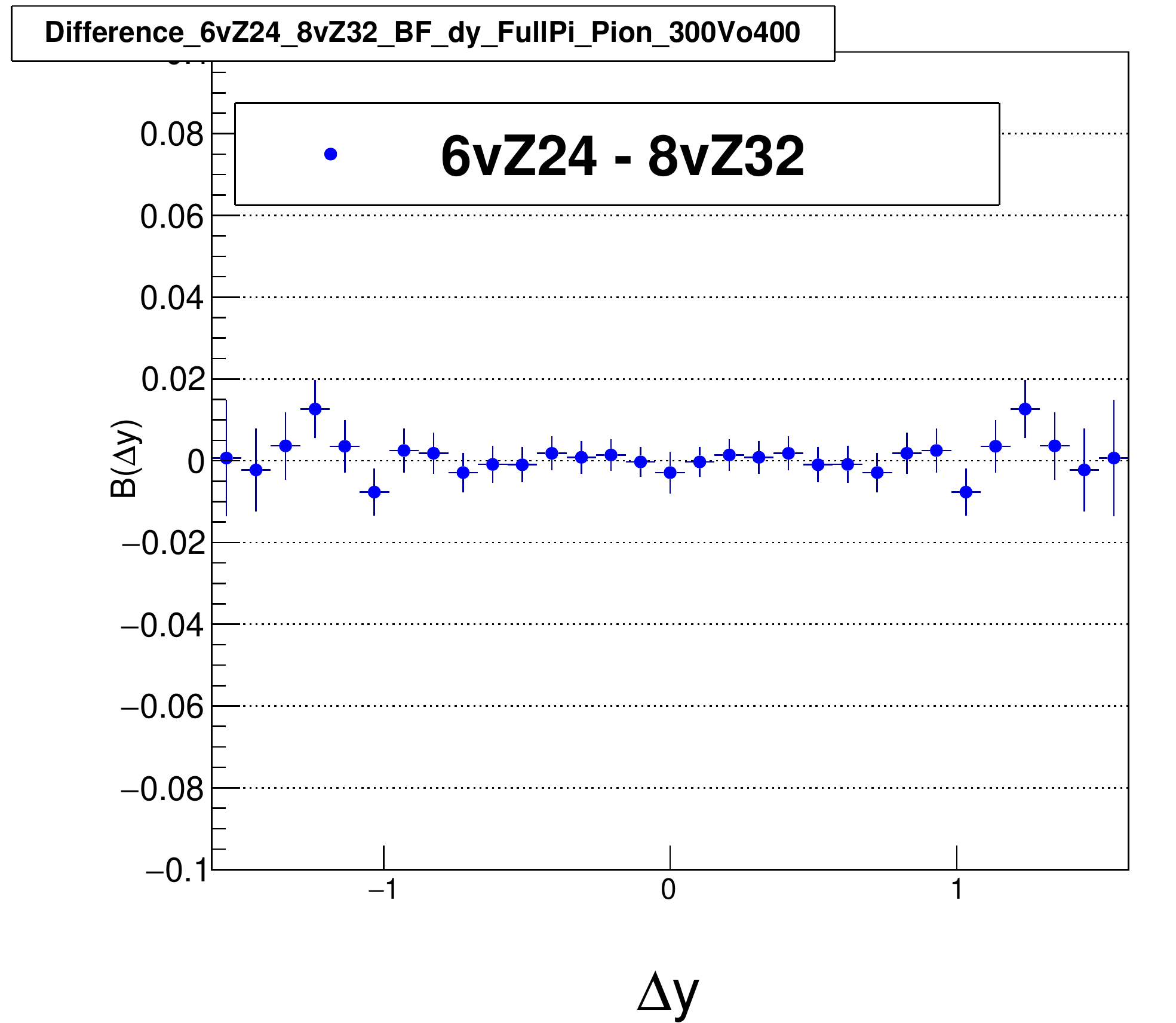}
  \includegraphics[width=0.32\linewidth]{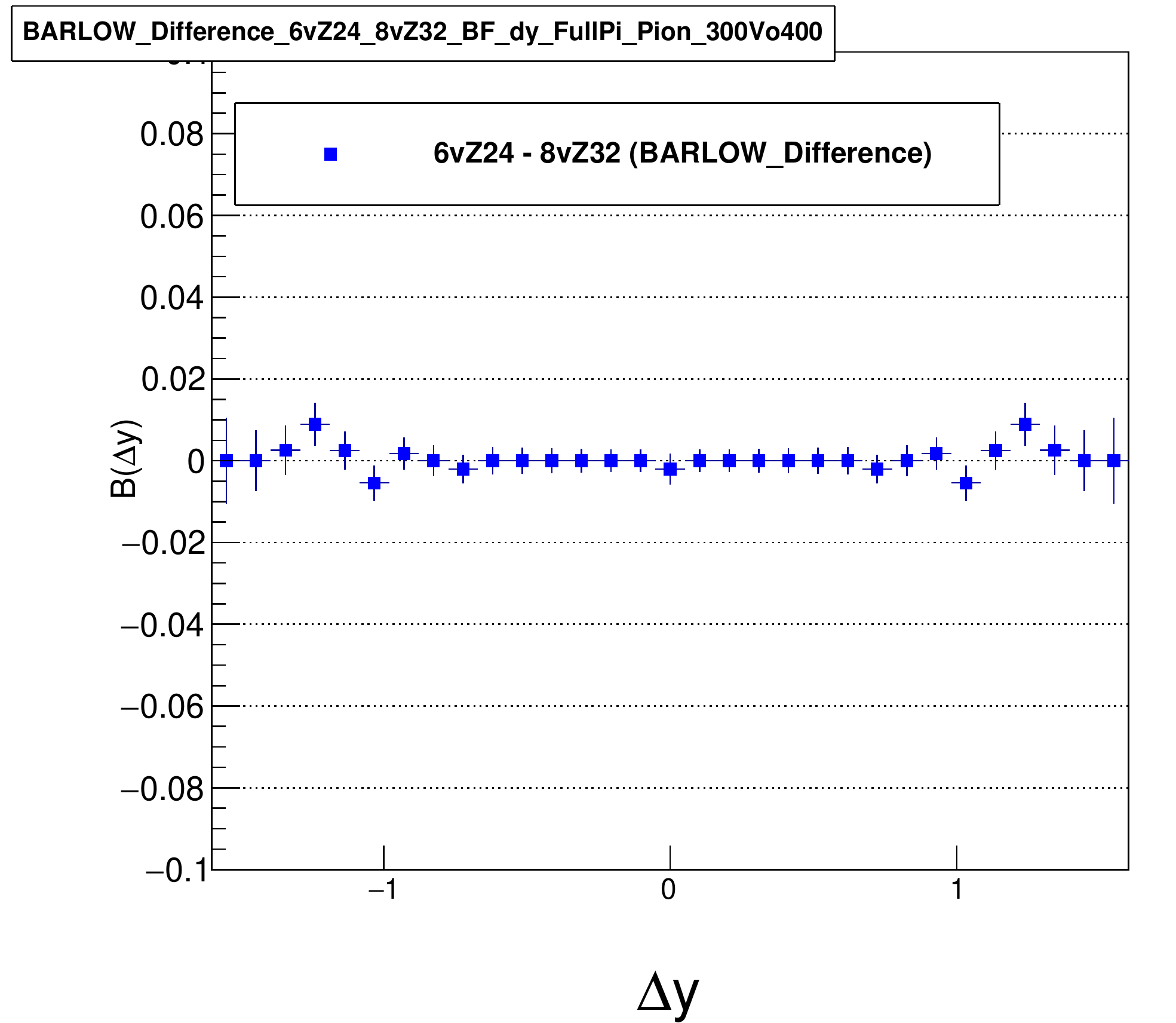}
  
  \includegraphics[width=0.32\linewidth]{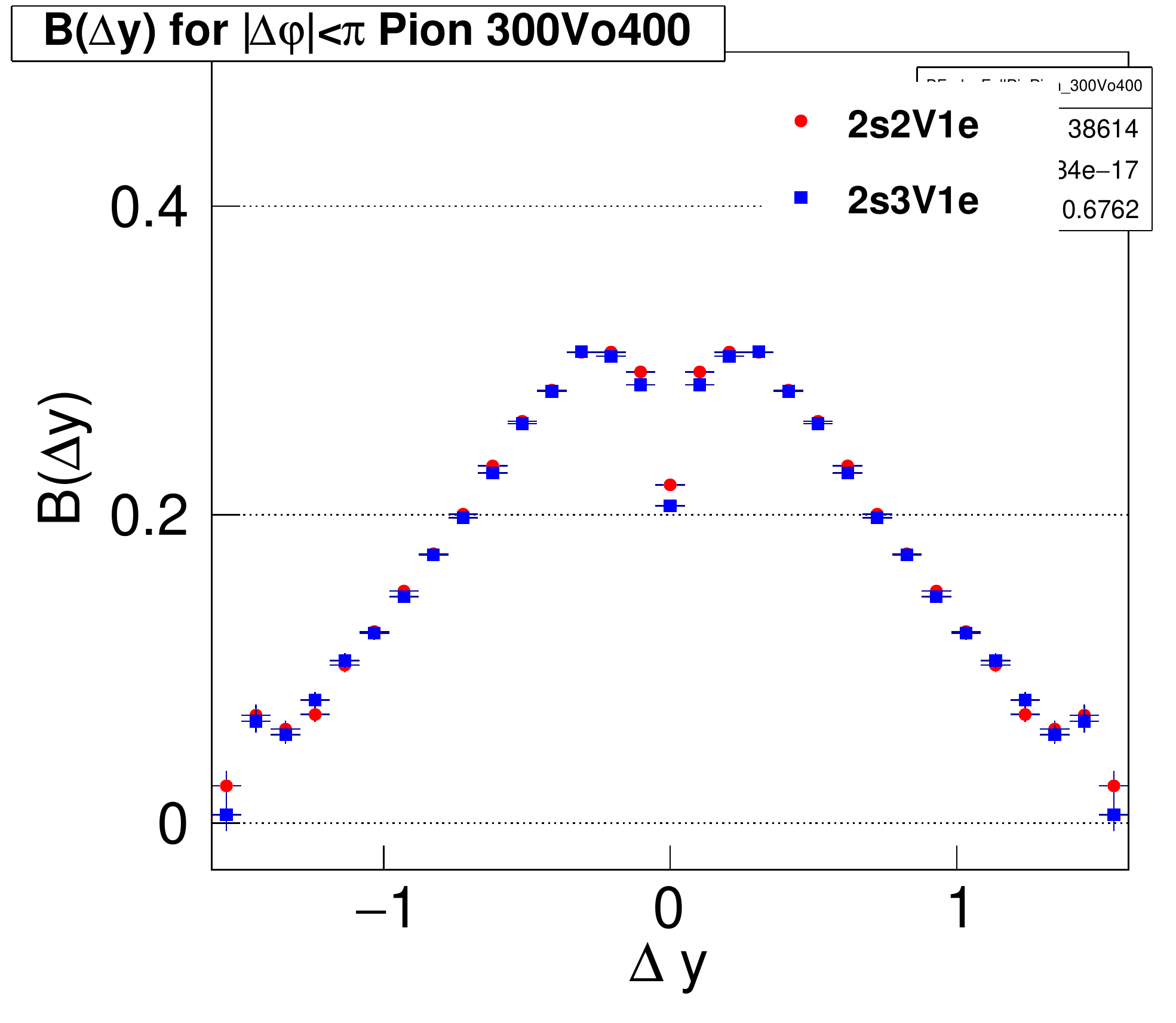}
  \includegraphics[width=0.32\linewidth]{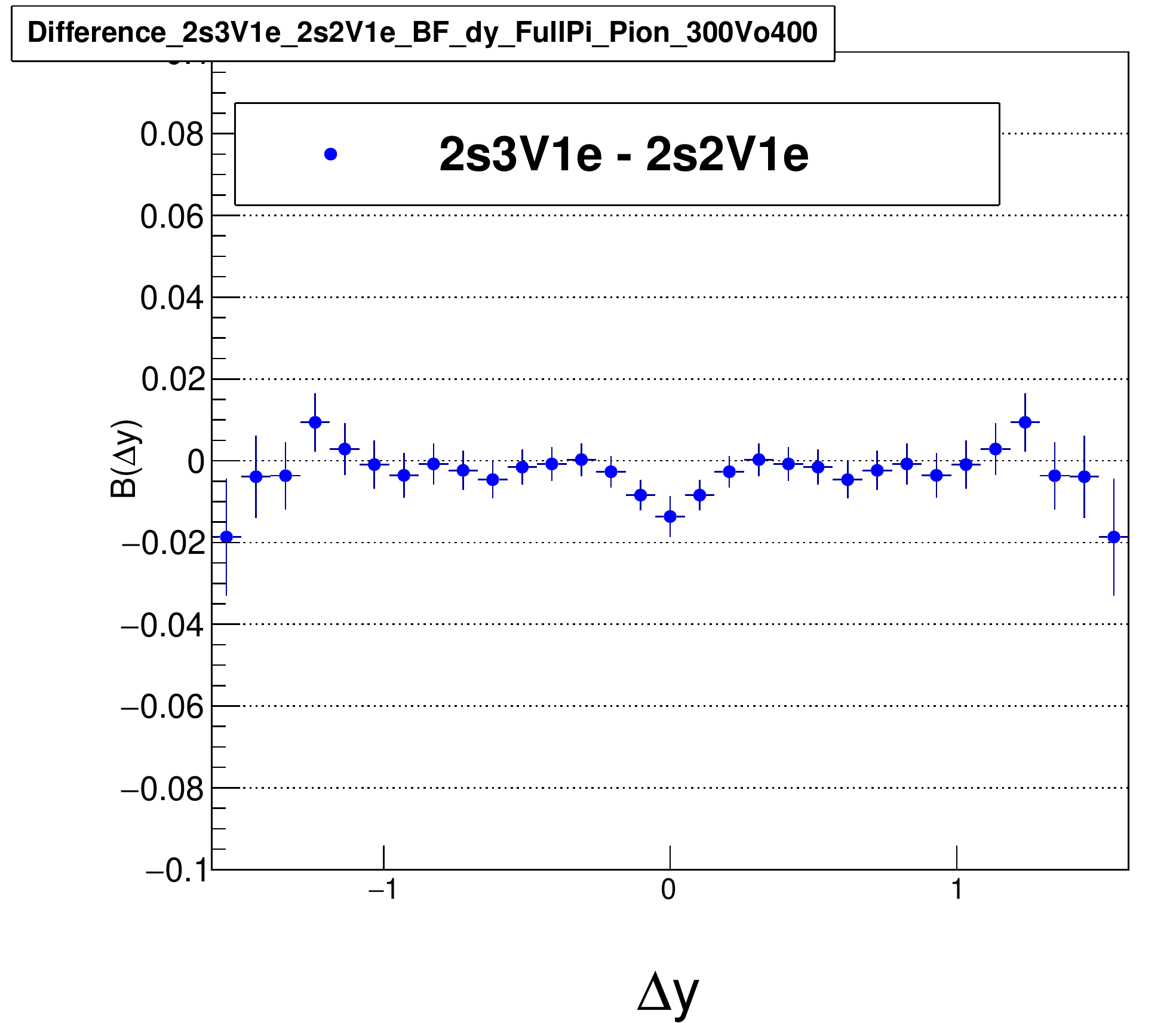}
  \includegraphics[width=0.32\linewidth]{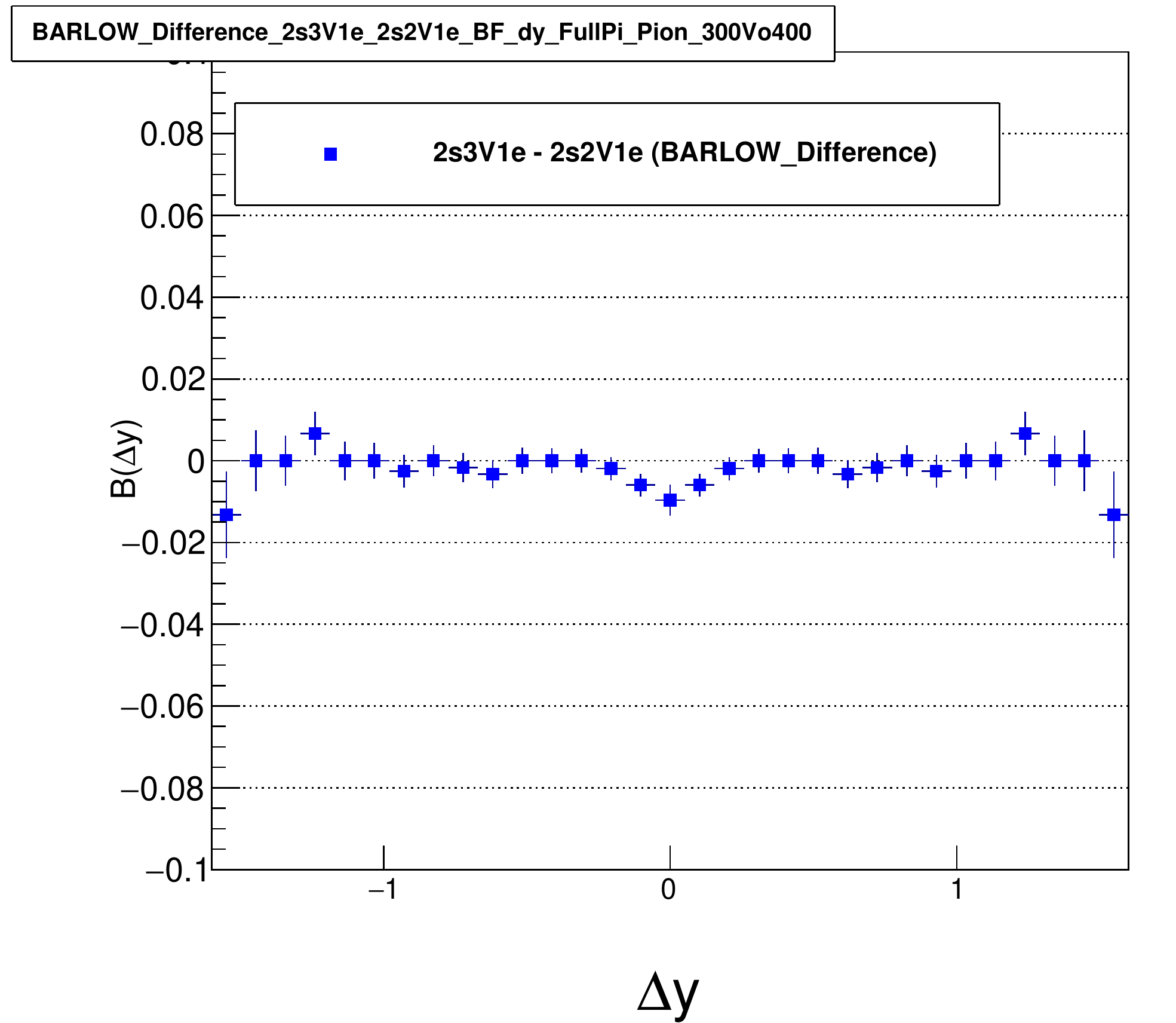}
  
  \includegraphics[width=0.32\linewidth]{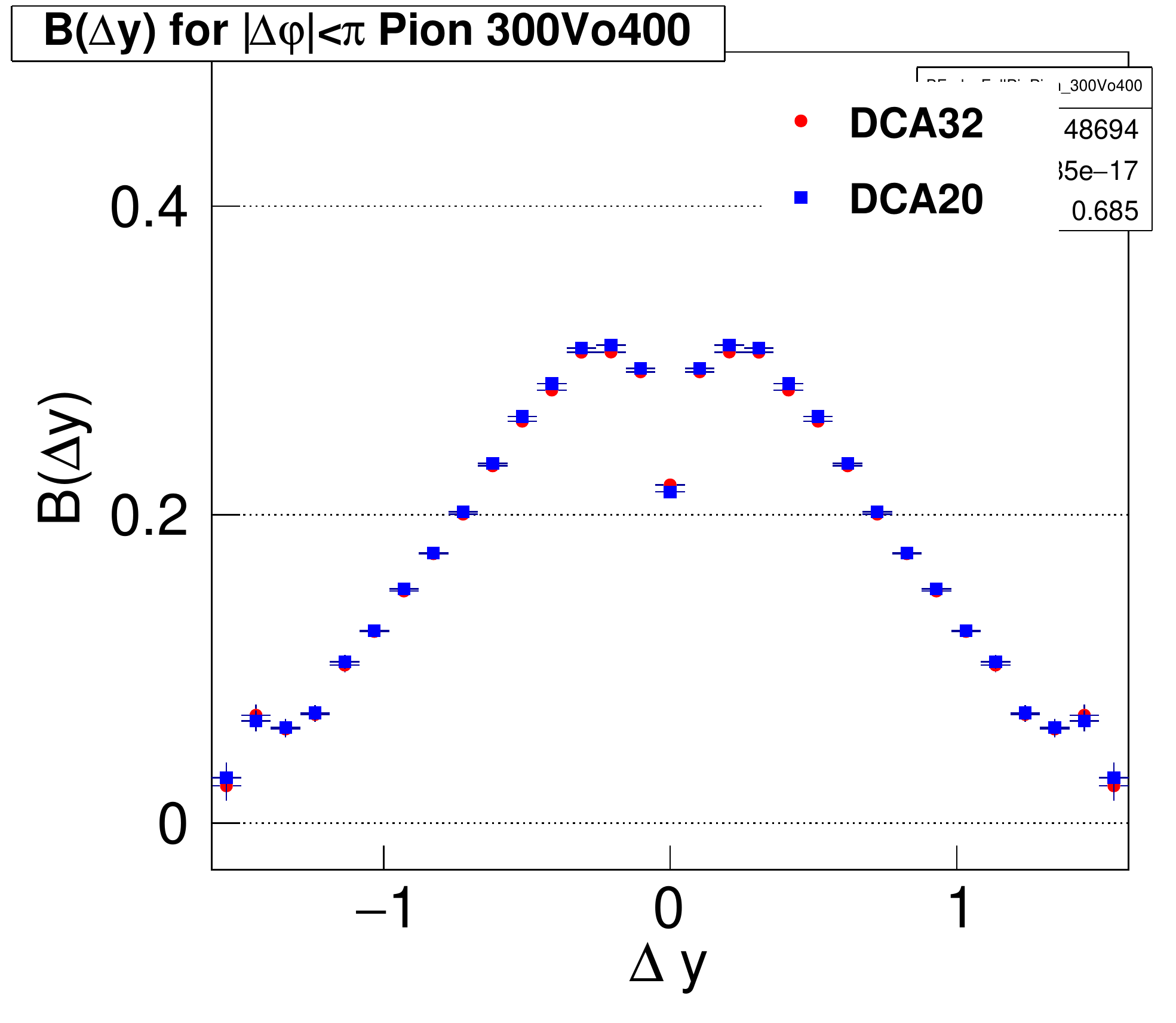}
  \includegraphics[width=0.32\linewidth]{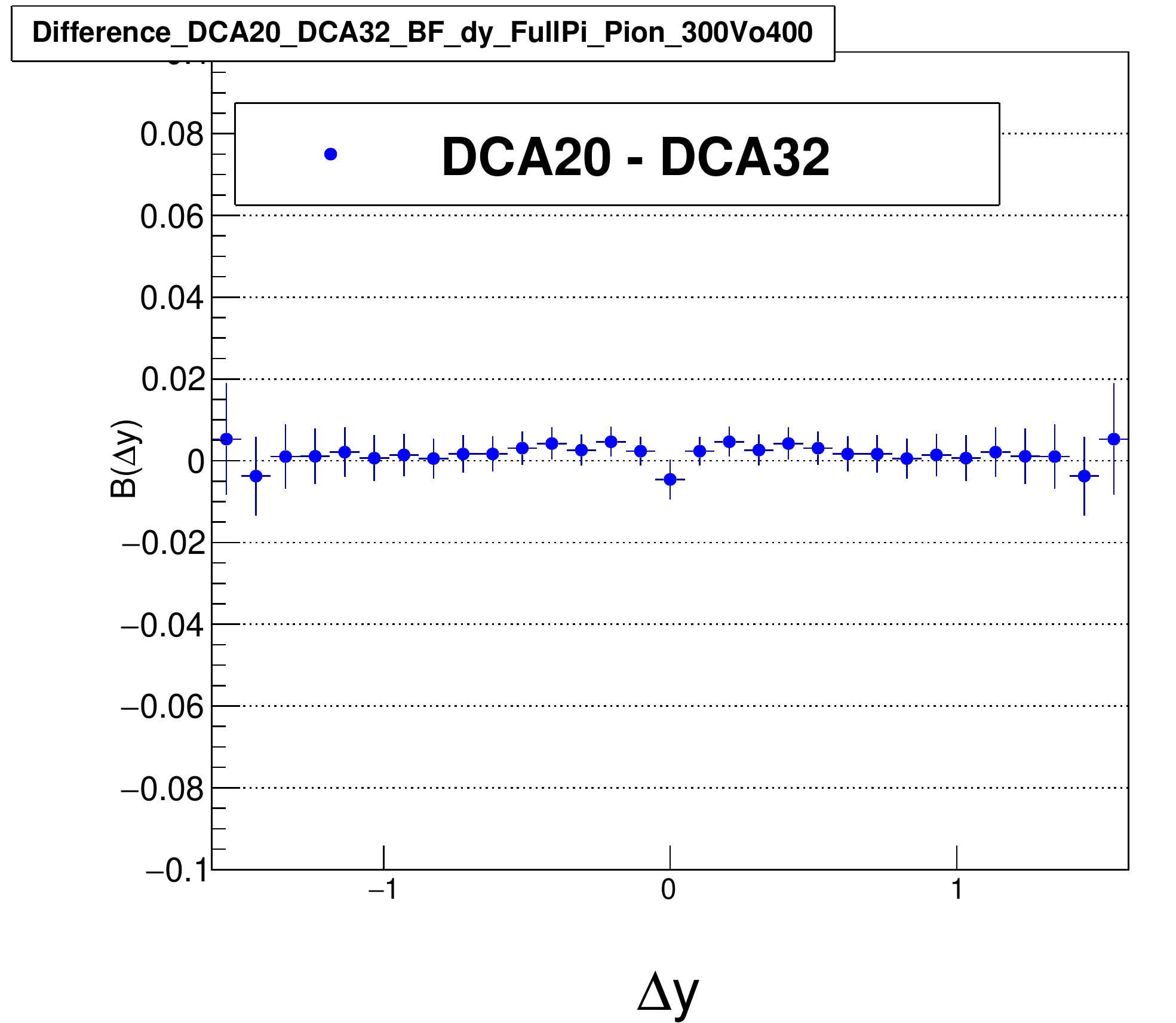}
  \includegraphics[width=0.32\linewidth]{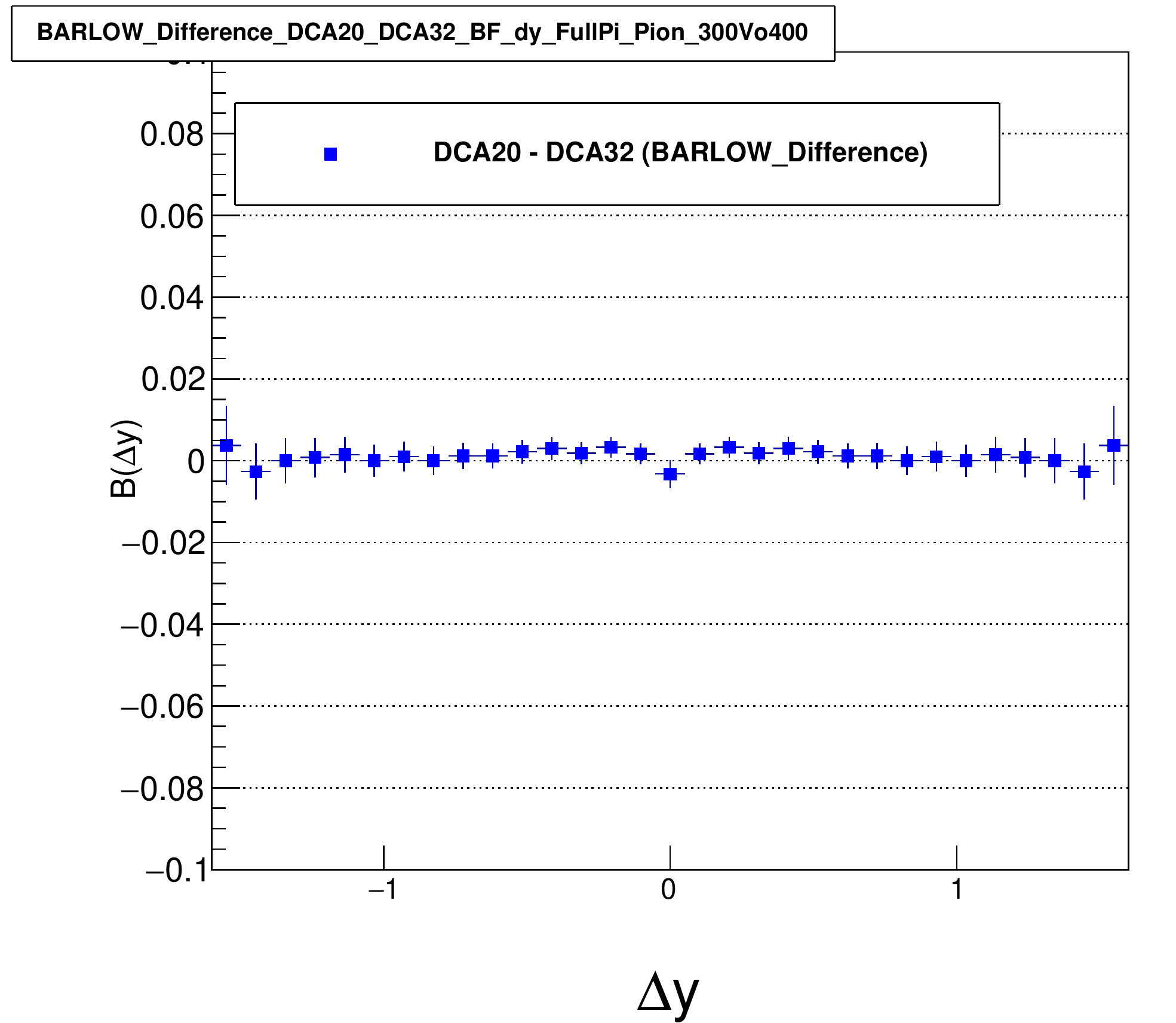}

  \caption{Systematic uncertainty contributions in $B^{\pi\pi}(\Delta y)$ from BField ($1^{st}$ row), $V_{z}$ ($2^{nd}$ row), PID ($3^{rd}$ row), and DCA ($4^{th}$ row). The comparisons between two sets of different cuts (left column), with their differences $d$ (middle column), and their differences after the Barlow check $D_{Barlow}$ (right column).}
  \label{fig:Sys_components_dy_projections_PionPion_300Vo400}
\end{figure}

\begin{figure}
\centering
  \includegraphics[width=0.32\linewidth]{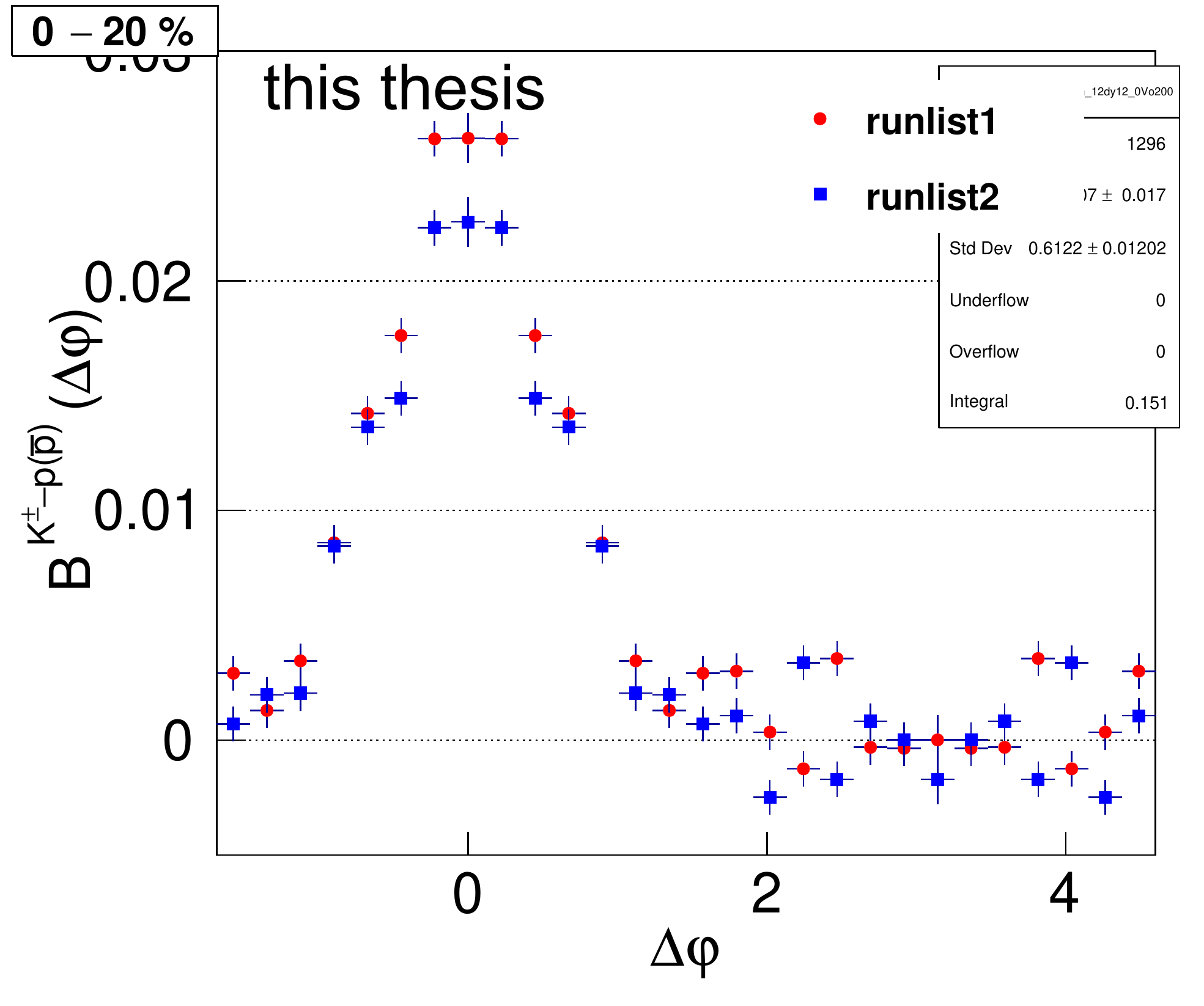}
  \includegraphics[width=0.32\linewidth]{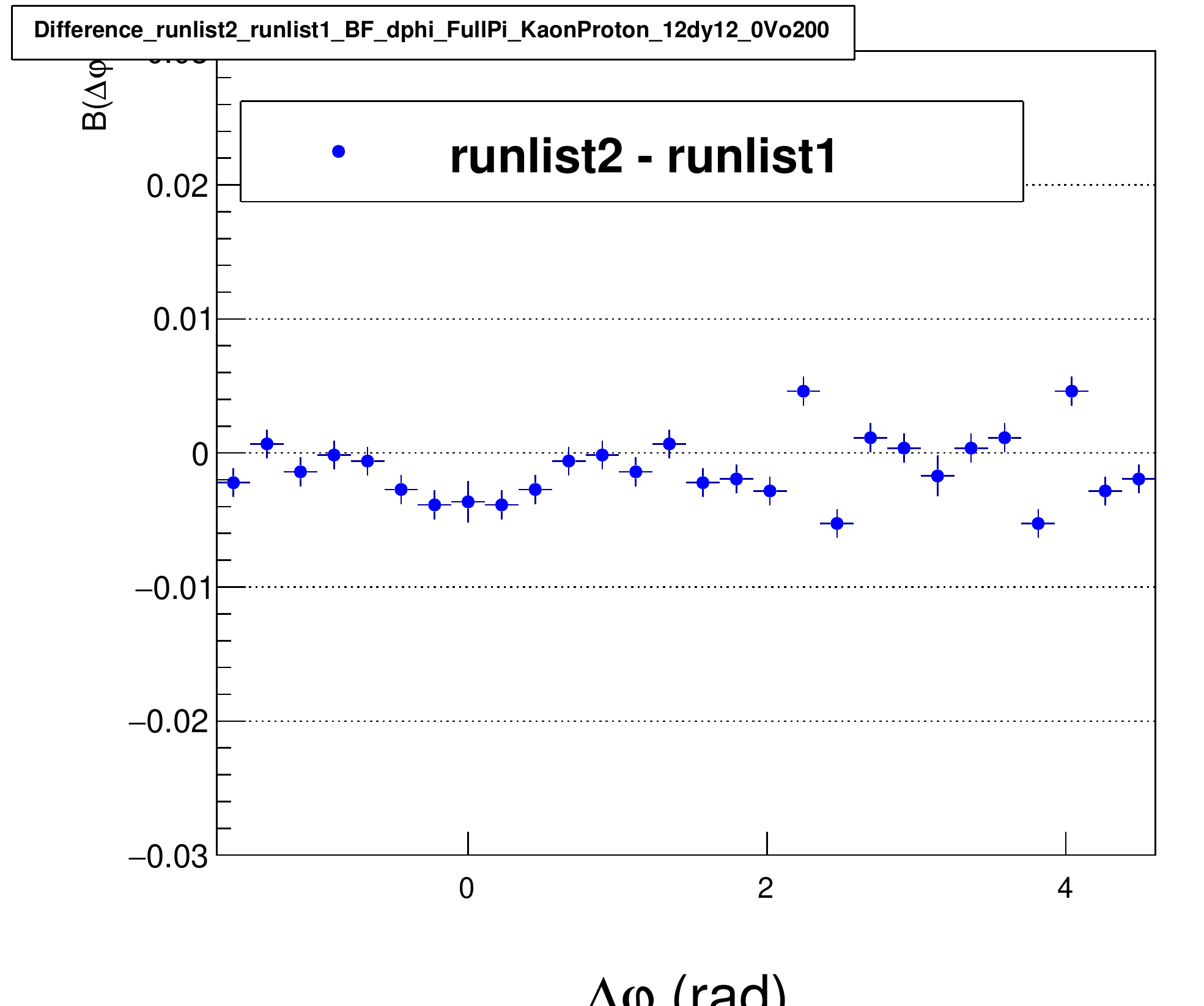}
  \includegraphics[width=0.32\linewidth]{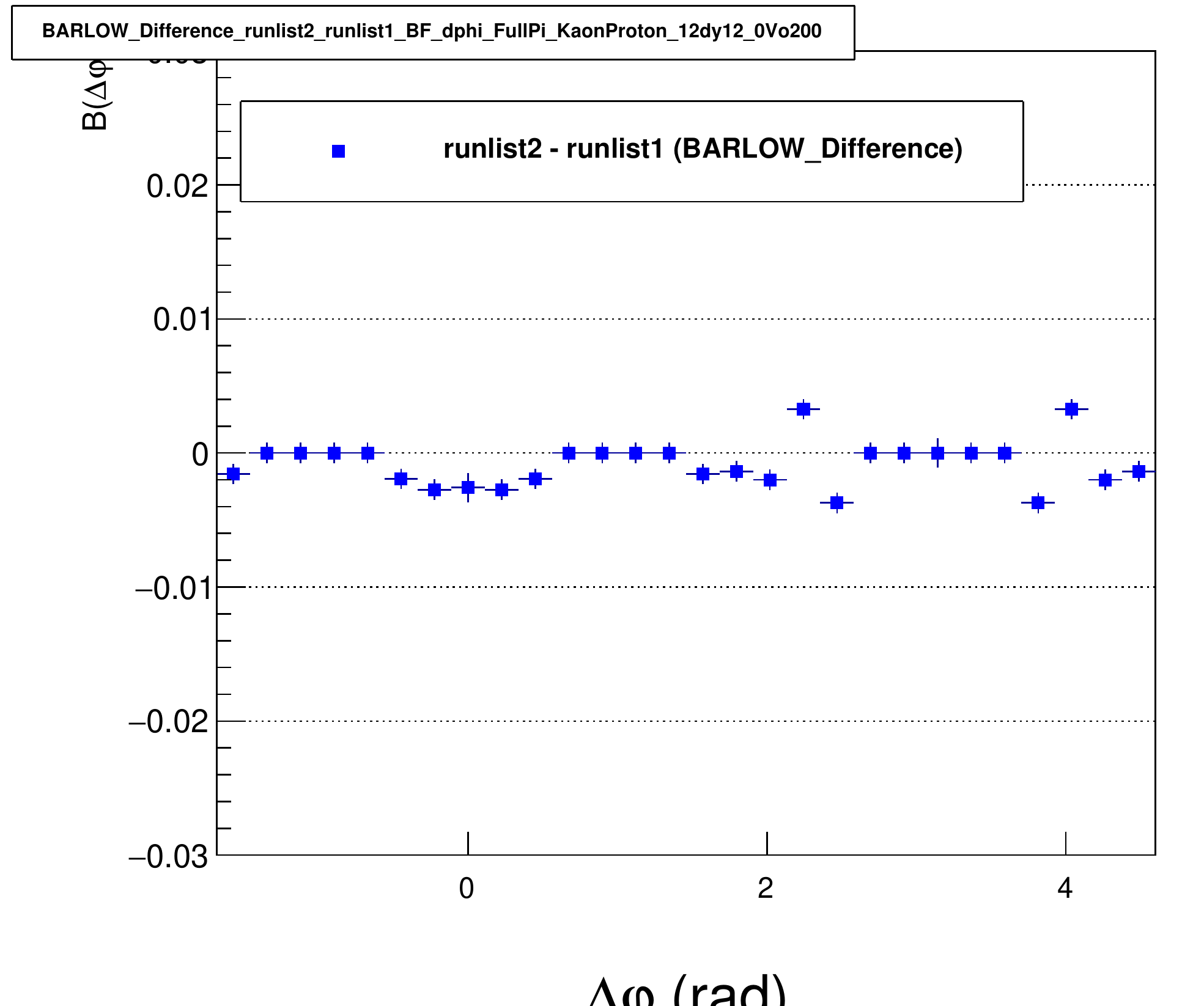}

  \includegraphics[width=0.32\linewidth]{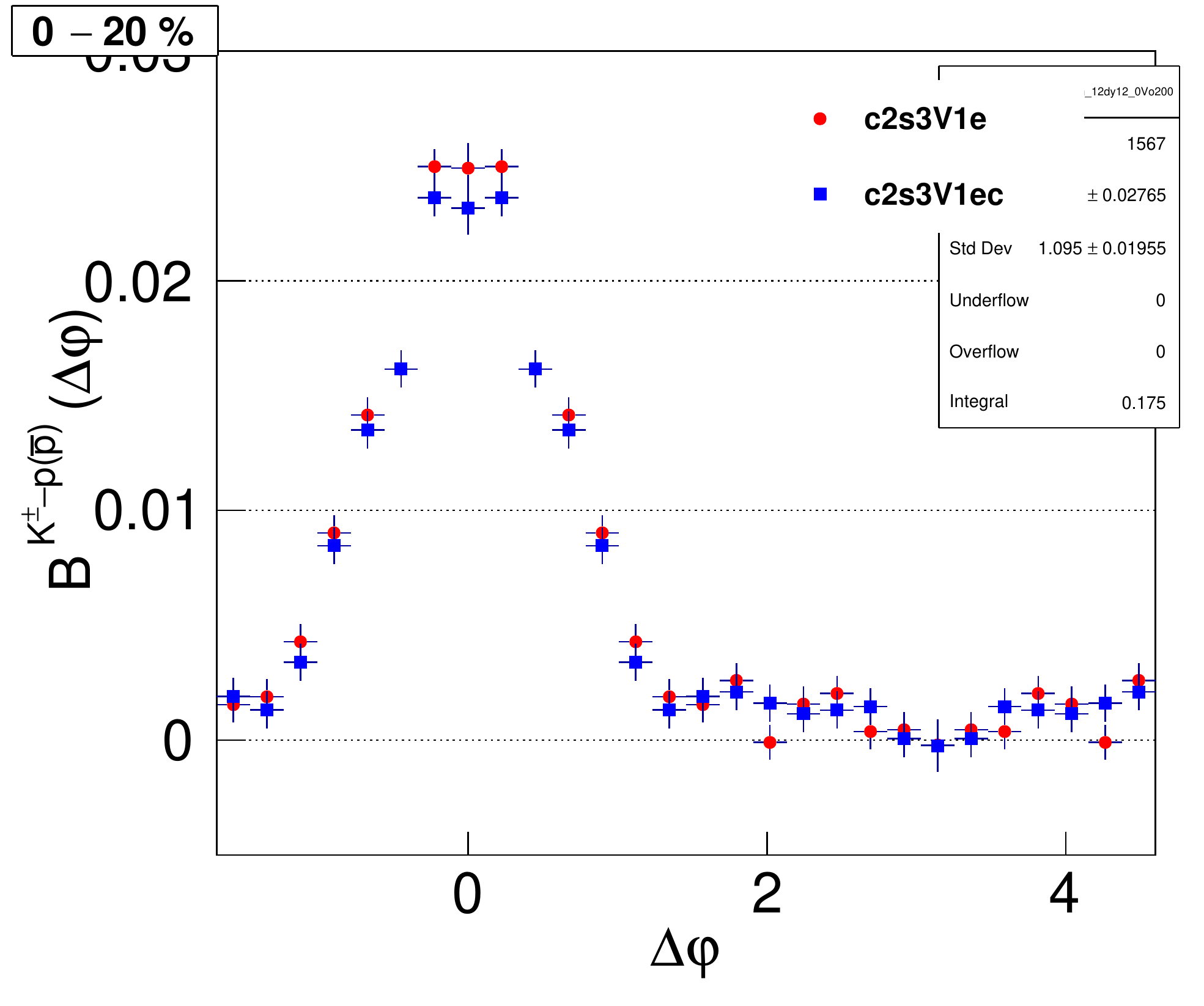}
  \includegraphics[width=0.32\linewidth]{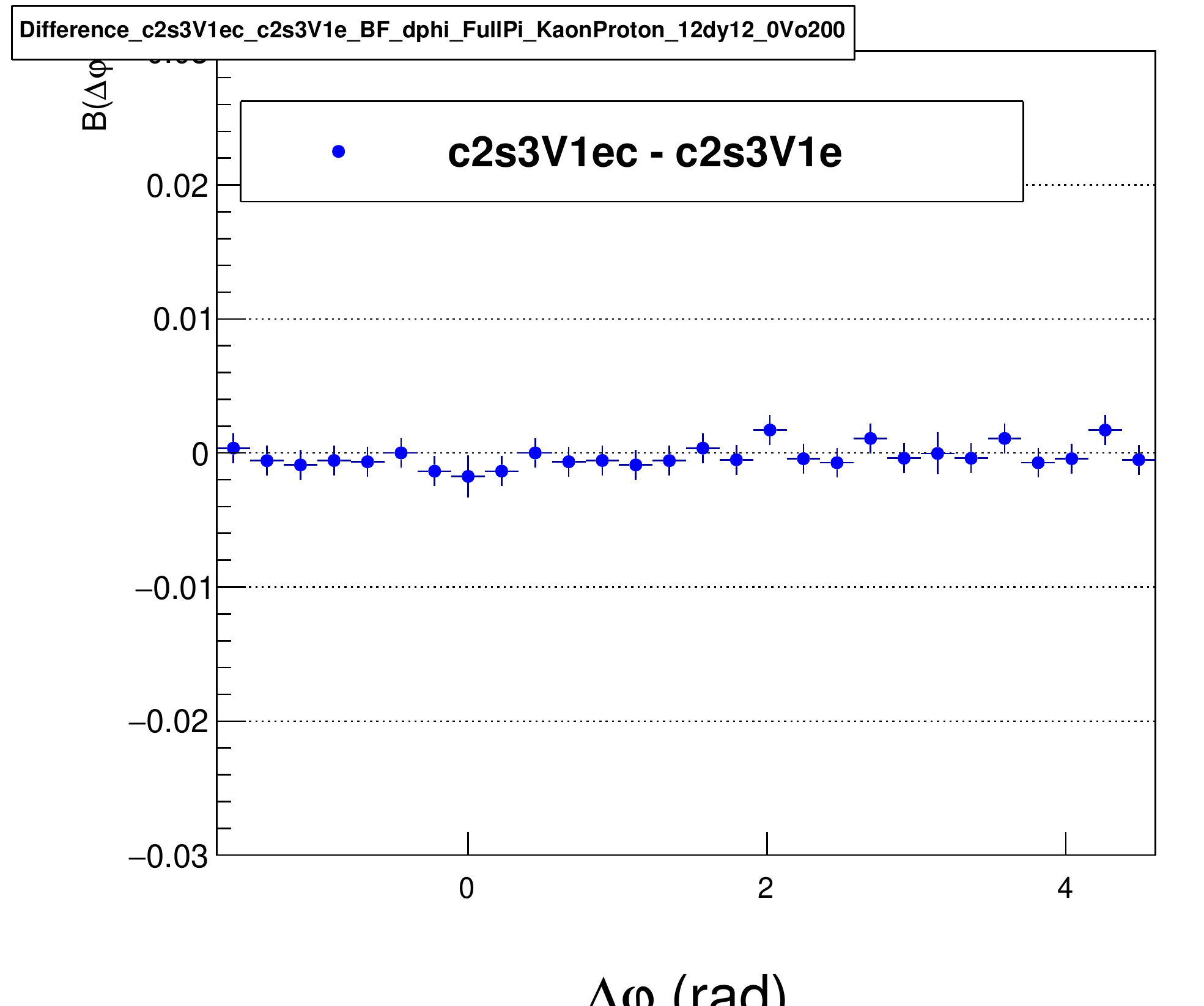}
  \includegraphics[width=0.32\linewidth]{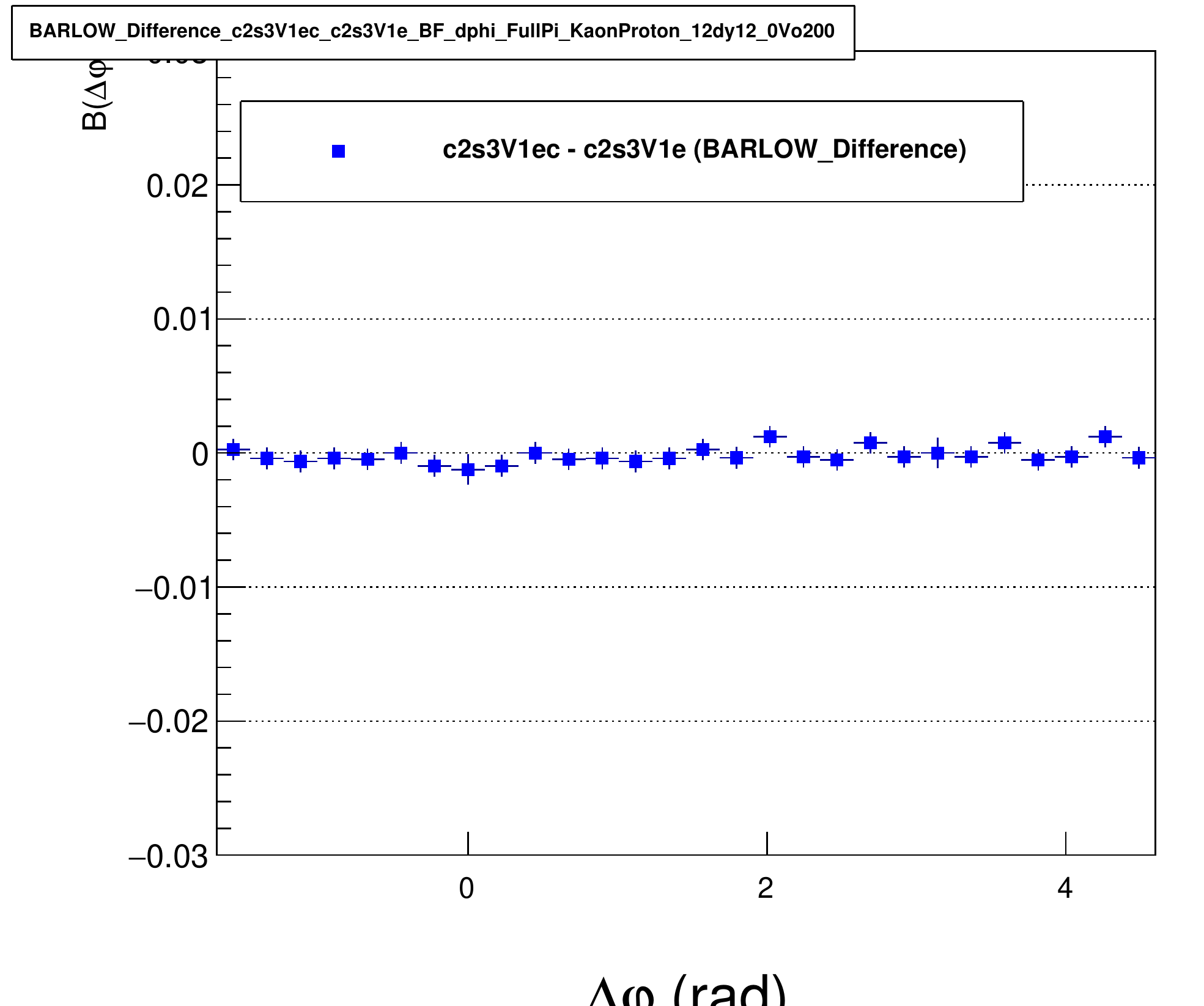}
  
  \includegraphics[width=0.32\linewidth]{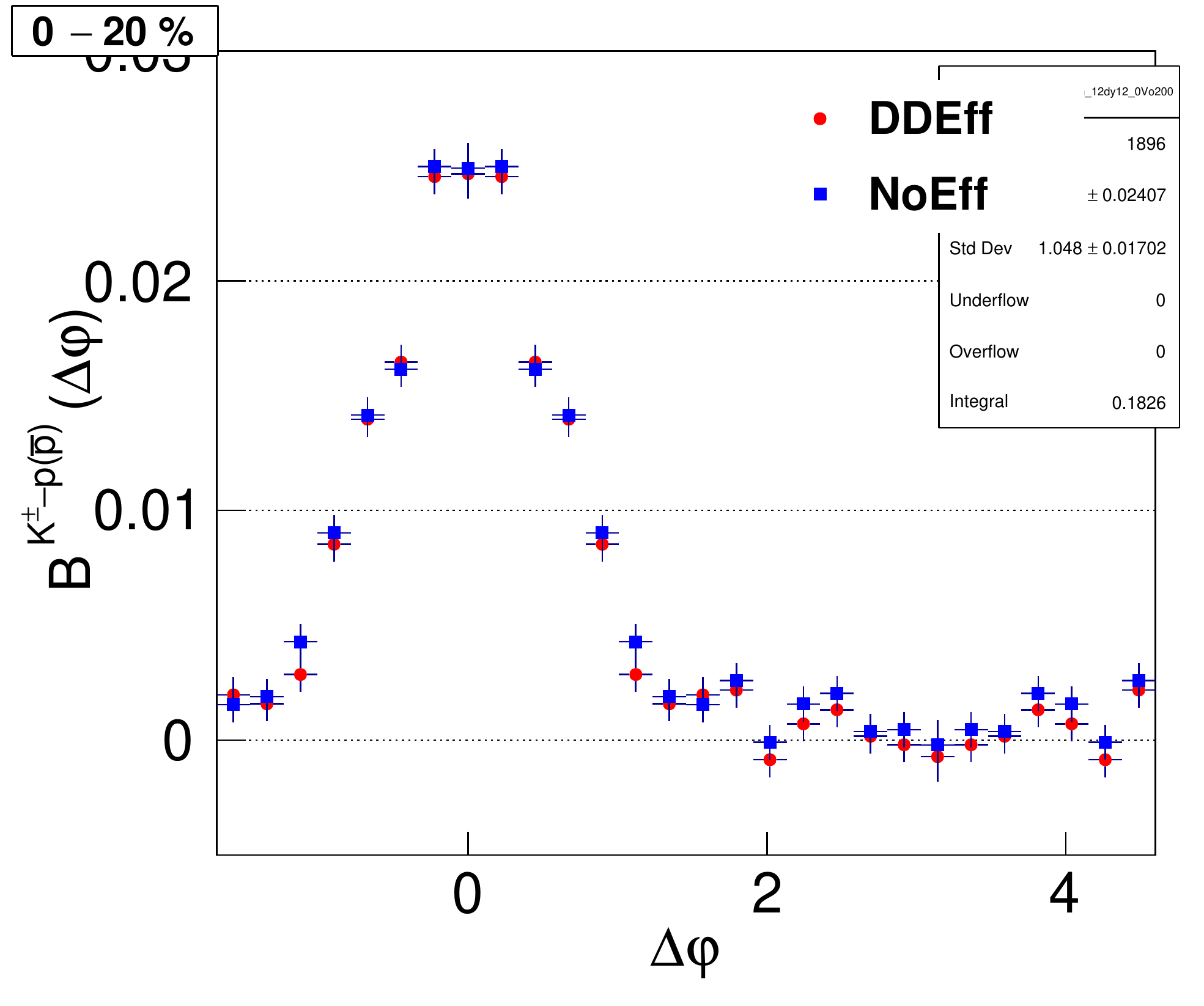}
  \includegraphics[width=0.32\linewidth]{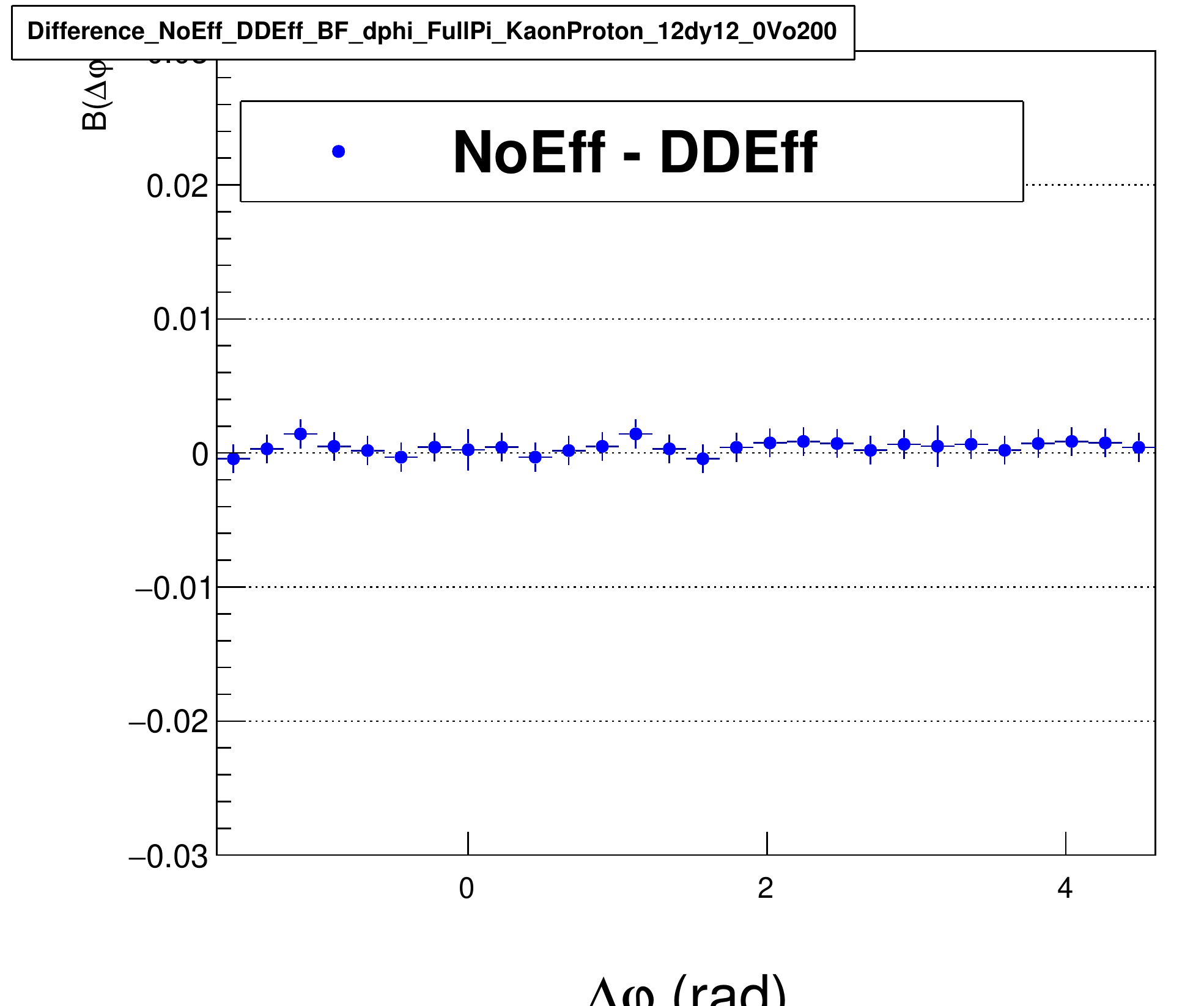}
  \includegraphics[width=0.32\linewidth]{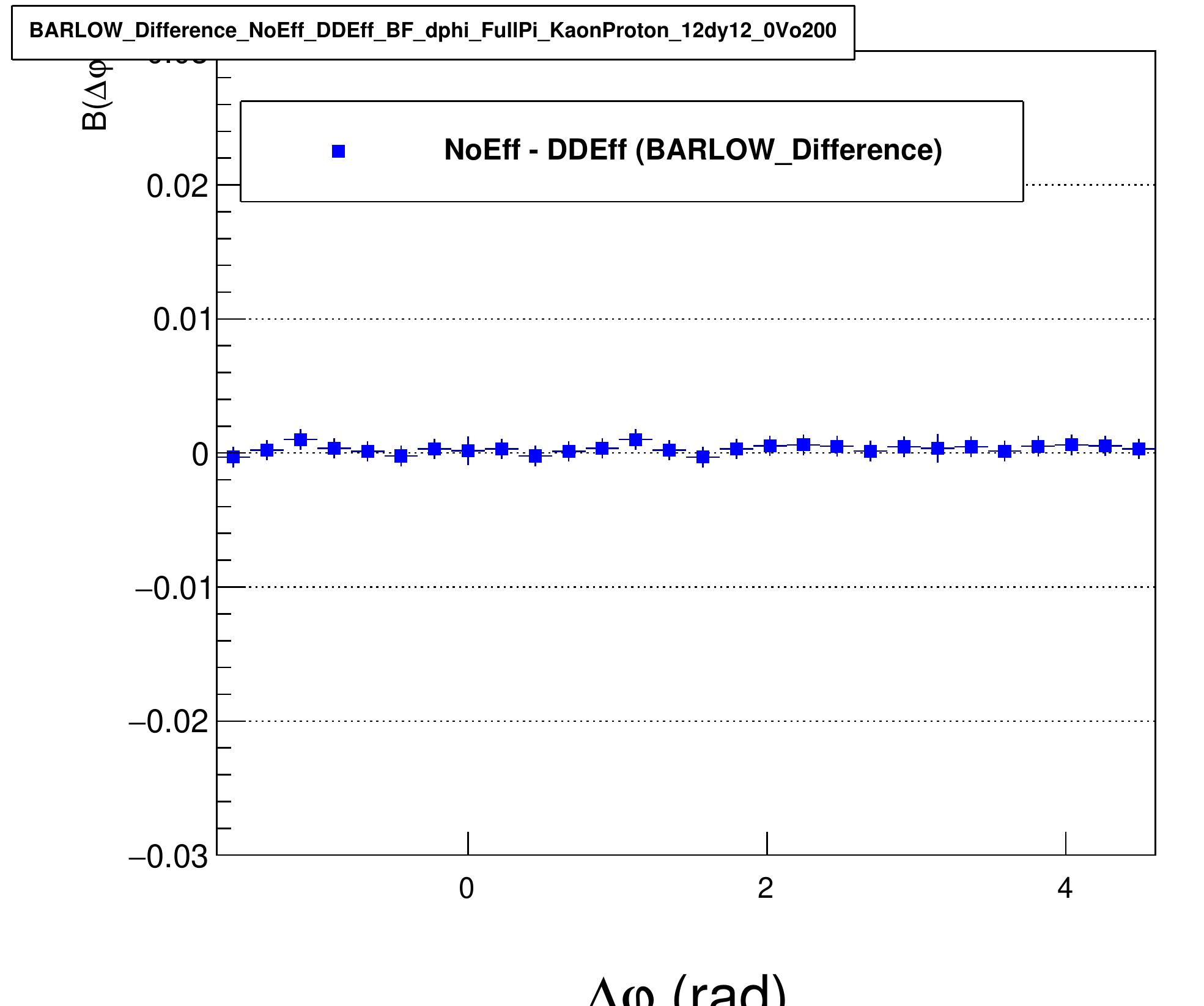}
  
  \includegraphics[width=0.32\linewidth]{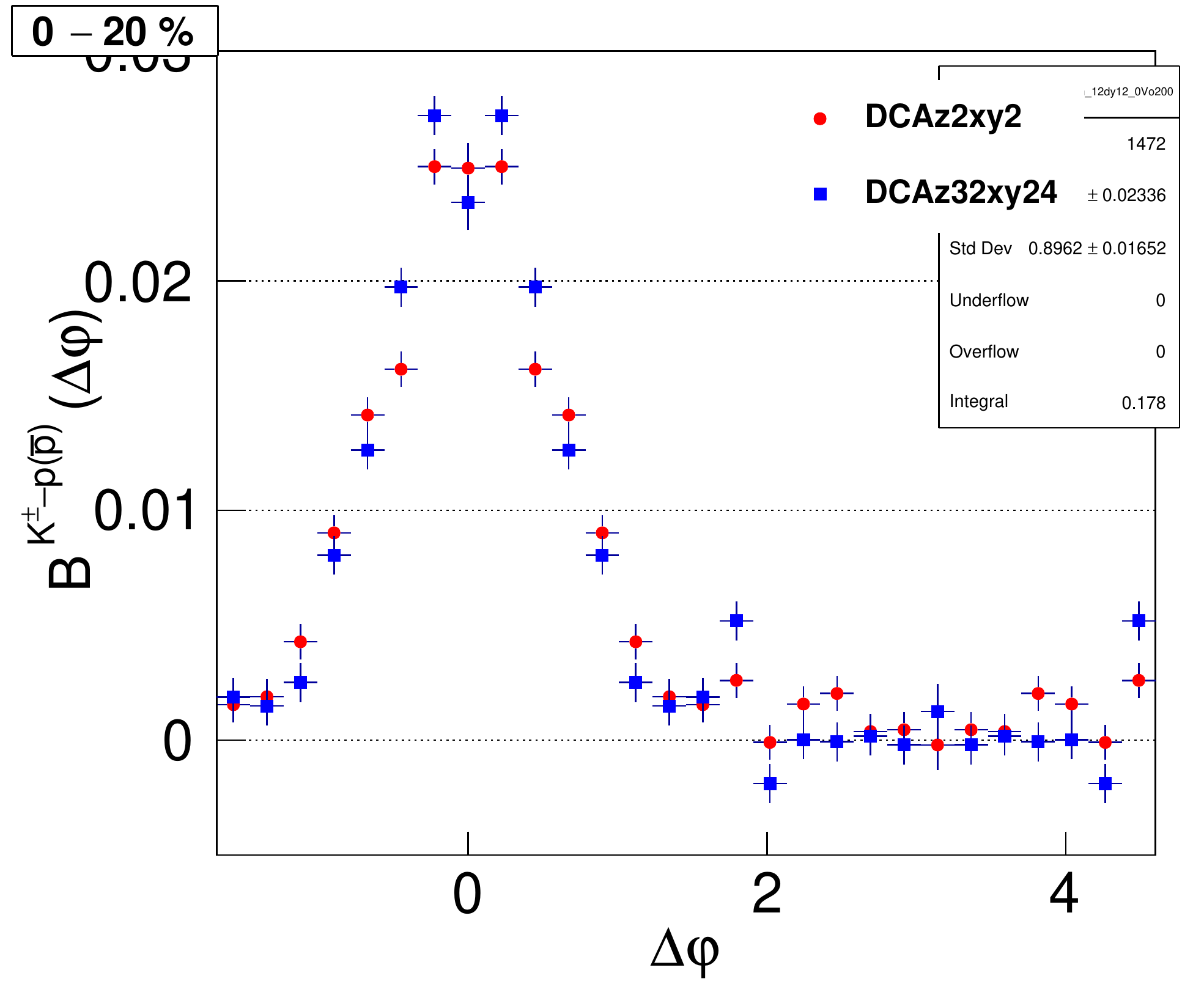}
  \includegraphics[width=0.32\linewidth]{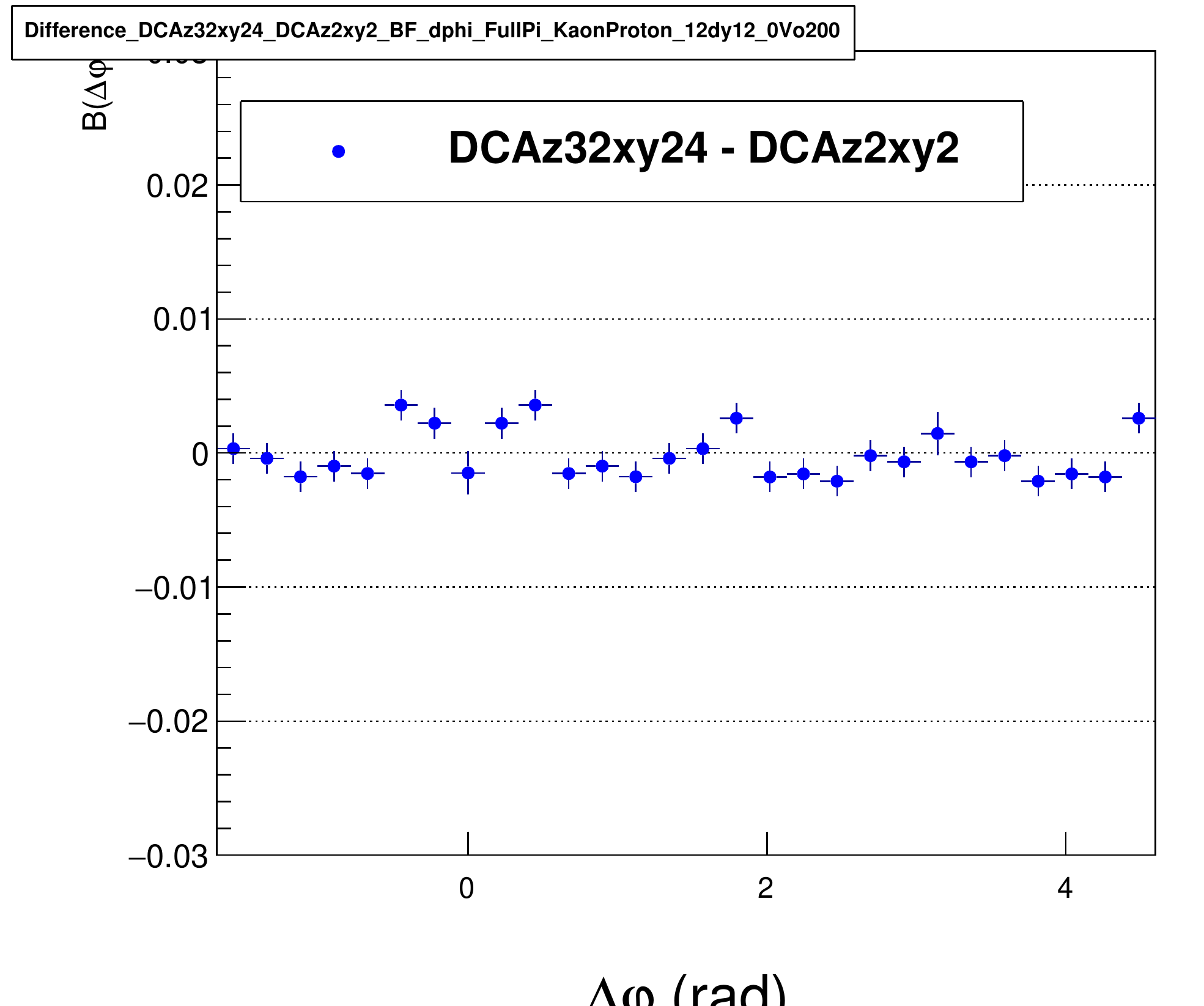}
  \includegraphics[width=0.32\linewidth]{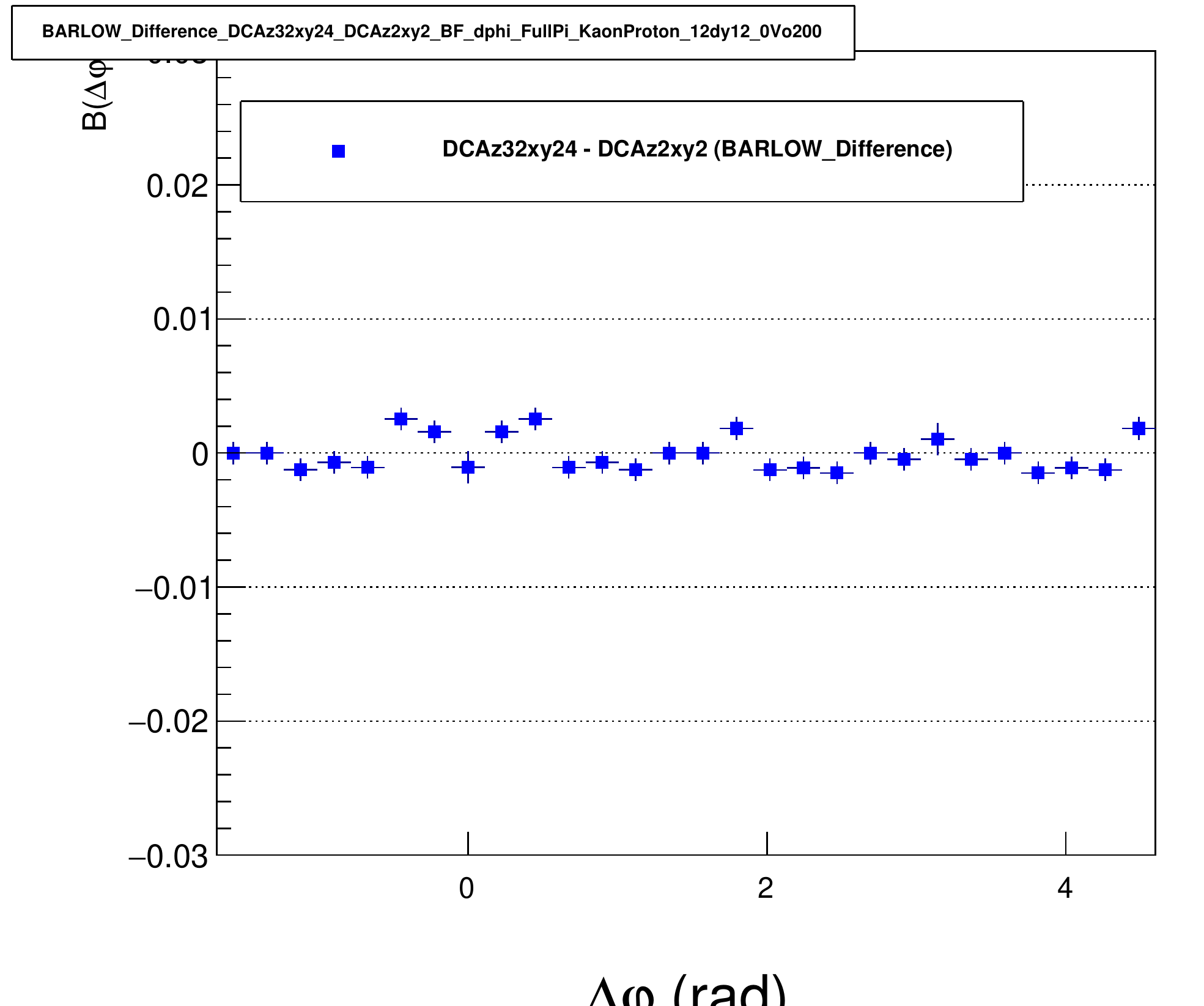}

  \caption{Systematic uncertainty contributions in $B^{Kp}(\Delta\varphi)$ from BField ($1^{st}$ row), PID ($2^{nd}$ row), additional $p_{\rm T}$-dependent efficiency corrections ($3^{rd}$ row), and DCA ($4^{th}$ row). The comparisons between two sets of different cuts (left column), with their differences $d$ (middle column), and their differences after the Barlow check $D_{Barlow}$ (right column).}
  \label{fig:Sys_components_dphi_projections_KaonProton}
\end{figure}

In this work, six possible sources of systematic uncertainties are investigated: magnetic field configurations, $V_{z}$ ranges, PID cuts, additional $p_{\rm T}$-dependent efficiency corrections, track DCA cuts, and TPC number of clusters.
The results of these studies are shown in Figures~\ref{fig:Sys_components_dy_projections_PionPion_300Vo400} -- 
%,~\ref{fig:Sys_components_dphi_KaonKaon_0Vo100},~\ref{fig:Sys_components_integral_ProtonProton},~\ref{fig:Sys_components_dy_widths_PionKaon},~\ref{fig:Sys_components_dphi_widths_PionProton},
 \ref{fig:Sys_components_dphi_projections_KaonProton} for all species pairs considered in this work.
Given these studies involve the variation of experimental conditions and cuts, we use the Barlow criterion~\cite{Barlow:2002yb} to determine the magnitude of uncertainties.

The   calculation of systematic uncertainties associated with variations of cuts proceeds as follows:

Step 1: For each component, we calculate the bin-by-bin differences ($d$) in BF $\Delta y$ and $\Delta\varphi$ projections along with their widths and integrals between two or more different cuts, as shown in the $2^{nd}$ column of Figs.~\ref{fig:Sys_components_dy_projections_PionPion_300Vo400} -- 
%~\ref{fig:Sys_components_dphi_KaonKaon_0Vo100},~\ref{fig:Sys_components_integral_ProtonProton},~\ref{fig:Sys_components_dy_widths_PionKaon},~\ref{fig:Sys_components_dphi_widths_PionProton},
~\ref{fig:Sys_components_dphi_projections_KaonProton} for all species pairs. \\
Step 2: The difference threshold is set to $D=d/\sqrt{2}$ according to Barlow, in cases where two extreme scenarios are compared. \\
Step 3: The difference threshold $D$ is subject to the Barlow criterion:
if $D>\sqrt{\sigma_{1}^{2} \pm \sigma_{2}^{2}}$ ($+$ for correlated samples; $-$ for uncorrelated samples; $\sigma_1$ and $\sigma_2$ stand for statistical uncertainties obtained with cut 1 and 2, respectively), then the Barlow criterion $D_{B}$  is set to  $D_{B}=D$, otherwise, it is set to   $D_{B}=0$, as shown in the $3^{rd}$ column of Figs.~\ref{fig:Sys_components_dy_projections_PionPion_300Vo400} --
%,~\ref{fig:Sys_components_dphi_KaonKaon_0Vo100},~\ref{fig:Sys_components_integral_ProtonProton},~\ref{fig:Sys_components_dy_widths_PionKaon},~\ref{fig:Sys_components_dphi_widths_PionProton},~
 \ref{fig:Sys_components_dphi_projections_KaonProton}. \\
Step 4: Systematic uncertainties on $R_{2}^{CD}$ are calculated assuming the six potential sources of uncertainties  are statistically independent, by taking a  sum in quadrature of the  non-vanishing contributions, as shown in Eq.(\ref{eq:sys_6_parts}).
\begin{equation}
\label{eq:sys_6_parts}
\sigma_{R_{2}^{CD}} = \left[ D_{B}^{2}(BField) + D_{B}^{2}(V_{z}) + D_{B}^{2}(PID).\right. \\ \nonumber
\left. + D_{B}^{2}(Eff) + D_{B}^{2}(DCA) + D_{B}^{2}(nCluster) \right]^{1/2}.
\end{equation}
Step 5: Calculate the average of systematic uncertainties of all the bins, denoted ${\bar\sigma_{R_{2}^{CD}}}$.\\
Step 6: For  $\Delta y$ and $\Delta\varphi$ projections of balance functions, and their integrals,  add, in quadrature,  systematic uncertainties of single particle densities of $\pi^{\pm}$, $K^{\pm}$ and $p/\bar{p}$ ~\cite{PhysRevC.88.044910}. Thus, the  systematic uncertainties on BF amplitudes are the quadratic sum of systematic uncertainties of CF ($\sigma_{R_{2}^{CD}}$) and the systematic uncertainties from the published single particle densities ($\sigma_{\rho}$), as shown in Eq.(\ref{eq:sys_density_plus_this_analysis}). 
The values of total systematic uncertainties ($\sigma_{BF}$) are shown in Tables~\ref{tab:Final_Systematic_Uncertainty_BF_PiPi} 
%~\ref{tab:Final_Systematic_Uncertainty_BF_PiK},~\ref{tab:Final_Systematic_Uncertainty_BF_PiPr},~\ref{tab:Final_Systematic_Uncertainty_BF_KPi},~\ref{tab:Final_Systematic_Uncertainty_BF_KK},~\ref{tab:Final_Systematic_Uncertainty_BF_KPr},~\ref{tab:Final_Systematic_Uncertainty_BF_PrPi},~\ref{tab:Final_Systematic_Uncertainty_BF_PrK},
-- \ref{tab:Final_Systematic_Uncertainty_BF_PrPr} for all species pairs.

\begin{equation}
\sigma_{BF}=\sqrt{{{\bar\sigma_{R_{2}^{CD}}}}^{2}+\sigma_{\rho}^{2}}
\label{eq:sys_density_plus_this_analysis}
\end{equation}
\begin{figure}
\centering
  \includegraphics[width=0.32\linewidth]{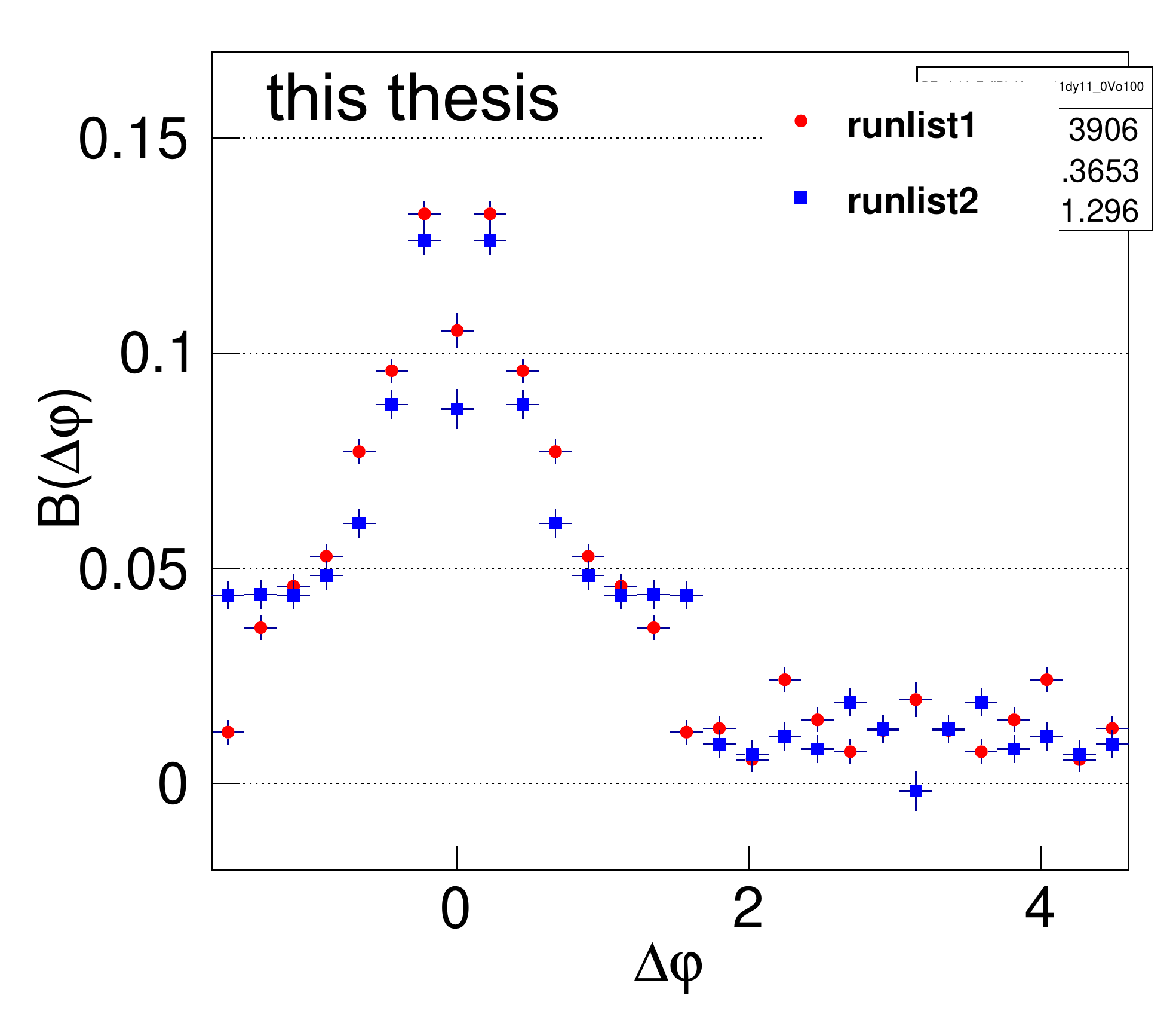}
  \includegraphics[width=0.32\linewidth]{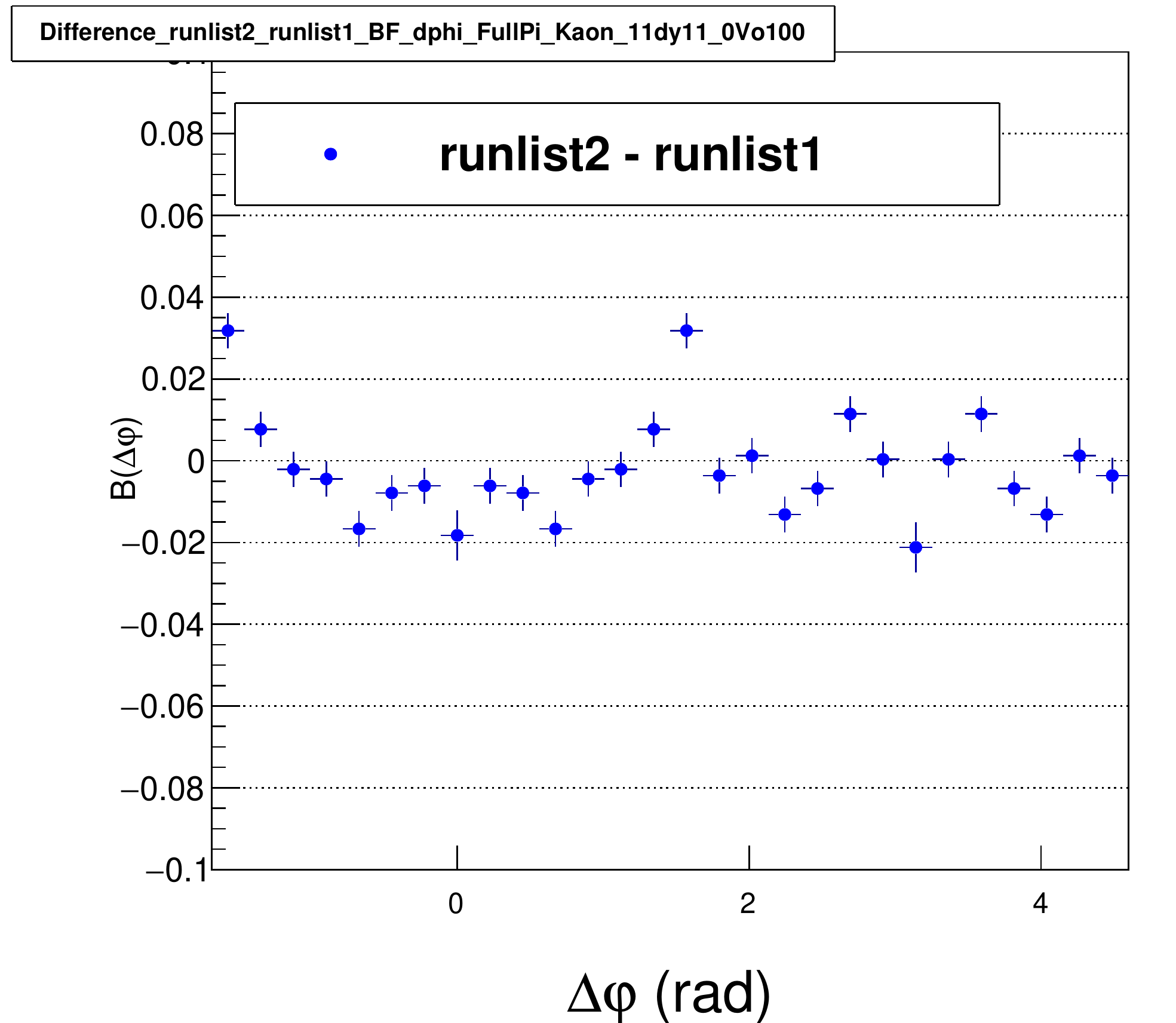}
  \includegraphics[width=0.32\linewidth]{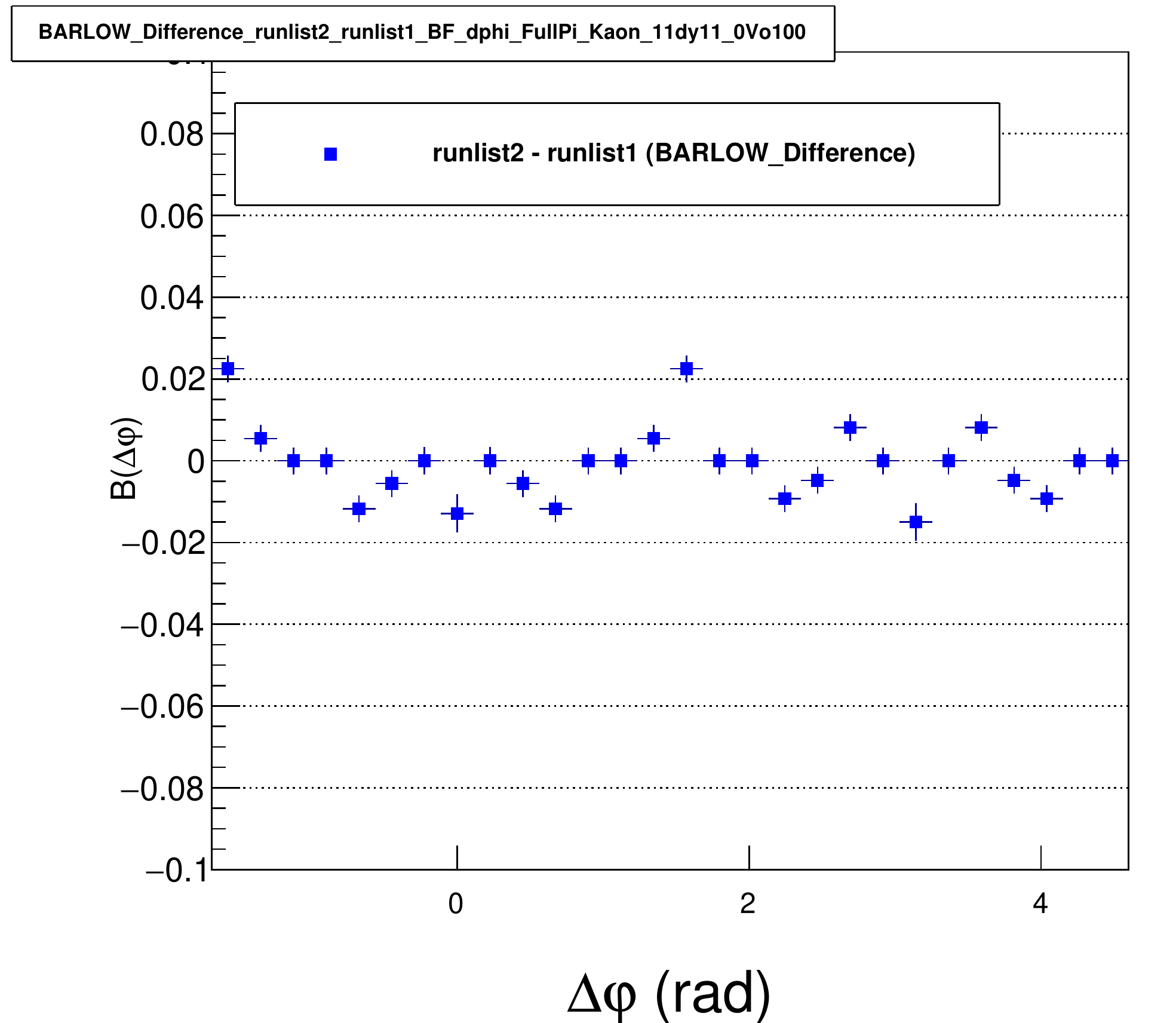}
  
  \includegraphics[width=0.32\linewidth]{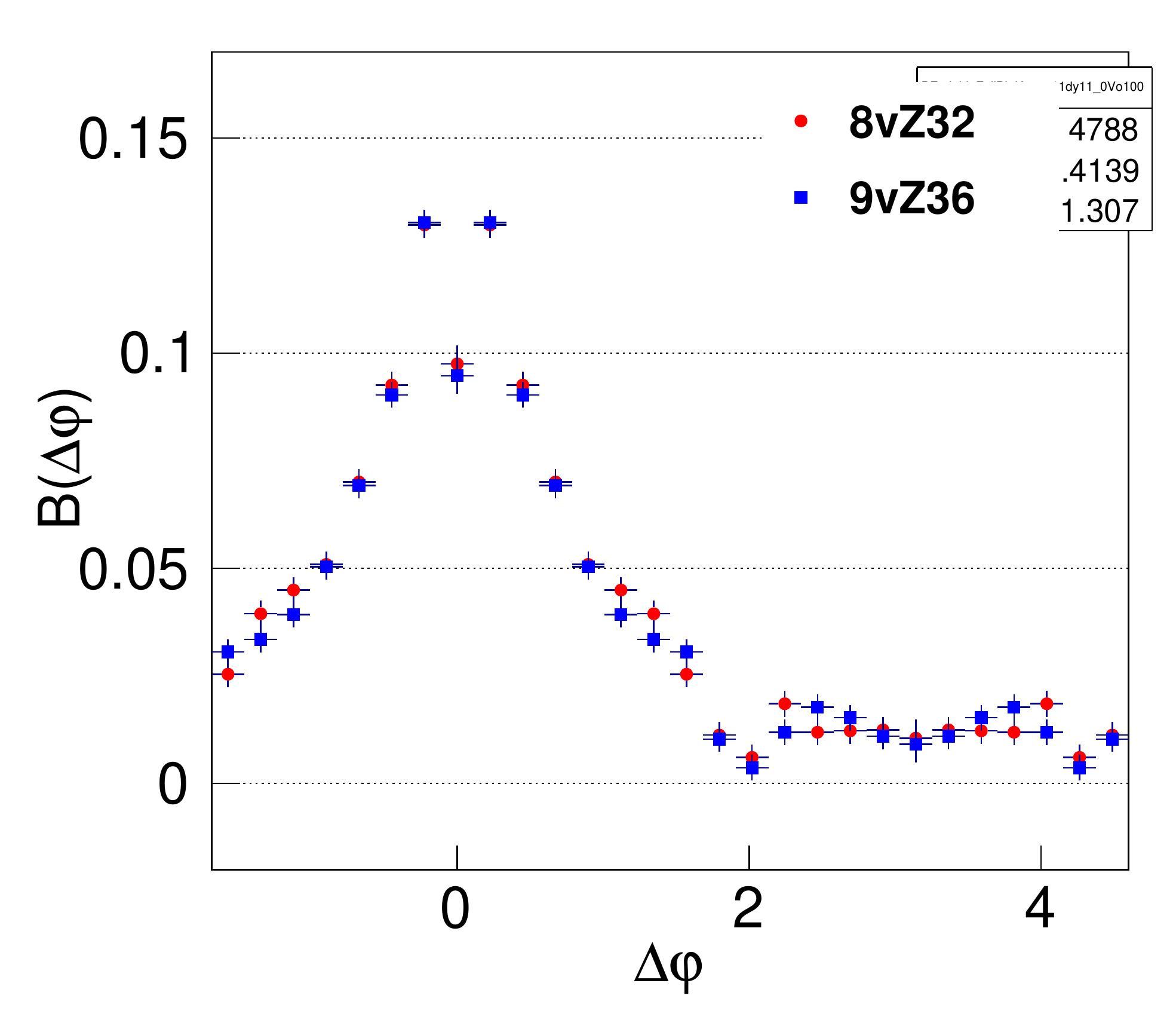}
  \includegraphics[width=0.32\linewidth]{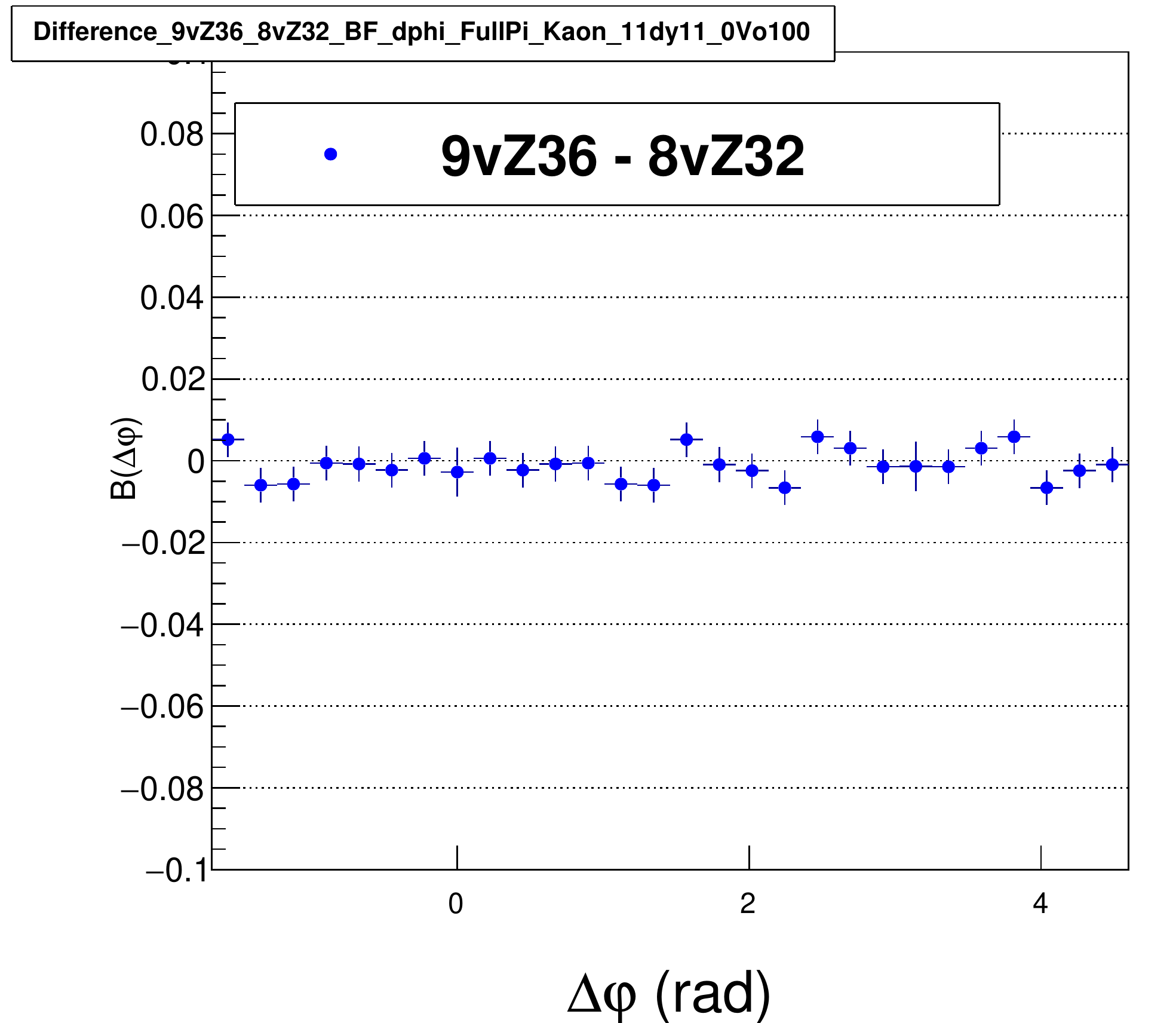}
  \includegraphics[width=0.32\linewidth]{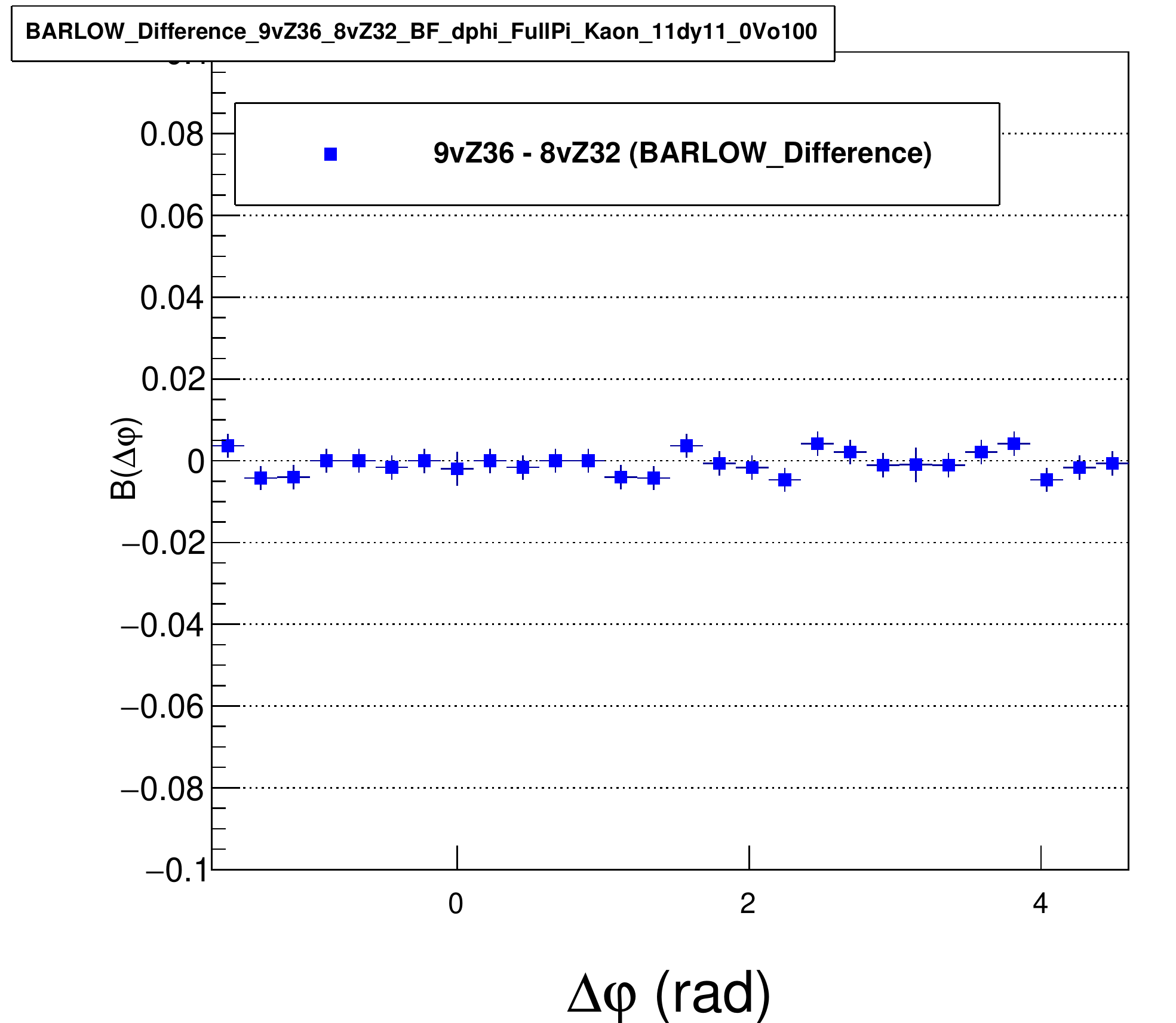}
  
  \includegraphics[width=0.32\linewidth]{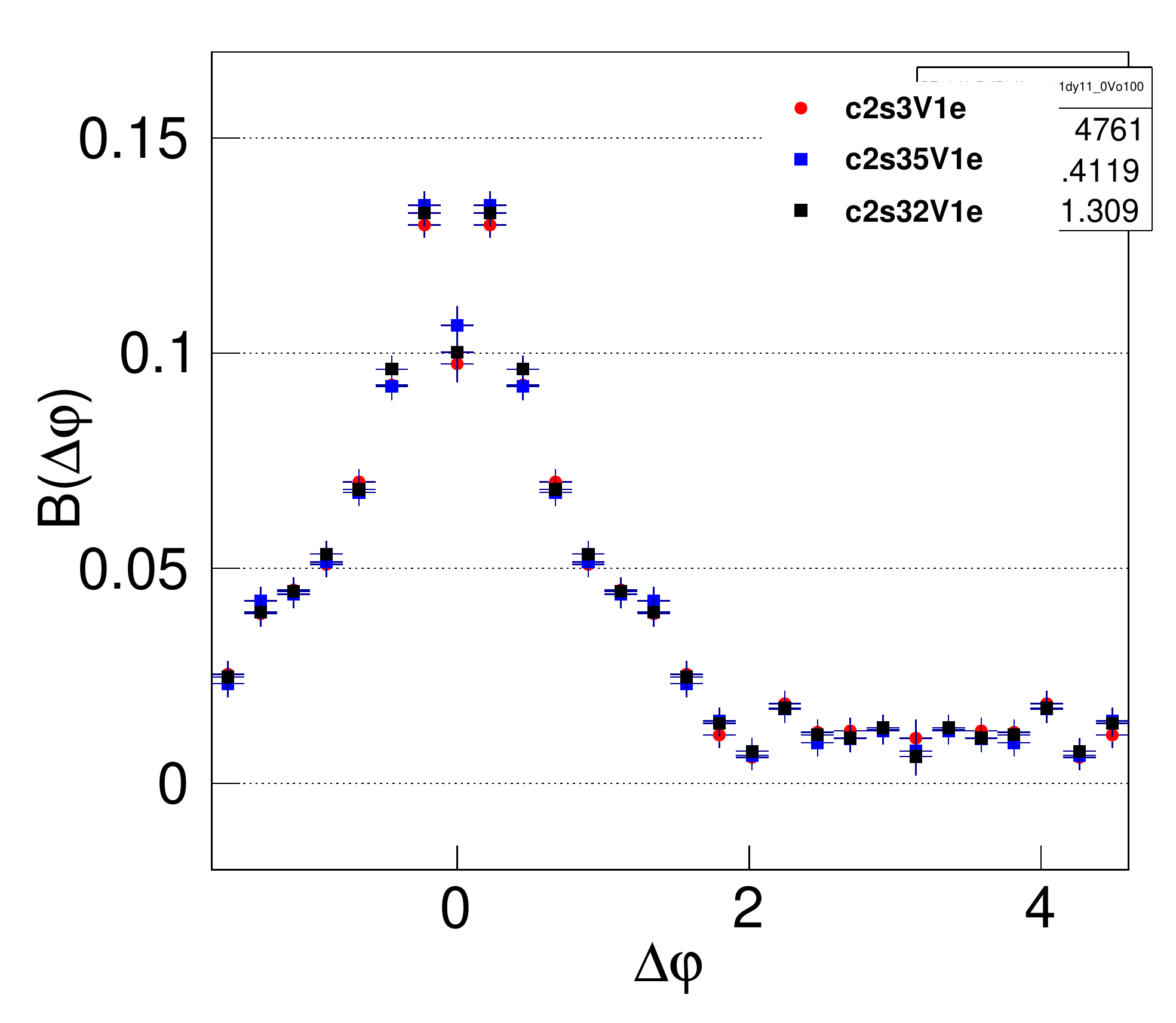}
  \includegraphics[width=0.32\linewidth]{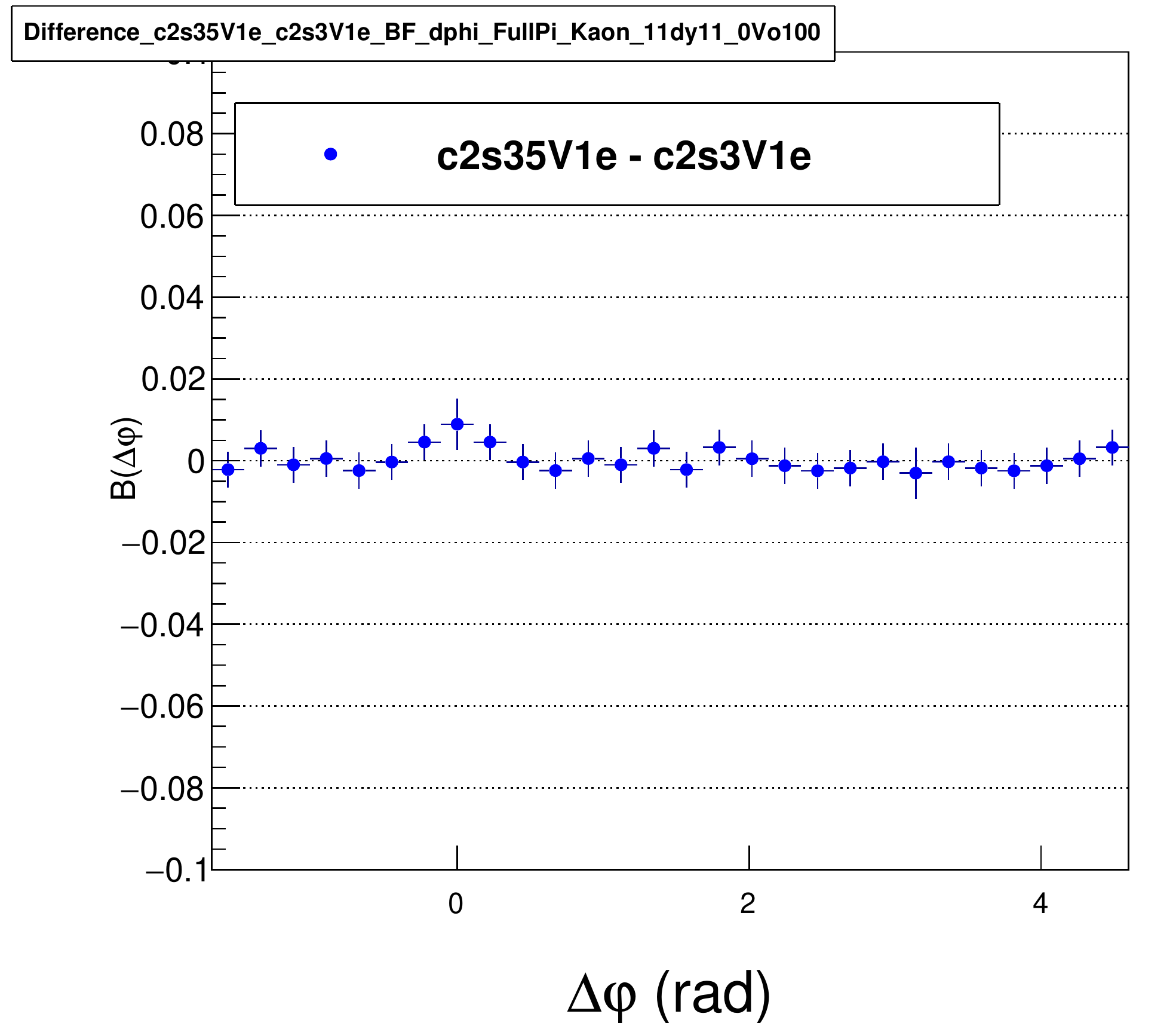}
  \includegraphics[width=0.32\linewidth]{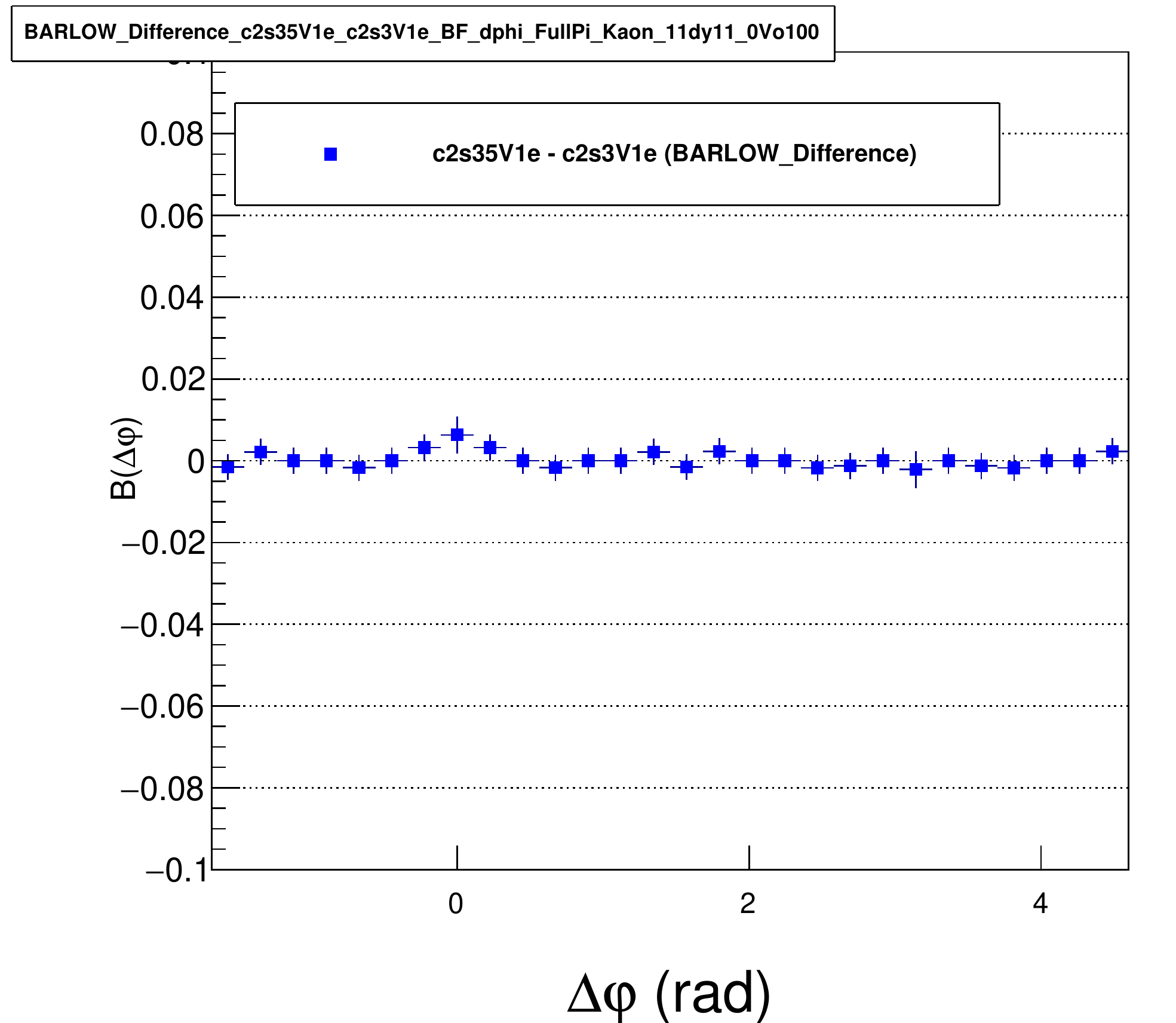}

  \includegraphics[width=0.32\linewidth]{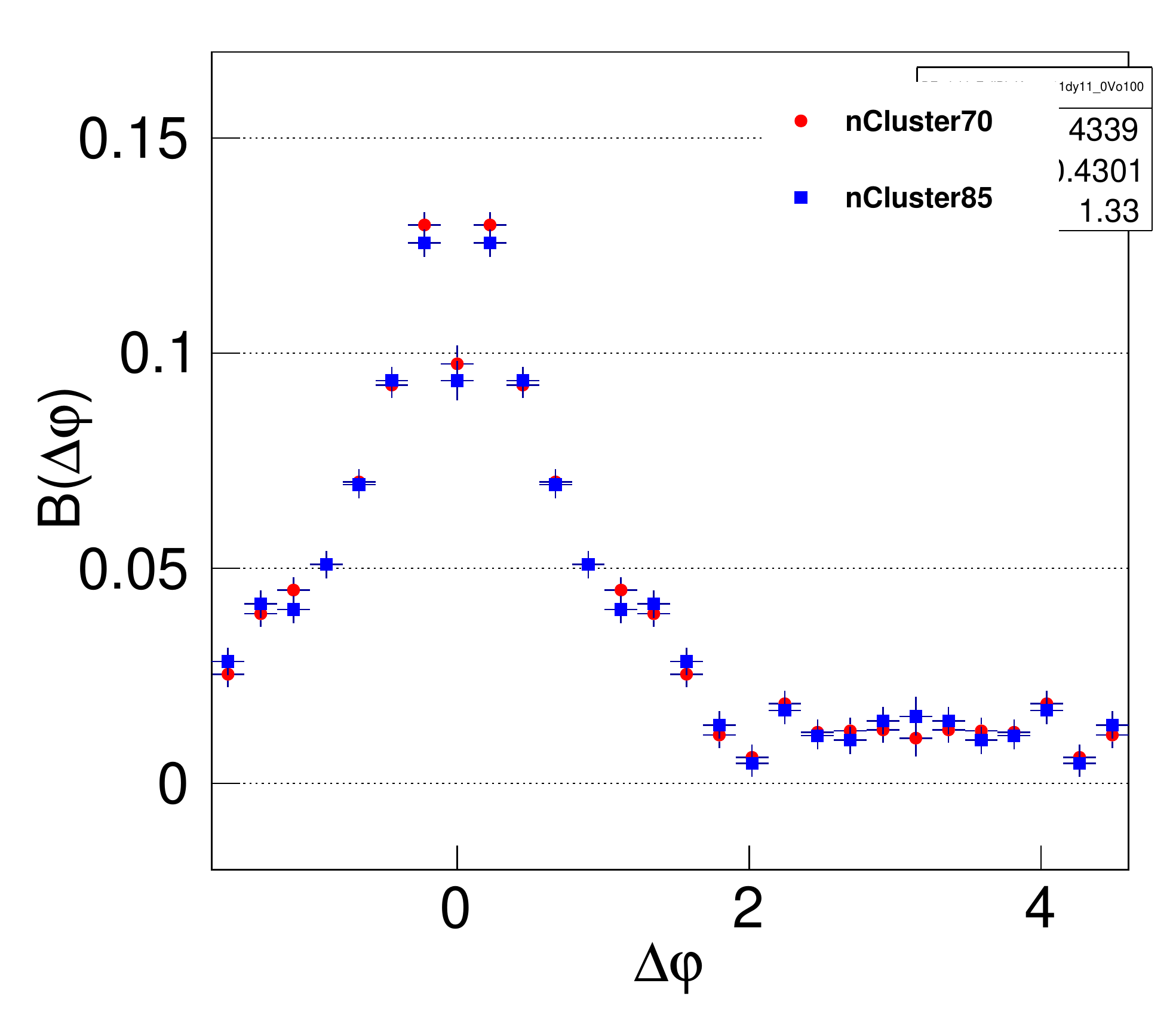}
  \includegraphics[width=0.32\linewidth]{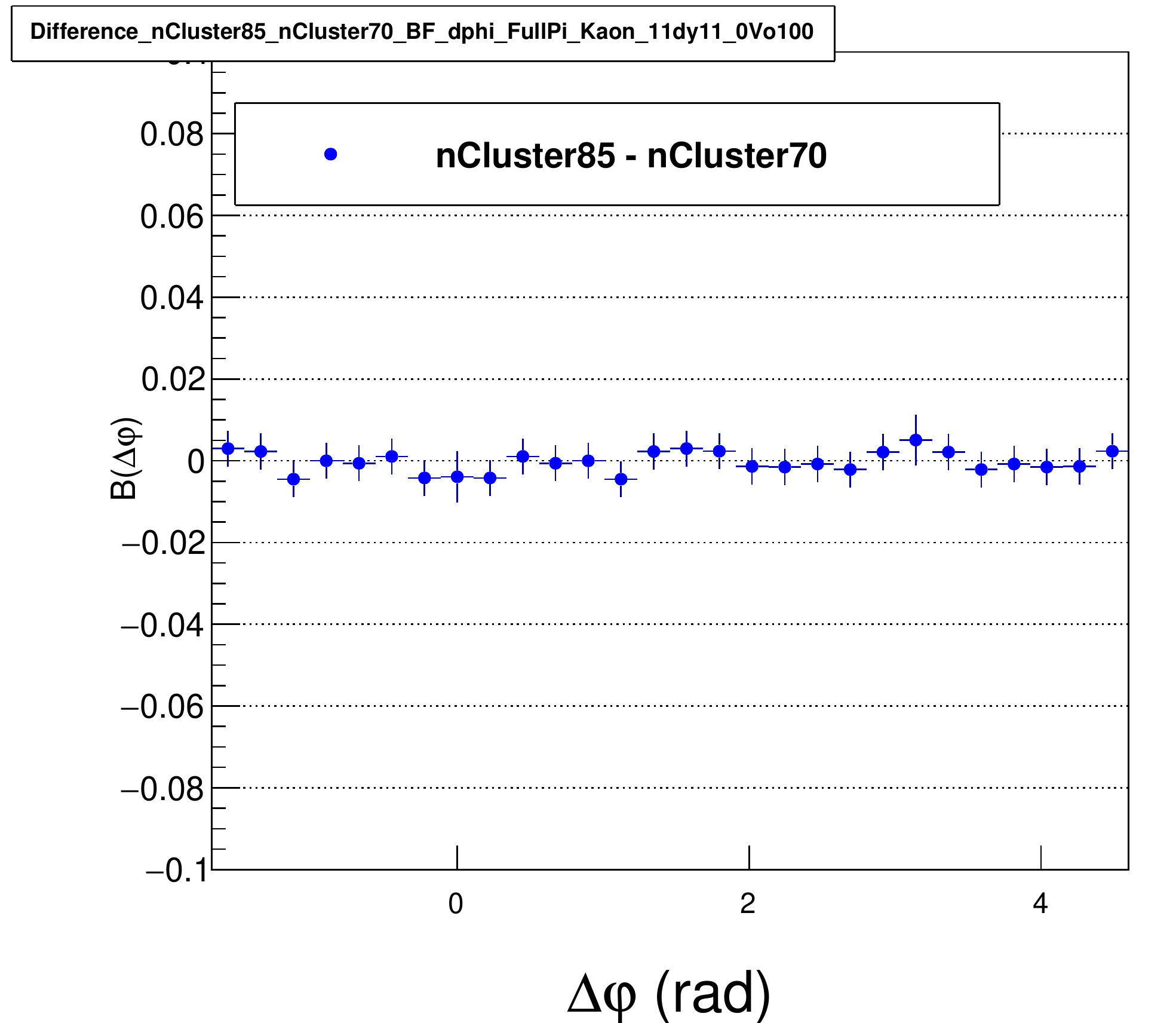}
  \includegraphics[width=0.32\linewidth]{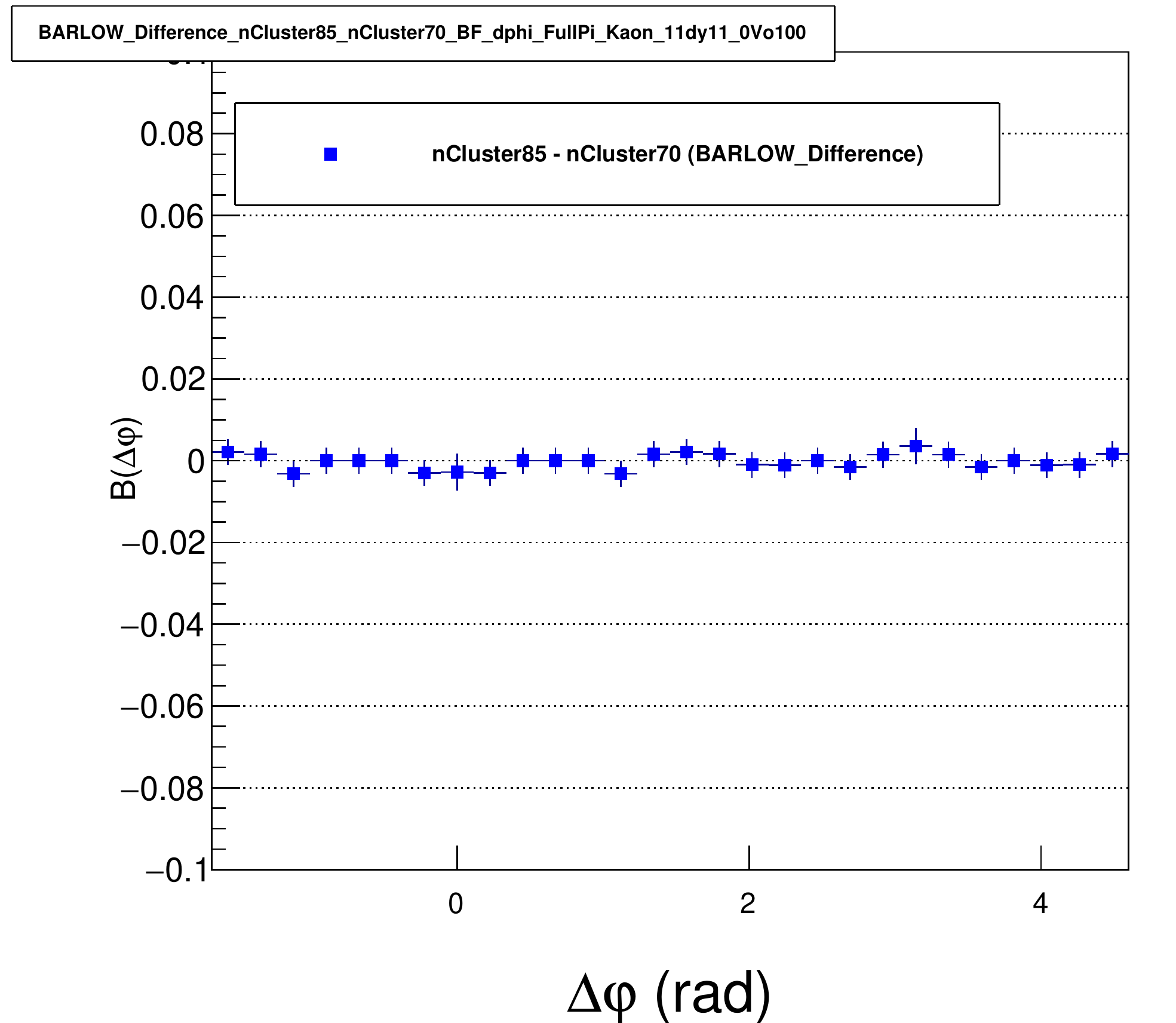}

  \caption{Systematic uncertainty contributions in $B^{KK}(\Delta\varphi)$ from BField ($1^{st}$ row), $V_{z}$ ($2^{nd}$ row), PID ($3^{rd}$ row), and nClusters ($4^{th}$ row). The comparisons between two sets of different cuts (left column), with their differences $d$ (middle column), and their differences after the Barlow check $D_{Barlow}$ (right column).}
  \label{fig:Sys_components_dphi_KaonKaon_0Vo100}
\end{figure}
\begin{figure}
\centering
  \includegraphics[width=0.32\linewidth]{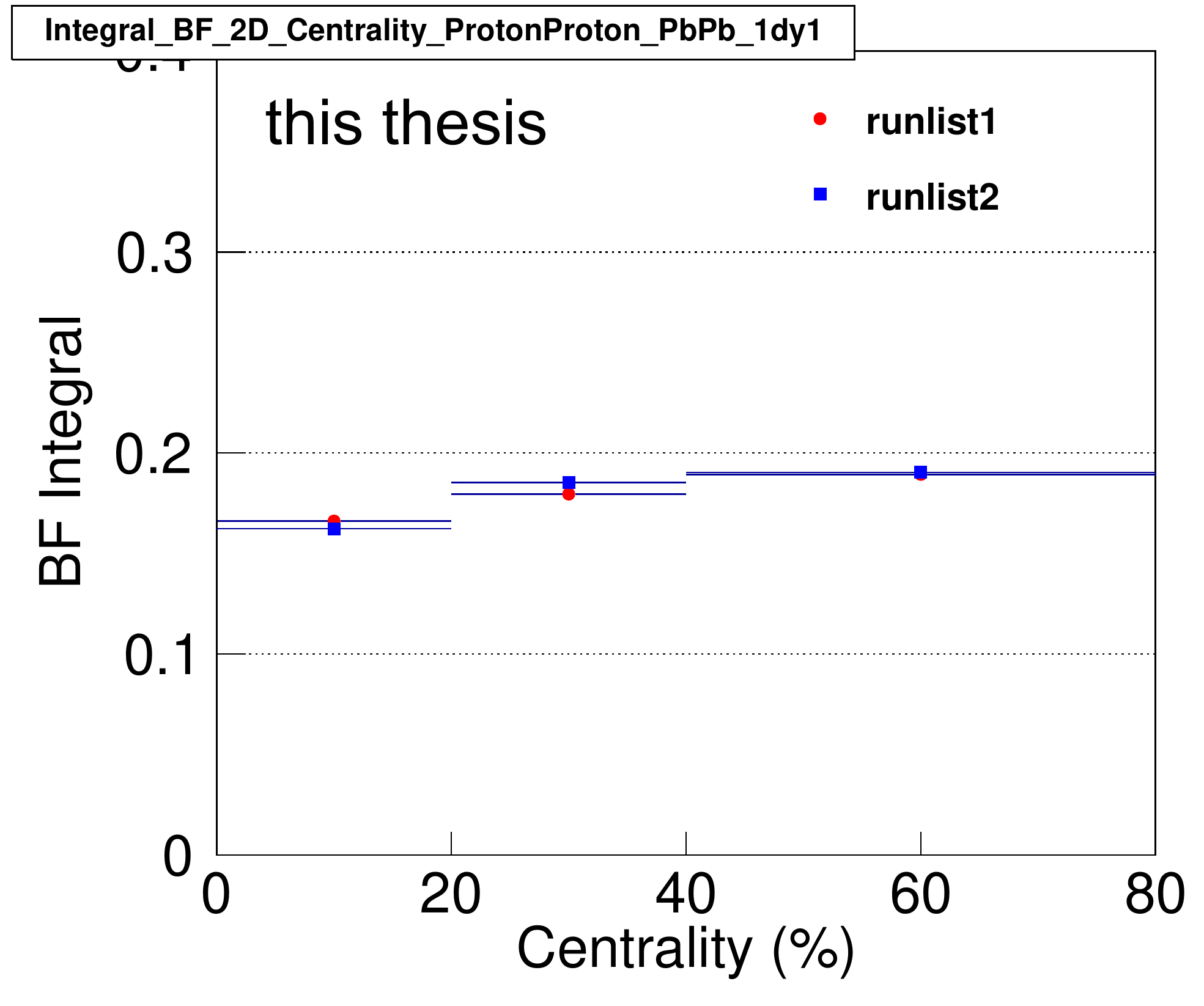}
  \includegraphics[width=0.32\linewidth]{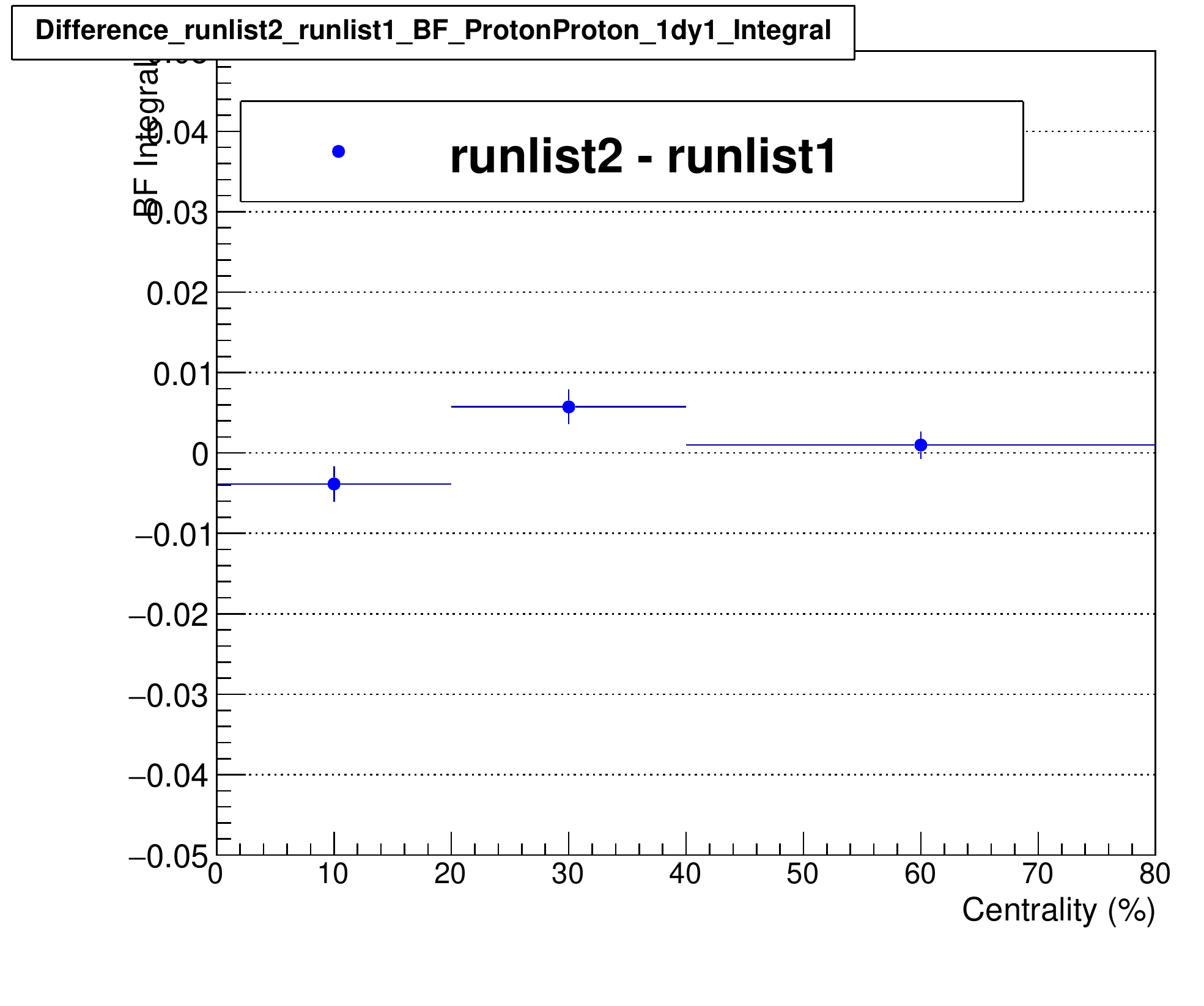}
  \includegraphics[width=0.32\linewidth]{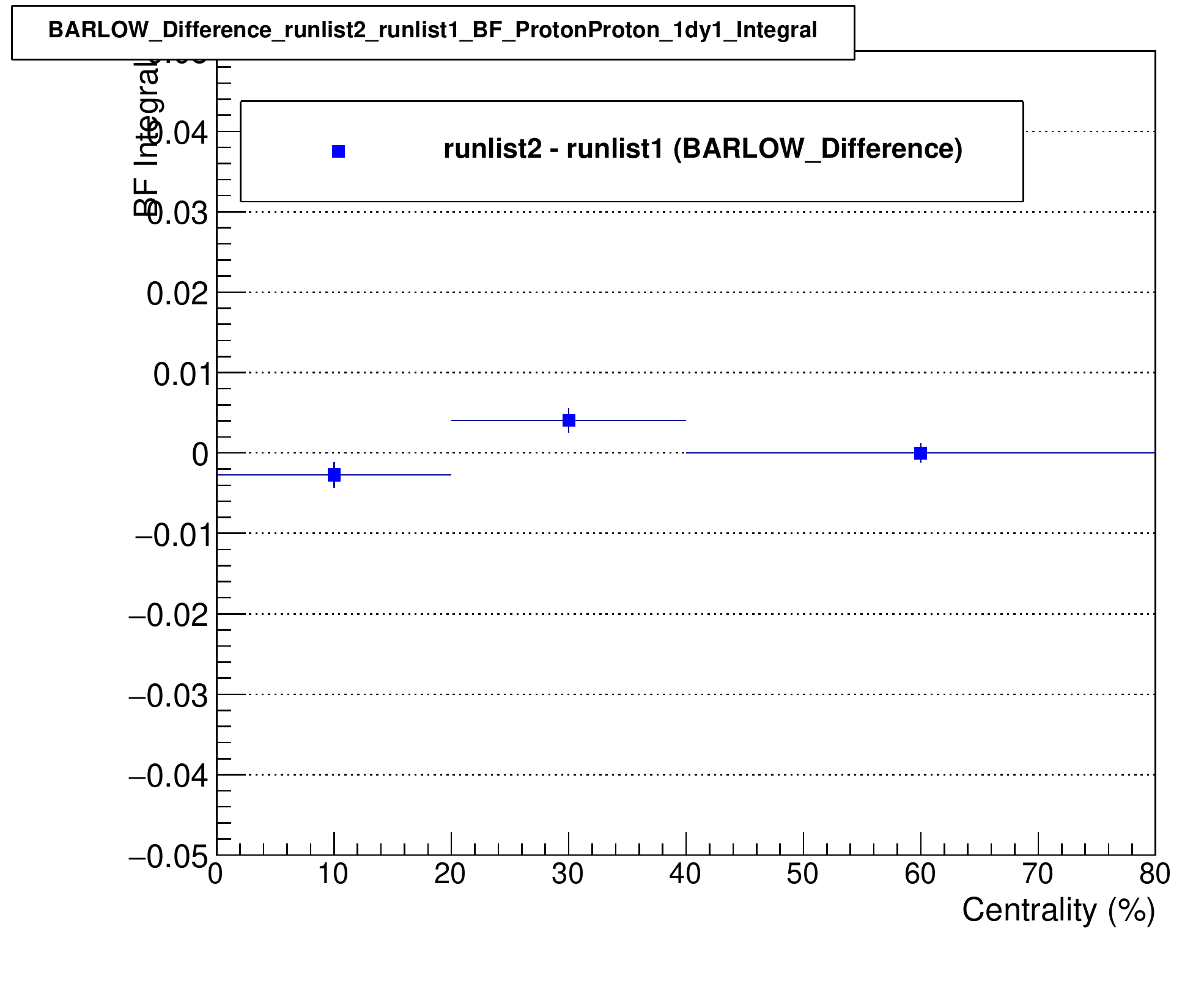}
  
  \includegraphics[width=0.32\linewidth]{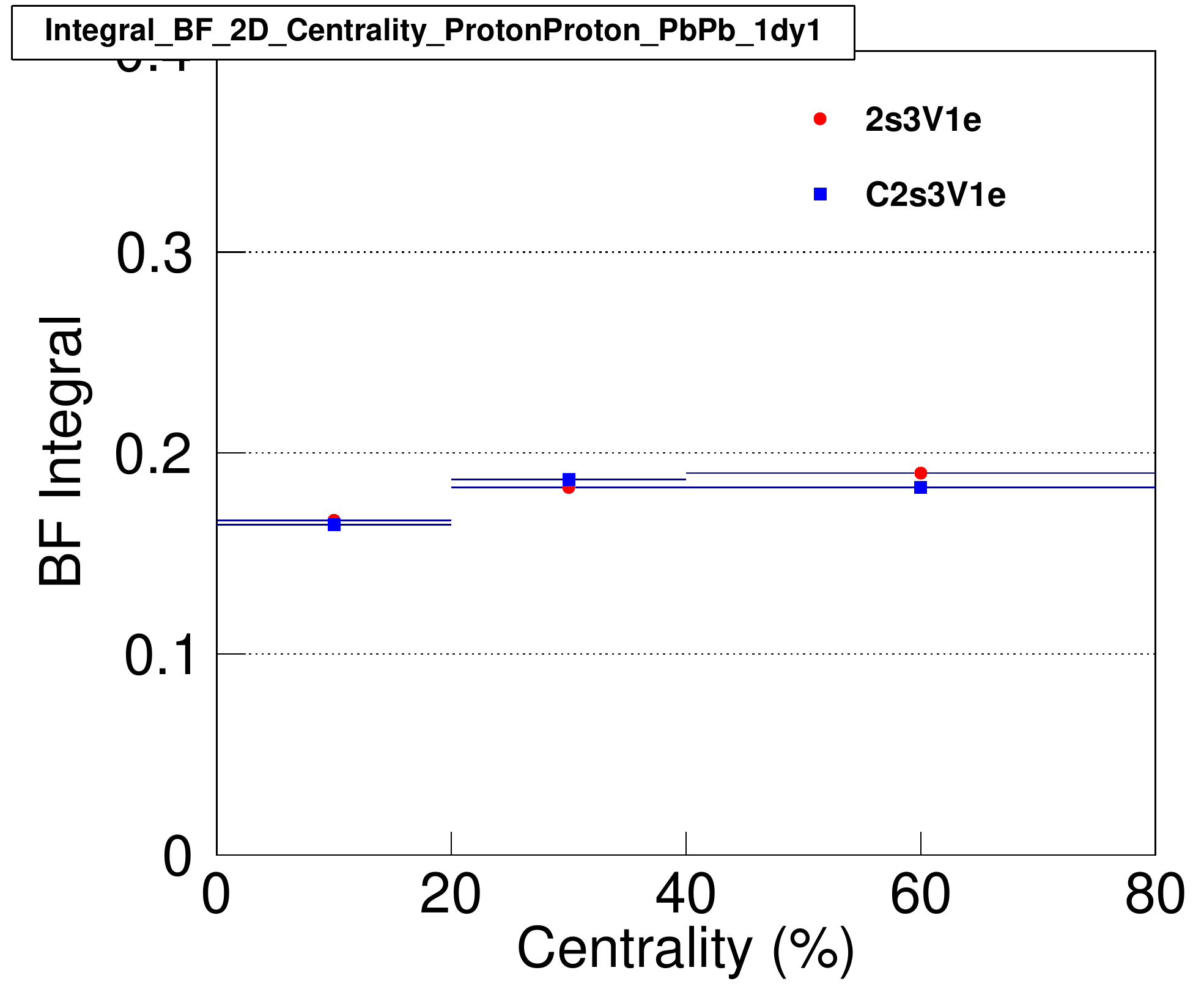}
  \includegraphics[width=0.32\linewidth]{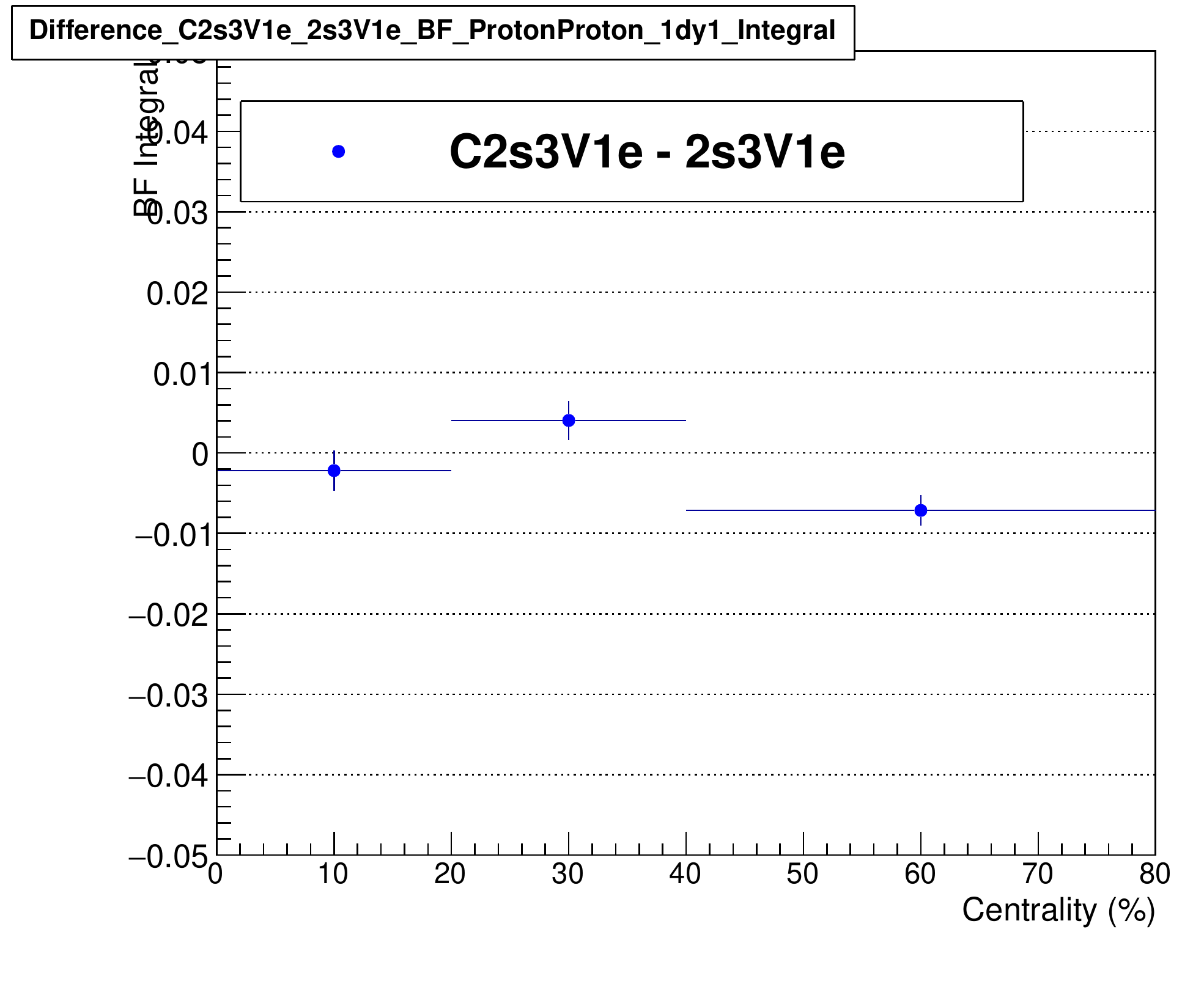}
  \includegraphics[width=0.32\linewidth]{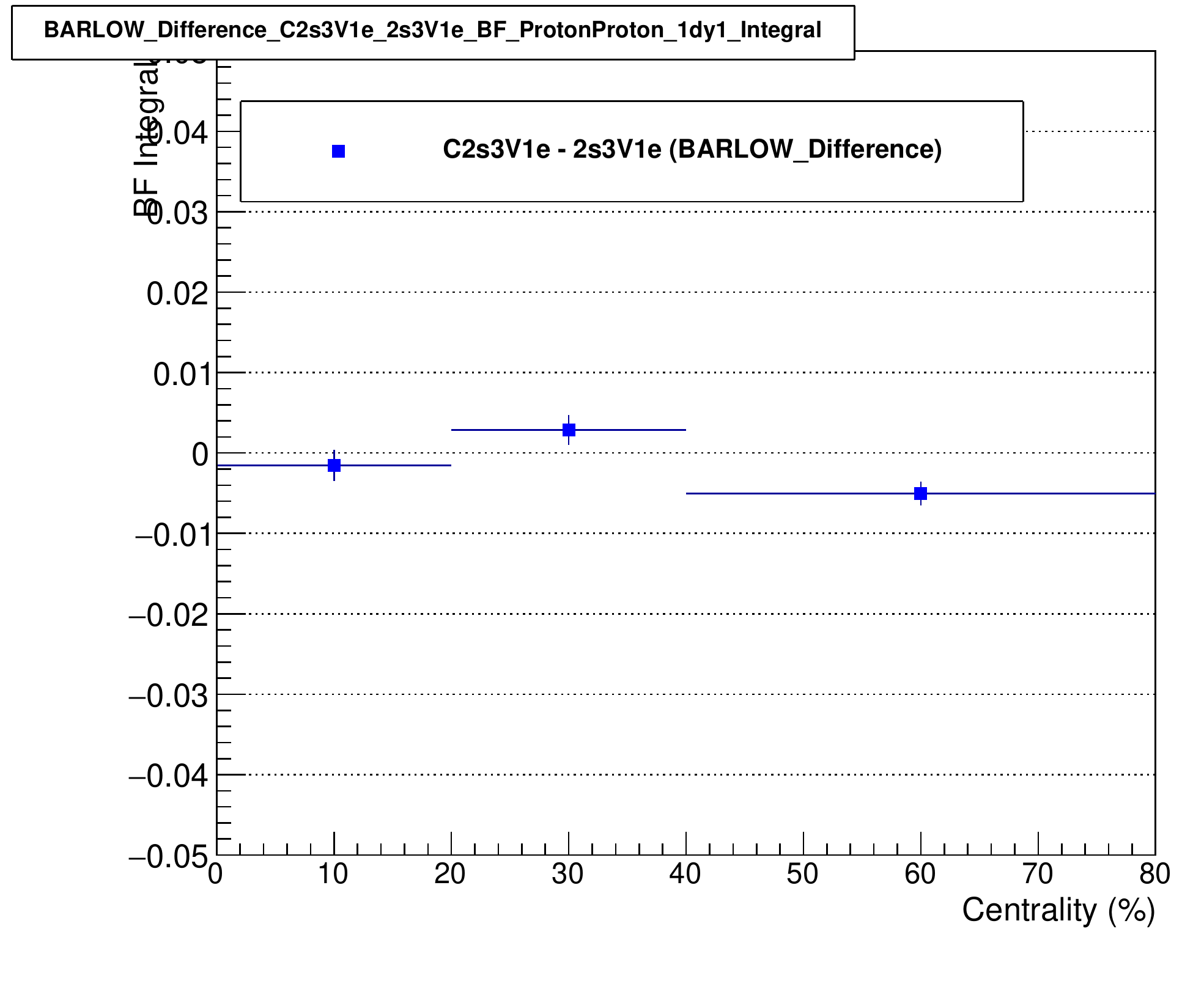}
  
  \includegraphics[width=0.32\linewidth]{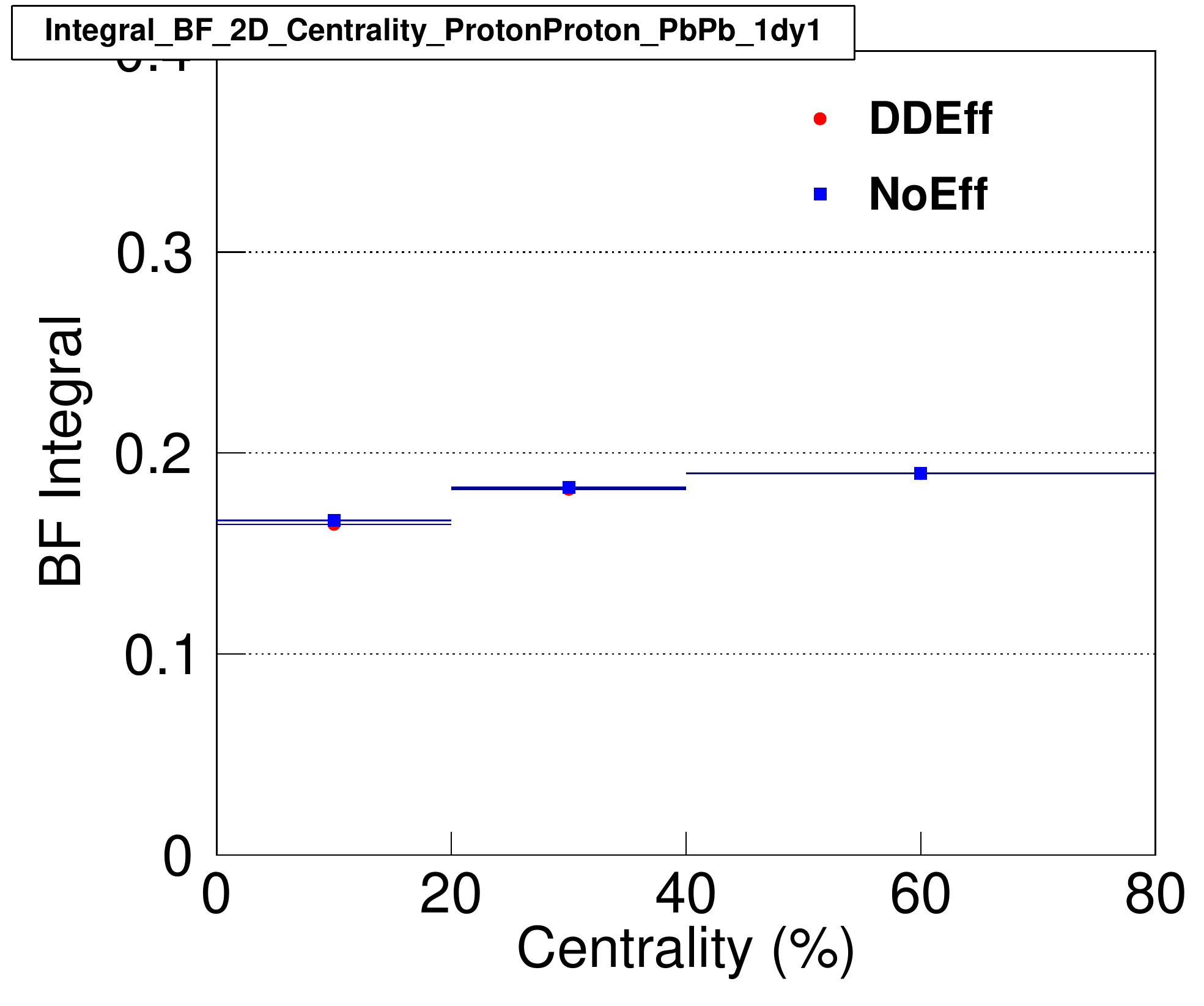}
  \includegraphics[width=0.32\linewidth]{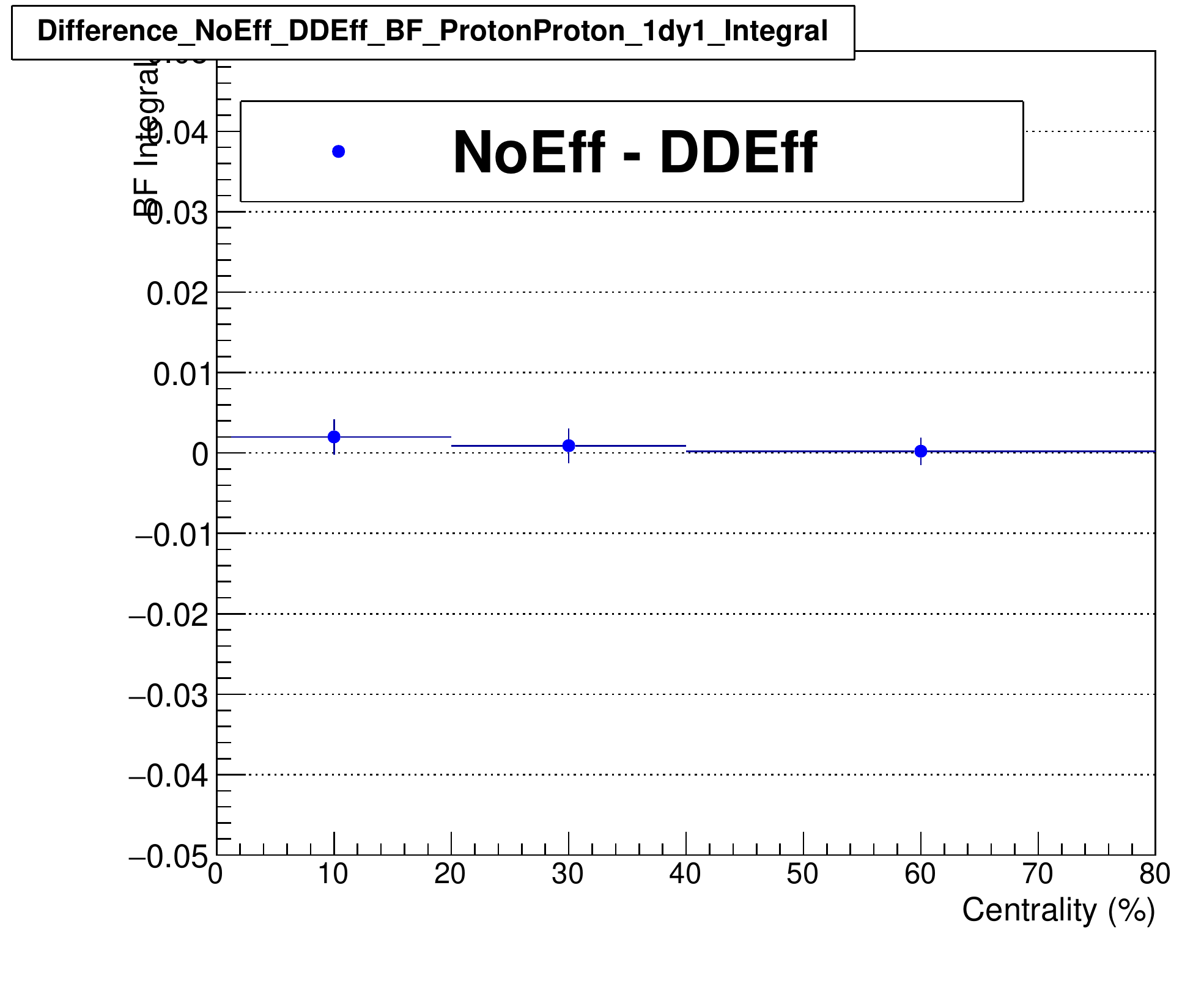}
  \includegraphics[width=0.32\linewidth]{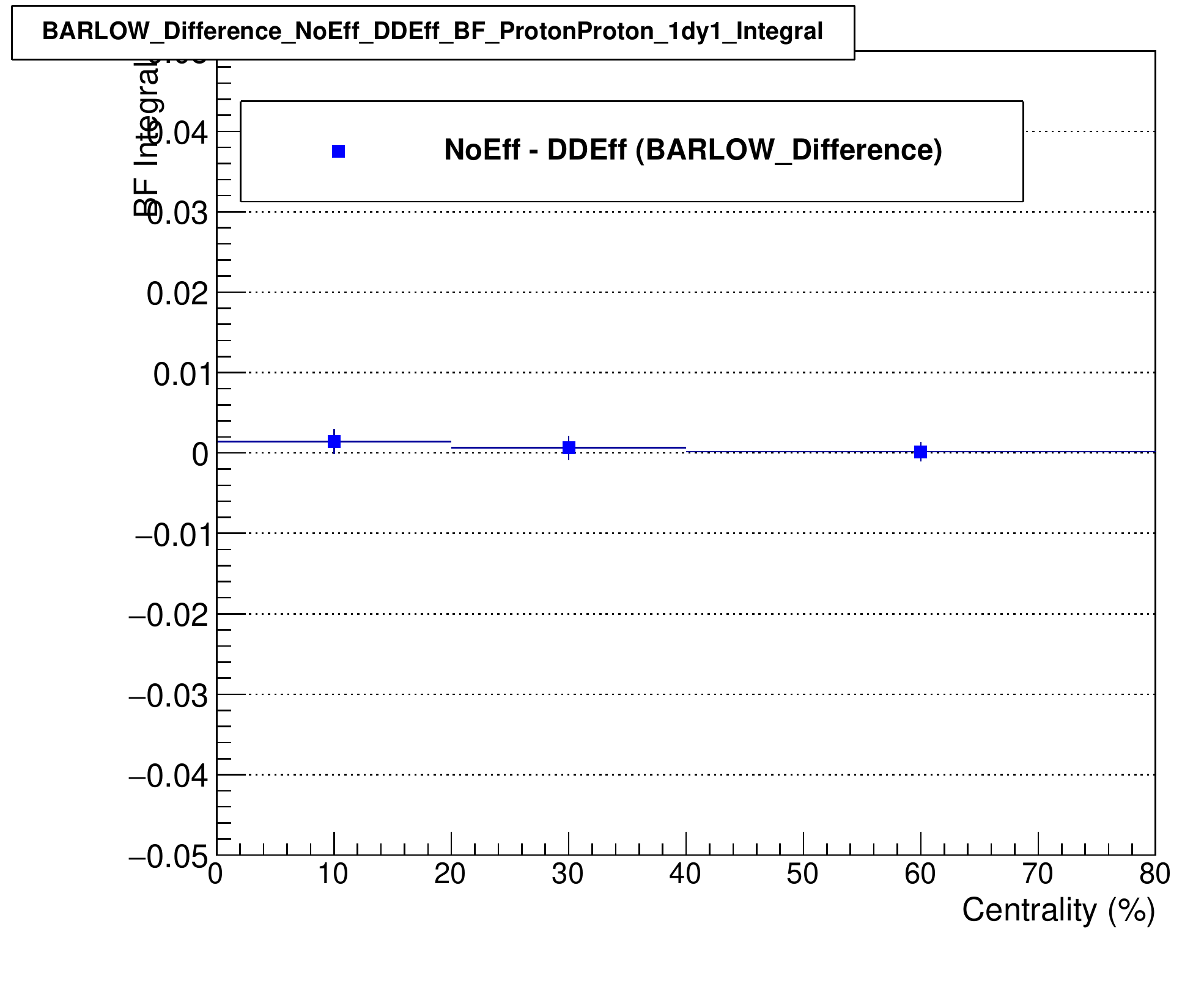}
  
  \includegraphics[width=0.32\linewidth]{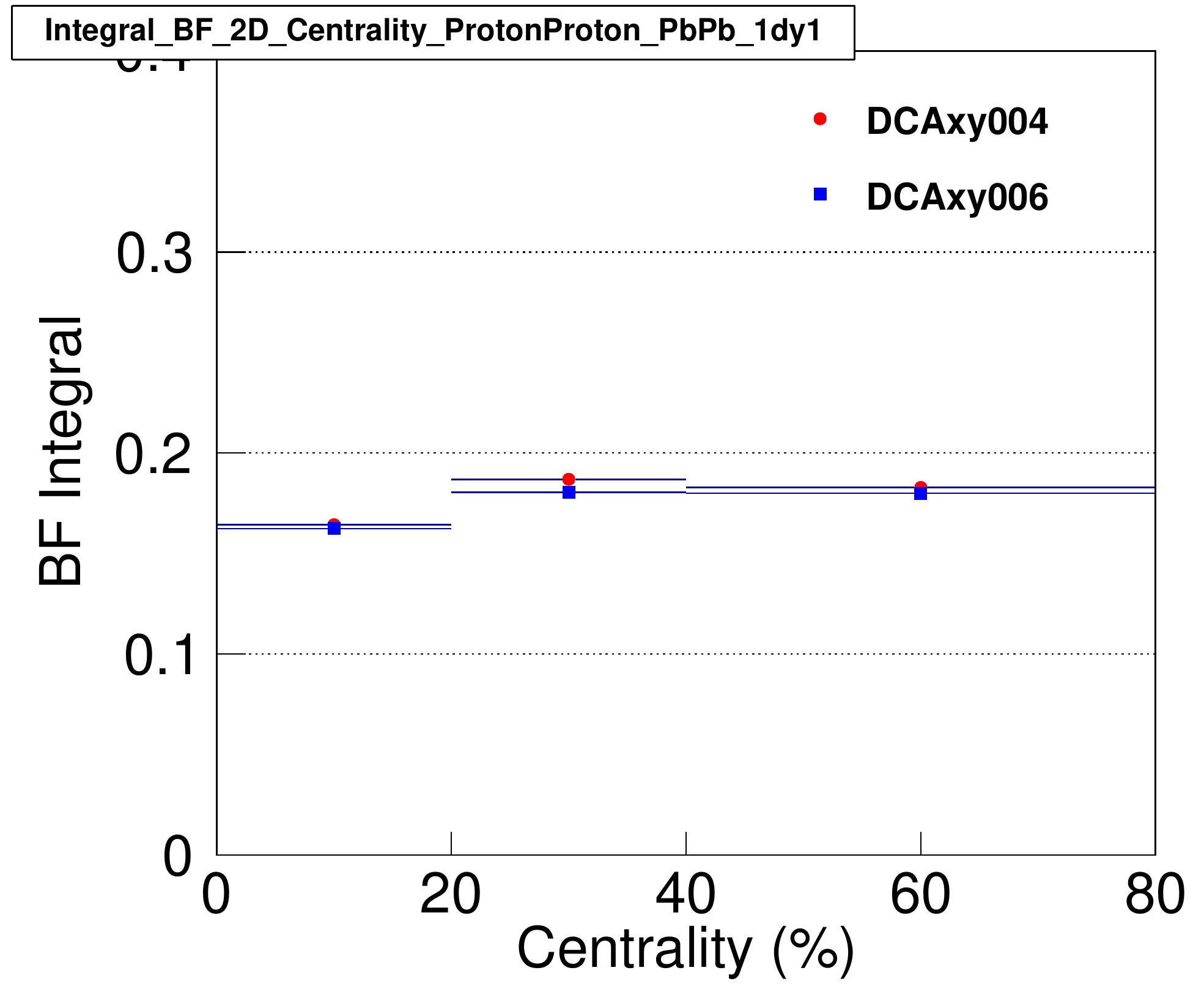}
  \includegraphics[width=0.32\linewidth]{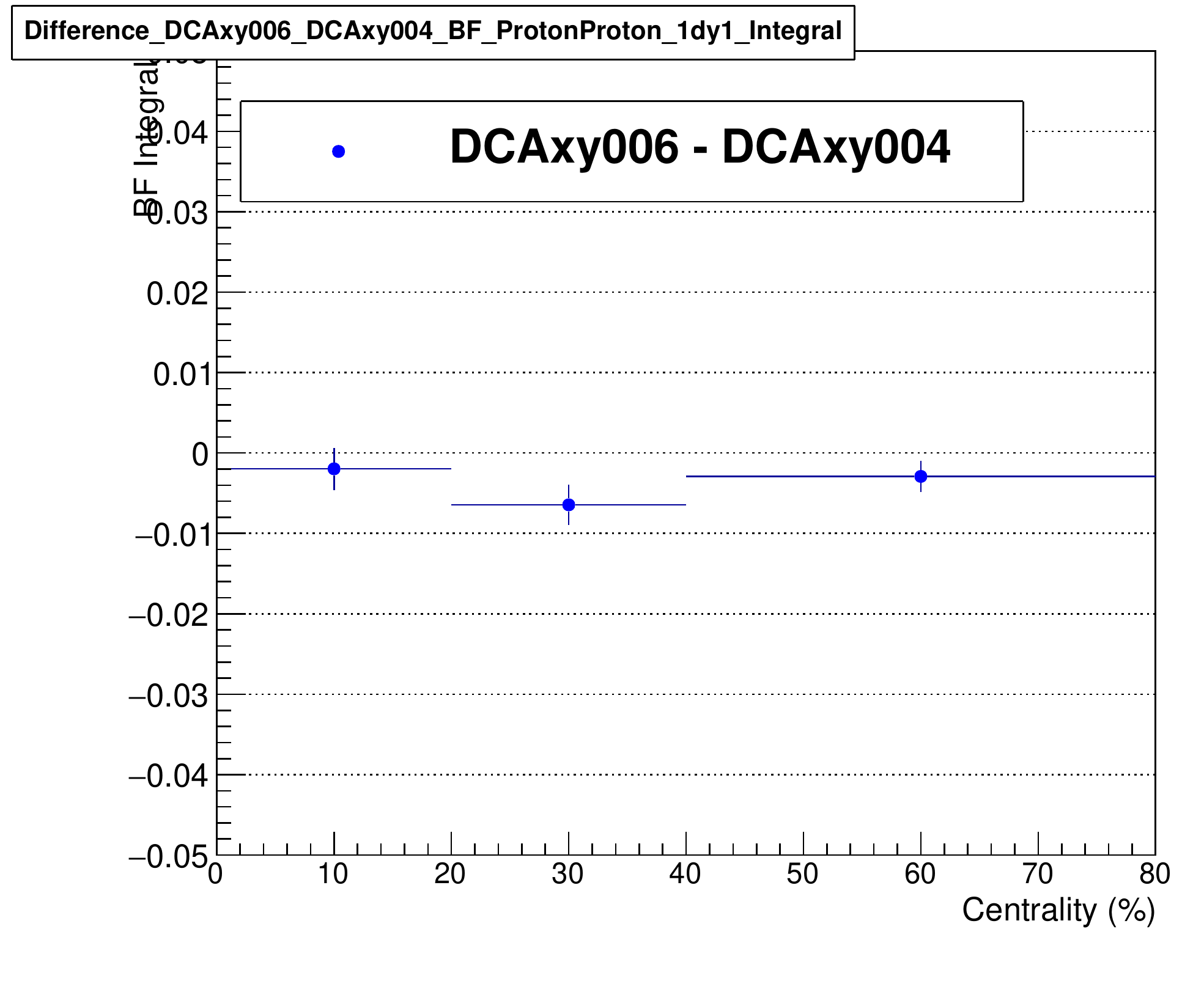}
  \includegraphics[width=0.32\linewidth]{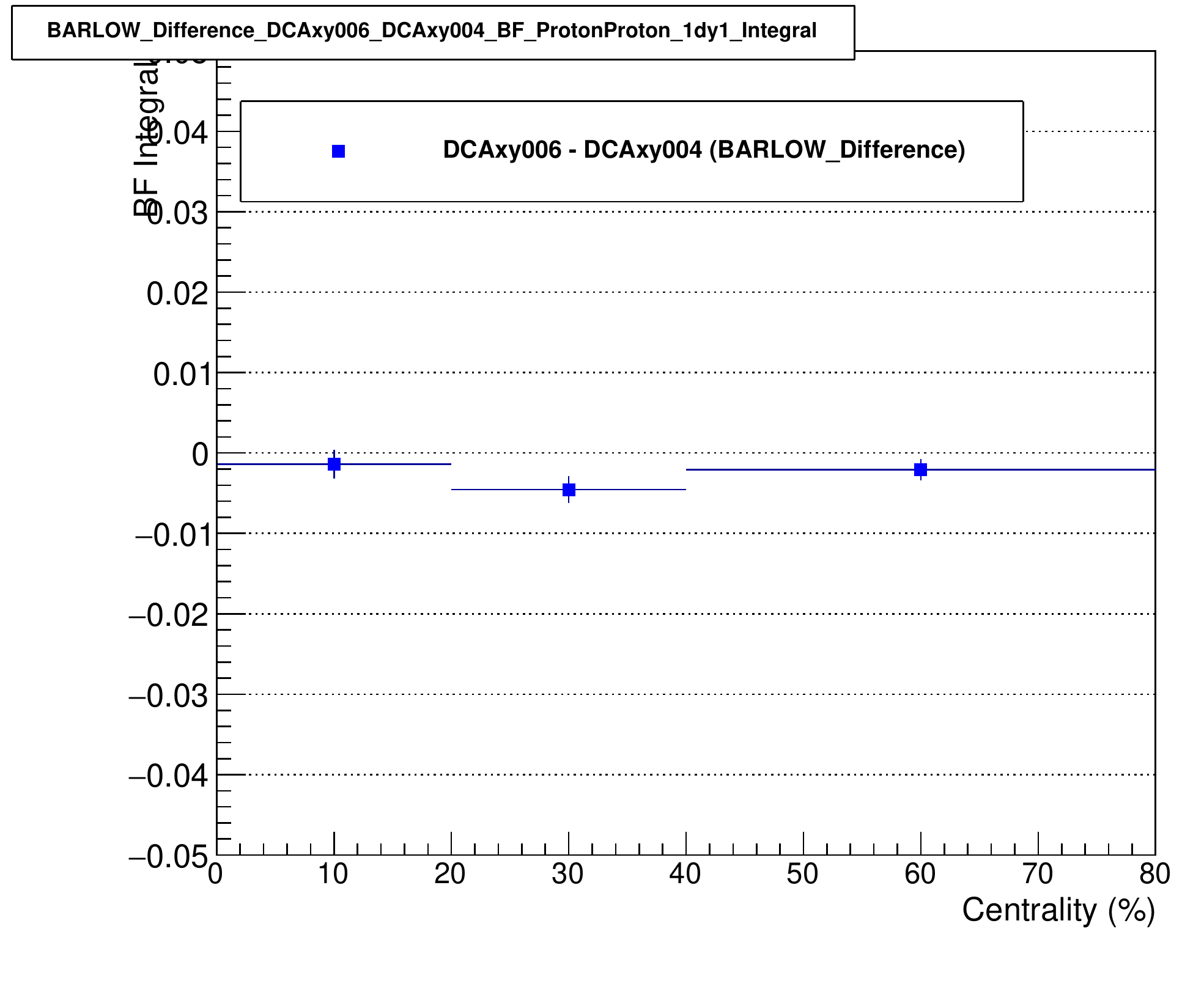}

  \caption{Systematic uncertainty contributions in $B^{pp}$ integrals from BField ($1^{st}$ row), PID ($2^{nd}$ row), additional $p_{\rm T}$-dependent efficiency corrections ($3^{rd}$ row), and DCA ($4^{th}$ row). The comparisons between two sets of different cuts (left column), with their differences $d$ (middle column), and their differences after the Barlow check $D_{Barlow}$ (right column).}
  \label{fig:Sys_components_integral_ProtonProton}
\end{figure}
\begin{figure}
\centering
  \includegraphics[width=0.32\linewidth]{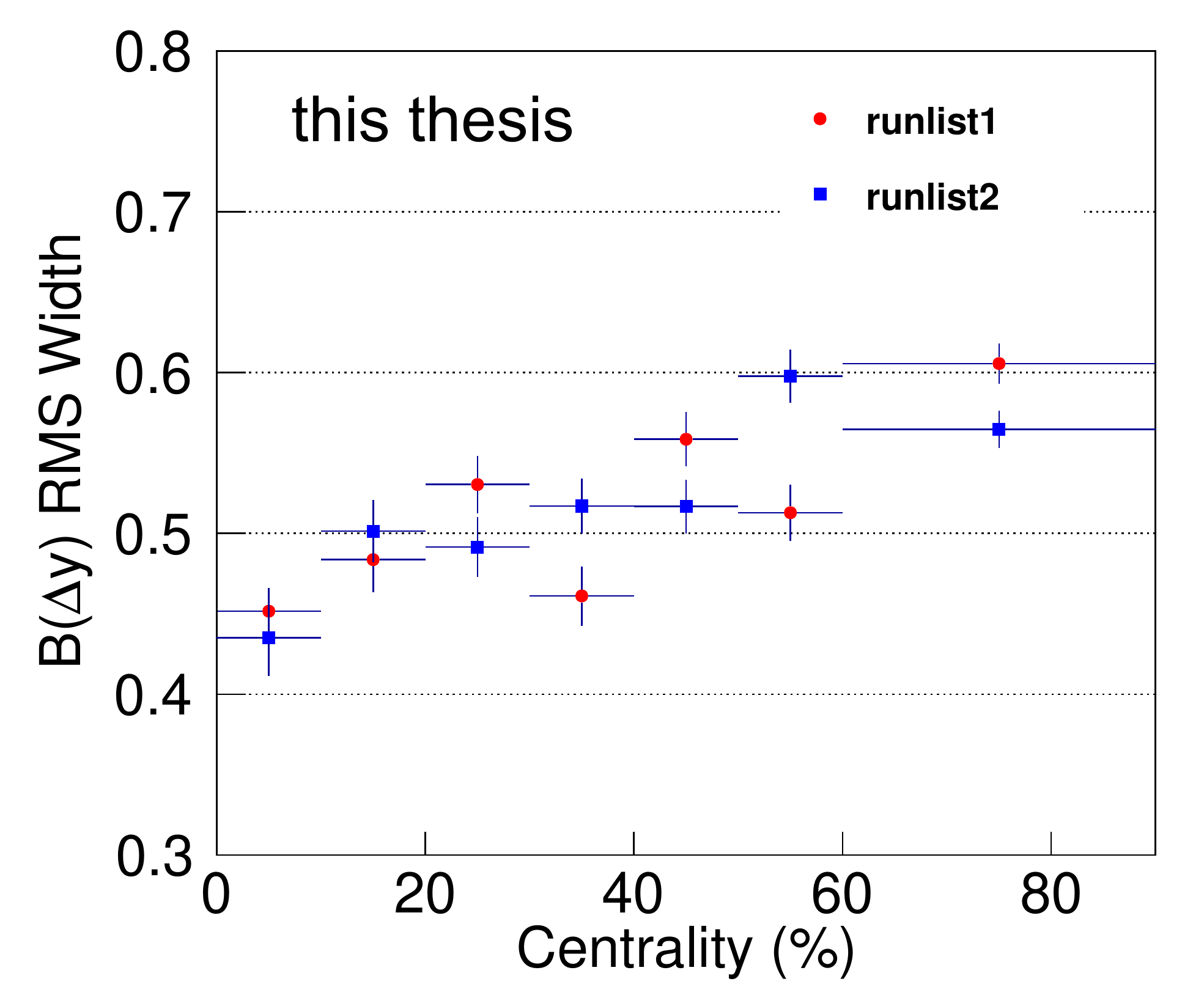}
  \includegraphics[width=0.32\linewidth]{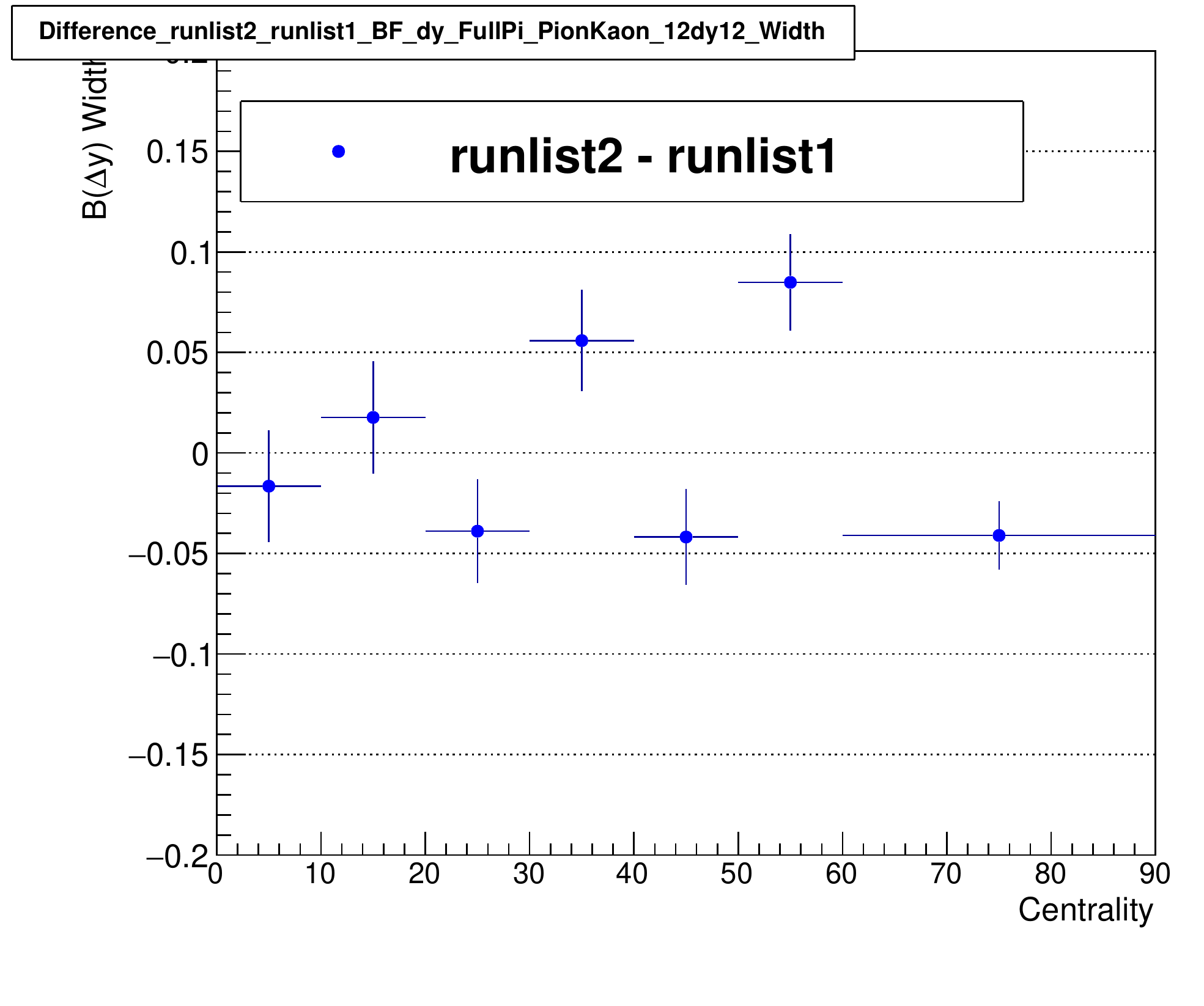}
  \includegraphics[width=0.32\linewidth]{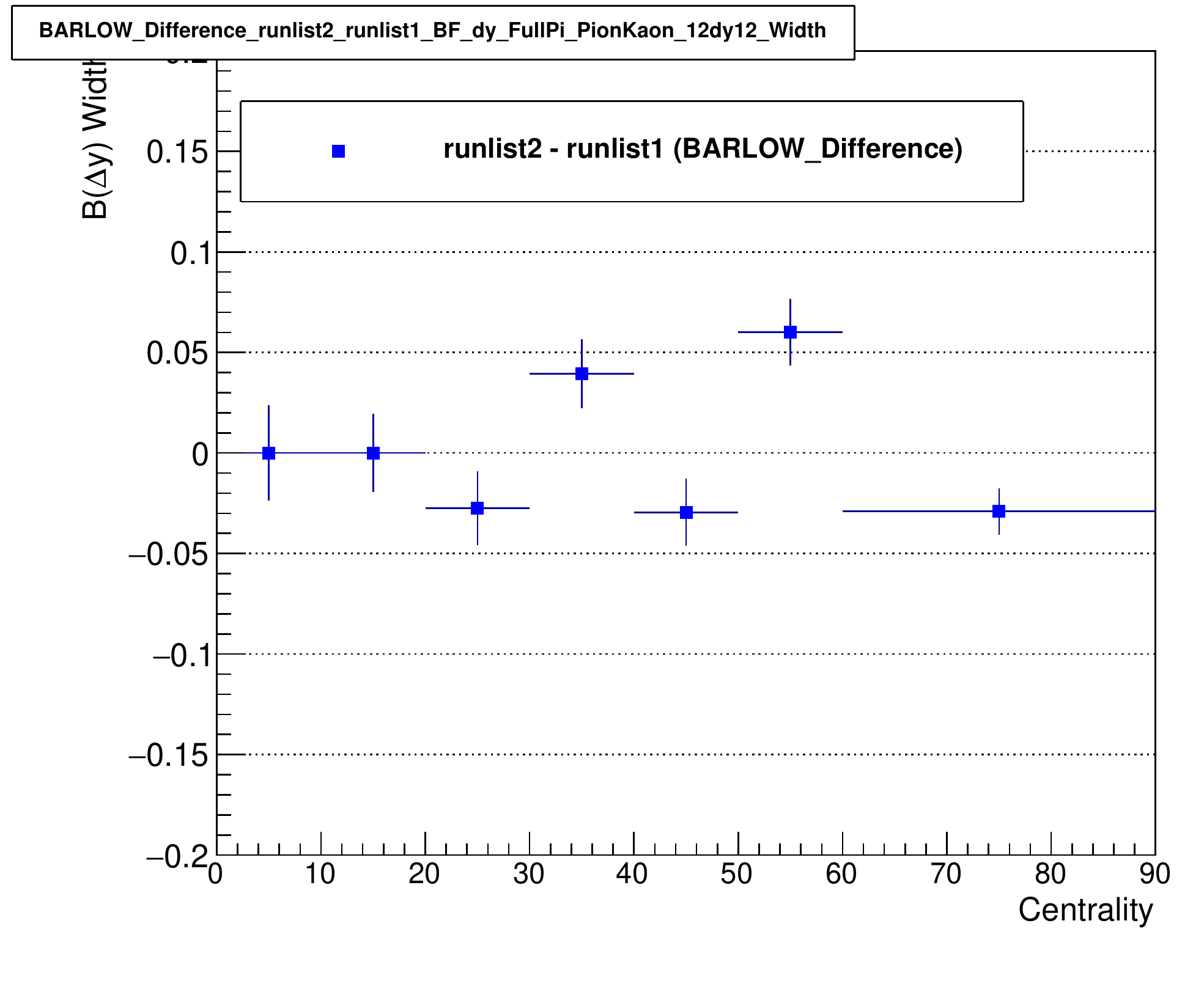}
  
  \includegraphics[width=0.32\linewidth]{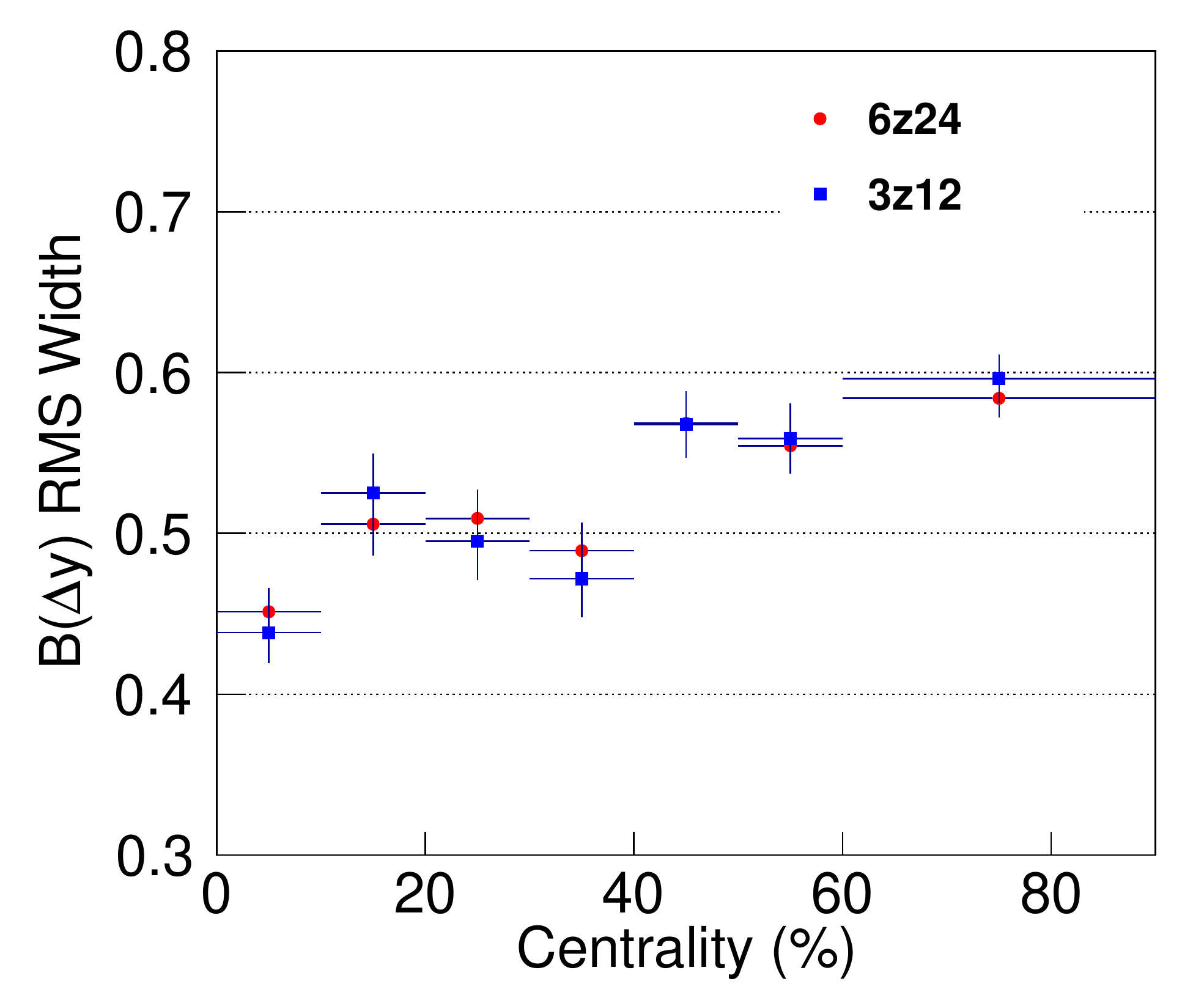}
  \includegraphics[width=0.32\linewidth]{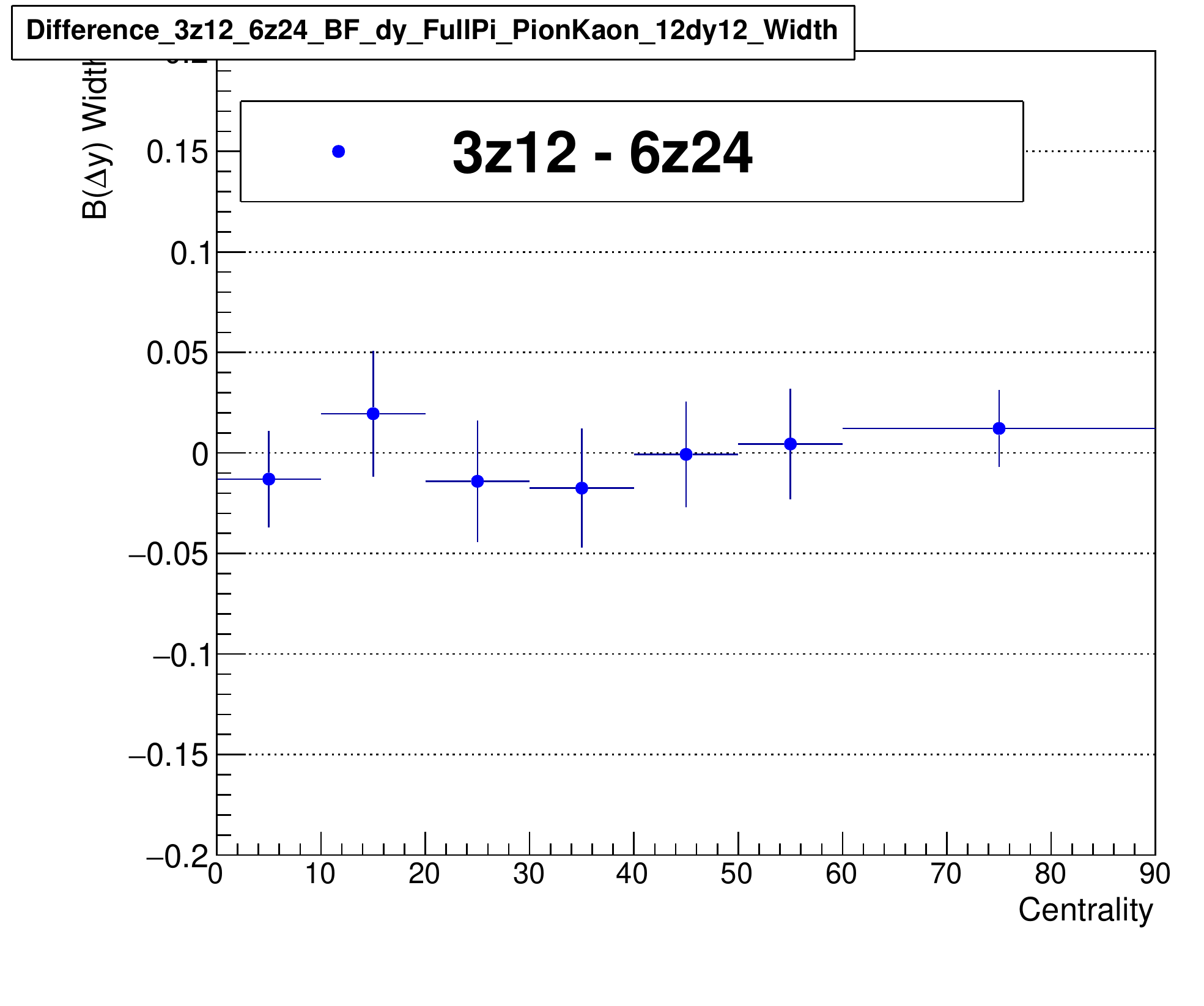}
  \includegraphics[width=0.32\linewidth]{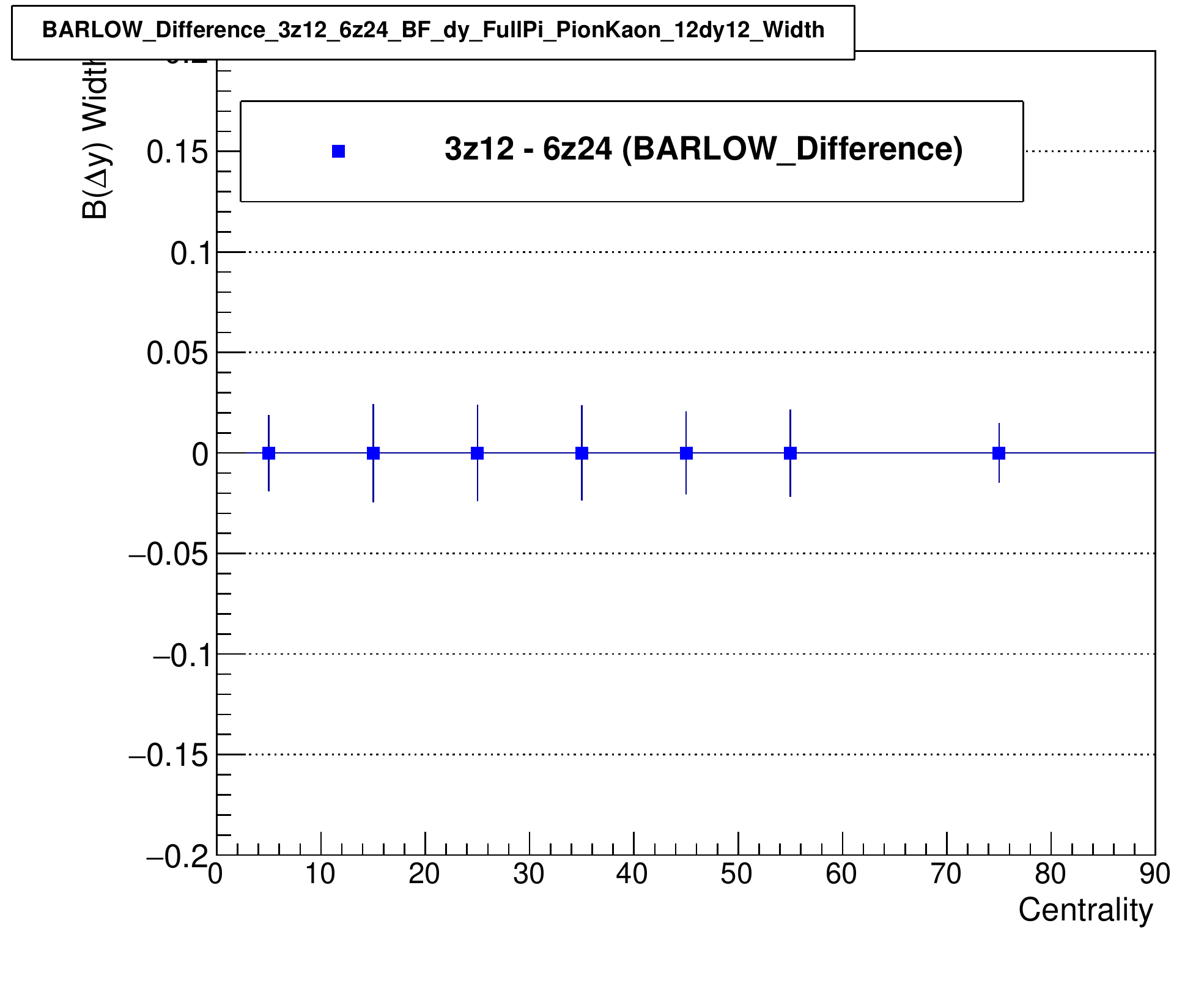}
  
  \includegraphics[width=0.32\linewidth]{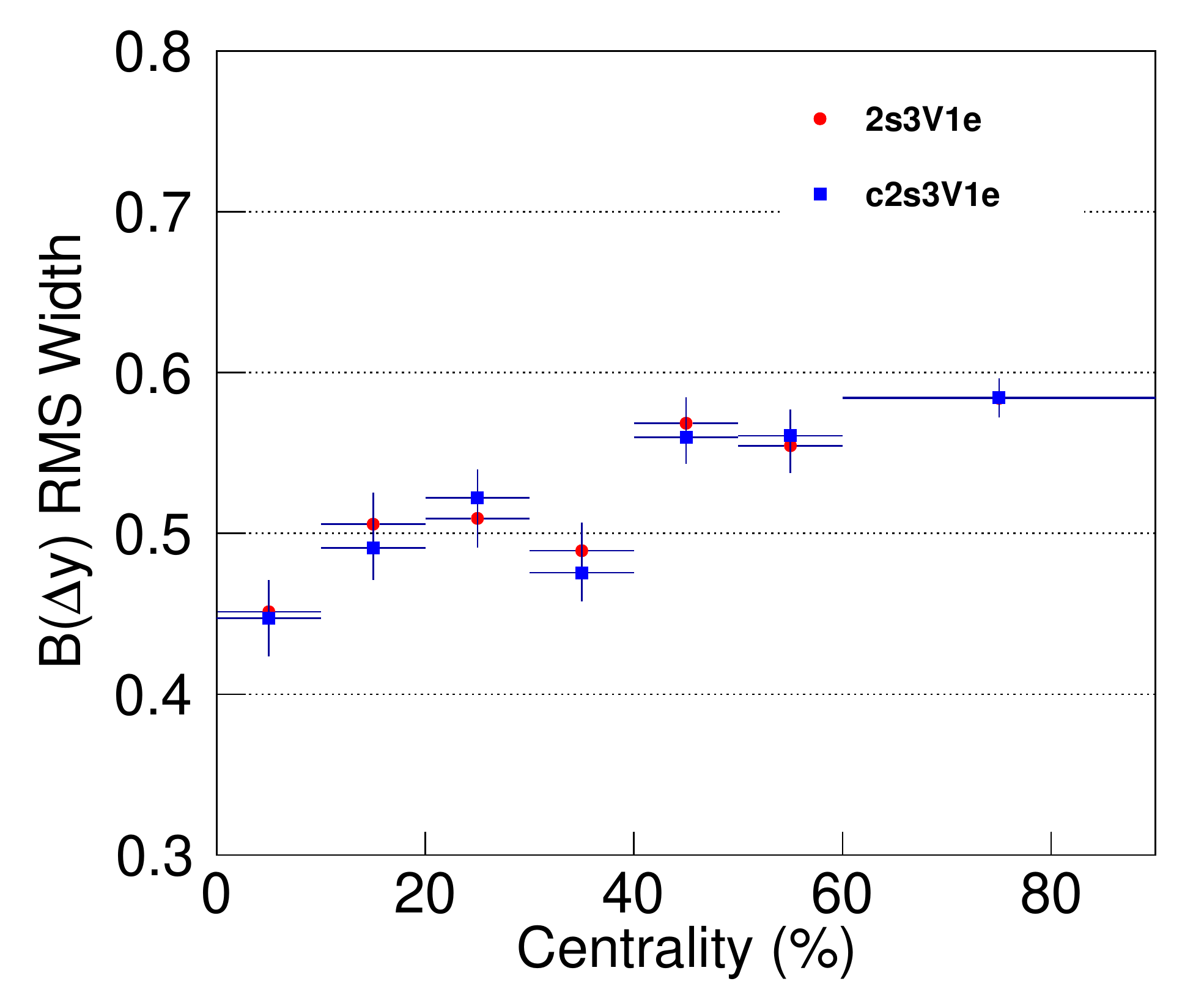}
  \includegraphics[width=0.32\linewidth]{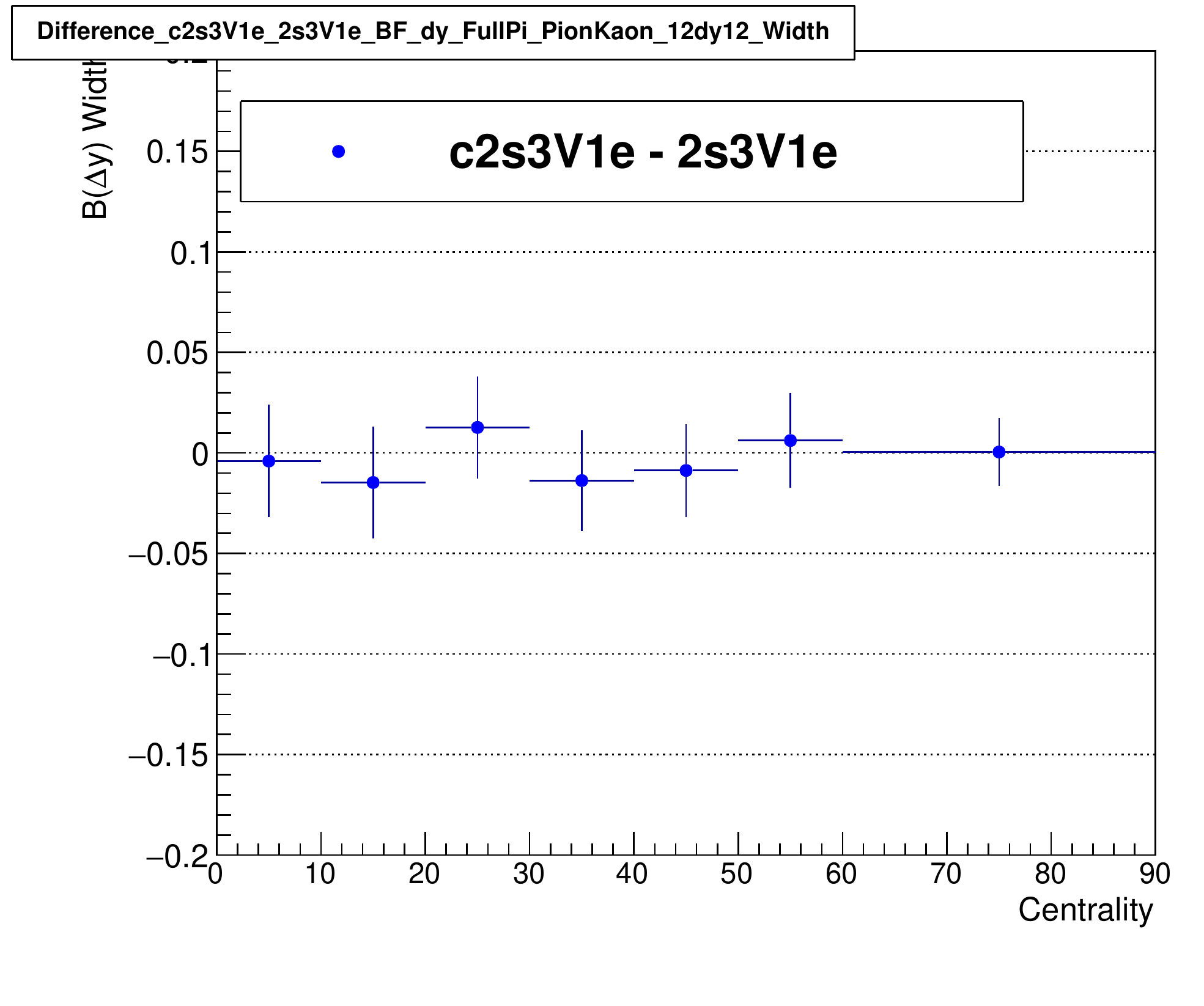}
  \includegraphics[width=0.32\linewidth]{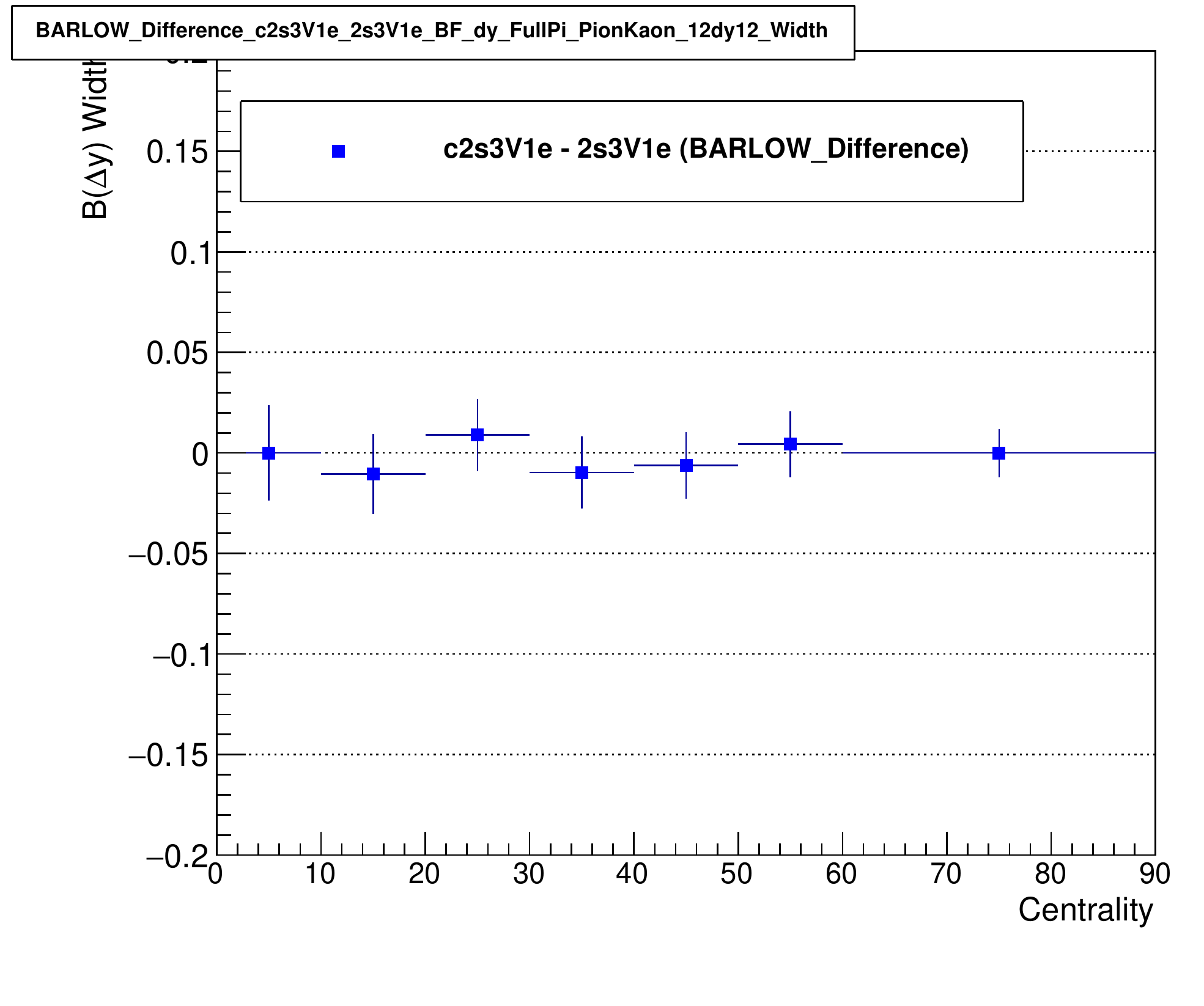}
  
  \includegraphics[width=0.32\linewidth]{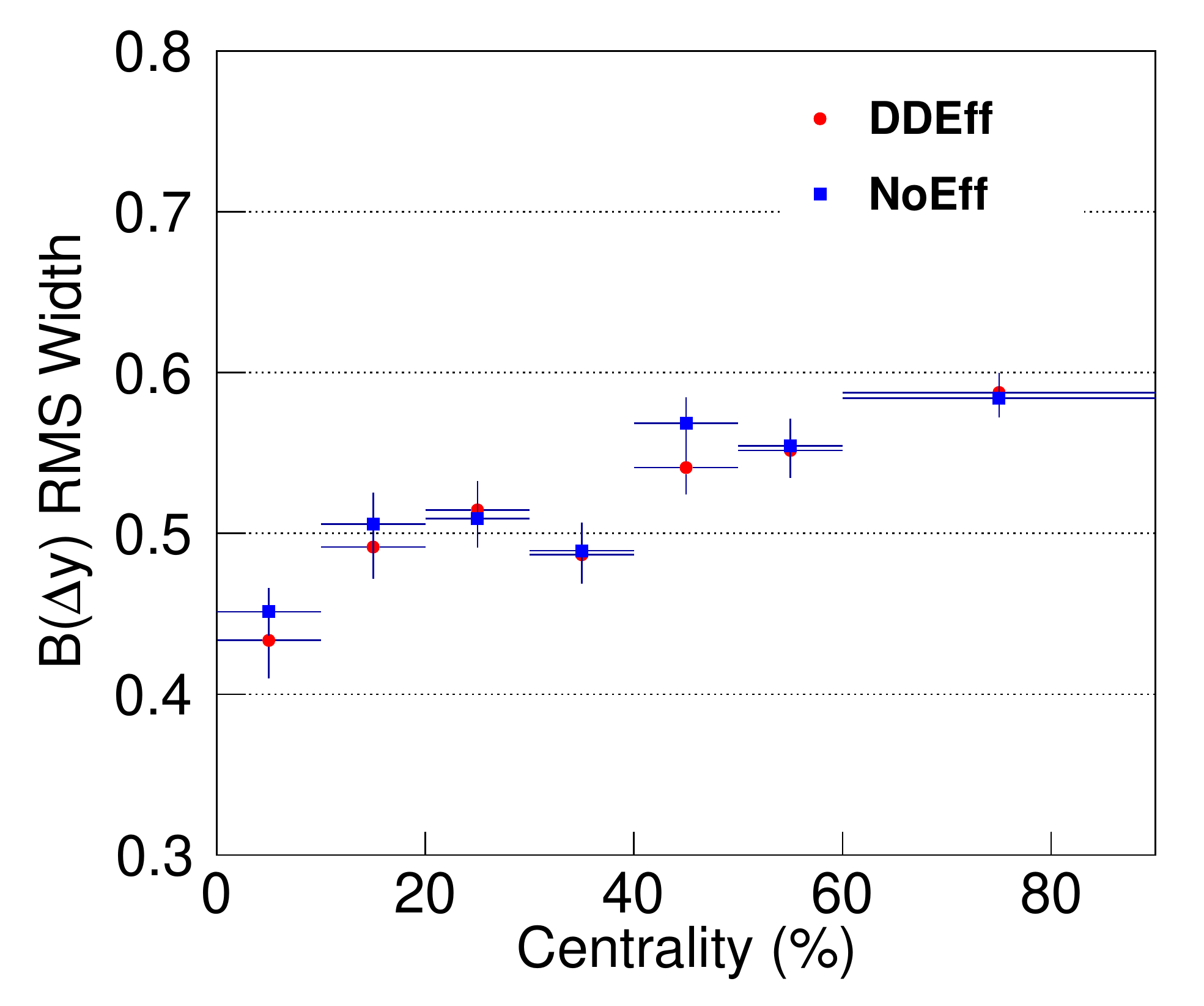}
  \includegraphics[width=0.32\linewidth]{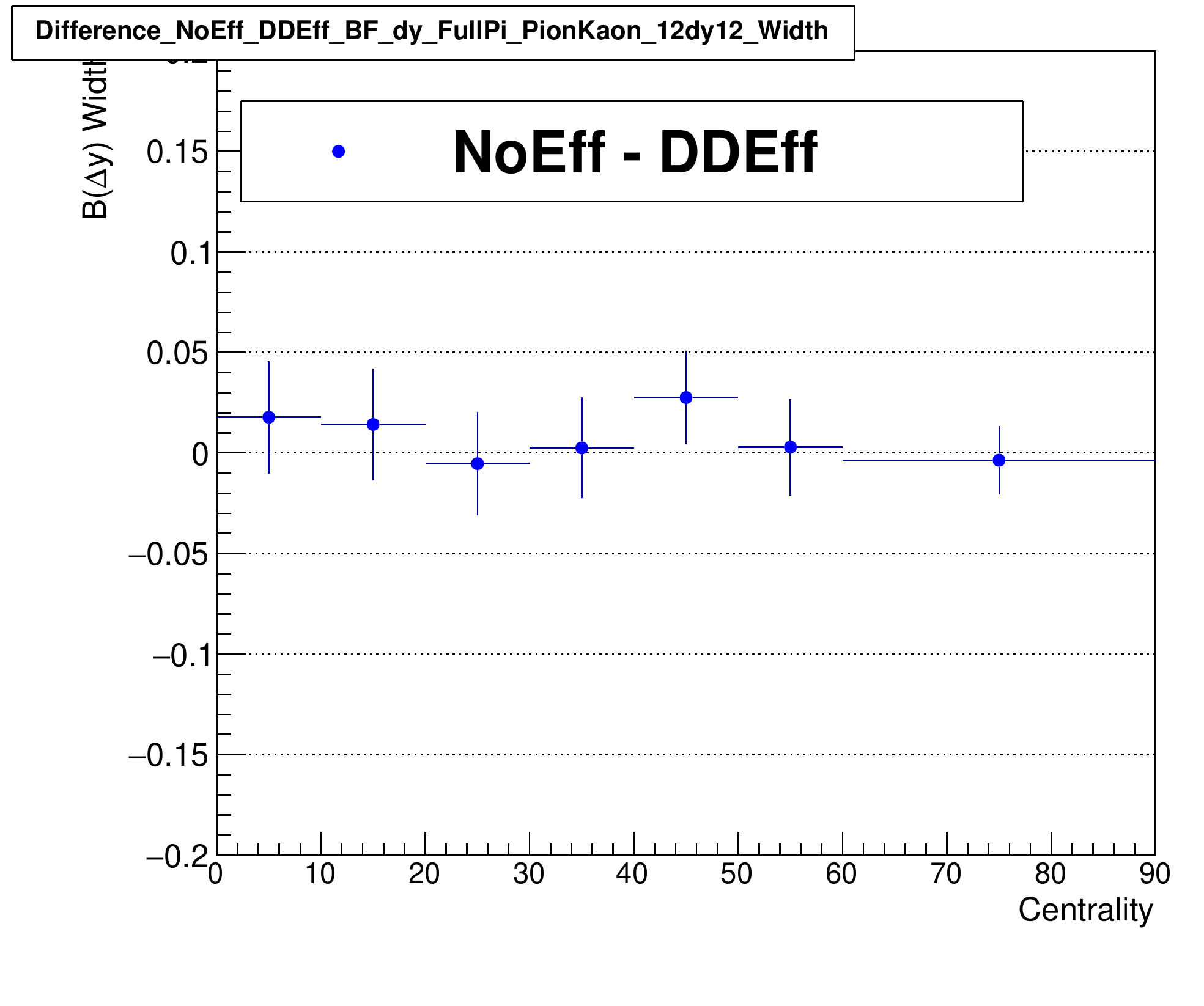}
  \includegraphics[width=0.32\linewidth]{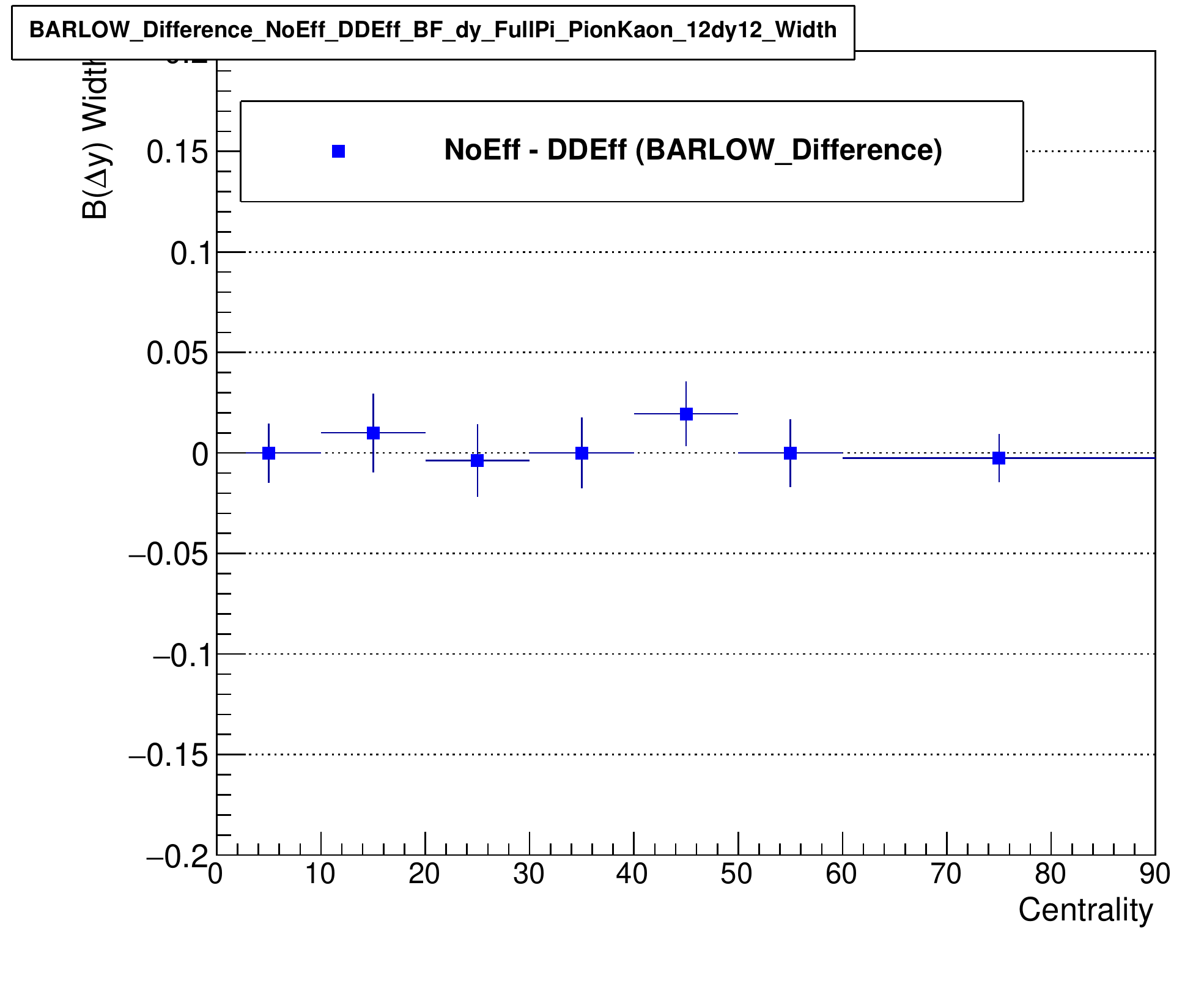}

  \caption{Systematic uncertainty contributions in $B^{\pi K}$ $\Delta y$ widths from BField ($1^{st}$ row), $V_{z}$ ($2^{nd}$ row), PID ($3^{rd}$ row), and additional $p_{\rm T}$-dependent efficiency corrections ($4^{th}$ row). The comparisons between two sets of different cuts (left column), with their differences $d$ (middle column), and their differences after the Barlow check $D_{Barlow}$ (right column).}
  \label{fig:Sys_components_dy_widths_PionKaon}
\end{figure}
\begin{figure}
\centering
  \includegraphics[width=0.32\linewidth]{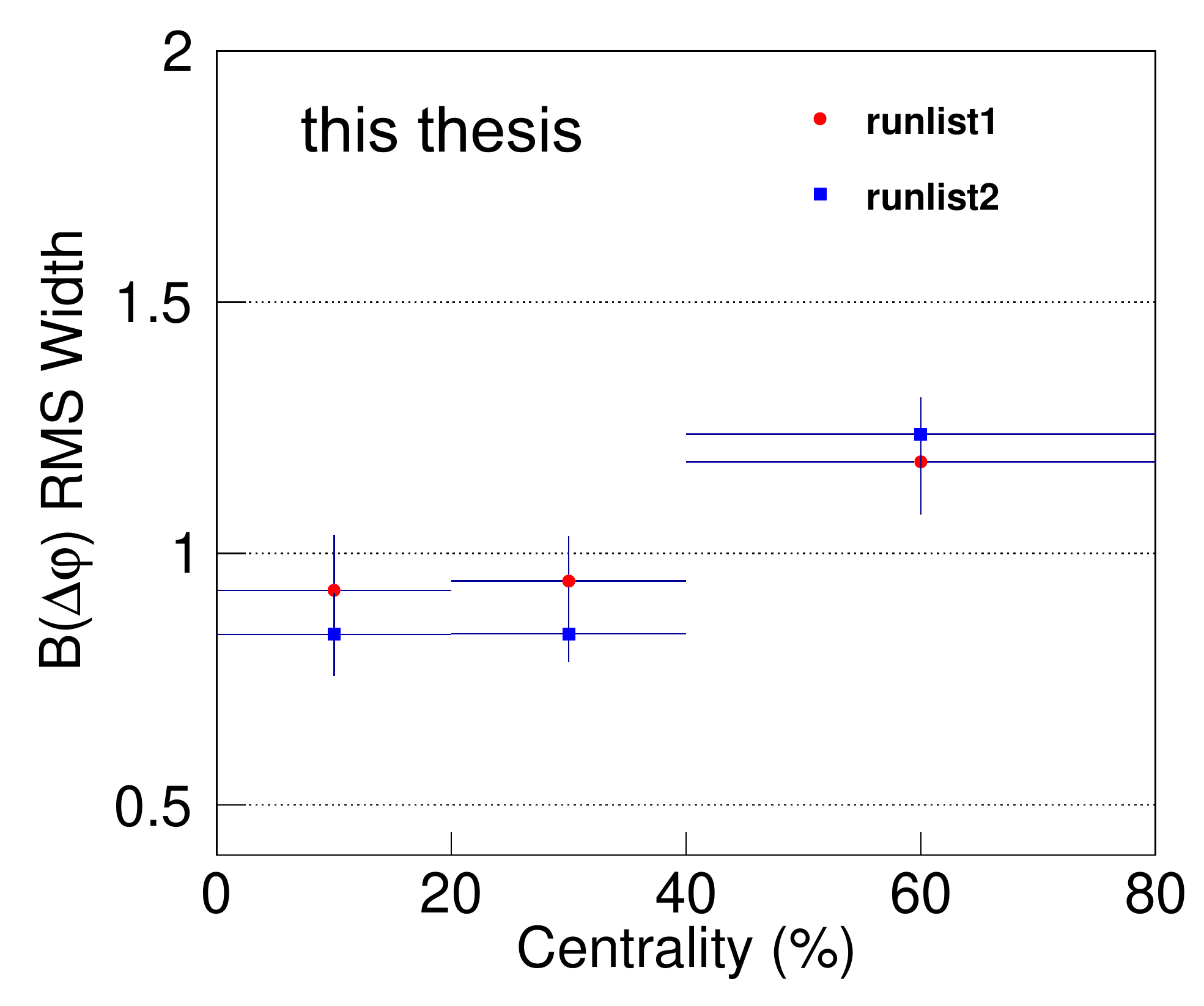}
  \includegraphics[width=0.32\linewidth]{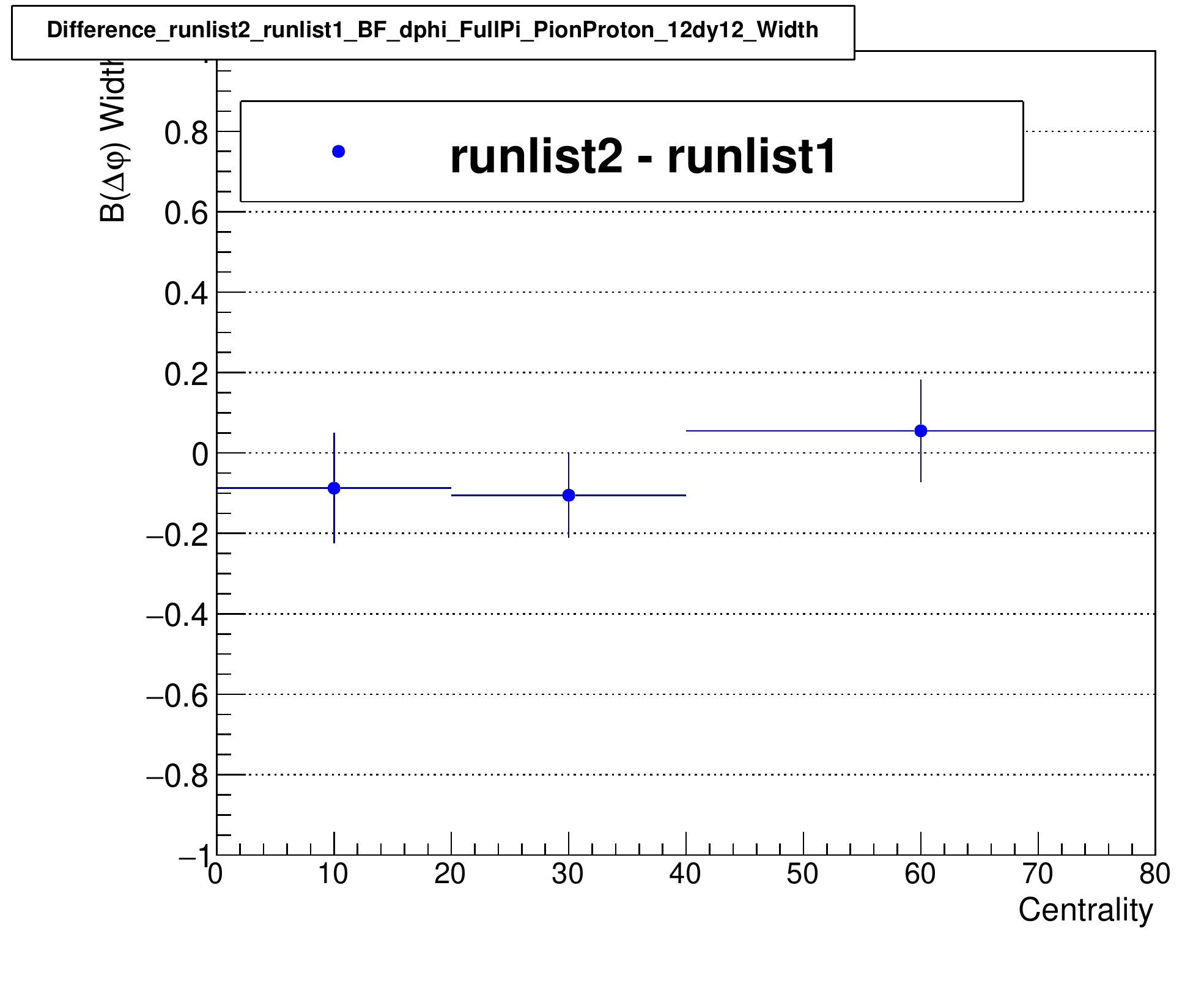}
  \includegraphics[width=0.32\linewidth]{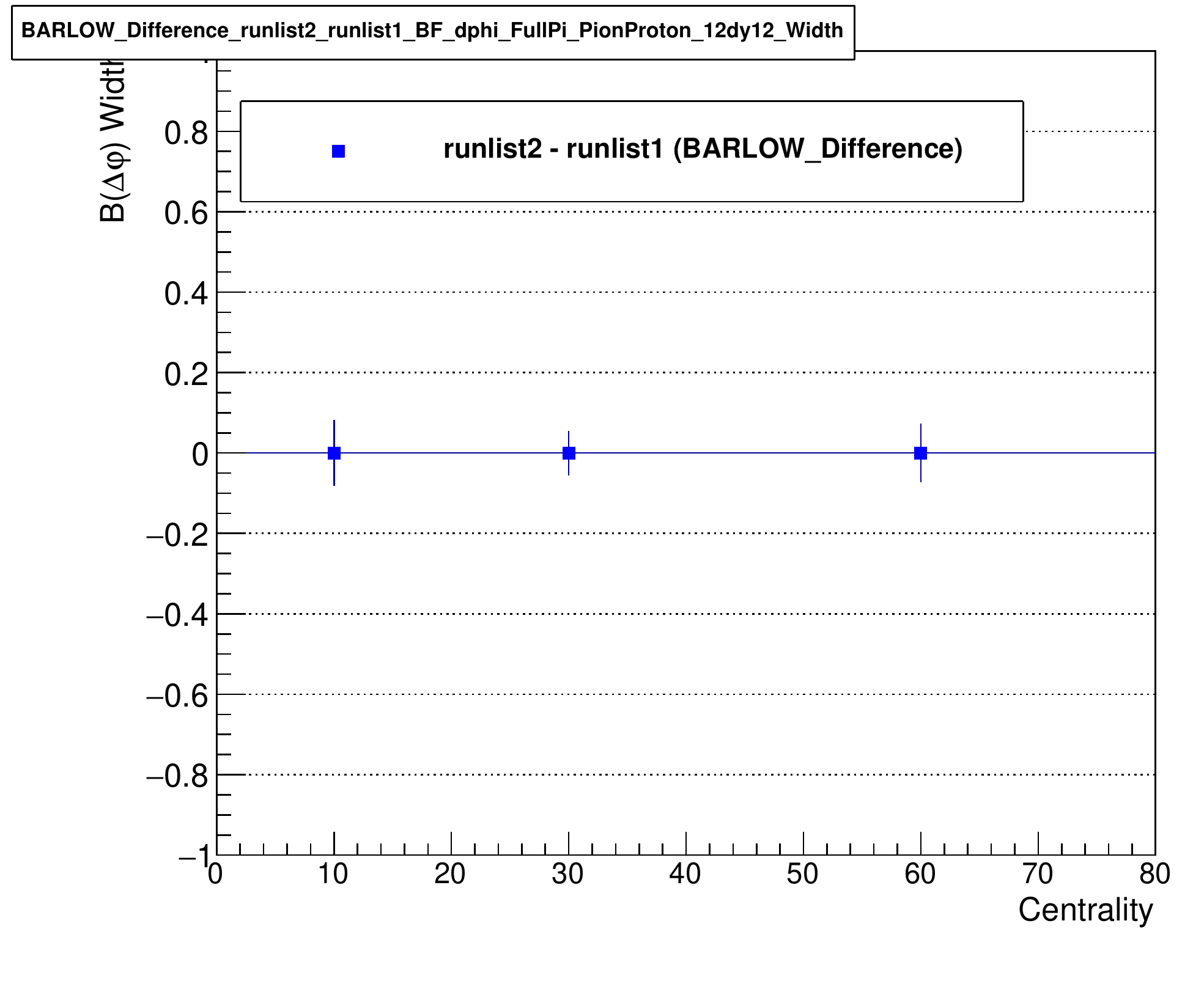}
  
  \includegraphics[width=0.32\linewidth]{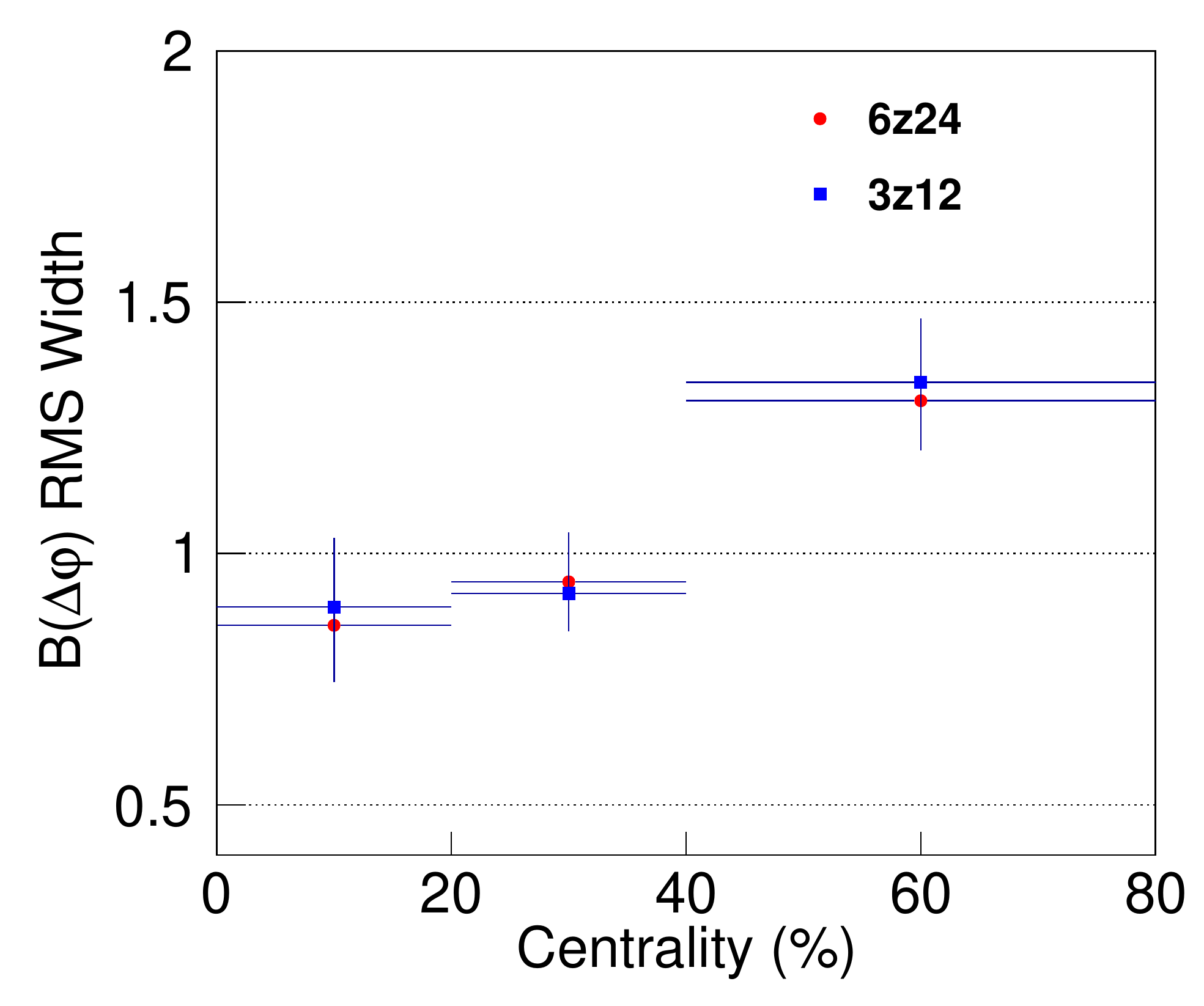}
  \includegraphics[width=0.32\linewidth]{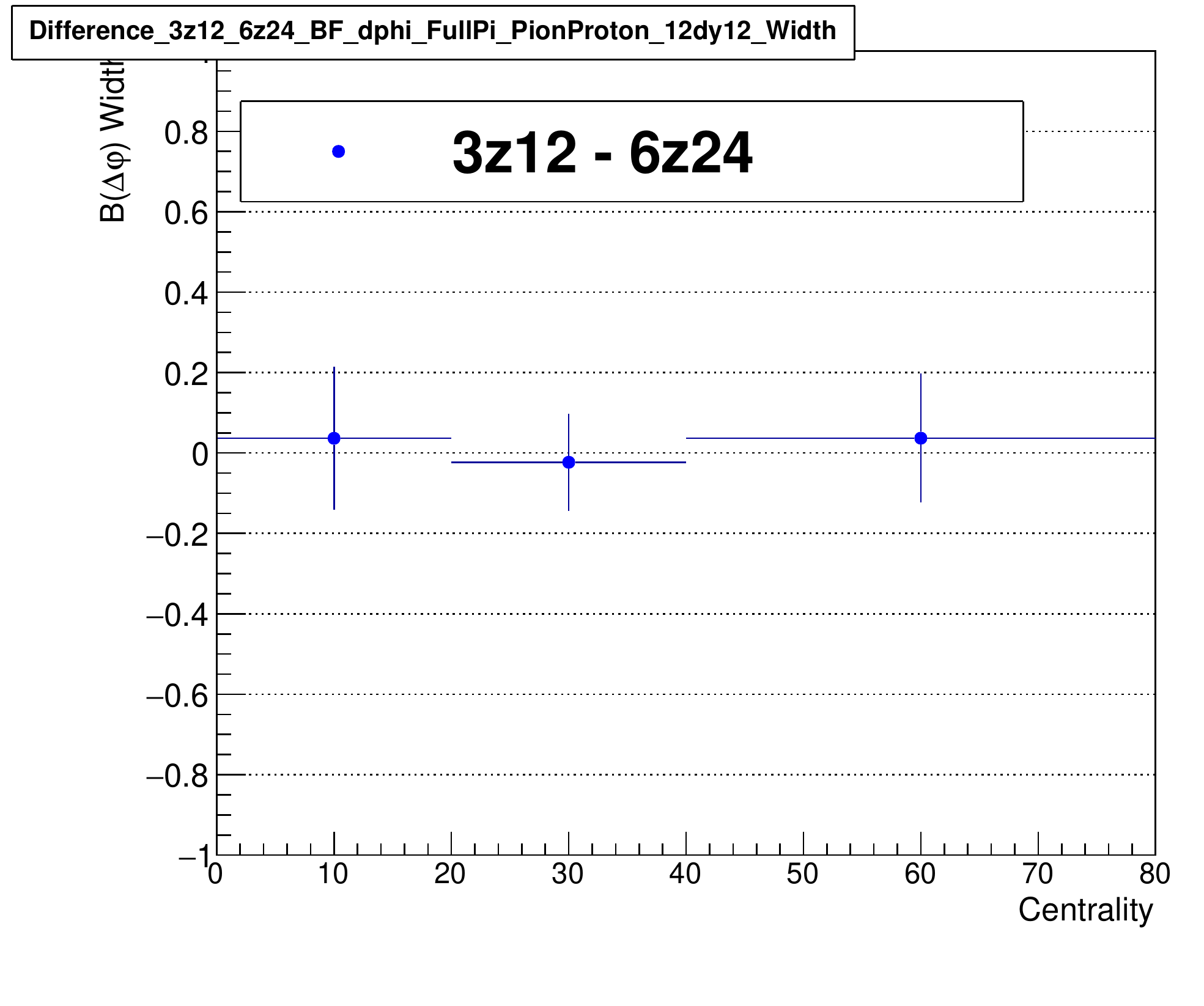}
  \includegraphics[width=0.32\linewidth]{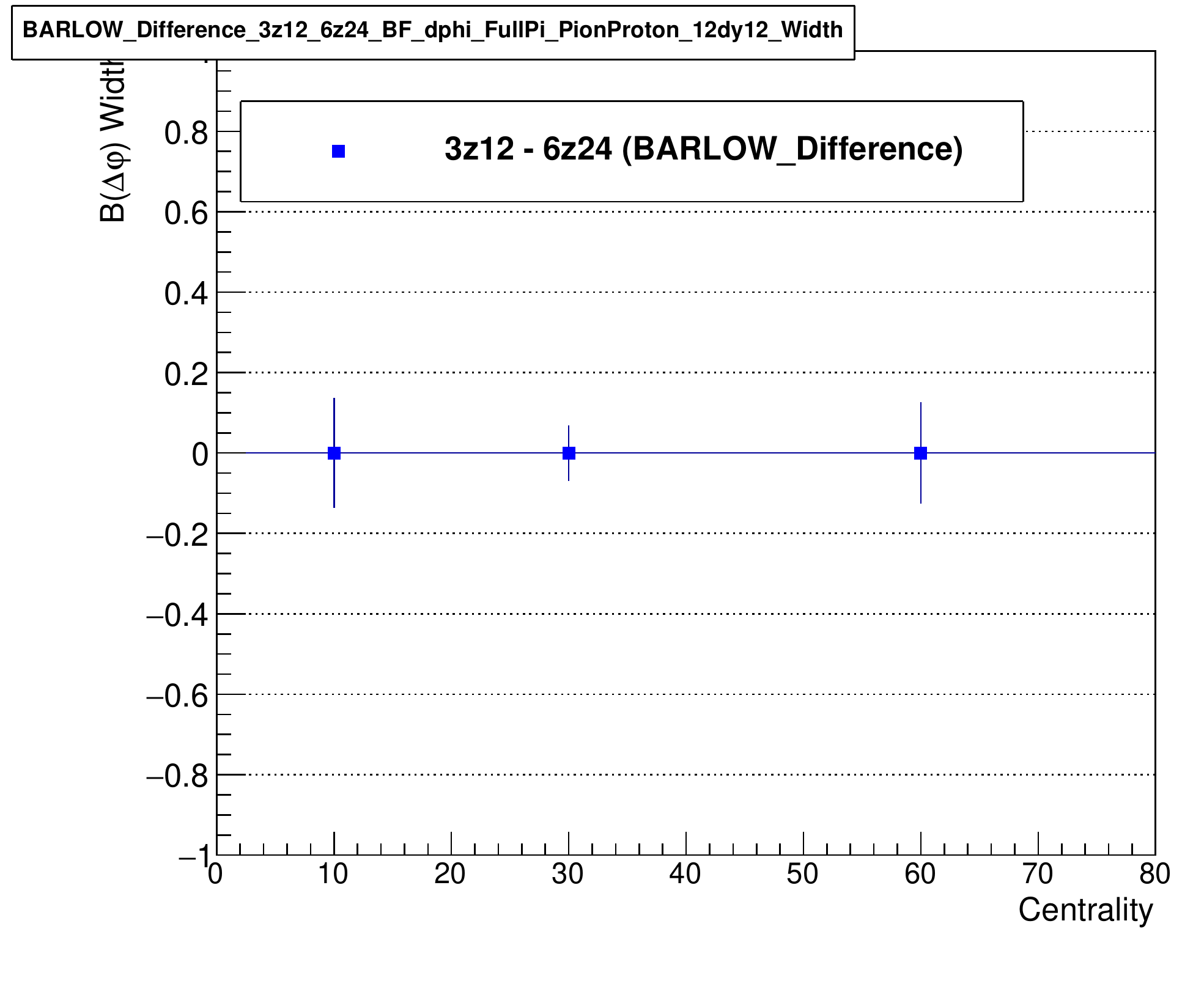}
  
  \includegraphics[width=0.32\linewidth]{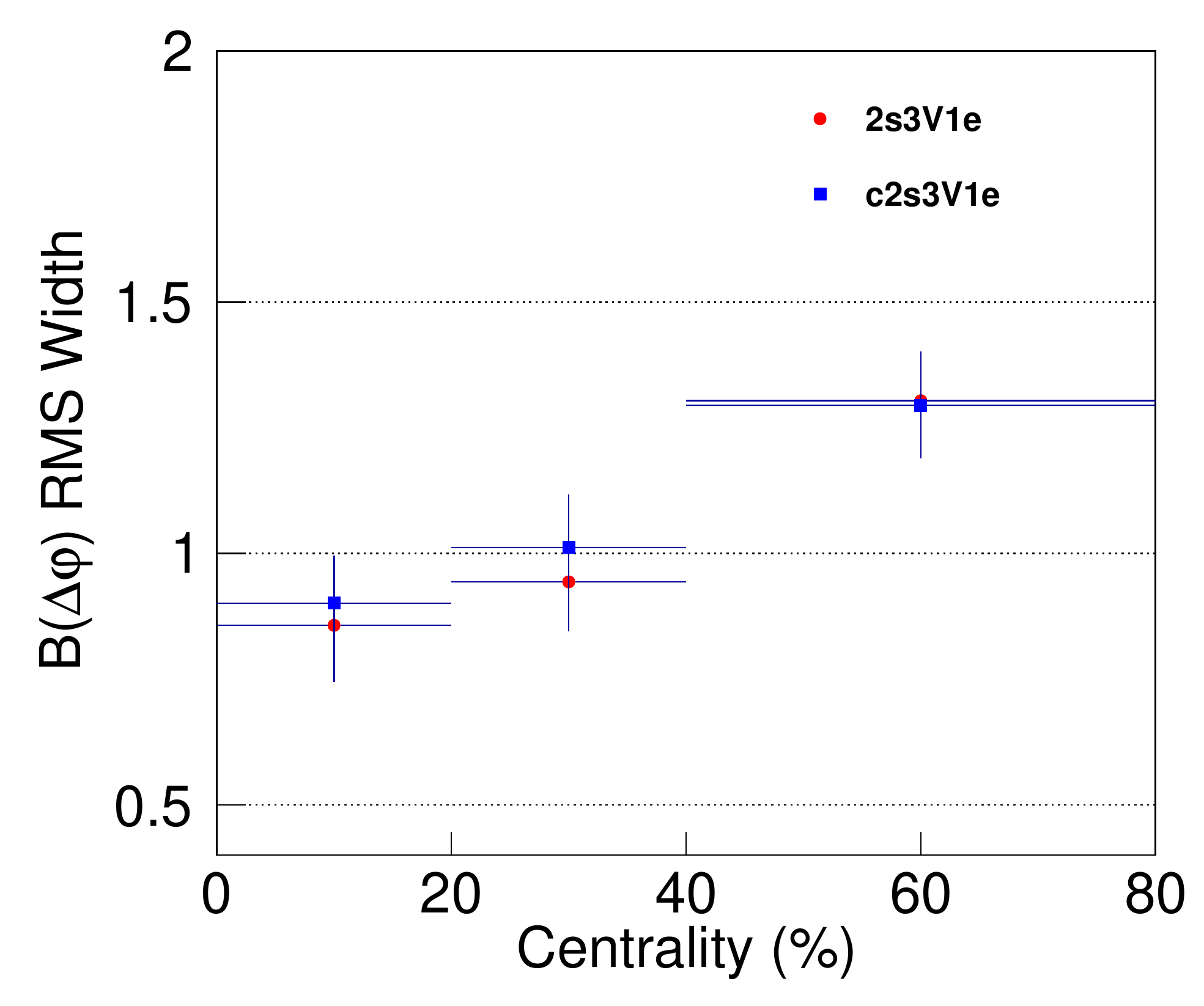}
  \includegraphics[width=0.32\linewidth]{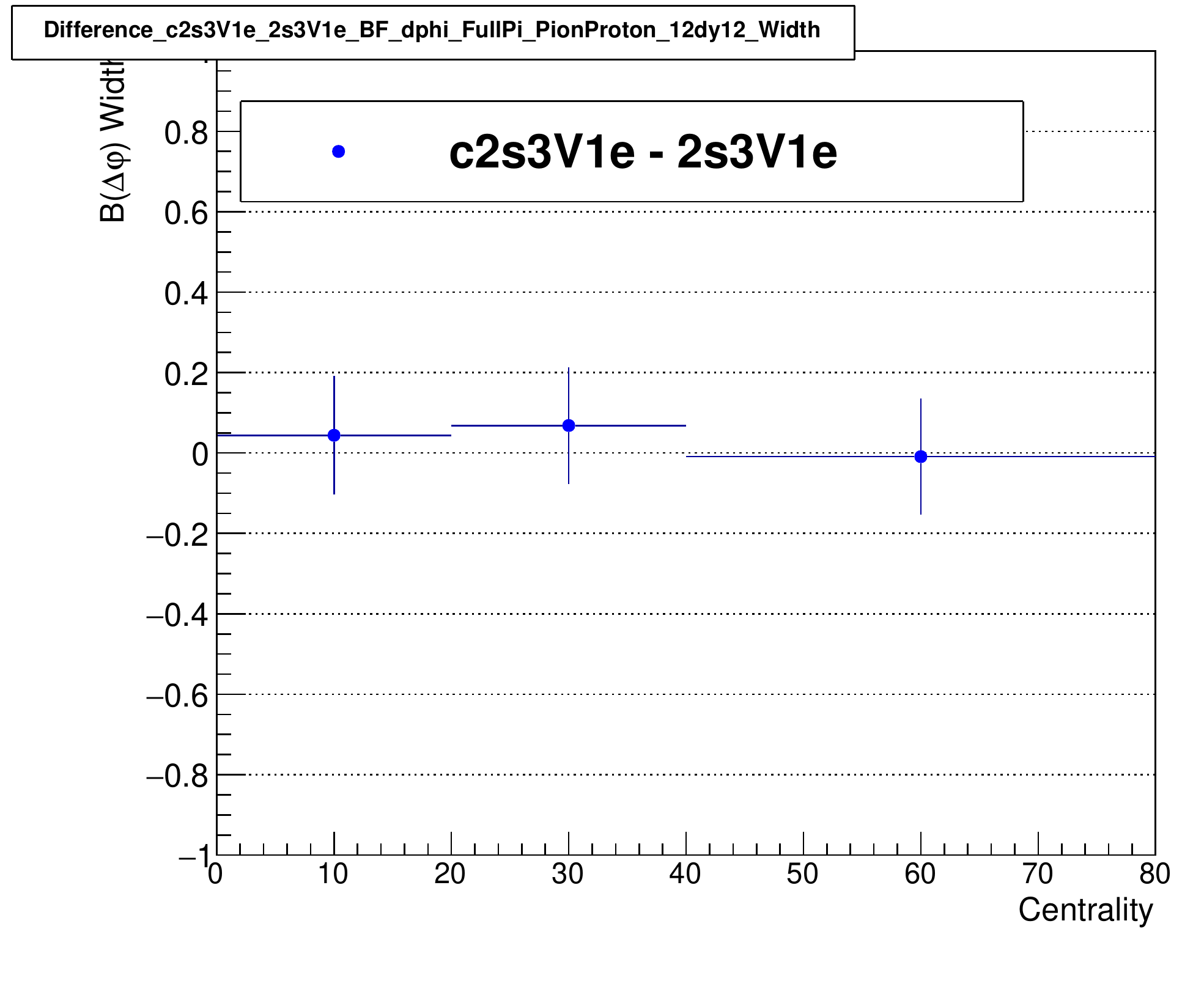}
  \includegraphics[width=0.32\linewidth]{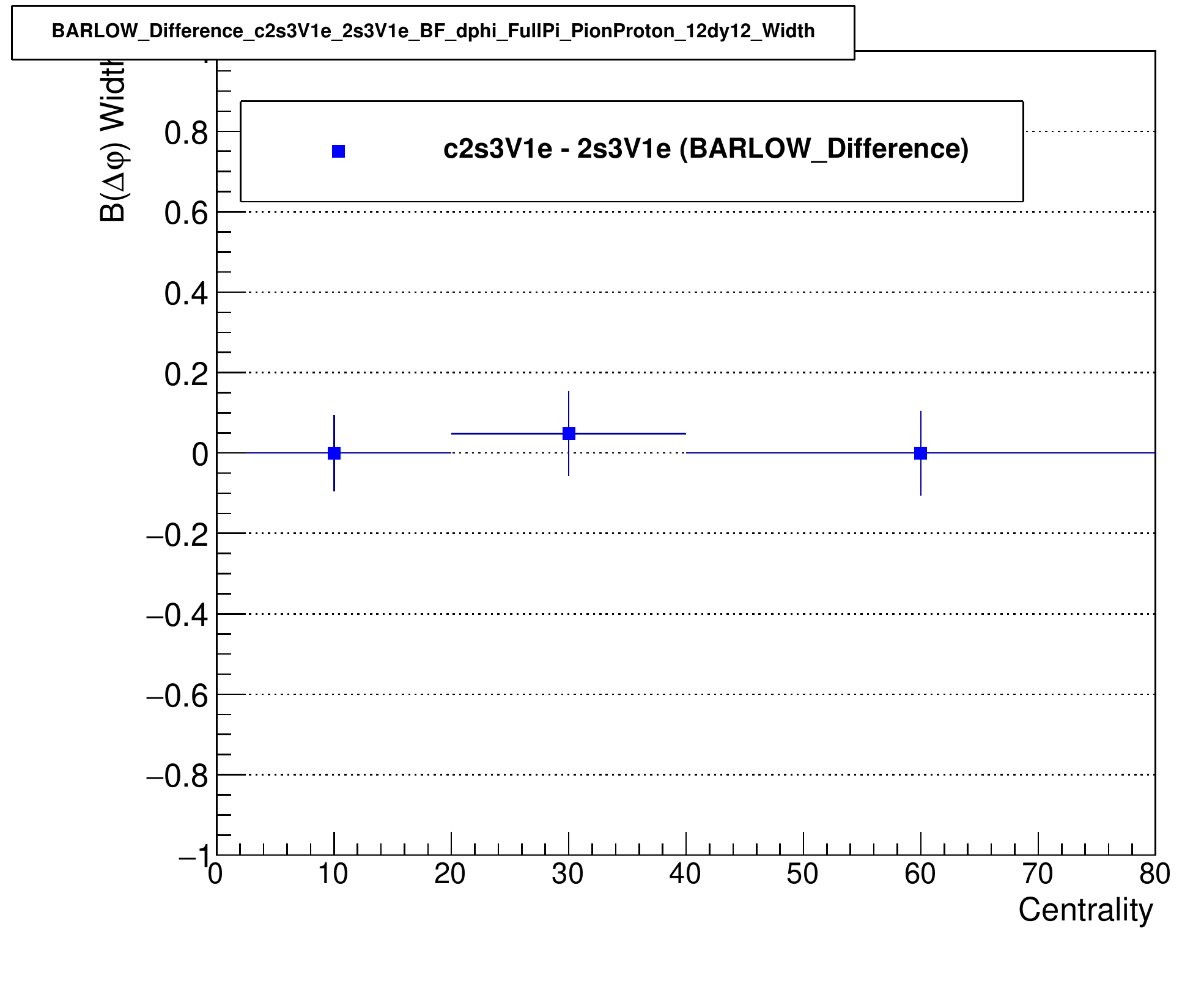}
  
  \includegraphics[width=0.32\linewidth]{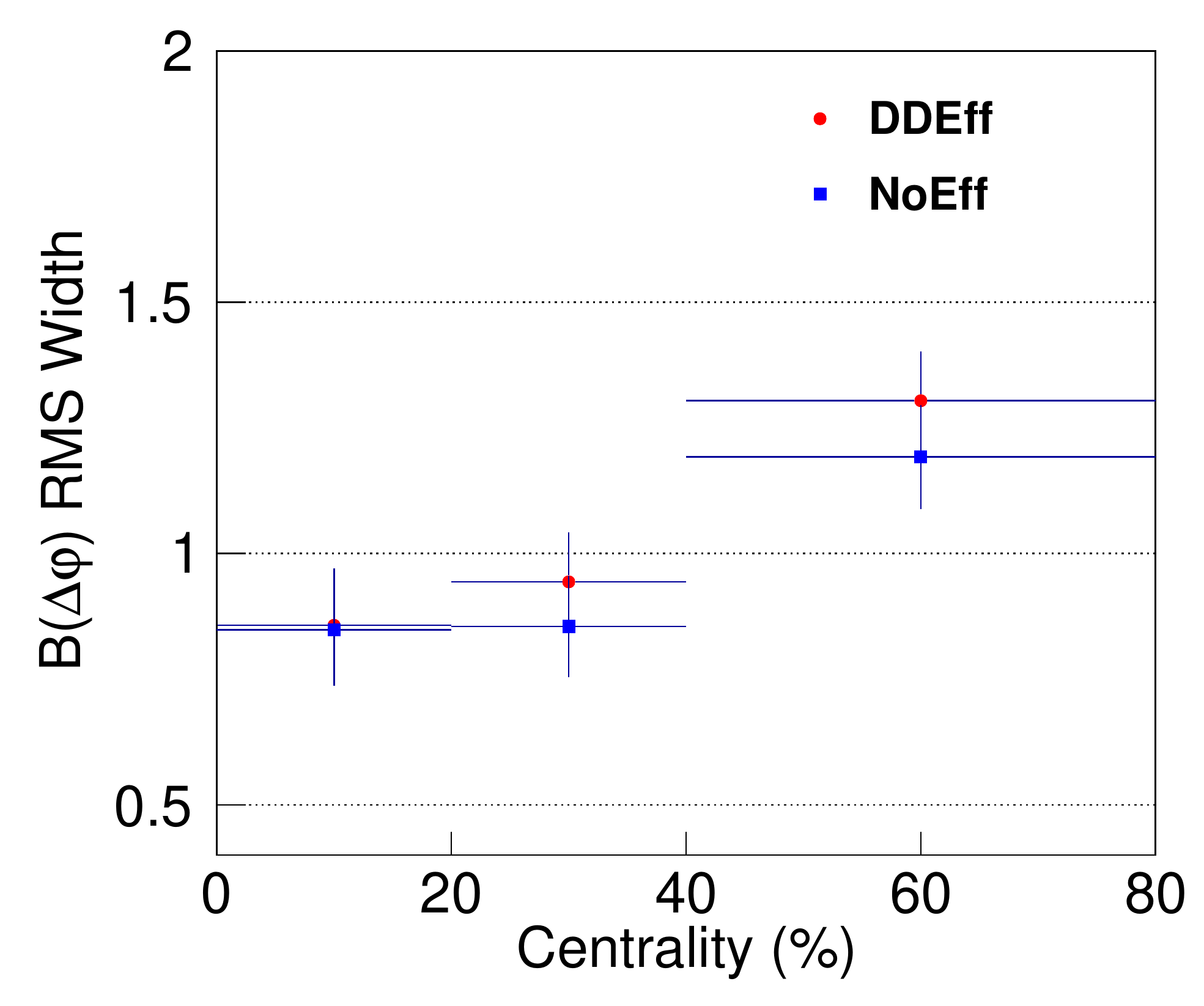}
  \includegraphics[width=0.32\linewidth]{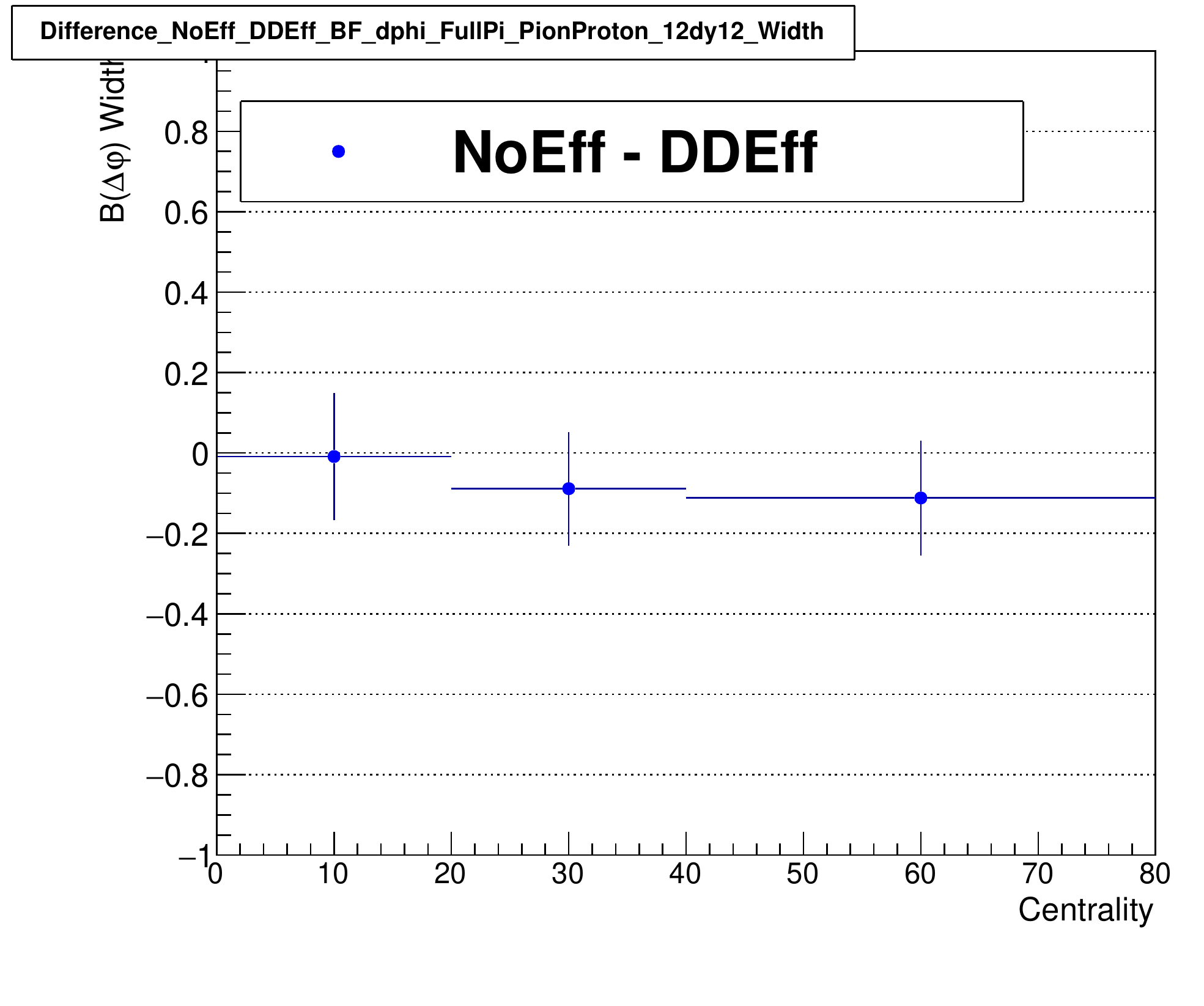}
  \includegraphics[width=0.32\linewidth]{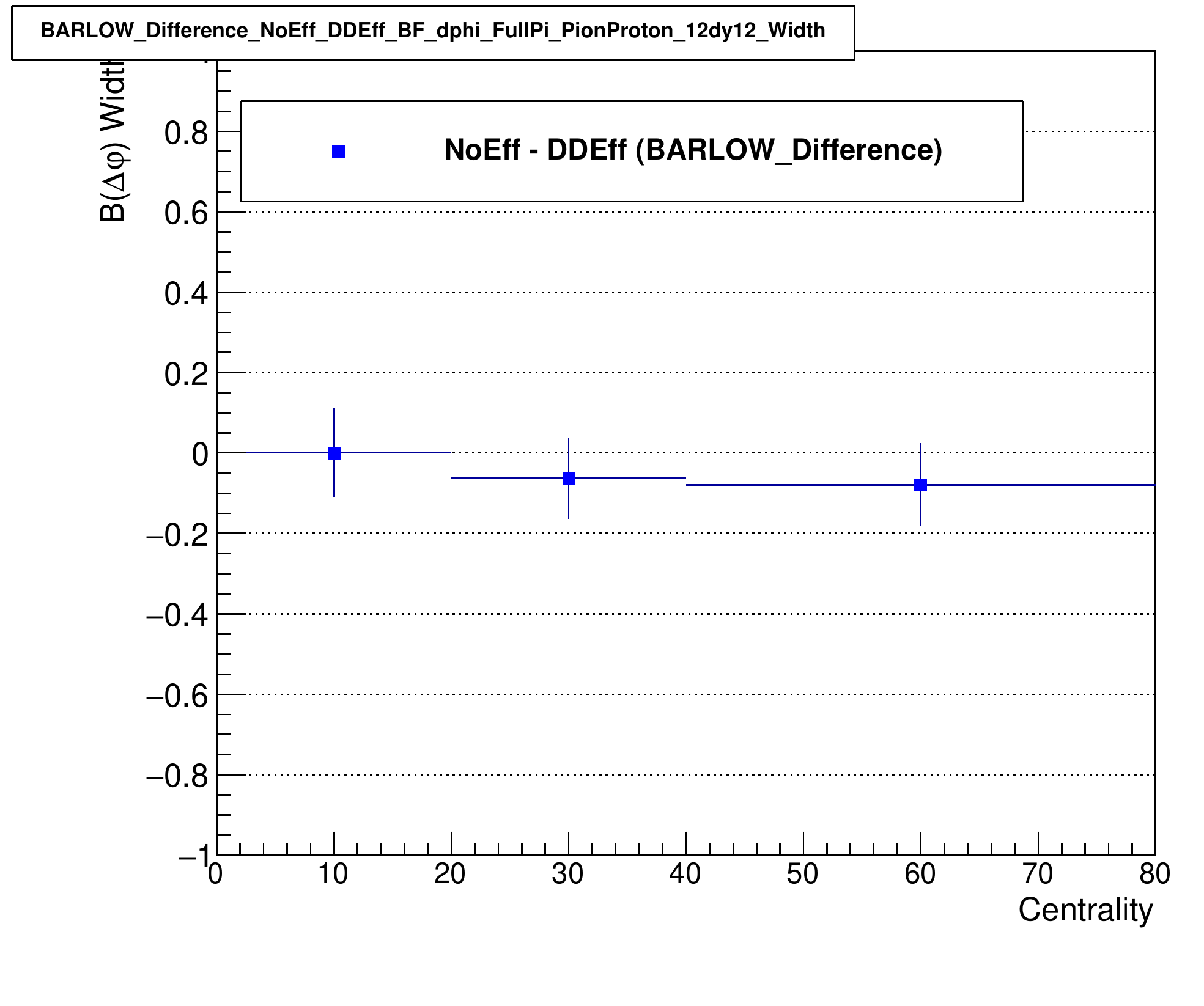}

  \caption{Systematic uncertainty contributions in $B^{\pi p}$ $\Delta\varphi$ widths from BField ($1^{st}$ row), $V_{z}$ ($2^{nd}$ row), PID ($3^{rd}$ row), and additional $p_{\rm T}$-dependent efficiency corrections ($4^{th}$ row). The comparisons between two sets of different cuts (left column), with their differences $d$ (middle column), and their differences after the Barlow check $D_{Barlow}$ (right column).}
  \label{fig:Sys_components_dphi_widths_PionProton}
\end{figure}
%

% With density contribution
\begin{table}[h!]
  \begin{center}
    \caption{Total Systematic Uncertainties ($\sigma_{BF}$) on $B^{\pi\pi}$.}
    \label{tab:Final_Systematic_Uncertainty_BF_PiPi}
    \begin{tabular}{cccccc}
      Centrality & $B(\Delta y)$ & $B(\Delta\varphi)$ & $B(\Delta y)$ Width & $B(\Delta\varphi)$ Width & BF Integral \\
      0-5\%       & 0.016             & 0.011                     & 0.010                       & 0.027                               & 0.030 \\
      5-10\%     & 0.015             & 0.010                     & 0.010                       & 0.027                               & 0.030 \\
      10-20\%   & 0.014             & 0.0097                   & 0.010                       & 0.027                               & 0.030 \\
      20-30\%   & 0.014             & 0.0090                   & 0.010                       & 0.027                               & 0.030 \\
      30-40\%   & 0.013             & 0.0086                   & 0.010                       & 0.027                               & 0.030 \\
      40-50\%   & 0.012             & 0.0088                   & 0.010                       & 0.027                               & 0.030 \\
      50-60\%   & 0.010             & 0.0085                   & 0.010                       & 0.027                               & 0.030 \\
      60-70\%   & 0.010             & 0.0084                   & 0.010                       & 0.027                               & 0.030 \\
      70-90\%   & 0.0094           & 0.0089                   & 0.010                       & 0.027                               & 0.030 \\
     \end{tabular}
  \end{center}
\end{table}
%//$$/$$/$$/$J/$$/$$/$$/$i/$$/$$/$$/$n/$$/$$/$$/$J/$$/$$/$$/$i/$$/$$/$$/$n/$$/$$/$$/$P/$$/$$/$$/$a/$$/$$/$$/$n/$$/$$/$$//%

% With density contribution
%MISSING DCA
\begin{table}[h!]
  \begin{center}
    \caption{Total Systematic Uncertainties ($\sigma_{BF}$) on $B^{\pi K}$.}
    \label{tab:Final_Systematic_Uncertainty_BF_PiK}
    \begin{tabular}{cccccc}
      Centrality & $B(\Delta y)$ & $B(\Delta\varphi)$ & $B(\Delta y)$ Width & $B(\Delta\varphi)$ Width & BF Integral \\
      0-10\%     & 0.0041           & 0.0020                   & 0.033                       & 0.079                               & 0.0072 \\
      10-20\%   & 0.0039           & 0.0022                   & 0.033                       & 0.079                               & 0.0072 \\
      20-30\%   & 0.0038           & 0.0021                   & 0.033                       & 0.079                               & 0.0072 \\
      30-40\%   & 0.0035           & 0.0025                   & 0.033                       & 0.079                               & 0.0072 \\
      40-50\%   & 0.0032           & 0.0023                   & 0.033                       & 0.079                               & 0.0072 \\
      50-60\%   & 0.0034           & 0.0021                   & 0.033                       & 0.079                               & 0.0072 \\
      60-90\%   & 0.0031           & 0.0017                   & 0.033                       & 0.079                               & 0.0072 \\
     \end{tabular}
  \end{center}
\end{table}
%//$$/$$/$$/$J/$$/$$/$$/$i/$$/$$/$$/$n/$$/$$/$$/$J/$$/$$/$$/$i/$$/$$/$$/$n/$$/$$/$$/$P/$$/$$/$$/$a/$$/$$/$$/$n/$$/$$/$$//

\begin{table}[h!]
  \begin{center}
    \caption{Total Systematic Uncertainties ($\sigma_{BF}$) on $B^{\pi p}$.}
    \label{tab:Final_Systematic_Uncertainty_BF_PiPr}
    \begin{tabular}{cccccc}
      Centrality & $B(\Delta y)$ & $B(\Delta\varphi)$ & $B(\Delta y)$ Width & $B(\Delta\varphi)$ Width & BF Integral \\
      0-20\%     & 0.0015           & 0.00066                 & 0.054                       & 0.053                               & 0.019 \\
      20-40\%   & 0.0012           & 0.00077                 & 0.054                       & 0.053                               & 0.019 \\
      40-80\%   & 0.00094         & 0.00063                 & 0.054                       & 0.053                               & 0.019 \\
     \end{tabular}
  \end{center}
\end{table}
%//$$/$$/$$/$J/$$/$$/$$/$i/$$/$$/$$/$n/$$/$$/$$/$J/$$/$$/$$/$i/$$/$$/$$/$n/$$/$$/$$/$P/$$/$$/$$/$a/$$/$$/$$/$n/$$/$$/$$//

% With density contribution
%MISSING DCA
\begin{table}[h!]
  \begin{center}
    \caption{Total Systematic Uncertainties ($\sigma_{BF}$) on $B^{K\pi}$.}
    \label{tab:Final_Systematic_Uncertainty_BF_KPi}
    \begin{tabular}{cccccc}
      Centrality & $B(\Delta y)$ & $B(\Delta\varphi)$ & $B(\Delta y)$ Width & $B(\Delta\varphi)$ Width & BF Integral \\
      0-10\%     & 0.015             & 0.0062                   & 0.033                       & 0.079                               & 0.022 \\
      10-20\%   & 0.014             & 0.0060                   & 0.033                       & 0.079                               & 0.022 \\
      20-30\%   & 0.012             & 0.0064                   & 0.033                       & 0.079                               & 0.022 \\
      30-40\%   & 0.013             & 0.0070                   & 0.033                       & 0.079                               & 0.022 \\
      40-50\%   & 0.012             & 0.0059                   & 0.033                       & 0.079                               & 0.022 \\
      50-60\%   & 0.012             & 0.0063                   & 0.033                       & 0.079                               & 0.022 \\
      60-90\%   & 0.011             & 0.0057                   & 0.033                       & 0.079                               & 0.022 \\
     \end{tabular}
  \end{center}
\end{table}
%//$$/$$/$$/$J/$$/$$/$$/$i/$$/$$/$$/$n/$$/$$/$$/$J/$$/$$/$$/$i/$$/$$/$$/$n/$$/$$/$$/$P/$$/$$/$$/$a/$$/$$/$$/$n/$$/$$/$$//

% With density contribution
\begin{table}[h!]
  \begin{center}
    \caption{Total Systematic Uncertainties ($\sigma_{BF}$) on $B^{KK}$.}
    \label{tab:Final_Systematic_Uncertainty_BF_KK}
    \begin{tabular}{cccccc}
      Centrality & $B(\Delta y)$ & $B(\Delta\varphi)$ & $B(\Delta y)$ Width & $B(\Delta\varphi)$ Width & BF Integral \\
      0-10\%     & 0.024             & 0.0095                   & 0.016                       & 0.060                               & 0.021 \\
      10-20\%   & 0.019             & 0.0095                   & 0.016                       & 0.060                               & 0.021 \\
      20-30\%   & 0.018             & 0.0094                   & 0.016                       & 0.060                               & 0.021 \\
      30-40\%   & 0.018             & 0.011                     & 0.016                       & 0.060                               & 0.021 \\
      40-60\%   & 0.022             & 0.0090                   & 0.016                       & 0.060                               & 0.021 \\
      60-90\%   & 0.015             & 0.0077                   & 0.016                       & 0.060                               & 0.021 \\
     \end{tabular}
  \end{center}
\end{table}
%//$$/$$/$$/$J/$$/$$/$$/$i/$$/$$/$$/$n/$$/$$/$$/$J/$$/$$/$$/$i/$$/$$/$$/$n/$$/$$/$$/$P/$$/$$/$$/$a/$$/$$/$$/$n/$$/$$/$$//

% With density contribution
\begin{table}[h!]
  \begin{center}
    \caption{Total Systematic Uncertainties ($\sigma_{BF}$) on $B^{Kp}$.}
    \label{tab:Final_Systematic_Uncertainty_BF_KPr}
    \begin{tabular}{cccccc}
      Centrality & $B(\Delta y)$ & $B(\Delta\varphi)$ & $B(\Delta y)$ Width & $B(\Delta\varphi)$ Width & BF Integral \\
      0-20\%     & 0.0046           & 0.0022                   & 0.068                       & 0.15                               & 0.028 \\
      20-40\%   & 0.0037           & 0.0019                   & 0.068                       & 0.15                               & 0.028 \\
      40-80\%   & 0.0047           & 0.0025                   & 0.068                       & 0.15                               & 0.028 \\
     \end{tabular}
  \end{center}
\end{table}
%//$$/$$/$$/$J/$$/$$/$$/$i/$$/$$/$$/$n/$$/$$/$$/$J/$$/$$/$$/$i/$$/$$/$$/$n/$$/$$/$$/$P/$$/$$/$$/$a/$$/$$/$$/$n/$$/$$/$$//

% With density contribution
\begin{table}[h!]
  \begin{center}
    \caption{Total Systematic Uncertainties ($\sigma_{BF}$) on $B^{p\pi}$.}
    \label{tab:Final_Systematic_Uncertainty_BF_PrPi}
    \begin{tabular}{cccccc}
      Centrality & $B(\Delta y)$ & $B(\Delta\varphi)$ & $B(\Delta y)$ Width & $B(\Delta\varphi)$ Width & BF Integral \\
      0-20\%     & 0.012           & 0.010                   & 0.054                       & 0.053                               & 0.023 \\
      20-40\%   & 0.011           & 0.010                   & 0.054                       & 0.053                               & 0.023 \\
      40-80\%   & 0.011           & 0.010                   & 0.054                       & 0.053                               & 0.023 \\
     \end{tabular}
  \end{center}
\end{table}
%//$$/$$/$$/$J/$$/$$/$$/$i/$$/$$/$$/$n/$$/$$/$$/$J/$$/$$/$$/$i/$$/$$/$$/$n/$$/$$/$$/$P/$$/$$/$$/$a/$$/$$/$$/$n/$$/$$/$$//

\begin{table}[h!]
  \begin{center}
    \caption{Total Systematic Uncertainties ($\sigma_{BF}$) on $B^{pK}$.}
    \label{tab:Final_Systematic_Uncertainty_BF_PrK}
    \begin{tabular}{cccccc}
      Centrality & $B(\Delta y)$ & $B(\Delta\varphi)$ & $B(\Delta y)$ Width & $B(\Delta\varphi)$ Width & BF Integral \\
      0-20\%     & 0.0092           & 0.0047                   & 0.068                       & 0.15                                 & 0.014 \\
      20-40\%   & 0.0075           & 0.0042                   & 0.068                       & 0.15                                 & 0.014 \\
      40-80\%   & 0.0090           & 0.0051                   & 0.068                       & 0.15                                 & 0.014 \\
     \end{tabular}
  \end{center}
\end{table}
%//$$/$$/$$/$J/$$/$$/$$/$i/$$/$$/$$/$n/$$/$$/$$/$J/$$/$$/$$/$i/$$/$$/$$/$n/$$/$$/$$/$P/$$/$$/$$/$a/$$/$$/$$/$n/$$/$$/$$//

% With density contribution + Vz ( Vz increased dy, dphi widths, and integral )
\begin{table}[h!]
  \begin{center}
    \caption{Total Systematic Uncertainties ($\sigma_{BF}$) on $B^{pp}$.}
    \label{tab:Final_Systematic_Uncertainty_BF_PrPr}
    \begin{tabular}{cccccc}
      Centrality & $B(\Delta y)$ & $B(\Delta\varphi)$ & $B(\Delta y)$ Width & $B(\Delta\varphi)$ Width & BF Integral \\
      0-20\%     & 0.011             & 0.0051                   & 0.0085                     & 0.067                               & 0.013 \\
      20-40\%   & 0.0088           & 0.0047                   & 0.0085                     & 0.067                               & 0.013 \\
      40-80\%   & 0.0098           & 0.0040                   & 0.0085                     & 0.067                               & 0.013 \\
     \end{tabular}
  \end{center}
\end{table}
%//$$/$$/$$/$J/$$/$$/$$/$i/$$/$$/$$/$n/$$/$$/$$/$J/$$/$$/$$/$i/$$/$$/$$/$n/$$/$$/$$/$P/$$/$$/$$/$a/$$/$$/$$/$n/$$/$$/$$//

\clearpage

	\chapter{Results}\label{chap:Results}

We have analyzed data from Pb--Pb collisions at $\sqrt{s_{_{\rm NN}}} =$2.76 TeV, acquired with the ALICE detector at the Large Hadron Collider, to measure balance functions (BF) of charged hadron pairs $(\pi,K,p)\otimes (\pi,K,p)$.
And it is worth to mention that the preliminary results of $B^{\pi\pi}$ and $B^{KK}$ were presented in an oral presentation at the 2018 Quark Matter conference and published in the proceedings~\cite{PAN2019315}.
The dataset on which this work is based is presented in Sec.~\ref{subsec:RealDataSamples}, whereas event and track selection criteria are described in Sec.~\ref{subsec:EventTrackSelection}.
The hadron identification method is introduced in Sec.~\ref{sec:Particle Identification}, and the detection efficiency correction and optimization methods are reported in Sec.~\ref{sec:Efficiency_Correction}.
Systematic uncertainties are determined according to the methods presented in Chapter~\ref{chap:SystematicUncertainties}.

In this Chapter, we present the main results of this dissertation. Two-dimensional  balance functions measured as functions of rapidity and azimuth differences are presented in Sec.~\ref{sec:2DResults}, while their projections are discussed in Sec.~\ref{sec:1DProjections}.
Measurements of the widths and integrals of these balance functions are considered in Sec.~\ref{sec:WidthsIntegrals}.

\section{Two-dimensional correlators and balance functions}
\label{sec:2DResults}

\begin{figure}
\centering
  \includegraphics[width=0.99\linewidth]{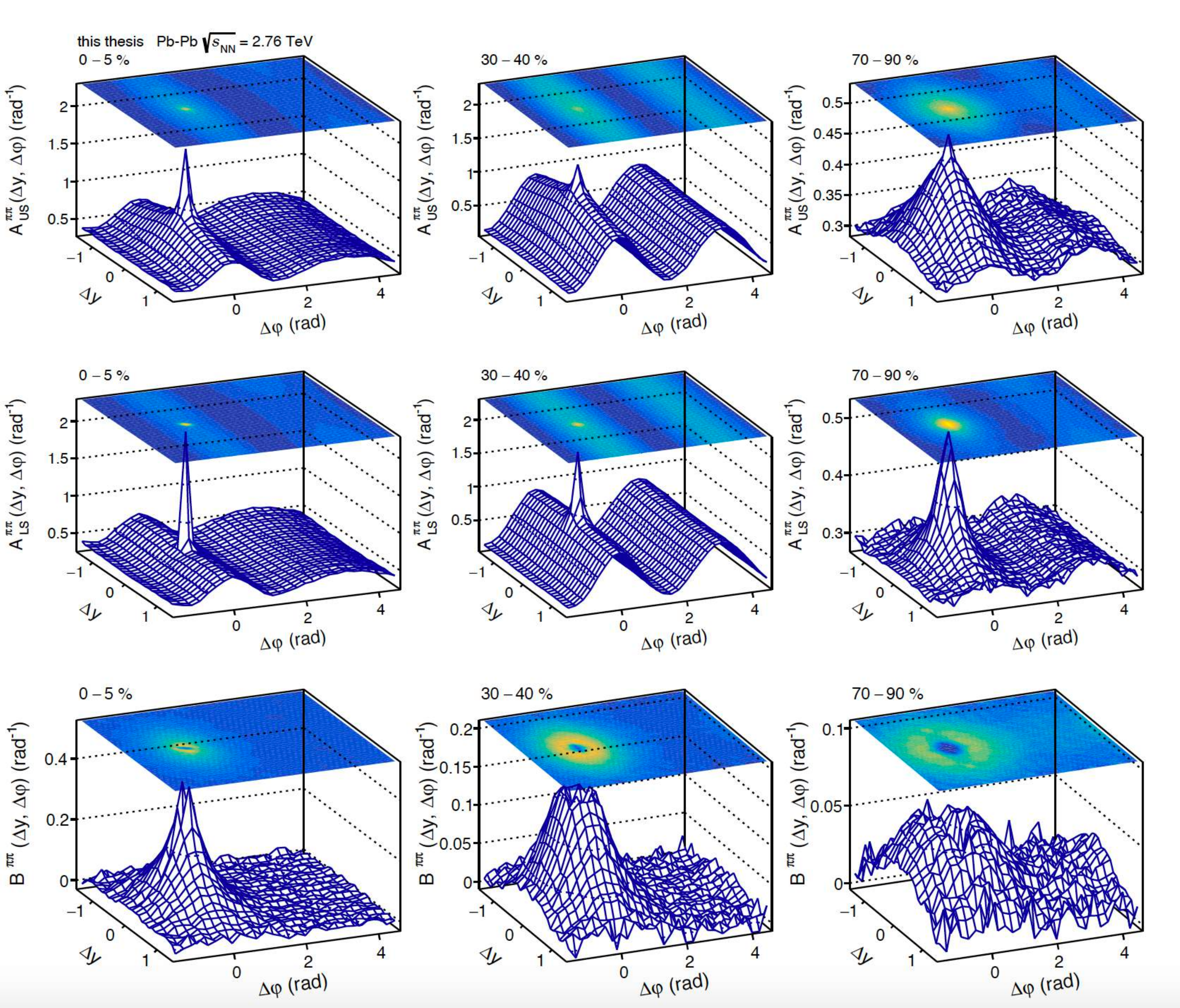}
  \caption{Correlation functions and balance functions of $\pi\pi$ pair for selected Pb--Pb collision centralities.
  Top row: unlike-sign correlation function $A_{US}^{\pi\pi}(\Delta y,\Delta\varphi)$;
  Middle row: like-sign correlation function $A_{LS}^{\pi\pi}(\Delta y,\Delta\varphi)$;  
  Bottom row: balance function $B^{\pi\pi}(\Delta y,\Delta\varphi)$.}
   \label{fig:PiPi_BF_2d}
\end{figure}

\begin{figure}
\centering
  \includegraphics[width=0.99\linewidth]{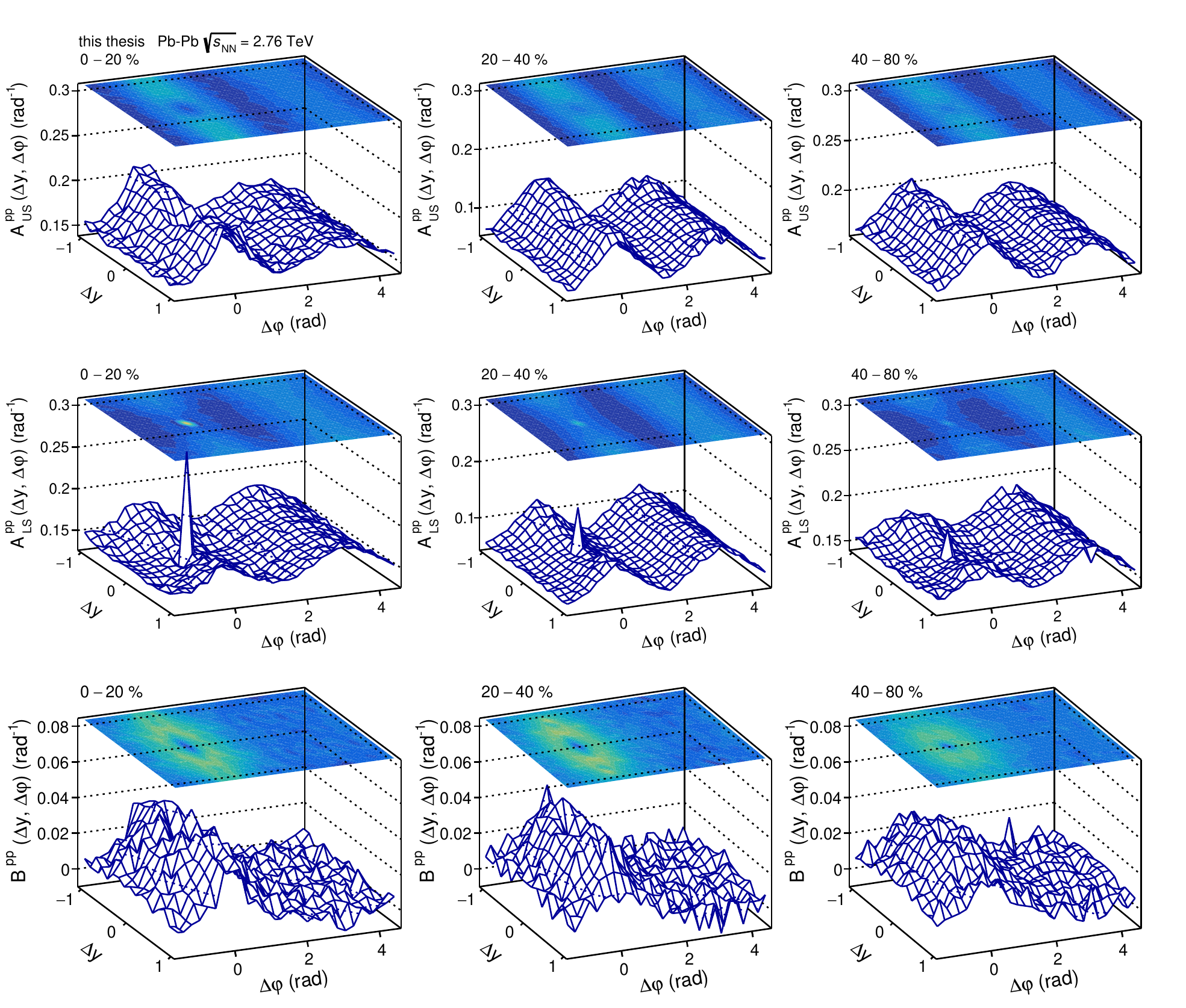}
  \caption{Correlation functions and balance functions of $pp$ pair for selected Pb--Pb collision centralities.
  Top row: unlike-sign correlation function $A_{US}^{pp}(\Delta y,\Delta\varphi)$;
  Middle row: like-sign correlation function $A_{LS}^{pp}(\Delta y,\Delta\varphi)$;  
  Bottom row: balance function $B^{pp}(\Delta y,\Delta\varphi)$.}
   \label{fig:PrPr_BF_2d}
\end{figure}

A complete set of two-dimensional   US and  LS correlators as well as  balance functions, plotted vs.   $\Delta y$ and $\Delta\varphi$, for selected collision centralities, 
is presented in Figs.~\ref{fig:PiPi_BF_2d} -- 
%,~\ref{fig:PiK_BF_2d}, ~\ref{fig:PiPr_BF_2d},~\ref{fig:KPi_BF_2d},~\ref{fig:KK_BF_2d}, ~\ref{fig:KPr_BF_2d}, ~\ref{fig:PrPi_BF_2d},~\ref{fig:PrK_BF_2d}, 
  \ref{fig:PrPr_BF_2d} for $\pi\pi$, $\pi K$, $\pi p$, $K\pi$, $KK$, $Kp$, $p\pi$, $pK$ and $pp$ species pairs.

\subsection{US and LS correlators}
\label{subsec:2DCFResults}

US and LS correlators of all the nine species pairs exhibit a collision centrality dependent $\Delta\varphi$ flow-like modulation dominated by the second harmonic cos(2$\Delta\varphi$).

The US correlators of all nine species pairs exhibit similar features but with varying degrees of importance and centrality dependence. Common features include a prominent near-side peak centered at $(\Delta y, \Delta\varphi) = (0,0)$, which varies in width and amplitude with centrality. For example, the $\pi\pi$ US correlator has a broad and strong near-side peak in 70-90\% collision centralities, but the peak becomes very narrow in central collisions.
Correlators of $\pi K$ and $\pi p$ pairs behave similarly, but their near-side peaks are not as prominent. The correlators of
$Kp$ and $pp$ US pairs are rather  different. Instead of a near-side peak, they feature a small (shallow and narrow) dip at the origin.

The LS correlators exhibit a mix of interesting features.
For example, $\pi\pi$ LS correlators feature a prominent near-side peak which may largely be associated to Hanbury-Brown and Twist (HBT) correlations. The widths of these correlation peaks decrease significantly with system size, whereas the cross-species pairs feature a depression near $(\Delta y, \Delta\varphi) = (0,0)$, which may  result in part from Coulomb effects and different source emission velocity and times.

\subsection{Balance Functions}
\label{subsec:2DBFResults}

All  nine species pair BFs exhibit common features: a prominent near-side peak centered at $(\Delta y, \Delta \varphi) = (0,0)$ and a relatively flat and featureless away-side.
The flat away-side stems for the fact that positive and negative particles of a given species feature essentially equal azimuthal anisotropy relative to the collision symmetry plane.
More importantly, it indicates the presence of a fast radial flow profile of the particle emitting sources~\cite{VOLOSHIN2006490}. 
For instance, emission from a quasi-thermal source at rest is expected to produce particles nearly isotropically. 
However, particles produced from such a source traveling at high speed in the lab frame are closely correlated, i.e., emitted at relatively small $\Delta y$ and $\Delta\varphi$.

One notes, however, that the different species pairs feature a mix of near-side peak shapes, widths, magnitudes, and dependences on collision centrality that indicate that they are subject to different charge balancing pair production and transport mechanisms, as well as final state effects. 
For instance, $B^{\pi\pi}$ exhibits a deep and narrow dip, within the near-side correlation peak, resulting from HBT effects, with a depth and width that vary inversely to the source size (and collision centrality). 
One observes that $B^{KK}$ exhibits much weaker HBT effects, whereas $B^{pp}$ also features a narrow dip centered at $(\Delta y, \Delta \varphi) = (0,0)$ within a somewhat elongated near-side peak that likely reflects annihilation of $p\bar{p}$ pairs. Produced protons and anti-protons are more likely to annihilate if emitted at small $\Delta y$, $\Delta\varphi$ and $\Delta p_{\rm T}$. Annihilation of $p\bar{p}$ pairs results in the production of several pions (on average), and should thus yield a depletion near the origin $(\Delta y, \Delta \varphi) = (0,0)$ of BFs.

\begin{figure}
\centering
  \includegraphics[width=0.99\linewidth]{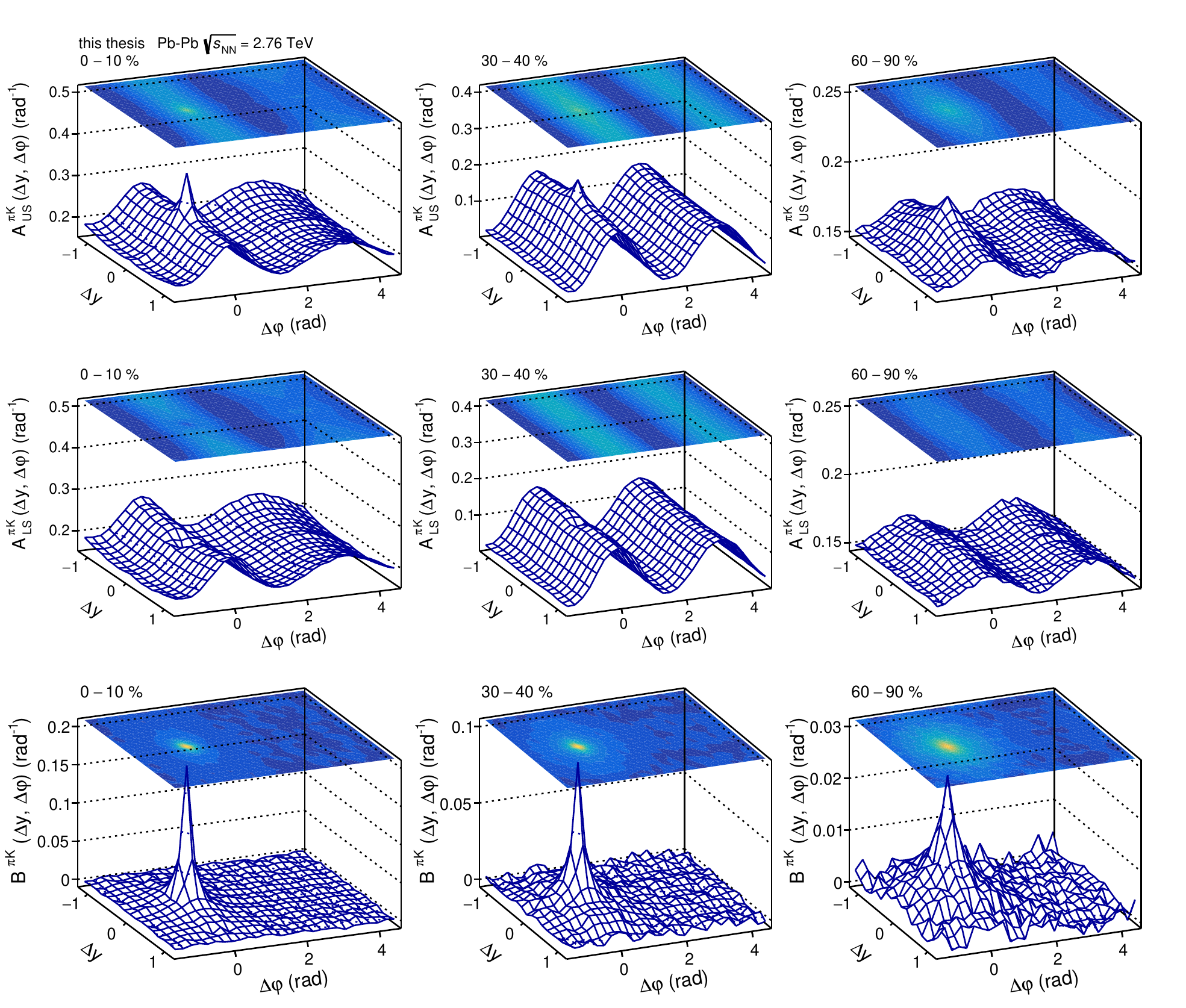}
  \caption{Correlation functions and balance functions of $\pi K$ pair for selected Pb--Pb collision centralities.
  Top row: unlike-sign correlation function $A_{US}^{\pi K}(\Delta y,\Delta\varphi)$;
  Middle row: like-sign correlation function $A_{LS}^{\pi K}(\Delta y,\Delta\varphi)$;  
  Bottom row: balance function $B^{\pi K}(\Delta y,\Delta\varphi)$.}
   \label{fig:PiK_BF_2d}
\end{figure}

\begin{figure}
\centering
  \includegraphics[width=0.99\linewidth]{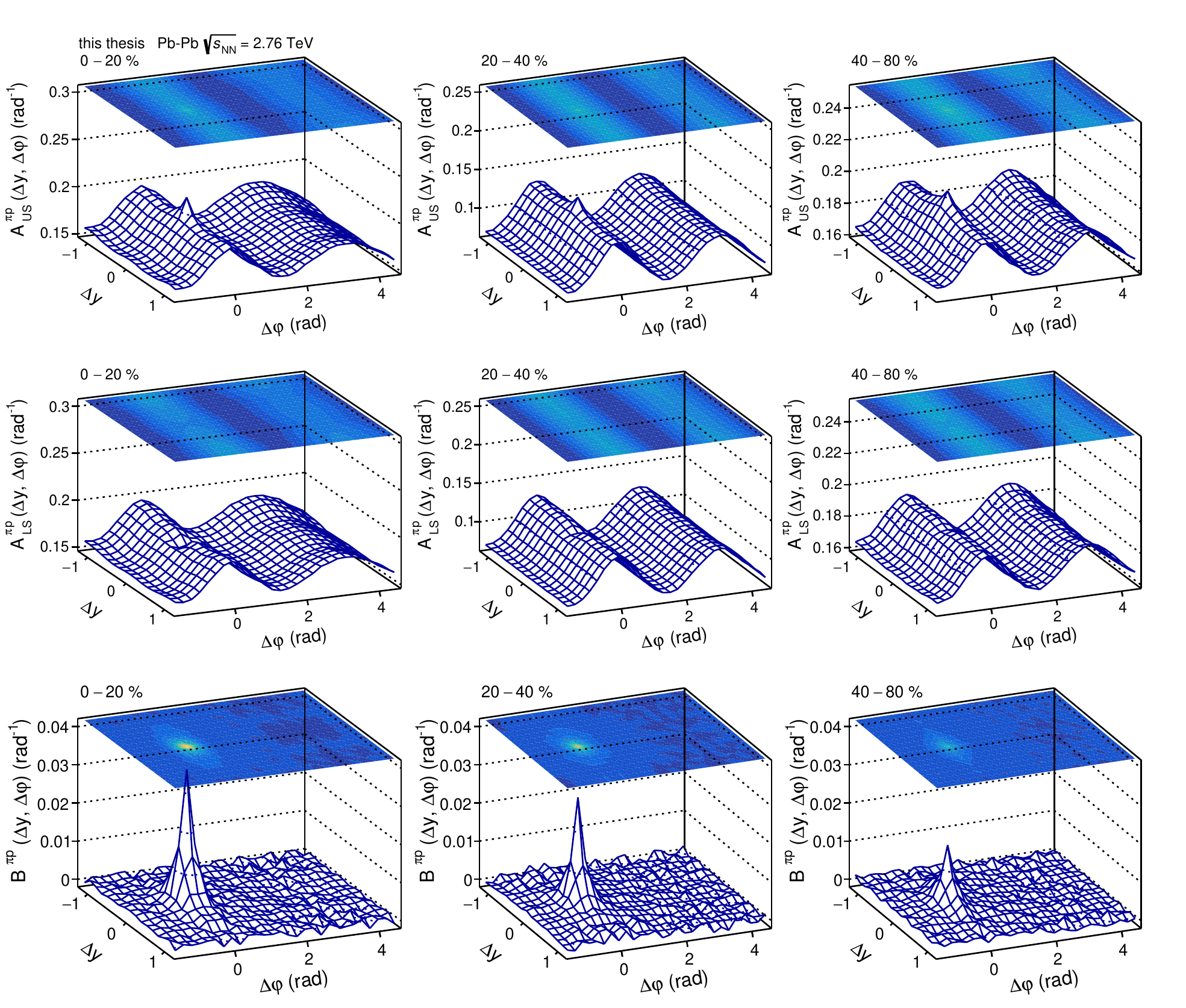}
  \caption{Correlation functions and balance functions of $\pi p$ pair for selected Pb--Pb collision centralities.
  Top row: unlike-sign correlation function $A_{US}^{\pi p}(\Delta y,\Delta\varphi)$;
  Middle row: like-sign correlation function $A_{LS}^{\pi p}(\Delta y,\Delta\varphi)$;  
  Bottom row: balance function $B^{\pi p}(\Delta y,\Delta\varphi)$.}
   \label{fig:PiPr_BF_2d}  
\end{figure}

\begin{figure}
\centering
  \includegraphics[width=0.99\linewidth]{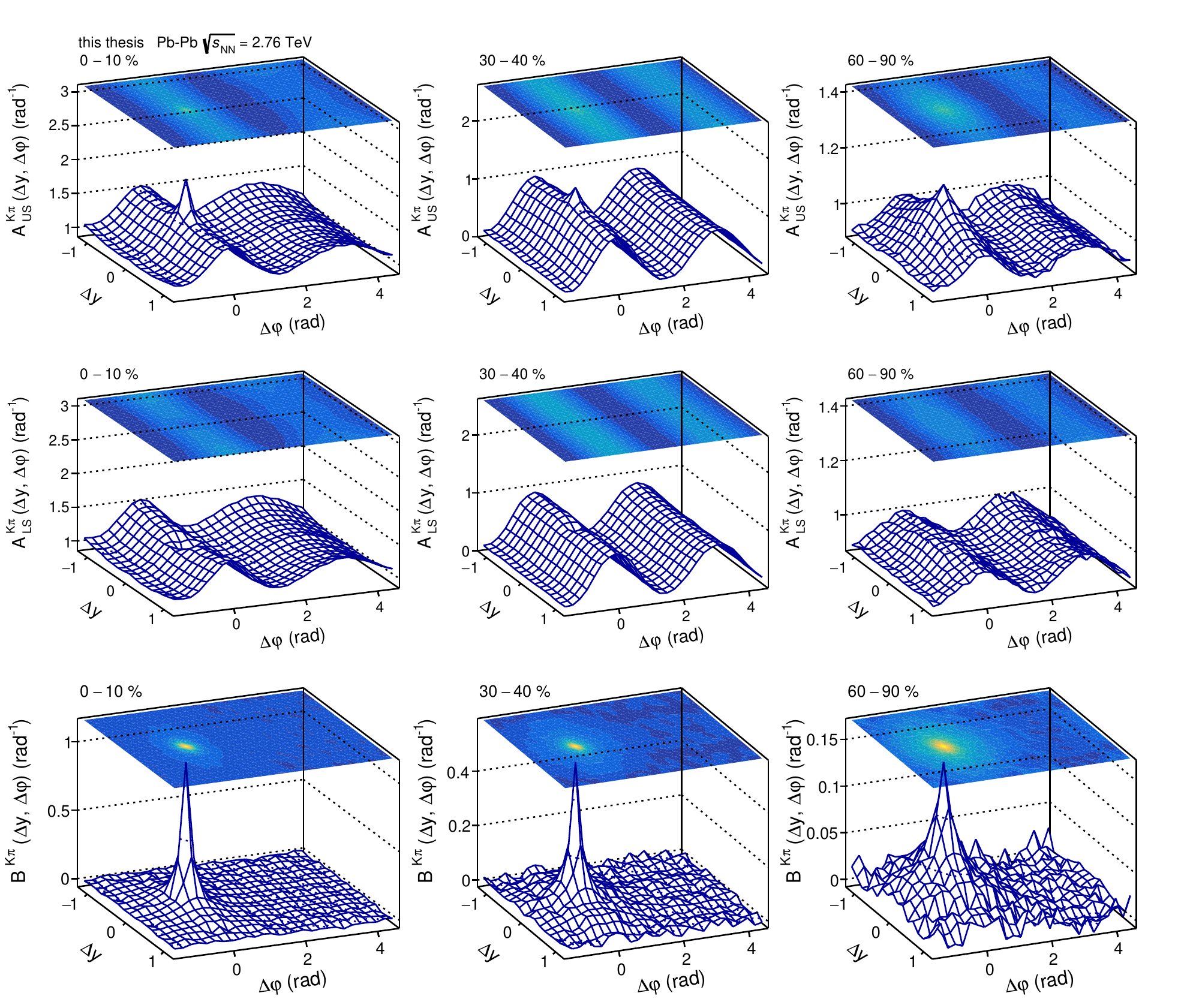}
  \caption{Correlation functions and balance functions of $K\pi$ pair for selected Pb--Pb collision centralities.
  Top row: unlike-sign correlation function $A_{US}^{K\pi}(\Delta y,\Delta\varphi)$;
  Middle row: like-sign correlation function $A_{LS}^{K\pi}(\Delta y,\Delta\varphi)$;  
  Bottom row: balance function $B^{K\pi}(\Delta y,\Delta\varphi)$.}
   \label{fig:KPi_BF_2d}
\end{figure}

\begin{figure}
\centering
  \includegraphics[width=0.99\linewidth]{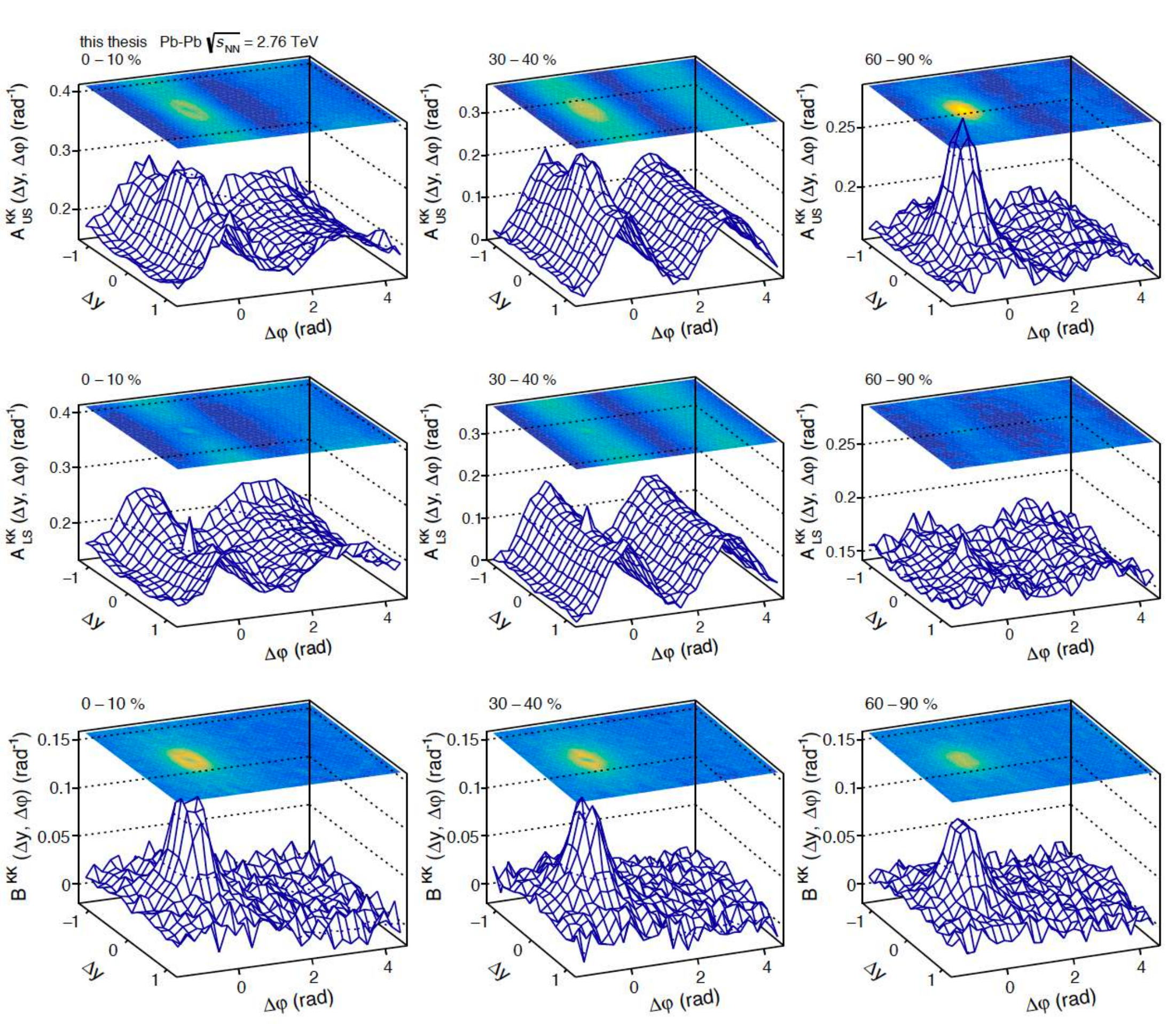}
  \caption{Correlation functions and balance functions of $KK$ pair for selected Pb--Pb collision centralities.
  Top row: unlike-sign correlation function $A_{US}^{KK}(\Delta y,\Delta\varphi)$;
  Middle row: like-sign correlation function $A_{LS}^{KK}(\Delta y,\Delta\varphi)$;  
  Bottom row: balance function $B^{KK}(\Delta y,\Delta\varphi)$.}
   \label{fig:KK_BF_2d}
\end{figure}

\begin{figure}
\centering
  \includegraphics[width=0.99\linewidth]{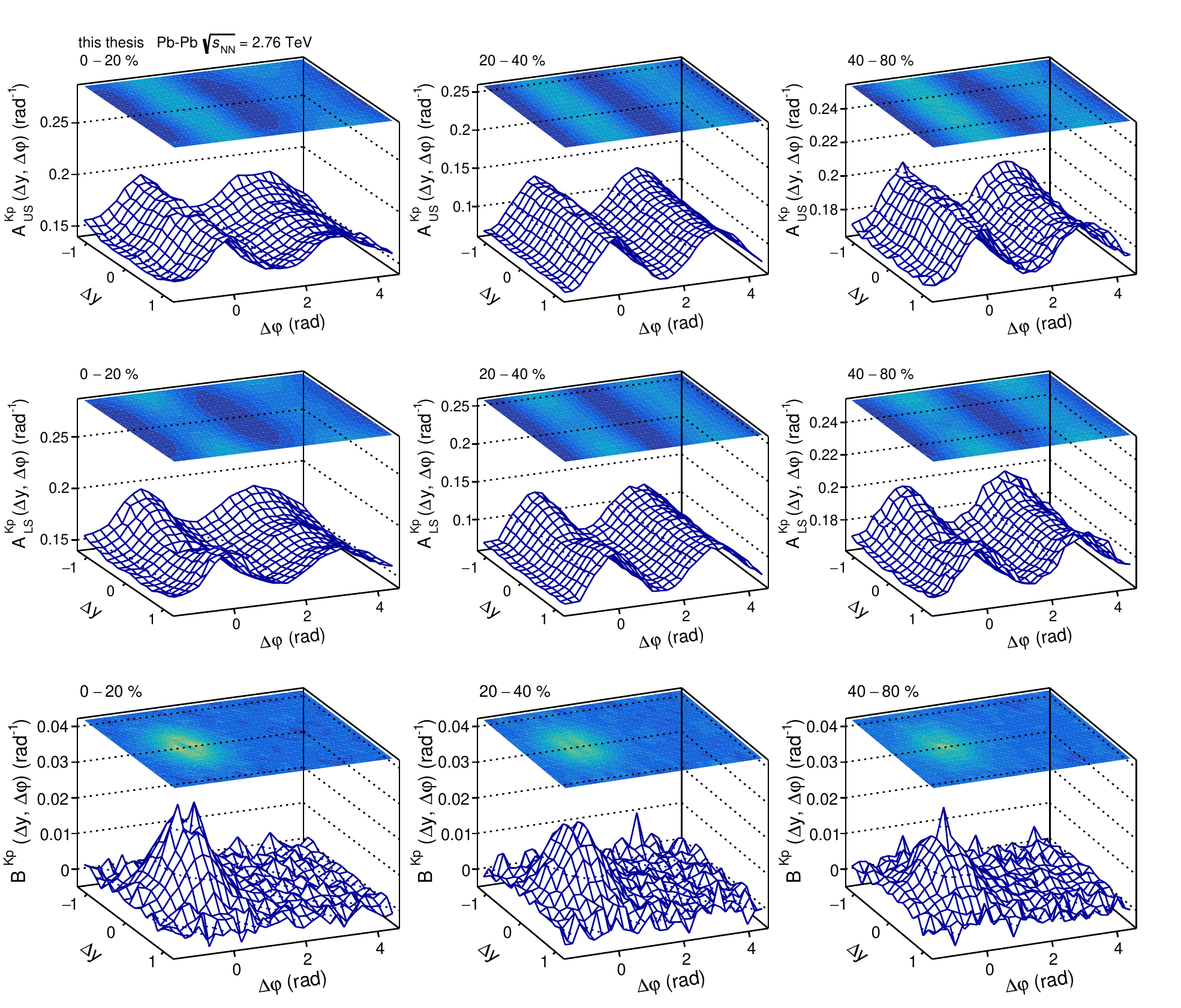}
  \caption{Correlation functions and balance functions of $Kp$ pair for selected Pb--Pb collision centralities.
  Top row: unlike-sign correlation function $A_{US}^{Kp}(\Delta y,\Delta\varphi)$;
  Middle row: like-sign correlation function $A_{LS}^{Kp}(\Delta y,\Delta\varphi)$;  
  Bottom row: balance function $B^{Kp}(\Delta y,\Delta\varphi)$.}
   \label{fig:KPr_BF_2d}  
\end{figure}

\begin{figure}
\centering
  \includegraphics[width=0.99\linewidth]{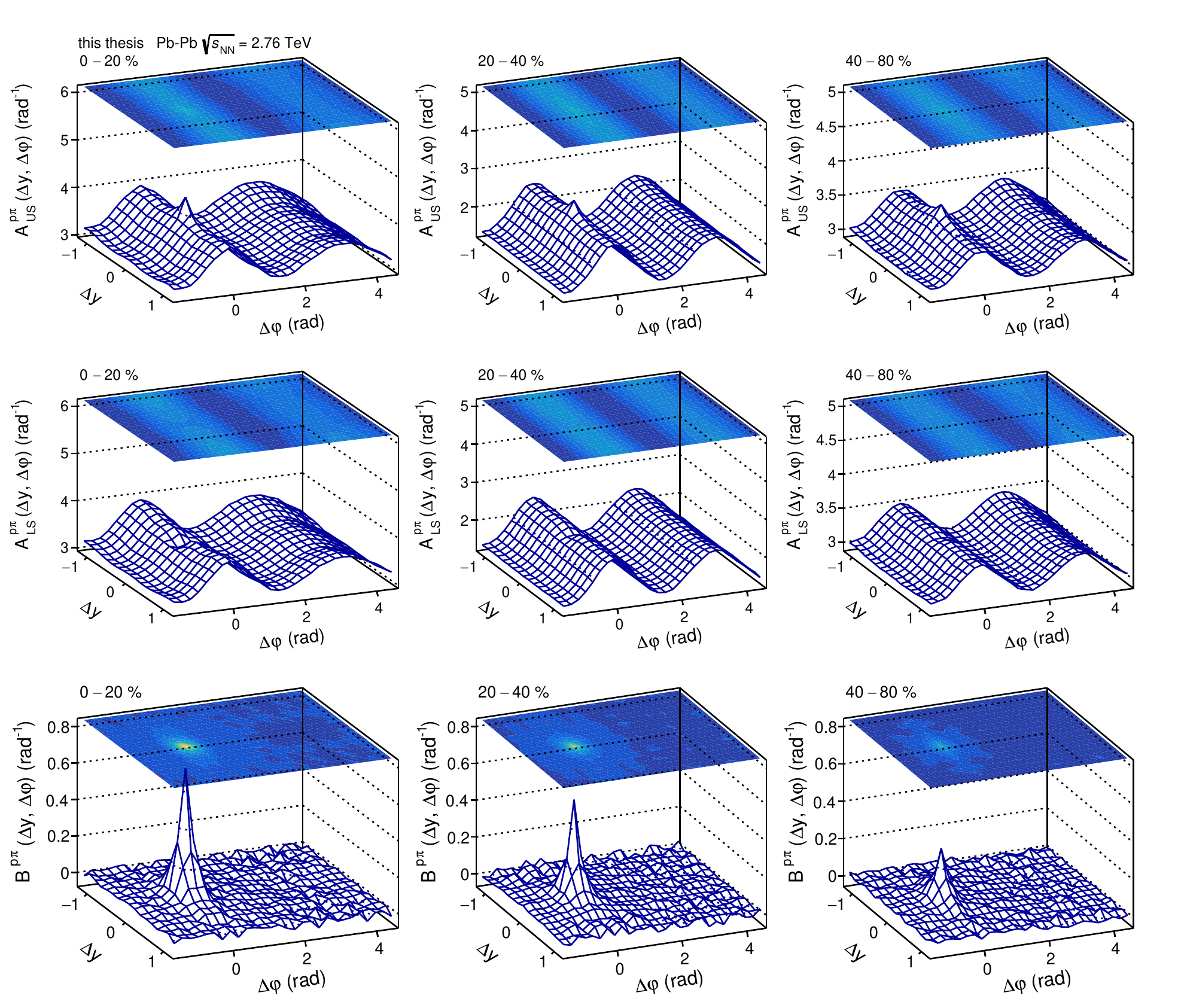}
  \caption{Correlation functions and balance functions of $p\pi$ pair for selected Pb--Pb collision centralities.
  Top row: unlike-sign correlation function $A_{US}^{p\pi}(\Delta y,\Delta\varphi)$;
  Middle row: like-sign correlation function $A_{LS}^{p\pi}(\Delta y,\Delta\varphi)$;  
  Bottom row: balance function $B^{p\pi}(\Delta y,\Delta\varphi)$.}
   \label{fig:PrPi_BF_2d}  
\end{figure}

\begin{figure}
\centering
  \includegraphics[width=0.99\linewidth]{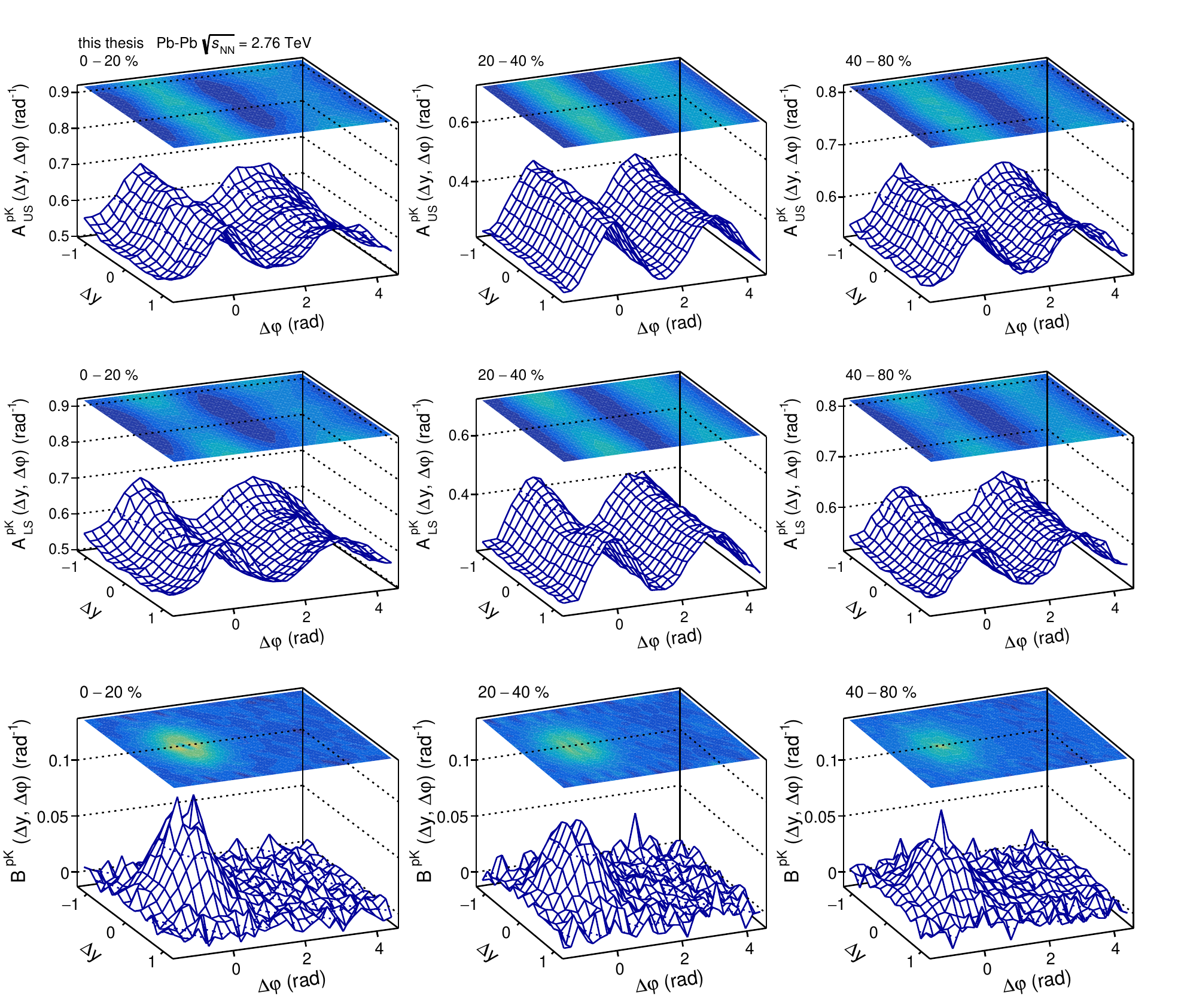}
  \caption{Correlation functions and balance functions of $pK$ pair for selected Pb--Pb collision centralities.
  Top row: unlike-sign correlation function $A_{LS}^{pK}(\Delta y,\Delta\varphi)$;
  Middle row: like-sign correlation function $A_{LS}^{pK}(\Delta y,\Delta\varphi)$;  
  Bottom row: balance function $B^{pK}(\Delta y,\Delta\varphi)$.}
   \label{fig:PrK_BF_2d}  
\end{figure}

\clearpage

%%%%%%%%%%%%%%%%%%%%%%%%%%%%%%%%%%%%%%%%%%%%%%%%%%%%%%%
% 1D BF Projections
%%%%%%%%%%%%%%%%%%%%%%%%%%%%%%%%%%%%%%%%%%%%%%%%%%%%%%%
\section{Balance Function Projections}
\label{sec:1DProjections}

\begin{figure}[h!]
 \includegraphics[width=0.99\linewidth]{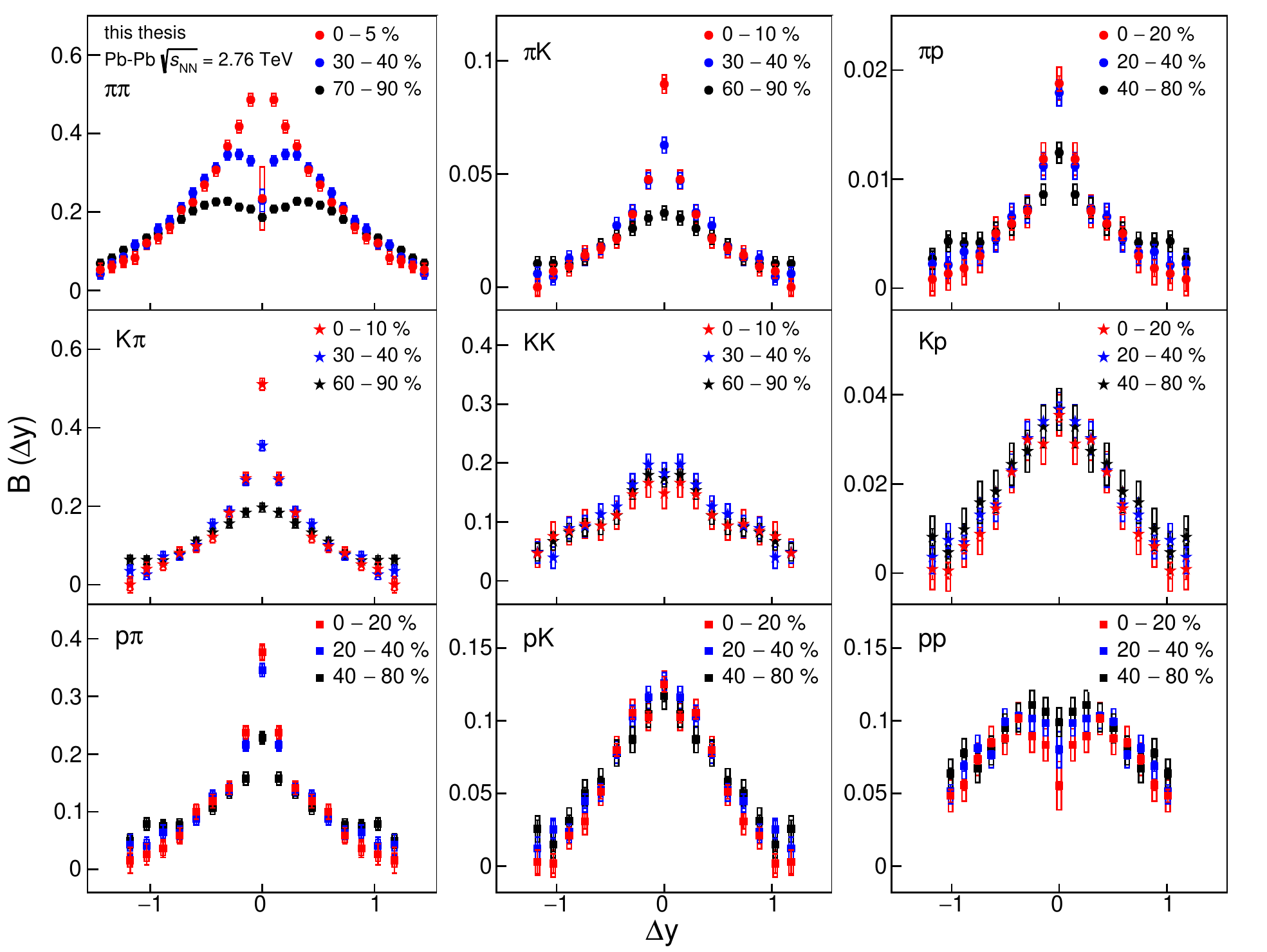}
 \caption{Longitudinal ($\Delta y$) projections of balance functions for $|\Delta\varphi|\le\pi$ of the full species matrix of $\pi^{\pm}$, $K^{\pm}$ and $p/\bar{p}$, with $\pi^{\pm}$, $K^{\pm}$ and $p/\bar{p}$ as reference particle in the $1^{st}$, $2^{nd}$ and $3^{rd}$ row, respectively. Systematic uncertainties are shown as boxes.}
  \label{fig:1d_BF_dy}
\end{figure}

We further examine the collision centrality evolution of the nine species pair BFs by plotting their projections onto the $\Delta y$ axis in Fig.~\ref{fig:1d_BF_dy}. 
We find that the $\Delta y$ projection shape and amplitude of $B^{\pi\pi}$ exhibit the strongest centrality dependence, whereas those of $B^{\pi K}$, $B^{\pi p}$, $B^{K\pi}$, $B^{p\pi}$, and $B^{pp}$ display significant albeit smaller dependence on centrality. 
The $\Delta y$ projections of $B^{KK}$, $B^{Kp}$, and $B^{pK}$, on the other hand, feature minimal centrality dependence, if any.

%

%The BF $\Delta y$ projections of the full species matrix of $\pi^{\pm}$, $K^{\pm}$ and $p/\bar{p}$, as shown in Fig.~\ref{fig:1d_BF_dy}, provide critical information on the QGP hadronization chemistry, including resonance decays, and general charge (electric charge, strangeness, baryon) production, which could be used to test the particle production theories including color tube fragmentations and quack coalescence. $B^{\pi\pi}$($\Delta y$) display a clear centrality dependence, while $B^{KK}$($\Delta y$) show almost no centrality dependence, which is qualitatively consistent with the two-wave quark production scenario~\cite{Pratt:2012dz}. $B^{pp}$ and cross-species pairs show moderate centrality dependence.

%
\begin{figure}[h!]
 \includegraphics[width=0.99\linewidth]{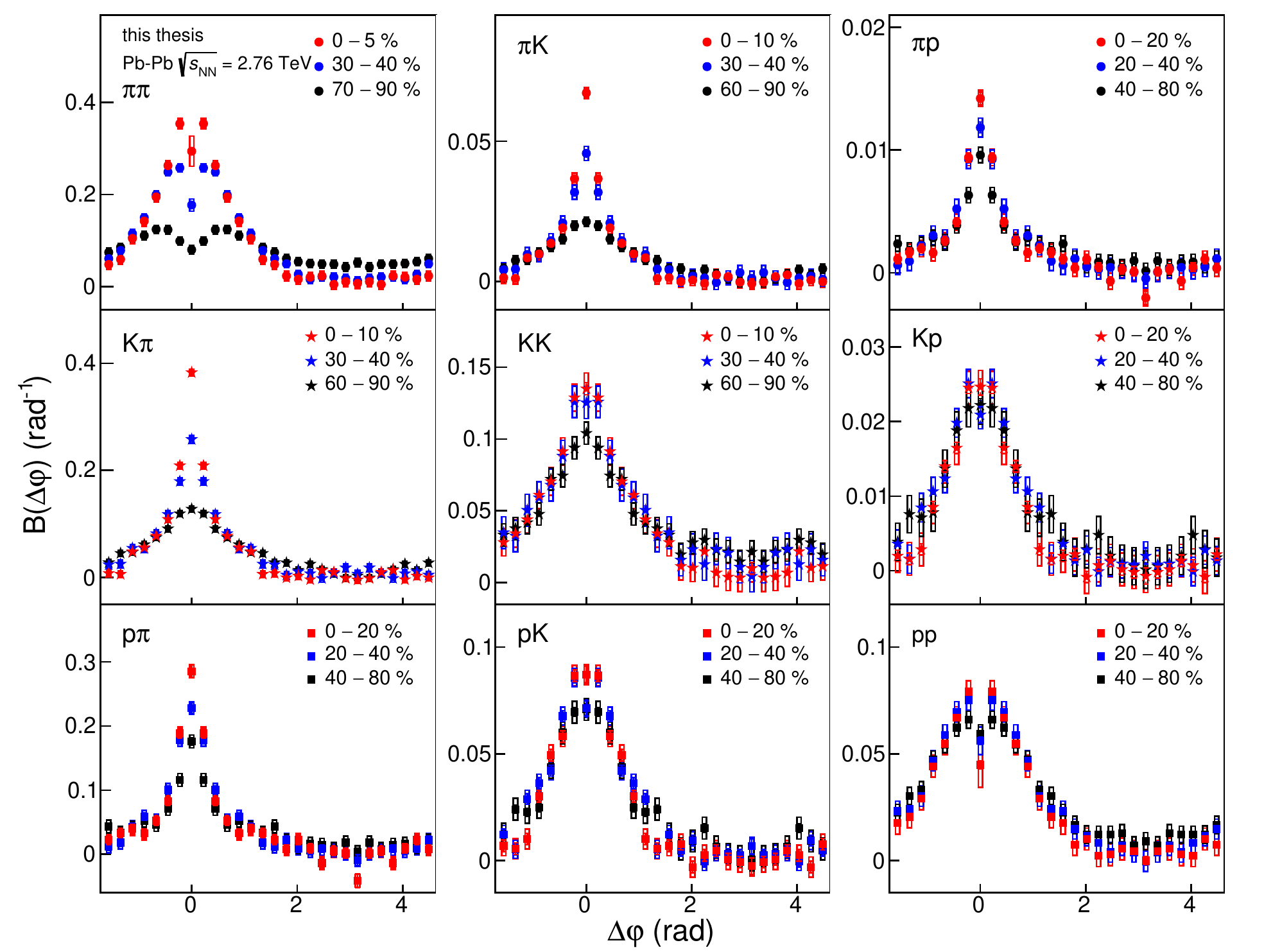}
 \caption{BF $\Delta\varphi$ projections of the full species matrix of $\pi^{\pm}$, $K^{\pm}$ and $p/\bar{p}$, with $\pi^{\pm}$, $K^{\pm}$ and $p/\bar{p}$ as the interest particle in the $1^{st}$, $2^{nd}$ and $3^{rd}$ row, respectively. Systematic uncertainties are shown as boxes.}
  \label{fig:1d_BF_dphi}
\end{figure}

The BF $\Delta\varphi$ projections of the full species matrix of $\pi^{\pm}$, $K^{\pm}$ and $p/\bar{p}$ are presented in Figure~\ref{fig:1d_BF_dphi}. We find that the $\Delta\varphi$ projection shape and amplitude of $B^{\pi\pi}$ exhibit the strongest centrality dependence, whereas those of $B^{\pi K}$, $B^{\pi p}$, $B^{K\pi}$ and $B^{p\pi}$, display significant albeit smaller dependence on centrality. The $\Delta\varphi$ projections of $B^{KK}$, $B^{Kp}$, $B^{pK}$, and $B^{pp}$ on the other hand, feature minimal centrality dependence, if any.

\clearpage

%%%%%%%%%%%%%%%%%%%%%%%%%%%%%%%%%%%%%%%%%%%%%%%%%%%%%%%
% BF Widths and Integrals
%%%%%%%%%%%%%%%%%%%%%%%%%%%%%%%%%%%%%%%%%%%%%%%%%%%%%%%
\section{BF RMS Widths and Integrals}
\label{sec:WidthsIntegrals}

\begin{figure}[h!]
 \includegraphics[width=0.6\linewidth]{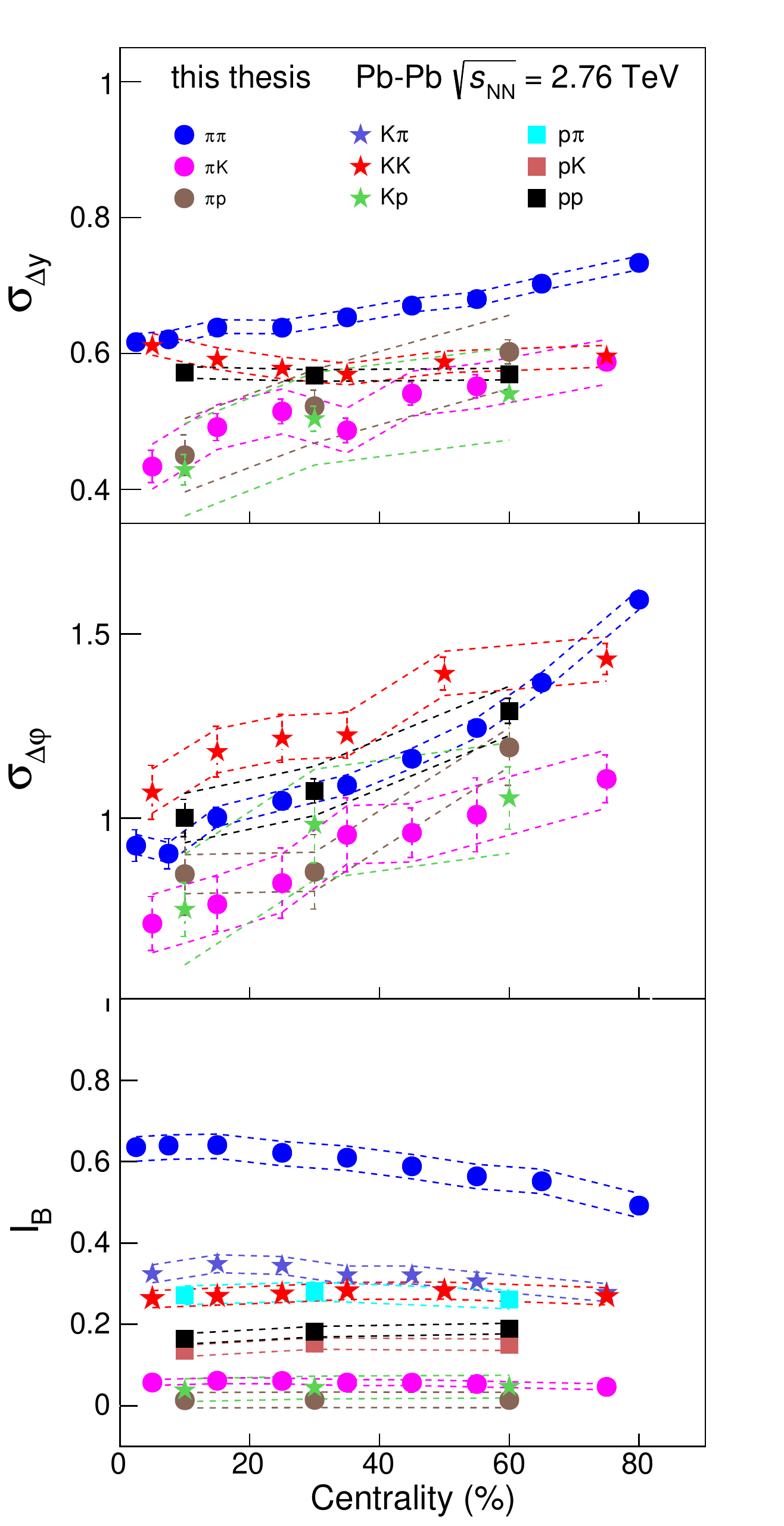}
 \caption{$\Delta y$ rms widths (top), $\Delta\varphi$ rms widths (middle), and integrals (bottom) of BFs of the full species matrix of $\pi^{\pm}$, $K^{\pm}$ and $p/\bar{p}$ as a function of collision centrality. For $\Delta y$ and $\Delta\varphi$ widths, $K\pi$, $p\pi$ and $pK$ have the same values with $\pi K$, $\pi p$, and $Kp$, respectively. The relative azimuthal angle range for all the species pairs is $|\Delta\varphi|\le\pi$. The relative rapidity range for all the species pairs is $|\Delta y|\le1.2$, with exceptions that for $\pi\pi$ it is $|\Delta y|\le1.4$, while for $pp$ it is $|\Delta y|\le1.0$. Systematic uncertainties are shown as dashed lines.}
  \label{fig:dy_dphi_widths_integral}
\end{figure}

We characterize the collision centrality evolution of the shape and strength of the BFs in terms of their longitudinal and azimuthal rms widths, as well as their integrals in Fig.~\ref{fig:dy_dphi_widths_integral}. One thing to note is that, the points of the $\Delta y$ and $\Delta\varphi$ projections are not included in the rms calculations, if they are smaller than 0. However, all the points of the $\Delta y$ and $\Delta\varphi$ projections are included in the BF integral calculations.

The BFs of all measured species combinations exhibit a strong narrowing in azimuth from peripheral to central collisions (e.g., 50\% for $\pi\pi$ pair). Such narrowing, already observed for unidentified charged hadrons at both RHIC and LHC energies, and for pions at RHIC, is commonly ascribed to strong radial collective motion with average velocities  that monotonically increase towards central collisions~\cite{VOLOSHIN2006490}. In the longitudinal direction, the BF of all  species pairs, except those of  $KK$ and $pp$, also exhibit significant narrowing from peripheral to central collision. We find that the longitudinal  projections  and  rms of $KK$ pairs  display no dependence on collision centrality. However, we remark  that while the longitudinal rms of $B^{pp}$ are essentially invariant, their projections  do change shape with collision centrality. This apparent contradiction results largely from the limited longitudinal acceptance of the $pp$ BF measurements.
% and the fact that, manifestly, only a rather limited fraction of $pp$  pairs can be captured within the measured $\Delta y$ range. 
It is also worth noticing that longitudinal rms values of cross-species pairs are rather similar and considerably smaller than those of $\pi\pi$ pair. By contrast, the azimuthal rms of all pairs  exhibit a wide spread of values, with $KK$ pairs featuring the largest values while $\pi K$ pairs feature the smallest. Qualitatively, one expects  that radial flow boosts should have a  larger impact on heavier pair progenitors (the objects that decay or produce the observed correlated pairs),  thereby yielding narrowest azimuthal BFs for those pairs. However, $\pi\pi$ and $pp$ feature similar rms widths in azimuth but rather different longitudinal widths therefore suggesting that other mechanisms are at play in determining the shape and widths of their respective balance functions.
The B($\Delta y$) width results indicate that the balancing pair production mechanisms of $K^{\pm}$ and $p/\bar{p}$ are rather different from those of $\pi^{\pm}$. On the one hand, the $KK$ results are qualitatively consistent with the two-wave quark production scenario, which stipulates that the production of up and down quarks dominates the late stage whereas strange quarks are predominantly produced during the early stages of collisions. On the other hand, the broad $pp$ BFs, whose widths are not fully captured in Fig.~\ref{fig:dy_dphi_widths_integral}, suggest that baryon anti-baryon pair creation might also occur rather early in A--A collisions.

%$\pi^{\pm}$, $K^{\pm}$ and $p/\bar{p}$

%For the B($\Delta y$) widths in Fig.~\ref{fig:dy_dphi_widths_integral}, $KK$ and $pp$ pairs are almost not dependent on collision centrality, while $\pi\pi$ and all the cross-species pairs show a narrowing trend towards central collisions. Furthermore, the B($\Delta y$) width values of all the cross-species pairs are very similar, which could be qualitatively explained by the strong radial flow effect. The B($\Delta y$) width results indicates the balancing pair production mechanisms of charged kaons and (anti-)protons are different from charged pions. These results are qualitatively consistent with the two-wave quark production scenario, which assumes up and down quarks are produced both at the beginning of collision and at hadronization, while strangeness quarks are only produced at the beginning of collision. In addition, the results indicate a difference on the balancing pair production mechanisms between mesons with different quark components and baryons. For the B($\Delta\varphi$) widths in Fig.~\ref{fig:dy_dphi_widths_integral}, even though different species pairs have different values, all the species pairs show a narrowing trend towards central collisions, which is qualitatively consistent with the radial flow effect.

Shifting our attention to the bottom panel of Figure~\ref{fig:dy_dphi_widths_integral}, which displays the collision centrality evolution of integrals, noted $I_B^{\alpha\beta}$, of BFs of all nine species pairs $\alpha\beta$, we note that  $I_B^{\pi\pi}$ exhibits a modest increasing trend towards central collisions whereas integrals of other species pairs  are essentially invariant.
By construction, integrals of $B^{\alpha\beta}(\Delta y,\Delta\varphi)$, reported for the first time in this work, measure the probability of observing a  charge balancing (i.e., an associated) particle  of species $\beta$ given a reference particle of species $\alpha$ has been observed. We may thus call $I_B^{\alpha\beta}$ hadron species pairing probability. Results shown in Fig.~\ref{fig:dy_dphi_widths_integral} are surprising on two accounts. First, the lack of collision centrality dependence of integrals $I_B^{\alpha\beta}$ observed for all species pairs, but one, imply hadron species pairing probabilities are invariant with collision centrality. Second, close examination of these hadron species pairing probabilities show they are rather different than inclusive probabilities of observing $\pi^{\pm}$, $K^{\pm}$, and $p/\bar p$ in Pb--Pb collisions. For instance, $I_B^{K\pi}$ is not larger than $I_B^{KK}$ by the $\pi/K$ ratio of inclusive single particle yields~\cite{PhysRevC.88.044910} and $I_B^{pp}$ is larger than $I_B^{pK}$ also in contrast to observed ratios of $K/p$ yield ratios. Hadron pairing probabilities are thus indeed very different than the relative probabilities of single hadrons.
%This should not be too much of a surprise, however, given the set of processes $P_2$ that lead to a specific balancing pair (e.g.,  $P_2: \rightarrow \alpha^{\pm} + \beta^{\mp} + X$) is, by construction, far smaller than the set of processes $P_1$ leading of a given particle species $\alpha$ (e.g., $P_1: \rightarrow \alpha^{\pm}  + X$). 
%It is remarkable, however, that the hadron species pairing probabilities $I_B^{\alpha\beta}$ exhibit essentially no collision centrality dependence while single particle yields are known to exhibit a significant dependence on collision centrality. 
Note that the observed rise of $I_B^{\pi\pi}$ in more central collisions may artificially result from increased kinematic focusing of pions with centrality in the  $p_{\rm T}$ and $\Delta y$ acceptance of this measurement. The higher velocity flow fields encountered in more central Pb--Pb collisions could indeed shift and focus the yield of associated pions. Why such a shift is not as important for other charge balancing pairs remains to be elucidated with a comprehensive model accounting for the  flow velocity profile and appropriate sets of charge conserving processes yielding balancing charges in the final state of collisions.  Recent deployment of hydrodynamics models feature the former but lack the latter. Further theoretical work is thus required to interpret the observed collision centrality dependence of the hadron species pairing probabilities displayed in Fig.~\ref{fig:dy_dphi_widths_integral}.

%Prior to this letter, various hadron spectra and ratio measurements have been carried out~\cite{ALICE:2013Pt}, which however only provided single particle production information, leaving particle pair production information absent. Thanks to the proper detector acceptance and efficiency corrections in this work, the balancing charge yield (i.e. balance function integral) was measured for the first time in relativistic heavy ion collisions shown in Fig.~\ref{fig:dy_dphi_widths_integral}, which provides critical information on balancing pair production probability of the full species matrix of charged pion, charged kaon and (anti-)proton. The balancing charge yields of all the species pairs show very little centrality dependence, except for $\pi\pi$ which displays an increasing trend towards central collisions. This could be partially explained by the fact that the $\pi\pi$ balance function leaks into larger $\Delta y$ range beyond the detector acceptance more for peripheral collisions than central collisions. The balancing charge yield results indicate the QGP hadronization chemistry does not change much with centrality. The fact that $I^{K\pi}$ is not larger than $I^{KK}$ by the amount of $\pi^{\pm}/K^{\pm}$ ratio~\cite{ALICE:2013Pt} and $I^{pp}$ have larger values than $I^{pK}$, shows the balancing pair production probabilities are very different than single hadron ratios.

\clearpage
	\chapter{Summary}\label{chap:Summary}

In summary, we presented first measurements of the collision centrality evolution of balance functions (BF) of the full species matrix of charged hadrons, $(\pi,K,p)\otimes (\pi,K,p)$ in Pb-Pb collisions at $\sqrt{s_{_{\rm NN}}} =$2.76 TeV. Measured as functions of particle pair separation in rapidity ($\Delta y$) and azimuth ($\Delta\varphi$), the BFs exhibit a common prominent near-side peak centered at $(\Delta y,\Delta\varphi)=(0,0)$. The peaks with different species pairs feature different shapes, amplitudes, and widths, and varied dependences on collision centrality. The BFs of all species pairs feature narrowing $\Delta\varphi$ rms widths in more central collisions, owing to the strong radial flow field present in central Pb--Pb collisions.  In the longitudinal direction, the rms widths of BFs of all species pairs narrow with centrality except for those of $KK$ and $pp$ pairs. The shape and width of $KK$ BFs are invariant with collision centrality, while the $pp$ BFs exhibit some shape dependence on collision centrality but essentially invariant rms values with the acceptance of the measurement. The observed centrality invariance of the $KK$ longitudinal and narrowing rms of other species  in the longitudinal direction are  qualitatively consistent with  effects associated with radial flow and the two-wave quark production scenario. However, a comprehensive model accounting for hadron chemistry at finite temperature, charge conserving pair creation, and strong radial flow fields is   required in order to interpret the data presented in greater detail.
	
	% insert appendices
	% Comment the following line if you don't have an appendix
	% for a new appendix, copy appendices/appendix_a.tex and add a new line here
%	\begin{appendices}
%	\include{appendices/appendix_header}
%	\include{appendices/appendix_a}
%	\end{appendices}

	\addcontentsline{toc}{chapter}{Bibliography}
  \bibliographystyle{unsrt}
  \bibliography{references}

	\newpage

	% add abstract
  \addcontentsline{toc}{chapter}{Abstract}        
	
\begin{center}
\textbf{ABSTRACT}

	\singlespacing
\textbf{Balance functions of charged hadron pairs $(\pi,K,p)\otimes (\pi,K,p)$ in Pb--Pb collisions at $\sqrt{s_{_{\rm NN}}} =$2.76 TeV}\\
	\doublespacing
	
	by\\
	
	\textbf{Jinjin Pan}\\
	\textbf{2019}\\
\end{center}
\begin{tabular}{ll}
\textbf{Advisor:} &Prof. Claude Pruneau\\
\textbf{Major:}   &Physics\\
\textbf{Degree:}  &Doctor of Philosophy
\end{tabular}
\bigskip

We present the first balance function (BF) measurement of charged hadron pairs $(\pi,K,p)\otimes (\pi,K,p)$ in Pb--Pb collisions at $\sqrt{s_{_{\rm NN}}} =$2.76 TeV. 
The BF measurements are carried out as two-dimensional (2D) differential correlators vs. the relative rapidity ($\Delta y$) and azimuthal angle ($\Delta\varphi$) of hadron pairs, and studied as a function of collision centrality. While the BF azimuthal widths of all pairs substantially decrease from peripheral to central collisions, the longitudinal widths exhibit mixed behaviors: BF of $\pi\pi$ and cross-species pairs narrow in more central collisions whereas those of $KK$ and $pp$ are found invariant with collision centrality. This dichotomy is qualitatively consistent with the presence of strong radial flow effects and the existence of two waves of quark production in relativistic heavy ion collisions. We additionally present first measurements of the BF integrals and find that hadron pairing probabilities are very different from single hadron ratios and feature minimal collision centrality dependence.
Overall, the results presented provide new and challenging constraints for theoretical models of hadron production and transport in relativistic heavy ion collisions.

	% add autobiographical statement
	\addcontentsline{toc}{chapter}{Autobiographical Statement}

\end{document}